\documentclass[12pt]{book}

\usepackage{amsfonts}
\usepackage{makeidx}

\newcommand{\cross }[1]{#1 \hspace{-0.6em} /}

\newtheorem{theorem}{Theorem}[chapter]

\newtheorem{assertion}[theorem]{Assertion}

\newtheorem{lemma}[theorem]{Lemma}

\newtheorem{postulate}[theorem]{Postulate}

\newtheorem{statement}[theorem]{Statement}

\newenvironment{proof}[1][Proof]{\noindent\textbf{#1.} }{\
\rule{0.5em}{0.5em}}

\makeindex
\usepackage{graphicx}

\begin{document}

\frontmatter
\title{RELATIVISTIC QUANTUM DYNAMICS}
\author{Eugene V. Stefanovich}
\date{2014}
\maketitle

\bigskip

\newpage

\begin{eqnarray*}
\mbox{ }
\end{eqnarray*}

\newpage

\vspace {2in}

Draft, 3rd edition

\begin{center}
\section* {RELATIVISTIC QUANTUM DYNAMICS:  }
\subsection* {A Non-Traditional Perspective on Space, Time, Particles, Fields, and
Action-at-a-Distance}

Eugene V. Stefanovich \footnote{e-mail: $eugene\_stefanovich@usa.net$

web address: 
$http://www.arxiv.org/abs/physics/0504062$}

\vspace {4in}

Mountain View, California \\

\bigskip

\end{center}

Copyright \copyright  2004 - 2014 Eugene V. Stefanovich

\newpage

\begin{eqnarray*}
\mbox{ }
\end{eqnarray*}

\newpage

\vspace {6in}

\Large

\textbf{                     To  Regina}

\normalsize

\newpage

\noindent \textbf{Abstract}

\bigskip

\noindent This book is an attempt to build a consistent relativistic
quantum theory of interacting particles. In the first part of the
book ``Quantum electrodynamics'' we follow rather traditional approach to particle physics. Our discussion proceeds
systematically from the principle of relativity and postulates of quantum
measurements to the renormalization in quantum electrodynamics. In
the second part of the book ``Quantum theory of particles'' this
traditional approach is reexamined. We find that formulas of special
relativity should be modified to take into account particle
interactions. We also suggest reinterpreting quantum field theory in
the language of physical ``dressed'' particles.  This formulation eliminates the need for renormalization and
opens up a new way for studying dynamical and bound state properties
of quantum interacting systems. The developed theory is applied to
realistic physical objects and processes including the energy spectrum of the hydrogen
atom,  the decay law of moving unstable particles, and the electric field of relativistic electron beams.
These results force us to take a fresh look at some core issues of
modern particle theories, in particular, the Minkowski space-time
unification, the role of quantum fields and renormalization as well as the
alleged impossibility of action-at-a-distance. A new perspective on
these issues is suggested. It can help to solve the old problem
of theoretical physics -- a consistent unification of
relativity and quantum mechanics.

\tableofcontents

\chapter{PREFACE}
\label{sc:preface}

Looking back at theoretical physics of the 20th century, we see two
monumental achievements that radically changed the way we understand
space, time, and matter -- the special theory of relativity and
quantum mechanics.  These theories extended our comprehension to those parts of the natural
world that are not normally accessible to human senses and
experience. Special relativistic descriptions encompassed observers and objects
moving with extremely high speeds and high energies. Quantum
mechanics was essential for understanding properties of matter at the
microscopic scale: nuclei, atoms, molecules, etc. In the 21st century the challenge
remains in the unification of these two theories, i.e., in the
theoretical description of energetic elementary particles and
their interactions.

It is commonly accepted that the most promising candidate for such
an unification is the local quantum field theory (QFT). \index{QFT (quantum field theory)}
Indeed, this theory achieved astonishing accuracy in calculations of
certain physical observables, such as scattering cross-sections and energy spectra. In
some instances, the discrepancies between
experiments and  predictions of quantum electrodynamics (QED) \index{QED (quantum electrodynamics)}
 are less than 0.000000001\%. It is difficult to find
such accuracy anywhere else in science! However, in spite of its
success, quantum field theory cannot be regarded as the ultimate
unification of relativity and quantum mechanics. Just too many
fundamental questions remain unanswered and too many serious
problems are left unsolved.

It is fair to say that everyone trying to learn QFT was struck by its detachment from
physically intuitive ideas and
enormous complexity. A successful physical theory is expected
to have, as much as possible, real-life counterparts for its mathematical constructs. This is often not
the case in QFT, where such physically transparent concepts of
quantum mechanics as the Hilbert space, wave functions, particle
observables, and Hamiltonian  were substituted
(though not completely discarded) by more formal and obscure notions
of quantum fields, ghosts, propagators, and Lagrangians. It was even
declared that the concept of a particle is not fundamental anymore
and must be abandoned in favor of the field description of nature:

\begin{quote}
\emph{In its mature form, the idea of quantum field theory is that
quantum fields are the basic ingredients of the universe and
particles are just bundles of energy and momentum of the fields.} S.
Weinberg \cite{weinberg_field}
\end{quote}

  The most notorious failure
of QFT is the problem of  ultraviolet
divergences: To obtain sensible  results from QFT calculations
one must drop certain infinite
terms. Although rules for doing such tricks are well-established, they
cannot be considered a part of a mathematically sound theory. As Dirac
remarked

\begin{quote}
\emph{This is just not sensible mathematics. Sensible mathematics involves
neglecting a quantity when it turns out to be small --
 not neglecting it because it is infinitely large and you do not want
it!} P. A. M. Dirac
\end{quote}

\noindent In modern QFT the problem of ultraviolet infinities is not
solved, it is ``swept under the rug.'' Even if the infinities in scattering amplitudes are ``renormalized'', one ends up with an ill-defined Hamiltonian, which is not suitable for describing the time evolution of states. The prevailing opinion is
that ultraviolet divergences are related to our lack of
understanding of physics at short distances. It is often argued that
QFT is a low energy (effective) approximation to some yet unknown truly
fundamental theory, and that in this final theory the small distance
or high energy (ultraviolet) mischiefs will be tamed somehow. There
are various guesses about what this ultimate theory may be.
 Some think that
future theory will reveal a non-trivial, probably discrete, or
non-commutative structure of space at distances comparable to the
Planck scale of $10^{-33}$ cm. Others hope that paradoxes will
go away if we replace point-like particles by tiny extended
objects, like strings.

Many researchers agree that the most fundamental obstacle on the way
forward is the deep contradiction between quantum theory and
Einstein's relativity theory (both special and general). In a more
general sense, the basic question is ``what is space and time?'' The
answers given by Einstein's theory of relativity and by quantum
mechanics are quite different. In special relativity, position and
time are treated on an equal footing, both of them being coordinates
in the 4-dimensional Minkowski space-time. However in quantum
mechanics position and time play very different roles. Position (as any other physical observable) is
an observable described  by an
Hermitian operator, whereas time is a numerical parameter, which
cannot be cast into the operator form without contradictions.

In our book we would like to take a fresh look at these issues. Two basic postulates of our approach are completely non-controversial. They are the \emph{principle of
relativity} (= the equivalence of all inertial frames of reference) and the
laws of \emph{quantum mechanics}. From the mathematical perspective, the former
postulate is embodied in the notion of the Poincar\'e group and the
latter postulate leads to the algebra of operators in the Hilbert
space. When combined, these two statements inevitably imply the idea
of unitary representations of the Poincar\'e group in the Hilbert
space as the major mathematical tool for the description of any isolated
physical system. One of our goals is to demonstrate that observable physics fits nicely into
this mathematical framework. We will also see that traditional theories sometimes deviate from these postulates, which often leads to unphysical conclusions and paradoxes. Our goal is to find, analyze, and correct these deviations.

Although the ideas presented here have rather general
nature, most calculations will be performed for systems of
charged particles and photons and electromagnetic forces acting
between them. Traditionally, these
systems were described by quantum electrodynamics (QED). However, our
approach will lead us to a different theory, which we call
\emph{relativistic quantum dynamics} or RQD. \index{relativistic
quantum dynamics} \index{RQD (relativistic quantum dynamics)} Our approach is exactly equivalent to the renormalized QED as long as
properties related to the $S$-matrix (scattering cross-sections,
lifetimes, energies of bound states, etc.) are concerned. However,
different results are expected for the time evolution and boost transformations in interacting systems.

RQD differs from the traditional approach in two important aspects:
the recognition of the \emph{dynamical character of boosts} and the
\emph{primary role of particles} rather than fields.

\bigskip
\textbf{The dynamical character of boosts.} Lorentz transformations for space-time coordinates of events are the most fundamental relationships in Einstein's special relativity. These formulas are usually derived for simple events associated either with light beams or with free (non-interacting) particles. Nevertheless, special relativity tacitly assumes that these Lorentz
formulas can be extended to all events with interacting
particles regardless of the interaction strength. We will show that this assumption is actually wrong. We
will derive boost transformations of particle observables by using
Wigner's theory of unitary representations of the Poincar\'e group
\cite{Wigner_unit} and Dirac's approach to relativistic interactions
\cite{Dirac}. It will then follow that boost transformations should
be interaction-dependent. Usual universal Lorentz transformations of special
relativity are thus only approximations. The Minkowski 4-dimensional space-time is an
approximate concept as well.

\bigskip

\textbf{Particles rather than fields.} Presently accepted quantum
field theories (e.g., the renormalized QED) have serious difficulties in describing the time
evolution of even simplest systems, such as vacuum and
single-particle states. Direct application of the QED time evolution
operator to these states leads to spontaneous creation of extra
(virtual) particles, which have not been observed in experiments. The
problem is that bare particles of QED have
rather remote relationship to physically observed electrons,
positrons, etc., while the rules connecting bare and physical
particles are not well established. We
 solve this problem by using the ``dressed particle'' formalism, which
is the cornerstone of our RQD approach.  The ``dressed'' Hamiltonian of RQD is
obtained by applying a unitary dressing transformation to the traditional QED
Hamiltonian. This transformation does not change the $S$-operator of QED,
therefore the perfect agreement with experimental data is preserved.
The RQD Hamiltonian describes electromagnetic phenomena in terms
of directly interacting physical particles (electrons, photons, etc.) without
reference to spurious bare and virtual particles. Quantum fields play only an auxiliary
role. In addition to accurate
scattering amplitudes, our approach allows us to obtain the time
evolution of interacting particles and offers a rigorous way to find
both energies and wave functions of bound states. All calculations
with the RQD Hamiltonian can be done by using standard recipes of
quantum mechanics without encountering embarrassing ultraviolet divergences and
without the need for artificial cutoffs, regularization,
renormalization, etc.

Of course, the idea of particles with action-at-a-distance forces is not new. The original Newtonian theory of gravity had exactly this form, and (quasi-)particle approaches are often used in modern theories. However, the consensus opinion is that such approaches can be only approximate, in particular, because instantaneous interactions are believed to violate important principles of relativistic invariance and causality. Textbooks try to convince us that these important principles can be reconciled with quantum postulates only in a theory based on local (quantum) fields with retarded interactions. In this book we are going to challenge this consensus and demonstrate that the particle picture and action-at-a-distance do not contradict relativity and causality.

Our central message can be summarized in few sentences

\begin{quote}
\emph{\textbf{The physical world is composed of point-like particles. They obey laws of quantum mechanics and interact with each other via instantaneous action-at-a-distance potentials, which depend on distances between the particles and on their momenta. These potentials may lead to the creation and annihilation of particles as well. This picture is in full agreement with principles of relativity and causality. In order to establish this agreement one should recognize that boost transformations of particle observables depend on interactions acting in the system. Thus special-relativistic formulas for Lorentz transformations are approximate. Exact relativistic theories of interacting particles should be formulated without reference to the unphysical 4D Minkowski space-time.}} \label{st:basic}
\end{quote}

\bigskip
 This book is divided into three
parts. Part I: \textbf{QUANTUM ELECTRODYNAMICS} comprises ten
chapters  \ref{ch:QM} - \ref{ch:renormalization}. In this part we avoid controversial issues and stick to traditionally accepted views on relativistic quantum theories, such as QFT. We specify our basic assumptions, notation, and terminology while trying to follow a logical path starting from basic postulates of relativity and probability and culminating in calculation of the renormalized $S$-matrix in QED.  The purpose of Part I is to set the stage for introducing our non-traditional particle-based approach in the second part of the book.

 In chapter \ref{ch:QM} \textbf{Quantum
mechanics} the basic laws of quantum mechanics are derived from
simple axioms of measurements (quantum logic).

In chapter
\ref{ch:relativity} \textbf{Poincar\'e group} we introduce the
Poincar\'e group as a set of transformations that relate different
(but equivalent) inertial reference frames.

Chapter
\ref{ch:QM-relativity} \textbf{Quantum mechanics and relativity}
unifies the two above pieces and establishes unitary representations of
the Poincar\'e group in the Hilbert space of states as the most
general mathematical description of any isolated physical system.

In chapter
\ref{ch:operators} \textbf{Operators of observables} we find the
correspondence between known physical observables (such as mass,
energy, momentum, spin, position, etc.) and concrete Hermitian
operators in the Hilbert space.

Chapter \ref{ch:single}
\textbf{Single particles} is devoted to Wigner's theory of
irreducible representations of the Poincar\'e group. It provides a
complete description of basic properties and dynamics of isolated
stable elementary particles.

In chapter \ref{ch:interaction}
\textbf{Interaction} we discuss
relativistically invariant interactions in multi-particle systems.

Chapter \ref{sc:scattering} \textbf{Scattering} is devoted to quantum-mechanical description of particle collisions.

In chapter
\ref{ch:fock-space} \textbf{Fock space} we consider the general class of systems in which particles can be
created and annihilated and their numbers are not conserved.

In chapter  \ref{ch:QED} \textbf{Quantum
electrodynamics} we apply all the above ideas to  systems of charged particles and photons in the
formalism of QED.

Chapter \ref{ch:renormalization} \textbf{Renormalization} concludes this first ``traditional'' part of the
book. This chapter discusses the appearance of ultraviolet divergences in QED and explains their elimination by means of counterterms added to the Hamiltonian.

Part II of the book \textbf{QUANTUM THEORY OF PARTICLES} (chapters
\ref{ch:rqd} - \ref{ch:summary}) examines the new particle-based RQD approach, its
connection to the traditional theory from part I, and its advantages. Our goal is to dispel the common prejudice against using particle interpretation in relativistic quantum theories. We show that the view of the world as consisting of point particles interacting via instantaneous direct potentials is not contradictory and is capable to explain physical phenomena just as well - or even better - as the mainstream field-based view.

This
``non-traditional'' part of the book begins with chapter \ref{ch:rqd}
\textbf{Dressed particle approach}, which provides a deeper analysis of
renormalization and the bare particle picture in quantum field
theories. The main ideas of our particle-based approach are formulated here and QED
is being rewritten in terms of creation and annihilation operators of physical particles, rather than
bare quantum fields.

In chapter \ref{sc:coulomb} \textbf{Coulomb potential and beyond} we derive the dressed interaction between charged particles and use it to calculate the spectrum of the hydrogen atom.

Chapter \ref{ch:decays} \textbf{Decays} deals with a rigorous description of unstable quantum systems. The special focus is on decays of fast moving
particles. Here we show that the usual Einstein's time dilation formula is not an accurate description of such phenomena. In principle, it should be possible to observe deviations from this formula in experiments, but, unfortunately, the required precision cannot be reached with the currently available technology.

The mathematics of particle decays is applied to radiative transitions in the hydrogen atom in chapter \ref{ch:hydrogen-revisited} \textbf{RQD in higher orders}. In this chapter we also discuss infrared divergences (and their cancelation) in high perturbation orders. In particular, we calculate the electron's anomalous magnetic moment and the Lamb shifts of atomic energy levels.

In chapter \ref{ch:theories}
\textbf{Classical electrodynamics} we show that classical electrodynamics can be reformulated as a Hamiltonian  theory of charged particles with action-at-a-distance forces. These forces depend not only on the distance between the charges, but also on their velocities and spins. In this formulation, electromagnetic fields and potentials are not present at all and Maxwell's equations do not play any role. This allows us to resolve a number of theoretical paradoxes and, at the same time, remain in agreement with experimental data. Even the famous Aharonov-Bohm experiment gets its explanation as an effect of inter-particle interactions on the phases of quantum wave packets -- i.e., without any involvement of electromagnetic potentials and non-trivial space topology.

We conclude our discussion of electromagnetic phenomena by the chapter \ref{ch:support} \textbf{Experimental support for RQD}, where we briefly describe several experiment supporting our idea about the instantaneous propagation of Coulomb and magnetic interactions.

In chapter \ref{sc:obs-interact} \textbf{Particles
and relativity} we discuss real and imaginary paradoxes usually
associated with the
particle interpretation of QFT. In particular, we discuss the superluminal spreading of
localized wave packets and the Currie-Jordan-Sudarshan ``no
interaction'' theorem. We show that superluminality and action-at-a-distance can coexist with causality if the relativistic invariance of interactions is properly understood.

The final small chapter \ref{ch:summary} \textbf{Conclusions} summarizes major results and conclusions of this work and briefly mentions possible directions for future investigations.

Some useful mathematical facts and more
technical derivations are collected in the  Part III:
\textbf{MATHEMATICAL APPENDICES}.

\bigskip

 Remarkably, the development of the new particle-based RQD  approach did not require introduction of radically new physical
ideas. Actually, all key ingredients of this study  were formulated
a long time ago, but for some reason they have not attracted the
attention they deserve. For example, the fact that either
translations or rotations or boosts must have dynamical dependence
on interactions was first established
 in Dirac's work \cite{Dirac}. These
ideas were further developed in ``direct interaction'' theories by
Bakamjian and Thomas \cite {Bakamjian_Thomas}, Foldy \cite{Foldy},
Sokolov \cite{Sokolov_Shatnii, Sokolov_Shatnii2}, Coester and
Polyzou \cite{Coester_Polyzou} and many others. The primary role of
particles in formulation of quantum field theories was emphasized in
an excellent book by Weinberg \cite{book}. The ``dressed particle''
approach was advocated by Greenberg and Schweber \cite{GS}.\footnote{A very similar unitary transformation technique was developed even earlier by Fr\"{o}hlich \cite{Frohlich, Frohlich1961} in the theory of electron-phonon interactions in solids.} First
indications that this approach can solve the problem of ultraviolet
divergences in QFT are contained in papers by Ruijgrok \cite{Ruij},
Shirokov and Vi\c sinesku \cite{Shirokov-72, Vishinesku}. The formulation of RQD  presented in this book just combined all these
good ideas into one comprehensive approach, which, we believe, is a
step toward a consistent unification of quantum mechanics and the
principle of relativity.

In this book we are using the
Heaviside-Lorentz system of units\footnote{see Appendix in
\cite{Jackson}} in which the Coulomb law has the form $V = q_1q_2/(4 \pi r)$ and the proton charge has the value of $e= 2 \sqrt{\pi}  \times 4.803 \times 10^{-10}$ statcoulomb. The speed of light is $c=2.998 \times 10^{10}$ cm/s; the Planck constant is $\hbar = 1.054 \times 10^{-27} erg \cdot s$, so the \emph{fine structure constant} is $\alpha \equiv e^2/(4 \pi \hbar c) \approx 1/137$. \index{fine structure constant}

The new material contained in this book was partially covered in
six  articles \cite{Stefanovich_qft, Stefanovich_decay,
Stefanovich_Mink, Stefanovich_ren, Stefanovich_dec, Ahar}.

I would like to express my gratitude to Peter Enders, Theo Ruijgrok
and Boris Zapol for reading this book and making valuable critical
comments and suggestions. I also would like to thank  Harvey R. Brown, Rainer Grobe, William
Klink, Vladimir Korda, Chris Oakley, Federico Piazza, Wayne Polyzou,
Alexander Shebeko and Mikhail Shirokov for enlightening
conversations as well as Bilge, Bernard Chaverondier,
 Wolfgang Engelhardt, Juan R. Gonz\'alez-\'Alvarez,  Bill Hobba, Igor
Khavkine, Mike Mowbray, Arnold Neumaier and Dan Solomon for online
discussions and fresh ideas that allowed me to improve the quality
of this manuscript over the years. These acknowledgements do not imply any direct or
indirect endorsements of my work by these distinguished researchers.
All errors and misconceptions contained in this book are mine and
only mine.

\chapter{INTRODUCTION }
\label{ch:intro}

\begin{quote}
\textit{It is wrong to think that the task of physics is to find out
how nature \emph{is}. Physics concerns what we can \emph{say} about
nature...}

\small
\hspace{1in}  Niels Bohr
\normalsize
\end{quote}

\vspace {0.5in}

In this Introduction, we will try to specify more exactly what is
the purpose of theoretical physics and what are the fundamental
 concepts and their relationships studied by this branch of
science. Some of the definitions and statements made here may look
self-evident or even trivial. Nevertheless, it seems important to
spell out these definitions explicitly, in order to avoid
misunderstandings in later parts of the book.

\begin{figure}[!t]
\includegraphics[width=13cm,height=7cm]{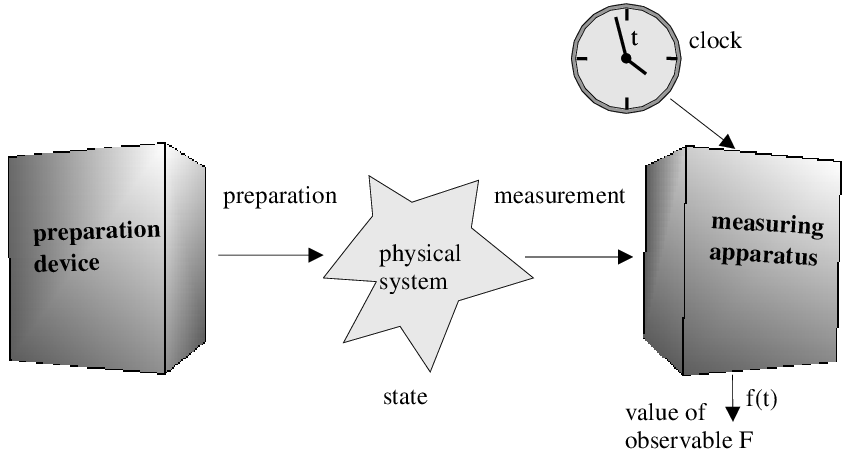} \caption{Schematic
representation of the preparation/measurement process.}
\label{fig:3.4}
\end{figure}

We obtain all information about the physical world from
\emph{measurements}, \index{measurement} and the fundamental
goal of theoretical physics is to describe and predict the results
of these measurements. The act of measurement
requires at least three objects (see Fig.
\ref{fig:3.4}): a \emph{preparation device}, \index{preparation
device} a \emph{physical system}, and a \emph{measuring apparatus}.
\index{measuring apparatus} The preparation device prepares the
physical system \index{physical system} in a certain
\emph{state}. \index{state} The state of the system has some
attributes or properties. If an attribute or property can be
assigned a numerical value it will be called \emph{observable} $F$.
\index{observable} The observables are measured by bringing the
system into contact with the measuring apparatus. The result of the
measurement is a numerical value of the observable, which is a real
number $f$. We assume that every measurement of $F$ yields
\emph{some} value $f$, so that there are no misfirings of the
measuring apparatus.

This was just a brief list of relevant notions. Let us now look at
all these ingredients in more detail.

\bigskip

\textbf{Physical system.} Loosely speaking, the physical system is any
object that can trigger a response (measurement) \index{measurement}
in the measuring apparatus. \index{measuring apparatus} As physical
system is the most basic concept in physics, it is difficult to give
a more precise definition. An intuitive understanding will be
sufficient for our purposes. For example, an electron, a hydrogen
atom, a book, a planet are all examples of physical systems.

Physical systems can be either \emph{elementary} (also called
\emph{particles}) \index{particle} or \emph{compound},
\index{compound system} i.e., consisting of two or more particles.

In this book we will limit our discussion to \emph{isolated
systems}, \index{isolated system} which do not interact with the
rest of the world or with any external potential.\footnote{Of course, the interaction with the measuring apparatus must be allowed, because this interaction is the only way to get objective information about the system. However, we reject the idea that the process of measurement should have a dynamical description in the theory. See subsection \ref{sc:measurement}.} By doing so, we
exclude some interesting physical systems and effects, like atoms
 in external electric and magnetic fields. However, this
does not limit the generality of our treatment. Indeed, one can
always combine the atom and the field-creating device into one
unified system that can be studied within the ``isolated system''
approach.

\bigskip

\textbf{States.}
 Any physical system may
exist in a variety of different states: \index{state} a book can be
on your desk or in the library; it can be open or closed; it can be
at rest or fly with a high speed. The distinction between different
systems and different states of the same system is sometimes far
from obvious. For example, a separated pair of particles (electron +
proton) does not look like the hydrogen atom. So, one
may conclude that these are two different systems. However, in
reality these are two different states of the same compound system.

\bigskip

\textbf{Preparation and measuring devices.} Generally, preparation
and measuring devices can be rather sophisticated, e.g.,
accelerators, bubble chambers, etc. It would be hopeless to include
in our theoretical framework a detailed description of
 their design and how they interact with the physical system. Instead, we
 will use an idealized representation of both the preparation and
measurement  acts. In particular, we will assume
that the measuring apparatus is a black box whose job is to  produce
just one real number - the value of some observable - upon
the act of measurement.

It is important to note that generally the measuring device can measure only \emph{one} observable. We will not assume that it is possible to measure several observables at once with the same device. For example, a particle's position and momentum cannot be obtained in one measurement.

We will also see that one preparation/measurement act is not sufficient for a full characterization of the studied physical system. Our preparation/measurement setup should be able to process multiple copies of the same system prepared in exactly the same conditions.\footnote{This is also called an \emph{ensemble}.} A striking property of nature is that in such repetitive measurements we are not guaranteed to obtain exactly the same results. We will see that in many cases results of measurements are subject to a random scatter. So, theoretical descriptions of states can be only probabilistic. This idea is the starting point of \emph{quantum mechanics}.

\bigskip

\textbf{Observables.} Theoretical physics is inclined to study
simplest physical systems and their most fundamental observable
properties, such as mass, velocity, spin, etc. We will assume exact
measurability of any observable. Of course, this claim  is an
idealization. Clearly, there are precision limits for all real
measuring apparatuses.  However, we will
suppose that with certain efforts one can always make more and more
accurate measurements,  so the precision is, in
principle, unlimited.\footnote{For example, it is impossible in
practice to measure location of the electron inside the hydrogen
atom. Nevertheless, we will assume that this can be done in our
idealized theory. Then each individual measurement of the electron's
position would yield a certain result $\mathbf{r}$. However, as will
be discussed in chapter \ref{ch:QM}, results of repetitive
measurements in the ensemble are generally non-reproducible and
random. So, in quantum mechanics the electron's state should be
describable by a probability distribution $|\psi(\mathbf{r})|^2$.}

 Some observables can take a value anywhere on the real axis
$\mathbb{R}$. The Cartesian components of position $R_x$, $R_y$, and
$R_z$ are good examples of such (unlimited range, continuous)
 observables. However there
are also observables for which this is not true and the allowed
values form only a subset of the real axis. Such a subset is called
the \emph{spectrum} \index{spectrum of observable} of the
observable. For example, it is known (see Chapter \ref{ch:single})
that each component of particle's velocity cannot exceed the speed
of light $c$, so the spectrum of the velocity components $V_x$,
$V_y$, and $V_z$ is $[-c,c]$. Both position and velocity have
\emph{continuous} \index{continuous spectrum} spectra. However,
there are many observables having a \emph{discrete} \index{discrete
spectrum} spectrum. For example, the number of particles in the
system (which is also a valid observable)
 can only take integer values 0, 1, 2, ... Later we will also meet observables whose
spectrum is a combination of discrete and continuous parts, e.g.,
the energy spectrum of the hydrogen atom. \index{hydrogen atom}

Clearly the measured values of observables must  depend on the kind
of the system being measured and on its state. The
measurement of any true observable must involve some kind of
interaction or contact between the observed system and the measuring
apparatus. We emphasize this fact because there are numerical
quantities in physics which are not associated with any physical
system and therefore they are not called observables. For example,
the number of space dimensions (3) is not an observable, because we
do not regard space as an example of a physical system.

\bigskip

\textbf{Time and clocks.} Another important physical quantity, that
does not belong to the class of observables, is time.  We cannot say
that time is a property of a physical system, because a
``measurement'' of time (looking at positions of the clock's arms) does
not involve any interaction with the physical system. One can
``measure'' time even in the absence of any physical system in our laboratory. To do
that, one just needs to have a clock, which is a necessary part of
any experimental setup and not a physical system by itself.\footnote{Of
course, one can decide to consider the laboratory
clock as a physical system and perform physical measurements on it. For example, one can investigate the quantum uncertainty in
the clock's arm position. However, then this particular clock is no
longer suitable for ``measuring'' time. Some other device must be
used for time-keeping purposes. } The clock assigns a time label (a
numerical parameter) to each measurement of true
observables, and this label does not depend on the state of the
observed system. The unique place of the clock and time in the
measurement process is indicated in Fig. \ref{fig:3.4}.

\bigskip

\textbf{Observers.}
 We will call
\emph{observer} \index{observer} $O$ a collection of measuring
apparatuses  (plus a specific device called \emph{clock}), which are designed to
measure all possible observables. \emph{Laboratory}
\index{laboratory} is a full experimental setup, i.e., a
preparation device   plus observer $O$
with all his measuring devices.

In this book we consider only \emph{inertial observers} (= inertial
frames of reference) \index{inertial frame of reference}
\index{inertial observer} or inertial laboratories. These are
observers that move uniformly without acceleration
 and rotation, i.e., observers whose velocity
and orientation of axes does not change with time. The importance of
choosing inertial observers will become clear in section
\ref{sc:inertial} where we will see that measurements
performed by these observers obey the principle of relativity.

The minimal set of measuring devices associated with an observer
include a yardstick for measuring distances, a clock for registering
time, a fixed point of origin and three mutually perpendicular axes
erected from this point. In addition to measuring properties of
physical systems, our observers can also see their fellow observers.
With the measuring kit described above each observer $O$ can
characterize another observer $O'$ by ten parameters $\{\vec{\phi},
\mathbf{v}, \mathbf{r}, t \}$. These parameters include i) the time shift $t$ between  $O$ and $O'$; ii) the position vector $\mathbf{r}$ that
connects the origin of $O$ with the origin of $O'$; iii) the rotation
angle\footnote{The vector parameterization of rotations is discussed
in Appendix \ref{ss:parameterization}.} $\vec{\phi}$ that relates
 orientations of axes in $O'$ to  orientations of axes in $O$ and iv) the velocity $\mathbf{v}$ of $O'$ relative to $O$.

It is convenient to introduce the notion of \emph{inertial
transformations} \index{inertial transformations of observers} of
observers and laboratories. Transformations of this kind include

\begin{itemize}
\item  rotations,
\item  space translations,
\item  time translations,
\item  changes of velocity or \emph{ boosts}. \index{boost} \label{page:parameters}
\end{itemize}

\noindent There are three independent rotations (around $x$, $y$, and $z$ axes), three independent translations and three independent
boosts. So, along with the time translations that makes 10 basic types of inertial
transformations. More general inertial transformations can be made
by performing two (or more) basic transformations in succession. We
will postulate that for any pair of inertial observers $O$ and $O'$
one can always find an inertial transformation $g$, such that $O'
=gO$. Conversely, application of any inertial transformation $g$
to any inertial observer $O$ leads to a different valid inertial observer
$O' = gO$. In chapter \ref{ch:relativity} we will make an important observation that transformations $g$ form a group.

An important comment should be made about the definition of
``observer'' used in this book. Usually, an observer is understood
as a person (or a measuring apparatus)
that exists and performs measurements for
infinitely long time. For example, it is common to
discus the time evolution of a physical system ``from the
point of view'' of this or that observer. However, this colloquial
definition does not fit our purposes. The problem  with this
definition is that it singles out time translations as being different
from space translations, rotations, and boosts. In this approach time translations become associated with the
observer herself rather than being treated equally with other inertial transformations between observers. The central idea of our approach to relativity is
to treat all ten types of inertial transformations on equal
footing. Therefore, we will use a  definition of
observer that is slightly different from the one described above. In our definition observers are ``short-living.'' They
exist and perform measurements in a short time interval and they can
see only a snapshot of the world around them. So, individual observers
cannot ``see'' the time evolution of a physical system. In our
approach the time evolution is described as a succession of
measurements performed by a series of instantaneous observers
related to each other by time translations. Then the colloquial
``observer'' is actually a continuous sequence of our ``short-living'' observers
displaced in time with respect to each other.

\bigskip

One of the most important tasks of physics is to establish the
relationship between measurements  performed by
two different observers on the same physical system. These
relationships will be referred to as \emph{inertial transformations
of observables}. \index{inertial transformations of observables} In
particular, if values of some observables measured by $O$ are known, and
the inertial transformation connecting $O$ with $O'$ is known as
well, then we should be able to figure out the values of
those observables from the point of view of $O'$. Probably the most important
and challenging task of this kind is the description of
\emph{dynamics}  or \emph{time evolution}. \index{time evolution}  In this case, observers $O'$ and $O$ are
connected by a time translation.

\textbf{The goals of physics.} The above discussion  can be summarized by indicating five essential
goals of theoretical physics:

\begin{itemize}
\item provide a classification of physical systems;
\item for each physical system give a list of observables and their spectra;
\item  for each physical system give a list of possible states;
\item for each state of the system describe the results of measurements of relevant
observables;
\item given one observer's description of the system's state find out how other observers see the same
state.\label{goals}
\end{itemize}

\mainmatter

\part{QUANTUM ELECTRODYNAMICS}

\chapter{QUANTUM MECHANICS}
\label{ch:QM}

\begin{quote}
\textit{The nature of light is a subject of no material importance to the
concerns of life or to the practice of the arts, but it is in many other respects extremely
interesting. }

\small
\hspace{1in} Thomas Young
 \normalsize
\end{quote}

\vspace {0.5in}

In this chapter we are going to discuss the most basic
inter-relationships between preparation devices,  physical systems, and measuring apparatuses (see Fig.
\ref{fig:3.4}). In particular, we will ask what kind of information
about the physical system can be obtained by the observer and how
this information depends on the state of the system?

Until the end of the 19th century these questions were answered by
classical mechanics which, basically, said that in each state the
physical system has a number of observables (e.g, position,
momentum, mass, etc) whose values can be measured simultaneously,
accurately, and reproducibly. These deterministic views were
held to be indisputable and self-evident not only in classical mechanics, but throughout
classical physics.

Dissatisfaction with the classical theory started to grow at the
end of the 19th century when this theory was found inapplicable to
microscopic phenomena, such as the radiation spectrum of heated
bodies,  discrete spectra of atoms, and the photo-electric
effect. Solutions for these and many other problems were found in
\emph{quantum mechanics} \index{quantum mechanics} whose creation
involved joint efforts and passionate debates of such outstanding
scientists as Bohr, Born, de Broglie, Dirac, Einstein, Fermi, Fock,
Heisenberg, Pauli, Planck, Schr\"odinger, Wigner, and many others.
The picture of the physical world emerged from these efforts was
weird, paradoxical, and completely different from the familiar
classical picture.
 However, despite this apparent weirdness, predictions of
quantum mechanics
 are being tested countless times everyday
in physical and chemical laboratories around the world and not a
single time were these ``weird'' predictions  found wrong. This
makes quantum mechanics the most successful and accurate physical
theory of all times.

There are dozens of good textbooks, which explain the laws and rules of quantum mechanics and how they can be used to perform calculations in each specific case. These laws and rules are not controversial and the reader of this book is supposed to be familiar with them. However, the deeper meaning and interpretation of the quantum formalism is still a subject of a fierce debate. Why does nature obey the rules of quantum mechanics? Why there are wave functions satisfying the linear superposition principle? Is it possible to change the rules (e.g., introduce some non-linearity) without finding ourselves in contradiction with experiments?  People are asking these questions more frequently in recent years as the search for quantum gravity has intensified, and one fashionable idea was that one should modify the rules quantum mechanics in order to reconcile them with general relativity.

In this chapter we will present a less-known viewpoint on theoretical origins of quantum laws. This approach says that the true laws of logic applicable to physical measurements are different from the classical laws of Aristotle and Boole. The familiar classical logic should be replaced by the so-called \emph{quantum logic}. \index{quantum logic} We will argue that the formalism of quantum mechanics (including vectors and Hermitian operators in the Hilbert space) follows almost inevitably from simple properties of measurements and quantum-logical relationships between them. These properties and relationships are so basic, that it seems impossible to modify them and thus to change quantum laws without destroying their internal consistency and their consistency with observations. In section \ref{sc:complete} we will also add some thoughts to the never-ending philosophical debate about interpretations of quantum mechanics.

\section{Why do we need quantum mechanics?}
\label{sc:thought}

The inadequacy of classical concepts is best seen by analyzing the
debate
 between the corpuscular and wave theories of
light.    Let us demonstrate the essence of this centuries-old
debate on an example of a  thought experiment with a pinhole camera.

\subsection{Corpuscular theory of light}
\label{ss:corpuscular}

You probably saw or heard about a simple device called \emph{camera
obscura} \index{camera obscura} or pinhole camera. You can easily
make this device yourself: Take a light-tight box,
 put a photographic plate inside  the box and make a small
hole in the center of the  side opposite to the photographic plate
(see Fig. \ref{fig:3.1}). The light passing  through the hole inside
the box creates a sharp inverted image on the photographic plate.
 You will get even
sharper image by decreasing the size of the hole, though the
 image will become dimmer, of course. This behavior
of light was well known for centuries (a drawing of the camera
obscura is present in Leonardo da Vinci's  papers). One of the
earliest scientific explanations of this and other  properties of
light (reflection, refraction, etc.) was suggested by Newton. In
modern language, his \emph{corpuscular theory} \index{corpuscular
theory} would explain the formation of the image like this:

\begin{figure}
\centering
\includegraphics{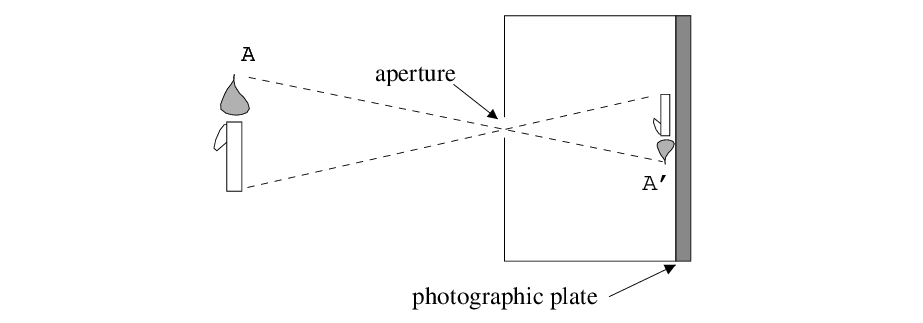} \caption{The image in the \emph{camera
obscura} with a pinhole aperture is created by straight light rays:
the image at the point $A'$ on the photographic plate is created only by
light rays emitted from the point $A$ and passed straight through the
hole.} \label{fig:3.1}
\end{figure}

\begin{quote}
\textbf{Corpuscular theory:}
 Light
is a flow of tiny particles (photons) propagating along straight classical
trajectories (light rays). Each particle in the ray carries a
certain amount of energy, which gets released upon impact in a very
small volume corresponding to one grain of the photographic emulsion
and produces a small dot image.
 When intensity
of the source is high, there are so many particles, that we cannot
distinguish their individual dots. All these dots merge into one
continuous image, and the intensity of the image is proportional to
the number of particles hitting the photographic plate during the
exposure time.
\end{quote}

\begin{figure}
 \includegraphics{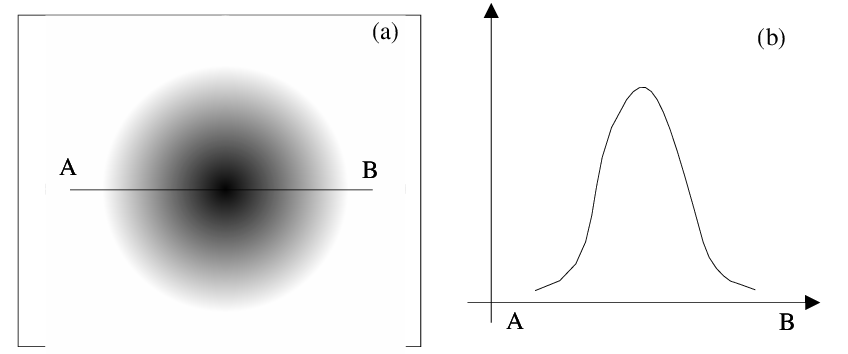} \caption{(a) Image in the
pinhole camera with a very small aperture; (b) the density of the
image along the line AB} \label{fig:3.2}
\end{figure}

Let us continue our experiment with the pinhole camera and decrease
the size of the hole even further. The corpuscular theory would
insist that the smaller size of the hole must result in a sharper image.
However this is not what experiment shows! For a very small hole the
image on the photographic plate will be blurred. If we further
decrease the size of the hole, the detailed picture will completely
disappear and the image will look like one large diffuse spot (see Fig.
\ref{fig:3.2}), independent on the form and shape of the light
source outside the camera. It appears as if light rays scatter in
all directions when they pass through a small aperture or near a
small object. This effect of the  light spreading is called
\emph{diffraction}, \index{diffraction} and it was discovered by
Grimaldi in the middle of the 17th century.

\begin{figure}
\includegraphics{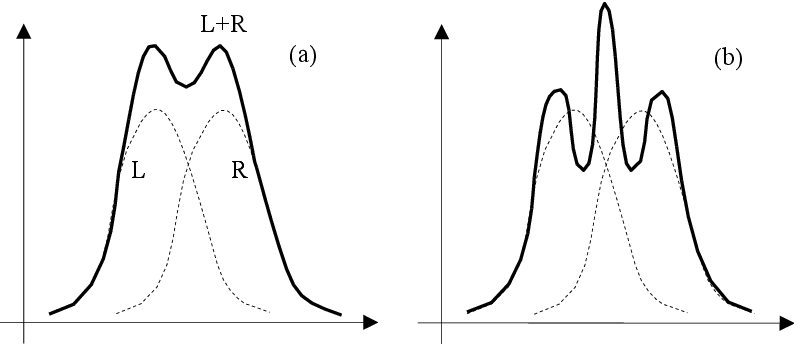} \caption{(a) The density
of the image in a two-hole camera according to na\"ive corpuscular
theory is a superposition of images created by the left (L) and
right (R) holes; (b) Actual interference picture: In some places the
density of the image is higher than L+R (constructive interference);
in other places the density is lower than L+R (destructive
interference). } \label{fig:3.3}
\end{figure}

Diffraction is rather difficult to reconcile with the
corpuscular theory. For example,  we can try to save this theory by
assuming  that light rays deviate from their straight paths due to
some interaction with the box material surrounding the hole. However
this is not a satisfactory explanation, because one can easily establish by experiment that the shape of the diffraction picture is completely independent on the type of material used to make the walls of the pinhole camera. The most striking evidence
of the fallacy of the na\"ive corpuscular theory is the effect of
light \emph{interference} \index{interference}
 discovered by Young in 1802 \cite{Young}.
To observe the  interference, we can slightly modify our pinhole
camera by making two small holes close to each other, so that their
diffraction spots on the photographic plate overlap. We already know
that when we leave the left hole open and close the right hole we
get a diffuse spot L (see Fig. \ref{fig:3.3}(a)). When we leave open
the right hole and close the left hole we get another spot R. Let us
try to predict what kind of image we will get if both holes are opened.

Following the corpuscular theory and simple logic we might conclude
that  particles  reaching the photographic plate are of two sorts:
those passed through the left hole and those passed through the
right hole.
 When the two holes are opened at the same time,  the density of the
``left hole`` particles should add to the density of the ``right
hole'' particles and the resulting image should be a superposition
L+R of the two images (full line in Fig. \ref{fig:3.3}(a)).
 Right? Wrong!
This seemingly reasonable explanation disagrees with experiment.
 The actual image  has
new features (brighter and darker regions) called the \emph{interference
picture}  (full line in  Fig. \ref{fig:3.3}(b)).

Can the corpuscular theory explain this strange interference pattern? We
could assume, for example,  that some kind of interaction between
light corpuscles is responsible for the interference, so that passages of different particles
through left and right holes are not independent events, and the law of addition
of probabilities does not hold for them. However, this
idea must be rejected because, as we will see later, the interference picture persists even if photons are released one-by-one, so that they cannot interact with each other.

\subsection{Wave theory of light}
\label{ss:wave}

The inability to explain such basic effects of light propagation as
diffraction and interference was a major embarrassment for the
Newtonian corpuscular theory. These effects as well as all other
properties of light known before quantum era (reflection,
refraction, polarization, etc.)
 were brilliantly explained by the wave
theory of light  advanced by Grimaldi, Huygens, Young, Fresnel, and
others.  The wave theory gradually replaced Newtonian corpuscles
 in the course of the 19th century. The idea of light as a  wave found its strongest support from
Maxwell's  electromagnetic theory which unified optics with electric
and magnetic phenomena.  Maxwell explained that  the light wave
is actually an oscillating field of electric
$\mathbf{E}(\mathbf{x},t)$ and magnetic $\mathbf{B}(\mathbf{x},t)$
vectors -- a sinusoidal wave propagating with the speed of light.
According to the Maxwell's theory, the energy of the wave and
consequently the intensity of light $I$, is proportional to the
square of the amplitude of the field vector oscillations, e.g., $I
\propto \mathbf{E}^2$. Then
formation of the photographic image can be explained as follows:

\begin{quote}
\textbf{Wave theory:} Light is a continuous wave or field
propagating in space in an
 undulatory fashion.
 When the light wave meets
molecules of the photo-emulsion, the charged parts of the molecules
start to oscillate under the influence of the light's electric and
magnetic field vectors. The portions of the photographic plate with higher field amplitudes
have more violent molecular oscillations and higher image densities.
\end{quote}

This provides a natural explanation for both diffraction and interference:  Diffraction simply means that light  waves can
deviate from straight paths and go around corners, just like
sound waves do.\footnote{Wavelengths corresponding to the visible light
are  between 0.4 micron for the violet light
and 0.7 micron for the red light.
 So for large
 obstacles or holes, the deviations from the straight path are very small and
  the  corpuscular theory of light works reasonably well. }  To
explain the interference, we just need to
 note that
 when two portions of the  wave pass through different holes and meet
 on the photographic plate, their electric vectors add up.
However intensities of the waves are not additive: $I \propto (\mathbf{E}_1 +
\mathbf{E}_2)^2 = \mathbf{E}_1^2 +
 2 \mathbf{E}_1 \cdot \mathbf{E}_2 + \mathbf{E}_2^2 \neq \mathbf{E}_1^2 +
\mathbf{E}_2^2 \propto I_1 + I_2 $. It follows from simple geometric
considerations  that in the two-hole configuration there are places
on the photographic plate where the two waves always come in phase
($\mathbf{E}_1(t) \uparrow \uparrow \mathbf{E}_2(t)$ and
$\mathbf{E}_1 \cdot \mathbf{E}_2 > 0$, which means \emph{constructive
interference}) and there are other places  where the two waves
always come with opposite phases ($\mathbf{E}_1(t) \uparrow \downarrow
\mathbf{E}_2(t)$ and $\mathbf{E}_1 \cdot \mathbf{E}_2 < 0$, i.e.,
\emph{destructive interference}).

\subsection{Low intensity light and other experiments}
\label{ss:low-intensity}

In the 19th century physics, the wave-particle debate was decided in
favor of the wave theory. However, further experimental evidence
showed that the victory
 was declared prematurely.
To see what goes wrong with the wave theory, let us continue our
thought experiment with the interference picture created by two
holes and  gradually tune
down the intensity of the light source. At first,
nothing interesting will happen: we will see that the density of the
image predictably decreases.
 However, after some point  we will
recognize that the photographic image is not uniform and continuous as before. It
consists of small blackened dots as if some grains of photo-emulsion
were exposed to light and others were not. This observation is very
difficult to reconcile with the  wave theory. How a continuous wave
can produce this dotty image?  However this is exactly what the
corpuscular theory would predict. Apparently the dots are created by
particles   hitting the
 photographic plate one-at-a-time.

A number of other effects were discovered at the end of the 19th
century and in the beginning of the 20th century, which could not be
explained by the wave theory of light. One of them was the
photo-electric effect: It was observed that when the light is shined
on a piece of metal, electrons can escape from the metal into the
vacuum. This observation was not  surprising  by itself. However it
was rather puzzling how the number of emitted electrons depended on
the frequency and the intensity of the incident light. It was found that
only light waves with frequencies above some threshold $\omega_0$
were capable of knocking out electrons from the metal. Radiation
with frequency below $\omega_0$ could not produce the electron
emission even if the light intensity  was high. According to the
wave theory, one could assume that the electrons
are knocked out of the metal due to some kind of force exerted on
them by electromagnetic vectors $\mathbf{E}, \mathbf{B}$ in the wave. A larger light intensity  (= larger wave amplitude = higher values of $\mathbf{E}$ and  $\mathbf{B}$) naturally means a larger force and a larger chance of
the electron emission. Then why the low frequency but high intensity
light could not do the job?

In 1905 Einstein  explained the photo-electric effect by bringing
back Newtonian corpuscles in the form of ``light quanta'' later
called \emph{photons}. \index{photon} He described the light as
\emph{``consisting of finite number of energy quanta which are
localized at points in space, which move without dividing and which
can only be produced and absorbed as complete units''}
\cite{Arons}. According to the Einstein's explanation,  each
photon carries the energy of $\hbar \omega$, where $\omega$ is the
frequency\footnote{$\omega$ is the so-called \emph{circular
frequency} \index{circular frequency} (measured in radians per
second) which is related to the usual frequency $\nu$ (measured in
oscillations per second) by the formula $\omega = 2 \pi \nu$.} of
the light wave, and $\hbar$ is the
 \emph{Planck constant}. \index{Planck constant} Each photon has a chance to collide with and pass its
energy
to just one electron in the metal. Only
 high energy photons (those corresponding to high frequency light) are capable
of passing enough energy to the electron to overcome certain energy
barrier\footnote{The barrier's energy is roughly proportional to the threshold frequency $E_b \approx \hbar \omega_0$.} $E_b$ between the metal and the vacuum. Low-frequency light has photons with low energy $\hbar \omega < E_b$. Then, no matter what is
the amplitude  (= the number of photons) of such light, its photons
 are just too ``weak'' to kick the  electrons
with sufficient energy.\footnote{Actually, the low-frequency light
may lead to the electron emission when two or more  low-energy
photons collide simultaneously with the same electron, but such
events have very low probability and become observable only at  very
high light intensities.} In the Compton's experiment (1923)
\index{Compton scattering} the interaction of light with electrons
could be studied with much greater detail. And indeed, this
interaction more resembled a collision of two particles rather than
shaking of the electron by a periodic electromagnetic wave.

These observations clearly lead us to the conclusion
 that light is a flow of particles after all. But what about the
interference? We already agreed that the corpuscular theory had no
logical explanation of this effect.

For example, in an interference experiment
conducted by
 Taylor  in 1909 \cite{Taylor}, the intensity of light was so low that
no more than one photon was present at any time instant, thus eliminating any possibility of the
photon-photon interaction and its effect on the interference picture. Another ``explanation''
that the photon somehow splits, passes through both holes, and then rejoins again
at the point of collision with the photographic plate does not stand
criticism as well: One photon
never creates two dots on the photographic plate, so it is unlikely
that the photon can split during propagation. Finally, can we assume that the
particle passing through the right hole somehow ``knows'' whether
the left hole is open or closed and adjusts its trajectory
accordingly? Of course, there is  some  effect
on the photon near the left hole  depending on whether the right
hole is open or not. However by all estimates this effect is
negligibly small.

So, young quantum theory had an
almost impossible task to reconcile two apparently contradicting
classes of experiments with light: Some experiments (diffraction,
interference) were easily explained with the wave theory,  while the
corpuscular theory had serious difficulties. Other experiments
(photo-electric effect, Compton scattering) could not be explained
from the wave properties and clearly showed that light consists of
particles. Just adding to the confusion,  de Broglie in 1924
advanced a hypothesis that such \emph{particle-wave duality}
\index{particle-wave duality} is not specific to photons. He
proposed that all particles of matter -- like electrons -- have
wave-like properties. This ``crazy'' idea was soon confirmed by
Davisson and Germer who observed the diffraction and interference of
electron beams in 1927.

Certainly, in the first quarter of the 20th century, physics faced
the greatest challenge in its history. This is how  Heisenberg
described the situation:

\bigskip

\begin{quote}
\emph{I remember discussions with Bohr which went through many hours
till very late at night and ended almost in despair; and when at the
end of the discussion I went alone for a walk in the neighboring
park I repeated to myself again and again the question: Can nature
possibly be as absurd as it seemed to us in those atomic
experiments?} W. Heisenberg \cite{Heisenberg}
\end{quote}

\section{Physical foundations of quantum mechanics}
\label{ss:measurements}

In the rest of this chapter we will introduce basic formalism of quantum theory. This theory rejects the duplicitous claim of particle-wave duality. It insists that matter and light are made of point-like particles (like photons and electrons) whose propagation is governed by non-classical rules of quantum mechanics.  In particular, these rules are responsible for the wave-like behavior of quantum particles, such as  in the double-hole experiment.  We will explain this experiment from the quantum-mechanical point of view in subsection \ref{ss:time-wave}.

In this section we will try to explain the main difference between
classical and quantum views of the world. To understand quantum
mechanics, we must accept that certain concepts,
 which were taken for granted in  classical
physics, cannot be applied to micro-objects like photons and
electrons. To see what is different, we should revisit basic ideas
about what is physical  system, how its states are prepared, and
how its observables are measured.

\subsection{Single-hole experiment}
\label{ss:ensembles}

The best way to understand the main idea of quantum mechanics is to
analyze the single-hole experiment discussed in the preceding
section. We have established that in the low-intensity regime, when
the source emits individual photons one-by-one, the image on the
screen consists of separate dots. We now accept this fact as a sufficient evidence that light is made of
small countable localizable particles, called photons.

However, the behavior of these particles is quite different from the
one expected in classical physics.  Classical physics is based on
one tacit axiom, which we formulate here as an Assertion\footnote{In
this book we will distinguish \emph{Postulates}, \index{postulate}
\emph{Statements}, and \emph{Assertions}. \index{assertion}
Postulates form a foundation of our theory. In most cases they
follow undoubtedly from experiments. Statements follow logically
from Postulates \index{statement} and we believe them to be true.
Assertions are commonly accepted in the literature, but they do not
have a place in the RQD
approach developed in this book.}

\begin{assertion} [determinism]
 If we prepare a physical system repeatedly in
the same state and measure the same observable, then each time we will get the same measurement
result. \label{assertionJ_c}
\end{assertion}

\noindent This seemingly obvious Assertion is violated in the single-hole
experiment. Indeed, suppose that the light source is monochromatic,
so that all photons reaching the hole have the same momentum and
energy. At the moment of passing through the hole the photons have rather
well-defined $x$, $y$, and $z$-components of position too. This
guarantees that at this moment all photons are prepared in (almost) the same
state, as required by the Assertion \ref{assertionJ_c}. We can make
this state to be defined even better  by reducing the size of the
aperture. Then according to the Assertion, each photon should
produce the same measurement result, i.e., each photon should
land at exactly the same point on the photographic plate. However, this is not what
happens in reality! The dots made by photons are scattered all over
the photographic plate. Moreover, the smaller is the aperture the wider is the
distribution of the dots. Results of measurements are not
reproducible even though preparation conditions are tightly
controlled!

Remarkably, it is not possible to find an ``ordinary'' explanation
of this extraordinary fact. For example, one can assume that
different photons passing through the hole are not exactly in the
same conditions. What if they interact with atoms surrounding the
hole and for each passing photon the configuration of the nearby atoms is
different? This explanation does not seem plausible, because one can repeat the
single-hole experiment with different materials (paper, steel, etc.)
without any visible difference. Moreover, the same diffraction
picture is observed if other particles (electrons, neutrons, $C_{60}
$ molecules, etc) are used instead of photons. It appears that there
are only two parameters, which determine the shape of the
diffraction spot -- the size of the aperture and the particle's
momentum. So, the explanation of this shape must be rather general
and should not depend on the nature of particles and on the material
surrounding the hole. Then we must accept a rather striking
conclusion: all these carefully prepared particles behave \emph{randomly}.
It is impossible to predict which point on the screen will be hit by
the next released particle.

\subsection{Ensembles and measurements in quantum mechanics}
\label{ss:quantum-case}

The main lesson of the single-hole experiment is that classical
Assertion \ref{assertionJ_c} is not true. Even if the system is prepared each time in the same state, we are not going to get
reproducible results in repeated measurements. Why does this happen? The honest answer is
that nobody knows. This is one of greatest mysteries of nature.  Quantum theory does not even attempt to explain the
physical origin of randomness in microsystems. This theory assumes
the randomness as given\footnote{In this book we claim that this randomness is absolute and fundamental; that it cannot be explained and does not even require an explanation. In section \ref{sc:complete} we will briefly discuss other approaches to this deep question. } and simply tries to formulate its
mathematical description. In order to move forward,
we should go beyond simple constatation of randomness in microscopic events and introduce
more precise  statements and new definitions.

 We will call \emph{experiment} \index{experiment}
the preparation of an \emph{ensemble} \index{ensemble} (= a set of
identical copies of the system prepared in the same conditions) and
performing measurements of the same observable in each member of the
ensemble.\footnote{\label{foot:repeat}It is worth noting here that
in this book we are not considering repeated measurements performed
on the same copy of the physical system. We will work under
assumption that after the measurement has been performed, the
copy of the system is discarded. Each measurement requires a fresh
copy of the physical system. This means, in particular, that we are
not interested in the state to which the system may have
``collapsed'' after the measurement. The description of repetitive quantum
measurements is an interesting subject, but it is beyond the scope
of this book.}

Suppose that we prepared an ensemble of $N$ identical copies of the
system and measured the same observable $N$ times. As we have established
above, we cannot say \emph{a priori} that all these measurements
will produce the same results.
 However, it seems reasonable to assume  the existence of ensembles in
which  measurements of \emph{one} observable can be repeated with
the same reproducible result infinite number of times. Indeed, there is no point to
talk about a physical quantity, if there are no ensembles in which this quantity can be
 measured with certainty. So, we begin our construction of the mathematical formalism of quantum mechanics by introducing the following Postulate

\begin{postulate} [partial determinism]
 For any observable $F$ and any value $f$ from
its spectrum, one can prepare an ensemble in such a state that
measurements of this observable are reproducible, i.e., always yield
the same value $f$. \label{postulateJ}
\end{postulate}

Note that in classical mechanics this Postulate follows immediately from Assertion \ref{assertionJ_c}.
The Postulate itself allows us to talk only about the reproducibility of
just \emph{one} observable in the ensemble, while  Assertion \ref{assertionJ_c} referred to \emph{all} observables. So, our quantum Postulate \ref{postulateJ} is a milder requirement than the classical Assertion \ref{assertionJ_c}.

Note also that Postulate \ref{postulateJ} permits existence of  certain groups of (\emph{compatible})
observables \index{compatible observables} whose measurements can be reproducible in a
given ensemble. For example, we will see in chapter \ref{ch:operators} that the three components of particle's momentum $(p_x, p_y, p_z)$ are compatible observables. The same is true for the three components of position $(r_x, r_y, r_z)$.
Thus, quantum mechanics says that with certain efforts we can
prepare an ensemble of particles in such a state that measurements
 of all 3 components of position yield the same
result each time, but then results for the components of momentum
would be randomly scattered. We can also prepare (another)
ensemble in a state with certain momentum, then the position will be
all over the place. However, we cannot prepare an ensemble in which the
uncertainties of both position $\Delta \mathbf{r} $ and momentum
$\Delta \mathbf{p}$ are zero.\footnote{See discussion of the
Heisenberg uncertainty relation in subsection
\ref{ss:heisenberg-unc}.}

\section{Lattice of propositions}
\label{sc:lattice}

Having described the fundamental Postulate \ref{postulateJ} of quantum mechanics in
the preceding section, we now need to turn it into a working
mathematical formalism. This is the goal of the present section and
next two sections. As we mentioned in the beginning of this chapter, our explanation of quantum mechanics will be based on the controversial but very powerful idea that microworld is governed by a non-classical logic.

When we say that measurements of observables in the quantum world  can be irreproducible we mean that this irreproducibility or randomness is of basic, fundamental nature. It is different from the randomness often observed  in the classical world,\footnote{e.g., when we roll a die} which is related merely to our  inability to prepare well-controlled ensembles, incomplete knowledge of initial conditions and inability to solve dynamical equations even if the initial conditions are known. So, classical randomness is of technical rather than fundamental nature. On the other hand, the ever-present quantum randomness cannot be reduced by a more thorough control of preparation conditions, lowering the temperature, etc. This means that quantum mechanics is bound to be a statistical theory based on postulates of probability.  However, this does not mean that quantum mechanics is a subset of classical statistical mechanics. Actually, the statistical theory underlying quantum mechanics is of a different more general non-classical kind.

We know that classical (Boolean) logic is at the core of classical probability theory. The latter theory assign probabilities (real numbers between 0 and 1) to logical propositions and tells us how these numbers are affected by logical operations (AND, OR, NOT, etc.) with propositions. Quite similarly, quantum probability theory is based on logic, but this time the logic is not Boolean. In quantum mechanics we are dealing with \emph{quantum logic} whose postulates differ slightly from the Boolean ones.
So, at the most fundamental level, quantum mechanics is built on two simple ideas: the prevalence of randomness in nature and the non-classical logical relationships between experimental propositions.

The initial idea that the fundamental difference between classical
and quantum mechanics lies in their different logical structures belongs to
Birkhoff and von Neumann. They suggested to substitute classical
logic of Aristotle and Boole by the \emph{quantum logic}.
\index{quantum logic} The formalism presented in this section
summarizes their seminal work \cite{Birkhoff} as well as further
research most notably by Mackey \cite{Mackey} and Piron \cite{Piron2, Piron}. Even for those who do not accept the necessity of such
radical change in our views on logic, the study of quantum logic may
provide a desirable bridge between intuitive concepts of classical
mechanics and abstract formalism of quantum theory.

\begin{quote}
\emph{In introductory quantum physics classes (especially in the
United States), students are informed \emph{ex cathedra} that the
state of a physical system is represented by a complex-valued wave
function $\psi$, that observables correspond to self-adjoint
operators, that the temporal evolution of the system is governed by
a Schr\"odinger equation and so on. Students are expected to accept
all this uncritically, as their professors probably did before them.
Any question of why is dismissed with an appeal to authority and an
injunction to wait and see how well it all works. Those students
whose curiosity precludes blind compliance with the gospel according
to Dirac and von Neumann are told that they have no feeling for
physics and that they would be better off studying mathematics or
philosophy. A happy alternative to teaching by dogma is provided by
basic quantum logic, which furnishes a sound and intellectually
satisfying background for the introduction of the standard notions
of elementary quantum mechanics.} D. J. Foulis \cite{Foulis}
\end{quote}

The main purpose of our sections \ref{sc:lattice} -
\ref{sc:quant-mech} is to demonstrate that postulates of quantum
mechanics are robust and inevitable consequences of the laws of
probability and simple properties of measurements. Basic axioms of quantum logic which are common
for both classical and quantum mechanics are presented in section
\ref{sc:lattice}. The close connection between the classical logic
and the phase space formalism of classical mechanics is discussed in
section \ref{sc:class-logic}. In section \ref{sc:quant-mech} we
will note a remarkable fact that the only difference between
classical and quantum logics (and, thus, between classical and
quantum physics in general) is in a single obscure
\emph{distributivity} postulate. \index{distributivity postulate}
This classical postulate must be replaced by the
\emph{orthomodularity} postulate
of quantum theory. In section \ref{ss:quant-obs} we will demonstrate how postulates of
quantum logic lead us (via Piron's theorem) to the standard
formalism of quantum mechanics with Hilbert spaces, Hermitian
operators, wave functions, etc.  In section \ref{sc:complete} we will briefly mention some interpretations of quantum mechanics and related philosophical issues.

\subsection{Propositions and states}
\label{ss:pure}

There are different types of observables in physics: position,
momentum, spin, mass, etc. As we discussed in Introduction, we are
not interested in the design of the apparatus measuring each
particular observable $F$. We can imagine such an apparatus as a
black box that produces a real number $f \in \mathbb{R}$ (the measured value of the
observable) each time it interacts with the physical system. So, the
act of measurement can be simply described by words ``the value of
observable $F$ was found to be $f$.''  However, for our purposes a
slightly different description of measurements will be more convenient.

With each subset\footnote{Subsets $X$ are not necessarily
limited to contiguous intervals in $\mathbb{R}$. All results remain
valid for more complex subsets of $\mathbb{R}$, such as unions of
any number of disjoint intervals.} $X$ of the real line $\mathbb{R}$
we can associate a \emph{proposition} \index{proposition} $x$ of the
type ``the value of observable $F$ is inside the subset $X$.'' If
the measured value $f$ has been found inside the subset $X$, then we say that
the proposition $x$ is \emph{true}. Otherwise we say that the
proposition $x$ is \emph{false}. At the first sight it seems that we
have not gained much by this reformulation. However, the true advantage is
that propositions introduced above can be regarded as key elements of mathematical
logic and probability theory. The powerful machinery of these
theories will be found very helpful in our analysis of quantum
measurements.  It is also useful to note that the ``yes-no''
propositions can be also regarded as special types of observables
whose spectrum consists of just two points: 1 (if the proposition is
true; the answer is ``yes'') and 0 (if the proposition is false; the answer is ``no'').

Yes-no propositions are not necessarily related to single observables. As
we will see later, there are sets of (compatible) observables $F_1, F_2,
\ldots, F_n$ that are simultaneously measurable with arbitrary
precision. For such sets of observables one can define propositions
corresponding to subsets in the ($n$-dimensional) common spectrum of
these observables.  For example, the proposition ``particle's
position is inside volume $V$'' is a statement about simultaneous
measurements of three observables - the $x, y$, and $z$ components
of the position vector. Experimentally, this proposition can be realized using
a Geiger counter occupying the volume $V$. The counter  clicks (the
proposition is true) if the particle  passes through the counter's
chamber and does not click (the proposition is false) if the
particle is outside of $V$.

In what follows we will denote by $\mathcal{L}$ the set of all propositions\footnote{$\mathcal{L}$ also called the \emph{propositional system}.} about the
physical system.
\index{propositional system} The set of all
possible states of the system will be denoted by $\Phi$. Our goal in this chapter is
to study the mathematical relationships between elements $x \in
\mathcal{L}$ and $\phi \in \Phi$ in these two sets.

The above discussion referred to a single measurement
 performed on one copy of the physical system.
Let us now prepare multiple copies (an ensemble) of the system and
perform  measurements  of the same proposition in all copies. As we discussed earlier, there is no guarantee that the results
of all these measurements will be the same. So, for some members in
the ensemble the proposition $x$ will be found 'true', while for
other members it will be 'false', even if every effort is made to
ensure that the state of the prepared system  is exactly the same in all cases.

 Using results
of measurements as described above, we can introduce a function
$(\phi|x)$ called the \emph{probability measure} \index{probability
measure} which assigns to each state $\phi$ and to each proposition
$x$ the probability of $x$ being true in the state $\phi$. The value
of this function (a real number between 0 and 1) is obtained by the
following recipe:

\begin{itemize}
\item[(i)] prepare a copy of the system in the state $\phi$;
\item[(ii)] determine whether $x$ is true or false;
\item[(iii)] repeat steps (i) and (ii) $N$ times, then

\begin{eqnarray*}
(\phi|x) = \lim_{N \to \infty} \frac{M}{N}
\end{eqnarray*}
\end{itemize}

\noindent where $M$ is the number of times when the proposition $x$
was found to be true.

Two states $\phi$ and $\psi$ of the same system are said to be equal
($\phi=\psi$) if for any proposition $x$ we have

\begin{eqnarray*}
(\phi|x) = (\psi|x)
\end{eqnarray*}

\noindent Indeed, there is no physical difference between these two states as
 all experiments yield the same results (=probabilities). For the same
reason we will say that two propositions $x$ and $y$ are identical
($x=y$) if for all states $\phi$ of the  system

\begin{eqnarray}
(\phi|x) = (\phi|y) \label{eq:(4.1)}
\end{eqnarray}

\noindent It follows from the above discussion that the probability
measure $(\phi|x)$ considered as a function on the set of all states
$\Phi$ is a unique representative of proposition $x$ (i.e., different
propositions have different probability measures on $\Phi$). So, we can gain some
insight into the properties of different propositions
 by studying properties of the
corresponding probability measures.

There are propositions which are always true independent on the
state of the system. For example, the proposition ``the value of the
observable $F_1$ is somewhere on the real line'' is always
true.\footnote{Measurements  of observables always yield \emph{some}
value, since we agreed in Introduction that an ideal measuring
apparatus  never misfires.} For any other
observable $F_2$, the proposition ``the value of the observable
$F_2$ is somewhere on the real line'' is also true for all states.
Therefore, according to (\ref{eq:(4.1)}), we will say that these two
propositions are identical. So, we can define a unique
\emph{maximal} \index{maximal proposition} proposition $\mathcal{I}
\in \mathcal{L}$ which is always true. Inversely, the proposition
``the value of observable is not on the real
line'' is always false and will be called the \emph{ minimal}
proposition. \index{minimal proposition} There is just one minimal
proposition $\emptyset$ in the set $\mathcal{L}$, and for each state
$\phi$ we can write

\begin{eqnarray}
(\phi|\mathcal{I}) &=& 1 \label{eq:maximum}\\
(\phi|\emptyset) &=& 0 \label{eq:minimum}
\end{eqnarray}

\subsection{Partial ordering}
\label{ss:operations}

Suppose that we found  two propositions $x$ and $y$, such that their
measures satisfy  $(\phi|x) \leq (\phi|y)$ for all states $\phi$. Then we will say that proposition  $x$ is \emph{less
than or equal to} \index{less than or equal to} proposition $y$ and
denote this relation by $x \leq y$. \index{$\leq$ ``less than or
equal to''} The meaning of this relation is obvious when  $x$ and $y$ are propositions about the same
observable and  $X$ and $Y$ are their corresponding subsets in $\mathbb{R}$. Then $x \leq y$ if the subset $X$ is inside the subset
$Y$, i.e., $X \subseteq Y$. In this case, the relation $x \leq y$ is
associated with logical \emph{implication}, \index{implication}
i.e., if $x$ is true then $y$ is definitely true as well; $x$
IMPLIES $y$; or ``IF $x$ THEN $y$''.
If $x \leq y$ and
$x \neq y$ we will say that $x$ is \emph{less than} \index{less
than} $y$ and denote this relationship by   \index{$<$ ``less than''} $ x < y$.

The implication relation $\leq$ has three obvious properties.

\begin{lemma} [reflectivity] \index{reflectivity}
 Any proposition implies itself: $x \leq x$.
  \label{lemmaK1}
\end{lemma}
\begin{proof}
This follows from the fact that for any $\phi$ it is true that $(\phi|x) \leq (\phi|x)$.
\end{proof}

\begin{lemma} [symmetry] \index{symmetry}
 If two propositions imply each other, then they are equal:   If $x \leq y$ and $y \leq x$ then $x = y$.
    \label{lemmaK2}
\end{lemma}
\begin{proof}
If two propositions $x$ and $y$ are less than or equal to each
other, then $(\phi|x) \leq (\phi|y)$ and $(\phi|y) \leq (\phi|x)$
for each state $\phi$,  which implies $(\phi|x) = (\phi|y)$ and,
according to (\ref{eq:(4.1)}), $x = y$.
\end{proof}

\begin{lemma} [transitivity] \index{transitivity}
 If $x \leq y$ and $y \leq z$, then  $x \leq z$.
\label{lemmaK3}
\end{lemma}
\begin{proof}
If $x \leq y$ and $y \leq z$, then $(\phi|x) \leq (\phi|y) \leq
(\phi|z)$ for every state $\phi$. Consequently, $(\phi|x) \leq
(\phi|z)$ for each state $\phi$ and $x \leq z$.
\end{proof}

\bigskip

 Properties \ref{lemmaK1}, \ref{lemmaK2}, and \ref{lemmaK3} tell us
that $\leq$
 is  a  \emph{partial ordering} \index{partial ordering}
relation. It is \emph{ordering} because it tells which proposition
is ``smaller'' and which is ``larger.'' The ordering is \emph{partial},
because it doesn't apply to all pairs of propositions.\footnote{There could
be propositions $x$ and $y$, such that for some states $(\phi|x) >
(\phi|y)$, while for other states $(\phi|x) < (\phi|y)$.} Thus, the
set $\mathcal{L}$ of all propositions is a \emph{partially ordered
set}. \index{partially ordered set} From equations (\ref{eq:maximum}) and
(\ref{eq:minimum}) we also conclude that

\begin{postulate} [definition of $\mathcal{I}$]
  $ x \leq \mathcal{I}$ for any $x \in \mathcal{L}$.
  \label{lemmaK4a}
  \end{postulate}

\begin{postulate} [definition of $\emptyset$]
 $\emptyset  \leq x$ for any $x \in \mathcal{L}$.
\label{lemmaK4b}
\end{postulate}

\subsection{Meet}
\label{ss:meet_join}

For two propositions $x$ and $y$, suppose that we found a third proposition
$z$
 such that

\begin{eqnarray}
z &\leq& x \label{eq:(4.4)} \\
z &\leq& y \label{eq:(4.5)}
\end{eqnarray}

\noindent There could be more than one proposition  satisfying these
properties. We will assume that we can always find one maximal
proposition $z$ in this set. This maximal proposition will be called a \emph{meet}
\index{meet} of $x$ and $y$ and denoted by $x \wedge y$.

 The existence of a unique
meet is obvious in the case when $x$ and $y$ are propositions about
the same observable, such that they correspond to two subsets of the real line
$\mathbb{R}$: $X$ and $Y$, respectively. Then the meet $z= x \wedge
y$ is a proposition
 corresponding to the intersection of these two subsets $Z = X \cap
 Y$.\footnote{If $X$ and $Y$ do not intersect, then $z= \emptyset$.}
 In this one-dimensional case the operation meet can be identified
 with the logical operation AND: the proposition $x \wedge y$ is true only
 when both $x$ AND $y$ are true.
\index{$\wedge$ meet}

The above definition of \emph{meet} can be formalized in the form of two postulates

\begin{postulate} [definition of $\wedge$]
 $x \wedge y \leq x$ and $x \wedge y \leq y$ for all $x$ and $y$.
\label{postulateK5a}
\end{postulate}

\begin{postulate} [definition of $\wedge$]
 If $z \leq x \mbox{  }$ and $ z \leq y$ then
$ z \leq x \wedge y$. \label{postulateK6a}
\end{postulate}

\noindent It seems reasonable to assume that the order in which  meet
operations are performed is not relevant

\begin{postulate} [commutativity of $\wedge$] \index{commutativity}
 $x \wedge y = y \wedge x$.
\label{postulateK7a}
\end{postulate}

\begin{postulate} [associativity of $\wedge$] \index{associativity}
  $(x \wedge y) \wedge z = x \wedge (y \wedge z)$.
\label{postulateK8a}
\end{postulate}

\subsection{Join}

Similar to our discussion of meet, we can assume that for any two propositions $x$ and $y$
there always exists a unique \emph{join} $x \vee y$, \index{$\vee$
join} \index{join} such that both $x$ and $y$ are less or equal than
$x \vee y$, and $x \vee y$ is the minimal proposition with such a
property.

In the case of propositions about the same observable, the join of
$x$ and $y$ is a proposition $z = x \vee y$ whose associated subset $Z$ of the real
line is a union of the subsets corresponding to $x$ and $y$: $Z = X
\cup Y$. The proposition $z$ is true when either $x$ OR $y$ is true.
So, the join can be identified with the logical operation
OR.\footnote{It is important to note that from $x
\vee y$ being true it does not necessarily follow that either $x$ or
$y$ are definitely true, as we had it in classical logic.}

 The formal version of the above definition of \emph{join} is

\begin{postulate} [definition of $\vee$ ]
 $ x \leq x \vee y$ and $y \leq x \vee y$.
\label{postulateK5b}
\end{postulate}

\begin{postulate} [definition of $\vee$ ]
 If $x \leq z$ and
$y \leq z$ then $ x \vee y \leq z $. \label{postulateK6b}
\end{postulate}

Similar to Postulates \ref{postulateK7a} and \ref{postulateK8a} we insist
that the order of join operations is irrelevant

\begin{postulate} [commutativity of $\vee$] \index{commutativity}
  $x \vee y = y \vee x$.
\label{postulateK7b}
\end{postulate}

\begin{postulate} [associativity of $\vee$] \index{associativity}
 $(x \vee y) \vee z = x \vee (y \vee z)$.
\label{postulateK8b}
\end{postulate}

The properties of propositions listed so far (partial ordering,
meet, and join) mean that the  set of propositions $\mathcal{L}$ is
what mathematicians call a \emph{complete lattice}. \index{lattice}

\subsection{Orthocomplement}
\label{ss:orthocomplement}

 There is one more
operation on the set of propositions that we need to consider. This
operation is called \emph{orthocomplement}. \index{$\perp$
orthocomplement}\index{orthocomplement} It has the
meaning of the logical negation (operation NOT). For any proposition $x$ its
orthocomplement is denoted by $x^{\perp}$. In the case of
a single observable, if proposition $x$
corresponds to the subset $X$ of the real line, then its
orthocomplement $x^{\perp}$ corresponds to the relative complement
of $X$ with respect to $\mathbb{R}$ (denoted by $\mathbb{R}
\setminus X$). When the value of observable $F$ is found inside $X$,
i.e., the value of $x$ is 1, we immediately know that the value  of
$x^{\perp}$ is zero. Inversely, if $x$ is false then $x^{\perp}$ is
necessarily true.

More formally, the orthocomplement $x^{\perp}$ is defined as a
proposition whose probability measure for each state $\phi$ is

\begin{eqnarray}
(\phi|x^{\perp}) = 1 - (\phi|x)
 \label{eq:orthocomp}
\end{eqnarray}

\begin{lemma} [non-contradiction] \index{non-contradiction}
  $x \wedge x^{\perp} = \emptyset$.
\label{postulateK9a}
\end{lemma}
\begin{proof}
Let us prove this Lemma in the case when $x$ is a proposition about
one observable $F$. Suppose that $x \wedge x^{\perp} = y \neq
\emptyset$, then, according to Postulate \ref{postulateJ}, there
exists a state $\phi$ such that $(\phi|y) =1$ and, from Postulate
\ref{postulateK5a} we obtain

\begin{eqnarray*}
y & \leq& x \\
y & \leq& x^{\perp} \\
1 = (\phi|y) &\leq& (\phi|x) \\
1 = (\phi|y) &\leq& (\phi|x^{\perp})
\end{eqnarray*}

\noindent It then follows that $(\phi|x) =1$ and $(\phi|x^{\perp})
=1$, which means that any measurement of the observable $F$ in the
state $\phi$ will result in a value inside both $X$ and $\mathbb{R}
\setminus X$ simultaneously, which is impossible. This contradiction
should convince us that $x \wedge x^{\perp} = \emptyset$.
\end{proof}

\begin{lemma} [double negation] \index{double negation}
   $(x^{\perp})^{ \perp} = x$.
\label{statementK9a}
\end{lemma}
\begin{proof}
 From equation (\ref{eq:orthocomp}) we can write for any state $\phi$

\begin{eqnarray*}
(\phi|(x^{\perp})^{ \perp}) = 1 - (\phi|x^{\perp}) =  1 - (1 -
(\phi|x))= (\phi|x)
\end{eqnarray*}
\end{proof}

\begin{lemma} [contraposition] \index{contraposition}
If $x \leq y$  then $y^{\perp} \leq x^{\perp}$. \label{postulateK11}
\end{lemma}
\begin{proof}
If $x \leq y$ then $(\phi|x) \leq (\phi|y)$ and $(1-(\phi|x)) \geq
(1- (\phi|y))$ for all states $\phi$. But according to our
definition (\ref{eq:orthocomp}), the two sides of the latter inequality
are probability measures for propositions $x^{\perp}$ and
$y^{\perp}$, respectively, which proves the Lemma.
\end{proof}

 Propositions $x$ and $y$ are said to be \emph{disjoint}
\index{disjoint propositions} if $x \leq y^{\perp}$ or,
equivalently, $y \leq x^{\perp}$.

 When $x$ and
 $y$ are disjoint propositions about the same observable, their
 corresponding subsets do not intersect: $X \cap Y =
\emptyset$. For such mutually exclusive propositions the probability
of either $x$ OR $y$ being true (i.e., the probability corresponding
to the proposition $x \vee y$) is the sum of probabilities for $x$
and $y$. It seems natural to generalize this property to all pairs
of disjoint propositions in $\mathcal{L}$:

\begin{postulate} [probabilities for mutually exclusive propositions]
If $x$ and $y$ are disjoint, then for any state $\phi$

\begin{eqnarray*}
(\phi|x \vee y) = (\phi|x) + (\phi|y)
\end{eqnarray*}

  \label{postulateL}
\end{postulate}

\noindent The following Lemma establishes that the join of any proposition
$x$ with its orthocomplement $x^{\perp}$ is always the trivial proposition.

\begin{lemma} [non-contradiction] \index{non-contradiction}
\label{postulateK9b} $x \vee x^{\perp} = \mathcal{I}$.
\end{lemma}
\begin{proof}
From Lemmas \ref{lemmaK1} and \ref{statementK9a} it follows that $x
\leq x = (x^{\perp})^{\perp}$, so that propositions $x$ and
$x^{\perp}$ are disjoint. Then, by Postulate \ref{postulateL}, for
any state $\phi$ we obtain

\begin{eqnarray*}
 (\phi|x \vee x^{\perp}) = (\phi|x)
+ (\phi| x^{\perp}) = (\phi|x) + (1 - (\phi|x)) = 1
\end{eqnarray*}

\noindent which proves the Lemma.
\end{proof}

Adding the orthocomplement to the properties of the propositional
system (complete lattice)
 $\mathcal{L}$, we conclude that $\mathcal{L}$  is an
\emph{orthocomplemented lattice}. \index{orthocomplemented lattice}
Axioms of orthocomplemented lattices are collected in the upper part
of Table \ref{table:2.1} for easy reference.

\newpage

\begin{table}[h]
\caption{Axioms of quantum  logic}
\begin{tabular*}{\textwidth}{@{\extracolsep{\fill}}ccc}
 \hline
Property     &       &  Postulate/Lemma    \cr \hline
   &   \multicolumn{2}{c}{ \textbf{Axioms of orthocomplemented lattices}}
  \cr
Reflectivity  & \ref{lemmaK1} & $x \leq x$  \cr Symmetry &
\ref{lemmaK2}  & $(x \leq y) \mbox{  } \& \mbox{  } (y \leq x) \mbox{ }
\Rightarrow \mbox{ } x = y $                  \cr Transitivity  &
\ref{lemmaK3} & $(x \leq y) \mbox{ } \& \mbox{  } (y \leq z) \mbox{ }
\Rightarrow \mbox{  } x \leq z$ \cr Definition of $\mathcal{I}$  & \ref{lemmaK4a} & $x \leq \mathcal{I}$ \cr
Definition of
$\emptyset$ & \ref{lemmaK4b}  & $\emptyset
\leq x$   \cr Definition of $\wedge$    &
\ref{postulateK5a}           & $x \wedge y \leq x$   \cr
Definition of $\wedge$     & \ref{postulateK6a} & $(z \leq
x) \mbox{  } \& \mbox{  } (z \leq y) \Rightarrow z \leq (x \wedge y)$ \cr
 Definition of  $\vee$   & \ref{postulateK5b} & $ x \leq x \vee y $  \cr
 Definition of  $\vee$   &  \ref{postulateK6b} & $(x \leq z) \mbox{  } \& \mbox{  }
(y \leq z) \Rightarrow (x \vee y) \leq z $ \cr Commutativity &
\ref{postulateK7a} & $x \vee y = y \vee x$   \cr
Commutativity &  \ref{postulateK7b} &
$x \wedge y = y \wedge x$   \cr Associativity  & \ref{postulateK8a}
& $(x \vee y) \vee z = x \vee (y \vee z)$   \cr
Associativity &  \ref{postulateK8b} &
$(x \wedge y) \wedge z = x \wedge (y \wedge z)$  \cr
Non-contradiction  & \ref{postulateK9a}  & $x \wedge x^{\perp} =
\emptyset$    \cr
Non-contradiction &  \ref{postulateK9b} & $x \vee x^{\perp} = \mathcal{I}$    \cr
Double negation & \ref{statementK9a} & $(x^{\perp})^{\perp} = x$ \cr
Contraposition      & \ref{postulateK11}        & $x \leq  y
\Rightarrow y^{\perp} \leq x^{\perp}$       \cr
Sum of probabilities & \ref{postulateL} & $(\phi|x \vee y) = (\phi|x) + (\phi|y)$ if $x \leq y^{\perp}$ \cr
Atomicity &
\ref{postulateM} & existence of elementary propositions \cr \hline
   &     \multicolumn{2}{c}{ \textbf{Additional assertions of classical
logic}}    \cr Distributivity & \ref{assertionK12a} & $ x \vee (y
\wedge z)  = (x \vee y) \wedge (x \vee z)$ \cr
 &  \ref{assertionK12b} & $ x \wedge (y \vee z) = (x \wedge y) \vee (x \wedge
z)  $
       \cr
\hline
   &    \multicolumn{2}{c}{ \textbf{Additional postulate of quantum logic}}
  \cr
Orthomodularity & \ref{postulateK13} & $ x \leq y  \Rightarrow   x
\leftrightarrow y  $     \cr
 \hline
\end{tabular*}
\label{table:2.1}
\end{table}

\newpage

\subsection{Atomic propositions}
\label{ss:atoms}

One says that proposition $y$ \emph{covers} \index{cover}
proposition $x$ if the following two statements are true:

\begin{itemize}
\item[1)] $x < y $
\item[2)] If $x \leq z \leq y$, then either $z=x$  or  $ z=y $
\end{itemize}

\noindent In simple words, this definition means that $x$ implies $y$ and there are no propositions ``intermediate''
between $x$ and $y$.

If $x$ is a proposition about a single observable corresponding to the
interval $X \subseteq \mathbb{R}$, then the interval corresponding
to the covering proposition $y$  can be obtained by adding just one
extra point to the interval $X$.

A proposition covering $\emptyset$ is called an \emph{atomic
proposition} or simply an \emph{atom}. \index{atom}\index{atomic
proposition} So, atoms are smallest non-vanishing propositions. They
unambiguously specify properties of the system in the most exact
way.

We will say that the atom $p$ is \emph{contained} in the proposition
$x$ if $p \leq x$. The existence of atoms is not a necessary property of mathematical lattices. Both atomic and non-atomic lattices can be studied. However, on physical grounds we will postulate that the lattice of propositions is atomic

\begin{postulate} [atomicity] The propositional system $\mathcal{L}$
 is an \emph{atomic
lattice}. \index{atomic lattice} This means that

1.  If $x \neq \emptyset $, then there exists at least one atom $p$
 contained in $x$.

2. Each proposition $x$ is a join of all atoms contained in it:
\begin{eqnarray*}
x = \vee_{p \leq x} p
\end{eqnarray*}

3.  If $p$ is an atom and $p \wedge x = \emptyset$, then $p \vee x$
covers $x$.

\label{postulateM}
\end{postulate}

\noindent There are three simple Lemmas that follow directly from
this Postulate.

\begin{lemma} \label{Lemma4.6} If $p$ is an atom and $x$ is any
proposition then either $p \wedge x = \emptyset $ or $p \wedge x = p
$.
\end{lemma}
\begin{proof}
We know that $ \emptyset \leq p \wedge x \leq p$ and  that $p$
covers $\emptyset$. Then, according to the definition of covering,
either $p \wedge x = \emptyset $ or $p \wedge x = p $.
\end{proof}
\bigskip

\begin{lemma} \label{Lemma4.7}  $x \leq y$ if and only if all atoms
contained in $x$ are contained in $y$ as well.
\end{lemma}
\begin{proof}
  If $x \leq y$ then for each atom  $p$ contained in $x$ we
have $p \leq x \leq y$ and $p \leq y $ by the transitivity property
\ref{lemmaK3}. To prove the inverse statement we notice that if we
assume that all atoms in $x$ are also contained in $y$ then by
Postulate \ref{postulateM}(2)

\begin{eqnarray*}
y &=& \vee_{p \leq y} p = (\vee_{p \leq x} p) \vee (\vee_{p \cross
\leq x} p) = x \vee (\vee_{p \cross \leq x} p)  \geq  x
\end{eqnarray*}

\end{proof}
\bigskip

\begin{lemma}  \label{Lemma4.8} The meet $x \wedge y$ of two propositions
$x$ and $y$ is a union of atoms contained in both $x$ and $y$.
\end{lemma}
\begin{proof}   If $p$ is an atom
contained in both $x$ and $y$ ($p\leq x$ and $p\leq y$), then $p\leq
x \wedge y$.
 Conversely, if $p\leq x \wedge y$, then $p\leq x$ and
$p\leq y$ by Lemma \ref{Lemma4.1}
\end{proof}
\bigskip

\section{Classical logic}
\label{sc:class-logic}

\subsection{Truth tables and distributive law}
\label{ss:truth}

This is not a surprise that the theory constructed above is  similar to
classical logic. Indeed if we make substitutions: 'less than or equal to' $\to$ IF...THEN...; join
$\to$ OR; meet $\to$ AND; and so on, as shown in  Table
\ref{table:2.2},
 then properties described in Postulates and Lemmas \ref{lemmaK1} - \ref{Lemma4.8}
 exactly match axioms
of classical Boolean logic. \index{Boolean logic} For example, the
transitivity property in Lemma \ref{lemmaK3} allows us to make
syllogisms, like the  one analyzed by Aristotle

\bigskip

\noindent \emph{If all humans are mortal,}

\noindent \emph{and all Greeks are humans,}

\noindent  \emph{then all Greeks are mortal.}

\bigskip

\begin{table}[h]
\caption{Four operations and two special elements of lattice theory
and logic}
\begin{tabular*}{\textwidth}{@{\extracolsep{\fill}}cccc}
\hline
 Name   &  Name         & Meaning          &  Symbol          \cr
 in lattice theory &  in logic     & in classical logic          &
\cr
 \hline
 less or equal  & implication  & IF $x$ THEN $y$ &  $x \leq y$ \cr
 meet   & injunction  & $x $ AND $y$      &  $x \wedge y$    \cr
 join  & disjunction  & $x $ OR $y$      &  $x \vee y$      \cr
 orthocomplement & negation       & NOT $x$           &  $x^{\perp}$     \cr
 maximal element & tautology   &  always true      &  $\mathcal{I}$    \cr
 minimal element & absurdity   &  always false      &  $\emptyset$ \cr
 \hline
\end{tabular*}
\label{table:2.2}
\end{table}

\noindent Lemma \ref{postulateK9a} tells that a proposition and its
negation cannot be true at the same time. Lemma \ref{postulateK9b}
is the famous \emph{tertium non datur} law of logic: either a
proposition or its negation is true with no third possibility.

Note, however, that properties \ref{lemmaK1} - \ref{Lemma4.8} are
not sufficient to build a complete theory of mathematical logic:
Boolean logic \index{Boolean logic} has two additional axioms, which have not been mentioned yet.
They are called \emph{distributive laws} \index{distributive laws}

\begin{assertion} [distributive law]
\label{assertionK12a} $x \vee (y \wedge z)  = (x \vee y) \wedge (x
\vee z)$. \end{assertion}

\begin{assertion} [distributive law]
\label{assertionK12b}  $ x \wedge (y \vee z) = (x \wedge y) \vee (x
\wedge z).  $
\end{assertion}

\noindent These laws, unlike axioms of orthocomplemented lattices,
cannot be justified by using our previous approach that relied on general probability measures $(\phi|x)$.
This is the reason why we call them Assertions. In the next section we will see that these Assertions are not valid in the quantum case. However, they
can be proven if we use
 the fundamental Assertion \ref{assertionJ_c} of classical
mechanics, which says that in classical pure states all measurements
yield the same results, i.e., reproducible. Then for a given
classical pure state $\phi$ each proposition $x$  is either always
true or always false and the probability measure can have only two
values: $(\phi|x)=1$ or $(\phi|x)=0$. Such classical probability
measure is called the \emph{truth function}. \index{truth function}
In the double-valued (true-false) Boolean logic, \index{Boolean
logic}\index{classical logic}  the job of performing logical
operations with propositions is greatly simplified by analyzing
their truth functions. For example, to show the equality of two
propositions it is sufficient to demonstrate that the values of
their truth functions are the same for all classical pure states.

Let us consider an example.  Given two propositions $x$ and $y$ and an arbitrary state $\phi$,
there are at most four possible values for the pair of their truth
functions $(\phi|x)$ and $(\phi|y)$: (1,1), (1,0), (0,1), and (0,0).
To analyze these possibilities it is convenient to put the values of
the truth functions in a \emph{truth table}. \index{truth table}
 Table \ref{table:2.3} is the truth table  for propositions $x$, $y$,  $x \wedge y$, $x
\vee y$, $x^{\perp}$ and $y^{\perp}$.\footnote{Here we assumed that
all these propositions are non-empty.} The first row in table
\ref{table:2.3} refers to all classical pure states in which both
propositions $x$ and   $y$ are false, i.e., $(\phi|x) = (\phi|y) = 0$. The other entries in this row tell that for such states $(\phi|x \wedge y) = (\phi|x \vee y) = 0$  and $( \phi | x^{\perp}) = ( \phi |y^{\perp}) =   0$. The second row refer to states
in which $x$ is false and $y$ is true, etc.

\begin{table}[h]
\caption{Truth table for basic logical operations}
\begin{tabular*}{\textwidth}{@{\extracolsep{\fill}}cccccc}
 \hline
 $x$ & $y$ & $x \wedge y $ & $x \vee  y$  & $x^{\perp}$  & $y^{\perp}$ \cr
\hline
 0  & 0 & 0 & 0  & 1 & 1   \cr
 0  & 1  & 0 & 1 & 1 & 0  \cr
 1  & 0  & 0 & 1 & 0 & 1  \cr
 1  & 1  & 1 & 1 & 0 & 0 \cr
 \hline
\end{tabular*}
\label{table:2.3}
\end{table}

\begin{table}[h]
\caption{Demonstration of the distributive law using truth table}
\begin{tabular*}{\textwidth}{@{\extracolsep{\fill}}cccccccc}
 \hline
 $x$ & $y$ & $z$ & $y \wedge z$ & $x \vee (y \wedge z)$ & $x \vee y$ &
$x \vee z$ & $(x \vee y) \wedge ( x \vee z)$ \cr
\hline
 0  & 0 & 0 & 0 & 0 & 0 & 0 & 0   \cr
 0  & 0 & 1 & 0 & 0 & 0 & 1 & 0   \cr
 0  & 1 & 0 & 0 & 0 & 1 & 0 & 0   \cr
 1  & 0 & 0 & 0 & 1 & 1 & 1 & 1   \cr
 0  & 1 & 1 & 1 & 1 & 1 & 1 & 1   \cr
 1  & 0 & 1 & 0 & 1 & 1 & 1 & 1   \cr
 1  & 1 & 0 & 0 & 1 & 1 & 1 & 1   \cr
 1  & 1 & 1 & 1 & 1 & 1 & 1 & 1   \cr
 \hline
\end{tabular*}
\label{table:2.4}
\end{table}

The truth table for three arbitrary non-empty propositions $x$, $y$, and $z$   is shown in Table \ref{table:2.4}.   This table has $2^3 = 8$ rows that correspond to
groups of classical states having different values of their truth functions on the propositions $x$, $y$ and $z$. In all these cases truth functions in
columns 5 and 8 are identical, which means that

\begin{eqnarray*}
x \vee (y \wedge z)  = (x \vee y) \wedge (x \vee z)
\end{eqnarray*}

\noindent and that the classical distributive law (Assertion
\ref{assertionK12a}) is valid. Assertion \ref{assertionK12b} can be derived in a similar
manner.

Thus we have shown that in the deterministic world of classical mechanics governed by the Assertion \ref{assertionJ_c}, the set
of propositions $\mathcal{L}$ is an orthocomplemented atomic lattice
where the distributive laws \ref{assertionK12a} and \ref{assertionK12b} hold true.
Such a lattice will be called a \emph{classical propositional
system} or, shortly, \emph{classical logic}. \index{classical logic}
A study of classical logics and its relationship to classical
mechanics is the topic of the present section.

\subsection{Atomic propositions in classical logic}
\label{ss:atoms-class}

Our next step is to demonstrate that classical logic provides the
entire mathematical framework of classical mechanics, i.e., the
description of observables and states in the \emph{phase space}.\footnote{An example of the phase space for a single classical particle will be presented in subsection \ref{ss:phase_space}.}
\index{phase space} First,  we prove four Lemmas.

\bigskip

\begin{lemma} \label{Lemma4.9} In classical logic, if $x < y$, then there
exists an atom  $p$ such that $p \wedge x = \emptyset$ and $p \leq y
$. \end{lemma}
\begin{proof}
Clearly, $ y \wedge x^{\perp} \neq \emptyset$, because otherwise we
would have

\begin{eqnarray*}
y &=& y \wedge \mathcal{I} = y \wedge (x \vee x^{\perp}) = (y \wedge
x) \vee (y \wedge x^{\perp}) = (y \wedge x) \vee \emptyset = y
\wedge x  \leq x
\end{eqnarray*}

\noindent and, by Lemma \ref{lemmaK2},  $x=y$ in contradiction with
the condition of the present Lemma. Since $ y \wedge x^{\perp}$ is non-zero,
then by Postulate \ref{postulateM}(1) there exists an atom
 $p$ such that $p \leq y \wedge x^{\perp}$. It then
follows that  $p \leq x^{\perp}$ and by Lemma \ref{Lemma4.3} $p
\wedge x \leq x^{\perp}\wedge x = \emptyset$.
\end{proof}
\bigskip

\begin{lemma}  \label{Lemma4.10} In classical logic, the orthocomplement
$x^{\perp} $ of a proposition $x$ (where $x \neq \mathcal{I}$) is a
join of all atoms not contained in $x$.
\end{lemma}
\begin{proof} First, it is clear that there should exist al least one
atom $p$ that is not contained in $x$. If it were not true, then we
would have $x = \mathcal{I}$ in contradiction to the condition of
the Lemma. Let us now prove that the atom $p$ is contained in
$x^{\perp}$. Indeed, using the distributive law \ref{assertionK12b}
we can write

\begin{eqnarray*}
p &=& p \wedge \mathcal{I} = p \wedge (x \vee x^{\perp})
= (p \wedge
x) \vee (p \wedge x^{\perp})
\end{eqnarray*}

\noindent According to Lemma \ref{Lemma4.6} we now have four
possibilities:

\begin{enumerate}
\item  $p \wedge x = \emptyset$ and $p \wedge x^{\perp} = \emptyset$; then
$p \leq x \wedge x^{\perp} = \emptyset$, which is impossible;

\item $p \wedge x =p$ and $p \wedge x^{\perp} = p$; then
$p = p \wedge p = (p \wedge x) \wedge (p \wedge x^{\perp}) = p
\wedge (x \wedge x^{\perp}) = p \wedge \emptyset = \emptyset$, which
is impossible;

\item  $p \wedge x =p$ and $p \wedge x^{\perp} = \emptyset$; from
Postulate \ref{postulateK5a} it follows that $p \leq x$, which
contradicts our assumption and should be dismissed;

\item  $p \wedge x =\emptyset$ and $p \wedge x^{\perp} = p$; from this we
have $ p \leq x^{\perp}$, i.e., $p$ is contained in $x^{\perp}$.
\end{enumerate}

\noindent This shows that all atoms not contained in $x$ are
contained in $x^{\perp}$. Further, Lemmas \ref{postulateK9a}
and \ref{Lemma4.8}  imply that all atoms contained in
$x^{\perp}$ are not contained in $x$. The statement of the Lemma
then follows from Postulate \ref{postulateM}(2).
\end{proof}
\bigskip

\begin{lemma}  \label{Lemma4.11} In classical logic,  two different atoms
$p$ and $q$ are always disjoint: $q \leq p^{\perp}$.
\end{lemma}
\begin{proof}
By Lemma \ref{Lemma4.10},  $ p^{\perp}$ is a join of all atoms
 different from $p$, including $q$, thus $q \leq
p^{\perp}$.
\end{proof}
\bigskip

\begin{lemma}  \label{Lemma4.12} In classical logic, the join $x \vee y$ of
two propositions $x$ and $y$ is a join of atoms contained in either
$x$ or $y$.
\end{lemma}
\begin{proof}
If $ p \leq x $ or $ p \leq y $ then
 $ p \leq x \vee y$.
Conversely, suppose that $ p \leq x
\vee y$ and $p \wedge x = \emptyset$, $p \wedge y = \emptyset$, then

\begin{eqnarray*}
p &=& p \wedge (x \vee y) = (p \wedge x) \vee (p \wedge y) =
\emptyset \vee \emptyset
= \emptyset
\end{eqnarray*}

\noindent which is absurd.
\end{proof}
\bigskip

Now we are ready to prove the important fact that in classical
mechanics (or in classical logic) propositions can be interpreted as subsets of a set $S$,
which is called the \emph{phase space}. \index{phase space}

\bigskip

\begin{theorem} \label{Theorem4.13}
For any classical logic $\mathcal{L}$, there exists a set $S$
 and an isomorphism $f(x)$ between
elements $x$ of $\mathcal{L}$ and subsets of the set $S$ such that

\begin{eqnarray}
x \leq y &\Leftrightarrow& f(x) \subseteq f(y) \label{eq:th4.13.1} \\
f(x \wedge y) &=& f(x) \cap f(y) \label{eq:th4.13.2} \\
f(x \vee y) &=& f(x) \cup f(y) \label{eq:th4.13.3} \\
f(x^{\perp}) &=& S \setminus f(x) \label{eq:th4.13.4}
\end{eqnarray}

\noindent where $\subseteq$, $\cap$, $\cup$ and $\setminus$ are usual
set-theoretical operations of inclusion, intersection, union and
relative complement.
\end{theorem}
\begin{proof}
 The statement of the theorem follows immediately  if we choose
$S$ to be the set of all atoms.  Then property (\ref{eq:th4.13.1})
follows from Lemma \ref{Lemma4.7}, equation (\ref{eq:th4.13.2}) follows
from Lemma \ref{Lemma4.8}. Lemmas \ref{Lemma4.12} and
\ref{Lemma4.10} imply equations (\ref{eq:th4.13.3}) and
(\ref{eq:th4.13.4}), respectively.
\end{proof}
\bigskip

\subsection{Atoms and pure classical states}
\label{ss:atoms-pure}

\begin{lemma} \label{Lemma4.14} In classical logic, if $p$ is an atom and
$\phi$ is a pure state such that $(\phi|p) = 1$,\footnote{such a
state always exists due to Postulate \ref{postulateJ}.} then for any
other atom $q \neq p$ we have  $(\phi |q) = 0$.
\end{lemma}
\begin{proof}
  According to Lemma \ref{Lemma4.11},  $q \leq p^{\perp}$ and due
to equation (\ref{eq:orthocomp})

\begin{eqnarray*}
(\phi|q) \leq  (\phi|p^{\perp}) = 1 - (\phi | p) = 0
\end{eqnarray*}
\end{proof}
\bigskip

\begin{lemma}  \label{Lemma4.15} In classical logic, if $p$ is an atom  and
$\phi$ and $\psi$ are two pure states such that $(\phi|p) = (\psi
|p) = 1$, then $\phi = \psi$.
\end{lemma}
\begin{proof}
  There are propositions of two kinds: those containing
the atom $p$ and those not containing the atom $p$. For any
proposition $x$ containing the atom $p$ we denote by $q$ the atoms
contained in $x$ and obtain using Postulate \ref{postulateM}(2),
Lemma \ref{Lemma4.11}, Postulate \ref{postulateL} and Lemma
\ref{Lemma4.14}

\begin{eqnarray*}
 (\phi |x)  &=& (\phi |\vee_{q \leq x} q)
 = \sum_{q \leq x} (\phi|q)
 =  (\phi|p)
 = 1
\end{eqnarray*}

\noindent The same equation holds for the state $\psi$. Similarly we
can show that for any proposition $y$ not containing the atom $p$

\begin{eqnarray*}
 (\phi |y) =  (\psi|y) = 0
\end{eqnarray*}

\noindent Since probability measures of $\phi$ and $\psi$ are the
same for all propositions, these two states are equal.
\end{proof}
\bigskip

\begin{theorem} \label{Theorem4.16} In classical logic, there is an
isomorphism between atoms $p$ and pure states $\phi_p$ such that

\begin{eqnarray}
(\phi_p|p) = 1 \label{eq:th4.16}
\end{eqnarray}

\end{theorem}
\begin{proof}
From Postulate \ref{postulateJ} we know that for each atom $p$
 there is a state $\phi_{p}$ in which equation
(\ref{eq:th4.16}) is valid. From Lemma \ref{Lemma4.14} this state is
unique. To prove the reverse statement we just need to show that for
each pure state $\phi_p$ there is a unique atom $p$ such that
$(\phi_p|p)=1$. Suppose that for each atom $p$  we have $(\phi_p|p)
= 0$. Then, taking into account that $\mathcal{I}$ is a join of all
atoms, that all atoms are mutually disjoint and using (\ref{eq:maximum}) and Postulate
\ref{postulateL}, we obtain

\begin{eqnarray*}
1 &=& (\phi_p |\mathcal{I}) = (\phi_p |\vee_{p \leq \mathcal{I}} p)
= \sum_{p \leq \mathcal{I}} (\phi_p|p)
= 0
\end{eqnarray*}

\noindent which is absurd. Therefore, for each state $\phi_p$ one
can always find at least one atom $p$ such that equation
(\ref{eq:th4.16}) is valid. Finally, we need to show that if $p$ and
$q$ are two such atoms, then $p = q$. This follows from the fact
that for each pure classical state the probability measures (or the
truth functions) corresponding to propositions $p$ and $q$ are
exactly the same. For the state $\phi_p$ the truth function is equal
to 1, for all other pure states the truth function is equal to 0.
\end{proof}
\bigskip

\subsection{Phase space of classical mechanics}
\label{ss:phase-space}

Now we are fully equipped to discuss the phase space representation
 in classical
mechanics. Suppose that the physical system under consideration has
observables $A,B,C, \ldots$ with corresponding spectra $S_A$, $S_B$,
$S_C$, ... According to Theorem \ref{Theorem4.16}, for each atom $p$
 of the propositional system we can find its
corresponding pure state $\phi_p$. All observables $A,B,C, \ldots$
have definite values in this state.\footnote{See Assertion \ref{assertionJ_c}} Therefore the state $\phi_p$ is
characterized by a set of real numbers $A_p,B_p,C_p, \ldots$ - the
values of observables. Let us suppose that the full set of
observables $\{A,B,C, \ldots \}$ contains a minimal subset of
observables $\{X,Y,Z, \ldots\}$ whose values $\{X_p,Y_p,Z_p,
\ldots\}$ uniquely enumerate all pure states $\phi_p$ and therefore
all atoms $p$. So, there is a one-to-one correspondence between
groups of numbers $\{X_p,Y_p,Z_p, \ldots\}$ and atoms $p$. Then the set of
all atoms can be identified with the direct product\footnote{\emph{Direct product} $A \times B$ of two sets \index{direct
product} $A$ and $B$ is a set of all ordered pairs $(x,y)$, where $x
\in A$ and $y \in B$.} of
spectra of this minimal set of observables $S = S_X \times S_Y \times
S_Z \times \ldots$. This direct product is called the \emph{phase
space} \index{phase space} of the system.
 The values
$\{X_s,Y_s,Z_s, \ldots\}$ of the independent observables $\{X,Y,Z,
\ldots\}$  in each point $s \in S$ provide the phase space with
``coordinates.'' Other (dependent) observables $A,B,C, \ldots$ can
be represented as real functions $A(s), B(s), C(s), \ldots$ on $S$
or as functions of independent observables $\{X,Y,Z, \ldots\}$.

In this representation, propositions can be viewed as subsets of the
phase space. Another way is to consider propositions as special
cases of observables (= real functions on $S$): The function
corresponding to the proposition about the subset $T$ of the phase
space is the \emph{characteristic function} \index{characteristic
function} of this subset

\begin{eqnarray}
\xi(s) = \left \{ \begin{array}{c}
 1, \mbox{ if } s \in T  \\
 0, \mbox{ if } s \notin T
\end{array} \right. \label{eq:character}
\end{eqnarray}

\noindent Atomic propositions
correspond to single-point subsets of the phase space $S$.  In subsection \ref{ss:phase_space} we will consider one massive particle and build an explicit and realistic example of the phase space for this physical system.

\subsection{Classical probability measures}

Probability measures have a simple interpretation in the classical
phase space. Each state $\phi$ (not necessarily a pure state)
defines probabilities $(\phi|p) $ for all atoms $p$ (= all points
$s$ in the phase space). Each proposition $x$ is a join of disjoint
atoms contained in $x$.

\begin{eqnarray*}
 x  = \vee_{q \leq x} q
\end{eqnarray*}

\noindent Then, by Postulate \ref{postulateM},
Lemma \ref{Lemma4.11}, and Postulate \ref{postulateL} the probability of the proposition $x$
being true in the state $\phi$ is

\begin{eqnarray}
(\phi| x)  &=& (\phi |\vee_{q \leq x} q) = \sum_{q \leq x} (\phi |q)
\label{eq:prob-sum}
\end{eqnarray}

\noindent So, the value of the probability measure for all
propositions $x$ is uniquely determined by its values on atoms.
 In many important cases, the phase space is continuous, and
instead of considering probabilities $(\phi |q)$ at points in the
phase space (= atoms) it is convenient to consider
\emph{probability densities} \index{probability density} which are
functions $\Phi(s)$ on the phase space such that

\begin{itemize}
\item[1)] $\Phi(s) \geq 0$;
\item[2)] $\int \limits_{S} \Phi(s) ds = 1$.
\end{itemize}

\noindent Then the  value of the probability measure $(\phi|x)$
 is obtained by
the integral

\begin{eqnarray*}
(\phi|x) = \int \limits_{X} \Phi(s) ds
\end{eqnarray*}

\noindent over the subset $X$ corresponding to the proposition $x$.

For a pure classical state $\phi$, the probability density is
represented by the delta function $\Phi(s) = \delta(s - s_0)$ localized at one point $s_0$ in the phase space. For such states, the value of the
probability measure in each proposition
$x$ can be either 0 or 1: $(\phi|x)=0$  if the point $s_0$ does not belong
to the subset $X$ corresponding to the proposition $x$ and $(\phi|x)=1$ otherwise\footnote{ States whose probability densities are nonzero at more than one
point in the phase space (i.e., the probability density is different from the delta function) are called
\emph{classical mixed states}. \index{classical mixed state} We will not discuss them in this book. }

\begin{eqnarray*}
(\phi|x)  = \int \limits_{X}  \delta(s - s_0) ds = \left \{
\begin{array}{c}
 1, \mbox{ if } s_0 \in X  \\
 0, \mbox{ if } s_0 \notin X
\end{array} \right. \label{eq:class-pure}
\end{eqnarray*}

\noindent This shows that for pure classical states the probability
measure degenerates into a two-valued truth function. This is in agreement
with our discussion in subsection \ref{ss:truth}.

As we discussed earlier, in classical pure states all observables have well defined values.  So, classical mechanics is
a fully deterministic theory in which one can, in principle, obtain
a full information about the system at any given time and, knowing
the rules of dynamics, predict exactly its development in the future.
This belief was best expressed by P.--S. Laplace:

\begin{quote}
\emph{An
intelligence that would know at a certain moment all the forces
existing in nature and the situations of the bodies that compose
nature and if it would be powerful enough to analyze all these
data, would be able to grasp in one formula the movements of the
biggest bodies of the Universe as well as of the lightest atom.}
\end{quote}

\section{Quantum logic}
\label{sc:quant-mech}

The above discussion of classical logic and phase
spaces relied heavily on the determinism (Assertion \ref{assertionJ_c}) of
classical mechanics and on the validity of distributive laws
(Assertions \ref{assertionK12a} and \ref{assertionK12b}). In quantum
mechanics we are not allowed to use these Assertions. This is how we are going to proceed in this section when building the formalism of \emph{quantum logic}. \index{quantum logic} By design, this theory is more general than the familiar classical logic and it includes the latter as a particular (limiting) case. As we will see in the rest of this chapter, quantum logic is a foundation of the entire mathematical formalism of quantum theory.

\subsection{Compatibility of propositions}
\label{ss:compatibility}

Propositions $x$ and $y$ are said to be \emph{compatible}
\index{compatible propositions}\index{compatibility} (denoted $x
\leftrightarrow y$) \index{$\leftrightarrow$ compatibility} if

\begin{eqnarray}
x   &=& (x \wedge y) \vee (x \wedge y^{\perp}) \label{eq:compat1} \\
y   &=& (x \wedge y) \vee (x^{\perp} \wedge y) \label{eq:compat2}
\end{eqnarray}

\noindent The notion of compatibility has a great importance for quantum
theory. In subsection \ref{ss:compatible} we will see that two propositions
can be
measured simultaneously if and only if they are compatible.

\bigskip

\begin{theorem}
\label{Theorem4.17} In an orthocomplemented lattice all propositions
are compatible if and only if the lattice is distributive.
\end{theorem}
\begin{proof}
If the lattice is distributive then for any two propositions $x$ and
$y$

\begin{eqnarray*}
(x \wedge y) \vee (x \wedge y^{\perp}) = x \wedge (y \vee y^{\perp})
= x \wedge \mathcal{I} = x
\end{eqnarray*}

\noindent and, changing places of $x$ and $y$

\begin{eqnarray*}
(x \wedge y) \vee (x^{\perp} \wedge y) = y
\end{eqnarray*}

\noindent These formulas coincide with our definitions of
compatibility (\ref{eq:compat1}) and (\ref{eq:compat2}), which proves
the direct statement of the theorem.

The proof of the inverse statement (compatibility $\to$
distributivity) is more lengthy. We assume that all propositions in
our lattice are compatible with each other and choose three
arbitrary propositions $x$, $y$, and $z$.  Now we are going to prove
that the distributive laws\footnote{Assertions \ref{assertionK12a}
and \ref{assertionK12b}}

 \begin{eqnarray}
(x \wedge z) \vee (y \wedge z) &=&   (x \vee y) \wedge  z
\label{eq:dist_law1} \\
(x \vee z) \wedge (y \vee z) &=&   (x \wedge y) \vee  z
\label{eq:dist_law2}
\end{eqnarray}

\noindent are valid. First we  prove that the following 7
propositions (some of them may be empty) are mutually disjoint (see
Fig. \ref{fig:4.1})

\begin{eqnarray*}
q_1 &=& x          \wedge y          \wedge z \\
q_2 &=& x^{\perp}  \wedge y          \wedge z \\
q_3 &=& x          \wedge y^{\perp}  \wedge z \\
q_4 &=& x          \wedge y          \wedge z^{\perp} \\
q_5 &=& x          \wedge y^{\perp}  \wedge z^{\perp} \\
q_6 &=& x^{\perp}  \wedge y          \wedge z^{\perp} \\
q_7 &=& x^{\perp}  \wedge y^{\perp}  \wedge z
\end{eqnarray*}

\begin{figure}
\centering
\includegraphics{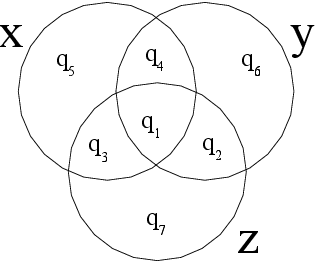}
\caption{To the proof of Theorem \ref{Theorem4.17}.}
\label{fig:4.1}
\end{figure}

\noindent For example, to show that propositions $q_3$ and $q_5$ are
disjoint we notice that $q_3 \leq z$ and $q_5 \leq z^{\perp}$ (by
Postulate  \ref{postulateK5a}). Then by Lemma \ref{postulateK11} $ z
\leq q_5 ^{\perp}$ and $ q_3 \leq z \leq q_5 ^{\perp}$. Therefore by
Lemma \ref{lemmaK3} $ q_3 \leq q_5 ^{\perp}$.

Since by our assumption both $x \wedge z$ and $x \wedge z^{\perp}$
are compatible with $y$, we obtain

\begin{eqnarray*}
x \wedge z &=& (x \wedge z \wedge y) \vee (x \wedge  z \wedge y^{\perp} ) =
q_1 \vee q_3\\
x \wedge z^{\perp} &=& (x \wedge  z^{\perp} \wedge y)
\vee (x \wedge  z^{\perp} \wedge y^{\perp}) = q_4 \vee q_5 \\
x &=& (x \wedge z) \vee (x \wedge z^{\perp}) = q_1 \vee q_3 \vee q_4 \vee q_5
\end{eqnarray*}

\noindent Similarly we show

\begin{eqnarray*}
y \wedge z &=& q_1 \vee q_2 \\
y &=& q_1 \vee q_2 \vee q_4 \vee q_6 \\
z &=& q_1 \vee q_2 \vee q_3 \vee q_7
\end{eqnarray*}

\noindent Then denoting $Q = q_1 \vee q_2 \vee q_3$ we obtain

\begin{eqnarray}
(x \wedge z) \vee (y \wedge z) &=& (q_1 \vee q_3) \vee (q_1 \vee q_2)
= q_1 \vee q_2 \vee q_3 = Q \label{eq:(4.23)}
\end{eqnarray}

\noindent From Postulate \ref{postulateK6a} and $y \vee x = Q \vee
q_4 \vee q_5 \vee q_6$ it follows that

\begin{eqnarray}
Q \leq (Q \vee q_7) \wedge (Q \vee q_4 \vee q_5 \vee q_6) = (x \vee y)
\wedge z
\label{eq:(4.24)}
\end{eqnarray}

\noindent On the other hand, from $q_4 \vee q_5 \vee q_6 \leq
q_7^{\perp}$, Lemma \ref{Lemma4.3} and the definition of
compatibility it follows that

\begin{eqnarray}
(x \vee y)  \wedge z =    (Q \vee q_4 \vee q_5 \vee q_6)\wedge  (Q \vee
q_7) \leq
  (Q \vee q_7^{\perp}) \wedge (Q \vee q_7) = Q
\label{eq:(4.25)}
\end{eqnarray}

\noindent Therefore, applying the  symmetry Lemma \ref{lemmaK2}
to equations (\ref{eq:(4.24)}) and (\ref{eq:(4.25)}), we obtain

\begin{eqnarray}
(x \vee y)  \wedge z =  Q \label{eq:(4.26)}
\end{eqnarray}

\noindent Comparing equations (\ref{eq:(4.23)}) and (\ref{eq:(4.26)}) we
see  that the distributive law (\ref{eq:dist_law1}) is valid. The
other distributive law (\ref{eq:dist_law2}) is obtained from equation
(\ref{eq:dist_law1}) by duality (see Appendix \ref{ss:theorems}).
\end{proof}
\bigskip

\noindent This theorem tells us that in classical mechanics all propositions are compatible. The presence of incompatible propositions is a characteristic feature of quantum theories.

\subsection{Logic of quantum mechanics}
\label{ss:qm-logic}

 In quantum mechanics we are not allowed to
use classical Assertion \ref{assertionJ_c} and  we must
abandon the distributive laws. However, in order to get a non-trivial
theory we need some substitutes for these two properties. This additional
postulate should be specific enough to yield sensible physics and
general enough  to be non-empty and to include the distributive law
as a particular case. The latter requirement is justified by our
desire to have classical mechanics as a particular case of the more
general quantum mechanics.

To find such a generalization we suggest the following arguments.
From Theorem \ref{Theorem4.17} we know that the  compatibility of
all propositions is a characteristic property of classical Boolean
lattices. We also mentioned that this property is equivalent to
simultaneous measurability of all propositions. We know that in quantum
mechanics not all propositions are simultaneously measurable,
therefore they cannot be compatible as well. This suggests that we
may try to find a generalization of classical theory by limiting the
set of propositions that are  mutually compatible. More specifically, we will postulate that two propositions are definitely
compatible if one implies the other and leave it to mathematics to
tell us about the compatibility of other pairs.

\begin{postulate} [orthomodularity]
 Propositions about physical systems obey
the orthomodular law: If $a$ implies $b$, then these two propositions are compatible

\begin{eqnarray}
 a \leq b  \Rightarrow  a   \leftrightarrow b.
\label{eq:orthomodularity}
\end{eqnarray} \label{postulateK13}
\end{postulate}

\noindent Orthocomplemented lattices with additional orthomodular
Postulate \ref{postulateK13} are called \emph{ orthomodular
lattices}. \index{orthomodularity}\index{orthomodular lattice}

Is there any deeper physical justification for the above postulate? As far as I know, there is none. The only justification is that the orthomodularity postulate really works, i.e., it results in the well-known mathematical structure of quantum mechanics, which has been thoroughly tested in experiments. In principle, one can try to introduce a different postulate to replace the classical distributivity relationships. If the resulting set of postulates turned out to be self-consistent, then one would obtain a non-classical theory that is also different from quantum mechanics. To the author's best knowledge, this line of reasoning has not been fruitful. So, in this book we will stick to orthomodular lattices and to traditional laws of quantum mechanics that follow from them.

Before proceeding further, we need to introduce important notions of the irreducibility of lattices and their rank.
The \emph{center}
\index{center of a lattice} of a lattice is the set of elements
compatible with all others. Obviously $\emptyset$ and $\mathcal{I}$
are in the center. A propositional system in which there are only
two elements in the center ($\emptyset$ and $\mathcal{I}$) is called
\emph{irreducible}. \index{lattice irreducible}\index{irreducible
lattice} Otherwise it is called \emph{reducible}. \index{lattice
reducible}\index{reducible lattice} Any Boolean lattice
\index{Boolean lattice} having more than two elements ($\emptyset$
and $\mathcal{I}$ are present in any lattice, of course) is
reducible and its center coincides with the entire lattice.
Orthomodular atomic irreducible lattices are called \emph{quantum
propositional systems} \index{quantum logic}\index{propositional
system quantum} or \emph{quantum logics}. The \emph{rank}
\index{rank of a lattice} of a propositional system is defined as
the maximum number of mutually disjoint atoms.
 For example, the rank of the classical propositional
system of one massive spinless particle described in subsection
\ref{ss:phase_space} is the ``number of points in the phase space
$\mathbb{R}^6$.''

The most fundamental conclusion from our discussion in this section
 is the following

\begin{statement} [quantum logic]
\label{statementN}
 Experimental propositions
form a quantum propositional system (=orthomodular atomic irreducible lattice).
\end{statement}

\noindent In principle, it should be possible to perform all
constructions and calculations in quantum theory by using the
formalism of orthomodular lattices based on just described
postulates. Such an approach would have certain advantages because
all its components have clear physical meaning: experimental propositions $x$ are
realizable in laboratories, and probabilities $(\phi|x)$ can be
directly measured in experiments. However, this approach meets
tremendous difficulties mainly because lattices are rather exotic
mathematical objects, and we lack intuition when dealing with lattice
operations.

We saw that in classical mechanics the happy alternative to the obscure
lattice theory is provided by Theorem  \ref{Theorem4.13} which
proves the isomorphism between the language of classical logic and the physically transparent
language of phase spaces. Is there a similar equivalence theorem in
the quantum case? To answer this question, we may notice that there
is a striking similarity between algebras of projections on closed
subspaces in a complex Hilbert space $\mathcal{H}$ (see Appendices
\ref{sc:hilbert-space} and \ref{sc:subspaces}) and quantum
propositional systems discussed above. In particular, if operations
between projections (or subspaces) in the Hilbert space are
translated to lattice operations according to Table
\ref{table:2.5},\footnote{We denote $Sp(A, B)$ the linear span \index{span} \index{$Sp(\ldots,\ldots)$, span of subspaces} of two subspaces $A$ and $B$ in the Hilbert space. (See Appendix \ref{ss:vector-space}.)  $A \cap B$ denotes the intersection of these subspaces. \index{$\cap$, intersection of subspaces} $A'$ is the orthogonal complement of $A$.} then all axioms of quantum logic can be directly
verified. For example, the validity of the Postulate
\ref{postulateK13} follows from Lemmas \ref{LemmaA.16} and
\ref{LemmaA.17}. Atoms \index{atom} can be identified with
one-dimensional subspaces or \emph{rays} in $\mathcal{H}$.
\index{ray} The irreducibility follows from Lemma \ref{LemmaG.6}.

\begin{table}[h]
\caption{Translation of terms, symbols, and operations used for
subspaces in the Hilbert space, projections on these subspaces, and propositions in
quantum logic.}
\begin{tabular*}{\textwidth}{@{\extracolsep{\fill}}ccc}
 \hline
Subspaces           & Projections   & Propositions  \cr \hline $X
\subseteq Y$     & $P_XP_Y = P_YP_X = P_X$ &  $x \leq y$  \cr $X
\cap Y$       & $P_{X \cap Y}  $&  $x \wedge y$    \cr $Sp(X, Y) $
& $P_{Sp(X, Y)}$ &  $ x \vee y$    \cr $X'$      & $1 - P_X$ &   $
x^{\perp}$   \cr $X$ and $Y$ are compatible   & $[P_X, P_Y] = 0$  &
$ x \leftrightarrow y$ \cr $X \perp Y$        & $P_XP_Y = P_YP_X =
0$   & $x \leq y^{\perp}$  \cr $\mathbf{0}$     &  $ 0$ &
$\emptyset$ \cr $\mathcal{H}$                &  $1$           &
$\mathcal{I}$ \cr $ray \mbox{ } x$ & $|x \rangle \langle x|$ & $x$
is an atom \cr
 \hline
\end{tabular*}
\label{table:2.5}
\end{table}

\subsection{Example: 3-dimensional Hilbert space}

\begin{figure}
\centering
 \includegraphics{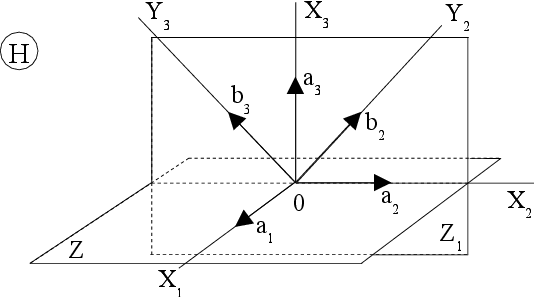} \caption{Subspaces in a 3-dimensional Hilbert space $\mathcal{H}$.} \label{fig:4.2}
\end{figure}

One can  verify directly that distributive laws
\ref{assertionK12a} and \ref{assertionK12b} are generally not valid for
subspaces in the Hilbert space $\mathcal{H}$. Consider, for example, the system of basis vectors and subspaces in a 3-dimensional Hilbert space $\mathcal{H}$ shown in Fig. \ref{fig:4.2}. The triples of vectors $(a_1, a_2, a_3)$ and $(a_1, b_2, b_3)$ form two orthogonal sets. They correspond to 1-dimensional subspaces $X_1, X_2, X_3, Y_2, Y_3$. In addition, two 2-dimensional subspaces $Z$ and $Z_1$ can be formed as $Z = Sp(X_1, X_2)$ and $Z_1 = Sp(X_2, X_3) = Sp(Y_2, X_3)$. These subspaces satisfy obvious relationships

\begin{eqnarray*}
Sp(X_2, X_3) &=& Sp(Y_2, X_3) = Sp(X_2, Y_2) = Z_1, \ \ \
Sp(X_1, X_2) = Z, \ \ \
X_3 \cap Y_2 = \mathbf{0} \\
X_2 \cap Y_2 &=& \mathbf{0}, \ \ \
X_1'  = Z_1, \ \ \
Z \cap Z_1 = X_2
\end{eqnarray*}

\noindent Then one can find a triple of subspaces for which the distributive laws in Assertions \ref{assertionK12a} and \ref{assertionK12b} are not satisfied

\begin{eqnarray*}
Sp(Y_2 , (X_3 \cap X_2)) &=& Sp(Y_2, \mathbf{0}) = Y_2 \neq Z_1 = Z_1 \cap Z_1 = Sp(Y_2, X_3) \cap Sp(Y_2, X_2) \\
Y_2 \cap Sp(X_3, X_2) &=& Y_2 \cap Z_1 = Y_2 \neq \mathbf{0} = Sp(\mathbf{0}, \mathbf{0}) = Sp((Y_2 \cap X_3), (Y_2 \cap X_2))
\end{eqnarray*}

\noindent This means that the logic represented by subspaces in the
Hilbert space is different from the classical Boolean logic.
\index{Boolean logic} However, the orthomodularity postulate is valid there. For example, $X_1 \subseteq Z$ so the condition in (\ref{eq:orthomodularity}) is satisfied and these two subspaces are compatible according to (\ref{eq:compat1}) - (\ref{eq:compat2})

\begin{eqnarray*}
Sp((Z \cap X_1), (Z \cap X_1')) &=& Sp(X_1, (Z \cap Z_1)) = Sp(X_1, X_2) = Z \\
Sp((Z \cap X_1), (Z' \cap X_1)) &=& Sp(X_1, (X_3 \cap X_1)) = Sp(X_1, \mathbf{0})  = X_1
\end{eqnarray*}

\subsection{Piron's theorem}

 Thus we have established that the set
of closed subspaces (or projections on these subspaces) in any complex Hilbert space
$\mathcal{H}$ is a representation of some quantum propositional
system. The next question is: can we find a Hilbert space
representation for \emph{each} quantum propositional system? The positive
answer to this question is given by the important \emph{Piron's theorem}
\index{Piron's theorem}
 \cite{Piron, Piron2}

\begin{theorem} [Piron]
 \label{Theorem.Piron}   Any irreducible quantum
propositional system (= orthomodular atomic irreducible lattice) $\mathcal{L}$ of rank 4 or higher is isomorphic to the lattice of closed
subspaces in a Hilbert space $\mathcal{H}$ such that the correspondences
shown in Table \ref{table:2.5} are true.
\end{theorem}

\noindent The proof of this theorem is beyond the
scope of our book. Two further remarks can be made regarding this theorem's statement.
First, all
propositional systems of interest to physics have infinite (even
uncountable) rank, so the condition ``rank $\geq 4$'' is not a
significant restriction. Second, the original Piron's theorem does not specify the nature of scalars in the Hilbert space. This theorem
leaves the freedom of choosing  any \emph{division ring with involutive
antiautomorphism} as the set of scalars in $\mathcal{H}$. We can
greatly reduce this freedom if we remember the important role played
by real numbers in physics.\footnote{e.g., values of observables are always in
$\mathbb{R}$} Therefore, it makes physical sense to consider only
those rings which include $\mathbb{R}$ as a subring. In 1877
Frobenius proved that there are only three such rings. They are
\emph{real numbers} $\mathbb{R}$, \emph{complex numbers}
$\mathbb{C}$, and \emph{quaternions} \index{quaternions}
$\mathbb{H}$. Although there is vast literature on real and,
especially, quaternionic quantum mechanics \cite{Stueckelberg,
quaternionic}, the relevance of these theories to physics remains
uncertain. Therefore, we will stick with complex numbers from now on.

 Piron's theorem forms the foundation of the mathematical
formalism of quantum physics. In particular, it allows us to express
the important notions of observables and states
 in the new language of Hilbert spaces. In this language pure quantum states are described by unit vectors in the Hilbert space. Observables are described by Hermitian operators in the same Hilbert space. These correspondences will be explained in the next
 section. Orthomodular lattices of quantum logic, phase spaces of
 classical mechanics, and Hilbert spaces of quantum mechanics are
 just different languages for describing relationships between
states, observables, and their measured values. Table
\ref{table:2.6} can be helpful for translation between these three
languages.

\begin{table}[h]
\caption{Glossary of terms used in general quantum logic, in
classical phase space, and in the Hilbert space of quantum
mechanics.}
\begin{tabular*}{\textwidth}{@{\extracolsep{\fill}}cccc}
 \hline
nature            & Quantum logic   & Phase space      & Hilbert
space  \cr \hline \hline Statement      & proposition & subset &
closed subspace   \cr \hline Unambiguous        & Atom         &
Point      & Ray           \cr statement                  & &
&     \cr \hline AND & meet                 & intersection &
intersection \cr \hline OR  & join              & union   & linear
span   \cr \hline NOT & orthocomplement    & relative   & orthogonal
\cr
                  &           & complement            & complement  \cr
\hline IF...THEN   & implication   &  inclusion  &  inclusion
 \cr
            &               &   of subsets    & of subspaces
\cr \hline Observable & proposition-valued   & real function &
Hermitian
 \cr
 &  measure  on $\mathcal{R}$ &  &
operator   \cr \hline jointly  & compatible   & all observables &
commuting   \cr measurable  &  propositions &   are  compatible &
operators  \cr observables &             &               &  \cr
\hline mutually exclusive  & disjoint   &
non-intersecting & orthogonal \cr statements   &  propositions      &
subsets & subspaces  \cr \hline
Pure state  &  Probability  & delta function & Ray   \cr
 &   measure &  &    \cr \hline
Mixed state &  Probability   & Probability   & Density   \cr
 &  measure  &     function  &  operator  \cr
 \hline
\end{tabular*}
\label{table:2.6}
\end{table}

\subsection{Should we abandon classical logic?}

In this section we have reached a seemingly paradoxical conclusion that one
cannot use classical logic as well as classical probability
theory for reasoning about
quantum systems. How could that be? Classical logic
is the foundation of the whole mathematics and the scientific method in general. All mathematical theorems are being
proved with the logic of Aristotle and Boole.\footnote{Note, however, attempts \cite{Dejonghe} to develop so-called ``quantum mathematics'' that is based on quantum logic.} Even theorems of
quantum mechanics are being proved using this logic. However, quantum mechanics insists that propositions about physical systems satisfy laws of the non-Boolean quantum logic.   The logic of classical distributive lattices is just an approximation. Isn't there a
contradiction? Not really.

It is still permissible to use classical logic  in quantum mechanical proofs because, thanks to the Piron's theorem, we have replaced real life
objects, such as experimental propositions and probability measures,
with abstract and artificial notions of Hilbert spaces, state
vectors, and Hermitian operators. These abstractions
is the price we pay for the privilege to keep using simple classical logic.
In principle, it should be possible to formulate entire quantum theory using the language of propositions, quantum logic, and probability measures. However, such an approach has not been developed yet.

\section{Quantum observables and states} \label{ss:quant-obs}

\subsection{Observables}
\label{ss:observables}

Each observable $F$ naturally defines a mapping (called a
\emph{proposition-valued measure}) \index{proposition-valued
measure} from the set of intervals of the real line $\mathbb{R}$ to
propositions $F_E$ in $\mathcal{L}$. These propositions can be described in words:
$F_E=$ ``the value of the observable $F$ is inside the interval $E$ of the
real line $\mathbb{R}$.'' We already discussed properties of
propositions about one observable. They can be summarized as
follows:

\begin{itemize}
\item The proposition corresponding to the intersection of intervals
$E_1$ and $E_2$ is the meet of propositions corresponding to these
intervals
\begin{eqnarray}
F_{E_1 \cap E_2} &=& F_{E_1} \wedge F_{E_2} \label{eq:obs1}
\end{eqnarray}
\item The proposition corresponding to the union of intervals
$E_1$ and $E_2$ is the join of propositions corresponding to these
intervals
\begin{eqnarray}
F_{E_1 \cup E_2} &=& F_{E_1} \vee F_{E_2} \label{eq:obs2}
\end{eqnarray}
\item The proposition corresponding to the complement of interval
$E$ is the orthocomplement of the proposition corresponding to $E$
\begin{eqnarray}
F_{\mathbb{R} \setminus E} &=& F_{E}^{\perp} \label{eq:obs3}
\end{eqnarray}
\item The minimal proposition corresponds to the empty subset of the
real line
\begin{eqnarray}
F_{\emptyset} &=& \emptyset \label{eq:obs4}
\end{eqnarray}
\item The maximal proposition corresponds to the real line itself.
\begin{eqnarray}
F_{\mathbb{R}} &=& \mathcal{I} \label{eq:obs5}
\end{eqnarray}
\end{itemize}

Intervals $E$ of the real line form a Boolean (distributive) lattice
with respect to set theoretical operations $\subseteq, \cap, \cup$, and $\setminus$. Due to the isomorphism (\ref{eq:obs1}) -
(\ref{eq:obs5}), the corresponding one-observable propositions $F_E$ also form a
Boolean lattice, which is a \emph{sublattice} \index{sublattice} of
our full propositional system. Therefore, according to Theorem
\ref{Theorem4.17}, all propositions about the same observable are
compatible. Due to the isomorphism
``propositions''$\leftrightarrow$``subspaces,'' we can use the same
notation $F_E$ for subspaces (projections) in $\mathcal{H}$
corresponding to intervals $E$. Then, according to Lemma
\ref{LemmaA.17}, all projections $F_E$, referring to one observable $F$, commute with each other.

Each point $f$ in the spectrum of observable $F$ is called an
\emph{eigenvalue} of this observable. \index{eigenvalue}
The subspace $F_{f} \subset \mathcal{H}$ corresponding to the
eigenvalue $f$ is called \emph{eigensubspace} \index{eigensubspace}
and projection $P_{f}$ onto this subspace is called a \emph{spectral
projection}. \index{spectral projection} Each vector in the
eigensubspace is called \emph{eigenvector}.\footnote{In the next subsection we will see that each vector in the Hilbert space defines a unique pure quantum state. States defined by the above eigenvectors will be called \emph{eigenstates} (of the given observable $F$).
\index{eigenvector} \index{eigenstate} Apparently, the observable $F$ has
definite values (=eigenvalues) in its eigenstates. This means that
eigenstates are examples of states whose existence was guaranteed by
Postulate \ref{postulateJ}.}

Consider two distinct eigenvalues $f$ and $g$ of observable $F$. The
corresponding intervals (=points) of the real line are disjoint.
Then  propositions $F_f$
and $F_g$ are disjoint too, and corresponding (eigen)subspaces are
orthogonal. The linear span of subspaces $F_{f}$, where $f$ runs
through entire spectrum of $F$, is the full Hilbert space
$\mathcal{H}$. Therefore, spectral projections of any observable
form a \emph{decomposition of unity}.\footnote{see Appendix
\ref{sc:projections}} \index{decomposition of unity} So, according
to discussion in Appendix \ref{ss:commuting}, we can associate an
Hermitian operator

\begin{eqnarray}
F = \sum \limits_{f} f P_{f} \label{eq:op-obs}
\end{eqnarray}

\noindent with each observable $F$. In what follows we will often
use terms ``observable'' and ``Hermitian operator'' as synonyms.

\subsection{States}
\label{ss:states}

As we discussed in subsection \ref{ss:pure}, each state $\phi$ of
the system defines a probability measure $(\phi|x)$ on propositions
in quantum logic $\mathcal{L}$. According to the isomorphism
``propositions $ \leftrightarrow$ subspaces'',\footnote{See Table \ref{table:2.6}.} the state $\phi$ also
defines a \emph{probability measure} \index{probability measure}
$(\phi|X)$ on subspaces $X$ in the Hilbert space $\mathcal{H}$.
This probability measure is a function from subspaces to the
interval $[0,1] \subseteq \mathbb{R}$ whose properties follow
directly from equations (\ref{eq:maximum}), (\ref{eq:minimum}), and
Postulate \ref{postulateL}

\begin{itemize}
\item The probability corresponding to the whole Hilbert space is 1
in all states
\begin{eqnarray}
(\phi|\mathcal{H}) &=& 1 \label{eq:prob1}
\end{eqnarray}
\item The probability corresponding to the empty subspace is 0 in
all states
\begin{eqnarray}
(\phi|\mathbf{0}) &=& 0 \label{eq:prob2}
\end{eqnarray}
\item The probability corresponding to the direct sum  of orthogonal
subspaces\footnote{Note that according to Table \ref{table:2.6} orthogonal subspaces correspond to disjoint propositions. For definition of the direct sum of subspaces see Appendix \ref{sc:projections}.} is the sum of probabilities for each subspace
\begin{eqnarray}
 (\phi|X \oplus Y) &=&
(\phi|X) + (\phi|Y) \mbox{, if } X \perp Y \label{eq:prob3}
\end{eqnarray}
\end{itemize}

\noindent The following important theorem provides a classification
of all such probability measures (= all states of the physical
system).

\begin{theorem} [Gleason \cite{Gleason}] \index{Gleason theorem}
\label{Theorem.Gleason}  If $(\phi|X)$ is a probability measure on
closed subspaces in the Hilbert space $\mathcal{H}$ with properties
(\ref{eq:prob1}) - (\ref{eq:prob3}), then there exists a
non-negative Hermitian operator $\rho$
\index{density matrix}\index{density operator} in $\mathcal{H}$ such
that

\begin{eqnarray}
Tr (\rho) = 1 \label{eq:(4.28)}
\end{eqnarray}

\noindent and for any subspace $X$ with projection $P_X$ the value
of the probability measure is

\begin{eqnarray}
(\phi|X) = Tr (P_X \rho)  \label{eq:(4.29)}
\end{eqnarray}

\end{theorem}

\bigskip

\noindent Just a few comments about the terminology and notation used here: First, a
Hermitian operator is called \emph{non-negative} if all its eigenvalues are greater
than or equal to zero. Second, the operator $\rho$  is usually
called the \emph{density operator} \index{density operator} or the \emph{density matrix}. \index{density matrix} Third, $Tr$ denotes \emph{trace}\footnote{Trace is, basically, the sum of all diagonal elements of a matrix. See Appendix \ref{ss:functions}.} of the matrix of the operator
$\rho$.  \index{trace}

The proof of Gleason's theorem is far from trivial, and we refer
interested reader to original works \cite {Gleason,
Richman_Bridges}. Here we will focus on the physical interpretation
of this result. First, we may notice that, according to the spectral
theorem \ref{spectral-th}, the operator $\rho$ can be always written
as

\begin{eqnarray}
\rho = \sum_{i} \rho_i | e_i \rangle \langle e_i | \label{eq:rho}
\end{eqnarray}

\noindent where $| e_i \rangle$ is an orthonormal basis
 in $\mathcal{H}$. Then the Gleason's theorem means that

\begin{eqnarray}
\rho_i &\geq& 0 \label{eq:(4.30)}\\
\sum_i \rho_i & =& 1 \label{eq:(4.31)}\\
0 &\leq& \rho_i \leq 1 \label{eq:(4.32)}
\end{eqnarray}

\noindent Among all states satisfying equation (\ref{eq:(4.30)}) -
(\ref{eq:(4.32)}) there are simple states for which just one
coefficient $\rho_i$ is non-zero. Then, from (\ref{eq:(4.31)}) it
follows that $\rho_i = 1$, $\rho_j = 0$ if $j \neq i$, and the
density operator degenerates to a projection onto the
one-dimensional subspace $|e_i \rangle \langle
e_i|$.\footnote{One-dimensional subspaces are also called
\emph{rays}. \index{ray}} Such states will be called \emph{pure
quantum}  \index{pure quantum state}
states. It is also common to describe a pure state
 by a unit vector from
its ray. Any unit vector from this ray represents the same state,
i.e., in the vector representation of states there is a freedom of
choosing an unimodular \emph{phase factor} of the state vector. In
what follows we will often use the terms ``pure quantum state'' and
``state vector'' as synonyms.

Mixed quantum states \index{mixed state} are
expressed as weighed sums of pure states whose coefficients $\rho_i$
in equation (\ref{eq:rho}) reflect the probabilities with which the pure
states enter in the  mixture. Therefore, in quantum mechanics there
are uncertainties of two types. The first type is the uncertainty
present in mixed states. This uncertainty is already familiar to us
from classical (statistical) physics. This uncertainty results from
our insufficient control of preparation conditions (like when a
bullet is fired from a shaky riffle). The second type of uncertainty is
present even in pure quantum states.  It does not have a counterpart in classical physics, and
it cannot be avoided by tightening the preparation conditions. This
uncertainty is a reflection of the mysterious unpredictability of
microscopic phenomena.

We will not discuss mixed quantum states in this book. So, we will deal only with uncertainties of the fundamental quantum type. Thus, when speaking about a quantum state $\phi$, we will always assume that there exists a corresponding state vector $|\phi \rangle$. As discussed above, this vector is not the unique representative of the state. Any vector $e^{i \alpha}|\phi \rangle$ that differs from $|\phi \rangle$ by a unimodular phase factor $e^{i \alpha}$ (where $\alpha \in \mathbb{R}$), is also a valid representative of the state $\phi$.

\subsection{Commuting and compatible observables}
\label{ss:compatible}

In subsection \ref{ss:compatibility} we defined the  notion of
compatible propositions.  In Lemma \ref{LemmaA.17} we showed
that the compatibility of propositions is equivalent to the
commutativity of corresponding projections. The importance of these
definitions for physics comes from the fact that for a pair of
compatible propositions (=projections=subspaces) there are states in
which both these propositions are certain, i.e., simultaneously
measurable. A similar statement can be made for two compatible
(=commuting) Hermitian operators of observables. According to
Theorem \ref{TheoremA.20}, such two operators have a common basis
 of eigenvectors (=eigenstates). In these eigenstates
both observables have definite (eigen)values.

We will assume  that for any physical system there always exists a
\emph{minimal} and \emph{complete} set of mutually compatible (= commuting) observables $F, G,
H, \ldots$.\footnote{A set $F, G, H, \ldots$ is called \emph{minimal} if no
observable from the set can be expressed as a function of other
observables from the same set. A set is complete if no more observables can be added to it. For example, a single massive particle has a mutually commuting set of observables $(P_x, P_y, P_z, S^2, S_z)$, where $\mathbf{P}$ and $\mathbf{S}$ are the momentum and spin operators, respectively. See section \ref{sc:massive}. Any function of observables from the
minimal commuting set also commutes with $F, G, H, \ldots$, and with
any other such function. } Then,  we should be able to build an orthonormal basis
 of common eigenvectors $|e_i \rangle$ such that each
basis vector is uniquely labeled by eigenvalues $f_i, g_i, h_i,
\ldots$ of operators $F, G, H, \ldots$, i.e., if $|e_i \rangle$ and
$|e_j \rangle$ are two eigenvectors then there is at least one
different number in the two sets of eigenvalues $\{f_i, g_i, h_i, \ldots \}$ and $\{f_j,
g_j, h_j, \ldots \}$.

Each state vector $| \phi \rangle$ can be represented as a linear
combination of these basis vectors

\begin{eqnarray}
| \phi \rangle = \sum_i \phi_i |e_i \rangle \label{eq:(4.33)}
\end{eqnarray}

\noindent where in the bra-ket notation\footnote{see Appendix
\ref{ss:bra-ket}}

\begin{eqnarray}
\phi_i = \langle e_i | \phi \rangle \label{eq:(4.33a)}
\end{eqnarray}

 The set of coefficients $\phi_i$
can be viewed as a function $\phi(f,g,h, \ldots)$ on the common
spectrum of observables $F, G, H, \ldots$. In this form, the
coefficients  $\phi_i$ are referred to as the \emph{wave function}
\index{wave function} of the state $| \phi \rangle$ in the
\emph{representation} defined by observables $F, G, H, \ldots$. When
the spectrum of operators $F, G, H, \ldots$ is continuous, the index
$i$ is, actually, a continuous variable.\footnote{Wave functions in the momentum and position representations for a single particle will be discussed in  section \ref{sc:representations}.}

\subsection{Expectation values}
\label{ss:expectation}

Equation (\ref{eq:op-obs}) defines a spectral decomposition for each
observable $F$, where index $f$ runs over all distinct eigenvalues
of $F$. Then for each pure state $|\phi \rangle$ we can find the
probability of measuring a value $f$ of the observable $F$ in this
state by using formula\footnote{This is simply the value of the
probability measure $(\phi| P_f)$ (see subsection \ref{ss:pure})
corresponding to the spectral projection $P_f$. One can also see that this formula is equivalent to the Gleason's expression (\ref{eq:(4.29)}).}

\begin{eqnarray}
\rho_{f} = \sum \limits_{i=1}^m |\langle  e_i^{f} | \phi \rangle |^2
\label{eq:(4.34)}
\end{eqnarray}

\noindent where $| e_i^{f} \rangle$ are basis vectors in the range
of the projection

\begin{eqnarray}
P_{f} \equiv \sum \limits_{i=1}^m |  e_i^{f} \rangle \langle e_i^{f} | \label{eq:1.39a}
\end{eqnarray}

\noindent and $m$ is the dimension of the
corresponding subspace.  Sometimes we also need to know the weighed average of
values $f$. This is called the \emph{expectation value}
\index{expectation value} of the observable $F$ in the state $|\phi
\rangle$ and denoted $\langle F \rangle$

\begin{eqnarray*}
\langle F \rangle \equiv \sum \limits_{f} \rho_{f} f
\end{eqnarray*}

\noindent  Substituting here equation (\ref{eq:(4.34)}) we obtain

\begin{eqnarray*}
\langle F \rangle &=& \sum \limits_{j=1}^n |\langle  e_j | \phi
\rangle |^2 f_{j}
\equiv \sum \limits_{j=1}^n |\phi_j|^2 f_{j}
\end{eqnarray*}

\noindent where the summation is carried out over the entire basis
 $|e_j \rangle $ of eigenvectors of the operator $F$
with eigenvalues $f_j$.  By using  decompositions (\ref{eq:(4.33)}), (\ref{eq:op-obs}), and (\ref{eq:1.39a}) we obtain a more compact formula for the
expectation value $\langle F \rangle$

\begin{eqnarray}
\langle \phi|F| \phi \rangle &=&
\left(\sum_i \phi_i^* \langle e_i |\right) \left(\sum_j  | e_j \rangle f_j \langle e_j |\right)
\left(\sum_k \phi_k |e_ k \rangle\right) \nonumber \\
&=&
\sum_{ijk} \phi_i^* f_j \phi_k \langle e_i |  e_j \rangle \langle e_j |
  e_k \rangle =
\sum_{ijk} \phi_i^* f_j \phi_k\delta_{ij} \delta_{jk} \nonumber =
\sum_{j} |\phi_j|^2 f_j \\
&=& \langle F \rangle \label{eq:fff}
\end{eqnarray}

\subsection{Basic rules of classical and quantum mechanics}
\label{ss:rules}

Results obtained in this chapter can be summarized as follows. If our system prepared in a pure state $\phi$ and we
want to calculate the probability $\rho$ for measuring the value of
the observable $F$ inside the interval $E \subseteq \mathbb{R}$, then we need to perform the
following steps:

\bigskip

\noindent \textbf{In classical mechanics:}

\begin{itemize}
\item [1.] Define the phase space $S$ of the  physical system;
\item [2.] Find a real function $f: S \to \mathbb{R}$ corresponding to the observable $F$;
\item [3.] Find the subset $U$ of $S$ corresponding to the subset $E$ of
the spectrum of the observable $F$ ($U$ is the set of all points $s
\in S$ such that $f(s) \in E$);
\item [4.] Find the point $s_{\phi} \in S$  representing the pure classical state
$\phi$;
\item [5.] The probability $\rho$ is equal to 1 if $s_{\phi} \in U$
and $\rho = 0 $ otherwise.
\end{itemize}

\noindent \textbf{In quantum mechanics:}

\begin{itemize}
\item [1.] Define the Hilbert space $\mathcal{H}$ of the  physical system;
\item [2.] Find the Hermitian operator $F$  in
$\mathcal{H}$ corresponding to the observable;
\item [3.] Find the eigenvalues and eigenvectors of the  operator $F$;
\item [4.] Find a spectral projection $P_E$ corresponding to the subset $E$
of the spectrum of the operator $F$.
\item [5.] Find the unit vector $| \phi \rangle$  (defined up to an arbitrary
unimodular factor) representing the  state of the system.
\item [6.] Use formula $\rho = \langle \phi | P_E | \phi \rangle$
\end{itemize}

\noindent At this point, there seems to be no connection between the
classical and quantum rules. However, we will see in section
\ref{sc:classical}  that in the macroscopic world with massive
objects and poor resolution of instruments, the classical rules
emerge as a decent approximation to the quantum ones.

\section{Interpretations of quantum mechanics}
\label{sc:complete}

In  sections \ref{sc:lattice} - \ref{ss:quant-obs} of this chapter
we focused on the mathematical formalism of quantum mechanics. Now
it is time to discuss the physical meaning and interpretation of
these formal rules.

\subsection{Quantum unpredictability in microscopic systems}
\label{ss:unpredictability}

Experiments with quantum microsystems have revealed one simple and
yet mysterious fact: if we prepare $N$ absolutely identical physical
systems in the same conditions and measure the same observable in
each of them, we may find $N$ different results.

Let us illustrate this experimental finding by few examples. We know
from experience that
 each photon passing through the hole in the \emph{camera obscura} \index{camera obscura} will
hit the photographic plate at some point on the photographic plate.
However, each new released photon will land at a different point.
Quantum mechanics allows us to calculate the probability density for
these points, but apart from that, the behavior of each individual
photon appears to be completely random. Quantum mechanics does not
even attempt to predict where each individual photon will hit the
target.

Another example of such an apparently random behavior is the decay
of unstable nuclei. The nucleus of $^{232}Th$ has the lifetime of 14
billion years. This means that in any sample containing thorium,
approximately half of all $^{232}Th$ nuclei will decay after 14
billion years. In principle, quantum mechanics can
 calculate the probability  of
the nuclear decay as a function of time by solving the corresponding
Schr\"odinger equation.\footnote {though our current knowledge of
nuclear forces is insufficient to make a reliable calculation of
that sort for thorium.} However, quantum mechanics cannot even
approximately guess when any given nucleus will decay. It could
happen today, or it could happen 100 billion years from now.

Although, such unpredictability is certainly a hallmark of
microscopic systems it would be wrong to think that it is not
affecting our macroscopic world.   Quite often the effect of random
microscopic processes can be amplified to produce a sizable equally
random macroscopic effect. One famous example of the amplification
of quantum uncertainties is the thought experiment with the
``Schr\"odinger cat'' \cite{cat}. \index{Schr\"odinger cat}

So, our world (even at the macroscopic scale) is full of truly
random events whose exact description and prediction is beyond
capabilities of modern science. Nobody knows why physical systems
have this random unpredictable behavior. Quantum mechanics simply
accepts this fact and does not attempt to explain it. Quantum
mechanics does not describe what actually happens; it describes the
full range of possibilities of what might have happened and the
probability of each possible outcome.  Each time nature chooses just one possibility from this
range, while obeying the probabilities predicted by quantum
mechanics. QM cannot say anything about which particular choice will
be made by nature. These choices are
completely random and beyond explanation by modern science. This
observation is a bit disturbing and embarrassing. Indeed, we have
real physically measurable effects (the actual choices made by
nature) for which we have no control and no power to predict the
outcome. These are facts without an explanation, effects without a
cause. It seems that microscopic particles obey some mysterious
random force. Then it is appropriate to ask what is the reason for
such stochastic behavior of micro-systems? Is it truly random or it
just seems to be random? If quantum mechanics cannot explain this
random behavior, maybe there is a deeper theory that can?

\subsection{Hidden variables}
\label{ss:hidden}

One school of thought attributes  the apparently random behavior of
microsystems to some yet unknown ``hidden'' variables, which are
currently beyond our observation and control. According to these
views, put somewhat simplistically, each photon in camera obscura
has a guiding mechanism which directs it to a certain predetermined
spot on the photographic plate. Each unstable nucleus has some
internal ``alarm clock'' ticking inside. The nucleus decays when the
alarm goes off. The behavior of quantum systems just appears to be
random to us because so far we don't have a clue about these
``guiding mechanisms'' and ``alarm  clocks.''

According to the ``hidden variables'' theory, quantum mechanics is
not the final word, and future theory will be able to
 fully describe the properties of individual systems and predict events without relying
on chance. There are two problems with this point of view. First, so
far nobody was able to build a convincing theory of hidden variables
and to predict (even approximately) outcomes of quantum measurements
beyond calculated probabilities. The second reason to reject the
``hidden variables'' argument is more formal.

The "hidden variables" theory says that the randomness of
micro-systems does not have any special quantum-mechanical origin.
It is the same classical pseudo-randomness as seen in the usual
coin-tossing or die-rolling. The "hidden variables" theory implies that rules of classical mechanics
apply to micro-systems just as well as to macro-systems. As we saw
in section \ref{ss:truth}, these rules are based on the classical
Assertion \ref{assertionJ_c} of
 determinism. Quantum mechanics simply discards this
unprovable Assertion and replaces it with a weaker Postulate
\ref{postulateJ}. So, quantum mechanics with its probabilities is a
more general mathematical framework, while classical mechanics with
its determinism can be represented as a particular case of this
framework. As Mittelstaedt put it \cite{Mittelstaedt}

\begin{quote}
\emph{...classical mechanics is loaded with metaphysical hypotheses
which clearly exceed our everyday experience. Since quantum
mechanics is based on strongly relaxed hypotheses of this kind,
classical mechanics is less intuitive and less plausible than
quantum mechanics. Hence classical mechanics, its language and its
logic cannot be the basis of an adequate interpretation of quantum
mechanics.} P. Mittelstaedt
\end{quote}

\subsection{Measurement problem}
\label{sc:measurement}

If we now accept the probabilistic quantum view of reality, we must
address some deep paradoxes. One disturbing ``paradox'' is that
in quantum mechanics the wave function of a physical system evolves
in time smoothly and unitarily according to the Schr\"odinger
equation (\ref{eq:psi-timex}) up until the
instant of measurement, at which point the wave function experiences
an unpredictable and abrupt collapse.

The seemingly puzzling part is that it is not clear
 how the wave function ``knows'' when it can undergo the continuous
evolution and when it should ``collapse.'' Where is
the boundary between the measuring device and the quantum system?
For example, it is customary to say that the photon is the quantum
system while the photographic plate is the measuring device. However,
we can adopt a different view and include the photographic plate
together with the photon in our quantum system. Then, we should, in
principle, describe both the photon and the photographic plate by a
joint wave function. When does this wave function collapse? Where is
the measuring apparatus \index{measuring apparatus} in this case?
Human's eye? Does it mean that while we are not looking, the entire
system (photon + photographic plate) remains in a superposition
state? Following this logic we may easily reach a seemingly absurd conclusion
that
 the ultimate measuring
device is human's brain, and all events remain potentialities until
they are registered by mind. For many physicists these
contradictions signify some troubling incompleteness of quantum
theory, its inability to describe the world ``as it is.''

In order to avoid the controversial wave function collapse, several
so-called ``interpretations'' of quantum mechanics were proposed. In
the de Broglie--Bohm's ``pilot wave'' interpretation \index{pilot
wave interpretation} it is postulated that the electron propagating in
the double-slit setup is actually a classical point particle, whose movement
is ``guided'' by a separate material ``wave'' that obeys the Schr\"odinger
equation. In the ``many worlds'' interpretation \index{many worlds
interpretation} it is assumed that at the instant of measurement
(when several outcomes are possible, according to quantum mechanics)
the world splits into several (or even infinite number of) copies,
so all outcomes are realized at once. We see only a single outcome
because we live just in one copy of the world and lack the ``bird
view'' of the many worlds reality.

Interpretations of quantum mechanics attempt to suggest some kinds
of physical mechanisms of the quantum system's behavior and the
measurement process. However, how we can be sure that these
mechanisms are correct? The only method of verification available in
physics is experiment, but suggested mechanisms are related to
things happening in the physical system \emph{while it is not
observed}. So, it is impossible to design experiments that would
(dis)prove interpretations of quantum mechanics. Being unaccessible
to experimental verification these interpretations should belong to
philosophy rather than physics.

Actually, the ``collapse'' or ``measurement'' paradoxes are not as
serious as they look. In author's view, their appearance is related
simply to our unrealistic expectations regarding the explanatory
power of physical theory. Intuitively we wish to have a physical
theory that encompasses all physical reality: the physical system,
the measuring apparatus, the observer, and the entire universe.
However this goal is perhaps too ambitious and misleading. Recall
that the goal of a physical theory declared in Introduction is to
provide a formalism that allows us to \emph{predict results of
experiments}.\footnote{More precisely, the theory should be able to
calculate probabilities of measurements.} In physics we do not want
and do not need to describe the whole world ``as it is.''  We should
be entirely satisfied if our theory allows us to calculate the
outcome of any conceivable measurement, which is a more modest task.

Thus it is not a surprise that certain aspects of reality  are beyond the reach of quantum theory. In this sense, quantum mechanics can be regarded as ``incomplete'' theory. However, this author believes that this incompleteness is not a problem, but a reflection of the fundamental unavoidable unpredictability of nature. Here the following quote from Einstein seems appropriate:

\begin{quote}
\emph{I now imagine a quantum theoretician who may even admit that
the quantum-theoretical description refers to ensembles of systems
and not to individual systems, but who, nevertheless, clings to the
idea that the type of description of the statistical quantum theory
will, in its essential features, be retained in the future. He may
argue as follows: True, I admit that the quantum-theoretical
description is an incomplete description of the individual system. I
even admit that a complete theoretical description is, in principle,
thinkable. But I consider it proven that the search for such a
complete description would be aimless. For the lawfulness of nature
is thus constructed that the laws can be completely and suitably
formulated within the framework of our incomplete description. To
this I can only reply as follows: Your point of view - taken as
theoretical possibility - is incontestable.} A. Einstein
\cite{incontestable}
\end{quote}

The quantum mechanical distinction between the observed system and the measuring apparatus is not as problematic as often claimed. This distinction is naturally present in every experiment. If properly asked, the experimentalist will always tell you which part of his setup is the observed system and which part is the measuring apparatus.\footnote{For
instance, in the above example of the double-slit experiment the photon is the physical system and
the photographic plate is the measuring apparatus. The one-photon
Hilbert space should be used for the quantum-mechanical analysis of
this experiment. If we like, we can consider the ``photon +
photographic plate'' as our physical system, but this would mean that
we have changed completely the experimental setup, the measuring apparatus, and the range of meaningful questions that can be asked and answered about the system. The new setup
should be described quantum-mechanically in a different Hilbert
space with different state vectors and different operators of
observables. } Therefore there is nothing wrong in applying different
descriptions to these two parts.
In quantum theory the state of the physical system is described as a
vector in the Hilbert space, and the measuring apparatus is described
as an Hermitian operator in the same Hilbert space. The measuring apparatus is not considered to be a
dynamical object. This means that there is no point to describe the
act of measurement as ``interaction'' between the physical system
and the measuring apparatus by means of some dynamical
theory.\footnote{as it was done,  e.g., in von Neumann's measurement theory.}

The ``collapse'' of the wave function
is not a dynamical process, it is just a part of a mathematical
formalism that allows us to fulfil to true task of any physical
theory - to predict outcomes of experiments. So there is no any
contradiction or paradox between the unitary time evolution of wave
functions and the abrupt ``collapse'' at the time of
measurement.

\subsection{Agnostic interpretation of quantum mechanics}

The things we said above can be  summarized in the following statements:

\begin{enumerate}
\item Quantum mechanics does not pretend to provide a description of the entire universe.
It only applies to the description of specific experiments in which
the physical system and the measuring apparatus are clearly
separated.
\item  Quantum mechanics does not provide a mechanism of what goes on
while the physical system is not observed or while it is measured.
Quantum mechanics is just a mathematical recipe for calculating
probabilities of experimental outcomes. The ingredients used in this
recipe (Hilbert space, state superpositions, wave functions,
Hermitian operators, etc.) have no direct relationship to things
observable in nature. They are just mathematical symbols.
\item We cannot measure all observables at once. A realistic experiment measures only one observable, or, in the best case, a few mutually compatible observables.
\item Nature is inherently probabilistic. There exists a certain level of random ``noise'' that leads to unpredictability of the results of measurements.
\item Logical propositions about measurements do not obey the set of classical Boolean axioms. The distributive law of logic is not valid and should be replaced by the orthomodular law.
\end{enumerate}

The most important philosophical lesson taught to us by quantum mechanics
is the unwillingness to speculate about things that are not
observable. Einstein was very displeased with this point of view. He
wrote:

\begin{quote}
\emph{I think that a particle must have a separate reality
independent of the measurements. That is an electron has spin,
location and so forth even when it is not being measured. I like to
think that the moon is there even if I am not looking at it.} A.
Einstein
\end{quote}

\noindent Actually, quantum mechanics does not make any claims about things that are not measured. So, we will also prefer to remain agnostic about non-observable features and refuse to use
them as a basis for building our theory.

\chapter{POINCAR\'E GROUP }

\label{ch:relativity}

\begin{quote}
\textit{There are more things in Heaven and on earth, dear Horacio,
than are dreamed of in your philosophy.}

\small
\hspace{1in} Hamlet
 \normalsize
\end{quote}

\vspace {0.5in}

In the preceding chapter we have learned that each physical system
can be described mathematically by a Hilbert space. Rays  in this space are in
one-to-one correspondence with (pure) states of the system.
Observables are described by Hermitian operators. This vague
description is not sufficient for a working theory. We are still
lacking precise classification of possible physical systems; we
still do not know which operators correspond to usual observables
like position, momentum, mass, energy, spin, etc. and how these
operators are related to each other; we still cannot tell how states
and observables evolve in time. Our theory is not complete.

It appears that many missing pieces mentioned above are supplied by
the \emph{principle of relativity} - which is one of the most
powerful ideas in physics. This principle has a very general
character. It works independent on what physical system, state or
observable is considered. Basically, this principle says that there is no preferred inertial reference frame (or observer or laboratory). All frames are equivalent if they are at rest or move uniformly without rotation or acceleration. Moreover, the principle of relativity recognizes certain (group)
properties of inertial transformations
between these frames or observers.  Our primary goal here is to establish that the group
of transformations between inertial observers
is the celebrated Poincar\'e group. In the rest of this book we will
have many opportunities to appreciate the fundamental importance of
this idea for relativistic physics.

One can notice that the principle of relativity discussed here is the same as the first postulate of Einstein's special relativity. In this book we will not need Einstein's second postulate, which claims the independence of the speed of light on the velocity of the source or observer. Actually, we will find out that by combining the first postulate, the Poincar\'e group idea, and laws of quantum mechanics we can obtain a complete working formalism of relativistic quantum theory. This will be done in chapter \ref{ch:QM-relativity}. We will also see in chapter \ref{ch:single} that the second postulate is redundant, because the speed of massless photons appears to be invariant anyway. Another distinctive feature of our relativistic approach is that we never assume the existence of the 4-dimensional Minkowski manifold, which unifies space and time. Time and position play very different roles in our theory. The significance of this idea will be discussed in chapter \ref{sc:obs-interact} in the second part of this book.

\section{Inertial observers}
\label{sc:inertial}

\subsection{Principle of relativity}
\label{ss:principle_of_relativity}

As has been said in Introduction, in this book we consider only
inertial laboratories. What is so special about them?
 The answer is that one can apply the powerful
 \emph{principle of relativity} \index{principle of relativity}
 to such laboratories.
The essence of this principle  was best explained by Galileo more
than 370 years ago \cite{Galileo}:

\begin{quotation}
\emph{Shut yourself up with some friend in the main cabin below
decks on some large ship and have with you there some flies,
butterflies and other small flying animals. Have a large bowl of
water with some fish in it; hang up a
 bottle that empties drop by drop into a wide vessel beneath it. With the ship
standing still, observe carefully how the little animals fly with
equal speed to all sides of the cabin. The fish swim indifferently
in all directions; the drops fall into the vessel beneath; and, in
throwing something to your friend, you need to throw it no more
 strongly in one direction than another, the distances being equal; jumping
with your feet together, you pass equal spaces in every direction.
When you have observed all of these things carefully (though there
is no doubt that when the ship is standing still everything must
happen this way), have the ship proceed with any speed you like, so
long as the motion is uniform and not fluctuating this way and that.
You will discover not the least change in all the effects named, nor
could you tell from any of them whether the ship was moving or
standing still. In jumping, you will pass on the floor the same
spaces as before, nor will you make larger jumps toward the stern
than towards the prow even though the ship is moving quite rapidly,
despite the fact that during the time that you are in the air the
floor under you will be going in a direction opposite to your jump.
In throwing something to your companion, you will need no more force
to get it to him whether he is in the direction of the
 bow or the stern, with yourself situated opposite. The droplets will fall as
before into the vessel beneath without dropping towards the stern,
although while the drops are in the air the ship runs many spans.
The fish in the water will swim towards the front of their bowl with
no more effort than toward the back and will go with equal ease to
bait placed anywhere around the edges of the bowl. Finally the
butterflies and flies will continue their flights indifferently
toward every side, nor will it ever happen that they are
concentrated toward the stern, as if tired out from keeping up with
the course of the ship, from which they will have been separated
during long intervals by keeping themselves in the air.}
\end{quotation}

\noindent These observations can be translated into a single statement
that all inertial laboratories
 cannot be
distinguished from the laboratory at rest by performing experiments
confined to those laboratories.  Any  experiment performed in one
laboratory,
  will yield exactly the same result as an identical
experiment in any other  laboratory. The results will be the same,
independent on how far apart the laboratories are and  what are
their relative orientations and velocities. Moreover, we may repeat
the same experiment at any time, tomorrow, or many years later,
still results will be the same. This allows us to formulate one of
the most important and deep postulates in physics

\begin{postulate} [the principle of relativity]
\index{principle of relativity} \textit{ In all inertial
laboratories, the laws of nature are the same: they do not change
with time, they do not depend on the position and orientation of the
laboratory in space and on its velocity. The laws of physics are
invariant against inertial transformations of laboratories.}
\label{postulateA}
\end{postulate}

\subsection{Inertial transformations}
\label{ss:inertial_transformations}

Our next goal is to study inertial transformations between
laboratories in more detail. To do this we do not need
to consider physical systems at all. It is sufficient to think about
a world inhabited only by laboratories. The only thing these laboratories
can do is to measure parameters\footnote{These parameters were explained on page \pageref{page:parameters} of the Introduction.} $\{ \vec{\phi}; \mathbf{v};
\mathbf{r}; t\}$ of their fellow laboratories. It appears that even in
this oversimplified world we can learn quite a few useful things
about properties of inertial laboratories and their relationships to
each other.

Let us first introduce a convenient labeling of inertial observers
and inertial transformations. We choose an arbitrary frame $O$ as
our reference observer, then other examples of observers  are

\begin{itemize}
\item[(i)] an observer $\{ \vec{0}; \mathbf{0}; \mathbf{0}; t_1 \} O$
 displaced in time by the
amount $t_1$;
\item[(ii)] an observer $\{\vec{0};  \mathbf{0}; \mathbf{r}_1; 0\} O$
 shifted in space by the
vector  $\mathbf{r}_1$;
\item[(iii)] an observer $\{ \vec{0}; \mathbf{v}_1; \mathbf{0}; 0 \} O $
moving
 with velocity $\mathbf{v}_1$;
\item[(iv)]
an observer $\{ \vec{\phi}_1; \mathbf{0}; \mathbf{0};  0\} O$
rotated by the vector $\vec{\phi}_1$.\footnote{The parameterization
of rotations by 3-vectors is discussed in Appendix
\ref{ss:parameterization}.}
\end{itemize}

\noindent  Suppose now that we have three different inertial
observers $O$, $O'$, and $O''$.
 There is an inertial transformation  $\{ \vec{\phi}_1;
\mathbf{v}_1; \mathbf{r}_1; t_1\}$ which connects $O$ and $O'$

\begin{eqnarray}
O' = \{ \vec{\phi}_1; \mathbf{v}_1; \mathbf{r}_1; t_1\} O
\label{eq:tran-1}
\end{eqnarray}

\noindent where parameters $\vec{\phi}_1$, $\mathbf{v}_1$,
$\mathbf{r}_1$, and $t_1$  are measured by the ruler and clock
belonging to the reference frame $O$ with respect to its basis
 vectors. Similarly,  there is an inertial
transformation that connects $O'$ and $O''$

\begin{eqnarray}
O'' = \{ \vec{\phi}_2; \mathbf{v}_2; \mathbf{r}_2; t_2\} O'
\label{eq:tran-2}
\end{eqnarray}

\noindent where parameters $\vec{\phi}_2$, $\mathbf{v}_2$,
$\mathbf{r}_2$, and $t_2$
 are defined with respect to the basis vectors, ruler and clock of
the observer $O'$. Finally, there is a transformation that connects
$O$ and $O''$

\begin{eqnarray}
O'' = \{ \vec{\phi}_3; \mathbf{v}_3; \mathbf{r}_3; t_3\} O
\label{eq:tran-3}
\end{eqnarray}

\noindent with all transformation parameters referring to $O$. We
can represent the transformation (\ref{eq:tran-3}) as a
\emph{composition}
 or \emph{product} of transformations  \index{composition of transformations}
(\ref{eq:tran-1}) and (\ref{eq:tran-2})

\begin{eqnarray}
\{ \vec{\phi}_3; \mathbf{v}_3; \mathbf{r}_3; t_3\} =  \{
\vec{\phi}_2; \mathbf{v}_2; \mathbf{r}_2; t_2\}  \{ \vec{\phi}_1;
\mathbf{v}_1; \mathbf{r}_1; t_1\} \label{eq:prod-inert}
\end{eqnarray}

\noindent Apparently, this product  has the property of
associativity.\footnote{see equation (\ref{eq:A.1})}
 Also, there exists a trivial (\emph{identity})
\index{identity transformation} transformation $\{ \vec{0};
\mathbf{0}; \mathbf{0}; 0\}$  that leaves all observers unchanged, and for each inertial transformation $ \{ \vec{\phi}; \mathbf{v};
\mathbf{r}; t\}$ there is an inverse transformation $ \{
\vec{\phi}; \mathbf{v}; \mathbf{r}; t\}^{-1}$ such that their
product is the identity transformation

\begin{eqnarray}
  \{ \vec{\phi}; \mathbf{v}; \mathbf{r}; t\}  \{ \vec{\phi};
\mathbf{v}; \mathbf{r}; t\}^{-1}  = \{ \vec{\phi}; \mathbf{v};
\mathbf{r}; t\}^{-1}  \{ \vec{\phi}; \mathbf{v}; \mathbf{r}; t\}  =
\{ \vec{0}; \mathbf{0}; \mathbf{0}; 0\}
 \label{eq:inv-inert}
\end{eqnarray}

\noindent In other words, the set of inertial transformations forms
a group (see Appendix \ref{ss:groups}). Moreover, since these transformations
smoothly depend on their parameters, this is a Lie group (see
Appendix \ref{ss:lie}).  The main goal of the
present chapter is to study the properties of this group in some
detail. In particular, we will need explicit formulas for the
composition and inversion laws.

First we notice that a general inertial transformation $\{
\vec{\phi}; \mathbf{v}; \mathbf{r}; t\}$ can be represented as a
product of basic transformations (i) - (iv). As these basic
transformations generally do not commute, we must agree on the
canonical order in this product. For our purposes the following
choice is convenient

\begin{eqnarray}
   \{ \vec{\phi}; \mathbf{v}; \mathbf{r};
t\}O = \{  \vec{\phi}; \mathbf{0};  \mathbf{0}; 0\} \{  \vec{0};
\mathbf{v}; \mathbf{0}; 0 \} \{ \vec{0}; \mathbf{0}; \mathbf{r}; 0\}
\{ \mathbf{0}; \mathbf{0}; \vec{0}; t \}O \label{eq:2.0a}
\end{eqnarray}

\noindent This means that in order to obtain observer $O' = \{
\vec{\phi}; \mathbf{v}; \mathbf{r}; t\}O $ we first shift observer
$O$ in time by the amount $t$,\footnote{Recall that observers considered in this book are \emph{instantaneous}, so the time flow is regarded as time translation of observers.} then shift the time-translated
observer by the vector $\mathbf{r}$, then give it velocity
$\mathbf{v}$, and finally rotate the obtained observer by the angle
$\vec{\phi}$.

\section{Galilei group}
\label{sc:galilei}

In this section we begin our study of the  group of inertial
transformations by considering a non-relativistic world in which
observers move with low speeds. This is a relatively easy task,
because in these derivations we can use our everyday experience and
``common sense.'' The relativistic group of transformations will be
approached in section \ref{sc:poincare} as a formal generalization
of the Galilei group derived here.

\subsection{Multiplication law of the Galilei group}
\label{ss:multiplication}

Let us first consider four examples of products (\ref{eq:prod-inert}) in
which $\{\vec{\phi}_1; \mathbf{v}_1; \mathbf{r}_1; t_1\} $ is a
general inertial transformation and $\{ \vec{\phi}_2; \mathbf{v}_2;
\mathbf{r}_2; t_2\}$ is one of the  basic transformations from the
list (i) - (iv). Applying a time translation to a general reference
frame $\{\vec{\phi}_1; \mathbf{v}_1; \mathbf{r}_1; t_1\} O$ will
change its time label and change its position in space according to
equation

\begin{eqnarray}
  \{ \vec{0}; \mathbf{0}; \mathbf{0};
t_2\}  \{ \vec{\phi}_1; \mathbf{v}_1; \mathbf{r}_1; t_1\}O = \{
\vec{\phi}_1; \mathbf{v}_1; \mathbf{r}_1 + \mathbf{v}_1 t_2; t_1 +
t_2\}O \label{eq:2.0b}
\end{eqnarray}

\noindent Space translations  affect the position

\begin{eqnarray}
  \{\vec{0}; \mathbf{0};  \mathbf{r}_2;
0\}  \{ \vec{\phi}_1; \mathbf{v}_1; \mathbf{r}_1; t_1\}O =  \{
\vec{\phi}_1; \mathbf{v}_1; \mathbf{r}_1 + \mathbf{r}_2; t_1\}O
\label{eq:2.0c}
\end{eqnarray}

\noindent Boosts  change the velocity \index{boost}

\begin{eqnarray}
  \{ \vec{0};  \mathbf{v}_2; \mathbf{0};
0\}  \{ \vec{\phi}_1; \mathbf{v}_1; \mathbf{r}_1; t_1\}O =  \{
\vec{\phi}_1; \mathbf{v}_1 + \mathbf{v}_2; \mathbf{r}_1; t_1\}O
\label{eq:2.0d}
\end{eqnarray}

\noindent Rotations  affect all vector parameters\footnote{For
definition of $3 \times 3$ rotation matrices $R_{\vec{\phi}}$ and function $\vec{\Phi}$
see Appendix \ref{ss:parameterization}.}

\begin{eqnarray}
  \{ \vec{\phi}_2; \mathbf{0}; \mathbf{0};
0\}  \{ \vec{\phi}_1; \mathbf{v}_1; \mathbf{r}_1; t_1\}O = \{
\vec{\Phi}(R_{\vec{\phi}_2}R_{\vec{\phi}_1});
R_{\vec{\phi}_2}\mathbf{v}_1; R_{\vec{\phi}_2}\mathbf{r}_1; t_1 \}O
\label{eq:2.0e}
\end{eqnarray}

\noindent Now we can calculate the product of two general inertial
transformations in (\ref{eq:prod-inert}) by using (\ref{eq:2.0a}) -
(\ref{eq:2.0e})\footnote{Note that sometimes the product of Galilei
transformations is written in other forms. See, for example,  section 3.2 in
\cite{Ballentine}, where the assumed
canonical order of factors was different from our  formula (\ref{eq:2.0a}). }

\begin{eqnarray}
&\mbox{ } &  \{ \vec{\phi}_2; \mathbf{v}_2; \mathbf{r}_2; t_2\}  \{
\vec{\phi}_1; \mathbf{v}_1;
\mathbf{r}_1; t_1\} \nonumber \\
&=& \{ \vec{\phi}_2;\mathbf{0}; \mathbf{0};  0\} \{ \vec{0};
\mathbf{v}_2; \mathbf{0};   0 \} \{ \vec{0}; \mathbf{0};
\mathbf{r}_2;   0\} \{ \vec{0}; \mathbf{0}; \mathbf{0}; t_2 \}
 \{ \vec{\phi}_1; \mathbf{v}_1;
\mathbf{r}_1; t_1\} \nonumber \\
&=& \{ \vec{\phi}_2;\mathbf{0}; \mathbf{0};  0\} \{ \vec{0};
\mathbf{v}_2; \mathbf{0};   0 \} \{ \vec{0}; \mathbf{0};
\mathbf{r}_2;   0\}
 \{ \vec{\phi}_1; \mathbf{v}_1;
\mathbf{r}_1 + \mathbf{v}_1 t_2; t_1 + t_2\} \nonumber \\
&=& \{ \vec{\phi}_2;\mathbf{0}; \mathbf{0};  0\} \{ \vec{0};
\mathbf{v}_2; \mathbf{0};   0 \}
 \{ \vec{\phi}_1; \mathbf{v}_1;
\mathbf{r}_1 + \mathbf{v}_1 t_2  + \mathbf{r}_2; t_1 + t_2\}
\nonumber
\\
&=& \{ \vec{\phi}_2;\mathbf{0}; \mathbf{0};  0\}
 \{ \vec{\phi}_1; \mathbf{v}_1 + \mathbf{v}_2;
\mathbf{r}_1 + \mathbf{v}_1 t_2  + \mathbf{r}_2; t_1 + t_2\}
\nonumber
\\
&=&
 \{\vec{\Phi}(R_{\vec{\phi}_2} R_{\vec{\phi}_1});
R_{\vec{\phi}_2}(\mathbf{v}_1 + \mathbf{v}_2);
R_{\vec{\phi}_2}(\mathbf{r}_1 + \mathbf{v}_1 t_2  + \mathbf{r}_2);
t_1 + t_2\} \label{eq:general-galilei}
\end{eqnarray}

\noindent  By direct substitution to equation (\ref{eq:inv-inert})
it is easy to check that the inverse of a general inertial
transformation $\{ \vec{\phi}; \mathbf{v}; \mathbf{r}; t\}$ is

\begin{eqnarray}
\{ \vec{\phi}; \mathbf{v}; \mathbf{r}; t\}^{-1}  =\{ -\vec{\phi}; -
\mathbf{v};
 -\mathbf{r}+
\mathbf{v} t ; -t\} \label{eq:inverse-galilei}
\end{eqnarray}

\noindent Equations (\ref{eq:general-galilei}) and
(\ref{eq:inverse-galilei}) are multiplication and inversion laws
which fully determine  the structure of the Lie group of inertial
transformations in non-relativistic physics. This group is called
the \emph{Galilei group}. \index{Galilei group}

\subsection{Lie algebra of the Galilei group}
\label{ss:lie-galilei}

In physical applications the Lie algebra of the group of inertial
transformations plays even greater role than the group itself.
According to our discussion in Appendix \ref{sc:lie-groups},
 we can obtain the basis $(\mathcal{H}, \vec{\mathcal{P}},
\vec{\mathcal{K}}, \vec{\mathcal{J}})$  in the
 Lie algebra of generators of the Galilei group by taking derivatives with respect to
parameters of one-parameter subgroups. For example, the generator of
time translations is

\begin{eqnarray*}
\mathcal{H} &=& \lim_{t \to 0}\frac{d}{dt} \{ \vec{0}; \mathbf{0};
\mathbf{0}; t\}
\end{eqnarray*}

\noindent For generators of space translations and boosts along the
$x$-axis we obtain

\begin{eqnarray*}
\mathcal{P}_x &=& \lim_{x \to 0} \frac{d}{dx} \{\vec{0};
\mathbf{0}; x,0,0 ;
0 \} \\
\mathcal{K}_x &=& \lim_{v \to 0} \frac{d}{dv} \{\vec{0}; v,0,0; \mathbf{0};
 0 \}
\end{eqnarray*}

\noindent The generator of rotations around the $x$-axis is

\begin{eqnarray*}
\mathcal{J}_x &=& \lim_{\phi \to 0} \frac{d}{d\phi} \{\phi, 0, 0;
\mathbf{0};
 \mathbf{0};
0 \}
\end{eqnarray*}

\noindent Similar formulas are valid  for $y$- and $z$-components.
According to (\ref{eq:A.28}) we can also express finite
transformations as exponents of generators

\begin{eqnarray}
\{ \vec{0}; \mathbf{0};  \mathbf{0}; t\} &=& e^{\mathcal{H}t}
\approx 1 +\mathcal{H}t
\label{eq:galilei_time} \\
\{\vec{0};  \mathbf{0}; \mathbf{r}; 0\} &=& e^{\vec{\mathcal{P}}
\mathbf{r}} \approx 1 + \vec{\mathcal{P}} \mathbf{r}
\label{eq:galilei_space}\\
 \{\vec{0};  \mathbf{v}; \mathbf{0}; 0\}
&=& e^{\vec{\mathcal{K}} \mathbf{v}} \approx 1 + \vec{\mathcal{K}}
\mathbf{v}
\label{eq:galilei_motion} \\
\{\vec{\phi}; \mathbf{0}; \mathbf{0}; 0 \} &=& e^{\vec{\mathcal{J}}
\vec{\phi}} \approx 1 + \vec{\mathcal{J}} \vec{\phi} \nonumber
\end{eqnarray}

\noindent Then each group element can be represented in its
canonical form (\ref{eq:2.0a}) as the following function of
parameters

\begin{eqnarray}
\{ \vec{\phi}; \mathbf{v};  \mathbf{r}; t \}& \equiv & \{
\vec{\phi}; \mathbf{0};  \mathbf{0}; 0\} \{  \vec{0};  \mathbf{v};
\mathbf{0}; 0 \} \{ \vec{0}; \mathbf{0}; \mathbf{r}; 0\}
\{ \mathbf{0}; \mathbf{0}; \vec{0};t \} \nonumber \\
&=& e^{\vec{\mathcal{J}} \vec{\phi}} e^{\vec{\mathcal{K}}
\mathbf{v}} e^{\vec{\mathcal{P}} \mathbf{r}}
 e^{\mathcal{H}t} \label{eq:galilei_general}
\end{eqnarray}

 Let us now
find the commutation relations between  generators, i.e., the
structure constants of the \emph{Galilei Lie algebra}.
\index{Galilei Lie algebra} Consider, for example, translations in
time and space. From equation (\ref{eq:general-galilei})  we have

\begin{eqnarray*}
\{\vec{0}; \mathbf{0}; \mathbf{0}; t\} \{\vec{0}; \mathbf{0}; x, 0,
0; 0\}
 = \{\vec{0}; \mathbf{0}; x, 0, 0;
0\} \{\vec{0}; \mathbf{0}; \mathbf{0}; t\}
\end{eqnarray*}

\noindent This implies

\begin{eqnarray*}
e^{ \mathcal{H}t}  e^{ \mathcal{P}_x x} &=& e^{ \mathcal{P}_x x}
e^{ \mathcal{H}t}  \\
1 &=&   e^{ \mathcal{P}_x x} e^{ \mathcal{H}t} e^{ -\mathcal{P}_x x}
e^{-\mathcal{H}t}
\end{eqnarray*}

\noindent Using equations (\ref{eq:galilei_time}) and
(\ref{eq:galilei_space}) for the exponents we can write to the first
order in $x$ and to the first order in $t$

\begin{eqnarray*}
1 &\approx& (1 + \mathcal{P}_x x)(1 +\mathcal{H}t) (1 -
\mathcal{P}_x x) (1
-\mathcal{H}t) \\
&\approx& 1 + \mathcal{P}_x \mathcal{H} xt - \mathcal{P}_x
\mathcal{H} xt - \mathcal{H}\mathcal{P}_x xt
 + \mathcal{P}_x\mathcal{H} xt \\
&=& 1  -  \mathcal{H}\mathcal{P}_x xt
 + \mathcal{P}_x\mathcal{H} xt \\
\end{eqnarray*}

\noindent hence

\begin{eqnarray*}
[\mathcal{P}_x, \mathcal{H}] \equiv \mathcal{P}_x \mathcal{H} -
 \mathcal{H} \mathcal{P}_x = 0
\end{eqnarray*}

\noindent So, generators of space and time translations have
vanishing Lie bracket. Similarly we obtain Lie brackets

\begin{eqnarray*}
[\mathcal{H}, \mathcal{P}_i] = [\mathcal{P}_i, \mathcal{P}_j] =
[\mathcal{K}_i, \mathcal{K}_j] = [\mathcal{K}_i, \mathcal{P}_j] = 0
\end{eqnarray*}

\noindent for any $i,j = x,y,z$ (or $i,j = 1,2,3$). The composition
of a time translation and a boost is more interesting since they do
not commute. We calculate from equation (\ref{eq:general-galilei})

\begin{eqnarray*}
e^{ \mathcal{K}_xv} e^{ \mathcal{H}t} e^{ -\mathcal{K}_xv} &=& \{
\vec{0}; v, 0, 0;\mathbf{0}, 0 \} \{\vec{0};
\mathbf{0}; \mathbf{0}; t \} \{ \vec{0}; -v, 0, 0;\mathbf{0}; 0 \}\\
&=& \{ \vec{0}; v, 0, 0;\mathbf{0}; 0 \} \{\vec{0};
-v, 0, 0;  -vt, 0 ,0; t \} \\
&=&
 \{\vec{0}; 0, 0, 0; -vt, 0 ,0;
 t \} \\
&=& e^{ \mathcal{H}t} e^{ -\mathcal{P}_xvt}
\end{eqnarray*}

\noindent Therefore, using equations (\ref{eq:galilei_time}),
(\ref{eq:galilei_motion}), and (\ref{eq:A.39}) we obtain

\begin{eqnarray*}
\mathcal{H}t + [\mathcal{K}_x, \mathcal{H}]vt &=& \mathcal{H}t -
\mathcal{P}_x vt \\
\ [\mathcal{K}_x, \mathcal{H}] &=& - \mathcal{P}_x
\end{eqnarray*}

\noindent Proceeding in a similar fashion for other pairs of
transformations we can obtain the full set of commutation relations for
the Lie algebra of the Galilei group.

\begin{eqnarray}
[\mathcal{J}_i, \mathcal{P}_j] &=& \sum_{k=1}^3 \epsilon_{ijk}
\mathcal{P}_k \label{eq:galilei_1}
\\
\mbox{ } [\mathcal{J}_i, \mathcal{J}_j] &=& \sum_{k=1}^3
\epsilon_{ijk} \mathcal{J}_k \label{eq:galilei_2}
\\
\mbox{ } [\mathcal{J}_i, \mathcal{K}_j] &=&  \sum_{k=1}^3
\epsilon_{ijk} \mathcal{K}_k \label{eq:galilei_3}
\\
\mbox{ } [\mathcal{J}_i,\mathcal{H}] &=&  0 \label{eq:galilei_4}
\\
\mbox{ }[\mathcal{P}_i, \mathcal{P}_j] &=&  [\mathcal{P}_i,
\mathcal{H}] = 0 \label{eq:galilei_5}
\\
\mbox{ }[\mathcal{K}_i, \mathcal{K}_j] &=& 0 \label{eq:galilei_6}
\\
\mbox{ }[\mathcal{K}_i, \mathcal{P}_j] &=& 0 \label{eq:galilei_7}
\\
\mbox{ } [\mathcal{K}_i, \mathcal{H}] &=&  -\mathcal{P}_i
\label{eq:galilei_8}
\end{eqnarray}

\noindent From these Lie brackets one can  identify several
important sub-algebras of the Galilei Lie algebra and, therefore,
subgroups of the Galilei group. In particular, there is an Abelian
subgroup of space and time translations (with generators
$\vec{\mathcal{P}}$ and $\mathcal{H}$, respectively), a subgroup of
rotations (with generators $\vec{\mathcal{J}}$), and an Abelian
subgroup of boosts (with generators $\vec{\mathcal{K}}$).

\subsection{Transformations of generators under rotations}
\label{ss:trans-rot}

Consider two reference frames $O$ and $O'$ connected to each other
by the group element $g$:

\begin{eqnarray*}
O' = g O
\end{eqnarray*}

\begin{figure}
\centering
\includegraphics {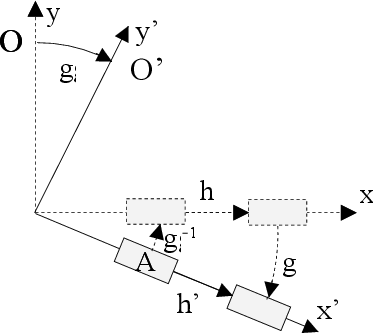} \caption{Connection
between similar transformations $h$ and $h'$ in different reference frames.  $g = \exp(J_z \phi)$ is a rotation
around the $z$-axis that is perpendicular to the page.}
\label{fig:2.1}
\end{figure}

\noindent Suppose that observer $O$ performs an (active) inertial
transformation with the group element $h$ (e.g., $h$ is a
translation along the $x$-axis). We want to find a transformation
$h'$ which is related to the observer $O'$ in the same way as $h$ is
related to $O$ (i.e., $h'$ is the translation along the $x'$-axis
belonging to the observer $O'$). As seen from the example in Fig.
\ref{fig:2.1}, the
transformation $h'$ of the object $A$ can be obtained by first going
from $O'$ to $O$, performing translation $h$ there, and then
returning back to the reference frame $O'$

\begin{eqnarray*}
h' = ghg^{-1}
\end{eqnarray*}

\noindent Similarly, if $\mathcal{A}$ is a generator of an inertial
transformation in the reference frame $O$, then

\begin{eqnarray}
\mathcal{A}' = g \mathcal{ A} g^{-1} \label{eq:generator-trans}
\end{eqnarray}

\noindent is ``the same'' generator in the reference frame $O' = g
O$.

Let us consider the effect of rotation around the $z$-axis on
generators of the Galilei group. We can write

\begin{eqnarray*}
\mathcal{A}_x' \equiv \mathcal{A}_x(\phi) &=& e^{\mathcal{J}_z
\phi}
\mathcal{A}_x e^{-\mathcal{J}_z  \phi} \\
\mathcal{A}_y' \equiv \mathcal{A}_y(\phi) &=& e^{\mathcal{J}_z
\phi}
\mathcal{A}_y e^{-\mathcal{J}_z  \phi} \\
\mathcal{A}_z' \equiv \mathcal{A}_z(\phi) &=& e^{\mathcal{J}_z \phi}
\mathcal{A}_z e^{-\mathcal{J}_z  \phi}
\end{eqnarray*}

\noindent where $\vec{\mathcal{A}}  $ is any of the generators
$\vec{\mathcal{P}}$, $\vec{\mathcal{J}}$ or $\vec{\mathcal{K}}$.
From Lie brackets (\ref{eq:galilei_1}) - (\ref{eq:galilei_3}) we
obtain

\begin{eqnarray}
\frac{\partial}{\partial \phi }\mathcal{A}_x(\phi) &=&
e^{\mathcal{J}_z  \phi}( \mathcal{J}_z \mathcal{A}_x - \mathcal{A}_x
\mathcal{J}_z) e^{-\mathcal{J}_z  \phi} = e^{\mathcal{J}_z
\phi}\mathcal{A}_y e^{-\mathcal{J}_z  \phi} =\mathcal{A}_y(\phi)
\label{eq:(2.16)} \\
\frac{\partial}{\partial \phi }\mathcal{A}_y(\phi) &=&
e^{\mathcal{J}_z  \phi}( \mathcal{J}_z \mathcal{A}_y - \mathcal{A}_y
\mathcal{J}_z) e^{-\mathcal{J}_z  \phi} = -e^{\mathcal{J}_z
\phi}\mathcal{A}_x e^{-\mathcal{J}_z  \phi} =-\mathcal{A}_x(\phi)
\nonumber \\
\frac{\partial}{\partial \phi }\mathcal{A}_z(\phi) &=&
e^{\mathcal{J}_z  \phi}( \mathcal{J}_z \mathcal{A}_z - \mathcal{A}_z
\mathcal{J}_z) e^{-\mathcal{J}_z  \phi} = 0 \label{eq:(2.16b)}
\end{eqnarray}

\noindent Taking a derivative of equation (\ref{eq:(2.16)}) by $\phi$ we
obtain a second order differential equation

\begin{eqnarray*}
\frac{\partial ^2}{\partial^2\phi }\mathcal{A}_x(\phi)  &=&
\frac{\partial}{\partial \phi }\mathcal{A}_y(\phi) = -
\mathcal{A}_x(\phi)
\end{eqnarray*}

\noindent with the general solution

\begin{eqnarray*}
\mathcal{A}_x(\phi) = \mathcal{B} \cos \phi + \mathcal{D} \sin \phi
\end{eqnarray*}

\noindent where $\mathcal{B}$ and $\mathcal{D}$ are arbitrary functions of generators. From the initial conditions  we obtain

\begin{eqnarray*}
 \mathcal{B} &=& \mathcal{A}_x(0) = \mathcal{A}_x \\
 \mathcal{D} &=& \frac{d}{d \phi}\mathcal{A}_x(\phi) \Bigl|_{\phi = 0} =
\mathcal{A}_y
\end{eqnarray*}

\noindent so that finally

\begin{eqnarray}
\mathcal{A}_x(\phi)&=& \mathcal{A}_x \cos \phi + \mathcal{A}_y \sin
\phi \label{eq:(2.17)}
\end{eqnarray}

\noindent Similar calculations show that

\begin{eqnarray}
\mathcal{A}_y(\phi)&=&  - \mathcal{A}_x \sin \phi + \mathcal{A}_y
\cos \phi
\label{eq:(2.18)} \\
\mathcal{A}_z(\phi)&=&  \mathcal{A}_z \label{eq:(2.18a)}
\end{eqnarray}

\noindent Comparing (\ref{eq:(2.17)}) - (\ref{eq:(2.18a)}) with equation
(\ref{eq:A.13}), we see that

\begin{eqnarray}
 \mathcal{A}'_i = e^{\mathcal{J}_z \phi}
\mathcal{A}_i e^{-\mathcal{J}_z \phi} = \sum_{j=1}^3(R_z)_{ij}
\mathcal{A}_j \label{eq:rotation}
\end{eqnarray}

\noindent where $R_z$ is the rotation matrix. As shown in equation
(\ref{eq:A.21}), we can also find the result of application of a general
rotation  $\{\vec{\phi}; \mathbf{0}; \mathbf{0}; 0  \}$ to
generators $\vec{\mathcal{A}}$

\begin{eqnarray*}
\vec{\mathcal{A}}' &=& e^{ \vec{\mathcal{J}}\vec{\phi}}
\vec{\mathcal{A}}  e^{ -\vec{\mathcal{J}}\vec{\phi}} \\
&=& \vec{\mathcal{A}} \cos \phi + \frac{\vec{\phi}}{\phi}
\left(\vec{\mathcal{A}} \cdot \frac{\vec{\phi}}{\phi}\right) (1 - \cos \phi) -
\left[\vec{\mathcal{A}} \times
\frac{\vec{\phi}}{\phi} \right] \sin \phi \\
& =& R_{\vec{\phi}} \vec{\mathcal{A}}
\end{eqnarray*}

\noindent This means that $\vec{\mathcal{P}}$, $\vec{\mathcal{J}}$, and $\vec{\mathcal{K}}$ are 3-vectors.\footnote{see Appendix
\ref{ss:scalars}}  The Lie bracket
(\ref{eq:galilei_4}) obviously means that $\mathcal{H}$ is a
3-scalar.

\begin{figure}
\centering
\includegraphics{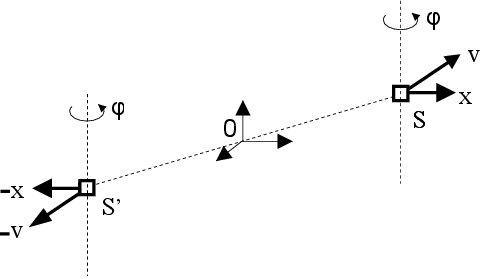}
\caption{Transformation of generators under space inversion.}
\label{fig:2.2}
\end{figure}

\subsection{Space  inversions}
\label{ss:inversions}

We will not consider physical consequences of discrete
transformations (inversion and time reversal) in this book. It is
physically impossible to prepare an exact mirror image or a
time-reversed image of a  laboratory, so the relativity postulate
has nothing to say about such transformations. Indeed, it has been
proven by experiment that these discrete symmetries are not exact.
Nevertheless, we will find it useful to know how generators behave
with respect to space inversions. Suppose we have a classical system
$S$ and its inversion image $S'$ (see Fig. \ref{fig:2.2}) with
respect to the origin 0. The question is: how the image $S'$ will
transform if we apply a certain inertial transformation to $S$?

Apparently, if we shift $S$ by vector $\mathbf{x}$, then $S'$ will
be shifted by $-\mathbf{x}$. This can be interpreted as the change
of sign of the  generator of translation $\vec{\mathcal{P}}$ under
inversion. The same with boost: the inverted image $S'$ acquires
velocity $-\mathbf{v}$ if the original was boosted by $\mathbf{v}$.
So, inversion changes the sign of the boost generator as well

\begin{eqnarray}
\vec{\mathcal{K}} \to -\vec{\mathcal{K}} \label{eq:boost-inversion}
\end{eqnarray}

\noindent Vectors, such as $\vec{\mathcal{P}}$ and
$\vec{\mathcal{K}}$, changing their sign after inversion are called
\emph{true vectors}. \index{true vector} However, the generator of
rotation $\vec{\mathcal{J}}$ is not a true vector. Indeed, if we
rotate $S$ by angle $\vec{\phi}$, then the image $S'$ is also
rotated by the same angle (see Fig. \ref{fig:2.2}). So,
$\vec{\mathcal{J}}$ does not change the sign after inversion. Such
vectors are called \emph{pseudovectors}. \index{pseudovector}
Similarly we can introduce the notions of true scalars/pseudoscalars
and true tensors/pseudotensors. It is conventional to define their
properties in a way opposite to those of true vectors/pseudovectors.
In particular, true scalars and true tensors (of rank 2) do not
change their signs after inversion. For example, $\mathcal{H}$ is a
true scalar. Pseudoscalars and rank-2 pseudotensors  do change
their signs after inversion. \index{pseudoscalar}
\index{pseudotensor}

\section{Poincar\'e group}
\label{sc:poincare}

It appears that the Galilei group described above is valid only for
observers moving with low speeds. In the general case a different
multiplication law should be used, and the group of inertial
transformations is, in fact, the Poincar\'e group (also known as the
\emph{inhomogeneous Lorentz group}). \index{inhomogeneous Lorentz
group} This is a very important lesson following from the theory of relativity
developed in the beginning of the 20th century by Einstein and Poincar\'e.

Derivation of the relativistic group of inertial transformations is
a difficult task, because we lack the experience of dealing with
fast-moving objects in our everyday life. So, we will use  more
formal mathematical arguments instead. In this section we will find
that there is almost a unique way to obtain the Lie algebra of
the Poincar\'e group by generalizing the commutation relations of
the Galilei Lie algebra (\ref{eq:galilei_1}) - (\ref{eq:galilei_8}),
so that they remain compatible with some simple physical
requirements.

\subsection{Lie algebra of the Poincar\'e group}
\label{sc:poincare-lie}

 We can be
confident about the validity of Galilei Lie brackets involving
generators of space-time translations and rotations, because
properties of these transformations have been verified in everyday
life and in physical experiments over a wide range of involved
parameters (distances, times, and angles).
 The situation  with respect to
boosts is quite different. Normally, we do not experience high
speeds in our life and we lack any physical intuition that was so
helpful in deriving the Galilei Lie algebra.
 Therefore the arguments that lead us to
the Lie brackets (\ref{eq:galilei_6}) - (\ref{eq:galilei_8}) involving boost generators may be not exact, and these formulas may be just
approximations that can be tolerated only for low-speed observers.
So, we will base our derivation of the relativistic group of
inertial transformations on the following ideas.

\begin{itemize}
\item [(I)] Just as in the non-relativistic world,
the set of inertial transformations should remain a 10-parameter Lie
group. However, Lie brackets in the exact (Poincar\'e) Lie algebra
are expected to be
 different from the Galilei Lie brackets
(\ref{eq:galilei_1}) - (\ref{eq:galilei_8}).
\item[(II)] The Galilei group does a good job in
describing the low-speed transformations, and the speed of light $c$
is a natural measure of speed.  Therefore we may guess that the
correct Lie brackets should include $c$ as a parameter, and they
must tend to the Galilei Lie brackets in the limit $c \to
\infty$.\footnote{Note that here we do not assume that $c$ is a
limiting speed or that the speed of light is invariant. These facts
will come out as a result of application of our approach to massive and massless particles in
chapter \ref{ch:single}.}
\item [(III)] We will assume that only
Lie brackets involving boosts may be subject to revision.
\item [(IV)] We will further
assume that relativistic generators of boosts $\vec{\mathcal{K}}$
still form components of a true vector,  so equations
(\ref{eq:galilei_3}) and (\ref{eq:boost-inversion})  remain valid.
\end{itemize}

\noindent Summarizing requirements (I) - (IV),  we can write the
following relativistic generalizations for the Lie brackets
(\ref{eq:galilei_6}) - (\ref{eq:galilei_8})

\begin{eqnarray}
[\mathcal{K}_i, \mathcal{P}_j] &=& \mathcal{U}_{ij} \label{eq:(2.21a)} \\
\mbox{ } [\mathcal{K}_i,
\mathcal{K}_j] &=& \mathcal{T}_{ij} \nonumber \\
\mbox{ } [\mathcal{K}_i, \mathcal{H}] &=&  -\mathcal{P}_i +
\mathcal{V}_{i} \label{eq:(2.21)}
\end{eqnarray}

\noindent where  $\mathcal{T}_{ij}$, $\mathcal{U}_{ij}$, and
$\mathcal{V}_{ij}$ are  some yet unknown linear combinations of
generators. The coefficients of these linear combinations
 must be selected in such a way
 that all Lie algebra properties\footnote{in particular, the Jacobi identity
(\ref{eq:jacobi})} are preserved. Let us try to satisfy these
conditions step by step.

First note that the Lie bracket $[\mathcal{K}_i, \mathcal{P}_j]$ is
a 3-tensor. Indeed, using equation (\ref{eq:rotation}) we obtain the
tensor transformation law (\ref{eq:tensor-tr})

\begin{eqnarray*}
e^{\vec{\mathcal{J}} \vec{\phi}} [\mathcal{K}_i,
\mathcal{P}_j]e^{-\vec{\mathcal{J}} \vec{\phi}}
 &=& \left[\sum_{k=1}^3 R_{ik}
(\vec{\phi})\mathcal{K}_k,
\sum_{k=1}^3 R_{jl} (\vec{\phi})\mathcal{P}_l \right] \\
&=& \sum_{kl=1}^3R_{ik}(\vec{\phi}) R_{jl} (\vec{\phi})
[\mathcal{K}_k, \mathcal{P}_l]
\end{eqnarray*}

\noindent  Since both $\vec{\mathcal{K}}$ and $\vec{\mathcal{P}}$
change their signs upon inversion,
 this a true tensor.
Therefore $\mathcal{U}_{ij}$ must be a true tensor as well. This
tensor should be constructed as a linear function of generators
among which we have a true scalar $\mathcal{H}$, a pseudovector
$\vec{\mathcal{J}}$, and two true vectors  $\vec{\mathcal{P}}$ and
$\vec{\mathcal{K}}$.
 According to our discussion in Appendix \ref{ss:invariant},
the only  way to make a true tensor from these ingredients
  is by using formulas in the first and third rows in table \ref{table:C.1}.
Therefore, the most general expression for the Lie bracket
(\ref{eq:(2.21a)}) is

\begin{eqnarray*}
[\mathcal{K}_i, \mathcal{P}_j] = -\beta \mathcal{H} \delta_{ij} +
\gamma \sum_{k=1}^3 \epsilon_{ijk}\mathcal{J}_k
\end{eqnarray*}

\noindent where $\beta$ and $\gamma$ are yet unspecified real
constants.

Similar arguments suggest that  $\mathcal{T}_{ij}$ is also a true
tensor. Due to the relationship

\begin{eqnarray*}
 [\mathcal{K}_i, \mathcal{K}_j]= - [\mathcal{K}_j,\mathcal{K}_i]
\end{eqnarray*}

\noindent this tensor must be antisymmetric with respect to indices $i$ and $j$. This excludes the term
proportional to $\delta_{ij}$, hence

\begin{eqnarray*}
[\mathcal{K}_i, \mathcal{K}_j] =  \alpha \sum_{k=1}^3
\epsilon_{ijk}\mathcal{J}_k,
\end{eqnarray*}

\noindent where $\alpha$ is, again, a yet undefined constant.

The quantity $ \mathcal{V}_{i}$ in equation (\ref{eq:(2.21)}) must be a
true vector, so, the most general form of the Lie bracket
(\ref{eq:(2.21)}) is

\begin{eqnarray*}
 [\mathcal{K}_i, \mathcal{H}] &=&  -(1 + \sigma )\mathcal{P}_i + \kappa
\mathcal{K}_i.
\end{eqnarray*}

So, we have reduced the task of generalization of Galilei Lie
brackets to finding just five real parameters $\alpha, \beta,
\gamma, \kappa$, and $\sigma$.  To proceed further, let us first use the
following Jacobi identity

\begin{eqnarray*}
0 &=&  [\mathcal{P}_x, [\mathcal{K}_x,\mathcal{H}]] +
[\mathcal{K}_x, [\mathcal{H}, \mathcal{P}_x]] + [\mathcal{H},
[\mathcal{P}_x,\mathcal{K}_x]] \\
&=&  \kappa  [\mathcal{P}_x, \mathcal{K}_x] \\
&=&
  \beta \kappa \mathcal{H}
\end{eqnarray*}

\noindent which implies

\begin{eqnarray}
 \beta \kappa = 0
\label{eq:(2.22)}
\end{eqnarray}

\noindent Similarly,

\begin{eqnarray*}
0 &=&  [\mathcal{K}_x, [\mathcal{K}_y,\mathcal{P}_y]]
+[\mathcal{K}_y, [\mathcal{P}_y, \mathcal{K}_x]] + [\mathcal{P}_y,
[\mathcal{K}_x,\mathcal{K}_y]] \\
&=& -\beta [\mathcal{K}_x,\mathcal{H}] -  \gamma [\mathcal{K}_y,
\mathcal{J}_z] + \alpha [\mathcal{P}_y, \mathcal{J}_z] \\
&=&   \beta (1 + \sigma) \mathcal{P}_x - \beta \kappa \mathcal{K}_x
-
\gamma \mathcal{K}_x + \alpha \mathcal{P}_x \\
&=& (\alpha + \beta + \beta \sigma) \mathcal{P}_x -
( \beta \kappa + \gamma) \mathcal{K}_x \\
&=& (\alpha + \beta + \beta \sigma) \mathcal{P}_x -
 \gamma \mathcal{K}_x
\end{eqnarray*}

\noindent implies

\begin{eqnarray}
\alpha &=& - \beta(1 +  \sigma)
\label{eq:(2.23)} \\
\gamma &=& 0 \label{eq:(2.24)}
\end{eqnarray}

\noindent The system of equations (\ref{eq:(2.22)}) -
(\ref{eq:(2.23)}) has two possible solutions  (in both cases
$\sigma$ remains undefined)

\begin{itemize}
\item[(i)] If $\beta \neq 0$, then
  $\alpha = - \beta(1 +  \sigma)$ and  $\kappa = 0$.
\item[(ii)]  If $\beta = 0$, then  $\alpha = 0$ and $\kappa$ is
arbitrary.
\end{itemize}

\noindent From the condition (II)   we know that parameters $\alpha,
\beta, \sigma, \kappa$ must depend on $c$ and tend to zero as $c \to
\infty$

\begin{eqnarray}
\lim_{c \to \infty} \kappa = \lim_{c \to \infty} \sigma = \lim_{c
\to \infty} \alpha = \lim_{c \to \infty} \beta = 0 \label{eq:(2.25)}
\end{eqnarray}

Additional insight into the values of these parameters may be
obtained by examining their dimensions. To keep the arguments of
exponents in (\ref{eq:galilei_general})  dimensionless we must
assume the following dimensions \index{dimension} (denoted by angle brackets) of
the generators

\begin{eqnarray*}
<\mathcal{H}> &=& <time>^{-1} \\
<\mathcal{P}> &=& <distance>^{-1} \\
<\mathcal{K}> &=& <speed>^{-1} \\
<\mathcal{J}> &=& <angle>^{-1} = dimensionless
\end{eqnarray*}

\noindent It then follows that

\begin{eqnarray*}
<\alpha> &=&  \frac{<\mathcal{K}>^2}{<\mathcal{J}>} =
<speed>^{-2} \\
<\beta> &=& \frac{<\mathcal{K}><\mathcal{P}>}{<\mathcal{H}>} =
<speed>^{-2} \\
<\kappa> &=& <\mathcal{H}> = <time>^{-1} \\
<\sigma> &=& dimensionless
\end{eqnarray*}

\noindent and we can satisfy condition (\ref{eq:(2.25)}) only by
setting  $\kappa = \sigma = 0$ (i.e., the choice (i) above) and
assuming
 $\beta = -\alpha   \propto c^{-2}$.
This approach does not specify the coefficient of proportionality
between $\beta$ (and $-\alpha$) and $c^{-2}$. To be in agreement
with experimental data we must choose this coefficient equal to 1.

\begin{eqnarray*}
\beta = -\alpha  = \frac{1}{c^2}
\end{eqnarray*}

\noindent Then the resulting Lie brackets are

\begin{eqnarray}
[\mathcal{J}_i, \mathcal{P}_j] &=& \sum_{k=1}^3 \epsilon_{ijk}
\mathcal{P}_k \label{eq:poincare_1}
\\
\mbox{ }[\mathcal{J}_i, \mathcal{J}_j] &=& \sum_{k=1}^3
\epsilon_{ijk} \mathcal{J}_k \label{eq:poincare_2}
\\
\mbox{ }[\mathcal{J}_i, \mathcal{K}_j] &=& \sum_{k=1}^3
\epsilon_{ijk} \mathcal{K}_k \label{eq:poincare_3}
\\
\mbox{ } [\mathcal{J}_i,\mathcal{H}]  &=& 0 \label{eq:poincare_4}
\\
\mbox{ } [\mathcal{P}_i, \mathcal{P}_j]  &=&  [\mathcal{P}_i,
\mathcal{H}] = 0 \label{eq:poincare_5}
\\
\mbox{ } [\mathcal{K}_i, \mathcal{K}_j] &=&  -\frac{1}{c^2}
\sum_{k=1}^3 \epsilon_{ijk}\mathcal{J}_k \label{eq:poincare_6}
\\
\mbox{ } [\mathcal{K}_i, \mathcal{P}_j] &=& -\frac{1}{c^2}
\mathcal{H} \delta_{ij} \label{eq:poincare_7}
\\
\mbox{ } [\mathcal{K}_i, \mathcal{H}] &=& -\mathcal{P}_i
\label{eq:poincare_8}
\end{eqnarray}

\noindent This set of Lie brackets is called the
 \emph{Poincar\'e Lie algebra} \index{Poincar\'e Lie algebra} and it differs from the Galilei algebra
(\ref{eq:galilei_1}) - (\ref{eq:galilei_8}) only by small terms on
the right hand sides of Lie brackets (\ref{eq:poincare_6}) and
(\ref{eq:poincare_7}).  The general element of the corresponding
\emph{Poincar\'e group} \index{Poincar\'e group} has the
form\footnote{Note that here we adhere to the conventional order of
basic transformations adopted in (\ref{eq:2.0a}); from right to
left: time translation $\to$ space translation $\to$ boost $\to$ rotation.}

\begin{eqnarray}
e^{\vec{\mathcal{J}} \vec{\phi}}
 e^{\vec{\mathcal{K}} c \vec{\theta}}
e^{\vec{\mathcal{P}} \mathbf{x}} e^{\mathcal{H}t}
\label{eq:poincare_exp}
\end{eqnarray}

\noindent In equation (\ref{eq:poincare_exp}) we denoted the parameter of
boost by $c \vec{\theta}$, where  $\theta = |\vec{\theta}|$ is a
dimensionless quantity  called \emph{rapidity}.  \index{rapidity} Its relationship to the
velocity of boost $\mathbf{v}$ is

\begin{eqnarray*}
\mathbf{v}( \vec{\theta}) &=&\frac{\vec{\theta}}{\theta} c  \tanh
\theta \\
\cosh \theta &=& (1-v^2/c^2)^{-1/2}
\end{eqnarray*}

\noindent The reason
for introducing this new quantity is that rapidities of successive
boosts in the same direction are additive, while velocities are not.\footnote{ see
equation (\ref{eq:6.4})}

In spite of their simplicity, equations (\ref{eq:poincare_1}) -
(\ref{eq:poincare_8}) are among the most important  equations in
physics, and they have such an abundance of experimental
confirmations that one cannot  doubt their validity. We therefore
accept that the Poincar\'e group is the true mathematical expression
of relationships between different inertial laboratories.

\begin{postulate} [the Poincar\'e group]
  Transformations
between inertial laboratories form the Poincar\'e group.
\label{postulateH}
\end{postulate}

\noindent Even a brief comparison of the Poincar\'e
(\ref{eq:poincare_1}) - (\ref{eq:poincare_8}) and Galilei
(\ref{eq:galilei_1}) - (\ref{eq:galilei_8}) Lie brackets reveals a
number of important new features in the relativistic theory.
 For example, due to the Lie bracket (\ref{eq:poincare_6}), boosts no longer form a
subgroup.
 However,
boosts together with  rotations do form a 6-dimensional subgroup of
the Poincar\'e group which  is called the \emph{Lorentz group}.
\index{Lorentz group}

\subsection{Transformations of translation generators under boosts}
\label{ss:4-vectors}

Poincar\'e Lie brackets allow us to  derive transformation
properties
 of generators $\vec{\mathcal{P}}$ and $\mathcal{H}$ with respect
to boosts. Using Equation (\ref{eq:generator-trans}) and Lie brackets
(\ref{eq:poincare_7}) - (\ref{eq:poincare_8}) we  find that if
$\mathcal{P}_x$ and $\mathcal{H}$ are generators in the reference
frame at rest $O$, then their counterparts $\mathcal{P}_x(\theta)$
and $\mathcal{H}(\theta)$ in the reference frame $O'$ moving along
the $x$-axis are

\begin{eqnarray*}
\mathcal{H}(\theta) &=& e^{\mathcal{K}_x c \theta} \mathcal{H}
e^{-\mathcal{K}_x c \theta} \\
\mathcal{P}_x(\theta) &=& e^{\mathcal{K}_x c \theta} \mathcal{P}_x
e^{-\mathcal{K}_x c \theta}
\end{eqnarray*}

\noindent Taking derivatives of these equations with respect to the
parameter $\theta$

\begin{eqnarray}
\frac{\partial}{\partial \theta }\mathcal{H}(\theta) &=& c
e^{\mathcal{K}_x c \theta}( \mathcal{K}_x\mathcal{H} - \mathcal{H}
\mathcal{K}_x) e^{-\mathcal{K}_x c \theta} \nonumber \\
&=& -c  e^{\mathcal{K}_x c \theta}\mathcal{P}_x
e^{-\mathcal{K}_x c \theta} = -c \mathcal{P}_x(\theta) \nonumber \\
\frac{\partial}{\partial \theta }\mathcal{P}_x(\theta) &=& c
e^{\mathcal{K}_x c\theta}( \mathcal{K}_x\mathcal{P}_x -
\mathcal{P}_x \mathcal{K}_x) e^{-\mathcal{K}_x c \theta} \nonumber \\
 &=& -\frac{1}{c} e^{\mathcal{K}_x c \theta}\mathcal{H}
 e^{-\mathcal{K}_x c \theta} = -\frac{1}{c}\mathcal{H}(\theta) \label{eq:(2.35)}
\end{eqnarray}

\noindent and taking a derivative of equation (\ref{eq:(2.35)}) again, we
obtain a differential equation

\begin{eqnarray*}
\frac{\partial ^2}{\partial^2\theta }\mathcal{P}_x(\theta) &=& -
\frac{1}{c} \frac{\partial}{\partial \theta }\mathcal{H}(\theta) =
\mathcal{P}_x(\theta)
\end{eqnarray*}

\noindent with the general solution

\begin{eqnarray*}
\mathcal{P}_x(\theta) = \mathcal{A} \cosh \theta + \mathcal{B} \sinh
\theta
\end{eqnarray*}

\noindent From the initial conditions
 we obtain

\begin{eqnarray*}
 \mathcal{A} &=& \mathcal{P}_x(0) = \mathcal{P}_x \\
 \mathcal{B} &=& \frac{\partial}{\partial \theta }
\mathcal{P}_x(\theta) \Bigl|_{\theta = 0} = -\frac{1}{c}\mathcal{H}
\end{eqnarray*}

\noindent and finally

\begin{eqnarray*}
 \mathcal{P}_x(\theta) &=& \mathcal{P}_x \cosh \theta  - \frac{\mathcal{H}}{c}
\sinh \theta
\end{eqnarray*}

\noindent Similar calculation shows that

\begin{eqnarray}
 \mathcal{H} (\theta) &=& \mathcal{H} \cosh \theta - c \mathcal{P}_x \sinh
\theta \label{eq:h-theta} \\
\mathcal{P}_y (\theta) &=& \mathcal{P}_y \nonumber \\
\mathcal{P}_z (\theta) &=& \mathcal{P}_z \nonumber
\end{eqnarray}

Similar to our discussion of rotations in subsection
\ref{ss:parameterization}, we can find the transformation of
$\vec{\mathcal{P}}$ and $\mathcal{H}$ corresponding to a general
boost vector $\vec{\theta}$ in the coordinate-independent form.
First we decompose $\vec{\mathcal{P}}$ into sum of two vectors
$\vec{\mathcal{P}} = \vec{\mathcal{P}}_{\parallel} +
\vec{\mathcal{P}}_{\perp}$. The vector
$\vec{\mathcal{P}}_{\parallel} = (\vec{\mathcal{P}} \cdot
\frac{\vec{\theta}}{\theta}) \frac{\vec{\theta}}{\theta}$  is
parallel to the direction of the boost,
 while vector  $\vec{\mathcal{P}}_{\perp} = \vec{\mathcal{P}} -
\vec{\mathcal{P}}_{\parallel}$ is perpendicular to that direction. The perpendicular part $\vec{\mathcal{P}}_{\perp}$
remains unchanged under the boost, while
$\vec{\mathcal{P}}_{\parallel}$ transforms according to $\exp(
\vec{\mathcal{K}} c \vec{\theta}) \vec{\mathcal{P}}_{\parallel}
\exp(- \vec{\mathcal{K}} c \vec{\theta}) =
\vec{\mathcal{P}}_{\parallel} \cosh \theta - c^{-1}\mathcal{H} \sinh
\theta \frac{\vec{\theta}}{\theta}$. Therefore

\begin{eqnarray}
\vec{\mathcal{P}}' = e^{ \vec{\mathcal{K}} c \vec{\theta}}
\vec{\mathcal{P}}  e^{- \vec{\mathcal{K}} c \vec{\theta}} &=&
\vec{\mathcal{P}} + \frac{\vec{\theta}}{\theta} \left[\left(\vec{\mathcal{P}}
\cdot \frac{\vec{\theta}}{\theta}\right) (\cosh \theta - 1) - \frac{1}{c}
\mathcal{H} \sinh \theta\right]
\label{eq:boost-momentum}\\
\mathcal{H}' = e^{ \vec{\mathcal{K}} c \vec{\theta}} \mathcal{H}
e^{- \vec{\mathcal{K}} c \vec{\theta}} &=&
 \mathcal{H} \cosh \theta - c \left(\vec{\mathcal{P}} \cdot
\frac{\vec{\theta}}{\theta}\right) \sinh \theta \label{eq:boost-energy}
\end{eqnarray}

\noindent It is clear from (\ref{eq:boost-momentum}) and
(\ref{eq:boost-energy}) that boosts perform  linear transformations
of components $c\vec{\mathcal{P}}$ and $\mathcal{H}$. These
transformations can be represented in a matrix form if four
generators
 $( \mathcal{H}, c\vec{\mathcal{P}})$ are arranged in  a
column 4-vector
\index{4-vector}

\begin{eqnarray*}
  \left[ \begin{array}{c}
   \mathcal{H}'  \\
   c\mathcal{P}_x' \\
   c\mathcal{P}_y' \\
   c\mathcal{P}_z'
\end{array} \right] = B(\vec{\theta})  \left[ \begin{array}{c}
   \mathcal{H} \\
   c\mathcal{P}_x \\
   c\mathcal{P}_y \\
   c\mathcal{P}_z
\end{array} \right].
\end{eqnarray*}

\noindent Explicit form of the matrix $B(\vec{\theta})$ can be found in equation (\ref{eq:boost-matrix}).

\chapter{QUANTUM MECHANICS AND RELATIVITY}
\label{ch:QM-relativity}

\begin{quote}
 \emph{I am ashamed to tell you to how many figures I carried these computations, having no other business at the time.}

\small
\hspace{1in} Isaac Newton
\normalsize
\end{quote}

\vspace {0.5in}

\noindent  Two preceding chapters discussed the
ideas of quantum mechanics and relativity separately. Now is the
time to unify them in one theory. The major contribution to such an
unification was made by Wigner who formulated and proved the famous
Wigner's theorem and developed the theory of unitary representations
of the Poincar\'e group in Hilbert spaces. This theory is the mathematical foundation of the entire relativistic quantum approach presented in this book.  Its discusion will occupy the present chapter as well as the two following chapters \ref{ch:operators} and \ref{ch:single}.

\section{Inertial transformations in quantum mechanics}
\label{sc:inertial-tr}

The relativity Postulate \ref{postulateA}  tells us that any
inertial laboratory $L$ is physically equivalent to any other
laboratory $L' = gL$ obtained from $L$ by applying an inertial
transformation $g$. This means that for identically arranged
experiments in these two laboratories the corresponding probability
measures $(\phi|X)$ are the same. As shown in Fig. \ref{fig:3.4},
laboratories are composed of two major parts: the preparation device
 $P$ and the observer $O$.  The inertial
transformation $g$ of the laboratory results in changes of both these part. The change of the preparation
device can be interpreted as a change of the state of the system. We
can formally denote this change by $\phi \to g \phi$. The change of
the observer (or measuring apparatus)
can be viewed as a change of the experimental proposition $X \to
gX$. Then, the mathematical expression of the relativity principle
is that for any $g$, $\phi$, and $X$

\begin{eqnarray}
(g \phi | g X) = (\phi | X) \label{eq:5.1}
\end{eqnarray}

\noindent In the rest of this chapter (and in chapters
\ref{ch:operators} -- \ref{ch:interaction}) we will develop a
mathematical formalism for representing transformations $g \phi$ and
$g X$ in the Hilbert space. This is the formalism of unitary
representations of the Poincar\'e group, which is a cornerstone of
any relativistic approach in quantum physics.

\subsection{Wigner's theorem}
\label{ss:wigner}

Let us first focus on inertial transformations of propositions $X
\to gX$.\footnote{We will turn to transformations of states $\phi
\to g \phi$ in the next subsection.}  The
experimental propositions attributed to the observer $O$ form a
propositional lattice $\mathcal{L}(\mathcal{H})$ which is realized
as a set of closed subspaces in the  Hilbert space $\mathcal{H}$.
Observer $O' = gO$  also represents her propositions as subspaces in the
same Hilbert space $\mathcal{H}$. As these two observers are
equivalent, we may expect that their propositional systems have
exactly the same mathematical structures, i.e., they are isomorphic.
This means that there exists a one-to-one mapping

\begin{eqnarray*}
K_g : \mathcal{L}(\mathcal{H})  \to \mathcal{L}(\mathcal{H})
\end{eqnarray*}

\noindent that  connects propositions of the observer $O$ with
propositions of the observer $O'$, such that all lattice relations
between propositions remain unchanged. In particular, we will
require that
 $K_g$ transforms atoms to atoms; $K_g$  maps minimal and
 maximal propositions of $O$ to the minimal and maximal propositions
 of $O'$, respectively

\begin{eqnarray}
K_g(\mathcal{I}) &=& \mathcal{I} \label{eq:5.2} \\
K_g(\emptyset) &=& \emptyset
\end{eqnarray}

\noindent  and for any $X, Y \in \mathcal{L}(\mathcal{H})$

\begin{eqnarray}
K_g(X \vee Y) &=& K_g(X) \vee K_g(Y)   \\
K_g(X \wedge Y) &=& K_g(X) \wedge K_g(Y)   \\
K_g(X^{\perp})   &=& K_g(X)^{\perp} \label{eq:5.6}
\end{eqnarray}

As discussed in subsection \ref{ss:qm-logic}, working with
propositions is rather inconvenient. It would be better to translate
conditions (\ref{eq:5.2}) - (\ref{eq:5.6}) into the language of
vectors in the Hilbert space. In other words, we would like to find
a vector-to-vector transformation $k_g : \mathcal{H} \to
\mathcal{H}$ which \emph{generates} the subspace-to-subspace
transformation $K_g$. More precisely, we demand that for each
subspace $X$, if $K_g(X) = Y$, then the generator $k_g$ maps all
vectors in $X$  into vectors in $Y$, so that $Sp(k_g(x)) = Y$,
where $x$ runs through all vectors in $X$.

The problem with finding generators $k_g$ is that  there are just
too many of them. For example, if  a ray $p$ goes to the ray
$K_g(p)$, then the generator $k_g$ must  map each  vector $|x
\rangle \in p$ somewhere inside $K_g(p)$, but the exact value of
$k_g |x \rangle $ remains undetermined. Actually, we can multiply
each image vector $k_g |x \rangle $ by an arbitrary nonzero factor
$\eta(|x \rangle )$ and still have a valid generator. Factors
$\eta(|x \rangle )$ can be chosen independently for each $|x \rangle
\in \mathcal{H}$. This freedom is very inconvenient from the
mathematical point of view.

This problem was solved by the
 celebrated Wigner's  theorem, \cite{Wigner} which states that  we can always
select factors  $\eta(|x \rangle )$ in such a way that the
vector-to-vector mapping $\eta(|x \rangle )k_g $ becomes either
unitary (linear) or antiunitary (antilinear).\footnote{See Appendix
\ref{ss:functions} for definitions of antilinear and antiunitary
operators.} \index{antiunitary operator}

\bigskip
\label{wigner-theo}
\begin{theorem} [Wigner]
\index{Wigner's theorem}  For any isomorphic mapping  $K_g$ of a
propositional lattice $\mathcal{L}(\mathcal{H})$ onto itself, one
can find either unitary or antiunitary transformation $k_g$ of
vectors in the Hilbert space $\mathcal{H}$, which  generates $K_g$. This transformation is defined uniquely up to an unimodular factor.
For a given $K_g$ only one of these two possibilities (unitary or antiunitary) is
realized.
\end{theorem}

\bigskip

\noindent In this formulation, Wigner's theorem has been proven in
ref. \cite{Uhlhorn} (see also \cite{AertsI}). The significance of
this theorem comes from the fact that there is a powerful
mathematical apparatus for working with unitary and antiunitary
transformations, so that their properties (and, thus, properties of
subspace transformations $K_g$) can be studied in great detail by familiar techniques of linear algebra in Hilbert spaces.

From our study of inertial transformations in chapter
\ref{ch:relativity}, we know that there is always a continuous path
from the identity transformation $e = \{\vec{0}, \mathbf{0},
\mathbf{0}, 0\}$ to any other element $g = \{\vec{\phi}, \mathbf{v},
\mathbf{r}, t\}$ in the Poincar\'e group. It is convenient to represent the identity
transformation $e$  by the
 identity operator  which is, of
course, unitary.  It also seems reasonable to demand that the mappings $g
\to K_g$ and $g \to k_g$ are continuous, so, the representative
$k_g$ cannot suddenly switch from unitary to antiunitary along the
path connecting $e$ with $g$. Then we can reject antiunitary
transformations as representatives of $K_g$.\footnote{ The
antiunitary operators \index{antiunitary operator} may still
represent discrete transformations, e.g., time inversion, but we
agreed not to discuss such transformations in this book, because
they do not correspond to exact symmetries.}

Although Wigner's theorem reduces the freedom of choosing
generators, it does not eliminate this freedom completely: Two
unitary transformations $k_g$ and $\beta k_g$ (where $\beta$ is any
unimodular constant) generate the same subspace mapping. Therefore,
for each  $K_g$ there is a set of generating unitary transformations
$U_g$ differing from each other by a multiplicative constant. Such a
set is called a \emph{ray} \index{ray} of  transformations $[U_g]$.

Results of this subsection can be summarized as follows: each inertial
 transformation $g$ of the observer can be represented by a unitary operator
$U_g$  in $\mathcal{H}$ defined up to an arbitrary unimodular
factor: ket vectors are  transformed according to $|x \rangle \to
U_g |x \rangle$, and bra vectors are transformed as $\langle x | \to
\langle x |U_g^{-1}$. If $X= \sum_i |e_i \rangle \langle
e_i|$\footnote{Here $|e_i \rangle $ is an orthonormal basis
 in the subspace $X$.}
is a projection (proposition) associated with the observer $O$, then
observer $O'=gO$ represents the same proposition by the projection

\begin{eqnarray*}
X' &=& \sum_i  U_g |e_i \rangle \langle e_i| U_g ^{-1}  =   U_g XU_g
^{-1}
\end{eqnarray*}

\noindent Similarly, if $F = \sum_i f_i |e_i \rangle \langle e_i| $
is an operator of observable associated with the observer $O$ then

\begin{eqnarray}
F' &=& \sum_i f_i U_g |e_i \rangle \langle e_i| U_g ^{-1} = U_g F
U_g ^{-1}  \label{eq:5.7}
\end{eqnarray}

\noindent  is
operator of the same observable from the point of view  of the observer
$O'=gO$.

\subsection{Inertial transformations of states}
\label{ss:inertial-states}

In the preceding subsection we analyzed the effect of an inertial
transformation $g$ on observers, measuring apparatuses,
 propositions, and observables. Now we
are going to examine the effect of $g$ on preparation devices
 and states. We will try to answer the
following question: if $| \Psi \rangle$ is a vector describing a
pure state prepared by
 the preparation device $P$, then which state vector $| \Psi' \rangle$
describes the state prepared by the transformed  preparation device
$P' = gP$?

To find the connection between $| \Psi \rangle$ and $| \Psi'
\rangle$ we will use the relativity principle. According to equation
(\ref{eq:5.1}), for every observable $F$, its expectation value (\ref{eq:fff})
should not change after inertial transformation of the entire
laboratory (= both the preparation device
and the observer). In the bra-ket notation, this condition can be written as

\begin{eqnarray}
\langle \Psi | F| \Psi \rangle &=&  \langle \Psi' | F'| \Psi'
\rangle =\langle \Psi' | U_g F U_g ^{-1} | \Psi' \rangle
\label{eq:5.8}
\end{eqnarray}

\noindent This equation should be valid for any choice of observable
 $F$. Let us choose $F = | \Psi \rangle \langle \Psi|$, i.e., the
projection onto the ray containing vector $| \Psi \rangle$. Then equation
(\ref{eq:5.8}) takes the form

\begin{eqnarray*}
\langle \Psi | \Psi \rangle \langle \Psi | \Psi \rangle  &=& \langle
\Psi' | U_g| \Psi \rangle \langle \Psi | U_g ^{-1} | \Psi' \rangle =
\langle \Psi' | U_g| \Psi \rangle \langle \Psi' |
U_g| \Psi \rangle^* \\
&=&  |\langle \Psi' |
U_g| \Psi \rangle|^2
\end{eqnarray*}

\noindent The left hand side of this equation is equal to 1.
So, for each $| \Psi \rangle$, the transformed vector $| \Psi'
\rangle$
is such that

\begin{eqnarray*}
  |\langle \Psi' | U_g| \Psi \rangle |^2  = 1
\end{eqnarray*}

\noindent Since both $U_g| \Psi\rangle $ and $| \Psi' \rangle$ are
unit vectors, we must have

\begin{eqnarray*}
  | \Psi' \rangle = \sigma(g) U_g| \Psi \rangle
\end{eqnarray*}

\noindent where $\sigma(g)$ is an unimodular factor.  Operator $U_g$
is defined up to a unimodular factor,\footnote{see subsection
\ref{ss:wigner}} therefore, we can absorb the factor $\sigma(g)$
into the uncertainty of $U_g$ and finally write the action of the
inertial transformation $g$  on states

\begin{eqnarray}
| \Psi \rangle &\to& | \Psi' \rangle = U_g | \Psi \rangle \label{eq:5.9}
\end{eqnarray}

\noindent Then, taking into account the transformation law for
observables (\ref{eq:5.7}) we can check that, in agreement with the
relativity principle (\ref{eq:5.8}), the expectation values remain the same in all
laboratories

\begin{eqnarray}
\langle F' \rangle &=& \langle \Psi' | F' | \Psi' \rangle = (\langle
\Psi |
U_g^{-1})(U_g F U_g^{-1})( U_g| \Psi \rangle ) = \langle \Psi | F | \Psi \rangle \nonumber \\
&=& \langle F \rangle \label{eq:exp-val-cons}
\end{eqnarray}

\subsection{Heisenberg and Schr\"odinger pictures}
\label{ss:heisenberg}

The conservation of expectation values (\ref{eq:exp-val-cons}) is
valid only in the case when inertial transformation $g$ is
applied to the laboratory as a whole. What would happen if only
observer or only preparation device  is
transformed?

Let us first consider inertial transformations of observers. If we
change the observer without changing the preparation device
 (=state) then operators of observables
change according to (\ref{eq:5.7}) while the state vector remains
the same $| \Psi \rangle$. As expected, this transformation changes
results of experiments. For example, the expectation values of
observable $F$ are generally different for different observers $O$
and $O' = gO$

\begin{eqnarray}
 \langle F' \rangle &=& \langle \Psi|(U_g F U_g ^{-1})| \Psi \rangle \neq  \langle
\Psi| F | \Psi \rangle = \langle F \rangle \label{eq:5.10}
\end{eqnarray}

\noindent On the other hand, if the inertial transformation is applied to
 the preparation device  and  the state of the system changes according to equation
(\ref{eq:5.9}), then the results of measurements are also affected

\begin{eqnarray}
 \langle F'' \rangle &=& (\langle \Psi|U_g^{-1})  F (U_g | \Psi \rangle )
 \neq \langle \Psi|  F  | \Psi \rangle = \langle F
\rangle\label{eq:5.11}
\end{eqnarray}

\noindent Formulas (\ref{eq:5.10})  and (\ref{eq:5.11}) play a
prominent  role because many problems in physics can be formulated
as questions about descriptions of systems affected by inertial transformations. An important example
 is \emph{dynamics}, \index{dynamics} i.e., the
time evolution of the system. In this case one considers time
translation elements of the Poincar\'e group $g = \{ \vec{0};
\mathbf{0};
 \mathbf{0}; t\}$. Then equations (\ref{eq:5.10})  and
(\ref{eq:5.11}) provide two equivalent descriptions of dynamics. Equation
(\ref{eq:5.10}) describes dynamics in
 the  \emph{Heisenberg picture}. \index{Heisenberg picture} In this picture the state vector of the system
 remains fixed
while operators of observables change with time.  Equation
(\ref{eq:5.11}) provides an alternative description of dynamics in
the \emph{Schr\"odinger picture}. \index{Schr\"odinger picture} In
this description, operators of observables are time-independent,
while the  state vector of the system  depends on time. These two
pictures are equivalent because according to (\ref{eq:5.1}) a shift
of the observer by $g$ (forward time translation) is equivalent to
the shift of the preparation device  by
$g^{-1}$ (backward time translation).

The notions of  Schr\"odinger  and Heisenberg
pictures can be applied not only to time translations. They can be
generalized to other types of inertial transformations; i.e., the transformation $g$ above can stand for space
translations, rotations,  boosts, or any combination of them.

\section{Unitary representations of the Poincar\'e group}
\label{sc:repres-inertial}

In the preceding section we discussed the representation of a single
inertial transformation $g$ by an isomorphism $K_g$ of the lattice
of propositions and by a ray of unitary operators $[U_g]$, which act
on states and/or observables in the Hilbert space. We know from
chapter \ref{ch:relativity} that inertial transformations form the
Poincar\'e group. Then subspace mappings $K_{g_1}$, $K_{g_2}$,
$K_{g_3}$, $\ldots$ corresponding to different group elements $g_1,
g_2, g_3, \ldots$ cannot be arbitrary. They must satisfy conditions

\begin{eqnarray}
K_{g_2} K_{g_1} &=& K_{g_2 g_1} \label{eq:5.14aa} \\
K_{g^{-1}} &=& K_{g} ^{-1} \label{eq:5.14bb} \\
K_{g_3} (K_{g_2}K_{g_1}) &=& K_{g_3} K_{g_2g_1} = K_{g_3(g_2g_1)} = K_{(g_3g_2)g_1} = (K_{g_3} K_{g_2} ) K_{g_1}
\label{eq:5.14}
\end{eqnarray}

\noindent which reflect group properties of inertial
transformations $g$. Our goal
in this section is to find out which conditions are imposed by
(\ref{eq:5.14aa}) - (\ref{eq:5.14}) on the set of unitary
representatives $U_g$ of the Poincar\'e group.

\subsection{Projective representations of groups}
\label{ss:projective}

For each group element $g$ let us choose an arbitrary unitary
representative $U_{g}$ in the ray $ [U_{g}]$. For example, let us
choose the representatives (also called generators) $ U_{g_1} \in
[U_{g_1}]$, $ U_{g_2} \in [U_{g_2}]$, and $ U_{g_2 g_1} \in [U_{g_2
g_1}]$. The product $U_{g_2} U_{g_1}$ should generate the mapping $
K_{g_2 g_1}$, therefore it can differ from our chosen representative
$ U_{g_2 g_1} $ by at most a unimodular constant $\alpha(g_2,g_1)$.
So, we can write for any two transformations $g_1$ and $g_2$

\begin{eqnarray}
U_{g_2} U_{g_1}  = \alpha(g_2,g_1) U_{g_2g_1} \label{eq:5.15}
\end{eqnarray}

\noindent The factors $\alpha$
have three properties. First, they are unimodular.

\begin{eqnarray}
|\alpha(g_2,g_1)| = 1 \label{eq:5.16}
\end{eqnarray}

\noindent Second, from the property (\ref{eq:A.2}) of the unit
element  we have for any $g$

\begin{eqnarray}
U_{g} U_{e}  &=& \alpha(g,e) U_{g} = U_g \label{eq:5.17} \\
 U_{e} U_{g} &=& \alpha(e,g) U_{g} = U_g \label{eq:5.18}
\end{eqnarray}

\noindent which implies

\begin{eqnarray}
\alpha(g, e) &=& \alpha( e, g) = 1 \label{eq:5.19}
\end{eqnarray}

\noindent Third, the  associative law (\ref{eq:5.14}) implies

\begin{eqnarray}
U_{g_3}(\alpha(g_2,g_1)  U_{g_2g_1}) &=&  (\alpha(g_3,g_2) U_{g_3g_2})
U_{g_1} \nonumber \\
\alpha(g_2,g_1) \alpha(g_3, g_2g_1) U_{g_3g_2g_1} &=& \alpha(g_3g_2,
g_1) \alpha(g_3,g_2) U_{g_3g_2g_1} \nonumber \\
\alpha(g_2,g_1) \alpha(g_3, g_2g_1)  &=& \alpha(g_3g_2, g_1)
\alpha(g_3,g_2)
\label{eq:5.20}
\end{eqnarray}

\noindent The mapping $U_g$ from group elements to unitary operators
in $\mathcal{H}$ is called a \emph{projective representation}
\index{projective representation} of the group if it satisfies equations
(\ref{eq:5.15}), (\ref{eq:5.16}), (\ref{eq:5.19}), and
(\ref{eq:5.20}).

 \subsection{Elimination of central charges in
the Poincar\'e algebra} \label{ss:central-charges}

In principle, we could keep the arbitrarily chosen  unitary
representatives of the subspace transformations $U_{g_1}$,
$U_{g_2}$, $\ldots$, as discussed above, and work with thus obtained
projective representation of the Poincar\'e group, but this would
result in a rather complicated mathematical formalism. The theory
would be significantly simpler if we could judiciously choose the
representatives\footnote{i.e., multiply our previously chosen unitary operators $U_g$ by
some unimodular factors $U_g \to \beta(g)U_g$} in such a way that the
factors $\alpha(g_2,g_1)$ in (\ref{eq:5.15}) are simplified or
eliminated altogether. Then we would have a much simpler \emph{linear}
unitary group representation (see Appendix \ref{sc:group-reps})
instead of the projective group representation. In this subsection
we are going to demonstrate that in any projective representation of
the Poincar\'e group such elimination of factors $\alpha(g_2,g_1)$
is indeed possible \cite{CJS}.

The proof of the last statement is significantly simplified if
conditions (\ref{eq:5.16}), (\ref{eq:5.19}), and (\ref{eq:5.20}) are
expressed in a Lie algebra notation. In the vicinity of the unit
element of the group we can use vectors $\vec{\zeta}$ from the
Poincar\'e Lie algebra to identify other group elements (see equation
(\ref{eq:A.28})), i.e.

\begin{eqnarray*}
g &=& e^{\vec{\zeta} } = \exp \left( \sum_{a=1}^{10} \zeta^a \vec{t}_a \right)
\end{eqnarray*}

\noindent where $t_a$ is the basis of the Poincar\'e Lie algebra
$(\mathcal{H}, \vec{\mathcal{P}}, \vec{\mathcal{K}},
\vec{\mathcal{J}})$ from subsection \ref{sc:poincare-lie}. Then we
can write unitary representatives $U_g$ of inertial transformations
$g$ in the form\footnote{Here we have used Stone's theorem
\ref{Theorem.Stone's}. \index{Stone's theorem} The nature of
one-parameter subgroups featuring in the theorem is rather obvious.
These are subgroups of similar transformations, i.e., a subgroup of
space translations, a subgroup of rotations about the $z$-axis,
etc.}

\begin{eqnarray}
U_{\vec{\zeta} } = \exp \left(-\frac{i}{\hbar} \sum_{a=1}^{10}
\zeta^a F_a \right) \label{eq:5.21}
\end{eqnarray}

\noindent where  $\hbar$ is a real constant which will be left
unspecified at this point,\footnote{We will identify $\hbar$ with
the Planck constant in subsection \ref{ss:energy}} and $F_a $ are
ten Hermitian operators in the Hilbert space $\mathcal{H}$ called
the \emph{generators} \index{generator} of
 the unitary projective representation.
 Then we can
write
equation (\ref{eq:5.15}) in the form

\begin{eqnarray}
U_{\vec{\zeta} } U_{ \vec{\xi} } =
 \alpha(\vec{\zeta}, \vec{\xi})U_{ \vec{\zeta}\vec{\xi} } \label{eq:5.22}
\end{eqnarray}

\noindent Since $\alpha$ is unimodular we can set
$\alpha(\vec{\zeta}, \vec{\xi})= \exp[i\kappa (\vec{\zeta},
\vec{\xi})]$, where $\kappa (\vec{\zeta}, \vec{\xi})$ is a real
function. Conditions (\ref{eq:5.19}) and (\ref{eq:5.20}) then
will be rewritten in terms of $\kappa$

\begin{eqnarray}
\kappa(\vec{\zeta},\vec{0}) &=& \kappa(\vec{0},\vec{\zeta}) = 0
\label{eq:5.23}\\
\kappa(\vec{\xi},\vec{\zeta})
+ \kappa(\vec{\chi}, \vec{\xi}\vec{\zeta})
&=& \kappa(\vec{\chi} \vec{\xi}, \vec{\zeta}) +
\kappa(\vec{\chi},\vec{\xi})
\label{eq:5.24}
\end{eqnarray}

\noindent Note that we can write the lowest order term in the Taylor
series for
 $\kappa$
near the  group identity element in the form\footnote{see also the third term on the right hand side of equation (\ref{eq:A.31})}

\begin{eqnarray}
\kappa(\vec{\zeta}, \vec{\xi}) = \sum_{ab=1}^{10}h_{ab}
\zeta^a \xi^b
\label{eq:5.25}
\end{eqnarray}

\noindent The constant term, the terms linear in $\zeta^a$ and
$\xi^b$, as well as the terms proportional to $\zeta^a \zeta^b$ and
$\xi^a \xi^b$ are absent on the right hand of (\ref{eq:5.25}) as a
consequence of the condition (\ref{eq:5.23}).

Using the same arguments as during our derivation of equation
(\ref{eq:A.33}), we can expand all terms in (\ref{eq:5.22}) around
$\vec{\zeta}=  \vec{\xi} = \vec{0}$

\begin{eqnarray*}
&\mbox{ }& \left(1 - \frac{i}{\hbar} \sum_{a=1}^{10} \xi^a F_a -
\frac{1}{2 \hbar^2} \sum_{bc=1}^{10} \xi^b \xi^c F_{bc} + \ldots
\right) \left(1 - \frac{i}{\hbar}\sum_{a=1}^{10} \zeta^a F_a -
 \frac{1}{2 \hbar^2}
\sum_{bc=1}^{10} \zeta^b  \zeta^c F_{bc} + \ldots \right) \\ &=&
\left(1 + i \sum_{ab=1}^{10} h_{ab}\zeta^a \xi^b + \ldots \right)
\Bigl[1 - \frac{i}{\hbar} \sum_{a=1}^{10} \left(\zeta^a + \xi^a +
\sum_{bc=1}^{10} f^a_{bc} \xi^b \zeta^c + \ldots \right)F_a \\
&-& \frac{1}{2 \hbar^2} \sum_{ab=1}^{10} (\zeta^a + \xi^a + \ldots)
(\zeta^b + \xi^b + \ldots) F_{ab} + \ldots \Bigr]
\end{eqnarray*}

\noindent Equating the coefficients multiplying products
$\xi^a\zeta^b$ on both sides, we obtain

\begin{eqnarray*}
-\frac{1}{2 \hbar^2} (F_{ab} + F_{ba}) = \frac{1}{\hbar^2} F_aF_b -
i h_{ab} + \frac{i}{\hbar} \sum_{c=1}^{10} f^c_{ab} F_c
\end{eqnarray*}

\noindent The left hand side of this equation is symmetric with
respect to interchange of indices $a \leftrightarrow b$. The same
should be true for the right hand side. From this condition we
obtain commutators of generators $F$

\begin{eqnarray}
 F_a F_b - F_b F_a  = i \hbar \sum_{c=1}^{10} C^c_{ab} F_c + E_{ab}
\label{eq:5.26}
\end{eqnarray}

\noindent where $C^c_{ab} \equiv f^c_{ab} - f^c_{ba}$ are familiar
structure constants \index{structure constants} of the Poincar\'e
Lie algebra (\ref{eq:poincare_1}) - (\ref{eq:poincare_8}) and
$E_{ab} = i\hbar^2 (h_{ab} - h_{ba})$ are imaginary constants, which
depend on our original choice of representatives $U_g$ in rays
$[U_g]$.\footnote{To be exact, we must write $E_{ab}$ on the right
hand side of equation (\ref{eq:5.26}) multiplied by the identity operator
$I$. However, we will omit the symbol $I$ here for brevity.} These
constants are called \emph{central charges}. \index{central charges}
Our main task in this subsection is to prove that representatives
$U_g$ can be chosen in such a way that $E_{ab}=0$, i.e., the central
charges get eliminated.

First we consider the original (arbitrary) set of representatives $U_g$. In
accordance with our notation in section \ref{sc:poincare} we will
use symbols

\begin{eqnarray}
(\tilde{H}, \tilde{\mathbf{P}}, \tilde{\mathbf{J}},
\tilde{\mathbf{K}}) \label{eq:geners-poinc}
\end{eqnarray}

\noindent to denote ten generators $F_a$ of an arbitrary projective
representation $U_g$.\footnote{Note that generators $(\mathcal{H}, \vec{\mathcal{P}}, \vec{\mathcal{K}},
\vec{\mathcal{J}})$ in subsection \ref{sc:poincare-lie} were abstract quantities that could be interpreted as ``derivatives of group transformations'', while generators $(\tilde{H}, \tilde{\mathbf{P}}, \tilde{\mathbf{J}},
\tilde{\mathbf{K}})$ here are Hermitian operators in the Hilbert space of states of our physical system.} These generators correspond to time
translation, space translation, rotations, and boosts, \index{boost
operator} respectively. Then using the structure constants
$C^a_{bc}$ of the Poincar\'e Lie algebra from equations
(\ref{eq:poincare_1}) - (\ref{eq:poincare_8}) we obtain the full
list of commutators (\ref{eq:5.26}).

\begin{eqnarray}
[\tilde{J}_i, \tilde{P}_j] &=& i\hbar \sum_{k=1}^3 \epsilon_{ijk}
\tilde{P}_k +  E^{(1)}_{ij} \label{eq:5.27} \\ \mbox{ }
 [\tilde{J}_i,
\tilde{J}_j] &=& i\hbar \sum_{k=1}^3 \epsilon_{ijk} (\tilde{J}_k  +
 iE^{(2)}_{k})
\label{eq:5.28}
\\ \mbox{ }
[\tilde{J}_i, \tilde{K}_j] &=& i\hbar \sum_{k=1}^3 \epsilon_{ijk}
\tilde{K}_k  + E^{(3)}_{ij} \label{eq:5.29}
\\ \mbox{ }
[\tilde{P}_i,\tilde{P}_j] &=&  E_{ij}^{(4)} \label{eq:5.30}
\\ \mbox{ }
[\tilde{J}_i,\tilde{H}] &=& E_i^{(5)} \label{eq:5.31}
\\ \mbox{ }
 [\tilde{P}_i, \tilde{H}] &=& E_i^{(6)}
\label{eq:5.32}
\\ \mbox{ }
[\tilde{K}_i, \tilde{K}_j] &=& -i\frac{\hbar}{c^2} \sum_{k=1}^3
\epsilon_{ijk} (\tilde{J}_k +  iE^{(7)}_{k}) \label{eq:5.33}
\\ \mbox{ }
[\tilde{K}_i, \tilde{P}_j] &=& -i\frac{\hbar}{c^2} \tilde{H}
\delta_{ij} + E^{(8)}_{ij}, \label{eq:5.34}
\\ \mbox{ }
 [\tilde{K}_i, \tilde{H}] &=& -i\hbar \tilde{P}_i + E^{(9)}_{i}
\label{eq:5.35}
\end{eqnarray}

\noindent Here we arranged $E_{ab}$ into nine sets of central
charges $E^{(1)} \ldots E^{(9)}$. In equation (\ref{eq:5.28}) and
(\ref{eq:5.33}) we took into account that their left hand sides are
 antisymmetric tensors.  So, the central charges must form antisymmetric tensors as
well, and, according to Table \ref{table:C.1}, they can be
represented as $
 -\hbar \sum_{k=1}^3 \epsilon_{ijk}
 E^{(2)}_k
$ and $
 \hbar c^{-2} \sum_{k=1}^3 \epsilon_{ijk}
  E^{(7)}_k$,
respectively, where $E_k^{(2)}$ and $E_k^{(7)}$ are 3-vectors.

Next we will use the requirement that commutators (\ref{eq:5.27}) -
(\ref{eq:5.35}) must satisfy the Jacobi identity.\footnote{equation
(\ref{eq:jacobi}), which is equivalent to the associativity
condition (\ref{eq:5.14}) or (\ref{eq:5.24})} This will allow us to make some simplifications.
For example, using\footnote{Here we used equation (\ref{eq:5.27}).} $\tilde{P}_3 = -\frac{i}{\hbar}[\tilde{J}_1,
\tilde{P}_2] + \frac{i}{\hbar} E_{12}^{(1)}$ and the fact that all constants $E$ commute with
 generators of the group, we obtain

\begin{eqnarray*}
[\tilde{P}_3,\tilde{P}_1] &=& -\frac{i}{\hbar}[([\tilde{J}_1,
\tilde{P}_2] + E^{(1)}_{12}), \tilde{P}_1] = -\frac{i}{\hbar}[[\tilde{J}_1, \tilde{P}_2], \tilde{P}_1] \\
 &=& -\frac{i}{\hbar}[[\tilde{P}_1, \tilde{P}_2], \tilde{J}_1] -
\frac{i}{\hbar} [[\tilde{J}_1, \tilde{P}_1], \tilde{P}_2] = -\frac{i}{\hbar} [E^{(4)}_{12}, \tilde{J}_1] - \frac{i}{\hbar}
[E^{(1)}_{11}, \tilde{P}_2] \\
&=& 0
\end{eqnarray*}

\noindent so $E^{(4)}_{31} = 0$. Similarly, we can show that $E_{ij}^{(4)} =
E^{(5)}_i =
E^{(6)}_i = 0$ for all values of indices $i,j = 1,2,3$.

Using the Jacobi identity we further obtain

\begin{eqnarray}
i \hbar[\tilde{J}_3, \tilde{P}_3] &=&  [[\tilde{J}_1, \tilde{J}_2],
\tilde{P}_3] = [[\tilde{P}_3, \tilde{J}_2], \tilde{J}_1] +
[[\tilde{J}_1,
\tilde{P}_3], \tilde{J}_2] \nonumber \\
&=&  i \hbar[\tilde{J}_1, \tilde{P}_1] + i \hbar [\tilde{J}_2,
\tilde{P}_2] \label{eq:5.36}
\end{eqnarray}

\noindent and, similarly,

\begin{equation}
i \hbar [\tilde{J}_1,\tilde{P}_1]  = i \hbar [\tilde{J}_2,
\tilde{P}_2] + i \hbar  [\tilde{J}_3, \tilde{P}_3] \label{eq:5.37}
\end{equation}

\noindent By adding equations (\ref{eq:5.36}) and (\ref{eq:5.37}) we see that

\begin{equation}
[\tilde{J}_2, \tilde{P}_2] = 0
\label{eq:5.38}
\end{equation}

\noindent Similarly, we obtain $ [\tilde{J}_1, \tilde{P}_1] = [\tilde{J}_3,
\tilde{P}_3] =
0$, which means that

\begin{equation}
E_{ii}^{(1)} = 0
\label{eq:5.39}
\end{equation}

\noindent Using the Jacobi identity again, we obtain

\begin{eqnarray*}
i \hbar [\tilde{J}_2,\tilde{P}_3] &=& [[\tilde{J}_3, \tilde{J}_1],
\tilde{P}_3]= [[\tilde{P}_3, \tilde{J}_1], \tilde{J}_3] +
[[\tilde{J}_3,
\tilde{P}_3], \tilde{J}_1] \\
&=& -i \hbar[\tilde{J}_3, \tilde{P}_2]
\end{eqnarray*}

\noindent This antisymmetry property is also true in the general case (for any
$i,j = 1,2,3; i\neq j$)

\begin{equation}
[\tilde{J}_i,\tilde{P}_j] =  -[\tilde{J}_j, \tilde{P}_i]
\label{eq:5.40}
\end{equation}

\noindent Putting together (\ref{eq:5.38}) and (\ref{eq:5.40}) we
see that tensor $[\tilde{J}_i,\tilde{P}_j]$ is antisymmetric.
 This implies that we can introduce a
vector $E_k^{(1)}$ such that

\begin{eqnarray}
E^{(1)}_{ij} &=& -\hbar \sum_{i=1}^3\epsilon_{ijk}
  E^{(1)}_k \nonumber \\
\ [\tilde{J}_i, \tilde{P}_j] &=& i\hbar \sum_{i=1}^3\epsilon_{ijk}
(\tilde{P}_k +  iE^{(1)}_k) \label{eq:5.41}
\end{eqnarray}

\noindent Similarly, we can show that $E^{(3)}_{ii} = 0$ and

\begin{eqnarray*}
[\tilde{J}_i, \tilde{K}_j] = i\hbar \sum_{i=1}^3 \epsilon_{ijk}
(\tilde{K}_k +  iE^{(3)}_k)
\end{eqnarray*}

Taking into account the above results, commutation relations
(\ref{eq:5.27})-(\ref{eq:5.35}) now take the form

\begin{eqnarray}
[\tilde{J}_i, \tilde{P}_j] &=& i\hbar \sum_{k=1}^3 \epsilon_{ijk}
(\tilde{P}_k +   iE^{(1)}_k) \label{eq:5.42}
\\ \mbox{ }
[\tilde{J}_i, \tilde{J}_j] &=& i\hbar \sum_{k=1}^3 \epsilon_{ijk}
(\tilde{J}_k  +
 iE^{(2)}_k)
\label{eq:5.43}
\\ \mbox{ }
[\tilde{J}_i, \tilde{K}_j] &=& i\hbar \sum_{k=1}^3 \epsilon_{ijk}
(\tilde{K}_k  +  iE^{(3)}_k) \label{eq:5.44}
\\ \mbox{ }
[\tilde{P}_i,\tilde{P}_j] &=&  [\tilde{J}_i,\tilde{H}] =
[\tilde{P}_i, \tilde{H}] = 0 \label{eq:5.45}
\\ \mbox{ }
[\tilde{K}_i, \tilde{K}_j] &=& -i\frac{\hbar}{c^2} \sum_{k=1}^3
\epsilon_{ijk} (\tilde{J}_k +  iE^{(7)}_k) \label{eq:5.46}
\\ \mbox{ }
[\tilde{K}_i, \tilde{P}_j] &=& -i\frac{\hbar}{c^2} \tilde{H}
\delta_{ij} + E^{(8)}_{ij}, \label{eq:5.47}
\\ \mbox{ }
 [\tilde{K}_i, \tilde{H}] &=& -i\hbar \tilde{P}_i + E^{(9)}_{i}
\label{eq:5.48}
\end{eqnarray}

\noindent where $E$ on the right hand sides are certain imaginary
constants. The next step in elimination of the central charges $E$
is to use the freedom of choosing unimodular factors $\beta(g)$ in
front of operators of the representation $U_g$: Two unitary
operators $U_{\vec{\zeta}}$ and $\beta(\vec{\zeta})U_{\vec{\zeta}}$
differing by a unimodular
 factor $\beta(\vec{\zeta})$ generate the same subspace
transformation $K_{\vec{\zeta}}$. Correspondingly, the choice of generators $F_a$ has some
degree
of arbitrariness as well. Since $\beta(\vec{\zeta})$ are unimodular,
we can write

\begin{eqnarray*}
\beta(\vec{\zeta}) = \exp(i \gamma(\vec{\zeta})) \approx 1 + i
\sum_{a=1}^{10}
R_a \zeta^a
\end{eqnarray*}

\noindent Therefore, in the first order, the presence of factors
$\beta(\vec{\zeta})$
results in adding some real constants $R_a$ to generators $F_a$.
  We would like to show
that by adding such constants we can make all central charges  equal
to zero.

Let us now add  constants $R$ to the generators $\tilde{P}_j$,
$\tilde{J}_j$, and $\tilde{K}_j$ and denote the
redefined generators as

\begin{eqnarray*}
P_j &=& \tilde{P}_j + R^{(1)}_j \\
J_j &=& \tilde{J}_j + R^{(2)}_j \\
K_j &=& \tilde{K}_j + R^{(3)}_j
\end{eqnarray*}

\noindent Then commutator (\ref{eq:5.43}) takes the form

\begin{eqnarray*}
[J_i, J_j] &=& [\tilde{J}_i + R^{(2)}_i, \tilde{J}_j + R^{(2)}_j] =
[\tilde{J}_i , \tilde{J}_j ] = i\hbar \sum_{k=1}^3 \epsilon_{ijk}
(\tilde{J}_k  +  iE^{(2)}_k)
\end{eqnarray*}

\noindent So, if we choose $R^{(2)}_k = iE^{(2)}_k$, then

\begin{eqnarray*}
[J_i, J_j] =
i\hbar \sum_{k=1}^3 \epsilon_{ijk} J_k
\end{eqnarray*}

\noindent and central charges are eliminated from this commutator.

Similarly, central charges can be eliminated from  commutators

\begin{eqnarray}
[J_i, P_j] &=&
i\hbar \sum_{k=1}^3\epsilon_{ijk} P_k \nonumber \\
\ [J_i, K_j] &=&
i\hbar \sum_{k=1}^3 \epsilon_{ijk} K_k
\label{eq:5.49}
\end{eqnarray}

\noindent by choosing $R^{(1)}_k = iE^{(1)}_k$ and $R^{(3)}_k =
iE^{(3)}_k$. From equation (\ref{eq:5.49}) we then obtain

\begin{eqnarray*}
[K_1,K_2] &=& - \frac{i}{\hbar}[[J_2, K_3], K_2]=  - \frac{i}{\hbar}
[[J_2, K_2], K_3]  - \frac{i}{\hbar} [[K_2, K_3], J_2]
\\
 &=&  - \frac{i}{\hbar} [-\frac{i \hbar}{c^2}(J_1 + iE_1^{(7)}), J_2]
 = -\frac{i \hbar}{c^2} J_3
\end{eqnarray*}

\noindent so, our choice of the constants $R^{(1)}_k$, $R^{(2)}_k$, and
$R^{(3)}_k$
eliminates the central charges $E_i^{(7)}$.

From equation (\ref{eq:5.49}) we also obtain

\begin{eqnarray*}
[K_3, \tilde{H}] &=&  - \frac{i}{\hbar} [[J_1, K_2], \tilde{H}] = -
\frac{i}{\hbar} [[\tilde{H}, K_2], J_1]  -
\frac{i}{\hbar} [[J_1,\tilde{ H}], K_2] \\
&=& -[J_1, P_2] = -i \hbar P_3
\end{eqnarray*}

\noindent which implies that the central charge $E^{(9)}$ is canceled as
well.
Finally

\begin{eqnarray*}
[K_1, P_2] &=& - \frac{i}{\hbar} [[J_2, K_3], P_2] =
-\frac{i}{\hbar}[[J_2, P_2], K_3] + \frac{i}{\hbar}[[K_3, P_2], J_3]
= 0 \\
\ [K_1, P_1] &=& - \frac{i}{\hbar} [[J_2, K_3], P_1] =
-\frac{i}{\hbar}[[J_2, P_1], K_3] + \frac{i}{\hbar}[[K_3, P_1],
J_3] \\
&=& [K_3, P_3]
\end{eqnarray*}

\noindent It then follows that $E^{(8)}_{ij} = 0$ if $i \neq j$ and we
can introduce a real scalar $E^{(8)}$ such that

\begin{eqnarray*}
E^{(8)}_{11} &=& E^{(8)}_{22}  = E^{(8)}_{33} \equiv - \frac{i
\hbar}{c^2}E^{(8)} \\
\ [K_i, P_i] &=& - \frac{i \hbar}{c^2} \delta_{ij} (\tilde{H}  + E^{(8)})
\end{eqnarray*}

\noindent Finally, by redefining the generator of time translations
$H = \tilde{H} + E^{(8)}$ we eliminate all central charges from
commutation relations of the Poincar\'e Lie algebra

\begin{eqnarray}
[J_i, P_j] &=& i\hbar \sum_{k=1}^3 \epsilon_{ijk} P_k
\label{eq:5.50}
\\
\mbox{ } [J_i, J_j] &=& i\hbar \sum_{k=1}^3 \epsilon_{ijk} J_k
\label{eq:5.51}
\\
\mbox{ } [J_i, K_j] &= &i\hbar \sum_{k=1}^3 \epsilon_{ijk} K_k
\label{eq:5.52}
\\
\mbox{ } [P_i,P_j] &=&  [J_i,H] = [P_i, H] = 0 \label{eq:5.53}
\\
 \mbox{ } [K_i, K_j] &=& -\frac{i\hbar}{c^2} \sum_{k=1}^3 \epsilon_{ijk} J_k
\label{eq:5.54}
\\
\mbox{ } [K_i, P_j] &=& -\frac{i\hbar}{c^2} H \delta_{ij} \label{eq:5.55}\\
 \mbox{ } [K_i, H] &=& -i\hbar P_i
\label{eq:5.56}
\end{eqnarray}

\noindent Thus Hermitian operators $H$, $\mathbf{P}$, $\mathbf{J}$, and $\mathbf{K}$ provide a representation of the Poincar\'e Lie
algebra, and the redefined unitary operators $\beta(g) U_g$ form a
\emph{unique} unitary representation of the Poincar\'e group that
corresponds to the given projective representation $ U_g$ in the
vicinity of the group identity. We have proven that projective
representations of the Poincar\'e group are equivalent to certain unitary representations,
which are much easier objects for study (see Appendix
\ref{sc:group-reps}).

Commutators (\ref{eq:5.50}) - (\ref{eq:5.56}) are probably the most
important equations of relativistic quantum theory. In the rest of
this book we will have many opportunities to appreciate  a deep
physical content of these formulas.

\subsection{Single-valued and double-valued representations}
\label{ss:single-valued}

In the preceding subsection we eliminated the phase factors
$\alpha(g_2, g_1)$ from equation (\ref{eq:5.15}) by resorting to Lie
algebra arguments. However, these arguments work only in the
vicinity of the group's unit element. There is a possibility that
non-trivial phase factors may reappear in the multiplication law
(\ref{eq:5.15}) when the group manifold has a non-trivial topology
and group elements are considered which are far from the unit
element.

In Appendix \ref{ss:double-valued} we established that this possibility is realized in the case of the rotation group.  This means that for quantum-mechanical applications we need to consider both single-valued and double-valued representations of this group.  Since the rotation group is a subgroup of the Poincar\'e
group, the same conclusion is relevant for the Poincar\'e group: both single-valued and double-valued unitary representations should be considered.\footnote{Equivalently, one can choose to consider all single-valued representations of the \emph{universal covering group} \index{universal covering group} of the Poincar\'e group.} In chapter \ref{ch:single} we will see that these two cases correspond to integer-spin and half-integer-spin systems, respectively.

\subsection{Fundamental statement of relativistic
quantum theory} \label{ss:fundamental}

The most important result of this chapter is the connection between
relativity and quantum mechanics summarized in the following
statement (see, e.g., \cite{book})

\begin{statement} [Unitary representations of the Poincar\'e
group] \label{statementO} In a relativistic quantum description of a
physical system, inertial transformations  are represented by
unitary operators which furnish a unitary (single- or double-valued)
representation of the Poincar\'e group in the Hilbert space of the
system.
\end{statement}

\noindent It is important to note that this statement is completely
general. The Hilbert space of \emph{any} isolated physical system
(no matter how complex) must carry a unitary representation of the
Poincar\'e group. Construction of Hilbert spaces and Poincar\'e
group representations in them is the major part of theoretical
description of physical systems. The rest of this book is primarily
devoted to performing these difficult tasks.

Basic inertial transformations from the Poincar\'e group are
represented in the Hilbert space by unitary operators:
$e^{-\frac{i}{\hbar}\mathbf{P}\mathbf{r}}$ for spatial translations,
$e^{-\frac{i}{\hbar}\mathbf{J} \vec{\phi}}$ for rotations,
$e^{-\frac{ic}{\hbar}\mathbf{K}  \vec{\theta}}$ for boosts,
\index{boost operator} and $e^{\frac{i}{\hbar}Ht}$ for time
translations,\footnote{The exponential form of the unitary group
representatives follows from equation (\ref{eq:5.21}). }
A general inertial transformation $g = \{\vec{\phi} ,
\mathbf{v}(\vec{\theta}), \mathbf{r}, t \}$  is represented by the
unitary operator\footnote{compare with equation
(\ref{eq:galilei_general})}

\begin{eqnarray}
U_g = e^{-\frac{i}{\hbar}\mathbf{J} \vec{\phi}}
e^{-\frac{ic}{\hbar}\mathbf{K}  \vec{\theta}}
e^{-\frac{i}{\hbar}\mathbf{P}\mathbf{r}} e^{\frac{i}{\hbar}Ht}
\label{eq:5.57}
\end{eqnarray}

\noindent We will frequently use notation

\begin{eqnarray}
U_g \equiv U(\vec{\phi}, \vec{\theta}; \mathbf{r}, t) \equiv U(\Lambda; \mathbf{r}, t) \label{eq:5.57a}
\end{eqnarray}

\noindent where $\Lambda$ is a Lorentz transformation of inertial frames that combines boost $\vec{\theta}$ and rotation $\vec{\phi}$. Then, in the Schr\"odinger picture\footnote{see subsection \ref{ss:heisenberg}} \index{Schr\"odinger picture} state vectors transform
between different inertial reference frames according to\footnote{We
will see in subsection \ref{ss:inertial-obs} that this is
\emph{active} transformation of states. In most physical
applications one is interested in \emph{passive} transformations of
states (i.e., how the same state is seen by two different observers), which are given by the inverse operator $U_g^{-1}$.}

\begin{eqnarray}
| \Psi' \rangle &=& U_g | \Psi \rangle \label{eq:5.58}
\end{eqnarray}

\noindent In the Heisenberg picture inertial transformations of
observables have the form \index{inertial transformations of
observables}

\begin{eqnarray}
 F' &=& U_g F U_g^{-1} \label{eq:5.59}
\end{eqnarray}

\noindent For example, the equation describing the time evolution of
the observable $F$ in the Heisenberg picture\footnote{see equation
(\ref{eq:A.39})} \index{Heisenberg picture}

\begin{eqnarray}
F(t)  &=& e^{\frac{i}{\hbar} Ht} F e^{-\frac{i}{\hbar} Ht}
\label{eq:5.60} \\
&=& F + \frac{i}{\hbar} [H,F]t - \frac{1}{2 \hbar^2} [H, [H,F]]t^2 +
\ldots
\label{eq:5.61}
\end{eqnarray}

\noindent can be also written  in
a differential form

\begin{eqnarray*}
\frac{dF(t)}{dt}  &=&  \frac{i}{\hbar} [H, F]
\end{eqnarray*}

\noindent which is the familiar \emph{Heisenberg equation}.
\index{Heisenberg equation}

Note also that analogous ``Heisenberg equations'' can be written for
transformations of observables with respect to space translations,
rotations, and boosts

\begin{eqnarray}
\frac{dF(\mathbf{r})}{d \mathbf{r}}  &=& - \frac{i}{\hbar} [ \mathbf{P},F] \nonumber \\
\frac{dF(\vec{\phi})}{d \vec{\phi}}  &=& - \frac{i}{\hbar} [
\mathbf{J},F] \nonumber \\
\frac{dF(\vec{\theta})}{d\vec{\theta}}  &=&   -
\frac{ic}{\hbar}[\mathbf{K}, F ] \label{eq:icKF}
\end{eqnarray}

We already mentioned that transformations of observables with
respect to inertial transformations of observers cover many
interesting problems in physics (the time evolution, boost
transformations, etc.). \index{Lorentz transformations} From the
above formulas we see that solution of these problems requires the
knowledge of commutators between observables $F$ and generators ($H, \mathbf{P}, \mathbf{J}$ and
$\mathbf{K}$) of
the relevant Poincar\'e group representation. In the next chapter we will discuss definitions of
various observables, their connections to Poincar\'e generators, and
their commutation relations.

\chapter{OPERATORS OF OBSERVABLES}
\label{ch:operators}

\begin{quote}
\textit{Throwing pebbles into the water, look at the ripples they
form on the surface, otherwise, such occupation becomes an idle
pastime.}

\small
\hspace{1in} Kozma Prutkov
\normalsize
\end{quote}

\vspace{0.5in}

\noindent In chapters \ref{ch:QM} and \ref{ch:QM-relativity} we
established that in quantum theory any physical system is described
by a complex Hilbert space $\mathcal{H}$, pure states are
represented by rays in $\mathcal{H}$,  observables are represented
by Hermitian operators in $\mathcal{H}$, and there is a unitary
representation $U_g$ of the Poincar\'e group in  $\mathcal{H}$ which
determines how state vectors and operators of observables change
when the preparation device  or the
measuring apparatus  undergoes an
inertial transformation. Our next goal is to clarify the structure
of the set of observables. In particular, we wish to find which
operators correspond to such familiar observables as velocity,
momentum, energy, mass, position, etc, what are their spectra and
what are the relationships between these operators? We will also find out how these observables change under inertial transformations from the Poincar\'e group. This implies that we will use the Heisenberg picture everywhere in this chapter.

We should stress that physical systems considered in this chapter are completely arbitrary: they can be either elementary particles or compound systems of many elementary particles or even systems (such as unstable particles) in which the number of particles is not precisely defined. The only significant requirement is that our system must be isolated, i.e., its interaction with the rest of the universe can be neglected.

In this chapter we will focus on observables whose operators can be
expressed as functions of  generators ($\mathbf{P}$, $\mathbf{J}$,
$\mathbf{K}$, $H$) of the Poincar\'e group representation $U_g$. In
chapter \ref{ch:fock-space}, we will meet other observables, such as
 the number of particles. They cannot be expressed through ten
generators of the Poincar\'e group.

\section{Basic observables}
\label{sc:basic-observables}

\subsection{Energy, momentum, and angular momentum}
\label{ss:energy}

The generators of the Poincar\'e group representation  in the
Hilbert space of any system are Hermitian operators $H$,
$\mathbf{P}$, $\mathbf{J}$, and $\mathbf{K}$, and we might suspect
that they are related to certain observables pertinent to this
system. What are these observables? In order to get a hint, let us
now postulate that the constant $\hbar$ introduced in equation (\ref{eq:5.21}) is the Planck constant \index{Planck
constant}

\begin{eqnarray}
\hbar =6.626\cdot 10^{-34} \frac{kg \cdot m^2}{s}
\label{Planck_constant}
\end{eqnarray}

\noindent whose dimension can be also expressed as $<\hbar> =
<mass><speed><distance>$. Then the dimensions of generators can be
found from the condition that the arguments of exponents in
(\ref{eq:5.57}) must be dimensionless

\begin{itemize}
\item  $<H> = \frac{<\hbar >}{<time>} = <mass><speed>^2$;
\item  $<\mathbf{P}> = \frac{<\hbar>}{<distance>} = <mass><speed>$;
\item  $<\mathbf{J}> = <\hbar> = <mass><speed><distance>$
\item  $<\mathbf{K}>  = \frac{<\hbar>}{<speed>} =
<mass><distance>$;
\end{itemize}

\noindent Based on these dimensions we can guess that we are dealing with
observables of \emph{ energy} \index{energy}\index{Hamiltonian} (or
\emph{Hamiltonian}) $H$, \emph{ momentum} $\mathbf{P}$,
\index{momentum} and \emph{ angular momentum} \index{angular
momentum} $\mathbf{J}$ of the system.\footnote{There is no common
observable directly associated with the boost generator
$\mathbf{K}$, but we will see later that $\mathbf{K}$ is intimately related to
system's  position and spin. \index{boost operator}} We will
call them \emph{basic observables}. \index{basic observables}
Operators
 $H$,  $\mathbf{P}$, and  $\mathbf{J}$ generate transformations of the system as a whole,
so we will assume that these are observables for the entire system,
\index{total observables} i.e., the \emph{total} energy, the
\emph{total} momentum, and the \emph{total} angular momentum.  Of
course, these dimensionality considerations are not a proof. The
justification of these choices will become more clear  later, after studying  properties of operators and relations between them.

Using this interpretation and commutators in the Poincar\'e Lie
algebra (\ref{eq:5.50}) - (\ref{eq:5.56}), we immediately obtain
commutation relations between operators of observables. Then we know
which pairs of observables can be simultaneously measured.\footnote{i.e., they have a common basis of eigenvectors, as explained in subsection \ref{ss:states} and Appendix \ref{ss:commuting}} For
example, we see from (\ref{eq:5.53}) that energy is simultaneously
measurable with the momentum and angular momentum.  From
(\ref{eq:5.51}) it is clear that different components of the angular
momentum cannot be measured simultaneously. These facts are
well-known in non-relativistic quantum mechanics. Now we have them
as direct consequences of the principle of relativity and the Poincar\'e group structure.

 From commutators (\ref{eq:5.50}) - (\ref{eq:5.56})  we can also find
formulas for transformations of operators $H$,  $\mathbf{P}$,
$\mathbf{J}$, and $\mathbf{K}$
 from one inertial frame to another.
For example, each vector observable $\mathbf{F}= \mathbf{P},
\mathbf{J}$ or $\mathbf{K}$ transforms  under rotations
as\footnote{see equation (\ref{eq:A.21})}

\begin{eqnarray}
\mathbf{F}(\vec{\phi})
&=& e^{-\frac{i}{\hbar} \mathbf{J}\vec{\phi}}
\mathbf{F}  e^{ \frac{i}{\hbar} \mathbf{J}  \vec{\phi}} =
\mathbf{F} \cos \phi + \frac{\vec{\phi}}{\phi} \left(\mathbf{F} \cdot
\frac{\vec{\phi}}{\phi} \right) (1 - \cos \phi) - \left[\mathbf{F} \times
\frac{\vec{\phi}}{\phi} \right] \sin \phi \nonumber \\
\label{eq:6.1}
\end{eqnarray}

\noindent The boost transformation law for generators of
translations is\footnote{see equations (\ref{eq:boost-momentum}) and
(\ref{eq:boost-energy})}

\begin{eqnarray}
\mathbf{P}(\theta)
&=& e^{-\frac{ic}{\hbar} \mathbf{K}  \vec{\theta}}
\mathbf{P}  e^{ \frac{ic}{\hbar} \mathbf{K}  \vec{\theta}} =
\mathbf{P} + \frac{\vec{\theta}}{\theta} \left(\mathbf{P} \cdot
\frac{\vec{\theta}}{\theta} \right) (\cosh \theta - 1) -
\frac{\vec{\theta}}{c \theta} H \sinh \theta \nonumber \\
\label{eq:6.2} \\
H(\theta) &=& e^{ -\frac{ic}{\hbar} \mathbf{K}  \vec{\theta}} H
e^{ \frac{ic}{\hbar} \mathbf{K}  \vec{\theta}} =
 H \cosh \theta - c \left(\mathbf{P} \cdot
\frac{\vec{\theta}}{\theta}\right) \sinh \theta
\label{eq:6.3}
\end{eqnarray}

\noindent  It also follows from (\ref{eq:5.53}) that energy $H$,
momentum $\mathbf{P}$, and angular momentum  $ \mathbf{J}$ do not
depend on time, i.e., they are \emph{conserved observables}.
\index{conserved observable}

\subsection{Operator of velocity}
\label{ss:velocity}

The operator of velocity is defined as\footnote{The ratio of
operators is well-defined here because $\mathbf{P}$ and $H$ commute
with each other.} (see, e.g., \cite{velocity, velocity2})

\begin{eqnarray}
\mathbf{V} \equiv \frac{\mathbf{P}c^2}{H}  \label{eq:op-of-vel}
\end{eqnarray}

\noindent  Denoting $\mathbf{V} (\theta) $ the velocity measured in the frame of reference moving with the speed $v =
c \tanh \theta$ along the $x$-axis, we obtain

\begin{eqnarray}
V_x(\theta) &=& e^{-\frac{ic}{\hbar} K_x  \theta} \frac{P_xc^2}{H}
e^{\frac{ic}{\hbar} K_x  \theta} = \frac{c^2P_x \cosh \theta - c H
\sinh \theta}{H \cosh \theta - c P_x \sinh \theta} \nonumber \\
&=&
\frac{c^2P_xH^{-1}  - c  \tanh \theta}{ 1 - c P_xH^{-1} \tanh
\theta}
= \frac{V_x -v}{ 1- V_x v/c^2 } \label{eq:6.4} \\
V_y(\theta) &=& \frac{V_y }{ (1 - \frac{V_x}{c} \tanh \theta) \cosh
\theta}
= \frac{V_y \sqrt{1 - v^2/c^2}}{ 1 - V_x v/c^2 }, \\
V_z(\theta) &=& \frac{V_z }{ (1 - \frac{V_x}{c} \tanh \theta) \cosh
\theta} \label{eq:6.5}
= \frac{V_z \sqrt{1 - v^2/c^2}}
{ 1 -V_x v/c^2  } \label{eq:6.6}
\end{eqnarray}

\noindent These formulas  coincide with  the usual relativistic
\emph{law of addition of velocities}. \index{law of addition of
velocities} In the  limit $c \to \infty$ they reduce to the familiar
non-relativistic form

\begin{eqnarray*}
V_x(v) &=& V_x - v \\
V_y(v) &=& V_y  \\
V_z(v) &=& V_z
\end{eqnarray*}

\section{Casimir operators}
\label{sc:casimir}

Observables  $H$,  $\mathbf{P}$, $\mathbf{V}$, and  $\mathbf{J}$
depend on the observer, so they do not represent intrinsic
fundamental properties of the system. For example, if a system has
momentum $\mathbf{p} \in \mathbb{R}^3$ in one frame of reference, then, according to (\ref{eq:6.2}),  there are
other (moving) frames of reference in which momentum takes different values. The measured momentum depends on
both the state of the system and the reference frame in which the
observation is made.  Are there observables which reflect some
\emph{intrinsic} \index{intrinsic properties} observer-independent
properties of the system? If there are such observables,  \index{Casimir operator} then their
operators (they are
called \emph{Casimir operators}) must commute with all generators of the Poincar\'e group.
 It can be shown that the Poincar\'e group has only two
independent Casimir operators  \cite{Fushchich}. Any other Casimir
operator of the Poincar\'e group is a function of these two. So,
there are two invariant physical properties of any physical system.
One such property is \emph{mass}, \index{mass} which is a measure of
the matter content in the system. The corresponding Casimir operator
will be considered in subsection \ref{ss:mass}. Another invariant
property is related to the speed of rotation of the system around
its own axis or \emph{spin}.\footnote{The
invariance of the absolute value of spin is evident for macroscopic
freely spinning objects. Indeed, no matter how we translate, rotate or
boost the frame of reference we cannot stop or reverse the spinning motion of
the observed system.
\label{footn}} \index{spin} The Casimir
 operator corresponding to this invariant property will be found in
subsection \ref{ss:pauli}.

\subsection{4-vectors}
\label{ss:4-vectors2}

Before addressing Casimir operators, let us introduce some
useful definitions. We will call a quadruple of operators
$(\mathcal{A}_0, \mathcal{A}_x, \mathcal{A}_y, \mathcal{A}_z)$ a
\emph{4-vector}\footnote{see also Appendix
\ref{ss:4-dim-rep}}\index{4-vector} if $(\mathcal{A}_x,
\mathcal{A}_y, \mathcal{A}_z)$ is a 3-vector,  $\mathcal{A}_0$ is a
3-scalar, and their commutators with the boost generators are

\begin{eqnarray}
[\mathcal{K}_i, \mathcal{A}_j] &=&  -\frac{i \hbar}{c}\mathcal{A}_0
\delta_{ij} \ \ \ \ \ (i, j = x,y,z) \label{eq:4-vector-a}
\\
\ [\vec{\mathcal{K}}, \mathcal{A}_0] &=& -\frac{i \hbar}{c} \vec{\mathcal{A}}
\label{eq:4-vector-b}
\end{eqnarray}

\noindent Then, it is easy to show that the \emph{4-square}
\index{4-square} $ \tilde{\mathcal{A}}^2 \equiv \mathcal{A}_x^2 + \mathcal{A}_y^2 +\mathcal{A}_z^2 - \mathcal{A}_0^2 $ of the  4-vector $ \tilde{\mathcal{A}}$
  is a
\emph{4-scalar}, \index{4-scalar} i.e.,  it commutes with both
rotations and boosts. For example

\begin{eqnarray*}
[K_x, \tilde{\mathcal{A}}^2] &=&  [K_x,   \mathcal{A}_x^2 + \mathcal{A}_y^2 + \mathcal{A}_z^2 - \mathcal{A}_0^2] \nonumber \\
&=& -\frac{i\hbar}{c} (\mathcal{A}_x\mathcal{A}_0 + \mathcal{A}_0 \mathcal{A}_x -\mathcal{A}_0\mathcal{A}_x - \mathcal{A}_x\mathcal{A}_0)= 0
\end{eqnarray*}

\noindent Therefore,  in order to find the Casimir operators of the
Poincar\'e group we should
 be looking for two functions of the Poincar\'e generators, which are 4-vectors
and, in addition, commute with $H$ and $\mathbf{P}$. Then 4-squares
of these 4-vectors are guaranteed to commute with all Poincar\'e
generators.

\subsection{Operator of mass}
\label{ss:mass}

It follows from (\ref{eq:6.1}) - (\ref{eq:6.3}) that four operators ($H, c\mathbf{P}$) satisfy all
conditions specified in subsection \ref{ss:4-vectors2} for
4-vectors.  Moreover, they commute with each other. These operators are called the
\emph{energy-momentum 4-vector}. \index{energy-momentum 4-vector}
Then
 we can construct the first Casimir invariant -- called the
\emph{mass operator} --\index{mass operator}\index{operator of mass}
as the 4-square  of this 4-vector

\begin{eqnarray}
M = + \frac{1}{c^2} \sqrt{H^2 - P^2 c^2} \label{eq:6.7}
\end{eqnarray}

\noindent The operator of mass must be  Hermitian, therefore we
demand that for any physical system $H^2 - P^2 c^2 \geq 0$, i.e.,
that the spectrum of operator $H^2 - P^2 c^2 $ does not contain
negative values. Honoring the fact that masses of all known physical
systems are non-negative we choose the positive value of the square
root in (\ref{eq:6.7}). Then the relationship between energy,
momentum, and mass takes the form

\begin{eqnarray}
H &=& +  \sqrt{P^2 c^2 + M^2 c^4} \label{eq:6.8}
\end{eqnarray}

\noindent In the non-relativistic limit ($c \to \infty$)  we obtain from equation
(\ref{eq:6.8})

\begin{eqnarray*}
H &\approx& M c^2 + \frac{P^2}{2M}
\end{eqnarray*}

\noindent which is the sum of the famous Einstein's \emph{rest mass
energy} \index{rest mass energy} $E= Mc^2$ and the usual kinetic
energy term $P^2/(2M) $.

\subsection{Pauli-Lubanski 4-vector}
\label{ss:pauli}

The second 4-vector commuting with $H$ and  $\mathbf{P}$ is the
\emph{Pauli-Lubanski operator} \index{Pauli-Lubanski operator} whose
components are defined as\footnote{These definitions involve
products of Hermitian commuting operators, therefore operators $W^0$
and $\mathbf{W}$ are guaranteed to be Hermitian.}

\begin{eqnarray}
W^0 &=& (\mathbf{P} \cdot \mathbf{J})
\label{eq:6.9} \\
\mathbf{W} &=&
\frac{1}{c} H\mathbf{J} - c[\mathbf{P} \times \mathbf{K}]
\label{eq:6.10}
\end{eqnarray}

\noindent Let us check that all required 4-vector properties are,
indeed, satisfied for ($W^0, \mathbf{W}$). We can immediately
observe that

\begin{eqnarray*}
[\mathbf{J},W^0] &=& 0
\end{eqnarray*}

\noindent so $W^0$ is a scalar. Moreover, $W^0$  changes its sign
after changing the sign of $\mathbf{P} $ so it is a
pseudoscalar.
\index{pseudoscalar} $\mathbf{W}$ is a pseudovector, because it does
not change its sign after changing the signs of $\mathbf{K} $ and
$\mathbf{P} $ and

\begin{eqnarray*}
[J_i, W_j] = i \hbar \sum_{k=1}^3 \epsilon_{ijk} W_k
\end{eqnarray*}

\noindent  Let us now check the commutators with boost generators

\begin{eqnarray}
[K_x, W^0] &=& [K_x, P_xJ_x + P_y J_y + P_z J_z] \nonumber\\
& =& -i \hbar \left(\frac{H
J_x}{c^2} - P_y K_z  +  P_z K_y \right)=-\frac{i \hbar}{c} W_x \label{eq:6.11} \\
\ [K_x, W_x] &=& \left[K_x, \frac{HJ_x}{c} - cP_y K_z  +  cP_zK_y \right] \nonumber\\
&=& \frac{i\hbar}{c}
( -P_x J_x - P_y J_y -
P_z J_z) = - \frac{i \hbar}{c} W^0 \label{eq:6.12} \\
\ [K_x, W_y] &=& \left[K_x, \frac{HJ_y}{c} - cP_zK_x  + cP_xK_z \right] \nonumber\\
&=& \frac{i\hbar}{c} (HK_z  - P_xJ_y  - HK_z  + P_xJ_y) = 0 \label{eq:6.13} \\
 \mbox{ }[K_x, W_z] &=& 0 \label{eq:6.14}
\end{eqnarray}

\noindent Putting equations (\ref{eq:6.11}) - (\ref{eq:6.14})
together we obtain the characteristic 4-vector relations
(\ref{eq:4-vector-a}) - (\ref{eq:4-vector-b})

\begin{eqnarray}
[\mathbf{K}, W^0] &=& -\frac{i\hbar}{c} \mathbf{ W} \label{eq:6.15}\\
\  [K_i,
W_j] &=& -\frac{i\hbar}{c} \delta_{ij} W^0 \label{eq:6.16}
\end{eqnarray}

Next we need to verify that commutators  with generators of
translations $(H, \mathbf{P})$ are all zero. First, for $W^0$ we obtain

\begin{eqnarray*}
[W^0, H] &=& [\mathbf{P} \cdot \mathbf{J}, H] = 0 \\ \mbox{ }
[W^0, P_x] &=& [J_x P_x + J_y P_y + J_z P_z, P_x]
= P_y [J_y, P_x] + P_z [J_z, P_x] \\
           &=& -i \hbar P_y P_z + i \hbar P_z P_y = 0
\end{eqnarray*}

\noindent For the vector part $\mathbf{W}$ we obtain

\begin{eqnarray*}
[\mathbf{W}, H] &=& -c[[\mathbf{P} \times \mathbf{K}], H]= -c
[[\mathbf{P}, H] \times \mathbf{K}] -c [\mathbf{P} \times [ \mathbf{K},
H]] = 0
\\ \mbox{ }
 [W_x, P_x] &=& \frac{1}{c}[H J_x, P_x] -
 c[[\mathbf{P} \times \mathbf{K}]_x, P_x] =
 -c[P_yK_z - P_zK_y, P_x] = 0 \\ \mbox{ }
 [W_x, P_y] &=& \frac{1}{c}[H J_x, P_y] -
 c[[\mathbf{P} \times \mathbf{K}]_x, P_y] =
\frac{i \hbar }{c} HP_z - c[P_yK_z - P_zK_y, P_y] \\
&=& \frac{i \hbar}{c}H P_z   - \frac{i \hbar}{c}H P_z  = 0
\end{eqnarray*}

\noindent
 This completes the proof that the 4-square of the
Pauli-Lubanski 4-vector

\begin{eqnarray*}
\Sigma^2 =  \mathbf{W} ^2 - W_0^2
\end{eqnarray*}

\noindent is a Casimir operator. Although operators $(W^0,
\mathbf{W})$ do not have direct physical interpretation, we will
find them very useful in the next section for deriving the operators
of position $\mathbf{R}$ and spin $\mathbf{S}$. For these
calculations we will need commutators between components of the
Pauli-Lubanski 4-vector.  For example,

\begin{eqnarray*}
[W_x, W_y] &=& \left[W_x, \frac{HJ_y}{c}
 + cP_xK_z - cP_z K_x \right] = i \hbar \left(\frac{H W_z}{c} - W^0 P_z \right) \\
\mbox{ } [W_0, W_x] &=&   \left[W_0, \frac{HJ_x}{c}  -
cP_yK_z + cP_z K_y \right] = -i \hbar P_y W_z  + i \hbar P_z W_y \\
&=& -i \hbar [\mathbf{P} \times \mathbf{W}]_x
\end{eqnarray*}

\noindent The above equations are easily generalized for all components

\begin{eqnarray}
[W_i, W_j] &=& \frac{i\hbar}{c}  \sum_{k=1}^3  \epsilon_{ijk} (H W_k
- cW^0 P_k) \label{eq:6.17} \\
\mbox{ }  [W_0, W_j] &=& - i\hbar  [\mathbf{P} \times \mathbf{W}]_j
\label{eq:6.18}
\end{eqnarray}

\section{Operators of spin and position}
\label{sc:spin}

Now we are ready to tackle the problem of finding expressions for spin
and position as functions of the Poincar\'e
group generators \cite{Pryce, Newton_Wigner, Berg,
Jordan}.

\subsection{Physical requirements}
\label{ss:requirements}

We will be looking for the total  \emph{spin operator} \index{spin
operator} $\mathbf{S}$ and the center-of-mass \emph{position
operator} \index{position operator} $\mathbf{R}$ which have the
following natural properties:

\begin{itemize}

\item[(I)]  Owing to the similarity between spin and angular momentum,\footnote{It is often stated that
spin is a purely quantum-mechanical observable which does not have a
classical counterpart. We do not share this point of view. From classical mechanics we know that the total angular
momentum of a body is a sum of two parts. The first part is the
angular momentum resulting from the linear movement of the body as a
whole with respect to the observer. The second part is related to
the rotation of the body around its own axis, or  spin. The only
significant difference between classical and quantum intrinsic
angular momenta (spins) is that the latter has a discrete spectrum,
while the former is continuous. In addition, components of the quantum spin operator do not commute with each other.} we  demand that $\mathbf{S}$
is a pseudovector
(just like $\mathbf{J}$)

\begin{eqnarray*}
[J_j, S_i] = i\hbar \sum_{k=1}^3 \epsilon_{ijk} S_k
\end{eqnarray*}

\item[(II)]  and  that components of
$\mathbf{S}$ satisfy the same commutation relations as components of
$\mathbf{J}$ (\ref{eq:5.51})

\begin{eqnarray}
[S_i, S_j] = i\hbar \sum_{k=1}^3 \epsilon_{ijk} S_k
\label{eq:6.19}
\end{eqnarray}

\item[(III)] We also demand that spin can be measured simultaneously with
momentum

\begin{eqnarray*}
[\mathbf{P}, \mathbf{S}] = 0
\end{eqnarray*}

\item[(IV)] and with position

\begin{eqnarray}
[\mathbf{R}, \mathbf{S}] = 0
\label{eq:6.20}
\end{eqnarray}

\item[(V)] From the physical meaning of $\mathbf{R}$ it follows that
space translations of the observer simply shift the values of
position.

\begin{eqnarray*}
e^{-\frac{i}{\hbar}P_x a} R_x e^{\frac{i}{\hbar}P_x a} &=& R_x - a \\
e^{-\frac{i}{\hbar}P_x a} R_y e^{\frac{i}{\hbar}P_x a} &=& R_y \\
e^{-\frac{i}{\hbar}P_x a} R_z e^{\frac{i}{\hbar}P_x a} &=& R_z
\end{eqnarray*}

This implies the following commutation relations\footnote{Note that in most textbooks this commutator is simply postulated. In our approach it is derived from the definition of $\mathbf{P}$ as the generator of space translations. }

\begin{eqnarray}
 [R_i, P_j] = i\hbar  \delta_{ij}
\label{eq:6.21}
\end{eqnarray}

\item[(VI)] Finally, we will assume that  position is a true
vector

\begin{eqnarray}
 [J_i, R_j] = i\hbar \sum_{k=1}^3 \epsilon_{ijk} R_k
\label{eq:6.22}
\end{eqnarray}

\end{itemize}

\subsection{Spin operator}
\label{ss:spin-oper}

Now we would like to make the following  guess about the form of the
spin operator\footnote{Note that operator $\mathbf{S} $ has the mass
operator $M$ in the denominator, so expressions (\ref{eq:6.23}) and
(\ref{eq:6.24}) have mathematical sense only for systems with
a strictly positive mass spectrum. \label{foot:positive}}

\begin{eqnarray}
\mathbf{S}
&=& \frac{\mathbf{W}}{Mc} - \frac{ W_0 \mathbf{P}}{M(Mc^2 + H)}
\label{eq:6.23}\\
&=& \frac{H \mathbf{J} }{Mc^2} - \frac{[\mathbf{P} \times
\mathbf{K}] } {M}  - \frac{\mathbf{P} (\mathbf{P}
\cdot \mathbf{J})}{(H+Mc^2)M}
\label{eq:6.24}
\end{eqnarray}

\noindent which is a pseudovector commuting with $\mathbf{P}$ as required by
the above conditions (I) and (III).
Next we are going to verify that
condition (II) is also valid for this operator.
 To calculate the commutators (\ref{eq:6.19}) between spin components we  denote

\begin{eqnarray}
 F \equiv - \frac{1}{M(Mc^2+H)}
\label{eq:6.25}
\end{eqnarray}

\noindent use  commutators (\ref{eq:6.17}) and (\ref{eq:6.18}), the equality

\begin{eqnarray}
(\mathbf{P} \cdot \mathbf{W}) &=& \frac{1}{c} H (\mathbf{P} \cdot
\mathbf{J}) = \frac{1}{c} H W_0 \label{eq:4.25a}
\end{eqnarray}

\noindent  and equation (\ref{eq:A.17}). Then

\begin{eqnarray*}
[S_x, S_y] &=& \left[FW_0P_x + \frac{W_x}{Mc}, FW_0P_y
+ \frac{W_y}{Mc}\right] \\
&=&i \hbar\left(- \frac{ FP_x [\mathbf{P} \times \mathbf{W}]_y}{Mc}
+ \frac{FP_y [\mathbf{P} \times
\mathbf{W}]_x}{Mc}  +  \frac{HW_z -c W_0 P_z}{M^2c^3} \right)\\
&=&i \hbar\left(-  \frac{F[\mathbf{P} \times [\mathbf{P} \times
\mathbf{W}]]_z}{Mc} + \frac{HW_z - cW_0 P_z}{M^2c^3} \right)\\
&=&i \hbar \left(-\frac{F (P_z (\mathbf{P} \cdot \mathbf{W})
- W_z P^2)}{Mc} +  \frac{HW_z - cW_0 P_z}{M^2c^3} \right) \\
&=&i \hbar \left(- \frac{F(P_z HW^0 c^{-1}
- W_z P^2)}{Mc} +  \frac{HW_z - cW_0 P_z}{M^2c^3} \right) \\
&=& i \hbar W_z  \left(\frac{P^2F}{Mc} + \frac{H}{M^2c^3} \right) +
i \hbar P_zW^0 \left(-\frac{HF}{M c^2}  - \frac{1}{M^2c^2} \right)
\end{eqnarray*}

\noindent For the expressions in parentheses we obtain

\begin{eqnarray*}
\frac{P^2F}{Mc} + \frac{H}{M^2c^3} &=& -\frac{P^2}{M^2c(Mc^2 + H)} +
\frac{H}{M^2 c^3} =
\frac{H(Mc^2 + H) - P^2 c^2}{M^2 c^3(Mc^2 + H)} \\
&=& \frac{H(Mc^2 + H) - (Mc^2 + H)(H- Mc^2 )}{M^2 c^3(Mc^2 + H)} =
\frac{1}{Mc} \\
-\frac{HF}{M c^2}  - \frac{1}{M^2c^2} &=&
\frac{H}{M^2 c^2 (Mc^2+H)} - \frac{1}{M^2 c^2 } \\
&=& \frac{H - (Mc^2+H)}{M^2 c^2 (Mc^2+H)} = -\frac{1}{M (Mc^2+H)} =
F
\end{eqnarray*}

\noindent Thus,  property (\ref{eq:6.19}) follows

\begin{eqnarray*}
[S_x, S_y] &=&i \hbar \left( \frac{W_z}{Mc} + F W^0 P_z\right) = i \hbar S_z
\end{eqnarray*}

 Let us now prove  that spin squared $\mathbf{S}^2$ is a function of
$M^2$ and $\Sigma^2$, i.e., a Casimir operator

\begin{eqnarray*}
 \mathbf{S}^2 &=& \left(\frac{\mathbf{W}}{Mc} + W_0 \mathbf{P}F \right)^2
 = \frac{W^2}{M^2c^2} + \frac{ 2W_0 F (\mathbf{P} \cdot
\mathbf{W})}{Mc}  +  W_0^2 P^2F^2 \\
&=&
\frac{W^2}{M^2c^2} +
W_0^2 F
 \left(\frac{2  H }{Mc^2} +   P^2F \right)= \frac{W^2}{M^2c^2} +  W_0^2 F
  \frac{2H (Mc^2+H) - P^2c^2}{Mc^2(Mc^2+H)} \\
  &=& \frac{W^2}{M^2c^2} - W_0^2
  \frac{  H^2 + 2HMc^2 + M^2 c^4}{M^2c^2(Mc^2+H)^2} = \frac{W^2-  W_0^2 }{M^2c^2} \\
   &=& \frac{\Sigma^2 }{M^2c^2}
\end{eqnarray*}

\noindent So far we guessed one particular form of the spin operator and
verified that the required properties are
satisfied. In subsection \ref{ss:unique-spin} we will
demonstrate that this  is the unique expression satisfying all
conditions from subsection \ref{ss:requirements}.

Sometimes it is convenient to use the operator of spin's projection on momentum
$(\mathbf{S} \cdot \mathbf{P})/P$ that is called \emph{helicity}. \index{helicity}
This operator is related to the 0-th component of the Pauli-Lubanski 4-vector

\begin{eqnarray}
(\mathbf{P} \cdot \mathbf{S}) = \frac{(\mathbf{P} \cdot \mathbf{J})
H}{Mc^2} -
\frac{P^2 (\mathbf{P} \cdot \mathbf{J})(H-Mc^2)}{P^2 Mc^2} = (\mathbf{P} \cdot \mathbf{J}) = W_0 \label{eq:helicity}
\end{eqnarray}

\subsection{Position operator}
\label{ss:position}

Now we are going to switch to the derivation of the position
operator. Here we will follow a  route similar to that for $\mathbf{S}$: we will first guess
the form of the operator $\mathbf{R}$ and then in subsection
\ref{ss:unique-position} we will prove that this is the unique
expression satisfying all requirements from subsection
\ref{ss:requirements}. Our guess for $\mathbf{R}$ is  the
\index{Newton-Wigner position operator}\index{position operator}
\emph{Newton-Wigner position operator}\footnote{Similarly to the
operator of spin, the
Newton-Wigner position operator is defined only for systems whose
mass spectrum is strictly positive. } \cite{Pryce, Newton_Wigner,
Berg, Jordan, Candlin}

\begin{eqnarray}
\mathbf{R} &=&
-\frac{c^2}{2}(H^{-1}\mathbf{K} + \mathbf{K}H^{-1}) -
\frac{c^2[\mathbf{P} \times \mathbf{S}]}{H(Mc^2+H)}
\label{eq:6.26}\\
&=&
- \frac{c^2} {H}\mathbf{K} -  \frac{i \hbar c^2 \mathbf{P}}{2H^2} -
\frac{c[\mathbf{P} \times \mathbf{W}]}{MH(Mc^2+H)}
\label{eq:6.27}
\end{eqnarray}

\noindent which is a  true vector
having properties (V) and (VI), e.g.,

\begin{eqnarray*}
[R_x, P_x] &=& -\frac{c^2}{2} [(H^{-1}K_x + K_xH^{-1}), P_x] =
\frac{i \hbar}{2}(H^{-1}H + HH^{-1}) = i \hbar
\\
\mbox{ }[R_x, P_y] &=& -\frac{c^2}{2} [(H^{-1}K_x + K_xH^{-1}), P_y]
= 0
\end{eqnarray*}

\noindent Let us now calculate\footnote{Note that $
[K_xP_y - K_y P_x, H] =
-i \hbar(P_x P_y - P_y P_x) = 0
$,
therefore $[\mathbf{K} \times \mathbf{P}]$ commutes with $H$, and operator
$ H^{-1}
[\mathbf{K} \times \mathbf{P}]$ is Hermitian.}

\begin{eqnarray*}
&\mbox{ }& \mathbf{J} - [\mathbf{R} \times \mathbf{P}] =
\mathbf{J} + \frac{c^2} {H} [\mathbf{K} \times \mathbf{P}]
+ \frac{c^2[[\mathbf{P} \times \mathbf{S}] \times \mathbf{P}] }{H(Mc^2+H)} \\
&=& \mathbf{J} + \frac{c^2} {H}
[\mathbf{K} \times \mathbf{P}]
- \frac{c^2 ( \mathbf{P} (\mathbf{P} \cdot \mathbf{S})
- \mathbf{S} P^2)}{H(Mc^2+H)} \\
&=& \mathbf{J} + \frac{c^2} {H}
[\mathbf{K} \times \mathbf{P}]
- \frac{ ( c^2\mathbf{P} (\mathbf{P} \cdot \mathbf{S})
- \mathbf{S} (H-Mc^2)(H+Mc^2))}{H(Mc^2+H)}  \\
&=& \mathbf{J} + \frac{c^2} {H}
[\mathbf{K} \times \mathbf{P}] + \mathbf{S}
- \frac{c^2 \mathbf{P} (\mathbf{P} \cdot \mathbf{S}) }{H(Mc^2+H)}
- \frac{Mc^2 }{H} \mathbf{S} \\
&=& \mathbf{J} + \frac{c^2} {H} [\mathbf{K} \times \mathbf{P}] + \mathbf{S}
- \frac{c^2 \mathbf{P} (\mathbf{P} \cdot \mathbf{S})}{H(Mc^2+H)}
-\mathbf{J} + \frac{c^2 \mathbf{P} (\mathbf{P} \cdot
\mathbf{J})}{H(Mc^2+H)}
+ \frac{c^2}{H} [\mathbf{P} \times \mathbf{K}]\\
&=& \mathbf{S}
\end{eqnarray*}

\noindent Therefore, just as  in classical physics, the total
angular momentum \index{angular momentum} is a sum of two parts: the \emph{orbital angular
momentum} \index{orbital angular momentum} $[\mathbf{R} \times
\mathbf{P}]$ and
 the \emph{intrinsic angular momentum}\index{intrinsic angular
momentum}\index{spin} or spin $\mathbf{S}$

\begin{eqnarray*}
\mathbf{J} = [\mathbf{R} \times \mathbf{P}] + \mathbf{S}
\end{eqnarray*}

\noindent Next we can check that condition (IV)  is satisfied as well, e.g.,

\begin{eqnarray*}
[S_x, R_y] &=& [J_x - [\mathbf{R} \times \mathbf{P}]_x, R_y] = i \hbar R_z - [P_yR_z - P_zR_y, R_y] \\
 &=& i \hbar R_z -i \hbar R_z = 0
\end{eqnarray*}

\bigskip

\begin{theorem}
\label{Theorem6.1} All components of the position operator commute
with each other:  $[R_i, R_j] = 0$.
\end{theorem}
\begin{proof}
   First, we calculate the commutator $[HR_x, HR_y] $
which is related to $[R_x, R_y] $ via formula\footnote{here we used (\ref{eq:A.37})}

\begin{eqnarray}
[HR_x, HR_y] &=& [HR_x, H] R_y + H[HR_x, R_y] \nonumber \\
&=& H [R_x, H] R_y + H[H, R_y]R_x +  H^2[R_x, R_y] \nonumber \\
&=& i \hbar c^2 (P_x R_y - R_yP_x) +  H^2[R_x, R_y] \nonumber \\
 &=& i \hbar
c^2 [\mathbf{P} \times \mathbf{R}]_z  +  H^2[R_x, R_y]
\nonumber \\
&=& -i \hbar c^2 J_z + i \hbar c^2 S_z+  H^2[R_x, R_y]
\label{eq:6.28}
\end{eqnarray}

\noindent Using formula (\ref{eq:6.27}) for the position operator, we obtain

\begin{eqnarray*}
&\ & [HR_x, HR_y] \\
&=& \left[-c^2 K_x -
\frac{i \hbar c^2P_x}{2H} + c F[\mathbf{P} \times \mathbf{W}]_x, -c^2 K_y -
 \frac{i \hbar c^2P_y}{2H} +c F[\mathbf{P} \times
\mathbf{W}]_y \right]
\end{eqnarray*}

\noindent Non-zero contributions to this commutator are

\begin{eqnarray}
[-c^2 K_x ,-c^2 K_y ] &=& c^4 [K_x , K_y ] = -i \hbar c^2 J_z \label{eq:KxKy1} \\
\mbox{ }  \left[ -  \frac{i \hbar c^2P_x}{2H} , -c^2 K_y
\right] &=& -\frac{i \hbar c^2}{2}   \left[ K_y,  \frac{P_x}{H}
\right] =  \frac{\hbar^2 c^4P_y P_x}{2H^2} \label{eq:KxKy2}
\\
\mbox{ }\left[-c^2 K_x , -
\frac{i \hbar c^2P_y}{2H}\right] &=& - \frac{\hbar^2 c^4P_y
P_x}{2H^2} \label{eq:KxKy3}
\end{eqnarray}

\begin{eqnarray*}
&\mbox{ }&[-c^2 K_x,  c F[\mathbf{P} \times
\mathbf{W}]_y] \nonumber \\
&=& \frac{c^3}{M} \left[ K_x,
\frac{P_z W_x - P_x W_z}{H+Mc^2} \right] \nonumber \\
&=& \frac{c^3}{M}\left(-\frac{P_z W_x - P_x W_z}{(H+Mc^2)^2} [K_x,
H] +\frac{P_z [ K_x, W_x] }{H+Mc^2} -\frac{ [ K_x, P_x] W_z}{H+Mc^2}
\right) \nonumber
\\
&=& i \hbar c^3(MF^2(P_z W_x - P_x W_z) P_x + FP_z W_0c^{-1} - FH
W_z c^{-2})
\end{eqnarray*}

\begin{eqnarray*}
 & \mbox{ }& [ c F[\mathbf{P} \times
\mathbf{W}]_x, -c^2 K_y ] \nonumber \\
&=&- c^3(- MF^2(P_y W_z - P_z W_y) [K_y, H]
 +F P_z [ K_y, W_y]  - F[ K_y, P_y] W_z) \nonumber
\\
&=& i \hbar c^3(MF^2(P_y W_z - P_z W_y) P_y + FP_z W_0 c^{-1} - F H
W_zc^{-2})
\end{eqnarray*}

\noindent Adding together two last results
and using (\ref{eq:4.25a}) we obtain

\begin{eqnarray}
&\mbox{ }& [-c^2 K_x, c F[\mathbf{P} \times \mathbf{W}]_y] + [ cF
[\mathbf{P} \times \mathbf{W}]_x, -c^2 K_y
] \nonumber\\
&=& i \hbar c^3(MF^2 [\mathbf{P} \times [\mathbf{P} \times
\mathbf{W}]]_z
+2FP_z W_0 c^{-1} - 2 FH W_zc^{-2}) \nonumber\\
&=& i \hbar c^3(MF^2(P_z (\mathbf{P} \cdot \mathbf{W}) - W_z P^2)
+ 2FP_z W_0c^{-1} -2 FH W_zc^{-2}) \nonumber\\
&=& i \hbar c^3(MF^2(P_z HW_0c^{-1} - W_z P^2 )
+2FP_z W_0 c^{-1} -2 FH W_zc^{-2}) \nonumber\\
&=& i \hbar c^2 MF^2P_z W_0 (H - 2(H+Mc^2)) +i \hbar c MF^2 W_z
(- (H-Mc^2)(H+Mc^2) \nonumber \\
&\mbox{ }&+ 2 H (H+Mc^2)) \nonumber\\
&=& i \hbar c^3P_z W_0 (Fc^{-1} - M^2 F^2 c  ) +\frac{i \hbar
cW_z}{M} \label{eq:6.32}
\end{eqnarray}

\noindent One more commutator is

\begin{eqnarray}
&\mbox{ }&  [c F[\mathbf{P} \times \mathbf{W}]_x,
c F[\mathbf{P} \times \mathbf{W}]_y] \nonumber\\
&=&
 c^2F^2[P_y W_z - P_z W_y, P_z W_x - P_x W_z] \nonumber\\
&=&
 c^2F^2(P_zP_y [W_z, W_x] - P_z^2 [W_y, W_x] + P_x
P_z [W_y, W_z]) \nonumber\\
&=&
 i \hbar cF^2(P_zP_y (HW_y - c W_0P_y) + P_z^2
(HW_z - c W_0P_z) + P_x
P_z (H W_x - c W_0 P_x)) \nonumber\\
&=&
 i \hbar cF^2(-W_0cP_z(P_x^2 + P_y^2 + P_z^2)
+ HP_z(P_x W_x + P_yW_y + P_zW_z)) \nonumber\\
&=&
 i \hbar c^2F^2(-W_0P_zP^2
+ H^2P_z W_0/c^2) \nonumber\\
&=&
 i \hbar F^2 W_0 (-P_z(H^2 -M^2 c^4)
+ H^2P_z )\nonumber\\
&=&
 \frac{i \hbar c^4 W_0 P_z  }{(H+Mc^2)^2 }
\label{eq:6.31}
\end{eqnarray}

\noindent Now we collect all terms  (\ref{eq:KxKy1}) - (\ref{eq:6.31}) and
 finally calculate

\begin{eqnarray*}
[HR_x, HR_y] &=& -i \hbar c^2 J_z + i \hbar c^3P_z W_0 (F -M^2F^2c )
+\frac{i \hbar cW_z}{M}
 + i \hbar c^4 M^2 F^2 W_0 P_z   \\
&=& -i \hbar c^2 J_z  + i \hbar c^2 \left(-\frac{P_z W_0
}{M(H+Mc^2)} +\frac{W_z}{Mc}\right) = -i \hbar c^2 J_z  + i \hbar
c^2 S_z
\end{eqnarray*}

\noindent Comparing this with equation (\ref{eq:6.28}) we obtain

\begin{eqnarray*}
 H^2[R_x, R_y] &=& 0
\end{eqnarray*}

\noindent Operator $H^2 = M^2c^4 + P^2c^2$ has no zero eigenvalues, because we have assumed that $M$ is strictly positive. Thus we get the desired result

\begin{eqnarray*}
\ [R_x, R_y] &=& 0
\end{eqnarray*}
\end{proof}

\subsection{Alternative set of basic operators}
\label{ss:alternative}

So far, our plan was to construct  operators of observables from
 10 basic generators \{$\mathbf{P}$, $\mathbf{J}$,
$\mathbf{K}$,  $H$\}. However, this set of operators is sometimes
difficult to use in calculations due to rather complicated
commutation relations in the Poincar\'e Lie algebra (\ref{eq:5.50})
- (\ref{eq:5.56}). For systems with a strictly positive spectrum of
the mass operator,  we may find it more convenient to use an
alternative set of basic operators \{$\mathbf{P}$, $\mathbf{R}$,
 $\mathbf{S}$,  $M$\} whose commutation relations are much simpler

\begin{eqnarray}
 [\mathbf{P}, M] &=& [\mathbf{R}, M] = [\mathbf{S}, M] = [R_i,
R_j]= [P_i, P_j] = 0 \label{eq:6.33} \\
\mbox{ } [R_i, P_j] &=& i \hbar \delta_{ij} \nonumber \\
\mbox{ } [\mathbf{P}, \mathbf{S}] &=& [\mathbf{R}, \mathbf{S}] =0
\nonumber
\\
\mbox{ } [S_i, S_j] &=& i \hbar \sum_{k=1}^{3}\epsilon_{ijk}S_k
\end{eqnarray}

\noindent  Summarizing our previous results, we can express
operators in this set through generators of the Poincar\'e group\footnote{Operator $\mathbf{P}$ is the same in both sets.}

\begin{eqnarray}
\mathbf{R} &=&
-\frac{c^2}{2}(H^{-1}\mathbf{K} + \mathbf{K}H^{-1}) -
\frac{c[\mathbf{P} \times \mathbf{W}]}{MH(Mc^2+H)}
\label{eq:6.35}\\
\mathbf{S} &=& \mathbf{J} -
[\mathbf{R} \times  \mathbf{P}]
\label{eq:6.36}\\
M &=& +\frac{1}{c^2} \sqrt{H^2 - P^2c^2} \label{eq:6.37}
\end{eqnarray}

\noindent Conversely, we can express generators of the Poincar\'e
group $\{ \mathbf{P}, \mathbf{K}, \mathbf{J}, H \}$ through
operators $\{ \mathbf{P}, \mathbf{R}, \mathbf{S}, M \}$.  For the
energy and angular momentum  we obtain

\begin{eqnarray}
H &=& +\sqrt{M^2c^4 + P^2 c^2}
\label{eq:6.38}\\
\mathbf{J} &=& [\mathbf{R} \times \mathbf{P}] + \mathbf{S}
\label{eq:6.39}
\end{eqnarray}

\noindent and the expression for the boost operator is \index{boost
operator}

\begin{eqnarray}
  &-&\frac{1}{2c^2}(\mathbf{R}H
+ H\mathbf{R}) - \frac{[\mathbf{P} \times \mathbf{S}]}{Mc^2+H}
\nonumber \\
=  &-&\frac{1}{2} \left(-\frac{1}{2}(H^{-1}\mathbf{K}H +
\mathbf{K}) -
\frac{[\mathbf{P} \times \mathbf{S}]}{Mc^2+H} \right) \nonumber\\
&-&\frac{1}{2} \left( -\frac{1}{2}(\mathbf{K} +
H\mathbf{K}H^{-1}) - \frac{[\mathbf{P} \times \mathbf{S}]}{Mc^2+H}
\right) - \frac{[\mathbf{P}
\times \mathbf{S}]}{Mc^2+H} \nonumber\\
= &\mbox{ }& \frac{1}{4} (H^{-1}\mathbf{K}H + \mathbf{K}+ \mathbf{K} +
H\mathbf{K}H^{-1}) \nonumber\\
= &\mbox{ }& \mathbf{K} -\frac{i \hbar}{4} (H^{-1}\mathbf{P} -
\mathbf{P}H^{-1}) \nonumber\\
= &\mbox{ }& \mathbf{K} \label{eq:6.40}
\end{eqnarray}

\noindent These two sets
 provide equivalent
descriptions of Poincar\'e invariant theories. Any function of
operators from the set $\{\mathbf{P}, \mathbf{J}, \mathbf{K}, H\}$
can be expressed as a function of operators from the set
$\{\mathbf{P}, \mathbf{R}, \mathbf{S}, M\}$, and \emph{vice versa}.
We will use this property in subsections \ref{ss:canonical-form},
\ref{ss:bakamjian}, and \ref{bakam-point}.

\subsection{Canonical form and ``power'' of operators}
\label{ss:canonical-form}

In this subsection,
we would like to mention
some  mathematical facts which will be helpful in further work.
When performing calculations with functions of Poincar\'e generators, we meet
a
problem that the same operator can be expressed in many
equivalent functional forms. For example, according to (\ref{eq:5.56})
 $K_xH$ and $HK_x -i\hbar P_x$
are two forms of the same operator.  To solve this non-uniqueness
problem, we will agree to write operator factors always in the
\emph{canonical form}, \index{canonical form of operator} i.e., from
left to right in the following order:\footnote{Since $H, P_x, P_y$, and $P_z$ commute with each other, the part of the operator
depending on these factors can be written as an ordinary function of
commuting arguments $C(P_x, P_y, P_z, H)$, whose order is
irrelevant. }

\begin{eqnarray}
C(P_x, P_y, P_z, H), J_x, J_y, J_z, K_x, K_y, K_z
\label{eq:6.41}
\end{eqnarray}

\noindent Consider, for example, the non-canonical product $K_y P_y J_x$. To bring it to the canonical
form,  we first move factor $P_y$ to the leftmost position using
(\ref{eq:5.55})

\begin{eqnarray*}
K_y P_y J_x &=& P_y K_y J_x + [K_y, P_y] J_x = P_y K_y J_x - \frac{i
\hbar}{c^2} H J_x
\end{eqnarray*}

\noindent The second term on the right hand side is already in the canonical
form, but the
first term is not. We need to switch factors  $J_x$
and $K_y$ there:

\begin{eqnarray}
K_y P_y J_x &=& P_y J_x K_y  + P_y  [K_y, J_x] - \frac{i \hbar}{c^2} H
J_x  \nonumber\\
&=& P_y J_x K_y  -i \hbar  P_y  K_z - \frac{i \hbar}{c^2} H J_x
\label{eq:6.43}
 \end{eqnarray}

\noindent Now all terms in (\ref{eq:6.43}) are in the canonical form.

The procedure for bringing a general operator to the canonical form
is not more difficult than in the above example. If we call the
original operator the \emph{primary term}, \index{primary term} then
this procedure can be formalized as the following sequence of steps:
 First we transform the primary term itself to the canonical
form. We do that by switching the order of pairs of neighboring
factors if they occur in the ``wrong'' order. Let us call them the
``left factor'' $L$ and the ``right factor'' $R$.
 If $R$
happens to commute with $L$, then such a change  has no other
effect. If  $R$ does not commute with $L$, then the result of the
switch is $LR \to RL + [L,R]$. This means that apart from switching $L \leftrightarrow R$
we must also add another \emph{secondary term} \index{secondary term} to
the original expression. The secondary term is obtained from  the primary term by
replacing the  product $L R$ with the commutator $[L,
R]$.\footnote{The second and third terms on the right hand side of
(\ref{eq:6.43}) are secondary.} At the end of the first step we have
all factors in the primary term in the canonical order. If during
this process all commutators $[L,R]$ were zero, then we are done. If
there were nonzero commutators, then we have a number of additional
secondary terms. In the general case, these terms are not yet in the
canonical form, and the above procedure should be repeated for them
resulting in \emph{tertiary}, \index{tertiary term}  etc. terms
until all terms are in the canonical order.

Then, for each operator there is a unique representation
  as a sum of  terms in the canonical form

\begin{eqnarray}
F = C^{00} + \sum_{i=1}^3 C^{10}_iJ_i +  \sum_{i=1}^3C^{01}_iK_i +
 \sum_{i,j=1}^3 C^{11}_{ij}J_iK_j +
\sum_{i,j=1; i \leq j}^3 C^{02}_{ij}K_iK_j
 + \ldots \nonumber \\
\label{eq:6.44}
\end{eqnarray}

\noindent where $C^{\alpha \beta} =C^{\alpha \beta}(P_x, P_y, P_z, H) $ are
 functions of translation generators.

We will also find useful the notion of \emph{power} \index{power
of operator} \index{\emph{pow}($\ldots$), power of operator} of terms in (\ref{eq:6.44}). We will denote $pow(A)$  the number of factors $J$ and/or $K$ in the
term $A$. For example, the first term on the right hand side of
(\ref{eq:6.44}) has power 0. The second and third terms have power
1, etc. The power of a general operator (\ref{eq:6.44})
 is defined as the maximum  power among terms in
$F$. For operators considered earlier in this chapter, we have

\begin{eqnarray*}
pow(H) &=&
pow(\mathbf{P}) = pow(\mathbf{V}) = 0 \\
pow(W^0) &=&
pow(\mathbf{W}) = pow(\mathbf{S}) = pow(\mathbf{R}) = 1
\end{eqnarray*}

\bigskip

\begin{lemma} \label{Lemma6.2} If $L$ and $R$ are operators from the list
(\ref{eq:6.41}) and $[L,R] \neq 0$, then

\begin{eqnarray}
pow([L,R]) = pow(L) + pow(R) -1  \label{eq:pow}
\end{eqnarray}
\end{lemma}
\begin{proof}
 The commutator $[L,R]$ is non-zero in two cases.

1. $pow(L) = 1$ and $pow(R) = 0$ (or, equivalently,  $pow(L) = 0$ and $pow(R)
= 1$).
From commutation relations (\ref{eq:5.50}), (\ref{eq:5.53}),
(\ref{eq:5.55}), and (\ref{eq:5.56}),
 it follows that non-vanishing
commutators between Lorentz generators and translation generators are
functions of translation generators, i.e., have zero power. The same
is true for commutators between Lorentz generators and arbitrary
functions of translation generators $C(P_x, P_y, P_z, H)$.

2. If $pow(L) = 1$ and $pow(R) = 1$, then $pow([L,R]) =1$ follows
directly from commutators (\ref{eq:5.51}), (\ref{eq:5.52}) and
(\ref{eq:5.54}).
\end{proof}
\bigskip

One can easily see that (\ref{eq:pow}) holds for more complex operators as well. For example, if $C$ and $D$ are two functions of $P_x, P_y, P_z, H$, then using (\ref{eq:A.37}) and $[C,D]=0$ we obtain

\begin{eqnarray*}
[CJ_x, DJ_y] = [CJ_x, D]J_y + D[CJ_x,J_y] = C[J_x, D]J_y + DC [J_x, J_y] + DJ_x[C,J_y]
\end{eqnarray*}

\noindent The power of the right hand side is 1, in agreement with (\ref{eq:pow}).
Similarly, one can see that $pow([CL, DR]) =1$ if $L$ and $R$ are any non-commuting components of $\mathbf{J}$ or $\mathbf{K}$.  This proves formula (\ref{eq:pow}) for all operators $L$ and $R$ having power 0 or 1. Let us now try to extend this result to general operators.

The primary term for the product of two terms $AB$  has exactly the
same number of Lorentz generators as the original operator, i.e.,
$pow(A) + pow(B) $.

\bigskip

\begin{lemma} \label{Lemma6.3} For two terms $A$ and $B$, either secondary
term in the product $AB$  is zero or its power is equal to
 pow(A) + pow(B) -1.
 \end{lemma}
\begin{proof}   Each secondary term results from replacing a
product of two generators $LR$ in the primary term with their
commutator $[L,R]$. According to Lemma \ref{Lemma6.2}, if $[L,R]
\neq 0$ such a replacement decreases the power of the term by one unit.
\end{proof}
\bigskip

\noindent  The
powers of tertiary and higher order terms are less than the power of
secondary terms. Therefore, for any product $AB$ its power is determined by the primary term only

\begin{eqnarray*}
pow(AB) &=& pow(BA) = pow(A) + pow(B)
\end{eqnarray*}

\noindent This implies

\bigskip

\begin{theorem}\footnote{This theorem was used by Berg in
ref. \cite{Berg}.}
 \label{Theorem6.4} For two non-commuting terms $A$ and $B$

\begin{eqnarray*}
pow([A,B]) = pow(A) + pow(B) -1
\end{eqnarray*}
\end{theorem}
\begin{proof}
  In the commutator $AB - BA$, the primary term of $AB$
cancels out the primary term of $BA$. If $[A,B] \neq 0$, then the
secondary terms  do not cancel.  Therefore, there is at least one
non-zero secondary term whose power is $pow(A) + pow(B) -1$
according to Lemma \ref{Lemma6.3}.
\end{proof}
\bigskip

Having at our disposal basic operators $\mathbf{P}$, $\mathbf{R}$,
$\mathbf{S}$, and $M$ we can form a number of Hermitian scalars,
vectors, and tensors, which are classified in  table \ref{table:4.1}
according to their true/pseudo character and power:

\begin{table}[h]
\caption{Scalar,  vector, and tensor functions of basic operators}
\begin{tabular*}{\textwidth}{@{\extracolsep{\fill}}cccc}
 \hline
              & power 0     & power 1                      & power 2 \cr
 \hline
 True scalar & $P^2$; $M$  & $\mathbf{P}\cdot \mathbf{R} + \mathbf{R}\cdot
\mathbf{P}$ & $R^2$; $S^2$  \cr
 Pseudoscalar &             & $\mathbf{P}\cdot \mathbf{S}$ & $\mathbf{R}\cdot
\mathbf{S}$  \cr
 True vector &$\mathbf{P}$ & $\mathbf{R}$; $[\mathbf{P}\times
\mathbf{S}]$ & $[\mathbf{R}\times \mathbf{S}]$ \cr
 Pseudovector &             & $\mathbf{S}$; $[\mathbf{P}\times \mathbf{R}]$ &
\cr
 True tensor & $P_iP_j$     & $\sum_{k=1}^3 \epsilon_{ijk}S_k$; $P_iR_j+R_jP_i$
& $S_iS_j+S_jS_i$; $R_iR_j$\cr
 Pseudotensor &$\sum_{k=1}^3 \epsilon_{ijk} P_k$ & $\sum_{k=1}^3
\epsilon_{ijk} R_k$; $P_iS_j$ & $R_iS_j$ \cr
 \hline
\end{tabular*}
\label{table:4.1}
\end{table}

\subsection{Uniqueness of the spin operator}
\label{ss:unique-spin}

 Let us
now prove that (\ref{eq:6.23}) is the unique spin operator
satisfying conditions (I) - (IV) from subsection
\ref{ss:requirements}. Suppose that there is another spin operator
$\mathbf{S}'$ satisfying the same conditions. Denoting the power of the
spin components  by $p = pow(S'_x) = pow(S'_y) = pow(S'_z) $ we
obtain from (\ref{eq:6.19}) and Theorem \ref{Theorem6.4}

\begin{eqnarray*}
pow([S'_x,S'_y]) &=& pow(S'_z) \\
2p -1 &=& p
\end{eqnarray*}

\noindent Therefore, the components of  $\mathbf{S}'$ must have power
1.
The most general form of a pseudovector operator having power 1
 can be deduced from Table \ref{table:4.1}

\begin{eqnarray*}
\mathbf{S}' = b(M, P^2) \mathbf{S} +
f(M, P^2) [\mathbf{P} \times \mathbf{R}] + e(M, P^2) (\mathbf{S} \cdot
\mathbf{P}) \mathbf{P}
\end{eqnarray*}

\noindent where $b$, $f$, and $e$ are arbitrary real functions.\footnote{These functions depend on scalars $P^2$ and $M$ in order to satisfy condition (I).} From
 condition (III) we obtain $f(M, P^2) =
0$. Comparing  commutator\footnote{Here we used equation $
[\mathbf{S} , (\mathbf{S} \cdot \mathbf{P})] = i \hbar [\mathbf{S}
\times \mathbf{P}]$. }

\begin{eqnarray*}
[S'_x , S'_y] &=& [bS_x + e(\mathbf{S} \cdot \mathbf{P})P_x,
bS_y + e(\mathbf{S} \cdot \mathbf{P})P_y] \\
&=& b^2 [S_x, S_y] - i \hbar eb P_x [\mathbf{S} \times \mathbf{P}]_y +
i \hbar  eb P_y [\mathbf{S} \times \mathbf{P}]_x \\
&=& i\hbar b^2 S_z - i \hbar eb (\mathbf{P} \times [\mathbf{S} \times
\mathbf{P}] )_z \\
&=& i\hbar (b^2 S_z -  eb P^2 S_z + eb (\mathbf{S} \cdot \mathbf{P})
P_z)
\end{eqnarray*}

\noindent  with the requirement (II)

\begin{eqnarray*}
[S'_x , S'_y] &=& i \hbar  S_z' = i \hbar(bS_z + e(\mathbf{S} \cdot
\mathbf{P})P_z)
\end{eqnarray*}

\noindent we  obtain the system of equations

\begin{eqnarray*}
b^2 -ebP^2  &=& b \\
eb &=& e
\end{eqnarray*}

\noindent whose non-trivial solution is $b=1$ and $e=0$. Therefore, the spin
operator is unique $\mathbf{S}' = \mathbf{S}$.

\subsection{Uniqueness of the position operator}
\label{ss:unique-position}

Assume that in addition to the Newton-Wigner position operator
$\mathbf{R}$ there is another position operator $\mathbf{R}'$
satisfying all properties (IV) - (VI). Then it follows from
condition (V) that $\mathbf{R}'$ has power 1. The most general true
vector with this property is

\begin{eqnarray*}
\mathbf{R}' = a(P^2, M) \mathbf{R}  +  d(P^2, M) [\mathbf{S} \times
\mathbf{P}]
 + g (P^2, M) \mathbf{P}
\end{eqnarray*}

\noindent where $a$, $d$, and $g$ are arbitrary real functions. From
 condition (IV) it follows, for example, that

\begin{eqnarray*}
0 &=& [R_x', S_y] = d(P^2, M) [S_yP_z - S_zP_y, S_y]  =i \hbar
d(P^2, M) P_yS_x
\end{eqnarray*}

\noindent  which implies that $d (P^2, M)=  0$.
From (V) we obtain

\begin{eqnarray*}
i \hbar &=& [R_x', P_x] = a(P^2, M) [R_x, P_x] = i \hbar a(P^2, M)
\end{eqnarray*}

\noindent and $a(P^2, M) = 1$. Therefore   the most
general form of the position operator is

\begin{eqnarray}
\mathbf{R}' = \mathbf{R}  + g (P^2, M) \mathbf{P}
\label{eq:6.45}
\end{eqnarray}

\noindent In Theorem \ref{theorem:lorentz-tr}  we will consider
boost transformations for times and positions of events in
non-interacting systems of particles.  If the
 term $g(P^2, M) \mathbf{P}$ in  (\ref{eq:6.45}) were non-zero, we would not get an
agreement with Lorentz transformations
known from Einstein's special relativity.\footnote{see
(\ref{eq:lor-transform-t}) - (\ref{eq:lor-transform-comp}) and Appendix \ref{sc:lorentz-time-pos2}} Therefore, we will assume that the factor
 $g(P^2, M)$  vanishes and $\mathbf{R}' = \mathbf{R}$. So, from now on, we
will use the Newton-Wigner operator $\mathbf{R}$ as the representative of the
position observable.

It follows from  commutator (\ref{eq:6.21}) that

\begin{eqnarray}
[R_x,P_x^n] &=& i \hbar n P_x^{n-1}
\label{eq:6.46}
\end{eqnarray}

\noindent so for any function $f(P_x)$

\begin{eqnarray}
[R_x,f(P_x)] &=& i \hbar \frac{ \partial f(P_x)}{\partial P_x} \label{eq:rxfpx}
\end{eqnarray}

\noindent For example,

\begin{eqnarray*}
[\mathbf{R}, H]
&=& [\mathbf{R}, \sqrt{P^2c^2 + M^2 c^4}]
= i \hbar \frac{ \partial \sqrt{P^2c^2 + M^2 c^4}}{\partial \mathbf{P}} \\
&=& \frac{i \hbar \mathbf{P} c^2}{\sqrt{P^2c^2 + M^2 c^4}} = i \hbar
\frac{ \mathbf{P}c^2}{ H} = i \hbar  \mathbf{V}
\end{eqnarray*}

\noindent where $\mathbf{V}$ is the velocity operator (\ref{eq:op-of-vel}). Therefore, as expected, for an observer shifted in time by
the amount $t$, the position of the physical system appears shifted
\index{velocity} by $\mathbf{V} t$:

\begin{eqnarray}
\mathbf{R}(t) &=& \exp \left(\frac{i}{\hbar} H t \right) \mathbf{R} \exp
\left(-\frac{i}{\hbar} Ht \right) \label{eq:4.49a} \\
 &=& \mathbf{R}
+\frac{i}{\hbar} [H, \mathbf{R}] t = \mathbf{R} + \mathbf{V} t
\label{eq:6.47}
\end{eqnarray}

\noindent Thus the center of mass $\mathbf{R}$ of any isolated physical system moves with constant velocity $\mathbf{V}$, as expected. This result is independent on the internal composition of the systemand on interactions between its different parts.

\subsection{Boost transformations of the position operator}
\label{ss:lorentz-position}

Let us now find how the vector of position (\ref{eq:6.26})
transforms with respect to boosts, i.e., we are looking  for the
connection between position observables in two inertial reference
frame moving with respect to each other. For simplicity, we consider
a massive system without spin, so that the center-of-mass position
in the reference frame at rest $O$ can be written as

\begin{eqnarray}
 \mathbf{R}  &=&
-\frac{c^2}{2}(\mathbf{K} H^{-1} + H^{-1} \mathbf{K}) \label{eq:r-nospin}
\end{eqnarray}

\noindent First, we need to determine boost transformations of
the boost operator itself. For example, the transformation of the
component $K_y$ with respect to the boost along the $x$-axis is
obtained by using equations (\ref{eq:A.39}), (\ref{eq:5.52}), and
(\ref{eq:5.54})

\begin{eqnarray*}
&\mbox{ }&  K_y(\theta) \nonumber \\ &=&
e^{ -\frac{ic}{\hbar} K_x \theta}
 K_y e^{ \frac{ic}{\hbar}K_x \theta} \nonumber\\
&=& K_y -\frac{ic\theta}{\hbar}  [K_x,K_y] - \frac{c^2
\theta^2}{2!\hbar^2}[K_x,[K_x,K_y]]
+ \frac{i c^3 \theta^3}{3! \hbar^3}[K_x,[K_x,[K_x,K_y]]] + \ldots  \nonumber\\
&=& K_y -\frac{\theta}{c}  J_z + \frac{\theta^2}{2!} K_y -
\frac{\theta^3}{3! c}  J_z \ldots
= K_y \cosh \theta -\frac{1}{c}
J_z \sinh \theta
\end{eqnarray*}

\noindent Then  the $y$-component of position in the
reference frame $O'$ moving along the $x$-axis is\footnote{Here we used (\ref{eq:6.3}).}

\begin{eqnarray}
 R_y(\theta)  &=&
e^{ -\frac{ic}{\hbar} K_x \theta}
 R_y e^{ \frac{ic}{\hbar}K_x \theta}
 = -\frac{c^2}{2} e^{ -\frac{ic}{\hbar} K_x \theta}
 (K_y H^{-1} + H^{-1} K_y) e^{ \frac{ic}{\hbar}K_x \theta} \nonumber \\
 &=&-\frac{c^2}{2}(K_y\cosh\theta - \frac{J_z}{c}
\sinh\theta)(H \cosh\theta- c P_x \sinh\theta)^{-1} \nonumber  \\
&\mbox{ }& -\frac{c^2}{2} (H \cosh\theta- c P_x \sinh\theta)^{-1} (K_y\cosh\theta - \frac{J_z}{c}
\sinh\theta)
\label{eq:7.46}
\end{eqnarray}

\noindent Similarly, for the $x$- and $z$-components

\begin{eqnarray}
 R_x(\theta)
&=& -\frac{c^2}{2}K_x(H \cosh\theta- c P_x \sinh\theta)^{-1}
\nonumber \\
&\ & -\frac{c^2}{2} (H \cosh\theta- c P_x \sinh\theta)^{-1}
K_x \label{eq:7.47a} \\
 R_z(\theta)
&=&-\frac{c^2}{2}(K_z\cosh\theta + \frac{J_y}{c}
\sinh\theta)(H \cosh\theta- c P_x \sinh\theta)^{-1} \nonumber \\
&\mbox { }& -\frac{c^2}{2}(H \cosh\theta- c P_x \sinh\theta)^{-1} (K_z\cosh\theta + \frac{J_y}{c}
\sinh\theta)
\label{eq:7.47}
\end{eqnarray}

\noindent These transformations do not resemble usual Lorentz
formulas from special relativity.\footnote{See also ref.
\cite{Monahan} and equations (\ref{eq:r-x-theta-t'}) - (\ref{eq:r-z-theta-t'}), which are classical ($\hbar \to 0$) limits of (\ref{eq:7.46}) - (\ref{eq:7.47}).} This is not surprising, because the Newton-Wigner
position operator does not constitute a 3-vector component of any
4-vector quantity.\footnote{In our formalism, there is no ``time operator''
which could serve as a 4th component of such a 4-vector. The
difference between special relativity and our approach to space and time will be discussed in chapter \ref{sc:obs-interact}.}

Furthermore, we can find the time dependence of the position
operator in the moving reference frame $O'$. We use label $t'$ to
indicate the time measured in the reference frame $O'$ by its own
clock and notice that the time translation generator $H'$ in $O'$ is
different from that in $O$

\begin{eqnarray}
H'  &=& e^{ -\frac{i}{\hbar} K_x c\theta}
 H e^{ \frac{i}{\hbar}K_x c\theta}
\label{eq:7.44}
\end{eqnarray}

\noindent Then we obtain

\begin{eqnarray}
 \mathbf{R}(\theta,t')  &=&
e^{ \frac{i}{\hbar} H't'}
 \mathbf{R}(\theta) e^{ -\frac{i}{\hbar}H't'}
= e^{ \frac{i}{\hbar} H' t'}  e^{-\frac{ic}{\hbar} K_x \theta}
\mathbf{R} e^{\frac{ic}{\hbar} K_x \theta} e^{ -\frac{i}{\hbar}H'
t'} \nonumber
\\
  &=&  \left(e^{-\frac{ic}{\hbar} K_x \theta}
e^{ \frac{i}{\hbar}Ht'} e^{\frac{ic}{\hbar} K_x\theta}\right)
e^{-\frac{ic}{\hbar} K_x \theta} \mathbf{R}  e^{\frac{ic}{\hbar}
K_x\theta} \left(e^{-\frac{ic}{\hbar} K_x\theta} e^{
-\frac{i}{\hbar}Ht'}
 e^{\frac{ic}{\hbar} K_x\theta}\right) \nonumber \\
&=&e^{-\frac{ic}{\hbar} K_x\theta}  e^{ \frac{i}{\hbar}Ht'}
\mathbf{R} e^{ -\frac{i}{\hbar} Ht'} e^{\frac{ic}{\hbar} K_x \theta}
=e^{-\frac{ic}{\hbar} K_x\theta} (\mathbf{R} + \mathbf{V}t' )
e^{\frac{ic}{\hbar} K_x
\theta} \nonumber \\
&=&
\mathbf{R}(\theta) + \mathbf{V} (\theta) t' \label{eq:7.43}
\end{eqnarray}

\noindent where  velocity
$\mathbf{V} (\theta)$ in the reference frame $O'$ is given by equations
 (\ref{eq:6.4}) -
(\ref{eq:6.6}). As expected, this time evolution in the moving frame can be obtained by a boost transformation from the trajectory (\ref{eq:6.47}) in the rest frame.

\chapter{SINGLE PARTICLES}
\label{ch:single}

\begin{quote}
\textit{The electron is as inexhaustible as the atom...}

\small
\hspace{1in} V. I. Lenin
\normalsize
\end{quote}

\vspace{0.5in}

\noindent  Our discussion in the preceding chapter could be
universally applied to any isolated physical system, be it
an electron or the Solar System. We have not specified how the
system was put together, and we considered only total observables
pertinent to the system as a whole. The results we obtained are not
surprising: the total energy, momentum, and angular momentum of our system are conserved, and the center of mass is moving with
a constant speed along a straight line (\ref{eq:6.47}). Although the
time evolutions of these total observables are rather uneventful, the
internal structure of complex (compound) physical systems may
undergo dramatic changes due to collisions, reactions, decays, etc.
The description of such transformations is the most interesting and
challenging part of physics. To address such problems, we need to
understand how complex physical systems are put together. The central
idea of this book is that all material objects are composed of
\emph{elementary particles} \index{elementary particle} i.e.,
localizable and countable systems lacking internal structure.\footnote{This is in contrast to the
wide-spread belief that the fundamental ingredients of nature are
continuous fields. See discussion in section \ref{ss:are-fields-meas}.} In this chapter
we will study these most fundamental ingredients of nature.

In subsection \ref{ss:fundamental} we have established  that the Hilbert space of any physical
system carries a unitary representation of the Poincar\'e group. Any
such representation can be decomposed
into a direct sum of irreducible representations.\footnote{See Appendix \ref{ss:unitary-reps}.} Elementary
particles are defined as physical
 systems for which this
sum has only one summand. Therefore, by definition, the Hilbert space
 of a stable elementary particle
  carries an
\emph{irreducible} \index{irreducible representation} unitary
representation of the Poincar\'e group. So, in a sense, elementary particles have
simplest non-decomposable spaces of states. The classification of
irreducible representations of the Poincar\'e group and their
Hilbert spaces was given by Wigner \cite{Wigner_unit}. From Schur's
first Lemma (Lemma \ref{Lemma.schur}) we know that in any irreducible
unitary representation of the Poincar\'e group, the two Casimir
operators $M$ and $\mathbf{S}^2$ act as multiplication by a
constant. So, all different irreducible representations and,
therefore, all elementary particles,
 can be classified according to the values of these two
constants - the mass and the spin squared. Of course, there are many
other parameters describing elementary particles, such as charge,
magnetic moment, strangeness, etc. But all of them are related to
the manner in which particles participate in interactions. In the
world where all interactions are ``turned off,'' particles have just
two intrinsic properties -- mass and spin.

There are only six known stable elementary particles for which the
classification  by mass and spin applies (see Table
\ref{table:5.1}). Some reservations should be made about  this
statement. First, for each particle in the table (except photons)
there is a corresponding antiparticle having the same mass and spin
but opposite values of the electric, baryon, and lepton
charges.\footnote{see subsection \ref{ss:conservation-laws} for conservation laws associated with the charges} So, if
we also count antiparticles, there are eleven different stable
particle species. Second, there are many more particles, like
\emph{muons}, \index{muon} \emph{pions}, \index{pion}
\emph{neutrons}, \index{neutron} etc., which are usually called
elementary but all of them are unstable and eventually decay into
particles shown in Table \ref{table:5.1}. This does not mean that
unstable particles are ``made of'' stable particles or that they are
less elementary. Simply, the stable particles shown in the table
have the lowest masses and there are no lighter species to which they
 could decay without violating conservation laws. Third,
 we  do not list in Table \ref{table:5.1}
 \emph{quarks}, \emph{gluons}, \emph{gravitons}, and other particles predicted
theoretically, but never directly observed in experiment.  Fourth,
strictly speaking, the photon is not a true elementary particle as
it is not described by an irreducible representation of the
Poincar\'e group. We will see in subsection \ref{ss:rep-poincare}
that the photon is described by a \emph{reducible} \index{reducible
representation} representation of the Poincar\'e group which is a
direct sum of two irreducible representations with helicities $+\hbar$ and
$-\hbar$. Fifth, neutrinos are not truly stable elementary particles. \index{neutrino}
According to recent experiments, three flavors of neutrinos  are
 oscillating between
each other over time. \index{neutrino oscillations} Finally, it may be true that protons are not
elementary particles as well. They are usually regarded as being composed
of quarks. This leaves us with just two truly stable, elementary, and directly  observable
particles, which are the electron and the positron.

\begin{table}[h]
\caption{Properties of stable elementary particles}
\begin{tabular*}{\textwidth}{@{\extracolsep{\fill}}ccc}
 \hline
Particle     & Mass     &  Spin/helicity     \cr \hline Electron
\index{electron} &  0.511 MeV/$c^2$    &  $\hbar/2$     \cr Proton &
938.3 MeV/$c^2$   & $\hbar/2$    \cr Electron neutrino  &  $<1$
eV/$c^2$ & $\hbar/2$  \cr Muon neutrino  &  $<1$ eV/$c^2$       &
$\hbar/2$  \cr Tau neutrino  &  $<1$ eV/$c^2$       & $\hbar/2$  \cr
 Photon  &  0       & $\pm \hbar$  \cr
\hline
\end{tabular*}
\label{table:5.1}
\end{table}

In the following we will denote $m$ the eigenvalue of the
mass operator in the Hilbert space of elementary particle
and consider separately two cases: massive
particles ($m>0$) and  massless particles ($m=0$).\footnote{Wigner's classification also permits irreducible representations with negative and imaginary values of $m$, but there is no evidence that such particles exist in nature. We skip their discussion in this book.}
\label{elem-part-end}

\section{Massive particles}
\label{sc:massive}

\subsection{Irreducible representations of the Poincar\'e group}
\label{ss:irrep-poinc}

The Hilbert space $\mathcal{H}$ of a massive elementary particle
carries
 an  unitary irreducible representation $U_g$ of the Poincar\'e group
characterized by a single positive eigenvalue $m $
of the mass operator $M$.  As discussed in subsection \ref{ss:position}, the
position
 operator  $\mathbf{R}$  is
well-defined in this case. Components of the position and
momentum operators satisfy commutation relations of the
6-dimensional \emph{Heisenberg Lie algebra}\footnote{see equations
(\ref{eq:5.53}), (\ref{eq:6.21}) and Theorem \ref{Theorem6.1}}
\index{Heisenberg Lie algebra}

\begin{eqnarray*}
[P_i, P_j] &=& [R_i, R_j] = 0
\\
\mbox{ } [R_i, P_j] &=& i \hbar \delta_{ij}
\end{eqnarray*}

\noindent Then, according to the Stone-von Neumann theorem
\ref{Theorem.Stone},  each component $P_x, P_y, P_z, R_x, R_y, R_z$
has continuous spectrum occupying entire real axis $\mathbb{R}$. The components of the momentum operator $P_x, P_y, P_z$ commute with each other. So, the spectrum of the vector operator $\mathbf{P}$  is the 3-dimensional linear space $\mathbb{R}^3$.\footnote{The same is true for the spectrum of the position operator $\mathbf{R}$.} Thus, there exists a  decomposition of unity associated with
the spectrum of $\mathbf{P}$ and  the Hilbert
space $\mathcal{H}$ can be represented as a direct sum of
corresponding eigensubspaces $\mathcal{H}_{\mathbf{p}}$ of $\mathbf{P}$

\begin{eqnarray*}
\mathcal{H} = \oplus_{\mathbf{p} \in \mathbb{R}^3} \mathcal{H}_{\mathbf{p}}
\end{eqnarray*}

\noindent This implies that the 1-particle Hilbert space $\mathcal{H}$ is infinite-dimensional. It can be said that the number of mutually orthogonal basis vectors in this space is no less than the ``number of distinct points in the infinite 3D space $\mathbb{R}^3$'', i.e., uncountable.

Let us first focus on the  subspace $\mathcal{H}_{\mathbf{0}}$ with
zero momentum. This subspace
is invariant with respect to rotations, because
 for any vector
$|\mathbf{0} \rangle$ from this subspace the result of rotation
$e^{-\frac{i}{\hbar} \mathbf{J} \vec{\phi}} |\mathbf{0} \rangle$
belongs to $\mathcal{H}_{\mathbf{0}}$\footnote{Here we used equation (\ref{eq:6.1}).}

\begin{eqnarray}
\mathbf{P} e^{-\frac{i}{\hbar} \mathbf{J} \vec{\phi}} |\mathbf{0}
\rangle &=&
e^{-\frac{i}{\hbar} \mathbf{J} \vec{\phi}}
e^{\frac{i}{\hbar} \mathbf{J} \vec{\phi}}\mathbf{P} e^{-\frac{i}{\hbar}
\mathbf{J} \vec{\phi}} |\mathbf{0}
\rangle \nonumber \\
&=& e^{-\frac{i}{\hbar} \mathbf{J} \vec{\phi}} \left(\left(\mathbf{P} \cdot
\frac{\vec{\phi}}{\phi}\right) \frac{\vec{\phi}}{\phi} (1 - \cos \phi) +
\mathbf{P} \cos \phi - \left[\mathbf{P} \times \frac{\vec{\phi}}{\phi}\right]
\sin \phi\right) |\mathbf{0}
\rangle \nonumber \\
&=& \mathbf{0} \label{eq:proof1}
\end{eqnarray}

\noindent This means that representation of the rotation subgroup defined in the full Hilbert space $\mathcal{H}$ induces a unitary
representation $V_g$ of
this subgroup in
 $\mathcal{H}_{\mathbf{0}}$.

The generators of rotations  are, of course,
represented by the angular momentum  vector $\mathbf{J}$ in $\mathcal{H}$. However,
in the subspace $\mathcal{H}_{\mathbf{0}}$, they can be equivalently
represented by the vector of spin  $\mathbf{S}$, because

\begin{eqnarray*}
S_z| \mathbf{0} \rangle  &=& J_z | \mathbf{0} \rangle- [\mathbf{R}
\times \mathbf{P}]_z| \mathbf{0} \rangle =  J_z | \mathbf{0}
\rangle- (R_x P_y - R_y P_x)| \mathbf{0} \rangle =  J_z | \mathbf{0}
\rangle
\end{eqnarray*}

We will show later that the representation of the full Poincar\'e group is
irreducible if and only if the  representation
 $V_g$ of
the rotation group in
 $\mathcal{H}_{\mathbf{0}}$ is irreducible.
 So, we will be interested only in such irreducible
representations $V_g$. The classification of unitary irreducible
representations of  the rotation group (single- and double-valued)\footnote{see
Appendix \ref{ss:rotation-reps}}
depends on one integer or half-integer parameter $s$ that we will identify with particle's \emph{spin}.
\index{spin} The trivial  one-dimensional
representation is characterized by spin zero ($s=0$) and corresponds
to a \emph{spinless} particle. The two-dimensional representation
corresponds to particles with spin one-half ($s=1/2$). The
3-dimensional representation corresponds to particles with spin one
($s=1$), etc.  Correspondingly, the dimension of the zero-momentum subspace $\mathcal{H}_{\mathbf{0}}$ will be 1,2,3, $\ldots$.

It is customary to choose a  basis of eigenvectors of $S_z$  in
$\mathcal{H}_{\mathbf{0}}$ and denote these vectors by $|
\mathbf{0}, \sigma \rangle$, i.e.,

\begin{eqnarray*}
\mathbf{P} | \mathbf{0},
\sigma \rangle &=& 0 \\
 H | \mathbf{0},
\sigma \rangle &=& mc^2 | \mathbf{0},
\sigma \rangle \\
 M | \mathbf{0},
\sigma \rangle &=& m | \mathbf{0},
\sigma \rangle \\
 S^2 | \mathbf{0}, \sigma  \rangle &=&
\hbar^2 s(s+1) | \mathbf{0}, \sigma  \rangle \\
S_z | \mathbf{0}, \sigma \rangle &=&
\hbar \sigma | \mathbf{0}, \sigma  \rangle
\end{eqnarray*}

\noindent where $\sigma = -s, -s+1, \ldots, s-1, s$. The action of a
rotation on these basis vectors is

\begin{eqnarray}
e^{-\frac{i}{\hbar}\mathbf{J} \vec{\phi} }| \mathbf{0}, \sigma  \rangle
= e^{-\frac{i}{\hbar}\mathbf{S} \vec{\phi} }| \mathbf{0}, \sigma \rangle &=&
 \sum_{\sigma' = -s}^{s} D^s _{\sigma' \sigma} (\vec{\phi})
| \mathbf{0}, \sigma' \rangle
\label{eq:7.1}
\end{eqnarray}

\noindent where $D^s$ are $(2s+1) \times (2s +1)$  matrices of the
representation $V_g$. This definition implies that\footnote{Here
$\vec{\phi}_1 \vec{\phi}_2$ denotes the composition of two rotations
parameterized by vectors $\vec{\phi}_1$ and $ \vec{\phi}_2$, respectively.}

\begin{eqnarray*}
&\mbox{ } & \sum_{\sigma' = -s}^{s} D^s _{\sigma' \sigma} (\vec{\phi}_1
\vec{\phi}_2)
| \mathbf{0}, \sigma' \rangle \\
&=& e^{-\frac{i}{\hbar}\mathbf{J} \vec{\phi}_1 }
e^{-\frac{i}{\hbar}\mathbf{J} \vec{\phi}_2 }| \mathbf{0}, \sigma
\rangle = e^{-\frac{i}{\hbar}\mathbf{J} \vec{\phi}_1 }
 \sum_{\sigma'' = -s}^{s} D^s _{\sigma'' \sigma} (\vec{\phi}_2)
| \mathbf{0}, \sigma'' \rangle \\
&=&
 \sum_{\sigma' =
-s}^{s} \left( \sum_{\sigma'' = -s}^{s} D^s _{\sigma' \sigma''}
(\vec{\phi}_1)  D^s _{\sigma'' \sigma} (\vec{\phi}_2) \right)| \mathbf{0},
\sigma' \rangle
\end{eqnarray*}

\noindent and

\begin{eqnarray*}
  D^s _{\sigma' \sigma} (\vec{\phi}_1 \vec{\phi}_2)
=   \sum_{\sigma'' = -s}^{s}
D^s _{\sigma' \sigma''} (\vec{\phi}_1)  D^s _{\sigma'' \sigma} (\vec{\phi}_2)
\end{eqnarray*}

\noindent which means that matrices $D^s$ furnish a representation
of the rotation group.

\subsection{Momentum-spin basis}
\label{ss:momentum-basis}

\begin{figure}
\centering
\includegraphics{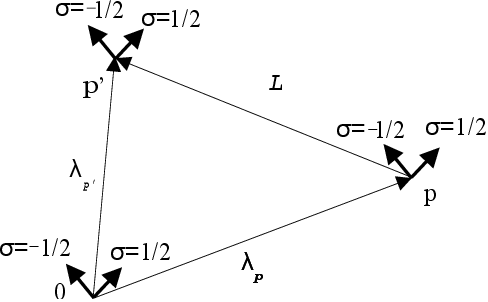} \caption{Construction of the
momentum-spin basis for a spin one-half particle. Spin eigenvectors
(with eigenvalues $\sigma=-1/2, 1/2$) at zero momentum are
propagated to non-zero momenta $\mathbf{p}$ and $\mathbf{p}'$ by
using pure boosts $\lambda_{\mathbf{p}}$ and
 $\lambda_{\mathbf{p}'}$, respectively. As discussed in subsection \ref{ss:momentum-poincare}, there is a unique pure boost
$L$ which connects momenta $\mathbf{p}$ and $\mathbf{p}'$.}
\label{fig:7.1}
\end{figure}

In the preceding subsection we constructed  basis vectors $|
\mathbf{0}, \sigma \rangle $
 in the subspace
$\mathcal{H}_{\mathbf{0}}$. We also need basis vectors $| \mathbf{p},
\sigma \rangle$ in other subspaces
$\mathcal{H}_{\mathbf{p}}$ with $\mathbf{p} \neq \mathbf{0}$. We will
build basis $| \mathbf{p},
\sigma \rangle$ by
 propagating state vectors $|\mathbf{0}, \sigma \rangle$ to other points in
the 3D momentum space using pure boost transformations.\footnote{Of
course, this choice is rather arbitrary. A different choice of
transformations connecting momenta $\mathbf{0}$ and $\mathbf{p}$
(e.g., boosts coupled with rotations) would result in a different but
equivalent basis set. However, once the basis set has been fixed,
all formulas should be written with respect to it.
\label{footnote:basis}} The unique pure boost, which transforms
momentum $\mathbf{0}$ to $\mathbf{p}$, will be denoted by $\lambda_{\mathbf{p}}$.\footnote{see Fig. \ref{fig:7.1}} The corresponding unitary operator in the Hilbert space is\footnote{for this notation see (\ref{eq:5.57a})}

\begin{eqnarray}
 U(\lambda_{\mathbf{p}};\mathbf{0},0) \equiv e^{-\frac{ic}{\hbar} \mathbf{K}
 \vec{\theta}_{\mathbf{p}}} \label{eq:unique-pure}
\end{eqnarray}

\noindent where

\begin{eqnarray}
\vec{\theta}_{\mathbf{p}} =
\frac{\mathbf{p}}{p} \sinh ^{-1} \frac{p}{mc}
\label{eq:7.2}
\end{eqnarray}

\noindent Therefore we can write

\begin{eqnarray}
 | \mathbf{p}, \sigma \rangle &=&
N(\mathbf{p}) U(\lambda_{\mathbf{p}};\mathbf{0},0) | 0, \sigma \rangle =
N(\mathbf{p}) e^{-\frac{ic}{\hbar} \mathbf{K}
\vec{\theta}_{\mathbf{p}}} | 0, \sigma \rangle \label{eq:7.3}
\end{eqnarray}

\noindent where
 $N(\mathbf{p})$ is
a normalization factor.  The explicit expression for
$N(\mathbf{p})$ will be given a bit later  in equation (\ref{eq:7.16}).

To verify that vector (\ref{eq:7.3}) is indeed an  eigenvector of
the momentum operator with eigenvalue $\mathbf{p}$ we use equation
(\ref{eq:6.2})

\begin{eqnarray}
 \mathbf{P} | \mathbf{p}, \sigma \rangle &=&  N(\mathbf{p})\mathbf{P}
e^{-\frac{ic}{\hbar} \mathbf{K} \vec{\theta}_{\mathbf{p}}} |
\mathbf{0}, \sigma \rangle  = N(\mathbf{p})e^{-\frac{ic}{\hbar}
\mathbf{K}  \vec{\theta}_{\mathbf{p}}} e^{\frac{ic}{\hbar}
\mathbf{K}  \vec{\theta}_{\mathbf{p}}} \mathbf{P}
e^{-\frac{ic}{\hbar} \mathbf{K}
 \vec{\theta}_{\mathbf{p}}} | \mathbf{0}, \sigma \rangle  \nonumber \\
&=& N(\mathbf{p})e^{-\frac{i}{\hbar} \mathbf{K} c
\vec{\theta}_{\mathbf{p}}}\frac{\vec{\theta}_{\mathbf{p}}}{c\theta_{\mathbf{p}}}
H \sinh \theta_{\mathbf{p}} | \mathbf{0}, \sigma \rangle
=
N(\mathbf{p})e^{-\frac{ic}{\hbar} \mathbf{K}
\vec{\theta}_{\mathbf{p}}}\frac{\vec{\theta}_{\mathbf{p}}}{\theta_{\mathbf{p}}}
mc \sinh \theta_{\mathbf{p}} | \mathbf{0}, \sigma \rangle  \nonumber \\
&=& N(\mathbf{p})\mathbf{p} e^{-\frac{ic}{\hbar} \mathbf{K}
\vec{\theta}_{\mathbf{p}}}
 | \mathbf{0}, \sigma \rangle  = \mathbf{p}
 | \mathbf{p}, \sigma \rangle \label{eq:proof2}
\end{eqnarray}

 Let us
now find the action of the spin component $S_z$ on the basis
 vectors $| \mathbf{p}, \sigma \rangle$\footnote{Here
we use (\ref{eq:6.23}) and take into account that $W_0 | \mathbf{0}, \sigma \rangle =
\mathbf{P} | \mathbf{0}, \sigma \rangle = 0$ and $H | \mathbf{0}, \sigma \rangle =
Mc^2 | \mathbf{0}, \sigma \rangle$. We also use boost
transformations (\ref{eq:6.2}) and (\ref{eq:6.3}) of the
energy-momentum 4-vector \index{energy-momentum 4-vector} $(H,
c\mathbf{P})$ and similar formulas for the Pauli-Lubanski 4-vector
$(W_0, \mathbf{W})$. For brevity, we denote $\theta_z$ the
$z$-component of the vector $\vec{\theta}_{\mathbf{p}}$ and $\theta$
its absolute value.}

\begin{eqnarray*}
&\ & S_z | \mathbf{p}, \sigma \rangle =
N(\mathbf{p})S_z
 e^{-\frac{ic}{\hbar} \mathbf{K}
 \vec{\theta}_{\mathbf{p}}} | \mathbf{0}, \sigma \rangle  \\
&=& N(\mathbf{p})e^{-\frac{ic}{\hbar} \mathbf{K}
\vec{\theta}_{\mathbf{p}}} e^{\frac{ic}{\hbar} \mathbf{K}
\vec{\theta}_{\mathbf{p}}} \left(\frac{W_z}{Mc} - \frac{ W_0
P_z}{M(Mc^2 + H)}\right)
 e^{-\frac{ic}{\hbar} \mathbf{K}
 \vec{\theta}_{\mathbf{p}}} | \mathbf{0}, \sigma \rangle  \\
&=& N(\mathbf{p})e^{-\frac{ic}{\hbar} \mathbf{K}
\vec{\theta}_{\mathbf{p}}} \Bigl(\frac{W_z + \frac{\theta_z}{\theta}
[(\mathbf{W} \cdot \frac{\vec{\theta}}{\theta}) (\cosh \theta - 1) -
W_0 \sinh
\theta]}{Mc} \\
&-& \frac{ \left( W_0 \cosh \theta -  (\mathbf{W} \cdot
\frac{\vec{\theta}}{\theta}) \sinh \theta \right) \left(P_z +
\frac{\theta_z}{\theta} [(\mathbf{P} \cdot
\frac{\vec{\theta}}{\theta}) (\cosh \theta - 1) - \frac{1}{c} H
\sinh \theta]\right)} {M(Mc^2 +
 H \cosh \theta - c (\mathbf{P} \cdot
\frac{\vec{\theta}}{\theta}) \sinh \theta)}\Bigr)
 | \mathbf{0}, \sigma \rangle  \\
&=& N(\mathbf{p})e^{-\frac{ic}{\hbar} \mathbf{K}
\vec{\theta}_{\mathbf{p}}} \Bigl(\frac{W_z + \frac{\theta_z}{\theta}
(\mathbf{W} \cdot
\frac{\vec{\theta}}{\theta}) (\cosh \theta - 1)}{Mc}
- \frac{  ( \mathbf{W} \cdot \frac{\vec{\theta}}{\theta}) \sinh
\theta ( \frac{\theta_z}{\theta}  Mc \sinh
\theta)} {M(Mc^2 +  Mc^2 \cosh \theta)}\Bigr)
 | \mathbf{0}, \sigma \rangle  \\
&=& N(\mathbf{p})e^{-\frac{ic}{\hbar} \mathbf{K}
\vec{\theta}_{\mathbf{p}}} \left(\frac{W_z }{Mc} +
\frac{\theta_z}{\theta} \left(\mathbf{W} \cdot
\frac{\vec{\theta}}{\theta}\right) \left(\frac{  \cosh \theta - 1}{Mc} - \frac{
\sinh^2 \theta } {Mc(1 +   \cosh \theta)}\right) \right)
 | \mathbf{0}, \sigma \rangle  \\
 &=& N(\mathbf{p})e^{-\frac{ic}{\hbar} \mathbf{K}
 \vec{\theta}_{\mathbf{p}}}
\frac{W_z}{Mc}
 | \mathbf{0}, \sigma \rangle = N(\mathbf{p})e^{-\frac{ic}{\hbar} \mathbf{K}
 \vec{\theta}_{\mathbf{p}}}
S_z
 | \mathbf{0}, \sigma \rangle  = N(\mathbf{p})e^{-\frac{ic}{\hbar} \mathbf{K}
 \vec{\theta}_{\mathbf{p}}}
\hbar \sigma
 | \mathbf{0}, \sigma \rangle  \\
&=&
\hbar \sigma
 | \mathbf{p}, \sigma \rangle  \\
\end{eqnarray*}

\noindent So, $| \mathbf{p}, \sigma \rangle$ are eigenvectors of the
momentum, energy, and $z$-component of spin\footnote{Note that
eigenvectors of the spin operator $\mathbf{S}$  were
obtained here by applying pure boosts to vectors at $\mathbf{p}=0$.
A different set of transformations connecting bases in
points $\mathbf{0}$ and $\mathbf{p}$ (see footnote on page
\pageref{footnote:basis}) would result in a different momentum-spin
basis $| \mathbf{p}, \sigma \rangle$ and in a different spin operator $\mathbf{S}$ (see \cite{RHD}).
Does this contradict our statement about the uniqueness of the spin
operator in subsection \ref{ss:unique-spin}? Not really. The point
is that the alternative spin operator $\mathbf{S}'$ (and the
corresponding alternative position operator $\mathbf{R}'$) will not
be expressed as a function of basic generators of the Poincar\'e
group. This condition was important for our proof of the uniqueness
of $\mathbf{S}$ (and $\mathbf{R}$) in section \ref{sc:spin}.}

\begin{eqnarray*}
\mathbf{P} | \mathbf{p},
\sigma \rangle &=& \mathbf{p} | \mathbf{p},
\sigma \rangle \\
 H | \mathbf{p},
\sigma \rangle &=& \omega_{\mathbf{p}}| \mathbf{p},
\sigma \rangle \\
M | \mathbf{p}, \sigma  \rangle &=&
m | \mathbf{p}, \sigma  \rangle \\
 S^2 | \mathbf{p}, \sigma  \rangle &=&
\hbar^2 s(s+1) | \mathbf{p}, \sigma  \rangle \\
S_z | \mathbf{p}, \sigma \rangle &=&
\hbar \sigma | \mathbf{p}, \sigma  \rangle
\end{eqnarray*}

\noindent where we denoted

\begin{eqnarray}
 \omega_{\mathbf{p}} \equiv \sqrt{m^2c^4 + p^2c^2}
 \label{eq:one-part-en}
\end{eqnarray}

\noindent the one-particle energy

\begin{figure}
\centering
 \includegraphics{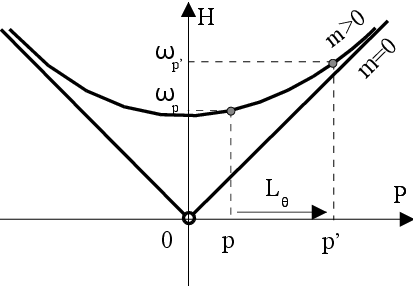} \caption{Mass hyperboloid
in the energy-momentum space for massive particles and the
zero-mass cone for $m=0$.} \label{fig:7.2}
\end{figure}

 The common spectrum of the energy-momentum eigenvalues $(\omega_{\mathbf{p}},
\mathbf{p})$ can be conveniently represented as points on the
\emph{mass hyperboloid} \index{mass hyperboloid} in the
4-dimensional energy-momentum space (see Fig. \ref{fig:7.2}). For
massive particles,
 the spectrum of the velocity operator  $\mathbf{V} = \mathbf{P}c^2/H$
is the interior of a 3-dimensional sphere $|\mathbf{v}| < c$ in the 4D energy-momentum space. This
spectrum does not include the surface of the sphere, therefore
massive particles cannot reach the speed of light.\footnote{In
quantum mechanics,
 the speed of propagation of particles is
not a well-defined concept. The value of particle's speed is
definite in states having certain momentum. However, such states are
described by infinitely extended  plane waves (\ref{eq:plane-wave}) and one cannot speak about particle propagation in such states. So,
strictly speaking, the speed of a particle cannot be obtained by
measuring its positions at two different time instants and dividing
the traveled distance by the time interval. This is a consequence of
the non-commutativity of the operators of position and velocity.}

\subsection{Action of Poincar\'e transformations}
\label{ss:momentum-poincare}

We can now define the action of transformations from the  Poincar\'e
group on the basis vectors $| \mathbf{p}, \sigma \rangle$
constructed above.\footnote{ We are working in the Schr\"odinger
picture here.} Translations act by simple multiplication

\begin{eqnarray}
e^{-\frac{i}{\hbar}\mathbf{P} \mathbf{a}}|\mathbf{p}, \sigma\rangle &=&
e^{-\frac{i}{\hbar}\mathbf{p}\mathbf{a}} |\mathbf{p} , \sigma\rangle
\label{eq:7.4} \\
e^{\frac{i}{\hbar} Ht}|\mathbf{p}, \sigma\rangle &=&
e^{\frac{i}{\hbar}\omega_{\mathbf{p}}t} | \mathbf{p}, \sigma\rangle
\label{eq:7.5}
\end{eqnarray}

\noindent Let us now apply rotation $e^{-\frac{i}{\hbar} \mathbf{J}
\vec{\phi}}$
to the vector $|\mathbf{p} , \sigma\rangle$ and use equations (\ref{eq:7.1}) and (\ref{eq:Rba})

\begin{eqnarray}
 e^{-\frac{i}{\hbar} \mathbf{J} \vec{\phi}}|\mathbf{p}, \sigma\rangle
&=& N(\mathbf{p})e^{-\frac{i}{\hbar} \mathbf{J} \vec{\phi}}
e^{-\frac{ic}{\hbar} \mathbf{K}
\vec{\theta}_{\mathbf{p}}} | 0, \sigma \rangle \nonumber \\
&=& N(\mathbf{p})e^{-\frac{i}{\hbar} \mathbf{J} \vec{\phi}}
e^{-\frac{ic}{\hbar} \mathbf{K}
\vec{\theta}_{\mathbf{p}}}  e^{\frac{i}{\hbar} \mathbf{J}
\vec{\phi}}  e^{-\frac{i}{\hbar} \mathbf{J} \vec{\phi}}| 0, \sigma
\rangle  \nonumber\\
&=& N(\mathbf{p}) e^{-\frac{ic}{\hbar} (R^{-1}_{\vec{\phi}}\mathbf{K}) \cdot
\vec{\theta}_{\mathbf{p}}}
  \sum_{\sigma'=-s}^s D_{\sigma' \sigma} (\vec{\phi}) | 0, \sigma' \rangle
\nonumber\\
&=& N(\mathbf{p}) e^{-\frac{ic}{\hbar} \mathbf{K} \cdot
 R_{\vec{\phi}} \vec{\theta}_{\mathbf{p}}}
  \sum_{\sigma'=-s}^s D_{\sigma' \sigma} (\vec{\phi})| 0, \sigma' \rangle
\nonumber \\
&=&
  \sum_{\sigma'=-s}^s D_{\sigma' \sigma} (\vec{\phi})
| R_{\vec{\phi}} \mathbf{p}, \sigma' \rangle
\label{eq:7.6}
\end{eqnarray}

\noindent This means that both momentum and spin of the particle are rotated
by
the angle $\vec{\phi}$, as expected.

Applying a boost $ L \equiv e^{-\frac{ic}{\hbar} \mathbf{K}
\vec{\theta}}$ to the vector $|\mathbf{p}, \sigma\rangle $ and
using (\ref{eq:7.3}) we obtain

\begin{eqnarray}
L |\mathbf{p}, \sigma\rangle &=& N(\mathbf{p}) L
U(\lambda_{\mathbf{p}};\mathbf{0},0)|\mathbf{0}, \sigma\rangle \label{eq:7.7}
\end{eqnarray}

\noindent The product of two boosts on the right hand side of equation (\ref{eq:7.7})
is a transformation from the Lorentz group, so
it can be represented in the  form (boost)$\times$(rotation)$=BQ$

\begin{eqnarray}
L
U(\lambda_{\mathbf{p}};\mathbf{0},0)
&=& B(\mathbf{p}, \vec{\theta}) Q(\mathbf{p}, \vec{\theta})
\label{eq:7.8}
\end{eqnarray}

\noindent Here $B$ and $Q$ are yet undefined quantum-mechanical operators, and now we are going to learn a bit more about them. Multiplying both sides of equation (\ref{eq:7.8}) by
$B^{-1}(\mathbf{p}, \vec{\theta})$, we obtain

\begin{eqnarray}
B^{-1}(\mathbf{p}, \vec{\theta}) L
U(\lambda_{\mathbf{p}};\mathbf{0},0) &=&
 Q(\mathbf{p}, \vec{\theta})
\label{eq:7.9}
\end{eqnarray}

\noindent Since operator $Q$ on the right hand side is a representative of a rotation, it keeps invariant
the subspace with zero momentum $\mathcal{H}_{\mathbf{0}}$.
Therefore, the sequence of boosts on the left hand side of equation
(\ref{eq:7.9}) when acting on a vector with zero momentum $|\mathbf{0},
\sigma \rangle$ returns this vector back to the zero momentum subspace. This fact is clearly seen in  Fig.
\ref{fig:7.1}: The zero momentum vector is mapped to a vector with
momentum $\mathbf{p}$ by the boost $\lambda_{\mathbf{p}}$.
Subsequent application of $L$ transforms this vector to another eigenstate of the momentum operator with eigenvalue $\mathbf{p}'$. This eigenvalue can be found easily by application of formula (\ref{eq:6.2})\footnote{For simplicity, in this derivation we omit spin indices.}

\begin{eqnarray*}
\mathbf{P} |\mathbf{p}' \rangle &=& \mathbf{P} L |\mathbf{p}\rangle  = \mathbf{P} e^{-\frac{ic}{\hbar}\mathbf{K} \vec{\theta}} |\mathbf{p} \rangle = e^{-\frac{ic}{\hbar}\mathbf{K} \vec{\theta}} e^{\frac{ic}{\hbar}\mathbf{K} \vec{\theta}} \mathbf{P} e^{-\frac{ic}{\hbar}\mathbf{K} \vec{\theta}} |\mathbf{p} \rangle \\
&=& e^{-\frac{ic}{\hbar}\mathbf{K} \vec{\theta}} \left( \mathbf{P} + \frac{\vec{\theta}}{\theta}
\left[\left(\mathbf{P} \cdot \frac{\vec{\theta}}{\theta}\right)
(\cosh \theta - 1) +
\frac{H}{c} \sinh \theta \right] \right) |\mathbf{p} \rangle \\
&=& \left( \mathbf{p} + \frac{\vec{\theta}}{\theta}
\left[\left(\mathbf{p} \cdot \frac{\vec{\theta}}{\theta}\right)
(\cosh \theta - 1) +
\frac{\omega_{\mathbf{p}}}{c} \sinh \theta \right] \right) |\mathbf{p}' \rangle
\end{eqnarray*}

\noindent So we conclude that

\begin{eqnarray}
\mathbf{p}'  & =& \mathbf{p} + \frac{\vec{\theta}}{\theta}
\left[\left(\mathbf{p} \cdot \frac{\vec{\theta}}{\theta}\right)
(\cosh \theta - 1) +
\frac{\omega_{\mathbf{p}}}{c} \sinh \theta \right] \equiv \Lambda \mathbf{p} \label{eq:p'}
\end{eqnarray}

\noindent It then follows that
$ B^{-1}(\mathbf{p}, \vec{\theta}) = \lambda^{-1}_{\mathbf{p}'}$ is the boost returning $\mathbf{p}'$ back to the zero-momentum vector.  For the  rotation on the right hand side of equation
(\ref{eq:7.9}) we will be using a special symbol

\begin{eqnarray}
U(R_{\vec{\phi}_W(\mathbf{p}, \Lambda)}; \mathbf{0}, 0) \equiv Q(\mathbf{p}, \vec{\theta}) = U^{-1}(\lambda_{\Lambda \mathbf{p}};\mathbf{0},0) L U(\lambda_{\mathbf{p}};\mathbf{0},0)  \label{eq:7.9a}
\end{eqnarray}

\noindent where $\vec{\phi}_W (\mathbf{p}, \Lambda)$ is called the \emph{Wigner angle}\footnote{Here function $\vec{\Phi}$ assigns a unique rotation angle to the given rotation matrix, as explained in Appendix \ref{ss:parameterization}.} \index{Wigner angle}

\begin{eqnarray}
\vec{\phi}_W(\mathbf{p}, \Lambda) = \vec{\Phi}(\lambda^{-1}_{\Lambda \mathbf{p}} \Lambda \lambda_{\mathbf{p}})  \label{eq:7.9x}
\end{eqnarray}

\noindent Explicit formulas for this angle
can be found, e.g., in ref. \cite{Ritus}. Then, substituting
(\ref{eq:7.9a}) in (\ref{eq:7.8}), we obtain

\begin{eqnarray}
 e^{-\frac{ic}{\hbar}\mathbf{K} \vec{\theta}}|\mathbf{p}, \sigma
\rangle
 &=&  N(\mathbf{p})
L U(\lambda_{\mathbf{p}};\mathbf{0},0)|\mathbf{0}, \sigma\rangle = N(\mathbf{p})
U(\lambda_{\Lambda \mathbf{p}};\mathbf{0},0) R_{\vec{\phi}_W(\mathbf{p}, \Lambda)}
|\mathbf{0}, \sigma \rangle \nonumber \\
 &=& N(\mathbf{p})
U(\lambda_{\Lambda \mathbf{p}};\mathbf{0},0)
\sum_{\sigma'=-s}^{s} D^s_{\sigma' \sigma}
(\vec{\phi}_W(\mathbf{p}, \Lambda))|\mathbf{0}, \sigma'
\rangle \nonumber \\
 &=&\frac{N(\mathbf{p})}{N(\Lambda \mathbf{p})}
\sum_{\sigma'=-s}^{s} D^s_{\sigma' \sigma}
(\vec{\phi}_W(\mathbf{p}, \Lambda))|\Lambda \mathbf{p}, \sigma'
\rangle
\label{eq:7.11}
\end{eqnarray}

Equations (\ref{eq:7.6}) and (\ref{eq:7.11}) show that rotations and
boosts are accompanied with turning the spin vector in each subspace
$\mathcal{H}_{\mathbf{p}}$ by rotation matrices $D^s$. If the
representation of the rotation group $D^s$ were reducible, then each
subspace $\mathcal{H}_{\mathbf{p}}$ would be represented as a direct
sum of irreducible components $\mathcal{H}^k_{\mathbf{p}}$

\begin{eqnarray*}
\mathcal{H}_{\mathbf{p}} = \oplus_k \mathcal{H}^k_{\mathbf{p}}
\end{eqnarray*}

\noindent and each subspace

\begin{eqnarray*}
\mathcal{H}^k =  \oplus_{\mathbf{p} \in R^3} \mathcal{H}^k_{\mathbf{p}}
\end{eqnarray*}

\noindent would be reducible with respect to the entire Poincar\'e
group. Therefore, in order to construct an irreducible
representation of the Poincar\'e group in $\mathcal{H}$, the
representation $D^s$  must be an irreducible unitary representation
of the rotation group, as was mentioned already in subsection
\ref{ss:irrep-poinc}.

In this book we will be interested in
describing interactions between electrons and protons, which are
massive particles with spin 1/2. Then the relevant representation
$D^s$ of the rotation group is the 2-dimensional representation
from Appendix \ref{ss:rotation-reps}.

Let us now recap the above construction of unitary irreducible
representations of the Poincar\'e group for massive
particles.\footnote{This construction is known as the \emph{induced
representation} method \index{induced representation method}
\cite{Mackey_unit}.} First we chose a \emph{standard momentum
vector} \index{standard momentum} $\mathbf{p} = \mathbf{0}$ and
found a \emph{little group}, \index{little group}   which was a
subgroup of the Lorentz group leaving this vector invariant. The
little group turned out to be the rotation group in our case. Then
we found that if the subspace $\mathcal{H}_{\mathbf{0}}$
corresponding to the standard vector carries an irreducible
representation of the little group, then the entire Hilbert space is
guaranteed to carry an irreducible representation of the Poincar\'e
group. In this representation, translations are represented by
multiplication (\ref{eq:7.4}) - (\ref{eq:7.5}), rotations and boosts
are represented by formulas (\ref{eq:7.6}) and (\ref{eq:7.11}),
respectively. It can be shown that a different choice of the
standard vector (i.e., $\mathbf{p} \neq \mathbf{0}$) in the spectrum of momentum would result in a
representation of the Poincar\'e group isomorphic to the one found
above.

\section{Momentum and position representations}
\label{sc:representations}

So far we discussed the action of inertial transformations on common
eigenvectors $| \mathbf{p}, \sigma \rangle$ of the operators
$\mathbf{P}$ and $S_z$. All other vectors in the Hilbert space
$\mathcal{H}$ can be represented as linear combinations of these
basis vectors, i.e., they can be represented as \emph{wave functions}
$\psi(\mathbf{p}, \sigma)$ in the momentum-spin representation.
Similarly one can construct the position space basis $| \mathbf{r}, \sigma \rangle$ from common
eigenvectors of the (commuting) Newton-Wigner position operator and
operator $S_z$. Then arbitrary states in $\mathcal{H}$ can be
represented in this basis by their position-spin wave functions
$\psi(\mathbf{r}, \sigma)$. In this section we will consider
wave function representations of states in greater detail. For
simplicity, we will omit the spin label and consider only spinless
particles. It is remarkable that formulas for the momentum-space and
position-space wave functions appear very similar to those in
non-relativistic quantum mechanics.

\subsection{Spectral decomposition of the identity operator}
\label{ss:spectral}

Two basis vectors with different momenta $| \mathbf{p} \rangle$ and
$| \mathbf{p}' \rangle$ are eigenvectors of the Hermitian operator $
\mathbf{P}$ with different eigenvalues, so they must be orthogonal

\begin{eqnarray*}
\langle\mathbf{p}| \mathbf{p}' \rangle = 0 \ \ \ \mbox{ if } \ \ \ \mathbf{p} \neq
\mathbf{p}'
\end{eqnarray*}

\noindent  If the spectrum of momentum values $\mathbf{p}$ were
discrete we could simply normalize the basis vectors to unity
$\langle\mathbf{p}| \mathbf{p} \rangle = 1$. However, this
normalization becomes problematic in the continuous momentum space. It will be more convenient to use non-normalizable eigenvectors of momentum.
We will call such eigenvectors $|\mathbf{p} \rangle$  \emph{improper}
\index{improper state} states and use them
 to  write  arbitrary ``proper''
normalizable state vectors $| \Psi \rangle$ as integrals

\begin{eqnarray}
 | \Psi \rangle = \int d \mathbf{p} \psi (\mathbf{p}) |\mathbf{p}
\rangle
\label{eq:7.12}
\end{eqnarray}

\noindent where $\psi (\mathbf{q})$ is called the \emph{wave
function in the momentum representation}. \index{wave function in
the momentum representation} It is convenient to demand, in analogy
with (\ref{eq:(4.33a)}), that normalizable wave functions $\psi
(\mathbf{q})$ are given by the inner product

\begin{eqnarray*}
 \psi(\mathbf{q}) = \langle \mathbf{q} |\Psi \rangle = \int
d\mathbf{p} \psi(\mathbf{p})
\langle \mathbf{q}
|\mathbf{p} \rangle
\end{eqnarray*}

\noindent This implies that the inner product of two basis
 vectors is given by the Dirac's delta function (see
Appendix \ref{sc:delta})\footnote{so that the norm of such improper vectors is, actually, infinite $\langle \mathbf{p} |\mathbf{p} \rangle = \infty$}

\begin{eqnarray}
 \langle \mathbf{q} |\mathbf{p} \rangle =  \delta (\mathbf{q}
-\mathbf{p})
\label{eq:7.13}
\end{eqnarray}

\noindent Then in analogy with equation (\ref{eq:A.58}) we can define the
decomposition of the identity operator

\begin{eqnarray}
I =\int d\mathbf{p}|\mathbf{p} \rangle \langle \mathbf{p}|
\label{eq:7.14}
\end{eqnarray}

\noindent Its action on any normalized state vector $| \Psi \rangle $
is trivial, as expected

\begin{eqnarray*}
 I | \Psi \rangle = \int d\mathbf{p}|\mathbf{p}
\rangle \langle \mathbf{p} |\Psi \rangle = \int
d\mathbf{p}|\mathbf{p} \rangle \psi( \mathbf{p} ) = | \Psi \rangle
\end{eqnarray*}

The identity  operator, of course, must be invariant with respect to
Poincar\'e transformations, i.e., we anticipate that

\begin{eqnarray*}
I = U(\Lambda; \mathbf{r}, t) IU^{-1}(\Lambda; \mathbf{r}, t)
\end{eqnarray*}

\noindent The invariance of $I$ with respect to translations follows
directly from equations (\ref{eq:7.4}) and (\ref{eq:7.5}). The invariance
with respect to rotations can be proven as follows

\begin{eqnarray*}
 I'
&=&  e^{-\frac{i}{\hbar}\mathbf{J} \vec{\phi}} I
e^{\frac{i}{\hbar}\mathbf{J} \vec{\phi}} =
e^{-\frac{i}{\hbar}\mathbf{J} \vec{\phi}} \left(\int d\mathbf{p}|
\mathbf{p} \rangle \langle \mathbf{p} | \right)
e^{\frac{i}{\hbar}\mathbf{J} \vec{\phi}}= \int
d\mathbf{p}|R_{\vec{\phi}}\mathbf{p} \rangle
\langle R_{\vec{\phi}} \mathbf{p} | \\
&=& \int d\mathbf{q} \det \left|\frac{d\mathbf{p}}{d\mathbf{q}}
\right| |\mathbf{q} \rangle \langle \mathbf{q} | = \int d\mathbf{q}
|\mathbf{q} \rangle \langle \mathbf{q} | = I
\end{eqnarray*}

\noindent where we used (\ref{eq:7.6}) and the fact that $\det
|d\mathbf{p}/d\mathbf{q}|= \det(R_{-\vec{\phi}}) = 1$ is the Jacobian of the transformation from
variables $\mathbf{p}$ to $ \mathbf{q} = R_{\vec{\phi}} \mathbf{p}
$.

  Let us consider more closely the invariance of $I$ with respect
to boosts. Using equation (\ref{eq:7.11}) we obtain

\begin{eqnarray}
 I'
&=&  e^{-\frac{ic}{\hbar}\mathbf{K}  \vec{\theta}} I
e^{\frac{ic}{\hbar}\mathbf{K} \vec{\theta}} =
e^{-\frac{ic}{\hbar}\mathbf{K} \vec{\theta}} \left(\int d\mathbf{p}|
\mathbf{p} \rangle \langle \mathbf{p} | \right)
e^{\frac{ic}{\hbar}\mathbf{K} \vec{\theta}}\nonumber \\
&=&
\int
d\mathbf{p}|\Lambda \mathbf{p} \rangle
\langle \Lambda \mathbf{p} |
\left|\frac{N(\mathbf{p})}{N(\Lambda \mathbf{p})} \right|^2 \nonumber \\
&=& \int d\mathbf{q}  \det \left|\frac{d \Lambda^{-1} \mathbf{q} }{
d \mathbf{q}} \right| | \mathbf{q} \rangle \langle \mathbf{q} |
\left|\frac{N(\Lambda^{-1}\mathbf{q})}{N(\mathbf{q})} \right|^2
\label{eq:7.15}
\end{eqnarray}

\noindent where $N(\mathbf{q})$ is the normalization factor
introduced in (\ref{eq:7.3}) and
 $\det |d \Lambda^{-1} \mathbf{q} / d\mathbf{q}|$ is the Jacobian of
the transformation from variables $\mathbf{p}$ to $ \mathbf{q}= \Lambda
\mathbf{p}$.  This Jacobian should not depend  on the direction of
the boost $\vec{\theta}$, and we can choose this direction along the
$z$-axis to simplify calculations. Then from (\ref{eq:p'}) we obtain

\begin{eqnarray}
\Lambda^{-1}   q_x&=& q_x \label{eq:labdax} \\
 \Lambda^{-1}  q_y&=& q_y  \label{eq:labday} \\
\Lambda^{-1} q_z &=& q_z \cosh \theta - \frac{1}{c}
\sqrt{m^2c^4 + q^2 c^2} \sinh
\theta \label{eq:labdaz} \\
\omega_{ \Lambda^{-1} \mathbf{q}} &=& \sqrt{m^2c^4 + c^2q_x^2 + c^2q_y^2
+ c^2 \left(q_z \cosh \theta - \frac{1}{c} \sqrt{m^2c^4 + q^2c^2} \sinh
\theta\right)^2} \nonumber \\
&=&  \omega_{ \mathbf{q}} \cosh \theta - cq_z \sinh \theta \nonumber
\end{eqnarray}

\noindent and\footnote{In particular, this means that inside 3D momentum integrals we
are allowed to use the equality

\begin{eqnarray}
\frac{d \mathbf{q}}{\omega_{\mathbf{q}}} = \frac{d (\Lambda
\mathbf{q})} {\omega_{\Lambda \mathbf{q}}} \label{eq:7.15x}
\end{eqnarray}

\noindent for any element $\Lambda$ of the Lorentz group, i.e., that
$d \mathbf{q}/\omega_{\mathbf{q}}$ is a ``Lorentz invariant measure.''  We will use this property quite often in our calculations.}

\begin{eqnarray}
\det \left|\frac{d \Lambda^{-1}\mathbf{q}}{ d\mathbf{q}} \right| &=&
\det \left[ \begin{array}{ccc}
 1 & 0 & 0   \\
  0 & 1 & 0   \\
 \frac{cq_x \sinh \theta}{\sqrt{m^2c^4 + q^2c^2} } &
\frac{cq_y \sinh \theta}{\sqrt{m^2c^4 + q^2c^2} } &
\cosh \theta - \frac{cq_z \sinh \theta}{\sqrt{m^2c^4 + q^2c^2} }
\end{array} \right] \nonumber \\
 &=& \cosh \theta - \frac{cq_z \sinh \theta}{\sqrt{m^2c^4 + q^2c^2}} = \frac{ \omega_{\Lambda^{-1} \mathbf{q} }}{\omega_{\mathbf{q}}}
\label{eq:7.15a}
\end{eqnarray}

\noindent Inserting this result  in equation (\ref{eq:7.15}) we obtain

\begin{eqnarray*}
I' &=& \int d\mathbf{p} \frac{\omega_{\Lambda^{-1} \mathbf{p}
}}{\omega_{\mathbf{p}}} | \mathbf{p} \rangle \langle \mathbf{p} |
\left| \frac{N(\Lambda^{-1} \mathbf{p})}{N(\mathbf{p})} \right|^2
\end{eqnarray*}

\noindent Thus, to ensure the invariance of $I$, we should
define our normalization factor as\footnote{We could also multiply this expression for
$N(\mathbf{p})$ by an arbitrary unimodular factor, but this would
not have any effect, because state vectors and their wave functions
are defined up to an unimodular factor anyway.}

\begin{eqnarray}
N(\mathbf{p}) = \sqrt{\frac{mc^2}{\omega_\mathbf{p}}}
\label{eq:7.16}
\end{eqnarray}

Putting together our results from  equations (\ref{eq:7.4}) -
(\ref{eq:7.6}), (\ref{eq:7.11}), and (\ref{eq:7.16}), we can find
the action of an arbitrary Poincar\'e group element on basis
 vectors $| \mathbf{p}, \sigma\rangle$.  Bearing in
mind that in a general Poincar\'e transformation\footnote{$\Lambda$ is a product of a rotation and a boost, as in equation (\ref{eq:LRB}).} $(\Lambda;
\mathbf{r}; t)$ we agreed\footnote{ see equation (\ref{eq:5.57})} first
to perform translations $(\mathbf{r}, t)$ and then
 boosts/rotations $\Lambda$,
we can find how this group element acts on an one-particle state\footnote{Here we restore spin indices and use active transformations of states, as explained in subsection \ref{ss:inertial-obs}. }

\begin{eqnarray}
&\ & U(\Lambda; \mathbf{r}, t)|\mathbf{p}, \sigma \rangle =
U(\Lambda; \mathbf{0}, 0) e^{-\frac{i}{\hbar}\mathbf{P}\mathbf{r}}
 e^{\frac{i}{\hbar}Ht}|\mathbf{p}, \sigma \rangle \nonumber \\
&=&\sqrt{\frac{\omega_{\Lambda \mathbf{p}}}
{\omega_{\mathbf{p}}}}e^{-\frac{i}{\hbar} \mathbf{p} \cdot \mathbf{r}
+\frac{i}{\hbar}\omega_{ \mathbf{p}}t}
\sum_{\sigma' = -s}^{s} D_{\sigma' \sigma} (\vec{\phi}_W(\mathbf{p},
\Lambda))|\Lambda
\mathbf{p}, \sigma' \rangle
\label{eq:7.17}
\end{eqnarray}

\subsection{Wave function in the momentum
representation}
\label{ss:wave-function}

 The inner product of two normalized vectors $| \Psi \rangle = \int d
\mathbf{p} \psi (\mathbf{p}) |\mathbf{p}\rangle $ and $| \Phi
\rangle = \int d \mathbf{p} \phi (\mathbf{p}) |\mathbf{p}\rangle $
can we written in terms of their wave functions

\begin{eqnarray}
 \langle \Psi | \Phi \rangle &=& \int d \mathbf{p} d
\mathbf{p}'
\psi^* (\mathbf{p}) \phi (\mathbf{p}')\langle \mathbf{p} |\mathbf{p}'\rangle
\nonumber \\
&=& \int d \mathbf{p} d
\mathbf{p}'
\psi^* (\mathbf{p}) \phi (\mathbf{p}')\delta (\mathbf{p}
-\mathbf{p}') \nonumber \\
&=& \int d \mathbf{p}
\psi^* (\mathbf{p}) \phi (\mathbf{p})
\label{eq:7.18}
\end{eqnarray}

\noindent So, for a  state vector $| \Psi \rangle$  with unit
normalization, its wave function $\psi (\mathbf{p})$ must satisfy
the condition

\begin{eqnarray*}
1 =  \langle \Psi | \Psi \rangle &=&
 \int d \mathbf{p}
|\psi (\mathbf{p})|^2
\end{eqnarray*}

\noindent  This wave function  has a direct probabilistic
interpretation, e.g., if $\Omega$ is a region in the momentum space,
then the integral $\int \limits_{\Omega} d\mathbf{p}
|\psi(\mathbf{p})|^2$ gives the probability of finding particle's
momentum inside this region.

Poincar\'e transformations  of the state vector $| \Psi \rangle$ can
be viewed as transformations of the corresponding momentum-space
wave function. For example, using equation (\ref{eq:7.17}) we obtain\footnote{Strictly speaking, operators always act on state vectors. When we apply operators to wave functions, as in (\ref{eq:x}), we will place a caret above the operator symbol.}

\begin{eqnarray}
e^{-\frac{ic}{\hbar}\hat{\mathbf{K}} \vec{\theta}} \psi(\mathbf{p})
&\equiv& \langle \mathbf{ p} | e^{-\frac{ic}{\hbar}\mathbf{K}
\vec{\theta}} | \Psi \rangle = \sqrt{\frac{\omega_{\mathbf{
\Lambda^{-1} p}}}{\omega_{\mathbf{p}}}}
\langle \Lambda^{-1}\mathbf{ p} |  \Psi \rangle \nonumber \\
&=& \sqrt{\frac{\omega_{\Lambda^{-1}\mathbf{ p
}}}{\omega_{\mathbf{p}}}} \psi( \Lambda^{-1}\mathbf{p}) \label{eq:x}
\end{eqnarray}

\noindent Then the boost invariance of the inner product (\ref{eq:7.18})
can be easily proven using property (\ref{eq:7.15x})

\begin{eqnarray*}
\langle \Phi'  | \Psi' \rangle
&=&
          \int d\mathbf{p}
\phi^*( \Lambda^{-1}\mathbf{ p}) \psi( \Lambda^{-1}\mathbf{ p})
\frac{\omega_{\Lambda^{-1}\mathbf{ p}}}{\omega_{\mathbf{p}}} \\
&=&
          \int  \frac{d  (\Lambda^{-1}\mathbf{ p})}{\omega
_{  \Lambda^{-1}\mathbf{ p}}}
\phi^*( \Lambda^{-1}\mathbf{ p}) \psi( \Lambda^{-1}\mathbf{ p})
\omega_{\Lambda^{-1}\mathbf{ p}}
\\
&=&
          \int  d  \mathbf{p}
\phi^*(\mathbf{ p}) \psi(\mathbf{p}) = \langle \Phi  | \Psi \rangle.
\end{eqnarray*}

The action of Poincar\'e generators and the Newton-Wigner position
operator
 on momentum-space wave functions of a massive spinless particle
can be derived from formula (\ref{eq:7.17})

\begin{eqnarray}
 \hat{P}_x \psi(\mathbf{p})  &=& i \hbar \lim_{a \to 0} \frac{d}{d
a}
e^{-\frac{i}{\hbar}\hat{P}_x a}
\psi(\mathbf{p}) = p_x \psi(\mathbf{p})
\label{eq:7.19}\\
  \hat{H}  \psi(\mathbf{p})  &=& -i \hbar \lim_{t \to 0} \frac{d}{d
t}
e^{\frac{i}{\hbar}\hat{H}t}
\psi(\mathbf{p}) =\omega_{\mathbf{p}} \psi(\mathbf{p})
\label{eq:7.20}\\
 \hat{K}_x \psi(\mathbf{p}) &=& \frac{i \hbar}{c} \lim_{\theta \to 0} \frac{d}{d
\theta}
e^{-\frac{ic}{\hbar}\hat{K}_x \theta}
\psi(\mathbf{p}) \nonumber \\
&=& \frac{i \hbar}{c} \lim_{\theta \to 0} \frac{d}{d \theta}
\sqrt{\frac{\sqrt{m^2c^4 + p^2 c^2} \cosh \theta - c p_x \sinh \theta}
{\sqrt{m^2c^4 + p^2 c^2}}} \times \nonumber \\
&\mbox{ }& \psi\left(p_x \cosh
\theta - \frac{1}{c} \sqrt{m^2c^4 + p^2 c^2} \sinh \theta, p_y, p_z \right) \nonumber
\\
&=& i \hbar \left(- \frac{\omega _{ \mathbf{ p}}}{c^2} \frac{d}{dp_x}
- \frac{p_x}{2\omega_{\mathbf{p}}}
 \right) \psi(\mathbf{p})
\label{eq:7.21} \\
\hat{R}_x \psi(\mathbf{p})  &=& -\frac{c^2}{2}(\hat{H}^{-1}\hat{K}_x + \hat{K}_x \hat{H}^{-1})
\psi(\mathbf{p}) \nonumber \\
 &=& -\frac{i \hbar}{2}\left(-\omega^{-1} _{ \mathbf{ p}} \omega _{ \mathbf{
p}} \frac{d}{dp_x} - \omega _{ \mathbf{p}}\frac{d}{dp_x}
 \omega^{-1} _{ \mathbf{ p}} - \frac{p_x c^2}{\omega^2_{\mathbf{p}}} \right)
\psi(\mathbf{p}) \nonumber \\
&=& i \hbar \frac{d}{dp_x} \psi(\mathbf{p}) \label{eq:7.22} \\
 \hat{J}_x \psi(\mathbf{p})  &=& (\hat{R}_y \hat{P}_z - \hat{R}_z \hat{P}_y) \psi(\mathbf{p})
 = i \hbar \left(p_z \frac{d}{dp_y} - p_y \frac{d}{dp_z} \right) \psi(\mathbf{p})
\label{eq:7.23}
\end{eqnarray}

 According to
equation (\ref{eq:7.22}), the exponent of the Newton-Wigner position
operator $e^{\frac{i}{\hbar}\mathbf{R b}}$ acts as a translation
operator in the momentum space: $e^{\frac{i}{\hbar}\mathbf{\hat{R} b}} \psi(\mathbf{p}) = \psi(\mathbf{p-b})$  This suggests the following useful representation for momentum eigenvectors

\begin{eqnarray}
 |\mathbf{p} \rangle & =&
 e^{\frac{i}{\hbar} \mathbf{R} \cdot \mathbf{p}}|\mathbf{0}
\rangle \label{eq:alpha}
\end{eqnarray}

\subsection{Position representation}
\label{ss:position-representation}

In the preceding subsection we considered particle's wave functions in
the momentum representation, i.e., with respect to common
eigenvectors of three commuting components of momentum $P_x, P_y$, and  $P_z$. Three components of the position operator $R_x, R_y$, and  $R_z$ also commute with each other,\footnote{see Theorem
\ref{Theorem6.1}} and their common eigenvectors $| \mathbf{r}
\rangle$ also form a basis in the Hilbert space $\mathcal{H}$ of one
massive spinless particle. In this section we will describe
particle's wave functions with respect to this basis set, i.e., in
the \emph{position representation}. \index{wave function in the
position representation}

First we can expand eigenvectors $|\mathbf{r}\rangle$ in the
momentum basis

\begin{eqnarray}
  | \mathbf{r} \rangle &=&
\int  d\mathbf{p}
\psi_{\mathbf{r}} (\mathbf{p}) |\mathbf{p} \rangle
\label{eq:7.24}
\end{eqnarray}

\noindent The momentum-space eigenfunctions are

\begin{eqnarray}
  \psi_{\mathbf{r}} (\mathbf{p}) &=& \langle \mathbf{p} | \mathbf{r}
\rangle = (2 \pi \hbar)^{-3/2}
 e^{-\frac{i}{\hbar}\mathbf{p}\mathbf{r}}
\label{eq:7.25}
\end{eqnarray}

\noindent as can be verified by substitution of (\ref{eq:7.22}) and
(\ref{eq:7.25}) to the eigenvalue equation

\begin{eqnarray*}
\hat{\mathbf{R}} \psi_{\mathbf{r}} (\mathbf{p}) &=& (2 \pi \hbar)^{-3/2}
\hat{\mathbf{R}}\mbox{ }
 e^{-\frac{i}{\hbar}\mathbf{p}\mathbf{r}}
= i \hbar (2 \pi \hbar)^{-3/2} \frac{d}{d \mathbf{p}} \mbox{ }
 e^{-\frac{i}{\hbar}\mathbf{p}\mathbf{r}}
\\
&=&
\mathbf{r} (2 \pi \hbar)^{-3/2}
   e^{-\frac{i}{\hbar}\mathbf{p}\mathbf{r}} = \mathbf{r} \psi_{\mathbf{r}} (\mathbf{p})
\end{eqnarray*}

\noindent  As operator  $\mathbf{R}$ is Hermitian, its eigenvectors
with different eigenvalues $\mathbf{r}$ and $\mathbf{r}'$ must be
orthogonal.  Indeed, using equation (\ref{eq:delta-rep}) we establish the
delta-function inner product

\begin{eqnarray}
\langle \mathbf{r'} | \mathbf{r} \rangle &=& (2\pi \hbar)^{-3} \int
d\mathbf{p}
 d\mathbf{p}'
e^{-\frac{i}{\hbar}\mathbf{p}\mathbf{r} +
\frac{i}{\hbar}\mathbf{p}'\mathbf{r'}} \langle \mathbf{p}' | \mathbf{p}
\rangle
\nonumber \\
&=& (2\pi \hbar)^{-3} \int
d\mathbf{p}
d\mathbf{p}'
e^{-\frac{i}{\hbar}\mathbf{p}\mathbf{r} +
\frac{i}{\hbar}\mathbf{p}'\mathbf{r'}}
\delta(\mathbf{p} - \mathbf{p}')
\nonumber \\
&=& (2\pi  \hbar)^{-3} \int d\mathbf{p}
e^{-\frac{i}{\hbar}\mathbf{p}(\mathbf{r} -\mathbf{r'})} = \delta
(\mathbf{r} -\mathbf{r'}) \label{eq:position-delta}
\end{eqnarray}

\noindent which means that $| \mathbf{r} \rangle$ are improper
states just as $| \mathbf{p} \rangle$ are. Similarly to
(\ref{eq:7.12}), a normalized state vector $| \Psi \rangle$ can be
represented as an integral over the position space

\begin{eqnarray*}
 |\Psi \rangle = \int d\mathbf{r} \psi(\mathbf{r})| \mathbf{r} \rangle
\end{eqnarray*}

\noindent where $\psi(\mathbf{r}) = \langle \mathbf{r} |\Psi
\rangle$ is the wave function of the state $| \Psi \rangle$ in the
position representation. The absolute square $|\psi(\mathbf{r})|^2$
of the wave function is the probability density for particle's
position. The inner product of two vectors $|\Psi \rangle $ and
$|\Phi \rangle $  can be expressed through their position-space wave
functions

\begin{eqnarray*}
 \langle \Phi |\Psi \rangle &=&
\int d\mathbf{r} d\mathbf{r}'  \phi^*(\mathbf{r})
\psi(\mathbf{r}') \langle \mathbf{r} | \mathbf{r}' \rangle \\
&=& \int d\mathbf{r} d\mathbf{r}'  \phi^*(\mathbf{r})
\psi(\mathbf{r}') \delta( \mathbf{r} - \mathbf{r}')  = \int
d\mathbf{r}   \phi^*(\mathbf{r}) \psi(\mathbf{r})
\end{eqnarray*}

Using Equations (\ref{eq:7.24}) and (\ref{eq:7.25}) we  find that the
position space wave function of a momentum eigenvector is the
usual plane wave

\begin{eqnarray}
  \psi_{\mathbf{p}} (\mathbf{r}) &=& \langle \mathbf{r} | \mathbf{p}
\rangle = (2 \pi \hbar)^{-3/2}
e^{\frac{i}{\hbar}\mathbf{p}\mathbf{r}} \label{eq:plane-wave}
\end{eqnarray}

\noindent As expected, eigenvectors of the position operator in its own representation are given by
delta-functions
(\ref{eq:position-delta}).\footnote{Note that position
eigenfunctions introduced in ref. \cite{Newton_Wigner} do not satisfy
this important requirement.}

 From (\ref{eq:7.14}) we can also obtain a position-space
representation of the identity operator

\begin{eqnarray*}
\int  d\mathbf{r} | \mathbf{r}  \rangle \langle \mathbf{r} |
&=& (2 \pi \hbar)^{-3}\int
d\mathbf{r} d\mathbf{p} d\mathbf{p}'
e^{-\frac{i}{\hbar}\mathbf{p}'\mathbf{r}} | \mathbf{p}  \rangle \langle
\mathbf{p}'|
e^{\frac{i}{\hbar}\mathbf{p}\mathbf{r}} \nonumber \\
&=& \int
d\mathbf{p}d\mathbf{p}'
 | \mathbf{p}  \rangle \langle \mathbf{p}'| \delta(\mathbf{p} - \mathbf{p}')
 = \int d \mathbf{p}
 | \mathbf{p}  \rangle \langle \mathbf{p}|
 =   I
\end{eqnarray*}

Similar to momentum-space formulas (\ref{eq:7.19}) - (\ref{eq:7.23})
we can represent generators of the Poincar\'e group
 by their actions on
position-space wave functions. For example, it follows from
(\ref{eq:7.4}), (\ref{eq:7.24}),
and (\ref{eq:7.25}) that

\begin{eqnarray*}
e^{-\frac{i}{\hbar}\mathbf{P} \mathbf{a}} \int d\mathbf{r}
\psi(\mathbf{r})|\mathbf{r} \rangle &=& \int d\mathbf{r}
\psi(\mathbf{r})|\mathbf{r+a} \rangle = \int d\mathbf{r}
\psi(\mathbf{r-a})|\mathbf{r} \rangle
\end{eqnarray*}

\noindent Therefore we can apply operators directly to wave functions

\begin{eqnarray}
e^{-\frac{i}{\hbar}\hat{\mathbf{P}} \mathbf{a}}
\psi(\mathbf{r}) &=&  \psi(\mathbf{r-a}) \nonumber \\
\hat{\mathbf{P}} \psi(\mathbf{r}) &=& i \hbar \lim_{\mathbf{a} \to 0}
\frac{d }{d \mathbf{a}} e^{-\frac{i}{\hbar}\hat{\mathbf{P}} \mathbf{a}}
\psi(\mathbf{r})  = i \hbar \lim_{\mathbf{a} \to 0} \frac{d }{d
\mathbf{a}} \psi(\mathbf{r-a})  = -i \hbar  \frac{d }{d \mathbf{r}}
\psi(\mathbf{r}) \nonumber \\
\label{eq:mom-oper}
\end{eqnarray}

\noindent Other operators in the position representation  have the
following forms\footnote{Here we used a formal notation for the
Laplacian operator

\begin{eqnarray*}
  \frac{d^2}{d\mathbf{r}^2}  \equiv
\frac{\partial^2}{\partial x^2} + \frac{\partial^2}{\partial y^2} +
\frac{\partial^2}{\partial z^2}
\end{eqnarray*}}

\begin{eqnarray*}
  \hat{H}  \psi(\mathbf{r})  &=& \sqrt{m^2c^4 -\hbar^2 c^2
\frac{d^2}{d\mathbf{r}^2}}
\psi(\mathbf{r}) \\
 \hat{J}_x \psi(\mathbf{r})  &=& -i \hbar \left(y \frac{d}{dz} - z \frac{d}{dy} \right)
\psi(\mathbf{r})  \\
 \hat{K}_x \psi(\mathbf{r}) &=& \frac{1}{2} \left( \sqrt{m^2c^4 -\hbar^2 c^2
\frac{d^2}{d\mathbf{r}^2}} x + x  \sqrt{m^2c^4 -\hbar^2 c^2
\frac{d^2}{d\mathbf{r}^2}} \right)
\psi(\mathbf{r})  \\
 \hat{\mathbf{R}} \psi(\mathbf{r})  &=&  \mathbf{r}\psi(\mathbf{r})  \\
\end{eqnarray*}

The switching between the position-space and momentum-space wave
functions of the same state are achieved by Fourier transformation
formulas. To see that, assume that the state $| \Psi \rangle$ has
a position-space wave function $\psi(\mathbf{r})$. Then using
(\ref{eq:7.24}) and (\ref{eq:7.25}) we obtain

\begin{eqnarray*}
| \Psi \rangle &=&  \int d\mathbf{r}
\psi(\mathbf{r}) | \mathbf{r} \rangle = (2 \pi \hbar)^{-3/2} \int d\mathbf{r}
\psi(\mathbf{r}) \int  d\mathbf{p}
 e^{-\frac{i}{\hbar}\mathbf{p}\mathbf{r}}
|\mathbf{p} \rangle \\
&=& (2 \pi \hbar)^{-3/2} \int  d\mathbf{p}
\left( \int d\mathbf{r}
\psi(\mathbf{r})
e^{-\frac{i}{\hbar}\mathbf{p}\mathbf{r}}\right) |\mathbf{p} \rangle
\end{eqnarray*}

\noindent and the corresponding  momentum-space wave function is

\begin{eqnarray}
\psi(\mathbf{p}) =  (2 \pi \hbar)^{-3/2}
 \int d\mathbf{r}
\psi(\mathbf{r}) e^{-\frac{i}{\hbar}\mathbf{p}\mathbf{r}}
\label{eq:7.26}
\end{eqnarray}

\noindent Inversely, if the momentum-space wave function is
$\psi(\mathbf{p})$, then the position-space wave function
is

\begin{eqnarray}
\psi(\mathbf{r}) = (2 \pi \hbar)^{-3/2} \int
d\mathbf{p}
e^{\frac{i}{\hbar}\mathbf{p}\mathbf{r}} \psi(\mathbf{p})
\label{eq:7.27}
\end{eqnarray}

\subsection{Inertial transformations of observables and states}
\label{ss:inertial-obs}

Here we would like to discuss how observables and states change
under inertial transformations of observers. We already touched this
issue in few places in the book, but it would be useful to summarize
the definitions and to clarify the physical meaning of
transformations. What do we mean exactly when expressing observables
and states in the reference frame $O'$ (primed) through observables
and states in the reference frame $O$ (unprimed)?

\begin{eqnarray}
F' &=& U_gFU_g^{-1} \label{eq:obs-trans} \\
| \Psi' \rangle &=& U_g |\Psi \rangle \label{eq:vec-trans}
\end{eqnarray}

\noindent where\footnote{see equation (\ref{eq:5.57})}

\begin{eqnarray*}
U_g = U_g(\Lambda; \mathbf{a}, t) = e^{-\frac{i}{\hbar}\mathbf{J} \vec{\phi}}
e^{-\frac{ic}{\hbar}\mathbf{K}  \vec{\theta}}
e^{-\frac{i}{\hbar}\mathbf{P}\mathbf{a}} e^{\frac{i}{\hbar}Ht}
\label{eq:5.57x}
\end{eqnarray*}

\noindent is the unitary representative of an inertial transformation
$g$ in the Hilbert space of the system.

\begin{figure}
\centering
 \includegraphics{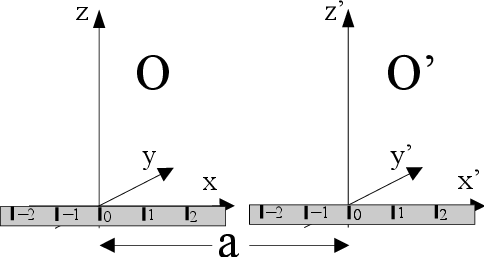} \caption{Rods for measuring the $x$-component of position
 in the reference frame
 $O$ and in the frame $O'$ displaced by the distance $a$.} \label{fig:5.2}
\end{figure}

Let us start with transformations of observables
(\ref{eq:obs-trans}). For definiteness we will assume that observable $F$ is
the $x$-component of position ($F=R_x$). This means that $F$ is a mathematical representation of a measuring rod at rest in the reference frame $O$, as shown in Fig. \ref{fig:5.2}.  The zero pointer on this rod coincides with the
origin of the coordinate system  $O$.   If we further assume that Poincar\'e group element $g$ is a
translation by the distance $a$ along the $x$-axis

\begin{eqnarray*}
U_g = U_g(1;a, 0, 0, 0) = e^{-\frac{i}{\hbar}P_xa}
\end{eqnarray*}

\noindent then the transformed observable

\begin{eqnarray}
R_x' = e^{-\frac{i}{\hbar}P_xa} R_x e^{\frac{i}{\hbar}P_xa}
\label{eq:x'x}
\end{eqnarray}

\noindent is the operator that describes measurements of position in
the reference frame $O'$ with respect to axes and measuring rods in
this frame.  This measuring rod $X'$
is shifted by the distance $a$ with respect to the rod $X$. The zero
pointer on $X'$ coincides with the origin of the coordinate system
$O'$. Of course, position measurements performed by $X$ and $X'$ on the same
particle yield different results and this difference is reflected in the difference of operators $R_x \neq R_x'$. For example, if the particle sits
in the origin of the reference frame $O$, then a measurement with the
rod $X$ yields position value $x=0$, but a measurement with the rod
$X'$ yields $x'=-a$. For this reason we say that observables $R_x$
and $R_x'$ are related by

\begin{eqnarray*}
R_x' = R_x-a
\end{eqnarray*}

\noindent Of course, the same relationship is obtained by a formal
application of (\ref{eq:x'x})

\begin{eqnarray*}
R_x' &=&  R_x - \frac{i}{\hbar}[P_x, R_x]a = R_x-a
\end{eqnarray*}

The position operator $R_x$ can be represented also through its
spectral decomposition

\begin{eqnarray*}
R_x &=& \int \limits_{-\infty}^{\infty} dx x |x \rangle \langle x|
\end{eqnarray*}

\noindent where $|x \rangle $ are eigenvectors (states) with
positions $x$. Then equation (\ref{eq:x'x}) can be rewritten as

\begin{eqnarray*}
R_x' &=& e^{-\frac{i}{\hbar}P_xa} \left( \int
\limits_{-\infty}^{\infty} dx x |x \rangle \langle x| \right)
e^{\frac{i}{\hbar}P_xa} \\
&=& \int \limits_{-\infty}^{\infty} dx x |x+a \rangle \langle x+a| =
\int \limits_{-\infty}^{\infty} dx (x-a) |x \rangle \langle x| =
R_x-a
\end{eqnarray*}

\noindent From this we see that the action of $U_g$ on state vectors\footnote{Here we switch to the Schr\"odinger representation, in which operators of observables are assumed to be fixed.}

\begin{eqnarray*}
 e^{-\frac{i}{\hbar}P_xa}  |x \rangle = |x+a \rangle
\end{eqnarray*}

\noindent should be interpreted as an \emph{active} shift of the
states, i.e., translation of the states by the distance $a$ in this
case.\footnote{One can also interpret
this as a result of the inertial transformation $g$ being applied to
the state preparation device. } For example, operator $e^{-\frac{i}{\hbar}P_xa}$ moves a particle localized in the origin of the frame $O$ ($x=0$) to the origin of the frame $O'$ ($x=a$). However, in many cases of practical interest we are not interested in active
transformations applied to states. More often we are interested in knowing
how the state of the system looks from the point of view of an
inertially transformed observer, i.e., we are interested in
\emph{passive} transformations of states. Apparently
such passive transformations should be represented by inverse
operators $U_g^{-1}$

\begin{eqnarray}
 |\Psi' \rangle &=& U_g^{-1}|\Psi \rangle \label{eq:passive}
\end{eqnarray}

\noindent This means, in particular, that if the vector $|\Psi
\rangle = |x \rangle$ describes a state of the particle located at
point $x$ from the point of view of the observer $O$ (measured by
the rod $X$), then \emph{the same} state is described by the vector

\begin{eqnarray}
|\Psi' \rangle &=& U_g^{-1}|\Psi \rangle = e^{\frac{i}{\hbar}P_xa}
|x \rangle = |x-a \rangle \label{eq:passive2}
\end{eqnarray}

\noindent from the point of view of the observer $O'$ (the position
is measured by the rod $X'$). A particle localized in the origin of the frame $O$ is described by the vector $|0 \rangle$ in that frame. From the point of view of observer $O'$ this same particle is described by the vector $|-a\rangle$.

We can also apply inertial transformations to wave functions instead of state vectors. For example, the state vector

\begin{eqnarray*}
|\Psi \rangle = \int d \mathbf{r} \psi(\mathbf{r}) |\mathbf{r}\rangle
\end{eqnarray*}

\noindent has wave function $\psi(x,y,z)$ in the position representation. When we shift the observer by the distance $a$ in  the positive direction along the $x$-axis, we should apply a passive transformation (\ref{eq:passive2}) to the state vector

\begin{eqnarray*}
|\Psi' \rangle = e^{\frac{i}{\hbar}P_xa}|\Psi \rangle = \int d \mathbf{r} \psi(x,y,z) |x-a, y, z\rangle = \int d \mathbf{r} \psi(x+a,y,z) |x, y, z\rangle
\end{eqnarray*}

\noindent This means that the passive transformation of the wave function has the form

\begin{eqnarray*}
 e^{\frac{i}{\hbar}\hat{P}_xa} \psi(x,y,z) = \psi(x+a,y,z)
\end{eqnarray*}

The above considerations find their most important applications in
the case when the inertial transformation $U_g$ is a time translation.
As we established in (\ref{eq:6.47}), the
position operator in time-translated reference frame $O''$ takes
the form

\begin{eqnarray*}
R_x'' &=& e^{\frac{i}{\hbar} Ht} R_x e^{-\frac{i}{\hbar}Ht} = R_x+
V_xt
\end{eqnarray*}

\noindent If we want to find how the state vector $|\Psi \rangle$
looks from the time translated reference frame $O''$, we should apply
the passive transformation (\ref{eq:passive})

\begin{eqnarray}
|\Psi'' \rangle &=& e^{-\frac{i}{\hbar} Ht} |\Psi \rangle
\label{eq:psi-time}
\end{eqnarray}

\noindent It is common to consider a continuous sequence of time
shifts parameterized by the value of time $t$ and to speak about
\emph{time evolution} of the state vector $|\Psi(t) \rangle $. Then
equation (\ref{eq:psi-time}) can be regarded as a solution of the
time-dependent Schr\"odinger equation \index{time dependent
Schr\"odinger equation}

\begin{eqnarray}
i \hbar \frac{d}{dt}|\Psi(t) \rangle &=& H |\Psi(t) \rangle
\label{eq:psi-timex}
\end{eqnarray}

\noindent In actual calculations it is more convenient to deal with
numerical functions (wave functions in a particular basis) rather than with abstract state vectors. To get this type of description, equation
(\ref{eq:psi-timex}) should be multiplied on the left by certain basis
bra-vectors. For example, if we multiply (\ref{eq:psi-timex}) by
position eigenvectors $\langle \mathbf{r} |$, we will obtain the
Schr\"odinger equation in the position representation

\begin{eqnarray}
i \hbar \frac{d}{dt}\langle \mathbf{r} |\Psi(t) \rangle &=& \langle
\mathbf{r} |H |\Psi(t) \rangle \nonumber \\
i \hbar \frac{d}{dt}\Psi( \mathbf{r},t) &=& \hat{H} \Psi(
\mathbf{r},t) \label{eq:schrod}
\end{eqnarray}

\noindent where $\Psi( \mathbf{r},t) \equiv \langle \mathbf{r}
|\Psi(t) \rangle$ is a wave function in the position
representation and the action of the Hamiltonian on this wave
function is denoted by $\hat{H} \Psi( \mathbf{r},t) \equiv \langle
\mathbf{r} |H |\Psi(t) \rangle$.

\section{Massless particles}
\label{sc:massless}

\subsection{Spectra of momentum,  energy, and velocity}
\label{ss:spectra}

In the case of massless particles ($m=0$), such as photons, the
method used in section \ref{sc:massive} to construct irreducible
unitary representations of the Poincar\'e group  does not work. Indeed, for
massless particles the
 position operator (\ref{eq:6.27}) cannot be defined.  Therefore we cannot
apply the Stone-von Neumann theorem \ref{Theorem.Stone} to figure
out the spectrum of the operator $\mathbf{P}$. To find the spectrum
of $\mathbf{P}$ for a single massless particle we will use another argument.

Let us choose a state of the massless particle with some
nonzero momentum   $\mathbf{p}$.\footnote{We assume that such a value exists
in the spectrum of the momentum operator $\mathbf{P}$.} There are
two kinds of inertial transformations that can affect this momentum value:
rotations and boosts. Any vector $\mathbf{p}'$ obtained from
$\mathbf{p}$ by rotations and boosts is also in the spectrum of
$\mathbf{P}$.\footnote{The proof of this statement is the same as in equations (\ref{eq:proof1}) and (\ref{eq:proof2}).}  So, we can use these transformations to explore the
spectrum of the momentum operator. Rotations generally change the
direction of the momentum vector, but preserve its length $p$, so
all rotation images of $\mathbf{p}$ form a surface of
a sphere with its center in the origin $\mathbf{0}$ and the radius of $p$. Boosts along the momentum vector
$\mathbf{p} $ do not change the direction of this vector, but do
change its length. To decrease the length of the momentum vector we
can use a boost vector $\vec{\theta}$ which points in the direction
opposite to $\mathbf{p} $, i.e.,
$\vec{\theta}/\theta = -\mathbf{p}/p$. Then, using formula
(\ref{eq:p'}) and equality\footnote{This equality follows from
(\ref{eq:one-part-en}) if $m=0$.}

\begin{equation}
\omega_{\mathbf{p}} = c p \label{eq:h-vs-p}
\end{equation}

\noindent  we can write

\begin{eqnarray}
\mathbf{p}' &=& \Lambda \mathbf{p}  = \mathbf{p} +
\frac{\mathbf{p}}{p}
[p (\cosh \theta - 1) - p \sinh \theta] =\mathbf{p} [\cosh \theta  -  \sinh \theta] \nonumber \\
& =&\mathbf{p}e^{-\theta} \label{eq:e-theta}
\end{eqnarray}

\noindent so the transformed momentum reaches zero only in the limit
$\theta \to \infty $. This means that  the point  $\mathbf{p} = 0$ cannot be reached from $\mathbf{p}$ by rotations and bosts. So, this point
does not belong to the spectrum of the momentum of any massless
particle\footnote{ The physical meaning of this result is clear
because there are no photons with zero momentum and energy. }. Then we see that for
massless particles the mass hyperboloid (\ref{eq:one-part-en}) degenerates to a cone
(\ref{eq:h-vs-p})
 with the point $\mathbf{p} = 0$ deleted (see Fig.
\ref{fig:7.2}). Therefore, the spectrum of velocity $\mathbf{V} =
\mathbf{P}c^2/H$ is the surface of a sphere $|\mathbf{v}| = c$. This
means that massless particles can move only with the speed of light
in any reference frame. This is the famous second postulate of
Einstein's special theory of relativity.

\begin{statement} [invariance of the speed of light]
The speed of massless particles (e.g., photons) is equal to $c$
independent on the velocity of the source and/or observer.
\label{statementB}
\end{statement}

\subsection{Representations of the little group}
\label{ss:little-group}

 Next we need to construct   unitary irreducible
representations of the Poincar\'e group for massless particles. To
do that
 we can slightly modify the method of induced representations used for massive particles in
section \ref{sc:massive}.

We already established that vector
$\mathbf{p} = (0,0,0)$ does not belong to the momentum spectrum
of a massless particle. So, unlike in the massive case, we cannot choose this vector  as the standard vector for
the construction of induced massless representations. We also mentioned that the choice of the
standard vector is arbitrary, and representations built on different
standard vectors are unitarily equivalent. Therefore, in the massless case we will choose
a different standard momentum  \index{standard momentum}

 \begin{eqnarray}
\mathbf{k}
= (0,0,1)
\label{eq:7.57}
\end{eqnarray}

\noindent
The next step is to find the little group corresponding to the vector $\mathbf{k}$, i.e., the subgroup of Lorentz transformations, which
leave this vector invariant. The energy-momentum 4-vector
corresponding to the standard vector (\ref{eq:7.57}) is $(ck,
c\mathbf{k}) = (c,0,0,c)$. Therefore, in the 4D notation from
Appendix \ref{ss:4-dim-rep}, the matrices $S$ of little group elements
must satisfy equation

\begin{eqnarray*}
S \left[ \begin{array}{c}
 c \\
 0 \\
 0 \\
 c \\
\end{array} \right]  =  \left[ \begin{array}{c}
 c \\
 0 \\
 0 \\
 c \\
\end{array} \right]
\end{eqnarray*}

\noindent Since the little group is a subgroup of the Lorentz group,
condition (\ref{eq:A.74}) must be fulfilled as well

\begin{eqnarray*}
S^T g S = g
\end{eqnarray*}

\noindent One can verify that the most general matrix $S$ with these
properties has the form \cite{Weinberg_1049}

\begin{eqnarray}
S(X_1, X_2, \theta) = \left[ \begin{array}{cccc}
1+ \frac{1}{2}(X_1^2 + X_2^2) & X_1 & X_2 & - \frac{1}{2}(X_1^2 + X_2^2) \\
X_1 \cos \theta + X_2 \sin \theta &\cos
\theta
 & \sin \theta & -X_1
\cos \theta -X_2 \sin \theta  \\
-X_1 \sin \theta +X_2 \cos \theta &  -\sin
\theta &  \cos \theta  & X_1
\sin \theta -X_2\cos \theta  \\
\frac{1}{2}(X_1^2 + X_2^2) & X_1 & X_2 & 1-\frac{1}{2}(X_1^2 + X_2^2)
\end{array} \right] \nonumber \\
\label{eq:7.58}
\end{eqnarray}

\noindent which depends on three independent real parameters $X_1$, $X_2$, and $\theta$.\footnote{So, just as in the massive case, the massless little group is a 3-dimensional subgroup of the Lorentz group. However, this subgroup is different from the rotation group of the massive case.}
The three generators of these transformations are obtained by
differentiation

\begin{eqnarray*}
T_1 &=&
\lim_{X_1, X_2, \theta \to 0} \frac{\partial}{\partial X_1}
S(X_1, X_2, \theta) = \left[ \begin{array}{cccc}
0 & 1 & 0 & 0 \\
1 & 0 & 0 & -1  \\
0 & 0 & 0 & 0  \\
0 & 1 & 0 & 0
\end{array} \right] = \mathcal{J}_x - c\mathcal{K}_y
\end{eqnarray*}

\begin{eqnarray*}
T_2 &=&
\lim_{X_1, X_2, \theta \to 0} \frac{\partial}{\partial X_2}
S(X_1, X_2, \theta) = \left[ \begin{array}{cccc}
0 & 0 & 1 & 0 \\
0 & 0 & 0 & 0  \\
1 & 0 & 0 & -1  \\
0 & 0 & 1 & 0
\end{array} \right] = \mathcal{J}_y + c\mathcal{K}_x
\end{eqnarray*}

\begin{eqnarray*}
R &=&
\lim_{X_1, X_2, \theta \to 0} \frac{\partial}{\partial \theta}
S(X_1, X_2, \theta) = \left[ \begin{array}{cccc}
0 & 0 & 0 & 0 \\
0 & 0 & 1 & 0  \\
0 & -1 & 0 & 0  \\
0 & 0 & 0 & 0
\end{array} \right] = \mathcal{J}_z
\end{eqnarray*}

\noindent where $\vec{\mathcal{J}}$ and $\vec{\mathcal{K}}$ are
Lorentz group generators (\ref{eq:A.77}) and (\ref{eq:A.78}). The
commutators are easily calculated

\begin{eqnarray*}
\ [T_1, T_2] &=&  0 \\
\  [R, T_2] &=& - T_1 \\
\ [R, T_1] &=&  T_2
\end{eqnarray*}

\noindent These are
commutation relations of the Lie algebra for the group of ``translations''
($T_1$ and $T_2$) and
rotations ($R$) in a 2D plane.

The next step is to find the full set of unitary irreducible
representations of the little group constructed above. We will do that by following the ``induced representation'' prescription outlined at the end of subsection \ref{ss:momentum-poincare}.
 First  we introduce three Hermitian operators
 $\vec{\Pi} = (\Pi_1, \Pi_2)$ and  $\Theta \equiv J_z$, which provide a representation of the  Lie
algebra generators $\mathbf{T}$ and $R$, respectively.\footnote{One can notice a formal analogy of operators $\vec{\Pi}$ and $\Theta$ with 2-dimensional ``momentum'' and ``angular momentum'', respectively.} So,
little group ``translations'' and rotations are represented in the subspace
$\mathcal{H}_{\mathbf{k}}$\footnote{This is the eigensubspace of the particle momentum operator $\mathbf{P}$, corresponding to the ``standard'' eigenvector $\mathbf{k}$.} by unitary operators $e^{-\frac{i}{\hbar} \Pi_1
x}$,
$e^{-\frac{i}{\hbar} \Pi_2 y}$ and $e^{-\frac{i}{\hbar} \Theta
\phi}$.

Next we should clarify the structure of the Hilbert subspace $\mathcal{H}_{\mathbf{k}}$, keeping in mind that this subspace should carry an irreducible representation of the little group. Suppose that the subspace $\mathcal{H}_{\mathbf{k}}$  contains a state
vector
$|\vec{\pi} \rangle$ which is an eigenvector of  $\vec{\Pi}$ with a
nonzero ``momentum'' $\vec{\pi} \neq 0$

\begin{eqnarray*}
\vec{\Pi} |\vec{\pi} \rangle = (\pi_1, \pi_2) |\vec{\pi} \rangle
\end{eqnarray*}

\noindent Then the rotated vector

\begin{eqnarray}
 e^{-\frac{i}{\hbar} \Theta
\phi} |\pi_1, \pi_2 \rangle = |\pi_1 \cos \phi + \pi_2 \sin \phi ,
\pi_1 \sin \phi - \pi_2 \cos \phi\rangle
\label{eq:7.59}
\end{eqnarray}

\noindent  also belongs to the subspace $\mathcal{H}_{\mathbf{k}}$. Vectors
(\ref{eq:7.59}) form a
 circle $\pi_1^2 + \pi_2^2 =
const$ in the 2D ``momentum'' plane. The linear span of these vectors form an infinite-dimensional Hilbert space. This means that
$\mathcal{H}_{\mathbf{k}}$, is infinite-dimensional. If we used this representation of the little group to build the unitary
irreducible representation of the full Poincar\'e group, we would obtain
massless particles having an infinite number of
internal (spin) degrees of freedom, or ``continuous spin.'' Such
particles have not been observed in nature, so we will not discuss
this possibility further. The only case having relevance to physics
is the ``zero-radius'' circle $\vec{\pi} = 0$. These vectors form a one-dimensional irreducible subspace
$\mathcal{H}_{\mathbf{k}}$, where ``translations''  are
represented trivially

\begin{eqnarray}
e^{-\frac{i}{\hbar} \vec{\Pi} \mathbf{r}} |\vec{\pi}  = \mathbf{0}
\rangle
= |\vec{\pi} = \mathbf{0} \rangle
\label{eq:trns-part}
\end{eqnarray}

\noindent and rotations around the $z$-axis are represented by
unimodular factors

\begin{eqnarray}
e^{-\frac{i}{\hbar} \Theta \phi} |\vec{\pi} = \mathbf{0}\rangle
&\equiv& e^{-\frac{i}{\hbar} J_z \phi} |\vec{\pi} =
\mathbf{0}\rangle
 = e^{i \tau \phi} |\vec{\pi} = \mathbf{0}\rangle
 \label{eq:unimod-rot}
\end{eqnarray}

\noindent The allowed values of the parameter $\tau $ can be
obtained from the fact that the representation must be either single- or
double-valued.\footnote{see Statement \ref{statementO}} Therefore,
the rotation through the angle $\phi= 2 \pi$ can be represented by
either 1 or -1, and $\tau$ must be either integer or half-integer
number

\begin{eqnarray}
\tau =\ldots, -1, -1/2, 0, 1/2, 1, \ldots \label{eq:tau}
\end{eqnarray}

We will refer
to the parameter $\tau $ as to \emph{helicity}.\footnote{Note that
$\hbar \tau $ is eigenvalue of the \emph{helicity operator} $ (\mathbf{J}
\cdot \mathbf{P})/P$. } \index{helicity} This parameter
distinguishes different massless unitary irreducible representations
of the Poincar\'e group, i.e., different types of elementary
massless particles.

\subsection{Massless representations of the Poincar\'e group}
\label{ss:rep-poincare}

In the preceding subsection we have built an unitary representation of the little group (which is a subgroup of the Poincar\'e group) in the 1-dimensional subspace $\mathcal{H}_{\mathbf{k}}$ of the standard momentum $\mathbf{k}= (0,0,1)$. In this subsection our goal is to build irreducible unitary representations of the full Poincar\'e group in the entire Hilbert space

\begin{eqnarray}
\mathcal{H} = \oplus_{\mathbf{p}} \mathcal{H}_{\mathbf{p}}
\end{eqnarray}

\noindent of a massless
particle with helicity $\tau$.

First we would like to build a basis in the Hilbert space $\mathcal{H}$. To do that we choose an arbitrary basis
vector $| \mathbf{k}, \tau \rangle$ in the subspace
$\mathcal{H}_{\mathbf{k}}$.\footnote{Here $\tau$ is a fixed half-integer number (\ref{eq:tau}) that specifies the helicity of our particle.} Similarly to what
we did in the massive case, we are going to propagate this basis vector to
other values of momentum $\mathbf{p}$ by using transformations from the Lorentz group. So,
we need to define elements $\lambda_{\mathbf{p}}$ of the Lorentz
group,
 which connect the standard momentum
$\mathbf{k}$ with all other momenta $\mathbf{p}$

\begin{eqnarray*}
\lambda_{\mathbf{p}} \mathbf{k} = \mathbf{p}
\end{eqnarray*}

\noindent Just as in the massive case, the choice of the set of transformations $\lambda_{\mathbf{p}}$ is not unique. However, one can show that representations of the Poincar\'e group constructed with differently chosen $\lambda_{\mathbf{p}}$ are unitarily equivalent. So, we are free to choose any set $\lambda_{\mathbf{p}}$, which makes our calculations more convenient. Our decision is to define $\lambda_{\mathbf{p}}$ as a Lorentz
boost along the z-axis

\begin{eqnarray}
B_{\mathbf{p}} = \left[ \begin{array}{cccc}
 \cosh \theta & 0 & 0 & \sinh \theta \\
 0 & 1  & 0 & 0 \\
 0 &  0 & 1  & 0 \\
 \sinh \theta &  0 & 0 & \cosh \theta
\end{array} \right]
\label{eq:7.60a}
\end{eqnarray}

\noindent followed by a
rotation  $R_{\mathbf{p}}$ which brings direction $\mathbf{k} =
(0,0,1) $ to $\frac{\mathbf{p}}{p}$

\begin{eqnarray}
\lambda_{\mathbf{p}} = R_{\mathbf{p}}B_{\mathbf{p}}
\label{eq:7.60}
\end{eqnarray}

\begin{figure}
\centering
\includegraphics {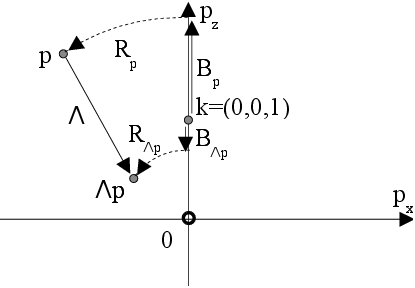} \caption{Each point $\mathbf{p}$
(except $\mathbf{p}=0$) in the momentum space of a massless particle
can be reached from the standard vector $\mathbf{k} = (0,0,1)$ by
applying a boost $B_{\mathbf{p}}$ along the $z$-axis followed by a
rotation $R_{\mathbf{p}}$. } \label{fig:7.4}
\end{figure}

\noindent (see Fig. \ref{fig:7.4}). The rapidity of the boost $\theta = log (p)$
is such that the length of $B_{\mathbf{p}} \mathbf{k}$ is equal to
$p$.\footnote{see formula (\ref{eq:e-theta})} The absolute value of the rotation angle in $R_{\mathbf{p}}$ is

\begin{eqnarray}
\cos \phi = \left(\mathbf{k} \cdot \frac{\mathbf{p}}{p} \right) =
p_z/p \label{eq:7.60b}
\end{eqnarray}

\noindent and the direction of $\vec{\phi}$ is

\begin{eqnarray*}
\frac{\vec{\phi}}{ \phi} &=& \left[\mathbf{k} \times
\frac{\mathbf{p}}{p} \right]/\sin \phi = (p_x^2 + p_y^2)^{-1/2}
(-p_y, p_x, 0) \label{eq:7.60c}
\end{eqnarray*}

\noindent The full basis $|\mathbf{p},\tau \rangle$ in $\mathcal{H}$ is now obtained by
propagating the basis vector $| \mathbf{k}, \tau \rangle$ to other fixed
momentum subspaces $\mathcal{H}_{\mathbf{p}}$\footnote{compare with equations
(\ref{eq:7.3}) and (\ref{eq:7.16})}

\begin{eqnarray}
|\mathbf{p}, \tau \rangle =
\frac{1}{\sqrt{|\mathbf{p}|}}U(\lambda_{\mathbf{p}}; \mathbf{0},
0)|\mathbf{k} ,\tau\rangle \label{eq:xx}
\end{eqnarray}

The next step is to consider how general elements of the
Poincar\'e group act on the basis vectors (\ref{eq:xx}). First we apply a general
transformation from the Lorentz subgroup $U(\Lambda; \mathbf{0},0)$
to an arbitrary basis vector $|\mathbf{p}, \tau\rangle $

\begin{eqnarray*}
 U(\Lambda; \mathbf{0},0)|\mathbf{p}, \tau\rangle
&=& U(\Lambda;\mathbf{0}, 0)
U(\lambda_{\mathbf{p}}; \mathbf{0},0) |\mathbf{k}, \tau \rangle \\
&=& U(\lambda _{\Lambda\mathbf{p}}; \mathbf{0},0)
U(\lambda ^{-1}_{\Lambda\mathbf{p}}; \mathbf{0},0)
U(\Lambda;\mathbf{0}, 0) U(\lambda_{\mathbf{p}};\mathbf{0}, 0)
|\mathbf{k}, \tau \rangle  \\
&=& U(\lambda _{\Lambda\mathbf{p}}; \mathbf{0},0)
U(\lambda ^{-1}_{\Lambda\mathbf{p}}
\Lambda \lambda_{\mathbf{p}}; \mathbf{0},0)
|\mathbf{k}, \tau \rangle  \\
&=& U(\lambda _{\Lambda\mathbf{p}}; \mathbf{0},0)
U(B^{-1}_{\Lambda\mathbf{p}} R^{-1}_{\Lambda\mathbf{p}}
\Lambda R_{\mathbf{p}} B_{\mathbf{p}};\mathbf{0}, 0)
|\mathbf{k}, \tau \rangle
\end{eqnarray*}

\noindent  The product of Lorentz group transformations $\lambda
^{-1}_{\Lambda\mathbf{p}} \Lambda \lambda_{\mathbf{p}} =
B^{-1}_{\Lambda\mathbf{p}} R^{-1}_{\Lambda\mathbf{p}} \Lambda
R_{\mathbf{p}} B_{\mathbf{p}}$ on the right hand side brings vector
$\mathbf{k}$ back to $\mathbf{k}$ (see Fig. \ref{fig:7.4}),
therefore this product is an element of the little group. The
``translation'' part of this element is irrelevant for us due to equation
(\ref{eq:trns-part}).  The relevant angle of rotation around the
$z$-axis is called  the \emph{Wigner angle} \index{Wigner angle}
$\phi_W (\mathbf{p}, \Lambda)$.\footnote{Explicit expressions for
the Wigner rotation angle can be found in \cite{Ritus, Caban}.}
According to equation (\ref{eq:unimod-rot}), this rotation acts as
multiplication by a unimodular factor

\begin{eqnarray*}
U(\lambda ^{-1}_{\Lambda\mathbf{p}} \Lambda \lambda_{\mathbf{p}};
\mathbf{0},0) |\mathbf{k}, \tau \rangle   = e^{i \tau \phi_W(\mathbf{p},
\Lambda)} |\mathbf{k}, \tau \rangle
\end{eqnarray*}

\noindent Thus, taking into account (\ref{eq:xx}) we can write for
arbitrary Lorentz transformation $\Lambda$

\begin{eqnarray*}
&\mbox{ }& U(\Lambda; \mathbf{0},0)|\mathbf{p}, \tau \rangle =
U(\lambda_{\Lambda\mathbf{p}}; \mathbf{0},0) e^{i \tau
\phi_W(\mathbf{p}, \Lambda)}|\mathbf{k}, \tau \rangle
=
\frac{\sqrt{|\Lambda \mathbf{p}|}}{\sqrt{|\mathbf{p}|}} e^{i \tau
\phi_W(\mathbf{p}, \Lambda)} |\Lambda\mathbf{p}, \tau \rangle
\end{eqnarray*}

\noindent For a general Poincar\'e group element we finally obtain a transformation that is similar to the massive case
result (\ref{eq:7.17})

\begin{eqnarray}
U(\Lambda; \mathbf{r}, t)|\mathbf{p}, \tau\rangle &=&
\sqrt{\frac{|\Lambda \mathbf{p}|}{|\mathbf{p}|}}
e^{-\frac{i}{\hbar} \mathbf{p} \cdot \mathbf{r}
+\frac{ic}{\hbar}| \mathbf{p}|t} e^{i \tau \phi_W(\mathbf{p}, \Lambda)}
 |\Lambda
\mathbf{p}, \tau\rangle
\label{eq:7.61}
\end{eqnarray}

As was mentioned in the beginning of this chapter, photons
\index{photon} are described  by a reducible representation of the
Poincar\'e group which is a direct sum of two irreducible
representations with   helicities $\tau=1$ and $\tau =-1$. In the
classical language these two irreducible components correspond to
the left and right circularly polarized light.

\subsection{Doppler effect and aberration}
\label{ss:doppler}

To illustrate results obtained in this section, here we are going to derive well-known formulas for the Doppler effect and the aberration of light. These formulas connects energies and propagation directions of photons viewed from reference frames in relative motion.

 We denote $H(0)$ the photon's energy and $\mathbf{P}(0)$ its
momentum in the reference frame $O$ at rest. We also denote $H(\theta)$ and
$\mathbf{P}(\theta)$  the photon's energy and momentum in the
reference frame $O'$ moving
 with velocity
$\mathbf{v} = c \frac{\vec{\theta}}{\theta} \tanh \theta$. Then  using (\ref{eq:6.3})  we obtain the usual  \emph{Doppler
effect} formula\footnote{The frequency of light is proportional to the photon's energy ($H = \hbar \omega$), so our formula (\ref{eq:doppler}) applies to frequencies as well.}\index{Doppler effect}

\begin{eqnarray}
H(\theta) &=& H(0) \cosh \theta - c\mathbf{P} (0) \cdot
\frac{\vec{\theta}}{\theta} \sinh \theta \nonumber \\
&=& H(0) \cosh \theta \left(1 - \frac{cP
(0)\mathbf{P} (0)}{H(0)P(0)} \cdot
\frac{\vec{\theta}}{\theta} \tanh \theta \right) \nonumber \\
&=& H(0) \cosh \theta \left(1 - \frac{v}{c}
\cos\phi_0  \right)
\label{eq:doppler}
\end{eqnarray}

\noindent where we  denoted $\phi_0 $ the angle between the direction of
photon's
propagation (seen in the reference frame $O$) and the direction of
movement of the reference frame $O'$ with respect to $O$

\begin{eqnarray}
\cos \phi_0 \equiv \frac{\mathbf{P} (0)}{P(0)} \cdot
\frac{\vec{\theta}}{\theta} \label{eq:phi0}
\end{eqnarray}

Sometimes the Doppler effect formula is written in another form
where the angle $\phi$ between the photon momentum and the reference
frame velocity is measured from the point of view of $O'$

\begin{eqnarray}
\cos \phi \equiv \frac{\mathbf{P} (\theta)}{P(\theta)} \cdot
\frac{\vec{\theta}}{\theta} \label{eq:phi1}
\end{eqnarray}

\noindent From (\ref{eq:6.3}) we can write

\begin{eqnarray*}
H(0) &=& H(\theta) \cosh \theta + c\mathbf{P} (\theta) \cdot
\frac{\vec{\theta}}{\theta} \sinh \theta \\
&=& H(\theta) \cosh \theta \left(1 +
\frac{cP(\theta) \mathbf{P} (\theta)}{H(\theta)P(\theta)}
\cdot
\frac{\vec{\theta}}{\theta} \tanh \theta \right) \\
&=& H(\theta) \cosh \theta \left(1 + \frac{v}{c}
\cos\phi  \right)
\end{eqnarray*}

\noindent Therefore

\begin{eqnarray}
H(\theta)
&=&\frac{ H(0)}{ \cosh \theta (1 + \frac{v}{c}
\cos\phi  )}
\label{eq:doppler2}
\end{eqnarray}

The difference between angles $\phi_0$ and $\phi$, i.e., the
dependence of the direction of light propagation on the observer is
known as the \emph{aberration} effect. \index{aberration} In order
to see the same star in the sky observers $O$ and $O'$ must point
their telescopes in different directions. These directions make
angles
 $\phi_0$ and $\phi$, respectively, with the direction
$\frac{\vec{\theta}}{\theta}$ of the relative velocity of $O$ and
$O'$. \index{aberration} The relationship between these angles  can be
found by taking the scalar product of both sides of (\ref{eq:6.2})
with $\frac{\vec{\theta}}{\theta}$ and taking into account equations
(\ref{eq:doppler}) - (\ref{eq:phi1}) and
$cP(\theta) = H(\theta)$

\begin{eqnarray*}
\cos \phi &=& \frac{P(0)}{P(\theta)} (\cosh \theta \cos \phi_0 -
\sinh \theta ) = \frac {\cosh \theta \cos \phi_0 - \sinh \theta }
{\cosh \theta (1 - \frac{v}{c}
\cos\phi_0  )} \\
&=& \frac
{\cos \phi_0 - v/c }
{1 - \frac{v}{c} \cos\phi_0}
\end{eqnarray*}

\begin{figure}
\centering
\includegraphics{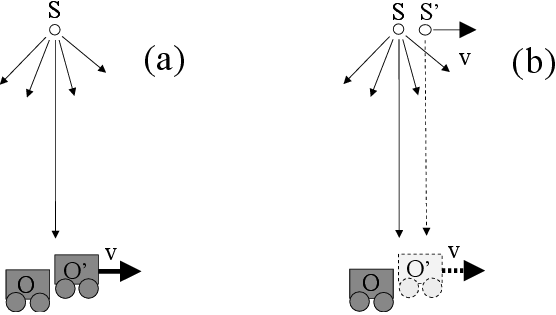} \caption{To the discussion of
the Doppler effect: (a) observer at rest $O$ and moving observer
$O'$ measure light from the same source (e.g., a star) $S$; (b) one
observer $O$ measures light from two sources $S$ and $S'$ that move
with respect to each other.} \label{fig:7.5}
\end{figure}

Our derivations above referred to the case when there was one source
of light and two observers moving with respect to each other (see
Fig. \ref{fig:7.5}(a)). However, this setup is not characteristic
for most astronomical observations of the Doppler effect. In these
observations one typically has one observer and two sources of light
(stars) that move with respect to each other with velocity
$\mathbf{v}$ (see Fig. \ref{fig:7.5}(b)). The aim of observations is to measure the energy (frequency) difference of photons emitted by the two stars. Let us assume that the
distance between the stars $S$ and $S'$ is much smaller than the distance from
the stars to the Earth. Photons emitted simultaneously by $S$ and
$S'$ move with the same speed $c$ and arrive to Earth at the same
time. Two stars are seen by $O$ in the same region of the sky
independent on the velocity $\mathbf{v}$. We also assume that
sources $S$ and $S'$ are identical, i.e., they emit photons of the
same energy in their respective reference frames. Furthermore, we assume
that the energy $h(0)$ of photons arriving from the source $S$ to
the observer $O$ is known. Our goal is to find the energy (denoted
by $h(\theta)$) of photons emitted by $S'$ from the point of view of
$O$. In order to do that, we introduce an imaginary observer $O'$
whose velocity $\mathbf{v}$ with respect to $O$ is the same as
velocity of $S'$ with respect to $S$ and apply the principle of
relativity. According to this principle, the energy of photons from
$S'$ registered by $O'$ is the same as the energy of photons from
$S$ registered by $O$, i.e., $h(0)$. Now, in order to find the
energy of photons from $S'$ seen by $O$ we can apply formula
(\ref{eq:doppler2}) with the opposite sign of velocity

\begin{eqnarray}
h(\theta) &=&\frac{ h(0)}{ \cosh \theta (1 - \frac{v}{c} \cos\phi )}
\label{eq:doppler4}
\end{eqnarray}

\noindent where $\phi$ is the angle between velocity $\mathbf{v}$ of
the star $S'$ and the direction
of
light arriving  from stars $S$ and $S'$ from the point of view of $O$.

Description of the Doppler effect from a different point of view will be found in subsection \ref{ss:doppler-effect}.

\chapter{INTERACTION}
\label{ch:interaction}

\begin{quote}
\textit{I myself, a professional mathematician, on re-reading
my own work find it strains my mental powers to recall
to mind from the figures the meanings of the
demonstrations, meanings which I myself originally put
into the figures  and the text from my mind. But
when I attempt to remedy the obscurity of the
material by putting in extra words, I see  myself falling
into the opposite fault of becoming chatty in something mathematical.}

\small
\hspace{1in} Johannes Kepler
\normalsize
\end{quote}

\vspace{0.5in}

\noindent In the preceding chapter we were concerned with isolated elementary
particles moving freely in space.  Starting from this chapter we
will focus on compound systems consisting of two or more particles.
In addition we will allow  a redistribution of energy and momentum
between different parts of the  system, in other words we will allow
\emph{interactions}. \index{interaction} In this chapter,
 our analysis will be limited to  cases in which  creation
and/or destruction of particles is not allowed, and
only few massive spinless  particles are present. Starting from
chapter \ref{ch:fock-space} we are going to
lift these restrictions and consider interacting systems in full generality.

\section{Hilbert space of a many-particle system}
\label{sc:many-particle}

In this section we will construct the Hilbert space of a compound
system. \index{compound system} In quantum mechanics textbooks it is tacitly assumed
that this space should be built as a tensor product of Hilbert
spaces of the components. Here we will show how this statement can
be proven from postulates of quantum logic.\footnote{see sections
\ref{sc:lattice} - \ref{sc:quant-mech}} For simplicity, we will start with the simplest case of a two-particle system.

\subsection{Tensor product theorem}
\label{ss:tensor-product}

Let  $\mathcal{L}_1$, $\mathcal{L}_2$, and $\mathcal{L}$ be quantum
propositional systems of particles 1, 2, and the compound system
1+2, respectively. It seems reasonable to assume that each
proposition about subsystem 1 (or 2) is still valid in the combined
system. So, propositions in $\mathcal{L}_1$ and $\mathcal{L}_2$ should be represented also as propositions  in
$\mathcal{L}$. Let us formulate this idea as a new postulate

\begin{postulate} [properties of compound systems]
\label{postulateP} If $\mathcal{L}_1$ and $\mathcal{L}_2$ are
quantum propositional systems describing two physical systems 1 and
2, and $\mathcal{L}$ is the quantum propositional system describing
the compound system 1+2, then there exist two mappings

\begin{eqnarray*}
f_1 :\mathcal{L}_1& \to& \mathcal{L} \\
f_2 :\mathcal{L}_2& \to& \mathcal{L}
\end{eqnarray*}

\noindent which satisfy the following conditions:

(I) The mappings $f_1$ and $f_2$ preserve all logical relationships between propositions, so
that

\begin{eqnarray*}
f_1 (\emptyset_{\mathcal{L}_1}) & =& \emptyset_{\mathcal{L}}  \\
f_1 (\mathcal{I}_{\mathcal{L}_1}) & =& \mathcal{I}_{\mathcal{L}}  \\
\end{eqnarray*}

\noindent and for  any two propositions $x,y  \in \mathcal{L}_1$

\begin{eqnarray*}
x\leq y  &\Leftrightarrow&   f_1(x) \leq f_1(x) \\
f_1 (x \wedge y)& =& f_1(x) \wedge f_1 (y)  \\
f_1 (x \vee y)& =& f_1(x) \vee f_1 (y)  \\
f_1 (x^{\perp})& =& (f_1(x))^{\perp}
\end{eqnarray*}

\noindent The same properties are valid for the mapping $f_2 :
\mathcal{L}_2
 \to \mathcal{L}$.

(II) The results of measurements on two subsystems are independent.
This means that in the compound system all propositions about
subsystem 1 are compatible with all propositions about subsystem 2:

\begin{eqnarray*}
f_1 (x_1) \leftrightarrow f_2 (x_2)
\end{eqnarray*}

\noindent where $x_1 \in \mathcal{L}_1$,  $x_2 \in \mathcal{L}_1$

(III) If we have full information about subsystems 1 and 2, then we
have full information about the combined system. Therefore, if $x_1
\in \mathcal{L}_1$ and $x_2 \in \mathcal{L}_2$ are atoms
 then the meet of their images $f_1 (x_1) \wedge f_2
(x_2) $ is also an atomic proposition in $\mathcal{L}$.
\label{postulate-compound}
\end{postulate}

The following theorem \cite{Matolcsi, Aerts} allows us to translate
the above properties of the
compound system from the language of quantum logic to the more
convenient language of
Hilbert spaces.

\bigskip

\begin{theorem} [Matolcsi]
\label{Theorem.Matolcsi} Suppose that $\mathcal{H}_1$,
$\mathcal{H}_2$, and $\mathcal{H}$ are three complex Hilbert spaces
corresponding to the propositional lattices $\mathcal{L}_1$,
$\mathcal{L}_2$, and $\mathcal{L}$, respectively. Suppose also that
$f_1$ and $f_2$ are two mappings satisfying all conditions from
Postulate \ref{postulateP}. Then the Hilbert space $\mathcal{H}$ of
the compound system is either one of the four \index{tensor product}
tensor products\footnote{For definition of the tensor product of two
Hilbert spaces see Appendix \ref{ss:tens-prod}. The star denotes a
dual Hilbert space \index{dual Hilbert space} as described in
Appendix \ref{ss:bra-ket}.} $\mathcal{H} = \mathcal{H}_1 \otimes
\mathcal{H}_2$, or $\mathcal{H} = \mathcal{H}_1^* \otimes
\mathcal{H}_2$, or
 $\mathcal{H} = \mathcal{H}_1
\otimes \mathcal{H}_2^*$, or $\mathcal{H} = \mathcal{H}_1^* \otimes
\mathcal{H}_2^*$.

\end{theorem}

\bigskip

\noindent The proof of this theorem is beyond the scope of our
book.

 So we have four ways to
couple two one-particle Hilbert spaces into one two-particle Hilbert
space. Quantum mechanics uses only the first possibility
$\mathcal{H} = \mathcal{H}_1 \otimes \mathcal{H}_2$.\footnote{It is
not yet clear what is the physical meaning of the other three
possibilities.} This means that if particle 1 is in a state $|1
\rangle \in \mathcal{H}_1$ and particle 2   is in a state $|2
\rangle \in \mathcal{H}_2$, then the state of the compound system is
described by the vector $|1 \rangle \otimes |2 \rangle \in
\mathcal{H}_1 \otimes \mathcal{H}_2$.

\subsection{Particle observables in a multiparticle system}
\label{ss:multi-part-obs}

The mappings $f_1$ and $f_2$ from Postulate \ref{postulateP} map
propositions (projections) from Hilbert spaces of individual
particles $\mathcal{H}_1$ and $\mathcal{H}_2$ into the Hilbert space
$\mathcal{H} = \mathcal{H}_1 \otimes \mathcal{H}_2$ of the compound
system. Therefore, they also map particle observables from
$\mathcal{H}_1$ and $\mathcal{H}_2$ to $\mathcal{H}$. For example,
consider an 1-particle observable $G^{(1)}$ that is represented in
the Hilbert space $\mathcal{H}_1$ by the Hermitian operator with a
spectral decomposition (\ref{eq:op-obs})

\begin{eqnarray*}
G^{(1)} = \sum_{g} g P^{(1)}_{g}
\end{eqnarray*}

\noindent Then the mapping $f_1$ transforms $G^{(1)}$ into a
Hermitian operator $f_1\left(G^{(1)}\right)$ in the Hilbert space $\mathcal{H}$
of the compound system

\begin{eqnarray*}
f_1(G^{(1)}) = \sum_{g} g f_1\left(P^{(1)}_{g}\right)
\end{eqnarray*}

\noindent which has the same spectrum $g$ as $G^{(1)}$. Thus we conclude that all
observables of individual particles, e.g., $\mathbf{P}_1$,
$\mathbf{R}_1$ in $\mathcal{H}_1$ and $\mathbf{P}_2$, $\mathbf{R}_2$
in $\mathcal{H}_2$ have well-defined meanings in the Hilbert space
$\mathcal{H}$ of the combined system.

  In what follows we will use small letters to
 denote observables of individual particles in
$\mathcal{H}$.\footnote{We will keep using capital letters for the
total observables ($H$, $\mathbf{P}$, $\mathbf{J}$, and
$\mathbf{K}$) of the compound system.} For example, the position and
momentum of the particle 1 in the two-particle system will be
denoted as $\mathbf{p}_1$ and $\mathbf{r}_1$.
 The operator of energy of the particle 1
will be written  as $h_1  = \sqrt{m_1^2c^4 + p_1^2 c^2}$,
etc. Similarly, observables of the particle 2 in $\mathcal{H}$ will
be denoted as $\mathbf{p}_2$, $\mathbf{r}_2$, and $h_2$. According
to Postulate \ref{postulate-compound}(II), spectral projections of
observables of different particles commute with each other in
$\mathcal{H}$. Therefore, observables of different particles commute
with each other as well.

Just as in the single-particle case discussed in chapter \ref{ch:single}, two-particle states can be also
described by wave functions. From the properties of the tensor
product of Hilbert spaces it can be derived that if
$\psi_1(\mathbf{r}_1) $ is the wave function of particle 1 and
$\psi_2(\mathbf{r}_2) $ is the wave function of particle 2, then the
wave function of the compound system is simply a product

\begin{eqnarray}
\psi(\mathbf{r}_1, \mathbf{r}_2) =
\psi_1(\mathbf{r}_1 ) \psi_2(\mathbf{r}_2)
\label{eq:8.1}
\end{eqnarray}

\noindent In this case, both particles 1 and 2 and the compound
system are in pure quantum states. However, the most general pure
2-particle state in $\mathcal{H}_1 \otimes \mathcal{H}_2$ is
described by a general (normalizable) function of two vector
variables $\psi(\mathbf{r}_1, \mathbf{r}_2)$ which is not
necessarily expressed in the product form (\ref{eq:8.1}). In this
case, individual subsystems are in mixed states: the results of
measurements performed on the particle 1 are correlated with the
results of measurements performed on the particle 2, even though the
particles do not interact with each other. The existence of such
\emph{entangled states} \index{entangled states} is a distinctive
feature of quantum mechanics, which is not present in the classical
world.

\subsection{Statistics}
\label{ss:statistics}

The above construction of the two-particle Hilbert space
$\mathcal{H}
 = \mathcal{H}_1 \otimes \mathcal{H}_2$ is valid when particles 1 and 2
belong to different species. If particles 1 and 2 are identical,
then there are vectors in $\mathcal{H}_1 \otimes \mathcal{H}_2$ that
do not correspond to physically realizable states, and the Hilbert
space of two-particle states is ``less'' than $\mathcal{H}_1 \otimes
\mathcal{H}_2$. Indeed, if two particles 1 and 2 are identical, then
no measurable quantity will change if these particles are interchanged.
Therefore, after permutation of two particles, the wave function
 may at
most acquire an insignificant unimodular phase factor $\beta$

\begin{eqnarray}
\psi(\mathbf{r}_2,\sigma_2;\mathbf{r}_1, \sigma_1) = \beta \psi(\mathbf{r}_1 , \sigma_1;
\mathbf{r}_2, \sigma_2) \label{eq:8.2}
\end{eqnarray}

\noindent If we swap the particles again then the original wave function must
be
restored

\begin{eqnarray*}
\psi(\mathbf{r}_1, \sigma_1; \mathbf{r}_2, \sigma_2) &=& \beta \psi(\mathbf{r}_2, \sigma_2;
\mathbf{r}_1, \sigma_1) = \beta^2 \psi(\mathbf{r}_1, \sigma_1; \mathbf{r}_2, \sigma_2)
\end{eqnarray*}

\noindent Therefore $\beta^2 = 1$, which implies
 that the factor $\beta$ for any physical state
$\psi(\mathbf{r}_1, \sigma_1; \mathbf{r}_2, \sigma_2)$ in $\mathcal{H}$ can be either 1
or -1. If a vector in $\mathcal{H}$ does not have this property, then this vector does not correspond to a physically realizable state. Thus the Hilbert space of physical states of two identical particles is only a subspace in $\mathcal{H}$.

Is it possible that in a system of two identical particles
one state $\phi_1(\mathbf{r}_1, \sigma_1; \mathbf{r}_2, \sigma_2)$ has factor $\beta $
equal to 1

\begin{eqnarray}
\phi_1(\mathbf{r}_1, \sigma_1; \mathbf{r}_2, \sigma_2) =  \phi_1(\mathbf{r}_2, \sigma_2;
\mathbf{r}_1, \sigma_1) \label{eq:beta1}
\end{eqnarray}

\noindent and another state $\phi_2(\mathbf{r}_1, \sigma_1; \mathbf{r}_2, \sigma_2)$ has
factor $\beta$ equal to -1?

\begin{eqnarray}
\phi_2(\mathbf{r}_1, \sigma_1; \mathbf{r}_2, \sigma_2) &=&  -\phi_2(\mathbf{r}_2, \sigma_2;
\mathbf{r}_1, \sigma_1) \label{eq:beta2}
\end{eqnarray}

\noindent If equations (\ref{eq:beta1}) and (\ref{eq:beta2}) were true,
then the linear combination of the states $\phi_1$ and $\phi_2$

\begin{eqnarray*}
\psi(\mathbf{r}_1, \sigma_1; \mathbf{r}_2, \sigma_2) &=& a \phi_1(\mathbf{r}_1, \sigma_1;
\mathbf{r}_2, \sigma_2) + b \phi_2(\mathbf{r}_1, \sigma_1; \mathbf{r}_2, \sigma_2)
\end{eqnarray*}

\noindent would not transform like (\ref{eq:8.2}) after permutation

\begin{eqnarray*}
\psi(\mathbf{r}_2, \sigma_2; \mathbf{r}_1, \sigma_1) &=&  a\phi_1(\mathbf{r}_2, \sigma_2;
\mathbf{r}_1, \sigma_1) +  b\phi_2(\mathbf{r}_2, \sigma_2; \mathbf{r}_1, \sigma_1) \\
&=&
a\phi_1(\mathbf{r}_1, \sigma_1;
\mathbf{r}_2, \sigma_2) -  b\phi_2(\mathbf{r}_1, \sigma_1; \mathbf{r}_2, \sigma_2) \\
&\neq& \pm \psi(\mathbf{r}_1, \sigma_1; \mathbf{r}_2, \sigma_2)
\end{eqnarray*}

\noindent It then follows that the factor $\beta$ must be the same
for all states in the Hilbert space $\mathcal{H}$ of the system of
two identical particles. This result implies that all particles in
nature are divided in two categories: \emph{bosons} \index{bosons}
and \emph{fermions}. \index{fermions}

For bosons $\beta =1$ and two-particle wave functions are
\emph{symmetric} \index{symmetric
wave function} with respect to permutations.
  Wave functions of two bosons form a
linear subspace $\mathcal{H}_1 \otimes_{sym} \mathcal{H}_2 \subseteq
\mathcal{H}_1 \otimes \mathcal{H}_2$. This means, in particular,
that two identical bosons may occupy the same quantum state, i.e.,
the wave function $\psi(\mathbf{r}_1, \sigma_1) \psi( \mathbf{r}_2, \sigma_2)$ belongs
to the bosonic subspace $\mathcal{H}_1 \otimes_{sym} \mathcal{H}_2$.

For fermions, $\beta =-1$ and two-particle wave functions are
\emph{antisymmetric} \index{antisymmetric wave function} with respect to
permutations of particle variables. The Hilbert space of two identical fermions is the
subspace of antisymmetric functions $\mathcal{H}_1 \otimes_{asym}
\mathcal{H}_2 \subseteq \mathcal{H}_1 \otimes \mathcal{H}_2$. This
means, in particular, that two identical fermions may not occupy the
same quantum state (this is called the \emph{Pauli exclusion
principle}),  \index{Pauli exclusion principle} i.e., the
wavefunction $\psi(\mathbf{r}_1, \sigma_1) \psi( \mathbf{r}_2, \sigma_2)$ does not belong
to the antisymmetric fermionic subspace $\mathcal{H}_1 \otimes_{asym}
\mathcal{H}_2$.

A remarkable \emph{spin-statistics theorem} \index{spin-statistics
theorem} has been proven in the framework of quantum field theory.
This theorem establishes (in full agreement with experiment) that
the symmetry of two-particle wave functions is related to their
spin: all particles with integer spin (e.g., photons) are bosons and
all particles with half-integer spin (e.g., neutrinos, electrons,
protons) are fermions.

All  results of this section  can be immediately generalized to the
case of $n$-particle system, where $n>2$. For example, the Hilbert
space of $n$ identical bosons is the symmetrized tensor product
$\mathcal{H} = \mathcal{H}_1 \otimes_{sym} \mathcal{H}_2
\otimes_{sym} \ldots \otimes_{sym}
 \mathcal{H}_n$, and the Hilbert space of $n$ identical fermions is the
antisymmetrized tensor
product $\mathcal{H} = \mathcal{H}_1 \otimes_{asym} \mathcal{H}_2
\otimes_{asym} \ldots \otimes_{asym}
 \mathcal{H}_n$.

\section{Relativistic Hamiltonian dynamics}
\label{sc:RHD}

To complete our description of the 2-particle system initiated in
the preceding section we  need to specify an unitary representation
$U_g$ of the Poincar\'e group  in the Hilbert space $\mathcal{H} =
\mathcal{H}_1 \otimes \mathcal{H}_2$.\footnote{For simplicity we
will assume that particles 1 and 2 are massive, spinless, and not
identical.} We already know from chapter \ref{ch:operators} that
generators of this representation (and some functions of generators) will
define total observables of the compound system. From subsection
\ref{ss:multi-part-obs} we also know how to define observables of
individual particles in $\mathcal{H}$. If we assume that
 total observables in $\mathcal{H}$ may be
expressed as functions of particle observables  $\mathbf{p}_1$,
$\mathbf{r}_1$, $\mathbf{p}_2$, and $\mathbf{r}_2$, then the
construction of $U_g$ is equivalent to finding 10 Hermitian operator
functions

\begin{eqnarray}
&\mbox{ }& H(\mathbf{p}_1, \mathbf{r}_1, \mathbf{p}_2,\mathbf{r}_2)
\label{eq:8.3}\\
&\mbox{ }& \mathbf{P}(\mathbf{p}_1, \mathbf{r}_1,
\mathbf{p}_2,\mathbf{r}_2)
\label{eq:8.4}\\
&\mbox{ }& \mathbf{J}(\mathbf{p}_1, \mathbf{r}_1,
\mathbf{p}_2,\mathbf{r}_2)
\label{eq:8.5}\\
&\mbox{ }& \mathbf{K}(\mathbf{p}_1, \mathbf{r}_1,
\mathbf{p}_2,\mathbf{r}_2)
\label{eq:8.6}
\end{eqnarray}

\noindent which  satisfy commutation relations of the Poincar\'e Lie
algebra (\ref{eq:5.50}) - (\ref{eq:5.56}). Even in the simplest two-particle
case this problem does not have a unique solution, and additional
physical principles should be applied to make sure that generators
(\ref{eq:8.3}) - (\ref{eq:8.6}) are selected in agreement with
observations. For a general multiparticle system, the construction
of the representation $U_g$ is the most difficult and the most
important part of relativistic quantum theories. A large portion of
the rest of this book is devoted to the analysis of different ways
to construct representation $U_g$. It is important to understand
that once this step is completed, we get everything we need for
 a full theoretical description of the physical system and for comparison of calculations with
experimental data.

\subsection{Non-interacting representation of the Poincar\'e
group} \label{ss:non-interacting}

There are infinitely many ways to define the
 representation $U_g$ of the Poincar\'e group, in the Hilbert space
$\mathcal{H} = \mathcal{H}_1 \otimes \mathcal{H}_2$.   Let us start
our analysis from one legitimate choice   which has a transparent
physical meaning. We  know from chapter \ref{ch:single} that one-particle Hilbert
spaces $\mathcal{H}_1 $ and $\mathcal{H}_2$ carry irreducible
unitary representations $U_g^1$ and $U_g^2$ of the Poincar\'e group.
Functions $f_1$ and $f_2$ defined in subsection
\ref{ss:tensor-product} allow us to map these representations to the
Hilbert space $\mathcal{H}$ of the compound system, i.e., to have
two representations of the Poincar\'e group $f_1(U_g^1)$ and $f_2(U_g^2)
$ in $\mathcal{H}$.\footnote{These representations are no longer irreducible, of course. For example, $f_1(U_g^1)$ is a direct sum of (infinitely) many irreducible representations isomorphic to $U_g^1$.} We can then define a new representation $U_g^0$
of the Poincar\'e group in $\mathcal{H}$ by making a (tensor
product) composition of $f_1(U_g^1)$ and $f_2(U_g^2) $. More
specifically, for any vector $|1 \rangle \otimes |2 \rangle \in
\mathcal{H}$ we define

\begin{eqnarray}
U_g^0 (|1 \rangle \otimes |2 \rangle) = f_1(U^1_g) |1 \rangle
\otimes  f_2(U^2_g) |2 \rangle \label{eq:non-int-rep}
\end{eqnarray}

\noindent and the   action of $U_g^0$ on other vectors in
$\mathcal{H}$ follows  by linearity. Representation
(\ref{eq:non-int-rep}) is called  the \emph{tensor product}
\index{tensor product} of unitary  representations $U^1_g$ and
$U^2_g$ and is denoted by $U_g^0 = U^1_g \otimes U^2_g$. Generators
of this representation are expressed as sums of one-particle
generators

\begin{eqnarray}
H_0 &=& h_1 + h_2
\label{eq:8.7}\\
\mathbf{P}_0 &=& \mathbf{p}_1 + \mathbf{p}_2
\label{eq:8.8}\\
\mathbf{J}_0 &=& \mathbf{j}_1 + \mathbf{j}_2
\label{eq:8.9}\\
\mathbf{K}_0 &=& \mathbf{k}_1 + \mathbf{k}_2
\label{eq:8.10}
\end{eqnarray}

\noindent The Poincar\'e commutation relations for generators
(\ref{eq:8.7}) - (\ref{eq:8.10}) follow immediately from the facts
that one-particle generators corresponding to particles 1 and 2
satisfy Poincar\'e commutation relations separately and that operators of different particles commute with
each other.

With definitions (\ref{eq:8.7}) - (\ref{eq:8.10}), inertial
transformations of particle observables with respect to the
representation $U_g^0$ are easy to find. For example, positions
of particles 1 and 2 change with time as

\begin{eqnarray*}
 \mathbf{r}_1(t) &=& e^{\frac{i}{\hbar}H_0 t } \mathbf{r}_1
e^{-\frac{i}{\hbar}H_0 t }  = e^{\frac{i}{\hbar}h_1  t }
\mathbf{r}_1
e^{-\frac{i}{\hbar}h_1 t } = \mathbf{r}_1 + \mathbf{v}_1 t \\
 \mathbf{r}_2(t)
&=& \mathbf{r}_2 + \mathbf{v}_2 t \\
\end{eqnarray*}

\noindent Comparing this with equation (\ref{eq:6.47}), we conclude that
all observables of particles 1 and 2 transform independently from
each other as if these particles were alone. So, the representation
(\ref{eq:8.7}) - (\ref{eq:8.10}) corresponds to the absence of
interaction
 and is called the
\emph{non-interacting} \index{non-interacting
representation} representation
of the Poincar\'e group.

\subsection{Dirac's forms of dynamics}
\label{ss:dirac-forms}

Obviously, the simple choice of  generators (\ref{eq:8.7}) -
(\ref{eq:8.10}) is not realistic, because particles in nature do
interact with each other. Therefore, to describe interactions in
multi-particle systems one should choose an
 \emph{interacting representation} \index{interacting representation}
 $U_g$ of the
Poincar\'e group in $\mathcal{H}$ which is different from  $U_g^0$.
First we write the generators $(H, \mathbf{P}, \mathbf{J},
\mathbf{K})$ of the desired representation $U_g$ in the most general
form where all generators are different from their non-interacting
counterparts by the presence of interaction terms denoted by
$V, \mathbf{U}, \mathbf{Y}, \mathbf{Z}$\footnote{ Our approach to the
 description of
interactions based on equations (\ref{eq:8.11}) - (\ref{eq:8.14}) and
their generalizations for multiparticle systems is called the
\emph{relativistic Hamiltonian dynamics} \cite{RHD}.
\index{relativistic Hamiltonian dynamics} For completeness, we
should mention that there is a number of other (non-Hamiltonian) methods for
describing interactions.
Overviews of these methods and further references can be found in
\cite{Keister, Van_Dam, covariant}. We will not discuss the
non-Hamiltonian approaches in this book.}

\begin{eqnarray}
H &=& H_0 + V( \mathbf{r}_1,
\mathbf{p}_1, \mathbf{r}_2, \mathbf{p}_2)
\label{eq:8.11}\\
\mathbf{P} &=& \mathbf{P}_0 +  \mathbf{U}( \mathbf{r}_1,
\mathbf{p}_1, \mathbf{r}_2, \mathbf{p}_2)
\label{eq:8.12}\\
\mathbf{J} &=& \mathbf{J}_0 + \mathbf{Y}( \mathbf{r}_1,
\mathbf{p}_1, \mathbf{r}_2, \mathbf{p}_2)
\label{eq:8.13}\\
\mathbf{K} &=& \mathbf{K}_0 +  \mathbf{Z}( \mathbf{r}_1,
\mathbf{p}_1, \mathbf{r}_2, \mathbf{p}_2)
\label{eq:8.14}
\end{eqnarray}

\noindent It may happen that some interaction operators on the right hand
sides of equations (\ref{eq:8.11}) - (\ref{eq:8.14}) are zero. Then these
generators and corresponding finite transformations
 coincide with generators and transformations of the non-interacting representation $U_g^0$. Such
generators and transformations will be referred to as \emph{kinematical}.
\index{kinematical inertial transformation} Generators which contain
interaction terms are called \emph{dynamical}. \index{dynamical
inertial transformation}

\begin{table}[h]
\caption{Comparison of three relativistic forms of dynamics}
\begin{tabular*}{\textwidth}{@{\extracolsep{\fill}}ccc}
 \hline
 Instant form & Point form & Front form \cr
\hline
\multicolumn{3}{c}{ \textbf{Kinematical generators}} \cr
 $P_{0x}$  & $K_{0x}$  & $P_{0x}$   \cr
 $P_{0y}$  & $K_{0y}$  & $P_{0y}$  \cr
 $P_{0z}$  & $K_{0z}$  & $\frac{1}{\sqrt{2}}(H_0 + P_{0z})$   \cr
 $J_{0x}$  & $J_{0x}$  & $\frac{1}{\sqrt{2}}(K_{0x} + J_{0y})$ \cr
 $J_{0y}$  & $J_{0y}$  & $\frac{1}{\sqrt{2}}(K_{0y} - J_{0x})$ \cr
 $J_{0z}$  & $J_{0z}$  & $J_{0z}$ \cr
   &   & $K_{0z}$ \cr
 \hline
\multicolumn{3}{c}{ \textbf{Dynamical generators}} \cr
 $H$ & $H$ & $\frac{1}{\sqrt{2}}(H-P_z)$   \cr
 $K_x$  & $P_x$  & $\frac{1}{\sqrt{2}}(K_x - J_y)$  \cr
 $K_y$  & $P_y$  & $\frac{1}{\sqrt{2}}(K_y + J_x)$   \cr
 $K_z$  & $P_z$  &  \cr
\hline
\end{tabular*}
\label{table:6.1}
\end{table}

The description of interaction by equations (\ref{eq:8.11}) -
(\ref{eq:8.14}) generalizes traditional classical
 \emph{non-relativistic Hamiltonian dynamics} in which
the only dynamical generator is the Hamiltonian $H$.
 To make sure that our theory reduces
to the familiar non-relativistic approach in the limit $c \to \infty$,
we will also assume that time translations are generated by a dynamical Hamiltonian $H=H_0 + V$ with a non-vanishing interaction $V$.
 The choice of other generators is restricted by the observation
that kinematical transformations should form a subgroup of the
Poincar\'e group, so that kinematical generators should form a
subalgebra of the Poincar\'e Lie algebra.\footnote{Indeed, if two
generators $A$ and $B$ do not contain interaction terms, then their
commutator $[A,B]$ should be interaction-free as well.} The set
$(\mathbf{P}, \mathbf{J}, \mathbf{K})$ does not form a subalgebra.
This explains why in the relativistic case we cannot introduce
interaction in the Hamiltonian alone.
 We must add interaction terms to some of the other
generators $\mathbf{P}$, $\mathbf{J}$, or $\mathbf{K}$ in order to
be consistent with relativity.  We will say that interacting
representations having different kinematical subgroups belong to
different \emph{forms of dynamics}. \index{forms of dynamics} In his
famous paper \cite{Dirac}, Dirac provided a classification of forms
of dynamics based on this principle. Table \ref{table:6.1} lists three Dirac's forms
of dynamics most frequently discussed in the literature.  In the case of the \emph{instant
form} \index{instant form} dynamics, the kinematical subgroup is the
subgroup of spatial translations and rotations. In the case of the
\emph{point form} \index{point form} dynamics the kinematical
subgroup is the Lorentz subgroup \cite{Thomas}. In both these
cases the kinematical subgroup has dimension 6. The \emph{front
form} \index{front form} dynamics has the largest number (7) of
kinematical generators.

\subsection{Total observables in a multiparticle system}
\label{ss:total-observables}

Once the interacting representation of the Poincar\'e group and its
generators $(H, \mathbf{P}, \mathbf{J}, \mathbf{K})$ are defined, we
immediately have expressions for \emph{total} observables of the
physical system considered as a whole. These are the \emph{total
energy} $H$, the \emph{total momentum} $\mathbf{P}$, and the
\emph{total angular momentum} $\mathbf{J}$.  Other total observables
of the system (the total mass $M$,  spin $\mathbf{S}$,
center-of-mass position $\mathbf{R}$, etc.) can be obtained as
functions of these generators by  formulas derived in chapter
\ref{ch:operators}.

Note also that inertial transformations of the total observables
$(H, \mathbf{P}, \mathbf{J}, \mathbf{K})$ coincide with those
presented in chapter \ref{ch:operators}. This is because total observables coincide with the Poincar\'e group generators, and this fact is independent on the interaction present in the system.
For example, the total
energy $H$ and the total momentum $\mathbf{P}$ form a 4-vector whose
boost transformations are always given by equation (\ref{eq:6.2}) -
(\ref{eq:6.3}). Boost transformations of the center-of-mass position
$\mathbf{R}$ are those derived in subsection \ref{ss:lorentz-position}.
Time translations result in a uniform movement of the center-of-mass
with constant velocity along a straight line (\ref{eq:6.47}). Thus
we conclude that inertial transformations of total observables are
completely independent on the form of dynamics and on the details of
interactions acting within the multiparticle system.

\section{Instant form of dynamics}
\label{sc:instant-form}

In sections \ref{sc:new-approach} and \ref{sc:discus} we will see that instant form of dynamics agrees with observations better than other forms. So, in this book we will prefer to use instant form interactions to describe realistic systems.

\subsection{General instant form interaction}
\label{sc:instant-general}

In the instant form we  can rewrite equations (\ref{eq:8.11}) -
(\ref{eq:8.14}) as

\begin{eqnarray}
 H &=& H_0 + V \label{eq:gen1}       \\
\mathbf{P} &=& \mathbf{P}_0 \\
\mathbf{J} &=& \mathbf{J}_0 \\
\mathbf{K} &=& \mathbf{K}_0 + \mathbf{Z} \label{eq:gen2}
\end{eqnarray}

\noindent  As we discussed in subsection \ref{ss:total-observables}, the observables $H,
\mathbf{P}, \mathbf{J}$, and $\mathbf{K}$ are total observables that
correspond to the compound system as a whole. The total momentum
$\mathbf{P}$ (\ref{eq:8.8}) and the total angular momentum  $\mathbf{J}$ (\ref{eq:8.9}) are
simply vector sums of the corresponding operators for individual
particles. The total energy $H$ and the boost operator $\mathbf{K}$
are written as sums of one-particle operators plus interaction
terms. The interaction term $V$ in $H$ is usually
called the \emph{potential energy} \index{potential energy}
operator. Similarly, we will call
  $ \mathbf{Z} $
the \emph{potential boost}. \index{potential boost} It is important to note that in an instant-form relativistic interacting system the potential boost operator cannot vanish. We will see later that the non-vanishing ``boost interaction'' has a profound effect on transformations of observables between moving reference frames. The potential boost $ \mathbf{Z} $ will play a crucial role in our non-traditional approach to relativity in the second part of this book.

Other total
observables (e.g., the total mass $M$, spin $\mathbf{S}$,
center-of-mass position $\mathbf{R}$, and its velocity
$\mathbf{V}$, etc.) are defined as functions of generators
(\ref{eq:gen1}) - (\ref{eq:gen2}) by formulas from chapter
\ref{ch:operators}. For interacting systems, these Hermitian operators are generally interaction-dependent as well.

According to the principle of relativity, ten operators (\ref{eq:gen1}) - (\ref{eq:gen2}) must obey Poincar\'e commutation relations (\ref{eq:5.50}) - (\ref{eq:5.56}). This requirement leads to the following equivalent relationships

\begin{eqnarray}
[\mathbf{J}_0, V] &=& [\mathbf{P}_0, V] = 0 \label{eq:8.17}
\\
\mbox{ } [Z_i, P_{0j}] &=& -\frac{i \hbar \delta_{ij}}{c^2} V
\label{eq:8.18}
\\
\mbox{ } [J_{0i}, Z_j] &=& i \hbar \sum_{k=1}^3\epsilon_{ijk} Z_k
\label{eq:8.19}
\\
\mbox{ } [K_{0i}, Z_j] + [Z_i, K_{0j}] + [Z_i, Z_j] &=& 0
\label{eq:8.20}
\\
\mbox{ } [\mathbf{Z}, H_0] + [\mathbf{K}_0, V] + [ \mathbf{Z}, V]
&=& 0 \label{eq:8.21}
\end{eqnarray}

\noindent So, the task of constructing a Poincar\'e invariant
theory of interacting particles has been  reduced to
 finding  a  non-trivial solution for the set of equations
(\ref{eq:8.17})
 - (\ref{eq:8.21}) with
respect to $V$ and $\mathbf{Z}$. These equations are necessary and
sufficient conditions for the relativistic  invariance of our theory.

\subsection{Bakamjian-Thomas construction}
\label{ss:bakamjian}

The set of  equations (\ref{eq:8.17}) - (\ref{eq:8.21}) is rather
complicated. The first non-trivial solution of these equations for
multiparticle systems was found by Bakamjian and Thomas
 \cite{Bakamjian_Thomas}. The idea of their approach was as follows.
 Instead of working with 10 generators $(\mathbf{P}, \mathbf{J}, \mathbf{K},
 H)$, it is convenient to use an alternative set of operators $ \{\mathbf{P}, \mathbf{R}, \mathbf{S}, M\}
 $ introduced in subsection \ref{ss:alternative}.
 Denote
$ \{\mathbf{P}_0, \mathbf{R}_0, \mathbf{S}_0, M_0\} $ and
$ \{\mathbf{P}_0, \mathbf{R}, \mathbf{S}, M\} $
the   sets of operators  obtained
 by using formulas (\ref{eq:6.35}) - (\ref{eq:6.37}) from the non-interacting
$(\mathbf{P}_0, \mathbf{J}_0, \mathbf{K}_0, H_0)$ and interacting
$(\mathbf{P}_0, \mathbf{J}_0, \mathbf{K}, H)$ generators,
respectively.  In a general instant form dynamics all three
operators $\mathbf{R}$, $\mathbf{S}$, and $M$ may contain
interaction terms. However, Bakamjian and Thomas decided to look for
 a simpler solution in
which  the  position operator remains
kinematical $\mathbf{R} = \mathbf{R}_0$. It then
immediately follows  that the spin operator is kinematical as well

\begin{eqnarray*}
\mathbf{S} &=& \mathbf{J} - [\mathbf{R} \times \mathbf{P}] =
\mathbf{J}_0 - [\mathbf{R}_0 \times \mathbf{P}_0] = \mathbf{S}_0
\end{eqnarray*}

\noindent  Then interaction term $N$ is present in the mass operator
only.

\begin{eqnarray*}
M = M_0 + N
\end{eqnarray*}

\noindent From commutators (\ref{eq:8.17}), the interaction  $N$
must satisfy

\begin{eqnarray}
 [\mathbf{P}_0, N] = [\mathbf{R}_0, N] = [\mathbf{J}_0, N] =0
 \label{eq:A}
\end{eqnarray}

\noindent So, we have reduced our task of solving
(\ref{eq:8.17}) - (\ref{eq:8.21}) to a simpler problem of finding
one  operator $N$ satisfying conditions (\ref{eq:A}). Indeed, by
knowing $N$ and non-interacting operators $M_0$, $\mathbf{P}_0$,
$\mathbf{R}_0$, $\mathbf{S}_0$, we can restore the interacting
generators using formulas (\ref{eq:6.38}) - (\ref{eq:6.40})

\begin{eqnarray}
\mathbf{P} &=& \mathbf{P}_0
\label{eq:8.22}\\
H &=& +\sqrt{M^2 c^4 + P_0^2 c^2}
\label{eq:8.23}\\
\mathbf{K} &=& -\frac{1}{2 c^2}(\mathbf{R}_0H + H\mathbf{R}_0) -
\frac{[\mathbf{P}_0 \times \mathbf{S}_0]}{Mc^2+H}
\label{eq:8.24}\\
\mathbf{J} &=& \mathbf{J}_0 = [\mathbf{R}_0 \times \mathbf{P}_0] +
\mathbf{S}_0 \label{eq:8.25}
\end{eqnarray}

Now let us turn to the construction of $N$ in the case of two
massive spinless particles. Suppose that we found two vector
operators $\vec{\pi}$ and $\vec{\rho}$
such that they form a 6-dimensional  Heisenberg Lie
\index{Heisenberg Lie algebra} algebra

\begin{eqnarray}
 [\pi_i, \rho_j] &=& i \hbar \delta_{ij}
\label{eq:8.26} \\
\mbox{ } [\pi_i, \pi_j] &=& [\rho_i, \rho_j] = 0 \label{eq:8.27}
\end{eqnarray}

\noindent  commuting with the center-of-mass position $\mathbf{R}_0$ and the
total
momentum
$\mathbf{P}_0$.

\begin{eqnarray}
 [\vec{\pi}, \mathbf{P}_0] = [\vec{\pi}, \mathbf{R}_0] =  [\vec{\rho},
\mathbf{P}_0]
=  [\vec{\rho}, \mathbf{R}_0] = 0
\label{eq:8.28}
\end{eqnarray}

\noindent Suppose also that these relative operators have the
following non-relativistic ($c \to \infty$) limits

\begin{eqnarray*}
 \vec{\pi} &\to& \mathbf{p}_1 - \mathbf{p}_2 \\
 \vec{\rho} &\to& \mathbf{r}_1 - \mathbf{r}_2
\end{eqnarray*}

\noindent Then observables $\vec{\pi}$ and $\vec{\rho}$ can be interpreted as \emph{relative momentum} \index{relative momentum} and \emph{relative position} \index{relative position} in the two-particle system, respectively. Moreover, any operator in the Hilbert space $\mathcal{H}$ can
be expressed either as a function of $(\mathbf{p}_1, \mathbf{r}_1,
\mathbf{p}_2, \mathbf{r}_2)$ or as a function of $(\mathbf{P}_0,
\mathbf{R}_0, \vec{\pi}, \vec{\rho})$. An interaction
operator $N$ satisfying conditions $ [N, \mathbf{P}_0] = [N,
\mathbf{R}_0] = 0$ can be expressed as a function of $\vec{\pi}$ and
$\vec{\rho}$ only. To satisfy the last condition $[N, \mathbf{J}_0]
= 0$ we will simply require
 $N$ to be an arbitrary function
of rotationally invariant combinations of the 2-particle relative observables

\begin{eqnarray}
 N = N (\pi^2, \rho^2, (\vec{\pi} \cdot \vec{\rho}))
\label{eq:8.29a}
\end{eqnarray}

\noindent In this \emph{ansatz}, the problem of building a
relativistically invariant interaction has reduced to finding
operators of relative positions $\vec{\rho}$ and momenta $\vec{\pi}$
satisfying equations (\ref{eq:8.26}) - (\ref{eq:8.28}). This problem has
been solved in a number of works \cite{Bakamjian_Thomas,
Barsella_Fabri, Osborn, Fong}. We will not need explicit formulas
for the operators of relative observables, so we will not reproduce
them here.

For systems of $n$ massive spinless particles ($n > 2$) similar
arguments apply, but
 instead of one pair of relative operators
$\vec{\pi}$ and $\vec{\rho}$ we will have $n-1$ pairs,

\begin{eqnarray}
\vec{\pi}_r, \vec{\rho}_r, \mbox{   } r=1,2, \ldots, n-1
\label{eq:8.29}
 \end{eqnarray}

\noindent These operators should form a $6(n-1)$-dimensional
Heisenberg algebra  commuting with $\mathbf{P}_0$ and
$\mathbf{R}_0$. Explicit expressions for $\vec{\pi}_r$ and
$\vec{\rho}_r$ were constructed, e.g., in ref. \cite{Chakrabarti}.
As soon as these expressions are found, we can build a
Bakamjian-Thomas interaction in an $n$-particle system by defining
the interaction $N$ as a function of
rotationally invariant combinations of relative operators
(\ref{eq:8.29})

\begin{eqnarray}
N = N(\pi_1^2, \rho_1^2, (\vec{\pi}_1 \cdot \vec{\rho}_1), \pi_2^2,
\rho_2^2, (\vec{\pi}_2 \cdot \vec{\rho}_2), (\vec{\pi}_1 \cdot
\vec{\rho}_2), (\vec{\pi}_2 \cdot \vec{\rho}_1), \ldots)
\label{eq:8.30}
 \end{eqnarray}
\label{bakamjian-end}

\subsection{Non-Bakamjian-Thomas instant forms of dynamics}
\label{ss:general-instant}

In the Bakamjian-Thomas construction, it was assumed that
$\mathbf{R} = \mathbf{R}_0$, but this limitation is rather
artificial and we will see later that realistic particle
interactions do not satisfy this condition. Any non-Bakamjian-Thomas variant of the
instant form dynamics has position operator $\mathbf{R}$  different
from the non-interacting Newton-Wigner position $\mathbf{R}_0$. Let
us now establish a connection between such a general instant form
interaction and the Bakamjian-Thomas form. We are going to
demonstrate that corresponding representations of the Poincar\'e
group are related by a unitary transformation.

Suppose that operators

\begin{eqnarray}
(\mathbf{P}_0,
\mathbf{J}_0, \mathbf{K}, H)
\label{eq:8.31}
\end{eqnarray}

\noindent  define a Bakamjian-Thomas dynamics. Let us now choose  an unitary operator
 $W$ commuting with $\mathbf{P_0}$ and
$\mathbf{J_0}$.\footnote{In the case of two massive spinless
particles such an operator must be a function of rotationally
invariant combinations of vectors $\mathbf{P_0}$, $\vec{\pi}$, and
$\vec{\rho}$.} and
apply this transformation
 to generators (\ref{eq:8.31}).

\begin{eqnarray}
\mathbf{J}_0 &=& W\mathbf{J}_0W^{-1}
\label{eq:8.32} \\
\mathbf{P}_0 &=& W\mathbf{P}_0W^{-1}
\label{eq:8.33}\\
\mathbf{K}' &=& W\mathbf{K}W^{-1}
\label{eq:8.34}\\
H' &=& WHW^{-1}
\label{eq:8.35}
\end{eqnarray}

\noindent  Since unitary transformations preserve commutators

\begin{eqnarray*}
W[A,B]W^{-1} = [WAW^{-1},WBW^{-1}]
\end{eqnarray*}

\noindent the transformed generators (\ref{eq:8.32}) -
(\ref{eq:8.35}) satisfy commutation relations of the Poincar\'e Lie
algebra in the instant form. However, generally, the new mass
operator $M' = c^{-2}\sqrt{(H')^2 - \mathbf{P}_0^2 c^2}$ does not
commute with $\mathbf{R}_0$, so (\ref{eq:8.32}) - (\ref{eq:8.35})
are not necessarily in the Bakamjian-Thomas form.

Thus we have a way to build a
non-Bakamjian-Thomas instant form representation $ (\mathbf{P}_0,
\mathbf{J}_0, \mathbf{K}', H')$ if a Bakamjian-Thomas representation
$ (\mathbf{P}_0, \mathbf{J}_0, \mathbf{K}, H)$ is given. However,
this construction does not answer the question if \emph{all} instant
form interactions can be connected to the Bakamjian-Thomas dynamics
by a unitary transformation? The answer to this question is ``yes'':
For any instant form interaction\footnote{Here it is convenient to use ``alternative sets'' of basic operators  introduced in subsection \ref{ss:alternative}.} $\{\mathbf{P}_0, \mathbf{R}',
\mathbf{S}', M' \}$ one can find a unitary operator $W$ which
transforms it to the Bakamjian-Thomas form \cite{Coester_Polyzou}

\begin{eqnarray}
W^{-1}\{\mathbf{P}_0, \mathbf{R}', \mathbf{S}', M' \} W =
\{\mathbf{P}_0, \mathbf{R}_0, \mathbf{S}_0, M \} \label{eq:WW}
\end{eqnarray}

\noindent To see that, let us consider the simplest two-particle
case. Operator

\begin{eqnarray*}
\mathbf{T} \equiv \mathbf{R}' - \mathbf{R}_0
\end{eqnarray*}

\noindent commutes with $\mathbf{P}_0$. Therefore, it can be written
as a function of $\mathbf{P}_0$ and relative operators $\vec{\pi}$
and  $\vec{\rho}$:  $\mathbf{T}(\mathbf{P}_0, \vec{\pi},
\vec{\rho})$. Then one can show that unitary operator\footnote{The
integral in (\ref{eq:integral2}) can be treated formally as an integral of
ordinary function (rather than operator) along the segment
$[\mathbf{0}, \mathbf{P}_0]$ in the 3D space of variable
$\mathbf{P}_0$ with arguments $\vec{\pi}$ and $\vec{\rho}$ being
fixed. }

\begin{eqnarray}
W &=& e^{\frac{i}{\hbar} \mathcal{W}} \nonumber \\
 \mathcal{W}  &=&
\int_{\mathbf{0}}^{\mathbf{P}_0} \mathbf{T}(\mathbf{P}, \vec{\pi},
\vec{\rho}) d \mathbf{P} \label{eq:integral2}
\end{eqnarray}

\noindent performs the desired transformation (\ref{eq:WW}). Indeed

\begin{eqnarray*}
 W ^{-1} \mathbf{P}_0 W  &=& \mathbf{P}_0 \\
 W ^{-1} \mathbf{J}_0 W  &=& \mathbf{J}_0
\end{eqnarray*}

\noindent  because $W$ is a scalar, which explicitly commutes with
$\mathbf{P}_0$. Operator $\mathcal{W}$ has the following commutators
with the center-of-mass position

\begin{eqnarray*}
 [\mathcal{W}, \mathbf{R}_0]  &=&
- \left[\mathbf{R_0}, \int_{\mathbf{0}}^{\mathbf{P}_0}
\mathbf{T}(\mathbf{P}, \vec{\pi}, \vec{\rho}) d \mathbf{P} \right] = -i \hbar \frac{\partial}{\partial \mathbf{P}_0}
\left(\int_{\mathbf{0}}^{\mathbf{P}_0} \mathbf{T}(\mathbf{P},
\vec{\pi},
\vec{\rho}) d \mathbf{P} \right) \\
&=& -i \hbar \mathbf{T}(\mathbf{P}_0, \vec{\pi},
\vec{\rho})  = -i \hbar (\mathbf{R}' - \mathbf{R}_0) \\
\mbox{ } [\mathcal{W}, [\mathcal{W}, \mathbf{R}_0]] &=& 0
\end{eqnarray*}

\noindent  Therefore

\begin{eqnarray*}
 W \mathbf{R}_0 W^{-1} &=&
 e^{\frac{i}{\hbar} \mathcal{W}} \mathbf{R}_0
e^{-\frac{i}{\hbar} \mathcal{W}} = \mathbf{R}_0 + \frac{i}{\hbar} [ \mathcal{W}, \mathbf{R}_0] -
\frac{1}{2! \hbar^2} [ \mathcal{W}, [ \mathcal{W}, \mathbf{R}_0]] + \ldots \\
&=& \mathbf{R}_0 + (\mathbf{R}' - \mathbf{R}_0)  = \mathbf{R}' \\
W^{-1} \mathbf{R}' W &=& \mathbf{R}_0 \\
 W ^{-1} \mathbf{S}' W & =&
W ^{-1} (\mathbf{J}_0 - \mathbf{R}' \times \mathbf{P}_0) W =
\mathbf{J}_0 - \mathbf{R}_0 \times \mathbf{P}_0 = \mathbf{S}_0
\end{eqnarray*}

\noindent Finally we can apply transformation $W$ to the mass
operator $M'$ and obtain operator

\begin{eqnarray*}
M  = W^{-1} M' W
\end{eqnarray*}

\noindent which commutes with both $\mathbf{R}_0$ and $\mathbf{P}_0$.
This demonstrates that operators on the right hand side of
(\ref{eq:WW}) describe a Bakamjian-Thomas instant form of dynamics.

\subsection{Cluster separability}
\label{ss:separability}

As we saw above, the requirement of Poincar\'e invariance imposes
rather loose conditions on interaction. Relativistic invariance can be satisfied in many different ways. However, there is another physical requirement
which limits the admissible form of interaction. We know from
experiment that all interactions between particles vanish  when
particles are separated by large distances.\footnote{We are not
considering here a hypothetical potential between quarks, which
supposedly grows as a linear function of the distance and results in
the confinement of quarks inside hadrons.} So, if in a 2-particle
system we remove particle 2 to infinity by using the space
translation operator $e^{\frac{i}{\hbar} \mathbf{p}_2 \mathbf{a}}$,
then interaction (\ref{eq:8.29a}) must tend to zero

\begin{eqnarray}
\lim _{\mathbf{a} \to \infty} e^{-\frac{i}{\hbar} \mathbf{p}_2
\mathbf{a}} N (\pi^2, \rho^2, (\vec{\pi} \cdot \vec{\rho}))
e^{\frac{i}{\hbar} \mathbf{p}_2 \mathbf{a}} = 0 \label{eq:8.36}
\end{eqnarray}

\noindent This condition is not difficult to satisfy in the
two-particle case. However, in the relativistic multi-particle case
the mathematical form of this condition becomes rather complicated.
This is because now there is more than one way to separate particles
in mutually non-interacting groups. The form of the $n$-particle
interaction (\ref{eq:8.30}) must ensure that each spatially
separated $m$-particle group ($m < n$) behaves as if it were alone.
This, in particular, implies that we cannot independently choose
interactions in systems with different number of particles. The
interaction in the $n$-particle sector of the theory must be
consistent with interactions in all $m$-particle sectors, where $m <
n$.

Interactions satisfying these conditions are called \emph{cluster separable}. \index{cluster separability} We will postulate that all interactions in nature have the property of separability.

\begin{postulate}[cluster separability of interactions]: All
interactions are cluster separable.  \label{postulateS} This means that  for any division of an
$n$-particle system ($n \geq 2$) into two spatially separated groups
(or \emph{clusters}) \index{cluster} of $l$ and $m$ particles
($l+m=n$)
\begin{enumerate}
\item  the interaction
separates too, i.e., the clusters move independent of each other;
\item  the interaction in each cluster is the same as in separate
$l$-particle and $m$-particle systems, respectively.
\end{enumerate}
\end{postulate}

A  counterexample of a non-separable interaction can be built in the
4-particle case. The interaction Hamiltonian

\begin{equation}
V=  \frac{1}{|\mathbf{r}_1 - \mathbf{r}_2| |\mathbf{r}_3 -
\mathbf{r}_4 |} \label{eq:non-separ}
\end{equation}

 \noindent has the property that no
matter how far two pairs of particles (1+2 and 3+4) are from each
other, the relative distance between 3 and 4 affects the force
acting between particles 1 and 2. Such infinite-range interactions
are not present in nature.

In the non-relativistic case the cluster separability is achieved
without much effort. For example,  the non-relativistic  Coulomb
potential energy in the system  of two charged  particles
is\footnote{Here we are interested just in the general functional
form of interaction, so we are not concerned with putting correct
factors in front of the potentials.}

\begin{eqnarray}
V_{12} &=& \frac{1}{|\vec{\rho}|} \equiv \frac{1}{|\mathbf{r}_1 -
\mathbf{r}_2|} \label{eq:1/r}
\end{eqnarray}

\noindent which clearly satisfies condition (\ref{eq:8.36}).
In the system of three charged particles 1, 2, and 3,
 the potential energy can be
written as a simple sum of two-particle terms

\begin{eqnarray}
V &=& V_{12} + V_{13} + V_{23} \nonumber \\
&=&\frac{1}{|\mathbf{r}_1 - \mathbf{r}_2|} +
\frac{1}{|\mathbf{r}_2 - \mathbf{r}_3|} + \frac{1}{|\mathbf{r}_1 -
\mathbf{r}_3|}
\label{eq:8.37}
\end{eqnarray}

\noindent The spatial separation between particle 3 and the cluster
of particles 1+2 can be increased by applying a large space
translation to the particle 3. In agreement with Postulate
\ref{postulateS}, such a translation  effectively cancels
 interaction between particles in
clusters 3 and 1+2, i.e.

\begin{eqnarray*}
&\mbox{ }& \lim _{\mathbf{a} \to \infty} e^{\frac{i}{\hbar} \mathbf{p}_3
\mathbf{a}}(V_{12} + V_{13} + V_{23}) e^{-\frac{i}{\hbar}
\mathbf{p}_3\mathbf{a}} \\
 &=& \lim _{\mathbf{a} \to \infty} \frac{1}{|\mathbf{r}_1 - \mathbf{r}_2|} +
\frac{1}{|\mathbf{r}_2 - \mathbf{r}_3 + \mathbf{a} |} + \frac{1}{|\mathbf{r}_1
-
\mathbf{r}_3 + \mathbf{a}|} \\
 &=& \frac{1}{|\mathbf{r}_1 - \mathbf{r}_2|}
\end{eqnarray*}

\noindent This is the same potential (\ref{eq:1/r}) as in an isolated 2-particle system. Therefore, both conditions (1) and (2) are satisfied and interaction (\ref{eq:8.37}) is cluster separable. As we will see below, in the relativistic case
construction of a general cluster-separable multi-particle
interaction is a more difficult task.

Let us now make some definitions which will be useful in discussions
of cluster separability.  A \emph{smooth} $m$-particle potential
\index{smooth potential}  $V^{(m)}$ is defined as operator that
depends on variables of $m$ particles and tends to zero if any
particle or a group of particles is removed to infinity.\footnote{In
section \ref{ss:diagrams-general} we will explain why we call such
potentials \emph{smooth}.} For example, the potential (\ref{eq:1/r}) is
smooth while (\ref{eq:non-separ}) is not.
 Generally, a cluster separable interaction in a $n$-particle system
can be written as a sum

\begin{eqnarray}
V = \sum_{\{2\}}V^{(2)} + \sum_{\{3\}}V^{(3)} + \ldots +  V^{(n)}
\label{eq:8.38}
\end{eqnarray}

\noindent where $\sum_{\{2\}} V^{(2)}$ is a sum of smooth
\emph{2-particle potentials} \index{2-particle potential} over all
pairs of particles; $\sum_{\{3\}}V^{(3)}$ is a sum of smooth
\emph{3-particle potentials} over all triples of particles, etc.
\index{3-particle potential} The example in equation
(\ref{eq:8.37}) is a sum of smooth 2-particle potentials.

\subsection{Non-separability of the Bakamjian-Thomas dynamics}
\label{ss:non-separability}

We expect that the property of cluster separability (Postulate \ref{postulateS}) must be valid for both potential
energy and potential boosts in realistic interacting systems. For example, in the relativistic case
of 3 massive spinless particles with interacting generators

\begin{eqnarray*}
H &=& H_0 + V(\mathbf{p}_1,\mathbf{r}_1;\mathbf{p}_2,\mathbf{r}_2;
\mathbf{p}_3,\mathbf{r}_3) \\
\mathbf{K} &=& \mathbf{K}_0 +
\mathbf{Z}(\mathbf{p}_1,\mathbf{r}_1;\mathbf{p}_2,\mathbf{r}_2;
\mathbf{p}_3,\mathbf{r}_3)
\end{eqnarray*}

\noindent  the cluster separability requires, in particular,  that

\begin{eqnarray}
 \lim _{\mathbf{a} \to \infty} e^{\frac{i}{\hbar} \mathbf{p}_3
\mathbf{a}}V(\mathbf{p}_1,\mathbf{r}_1;\mathbf{p}_2,\mathbf{r}_2;
\mathbf{p}_3,\mathbf{r}_3) e^{-i\frac{i}{\hbar}\mathbf{p}_3\mathbf{a}}
 &=& V_{12}(\mathbf{p}_1,\mathbf{r}_1;\mathbf{p}_2,\mathbf{r}_2)
\label{eq:8.39}\\
\lim _{\mathbf{a} \to \infty} e^{\frac{i}{\hbar} \mathbf{p}_3
\mathbf{a}} \mathbf{Z}(\mathbf{p}_1,\mathbf{r}_1;\mathbf{p}_2,\mathbf{r}_2;
\mathbf{p}_3,\mathbf{r}_3) e^{-\frac{i}{\hbar} \mathbf{p}_3\mathbf{a}}
 &=&
\mathbf{Z}_{12}(\mathbf{p}_1,\mathbf{r}_1;\mathbf{p}_2,\mathbf{r}_2)
\label{eq:8.40}
\end{eqnarray}

\noindent where $V_{12}$ and $\mathbf{Z}_{12}$ are interaction operators for
the
2-particle system.

Let us see if these principles can be satisfied by Bakamjian-Thomas interactions. In this case the potential energy  is

\begin{eqnarray*}
V &=& H - H_0 \\
&=& \sqrt{(\mathbf{p}_1 + \mathbf{p}_2 + \mathbf{p}_3)^2 c^2 + (M_0
+ N(\mathbf{p}_1,\mathbf{r}_1;\mathbf{p}_2,\mathbf{r}_2;
\mathbf{p}_3,\mathbf{r}_3))^2 c^4} \\
&\ & -\sqrt{(\mathbf{p}_1 + \mathbf{p}_2 + \mathbf{p}_3)^2 c^2 + M_0^2
c^4}
\end{eqnarray*}

\noindent By removing particle 3 to infinity we obtain

\begin{eqnarray}
&\mbox{ }& \lim _{\mathbf{a} \to \infty} e^{\frac{i}{\hbar} \mathbf{p}_3
\mathbf{a}}V(\mathbf{p}_1,\mathbf{r}_1;\mathbf{p}_2,\mathbf{r}_2;
\mathbf{p}_3,\mathbf{r}_3) e^{-\frac{i}{\hbar} \mathbf{p}_3\mathbf{a}}
\nonumber \\
 &=& \sqrt{(\mathbf{p}_1 + \mathbf{p}_2 + \mathbf{p}_3)^2 c^2
+ (M_0 + N(\mathbf{p}_1,\mathbf{r}_1;\mathbf{p}_2,\mathbf{r}_2;
\mathbf{p}_3,\infty))^2 c^4} \nonumber\\
&\ &
-\sqrt{(\mathbf{p}_1 + \mathbf{p}_2 + \mathbf{p}_3)^2 c^2
+ M_0^2 c^4}
\label{eq:8.41}
\end{eqnarray}

\noindent According to (\ref{eq:8.39}) we should require that the right hand
side of
equation (\ref{eq:8.41})
depends only on
 variables pertinent to particles 1 and 2. Then we must set

\begin{eqnarray*}
 N(\mathbf{p}_1,\mathbf{r}_1;\mathbf{p}_2,\mathbf{r}_2;
\mathbf{p}_3,\infty) = 0
\end{eqnarray*}

\noindent which also means that

\begin{eqnarray*}
 V(\mathbf{p}_1,\mathbf{r}_1;\mathbf{p}_2,\mathbf{r}_2;
\mathbf{p}_3,\infty) &=&  V_{12}(\mathbf{p}_1,\mathbf{r}_1;
\mathbf{p}_2,\mathbf{r}_2) = 0
\end{eqnarray*}

\noindent and interaction in the 2-particle sector 1+2 vanishes. Similarly, we can show that  interaction $V$  tends to zero when
either  particle 1 or particle 2 is removed to infinity. Therefore,
 $V$ is a smooth 3-particle potential, \index{3-particle potential}
and there is no interaction in any
2-particle subsystem: the interaction turns on only if there are
three or more particles close to each other. This is clearly
unphysical. So, we conclude that the Bakamjian-Thomas construction
cannot describe a non-trivial cluster-separable interaction in
many-particle systems (see also \cite{Mutze}).

\subsection{Cluster separable
3-particle interaction}
\label{ss:3-particle}

The problem of constructing relativistic cluster separable
many-particle interactions can be solved  by allowing
non-Bakamjian-Thomas instant form interactions.
 Our goal here is to construct the interacting
Hamiltonian $H$ and boost $\mathbf{K}$ operators
 in the Hilbert space $\mathcal{H} = \mathcal{H}_1 \otimes \mathcal{H}_2
\otimes \mathcal{H}_3$ of a 3-particle system so that interaction satisfies the separability
Postulate \ref{postulateS}, i.e., it reduces to a non-trivial
2-particle interaction when one of the particles is removed to
infinity. In this construction we follow ref.
\cite{Coester_Polyzou}.

Let us assume that  2-particle potentials $V_{ij}$ and
$\mathbf{Z}_{ij}$, $i,j =1,2,3$ resulting from removing particle $k
\neq i,j$ to infinity are known. They depend on variables of the
$i$-th and $j$-th particles only. For example, when particle 3 is
removed to infinity, the interacting operators take the form\footnote{Here we used (\ref{eq:6.26}) and took into account that $[\mathbf{P}_0 \times \mathbf{S}_{12}] = [\mathbf{P}_0 \times \mathbf{W}_{12}]/(M_{12}c)$. Similar equations result
from the removal of particles 1 or 2 to infinity. They are obtained
from (\ref{eq:8.42}) - (\ref{eq:8.45}) by permutation of indices
(1,2,3).}

\begin{eqnarray}
\lim _{\mathbf{a} \to \infty} e^{\frac{i}{\hbar}\mathbf{p}_3
\mathbf{a}}
H e^{-\frac{i}{\hbar}\mathbf{p}_3 \mathbf{a}} &=& H_0 +
V_{12} \equiv H_{12}
\label{eq:8.42}\\
\lim _{\mathbf{a} \to \infty}e^{\frac{i}{\hbar}\mathbf{p}_3
\mathbf{a}}\mathbf{K} e^{-\frac{i}{\hbar}\mathbf{p}_3 \mathbf{a}} &=&
\mathbf{K}_0 +
\mathbf{Z}_{12} \equiv \mathbf{K}_{12}
\label{eq:8.43}\\
\lim _{\mathbf{a} \to \infty} e^{\frac{i}{\hbar}\mathbf{p}_3
\mathbf{a}} M e^{-\frac{i}{\hbar}\mathbf{p}_3 \mathbf{a}} &=&
\frac{1}{c^2} \sqrt{ H^2_{12} - P_0^2 c^2} \equiv M_{12}
\label{eq:8.44}\\
\lim _{\mathbf{a} \to \infty} e^{\frac{i}{\hbar}\mathbf{p}_3
\mathbf{a}}
\mathbf{R} e^{-\frac{i}{\hbar}\mathbf{p}_3 \mathbf{a}}
&=& -\frac{c^2}{2}(\mathbf{K}_{12} H_{12} + H_{12}
\mathbf{K}_{12}) - \frac{c [\mathbf{P}_0 \times
\mathbf{W}_{12}]}{M_{12}H_{12}(M_{12}c^2 +H_{12})} \nonumber \\
&\equiv&
\mathbf{R}_{12}
\label{eq:8.45}
\end{eqnarray}

\noindent where operators $H_{12}$, $\mathbf{K}_{12}$, $M_{12}$, and
$\mathbf{R}_{12}$ (energy, boost, mass, and center-of-mass position,
respectively) will be considered as given. Now we want to combine the two-particle potentials $V_{ij}$
and $\mathbf{Z}_{ij}$ together in a cluster-separable 3-particle
interaction in analogy with (\ref{eq:8.37}). It appears that we
cannot form the interactions $V$ and $\mathbf{Z}$ in the 3-particle
system simply as a sum of 2-particle potentials. One can verify that
such a definition would violate Poincar\'e commutators. Therefore

\begin{eqnarray*}
V &\neq& V_{12} + V_{23} + V_{13} \\
\mathbf{Z} &\neq& \mathbf{Z}_{12} + \mathbf{Z}_{23} + \mathbf{Z}_{13}
\end{eqnarray*}

\noindent  and the relativistic ``addition of interactions'' \index{addition of interactions} should
be more complicated.

 When particles 1 and 2 are split apart, operators
$V_{12}$ and
$\mathbf{Z}_{12}$ must tend to zero, therefore

\begin{eqnarray}
\lim _{\mathbf{a} \to \infty}
e^{\frac{i}{\hbar}\mathbf{p}_1 \mathbf{a}} M_{12} e^{-\frac{i}{\hbar}
\mathbf{p}_1 \mathbf{a}} &=&
M_0 \label{eq:6.57a}\\
\lim _{\mathbf{a} \to \infty} e^{\frac{i}{\hbar}\mathbf{p}_2
\mathbf{a}}
M_{12} e^{-\frac{i}{\hbar}\mathbf{p}_2 \mathbf{a}} &=&
M_0 \label{eq:6.57b} \\
\lim _{\mathbf{a} \to \infty} e^{\frac{i}{\hbar}\mathbf{p}_3
\mathbf{a}}
M_{12} e^{-\frac{i}{\hbar}\mathbf{p}_3 \mathbf{a}} &=&
M_{12} \label{eq:6.57c}
\end{eqnarray}

\noindent The Hamiltonian $H_{12}$ and boost $\mathbf{K}_{12}$
define an instant form representation $U_{12}$ of the Poincar\'e
group in the 3-particle Hilbert space $\mathcal{H}$. The corresponding position
operator (\ref{eq:8.45}) is generally different from the
non-interacting Newton-Wigner position operator

\begin{eqnarray}
\mathbf{R}_0 = -\frac{c^2}{2}(\mathbf{K}_0 H_0 + H_0
\mathbf{K}_0) - \frac{c [\mathbf{P}_0 \times
\mathbf{W}_0]}{M_0H_0(M_0+H_0)}
\label{eq:newt-wig}
\end{eqnarray}

\noindent which is characteristic for the Bakamjian-Thomas form of
dynamics. However, we can unitarily transform the representation
$U_{12}$, so that it acquires a Bakamjian-Thomas form with operators
$\mathbf{R}_0, \overline{H}_ {12}, \overline{\mathbf{K}}_ {12},
\overline{M}_ {12} $.\footnote{see subsection
\ref{ss:general-instant}} Let us denote such an unitary
transformation operator by $B_{12}$. We can repeat the same steps
for two other pairs of particles 1+3 and 2+3 and write in the
general case $i,j = 1,2,3;$ $i \neq j$

\begin{eqnarray*}
B_{ij} \mathbf{R}_ {ij} B^{-1}_{ij} &=& \mathbf{R}_0 \\
B_{ij} H_ {ij} B^{-1}_{ij} &=& \overline{H}_ {ij} \\
B_{ij} \mathbf{K}_ {ij} B^{-1}_{ij} &=& \overline{\mathbf{K}}_ {ij} \\
B_{ij} M_ {ij} B^{-1}_{ij} &=& \overline{M}_ {ij}
\end{eqnarray*}

\noindent Operators $B_{ij} = \{B_{12}, B_{13}, B_{23} \}$ commute
with $\mathbf{P}_0$ and $\mathbf{J}_0$. Since representation
$U_{ij}$ becomes non-interacting when the distance between particles
 $i$ and $j$ tends to infinity, we can write

\begin{eqnarray}
\lim _{\mathbf{a} \to \infty} e^{\frac{i}{\hbar}\mathbf{p}_3
\mathbf{a}} B_{13} e^{-\frac{i}{\hbar}\mathbf{p}_3 \mathbf{a}} &=& 1
\label{eq:8.46}\\
\lim _{\mathbf{a} \to \infty} e^{\frac{i}{\hbar}\mathbf{p}_3
\mathbf{a}} B_{23}
e^{-\frac{i}{\hbar}\mathbf{p}_3 \mathbf{a}} &=& 1
\label{eq:8.47}\\
\lim _{\mathbf{a} \to \infty} e^{\frac{i}{\hbar}\mathbf{p}_3
\mathbf{a}} B_{12}
e^{-\frac{i}{\hbar}\mathbf{p}_3 \mathbf{a}} &=& B_{12}
\label{eq:8.48}
\end{eqnarray}

\noindent The transformed Hamiltonians $\overline{H}_ {ij}$ and boosts
$\overline{\mathbf{K}}_ {ij}$ define Bakamjian-Thomas representations and their mass
operators  $\overline{M}_{ij}$
now commute with $\mathbf{R}_0 $.
So, we can  add $\overline{M}_{ij}$  together to build a new mass operator

\begin{eqnarray*}
\overline{M} &=&
\overline{M}_{12} + \overline{M}_{13} + \overline{M}_{23} - 2 M_0\\
&=& B_{12} M_{12} B_{12}^{-1} + B_{13} M_{13} B_{13}^{-1} +
B_{23} M_{23} B_{23}^{-1} - 2 M_0
\end{eqnarray*}

\noindent which also commutes with $\mathbf{R}_0$. Using this mass operator,
we can build another
Bakamjian-Thomas representation with generators

\begin{eqnarray}
\overline{H}&=& \sqrt{P_0^2 + \overline{M}^2}
\label{eq:8.49}\\
\overline{\mathbf{K}}&=& -\frac{1}{2c^2}(\mathbf{R}_0 \overline{H}
+ \overline{H}\mathbf{R}_0) - \frac{c [\mathbf{P}_0
\times \mathbf{W}_0]}{\overline{M}
\overline{H}(\overline{M}c^2+\overline{H})}
\label{eq:8.50}
\end{eqnarray}

\noindent This representation has interactions between all particles,
however,
it does not satisfy the cluster property
yet. For example, by removing particle 3 to infinity we do not obtain
the interaction $M_{12}$ characteristic for the subsystem of two particles 1
and
2. Instead, we obtain a unitary transform of $M_{12}$\footnote{Here we used (\ref{eq:6.57a}) - (\ref{eq:6.57c}) and (\ref{eq:8.46}) - (\ref{eq:8.48}).}

\begin{eqnarray}
&\mbox { }& \lim _{\mathbf{a} \to \infty}
e^{\frac{i}{\hbar}\mathbf{p}_3 \mathbf{a}} \overline{M}
e^{-\frac{i}{\hbar}\mathbf{p}_3 \mathbf{a}}
\nonumber\\
&=& \lim _{\mathbf{a} \to \infty}  e^{\frac{i}{\hbar}\mathbf{p}_3 \mathbf{a}}
(B_{12} M_{12} B_{12}^{-1} + B_{13} M_{13} B_{13}^{-1} +
B_{23} M_{23} B_{23}^{-1} - 2 M_0) e^{-\frac{i}{\hbar}\mathbf{p}_3
\mathbf{a}} \nonumber\\
&=&  B_{12} M_{12} B_{12}^{-1} - 2M_0 +  \lim _{\mathbf{a} \to \infty}
(e^{\frac{i}{\hbar}\mathbf{p}_3
\mathbf{a}}
M_{13} e^{-\frac{i}{\hbar}\mathbf{p}_3 \mathbf{a}} +
e^{\frac{i}{\hbar}\mathbf{p}_3 \mathbf{a}} M_{23}
e^{-\frac{i}{\hbar}\mathbf{p}_3
 \mathbf{a}}) \nonumber\\
&=&  B_{12} M_{12} B_{12}^{-1} - 2 M_0 + 2 M_0 = B_{12} M_{12}
B_{12}^{-1} \label{eq:8.51}
\end{eqnarray}

\noindent To fix this deficiency, let us perform a unitary
transformation of the representation (\ref{eq:8.49}) - (\ref{eq:8.50}) with
operator $B$\footnote{which
must commute with $\mathbf{P}_0$ and $\mathbf{J}_0$, of course, to
preserve the instant form of interaction}

\begin{eqnarray}
H &=& B^{-1}\overline{H} B
\label{eq:8.52}\\
\mathbf{K} &=& B^{-1} \overline{\mathbf{K}} B
\label{eq:8.53}\\
M &=& B^{-1}\overline{M} B
\label{eq:8.54}
\end{eqnarray}

\noindent We choose the transformation $B$ from the requirement that
it must  cancel factors $B_{ij}$ and $ B_{ij}^{-1}$ in equations like
(\ref{eq:8.51}) as particle $k$ is removed to infinity. In other words,
$B$ can be any unitary operator, which has the following limits

\begin{eqnarray}
\lim _{\mathbf{a} \to \infty} e^{\frac{i}{\hbar}\mathbf{p}_3
\mathbf{a}}
B e^{-\frac{i}{\hbar}\mathbf{p}_3 \mathbf{a}} &=& B_{12}
\label{eq:8.55}\\
\lim _{\mathbf{a} \to \infty} e^{\frac{i}{\hbar}\mathbf{p}_2
\mathbf{a}} B
e^{-\frac{i}{\hbar}\mathbf{p}_2 \mathbf{a}} &=& B_{13}
\label{eq:8.56}\\
\lim _{\mathbf{a} \to \infty} e^{\frac{i}{\hbar}\mathbf{p}_1
\mathbf{a}}
B e^{-\frac{i}{\hbar}\mathbf{p}_1 \mathbf{a}} &=& B_{23}
\label{eq:8.57}
\end{eqnarray}

\noindent One can check that one possible
choice of $B$ is

\begin{eqnarray*}
B = \exp(\ln B_{12} + \ln B_{13} + \ln B_{23})
\end{eqnarray*}

\noindent Indeed, using equations (\ref{eq:8.46}) - (\ref{eq:8.48}) we obtain

\begin{eqnarray*}
&\mathbf{ } & \lim _{\mathbf{a} \to \infty}
e^{\frac{i}{\hbar}\mathbf{p}_3 \mathbf{a}} B
e^{-\frac{i}{\hbar}\mathbf{p}_3 \mathbf{a}} = \lim _{\mathbf{a} \to \infty}  e^{\frac{i}{\hbar}\mathbf{p}_3
\mathbf{a}}
\exp(\ln B_{12} + \ln B_{13} + \ln B_{23})
e^{-\frac{i}{\hbar}\mathbf{p}_3 \mathbf{a}} \\
 &=&
\exp(\ln B_{12}) = B_{12}
\end{eqnarray*}

\noindent Then, it is easy to show that the interacting
representation of the Poincar\'e group generated by operators
(\ref{eq:8.52}) and (\ref{eq:8.53}) satisfies cluster separability
properties (\ref{eq:8.42}) - (\ref{eq:8.45}). For example,

\begin{eqnarray*}
\lim _{\mathbf{a} \to \infty}  e^{\frac{i}{\hbar}\mathbf{p}_3 \mathbf{a}} H
e^{-\frac{i}{\hbar}\mathbf{p}_3 \mathbf{a}}
 &=& \lim _{\mathbf{a} \to \infty}  e^{\frac{i}{\hbar}\mathbf{p}_3
\mathbf{a}}
B^{-1}\overline{H} B
e^{-\frac{i}{\hbar}\mathbf{p}_3 \mathbf{a}} \\
 &=& \lim _{\mathbf{a} \to \infty} B_{12} ^{-1}
 e^{\frac{i}{\hbar}\mathbf{p}_3 \mathbf{a}}
\sqrt{P_0^2c^2 + \overline{M}^2 c^4}
e^{-\frac{i}{\hbar}\mathbf{p}_3 \mathbf{a}} B_{12} \\
 &=& B_{12} ^{-1}
\sqrt{P_0^2 c^2 + (B_{12}M_{12}B_{12} ^{-1})^2 c^4}
 B_{12} \\
 &=&
\sqrt{P_0^2c^2 +  M_{12}^2 c^4} = H_{12}
\end{eqnarray*}

\noindent Generally, operator $B$ does not commute with the
Newton-Wigner position operator (\ref{eq:newt-wig}). Therefore, the
mass operator (\ref{eq:8.54}) also does not commute with
$\mathbf{R}_0$, and
the representation generated by operators $(\mathbf{P}_0,
\mathbf{J}_0, \mathbf{K}, H)$ does not belong to the Bakamjian-Thomas form. This is consistent with our conclusion in subsection \ref{ss:non-separability} that Bakamjian-Thomas dynamics cannot be made cluster-separable.

Obviously, the above method of constructing relativistic cluster-separable interactions is very cumbersome. Moreover, its applicability is limited to interactions that conserve the number of particles. In chapters \ref{ch:QED} and \ref{ch:rqd} we will consider a  more general approach that seems to be more relevant to interactions occurring in nature.  This construction will be based on the idea of quantum fields

\section{Bound states and time evolution}
\label{sc:bound-states}

 We already mentioned  that
the knowledge of the Poincar\'e group representation $U_g$ in the Hilbert space $\mathcal{H}$ of
a multiparticle system is sufficient for getting any desired
physical information about the system. In this section, we would
like to make this statement more concrete by examining two types of
information, which can be compared with experiment: the mass and
energy spectra of the system and the time evolution of its observables. In the next
section we will discuss scattering experiments, which
are currently the most informative way of studying microscopic
systems.

\subsection{Mass and energy spectra}
\label{ss:mass-spectrum}

The mass operator of a  non-interacting 2-particle system is

\begin{eqnarray}
M_0 &=& +\frac{1}{c^2} \sqrt{H_0^2 - P_0^2 c^2} = +\frac{1}{c^2} \sqrt{(h_1 + h_2)^2 - (\mathbf{p}_1 +
\mathbf{p}_2)^2 c^2} \nonumber \\
&=& +\frac{1}{c^2} \sqrt{\left(\sqrt{m_1^2 c^4 + p_1^2 c^2} +
\sqrt{m_2^2 c^4 + p_2^2 c^2}\right)^2 - (\mathbf{p}_1 +
\mathbf{p}_2)^2 c^2} \label{eq:m00}
\end{eqnarray}

\begin{figure}
\centering
\includegraphics{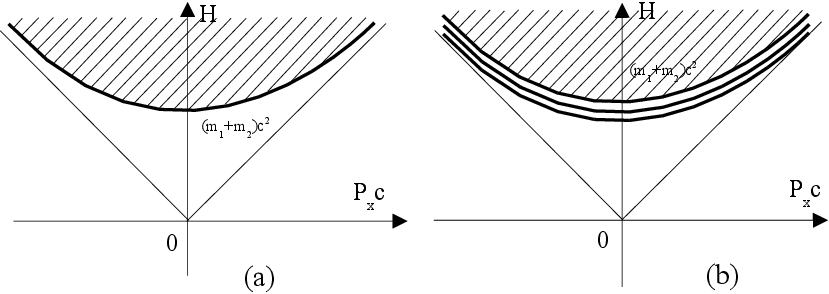} \caption{Typical momentum-energy spectrum
of (a) non-interacting  and (b) interacting  two-particle system. }
\label{fig:8.1}
\end{figure}

\noindent As particles' momenta can have any value in the 3D momentum space,
the eigenvalues $m$ of the mass operator have continuous spectrum in the
range

\begin{eqnarray}
m_1 + m_2 \leq  m  < \infty
\label{eq:8.58}
\end{eqnarray}

\noindent where the minimum value of mass $m_1 + m_2$ is obtained
from (\ref{eq:m00}) when both particles are at rest $\mathbf{p}_1 =
\mathbf{p}_2 = 0$. It then follows that the common spectrum of
mutually commuting operators  $\mathbf{P}_0$ and

\begin{eqnarray*}
H_0 = +\sqrt{M_0^2 c^4 + P_0^2 c^2}
\end{eqnarray*}

\noindent is  the union of mass hyperboloids\footnote{with masses in
the interval (\ref{eq:8.58})} in the 4-dimensional momentum-energy
space.  This spectrum is  shown by the hatched region in Fig.
\ref{fig:8.1}(a).

In the presence of interaction, the eigenvalues $\mu_n$ of the mass
operator $M = M_0 + N$ can be found by solving the \emph{stationary}
\index{stationary Schr\"odinger equation} Schr\"odinger equation

\begin{eqnarray}
   M |\Psi_n \rangle = \mu_n | \Psi_n \rangle
   \label{eq:mass-schr}
\end{eqnarray}

\noindent It is well-known  that in the presence of sufficiently weak interaction $N$, the spectrum of $M$ will not be perturbed much. For example, for a weak attractive $N$, new discrete eigenvalues in the mass spectrum may
split off below the threshold $m_1 + m_2$. The eigenvectors of the
interacting mass operator with eigenvalues $\mu_n < m_1 + m_2$ are
called \emph{bound states}. \index{bound states} The mass
eigenvalues $\mu_n$ are highly degenerate. For example, if $|\Psi_n
\rangle$ is an eigenvector corresponding to $\mu_n$, then for any
Poincar\'e group element $g$  the vector $U_g |\Psi_n \rangle$ is
also an eigenvector with the same mass eigenvalue.\footnote{This means that eigensubspaces with fixed mass $\mu_n$ are invariant with respect to Poincar\'e group actions.} To remove this
degeneracy (at least partially) one can consider operators
$\mathbf{P}_0$ and $H$, which commute with each other and with $M$,
so that they define a basis  of common eigenvectors

\begin{eqnarray*}
   M |\Psi_{\mathbf{p},n} \rangle &=& \mu_n | \Psi_{\mathbf{p},n}
   \rangle \\
 \mathbf{P}_0 |\Psi_{\mathbf{p},n}\rangle &=& \mathbf{p} | \Psi_{\mathbf{p},n}
   \rangle \\
   H |\Psi_{\mathbf{p},n}\rangle &=& \sqrt{M^2c^4 + P^2c^2} |\Psi_{\mathbf{p},n}\rangle
   = \sqrt{\mu_n^2c^4 + p^2c^2} | \Psi_{\mathbf{p},n}
   \rangle
\end{eqnarray*}

\noindent Then sets of common eigenvalues of $\mathbf{P}_0$ and $H$
with fixed $\mu_n <m_1+m_2$ form  hyperboloids

\begin{eqnarray*}
 h_n &=& \sqrt{\mu_n^2c^4 + p^2c^2}
\end{eqnarray*}

\noindent which are shown in Fig. \ref{fig:8.1}(b) below the
continuous part of the common spectrum of $\mathbf{P}_0$ and $H$. An
example of a bound system whose mass spectrum has both continuous
and discrete parts -- the hydrogen atom -- will be considered in greater
detail in section \ref{ss:hydrogen-nonr}. \index{hydrogen atom}

\subsection{Doppler effect revisited}
\label{ss:doppler-effect}

In our discussion of the Doppler effect in subsection
\ref{ss:doppler} we were interested in the energy of free photons
measured by moving observers or emitted by moving sources. To this end we
applied a boost transformation (\ref{eq:doppler}) to the energy $E$ of a free massless
photon. It is instructive to look at this problem from another point
of view. Photons are usually emitted by compound massive physical
systems (atoms, molecules, nuclei, etc.) in transitions between two
discrete energy levels $E_2$ and $E_1$, so that the photon's energy
is found simply from the energy conservation law\footnote{The transition energy $E$ is actually not well-defined,
because the excited state 2 is not a stationary state. (See section \ref{sc:general-decay}.) Therefore our
discussion in this subsection is valid only approximately for
long-living states 2, for which the uncertainty of energy can be
neglected. }

\begin{eqnarray*}
E = E_2 - E_1
\end{eqnarray*}

When the source is moving with respect to the observer (or observer
is moving with respect to the source), the energies of levels 1 and
2  experience inertial transformations given by formula
(\ref{eq:6.3}). Therefore, to check our theory for consistency, we would like to prove that
the Doppler shift calculated with this formula is  the same
as that obtained in subsection \ref{ss:doppler}.

\begin{figure}
\centering
\includegraphics[width=11cm,height=6cm]{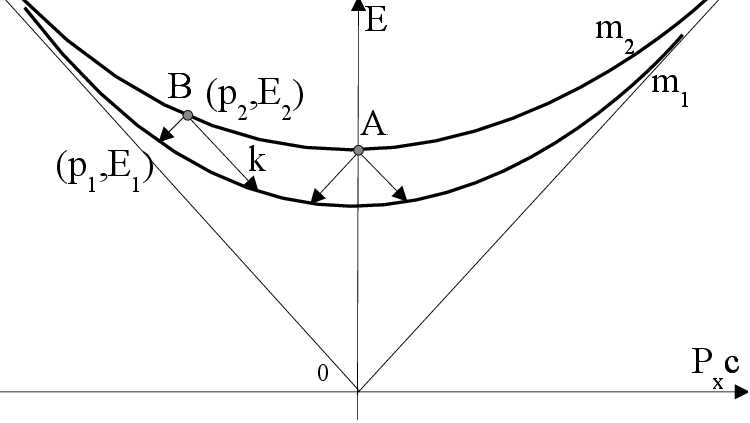} \caption{Energy level diagram for a bound
system with the ground state of mass $m_1$ and the excited state of
mass $m_2$. If the system is at rest, its excited state is
represented by point A. Note that the energy of emitted photons
(arrows) is less than $(m_2 - m_1)c^2$. A moving excited state with
momentum $\mathbf{p}_2$ is represented by point B. The energies and
momenta $\mathbf{k}$ of emitted photons depend on the angle between
$\mathbf{k}$ and $\mathbf{p}_2$.} \label{fig:8.2}
\end{figure}

Suppose that the compound system has two bound states characterized
by mass eigenvalues $m_1$ and $m_2 > m_1$ (see Fig. \ref{fig:8.2}).
Suppose also that initially the system is in the excited state with mass
$m_2$, total momentum $\mathbf{p}_2$, and energy $E_2 = \sqrt{m_2^2
c^4 + p_2^2c^2}$. In the final state we have the same system with
a lower mass $m_1$, different total momentum $\mathbf{p}_1$, and
energy $E_1 = \sqrt{m_1^2 c^4 + p_1^2c^2}$. In addition, there is a photon with
momentum $\mathbf{k}$ and energy $ck$. From the  momentum and energy conservation
laws we can write

\begin{eqnarray*}
\mathbf{p}_2 &=& \mathbf{p}_1 + \mathbf{k} \nonumber \\
E_2 &=& E_1 + c k \nonumber \\
\sqrt{m_2^2 c^4 + p_2^2c^2} &=& \sqrt{m_1^2 c^4 +
p_1^2c^2} + c k \nonumber \\
&=& \sqrt{m_1^2 c^4 + (\mathbf{p}_2 - \mathbf{k})^2c^2} + c k
\label{eq:p2m2k} \nonumber
\end{eqnarray*}

\noindent Taking squares of both sides of the last equality,
 we obtain

\begin{eqnarray*}
  k \sqrt{m_1^2 c^2 + (\mathbf{p}_2 -
\mathbf{k})^2} &=& \frac{1}{2} \mu^2c^2 +  p_2 k \cos \phi - k^2
\end{eqnarray*}

\noindent where $\mu^2 \equiv m_2^2 - m_1^2$ and  $\phi$ is the
angle between vectors $\mathbf{p}_2$ and $\mathbf{k}$.\footnote{
Note also that vector $\mathbf{k}$ points from the light
emitting system to the observer, so the angle $\phi$ can be
interpreted as the angle between the velocity of the source and the
line of sight, which is equivalent to the definition of $\phi$ in
subsection \ref{ss:doppler}.} Taking squares of both sides again we obtain
 a quadratic equation

\begin{eqnarray*}
k^2(  m_2^2 c^2 + p_2^2 - p_2^2 \cos^2 \phi) -  k\mu^2 c^2  p_2 \cos
\phi - \frac{1}{4}\mu^4c^4 =0
\end{eqnarray*}

\noindent with the solution\footnote{Only positive sign of the
square root leads to a physical solution with positive $k$}

\begin{eqnarray*}
k &=&  \frac{\mu^2c^2 }{2 m_2^2 c^2 + 2p_2^2 \sin^2 \phi} \Bigl( p_2
\cos \phi + \sqrt{ m_2^2 c^2 +
p_2^2}\Bigr) \\
\end{eqnarray*}

\noindent Introducing the rapidity $\theta$ of the initial state, we
obtain $p_2 = m_2c \sinh \theta$, $\sqrt{ m_2^2 c^2 +
p_2^2} = m_2c \cosh \theta $ and

\begin{eqnarray*}
k
&=&  \frac{ \mu^2 c(\sinh \theta \cos \phi + \cosh \theta)}
{2m_2(\cosh^2 \theta - \sinh^2 \theta\cos^2 \phi)} =  \frac{ \mu^2 c} {2m_2 \cosh \theta (1- \frac{v}{c}\cos \phi)}
\end{eqnarray*}

\noindent This formula gives the energy of the photon emitted by a
system moving with the speed $v = c \tanh \theta$

\begin{eqnarray*}
E(\theta, \phi)  &\equiv& ck  = \frac{ E(0)} {\cosh \theta (1-
\frac{v}{c}\cos \phi)}
\end{eqnarray*}

\noindent where

\begin{eqnarray*}
E(0) &=&  \frac{\mu^2 c^2}{2m_2}
\end{eqnarray*}

\noindent is the energy of the photon emitted by a source at rest.
This agrees with our earlier result (\ref{eq:doppler4}).

\subsection{Time evolution}
\label{ss:scattering-evol}

In addition to stationary energy spectra discussed above,
 we are often interested in the time evolution
of a compound system. This includes  reactions, scattering, decays,
etc. As we discussed in subsection \ref{ss:inertial-obs}, in quantum
theory the time evolution of states from (earlier) time $t'$ to (later) time $t$ is
described by the \emph{time evolution operator} \index{time
evolution operator}

\begin{eqnarray}
U(t \gets t') =  e^{-\frac{i}{\hbar}H(t-t')} \label{eq:8.59}
\end{eqnarray}

\noindent This operator has the following useful properties

\begin{eqnarray}
U(t \gets t') &=& e^{-\frac{i}{\hbar}H(t -t_1)}
e^{-\frac{i}{\hbar}H(t_1-t')} = U(t \gets t_1) U(t_1 \gets t')
\label{eq:property-A}  \\
U(t \gets t') &=& U^{-1}(t' \gets t) \label{eq:property-B}
\end{eqnarray}

\noindent for any $t, t', t_1$.

In the Schr\"odinger picture, the time evolution of a state vector
is given by (\ref{eq:psi-time})

\begin{eqnarray}
|\Psi (t) \rangle = U(t \gets t')| \Psi (t') \rangle =
e^{-\frac{i}{\hbar}H(t -t')}| \Psi (t') \rangle \label{eq:8.60}
\end{eqnarray}

\noindent $|\Psi (t) \rangle$ is also a solution of the \emph{time
dependent} \index{time dependent Schr\"odinger equation}
Schr\"odinger equation

\begin{eqnarray}
i \hbar \frac{d}{dt}|\Psi (t) \rangle &=& i \hbar
\frac{d}{dt}e^{-\frac{i}{\hbar}H(t-t')}|\Psi (t') \rangle
= He^{-\frac{i}{\hbar}H(t-t')}|\Psi (t') \rangle \nonumber \\
&=& H |\Psi (t) \rangle \label{eq:time-dep}
\end{eqnarray}

\noindent In spite of simple appearance of formula (\ref{eq:8.60}),
the evaluation of the exponents of the Hamilton operator is an
extremely difficult task. In rare cases when all eigenvalues $E_n$
and eigenvectors $| \Psi \rangle_n$  of the Hamiltonian are known

\begin{eqnarray*}
H | \Psi \rangle_n = E_n | \Psi \rangle_n
\end{eqnarray*}

\noindent the initial state can be represented as a sum (and/or
integral) of basis eigenvectors

\begin{eqnarray*}
| \Psi (0) \rangle = \sum_n C_n | \Psi \rangle_n
\end{eqnarray*}

\noindent and the time evolution can be calculated as

\begin{eqnarray}
|\Psi (t) \rangle &=& e^{-\frac{i}{\hbar}Ht} | \Psi (0) \rangle =
e^{-\frac{i}{\hbar}Ht}\sum_n C_n | \Psi \rangle_n
= \sum_n C_n
e^{-\frac{i}{\hbar}E_n t} | \Psi \rangle_n \label{eq:8.61a}
\end{eqnarray}

There is another useful formula for the state vector's time
evolution in a theory with Hamiltonian $H = H_0 + V$. Denoting

\begin{eqnarray*}
 V(t) = e^{\frac{i}{\hbar}H_0 (t-t_0)} V e^{-\frac{i}{\hbar}H_0 (t-t_0)}
\end{eqnarray*}

\noindent it is easy to verify that the time-dependent state
vector\footnote{Note that time integration variables satisfy
inequalities $t \geq t' \geq t'' \geq \ldots \geq t_0$.}

\begin{eqnarray}
|\Psi (t) \rangle &=&e^{-\frac{i}{\hbar}H_0
(t-t_0)}\left(1-\frac{i}{\hbar} \int_{t_0}^{t} V(t')dt'
-\frac{1}{\hbar^2} \int_{t_0}^{t} V(t')dt' \int_{t_0}^{t'}
V(t'')dt'' + \ldots \right) |\Psi (t_0) \rangle \nonumber \\
\label{eq:time-evoll}
\end{eqnarray}

\noindent satisfies the Schr\"odinger equation (\ref{eq:time-dep})
with the additional condition that at $t=t_0$ the solution coincides
with the given initial state $|\Psi (t_0) \rangle$. Indeed

\begin{eqnarray*}
&\ & i \hbar \frac{d}{dt}|\Psi (t) \rangle \\
&=& i \hbar \frac{d}{dt}e^{-\frac{i}{\hbar}H_0
(t-t_0)}\left(1-\frac{i}{\hbar} \int_{t_0}^{t} V(t')dt'
-\frac{1}{\hbar^2} \int_{t_0}^{t} V(t')dt' \int_{t_0}^{t'}
V(t'')dt'' + \ldots \right) |\Psi (t_0) \rangle \\
&=& H_0 e^{-\frac{i}{\hbar}H_0 (t-t_0)} \left(1- \frac{i}{\hbar}\int_{t_0}^{t}
V(t')dt' -\frac{1}{\hbar^2} \int_{t_0}^{t} V(t')dt' \int_{t_0}^{t'}
V(t'')dt'' + \ldots \right)
|\Psi (t_0) \rangle \\
&\ & +e^{-\frac{i}{\hbar}H_0 (t-t_0)}
 V(t)\left(1 -\frac{i}{\hbar}
\int_{t_0}^{t} V(t'')dt'' + \ldots \right) |\Psi (t_0) \rangle \\
&=& (H_0 + V) |\Psi (t) \rangle
\end{eqnarray*}

\noindent Perturbative formula (\ref{eq:time-evoll}) will be found useful in our
discussion of scattering in subsection \ref{ss:perturbation}.

 Unfortunately, the above methods for calculating the time evolution of quantum systems have very limited practical value: The full spectrum of eigenvalues and eigenvectors of the
interacting Hamiltonian $H$\footnote{which are required for formula (\ref{eq:8.61a})} can be found only for very simple
models. The convergence of the perturbative expansion (\ref{eq:time-evoll}) is usually rather poor. So, calculations of the time evolution in quantum mechanics are rather challenging. There are, however, two areas where we can make further progress in solving this problem. First, in most circumstances, quantum effects are too small to be observable. So, it is important to understand how solutions of the time dependent Schr\"odinger equation correspond to classical trajectories of particles that we see in everyday life. The classical limit of quantum mechanics will be tackled in section \ref{sc:classical}. Second, there is an important class of \emph{scattering experiments}, which do not require a detailed description of the time evolution of quantum states. The powerful formalism of scattering theory will be discussed in chapter \ref{sc:scattering}.

\section{Classical Hamiltonian dynamics} \label{sc:classical}

There are many studies devoted to the so-called problem of \emph{quantization}.  This means that given a classical theory\footnote{e.g., classical mechanics or classical field theory} one is trying to develop a corresponding quantum analog. However, as the world is fundamentally quantum, and its classical description is just a rough approximation, this line of research is not well justified. In our opinion, it seems more logical to go in the opposite direction: to build an (approximate) classical theory starting from its (exact) quantum analog.

In section \ref{ss:qm-logic} we have established  that  distributive
(classical) propositional systems are particular cases of
orthomodular (quantum) propositional systems. Therefore, we may
expect that quantum mechanics includes classical mechanics as a
particular case. However, it is not obvious how exactly the phase
space of classical mechanics is related to the quantum Hilbert space. We  would like to analyze this
relationship in the present section. For simplicity, we will use as an
example a system of spinless particles with non-zero masses $m_i > 0$.
For classical treatment of massless particles, e.g., photons, see subsection \ref{ss:way-forward}.

\subsection{Quasiclassical states}
\label{ss:limit}

 In the macroscopic world
 we do not meet localized eigenvectors $| \mathbf{r} \rangle$ of the position operator.
 According to equation (\ref{eq:7.25}), such states have infinite
uncertainty of momentum which is rather unusual. Similarly, we do
not meet states with sharply defined momentum. Such states are
delocalized over large distances (\ref{eq:plane-wave}). The reason
why such states are not commonly seen\footnote{Spatially delocalized states of particles play a role in such low-temperature effects as superconductivity and superfluidity.} is not well understood yet. The most
plausible hypothesis is that eigenstates of the position or
eigenstates of the momentum are susceptible to small perturbations
(e.g., due to temperature or external radiation) and rapidly
transform to more robust \emph{wave packets} \index{wave packet} or
\emph{quasiclassical states} \index{quasiclassical state} in which both position
and momentum have good, but not perfect localization.

So, when discussing the classical limit of quantum mechanics, we
will not consider general states allowed by quantum mechanics. We
will limit our attention only to the class of particle states $|
\Psi_{\mathbf{r}_0, \mathbf{p}_0} \rangle $ that we will call
quasiclassical. Wave functions of these states are assumed to be
well-localized in both position and momentum representations around the points  $\mathbf{r}_0$ and $\mathbf{p}_0$, respectively. Without loss of generality such wave functions in the position representation can
be written  as

\begin{eqnarray}
\psi_{\mathbf{r}_0, \mathbf{p}_0} (\mathbf{r}) \equiv \langle \mathbf{r}
| \Psi_{\mathbf{r}_0, \mathbf{p}_0} \rangle &=& \eta
(\mathbf{r}- \mathbf{r}_0) e^{\frac{i}{\hbar}\phi} e^{\frac{i}{\hbar}\mathbf{p}_0 (\mathbf{r} -\mathbf{r}_0)}
\label{eq:7.31}
\end{eqnarray}

\noindent where  $\eta (\mathbf{r}- \mathbf{r}_0)$ is a real smooth
(non-oscillating) function with a maximum near the expectation value of position
$\mathbf{r}_0$ and $\phi$ is a real phase.\footnote{such that $e^{\frac{i}{\hbar}\phi}$ is a unimodular phase factor: $|e^{\frac{i}{\hbar}\phi}|=1$. The introduction of this factor seems redundant here, because any wave function is defined up to a multiplier, anyway. However, we will find the factor $e^{\frac{i}{\hbar}\phi}$ important in our discussions of the interference effect in subsection \ref{ss:time-wave} and in section \ref{sc:a-b-effect}.} The last factor in (\ref{eq:7.31})  ensures that the expectation value of momentum is $\mathbf{p}_0$.\footnote{compare with the form (\ref{eq:plane-wave}) of momentum eigenfunctions in the position space } As we will see later, in order to discuss the
classical limit of quantum mechanics the exact choice of the
function $\eta (\mathbf{r}- \mathbf{r}_0)$ is not important. For
example, it is convenient to choose it in the form of a Gaussian

\begin{eqnarray}
 \psi_{\mathbf{r}_0, \mathbf{p}_0}(\mathbf{r})   = N e^{-( \mathbf{r}-
\mathbf{r}_0)^2/d^2}
 e^{\frac{i}{\hbar}\mathbf{p}_0\mathbf{r}}
 \label{eq:7.31a}
\end{eqnarray}

\noindent where $\phi=0$, parameter $d$ controls the degree of localization,
and $N$ is a coefficient required for the proper normalization

\begin{eqnarray*}
\int d \mathbf{r}| \psi_{\mathbf{r}_0, \mathbf{p}_0}(\mathbf{r})|^2
= 1
\end{eqnarray*}

\noindent The exact magnitude of this coefficient is not important
for our discussion, so we will not calculate it here.

\subsection{Heisenberg uncertainty relation}
\label{ss:heisenberg-unc}

Wave functions like (\ref{eq:7.31a}) cannot possess both sharp
position and sharp momentum at the same time. They are always
characterized by a non-vanishing uncertainty of position $\Delta r > 0$ and
a non-vanishing uncertainty of momentum $\Delta p > 0$. These uncertainties are
roughly inversely proportional to each other. To see the nature of
this inverse proportionality, we assume, for simplicity, that the
particle is at rest in the origin, i.e., $\mathbf{r}_0 =\mathbf{p}_0
= 0$. Then the position-space wave function is

\begin{eqnarray}
 \psi_{\mathbf{0}, \mathbf{0}}(\mathbf{r})   = N e^{-
 \mathbf{r}^2/d^2} \label{eq:position-wave-packet}
\end{eqnarray}

\noindent and its counterpart in the momentum space is\footnote{Here we used equations
(\ref{eq:7.26}) and (\ref{eq:A.96}).}

\begin{eqnarray}
\psi_{\mathbf{0}, \mathbf{0}}(\mathbf{p}) &=&  (2 \pi \hbar)^{-3/2}
N
 \int d\mathbf{r}
 e^{-\mathbf{r}^2/d^2}
e^{-\frac{i}{\hbar}\mathbf{p}\mathbf{r}} \nonumber \\
&=& (2 \hbar)^{-3/2} Nd^3 e^{-p^2d^2/(4 \hbar^2)} \label{eq:7.32}
\end{eqnarray}

\noindent The product of the uncertainties of the
momentum-space ($\Delta p \approx \frac{2 \hbar}{d} $) and
position-space ($\Delta r \approx d$) wave functions is independent on the parameter $d$

\begin{eqnarray}
\Delta r \Delta p \approx 2 \hbar \label{eq:heisenberg2}
\end{eqnarray}

\noindent This is an example of the \emph{Heisenberg
uncertainty relation}, \index{Heisenberg uncertainty relation} which
tells us that for all quantum states the above uncertainties must satisfy the famous inequality

\begin{eqnarray}
\Delta r \Delta p \geq
 \hbar/2
\label{eq:heisenberg}
\end{eqnarray}

\subsection{Spreading of quasiclassical wave packets}
\label{ss:wave-packets}

Suppose that at time $t=0$ the particle was prepared in the state
with well-localized wave function (\ref{eq:position-wave-packet}),
i.e., the uncertainty of position $\Delta r \approx d$ is small. The
corresponding time-dependent wave function in the momentum
representation is

\begin{eqnarray*}
  \psi (\mathbf{p}, t)
&=& e^{-\frac{i}{\hbar}\hat{H}t} \psi_{\mathbf{0}, \mathbf{0}} (\mathbf{p},
0)
\\
 &=&  \frac{Nd^3}{(2 \hbar)^{3/2}}
e^{-p^2d^2/(4 \hbar^2)}
 e^{-\frac{it}{\hbar}\sqrt{m^2c^4 + p^2c^2} }
\end{eqnarray*}

\noindent whose  position representation counterpart is\footnote{Due to the factor $ e^{-p^2d^2/(4
\hbar^2)}$, only small values of momentum contribute to the
integral, so we can use the non-relativistic approximation
$\sqrt{m^2c^4 + p^2c^2} \approx mc^2 + \frac{p^2}{2m}$ and equation
(\ref{eq:A.96}).}

\begin{eqnarray*}
  \psi (\mathbf{r}, t)
 &=&  \frac{Nd^3}{(4 \pi \hbar^2)^{3/2} }
 \int d\mathbf{p}
 e^{-p^2d^2/(4 \hbar^2)} e^{\frac{i}{\hbar}\mathbf{p} \mathbf{r}}
 e^{-\frac{it}{\hbar}\sqrt{m^2c^4 + q^2c^2} } \\
 &\approx& \frac{Nd^3}{(4 \pi \hbar^2)^{3/2} }
e^{-\frac{i}{\hbar}mc^2 t}  \int d\mathbf{p}
 \exp \left(- p^2 \left(\frac{d^2}{4  \hbar^2} + \frac{it}{2 \hbar m}\right)
+ \frac{i}{\hbar} \mathbf{p} \mathbf{r}\right) \\
 &=&  N
\left(\frac{d^2 m}{ d^2 m + 2i \hbar t}\right)^{3/2}
e^{-\frac{i}{\hbar}mc^2 t}
 \exp\left(- \frac{  m r^2}{d^2m + 2 i \hbar   t}\right)
\end{eqnarray*}

\noindent and the probability density is

\begin{eqnarray*}
\rho (\mathbf{r}, t) &=& |\psi(\mathbf{r}, t)|^2 \\
 &=&  |N|^2
\left(\frac{d^4 m^2}{ d^4m^2 + 4 \hbar^2 t^2}\right)^{3/2}
 \exp\left(- \frac{ 2  r^2d^2 m^2}{d^4m^2 + 4  \hbar^2t^2}\right) \\
\end{eqnarray*}

\noindent The size of the wave packet at large times $t \to \infty$
 is easily found as

\begin{eqnarray*}
\Delta r(t) \approx \sqrt{\frac{d^4m^2 + 4  \hbar^2t^2}{ d^2
m^2}} \approx \frac{2  \hbar t}{  d m}
\end{eqnarray*}

\noindent So, the position-space wave packet is spreading out, and
the speed of spreading $v_s$ is directly proportional to the
uncertainty of velocity in the initially
prepared state\footnote{Here we  used equality (\ref{eq:heisenberg2}).}

\begin{eqnarray}
 v_s \approx \frac{2 \hbar }{d m}  \approx \frac{\Delta p}{m}
\label{eq:7.33}
\end{eqnarray}

\noindent One can verify that at large times this speed does not
depend on the shape of the
 initial wave packet. The important parameters are the size $d$ of this wave
packet and the  particle's mass $m$.

A simple estimate demonstrates that for macroscopic objects this
spreading phenomenon can be safely neglected. For example,
for  a particle of mass $m=$ 1 mg and the initial position
uncertainty of $d=$ 1 micron, the time needed for the wave function
to spread to 1 cm is more than $10^{11}$ years. Therefore, for
quasiclassical states of macroscopic particles with sufficiently high masses, their positions and
momenta are well defined at all times and their time evolution can be described by a classical  \emph{trajectory} pretty well. \index{trajectory} So, in these conditions one can safely replace quantum mechanics with its classical counterpart.

\subsection{Phase space}
\label{ss:phase_space}

Let us now see how rules of classical Hamiltonian mechanics follow from the quantum Schr\"odinger equation.

In subsection \ref{ss:limit} we have established the general form (\ref{eq:7.31}) of quasiclassical
wave packets. In most circumstances the resolution of
measuring instruments is poor, i.e., much poorer than the quantum of action $\hbar$ \cite{Kofler}. Then the shape of the envelope function
$\eta(\mathbf{r} - \mathbf{r}_0)$ cannot be discerned.  All quantum states (\ref{eq:7.31}) with  different shapes of the function
$\eta(\mathbf{r} - \mathbf{r}_0)$ can now be treated as the same
classical state. So,
each classical state is fully characterized by two parameters: the
average position of the packet $\mathbf{r}_0$ and the average momentum
$\mathbf{p}_0$.
 These states are approximate
 eigenstates of both position and momentum operators
simultaneously:

\begin{eqnarray}
 \mathbf{R} | \Psi_{\mathbf{r}_0, \mathbf{p}_0} \rangle &\approx&
\mathbf{r}_0 | \Psi_{\mathbf{r}_0, \mathbf{p}_0} \rangle
\label{eq:7.34} \\
 \mathbf{P} | \Psi_{\mathbf{r}_0, \mathbf{p}_0} \rangle &\approx& \mathbf{p}_0
| \Psi_{\mathbf{r}_0, \mathbf{p}_0} \rangle \label{eq:7.35}
\end{eqnarray}

\noindent All such equivalent states can be represented by one point
$(\mathbf{r}_0, \mathbf{p}_0)$  in a 6-dimensional manifold $\mathbb{R}^6$ with
coordinates $r_x, r_y, r_z, p_x, p_y, p_z$.  This is the one-particle classical
\emph{phase space} \index{phase space} that was discussed from a logico-probabilistic point of view in subsection \ref{ss:phase-space}.

We can continue this line of reasoning and translate other quantum notions to the classical language as well. For example, we know that any 1-particle quantum observable $F$ can be expressed as a function
of the particle position $\mathbf{R}$, momentum $\mathbf{P}$, and
mass $M$.\footnote{See subsection \ref{ss:alternative}. Recall that
in this section we are talking only about spinless particles. So, we set $\mathbf{S}=0$.} The
eigenvalue of $M$ is just a constant. Therefore, in the classical
phase space picture, all observables (the energy, angular momentum,
 velocity, etc.) are represented as real functions $f(\mathbf{p}, \mathbf{r})$
on the phase space.

For example, consider a logical proposition $F$. As we have established in chapter \ref{ch:QM}, logical propositions  form a special class of observables (or functions on the phase space), whose spectrum consists of only two points 0 and 1. Thus, the phase space function $f(\mathbf{p}, \mathbf{r})$ that corresponds to the proposition $F$, defines a subset of the phase space -- the set of points where $f(\mathbf{p}, \mathbf{r})=1$.

 Let us consider two examples of
propositions/subsets in $\mathbb{R}^6$. The proposition $R$ = ``position of
the particle is exactly $\mathbf{r}_0$'' is represented in the phase space
by a 3-dimensional hyperplane with fixed position $\mathbf{r} =
\mathbf{r}_0$ and arbitrary momentum $\mathbf{p}$. The proposition
$P$ = ``momentum of the particle is exactly $\mathbf{p}_0$'' is represented
by another 3-dimensional hyperplane in which the value of momentum
is fixed, while position is arbitrary. The meet of these two
propositions is represented by the intersection of the two hyperplanes
$s = R \cap P$ which is a point $s =(\mathbf{r}_0, \mathbf{p}_0)$ in
the phase space and an atom in the classical propositional system.
In the classical case such an intersection always exists. Thus, there exist states in which both position and momentum are measurable simultaneously with absolute certainty.
However, this is not true in the quantum case. As we saw in subsection \ref{ss:heisenberg-unc}, quantum propositions about position $R$ and momentum $P$ can have a non-empty meet only if they are associated with uncertainties (intervals) $\Delta r$ and $\Delta p$, which satisfying the Heisenberg uncertainty relationship (\ref{eq:heisenberg}).

Similar to the one-particle case, we can introduce a $6N$-dimensional phase space for
any system of $N$ particles. This phase space is a classical replacement for
the quantum-mechanical $N$-particle Hilbert space, as we discussed
in subsection \ref{ss:phase-space}.

\subsection{Poisson brackets}
\label{ss:classical-observables}

 It follows from
(\ref{eq:7.34}) and (\ref{eq:7.35}) that quasiclassical states $| \Psi_{\mathbf{r}_0, \mathbf{p}_0} \rangle$ are
approximate eigenstates of any classical observable

\begin{eqnarray}
 f(\mathbf{R},\mathbf{P}) | \Psi_{\mathbf{r}_0, \mathbf{p}_0} \rangle
&\approx& f(\mathbf{r}_0, \mathbf{p}_0) | \Psi_{\mathbf{r}_0,
\mathbf{p}_0} \rangle \label{eq:frp}
\end{eqnarray}

\noindent The expectation value of observable
$f(\mathbf{R},\mathbf{P})$ in the quasiclassical state $|
\Psi_{\mathbf{r}_0, \mathbf{p}_0} \rangle $ is just the value of the
corresponding function $f(\mathbf{r}_0, \mathbf{p}_0)$

\begin{eqnarray*}
 \langle f(\mathbf{R},\mathbf{P}) \rangle
&=& f(\mathbf{r}_0, \mathbf{p}_0)
\end{eqnarray*}

\noindent and the expectation value of a product of two such
observables is equal to the product of expectation values

\begin{eqnarray}
 \langle f(\mathbf{R},\mathbf{P}) g(\mathbf{R},\mathbf{P}) \rangle
&=& f(\mathbf{r}_0, \mathbf{p}_0) g(\mathbf{r}_0, \mathbf{p}_0)  =
\langle f(\mathbf{R},\mathbf{P}) \rangle \langle
g(\mathbf{R},\mathbf{P}) \rangle \label{eq:6.126a}
\end{eqnarray}

 According to (\ref{eq:5.50}) - (\ref{eq:5.56}),
commutators of observables are proportional to $\hbar$, so in the classical  limit $\hbar \to 0$ all
operators of observables commute with each other.\footnote{This is also clear from (\ref{eq:6.126a}) as $\langle f(\mathbf{R},\mathbf{P}) g(\mathbf{R},\mathbf{P}) \rangle = \langle g(\mathbf{R},\mathbf{P}) f(\mathbf{R},\mathbf{P}) \rangle$.} There are two
important roles played by commutators in quantum
mechanics. First, the commutator of two observables determines
whether these observables can be measured simultaneously, i.e., whether
there exist states in which both observables have well-defined
values. Vanishing commutators of classical observables imply that
all such observables can be measured simultaneously.  Second, commutators of observables
with generators of the Poincar\'e group determine how these observables transform from one reference frame to another.
One example of such a transformation is the time translation in
(\ref{eq:5.61}). However, the zero classical limit of these
commutators as $\hbar \to 0$ does not mean that
 $t$-dependent terms on the right hand side of equation (\ref{eq:5.61}) become
zero, and that the time evolution stops in this limit. The
right hand side of (\ref{eq:5.61}) does not vanish even in the
classical limit, because the commutators in $n$-th order terms are
multiplied by large factors $(-i/\hbar)^n$. In the limit $\hbar \to 0$
we obtain

\begin{eqnarray}
F(t) = F - [H,F]_Pt + \frac{1}{2} [H, [H, F]_P]_P t^2 + \ldots
\label{eq:7.36}
\end{eqnarray}

\noindent where

\begin{eqnarray}
 [f,g]_P \equiv \lim _{\hbar \to 0} \frac{-i}{\hbar} [f(\mathbf{R},
\mathbf{P}), g(\mathbf{R}, \mathbf{P})] \label{eq:poiss-br}
\end{eqnarray}

\noindent  is called the \emph{Poisson bracket}. \index{Poisson bracket}
\index{$[\ldots,\ldots]_P$ Poisson bracket} So, even though
commutators of observables are effectively zero in classical
mechanics, we can still use non-vanishing Poisson brackets when
calculating the action of inertial transformations on
observables.

Now we are going to derive a useful explicit formula for the Poisson bracket (\ref{eq:poiss-br}). The exact commutator of two quantum mechanical operators
 $f(\mathbf{R}, \mathbf{P})$ and $g(\mathbf{R},
\mathbf{P})$ can be written generally as a series in powers of
$\hbar$

\begin{eqnarray*}
 [f,g] = i\hbar k_1 + i\hbar^2 k_2 + i\hbar^3 k_3 \ldots
\end{eqnarray*}

\noindent where $k_i$ are Hermitian operators. From equation
(\ref{eq:poiss-br}) it is clear that the Poisson bracket is equal to
the coefficient of the dominant term of the first order in
$\hbar$

\begin{eqnarray*}
 [f,g]_P = k_1
\end{eqnarray*}

\noindent As a consequence, the classical Poisson bracket $[f,g]_P$
is much easier to calculate than the full quantum commutator $[f,
g]$. The following theorem demonstrates that calculation of the
Poisson bracket can be reduced to simple differentiation.

\bigskip

\begin{theorem}
\label{Theorem7.1} If $f(\mathbf{R}, \mathbf{P})$ and $g(\mathbf{R},
\mathbf{P})$ are two observables of a massive spinless particle,
then\footnote{Equation (\ref{eq:7.37}) is the \emph{definition} of the Poisson bracket usually presented
in classical mechanics textbooks without proper justification. Here we are deriving this formula from quantum-mechanical commutators.}

\begin{eqnarray}
[f(\mathbf{R}, \mathbf{P}), g(\mathbf{R}, \mathbf{P})]_P   =
\frac{\partial f}{ \partial\mathbf{R} } \cdot  \frac{\partial g}{
\partial\mathbf{P} } - \frac{\partial f}{ \partial\mathbf{P} } \cdot
\frac{\partial g}{ \partial\mathbf{R} } \label{eq:7.37}
\end{eqnarray}
 \end{theorem}
\begin{proof}   Consider for simplicity the one-dimensional case (the 3D proof is similar) in
which the desired result (\ref{eq:7.37}) becomes

\begin{eqnarray}
\lim _{\hbar \to 0} \frac{-i}{\hbar} [f(R,P), g(R,P)] =
\frac{\partial f}{ \partial R } \cdot  \frac{\partial g}{ \partial P
} - \frac{\partial f}{ \partial P } \cdot
 \frac{\partial
g}{ \partial R} \label{eq:7.38}
\end{eqnarray}

\noindent First, functions $f(R,P)$ and $g(R,P)$ can be represented
by their Taylor expansions around the origin ($r=0$, $p=0$) in the
phase space, e.g.,

\begin{eqnarray*}
f(R,P) &=& C_{00} + C_{10} R + C_{01} P + C_{11} RP +  C_{20} R^2 +
C_{02} P^2 + C_{21} R^2P +  \ldots\\
g(R,P) &=& D_{00} + D_{10} R + D_{01} P + D_{11} RP +  D_{20} R^2 +
D_{02} P^2 + D_{21} R^2P +  \ldots
\end{eqnarray*}

\noindent where $C_{ij}$ and $D_{ij}$ are numerical coefficients, and we agreed to write factors $R$ to the left from factors $P$.
Then it is sufficient to prove formula (\ref{eq:7.38}) for $f$ and
$g$ being monoms of the form $R^nP^m$. In particular, we would like
to prove that

\begin{eqnarray}
[R^nP^m, R^qP^s]_P &=& \frac{\partial (R^nP^m)}{\partial R}
\frac{\partial (R^qP^s)}{\partial P} -
\frac{\partial (R^nP^m)}{\partial P} \frac{\partial (R^qP^s)}{\partial R} \nonumber \\
&=& ns R^{n+q-1} P^{m+s-1} - mq R^{n+q-1} P^{m+s-1} \nonumber  \\
&=& (ns - mq) R^{n+q-1} P^{m+s-1} \label{eq:poiss-brac1}
\end{eqnarray}

\noindent for all non-negative integers $n, m, q, s \geq 0$.  This result definitely holds if $f$ and $g$ are linear in $R$ and
$P$, i.e., when $n, m, q, s$ are either 0 or 1. For example, in the
case $n=1, m=0, q=0, s=1$ formula (\ref{eq:poiss-brac1}) yields

\begin{eqnarray*}
[R, P]_P &=& 1
\end{eqnarray*}

\noindent which agrees with definition (\ref{eq:poiss-br}) and with
quantum result (\ref{eq:6.21}).

To prove (\ref{eq:poiss-brac1}) for higher powers we will use
mathematical induction. Suppose that we established the validity of
(\ref{eq:poiss-brac1}) for a set of powers $n, m, q, s$ as well as
for any set of lower powers $n', m', q', s'$, where  $n'\leq n$,
$m'\leq m$, $q'\leq q$, $s'\leq s$. The proof by induction now
requires us to establish the validity of the following equations

\begin{eqnarray*}
[R^nP^m, R^{q+1}P^s]_P &=& (ns - mq -m) R^{n+q} P^{m+s-1} \\
\mbox{ } [R^nP^m, R^qP^{s+1}]_P &=& (ns - mq +n) R^{n+q-1} P^{m+s} \\
\mbox{ } [R^{n+1}P^m, R^qP^s]_P &=& (ns - mq +s) R^{n+q} P^{m+s-1} \\
\mbox{ } [R^nP^{m+1}, R^qP^s]_P &=& (ns - mq -q) R^{n+q-1} P^{m+s} \\
\end{eqnarray*}

\noindent Let us prove only the first equation. Three others are
proved similarly.  Using
 equations (\ref{eq:6.46}), (\ref{eq:poiss-brac1}), and (\ref{eq:A.37}) we, indeed,  obtain

\begin{eqnarray*}
[R^nP^m, R^{q+1}P^s]_P &=&  - \lim _{\hbar \to 0} \frac{i}{\hbar}[R^nP^m, R^{q+1}P^s] \\
&=&  - \lim _{\hbar \to 0} \frac{i}{\hbar} [R^nP^m,R] R^{q}P^s  -
\lim _{\hbar \to 0} \frac{i}{\hbar} R [R^nP^m,
R^{q}P^s] \\
&=&  [R^nP^m,R]_P R^{q}P^s  + R [R^nP^m,
R^{q}P^s]_P \\
&=& - m R^{n+q} P^{m+s-1} + (ns-mq) R^{n+q} P^{m+s-1}  \\
&=& (ns - mq-m) R^{n+q} P^{m+s-1}
\end{eqnarray*}

\noindent Therefore, by induction, equation (\ref{eq:7.38}) holds for all
values of $n,m,q,s \geq 0$ and for all smooth functions $f(R,P)$ and  $g(R,P)$.
\end{proof}
\bigskip

Since Poisson brackets are obtained from commutators (\ref{eq:poiss-br}), all properties
of commutators from Appendix \ref{ss:lie-algebras} remain valid for Poisson brackets as well.

For a concrete example, let us apply the above formalism of Poisson brackets to the time
evolution. We can use formulas (\ref{eq:7.36}) and (\ref{eq:7.37})
in the case when $F$ is either position or momentum and obtain\footnote{Here we used equation (\ref{eq:rxfpx}) and a similar formula $[P_x,f(R_x)] = -i \hbar  \partial f(R_x)/\partial R_x$.}

\begin{eqnarray}
\frac{d\mathbf{P}(t)}{dt} &=&  -
[H(\mathbf{R},\mathbf{P}),\mathbf{P} ]_P = -\frac{\partial
H(\mathbf{R},\mathbf{P})}{\partial \mathbf{R}} \label{eq:7.40} \\
\frac{d\mathbf{R}(t)}{dt} &=& - [H(\mathbf{R},\mathbf{P}),\mathbf{R}
]_P = \frac{\partial H(\mathbf{R},\mathbf{P})}{\partial \mathbf{P}}
\label{eq:7.39}
\end{eqnarray}

\noindent where one recognizes  the classical \emph{Hamilton's
equations of motion}. \index{Hamilton's equations of motion}

\subsection{Time evolution of wave packets}
\label{ss:time-wave}

Our result (\ref{eq:7.40}) - (\ref{eq:7.39})
means, in particular, that trajectories of centers of quasiclassical wave packets
are exactly the same as those predicted by classical Hamiltonian
mechanics. Here we would like to demonstrate how this conclusion follows from approximate solutions of the Schr\"odinger equation.

Earlier in this section we have established that in many cases the
spreading of a quasiclassical wave packet can be ignored and that
the center of the packet moves along a well-defined trajectory
$(\mathbf{r}_0(t), \mathbf{p}_0(t))$. This means that the shape of the wave packet is described by the function $\eta (\mathbf{r} - \mathbf{r}_0(t))$, which remains localized around the path $\mathbf{r} = \mathbf{r}_0(t)$. This also means that the
exponential $\mathbf{r}$-dependent factor in (\ref{eq:7.31}) is
approximately $\exp(\frac{i}{\hbar}\mathbf{p}_0(t)\mathbf{r})$. For
generality, we also need to assume that the phase $\phi(t)$ is time-dependent too. Then the time-dependent quasiclassical wave packet
is described by the following \emph{ansatz}

\begin{eqnarray}
 \Psi(\mathbf{r},t) &=& \eta (\mathbf{r} - \mathbf{r}_0(t))\exp \left(\frac{i}{\hbar}A(t)\right)
 \label{eq:ansatz2}\\
 A(t) &\equiv& \mathbf{p}_0(t)(\mathbf{r} - \mathbf{r}_0(t))+ \phi(t)
 \label{eq:ansatz}
\end{eqnarray}

\noindent where $\mathbf{r}_0(t)$, $\mathbf{p}_0(t)$, $\phi(t)$ are
yet undetermined numerical functions. In order to find these functions we insert
(\ref{eq:ansatz2}) - (\ref{eq:ansatz}) in the Schr\"odinger equation
(\ref{eq:schrod})

\begin{eqnarray}
i \hbar \frac{\partial \Psi(\mathbf{r},t)}{\partial t} +
\frac{\hbar^2}{2m} \frac{\partial^2 \Psi(\mathbf{r},t)}{\partial
\mathbf{r}^2} - V(\mathbf{r}) \Psi(\mathbf{r},t) = 0
\label{eq:schro}
\end{eqnarray}

\noindent which is valid for the position-space wave function
$\Psi(\mathbf{r},t)$ of a single particle moving in an external
potential $V(\mathbf{r})$.\footnote{Here we make several assumptions
and approximations to simplify our calculations. First, we
consider a particle moving in a fixed potential. This is not an
isolated system, which is a subject of most discussions in the book.
Nevertheless, this is still a good approximation in the case when
the object creating the potential $V(\mathbf{r})$ is heavy, so that
it can be considered fixed. Second, the position-dependent potential
$V(\mathbf{r})$ does not depend on the particle's momentum. Third,
we are working in the non-relativistic approximation.}
\index{time dependent Schr\"odinger equation} Omitting for brevity
time arguments and denoting time derivatives by dots we can write

\begin{eqnarray*}
i \hbar \frac{\partial \Psi(\mathbf{r})}{\partial t} &=& \left(-i
\hbar \left(\frac{\partial \eta}{\partial \mathbf{r}} \cdot
\dot{\mathbf{r}_0}\right)  - (\dot{\mathbf{p}_0} \cdot
\mathbf{r})\eta + (\dot{\mathbf{p}_0} \cdot\mathbf{r}_0)\eta +
(\mathbf{p}_0 \cdot \dot{\mathbf{r}_0})\eta - \dot{\phi} \eta
\right)
\exp\left(\frac{i}{\hbar}A\right) \\
\frac{\hbar^2}{2m} \frac{\partial^2 \Psi(\mathbf{r})}{\partial
\mathbf{r}^2} &=&  \frac{\hbar^2}{2m} \frac{\partial }{\partial \mathbf{r}}
\left(\frac{\partial \eta}{\partial \mathbf{r}}
\exp\left(\frac{i}{\hbar}A \right)
+ \frac{i}{\hbar} \mathbf{p}_0 \eta \exp\left(\frac{i}{\hbar}A \right)\right) \\
 &=& \frac{\hbar^2}{2m} \left(\frac{\partial^2 \eta}{\partial
\mathbf{r}^2} + \frac{2i}{\hbar} \left(\frac{\partial \eta}{\partial
\mathbf{r}} \cdot \mathbf{p}_0\right)
 - \frac{p_0^2 \eta}{\hbar^2 }
  \right) \exp\left(\frac{i}{\hbar}A\right)\\
 - V(\mathbf{r}) \Psi(\mathbf{r}) &=&  \left( - V(\mathbf{r}_0) \eta
 - \frac{\partial V(\mathbf{r})} {\partial \mathbf{r}}\Bigr|_{\mathbf{r}= \mathbf{r}_0}
 (\mathbf{r}- \mathbf{r}_0) \eta \right) \exp\left(\frac{i}{\hbar}A\right)
\end{eqnarray*}

\noindent There are three kinds of terms on the left hand side of (\ref{eq:schro}): those
proportional to $\hbar^0, \hbar^1$, and $\hbar^2$. They must vanish
independently. The $\hbar^2$-dependent terms are too small; they are beyond the
accuracy of the quasiclassical approximation and can be ignored. The
terms that are first order in $\hbar$ result in equation $\dot{\mathbf{r}}_0 = \frac{ \mathbf{p}_0}{m}$, which is the usual relationship between velocity and
momentum for momentum-independent potentials.\footnote{This is also
the 2nd Hamilton's equation (\ref{eq:7.39}).} $\hbar^0$ terms lead
to equation

\begin{eqnarray} 0&=& -(\dot{\mathbf{p}}_0 \cdot \mathbf{r}) +
(\dot{\mathbf{p}}_0 \cdot \mathbf{r}_0) + (\mathbf{p}_0 \cdot
\dot{\mathbf{r}}_0) - \dot{\phi} - \frac{p_0^2}{2m} -
V(\mathbf{r}_0)
 - \frac{\partial V(\mathbf{r})} {\partial \mathbf{r}}\Bigr|_{\mathbf{r}= \mathbf{r}_0}
 (\mathbf{r}- \mathbf{r}_0) \nonumber \\
 \label{eq:hbar0}
\end{eqnarray}

\noindent The function $\dot{\mathbf{p}}_0(t)$ can be determined
from the first Hamilton's equation (\ref{eq:7.40})

\begin{eqnarray*}
\dot{\mathbf{p}}_0 =
 - \frac{\partial V(\mathbf{r})} {\partial \mathbf{r}}\Bigr|_{\mathbf{r}= \mathbf{r}_0} \label{eq:2ndHam}
\end{eqnarray*}

\noindent So,  we can rewrite (\ref{eq:hbar0}) as an equation for
the last undetermined (phase) function $\phi(t)$

\begin{eqnarray*}
 \frac{\partial \phi} {\partial t}
   &=&   \frac{ p_0^2(t)}{2m}  - V(\mathbf{r}_0(t))
\end{eqnarray*}

\noindent The solution of this equation for a particle propagating in
the time interval $[t_0, t]$ is given by the so-called \emph{action
integral} \index{action integral}\footnote{Note that the integrand has the form (``kinetic energy'' - ``potential energy''), which is known in classical mechanics as the \emph{Lagrangian}. \index{Lagrangian}}

\begin{eqnarray}
 \phi( t)
   &=& \phi( t_0) + \int \limits_{t_0}^t dt' \left[\frac{ p_0^2(t')}{2m}  - V(\mathbf{r}_0(t'))
   \right] \label{eq:phase}
\end{eqnarray}

 From the above discussion we conclude that the center of a quasiclassical wave
packet moves along a trajectory determined by classical Hamilton's equations
of motion (\ref{eq:7.40}) - (\ref{eq:7.39}).  In addition, there is a genuine quantum
effect: the change of the overall phase of the wave packet according
to equation (\ref{eq:phase}).

\begin{figure}
\includegraphics{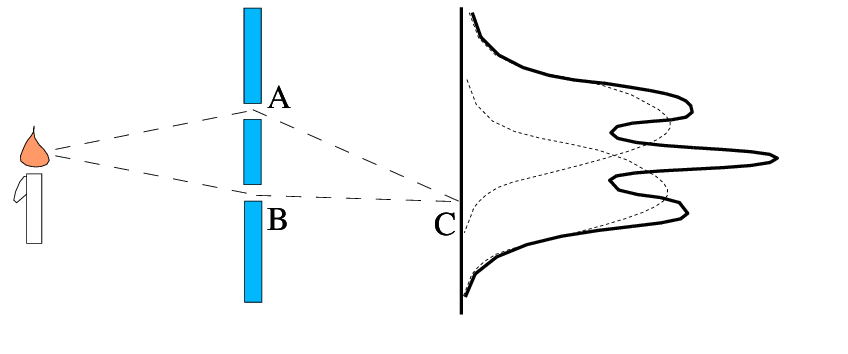} \caption{Interference picture in the two-slit experiment. Two dotted bell-shaped curves on the right show the image density profiles when one of the two slits is closed. The thick full line is the interference pattern when both slits are opened. Compare with Fig. \ref{fig:3.3}(b).} \label{fig:6.2}
\end{figure}

This phase change explains the double-slit (or double-hole) interference effect discussed in
section \ref{sc:thought}. Suppose that a monochromatic source emits electrons,\footnote{The explanation of the photon interference is similar.} which pass through two slits and form an image on the scintillating screen, as shown in Fig. \ref{fig:6.2}. The electron wave packets can reach the point $C$ on the screen by two alternative ways: either through slit $A$ or through slit $B$. Both kinds of packets contribute to the wave function at point $C$. Their complex phase factors $\exp(\frac{i}{\hbar} \phi(t))$ should be added together when calculating the probability amplitude for finding an electron at point $C$. In this particular case, the calculation of phase factors is especially simple, because there is no external potential ($V(\mathbf{r}) =
0$). The momentum (and speed) of each wave packet remains constant
($p_0^2(t) = const$), so that the action integral (\ref{eq:phase})
is proportional to the distance traveled by the wave packet from the slit to the screen. This means that the character of interference (constructive or
destructive phase shift) at point $C$ is fully determined by the difference between two
traveling distances $AC$ and $BC$.

Other
experimental manifestations of the phase formula (\ref{eq:phase})  will be discussed in
section \ref{sc:a-b-effect}.

\chapter{SCATTERING} \label{sc:scattering}

\begin{quote}
\textit{Physics is becoming so unbelievably complex that it is
taking longer and longer to train a physicist. It is
taking so long, in fact, to train a physicist to the place
where he understands the nature of physical problems
that he is already too old to solve them.}

\small
\hspace{1in} Eugene P. Wigner
\normalsize
\end{quote}

\vspace{0.5in}

As we discussed at the end of section \ref{sc:bound-states}, it is very difficult to solve the time-dependent Schr\"odinger equation (\ref{eq:time-dep}) even for simplest models. However, nature gives us a lucky break: there is a very
important class of experiments for which the description of dynamics
by equation (\ref{eq:time-dep}) is not needed; this
description is just too detailed. We are talking about  scattering experiments here.
They are performed by preparing free particles,\footnote{or their bound
states, like hydrogen atoms or deuterons} bringing them into
collision and studying the properties of free particles or stable bound
states leaving the region of the collision. In these experiments, often
it is not possible to observe the time evolution during interaction:
particle reactions occur almost instantaneously, and we can only
register the reactants and products moving freely before and
after the collision. In such situations the theory is not required
to describe the actual evolution of particles during the short time
interval of collision. It is sufficient to provide a mapping of free
states before the collision onto free states after the
collision. This mapping is provided by the $S$-operator, which we are
going to discuss in this chapter.

\section{Scattering operators} \label{ss:scattering}

\subsection{$S$-operator} \label{sc:scatter}

Let us consider a scattering experiment in which free states of
reactants  are prepared at time $t = -\infty$. The
collision occurs during a short time interval $[\eta', \eta]$ around
time zero.\footnote{The short interaction time can be guaranteed if three conditions are met: First, the interaction between particles is short-range or, more generally, cluster separable. Second, states of particles are describable by localized wave packets, such as those in subsection \ref{ss:limit}. Third, particles' velocities (or momenta) are sufficiently high.} The free states of the products are
registered at time $t = \infty$, so that inequalities $-\infty  \ll
\eta' < 0 < \eta \ll \infty $ hold. Here we assume that the two colliding particles do
not form bound states neither before nor after the collision.
Therefore, at asymptotic times the exact evolution is well
approximated by non-interacting time evolution operators $U_0(\eta' \gets -\infty)$
and $U_0(\infty \gets \eta)$, respectively.\footnote{Here we denoted $U_0(t \gets t') \equiv
\exp(-\frac{i}{\hbar}H_0(t-t'))$ the time evolution operator associated with the non-interacting Hamiltonian $H_0$.} Then we can write the time evolution operator from
the infinite past to the infinite future\footnote{Here we used
properties (\ref{eq:property-A}) and (\ref{eq:property-B}).}

\begin{eqnarray}
&\mbox{ } &U(\infty \gets -\infty) \nonumber \\  &\approx&
U_0(\infty \gets \eta) U(\eta \gets
\eta') U_0(\eta' \gets -\infty ) \nonumber\\
 &=&  U_0(\infty \gets \eta) U_0(\eta \gets 0)
[U_0(0 \gets \eta)  U(\eta \gets  \eta')
 U_0( \eta' \gets 0)]
U_0(0 \gets \eta')  U_0( \eta' \gets -\infty) \nonumber\\
&=&  U_0(\infty \gets 0) S_{\eta, \eta'} U_0(0 \gets -\infty)
\label{eq:8.63}
\end{eqnarray}

\noindent where

\begin{eqnarray}
S_{\eta, \eta'}
 &\equiv&
U_0(0 \gets \eta)  U(\eta \gets  \eta')
 U_0( \eta' \gets 0)
\label{eq:8.64}
\end{eqnarray}

\noindent Equation (\ref{eq:8.63})  means that a simplified description
of the time evolution in scattering events is possible in which the
evolution is free at all times except sudden change at $t=0$
described by the unitary operator $S_{\eta, \eta'}$:
 Approximation (\ref{eq:8.63})  becomes more
accurate if we increase the time interval $[\eta', \eta]$ during which
the exact time evolution
is taken into account, i.e., $\eta' \to -\infty, \eta \to \infty$.\footnote{provided that the right hand side of (\ref{eq:8.64}) converges in these limits. The issue of convergence is discussed in subsection \ref{ss:adiaba}.}
 Therefore, the exact formula for the time evolution from $- \infty$ to
$\infty$ can be written as

\begin{eqnarray}
U(\infty \gets -\infty)
&=&  U_0(\infty \gets 0)
S U_0(0 \gets -\infty)
\label{eq:8.65}
\end{eqnarray}

\noindent where the $S$\emph{-operator} \index{S-operator} (or
\emph{scattering operator}) is defined by formula

\begin{eqnarray}
S &=& \lim_{\eta' \to -\infty, \eta \to  \infty} S_{\eta, \eta'} =
\lim_{\eta' \to -\infty, \eta \to  \infty} U_0(0 \gets \eta) U(\eta
\gets
 \eta')
 U_0( \eta' \gets 0) \nonumber  \\
&=& \lim_{\eta' \to -\infty, \eta \to  \infty} e^{\frac{i}{\hbar}H_0
\eta} e^{-\frac{i}{\hbar}H (\eta - \eta')}
e^{-\frac{i}{\hbar}H_0 \eta'} \label{eq:8.73a} \\
  &=& \lim_{\eta \to \infty} S(\eta)  \nonumber \\
S(\eta) &\equiv& \lim_{\eta' \to - \infty} e^{\frac{i}{\hbar}H_0 \eta}
e^{-\frac{i}{\hbar}H(\eta-\eta')} e^{-\frac{i}{\hbar}H_0\eta'}
\label{eq:seta}
\end{eqnarray}

\begin{figure}
\centering
\includegraphics{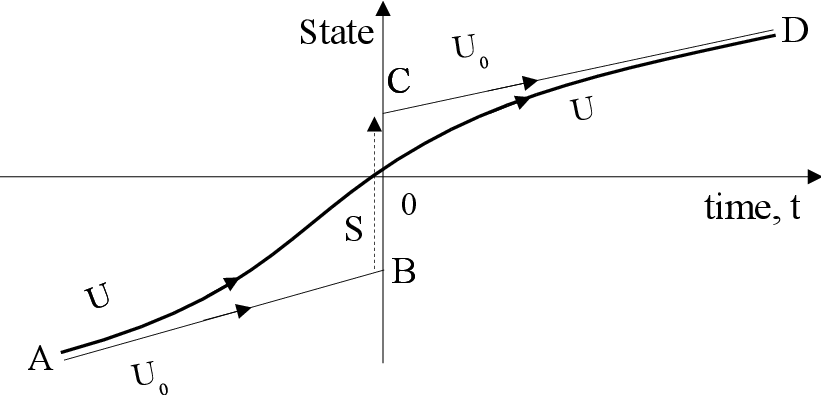} \caption{A schematic representation of
the scattering process.} \label{fig:6.1}
\end{figure}

A better understanding of how scattering theory describes time
evolution can be obtained from Fig. \ref{fig:6.1}. In this figure we
plot the state of the scattering system (represented abstractly as a
point on the vertical axis) as a function of time (the horizontal
axis). The exact evolution of the state is governed by the full time
evolution operator $U$ and is shown by the thick line $A \to D$. In
asymptotic regions (when $t$ is large negative or large positive)
the interaction between parts of the scattering system is weak. In these regions exact time evolutions can be well approximated by free time
evolutions. These free ``trajectories'' are governed by the operator $U_0$ and shown in the figure
by two thin straight lines with arrows: one for large positive times $C \to D$ and
another for large negative times $A \to B$.  The thick line (the exact interacting time evolution)
asymptotically approaches thin lines (free time evolutions) in the remote past (around $A$)
and in the remote future (around $D$). The past and future free
evolutions can be extrapolated to time $t=0$, and there is a gap
$B-C$ between these extrapolated states. The $S$-operator (which connects
states $B$ and $C$ as shown by
the dashed arrow) is designed to bridge this gap. This operator provides a
mapping between  free states extrapolated to time $t=0$. Thus, in scattering theory the exact time
evolution $A \to D$ is approximated by three steps: the system first evolves freely until time $t=0$,
i.e., from $A$ to $B$. Then there is a sudden jump $B \to C$
represented by the $S$-operator. Finally, the time evolution is free
again $C \to D$. As seen from the figure, this description of the
scattering process is perfectly exact, as long as we are interested
only in the mapping from asymptotic states in the remote past $A$ to
asymptotic states in the remote future $D$. However, it is also
clear that scattering theory does not provide a good description of
the time evolution in the interacting region around $t=0$. In the scattering operator $S$ the information about particle
interactions enters integrated over the infinite time interval $t
\in (-\infty, \infty)$. In order to describe the time evolution in
the interaction region ($t \approx 0$) the $S$-matrix approach is
not suitable. The full interacting time evolution operator $U$ is
needed for this purpose.

 In applications we are mostly interested in matrix elements of
the $S$-operator

\begin{eqnarray}
S_{i \to f} = \langle f | S | i \rangle \label{eq:8.62}
\end{eqnarray}

\noindent where $| i \rangle$ is a state of non-interacting initial
particles, and $|f \rangle$ is a state of non-interacting final
particles. Such matrix elements   are called the \emph{$S$-matrix}.
\index{S-matrix} Formulas relating the $S$-matrix to observable
quantities, such as scattering cross-sections,
 can be found in any textbook on scattering theory.

 An important property of the $S$-operator is its
``Poincar\'e invariance,'' i.e., zero commutators with  generators
of the non-interacting representation of the Poincar\'e group
\cite{book, Kazes}

\begin{eqnarray}
[S,H_0] = [S, \mathbf{P}_0] = [S, \mathbf{J}_0] = [S, \mathbf{K}_0] = 0
\label{eq:S-rel-inv}
\end{eqnarray}

\noindent The vanishing commutator $[S,H_0] =0$ implies that in (\ref{eq:8.65}) one can
change places of $U_0$ and $S$, so that the interacting time
evolution operator can be written as the full free time evolution
operator times the $S$-operator

\begin{eqnarray}
U(\infty \gets -\infty) &=&   S U_0(\infty \gets -\infty) =
U_0(\infty \gets -\infty) S \label{eq:8.65a}
\end{eqnarray}

\subsection{$S$-operator in perturbation theory}
\label{ss:perturbation}

 There are various techniques available for
calculations of the $S$-operator. Currently, the perturbation theory
is the most powerful and effective one. To derive the perturbation
expansion for the $S$-operator, first note that operator $S(t)$
in (\ref{eq:seta}) satisfies equation

\begin{eqnarray}
 &\ & \frac{d}{dt} S(t) \nonumber \\ &=& \frac{d}{dt} \lim_{t' \to - \infty}
e^{\frac{i}{\hbar}H_0t} e^{-\frac{i}{\hbar}H(t-t')}
e^{-\frac{i}{\hbar}H_0t'} \nonumber\\
&=&  \lim_{t' \to - \infty} \left(e^{\frac{i}{\hbar}H_0t}
\left(\frac{i}{\hbar}H_0\right) e^{-\frac{i}{\hbar}H(t-t')}
e^{-\frac{i}{\hbar}H_0t'} + e^{\frac{i}{\hbar}H_0t}
\left(-\frac{i}{\hbar}H\right) e^{-\frac{i}{\hbar}H(t-t')}
e^{-\frac{i}{\hbar}H_0t'}\right)\nonumber\\
&=& -\frac{i}{\hbar}  \lim_{t' \to - \infty}
e^{\frac{i}{\hbar}H_0t} (H- H_0) e^{-\frac{i}{\hbar}H(t-t')}
e^{-\frac{i}{\hbar}H_0t'} \nonumber\\
&=& -\frac{i}{\hbar}  \lim_{t' \to - \infty}
e^{\frac{i}{\hbar}H_0t} V e^{-\frac{i}{\hbar}H(t-t')}
e^{-\frac{i}{\hbar}H_0t'} \nonumber\\
&=& -\frac{i}{\hbar}  \lim_{t' \to - \infty}
e^{\frac{i}{\hbar}H_0t} V e^{-\frac{i}{\hbar}H_0t}
e^{\frac{i}{\hbar}H_0t}e^{-\frac{i}{\hbar}H(t-t')}
e^{-\frac{i}{\hbar}H_0t'} \nonumber\\
&=&  -\frac{i}{\hbar}  \lim_{t' \to - \infty}  V(t)
e^{\frac{i}{\hbar}H_0t}e^{-\frac{i}{\hbar}H(t-t')}
e^{-\frac{i}{\hbar}H_0t'} \nonumber \\
&=& -\frac{i}{\hbar} V(t) S(t)
\label{eq:8.66}
\end{eqnarray}

\noindent where we denoted\footnote{Note that the $t$-dependence of
$V(t)$ does not mean that we are considering time-dependent
interactions. The argument $t$ has very little to do with actual
\emph{time dependence} of operators in the Heisenberg representation, which must be
generated by the full interacting Hamiltonian $H$ and not by the
free Hamiltonian $H_0$ as in equation (\ref{eq:8.67}). In such cases we will use the term ``$t$-dependence'' instead of ``time dependence''. }

\begin{eqnarray}
V(t) = e^{\frac{i}{\hbar}H_0t} V e^{-\frac{i}{\hbar}H_0t}
\label{eq:8.67}
\end{eqnarray}

One can directly check that solution of equation (\ref{eq:8.66})
 with the natural initial condition $S( -\infty) = 1$  is
given by the ``old-fashioned'' perturbation expansion

\begin{eqnarray*}
    S(t) &=& 1 - \frac{i}{\hbar}\int_{-\infty}^{t} V(t')\,dt'
          - \frac{1}{\hbar^2} \int_{-\infty}^{t} V(t')\,dt'
\int_{-\infty}^{t'} V(t'')\,dt''
          + \ldots,  \label{eq:old-fash}
\end{eqnarray*}

\noindent Therefore, the $S$-operator can be calculated by putting $t=+\infty$ as the upper limit of $t$-integrals

\begin{eqnarray}
    S &=& 1 - \frac{i}{\hbar}\int_{-\infty}^{+\infty} V(t)\,dt
          - \frac{1}{\hbar^2} \int_{-\infty}^{+\infty} V(t)\,dt
\int_{-\infty}^{t} V(t')\,dt'
          + \ldots
\label{eq:8.68}
\end{eqnarray}

\noindent This formula can be also derived from equation (\ref{eq:time-evoll})  in the case when the initial time $t_0=  -\infty$ is
in the remote past and the final
time $t=+\infty$ is in the distant future

\begin{eqnarray*}
&\ &|\Psi (+\infty) \rangle \\
&=&\lim_{t \to +\infty}e^{-\frac{i}{\hbar}H_0
(t-t_0)}\Bigl(1-\frac{i}{\hbar} \int_{-\infty}^{\infty}
e^{\frac{i}{\hbar}H_0 (t'-t_0)} V e^{-\frac{i}{\hbar}H_0 (t'-t_0)}dt' \\
&\ & -\frac{1}{\hbar^2} \int_{-\infty}^{\infty} e^{\frac{i}{\hbar}H_0
(t'-t_0)} V e^{-\frac{i}{\hbar}H_0 (t'-t_0)}dt' \int_{-\infty}^{t'}
e^{\frac{i}{\hbar}H_0 (t''-t_0)} V e^{-\frac{i}{\hbar}H_0 (t''-t_0)}dt''
+ \ldots \Bigr) |\Psi (-\infty) \rangle \nonumber
\end{eqnarray*}

\noindent Next we shift integration variables $ t'-t_0 \to t'$ and $ t''-t_0 \to t''$, so that\footnote{Note that the trick of ``adiabatic
switching'' described in the next subsection (and tacitly assumed to be working here) allows us to keep
unchanged the infinite limits ($-\infty $ and $\infty$) of integrals.}

\begin{eqnarray*}
&\ &|\Psi (+\infty) \rangle \\
&=&\lim_{t \to +\infty} e^{-\frac{i}{\hbar}H_0
(t-t_0)}\Bigl(1-\frac{i}{\hbar} \int_{-\infty}^{\infty}
e^{\frac{i}{\hbar}H_0 t'} V e^{-\frac{i}{\hbar}H_0 t'}dt' \\
&\ &-\frac{1}{\hbar^2} \int_{-\infty}^{\infty} e^{\frac{i}{\hbar}H_0
t'} V e^{-\frac{i}{\hbar}H_0 t'}dt' \int_{-\infty}^{t'}
e^{\frac{i}{\hbar}H_0 t''} V e^{-\frac{i}{\hbar}H_0 t''}dt'' +
\ldots \Bigr) |\Psi (-\infty) \rangle \nonumber
\end{eqnarray*}

\noindent Comparing this formula with representation
(\ref{eq:8.65a}) of the time evolution operator we conclude that the
$S$-factor is the same as (\ref{eq:8.68}).

We will avoid discussion of (non-trivial) convergence properties
of the series on the right hand side of equation (\ref{eq:8.68}).
Throughout this book we will tacitly assume that all relevant
perturbation series do converge.

We will often use the following convenient shorthand notation for
$t$-integrals \index{$\underline{expr}$} \index{$\underbrace{expr}$}

\begin{eqnarray}
 \underline{Y(t)} &\equiv&
-\frac{i}{\hbar} \int_{-\infty }^{t} Y(t') d t' \label{eq:underline} \\
 \underbrace{Y(t)}&\equiv&
-\frac{i}{\hbar} \int_{-\infty }^{+\infty} Y(t') d t' = \underline{Y(\infty)}
\label{eq:underbrace}
\end{eqnarray}

\noindent In this notation the  perturbation expansion of the
$S$-operator (\ref{eq:8.68})  can be written compactly as

\begin{eqnarray}
S   &=&  1 +
\underbrace{\Sigma(t)}
\label{eq:8.69} \\
   \Sigma(t)  &=&
            V(t)
          + V(t) \underline{V(t')}
          + V(t) \underline{V(t') \underline{V(t'')}}
          + V(t) \underline{V(t') \underline{V(t'')
\underline{V(t''')}}}
          +   \ldots \nonumber \\
\label{eq:8.70}
\end{eqnarray}

Formula (\ref{eq:8.68}) is not the only way to write the
perturbation expansion for the $S$-operator and, perhaps, not the
most convenient one. In most books on quantum field theory the
covariant Feynman--Dyson perturbation expansion \cite{book} is used,
which involves a time ordering of operators in the
integrands\footnote{When applied to a product of several $t$-dependent bosonic operators, the time ordering \index{time ordering} symbol
$T$ changes the order of operators in such a way that the $t$ label increases from
right to left, e.g.

\begin{eqnarray}
T[A(t_1)B(t_2)]  =  \left \{
\begin{array}{c}
A(t_1)B(t_2), \mbox{  } if \mbox{  } t_1 > t_2 \\
B(t_2)A(t_1), \mbox{  } if \mbox {  } t_1 < t_2
\end{array}\right. \label{eq:time-order}
\end{eqnarray}

\noindent For the time ordered product of fermionic (anticommuting) operators see equation (\ref{eq:J.79}).}

\begin{eqnarray}
 S &=& 1 - \frac{i}{\hbar}\int \limits_{-\infty }^{+\infty} dt_1 V(t_1)
 - \frac{1}{2!\hbar^2} \int \limits_{-\infty }^{+\infty} dt_1 dt_2 T[V(t_1)
 V(t_2)] \nonumber \\
 &\ & +\frac{i}{3! \hbar^3} \int \limits_{-\infty }^{+\infty} dt_1 dt_2 dt_3T[V(t_1) V(t_2)
 V(t_3)] \nonumber \\
  &\ & +\frac{1}{4!\hbar^4} \int \limits_{-\infty }^{+\infty} dt_1 dt_2 dt_3 dt_4T[V(t_1) V(t_2) V(t_3)
  V(t_4)] + \ldots \label{eq:F-D}
\end{eqnarray}

\noindent For our purposes  in chapter \ref{ch:rqd} we
found more useful yet another equivalent perturbative expression
suggested by Magnus \cite{Magnus, Magnus2, Blanes08}

\begin{eqnarray}
   S = \exp(\underbrace{F(t)}) \label{eq:8.71}
\end{eqnarray}

\noindent where
  Hermitian operator $F(t)$ will be referred to as the
\emph{scattering phase} \index{scattering phase operator} operator.
It is represented as a series of multiple commutators with $t$-integrals

\begin{eqnarray}
   F(t)  &=&  V(t)
   - \frac{1}{2 }[\underline{V(t')},V(t)]
   +\frac{1}{6}[\underline{\underline{V(t'')},[V(t')},V(t)]] \nonumber  \\
 & \ & +\frac{1}{6}[\underline{[\underline{V(t'')},V(t')]},V(t)]
-\frac{1}{12}[\underline{\underline{\underline{V(t''')},[[V(t'')},V(t')]},V(t)]]
\nonumber \\
&\ & -\frac{1}{12}[\underline{[\underline{\underline{V(t''')},[V(t'')},V(t')]]},V(t)]
\nonumber \\
  &\ & -\frac{1}{12}[\underline{\underline{[\underline{V(t''')},V(t'')]},[V(t')},V(t)]]
   +\ldots
\label{eq:8.72}
\end{eqnarray}

\noindent One important advantage of this representation is that
expression (\ref{eq:8.71}) for the $S$-operator is manifestly
unitary in each perturbation order. The
three perturbative expansions (old-fashioned, Feynman--Dyson, and
Magnus) are equivalent in the sense that they converge to the same
result if all perturbation orders are added up to infinity. However,
in each fixed order $n$ the three types of terms can be different.\footnote{the
difference being of the order $n+1$ or higher}

We will often drop $t$-arguments in operator expressions. Then formulas for the $S$-operator (\ref{eq:8.69}) and (\ref{eq:8.71}) simplify

\begin{eqnarray}
S &=& \exp(\underbrace{F}) = 1 + \underbrace{\Sigma} \label{eq:7.63a}\\
F &=& V - \frac{1}{2}[\underline{V}, V] + \ldots \label{eq:7.63b} \\
\Sigma &=& V + V \underline{V} + V\underline{V \underline{V}} + \ldots \label{eq:7.63c}
\end{eqnarray}

\subsection{Adiabatic switching of interaction}
\label{ss:adiaba}

In formulas for scattering operators (\ref{eq:8.70}) and
(\ref{eq:8.72}) we meet $t$-integrals $\underline{V(t)}$.  A
straightforward calculation of such integrals gives rather
discouraging result. Let us introduce a complete basis $|n \rangle$
 of eigenvectors of the free Hamiltonian

\begin{eqnarray}
H_0 |n\rangle &=& E_n|n\rangle \label{eq:basis1} \\
\sum_n|n \rangle \langle n | &=& 1 \label{eq:basis2}
\end{eqnarray}

\noindent and calculate matrix elements of $\underline{V(t)}$ in
this basis

\begin{eqnarray}
\langle n|\underline{V(t)} | m \rangle &\equiv& -\frac{i}{\hbar}
\int \limits_{-\infty}^{t} \langle n| e^{\frac{i}{\hbar}H_0t'} V
e^{\frac{i}{\hbar}H_0t'}  | m \rangle dt'
= -\frac{i}{\hbar} V_{nm} \int \limits_{-\infty}^{t}
e^{\frac{i}{\hbar}(E_n-E_m)t'}
dt' \nonumber \\
 &=& - V_{nm} \left(\frac{
e^{\frac{i}{\hbar}(E_n-E_m)t}}{E_n-E_m} - \frac{
e^{\frac{i}{\hbar}(E_n-E_m)(-\infty)}}{ E_n-E_m} \right)
\label{eq:9.54}
\end{eqnarray}

\noindent What shall we do with the meaningless term containing $(-\infty)$ on the right
hand side?

 This term can be made harmless if we take into account an
important fact that the $S$-operator cannot be applied to all states in
the Hilbert space. It can be applied only to \emph{scattering states} $| \Psi \rangle$
\index{scattering states} in which free particles are far from each
other in asymptotic limits $t \to \pm \infty$. Then the time evolution of
these states coincides with the free evolution in the distant past
and distant future,\footnote{Of
course, interaction $V$ must be cluster separable to ensure
that.} and the full time evolution in the infinite time interval is given exactly by formula (\ref{eq:8.65a}).  Certainly, the above assumptions cannot be applied
to all states in the Hilbert space.
 For example, the time evolution of  bound states of
the interacting Hamiltonian $H$, does not resemble the free
 evolution at any time. It appears that if we exclude such bound states from consideration and limit our
application of the $S$-operator and $t$-integrals (\ref{eq:9.54})
only to scattering states consisting  of
one-particle wave packets with good localization in both position
and momentum spaces, then no ambiguity arises.

For scattering states the interaction operator is effectively zero in asymptotic regimes, so we can write

\begin{eqnarray}
 \lim_{t \to \pm \infty}  V e^{-\frac{i}{\hbar}H_0t}|\Psi \rangle &=& 0
\nonumber \\
\lim_{t \to \pm \infty}V(t)|\Psi \rangle &=& 0 \label{eq:9.55}
\end{eqnarray}

\noindent One approach to the  exact treatment of scattering is to explicitly
consider only wave packets described above.\footnote{see, e.g., \cite{Goldberger}} Then the cluster
separability of $V$ will ensure  the correct asymptotic behavior of the colliding wave
packets and the validity of equation (\ref{eq:9.55}).
 However, such an
approach is rather complicated, and we would like to stay away from
working with wave packets.

There is another way to achieve the same goal by using a trick
called the \emph{adiabatic switching} \index{adiabatic switching} of
 interaction. The trick is to add
  the property (\ref{eq:9.55}) to the interaction operator ``by
hand.'' To do that we multiply $V(t)$ by a numerical  function of
$t$ which slowly grows from the value of zero  at $t = - \infty$ to
the value of one at $t \approx 0$ (turning the interaction ``on'')
and then slowly decreases back to zero at $t = \infty$ (turning the
interaction ``off''). For example, it is convenient to choose

\begin{eqnarray}
V(t) = e^{\frac{i}{\hbar}H_0t} V
e^{-\frac{i}{\hbar}H_0t}e^{-\epsilon |t|} \label{eq:adiaba}
\end{eqnarray}

\noindent If the parameter $\epsilon$ is small and positive, such a
modification would not affect the movement of  wave
packets and the $S$-matrix. At the end of calculations we will take
the limit $\epsilon \to +0$. Then the $t$-integral (\ref{eq:9.54})
takes the form

\begin{eqnarray}
\langle n|\underline{V(t)} | m \rangle &\approx&
 - V_{nm} \left(\frac{
e^{\frac{i}{\hbar}(E_n-E_m)t - \epsilon |t|}}{E_n-E_m} - \frac{
e^{\frac{i}{\hbar}(E_n-E_m)(-\infty) - \epsilon (\infty)}}{ E_n-E_m}
\right) \nonumber \\
&\longrightarrow&  - V_{nm} \frac{ e^{\frac{i}{\hbar}(E_n-E_m)t }}{E_n-E_m}
\label{eq:9.56}
\end{eqnarray}

\noindent so the embarrassing expression $e^{i \infty}$ does not
appear.

\subsection{$T$-matrix}
\label{ss:t-matrix}

In this subsection we will introduce the idea of the $T$-matrix, which is often useful in scattering calculations.\footnote{I am indebted to Cao
Bin for numerous discussions which resulted in writing this
subsection.} Let us  calculate matrix elements of the
$S$-operator (\ref{eq:8.68}) in the basis of eigenvectors of the free Hamiltonian
(\ref{eq:basis1}) - (\ref{eq:basis2})\footnote{Summation over indices
$k$ and $l$ is implied. Formula (\ref{eq:9.56}) is used for
$t$-integrals.}

\begin{eqnarray}
&\ &\langle n|S|m \rangle \nonumber \\
&=&\delta_{nm} -\frac{i}{\hbar} \int_{-\infty}^{\infty} \langle n|
e^{\frac{i}{\hbar}H_0 t'}  V
 e^{-\frac{i}{\hbar}H_0 t'} |m \rangle dt' \nonumber \\
&\ &-\frac{1}{\hbar^2} \int_{-\infty}^{\infty} \langle n|
e^{\frac{i}{\hbar}H_0 t'}  V e^{-\frac{i}{\hbar}H_0 t'} |k \rangle
dt' \int_{-\infty}^{t'} \langle k | e^{\frac{i}{\hbar}H_0 t''} V
e^{-\frac{i}{\hbar}H_0
t''} |m \rangle dt'' + \ldots \nonumber \\
 &=&\delta_{nm} -\frac{i}{\hbar}
\int_{-\infty}^{\infty}  e^{\frac{i}{\hbar}(E_n - E_m) t'}   V_{nm}
  dt' \nonumber \\
&\ &-\frac{1}{\hbar^2} \int_{-\infty}^{\infty} e^{\frac{i}{\hbar}(E_n
- E_k) t'}   V_{nk}    dt' \int_{-\infty}^{t'}
e^{\frac{i}{\hbar}(E_k - E_m) t''} V_{km}  dt'' + \ldots \nonumber \\
 &=&\delta_{nm} -2 \pi i
\delta(E_n - E_m)    V_{nm} +\frac{i}{\hbar} \int_{-\infty}^{\infty}
e^{\frac{i}{\hbar}(E_n - E_k) t'}      dt'
\frac{e^{\frac{i}{\hbar}(E_k - E_m) t'}}{E_m - E_k} V_{nk} V_{km}   + \ldots \nonumber  \\
 &=&\delta_{nm} -2 \pi i
\delta(E_n - E_m)    V_{nm} +2\pi i \delta(E_n - E_m)
\frac{1}{E_m - E_k} V_{nk} V_{km}   + \ldots \nonumber \\
 &=&\delta_{nm} -2 \pi i
\delta(E_n - E_m) V_{nk} \times \nonumber \\
&\ &\left(  \delta_{km} + \frac{ 1}{E_m -
E_k}V_{km} + \frac{ 1}{E_m - E_k}V_{kl} \frac{ 1}{E_m - E_l} V_{lm}
+ \ldots \right) \label{eq:Snm}
\end{eqnarray}

\noindent The first term is a unit matrix that describes free propagation. The matrix in the second term is called the
$T$-\emph{matrix} (or \emph{transition matrix}). \index{T-matrix}

\begin{eqnarray*}
T_{nm}
 &\equiv& V_{nk} \left(  \delta_{km} + \frac{ 1}{E_m -
E_k}V_{km} + \frac{ 1}{E_m - E_k}V_{kl} \frac{ 1}{E_m - E_l} V_{lm}
+ \ldots \right) \\
&=& \langle n| V\left(1 + \frac{1}{ E_m - H_0}V + \frac{1}{E_m -H_0
}V\frac{1}{E_m -H_0 }V + \ldots \right) |m \rangle
\end{eqnarray*}

\noindent The series in the parentheses can be summed up using the
standard formula $(1-x)^{-1} = 1 + x + x^2 +\ldots$

\begin{eqnarray}
T_{nm} &=& \langle n| V \frac{1}{1 - (E_m - H_0 )^{-1}V}  |m \rangle
\label{eq:Tnm2} \\
&=& \langle n| V (E_m - H_0)(E_m - H_0  - V)^{-1} |m \rangle \nonumber \\
&=& \langle n| V (E_m - H_0)(E_m - H)^{-1} |m \rangle \label{eq:Tnm}
\end{eqnarray}

\noindent The beauty of this result is that it provides a
non-perturbative closed-form expression for the $S$-operator and
scattering amplitudes. Expression (\ref{eq:Tnm2}) has been used in
numerical scattering calculations \cite{Rescigno}. See also a related method of ``inverse matrix'' in \cite{Brown-Jackson, Korchin, Dubovyk}.

According to (\ref{eq:Snm}) and (\ref{eq:Tnm}), the $S$-matrix can
be represented as a function of energy ($E = E_m = E_n$)

\begin{eqnarray}
S_{nm}(E) &=&\delta_{nm} -2 \pi i
\delta(E_n - E_m) T_{nm} \nonumber \\
 &=&\delta_{nm} -2 \pi i
\delta(E_n - E_m) \langle n| V (E - H_0)(E - H)^{-1} |m \rangle
\label{eq:SnmE}
\end{eqnarray}

\noindent Let us analyze the structure of this expression in more
detail. As we discussed in subsection \ref{ss:adiaba}, the
$S$-matrix is defined only for states that behave asymptotically as
free states. For such states, the energy $E$ is not lower than the
energy of separated reactants $E_0$. In the center-of-mass reference frame
this threshold is the sum of rest energies of all $N$
particles participating in the scattering event

\begin{eqnarray*}
E_0  &=&\sum_{a=1}^{N} m_ac^2
\end{eqnarray*}

\noindent Therefore

\begin{eqnarray}
E \in [E_0, \infty) \label{eq:eine}
\end{eqnarray}

Let us pick a value $E$ in this interval. Then $T_{nm}$ in
(\ref{eq:SnmE}) can be calculated as a matrix of an energy-dependent
$T$-operator $T(E)$

\begin{eqnarray*}
T_{nm} = \langle n| V (E -H_0)(E - H)^{-1} |m \rangle = \langle n|
T(E) |m \rangle
\end{eqnarray*}

\noindent The product $\delta(E_n - E)T_{nm}$  can be interpreted as
a matrix, which is zero everywhere except the diagonal sub-block
corresponding to the eigenvalue $E$ of $H_0$. If we denote $P_E $
the projection on this eigensubspace, then $\delta(E_n -
E)T(E) = P_E T(E) P_E$ and the full $S$-operator (for all values of
$E$) can be written in a basis-independent form

\begin{eqnarray}
S = 1 -2 \pi i \sum_E P_E V (E -H_0)(E - H)^{-1} P_E \label{eq:SofE}
\end{eqnarray}

\noindent The corresponding $S$-matrix has a block-diagonal form,
i.e., the matrix element $S_{nm}$ is non-zero only if the indices $n$
and $m$ satisfy condition $E_m = E_n$. This implies that the
$S$-operator commutes with the free Hamiltonian $H_0$.\footnote{see also equation (\ref{eq:S-rel-inv})} Exact formula (\ref{eq:SofE}) is not very useful in practical scattering calculations. However, it helps to derive some interesting results, such as the connection between poles of the $S$-matrix and energies of bound states discussed in the next subsection.

\subsection{$S$-matrix and bound states}
\label{ss:s-matr-bound}

There are good reasons to believe that the $S$-operator is an
analytical function of energy $E$.
 So, it is interesting to find where
the poles of this function are located. If operator $V$ is
non-singular, then poles of $S$ coincide with those values of $E$
at which operator $(E - H_0)(E - H)^{-1}$ in (\ref{eq:SofE}) is singular. In other
words, these are values $E$ for which the denominator has zero
eigenvalue. Thus poles $E_{\alpha}$ can be found as solutions of the
eigenvalue equation

\begin{eqnarray*}
H|\Psi_{\alpha} \rangle&=&  E_{\alpha}|\Psi_{\alpha} \rangle
\end{eqnarray*}

\noindent Obviously, this is equivalent to the stationary
Schr\"odinger equation for bound states.  This means that
there exists a correspondence between poles of the $S$-operator (or
$T$-operator) and bound state energies $E_{\alpha}$ of the full Hamiltonian $H$.  Earlier we
found that operators $S(E)$ and $T(E)$ are defined only for energies
$E$ in the interval (\ref{eq:eine}). Energies of bound states are always lower than $E_0$, i.e., they are
outside the domain of validity of operators $S(E)$ and $T(E)$. Therefore, the above correspondence (poles of
the $S$-operator)$\leftrightarrow$(energies of bound states) can be
established only in terms of analytic continuation of the
$S$-operator from its natural energy range $E \geq E_0$ to energy values below
$E_0$.

It is important to stress that the above possibility to find energies of bound states $E_{\alpha}$ from the $S$-matrix does not mean that state vectors
of bound states $|\Psi_{\alpha} \rangle$ can be found as well. All
bound states are eigenstates of just one (infinite) eigenvalue of the
$T$-matrix. Therefore, the maximum we can do is to find the entire
subspace spanning all bound state vectors. This ambiguity is closely related to the scattering equivalence of Hamiltonians discussed in the next section.

\section{Scattering equivalence}
\label{sc:scatt-equiv}

Results from the preceding section allow us to conclude that even
full knowledge of the $S$-operator does not permit us to obtain the
unique corresponding Hamiltonian $H$. In other words, many different
Hamiltonians may have identical scattering properties. In this
section, we will discuss in more detail this one-to-many
relationship between $S$-operators and Hamiltonians.

\subsection{Scattering equivalence of Hamiltonians}
\label{ss:scatt-equiv}

The $S$-operator and the Hamiltonian provide two different ways to
describe dynamics. The Hamiltonian completely describes the time
evolution for all time intervals, large or small. On the other hand,
the $S$-operator represents time evolution in the ``integrated'' form,
i.e., knowing the state of the system in the remote past
$|\Psi(-\infty) \rangle$, the free Hamiltonian $H_0$, and the scattering operator $S$, we can find the evolved state in the
distant future\footnote{see equation (\ref{eq:8.65a})}

\begin{eqnarray*}
|\Psi(\infty) \rangle &=&
 U(\infty \gets -\infty) |\Psi(-\infty) \rangle \\
&=&  U_0(\infty \gets -\infty) S  |\Psi(-\infty) \rangle
\end{eqnarray*}

\noindent Calculations of the $S$-operator are much easier than
those of the detailed time evolution and yet they fully satisfy the
needs of current experiments in high energy physics. In particular, the knowledge of the $S$-operator is sufficient to
calculate scattering cross-sections as well as energies and
lifetimes of stable and metastable bound states.\footnote{The two
latter  quantities are represented by positions of poles of the
$S$-operator on the complex energy plane, as discussed in the preceding subsection.} This situation
created an impression that a comprehensive theory can be constructed
which uses the $S$-operator as the fundamental quantity rather than
the Hamiltonian and wave functions. However,  in order to
describe the detailed time evolution of all states and wavefunctions of bound states, the
knowledge of the $S$-operator is not enough: the full interacting
Hamiltonian $H$ is needed. Therefore the $S$-operator
description is not complete, and such a theory is applicable
only to a limited class of experiments.

Knowing the full interacting Hamiltonian $H$, we can calculate the
$S$-operator by formulas (\ref{eq:8.69}), (\ref{eq:F-D}), or  (\ref{eq:8.71}). However, the inverse statement is not
true: the same $S$-operator can be obtained from many different
Hamiltonians. Suppose that two Hamiltonians $ H $ and $ H' $ are
related to each other by a unitary transformation $e^{i\Phi}$

\begin{eqnarray*}
H' &=& e^{i\Phi} H e^{-i\Phi}
\end{eqnarray*}

\noindent Then they yield  the same scattering (and Hamiltonians $H$
and $H'$ are called \emph{scattering equivalent}) \index{scattering
equivalence} as long as condition

\begin{eqnarray}
\lim _{t \to  \pm \infty}  e^{\frac{i}{\hbar}H_0 t} \Phi
e^{-\frac{i}{\hbar}H_0t}= 0 \label{eq:8.75}
\end{eqnarray}

\noindent is satisfied.\footnote{A rather general class of operators
$\Phi$ that satisfy this condition will be found in Theorem
\ref{theorem:non-singular} from subsection \ref{ss:scatt-ham}.}
Indeed, in the limit $\eta \to +\infty, \eta' \to -\infty$ we obtain
from (\ref{eq:seta}) \cite{Ekstein}

\begin{eqnarray}
S' &=& \lim_{\eta' \to -\infty, \eta \to  \infty}
 e^{\frac{i}{\hbar}H_0\eta}  e^{-\frac{i}{\hbar}H'(\eta - \eta')}
e^{-\frac{i}{\hbar}H_0 \eta'} \nonumber \\
&=& \lim_{\eta' \to -\infty, \eta \to  \infty}
 e^{\frac{i}{\hbar}H_0\eta} \left(e^{i\Phi} e^{-\frac{i}{\hbar}H(\eta - \eta')}
e^{-i\Phi}\right) e^{-\frac{i}{\hbar}H_0 \eta'}
\nonumber \\
&=&\lim_{\eta' \to -\infty, \eta \to  \infty}
\left(e^{\frac{i}{\hbar}H_0\eta}
   e^{i\Phi} e^{-\frac{i}{\hbar}H_0\eta}\right) e^{\frac{i}{\hbar}H_0\eta}
e^{-\frac{i}{\hbar}H(\eta - \eta')} e^{-\frac{i}{\hbar}H_0 \eta'}
\left(e^{\frac{i}{\hbar}H_0 \eta'} e^{-i\Phi} e^{-\frac{i}{\hbar}H_0
\eta'}\right)
\nonumber \\
&=& \lim_{\eta' \to -\infty, \eta \to  \infty}
e^{\frac{i}{\hbar}H_0\eta} e^{-\frac{i}{\hbar}H(\eta - \eta')}
e^{-\frac{i}{\hbar}H_0\eta'} \nonumber \\
     &=& S \label{eq:6.101a}
\end{eqnarray}

Note that due to Lemma \ref{LemmaA.12}, the energy spectra of two
scattering equivalent Hamiltonians $H$ and $H'$ are identical.
However, their eigenvectors are different, and corresponding
descriptions of dynamics (e.g., via equation (\ref{eq:8.61a})) are
different too. Therefore scattering-equivalent theories may not be
\emph{physically} equivalent. \index{physical equivalence}

\subsection{Bakamjian's construction of the point form dynamics}
\label{bakam-point}

So far we have been working within the instant form of Dirac's relativistic dynamics. It appears, however, that the above conclusions about scattering-equivalent Hamiltonians can be made more general, in the sense that even two different forms of dynamics (e.g., the instant form and the point form) can have the same $S$-operators. This will be discussed in subsection \ref{ss:form-equiv-point}. To prepare for this discussion, here we will construct a particular version of the point form dynamics using the Bakamjian's prescription \cite{Bakamjian}.
This method bears some resemblance to the Bakamjian-Thomas approach described in subsection \ref{ss:bakamjian}.

We start from
non-interacting operators of mass $M_0$, linear momentum
$\mathbf{P}_0$, angular momentum  $\mathbf{J}_0$, position
$\mathbf{R}_0$, and spin $\mathbf{S}_0 = \mathbf{J}_0 -
[\mathbf{R}_0 \times \mathbf{P}_0]$. Next we introduce two new operators

\begin{eqnarray*}
\mathbf{Q}_0 &\equiv& \frac{\mathbf{P}_0}{ M_0 c^2} \\
\mathbf{X}_0 &\equiv& M_0 c^2 \mathbf{R}_0
\end{eqnarray*}

\noindent which  satisfy canonical commutation relations

\begin{eqnarray*}
[X_{0i}, Q_{0j}] = i \hbar \delta_{ij}.
\end{eqnarray*}

Then we can express generators of the non-interacting representation
of the Poincar\'e group in the following form

\begin{eqnarray*}
\mathbf{P}_0 &=& M_0 c^2\mathbf{Q}_0   \\
\mathbf{J}_0 &=&
[\mathbf{X}_0 \times \mathbf{Q}_0] + \mathbf{S}_0 \\
\mathbf{K}_0 &=& -\frac{1}{2}\left(\sqrt{1+ c^2Q_0^2} \mathbf{X}_0 +
\mathbf{X}_0 \sqrt{1+ c^2Q_0^2}\right) - \frac{[\mathbf{Q}_0 \times
\mathbf{S}_0]}{1 + \sqrt{1+
 c^2Q_0^2}} \\
H_0 &=& M_0c^2 \sqrt{1+ c^2Q_0^2}.
\end{eqnarray*}

\noindent A point form interaction can be now introduced by
modifying the mass operator $M_0 \to M$ provided that the following
conditions are satisfied\footnote{Just as in subsection
\ref{ss:bakamjian}, these conditions can be satisfied by defining
$M = M_0 + L$, where $L$ is a rotationally-invariant function of relative position and momentum operators
commuting with $\mathbf{Q}_0$ and $\mathbf{X}_0$.}

\begin{eqnarray*}
[M, \mathbf{Q}_0] = [M, \mathbf{X}_0] = [M, \mathbf{S}_0] = 0
\end{eqnarray*}

\noindent These conditions, in particular, guarantee that $M$ is
Lorentz invariant

\begin{eqnarray*}
[M, \mathbf{K}_0] = [M, \mathbf{J}_0] = 0.
\end{eqnarray*}

\noindent This modification of the  mass operator introduces
interaction in  generators of the translation subgroup

\begin{eqnarray}
\mathbf{P}  &=& M c^2 \mathbf{Q}_0 \nonumber \\
H &=& M c^2 \sqrt {1 + c^2Q_0^2} \label{eq:6.3.5}
\end{eqnarray}

\noindent while Lorentz subgroup generators $\mathbf{K}_0$ and
$\mathbf{J}_0$ remain interaction-free.

\subsection{Scattering equivalence of forms of dynamics}
\label{ss:form-equiv-point}

The $S$-matrix equivalence of Hamiltonians established in subsection \ref{ss:scatt-equiv} remains valid even if the transformation
$e^{i\Phi}$ changes the relativistic form of dynamics \cite{
Sokolov_Shatnii, Sokolov_Shatnii2}. Here we would like to
demonstrate this equivalence on an example of Dirac's point and
instant forms of dynamics \cite{Sokolov_Shatnii}. We will use
definitions and notation from subsections \ref{bakam-point} and \ref{ss:adiaba}.

First we  assume that a Bakamjian's
point form representation of the Poincar\'e group is given, which is
built on operators

\begin{eqnarray*}
M &\neq& M_0 \\
\mathbf{P} &=& \mathbf{Q}_0 Mc^2 \\
\mathbf{J} &=& \mathbf{J}_0 \\
\mathbf{R} &=& \frac{\mathbf{X}_0}{ Mc^2}
\end{eqnarray*}

\noindent Then we introduce the unitary operator

\begin{eqnarray*}
\Theta = \zeta_0  \zeta^{-1}
\end{eqnarray*}

\noindent where

\begin{eqnarray*}
\zeta_0 &=& \exp(-i \ln (M_0c^2) Z_0) \\
\zeta &=& \exp(-i \ln( Mc^2) Z_0) \\
Z_0 &=& \frac{1}{2 \hbar} (\mathbf{Q}_0 \cdot \mathbf{X}_0 +
\mathbf{X}_0 \cdot \mathbf{Q}_0)
\end{eqnarray*}

\noindent Our goal here is to demonstrate that the set of operators $\Theta
M \Theta^{-1}$, $\Theta \mathbf{P} \Theta^{-1}$, $\Theta \mathbf{J}
\Theta^{-1}$, and $\Theta \mathbf{R} \Theta^{-1}$ generates a
representation of the Poincar\'e group in the Bakamjian-Thomas instant form of dynamics.
Moreover, the $S$-operators computed with the point-form Hamiltonian
$H =\sqrt{M^2c^4 + P^2c^2}$ and the instant form Hamiltonian
$H'=\Theta H \Theta^{-1}$ are the same.

Let us denote

\begin{eqnarray*}
\mathbf{Q}_0(b) = e^{i b  Z_0} \mathbf{Q}_0 e^{-i  b
 Z_0}, \mbox{ } b \in \mathbb{R}
\end{eqnarray*}

\noindent From the commutator

\begin{eqnarray*}
[Z_0, \mathbf{Q}_0] &=& i \mathbf{Q}_0
\end{eqnarray*}

\noindent it follows that

\begin{eqnarray*}
\frac{d}{db} \mathbf{Q}_0(b) &=& i [Z_0, \mathbf{Q}_0] = -
\mathbf{Q}_0 \\
\mathbf{Q}_0(b) &=& e^{- b}\mathbf{Q}_0
\end{eqnarray*}

\noindent This formula remains valid even if $b$ is a Hermitian
operator commuting with both $\mathbf{Q}_0$ and $\mathbf{X}_0$. For
example, if $b = \ln(M_0c^2)$, then

\begin{eqnarray*}
 e^{i  \ln(M_0c^2) Z_0} \mathbf{Q}_0 e^{-i
\ln(M_0c^2) Z_0} &=& e^{- \ln (M_0c^2)}\mathbf{Q}_0 =M_0^{-1}
c^{-2} \mathbf{Q}_0
\end{eqnarray*}

\noindent Similarly, one can prove

\begin{eqnarray*}
 e^{i  \ln(Mc^2) Z_0} \mathbf{Q}_0 e^{-i
\ln(Mc^2) Z_0} &=& M^{-1}c^{-2} \mathbf{Q}_0 \\
 e^{i  \ln(M_0c^2) Z_0} \mathbf{X}_0 e^{-i
\ln(M_0c^2) Z_0} &=& M_0c^2 \mathbf{X}_0 \\
 e^{i  \ln(Mc^2) Z_0} \mathbf{X}_0 e^{-i
\ln(Mc^2) Z_0} &=& Mc^2 \mathbf{X}_0
\end{eqnarray*}

\noindent which imply

\begin{eqnarray*}
\Theta \mathbf{P} \Theta^{-1} &=& e^{-i \ln (M_0c^2) Z_0} e^{i \ln
(Mc^2) Z_0}
\mathbf{Q}_0 Mc^2  e^{-i \ln (Mc^2) Z_0} e^{i \ln (M_0c^2) Z_0} \\
&=& e^{-i \ln (M_0c^2) Z_0}  \mathbf{Q}_0 e^{i \ln (M_0c^2) Z_0} =
\mathbf{Q}_0
M_0c^2 =\mathbf{P}_0 \\
\Theta \mathbf{J}_0 \Theta^{-1} &=&
\mathbf{J}_0 \\
\Theta \mathbf{R} \Theta^{-1} &=& e^{-i \ln (M_0c^2) Z_0} e^{i \ln
(Mc^2) Z_0}
\mathbf{X}_0 M^{-1}c^{-2}  e^{-i \ln (Mc^2) Z_0} e^{i \ln (M_0c^2) Z_0} \\
&=& e^{-i \ln (M_0c^2) Z_0}  \mathbf{X}_0 e^{i \ln (M_0c^2) Z_0} =
\mathbf{X}_0 M_0^{-1} c^{-2}= \mathbf{R}_0
\end{eqnarray*}

\noindent From these formulas it is clear that the transformed
dynamics corresponds to the Bakamjian-Thomas instant form.

Let us now demonstrate that the scattering operator $S$  computed with the
point form Hamiltonian $H$ (\ref{eq:6.3.5}) is the same as  $S'$ computed with
 the instant form Hamiltonian $H' =
\Theta H \Theta^{-1}$. Note that we can write equation (\ref{eq:8.73a})
as

\begin{eqnarray*}
S = \Omega^{+}(H, H_0) \Omega^{-}( H, H_0)
\end{eqnarray*}

\noindent where operators

\begin{eqnarray*}
\Omega^{\pm}(H, H_0) \equiv \lim_{t \to \pm \infty}
e^{\frac{i}{\hbar}H_0 t} e^{-\frac{i}{\hbar}Ht}
\end{eqnarray*}

\noindent are called M\o ller \emph{wave operators}. \index{M\o ller
wave operator} Now we can use
 the Birman-Kato invariance principle \index{Birman-Kato invariance principle} \cite{Dollard} which states
that $\Omega^{\pm}(H, H_0) = \Omega^{\pm}(f(H), f(H_0))$ where $f$
is any smooth function with positive derivative. Using the following
connection between the point form mass operator $M$ and the instant
form mass operator $M'$

\begin{eqnarray*}
M &=& \zeta^{-1}  M \zeta = \zeta^{-1} \Theta^{-1} M' \Theta \zeta =
\zeta^{-1} \zeta \zeta_0^{-1}  M' \zeta_0 \zeta^{-1} \zeta =
\zeta_0^{-1}  M' \zeta_0
\end{eqnarray*}

\noindent we obtain

\begin{eqnarray*}
\Omega^{\pm}(H, H_0) &\equiv& \Omega^{\pm}\left( M c^2\sqrt{1+
c^2\mathbf{Q}_0^2},
M_0c^2 \sqrt{1+ c^2\mathbf{Q}_0^2}\right) \\
&=&\Omega^{\pm}(Mc^2, M_0c^2)
= \Omega^{\pm}( \zeta_0^{-1} M' \zeta_0 c^2 , M_0c^2) \\
 &=& \zeta_0
^{-1}  \Omega^{\pm}(M'c^2, M_0c^2) \zeta_0 \\
 &=& \zeta_0
^{-1}  \Omega^{\pm} \left(\sqrt{(M')^2 c^4 + \mathbf{P}_0c^2},
\sqrt{M_0^2 c^4 + \mathbf{P}_0c^2}\right) \zeta_0 \\
 &=& \zeta_0
^{-1}  \Omega^{\pm}(H', H_0) \zeta_0 \\
 \\
S' &=& \Omega^{+}(H', H_0) \Omega^{-}( H', H_0) \\
&=& \zeta_0
  \Omega^{+}(H, H_0) \zeta_0^{-1}
\zeta_0  \Omega^{-}(H, H_0) \zeta_0^{-1} \\
&=& \zeta_0
  \Omega^{+}(H, H_0)   \Omega^{-}(H, H_0) \zeta_0^{-1}
  =\zeta_0   S \zeta_0^{-1}
\end{eqnarray*}

\noindent but $S$ commutes with free generators (\ref{eq:S-rel-inv})
and hence commutes with $\zeta_0$, which implies that $S' = S$ and
transformation $\Theta$ conserves the $S$-matrix.

In addition to the scattering equivalence of the instant and point
forms proved above, Sokolov and Shatnii \cite{Sokolov_Shatnii,
Sokolov_Shatnii2} established the mutual scattering equivalence of all three major forms
of dynamics  - the instant, point, and front forms. Then, it seems
reasonable to assume that the same $S$-operator can be obtained in
any form of dynamics.

The scattering equivalence of the $S$-operator is of great help in
practical calculations. If we are interested only in scattering
properties, energies, and lifetimes of bound states, then we can choose the
most convenient Hamiltonian and the most
convenient form of dynamics. However, as we have mentioned already, the
scattering equivalence of Hamiltonians and forms of dynamics does
not mean their complete physical equivalence. We will see in
subsection \ref{ss:inequivalence} that the instant form of dynamics
should be preferred in those cases  when desired physical properties\footnote{e.g., the detailed time evolution}
cannot be described by the $S$-operator, but require knowledge of the full interacting Hamiltonian.

\chapter{FOCK SPACE}
\label{ch:fock-space}

\vspace{0.5in}

\begin{quote}
\textit{This subject has been thoroughly worked out
and is now understood. A thesis on this topic,
even a correct one, will not get you a job. }

\small
\hspace{1in} R.F. Streater
\normalsize
\end{quote}

\vspace{0.5in}

In
chapter \ref{ch:interaction} we discussed interacting quantum theories
in  the Hilbert space with a fixed
particle content. These theories were fundamentally incomplete, because they could not
describe many physical processes that can change  particle types
and/or numbers. Familiar examples of such processes include the
emission and absorption of light (photons) by atoms and nuclei,
decays, neutrino oscillations,   etc. The persistence of particle
creation and destruction processes at high energies
 follows from the famous Einstein's formula $E = mc^2$. This
formula, in particular, implies that if a system of particles has
sufficient energy $E$ of their relative motion, then this energy can
be converted to the mass $m$ of newly created particles.
Generally, there is no limit on how many particles can be created in
collisions, so the Hilbert space of any realistic quantum mechanical
system should include states with arbitrary numbers (from zero to infinity) of particles
 of all types. Such a Hilbert space is called
the \emph{Fock space}. \index{Fock space}

Our primary goal in this chapter (and for the most part in the rest of this book) is to understand electromagnetic interactions between five particle species:
electrons $e^-$, positrons $e^+$, protons $p^+$, antiprotons $p^-$, and photons $\gamma$ within, allegedly, the most successful physical theory -- \emph{quantum
electrodynamics} (QED). \index{quantum electrodynamics}

\section{Annihilation and creation operators}
\label{sc:annih-operators}

In this section we are going to build the Fock space $\mathcal{H}$
of QED  and introduce creation and annihilation operators, which
provide a very convenient notation for working with
operators in $\mathcal{H}$.

\subsection{Sectors with fixed numbers of particles}
\label{ss:sectors}

The numbers of particles of any type are readily measured in
experiments, so we can introduce 5 new observables in our theory: the
numbers of electrons ($N_{el}$), positrons ($N_{po}$), protons
($N_{pr}$), antiprotons ($N_{an}$), and photons ($N_{ph}$).
According to general rules of quantum mechanics, these observables
must be represented by five Hermitian operators in the Hilbert
(Fock) space $\mathcal{H}$. Apparently, the allowed values (the
spectrum) for the number of particles of each type are non-negative
integers (0,1,2,...). We assume that these observables can be
measured simultaneously, therefore the corresponding operators
commute with each other and  have common spectrum.
  So, the Fock space $\mathcal{H}$ separates
into a direct sum of corresponding orthogonal eigensubspaces or
\emph{sectors} \index{sector} $\mathcal{H} (i,j,k,l,m)$ with $i$
electrons, $j$ positrons, $k$ protons, $l$ antiprotons, and $m$
photons

\begin{eqnarray}
\mathcal{H} =\oplus _{ijklm = 0}^{\infty} \mathcal{H} (i,j,k,l,m)
\label{eq:fock-space}
\end{eqnarray}

\noindent  where

\begin{eqnarray*}
N_{el} \mathcal{H}(i,j,k,l,m) &=& i \mathcal{H}(i,j,k,l,m) \\
N_{po} \mathcal{H}(i,j,k,l,m) &=& j \mathcal{H}(i,j,k,l,m) \\
N_{pr} \mathcal{H}(i,j,k,l,m) &=& k \mathcal{H}(i,j,k,l,m) \\
N_{an} \mathcal{H}(i,j,k,l,m) &=& l \mathcal{H}(i,j,k,l,m) \\
N_{ph} \mathcal{H}(i,j,k,l,m) &=& m \mathcal{H}(i,j,k,l,m)
\end{eqnarray*}

The one-dimensional subspace with no particles $\mathcal{H}
(0,0,0,0,0)  $ is called the \emph{vacuum subspace}. \index{vacuum
subspace} The \emph{vacuum vector} \index{vacuum vector} $| 0
\rangle$ is then defined as a vector in this subspace, up to an
insignificant phase factor. The one-particle sectors are built using
prescriptions from chapter \ref{ch:single}.  The subspaces
$\mathcal{H} (1,0,0,0,0)$ and $\mathcal{H} (0,1,0,0,0)$ correspond
to one electron and one positron, respectively. They are subspaces
of unitary irreducible representations of the Poincar\'e group
characterized by the mass $m = 0.511$ MeV/$c^2$ and spin 1/2 (see
Table \ref{table:5.1}). The subspaces $\mathcal{H} (0,0,1,0,0)$ and
$\mathcal{H} (0,0,0,1,0)$ correspond to one proton and one
antiproton, respectively. They have mass $M = 938.3$ MeV/$c^2$ and
spin 1/2. The subspace $\mathcal{H} (0,0,0,0,1)$ correspond to one
photon. It is characterized by zero mass and it is a direct sum of
two irreducible subspaces with helicities 1 and -1.\footnote{see subsection
\ref{ss:rep-poincare}}

Sectors with two or more particles are constructed as
(anti)symmetrized tensor products of one-particle
sectors.\footnote{See  section \ref{sc:many-particle}. Note that electrons and protons are fermions, while photons are bosons.}
For example, if we denote $\mathcal{H}_{el}$ the one-electron
Hilbert space and $\mathcal{H}_{ph}$ the one-photon Hilbert space,
then sectors having only electrons and photons can be written as

\begin{eqnarray}
\mathcal{H} (0,0,0,0,0) &=& | 0 \rangle
\label{eq:9.1} \\
\mathcal{H} (1,0,0,0,0) &=& \mathcal{H}_{el}
\label{eq:9.2}\\
\mathcal{H} (0,0,0,0,1) &=& \mathcal{H}_{ph}
\label{eq:9.3}\\
\mathcal{H} (1,0,0,0,1) &=& \mathcal{H}_{el} \otimes \mathcal{H}_{ph}
\label{eq:9.4}\\
\mathcal{H} (2,0,0,0,0) &=& \mathcal{H}_{el} \otimes_{asym} \mathcal{H}_{el}
\label{eq:9.5}\\
\mathcal{H} (0,0,0,0,2) &=& \mathcal{H}_{ph} \otimes_{sym} \mathcal{H}_{ph}
\label{eq:9.6}\\
\mathcal{H} (1,0,0,0,2) &=& \mathcal{H}_{el} \otimes
(\mathcal{H}_{ph} \otimes_{sym} \mathcal{H}_{ph})
\label{eq:9.7}\\
\mathcal{H} (2,0,0,0,1) &=& \mathcal{H}_{ph} \otimes
(\mathcal{H}_{el} \otimes_{asym} \mathcal{H}_{el})
\label{eq:9.8}\\
\mathcal{H} (2,0,0,0,2) &=& (\mathcal{H}_{ph} \otimes_{sym} \mathcal{H}_{ph})
\otimes
(\mathcal{H}_{el} \otimes_{asym} \mathcal{H}_{el})
\label{eq:9.9} \\
&\ldots& \nonumber
\end{eqnarray}

\noindent  In each sector of the Fock space we can define
observables of individual particles, e.g., position momentum, spin,
etc., as described in subsection \ref{ss:multi-part-obs}.

For example, in each
(massive) 1-particle subspace of the Fock space there is a
Newton-Wigner operator that describes position measurements on this
particle. In 2-particle sectors we can define two different position
operators for each of the two particles. In addition, we can
also define the ``center-of-mass'' position operators for the
2-particle system in a usual way. Similar position operators exist
in each $N$-particle sector.

Then, in
each sector we can select a basis of common eigenvectors of a full
set of mutually commuting one-particle observables. A general state $| \Psi \rangle$ in the Fock space may have components in all sectors.\footnote{\emph{Superselection rules} \index{superselection rules} forbid linear combinations of states with, e.g., different charges. We will not discuss such rules here.} Thus the number of particles in $| \Psi \rangle$ may be not well-defined.

 For future discussions  it will be
 convenient to use the basis in
which momenta and $z$-components of the spin $\sigma$ of massive
particles (or helicity $\tau$ of massless particles) are diagonal.
For example, basis vectors in the two-electron sector
$\mathcal{H}_{el} \otimes_{asym} \mathcal{H}_{el}$ are denoted by $
|\mathbf{p}_1 \sigma_1;\mathbf{p}_2 \sigma_2 \rangle $. This
allows us to define in each sector multi-particle wave functions in the momentum-spin representation.

\subsection{Non-interacting representation of the Poincar\'e group}
\label{ss:non-int-fock}

 The above construction provides us with
the Hilbert (Fock) space $\mathcal{H}$ where multiparticle states
and observables of our theory reside and where a convenient
orthonormal basis set is defined. To complete the formalism we need
to build  a realistic interacting representation of the Poincar\'e
group in $\mathcal{H}$. Let us first fulfill an easier task and
construct the non-interacting representation $U_g^0$ of the
Poincar\'e group in the Fock space $\mathcal{H}$.

From subsection \ref{ss:non-interacting}, we already know how to build a
non-interacting representation of the
Poincar\'e group in each individual sector of $\mathcal{H}$.  This can be done
 by
making tensor products (with proper (anti)symmetrization) of single-particle irreducible representations  $U_g^{el}, U_g^{ph}$, etc.
Then the \emph{non-interacting} representation
\index{non-interacting representation} of the Poincar\'e group in
the entire Fock space can be constructed as a direct sum of such sector
representations. In agreement with the sector decomposition (\ref{eq:9.1}) - (\ref{eq:9.9}) we can write

\begin{eqnarray}
U_g^0 = 1 \oplus U_g^{el} \oplus U_g^{ph} \oplus  (U_g^{el} \otimes
U_g^{ph})  \oplus (U_g^{el} \otimes_{asym} U_g^{el}) \ldots
\label{eq:utensor-product}
\end{eqnarray}

\noindent Generators of this representation   will be denoted as
$(H_0, \mathbf{P}_0, \mathbf{J}_0, \mathbf{K}_0)$. In each sector
these generators are simply sums of one-particle generators.\footnote{For example, in each 2-particle sector equations (\ref{eq:8.7}) - (\ref{eq:8.10}) are valid.} As usual, we assume that operators $H_0$,
$\mathbf{P}_0$, and $\mathbf{J}_0$ describe the \emph{total} energy,
linear momentum, and angular momentum  of the non-interacting
system, respectively.

Here we immediately face a serious problem. For example, according to
(\ref{eq:utensor-product}),
the free Hamiltonian can be represented as a direct sum of sector components

\begin{eqnarray*}
H_0 = |0 \rangle \oplus H_0(1,0,0,0,0) \oplus H_0(0,0,0,0,1)
\oplus H_0(1,0,0,0,1) \oplus \ldots
\end{eqnarray*}

\noindent It is tempting to use notation from section \ref{sc:RHD}
and express each sector Hamiltonian using
observables of individual particles there: $\mathbf{p}_1$, $\mathbf{p}_2$,
etc. For example, in the sector
 $\mathcal{H}(1,0,0,0,0)$, the free Hamiltonian is

\begin{eqnarray}
H_0(1,0,0,0,0) =  \sqrt{m^2c^4 + p^2c^2} \label{eq:9.32}
\end{eqnarray}

\noindent while in the
sector $\mathcal{H}(2,0,0,0,2)$ the Hamiltonian is\footnote{Two photons
are denoted by indices 1 and 2 and two electrons are denoted by
indices 3 and 4}

\begin{eqnarray}
H_0(2,0,0,0,2) = p_1c + p_2c + \sqrt{m^2c^4 + p_3^2c^2} +
\sqrt{m^2c^4 + p_4^2c^2} \label{eq:9.33}
\end{eqnarray}

\noindent Clearly, this notation is very cumbersome because it does
not provide a unique expression for the operator $H_0$ in the entire
Fock space. Moreover, it is not clear at all how one can use one-particle observables to express operators changing the number of particles,
i.e., moving state vectors across sector boundaries. We need to find
a better and simpler way to write operators in the Fock space. This
task is accomplished by introduction of annihilation and creation
operators in the rest of this section.

\subsection{Creation and annihilation
operators. Fermions} \label{ss:discrete-momentum}

First,  it is instructive to consider the case of the discrete
spectrum of momentum. This can be achieved by using the standard
trick of putting the system in a box or applying periodic boundary
conditions. Then eigenvalues of the momentum operator form a
discrete 3D lattice $\mathbf{p}_i$, and the usual continuous
momentum spectrum can be obtained as a limit when the size of the
box tends to infinity.

Let us examine the case of electrons. We define the (linear)
\emph{creation operator} \index{creation operator}
$a^{\dag}_{\mathbf{p}, \sigma}$ for the electron with momentum
$\mathbf{p}$ and spin projection  $\sigma$  by  its action on basis
 vectors with $n$ electrons

\begin{eqnarray}
|\mathbf{p}_1, \sigma_1;\mathbf{p}_2, \sigma_2; \ldots
;\mathbf{p}_n, \sigma_n \rangle \label{eq:9.10}
\end{eqnarray}

\noindent We need to distinguish two cases. The first case is when
the one-particle state $(\mathbf{p}, \sigma)$ created by $a^{\dag}_{\mathbf{p}, \sigma}$ is among the states
listed in (\ref{eq:9.10}), for example $(\mathbf{p}, \sigma) =
(\mathbf{p}_i, \sigma_i)$. Since electrons are fermions and two
fermions cannot occupy the same state due to the Pauli exclusion
principle, \index{Pauli exclusion principle} this action leads to a
zero result, i.e.

\begin{eqnarray}
a^{\dag}_{\mathbf{p}, \sigma} |\mathbf{p}_1, \sigma_1;
\mathbf{p}_2, \sigma_2; \ldots ; \mathbf{p}_n, \sigma_n \rangle = 0
\label{eq:9.11}
\end{eqnarray}

\noindent The second case is when the created
one-particle state $(\mathbf{p}, \sigma)$ is not among the states
listed in (\ref{eq:9.10}). Then
 the
creation operator $a^{\dag}_{\mathbf{p}, \sigma}$ just adds one
electron in the state  $(\mathbf{p}, \sigma )$ to the beginning of
the list of particles

\begin{eqnarray}
a^{\dag}_{\mathbf{p}, \sigma} |\mathbf{p}_1, \sigma_1;\mathbf{p}_2,
\sigma_2; \ldots ; \mathbf{p}_n, \sigma_n \rangle = |\mathbf{p},
\sigma; \mathbf{p}_1, \sigma_1 ; \mathbf{p}_2, \sigma_2; \ldots
;\mathbf{p}_n, \sigma_n \rangle \label{eq:9.12}
\end{eqnarray}

\noindent Operator $a^{\dag}_{\mathbf{p}, \sigma} $ has transformed
a state with $n$ electrons  to a state with $n+1$ electrons.
Applying multiple creation operators to the vacuum state $|0
\rangle$ we can construct all basis vectors in the Fock space. For
example,

\begin{eqnarray*}
a^{\dag}_{\mathbf{p}_1, \sigma_1} a^{\dag}_{\mathbf{p}_2, \sigma_2}
|0 \rangle = |\mathbf{p}_1, \sigma_1; \mathbf{p}_2, \sigma_2\rangle
\end{eqnarray*}

\noindent is a basis vector in the 2-electron sector.

We define the electron \emph{annihilation operator}
\index{annihilation operator} $a_{\mathbf{p}, \sigma}$ as an operator
adjoint to the creation operator $a^{\dag}_{\mathbf{p}, \sigma}$. It
can be proven \cite{book} that the action of $a_{\mathbf{p},
\sigma}$
 on the $n$-electron state (\ref{eq:9.10}) is the following: If the electron
state with parameters $(\mathbf{p}, \sigma)$ was already occupied,
e.g. $(\mathbf{p},\sigma) = (\mathbf{p}_i,\sigma_i)$ then this state
is ``annihilated,'' and the number of particles in the system is
reduced by one

\begin{eqnarray}
&\mbox{ }& a_{\mathbf{p}, \sigma} |\mathbf{p}_1, \sigma_1; \ldots
;\mathbf{p}_{i-1}, \sigma_{i-1}; \mathbf{p}_i, \sigma_i;
\mathbf{p}_{i+1}, \sigma_{i+1}; \ldots ;\mathbf{p}_n,
\sigma_n \rangle \nonumber\\
&=& (-1)^P|\mathbf{p}_1, \sigma_1; \ldots ;\mathbf{p}_{i-1},
\sigma_{i-1}; \mathbf{p}_{i+1}, \sigma_{i+1}; \ldots ;\mathbf{p}_n,
\sigma_n \rangle \label{eq:9.13}
\end{eqnarray}

\noindent where $P$ is the number of permutations of particles
required to bring the one-particle $i$ to the first place in the
list. If the state $(\mathbf{p}, \sigma)$ is not present in the
list, i.e., $(\mathbf{p}, \sigma) \neq (\mathbf{p}_i, \sigma_i)$ for
each $i$, then

\begin{eqnarray}
a_{\mathbf{p}, \sigma} |\mathbf{p}_1, \sigma_1; \mathbf{p}_2,
\sigma_2; \ldots ;\mathbf{p}_n, \sigma_n \rangle = 0 \label{eq:9.14}
\end{eqnarray}

\noindent Annihilation operators always yield zero when acting on
the vacuum state

\begin{eqnarray*}
a_{\mathbf{p}, \sigma} |0 \rangle = 0
\end{eqnarray*}

The above formulas fully define the action of creation and
annihilation operators on basis vectors in purely electronic
sectors. These rules are easily generalized to all states: they
 do not change if other particles  are present and they can be
 extended to  linear combinations of the basis vectors
by linearity. Creation and annihilation operators for other fermions
-- positrons, protons and antiprotons -- are constructed similarly.

For brevity we will refer to creation and annihilation operators
collectively as to \emph{particle operators}. \index{particle
operators} This will distinguish them from operators of momentum,
position, energy, etc. of individual particles which will be called
\emph{particle observables}. \index{particle observables} Let us
emphasize that creation and annihilation operators are not intended
to directly describe any real physical process and they do not correspond to
physical observables. They are just formal mathematical objects that
simplify our notation for other operators having more direct
physical meaning. We will see how operators of observables are built
from particle operators later in this book, e.g., in subsection
\ref{ss:non-int-rep}.

\subsection{Anticommutators of particle operators}
\label{ss:commutators}

In practical calculations one often uses \emph{anticommutators} of fermion operators. \index{$\{\ldots,
\ldots\}$ anticommutator} First consider the case of unequal particle states $(\mathbf{p}, \sigma) \neq (\mathbf{p}',
\sigma')$

\begin{eqnarray*}
\{ a_{\mathbf{p}', \sigma'}, a^{\dag}_{\mathbf{p}, \sigma} \} \equiv
a^{\dag}_{\mathbf{p}, \sigma} a_{\mathbf{p}', \sigma'} +
 a_{\mathbf{p}', \sigma'} a^{\dag}_{\mathbf{p}, \sigma}
\end{eqnarray*}

\noindent  Then, acting on a state $|\mathbf{p}'', \sigma'' \rangle $,
which is different from both $|\mathbf{p}, \sigma \rangle $ and
$|\mathbf{p}', \sigma' \rangle $, we obtain

\begin{eqnarray*}
(a^{\dag}_{\mathbf{p}, \sigma} a_{\mathbf{p}', \sigma'} +
 a_{\mathbf{p}', \sigma'} a^{\dag}_{\mathbf{p}, \sigma})
|\mathbf{p}'', \sigma'' \rangle &=& a_{\mathbf{p}', \sigma'}
|\mathbf{p}, \sigma; \mathbf{p}'', \sigma'' \rangle = 0
\end{eqnarray*}

\noindent Similarly,  we obtain

\begin{eqnarray*}
(a^{\dag}_{\mathbf{p}, \sigma} a_{\mathbf{p}', \sigma'} +
 a_{\mathbf{p}', \sigma'} a^{\dag}_{\mathbf{p}, \sigma})
|\mathbf{p}, \sigma \rangle &=&
 0 \\
(a^{\dag}_{\mathbf{p}, \sigma} a_{\mathbf{p}', \sigma'} +
 a_{\mathbf{p}', \sigma'} a^{\dag}_{\mathbf{p}, \sigma})
|\mathbf{p}', \sigma' \rangle &=&
 a^{\dag}_{\mathbf{p}, \sigma} |0 \rangle  +
 a_{\mathbf{p}', \sigma'}
|\mathbf{p}, \sigma;\mathbf{p}', \sigma' \rangle \\
&=&
  |\mathbf{p}, \sigma \rangle  -
|\mathbf{p}, \sigma \rangle = 0
\end{eqnarray*}

\noindent One can easily demonstrate that the result is still zero
when acting on zero-, two-, three-, etc. particle states as well as
on their linear combinations. So, we conclude that in the entire Fock space

\begin{eqnarray*}
\{  a_{\mathbf{p}', \sigma'}, a^{\dag}_{\mathbf{p}, \sigma} \} = 0
\mbox {, if $(\mathbf{p}, \sigma) \neq (\mathbf{p}', \sigma') $ }
\end{eqnarray*}

\noindent Similarly, in the case $(\mathbf{p}, \sigma) =
(\mathbf{p}', \sigma')$ we  obtain

\begin{eqnarray*}
\{ a^{\dag}_{\mathbf{p}, \sigma}, a_{\mathbf{p}, \sigma} \} = 1
\end{eqnarray*}

\noindent Therefore for all values of $\mathbf{p}$, $\mathbf{p}'$,
$\sigma$, and $\sigma'$ we have

\begin{eqnarray}
\{ a^{\dag}_{\mathbf{p}, \sigma}, a_{\mathbf{p}', \sigma'} \} =
\delta_{\mathbf{p}, \mathbf{p}'} \delta_{\sigma, \sigma'}
\label{eq:9.18}
\end{eqnarray}

\noindent Using similar arguments one can show that

\begin{eqnarray*}
\{ a^{\dag}_{\mathbf{p}, \sigma}, a^{\dag}_{\mathbf{p}', \sigma'} \}
&=&  \{ a_{\mathbf{p}, \sigma}, a_{\mathbf{p}',
\sigma'} \} =0 \label{eq:9.20}
\end{eqnarray*}

\subsection{Creation and annihilation
operators. Photons} \label{ss:discrete-momentum-phot}

For photons, which are bosons, the properties of creation and
annihilation operators \index{annihilation operator}\index{creation operator}
are slightly different from those characteristic for fermion operators described
above. Two or more bosons may coexist in the same state. Therefore,
we define the action of the photon creation operator
$c^{\dag}_{\mathbf{p}, \tau} $ on a many-photon state as

\begin{eqnarray*}
c^{\dag}_{\mathbf{p}, \tau} |\mathbf{p}_1, \tau_1; \mathbf{p}_2,
\tau_2; \ldots ; \mathbf{p}_n, \tau_n \rangle = |\mathbf{p}, \tau;
\mathbf{p}_1, \tau_1; \mathbf{p}_2, \tau_2; \ldots ;\mathbf{p}_n,
\tau_n \rangle \label{eq:9.15}
\end{eqnarray*}

\noindent independent of whether or not the state $(\mathbf{p},
\tau)$ already existed. Just as in the case of fermions, boson annihilation operators $c_{\mathbf{p}, \tau}$  are adjoint to boson creation operators. The photon
annihilation operator $c_{\mathbf{p}, \tau}$ destroys a multi-photon state completely

\begin{eqnarray*}
c_{\mathbf{p}, \tau} |\mathbf{p}_1, \tau_1; \mathbf{p}_2, \tau_2;
\ldots ;\mathbf{p}_n, \tau_n \rangle = 0 \label{eq:9.16}
\end{eqnarray*}

\noindent if the annihilated 1-photon state $(\mathbf{p}, \tau)$ was  not present there. If the photon $(\mathbf{p}, \tau)$  was present in the $n$-photon state, then the annihilation operator $c_{\mathbf{p}, \tau}$ simply destroys this component, thus resulting in a $(n-1)$-photon state

\begin{eqnarray*}
&\mbox{ }& c_{\mathbf{p}_i, \tau_i} |\mathbf{p}_1, \tau_1; \ldots
;\mathbf{p}_{i-1}, \tau_{i-1}; \mathbf{p}_i, \tau_i;
\mathbf{p}_{i+1}, \tau_{i+1}; \ldots ;\mathbf{p}_n,
\tau_n \rangle \nonumber\\
&=& |\mathbf{p}_1, \tau_1; \ldots ;\mathbf{p}_{i-1}, \tau_{i-1};
\mathbf{p}_{i+1}, \tau_{i+1}; \ldots ;\mathbf{p}_n, \tau_n \rangle
\label{eq:9.17}
\end{eqnarray*}

The above formulas can be immediately  extended to states where, in addition
to photons, other particles are also present. The actions of creation
and annihilation operators  on linear combinations of basis
 vectors are obtained by  linearity. Then similar to subsection \ref{ss:commutators}, we obtain the following
\emph{commutation} relations for photon creation and annihilation
operators \index{commutation relations}

\begin{eqnarray*}
[ c_{\mathbf{p}, \tau}, c^{\dag}_{\mathbf{p}', \tau'} ] &=&
\delta_{\mathbf{p}, \mathbf{p}'} \delta_{\tau, \tau'}
\label{eq:9.21} \\
\mbox{ }  [ c_{\mathbf{p}, \tau}, c_{\mathbf{p}', \tau'} ]
&=&
 \mbox{ } [ c^{\dag}_{\mathbf{p}, \tau},
c^{\dag}_{\mathbf{p}', \tau'} ] = 0 \label{eq:9.23}
\end{eqnarray*}

\subsection{Particle number operators} \label{ss:particle-number}

With the help of particle creation and annihilation operators
 we can
now build explicit expressions for various observables in the Fock
space. Consider, for example, the product of two photon operators

\begin{eqnarray}
N_{\mathbf{p},\tau} =
 c^{\dag}_{\mathbf{p},\tau} c
_{\mathbf{p},\tau}
\label{eq:9.24}
\end{eqnarray}

\noindent Acting on a state with two photons with quantum numbers
$(\mathbf{p}, \tau ) $ this operator yields

\begin{eqnarray*}
&\ &N_{\mathbf{p},\tau} |\mathbf{p},\tau; \mathbf{p},\tau \rangle =
N_{\mathbf{p},\tau}
c^{\dag}_{\mathbf{p},\tau}c^{\dag}_{\mathbf{p},\tau} |0 \rangle =
 c^{\dag}_{\mathbf{p},\tau} c
_{\mathbf{p},\tau} c^{\dag}_{\mathbf{p},\tau}c^{\dag}_{\mathbf{p},\tau} |0
\rangle \\
&=&
 c^{\dag}_{\mathbf{p},\tau} c^{\dag}_{\mathbf{p},\tau} c
_{\mathbf{p},\tau} c^{\dag}_{\mathbf{p},\tau} |0 \rangle +
c^{\dag}_{\mathbf{p},\tau}  c^{\dag}_{\mathbf{p},\tau} |0 \rangle=
 c^{\dag}_{\mathbf{p},\tau} c^{\dag}_{\mathbf{p},\tau}
c^{\dag}_{\mathbf{p},\tau}
c
_{\mathbf{p},\tau}  |0
\rangle +  2 c^{\dag}_{\mathbf{p},\tau}  c^{\dag}_{\mathbf{p},\tau} |0
\rangle\\
&=& 2  |\mathbf{p},\tau; \mathbf{p},\tau \rangle
\end{eqnarray*}

\noindent while acting on the state
$|\mathbf{p},\tau;\mathbf{p}',\tau' \rangle$ we obtain

\begin{eqnarray*}
&\ &N_{\mathbf{p},\tau} |\mathbf{p},\tau; \mathbf{p}',\tau' \rangle
= N_{\mathbf{p},\tau}
c^{\dag}_{\mathbf{p},\tau}c^{\dag}_{\mathbf{p}',\tau'} |0 \rangle =
 c^{\dag}_{\mathbf{p},\tau} c
_{\mathbf{p},\tau}
c^{\dag}_{\mathbf{p},\tau}c^{\dag}_{\mathbf{p}',\tau'} |0 \rangle
\\
&=&
 c^{\dag}_{\mathbf{p},\tau} c^{\dag}_{\mathbf{p},\tau} c
_{\mathbf{p},\tau} c^{\dag}_{\mathbf{p}',\tau'} |0 \rangle +
c^{\dag}_{\mathbf{p},\tau}  c^{\dag}_{\mathbf{p}',\tau'} |0 \rangle=
 c^{\dag}_{\mathbf{p},\tau} c^{\dag}_{\mathbf{p},\tau}
c^{\dag}_{\mathbf{p}'\tau'}
c
_{\mathbf{p},\tau}  |0
\rangle +   c^{\dag}_{\mathbf{p},\tau}  c^{\dag}_{\mathbf{p}',\tau'} |0
\rangle\\
&=&   |\mathbf{p},\tau; \mathbf{p}',\tau' \rangle
\end{eqnarray*}

\noindent These examples should convince us  that
 operator $N_{\mathbf{p},\tau} $ works as a counter of the number of photons
with quantum numbers  $(\mathbf{p},\tau ) $.

\subsection{Continuous spectrum of momentum} \label{ss:continuous-momentum}

Properties of creation and annihilation operators presented in
preceding subsections were derived for the case of discrete spectrum
of momentum. In reality the spectrum of momentum is continuous, and
the above results should be modified by taking the ``large box''
limit. We can guess that in this limit equation (\ref{eq:9.18})
transforms to

\begin{eqnarray}
\{ a_{\mathbf{p}', \sigma'}, a^{\dag}_{\mathbf{p}, \sigma}  \} =
\delta_{\sigma, \sigma'} \delta(\mathbf{p} - \mathbf{p}')
\label{eq:9.25}
\end{eqnarray}

\noindent The following chain of formulas

\begin{eqnarray*}
&\ &\delta_{\sigma, \sigma'} \delta (\mathbf{p} - \mathbf{p}') \\
 &=&
\langle
\mathbf{p}, \sigma| \mathbf{p}', \sigma' \rangle = \langle 0 |
a_{\mathbf{p}, \sigma} a^{\dag}_{\mathbf{p}', \sigma'} | 0 \rangle = -\langle 0 |
a^{\dag}_{\mathbf{p}', \sigma'} a_{\mathbf{p}, \sigma}  | 0 \rangle
+ \delta_{\sigma, \sigma'} \delta(\mathbf{p} -
\mathbf{p}') \\
&=&  \delta_{\sigma, \sigma'} \delta(\mathbf{p} -
\mathbf{p}')
\end{eqnarray*}

\noindent confirms that our choice (\ref{eq:9.25}) is consistent
with the normalization of momentum eigenvectors (\ref{eq:7.13}).

The same arguments now can be applied to positrons (operators
$b_{\mathbf{p}, \sigma}$ and $b^{\dag}_{\mathbf{p}, \sigma}$),
protons ($d_{\mathbf{p}, \sigma}$ and $d^{\dag}_{\mathbf{p},
\sigma}$), antiprotons ($f_{\mathbf{p}, \sigma}$ and
$f^{\dag}_{\mathbf{p}, \sigma}$) and photons ($c_{\mathbf{p},
\tau}$ and $c^{\dag}_{\mathbf{p}, \tau}$). So, finally, we obtain
the full set of anticommutation and commutation relations pertinent
to QED

\begin{eqnarray}
 \{a_{\mathbf{p}, \sigma},a^{\dag}_{\mathbf{p}',
\sigma'}\} &=&\{b_{\mathbf{p}, \sigma},b^{\dag}_{\mathbf{p}',
\sigma'}\} = \{d_{\mathbf{p}, \sigma},d^{\dag}_{\mathbf{p}',
\sigma'}\}  = \{f_{\mathbf{p}, \sigma},f^{\dag}_{\mathbf{p}',
\sigma'}\} \nonumber\\
 &=& \delta (\mathbf{p}-\mathbf{p}')
\delta_{\sigma \sigma'}
\label{eq:9.26} \\
 \{a_{\mathbf{p}, \sigma},a_{\mathbf{p}', \sigma'}\}
&=&\{b_{\mathbf{p}, \sigma},b_{\mathbf{p}', \sigma'}\} =
\{d_{\mathbf{p}, \sigma},d_{\mathbf{p}', \sigma'}\} =
\{f_{\mathbf{p},
\sigma},f_{\mathbf{p}', \sigma'}\} \nonumber \\
&=&\{a^{\dag}_{\mathbf{p}, \sigma},a^{\dag}_{\mathbf{p}', \sigma'}\}
=\{b^{\dag}_{\mathbf{p}, \sigma},b^{\dag}_{\mathbf{p}', \sigma'}\} =
\{d^{\dag}_{\mathbf{p}, \sigma},d^{\dag}_{\mathbf{p}', \sigma'}\} \nonumber \\
&=& \{f^{\dag}_{\mathbf{p}, \sigma},f^{\dag}_{\mathbf{p}',
\sigma'}\}
= 0 \label{eq:9.27} \\
\   [c_{\mathbf{p}, \tau},c^{\dag}_{\mathbf{p}', \tau'}]
 &=& \delta
(\mathbf{p}-\mathbf{p}') \delta_{\tau \tau'} \label{eq:9.28} \\
 \ [c^{\dag}_{\mathbf{p}, \tau},c^{\dag}_{\mathbf{p}', \tau'}]
 &=&
 [c_{\mathbf{p}, \tau},c_{\mathbf{p}', \tau'}]
 = 0
\label{eq:9.29}
\end{eqnarray}

\noindent Commutators of operators related to different particles are always
zero.

In the continuous momentum
limit, the analog of the particle counter operator (\ref{eq:9.24})

\begin{eqnarray}
\rho_{\mathbf{p},\tau} =
 c^{\dag}_{\mathbf{p},\tau} c
_{\mathbf{p},\tau}
\label{eq:9.30}
\end{eqnarray}

\noindent can be interpreted as the \emph{density} of photons with helicity $\tau$ at momentum
$\mathbf{p}$. By summing over photon polarizations and integrating
density (\ref{eq:9.30}) over entire momentum space we can define an operator for the total
number of photons in the system

\begin{eqnarray*}
N_{ph}
&=& \sum_{\tau} \int d \mathbf{p} c^{\dag} _{\mathbf{p},
\tau}
c _{\mathbf{p}, \tau}
\end{eqnarray*}

\noindent We can also write down similar operator expressions for
the numbers of other particles. For example

\begin{eqnarray}
N_{el}
&=& \sum_{\sigma} \int d \mathbf{p} a^{\dag} _{\mathbf{p},
\sigma}
a _{\mathbf{p}, \sigma} \label{eq:electron_number}
\end{eqnarray}

\noindent is the electron number operator.  Then we conclude that operator

\begin{eqnarray}
N
  &=& N_{el} + N_{po} + N_{pr} + N_{an} + N_{ph}
\label{eq:9.31}
\end{eqnarray}

\noindent corresponds to the total number of all particles in the
system.

\subsection{Generators of the non-interacting representation}
\label{ss:non-int-rep}

Now we can  fully appreciate the benefits of introducing annihilation
and creation operators. The
expression for
the non-interacting
Hamiltonian $H_0$ can be simply obtained from the
particle number operator (\ref{eq:9.31}) by multiplying the integrands
(particle
densities in the momentum space) by  energies of free particles

\begin{eqnarray}
    H_0 &=& \int d\mathbf{p} \omega_{\mathbf{p}} \sum_{\sigma = \pm 1/2}
          [a^{\dag}_{\mathbf{p},\sigma} a _{\mathbf{p},\sigma}
        + b^{\dag}_{\mathbf{p},\sigma}b_{\mathbf{p},\sigma}] +\int d\mathbf{p} \Omega_{\mathbf{p}} \sum_{\sigma = \pm 1/2}
          [d^{\dag}_{\mathbf{p},\sigma} d _{\mathbf{p},\sigma}
        + f^{\dag}_{\mathbf{p},\sigma} f _{\mathbf{p},\sigma}] \nonumber \\
        &\ & + c \int d\mathbf{p} p \sum_{\tau = \pm 1}
          c^{\dag}_{\mathbf{p},\tau} c _{\mathbf{p},\tau}
\label{eq:9.34}
\end{eqnarray}

\noindent where we denoted $\omega_{\mathbf{p}} = \sqrt{m^2 c^4 +
p^2 c^2}$ the energy of free electrons and positrons,
$\Omega_{\mathbf{p}} = \sqrt{M^2 c^4 + p^2 c^2}$ is the energy of free
protons and antiprotons, and $cp$ is the energy of free photons.  One can easily verify that
$H_0$ in (\ref{eq:9.34}) acts on states in the sector
$\mathcal{H}(1,0,0,0,0)$ just as equation (\ref{eq:9.32}) and it acts on
states in the sector $\mathcal{H}(2,0,0,0,2)$ exactly as equation
(\ref{eq:9.33}). So, we have obtained a single expression which
works equally well in all sectors of the Fock space.  Similar
arguments demonstrate that operator

\begin{eqnarray}
\mathbf{P}_0& =& \int d\mathbf{p}
\mathbf{p} \sum_{\sigma = \pm 1/2}
          [a^{\dag}_{\mathbf{p},\sigma} a _{\mathbf{p},\sigma}
        + b^{\dag}_{\mathbf{p},\sigma}b_{\mathbf{p},\sigma}] + \int d\mathbf{p} \mathbf{p} \sum_{\sigma = \pm 1/2}
          [d^{\dag}_{\mathbf{p},\sigma} d _{\mathbf{p},\sigma}
        + f^{\dag}_{\mathbf{p},\sigma} f _{\mathbf{p},\sigma}] \nonumber \\
        &\ & +\int d\mathbf{p} \mathbf{p} \sum_{\tau = \pm 1}
          c^{\dag}_{\mathbf{p},\tau} c _{\mathbf{p},\tau}
\label{eq:9.35}
\end{eqnarray}

\noindent can be regarded as the total momentum operator in QED.

Expressions for the generators $\mathbf{J}_0$ and $\mathbf{K}_0$ are
more complicated as they involve derivatives of particle operators.
Let us illustrate their derivation on an example of a massive
spinless particle. Consider the action of a space rotation
$e^{-\frac{i}{\hbar}J_{0z} \phi}$ on the one-particle state
$|\mathbf{p} \rangle$\footnote{See equation (\ref{eq:7.17}). }

\begin{eqnarray*}
e^{-\frac{i}{\hbar}J_{0z} \phi } |\mathbf{p} \rangle &=&
|p_x \cos \phi + p_y \sin \phi, p_y \cos \phi - p_x \sin \phi, p_z \rangle
\end{eqnarray*}

\noindent This action can be represented as annihilation of the
state $|\mathbf{p} \rangle = |p_x, p_y, p_z \rangle$ followed by
creation of the state $ |p_x \cos \phi + p_y \sin \phi, p_y \cos
\phi - p_x \sin \phi, p_z \rangle$, i.e., if
$\alpha^{\dag}_{\mathbf{p}}$ and $\alpha_{\mathbf{p}}$ are,
respectively, creation and annihilation operators for the particle,
then

\begin{eqnarray*}
e^{-\frac{i}{\hbar}J_{0z} \phi } |p_x, p_y, p_z \rangle &=&
\alpha^{\dag}_{p_x \cos \phi + p_y \sin \phi, p_y \cos \phi - p_x
\sin \phi, p_z} \alpha_{p_x, p_y, p_z}|p_x, p_y, p_z \rangle \\
&=& \alpha^{\dag}_{R_z(\phi) \mathbf{p}} \alpha_{\mathbf{p}}|p_x,
p_y, p_z \rangle
\end{eqnarray*}

\noindent Therefore, for arbitrary 1-particle state, the operator of
finite rotation takes the form

\begin{eqnarray}
e^{-\frac{i}{\hbar}J_{0z} \phi }  &=&
\int d \mathbf{p} \alpha^{\dag}_{R_z(\phi) \mathbf{p}}
\alpha_{\mathbf{p}}
\label{eq:9.34a}
\end{eqnarray}

\noindent  It is easy to show that the same form is valid everywhere
on the Fock space. An explicit expression for the generator $J_{0z}$
can be obtained now by taking a derivative of  (\ref{eq:9.34a}) with
respect to $\phi$

\begin{eqnarray}
J_{0z} &=& i \hbar \lim_{\phi \to 0} \frac{d}{d \phi}
e^{-\frac{i}{\hbar}J_{0z} \phi } = i \hbar \lim_{\phi \to 0}
\frac{d}{d \phi} \int d \mathbf{p} \alpha^{\dag}_{R_z(\phi)
\mathbf{p}}
\alpha_{\mathbf{p}} \nonumber  \\
&=& i \hbar \int d \mathbf{p} \left(p_y \frac{\partial
\alpha^{\dag}_{\mathbf{p}}}{\partial p_x} - p_x \frac{\partial
\alpha^{\dag}_{\mathbf{p}}}{\partial p_y}\right) \alpha_{\mathbf{p}}
\label{eq:9.36a}
\end{eqnarray}

\noindent The action of a boost along the $z$-axis is obtained from (\ref{eq:7.17})

\begin{eqnarray}
e^{-\frac{ic}{\hbar}K_{0z} \theta } |\mathbf{p} \rangle &=&
\sqrt{\frac{\omega_{\mathbf{p}} \cosh \theta + cp_z \sinh \theta
}{\omega_{\mathbf{p}}}}|p_x , p_y , p_z \cosh \theta +
\omega_{\mathbf{p}} \cosh \theta \rangle \nonumber \\
\label{eq:boosted}
\end{eqnarray}

\noindent This transformation can be represented as annihilation of the state
$|\mathbf{p} \rangle = |p_x, p_y, p_z \rangle$ followed by creation
of the state (\ref{eq:boosted})

\begin{eqnarray*}
e^{-\frac{ic}{\hbar}K_{0z} \theta } |\mathbf{p} \rangle &=&
\sqrt{\frac{\omega_{\mathbf{p}} \cosh \theta + cp_z \sinh \theta
}{\omega_{\mathbf{p}}}} \alpha^{\dag}_{p_x , p_y , p_z \cosh \theta
+ \omega_{\mathbf{p}} \cosh \theta} \alpha_{p_x, p_y, p_z}|p_x, p_y,
p_z \rangle
\end{eqnarray*}

\noindent Therefore, for arbitrary 1-particle state in the Fock space, the
operator of a finite boost takes the form

\begin{eqnarray}
e^{-\frac{ic}{\hbar}K_{0z} \theta }  &=& \int d \mathbf{p}
\sqrt{\frac{\omega_{\Lambda \mathbf{p}}  }{\omega_{\mathbf{p}}}}
\alpha^{\dag}_{\Lambda \mathbf{p}} \alpha_{\mathbf{p}}
\label{eq:9.34b}
\end{eqnarray}

\noindent An explicit expression for the generator $K_{0z}$ can be
now obtained by taking a derivative of (\ref{eq:9.34b}) with respect
to $\theta$

\begin{eqnarray}
K_{0z} &=& \frac{i \hbar}{c} \lim_{\theta \to 0} \frac{d}{d \theta}
e^{-\frac{ic}{\hbar}K_{0z} \theta } \nonumber  \\
&=& \frac{i \hbar}{c} \lim_{\theta \to 0} \frac{d}{d \theta} \int d
\mathbf{p} \sqrt{\frac{\omega_{\mathbf{p}} \cosh \theta + cp_z \sinh
\theta }{\omega_{\mathbf{p}}}} \alpha^{\dag}_{p_x , p_y , p_z \cosh
\theta + c^{-1}\omega_{\mathbf{p}} \sinh \theta}
\alpha_{\mathbf{p}} \nonumber  \\
&=& i \hbar  \int d \mathbf{p} \left(\frac{p_z}{2
\omega_{\mathbf{p}}} \alpha^{\dag}_{\mathbf{p}} \alpha_{\mathbf{p}}
+
        \frac{\omega_{\mathbf{p}}}{c^2} \frac{\partial \alpha^{\dag}_{\mathbf{p}}}{\partial p_z}
         \alpha_{\mathbf{p}} \right)
\label{eq:9.36b}
\end{eqnarray}

\noindent Similar derivations can be done for other components of
$\mathbf{J}_0$ and $\mathbf{K}_0$.  \label{non-int-rep-end}

\subsection{Poincar\'e transformations of particle operators}
\label{ss:transf-laws}

From transformations (\ref{eq:7.17})  of 1-particle state vectors
with respect to  the non-interacting
representation

\begin{eqnarray*}
U_0(\Lambda; \mathbf{r}, t) \equiv e^{-\frac{i}{\hbar}\mathbf{J}_0 \vec{\phi}}
e^{-\frac{ic}{\hbar}\mathbf{K}_0 \vec{\theta}}
e^{-\frac{i}{\hbar}\mathbf{P}_0 \mathbf{r}}
e^{\frac{i}{\hbar}H_0t}
\end{eqnarray*}

\noindent in the Fock space, we can
find corresponding Poincar\'e transformations of creation-annihilation operators.
 For electron creation
operators we obtain\footnote{Here we took into account that the vacuum vector is invariant with respect to $U_0$.}

\begin{eqnarray*}
U_0(\Lambda; \mathbf{r}, t) a^{\dag}_{\mathbf{p}, \sigma}
U_0^{-1}(\Lambda; \mathbf{r}, t) | 0 \rangle &=& U_0(\Lambda;
\mathbf{r}, t) a^{\dag}_{\mathbf{p}, \sigma}  | 0 \rangle =
U_0(\Lambda; \mathbf{r}, t)  | \mathbf{p}, \sigma\rangle \\&=&
\sqrt{\frac{\omega_{\Lambda \mathbf{p}}}{\omega_{\mathbf{p}}}}
e^{-\frac{i}{\hbar}  \mathbf{p} \cdot \mathbf{r}
+\frac{i}{\hbar}\omega_{ \mathbf{p}}t} \sum_{\sigma'}
D^{1/2}_{\sigma' \sigma}
(\vec{\phi}_W(\mathbf{p}, \Lambda)) |\Lambda \mathbf{p}, \sigma' \rangle \\
\end{eqnarray*}

\noindent Therefore\footnote{Here $^*$ and $^{\dag}$ denote complex
conjugation and Hermitian conjugation, respectively. We also use the
property $D^T(\vec{\phi}) = (D^{\dag}(\vec{\phi}))^* =
(D^{-1}(\vec{\phi}))^* = D^*(-\vec{\phi})$ which is valid for the
unitary representation $D(\vec{\phi})$ of the rotation group.}

\begin{eqnarray}
&\mbox{ }& U_0(\Lambda; \mathbf{r}, t) a^{\dag}_{\mathbf{p}, \sigma}
U_0^{-1}(\Lambda; \mathbf{r}, t) = \sqrt{\frac{\omega_{\Lambda
\mathbf{p}}}{\omega_{\mathbf{p}}}} e^{-\frac{i}{\hbar}\mathbf{p}
\cdot \mathbf{r} +\frac{i}{\hbar}\omega_{\mathbf{p}}  t}
\sum_{\sigma'} D^{1/2}_{\sigma' \sigma}(\vec{\phi}_W(\mathbf{p},
\Lambda)) a^{\dag}_{\Lambda\mathbf{p},
\sigma'} \nonumber \\
&=& \sqrt{\frac{\omega_{\Lambda \mathbf{p}}}{\omega_{\mathbf{p}}}}
e^{-\frac{i}{\hbar}\mathbf{p}  \cdot \mathbf{r}
+\frac{i}{\hbar}\omega_{\mathbf{p}}  t} \sum_{\sigma'}
(D^{1/2})^*_{\sigma \sigma'}(-\vec{\phi}_W(\mathbf{p}, \Lambda))
a^{\dag}_{\Lambda\mathbf{p},
\sigma'}
\label{eq:9.36}
\end{eqnarray}

\noindent Similarly, we obtain the transformation law for
annihilation operators

\begin{eqnarray}
 U_0(\Lambda; \mathbf{r}, t) a_{\mathbf{p}, \sigma}
U_0^{-1}(\Lambda; \mathbf{r}, t) = \sqrt{\frac{\omega_{\Lambda \mathbf{p}}}{\omega_{\mathbf{p}}}}
e^{\frac{i}{\hbar}\mathbf{p} \cdot \mathbf{r}
-\frac{i}{\hbar}\omega_{\mathbf{p}}  t} \sum_{\sigma'}
D^{1/2}_{\sigma \sigma'}(-\vec{\phi}_W(\mathbf{p}, \Lambda))
a_{\Lambda\mathbf{p}, \sigma'} \nonumber \\
\label{eq:9.37}
\end{eqnarray}

Transformation laws for photon operators are obtained from equation
(\ref{eq:7.61})

\begin{eqnarray}
U_0 (\Lambda; \mathbf{r}, t) c^{\dag}_{\mathbf{p},\tau} U^{-1}_0
(\Lambda; \mathbf{r}, t) &=& \sqrt{\frac{|\Lambda \mathbf{p}|}{p}}
e^{-\frac{i}{\hbar}(\mathbf{p}  \cdot \mathbf{r}) +\frac{ic}{\hbar}p
t}
 e^{ i \tau \phi_W(\mathbf{p}, \Lambda)} c^{\dag}_{\Lambda
\mathbf{p},\tau} \label{eq:9.38} \\
 U_0 (\Lambda; \mathbf{r}, t)
c_{\mathbf{p},\tau}U^{-1}_0 (\Lambda; \mathbf{r}, t) &=&
\sqrt{\frac{|\Lambda \mathbf{p}|}{p}} e^{\frac{i}{\hbar}(\mathbf{p}
\cdot \mathbf{r}) -\frac{ic}{\hbar}p  t} e^{-i \tau
\phi_W(\mathbf{p}, \Lambda)} c_{\Lambda \mathbf{p},\tau}
\label{eq:9.39}
\end{eqnarray}

\section{Interaction potentials}
\label{sc:potentials}

Our primary goal in the rest of this first part of the book
is to learn how  to calculate the $S$-operator in QED, which is the
quantity most readily comparable with experiment. Equations in subsection \ref{ss:perturbation} tell us that in order
to do that we need to know the non-interacting part $H_0$ and the
interacting part $V$ of the full Hamiltonian

\begin{eqnarray*}
H = H_0 +V
\end{eqnarray*}

\noindent The non-interacting Hamiltonian $H_0$ has been constructed in equation
(\ref{eq:9.34}).
 The interaction energy $V$ (and the corresponding interaction boost
$\mathbf{Z}$) in  QED will be explicitly written only in section
\ref{sc:inter-qed}. Until then, we are going to study rather general
properties of interactions and $S$-operators in the Fock space. We
will try to use some physical principles to narrow down the allowed
form of the operator $V$.

Note that in our approach we assume that interaction does not have any effect on the structure of the Fock space. All properties of this space defined in the non-interacting case remain valid in the presence of interaction: the inner product, the orthogonality of $n$-particle sectors, the existence of particle number operators, etc. In this respect our theory is different from axiomatic or constructive quantum field theories, in which the Hilbert space has a non-Fock structure, which depends on interactions.

\subsection{Conservation laws}
\label{ss:conservation-laws}

From experiment we know that  interaction $V$ between charged particles obeys some
 several important restrictions called
\emph{conservation laws}. \index{conservation law} An observable $F$
is called \emph{conserved} \index{conserved observable} if it
remains unchanged in the course of time evolution

\begin{eqnarray*}
F(t) &\equiv& e^{\frac{i}{\hbar} Ht} F(0) e^{-\frac{i}{\hbar} Ht} =
F(0)
\end{eqnarray*}

\noindent It then follows that conserved observables commute with
the Hamiltonian $[F, H] = [F, H_0 + V] = 0$, which imposes some
restrictions on the interaction operator $V$. For example, the
conservation of the total momentum and the total angular momentum
implies that

\begin{eqnarray}
[V, \mathbf{P}_0]  &=& 0 \label{eq:9.40} \\
\mbox{ } [V, \mathbf{J}_0] &=& 0 \label{eq:9.41}
\end{eqnarray}

\noindent These commutators are automatically satisfied in the
instant form of dynamics (\ref{eq:8.17}) adopted in our study. It is also well-established that all interactions
conserve the \emph{lepton number} \index{lepton number} (the number
of electrons minus the number of positrons, in our case). Therefore,
$H = H_0 + V$ must
 commute with the lepton number operator

\begin{eqnarray}
L &=&  N_{el} - N_{po} = \sum_{\sigma} \int
d\mathbf{p}
( a^{\dag}_{\mathbf{p},\sigma} a
_{\mathbf{p},\sigma} -
b^{\dag}_{\mathbf{p},\sigma}b_{\mathbf{p},\sigma})
\label{eq:9.42}
\end{eqnarray}

\noindent Since $H_0$ already commutes with $L$, we obtain

\begin{eqnarray}
[V,L]  =  0
\label{eq:9.43}
\end{eqnarray}

\noindent Moreover, all known interactions conserve the \emph{baryon
number} \index{baryon number} (=the number of protons minus the
number of antiprotons in our case). So, $V$ must also commute with
the baryon number operator

\begin{eqnarray}
B &=& N_{pr} - N_{an} = \sum_{\sigma} \int d\mathbf{p} (
d^{\dag}_{\mathbf{p},\sigma} d _{\mathbf{p},\sigma} -
f^{\dag}_{\mathbf{p},\sigma}f_{\mathbf{p},\sigma}) \label{eq:9.44}
\\
\mbox{ } [V,B]  &=& 0 \label{eq:9.45}
\end{eqnarray}

\noindent  Taking into account that electrons have charge $-e$,
protons have charge $e$ and antiparticles have charges opposite to
those of particles, we can introduce the \emph{electric charge}
operator \index{electric charge}

\begin{eqnarray}
Q& =& e(B-L) \nonumber \\
&=&  e\sum_{\sigma}\int d\mathbf{p} ( b^{\dag}_{\mathbf{p},\sigma} b
_{\mathbf{p},\sigma} -
a^{\dag}_{\mathbf{p},\sigma}a_{\mathbf{p},\sigma}
+d^{\dag}_{\mathbf{p},\sigma} d _{\mathbf{p},\sigma} -
f^{\dag}_{\mathbf{p},\sigma}f_{\mathbf{p},\sigma})   \label{electric
charge}
\end{eqnarray}

\noindent and obtain the \emph{charge conservation law}
\index{charge conservation law}

\begin{eqnarray}
[H, Q] &=& [V,Q] = e[V, B-L] = 0 \label{eq:9.46}
\end{eqnarray}

\noindent from equations (\ref{eq:9.43}) and (\ref{eq:9.45}).

As we saw above, both operators $H_0$ and $V$ in QED commute with
$\mathbf{P}_0$, $\mathbf{J}_0$, $L$, $B$, and $Q$. Then from subsection
 \ref{ss:perturbation} it follows that scattering operators $F$, $\Sigma$, and $S$
also commute with $\mathbf{P}_0$, $\mathbf{J}_0$, $L$, $B$, and $Q$,
which means that corresponding observables (total momentum, total angular momentum,
lepton number, baryon number, and electric charge, respectively) are conserved in
scattering events. Although, separate numbers of particles of
individual species, i.e., electrons, or protons may not be
conserved, the above
 conservation laws require that charged  particles may be created or
annihilated only together with their antiparticles, i.e., in pairs.
Creation of pairs is suppressed in low energy reactions as such
processes require additional energy of $2m_{el} c^2 = 2 \times 0.51
MeV = 1.02 MeV$ for an electron-positron pair and $2m_{pr}c^2 =
1876.6 MeV$ for an proton-antiproton pair. These high-energy processes
can be safely neglected in
classical electrodynamics. However, even in this low-energy theory one cannot disregard the emission of photons. Since photons have zero mass, the energetic threshold for the photon emission  is zero.  Moreover, photons have zero electric charge, lepton, and baryon numbers, so there
are no conservation laws that would restrict their creation and annihilation. Photons can be created and destroyed in
any quantities. \label{conservation-end}

\subsection{Normal ordering}
\label{ss:normal-ordering}

In the next subsection we are going to express operators in the Fock
space as polynomials in particle creation and annihilation
operators. But first we need to overcome  one notational problem
 related to the
non-commutativity of particle operators: two different polynomials
may, actually, represent the same operator. To have a unique
polynomial representative for each operator, we will agree always to
write products of operators in the \emph{normal order},
\index{normal order} i.e., creation operators to the left from
annihilation operators. Among creation (annihilation) operators we
will enforce a certain order based on particle species: We will
write particle operators in the order proton - antiproton - electron
- positron - photon from left to right. With these rules and with
(anti)commutation relations (\ref{eq:9.26}) - (\ref{eq:9.29}) one can
always convert a product of particle operators to the normally
ordered form. This is illustrated by the following  example

\begin{eqnarray*}
a_{\mathbf{p}', \sigma'} c_{\mathbf{q}', \tau'}
a^{\dag}_{\mathbf{p}, \sigma}c^{\dag}_{\mathbf{q}, \tau}
 &=&
a_{\mathbf{p}', \sigma'} a^{\dag}_{\mathbf{p}, \sigma}
c_{\mathbf{q}', \tau'} c^{\dag}_{\mathbf{q}, \tau}  \\
&=&
 (a^{\dag}_{\mathbf{p}, \sigma} a_{\mathbf{p}', \sigma'}  +
\delta(\mathbf{p} - \mathbf{p}') \delta_{\sigma, \sigma'})
(-c^{\dag}_{\mathbf{q}, \tau} c_{\mathbf{q}', \tau'} +
 \delta(\mathbf{q} - \mathbf{q}') \delta_{\tau, \tau'}) )\\
&=&
-a^{\dag}_{\mathbf{p}, \sigma} c^{\dag}_{\mathbf{q}, \tau} a_{\mathbf{p}',
\sigma'}
 c_{\mathbf{q}', \tau'}
+   a^{\dag}_{\mathbf{p}, \sigma} a_{\mathbf{p}', \sigma'}
   \delta(\mathbf{q} - \mathbf{q}') \delta_{\tau, \tau'} \\
&-&  c^{\dag}_{\mathbf{q}, \tau} c_{\mathbf{q}', \tau'}
  \delta(\mathbf{p} - \mathbf{p}') \delta_{\sigma, \sigma'}
+  \delta(\mathbf{p} - \mathbf{p}') \delta_{\sigma, \sigma'}
   \delta(\mathbf{q} - \mathbf{q}') \delta_{\tau, \tau'}
\end{eqnarray*}

\noindent where the right hand side is in the normal order.

\subsection{General form of interaction operators}
\label{ss:inter-oper}

 A well-known theorem (see \cite{book}, p. 175) states that in the Fock space any
operator $V$  satisfying conservation laws (\ref{eq:9.40}) -
(\ref{eq:9.41})

 \begin{eqnarray}
[V, \mathbf{P}_0] = [V, \mathbf{J}_0]  = 0
\label{eq:9.47}
\end{eqnarray}

\noindent   can be written as a polynomial in particle creation and
annihilation operators\footnote{Here symbols $\alpha^{\dag}$ and $\alpha$ refer
to generic creation and annihilation operators without specifying
the type of the particle. Although this form does not involve
derivatives of particle operators, it still can be used to represent
operators like (\ref{eq:9.36a}) and (\ref{eq:9.36b}) if derivatives are approximated
by finite differences.}

\begin{eqnarray}
 V &=& \sum_{N=0}^{\infty} \sum_{M=0}^{\infty} V_{NM}
\label{eq:9.48} \\
 V_{NM}     &=&
 \sum_{\{\eta, \eta' \}}\int   d\mathbf{q}'_1 \ldots d\mathbf{q}'_N
d\mathbf{q}_1 \ldots d\mathbf{q}_M
   D_{NM}(\mathbf{q}'_1 \eta'_1;  \ldots ;\mathbf{q}'_N \eta'_N;
\mathbf{q}_1 \eta_1; \ldots
;\mathbf{q}_M \eta_M)
  \times \nonumber \\
    &\mbox{ }&  \delta
\left(\sum_{i=1}^{N} \mathbf{q'}_i - \sum_{j=1}^{M} \mathbf{q}_j
\right) \alpha^{\dag}_{\mathbf{q}'_1, \eta'_1} \ldots
\alpha^{\dag}_{\mathbf{q}'_N, \eta'_N} \alpha_{ \mathbf{q}_1,
\eta_1} \ldots \alpha_{\mathbf{q}_M, \eta_M} \label{eq:9.49}
\end{eqnarray}

\noindent where the  summation is carried over all spin/helicity
indices $\eta, \eta'$ of creation and annihilation operators, and the
integration is carried  over all particle momenta.
 Individual terms $V_{NM}$ in the expansion (\ref{eq:9.48}) of the interaction
Hamiltonian will be called \index{potential} \emph{potentials}. Each
potential is a normally ordered product of $N$ creation operators
$\alpha^{\dag}$ and $M$ annihilation operators
$\alpha$. The pair of integers $(N,M)$ will be
referred to as the \emph {index} \index{index of potential} of the
potential $V_{NM}$. A potential is called \emph{bosonic}
\index{bosonic operator} if it has an even number of fermion
particle operators $N_f+ M_f$. Conservation laws (\ref{eq:9.43}), (\ref{eq:9.45}) and
(\ref{eq:9.46})

 \begin{eqnarray}
 [V,L] = [V,B] = [V,Q] = 0
\label{eq:9.47a}
\end{eqnarray}

\noindent imply that all potentials in QED must be bosonic.

 $D_{NM}$ is a numerical
\emph{coefficient function} \index{coefficient function} which
depends on momenta and spin projections (or helicities) of all
created and annihilated particles.  In order to satisfy $[V,
\mathbf{J}_0] = 0$, the function $D_{NM}$ must be rotationally
invariant.    Translational invariance of
(\ref{eq:9.49}) is guaranteed by the momentum delta function

\begin{eqnarray*}
\delta \left(\sum_{i=1}^{N} \mathbf{q'}_i - \sum_{j=1}^{M}
\mathbf{q}_j \right)
\end{eqnarray*}

\noindent which expresses the conservation of momentum: the sum of
momenta of annihilated particles is equal to the sum of momenta of
created particles.

The interaction Hamiltonian $V$ enters in formulas  (\ref{eq:8.70}), (\ref{eq:F-D}), and (\ref{eq:8.72}) for the $S$-operator in the $t$-dependent form

 \begin{eqnarray}
V(t) = e^{\frac{i}{\hbar}H_0t} V e^{-\frac{i}{\hbar}H_0t}
\label{eq:9.50}
\end{eqnarray}

\noindent  Operators with $t$-dependence determined by the free
Hamiltonian $H_0$, as in equation (\ref{eq:9.50}), and satisfying
conservation laws (\ref{eq:9.47}), (\ref{eq:9.47a})
 will be called
\emph{regular}. \index{regular operator} Such operators will play an
important role in our calculations of the $S$-operator below. In
what follows, when we write a regular operator $V$ without its
$t$-argument, this means that either this operator is
$t$-independent, i.e., it commutes with $H_0$, or that we take its
value at $t=0$.

One final notational remark. If potential $V_{NM}$ has coefficient
function $D_{NM}$, we introduce  notation \index{$\circ$} $V_{NM}
\circ \zeta$ for the operator whose coefficient function $D_{NM}'$
is a product of $D_{NM}$ and a numerical function $\zeta$ of the same
arguments

 \begin{eqnarray*}
&\mbox { }& D_{NM}'(\mathbf{q}'_1 \eta'_1;  \ldots ;\mathbf{q}'_N \eta'_N;
\mathbf{q}_1 \eta_1; \ldots
;\mathbf{q}_M \eta_M) \\
&=& D_{NM}(\mathbf{q}'_1 \eta'_1;  \ldots ;\mathbf{q}'_N \eta'_N;
\mathbf{q}_1 \eta_1; \ldots
;\mathbf{q}_M \eta_M) \zeta(\mathbf{q}'_1 \eta'_1;  \ldots ;\mathbf{q}'_N
\eta'_N;
\mathbf{q}_1 \eta_1; \ldots
;\mathbf{q}_M \eta_M)
\end{eqnarray*}

\noindent  Then, substituting (\ref{eq:9.49}) in (\ref{eq:9.50}) and
using (\ref{eq:9.36}) - (\ref{eq:9.39}), the $t$-dependent form of
any regular potential $V_{NM}(t)$ can be written as

\begin{eqnarray*}
V_{NM}(t) &=& e^{\frac{i}{\hbar}H_0t} V_{NM}
e^{-\frac{i}{\hbar}H_0t} = V_{NM} \circ e^{\frac{i}{\hbar}E_{NM}t}
\end{eqnarray*}

\noindent  where

\begin{eqnarray}
 E_{NM}(\mathbf{q}'_1, \ldots, \mathbf{q}'_N, \mathbf{q}_1,  \ldots,
\mathbf{q}_M)
 \equiv \sum_{i=1}^N \omega_{\mathbf{q}'_i}
 - \sum_{j=1}^M \omega_{\mathbf{q}_j}
\label{eq:9.51}
\end{eqnarray}

\noindent is the difference of energies of particles created and
destroyed by $V_{NM}$, which is called the \emph{energy function}
\index{energy function} of the term $V_{NM}$. We can also extend
this notation to a general sum of potentials $V_{NM}$

\begin{eqnarray*}
V(t) &=& e^{\frac{i}{\hbar}H_0t} V e^{-\frac{i}{\hbar}H_0t} = V
\circ e^{\frac{i}{\hbar}E_Vt}
\end{eqnarray*}

\noindent which means that for each potential $V_{NM}(t)$ in the sum
$V(t)$, the
argument of the $t$-exponent contains the
corresponding energy function $E_{NM}$.
In this notation we can conveniently write

\begin{eqnarray}
\frac{d}{dt}V(t) &=& V(t) \circ \left( \frac{i}{\hbar} E_{V}\right) \nonumber \\
 \underbrace{V(t)} &\equiv& -\frac{i}{\hbar}\int _{-\infty}^{\infty} V(t) dt
=   -2 \pi i  V \circ \delta(E_{V}) \label{eq:9.52}
\end{eqnarray}

\noindent Equation (\ref{eq:9.52}) means that each term in $ \underbrace{V(t)}$
is non-zero only on the hypersurface
 of solutions of the equation

\begin{eqnarray}
 E_{NM}(\mathbf{q}'_1, \ldots, \mathbf{q}'_N, \mathbf{q}_1,  \ldots,
\mathbf{q}_M) =  0 \label{eq:en-shell}
\end{eqnarray}

\noindent (if such solutions exist). This hypersurface in the
$3(N+M)$ dimensional momentum space is called the \emph{energy shell} \index{energy
shell} of the potential $V$. We will also say that $
\underbrace{V(t)}$ in equation (\ref{eq:9.52}) is zero outside the energy
shell of $V$. Note that the scattering operator (\ref{eq:8.69}) $S =
1 + \underbrace{\Sigma(t)}$ is different from 1 only on the energy
shell, i.e., where the energy conservation condition
(\ref{eq:en-shell}) is satisfied.

\subsection{Five types of regular potentials}
\label{ss:five-types}

\begin{figure}
\centering
\includegraphics{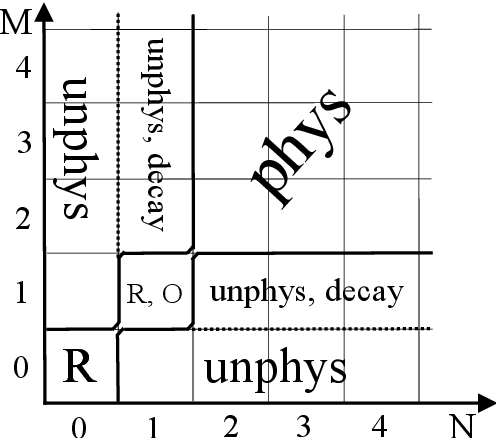} \caption{Positions of different operator
types in the ``index space'' $(N,M)$. $N$ and $M$ are numbers of creation and annihilation operators, respectively. R = ``renorm,'' O =
``oscillation.''} \label{fig:7.6}
\end{figure}

Here we would like to introduce a classification of regular
potentials (\ref{eq:9.49})  by dividing them into five groups
depending on their index $(N,M)$. We will call these types of
operators \emph{renorm}, \emph{oscillation}, \emph{decay},
\emph{phys}, and \emph{unphys}.\footnote{The correlation between
potential's index $(N,M)$ and its type is shown in Fig.
\ref{fig:7.6}. There is no established terminology for the types of
potentials. In the literature, our \emph{phys} operators are sometimes
called \emph{good}; \emph{unphys} operators may be called \emph{bad} or
\emph{virtual}.} The rationale for introducing this classification
and nomenclature will become clear in chapters
\ref{ch:renormalization} and \ref{ch:rqd} where we will examine
renormalization and the ``dressed particle'' approach in quantum
field theory.

\bigskip

\textbf{Renorm potentials} \index{renorm potential} have either
index (0,0) (such operator is simply a numerical constant) or index
(1,1) in which case both created and annihilated particles are
required to have the same mass.\footnote{Actually, in most cases the same particle type is created and annihilated, so renorm operators have the form $\alpha^{\dag}\alpha$.}  In QED the most general form of a \emph{renorm}
potential obeying conservation laws is the sum of a numerical
constant $C$ and (1,1) terms corresponding to each particle
type.\footnote{Here we write just the operator structure of $R$
omitting all numerical factors, indices, integration and summation
signs. Note also that terms like $a^{\dag} b$ or $d^{\dag} f $ are
forbidden by the charge conservation law (\ref{eq:9.46}).}

\begin{eqnarray}
R &=&
a^{\dag} a + b^{\dag} b  +  d^{\dag} d + f^{\dag} f  + c^{\dag} c +C
\label{eq:9.53}
\end{eqnarray}

\noindent  \emph{Renorm}  potentials are characterized
by the property that the energy function (\ref{eq:9.51}) is
identically zero.
 So, \emph{renorm}  potentials always have energy shell where they do
not vanish. \emph{Renorm}  potentials commute with $H_0$, therefore regular
\emph{renorm}  operators\footnote{whose $t$-dependence is determined by
(\ref{eq:9.50})} do not depend on $t$.  The free Hamiltonian (\ref{eq:9.34}) and  the
total momentum (\ref{eq:9.35}) are examples of  \emph{renorm}  operators,
i.e., sums of \emph{renorm}  potentials.

\bigskip

\textbf{Oscillation  potentials} \index{oscillation potential} have
index $(1,1)$. In contrast to \emph{renorm}  potentials with index (1,1),
\emph{oscillation} potentials destroy and create different particle species
having different masses. For this reason, the energy function
(\ref{eq:9.51}) of an oscillation potential
 never turns to zero, so there is no energy shell.
In QED there can be no oscillation potentials, because they would
violate either lepton number or baryon number \index{baryon number}
conservation law. However, there are particles in nature, such as
kaons and neutrinos, for which oscillation interactions play a
significant role. These interactions are responsible for
time-dependent oscillations between different particle species
\cite{oscillations}.  \index{neutrino oscillations}

\bigskip

\textbf{Decay  potentials} \index{decay potential} satisfy two
conditions:

\begin{enumerate}
\item they  have indices $(1,N)$ or
 $(N,1)$ with $N \geq 2$;
\item    they  have a
non-empty energy shell on which the coefficient function does not vanish;
\end{enumerate}

\noindent These potentials describe decay processes
$1 \to N$\footnote{as well as
reverse processes $N \to 1$} in which one particle decays into $N$ decay products, so that
the energy conservation condition is satisfied. There are no decay terms in the QED Hamiltonian and in the
corresponding $S$-matrix: decays of electrons, protons, or photons
would violate conservation laws.\footnote{Exceptions to this rule
are given by operators describing the decay of a photon into odd
number of photons, e.g., $c^{\dag}_{\mathbf{k}_1,\tau_1}c^{\dag}_{\mathbf{k}_2,\tau_2}c^{\dag}_{\mathbf{k}_3,\tau_3}c_
{\mathbf{k}_1 + \mathbf{k}_2 + \mathbf{k}_3,\tau_4}$.
This potential obeys all conservation laws if momenta of
involved photons  are collinear and $k_1 + k_2 + k_3 - |\mathbf{k}_1
+ \mathbf{k}_2 + \mathbf{k}_3| = 0$.
 However, it was shown in \cite{photon_decay} that
such contributions to the $S$-operator
 are zero on the energy shell, so photon decays are forbidden in QED.} Nevertheless,
particle decays play important roles in other areas of high energy
physics, and they  will be considered in chapter \ref{ch:decays}.

\bigskip

\textbf{Phys  potentials} \index{phys operator} have at least two
creation operators and at least two destruction operators (index
$(N,M)$ with $N \geq 2 $ and $ M \geq 2$). For \emph{phys} potentials the
energy shell always exists. For example, in the case of a \emph{phys}
potential $ d^{\dag}_{\mathbf{p} + \mathbf{k}, \rho}
f^{\dag}_{\mathbf{q} - \mathbf{k}, \sigma} a_{\mathbf{p}, \tau}
b_{\mathbf{q}, \eta} $
 the energy shell is determined by the solution of equation
$ \Omega_{\mathbf{p} + \mathbf{k}} + \Omega_{\mathbf{q} -
\mathbf{k}} = \omega_{\mathbf{p}} + \omega_{\mathbf{q}} $ which is
not empty.

\begin{table}[h]
\caption{Types of potentials in the Fock space.}
\begin{tabular*}{\textwidth}{@{\extracolsep{\fill}}cccc}
\hline Potential            & Index of potential & Energy shell &
Examples \cr
 & $(N,M)$      & exists?   &                  \cr
\hline
Renorm   & $(0,0)$,$(1,1)$ & yes    &  $a^{\dag}_{\mathbf{p}}
a_{\mathbf{p}}$ \cr
Oscillation   & $(1,1)$    & no &  forbidden in QED \cr
Unphys   &  $(0,N \geq 1)$,$(N \geq 1,0)$ & no  &
$a^{\dag}_{\mathbf{p}} b^{\dag}_{-\mathbf{p-k}} c^{\dag}_{\mathbf{k}} $ \cr
Unphys   & $(1,N \geq 2)$,$(N \geq 2,1)$ & no   &
$a^{\dag}_{\mathbf{p}} a_{\mathbf{p-k}} c_{\mathbf{k}}$ \cr
Decay   &  $(1,N \geq 2)$,$(N \geq 2,1)$ &yes &
 forbidden in QED \cr
Phys   &  $(N \geq 2 ,M \geq 2)$ & yes  &
$d^{\dag}_{\mathbf{q+k}} a^{\dag}_{\mathbf{p-k}} d_{\mathbf{q}} a_{\mathbf{p}}
$ \cr
\hline
\end{tabular*}
\label{table:7.1}
\end{table}

All regular operators not mentioned above belong to the class of

\noindent \textbf{Unphys potentials.} \index{unphys operator} They come in two subclasses with following indices

\begin{enumerate}
\item   $(0,N)$, or $(N,0)$, where $N \geq 1$.
Obviously, there is no energy shell
in this case.
\item  $(1,M)$ or
$(M,1)$, where $M \geq 2$. This is the same condition as for decay
potentials, however, in contrast to decay potentials, for \emph{unphys}
potentials it is required that either the energy shell does not exist
or the coefficient function vanishes on the energy shell.
\end{enumerate}

\noindent An example of an \emph{unphys} potential satisfying condition 2. is

\begin{equation}
 a^{\dag}_{\mathbf{p-k}, \sigma}  c^{\dag}_{\mathbf{k},\tau} a_{\mathbf{p}, \rho} \label{eq:unphys}
\end{equation}

\noindent Its energy shell equation is $\omega_{ \mathbf{p-k}} + ck =  \omega_{ \mathbf{p}} $ whose only solution is $\mathbf{k}=\mathbf{0}$. However zero
 vector  is excluded from
the photon momentum spectrum,\footnote{see subsection \ref{ss:spectra}}
 so the energy shell  of the potential (\ref{eq:unphys}) is empty. This means
 that a free electron cannot decay into the pair electron+photon without
 violating energy-momentum conservation laws.

Properties of potentials discussed above are summarized in Table
\ref{table:7.1}.
  These five types of potentials  exhaust all possibilities, therefore
 any regular operator $V$ must have a unique decomposition

\begin{eqnarray*}
 V =  V^{ren} + V^{unp} + V^{dec} + V^{ph} + V^{osc}
\end{eqnarray*}

\noindent As mentioned above, in QED interaction,  oscillation and
decay contributions are absent. So, everywhere in this
book\footnote{except chapter
\ref{ch:decays} where we discuss decays} we will assume that
the most general interaction operator is a  sum of \emph{renorm},  \emph{unphys}, and \emph{phys} potentials

\begin{eqnarray*}
 V =  V^{ren} + V^{unp} + V^{ph}
\end{eqnarray*}

Now we need to learn how to perform various operations with these
three classes of potentials, i.e., how to evaluate products, commutators, and
$t$-integrals required for calculations of scattering operators in
(\ref{eq:8.69})  - (\ref{eq:8.72}).

\subsection{Products and commutators of potentials}
\label{ss:products}

\begin{lemma} \label{Lemma9.3} The
 product of two (or any number of) regular operators is regular.
\end{lemma}
\begin{proof}
   If operators $A(t)$ and $B(t)$ are regular, then

\begin{eqnarray*}
A(t) &=& e^{\frac{i}{\hbar}H_0t} A e^{-\frac{i}{\hbar}H_0t} \\
B(t) &=& e^{\frac{i}{\hbar}H_0t} B e^{-\frac{i}{\hbar}H_0t}
\end{eqnarray*}

\noindent and their product $C(t) = A(t)B(t)$  has $t$-dependence

 \begin{eqnarray*}
C(t) &=& e^{\frac{i}{\hbar}H_0t} A e^{-\frac{i}{\hbar}H_0t}
 e^{\frac{i}{\hbar}H_0t} B e^{-\frac{i}{\hbar}H_0t} \\
&=& e^{\frac{i}{\hbar}H_0t} A  B e^{-\frac{i}{\hbar}H_0t}
\end{eqnarray*}

\noindent characteristic for regular operators. The conservation
laws (\ref{eq:9.47}), (\ref{eq:9.47a}) are valid for the product $AB$ if they are
valid for $A$ and $B$ separately. Therefore $C(t)$ is regular.
\end{proof}

\bigskip

\begin{lemma} \label{Lemma9.4} A Hermitian operator $A$ is \emph{phys} if and only
if it yields zero when acting on the vacuum $|0 \rangle$ and
one-particle states $|1 \rangle \equiv \alpha^{\dag}|0 \rangle$\footnote{Here $\alpha$ denotes
any one of the five particle operators ($a, b, d, f, c$) relevant for QED. Momentum and spin labels are omitted for simplicity.}

\begin{eqnarray}
 A |0 \rangle &=& 0
\label{eq:9.62}\\
A |1 \rangle &=& A \alpha^{\dag}|0 \rangle = 0
\label{eq:9.63}
\end{eqnarray}
\end{lemma}
\begin{proof}    Normally ordered \emph{phys} operators have two annihilation
operators on the right, so equations (\ref{eq:9.62}) and (\ref{eq:9.63})
are satisfied. Let us now prove the inverse statement. \emph{Renorm}
operators cannot satisfy (\ref{eq:9.62}) and (\ref{eq:9.63}) because
they conserve the number of particles. \emph{Unphys} operators $(1,N)$ can
satisfy equations (\ref{eq:9.62}) and (\ref{eq:9.63}), e.g.,

\begin{eqnarray*}
 \alpha_1^{\dag} \alpha_2 \alpha_3 |0 \rangle &=& 0 \\
\alpha_1^{\dag} \alpha_2 \alpha_3 |1 \rangle &=& 0.
\end{eqnarray*}

\noindent However, for Hermiticity, such operators should be always
present in pairs with $(N,1)$ operators $\alpha_2^{\dag}
\alpha_3^{\dag} \alpha_1$. Then, there exists at least one
one-particle state $|1 \rangle$ for which equation (\ref{eq:9.63}) is not
valid, e.g.,

\begin{eqnarray*}
\alpha_3^{\dag} \alpha_2^{\dag} \alpha_1   |1 \rangle = \alpha_3^{\dag} \alpha_2^{\dag}    |0 \rangle
& \neq& 0
\end{eqnarray*}

\noindent A similar argument is valid for \emph{unphys} operators having index
$(0,N)$.
Therefore, the only remaining possibility for $A$ is to be \emph{phys}.
\end{proof}
\bigskip

\begin{lemma} \label{Lemma9.5} Product and commutator of any two \emph{phys} operators
$A$ and $B$ is \emph{phys}.
\end{lemma}
\begin{proof}    By Lemma \ref{Lemma9.4} if $A$ and $B$ are
\emph{phys}, then

\begin{eqnarray*}
A |0 \rangle = B |0 \rangle = A |1 \rangle = B |1 \rangle = 0.
\end{eqnarray*}

\noindent Then the same conditions are true for the Hermitian
combinations $i(AB-BA)$ and $AB+BA$. Therefore, the commutator
$[A,B]$ and anticommutator \index{anticommutator} $AB +BA$ are \emph{phys} and

\begin{eqnarray*}
AB = \frac{1}{2} (AB+BA ) + \frac{1}{2}[A,B]
\end{eqnarray*}

\noindent is \emph{phys} as well.
\end{proof}
\bigskip

\begin{lemma} \label{Lemma9.6} If $R$ is a \emph{renorm}  operator  and $[A,R] \neq
0$, then operator $[A,R]$ has the same type (i.e., \emph{renorm},  \emph{phys}, or
\emph{unphys}) as $A$.
\end{lemma}
\begin{proof}    The general form of a \emph{renorm}
operator is given in equation (\ref{eq:9.53}). Let us consider just one
term in that sum

\begin{eqnarray*}
R = \int d \mathbf{p} f(\mathbf{p}) \alpha^{\dag}_{\mathbf{p}}
\alpha_{\mathbf{p}}
\end{eqnarray*}

\noindent We calculate the commutator $[A,R] = AR - RA$ by
moving  the factor $R$ in the
 term $AR$ step-by-step to the leftmost position. If the
product $\alpha^{\dag}_{\mathbf{p}} \alpha_{\mathbf{p}}$ (from $R$)
changes places with a particle operator (from $A$) different from
$\alpha^{\dag}$ or $ \alpha$ then nothing happens. If the product
$\alpha^{\dag} \alpha$ changes places with a creation operator
$\alpha^{\dag}_{\mathbf{q}} $ (from $A$)  then, as discussed in
subsection \ref{ss:normal-ordering}, a secondary term should be
added which, instead of $\alpha^{\dag}_{\mathbf{q}} $ contains the
commutator\footnote{The upper sign is for bosons and the lower sign
is for fermions.}

 \begin{eqnarray*}
&\mbox{ }& \alpha^{\dag}_{\mathbf{q}} \left(\int d\mathbf{p}
f(\mathbf{p}) \alpha^{\dag}_{\mathbf{p}} \alpha_{\mathbf{p}}\right)
-
 \left(\int d\mathbf{p} f(\mathbf{p})  \alpha^{\dag}_{\mathbf{p}}
\alpha_{\mathbf{p}}\right) \alpha^{\dag}_{\mathbf{q}} \\
&=& \pm \int
d\mathbf{p} f(\mathbf{p})   \alpha^{\dag}_{\mathbf{p}}
\alpha^{\dag}_{\mathbf{q}} \alpha_{\mathbf{p}}  -
 \int d\mathbf{p} f(\mathbf{p})  \alpha^{\dag}_{\mathbf{p}}
\alpha_{\mathbf{p}} \alpha^{\dag}_{\mathbf{q}} \\
&=&   \int d\mathbf{p} f(\mathbf{p})
\alpha^{\dag}_{\mathbf{p}}\alpha_{\mathbf{p}}  \alpha^{\dag}_{\mathbf{q}}
\pm \int d\mathbf{p} f(\mathbf{p}) \alpha^{\dag}_{\mathbf{p}} \delta(\mathbf{p} -
\mathbf{q})
 -  \int d\mathbf{p} f(\mathbf{p})  \alpha^{\dag}_{\mathbf{p}}
\alpha_{\mathbf{p}} \alpha^{\dag}_{\mathbf{q}} \\
&=&
\pm  f(\mathbf{q})
\alpha^{\dag}_{\mathbf{q}}
\end{eqnarray*}

\noindent This commutator is proportional to
$\alpha^{\dag}_{\mathbf{q}} $, so the secondary term has the same
operator structure as the primary term and it is already in the
normal order, so no tertiary terms need to be created. If the
product $\alpha^{\dag} \alpha$ changes places with an annihilation
operator $\alpha_{\mathbf{q}} $ then the commutator $\mp
f(\mathbf{q})  \alpha_{\mathbf{q}} $ is
proportional to the annihilation operator. If there are many
$\alpha^{\dag}$ and $\alpha$ operators in $A$ having non-vanishing
commutators with $R$, then each one of them results in one
additional term whose type remains the same as in the original
operator $A$.
\end{proof}
\bigskip

\begin{lemma}
A commutator $[P,U]$ of a Hermitian \emph{phys} operator $P$ and a
Hermitian \emph{unphys} operator $U$
 can be either \emph{phys} or \emph{unphys}, but not \emph{renorm}.
 \end{lemma}
\begin{proof}
 Acting by $[P,U] $ on a one-particle state $| 1 \rangle $, we
obtain

\begin{eqnarray*}
 [P,U] | 1 \rangle &=& (PU-UP) | 1 \rangle =
 PU | 1 \rangle
\end{eqnarray*}

\noindent If $U$ is Hermitian then the state $U | 1 \rangle$ has at
least two particles (see proof of Lemma \ref{Lemma9.4}) and the
same is true for the state $PU | 1 \rangle$. Therefore, $[P,U]$
creates several particles when acting on a one-particle state, which
is impossible if  $[P,U]$ were \emph{renorm}.
\end{proof}

Finally, there are no limitations on the  type of commutator of two
\emph{unphys} operators $[U,U']$. It can be a superposition of \emph{phys},
\emph{unphys}, and \emph{renorm}  terms.

The above results are summarized in Table \ref{table:7.2}.

\begin{table}[h]
\caption{Operations with regular operators in the Fock space.
(Notation: P=phys, U=unphys,  R=renorm, NR=non-regular.)}
\begin{tabular*}{\textwidth}{@{\extracolsep{\fill}}ccccccc}
\hline
 Type of operator            &  & &  & & & \cr
 $A$  & $[A,P] $  &  $[A,U]$    &   $[A,R]$ &   $\frac{dA}{dt}$ &
$\underline{A}$ & $\underbrace{A}$ \cr
                  &       &         &        &          &         &  \cr
\hline
 P                & P     & P+U   &  P    & P     &    P    & P \cr
 U                & P+U & P+U+R &  U  &  U        &
U    & 0 \cr
 R                & P     &  U      &  R   &  0        &    NR   &
$\infty$ \cr \hline
\end{tabular*}
\label{table:7.2}
\end{table}

\subsection{More about $t$-integrals}
\label{ss:adiabatic}

\bigskip

\begin{lemma} \label{Lemma9.1} A $t$-derivative of a regular operator
$A(t)$ is regular and has zero \emph{renorm}  part.
\end{lemma}
\begin{proof}
    The derivative of a regular operator has $t$-dependence that is
characteristic for regular operators:

\begin{eqnarray}
\frac{d}{dt}A(t) &=& \frac{d}{dt} e^{\frac{i}{\hbar}H_0t} A
e^{-\frac{i}{\hbar}H_0t} = \frac{i}{\hbar} e^{\frac{i}{\hbar}H_0t} [H_0,A]
e^{-\frac{i}{\hbar}H_0t} \nonumber \\
&=& \frac{i}{\hbar}  [H_0,A(t)]  \label{eq:da/dt}
\end{eqnarray}

\noindent In addition, it is easy to check that the derivative obeys
all conservation laws (\ref{eq:9.47}), (\ref{eq:9.47a}). Therefore, it is regular.

 Suppose that $\frac{d}{dt}A(t) $
has a non-zero \emph{renorm}  part $R$. Then $R$ is $t$-independent and
originates from a derivative of the term $Rt + S$ in $A(t)$, where
$S$ is $t$-independent. Since $A(t)$ is regular, its \emph{renorm}  part
must be $t$-independent, therefore $R=0$.
\end{proof}
\bigskip

 From formula (\ref{eq:9.56}) we conclude that $t$-integrals of regular \emph{phys} and \emph{unphys}
operators are regular\footnote{Here we assume the adiabatic switching of interaction
(\ref{eq:adiaba}) and use integral notation (\ref{eq:underline}) - (\ref{eq:underbrace}).}

\begin{eqnarray}
\underline{V(t)} &=& V(t) \circ \frac{-1}{E_{V}} \label{eq:underline2}
\end{eqnarray}

\noindent However, this property does not hold for $t$-integrals of \emph{renorm}
operators. These operators are $t$-independent, therefore

\begin{eqnarray}
\underline{V^{ren}} &=& \lim _{\epsilon \to +0} V^{ren} \circ  \frac
{i e^{-\epsilon t}} {\epsilon \hbar}  = \lim _{\epsilon \to +0}
V^{ren}  \circ \frac{i}{\epsilon \hbar} - V^{ren} \circ
\frac{it}{\hbar} + \ldots
\label{eq:9.57}\\
\underbrace{V^{ren}}  &=& \infty \label{eq:9.58}
 \end{eqnarray}

\noindent Thus, \emph{renorm}  operators are different from others in the
sense that their $t$-integrals (\ref{eq:9.57}) are infinite and
non-regular. Infinite-limit $t$-integrals (\ref{eq:9.58}) are infinite, unless $V^{ren}=0$.\footnote{
This fact does not limit the applicability of our theory, because,
as we will see in subsection \ref{ss:mass-renorm},  properly
renormalized expressions for scattering operators  should not contain \emph{renorm}  terms and pathological expressions like (\ref{eq:9.57}) - (\ref{eq:9.58}).}

Since for any \emph{unphys} operator $V^{unp}$ either the energy shell does
not exist or the coefficient function is zero on the energy shell,
we conclude from equation (\ref{eq:9.52}) that

\begin{eqnarray}
\underbrace{V^{unp}} = 0 \label{eq:9.61}
\end{eqnarray}

\noindent From equations (\ref{eq:8.70}) and (\ref{eq:8.72}) it is then
clear that \emph{unphys} terms in $F$ and $\Sigma$ do not make
contributions to the $S$-operator. \label{ss:adiabatic-end} Results obtained so far in this subsection are presented in last three columns of Table \ref{table:7.2}.

\subsection{Solution of one commutator equation}
\label{ss:solution}

In section \ref{sc:dressed} we will find it necessary to solve equation of the type

\begin{eqnarray}
i[H_0, A] = V \label{eq:H0AV}
\end{eqnarray}

\noindent where $H_0$ is the free Hamiltonian, $V$ is a given regular Hermitian operator in the Fock space and $A$ is the desired solution (an unknown Hermitian operator). What can we say about the solution of this commutator equation? Let us first multiply both sides of (\ref{eq:H0AV}) by the usual $t$-exponents $ e^{\frac{i}{\hbar}H_0t} \ldots
e^{-\frac{i}{\hbar}H_0t}$

\begin{eqnarray*}
i[H_0, A(t)] = V(t) \label{eq:H0AV2}
\end{eqnarray*}

\noindent Using (\ref{eq:da/dt}) this can be rewritten as

\begin{eqnarray}
\hbar \frac{d}{dt} A(t) = V(t) \label{eq:H0AV3}
\end{eqnarray}

Note that it follows from Lemma \ref{Lemma9.1} that the operator on the right hand side of (\ref{eq:H0AV}) cannot contain \emph{renorm}  terms. Indeed, according to (\ref{eq:H0AV3}), the left hand side of (\ref{eq:H0AV}) is a $t$-derivative, which cannot be \emph{renorm}.  Luckily, for our purposes in this book we will never meet equations of the above type with \emph{renorm}  right hand sides. Therefore, here we will assume  $V^{ren} = 0$.

Next we assume that the usual ``adiabatic switching'' (\ref{eq:adiaba}) works, so that $V(-\infty) = 0$ and the same property is valid for the solution $A(t)$

\begin{eqnarray}
 A(-\infty) = 0 \label{eq:H0AV4}
\end{eqnarray}

\noindent Then equation (\ref{eq:H0AV3}) with the initial condition (\ref{eq:H0AV4}) has a simple solution

\begin{eqnarray}
 A(t) = \frac{1}{\hbar} \int \limits_{-\infty}^{t} V(t') dt' = i  \underline{V(t)}
\end{eqnarray}

\noindent In order to get a $t$-independent solution of our original equation (\ref{eq:H0AV}), we can simply set $t=0$ and obtain

\begin{eqnarray}
 A \equiv A(0) = i  \underline{V} \equiv -i V \circ \frac{1}{E_V} \label{eq:soluti}
\end{eqnarray}

\subsection{Two-particle potentials}
\label{ss:2-particle}

Our next goal is to express $n$-particle potentials ($n \geq 2$)\footnote{studied in subsection \ref{ss:separability}} using the formalism of
annihilation and creation operators. These potentials conserve the
number and types of particles, so they must have equal numbers of
creation and annihilation operators  $(N = M, N \geq 2, M \geq 2)$.
Therefore, they must be of the \emph{phys} type.

As an example, let us consider a two-particle subspace $\mathcal{H}(1,0,1,0,0)$ of the
Fock space. This subspace describes states of the system consisting
of one electron and one proton. A general \emph{phys} operator leaving this
subspace invariant must have $N=2$, $M=2$ and, according to equation
(\ref{eq:9.49}), it can be written as\footnote{In this subsection we
use variables $\mathbf{p}$ and $\mathbf{q}$
 to denote momenta of the proton and
the electron, respectively. We also omit spin indices for brevity.}

\begin{eqnarray}
\hat{V}&=& \int d \mathbf{p} d \mathbf{q} d \mathbf{p}' d \mathbf{q}'
D_{22}(\mathbf{p},\mathbf{q},\mathbf{p}', \mathbf{q}'  )
\delta(\mathbf{p}+\mathbf{q}-\mathbf{p}'- \mathbf{q}')
 d^{\dag}_{\mathbf{p} }
a^{\dag}_{\mathbf{q} } d_{\mathbf{p}' }a_{\mathbf{q}' } \nonumber\\
&=&
 \int d \mathbf{p} d \mathbf{q} d \mathbf{p}'
D_{22}(\mathbf{p},\mathbf{q},\mathbf{p}', \mathbf{p+q-p}'  )
 d^{\dag}_{\mathbf{p} }
a^{\dag}_{\mathbf{q} } d_{\mathbf{p}' }a_{\mathbf{p+q-p}' }\nonumber \\
&=& \int d \mathbf{p} d \mathbf{q} d \mathbf{k} V
(\mathbf{p},\mathbf{q},\mathbf{k} )
 d^{\dag}_{\mathbf{p} }
a^{\dag}_{\mathbf{q} } d_{\mathbf{p-k} }a_{\mathbf{q+k} }
\label{eq:9.64}
\end{eqnarray}

\noindent where we denoted $\mathbf{k} = \mathbf{p-p'}$ the
``transferred momentum'' and

\begin{eqnarray*}
V (\mathbf{p},\mathbf{q},\mathbf{k} ) \equiv
D_{22}(\mathbf{p},\mathbf{q},\mathbf{p-k}, \mathbf{q+k}  )
\end{eqnarray*}

\noindent Acting by this operator on  an arbitrary state $|\Psi
\rangle$ of the two-particle system

\begin{eqnarray}
  |\Psi \rangle =  \int d\mathbf{p}'' d\mathbf{q}''
\Psi(\mathbf{p}'',  \mathbf{q}'')
  d^{\dag}_{\mathbf{p}''}a^{\dag}_{\mathbf{q}'' } |0 \rangle
\label{eq:9.65}
\end{eqnarray}

\noindent  we obtain

\begin{eqnarray}
 &\mbox{ }& \hat{V} |\Psi\rangle \nonumber\\ &=&
\int d \mathbf{p} d \mathbf{q} d \mathbf{k}
V(\mathbf{p},\mathbf{q},\mathbf{k} )
 d^{\dag}_{\mathbf{p} }
a^{\dag}_{\mathbf{q} } d_{\mathbf{p-k} }a_{\mathbf{q+k}}
 \int d\mathbf{p}'' d\mathbf{q}''
\Psi(\mathbf{p}'',  \mathbf{q}'')
  d^{\dag}_{\mathbf{p}'' }a^{\dag}_{\mathbf{q}'' } |0 \rangle \nonumber\\
&=& \int  d \mathbf{p} d \mathbf{q} d \mathbf{k} V
(\mathbf{p},\mathbf{q},\mathbf{k} ) \int d\mathbf{p}'' d\mathbf{q}''
 \Psi (\mathbf{p}'',  \mathbf{q}'')
\delta(\mathbf{p-k} - \mathbf{p}'') \delta(\mathbf{q+k} -
\mathbf{q}'')
 d^{\dag}_{\mathbf{p} }
a^{\dag}_{\mathbf{q}  }  |0 \rangle \nonumber\\
&=& \int  d \mathbf{p} d \mathbf{q} \Bigl( \int  d \mathbf{k} V
(\mathbf{p},\mathbf{q},\mathbf{k})
 \Psi (\mathbf{p-k},  \mathbf{q+k})\Bigr)
 d^{\dag}_{\mathbf{p} }
a^{\dag}_{\mathbf{q} }  |0 \rangle \label{eq:9.66}
\end{eqnarray}

\noindent Comparing this with (\ref{eq:9.65}) we see that the
momentum-space wave function $\Psi(\mathbf{p},  \mathbf{q} )$ has
been transformed by the action of $\hat{V}$ to the new wave function

\begin{eqnarray*}
\Psi'(\mathbf{p}, \mathbf{q} ) &\equiv& \hat{V} \Psi(\mathbf{p},  \mathbf{q} )
= \int   d \mathbf{k}
 V
(\mathbf{p},\mathbf{q},\mathbf{k} )
 \Psi (\mathbf{p-k},\mathbf{q+k} )
\end{eqnarray*}

\noindent This is the most general linear transformation of a
two-particle wave function which conserves the total momentum.

For comparison with traditionally used inter-particle potentials, it
is more convenient to have expression for operator $\hat{V}$ in the position space.
We can write\footnote{Here $\mathbf{x}$ and $\mathbf{y}$ are
positions of the proton and the electron, respectively; and we use
(\ref{eq:7.27}) to change from the momentum representation to the  position
representation.}

\begin{eqnarray}
&\mbox{ }& \Psi'(\mathbf{x},  \mathbf{y} ) \nonumber \\
&\equiv&
\hat{V} \Psi(\mathbf{x},  \mathbf{y} ) = \frac{1}{(2 \pi \hbar )^3} \int
e^{\frac{i}{\hbar}\mathbf{p}\mathbf{x} +
\frac{i}{\hbar}\mathbf{q}\mathbf{y}} d\mathbf{p} d\mathbf{q}
\Psi'(\mathbf{p}, \mathbf{q} )  \nonumber \\
&= &\frac{1}{(2 \pi \hbar)^3} \int
e^{\frac{i}{\hbar}\mathbf{p}\mathbf{x} +
\frac{i}{\hbar}\mathbf{q}\mathbf{y}} d\mathbf{p} d\mathbf{q} \left(\int
  d \mathbf{k}
V(\mathbf{p}, \mathbf{q}, \mathbf{k})
\Psi(\mathbf{p-k}, \mathbf{q+k} ) \right) \nonumber \\
& =& \frac{1}{(2\pi \hbar )^3} \int
e^{\frac{i}{\hbar}(\mathbf{p+k})\mathbf{x} +
\frac{i}{\hbar}(\mathbf{q-k})\mathbf{y}} d\mathbf{p} d\mathbf{q}
\int   d \mathbf{k} V(\mathbf{p+k}, \mathbf{q-k}, \mathbf{k})
\Psi(\mathbf{p}, \mathbf{q} )  \nonumber \\
 &=&  \int    d \mathbf{k}
e^{\frac{i}{\hbar}\mathbf{k}(\mathbf{x-y})} V(\mathbf{p+k},
\mathbf{q-k}, \mathbf{k} ) \left[\frac{1}{(2 \pi \hbar)^{3}} \int
d\mathbf{p} d\mathbf{q} e^{\frac{i}{\hbar}\mathbf{p}\mathbf{x} +
\frac{i}{\hbar}\mathbf{q}\mathbf{y}} \Psi(\mathbf{p}, \mathbf{q})
\right] \nonumber
\\ \label{eq:pot-act}
\end{eqnarray}

\noindent where expression in square brackets is recognized as the
original position-space wave function

\begin{eqnarray}
\Psi(\mathbf{x}, \mathbf{y}) = \frac{1}{(2 \pi \hbar)^3} \int d\mathbf{p} d\mathbf{q}
e^{\frac{i}{\hbar} \mathbf{p}\mathbf{x} +
\frac{i}{\hbar}\mathbf{q}\mathbf{y}}
\Psi(\mathbf{p}, \mathbf{q}) \label{eq:psixy}
\end{eqnarray}

\noindent and the rest is an operator acting on this wave function. This
operator acquires especially simple form if we assume that
$V(\mathbf{p}, \mathbf{q}, \mathbf{k})$ does not depend on
$\mathbf{p}$ and $\mathbf{q}$

\begin{eqnarray*}
V(\mathbf{p}, \mathbf{q}, \mathbf{k}) = v(\mathbf{k})
\end{eqnarray*}

\noindent Then

\begin{eqnarray}
 \hat{V} \Psi(\mathbf{x}, \mathbf{y}) =  \int    d \mathbf{k}
e^{\frac{i}{\hbar}\mathbf{k}(\mathbf{x-y})} v(\mathbf{k})
\Psi(\mathbf{x}, \mathbf{y}) =
w(\mathbf{x-y}) \Psi(\mathbf{x}, \mathbf{y} ) \label{eq:9.67}
\end{eqnarray}

\noindent where

\begin{eqnarray*}
w(\mathbf{r}) = \int d \mathbf{k}
e^{\frac{i}{\hbar}\mathbf{k}\mathbf{r}} v(\mathbf{k})
\end{eqnarray*}

 \noindent is the Fourier transform of $v(\mathbf{k}) $.
We see that interaction (\ref{eq:9.64}) acts as multiplication by
the function $w(\mathbf{r})$  in the position space. So, it is a
usual position-dependent potential. Note that the requirement of the
total momentum conservation implies automatically that this potential
depends on the relative position $\mathbf{r} \equiv \mathbf{x-y}$.

As an example consider interaction operator of the form (\ref{eq:9.64})

\begin{eqnarray}
\hat{V}
&=& \frac{q_1q_2}{(2 \pi)^3 \hbar}\int  \frac{d \mathbf{p} d \mathbf{q} d \mathbf{k}}{k^2}
 d^{\dag}_{\mathbf{p} }
a^{\dag}_{\mathbf{q} } d_{\mathbf{p-k} }a_{\mathbf{q+k} } \label{eq:Coulomb2}
\end{eqnarray}

\noindent where constants $q_1$ and $q_2$ can be interpreted as charges of the two particles and $v(\mathbf{k}) = q_1q_2/(8 \pi^3 \hbar k^2)$. Then the position-space interaction is the usual Coulomb potential\footnote{see equation (\ref{eq:A.90})}

\begin{eqnarray}
w(\mathbf{r}) = \frac{q_1q_2}{(2 \pi)^3 \hbar} \int \frac{d \mathbf{k}}{k^2}
e^{\frac{i}{\hbar}\mathbf{k}\mathbf{r}} =\frac{q_1q_2}{4 \pi r} \label{eq:wmatr}
\end{eqnarray}

Let us now consider the general case (\ref{eq:pot-act}). Without
loss of generality we can represent function $V(\mathbf{p+k},
\mathbf{q-k}, \mathbf{k} )$ as a series\footnote{For example, a
series of this form can be obtained by writing a Taylor expansion
with respect to the variable $\mathbf{k}$ with $\chi_j$ being the
coefficients depending on $\mathbf{p}$ and $\mathbf{q}$.}

\begin{eqnarray*}
V(\mathbf{p+k}, \mathbf{q-k}, \mathbf{k} ) = \sum_j \chi_j
(\mathbf{p}, \mathbf{q})  v_j (\mathbf{k})
\end{eqnarray*}

\noindent Then we obtain

\begin{eqnarray}
&\mbox{ }& \hat{V} \Psi(\mathbf{x}, \mathbf{y}) \nonumber\\
 &=& \sum_j \int    d \mathbf{k}
e^{\frac{i}{\hbar}\mathbf{k}(\mathbf{x-y})} v_j (\mathbf{k})
\left[\frac{1}{(2 \pi \hbar)^{3}} \int d\mathbf{p} d\mathbf{q}
\chi_j (\mathbf{p}, \mathbf{q}) e^{\frac{i}{\hbar}
\mathbf{p}\mathbf{x} + \frac{i}{\hbar}\mathbf{q}\mathbf{y}}
\Psi(\mathbf{p}, \mathbf{q} )\right] \nonumber\\
 &=& \sum_j
w_j (\mathbf{x-y}) \chi_j (\hat{\mathbf{p}}, \hat{\mathbf{q}})
\left[\frac{1}{(2 \pi \hbar)^{3}} \int d\mathbf{p} d\mathbf{q}
e^{\frac{i}{\hbar} \mathbf{p}\mathbf{x} +
\frac{i}{\hbar}\mathbf{q}\mathbf{y}}
\Psi(\mathbf{p}, \mathbf{q} )\right] \nonumber\\
&=& \sum_j w_j (\mathbf{x-y}) \chi_j (\hat{\mathbf{p}},
\hat{\mathbf{q}}) \Psi(\mathbf{x}, \mathbf{y} )
\label{eq:pos-rep-pot}
\end{eqnarray}

\noindent where $\hat{\mathbf{p}} = -i\hbar (d/d\mathbf{x})$ and
$\hat{\mathbf{q}} = -i\hbar (d/d\mathbf{y})$ are differential
operators, i.e.,  position-space representations (\ref{eq:mom-oper})
of the momentum operators of the two particles. Expression
(\ref{eq:pos-rep-pot}) then demonstrates that interaction
$d^{\dag}a^{\dag} da$ can be always represented as a general
2-particle potential \index{2-particle potential} depending on the
distance between particles and their momenta. We will use equation
(\ref{eq:pos-rep-pot}) in our derivation of 2-particle RQD
potentials in subsection \ref{ss:position-breit}.

\subsection{Cluster separability in the Fock space}
\label{ss:clus-sep}

We know that a cluster
separable interaction potential can be constructed as a sum of
smooth potentials (\ref{eq:8.38}) depending on particle observables (positions, momenta, and spins). However, this notation is very
inconvenient to use in the Fock space, because such sums have rather
different forms in different Fock sectors. For example, the Coulomb
interaction has the form (\ref{eq:1/r}) in the 2-particle sector and
the form (\ref{eq:8.37}) in the 3-particle sector. This notational difference for inter-particle interactions is very inconvenient. It would be more preferable to have a unique formula, which remains valid in all $N$-particle sectors. Fortunately, it is easy to satisfy cluster separability within our standard notation (\ref{eq:9.48}) - (\ref{eq:9.49}). We just need to make sure that factors $D_{NM}$ are smooth
functions of particle momenta.\footnote{see section 4 in \cite{book}.}

Let us verify this statement on a simple example. We are going to
find out how the 2-particle potential (\ref{eq:9.64})\footnote{As a concrete example, it is instructive to choose the Coulomb interaction (\ref{eq:Coulomb2}) in these calculations.} acts in the
3-particle (one proton and two electrons) sector of the Fock space
$\mathcal{H}(2, 0, 1, 0, 0)$ where state vectors have the form

\begin{eqnarray}
| \Psi \rangle &=&  \int d \mathbf{p} d \mathbf{q}_1 d \mathbf{q}_2
\psi (\mathbf{p},\mathbf{q}_1,\mathbf{q}_2 )
 d^{\dag}_{\mathbf{p} }
a^{\dag}_{\mathbf{q}_1 } a^{\dag}_{\mathbf{q}_2 } |0 \rangle
\label{eq:9.68a}
\end{eqnarray}

\noindent Applying operator (\ref{eq:9.64}) to this state vector we
obtain

\begin{eqnarray}
&\mbox{ }& \hat{V}| \Psi \rangle \nonumber \\
&=& \int d \mathbf{p'} d \mathbf{q'} d \mathbf{k} \int d \mathbf{p}
d \mathbf{q}_1 d \mathbf{q}_2 V (\mathbf{p'},\mathbf{q'},\mathbf{k}
) \psi (\mathbf{p},\mathbf{q}_1,\mathbf{q}_2 )
 d^{\dag}_{\mathbf{p'} }
a^{\dag}_{\mathbf{q'} } d_{\mathbf{p'-k} }a_{\mathbf{q'+k} }
 d^{\dag}_{\mathbf{p} }
a^{\dag}_{\mathbf{q}_1 } a^{\dag}_{\mathbf{q}_2 } |0 \rangle \nonumber \\
\label{eq:7.83x}
\end{eqnarray}

\noindent The product of particle operators acting on the vacuum
state can be normally ordered as

\begin{eqnarray*}
&\mbox{ }&  d^{\dag}_{\mathbf{p'} } a^{\dag}_{\mathbf{q'} }
d_{\mathbf{p'-k} }a_{\mathbf{q'+k} }
 d^{\dag}_{\mathbf{p} }
a^{\dag}_{\mathbf{q}_1 } a^{\dag}_{\mathbf{q}_2 } |0\rangle \nonumber \\
&=& - d^{\dag}_{\mathbf{p'} } a^{\dag}_{\mathbf{q'} }
d^{\dag}_{\mathbf{p} }  a^{\dag}_{\mathbf{q}_1 }
a^{\dag}_{\mathbf{q}_2 } d_{\mathbf{p'-k} } a_{\mathbf{q'+k} }
|0\rangle  + \delta(\mathbf{p'-k -p} )d^{\dag}_{\mathbf{p'} } a^{\dag}_{\mathbf{q'} }
a_{\mathbf{q'+k} } a^{\dag}_{\mathbf{q}_1
} a^{\dag}_{\mathbf{q}_2 }  |0\rangle \nonumber \\
&\ & -\delta(\mathbf{q}_1 - \mathbf{q'-k}) d^{\dag}_{\mathbf{p'} } a^{\dag}_{\mathbf{q'} }
d^{\dag}_{\mathbf{p} }   a^{\dag}_{\mathbf{q}_2 } d_{\mathbf{p'-k} }
|0\rangle + \delta(\mathbf{q}_2 -
\mathbf{q'-k} )
d^{\dag}_{\mathbf{p'} } a^{\dag}_{\mathbf{q'} } d^{\dag}_{\mathbf{p}
} a^{\dag}_{\mathbf{q}_1 } d_{\mathbf{p'-k} }  |0\rangle \nonumber \\
&=& \delta(\mathbf{p'-k -p} ) \delta(\mathbf{q}_1 -\mathbf{q'-k} )
 d^{\dag}_{\mathbf{p'} } a^{\dag}_{\mathbf{q'} }
  a^{\dag}_{\mathbf{q}_2}
   |0 \rangle - \delta(\mathbf{q}_2 - \mathbf{q'-k} ) \delta(\mathbf{p'-k -p}
 ) d^{\dag}_{\mathbf{p'} } a^{\dag}_{\mathbf{q'} }
 a^{\dag}_{\mathbf{q}_1 }  |0 \rangle
\end{eqnarray*}

\noindent Inserting this result in (\ref{eq:7.83x}) we obtain

\begin{eqnarray}
&\mbox{ }& \hat{V}| \Psi \rangle \nonumber \\
&=& \int   d \mathbf{k} d \mathbf{p} d \mathbf{q}_1 d \mathbf{q}_2 V
(\mathbf{k+p},\mathbf{q}_1 - \mathbf{k},\mathbf{k} ) \psi
(\mathbf{p},\mathbf{q}_1,\mathbf{q}_2 )
 d^{\dag}_{\mathbf{k+p} } a^{\dag}_{\mathbf{q}_1 - \mathbf{k}}
  a^{\dag}_{\mathbf{q}_2} |0 \rangle \nonumber \\
  &\ & -\int  d \mathbf{k} d \mathbf{p}
d \mathbf{q}_1 d \mathbf{q}_2 V (\mathbf{k+p},\mathbf{q}_2 -
\mathbf{k},\mathbf{k} ) \psi (\mathbf{p},\mathbf{q}_1,\mathbf{q}_2 )
 d^{\dag}_{\mathbf{k+p} } a^{\dag}_{\mathbf{q}_2 - \mathbf{k}}
 a^{\dag}_{\mathbf{q}_1 }   |0 \rangle
 \nonumber \\
  &=& \int d \mathbf{p} d \mathbf{q}_1 d \mathbf{q}_2 \Bigl( \int   d \mathbf{k}
    V
(\mathbf{p},\mathbf{q}_1,\mathbf{k} ) \psi
(\mathbf{p-k},\mathbf{q}_1 + \mathbf{k},\mathbf{q}_2 ) \nonumber \\
  &\ & +\int   d \mathbf{k} V (\mathbf{p},\mathbf{q}_2 ,\mathbf{k}
) \psi (\mathbf{p-k},\mathbf{q}_1,\mathbf{q}_2 +\mathbf{k}) \Bigr)
 d^{\dag}_{\mathbf{p} }
 a^{\dag}_{\mathbf{q}_1 } a^{\dag}_{\mathbf{q}_2 }  |0 \rangle
 \label{eq:v-psi}
\end{eqnarray}

\noindent Comparing this with equation (\ref{eq:9.66}) we see that, as
expected from the condition of cluster separability, the
two-particle interaction in the three-particle sector  separates in
two terms. One term acts on the pair of variables $(\mathbf{p},
\mathbf{q}_1)$. The other term acts on variables $(\mathbf{p},
\mathbf{q}_2)$.

Removing the electron 2 to infinity is equivalent to multiplying the
momentum-space wave function $\psi (\mathbf{p},\mathbf{q}_1,\mathbf{q}_2 )$ by
$\exp(\frac{i}{\hbar} \mathbf{q}_2\mathbf{a})$ where $\mathbf{a} \to
\infty$. The action of $V$ on such a wave function\footnote{i.e., the term
in parentheses in (\ref{eq:v-psi})} is

\begin{eqnarray*}
&\mbox{ }& \lim _{\mathbf{a} \to \infty} \Bigl[\int   d \mathbf{k} V
(\mathbf{p},\mathbf{q}_1,\mathbf{k} ) \psi
(\mathbf{p-k},\mathbf{q}_1 + \mathbf{k},\mathbf{q}_2 )
e^{\frac{i}{\hbar} \mathbf{q}_2\mathbf{a}}
 \nonumber \\
  &+& \int  d \mathbf{k}  V (\mathbf{p},\mathbf{q}_2 ,\mathbf{k}
) \psi (\mathbf{p-k},\mathbf{q}_1,\mathbf{q}_2 +\mathbf{k} )
e^{\frac{i}{\hbar} (\mathbf{q}_2 + \mathbf{k} )\mathbf{a}} \Bigr]
 \nonumber
\end{eqnarray*}

\noindent  In the limit $\mathbf{a} \to \infty $ the exponent in the
integrand of the second term is a rapidly oscillating function of
$\mathbf{k}$. If the coefficient function $V
(\mathbf{p},\mathbf{q},\mathbf{k} )$ is a smooth function of
$\mathbf{k}$ then
 the integral on $\mathbf{k}$ is zero
due to the Riemann-Lebesgue lemma \ref{lemma:Rim-Leb}. Therefore,
only the interaction proton - electron(1) does not vanish

\begin{eqnarray*}
&\mbox{ }& \lim _{\mathbf{a} \to \infty} \hat{V} e^{\frac{i}{\hbar}
\hat{\mathbf{q}}_2\mathbf{a}}| \Psi \rangle \nonumber \\
 &=& \lim _{\mathbf{a} \to \infty} \int   d \mathbf{p} d \mathbf{q}_1 d \mathbf{q}_2
 \left( \int   d \mathbf{k} V
(\mathbf{p},\mathbf{q}_1,\mathbf{k} ) \psi
(\mathbf{p-k},\mathbf{q}_1 + \mathbf{k},\mathbf{q}_2 ) \right)
e^{\frac{i}{\hbar} \mathbf{q}_2\mathbf{a}}
 d^{\dag}_{\mathbf{p} } a^{\dag}_{\mathbf{q}_1 }
  a^{\dag}_{\mathbf{q}_2} |0 \rangle \nonumber \\
 &=& \left(\lim _{\mathbf{a} \to \infty} \int    d \mathbf{q}_2 e^{\frac{i}{\hbar} \mathbf{q}_2\mathbf{a}} a^{\dag}_{\mathbf{q}_2} |0 \rangle \right)
 \left( \int   d \mathbf{p} d \mathbf{q}_1 d \mathbf{k} V
(\mathbf{p},\mathbf{q}_1,\mathbf{k} ) \psi
(\mathbf{p-k},\mathbf{q}_1 + \mathbf{k},\mathbf{q}_2 )
 d^{\dag}_{\mathbf{p} } a^{\dag}_{\mathbf{q}_1 } |0 \rangle\right)
   \nonumber
\end{eqnarray*}

\noindent Comparing this with equation (\ref{eq:9.66}), we see that the spatial translation of the electron 2 results in a state in which the remote electron 2 coexists with the interacting subsystem ``proton + electron 1''. This demonstrates that $\hat{V}$ is a cluster separable
potential.

For general potentials (\ref{eq:9.48}) - (\ref{eq:9.49}) with smooth coefficient functions the above arguments can be repeated: If
some particles are removed to infinity such potentials automatically
separate into sums of smooth terms, as required by cluster separability. Therefore,

\begin{statement} [cluster separability]
\label{statementT} \index{cluster separability} The cluster
separability of the general interaction (\ref{eq:9.48}) is guaranteed if
coefficient functions $D_{NM}$
of all interaction potentials $V_{NM}$ are  smooth functions of
momenta. \index{smooth potential}
\end{statement}

\noindent The power of this statement is that when expressing
interacting potentials through particle operators in the momentum
representation (as in (\ref{eq:9.49})) we have a very simple criterion of cluster
separability: the coefficient functions must be smooth, i.e., they
should not contain singular factors, like delta
functions.\footnote{This is the reason why cluster separable potentials were
called \emph{smooth} in subsection \ref{ss:separability}.} This demonstrates
the great advantage of writing interactions in terms of particle
(creation and annihilation) operators (\ref{eq:9.49}) instead of particle (position
and momentum) observables, as in section
\ref{sc:instant-form}.\footnote{
Recall that in subsection \ref{ss:3-particle} it was a very
non-trivial matter to ensure the cluster separability for
interaction potentials written in terms of particle observables even
in a simplest 3-particle system.}

\section{A toy model theory}
\label{sc:model-theory}

 Before
considering real QED interactions in the next chapter, in this section we are going to perform
a warm-up exercise. We will introduce a simple yet quite realistic
model
 theory with variable number of particles in the Fock space.
In this theory, the perturbation expansion of the $S$-operator can
be evaluated with minimal efforts, in particular, with the help of a
convenient \emph{diagram technique}.

\subsection{Fock space and Hamiltonian}
\label{ss:fock-space-ham}

The toy model introduced here is a rough approximation to QED.  No particle-antiparticle pair creation is allowed.
So, for simplicity we will ignore all other types of particles except electrons and photons.  The relevant part of the Fock space is a direct sum of electron-photon sectors like
those described in formulas (\ref{eq:9.1}) - (\ref{eq:9.9}). We will
also assume that
 interaction  does not affect the electron
spin and photon polarization degrees of freedom, so the
corresponding labels will be omitted. Then relevant
(anti)commutation relations of particle operators can be taken
from (\ref{eq:9.26}) - (\ref{eq:9.29})

\begin{eqnarray}
\{a_{\mathbf{p}},a^{\dag}_{\mathbf{p}'}\} &=& \delta
(\mathbf{p}-\mathbf{p}') \label{eq:aa} \\
\mbox{ } [c_{\mathbf{p}},c^{\dag}_{\mathbf{p}'}]
&=&  \delta (\mathbf{p}-\mathbf{p}') \label{eq:cc}\\
  \{a_{\mathbf{p}},a_{\mathbf{p}'}\}
&=& \{a^{\dag}_{\mathbf{p}},a^{\dag}_{\mathbf{p}'}\} = 0
\label{eq:aa2} \\
\mbox{ } [c_{\mathbf{p}},c_{\mathbf{p}'}] &=&
[c^{\dag}_{\mathbf{p}},c^{\dag}_{\mathbf{p}'}] = 0 \label{eq:cc2} \\
 \mbox{ } [a^{\dag}_{\mathbf{p}},c^{\dag}_{\mathbf{p}'}]&=&
[a^{\dag}_{\mathbf{p}},c_{\mathbf{p}'}] =
[a_{\mathbf{p}},c^{\dag}_{\mathbf{p}'}] =
[a_{\mathbf{p}},c_{\mathbf{p}'}] = 0 \nonumber
\end{eqnarray}

\noindent The full Hamiltonian

\begin{eqnarray}
H  = H_0 + V_1 \label{eq:full-ham}
\end{eqnarray}

\noindent as usual, is the sum of  the free Hamiltonian

\begin{eqnarray*}
    H_0 &=&  \int d\mathbf{p} \omega_{\mathbf{p} }
  a^{\dag}_{\mathbf{p}}a_{\mathbf{p}}
+ c \int d\mathbf{k} k   c^{\dag}_{\mathbf{k}} c_{\mathbf{k}}
\end{eqnarray*}

\noindent  and interaction, which we choose in the following \emph{unphys}
form

\begin{eqnarray}
    V_1 &=&  \frac{e c \hbar}{ (2\pi \hbar)^{3/2}} \int
     \frac{d\mathbf{p}d\mathbf{k}}{\sqrt{ck}}
a^{\dag}_{\mathbf{p}}c^{\dag}_{\mathbf{k}}a_{\mathbf{p+k}} + \frac{e
c \hbar}{ (2\pi \hbar)^{3/2}} \int
    \frac{d\mathbf{p}d\mathbf{k}}{\sqrt{ck}}
a^{\dag}_{\mathbf{p}}a_{\mathbf{p-k}}
     c_{\mathbf{k}}
\label{eq:9.69}
\end{eqnarray}

\noindent The \emph{coupling constant} $e$ \index{coupling constant} is
equal to the absolute value of the electron's
charge. Here and in what follows the \emph{perturbation order}
\index{perturbation order} of an operator (= the power of the
coupling constant $e$ in the operator) is shown by the subscript. For example, the
free Hamiltonian $H_0$ does not depend on $e$, so it is of zero
perturbation order; the perturbation order of $V_1$ is one, etc.

 The above theory
satisfies  conservation laws

\begin{eqnarray*}
[H, \mathbf{P}_0] = [H, \mathbf{J}_0] = [H,Q] = 0
\end{eqnarray*}

\noindent where operators $\mathbf{P}_0$, $\mathbf{J}_0$, and $Q$ refer to the total momentum, total angular momentum
operator and total electric charge, respectively

\begin{eqnarray*}
\mathbf{P}_0 &=&  \int d\mathbf{p} \mathbf{p}
 ( a^{\dag}_{\mathbf{p}}a_{\mathbf{p}}
+   c^{\dag}_{\mathbf{p}} c_{\mathbf{p}}) \\
Q &=& -e\int d\mathbf{p}
 a^{\dag}_{\mathbf{p}}a_{\mathbf{p}}
\end{eqnarray*}

\noindent The number of electrons is conserved, due to the conservation of charge, but the number of photons can vary without limitations. So, this theory can
describe important processes of the photon emission and
absorption.

However, our toy model has a major drawback: This model is not Poincar\'e invariant. This means that we cannot construct an interacting boost operator $\mathbf{K}$
such that the Poincar\'e commutation relations with $H$,
$\mathbf{P}_0$, and $ \mathbf{J}_0$ are satisfied. In this section
we will tolerate the lack of invariance, but in chapter \ref{ch:QED} we
will show how the Poincar\'e invariance can be satisfied  in a
more comprehensive theory (QED) which includes both particles and
antiparticles.

\subsection{Drawing a diagram in the toy model}
\label{ss:draw-diagram}

\begin{figure}
\centering
\includegraphics{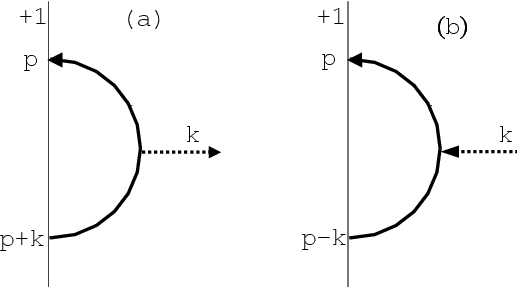} \caption{Diagram representation of the
interaction operator $V_1$.} \label{fig:9.1}
\end{figure}

In this subsection we would like to introduce a diagram
technique which would greatly facilitate perturbative calculations of
scattering operators (\ref{eq:7.63a}) - (\ref{eq:7.63c}). Let us
graphically represent each term in the interaction potential
(\ref{eq:9.69}) as a \emph{vertex} \index{vertex} (see Fig.
\ref{fig:9.1}). \index{diagram} Each particle operator in
$V_1$ is represented as an oriented \emph{line}
\index{line in diagram} or arrow. The line corresponding to an
annihilation operator  enters the vertex and the line corresponding
to a creation operator  leaves the vertex. Electron lines are shown by full arcs
 and photon lines are shown by dashed arrows on the diagram. Each line is marked
with the momentum label of the corresponding particle operator.
 Free ends of the electron lines
are attached to  the vertical electron ``order bar'' on the left
hand side of the diagram.  The order of these \emph{external lines}
\index{line external} (from bottom to top of the order bar)
corresponds to the order of electron particle operators in the
potential (from right to left).    An additional numerical factor is
indicated in the upper left corner of the diagram.

\begin{figure}
\centering
\includegraphics{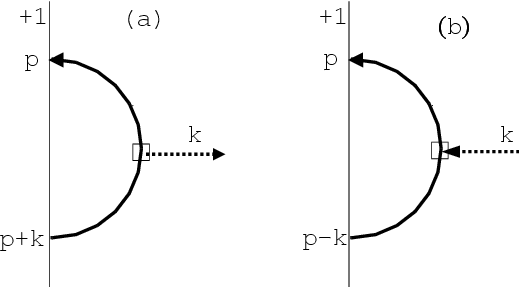} \caption{Diagram representation of the $t$-integral
$\underline{V_1(t)}$.} \label{fig:9.2}
\end{figure}

The $t$-integral

\begin{eqnarray}
\underline{V_1} &=& -\frac{e c \hbar }{(2\pi
\hbar)^{3/2}} \int
    \frac{d\mathbf{p}d\mathbf{k}}{\sqrt{ck}}
\frac{a^{\dag}_{\mathbf{p}}c^{\dag}_{\mathbf{k}}a_{\mathbf{p+k}}}{\omega_{ \mathbf{p}} + ck - \omega_{ \mathbf{p+k}}}
\nonumber\\
&\ & -\frac{e  c \hbar }{(2\pi \hbar)^{3/2}} \int
     \frac{d\mathbf{p}d\mathbf{k}}{\sqrt{ck}}
\frac{a^{\dag}_{\mathbf{p}}a_{\mathbf{p-k}}
     c_{\mathbf{k}}}{\omega_{ \mathbf{p}} - ck - \omega_{ \mathbf{p-k}}}
\label{eq:9.70}
\end{eqnarray}

\noindent  differs from $V_1$ only by
the factor $- E_{V_1}^{-1}$ (see equation (\ref{eq:underline2})), which is
represented in the diagram  \ref{fig:9.2} by drawing a box that crosses all
external lines. A line entering (leaving) the box contributes its
energy with the negative (positive) sign to the energy function
$E_{V_1}$.

\begin{figure}
\centering
\includegraphics{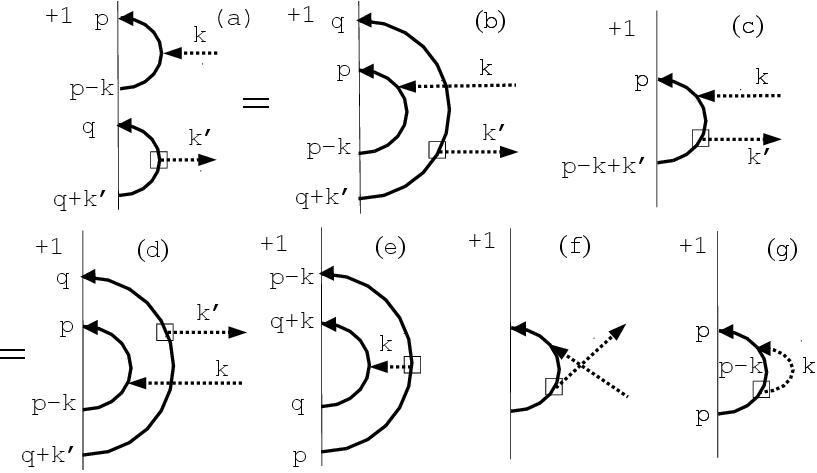} \caption{The normal product of operators
in Fig. \ref{fig:9.1}(b) and \ref{fig:9.2}(a).} \label{fig:9.3}
\end{figure}

In order to calculate the $\Sigma$-operator (\ref{eq:7.63c})  in the 2nd perturbation order we need products
$V_1\underline{V_1}$. The diagram corresponding to the product
of two diagrams $AB$ is obtained
  by simply placing the diagram $B$ below the diagram $A$ and attaching
external electron lines of both diagrams to the same order bar. For
example, the diagram for the product of the second term in
(\ref{eq:9.69}) (Fig. \ref{fig:9.1}(b)) and the first term in
(\ref{eq:9.70}) (Fig. \ref{fig:9.2}(a))

\begin{eqnarray}
V_1 \underline{V_1} &\propto&
(a^{\dag}_{\mathbf{p}}a_{\mathbf{p-k}} c_{\mathbf{k}})
(a^{\dag}_{\mathbf{q}} c^{\dag}_{\mathbf{k}'} a_{\mathbf{q+k'}}  ) +
\ldots
\label{eq:9.71}
\end{eqnarray}

\noindent  is shown in Fig. \ref{fig:9.3}(a).\footnote{By
convention, we will place free ends of photon external lines on the
right hand side of the diagram. The order of these free ends (from
top to bottom of the diagram) will correspond to the order of photon
particle operators in the expression (from left to right). For
example, in Fig. \ref{fig:9.3}(a) the incoming photon line is above
the outgoing photon line, which corresponds to the order $cc^{\dag}$
of photon operators in (\ref{eq:9.71}). } This product should be
further converted to the normal form, i.e., all creation operators should be moved to the left. On the diagram, the movement of creation operators from
right to left is represented by the movement of free outward pointing arrows upward, so that at the end of this process all outgoing
 lines are positioned below incoming lines. Due to anticommutation relations (\ref{eq:aa}) and
(\ref{eq:aa2}), each exchange of positions of electron particle
operators (full lines in the diagram) changes the total sign of the
expression.  Each permutation of annihilation and creation
operators (incoming and outgoing lines, respectively) of similar
particles
 creates an additional  expression, which corresponds to delta functions on the right hand sides of (\ref{eq:aa}) and (\ref{eq:cc}). We represent these additional terms
by diagrams in which the swapped lines are joined together forming
\emph{internal lines} that directly connects two vertices. \index{internal line}

Applying these rules to (\ref{eq:9.71}),  we
first move the photon operators  to rightmost
positions, move the operator $a^{\dag}_{\mathbf{q}}$ to the leftmost
position, and add another term due to the anticommutator
$\{a_{\mathbf{p-k}}, a^{\dag}_{\mathbf{q}} \} = \delta (\mathbf{q -
p+k})$.

\begin{eqnarray}
V_1 \underline{V_1} &\propto&
 a^{\dag}_{\mathbf{q}} a^{\dag}_{\mathbf{p}}a_{\mathbf{p-k}}
  a_{\mathbf{q+k'}} c_{\mathbf{k}}  c^{\dag}_{\mathbf{k}'}
+ \delta(\mathbf{q} - \mathbf{p+k})a^{\dag}_{\mathbf{p}}
  a_{\mathbf{q+k'}} c_{\mathbf{k}}  c^{\dag}_{\mathbf{k}'} \nonumber\\
&=& a^{\dag}_{\mathbf{q}} a^{\dag}_{\mathbf{p}}a_{\mathbf{p-k}}
  a_{\mathbf{q+k'}} c_{\mathbf{k}}  c^{\dag}_{\mathbf{k}'}
+ a^{\dag}_{\mathbf{p}}
  a_{\mathbf{p-k+k'}} c_{\mathbf{k}}  c^{\dag}_{\mathbf{k}'} + \ldots
\label{eq:9.72}
\end{eqnarray}

Expression (\ref{eq:9.72}) is represented by two diagrams \ref{fig:9.3}(b) and
\ref{fig:9.3}(c). In the diagram \ref{fig:9.3}(b) the electron line marked $\mathbf{q}$
has been moved to the top of the electron order bar. In the diagram
\ref{fig:9.3}(c) the product $\delta(\mathbf{q} - \mathbf{p+k})$ and the
integration by $\mathbf{q}$ are represented by  joining or
\emph{pairing} \index{pairing} the
 incoming electron line carrying momentum
$\mathbf{p-k}$ with the outgoing electron line carrying momentum
$\mathbf{q}$.  This produces an internal electron line carrying
momentum $\mathbf{p-k}$ between two vertices.

 In the expression (\ref{eq:9.72}),
 electron operators are in the normal order, however, photon operators
are not there yet.
The next step is to change the order of photon operators

\begin{eqnarray}
V_1 \underline{V_1} &\propto&
 a^{\dag}_{\mathbf{q}} a^{\dag}_{\mathbf{p}} c^{\dag}_{\mathbf{k}'} a_{\mathbf{p-k}}
  a_{\mathbf{q+k'}}  c_{\mathbf{k}}
+   a^{\dag}_{\mathbf{q}} a^{\dag}_{\mathbf{p}}a_{\mathbf{p-k}}
  a_{\mathbf{q+k'}} \delta(\mathbf{k}'-\mathbf{k}) \nonumber \\
&\ &  +a^{\dag}_{\mathbf{p}} c^{\dag}_{\mathbf{k}'}
  a_{\mathbf{p-k+k'}}  c_{\mathbf{k}}
+  a^{\dag}_{\mathbf{p}}
  a_{\mathbf{p-k+k'}}\delta(\mathbf{k}'-\mathbf{k})  + \ldots \nonumber \\
&=&  a^{\dag}_{\mathbf{q}} a^{\dag}_{\mathbf{p}} c^{\dag}_{\mathbf{k}'} a_{\mathbf{p-k}}
  a_{\mathbf{q+k'}}  c_{\mathbf{k}}
+   a^{\dag}_{\mathbf{q}} a^{\dag}_{\mathbf{p}}a_{\mathbf{p-k}}
  a_{\mathbf{q+k}}  \nonumber \\
&\ & + a^{\dag}_{\mathbf{p}} c^{\dag}_{\mathbf{k}'}
  a_{\mathbf{p-k+k'}}  c_{\mathbf{k}}
+  a^{\dag}_{\mathbf{p}}
  a_{\mathbf{p}}  + \ldots \label{eq:norm-order1}
\end{eqnarray}

\noindent   The normal ordering of photon operators in \ref{fig:9.3}(b) yields
diagrams \ref{fig:9.3}(d) and \ref{fig:9.3}(e) according to equation
(\ref{eq:cc}).  Diagrams \ref{fig:9.3}(f) and \ref{fig:9.3}(g)
are obtained from \ref{fig:9.3}(c) in a similar way.

\subsection{Reading a diagram in the toy model}
\label{ss:read-diagram}

With the above diagram rules and some practice,  one can perform
 calculations of scattering operators  (\ref{eq:7.63b}) and (\ref{eq:7.63c})
much easier than in the usual algebraic way.
During these
diagram manipulations we, actually, do not need to keep track of the momentum labels of
lines.
The
algebraic expression of the result can be easily restored from an
unlabeled diagram by following these steps:

\begin{itemize}
\item[(I)] Assign a distinct momentum label to
each external line, except one, whose momentum is obtained from the
(momentum conservation) condition that the sum of all incoming
external momenta minus the sum of all outgoing external momenta is
zero.
\item[(II)] Assign momentum labels to internal lines so
that the momentum conservation law is satisfied at each vertex: The
sum of momenta of lines entering the vertex is equal to the sum of
momenta of outgoing lines. If there are \emph{loops}, \index{loop}
one needs to introduce new independent \emph{loop
momenta}\footnote{see diagram \ref{fig:9.3}(g) in which $\mathbf{k}$
is the loop momentum.} \index{loop momentum}
\item[(III)]  Read external lines  from top to bottom of the
diagram and write  corresponding particle operators  from left to
right.\footnote{Incoming lines correspond to annihilation operators; outgoing lines correspond to creation operators.} Do it first for electron lines and then for photon lines.
\item[(IV)]  For
each box, write a factor $  (E_f - E_i)^{-1}$, where $E_f$ is the
sum of energies of particles going out of the box and $E_i$ is the
sum of energies of particles coming into the box.
\item[(V)] For each vertex introduce a  factor
$\frac{ e  c \hbar}{\sqrt{(2 \pi \hbar)^3 ck}}$, where
$\mathbf{k}$ is the momentum of the photon line attached to the
vertex.
\item[(VI)] Integrate the obtained expression by all independent external
momenta and loop momenta.
\end{itemize}

\subsection{Electron-electron scattering}
\label{ss:e-e-scatt}

Let us now try to extract  some physical information from the above
theory. We will  calculate low order terms in the perturbation
expansion (\ref{eq:7.63c}) for the $\Sigma$-operator

\begin{eqnarray}
\Sigma_1 &=&    V_1
\label{eq:9.73}  \\
\Sigma_2 &=&    (V _1 \underline{V _1})^{unp}
  + (V _1 \underline{V _1})^{ph}
  + (V _1\underline{V _1})^{ren}
\label{eq:9.74}
\end{eqnarray}

\noindent To obtain corresponding contributions to the
$S$-operator we need to take $t$-integrals

\begin{eqnarray*}
S = 1 + \underbrace{\Sigma_1 } + \underbrace{\Sigma_2 } + \ldots
\end{eqnarray*}

\noindent Note that the right hand side of (\ref{eq:9.73}) and the
first term on the right hand side of (\ref{eq:9.74}) are \emph{unphys}, so,
due to equation (\ref{eq:9.61}), they do not contribute to the
$S$-operator. For now, we also ignore the contribution of the \emph{renorm}
term in (\ref{eq:9.74}).\footnote{In fact, in a more consistent theory the \emph{renorm} term $(V_1 \underline{V _1 })^{ren}$ should have been canceled by a renormalization counterterm, as will be discussed in chapter
\ref{ch:renormalization}. \label{foot:renorm}} Then we
obtain in the 2nd perturbation order

\begin{eqnarray}
S_2 = \underbrace{(V _1  \underline{V _1 })^{ph}} + \ldots
\label{eq:9.75}
\end{eqnarray}

\noindent Operator $V_1 \underline{V}_1$ has several terms
corresponding to different scattering processes. Some of them were
calculated in subsection \ref{ss:draw-diagram}. For example,
terms of the type $a^{\dag} c^{\dag}ac $ (see Figs. \ref{fig:9.3}(c) and (f))
annihilate an electron and a photon in the initial state and
recreate them (with different momenta) in the final state. So, these
terms describe the electron-photon (Compton) scattering.
\index{Compton scattering} Let us consider in more detail the
electron-electron scattering term $a^{\dag} a^{\dag}aa $  described
by the diagram in Fig. \ref{fig:9.3}(e).  According to the rules
(I) - (VII), this diagram can be written algebraically as

\begin{eqnarray*}
 Fig. \ref{fig:9.3}(e)
&=&  \frac{\hbar^2 e^2 c}{ (2 \pi \hbar)^3 }
  \int
 \frac{  d \mathbf{p} d \mathbf{q}  d \mathbf{k}} {k(ck +
\omega_{\mathbf{p-k}} - \omega_{\mathbf{p}})}
 a^{\dag}_{\mathbf{p-k}}
a^{\dag}_{\mathbf{q+k}} a_{\mathbf{p}} a_{\mathbf{q}}
\end{eqnarray*}

\noindent The $t$-integral of this expression is

\begin{eqnarray}
&\mbox{ }&  \underbrace{Fig. \ref{fig:9.3}(e) } \nonumber \\
&=&
 -\frac{2 \pi i e^2 \hbar^2 c}{(2 \pi \hbar)^3} \int d \mathbf{p} d \mathbf{q}
d \mathbf{k}
 \frac{\delta (\omega_{\mathbf{p-k}} + \omega_{\mathbf{q+k}}
-\omega_{\mathbf{q}}- \omega_{\mathbf{p}})} {k(ck +
\omega_{\mathbf{p-k}} - \omega_{\mathbf{p}})}
 a^{\dag}_{\mathbf{p-k}}
a^{\dag}_{\mathbf{q+k}} a_{\mathbf{p}} a_{\mathbf{q}} \nonumber \\
\label{eq:9.76}
\end{eqnarray}

\noindent  The delta function in (\ref{eq:9.76}) expresses the conservation
of
energy in the scattering process. We will also say that expression
(\ref{eq:9.76})
is non-zero only on the \emph{energy shell}, which is defined as a solution of the
equation

\begin{eqnarray*}
 \omega_{\mathbf{p-k}} + \omega_{\mathbf{q+k}}
= \omega_{\mathbf{q}} + \omega_{\mathbf{p}}
\end{eqnarray*}

   In the non-relativistic approximation ($p, q \ll
mc$)

\begin{eqnarray}
\omega_{\mathbf{p}} &\equiv& \sqrt{p^2 c^2 + m^2 c^4}  \approx mc^2
+ \frac{p^2}{2m} \label{eq:omega-non-rel}
\end{eqnarray}

\noindent Then in the limit of small momentum transfer ($k \ll mc$)
the denominator in (\ref{eq:9.76})
 can
be approximated as

\begin{eqnarray}
k\left(ck + \omega_{\mathbf{p-k}} - \omega_{\mathbf{p}}\right)
&\approx& k\left(ck + mc^2 + \frac{(\mathbf{p-k})^2}{2m} - mc^2 -
\frac{p^2}{2m}\right)
\nonumber
\\ &\approx& ck^2 \label{eq:9.77}
\end{eqnarray}

\noindent Substituting this result in (\ref{eq:9.76}), we obtain the
second order contribution to the $S$-operator

\begin{eqnarray}
S_2[a^{\dag}a^{\dag} aa] \approx  -\frac{i e^2 }{ 4 \pi^2 \hbar}
 \int d \mathbf{p} d \mathbf{q}  d \mathbf{k}
 \frac{\delta (\omega_{\mathbf{p-k}} + \omega_{\mathbf{q+k}}
-\omega_{\mathbf{q}}- \omega_{\mathbf{p}})} {k^2}
 a^{\dag}_{\mathbf{p-k}}
a^{\dag}_{\mathbf{q+k}} a_{\mathbf{q}} a_{\mathbf{p}} \nonumber \\
\label{eq:9.78}
\end{eqnarray}

\subsection{Effective potential}
\label{ss:effective}

We have discussed in subsection \ref{ss:scatt-equiv} that the same
scattering matrix can correspond to different total Hamiltonians
with different interaction potentials.  Here we would like to demonstrate this idea by
constructing the following 2nd order \emph{effective interaction} between our model
electrons

\begin{eqnarray}
V_{2eff}  =  \frac{e^2 }{ (2 \pi)^3 \hbar}
 \int
 \frac{d \mathbf{p} d \mathbf{q}  d \mathbf{k}} {k^2}
 a^{\dag}_{\mathbf{p-k}}
a^{\dag}_{\mathbf{q+k}} a_{\mathbf{q}} a_{\mathbf{p}}
\label{eq:9.78x}
\end{eqnarray}

\noindent With this interaction the 2nd order amplitude for
electron-electron scattering is obtained by the usual formula
(\ref{eq:7.63a}), keeping only the first term in $F$

\begin{eqnarray*}
S_{2eff} &=&
\underbrace{V_{2eff}} \nonumber \\
&=& -\frac{i e^2 }{ 4 \pi^2 \hbar}
 \int d \mathbf{p} d \mathbf{q}  d \mathbf{k}
 \frac{\delta (\omega_{\mathbf{p-k}} + \omega_{\mathbf{q+k}}
-\omega_{\mathbf{q}}- \omega_{\mathbf{p}})} {k^2}
 a^{\dag}_{\mathbf{p-k}}
a^{\dag}_{\mathbf{q+k}} a_{\mathbf{q}} a_{\mathbf{p}}
\end{eqnarray*}

\noindent This is the same result as (\ref{eq:9.78}), in spite of
the fact that the new \emph{effective} Hamiltonian $H_0 + V_{2eff}$ is completely different
from the original Hamiltonian (\ref{eq:full-ham}). In
particular, it is important to note that interaction
(\ref{eq:9.78x}) is \emph{phys}, while (\ref{eq:9.69}) is \emph{unphys}. The replacement of an \emph{unphys} interaction with a scattering-equivalent effective \emph{phys} potential is the central idea of the ``dressed particle'' approach to quantum field theory that will be introduced in chapter \ref{ch:rqd}.

From equations (\ref{eq:9.67}) and (\ref{eq:A.90}) it follows that   interaction (\ref{eq:9.78x}) corresponds to the
ordinary position-space Coulomb potential\footnote{see also equation (\ref{eq:wmatr})} \index{Coulomb
potential}

\begin{eqnarray}
w(\mathbf{r}) &=& \frac{e^2 }{ (2 \pi)^3 \hbar} \int d\mathbf{k}
\frac{e^{\frac{i}{\hbar}\mathbf{k}\mathbf{r}}}{k^2} = \frac{e^2}{4
\pi r} \label{eq:coulomb}
\end{eqnarray}

\noindent So, our toy model is quite realistic.

\section{Diagrams in a general theory} \label{ss:diagrams-general}

\subsection{Properties of products and commutators}
\label{ss:diagrams}

The diagrammatic approach developed for the toy model above can be
easily extended to  interactions in the general form
(\ref{eq:9.48}): Each potential
 $V_{NM}$ with $N$ creation operators and $M$ annihilation operators
  can be represented by a vertex
 with $N$ outgoing and $M$ incoming lines.
  In calculations of scattering operators
(\ref{eq:7.63b}) and (\ref{eq:7.63c}) we  meet products of such
potentials.\footnote{$\mathcal{V}$ is the number of potentials in
the product.}

 \begin{eqnarray}
Y = V^{(1)}V^{(2)} \ldots V^{(\mathcal{V})} \label{eq:9.80}
\end{eqnarray}

\noindent As explained in subsection \ref{ss:normal-ordering}, we
should bring these products to the normal order. The normal ordering
transforms (\ref{eq:9.80}) into a sum of terms $y^{(j)}$

 \begin{eqnarray}
Y = \sum_{j} y^{(j)}
\label{eq:9.81}
\end{eqnarray}

\noindent each of which can be described by a
  diagram with $\mathcal{V}$ vertices.

Each potential $V^{(i)}$ in the product (\ref{eq:9.80}) is of the standard form (\ref{eq:9.49}) has $N^{(i)}
$ creation operators, $M^{(i)} $ annihilation operators, and $N^{(i)}
+ M^{(i)} $ momentum integrals. Then each term  $y^{(j)}$ in the expansion
(\ref{eq:9.81}) has

\begin{equation}
\mathcal{N} = \sum_{i=1}^{\mathcal{V}} (N^{(i)} + M^{(i)})
\label{eq:n-lines}
\end{equation}

\noindent  integrals and independent momentum integration variables. This
term also has a product of $\mathcal{V}$ delta functions, which express the
conservation of the total momentum in each of the factors $V^{(i)}$.
In the process of normal ordering of the product (\ref{eq:9.80}), a certain number
of pairs of external lines in the factors $V^{(i)}$
 have to be joined
together to make  internal lines and to introduce additional
 delta-functions. Let us denote by $\mathcal{I}$ the number of such delta functions in $y^{(j)}$. Then the total number of delta
functions in $y^{(j)}$ is

\begin{equation}
N_{\delta} = \mathcal{V} + \mathcal{I} \label{eq:n-delta}
\end{equation}

\noindent and the number of external lines is

\begin{equation}
\mathcal{E} = \mathcal{N} - 2\mathcal{I} \label{eq:n-ext}
\end{equation}

\noindent The terms
 $y^{(j)}$  in the normally ordered product (\ref{eq:9.81}) can be either \emph{disconnected}
 \index{disconnected diagram}  or \emph{connected}.
\index{connected diagram} In the latter case there is a continuous
sequence (path) of internal lines connecting  any two vertices. In the former case such a path does not exist, and the diagram splits into several separated pieces.

For illustration consider a product  of two potentials  $V^{(1)} = V_1 $ and $
V^{(2)} = \underline{V}_1$ from our example (\ref{eq:9.71}). The expansion of this product into a sum of normally ordered terms is shown diagrammaticaly in Figures \ref{fig:9.3}(d) - (g)

\begin{eqnarray}
V^{(1)} V^{(2)} = \sum_{j} y^{(j)} \label{eq:9.83a}
\end{eqnarray}

\noindent There is only one disconnected
term \ref{fig:9.3}(d) in the sum on the right hand side. Let us denote this term
$y^{(0)} \equiv (V^{(1)} V^{(2)})_{disc}$. This is the term in which
the factors from the original product are simply rearranged and no
pairings are introduced. All other terms
$y^{(1)}, y^{(2)}, \ldots $ on the right hand side of
(\ref{eq:9.83a})(d) are connected, because they have at least one
pairing. These pairings are  represented in diagrams \ref{fig:9.3}(e) - (g) by one or more internal lines
connecting vertices $V^{(1)}$ and $ V^{(2)} $.

\begin{lemma}
The disconnected part of a product of two connected bosonic
operators\footnote{As discussed in subsection \ref{ss:inter-oper},
all potentials considered in this book are bosonic.} does not depend
on the order of the product

\begin{eqnarray}
(V^{(1)} V^{(2)})_{disc} =  (V^{(2)}V^{(1)})_{disc} \label{eq:9.84}
\end{eqnarray} \label{Lemma-bosonic}
\end{lemma}
\begin{proof}
Operators $V^{(1)} V^{(2)}$ and $V^{(2)} V^{(1)}$  differ only by
the order of particle operators. So, after all particle operators
are brought to the normal order in $(V^{(1)} V^{(2)})_{disc}$ and
$(V^{(2)}V^{(1)})_{disc}$, they may differ, at most, by a sign. So, our goal is to show that this sign is plus $(+)$. Any reordering of boson particle operators does not affect the sign of an expression, so for our proof
we do not need to pay attention to creation and annihilation operators of bosons in the two factors. Let us now focus only on fermion particle operators
in $V^{(1)}$ and $V^{(2)}$. For simplicity, we will assume that only
electron and/or positron particle operators are present in $V^{(1)}$
and $V^{(2)}$. The inclusion of the proton and antiproton operators
will not change anything in this proof, except its length. For the
two factors $V^{(i)}$ (where $i =1,2$)  let us denote $N_e^{(i)}$
the numbers of electron creation operators, $N_p^{(i)}$ the numbers
of positron creation operators, $M_e^{(i)}$ the numbers of electron
annihilation operators, and $M_p^{(i)}$ the numbers of positron
annihilation operators. Taking into account that $V^{(i)}$ are
assumed to be normally ordered, we may formally write

\begin{eqnarray*}
V^{(1)} &\propto& [N_e^{(1)}] [ N_p^{(1)}] [ M_e^{(1)}][ M_p^{(1)} ] \\
V^{(2)} &\propto& [ N_e^{(2)}] [ N_p^{(2)}] [ M_e^{(2)}][ M_p^{(2)}
]
\end{eqnarray*}

\noindent where the bracket $[N_e^{(1)}]$ denotes the product of
$N_e^{(1)}$ electron creation operators from the term $V^{(1)}$, the
bracket $[N_p^{(1)}]$ denotes the product of $N_p^{(1)}$ positron
creation operators from the term $V^{(1)}$, etc. Then

\begin{eqnarray}
V^{(1)}V^{(2)} &\propto& [ N_e^{(1)}] [ N_p^{(1)}] [ M_e^{(1)}][
M_p^{(1)} ] [ N_e^{(2)}] [ N_p^{(2)}] [ M_e^{(2)}][ M_p^{(2)} ]
\label{eq:9.85}\\
V^{(2)} V^{(1)}  &\propto& [ N_e^{(2)}] [ N_p^{(2)}] [ M_e^{(2)}][
M_p^{(2)} ] [ N_e^{(1)}] [ N_p^{(1)}] [ M_e^{(1)}][ M_p^{(1)} ]
\label{eq:9.86}
\end{eqnarray}

\noindent Let us now bring particle operators on the right hand
side of (\ref{eq:9.86}) to the same order as on the right hand side
of (\ref{eq:9.85}). First we move $N_e^{(1)}$ electron creation
operators to the leftmost position in the product. This involves
$N_e^{(1)}M_e^{(2)}$ permutations with electron annihilation
operators from the factor $V^{(2)}$ and $N_e^{(1)}N_e^{(2)}$
permutations with electron creation operators from the factor
$V^{(2)}$. Each of these permutations changes the sign of the
disconnected term, so the acquired factor is $(-1)
^{N_e^{(1)}(N_e^{(2)} +M_e^{(2)})}$.

Next we need to move the $[ N_p^{(1)}]$ factor to the second
position from the left. The factor acquired after this move is $(-1)
^{N_p^{(1)}(N_p^{(2)} +M_p^{(2)})}$. Then we move the factors $[
M_e^{(1)}]$ and $[ M_p^{(1)}]$  to the third and fourth places in
the product, respectively. Finally, the total factor acquired by the
expression $(V^{(2)} V^{(1)})_{disc}$ after all its terms are
rearranged in the same order as in $(V^{(1)} V^{(2)})_{disc}$ is

\begin{eqnarray}
f &=& (-1)^{K_e^{(1)}K_e^{(2)}   + K_p^{(1)}K_p^{(2)}}
\label{eq:power-1}
\end{eqnarray}

\noindent where we denoted

\begin{eqnarray*}
K_e^{(i)} &\equiv& N_e^{(i)} + M_e^{(i)} \\
K_p^{(i)} &\equiv& N_p^{(i)} + M_p^{(i)} \\
\end{eqnarray*}

\noindent the total (= creation + annihilation) numbers of electron
and positron operators, respectively, in the factor $V^{(i)} $. Next, let us show that the power of (-1) in (\ref{eq:power-1}) is
even, so that $f = 1$. Indeed, consider the case when $K_e^{(1)}$ is
even and $K_e^{(2)}$ is odd. Then the product $K_e^{(1)}K_e^{(2)}$
is odd. From the bosonic character of $V^{(1)}$ and $V^{(2)}$ it
follows that $K_e^{(1)}+ K_p^{(1)} $ and  $K_e^{(2)}+ K_p^{(2)}$ are
even numbers. Therefore $K_p^{(1)}$ is odd and $K_p^{(2)}$ is even,
so that the product $K_p^{(1)}K_p^{(2)}$ is odd and the total power
of (-1) in (\ref{eq:power-1}) is even.

The same result is obtained for any other assumption about the
even/odd character of $K_e^{(1)}$ and $K_e^{(2)}$. This proves
(\ref{eq:9.84}).
\end{proof}

\bigskip
\begin{theorem}
\label{Theorem9.9} A multiple commutator of bosonic
potentials is connected.
\end{theorem}
\begin{proof}
    Let us first consider a single
commutator of two potentials  $V^{(1)}$ and $ V^{(2)}$.

\begin{eqnarray}
V^{(1)} V^{(2)} -  V^{(2)}V^{(1)} \label{eq:9.83}
\end{eqnarray}

\noindent  According to Lemma \ref{Lemma-bosonic}, the disconnected
terms $(V^{(1)} V^{(2)})_{disc}$  and $(V^{(2)} V^{(1)})_{disc}$ in
the commutator (\ref{eq:9.83}) are canceled. All other terms in the commutator
are
 connected. This proves the theorem for a single commutator (\ref{eq:9.83}).
Since this commutator is also bosonic, repeating the above
arguments by induction, we conclude that all multiple commutators of bosonic
operators  are connected.
\end{proof}
\bigskip

\begin{lemma}
In a connected diagram the number of independent loops is\footnote{Here $\mathcal{V}$ is the number of vertices  and $\mathcal{I}$ is the number of internal lines in the diagram.}

\begin{equation}
\mathcal{L} = \mathcal{I}- \mathcal{V}+1 \label{eq:n-loops}
\end{equation}

\end{lemma}
\begin{proof}
 If
there are $\mathcal{V}$ vertices, they can be connected together
without making loops by $\mathcal{V}-1$ internal lines. Each
additional internal line will make one independent loop. Therefore,
the total number of independent loops is
$\mathcal{I}-(\mathcal{V}-1)$.
\end{proof}
\bigskip

\begin{figure}
\centering
\includegraphics{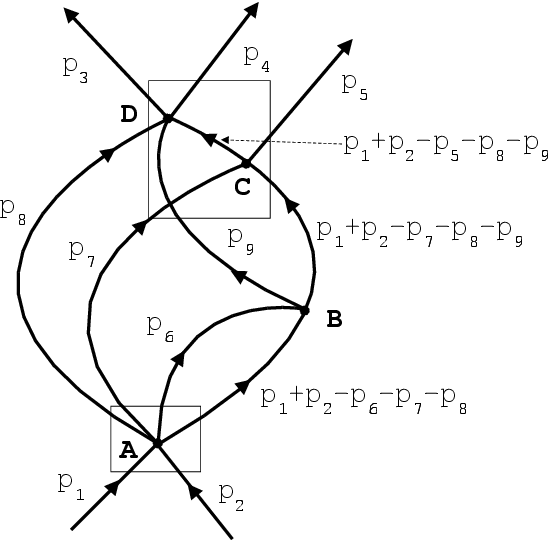} \caption{ A diagram representing one term
in the 4-order product of a hypothetical theory. Here we do not draw
the order bars as in subsection \ref{ss:draw-diagram}. However, we draw all outgoing lines on the top of the diagram and all incoming lines at the bottom to indicate that the diagram is
normally ordered. Note that all internal lines are oriented upwards
because all
 paired operators (i.e., those operators whose order should be changed by the normal ordering procedure)
 in the product (\ref{eq:9.80}) always occur in the order $\alpha \alpha^{\dag}$.}
 \label{fig:9.4}
\end{figure}

 An example of  a connected
diagram is shown in   Fig. \ref{fig:9.4}.  This diagram has
$\mathcal{V}=4$ vertices, $\mathcal{E}=5$ external lines,
$\mathcal{I}=7$ internal lines, and $\mathcal{L}= 4$ independent
loops. This diagram describes a nine-fold momentum integral.  Five integration momentum variables  correspond to
 external lines in the diagram: These are two incoming momenta $\mathbf{p}_1$, $\mathbf{p}_2$, and three outgoing momenta
$\mathbf{p}_{3}$,
 $\mathbf{p}_{4}$ and $\mathbf{p}_{5}$. These five integrals/variables are a part of the general expression for
the potential (\ref{eq:9.49}). Four additional integrals are performed by loop momenta
$\mathbf{p}_6$,
$\mathbf{p}_7$, $\mathbf{p}_{8}$, and $\mathbf{p}_{9}$. These
integrals can be absorbed  in the definition of
 the coefficient
function

\begin{eqnarray}
&\mbox{ }& D_{3,2}(\mathbf{p}_3, \mathbf{p}_4,
\mathbf{p}_5;\mathbf{p}_1, \mathbf{p}_2) = \nonumber \\
&\mbox{ }&
\int d\mathbf{p}_6 d\mathbf{p}_7 d\mathbf{p}_8 d\mathbf{p}_9
D_A(\mathbf{p}_6, \mathbf{p}_7, \mathbf{p}_8,
\mathbf{p}_1 + \mathbf{p}_2 - \mathbf{p}_6 - \mathbf{p}_7 -
\mathbf{p}_8; \mathbf{p}_1, \mathbf{p}_2) \times \nonumber \\
&\mbox{ }& D_B(\mathbf{p}_9, \mathbf{p}_1 + \mathbf{p}_2 -
\mathbf{p}_7
- \mathbf{p}_8 - \mathbf{p}_9; \mathbf{p}_6, \mathbf{p}_1 +
\mathbf{p}_2
- \mathbf{p}_6 - \mathbf{p}_7 - \mathbf{p}_8)  \times \nonumber \\
&\mbox{ }& D_C(\mathbf{p}_5, \mathbf{p}_1 + \mathbf{p}_2 -
\mathbf{p}_5
- \mathbf{p}_8 - \mathbf{p}_9; \mathbf{p}_7, \mathbf{p}_1 +
\mathbf{p}_2
- \mathbf{p}_7 - \mathbf{p}_8 - \mathbf{p}_9)
\times \nonumber \\
&\mbox{ }& D_D(\mathbf{p}_3,\mathbf{p}_4;\mathbf{p}_8,\mathbf{p}_9,
\mathbf{p}_1+\mathbf{p}_2-\mathbf{p}_5-\mathbf{p}_8-\mathbf{p}_9)
\frac{1}{E_A (E_C + E_D)}
\label{eq:9.82}
\end{eqnarray}

\noindent where $E_A$, $E_C$, and $E_D$ are energy functions of the
corresponding vertices and $D_A, D_B, D_C, D_D$ are coefficient functions at the vertices.

\subsection{Cluster separability of the $S$-operator}
\label{ss:linkedness}

In agreement with Postulate \ref{postulateS} (general cluster separability)
and Statement \ref{statementT} (cluster separability of smooth
potentials), interactions considered in this book are cluster-separable. In this subsection we are going to prove that the $S$-operator calculated with such interactions is always cluster-separable too. Physically, this means that if a multiparticle scattering system is separated into two (or more) distant subsystems, then the result of scattering in each subsystem will not depend on what is going on in other subsystem(s).

From perturbation formulas for the $S$-operator in subsection \ref{ss:perturbation} we know that, generally,  $S$ is a sum of products of interaction potentials like (\ref{eq:9.80}).
Mathematically, the cluster separability of interactions means  that coefficient functions of interaction
potentials $V^{(i)}$ in the product (\ref{eq:9.80}) are smooth. If we could show that the product (\ref{eq:9.80}) itself  is a sum of smooth operators, then the desired cluster separability of the $S$-operator would follow directly from  Statement \ref{statementT}.
The question about the smoothness of (\ref{eq:9.80}) is  not trivial, because bringing such a product
 to the normal order involves permutations of particle operators that produce singular
delta functions.

The following theorem establishes an important connection between
the smoothness of terms on the right hand side of (\ref{eq:9.81})
and the connectivity of corresponding diagrams.

\bigskip

\begin{theorem}
 \label{Theorem9.8} Each term $y^{(j)}$ in the expansion (\ref{eq:9.81})
of the product of smooth
 potentials is smooth if and only if it
is represented by a connected diagram.
\end{theorem}
\begin{proof} Let us first assume that $y^{(j)}$ is represented by a connected diagram. We will establish the smoothness of the term  $y^{(j)}$
by proving that it can be represented in the general form
(\ref{eq:9.49}) in which  the integrand contains only one delta
function required by the momentum conservation condition and the
coefficient function $D_{NM}$ is smooth.\footnote{This means that all other delta functions can be integrated out. Note also that singularities present in $y^{(j)}$  due to energy
denominators resulting from $t$-integrals (\ref{eq:underline2}) can be
made harmless by employing the ``adiabatic switching''  trick from
subsection \ref{ss:adiaba}. This trick essentially results in
adding small imaginary contributions to each denominator, which
remove the singularities. \label{foot:harmless}} From equation
(\ref{eq:n-lines}), the original number of integrals in $y^{(j)}$ is
$\mathcal{N}$. Integrals corresponding to $\mathcal{E}$ external
lines are parts of the general form (\ref{eq:9.49}), and  integrals
corresponding to $\mathcal{L}$ loops can be absorbed into the
definition of the coefficient function of $y^{(j)}$. The number of
remaining integrals is then obtained from (\ref{eq:n-ext}) and (\ref{eq:n-loops})

\begin{eqnarray}
\mathcal{N}' &=& \mathcal{N} - \mathcal{E} - \mathcal{L} =
\mathcal{I}+ \mathcal{V}-1 \label{eq:n-prime}
\end{eqnarray}

\noindent This is  just enough integrals to cancel all momentum
delta functions (\ref{eq:n-delta}) except one, which proves that the
term $y^{(i)}$ is smooth.

Inversely, suppose that the term $y^{(j)}$ is represented by a
disconnected diagram with $\mathcal{V}$ vertices and $\mathcal{I}$
internal lines. Then the number of independent loops $\mathcal{L}$
is greater than the value $\mathcal{I}- \mathcal{V}+1$
characteristic for connected diagrams. Then the number of
integrations $\mathcal{N}'$ in equation (\ref{eq:n-prime}) is less than
$\mathcal{I}+ \mathcal{V}-1$ and the number of delta functions
remaining in the integrand $\mathcal{N}' - \mathcal{N}_{\delta}$ is
greater than 1. This means that the term $y^{(j)}$ is represented by
expression (\ref{eq:9.49}) whose coefficient function is singular,
therefore the corresponding operator is not smooth.
\end{proof}
\bigskip

Theorem \ref{Theorem9.8} establishes that smooth operators are
represented by connected diagrams and \emph{vice versa}. In what
follows, we will use the terms \emph{smooth} \index{smooth operator}
\index{smooth potential} and \emph{connected} \index{connected
operator} \index{connected diagram} as synonyms, when applied to
operators.

Putting together Theorems \ref{Theorem9.9} and \ref{Theorem9.8} we
immediately obtain the following important

\begin{theorem}
All terms in a normally ordered multiple commutator of smooth
bosonic potentials are smooth. \label{theorem:linked-comm}
\end{theorem}

This theorem allows us to apply the property of cluster separability
to the $S$-operator. Let us write the $S$-operator in the form (\ref{eq:7.63a})

\begin{eqnarray}
S =  e^{ \underbrace{F}} \label{eq:9.87}
\end{eqnarray}

\noindent where $F$ is a series of multiple commutators
(\ref{eq:7.63b}) of smooth bosonic potentials in $V$. According to Theorem
\ref{theorem:linked-comm}, operators $F$ and $\underbrace{F}$ are
also smooth. Then, according to Statement \ref{statementT}, operator $\underbrace{F}$ is cluster
separable, and if all particles are divided into two spatially
separated groups 1 and 2, the argument of the exponent in
(\ref{eq:9.87}) takes the form of a sum

 \begin{eqnarray*}
\underbrace{F} \to \underbrace{F^{(1)}} + \underbrace{F^{(2)}}
\end{eqnarray*}

\noindent where $\underbrace{F^{(1)}}$ acts only on variables in the
group 1, and $\underbrace{F^{(2)}}$ acts only on variables in the
group 2. So, these two operators commute with each other, and  the
$S$-operator separates into the  product of two independent factors

 \begin{eqnarray*}
S &\to& \exp(\underbrace{F^{(1)}} + \underbrace{F^{(2)}}) =
\exp(\underbrace{F^{(1)}}) \exp( \underbrace{F^{(2)}}) = S^{(1)}S^{(2)}
\end{eqnarray*}

\noindent This relationship expresses the cluster separability of the
$S$-operator and the $S$-matrix:
The total scattering amplitude for
spatially separated events is given by the product of individual
amplitudes.

\subsection{Divergence of loop integrals}
\label{ss:convergence}

In the preceding subsection we showed that $S$-operator terms described by
connected diagrams are smooth. However, such terms involve loop
integrals, and generally there is no guarantee that these integrals
converge. This problem is evident in  our toy model: the loop
integral by $\mathbf{k}$ in diagram \ref{fig:9.3}(g) is
divergent

\begin{eqnarray}
(V_1  \underline{V_1 })^{ren} = -\frac{e^2 \hbar^2 c}{(2 \pi \hbar)^3}
\int
d\mathbf{p} d\mathbf{k}
 \frac{a^{\dag}_{\mathbf{p}}a_{\mathbf{p}}}
{(\omega_{\mathbf{p}- \mathbf{k}} -\omega_{\mathbf{p}}  + ck) k}
\label{eq:9.88}
\end{eqnarray}

\noindent Substituting this result to the right hand side of
(\ref{eq:9.74}) we see that the $S$-operator in the second order
 $\underbrace{\Sigma_2 }$ is
infinite, which makes it meaningless and unacceptable.

The appearance of divergences in perturbative formulas for the $S$-operator  is a commonplace in quantum field theories. So, we
need to  understand this phenomenon better. In this subsection, we
will
formulate a sufficient condition under which loop integrals are
convergent. We will find this result useful in our discussion of
the renormalization of QED in chapter \ref{ch:renormalization} and
in our construction of a divergence-free theory in section
\ref{sc:dressed}.

Let us consider, for example, the diagram in Fig. \ref{fig:9.4}.
There are three different reasons why loop integrals may diverge
there: \index{loop}

\begin{itemize}
\item[(I)] The coefficient functions $D_A$, $D_B$, $\ldots$ of interaction vertices in
(\ref{eq:9.82}) may
contain singularities. One example is interaction (\ref{eq:9.69}), which is singular at $\mathbf{k} = \mathbf{0}$. Such
singularities are usually related to the vanishing photon mass. They correspond to the so-called \emph{infrared divergences} of loop integrals. We will discuss them in greater detail in chapters \ref{ch:renormalization} and \ref{ch:hydrogen-revisited}.
\item[(II)] There can be also singularities due to zeroes in energy
denominators
$E_A$ and $E_C + E_D$. The energy denominators
may be rendered finite and harmless if we use the adiabatic switching
prescription from subsection \ref{ss:adiaba}.\footnote{see also footnote on page \pageref{foot:harmless}}
\item[(III)] The coefficient functions $D_A$, $D_B$, $\ldots$ may
not decay fast enough at large values of loop momenta, so that the
integrals may be divergent due to the infinite integration range. These  \emph{ultraviolet
divergences} \index{ultraviolet divergences} present
more serious problems, which we are going to discuss here in some detail.
\end{itemize}

\noindent    In particular, we would like to prove the
following

\bigskip

\begin{theorem} \label{Theorem9.10} If coefficient functions of potentials
decay sufficiently rapidly (e.g., exponentially) when arguments move
away from the energy shell, then all loop integrals converge.
\end{theorem}
\begin{proof}   [Idea of the proof] Equation (\ref{eq:9.82}) is
an integral in a 12-dimensional space of 4 loop momenta
$\mathbf{p}_{6}$, $\mathbf{p}_{7}$, $\mathbf{p}_{8}$, and
$\mathbf{p}_{9}$. Let us denote this space $\Xi$.
 Consider for example the dependence of the integrand in (\ref{eq:9.82}) on
 the loop momentum $\mathbf{p}_9$
as $\mathbf{p}_9 \to \infty$  and all other momenta fixed.
 Note that we have chosen integration variables
in Fig. \ref{fig:9.4} in such a way that each loop momentum is
present only in the internal lines forming the corresponding loop,
e.g., momentum $\mathbf{p}_{9}$ is confined to the loop $BDCB$, and
the energy function $E_A$ of the vertex $A$ does not depend on
$\mathbf{p}_9$. Such a selection of integration variables can be
done for any arbitrary diagram. Taking into account that  at large
values of momentum $\omega_{\mathbf{p}} \approx cp $,
  we obtain in the limit $\mathbf{p}_9 \to \infty$

\begin{eqnarray*}
E_A &\to& const, \\
 E_B &=&
\omega_{\mathbf{p}_1+\mathbf{p}_2-\mathbf{p}_7-\mathbf{p}_8-\mathbf{p}_9}
+ \omega_{\mathbf{p}_9} - \omega_{\mathbf{p}_6} -
\omega_{\mathbf{p}_1+\mathbf{p}_2-\mathbf{p}_6-\mathbf{p}_7-\mathbf{p}_8}
\\ &\approx& 2cp_9 \to \infty, \\ E_C &=&
\omega_{\mathbf{p}_1+\mathbf{p}_2-\mathbf{p}_5-\mathbf{p}_8-\mathbf{p}_9}
+ \omega_{\mathbf{p}_5} - \omega_{\mathbf{p}_7} -
\omega_{\mathbf{p}_1+\mathbf{p}_2-\mathbf{p}_7-\mathbf{p}_8-\mathbf{p}_9}
\\ &\to& const, \\ E_D &=& \omega_{\mathbf{p}_3} +
\omega_{\mathbf{p}_4} - \omega_{\mathbf{p}_8} -
\omega_{\mathbf{p}_9} -
\omega_{\mathbf{p}_1+\mathbf{p}_2-\mathbf{p}_5-\mathbf{p}_8-\mathbf{p}_9}
\\
&\approx& - 2cp_9 \to \infty.
\end{eqnarray*}

\noindent So, in this limit, according to the condition of the
theorem, the coefficient functions at vertices $B$ and $D$ tend to
zero rapidly, e.g., exponentially. So, the loop integral on $\mathbf{p}_9$ converges. In order to prove the convergence
of all four loop integrals, we need to make sure that the same rapid
decay is characteristic for all directions in the space $\Xi$. Here are some arguments that this is, indeed, true.

The above analysis is applicable to all loop variables: Any loop has
a \emph{bottom vertex} (vertex $B$ in our example), a \emph{top
vertex} (vertex $D$ in our example) and possibly a number of
\emph{intermediate vertices} (vertex $C$ in our example). As the
loop momentum goes to infinity, energy functions of the top and
bottom vertices tend to infinity, i.e., move away from the energy
shell. This ensures a fast (e.g., exponential) decay of the corresponding coefficient
function.

Now we can take an arbitrary direction to infinity in the space
$\Xi$. Along this direction, there is at least one loop momentum
which goes to infinity. Then there is at least one energy function
($E_A$, $E_B$, $E_C$, or $E_D$) which grows linearly, while others
stay constant (in the worst case). Therefore, according to the
condition of the theorem, the integrand decreases rapidly (e.g.,
exponentially) along this direction. Thus the integrand
rapidly tends to zero in all directions in $\Xi$, and
 integral (\ref{eq:9.82}) converges.
\end{proof}
\bigskip

In chapter \ref{ch:renormalization} we will see that in realistic theories, like
QED, the asymptotic decay of the coefficient functions of potentials
at large momenta is not fast
 enough, so Theorem \ref{Theorem9.10} is not applicable, and loop
integrals usually diverge.
 A detailed discussion of such divergences in quantum field theory and
their elimination  will be presented in chapter
\ref{ch:renormalization}, in section \ref{sc:dressed}, and in chapter \ref{ss:hydrogen}.

\chapter{QUANTUM ELECTRODYNAMICS}
\label{ch:QED}

\begin{quote}
\textit{If it turned out that some physical system could not
be described by a quantum field theory, it would
be a sensation; if it turned out that the system
did not obey the rules of quantum mechanics and
relativity, it would be a cataclysm.}

\small
\hspace{1in} Steven Weinberg
\normalsize
\end{quote}

\vspace{0.5in}

So far we have developed a general formalism of quantum theory in the Fock space. We
emphasized that any such theory must obey, at least, three important
requirements:

\begin{itemize}
\item the theory must be relativistically invariant, in the instant form of dynamics;
\item the interaction must be cluster separable;
\item the theory must allow for processes involving creation and
annihilation of particles.
\end{itemize}

We have considered a few model examples, but they were purely academic
and not directly relevant to real systems observed in nature. The
reason for such inadequacy was that our models failed to satisfy all
three requirements mentioned above simultaneously.

For example, in subsection \ref{ss:3-particle} we constructed an
interacting model that explicitly satisfied the requirement of
relativistic invariance. We also managed to ensure that the model is
cluster separable in the 3-particle sector. In principle,  by following this approach one can build
cluster-separable interactions in all $n$-particle sectors. There is
even a possibility for describing systems with variable number of
particles \cite{Polyzou-production}. However, the resulting
formalism is very cumbersome, and it can be applied only to model
systems.

In section \ref{sc:model-theory} we considered another toy model of
interacting particles, which was based on the formalism of creation
and annihilation operators. The great advantage of this formalism
was that the cluster separability condition could be conveniently
expressed in terms of smoothness of
interaction potentials.\footnote{see Statement \ref{statementT}} The
processes of particle creation and annihilation were easily
described as well. However, the difficult part was to ensure the
relativistic invariance. In our toy model we have not even tried to
make the theory relativistic.

Fortunately, there is a class of theories, which allow one
to satisfy all three conditions listed above. What is even more
important, these theories are directly applicable to realistic physical
systems and allow one to achieve an impressive agreement with
experiments. These are \emph{quantum field theories} (QFT).
\index{quantum field theory} A particular version of QFT for
describing interactions between electrically charged particles and
photons is called \emph{quantum electrodynamics} \index{quantum
electrodynamics} (QED). This is the topic of our discussion in the present
chapter.

In section \ref{sc:inter-qed} we will write down interaction terms
$V$ (potential energy) and $\mathbf{Z}$ (potential boost) in QED. The relativistic invariance of this approach will be proven in Appendix \ref{ss:relat-invar}. In
section \ref{ss:s-oper-qed} $S$-matrix elements will be calculated
in the lowest non-trivial order of perturbation theory.

\section{Interaction in QED}
\label{sc:inter-qed}

Our goal in this section is to build a realistic interacting representation
$U(\Lambda, a)$ of the Poincar\'e group in the Fock space
(\ref{eq:fock-space}). In this book we do not pretend to derive QED
interactions from first principles. We  simply borrow from the
traditional approach the form of four interacting Poincar\'e  generators
 $H$ and $\mathbf{K}$ in terms of quantum fields for
electrons/positrons $\psi_{\alpha}(\tilde{x})$, protons/antiprotons\footnote{Throughout this book, protons and antiprotons are treated as simple point charges. Their internal structures are disregarded as well as their participation in strong nuclear interactions.}
$\Psi_{\alpha}(\tilde{x})$, and photons  $A_{\mu}(\tilde{x})$. Definitions of quantum fields are given in  Appendices
\ref{sc:fermions} and \ref{sc:photons}.

At this point we do not offer any physical interpretation of quantum
fields. For us they are just abstract multicomponent functions from
the 4-dimensional Minkowski space-time $\mathcal{M}$ to operators in the Fock
space. In our approach, the only role of quantum fields is to provide convenient
``building blocks'' for the construction of Poincar\'e invariant
interactions $V$ and $\mathbf{Z}$. This attitude was inspired by a non-traditional way of
looking at quantum fields presented in Weinberg's book \cite{book}.
Also, we are not identifying coordinates $\mathbf{x}$ and $t$ in
$\mathcal{M}$ with positions and times of events measured in real
experiments. The space-time $\mathcal{M}$ will be understood as an
abstract 4-dimensional manifold with a pseudo-Euclidean metric. In
section \ref{ss:are-fields-meas} we will discuss in more detail the
meaning of quantum fields and  their
arguments $\tilde{x} \equiv (\mathbf{x}, t)$.

\subsection{Construction of simple quantum field theories} \label{ss:weinberg}

Before approaching QED we will do a warm-up exercise and build a class of simpler QFT theories, which would allow us to  demonstrate many important features characteristic for all QFT models. In simple QFT theories, relativistic interactions are constructed in
three steps \cite{book, Weinberg_lecture}:

\noindent \textbf{Step 1.} For each particle type\footnote{A
particle and its antiparticle are assumed to belong to the same
particle type.}
 participating in the theory we construct
a \emph{quantum field} \index{quantum field}  which is a
multicomponent operator-valued function\footnote{This means that for each value of its arguments $( \mathbf{x}, t)$ and index $i$, the symbol $\phi$ denotes an operator acting in the Fock space. } $\phi_i( \mathbf{x}, t)$
defined on an abstract Minkowski space-time $\mathcal{M}$\footnote{see
Appendix \ref{ss:4-dim-rep}} and satisfying  following
conditions:

\begin{itemize}

\item[(I)] Operator $\phi_i( \mathbf{x}, t)$  contains only  terms linear in creation or annihilation
operators of the particle and its antiparticle.

\item[(II)]
Quantum fields are supposed to have simple transformation laws

\begin{eqnarray}
U_0(\Lambda; \tilde{a}) \phi_i(\tilde{x}) U_0^{-1}(\Lambda; \tilde{a}) = \sum_j
D_{ij}(\Lambda^{-1}) \phi_j(\Lambda( \tilde{x} + \tilde{a})) \label{eq:10.1}
\end{eqnarray}

with respect to the non-interacting representation\footnote{see subsection
\ref{ss:non-int-rep}} $U_0(\Lambda; \tilde{a})$ of the Poincar\'e
group in the Fock space,  where $\Lambda$ is a boost/rotation, $a$ is a
space-time translation and $D_{ij}$ is a finite-dimensional
representation\footnote{The representation $D_{ij}$ is definitely
non-unitary, because the Lorentz group is non-compact and it is
known that non-compact groups cannot have finite-dimensional unitary
representations.} of the Lorentz group.

\item[(III)] Quantum fields turn to zero at $\mathbf{x}$-infinity, i.e.

\begin{eqnarray}
\lim_{|\mathbf{x}| \to \infty} \phi_i (\mathbf{x},t) =
 0
\label{eq:10.2}
\end{eqnarray}

\item[(IV)] Fermionic fields (i.e., fields for particles with half-integer spin)
 $\phi_i(\tilde{x})$ and $\phi_j(\tilde{x}')$
are required to \emph{anticommute} if $(\tilde{x}-\tilde{x}')$ is a space-like
4-vector,  or equivalently

\begin{eqnarray}
\{ \phi_i(\mathbf{x},t),  \phi_j(\mathbf{y},t) \} &=& 0 \mbox{   }
if \mbox{   } \mathbf{x} \neq \mathbf{y} \label{eq:10.3}
\end{eqnarray}

Fermionic quantum fields for electrons-positrons $\psi_{\alpha}(\tilde{x})$ and
protons-antiprotons $\Psi_{\alpha}(\tilde{x})$ are constructed and analyzed
in Appendix \ref{sc:fermions}.

\item[(V)] Bosonic  fields (i.e., fields for particles with integer spin or
helicity) at points $\tilde{x}$ and $\tilde{x}'$ are required to \emph{commute}  if
$(\tilde{x}-\tilde{x}')$ is a space-like 4-vector, or equivalently

\begin{eqnarray}
[ \phi_i(\mathbf{x},t),  \phi_j(\mathbf{y},t)] &=& 0 \mbox{   } if
\mbox{   }  \mathbf{x} \neq \mathbf{y} \label{eq:10.4}
\end{eqnarray}

Bosonic quantum field for photons $A_{\mu}(\tilde{x})$ is discussed in Appendix
\ref{sc:photons}.

\end{itemize}

\noindent \textbf{Step 2.} Having at our disposal quantum fields
$\phi_i(\tilde{x}), \psi_{j}(\tilde{x}), \chi_k (\tilde{x}), \ldots$
 for all particles we can build the \emph{potential energy density}
\index{potential energy density}

\begin{eqnarray}
V(\tilde{x}) \equiv V(\mathbf{x}, t) = \sum_n V_n(\mathbf{x}, t) \label{eq:10.4a}
\end{eqnarray}

\noindent in the form of a polynomial where each term is a product of
fields at the same $(\mathbf{x}, t)$ point

\begin{eqnarray}
V_n(\mathbf{x}, t) = \sum_{i,j,k, \ldots} G^n_{ijk \ldots}
\phi_i(\mathbf{x}, t) \psi_{j}(\mathbf{x}, t) \chi_k (\mathbf{x}, t)
\ldots \label{eq:10.5}
\end{eqnarray}

\noindent and coefficients $G^n_{ijk \ldots}$ are such that $V(\tilde{x})$

\begin{itemize}
\item[(I)] is a bosonic\footnote{i.e., there is an even number of fermionic
fields in each product in (\ref{eq:10.5})} Hermitian  operator
function on the space-time $\mathcal{M}$;

\item[(II)]  transforms
as a scalar with respect to the non-interacting representation of
the Poincar\'e group:

\begin{eqnarray}
U_0 (\Lambda; \tilde{a}) V(\tilde{x}) U_0^{-1} (\Lambda; \tilde{a}) = V(\Lambda \tilde{x} + \Lambda\tilde{a}) \label{eq:as-a-scalar}
\end{eqnarray}

\end{itemize}

From properties (\ref{eq:10.3}) - (\ref{eq:10.4}) and the bosonic
character of $V(\tilde{x})$ it is easy to prove that $V(\tilde{x})$ commutes with
itself at space-like separations, e.g.,

\begin{eqnarray}
[V(\mathbf{x}, t), V(\mathbf{y}, t)]  & =& 0 \mbox{   } if \mbox{ }
 \mathbf{x} \neq \mathbf{y}
\label{eq:10.6}
\end{eqnarray}

\noindent \textbf{Step 3.} Instant-form interactions in the Hamiltonian
and boost operator are obtained by integrating the potential energy
density (\ref{eq:10.4a}) on $\mathbf{x}$ and setting $t=0$

\begin{eqnarray}
H &=& H_0 + V =  H_0 + \int d\mathbf{x} V( \mathbf{x},0)
\label{eq:10.7} \\
\mathbf{K} &=& \mathbf{K}_0 + \mathbf{Z} =  \mathbf{K}_0  +
  \frac{1}{c^2}\int d\mathbf{x} \mathbf{x} V( \mathbf{x},0)
\label{eq:10.8}
\end{eqnarray}

 Non-interacting generators $H_0$, $\mathbf{P}_0$, $\mathbf{J}_0$, and $\mathbf{K}_0$ can be found in (\ref{eq:9.34}), (\ref{eq:9.35}), (\ref{eq:9.36a}), and (\ref{eq:9.36b}), respectively.  With these definitions,  the commutation relations of the
Poincar\'e Lie algebra are proved in Appendix \ref{ss:relat-invar-simple}.  Coefficient
functions of potentials in (\ref{eq:10.7}) are  smooth, so the cluster separability is established by reference to Statement \ref{statementT}. The terms that change the number of particles can be easily accomodated within the above 3-step approach. Then all three conditions listed in the beginning of this chapter are readily satisfied. This explains why quantum field theories are so useful for describing realistic physical systems.

\subsection{Interaction operators in QED}
\label{ss:interaction-qed}

 Unfortunately, formulas (\ref{eq:10.7}) and
(\ref{eq:10.8}) work only for simplest QFT models. More interesting
cases, such as QED,
 require
some modifications in this scheme. In particular,  the presence of
the additional term $ \Omega_{\mu} (\tilde{x}, \Lambda)$ in the
transformation law of the photon field (\ref{eq:10.35a}) does not
allow us to define the boost interaction in QFT by simple formula
(\ref{eq:10.8}). Let us now postulate QED interaction operators without proof.
\label{sc:qft-end}

 The total Hamiltonian of QED has the usual form

\begin{eqnarray}
   H = H_0 + V
\label{eq:11.4}
\end{eqnarray}

\noindent where the non-interacting Hamiltonian $H_0$ is that from
equation (\ref{eq:9.34}) and  interaction is composed of two
terms\footnote{Here and in what follows we denote the power of the
coupling constant $e$ (the perturbation order of an operator) by a
subscript, i.e., $H_0$ is zero order, $V_1$ is first order, $V_2$ is
second order, etc. }

\begin{eqnarray}
    V
 &=& V_1  + V_2
\label{eq:11.5}
\end{eqnarray}

\noindent The first order interaction is a pseudoscalar
product of two 4-component quantities. One of them is the 4-vector fermion current
operator $\tilde{j}(\tilde{x})$ defined in Appendix \ref{sc:current-dens}. The other is the photon
quantum field $\tilde{A}(\tilde{x})$\footnote{Here we mark the photon quantum field
$\tilde{A}$ by tilde as if it were a 4-vector. However, as shown in Appendix
\ref{ss:lorentz-photon}, the components of $\tilde{A}$ do
not transform by 4-vector rules. The last equality in (\ref{eq:11.6}) follows from equation
(\ref{eq:K13.a}).}

\begin{eqnarray}
V_1  &=& \frac{1}{c}  \int d\mathbf{x} \tilde{j}(\mathbf{x},0)
\cdot
 \tilde{A}( \mathbf{x}, 0)
 \equiv \frac{1}{c}\int d\mathbf{x} j_{\mu}( \mathbf{x},0) A^{\mu}( \mathbf{x},0) \nonumber \\
 &=& \frac{1}{c} \int d\mathbf{x} \mathbf{j}(
\mathbf{x},0) \cdot \mathbf{ A}( \mathbf{x},0) \label{eq:11.6}
\end{eqnarray}

\noindent The second order interaction  is

\begin{eqnarray}
 V_2 = \frac{1}{2c^2}\int d\mathbf{x} d\mathbf{y} \frac{j_0( \mathbf{x},0) j_0( \mathbf{y},0)}{8 \pi |\mathbf{x}-\mathbf{y}|}
\label{eq:11.7}
\end{eqnarray}

\noindent Interaction in the boost operator

\begin{eqnarray}
\mathbf{K}   = \mathbf{K}_0  + \mathbf{Z}
\label{eq:interact-boost}
\end{eqnarray}

\noindent is defined as

\begin{eqnarray}
 \mathbf{Z}
&=& \frac{1}{c^3}\int d\mathbf{x} \mathbf{x} \mathbf{j}(
\mathbf{x},0) \mathbf{ A}( \mathbf{x},0) +
       \frac{1}{2c^4} \int d\mathbf{x} d\mathbf{y} \frac{\mathbf{x} j_0( \mathbf{x},0) j_0( \mathbf{y},0)}{8 \pi |\mathbf{x}-\mathbf{y}|} \nonumber \\
 &\ & +\frac{1}{c^3} \int d\mathbf{x} j_0( \mathbf{x},0)
\mathbf{C} ( \mathbf{x},0) \label{eq:11.8}
\end{eqnarray}

\noindent where components of the operator function $\mathbf{C} (
\mathbf{x},t)$ are given by equation (\ref{eq:11.9}).

 The above operators of energy $H$ and  boost  $\mathbf{K}$ are
those usually written in the \emph{Coulomb gauge} \index{Coulomb
gauge} version of QED \cite{book, Weinberg_lecture}. In Appendix
\ref{ss:relat-invar} we prove that this theory is
Poincar\'e invariant. For this proof it is convenient to represent
interaction operators of QED in terms of quantum fields, as above.
However, for some calculations in this book it will be more
useful to express interaction $V$ through particle
creation and annihilation operators, as in chapter
\ref{ch:fock-space}. To do that, we just need to insert field expansions
(\ref{eq:10.10}) and (\ref{eq:10.26}) in equations (\ref{eq:11.6})
and (\ref{eq:11.7}). The resulting expressions are rather long and
cumbersome, so this derivation has been moved to Appendix
\ref{ss:int-part-oper}.

\section{$S$-operator in QED}
\label{ss:s-oper-qed}

Having at our disposal all 10 generators of the Poincar\'e group representation in the Fock space, in principle, we should be able to calculate all physical quantities related to systems of charged particles and photons. However, this statement is overly optimistic. In chapters \ref{ch:renormalization} and \ref{ch:rqd} we will see that the theory outlined above has serious problems and internal contradictions. In fact, this theory allows one to calculate only simplest physical properties only in low perturbation orders.  An example of such a calculation will be given in this section: Here we will  calculate the $S$-operator for the proton-electron
scattering in the 2nd perturbation order.

\subsection{$S$-operator in the second order }
\label{ss:2-nd-order}

We are interested in $S$-operator terms of the type $d^{\dag}a^{\dag}da$. It
will be convenient to start this calculation from expanding the
phase operator (\ref{eq:7.63b}) in powers of the coupling constant

\begin{eqnarray}
F  &=& F_1  + F_2  + \ldots \nonumber \\
F_1  &=&  V_1  \nonumber \\
F_2  &=&  V_2  - \frac{1}{2} [\underline{V_1 }, V_1 ]
\label{eq:11.29} \\
&\ldots& \nonumber
 \end{eqnarray}

\noindent Taking into account that operator $V_1 $ is \emph{unphys}, so
that
 $\underbrace{F_1 } =
\underbrace{V_1 } =  0$,\footnote{see equation (\ref{eq:9.61})} we obtain the following perturbation
expansion

\begin{eqnarray}
S &=& e^{\underbrace{F }} \nonumber \\
&=& 1 + \underbrace{F } + \frac{1}{2!}
\underbrace{F }\underbrace{F }+ \ldots \nonumber \\
  &=& 1 +  \underbrace{F_1 } + \underbrace{F_2 } + \frac{1}{2!}
\underbrace{F_1 } \underbrace{F_1 } +  \frac{1}{2!}
\underbrace{F_2 } \underbrace{F_1 } + \frac{1}{2!}
\underbrace{F_1 } \underbrace{F_2 } +
\underbrace{F_3 }  +\ldots \nonumber \\
&=& 1 + \underbrace{F_2 }  + \underbrace{F_3 } \ldots \nonumber \\
&=& 1 + \underbrace{V_2 }
-\frac{1}{2}\underbrace{[\underline{V_1 }, V_1 ]} +
\underbrace{F_3 } \ldots \label{eq:11.34}
 \end{eqnarray}

 Let us first evaluate expression
$-\frac{1}{2}[\underline{V}_1, V_1]$ in (\ref{eq:11.34}). Only the four first terms in equation (\ref{eq:11.32}) for $V_1$ are relevant for this calculation:\footnote{Operators $A$, $C$, $D$ are defined in (\ref{eq:11.25}) -
(\ref{eq:11.28a}) and (\ref{eq:K.6}).}

\begin{eqnarray*}
V_1
&=& -\frac{e }{ (2\pi \hbar)^{3/2} }  \int
 d\mathbf{k}
d\mathbf{p}
 \overline{A}^{\dag}_{ \alpha}(\mathbf{p+k}) A _{
\beta}(\mathbf{p})C_{\alpha \beta} (\mathbf{k})
 \nonumber\\
&\ & -\frac{ e } { (2\pi \hbar)^{3/2} } \int d\mathbf{k} d\mathbf{p}
 \overline{A}^{\dag}_{ \alpha}(\mathbf{p-k}) A _{
\beta}(\mathbf{p})C^{\dag}_{\alpha \beta}(\mathbf{k})
 \nonumber\\
&\ & +\frac{ e  } { (2\pi \hbar)^{3/2} } \int d\mathbf{k} d\mathbf{p}
 \overline{D}^{\dag}_{ \alpha}(\mathbf{p+k}) D _{
\beta}(\mathbf{p})C_{\alpha \beta} (\mathbf{k})
 \nonumber\\
&\ & +\frac{ e } { (2\pi \hbar)^{3/2} } \int d\mathbf{k} d\mathbf{p}
 \overline{D}^{\dag}_{ \alpha}(\mathbf{p-k}) D _{
\beta}(\mathbf{p})C^{\dag}_{\alpha \beta}(\mathbf{k})
 +\ldots \nonumber
\end{eqnarray*}

\noindent According to (\ref{eq:underline2}), the corresponding terms in
$\underline{V_1} $ are

\begin{eqnarray}
&\mbox{ } &\underline{ V}_1  \nonumber \\
 &=& \frac{ e } { (2\pi \hbar)^{3/2} } \int
 d\mathbf{k}
d\mathbf{p}
 \overline{A}^{\dag}_{ \alpha}(\mathbf{p+k}) A _{
\beta}(\mathbf{p})C_{\alpha \beta} (\mathbf{k})
\frac{1} {\omega_{\mathbf{p+k}} -
\omega_{\mathbf{p}}-
ck} \nonumber\\
& \ & +\frac{ e } { (2\pi \hbar)^{3/2} } \int d\mathbf{k} d\mathbf{p}
 \overline{A}^{\dag}_{ \alpha}(\mathbf{p-k}) A _{
\beta}(\mathbf{p})C^{\dag}_{\alpha \beta}(\mathbf{k})
\frac{1}
{\omega_{\mathbf{p-k}} - \omega_{\mathbf{p}} + ck} \nonumber\\
&\ & -\frac{ e } { (2\pi \hbar)^{3/2} } \int d\mathbf{k} d\mathbf{p}
 \overline{D}^{\dag}_{ \alpha}(\mathbf{p+k}) D _{
\beta}(\mathbf{p})C_{\alpha \beta} (\mathbf{k})
\frac{1} {\Omega_{\mathbf{p+k}} -
\Omega_{\mathbf{p}}-
ck} \nonumber\\
& \ & -\frac{ e  } { (2\pi \hbar)^{3/2} } \int d\mathbf{k} d\mathbf{p}
 \overline{D}^{\dag}_{ \alpha}(\mathbf{p-k}) D _{
\beta}(\mathbf{p})C^{\dag}_{\alpha \beta}(\mathbf{k})
\frac{1} {\Omega_{\mathbf{p-k}} - \Omega_{\mathbf{p}} + ck}
\nonumber \\
&\ & +\ldots \label{eq:11.35}
\end{eqnarray}

In order to obtain terms of the type $D^{\dag} A^{\dag}DA$ in the
expression $[\underline{V}_1, V_1]$,  we need to consider four
 commutators: the
1st term in $\underline{V}_1$ commuting with the 4th term in $V_1$,
the  2nd term in $\underline{V}_1$ commuting with the 3rd term in
$V_1$, the 3rd term in $\underline{V}_1$ commuting with the 2nd term
in $V_1$ and the 4th term in $\underline{V}_1$ commuting with the
1st term in $V_1$. Using commutator (\ref{eq:K.5a}) we then obtain

\begin{eqnarray}
&\mbox { }& -\frac{1}{2} [\underline{V_1}, V_1]  \nonumber \\
&=&  - \frac{e^2 }{2(2\pi \hbar)^{3}} \int
 d\mathbf{k}
d\mathbf{p}
 d\mathbf{k}'
d\mathbf{p}'
 \overline{A}^{\dag}_{ \alpha}(\mathbf{p+k}) A _{
\beta}(\mathbf{p})  \overline{D}^{\dag}_{ \gamma}(\mathbf{p'-k}') D
_{
\delta}(\mathbf{p'}) \times \nonumber \\
&\mbox { }& [C_{\alpha \beta} (\mathbf{k}), C^{\dag}_{\gamma
\delta}(\mathbf{k}')] \frac{1} {\omega_{\mathbf{p+k}} - \omega_{\mathbf{p}}-
ck} \nonumber \\
&-&  \frac{e^2 }{2(2\pi \hbar)^{3}} \int
 d\mathbf{k}
d\mathbf{p}
 d\mathbf{k}'
d\mathbf{p}'
 \overline{A}^{\dag}_{ \alpha}(\mathbf{p-k}) A _{
\beta}(\mathbf{p})  \overline{D}^{\dag}_{ \gamma}(\mathbf{p'+k}') D
_{
\delta}(\mathbf{p'}) \times \nonumber \\
&\mbox { }& [C^{\dag}_{\alpha \beta} (\mathbf{k}), C_{\gamma
\delta}(\mathbf{k}')] \frac{1} {\omega_{\mathbf{p-k}} - \omega_{\mathbf{p}}+
ck} \nonumber \\
&-&  \frac{e^2 }{2(2\pi \hbar)^{3}} \int
 d\mathbf{k}
d\mathbf{p}
 d\mathbf{k}'
d\mathbf{p}'
 \overline{D}^{\dag}_{ \alpha}(\mathbf{p+k}) D _{
\beta}(\mathbf{p})  \overline{A}^{\dag}_{ \gamma}(\mathbf{p'-k}') A
_{
\delta}(\mathbf{p'}) \times \nonumber \\
&\mbox { }& [C_{\alpha \beta} (\mathbf{k}), C^{\dag}_{\gamma
\delta}(\mathbf{k}')]
\frac{1} {\Omega_{\mathbf{p+k}} - \Omega_{\mathbf{p}}-
ck} \nonumber \\
&-&  \frac{e^2 }{2(2\pi \hbar)^{3}} \int
 d\mathbf{k}
d\mathbf{p}
 d\mathbf{k}'
d\mathbf{p}'
 \overline{D}^{\dag}_{ \alpha}(\mathbf{p-k}) D _{
\beta}(\mathbf{p})  \overline{A}^{\dag}_{ \gamma}(\mathbf{p'+k}') A
_{
\delta}(\mathbf{p'}) \times  \nonumber \\
&\mbox { }& [C^{\dag}_{\alpha \beta} (\mathbf{k}), C_{\gamma
\delta}(\mathbf{k}')]
\frac{1} {\Omega_{\mathbf{p-k}} - \Omega_{\mathbf{p}}+
ck}  + \ldots \nonumber \\
&=& \frac{e^2 \hbar^2 c}{4(2\pi \hbar)^{3}}\int \frac{ d\mathbf{k}
d\mathbf{p} d\mathbf{q}}{k}
 \gamma^{\mu}_{ \alpha \beta}
 \gamma^{\nu}_{\gamma \delta} h_{\mu \nu} (\mathbf{k}) \times \nonumber \\
\Bigr(&-& \overline{D}^{\dag}_{ \gamma}(\mathbf{p-k}) D _{
\delta}(\mathbf{p}) \overline{A}^{\dag}_{ \alpha}(\mathbf{q+k}) A _{
\beta}(\mathbf{q}) \frac{1}
{\omega_{\mathbf{q+k}} - \omega_{\mathbf{q}}-
ck} \nonumber \\
&+& \overline{D}^{\dag}_{ \gamma}(\mathbf{p+k}) D _{
\delta}(\mathbf{p})
 \overline{A}^{\dag}_{ \alpha}(\mathbf{q-k}) A _{
\beta}(\mathbf{q})
 \frac{1}
{\omega_{\mathbf{q-k}} - \omega_{\mathbf{q}}+
ck} \nonumber \\
&-&
 \overline{D}^{\dag}_{ \alpha}(\mathbf{p+k}) D _{
\beta}(\mathbf{p})  \overline{A}^{\dag}_{ \gamma}(\mathbf{q-k}) A _{
\delta}(\mathbf{q}) \frac{1} {\Omega_{\mathbf{p+k}} - \Omega_{\mathbf{p}}-
ck} \nonumber \\
&+&
 \overline{D}^{\dag}_{ \alpha}(\mathbf{p-k}) D _{
\beta}(\mathbf{p})  \overline{A}^{\dag}_{ \gamma}(\mathbf{q+k}) A _{
\delta}(\mathbf{q})
 \frac{1}
{\Omega_{\mathbf{p-k}} - \Omega_{\mathbf{p}}+
ck}  + \ldots \Bigr) \nonumber \\
&=& \frac{e^2 \hbar^2 c}{4(2\pi \hbar)^{3}}\int \frac{ d\mathbf{k}
d\mathbf{p} d\mathbf{q}}{k}
 \gamma^{\mu}_{ \alpha \beta}
 \gamma^{\nu}_{\gamma \delta} h_{\mu \nu}(\mathbf{k}) \times \nonumber \\
\Bigl(&-& \overline{D}^{\dag}_{ \alpha}(\mathbf{p-k}) D _{
\beta}(\mathbf{p}) \overline{A}^{\dag}_{ \gamma}(\mathbf{q+k}) A _{
\delta}(\mathbf{q}) \frac{1} {\omega_{\mathbf{q+k}} - \omega_{\mathbf{q}}-
ck} \nonumber \\
&+& \overline{D}^{\dag}_{ \alpha}(\mathbf{p-k}) D _{
\beta}(\mathbf{p})
 \overline{A}^{\dag}_{ \gamma}(\mathbf{q+k}) A _{
\delta}(\mathbf{q})
 \frac{1}
{\omega_{\mathbf{q+k}} - \omega_{\mathbf{q}}+
ck} \nonumber \\
&-&
 \overline{D}^{\dag}_{ \alpha}(\mathbf{p-k}) D _{
\beta}(\mathbf{p})  \overline{A}^{\dag}_{ \gamma}(\mathbf{q+k}) A _{
\delta}(\mathbf{q}) \frac{1} {\Omega_{\mathbf{p-k}} - \Omega_{\mathbf{p}}-
ck} \nonumber \\
&+&
 \overline{D}^{\dag}_{ \alpha}(\mathbf{p-k}) D _{
\beta}(\mathbf{p})  \overline{A}^{\dag}_{ \gamma}(\mathbf{q+k}) A _{
\delta}(\mathbf{q})
 \frac{1}
{\Omega_{\mathbf{p-k}} - \Omega_{\mathbf{p}}+ ck}  +
\ldots \Bigr) \nonumber \\
&=& -\frac{e^2 \hbar^2 c^2}{2(2\pi \hbar)^{3}}  \int
 d\mathbf{k}
d\mathbf{p} d\mathbf{q}
 \gamma^{\mu}_{ \alpha \beta}
 \gamma^{\nu}_{\gamma \delta}  \frac{h_{\mu \nu}(\mathbf{k}) }
{(\tilde{q}+\tilde{k} \div \tilde{q})^2 } \times \nonumber \\
&\mbox{ } &  \overline{D}^{\dag}_{ \alpha}(\mathbf{p-k})
\overline{A}^{\dag}_{ \gamma}(\mathbf{q+k}) D _{ \beta}(\mathbf{p})
A _{ \delta}(\mathbf{q}) \nonumber \\
&\ & -\frac{e^2 \hbar^2 c^2}{2(2\pi \hbar)^{3}}  \int
 d\mathbf{k}
d\mathbf{p} d\mathbf{q}
 \gamma^{\mu}_{ \alpha \beta}
 \gamma^{\nu}_{\gamma \delta} \frac{h_{\mu \nu}(\mathbf{k})}{(\tilde{P}-\tilde{K}
 \div \tilde{P})^2} \times \nonumber \\
&\mbox{ } &  \overline{D}^{\dag}_{ \alpha}(\mathbf{p-k})
\overline{A}^{\dag}_{ \gamma}(\mathbf{q+k}) D _{ \beta}(\mathbf{p})
A _{ \delta}(\mathbf{q}) \label{eq:com-v1v1}
\end{eqnarray}

\noindent where we denoted\footnote{ Recall that in Appendix \ref{ss:4-dim-rep} we agreed to denote 4-vectors by the tilde. Then
$(\tilde{p} \div \tilde{q})^2$ is a
4-square of the difference between 4-vectors $\tilde{p}$ and
$\tilde{q}$, i.e., $(\tilde{p} \div \tilde{q})^2 =
(\tilde{p}-\tilde{q})_{\mu}(\tilde{p}-\tilde{q})^{\mu}.$ Thus, for
example, $(\tilde{q}+\tilde{k}\div \tilde{q})^2 =
(\omega_{\mathbf{q+k}} - \omega_{\mathbf{q}})^2 - c^2 k^2$.}
\index{$\div$}

\begin{eqnarray}
(\tilde{p} \div \tilde{q})^2 &\equiv& (\omega_{\mathbf{p}} -
\omega_{\mathbf{q}})^2 -
c^2 (\mathbf{p-q})^2 \label{eq:pdivq}\\
(\tilde{P} \div \tilde{Q})^2 &\equiv& (\Omega_{\mathbf{p}} -
\Omega_{\mathbf{q}})^2 - c^2 (\mathbf{p-q})^2 \label{eq:PdivQ}
\end{eqnarray}

\noindent Next take into account that we need to know our $S$-operator only in the vicinity of the energy shell where

\begin{eqnarray}
\Omega_{\mathbf{p-k}}- \Omega_{\mathbf{p}} &=& \omega_{\mathbf{q}} -
\omega_{\mathbf{q+k}} \nonumber \\
(\tilde{P}-\tilde{K} \div \tilde{P})^2 &=& (\tilde{q}+\tilde{k} \div
\tilde{q})^2 \label{eq:PK2}
\end{eqnarray}

\noindent Also use notation (\ref{eq:uz}) - (\ref{eq:vz}) in which

\begin{eqnarray*}
\overline{A}^{\dag}(\mathbf{q+k}) \gamma^{\nu}
A(\mathbf{q}) &=& \frac{mc^2}{\sqrt{\omega_{\mathbf{q+k}}\omega_{\mathbf{q}}}} \sum_{\sigma \sigma'} U^{\nu}(\mathbf{q+k}, \sigma; \mathbf{q}, \sigma') a^{\dag}_{\mathbf{q+k}, \sigma}a_{\mathbf{q}, \sigma'}\\
 \overline{D}^{\dag}(\mathbf{p-k}) \gamma^{\mu} D _{ \beta}(\mathbf{p}) &=& \frac{Mc^2}{\sqrt{\Omega_{\mathbf{p-k}}\Omega_{\mathbf{p}}}} \sum_{\tau \tau'} W^{\mu}(\mathbf{p-k}, \tau; \mathbf{p}, \tau')
 d^{\dag}_{\mathbf{p-k}, \tau}d_{\mathbf{p}, \tau'}
\end{eqnarray*}

\noindent and equation (\ref{eq:10.29b}). Then

\begin{eqnarray*}
&\mbox { }& -\frac{1}{2} [\underline{V_1}, V_1]  \nonumber \\
&=&  -\frac{e^2 \hbar^2 c^2}{(2\pi \hbar)^{3}} \frac{Mmc^4}{\sqrt{\omega_{\mathbf{q+k}}\omega_{\mathbf{q}}}\sqrt{\Omega_{\mathbf{p-k}}\Omega_{\mathbf{p}}}} \sum_{\sigma, \tau, \sigma', \tau'}  \times \\
&\ &\int
 d\mathbf{k}
d\mathbf{p} d\mathbf{q}
  \frac{h_{\mu \nu}(\mathbf{k}) U^{\nu}(\mathbf{q+k}, \sigma; \mathbf{q}, \sigma')W^{\mu}(\mathbf{p-k}, \tau; \mathbf{p}, \tau')}
{(\tilde{q}+\tilde{k} \div \tilde{q})^2 }
  d^{\dag}_{\mathbf{p-k}, \tau} a^{\dag}_{\mathbf{q+k}, \sigma} d_{\mathbf{p}, \tau'} a_{\mathbf{q}, \sigma'} \\
&=& -\frac{e^2 \hbar^2 c^2}{(2\pi \hbar)^{3}} \frac{Mmc^4}{\sqrt{\omega_{\mathbf{q+k}}\omega_{\mathbf{q}}}\sqrt{\Omega_{\mathbf{p-k}}\Omega_{\mathbf{p}}}} \sum_{\sigma, \tau \sigma', \tau'}  \times \\
&\ &\int
 d\mathbf{k}
d\mathbf{p} d\mathbf{q}
\Bigl[  \frac{(\mathbf{U}(\mathbf{q+k}, \sigma; \mathbf{q}, \sigma') \cdot \mathbf{W}(\mathbf{p-k}, \tau; \mathbf{p}, \tau')}
{(\tilde{q}+\tilde{k} \div \tilde{q})^2 } \\
&-& \frac{(\mathbf{k} \cdot \mathbf{U}(\mathbf{q+k}, \sigma; \mathbf{q}, \sigma'))(\mathbf{k} \cdot \mathbf{W}(\mathbf{p-k}, \tau; \mathbf{p}, \tau')}
{k^2(\tilde{q}+\tilde{k} \div \tilde{q})^2 } \Bigr]
  d^{\dag}_{\mathbf{p-k}, \tau} a^{\dag}_{\mathbf{q+k}, \sigma} d_{\mathbf{p}, \tau'} a_{\mathbf{q}, \sigma'}
\end{eqnarray*}

Combining this
expression with the term $D^{\dag}A^{\dag}DA$ in
$V_2$,\footnote{ the third term in equation (\ref{eq:v2-final-ph})} we
see that operator $F_2$ in (\ref{eq:11.29}) takes  the
form

\begin{eqnarray}
 F_2
&=& \frac{e^2 \hbar^2 c^2}{(2\pi \hbar)^{3}} \frac{Mmc^4}{\sqrt{\omega_{\mathbf{q+k}}\omega_{\mathbf{q}}\Omega_{\mathbf{p-k}}\Omega_{\mathbf{p}}}} \sum_{\sigma, \tau, \sigma', \tau'}  \int
 d\mathbf{k}
d\mathbf{p} d\mathbf{q} \times \nonumber \\
&\ &
\Bigl[  -\frac{\mathbf{U}(\mathbf{q+k}, \sigma; \mathbf{q}, \sigma') \cdot \mathbf{W}(\mathbf{p-k}, \tau; \mathbf{p}, \tau')}
{(\tilde{q}+\tilde{k} \div \tilde{q})^2 } \nonumber \\
&+& \frac{(\mathbf{k} \cdot \mathbf{U}(\mathbf{q+k}, \sigma; \mathbf{q}, \sigma'))(\mathbf{k} \cdot \mathbf{W}(\mathbf{p-k}, \tau; \mathbf{p}, \tau'))}
{k^2(\tilde{q}+\tilde{k} \div \tilde{q})^2 } \nonumber \\
&-& \frac{U^0(\mathbf{q+k}, \sigma; \mathbf{q}, \sigma') W^0(\mathbf{p-k}, \tau; \mathbf{p}, \tau')}
{c^2k^2 } \Bigr]
  d^{\dag}_{\mathbf{p-k},\tau}a^{\dag}_{\mathbf{q+k}, \sigma} d_{\mathbf{p}, \tau'} a_{\mathbf{q}, \sigma'} \label{eq:F2xy}
\end{eqnarray}

\noindent From (\ref{eq:kk})  we further obtain

\begin{eqnarray*}
(\mathbf{k} \cdot \mathbf{W}(\mathbf{p}-\mathbf{k}, \tau; \mathbf{p}, \tau')) &=& -\frac{\Omega_{\mathbf{p}-\mathbf{k}}- \Omega_{\mathbf{p}}}{c} W^0(\mathbf{p}-\mathbf{k},\tau; \mathbf{p},\tau') \\
&=& \frac{\omega_{\mathbf{q}+\mathbf{k}}- \omega_{\mathbf{q}}}{c} W^0(\mathbf{p}-\mathbf{k},\tau; \mathbf{p}, \tau')  \nonumber \\
(\mathbf{k} \cdot \mathbf{U}(\mathbf{q}+ \mathbf{k},\sigma;
\mathbf{q}, \sigma')) &=& \frac{\omega_{\mathbf{q}+\mathbf{k}}- \omega_{\mathbf{q}}}{c} U^0(\mathbf{q}+ \mathbf{k},\sigma;
\mathbf{q},\sigma')
\end{eqnarray*}

\noindent and

\begin{eqnarray*}
 \frac{(\mathbf{k} \cdot \mathbf{U})(\mathbf{k} \cdot \mathbf{W})}
{k^2(\tilde{q}+\tilde{k} \div \tilde{q})^2 }
- \frac{U^0 W^0}
{c^2k^2 }
&=& \frac{(\omega_{\mathbf{q+k}}- \omega_{\mathbf{q}})^2 U^0W^0}
{c^2k^2(\tilde{q}+\tilde{k} \div \tilde{q})^2 }
- \frac{[(\omega_{\mathbf{q+k}}- \omega_{\mathbf{q}})^2 - c^2 k^2] U^0 W^0}
{c^2k^2 (\tilde{q}+\tilde{k} \div \tilde{q})^2}  \\
&=&  \frac{ U^0 W^0}
{(\tilde{q}+\tilde{k} \div \tilde{q})^2}  \\
\end{eqnarray*}

\noindent so that our final expression for the $F$-operator is

\begin{eqnarray}
 F_2
&=& \frac{e^2 \hbar^2 c^2}{(2\pi \hbar)^{3}} \frac{Mmc^4}{\sqrt{\omega_{\mathbf{q+k}}\omega_{\mathbf{q}}\Omega_{\mathbf{p-k}}\Omega_{\mathbf{p}}}} \sum_{\sigma, \tau, \sigma', \tau'}  \int
 d\mathbf{k}
d\mathbf{p} d\mathbf{q} \times \nonumber \\
&\ &
\frac{U^0(\mathbf{q+k}, \sigma; \mathbf{q}, \sigma') W^0(\mathbf{p-k}, \tau; \mathbf{p}, \tau') - (\mathbf{U}(\mathbf{q+k}, \sigma; \mathbf{q}, \sigma') \cdot \mathbf{W}(\mathbf{p-k}, \tau; \mathbf{p}, \tau')}
{(\tilde{q}+\tilde{k} \div \tilde{q})^2 } \times \nonumber \\
&\ & d^{\dag}_{\mathbf{p-k}, \tau} a^{\dag}_{\mathbf{q+k}, \sigma} d_{\mathbf{p}, \tau'} a_{\mathbf{q}, \sigma'} \nonumber  \\ \nonumber  \\
 &=& \frac{e^2 \hbar^2 c^2}{(2\pi \hbar)^{3}} \frac{Mmc^4}{\sqrt{\omega_{\mathbf{q+k}}\omega_{\mathbf{q}}\Omega_{\mathbf{p-k}}\Omega_{\mathbf{p}}}} \sum_{\sigma, \tau, \sigma', \tau'}  \int
 d\mathbf{k}
d\mathbf{p} d\mathbf{q} \times \nonumber \\
&\ &
\frac{U_{\mu}(\mathbf{q+k}, \sigma; \mathbf{q}, \sigma') W^{\mu}(\mathbf{p-k}, \tau; \mathbf{p}, \tau') }
{(\tilde{q}+\tilde{k} \div \tilde{q})^2 }  d^{\dag}_{\mathbf{p-k}, \tau} a^{\dag}_{\mathbf{q+k}, \sigma} d_{\mathbf{p}, \tau'} a_{\mathbf{q}, \sigma'} \label{eq:s-oper-2nd}
\end{eqnarray}

\noindent Now we insert this result in formula  (\ref{eq:11.34})
for the $S$-operator. According to (\ref{eq:9.52}), in order to perform the integration on
$t$ from $-\infty$ to $\infty$, we just need to multiply the coefficient function
 by the factor $-2\pi i \delta(E(\mathbf{p},
\mathbf{q}, \mathbf{k}))$, where

\begin{eqnarray*}
E(\mathbf{p}, \mathbf{q}, \mathbf{k}) = \omega_{\mathbf{q+k}} -
\omega_{\mathbf{q}}+ \Omega_{\mathbf{p-k}} - \Omega_{\mathbf{p}}
\end{eqnarray*}

\noindent is the energy function. This makes the $S$-operator non-trivial only on the
energy shell $E(\mathbf{p}, \mathbf{q}, \mathbf{k}) = 0$. Finally, we can represent the 2nd order scattering operator
in the general form (\ref{eq:9.49})

\begin{eqnarray}
S_2[d^{\dag}a^{\dag}da]&=& \sum_{\sigma, \tau, \sigma', \tau'}\int d
\mathbf{p} d \mathbf{q} d \mathbf{p'} d \mathbf{q'}
s_2(\mathbf{p},\mathbf{q},\mathbf{p}', \mathbf{q}' ; \sigma, \tau,
\sigma', \tau')
 d^{\dag}_{\mathbf{p}, \tau }
a^{\dag}_{\mathbf{q}, \sigma } d_{\mathbf{p}', \tau'
}a_{\mathbf{q}', \sigma' } \nonumber \\
\label{eq:S2general}
\end{eqnarray}

\noindent with the coefficient function

\begin{eqnarray}
&\mbox{ }& s_2(\mathbf{p},\mathbf{q},\mathbf{p}', \mathbf{q}';
\tau, \sigma, \tau', \sigma' ) \nonumber \\
  &=&    -\frac{ie^2c^2 mMc^4 \delta^4(\tilde{p}+\tilde{q}-\tilde{p}'-\tilde{q}')U^{\mu}
(\mathbf{q}, \sigma; \mathbf{q}', \sigma')W_{\mu} (\mathbf{p}, \tau;
\mathbf{p}', \tau')}{4 \pi^2\hbar
\sqrt{\omega_{\mathbf{q}}\omega_{\mathbf{q'}}\Omega_{\mathbf{p}}\Omega_{\mathbf{p'}}}
(\tilde{q}-\tilde{q}')^2 }
\label{eq:S_2}
\end{eqnarray}

\noindent where the 4-dimensional delta function\footnote{see Appendix \ref{ss:4dimdellta}}

\begin{eqnarray}
 \delta^4(\tilde{p}+\tilde{q}-\tilde{p}'-\tilde{q}')  \equiv \delta(\mathbf{p}+\mathbf{q}-\mathbf{p}'-
\mathbf{q}')
\delta(\Omega_{\mathbf{p}}+\omega_{\mathbf{q}}-\Omega_{\mathbf{p}'}-
\omega_{\mathbf{q}'}) \label{eq:en-delta}
\end{eqnarray}

\noindent guarantees the conservation of both momentum and energy in the collision process.

Note that  $s_2(\mathbf{p},\mathbf{q},\mathbf{p}', \mathbf{q}' ; \tau, \sigma, \tau', \sigma')$ in (\ref{eq:S_2}) is indeed a matrix element of the $S$-operator between two 2-particle states

\begin{eqnarray}
&\mbox{ }& \langle 0 | a_{\mathbf{q}, \sigma } d_{\mathbf{p}, \tau}
S_2[d^{\dag}a^{\dag}da] d^{\dag}_{\mathbf{p'}, \tau'}
a^{\dag}_{\mathbf{q'}, \sigma'} |0
\rangle \nonumber  \\
&=& \sum_{\pi, \rho, \pi ', \rho '} \langle 0 | a_{\mathbf{q}, \sigma
} d_{\mathbf{p}, \tau} \int d \mathbf{s} d \mathbf{t} d
\mathbf{s'} d \mathbf{t'}  d^{\dag}_{\mathbf{s}, \pi }
a^{\dag}_{\mathbf{t}, \rho } s_2(\mathbf{s},\mathbf{t},\mathbf{s}',
\mathbf{t}'; \pi, \rho, \pi ', \rho ' ) \times
\nonumber  \\
&\mbox{ }&
 d_{\mathbf{s}', \pi' }a_{\mathbf{t}',
\rho' } d^{\dag}_{\mathbf{p'}, \tau'} a^{\dag}_{\mathbf{q'},
\sigma'} |0 \rangle \nonumber  \\
&=& \sum_{\pi, \rho, \pi ', \rho '}  \int d \mathbf{s} d \mathbf{t}
d \mathbf{s'} d \mathbf{t'} s_2(\mathbf{s},\mathbf{t},\mathbf{s}',
\mathbf{t}'; \pi, \rho, \pi ', \rho ' ) \times \nonumber  \\
&\mbox{ }&
 \delta(\mathbf{s-p}) \delta_{ \pi, \tau }
\delta(\mathbf{t-q})\delta_{ \rho, \sigma } \delta(\mathbf{s'-p'})
\delta_{ \pi', \tau'} \delta(\mathbf{t' - q'}) \delta_{ \rho',
\sigma'} \nonumber  \\
&=& s_2(\mathbf{p},\mathbf{q},\mathbf{p}', \mathbf{q}'; \tau,
\sigma, \tau', \sigma' ) \label{eq:s2pq}
\end{eqnarray}

\subsection{Lorentz invariance of the $S$-operator}
\label{ss:Lor-S}

In this subsection we would like to verify the Lorentz invariance of the $S$-operator calculated above. More specifically, in accordance with (\ref{eq:S-rel-inv}) we would like to check that

\begin{eqnarray*}
e^{-\frac{ic}{\hbar}\mathbf{K}_0 \vec{\theta}} S_2[d^{\dag}a^{\dag}da] e^{\frac{ic}{\hbar}\mathbf{K}_0 \vec{\theta}} =
 S_2[d^{\dag}a^{\dag}da]
\end{eqnarray*}

\noindent On the left hand side of this equality we apply boosts to particle operators using formulas (\ref{eq:9.36}) - (\ref{eq:9.37})

\begin{eqnarray*}
&\ & \sum_{\sigma, \tau, \sigma', \tau'}\int d
\mathbf{p} d \mathbf{q} d \mathbf{p'} d \mathbf{q'}
s_2(\mathbf{p},\mathbf{q},\mathbf{p}', \mathbf{q}' ; \tau, \sigma,
\tau', \sigma') e^{\frac{ic}{\hbar}\mathbf{K}_0 \vec{\theta}}
 d^{\dag}_{\mathbf{p}, \tau}
a^{\dag}_{\mathbf{q}, \sigma} d_{\mathbf{p}', \tau'
}a_{\mathbf{q}', \sigma' }   e^{-\frac{ic}{\hbar}\mathbf{K}_0 \vec{\theta}} \\
&=& \sum_{\sigma, \tau, \sigma', \tau'}\int d
\mathbf{p} d \mathbf{q} d \mathbf{p'} d \mathbf{q'}
s_2(\mathbf{p},\mathbf{q},\mathbf{p}', \mathbf{q}' ; \tau, \sigma,
 \tau', \sigma') \times \\
&\ & \frac{\sqrt{\Omega_{\Lambda \mathbf{p}}\Omega_{\mathbf{p}}}}{\Omega_{\mathbf{p}}} \sum_{\rho}
(D^{1/2})^*_{\tau \rho}(-\vec{\phi}_W(\mathbf{p}, \Lambda))
d^{\dag}_{\Lambda\mathbf{p},
\rho}  \frac{\sqrt{\omega_{\lambda \mathbf{q}}\omega_{\mathbf{q}}}}{\omega_{\mathbf{q}}}
 \sum_{\pi}
(D^{1/2})^*_{\sigma \pi}(-\vec{\phi}_W(\mathbf{q}, \lambda))
a^{\dag}_{\lambda\mathbf{q}, \pi} \times \\
&\ &  \frac{\sqrt{\Omega_{\Lambda \mathbf{p}'}\Omega_{\mathbf{p}'}}}{\Omega_{\mathbf{p}'}}
 \sum_{\eta}
D^{1/2}_{\tau' \eta}(-\vec{\phi}_W(\mathbf{p}', \Lambda))
d_{\Lambda\mathbf{p}', \eta} \frac{\sqrt{\omega_{\lambda \mathbf{q}'}\omega_{\mathbf{q}'}}}{\omega_{\mathbf{q}'}}
 \sum_{\chi}
D^{1/2}_{\sigma' \chi}(-\vec{\phi}_W(\mathbf{q}', \lambda))
a_{\lambda\mathbf{q}', \chi}
\end{eqnarray*}

\noindent where $\omega_{\mathbf{q}} = \sqrt{m^2c^4 + q^2c^2}$, $\lambda\mathbf{q}$ is the boost transformation of the electron's momentum,\footnote{see formula (\ref{eq:p'})} and $\vec{\phi}_W(\mathbf{q}, \lambda)$ is the corresponding Wigner angle (\ref{eq:7.9x}). The analogous  proton-related quantities $\Omega_{\mathbf{q}} $, $\Lambda\mathbf{q}$, and $\vec{\phi}_W(\mathbf{q}, \Lambda)$ are obtained by replacing the electron mass $m$ with the proton mass $M$. Changing summation and integration variables and denoting $R(\mathbf{p}, \lambda)\tau$ and $R(\mathbf{p}, \Lambda)\sigma$ the results of Wigner rotations on the electron and proton spin components, respectively,  we obtain\footnote{We also used equation (\ref{eq:7.15x}) here.}

\begin{eqnarray*}
&=& \sum_{\sigma, \tau, \sigma', \tau'}\int d
(\Lambda^{-1}\mathbf{ p}) d (\lambda^{-1}\mathbf{ q}) d (\Lambda^{-1}\mathbf{p'}) d (\lambda^{-1}\mathbf{ q'})
\frac{\sqrt{\Omega_{\mathbf{p}}\Omega_{\Lambda^{-1} \mathbf{p}}
\omega_{\mathbf{q}}\omega_{\lambda^{-1}\mathbf{q}}\Omega_{\mathbf{p}'}\Omega_{\Lambda^{-1}\mathbf{p}'}
\omega_{\mathbf{q}'}\omega_{\Lambda^{-1}\mathbf{q}'}}}
{\Omega_{\Lambda^{-1}\mathbf{p}}\omega_{\lambda^{-1}\mathbf{q}}\Omega_{\Lambda^{-1}\mathbf{p}'} \omega_{\Lambda^{-1}\mathbf{q}'}} \times  \\
&\ &s_2(\Lambda^{-1} \mathbf{p}, \lambda^{-1}\mathbf{q}, \Lambda^{-1}\mathbf{p}', \lambda^{-1}\mathbf{q}'; R^{-1}(\mathbf{p}, \Lambda)\tau, R^{-1}(\mathbf{q}, \lambda)\sigma,
R(\mathbf{p}', \Lambda)\tau', R(\mathbf{q}', \lambda)\sigma') \times \\
&\ &  d^{\dag}_{\mathbf{p}, \tau}a^{\dag}_{\mathbf{q}, \sigma}
d_{\mathbf{p}', \tau'}
a_{\mathbf{q}', \sigma'} \\
&=& \sum_{\sigma, \tau, \sigma', \tau'}\int d
\mathbf{ p} d \mathbf{ q} d \mathbf{p'} d \mathbf{ q'}
\sqrt{\frac{\Omega_{\Lambda^{-1} \mathbf{p}}\omega_{\lambda^{-1} \mathbf{q}} \Omega_{\Lambda^{-1} \mathbf{p}'} \omega_{\lambda^{-1}\mathbf{q}'} }{\Omega_{\mathbf{p}} \omega_{ \mathbf{q}}\Omega_{ \mathbf{p}'}\omega_{ \mathbf{q}'}}} \times  \\
&\ &s_2(\Lambda^{-1} \mathbf{p}, \lambda^{-1}\mathbf{q}, \Lambda^{-1}\mathbf{p}', \lambda^{-1}\mathbf{q}'; R^{-1}(\mathbf{p}, \Lambda)\tau, R^{-1}(\mathbf{q}, \lambda)\sigma,
R(\mathbf{p}', \Lambda)\tau', R(\mathbf{q}', \lambda)\sigma') \times \\
&\ &  d^{\dag}_{\mathbf{p}, \tau}
a^{\dag}_{\mathbf{q}, \sigma}
d_{\mathbf{p}', \tau'}
a_{\mathbf{q}', \sigma'}
\end{eqnarray*}

\noindent Comparing this result with (\ref{eq:S2general}) we conclude that the Lorentz invariance condition will be satisfied if the coefficient function can be written in the form

\begin{eqnarray*}
&\ &s_2( \mathbf{p}, \mathbf{q}, \mathbf{p}', \mathbf{q}'; \tau, \sigma,
\tau', \sigma') \equiv
\frac{Mmc^4 }{\sqrt{\Omega_{\mathbf{p}} \omega_{ \mathbf{q}}\Omega_{ \mathbf{p}'}\omega_{ \mathbf{q}'}}} \mathcal{S}_2( \mathbf{p}, \mathbf{q}, \mathbf{p}', \mathbf{q}'; \tau, \sigma,
\tau', \sigma')
\end{eqnarray*}

\noindent and the new function $\mathcal{S}_2$ satisfies a simpler invariance condition

\begin{eqnarray}
&\ &\mathcal{S}_2( \mathbf{p}, \mathbf{q}, \mathbf{p}', \mathbf{q}'; \tau, \sigma,
\tau', \sigma') \nonumber \\
&=&
\mathcal{S}_2(\Lambda^{-1} \mathbf{p}, \lambda^{-1}\mathbf{q}, \Lambda^{-1}\mathbf{p}', \lambda^{-1}\mathbf{q}'; R^{-1}(\mathbf{p}, \Lambda)\tau, R^{-1}(\mathbf{q}, \lambda)\sigma,
R(\mathbf{p}', \Lambda)\tau', R(\mathbf{q}', \lambda)\sigma') \nonumber \\
\label{eq:S-lambda}
\end{eqnarray}

\noindent In our case (\ref{eq:S_2})

\begin{eqnarray*}
 \mathcal{S}_2  &=&   \frac{ie^2c^2 }{4 \pi^2 \hbar}
\delta^4(\tilde{p}+\tilde{q}-\tilde{p}'-\tilde{q}')
\frac{U^{\mu}(\mathbf{q}, \sigma;\mathbf{q'}, \sigma')
W_{\mu}(\mathbf{p}, \tau;\mathbf{p'},
\tau')}{(\tilde{q}-\tilde{q}')^2}
\end{eqnarray*}

\noindent As shown in Appendix  \ref{ss:Umuwmu}, the quantities $U^{\mu}$ and $W^{\mu}$ transform as 4-vectors under the change of arguments indicated on the right hand side of (\ref{eq:S-lambda}). So the 4-product $U^{\mu}W_{\mu}$ stays invariant. The 4-square $(\tilde{q}-\tilde{q}')^2$ is invariant as well. This proves Lorentz invariance of the 2nd order contribution (\ref{eq:S_2}) to the $S$-operator.

\subsection{$S_2$ in Feynman-Dyson perturbation theory}
\label{ss:Fey}

The relativistic invariance of the $S_2$ operator looks almost accidental in our approach. Indeed, we have used the interacting Hamiltonian $V_1 + V_2$, which did not have simple transformation properties with respect to boosts. We have also used a non-covariant form
(\ref{eq:10.29b}) of the matrix $h_{\mu \nu}(\mathbf{k})$ and
saw a lucky cancelation of non-covariant terms. However, as discussed in section 8.5 of
\cite{book}, this cancelation is actually not accidental. In fact, it is expected to
occur for all processes in all perturbation orders, so that $S$-matrix elements become explicitly Lorentz invariant. This observation opens up a
possibility to perform $S$-matrix calculations much more easily than
it has been done above, while maintaining the
manifest covariance at each calculation step. Such a possibility is realized
in the \emph{Feynman-Dyson perturbation theory}, \index{Feynman-Dyson perturbation theory} which is the method of
choice for $S$-matrix calculations in QFT.

The prescription used in the Feynman-Dyson approach has three
differences with respect to our subsection \ref{ss:2-nd-order} \cite{book,Weinberg_1049}.

\begin{enumerate}
\item Drop the 2nd order
interaction $V_2 $ in (\ref{eq:11.5}), so that the full
interaction operator is given simply by $V_1 $ in
(\ref{eq:11.6})\footnote{So, we will call $V_1$ the
\emph{Feynman-Dyson interaction operator} \index{Feynman-Dyson
interaction operator} to distinguish it from the \emph{Hamiltonian
interaction operator} \index{Hamiltonian interaction operator}
$V_1+V_2$.}

\begin{eqnarray}
V_1
&=&  \frac{1}{c}\int d\mathbf{x} j_{\mu}( \mathbf{x},0) A^{\mu}( \mathbf{x},0) \nonumber  \\
&=& \int d\mathbf{x} (-e \overline{\psi}( \mathbf{x},0) \gamma^{\mu}
\psi( \mathbf{x},0) + e \overline{\Psi}( \mathbf{x},0)  \gamma^{\mu} \Psi
( \mathbf{x},0)) A^{\mu}( \mathbf{x},0) \nonumber \\
  \label{eq:11.6a}
\end{eqnarray}

\item In calculations involving photon fields\footnote{commutators (\ref{eq:K.5a}) and photon propagators (\ref{eq:K.17a})} use the covariant expression ($g_{\mu \nu}$) for the matrix $h_{\mu \nu}(\mathbf{k})$ instead of our formula (\ref{eq:10.29b}).
\item Calculate the $S$-operator with the help of the Feynman-Dyson perturbation formula (\ref{eq:F-D}).
\end{enumerate}

We will omit a proof that this approach works in all orders. Let us simply show how it applies to our example. Here
we will repeat calculation of the 2nd order $S$-matrix element $S_2$ in the Feynman-Dyson formulation.
We use
formulas (\ref{eq:F-D}), (\ref{eq:s2pq}), and (\ref{eq:11.1})

\begin{eqnarray}
&\mbox{ }& s_2(\mathbf{p},\mathbf{q},\mathbf{p}', \mathbf{q}';
\tau, \sigma, \tau', \sigma' ) \nonumber \\
&=&  \langle 0 | a_{\mathbf{q}, \sigma } d_{\mathbf{p}, \tau} S_2
d^{\dag}_{\mathbf{p'}, \tau'}
a^{\dag}_{\mathbf{q'}, \sigma'} |0 \rangle \nonumber \\
 &=&  -\langle 0 | a_{\mathbf{q}, \sigma }
d_{\mathbf{p}, \tau} \left(\frac{1}{2! \hbar^2} \int
\limits_{-\infty }^{+\infty} dt_1 dt_2 T[V_1(t_1)
 V_1(t_2)] \right) d^{\dag}_{\mathbf{p'}, \tau'}
a^{\dag}_{\mathbf{q'}, \sigma'} |0 \rangle \nonumber \\
  &=& -\frac{1}{2!\hbar^2c^2} \int d^4x_1 d^4x_2    \langle 0 | a_{\mathbf{q}, \sigma } d_{\mathbf{p},
\tau}T[(J^{\mu}(\tilde{x}_1)A_{\mu} (\tilde{x}_1) +
\mathcal{J}^{\mu}(\tilde{x}_1)A_{\mu}
(\tilde{x}_1)) \times \nonumber \\
&\ & (J^{\mu}(\tilde{x}_2)A_{\mu} (\tilde{x}_2) +
\mathcal{J}^{\mu}(\tilde{x}_2)A_{\mu} (\tilde{x}_2)) ]
d^{\dag}_{\mathbf{p'}, \tau'} a^{\dag}_{\mathbf{q'}, \sigma'}
|0 \rangle \nonumber  \\
  &=& -\frac{1}{2!\hbar^2 c^2} \int d^4x_1 d^4x_2
   \langle 0 | a_{\mathbf{q}, \sigma } d_{\mathbf{p}, \tau}
\Bigl(T[J^{\mu}(\tilde{x}_1)A_{\mu} (\tilde{x}_1)
\mathcal{J}^{\mu}(\tilde{x}_2)A_{\mu}
(\tilde{x}_2) ] \nonumber \\
&\ & +T[ \mathcal{J}^{\mu}(\tilde{x}_1)A_{\mu} (\tilde{x}_1)
J^{\mu}(\tilde{x}_2)A_{\mu} (\tilde{x}_2)
]\Bigr)d^{\dag}_{\mathbf{p'}, \tau'} a^{\dag}_{\mathbf{q'}, \sigma'}
|0 \rangle \nonumber  \\
  &=& \frac{e^2}{\hbar^2}   \int d^4x_1 d^4x_2 \times \nonumber  \\
&\mbox{ }& \langle 0 | a_{\mathbf{q}, \sigma } d_{\mathbf{p}, \tau}
T[\overline{\psi}(\tilde{x}_1) \gamma^{\mu} \psi
(\tilde{x}_1)A_{\mu} (\tilde{x}_1) \overline{\Psi}(\tilde{x}_2)
\gamma^{\nu} \Psi (\tilde{x}_2) A_{\nu} (\tilde{x}_2) ]
d^{\dag}_{\mathbf{p'}, \tau'} a^{\dag}_{\mathbf{q'}, \sigma'} |0
\rangle \nonumber \\
\label{eq:t-order}
\end{eqnarray}

\noindent If the operator sandwiched between vacuum vectors $\langle 0 |
\ldots | 0\rangle$ is converted to the normal order, then all its
terms will not contribute to the matrix element, except the
$c$-number term. In order to provide such a $c$-number term, the
operator under the $T$-symbol should have the structure $d^{\dag}
a^{\dag} da$. From expressions (\ref{eq:10.10}) and
(\ref{eq:10.10a}) for quantum fields $\psi$ and $\Psi$ we conclude
that operator $d^{\dag}$ (with a corresponding numerical factor) may
come only from the factor $\overline{\Psi}$, operator $a^{\dag}$ comes from
$\overline{\psi}$ and operators $d$ and $a$ come from factors $\Psi$ and $\psi$, respectively. In the process of bringing the full operator
to the normal order, creation (annihilation) operators inside the
$T$-symbol change places with corresponding annihilation (creation)
operators outside this symbol. This leaves expressions like (momentum delta function) $\times$
(Kronecker delta symbol of spin labels) $\times$ (numerical factor).
The delta function and the Kronecker delta disappear after integration
(summation) and only the numerical factor is left. For example, the
electron creation operator from the factors
$\overline{\psi}_{\alpha}$ being coupled with the annihilation
operator $a_{\mathbf{q}, \sigma}$ results in the numerical factor

\begin{eqnarray}
\sqrt{\frac{mc^2}{(2 \pi \hbar)^3 \omega_{\mathbf{q}}}}
\exp\left(\frac{i}{\hbar}\tilde{q} \cdot \tilde{x}_i \right)
\overline{u}_{\alpha}(\mathbf{q}, \sigma) \label{eq:u-factor}
\end{eqnarray}

\noindent After these routine manipulations the coefficient function
takes the form\footnote{where the matrix element $\langle 0|
T[A^{\mu} (\tilde{x}_1) A^{\nu} (\tilde{x}_2) ]|0 \rangle$ (the
\emph{photon propagator}) \index{photon propagator} was taken from
(\ref{eq:photon-prop})}

\begin{eqnarray}
&\mbox{ }& s_2(\mathbf{p},\mathbf{q},\mathbf{p}', \mathbf{q}';
\tau, \sigma, \tau',  \sigma' ) \nonumber  \\
  &\approx&  \int d^4x_1 d^4x_2 \frac{e^2  mMc^4}{\hbar^2(2 \pi \hbar)^6
\sqrt{\omega_{\mathbf{q}}\omega_{\mathbf{q'}}\Omega_{\mathbf{p}}\Omega_{\mathbf{p'}}}}  \times \nonumber  \\
&\mbox{ }& \exp\left(\frac{i}{\hbar}\tilde{q} \cdot
\tilde{x}_1\right) \exp\left(-\frac{i}{\hbar}\tilde{q}' \cdot
\tilde{x}_1\right) \exp\left(\frac{i}{\hbar}\tilde{p} \cdot
\tilde{x}_2\right) \exp\left(-\frac{i}{\hbar}\tilde{p}'
\cdot \tilde{x}_2\right) \times \nonumber  \\
&\mbox{ }&  \overline{u}(\mathbf{q}, \sigma) \gamma_{\mu}
u(\mathbf{q'}, \sigma')
 \overline{w}(\mathbf{p}, \tau) \gamma_{\nu} w(\mathbf{p'},
 \tau')\langle 0| T[A^{\mu} (\tilde{x}_1) A^{\nu} (\tilde{x}_2) ]|0 \rangle \nonumber   \\
  &=&  \frac{1}{2 \pi i} \int d^4x_1 d^4x_2 d^4s \frac{e^2c^2  mMc^4}{(2 \pi \hbar)^9
\sqrt{\omega_{\mathbf{q}}\omega_{\mathbf{q'}}\Omega_{\mathbf{p}}\Omega_{\mathbf{p'}}}}  \times \nonumber   \\
&\mbox{ }& \exp\left(\frac{i}{\hbar}(\tilde{q}-\tilde{q}'-\tilde{s})
\cdot \tilde{x}_1\right)
\exp\left(\frac{i}{\hbar}(\tilde{p}-\tilde{p}'+\tilde{s}) \cdot \tilde{x}_2\right) \times \nonumber   \\
&\mbox{ }&  \overline{u}_a(\mathbf{q}, \sigma) \gamma^{ab}_{\mu}
u_b(\mathbf{q'}, \sigma') \frac{g_{\mu \nu}}{\tilde{s}^2}
 \overline{w}_c(\mathbf{p}, \tau) \gamma^{cd}_{\nu} w_d(\mathbf{p'}, \tau') \nonumber  \\
  &=&   -\frac{ie^2c^2   mMc^4}{4 \pi^2 \hbar
\sqrt{\omega_{\mathbf{q}}\omega_{\mathbf{q'}}\Omega_{\mathbf{p}}\Omega_{\mathbf{p'}}}}
\delta^4(\tilde{p}+\tilde{q}-\tilde{p}'-\tilde{q}') \times \nonumber    \\
&\mbox{ }&  \overline{u}_a(\mathbf{q}, \sigma) \gamma^{ab}_{\mu}
u_b(\mathbf{q'}, \sigma') \frac{g_{\mu \nu}}{(\tilde{q}-\tilde{q}')^2}
 \overline{w}_c(\mathbf{p}, \tau) \gamma^{cd}_{\nu} w_d(\mathbf{p'},
 \tau') \label{eq:2nd-order} \\
   &=&   -\frac{ie^2c^2   mMc^4 \delta^4(\tilde{p}+\tilde{q}-\tilde{p}'-\tilde{q}')U^{\mu}(\mathbf{q}, \sigma;\mathbf{q'}, \sigma')
W_{\mu}(\mathbf{p}, \tau;\mathbf{p'},
\tau')}{4 \pi^2 \hbar
\sqrt{\omega_{\mathbf{q}}\omega_{\mathbf{q'}}\Omega_{\mathbf{p}}\Omega_{\mathbf{p'}}}(\tilde{q}-\tilde{q}')^2}
\label{eq:8.31a}
\end{eqnarray}

\noindent which, as expected, is exactly the same as in the
non-covariant approach (\ref{eq:S_2}).

Using results from Appendix \ref{ss:non-rel} and assuming that the proton is very heavy ($M \to \infty$), we obtain this coefficient function in the $(v/c)^2$ approximation\footnote{We omit the energy delta function and introduce the vector of transferred momentum $\mathbf{k} = \mathbf{q}' - \mathbf{q} = \mathbf{p} - \mathbf{p}'$. }

\begin{eqnarray*}
&\mbox{ }&  s_2(\mathbf{p},\mathbf{q},\mathbf{p}', \mathbf{q}';
\tau, \sigma, \tau',  \sigma' )   \nonumber \\
&\approx& \frac{
ie^2 \delta_{\tau, \tau'}}{ (2\pi)^2 \hbar}    \frac{ mc^2}{
\sqrt{\omega_{\mathbf{q}}\omega_{\mathbf{q'}}}} \frac{1}{k^2}
\ U^{0}(\mathbf{q+k}, \sigma; \mathbf{q},\sigma')  \\
&\approx& \frac{
ie^2 \delta_{\tau, \tau'}}{ (2\pi)^2 \hbar}  \frac{1}{k^2} \left(1 - \frac{q^2}{2m^2c^2} -\frac{\mathbf{qk}}{2m^2c^2} -\frac{k^2}{4m^2c^2} \right)
\chi _{\sigma}^{\dag} \left(1 + \frac{(2 \mathbf{q} + \mathbf{k})^2}{8m^2c^2} +\frac{i\vec{\sigma}_{el} \cdot [\mathbf{k} \times \mathbf{q}]}{4m^2c^2}  \right) \chi_{\sigma'} \\
&\approx& \frac{ie^2 \delta_{\tau, \tau'}\delta_{\sigma, \sigma'}}{ (2\pi)^2 \hbar}  \frac{1}{k^2} \left(1 - \frac{q^2}{2m^2c^2} -\frac{\mathbf{qk}}{2m^2c^2} -\frac{k^2}{4m^2c^2} + \frac{q^2}{2m^2c^2} + \frac{\mathbf{q}\mathbf{k}}{2m^2c^2}  + \frac{k^2}{8m^2c^2} \right)  \\
&+& \frac{ie^2  \delta_{\tau, \tau'}}{ (2\pi)^2 \hbar}  \frac{1}{k^2} \chi _{\sigma}^{\dag} \frac{i\vec{\sigma}_{el} \cdot [\mathbf{k} \times \mathbf{q}]}{4m^2c^2}  \chi_{\sigma'} \\
&=&  \frac{ie^2 \delta_{\tau, \tau'}\delta_{\sigma, \sigma'}}{ (2\pi)^2 \hbar} \left( \frac{1}{k^2}   -\frac{1}{8m^2c^2}  \right)  - \frac{\alpha  \delta_{\tau, \tau'}}{ 4 \pi m^2c}  \chi _{\sigma}^{\dag} \frac{\vec{\sigma}_{el} \cdot [\mathbf{k} \times \mathbf{q}]}{ k^2}  \chi_{\sigma'}
\end{eqnarray*}

\noindent   In the extreme non-relativistic
approximation, we can ignore terms with $c$ in denominators and obtain

\begin{eqnarray}
 S_2[d^{\dag}a^{\dag}da]
 & =&   \frac{ ie^2}{ (2\pi)^2 \hbar} \sum_{\sigma  \tau }\int d\mathbf{p}
d\mathbf{q} d\mathbf{k} \frac{\delta(E(\mathbf{p}, \mathbf{q},
\mathbf{k}))}{k^2}
 d^{\dag}_{\mathbf{p-k}, \tau} d_{\mathbf{p}, \tau}
 a^{\dag}_{\mathbf{q+k}, \sigma} a_{\mathbf{q}, \sigma} \nonumber \\
 \label{eq:s-matr-2nd}
\end{eqnarray}

\noindent This is consistent with  our toy model (\ref{eq:9.78}).
The difference in sign is related to the fact that equation
(\ref{eq:9.78}) describes scattering of two electrons having the
same charge and, therefore, repelling each other, while equation
(\ref{eq:s-matr-2nd}) refers to a Coulomb-type attractive electron-proton
potential.\footnote{see subsection \ref{ss:effective}}

\subsection{Feynman diagrams}
\label{ss:Fdiagrams}

\begin{figure}
\centering
\includegraphics{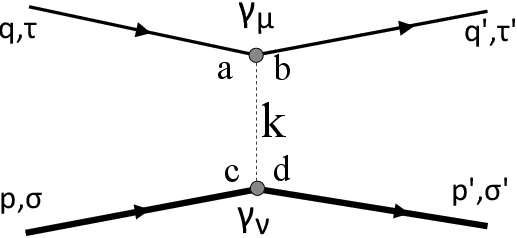} \caption{Feynman diagram for the
electron-proton scattering in the 2nd perturbation order. }
\label{fig:9.6}
\end{figure}

Expression (\ref{eq:2nd-order}) for the scattering amplitude can be
conveniently represented by the \emph{Feynman diagram}
 shown in Fig. \ref{fig:9.6}\footnote{Note
that in spite of some similarities, Feynman diagrams are
fundamentally different from diagrams considered in sections
\ref{sc:model-theory} and \ref{ss:diagrams-general}. Feynman
diagrams describe expressions obtained in the Feynman-Dyson
perturbation theory, while diagrams from sections
\ref{sc:model-theory} and \ref{ss:diagrams-general} are for the use
in the ``old-fashioned'' (non-covariant) perturbation theory. In fact, one Feynman-Dyson diagram is equivalent to a sum of many ``old-fashioned'' diagrams. This explains the convenience and popularity of the canonical perturbation theory.}
\index{Feynman diagram} Here we would like to formulate general
rules for drawing and interpreting Feynman diagrams in QED. They are
called \emph{Feynman rules}. \index{Feynman rules}

The initial state of the electron with momentum $\mathbf{q}$ and
spin component $\tau$ is represented by the factor\footnote{This is
 (\ref{eq:u-factor}) with the exponential
factor dropped. Exponential factors coming from different sources
will be collected and analyzed together later. }

\begin{eqnarray}
\sqrt{\frac{mc^2}{(2 \pi \hbar)^3 \omega_{\mathbf{q}}}}
\overline{u}_{a}(\mathbf{q}, \sigma) \label{eq:el-incom}
\end{eqnarray}

\noindent in (\ref{eq:2nd-order}) and by a thin incoming arrow in
the diagram \ref{fig:9.6}. Similarly, the final state of the
electron is represented by the factor

\begin{eqnarray}
\sqrt{\frac{mc^2}{(2 \pi \hbar)^3 \omega_{\mathbf{q}'}}}
u_{b}(\mathbf{q}', \sigma') \label{eq:el-outgoing}
\end{eqnarray}

\noindent in (\ref{eq:2nd-order}) and by a thin outgoing arrow in
the diagram \ref{fig:9.6}. The incoming and outgoing proton lines
are represented by thick arrows in the diagram and by factors

\begin{eqnarray}
\sqrt{\frac{Mc^2}{(2 \pi \hbar)^3 \Omega_{\mathbf{p}}}}
\overline{w}_{c}(\mathbf{p}, \tau), \ \ \ \ \ \ \
\sqrt{\frac{Mc^2}{(2 \pi \hbar)^3 \Omega_{\mathbf{p}'}}}
w_{d}(\mathbf{p}', \tau') \label{eq:pr-outgoing}
\end{eqnarray}

\noindent respectively. Similarly incoming and outgoing external
photon lines\footnote{See graph \ref{fig:compton}, where external photon lines are shown by dashed arrows.} with momentum $\mathbf{p}$ and helicity $\rho$ correspond to factors

\begin{eqnarray}
\frac{\hbar \sqrt{c}}{\sqrt{2p(2 \pi \hbar)^3}}e_{\mu}(\mathbf{p},
\rho), \ \ \ \ \ \ \  \frac{\hbar \sqrt{c}}{\sqrt{2p(2 \pi
\hbar)^3}}e^*_{\mu}(\mathbf{p}, \rho) \label{eq:free-phot}
\end{eqnarray}

\noindent respectively.

An internal photon line carrying 4-momentum $\tilde{k}$ corresponds to the
photon propagator in (\ref{eq:2nd-order})\footnote{Compare with
(\ref{eq:photon-prop}). Here we omit the integration sign and the
exponential factor, which will be tackled later.}

\begin{eqnarray}
 \frac{\hbar^2c^2g^{\mu \nu}}{2 \pi i (2 \pi \hbar)^3 \tilde{k}^2 }
 \label{eq:phot-propa}
\end{eqnarray}

\noindent It is shown by the dashed line in the diagram \ref{fig:9.6}. An internal electron line\footnote{For brevity, we call
it ``internal electron lines.'' However, in fact, the
corresponding propagator has contributions from both electron and
positron operators.} connects two vertices and corresponds to the
electron propagator\footnote{Compare with (\ref{eq:elec-prop}). }

\begin{eqnarray}
 \frac{(\cross{k} + mc^2)_{bc}}{2 \pi i (2 \pi \hbar)^3(\tilde{k}^2 - m^2
 c^4)} \label{eq:el-propa}
\end{eqnarray}

\noindent Similarly, an internal proton line is associated with the
factor

\begin{eqnarray}
 \frac{(\cross{k} + Mc^2)_{bc}}{2 \pi i (2 \pi \hbar)^3(\tilde{k}^2 - M^2
 c^4)} \label{eq:pr-propa}
\end{eqnarray}

The number of vertices ($\mathcal{V}$) in the graph is the same as
the order of perturbation theory (=2 in the diagram \ref{fig:9.6}). Each vertex is
associated with the factor

\begin{eqnarray*}
\frac{i (2 \pi \hbar)^4 e \gamma^{\mu}_{ab}}{\hbar}
\end{eqnarray*}

\noindent This factor has three summation indices. Two
 bispinor indices $a$ and $b$ are coupled to indices of fermion factors
(\ref{eq:el-incom}) - (\ref{eq:pr-outgoing}) or (\ref{eq:el-propa})
- (\ref{eq:pr-propa}). Thus, in the diagram, each vertex is
connected to two fermion lines (either external or internal). The
4-vector index $\mu$ is coupled to indices of either free photon
factor (\ref{eq:free-phot}) or photon propagator
(\ref{eq:phot-propa}). Correspondingly, each vertex connects to one
photon line (either external or internal).  So, we conclude that each contribution to the QED
S-matrix can be represented simply by drawing a
connected\footnote{As discussed in subsection \ref{ss:linkedness},
we should not consider disconnected diagrams, because they
correspond to non-interesting spatially separated scattering events.}
diagram, whose edges and vertices respect the above connectivity
rules.

Let us now discuss exponential factors and integrations. Each
interaction vertex is associated with a 4-integral $\int d^4x$. Each
incoming external particle line (attached to the vertex with
integration variable $\tilde{x}'$) is associated with exponential
factor $\exp(\frac{i}{\hbar}\tilde{p} \cdot \tilde{x}')$. Each
outgoing external particle line (attached to the vertex with
integration variable $\tilde{x}$) is associated with exponential
factor $\exp(-\frac{i}{\hbar}\tilde{p} \cdot \tilde{x})$. Each
internal line carrying 4-momentum $\tilde{p}$ and connecting
vertices marked by $\tilde{x}$ and $\tilde{x}'$ provides exponential
factor $\exp(\frac{i}{\hbar}\tilde{p} \cdot
(\tilde{x}-\tilde{x}'))$. So, the full exponential factor that
depends on $\tilde{x}$ is $\exp(\frac{i}{\hbar} (\tilde{p}_1 +
\tilde{p}_2 + \tilde{p}_3) \cdot \tilde{x})$, where $\tilde{p}_i$
are 4-momenta (with appropriate signs) of the three lines attached
to the vertex. Integrating on $d^4x$ we obtain the 4-momentum
$\delta$-function $(2 \pi \hbar)^4 \delta^4(\tilde{p}_1 +
\tilde{p}_2 + \tilde{p}_3)$, which expresses ``conservation'' of
the 4-momentum\footnote{We put the word ``conservation'' in quotes, because the 4-momenta of ``virtual particles'' involved here are just integration variables. They cannot be measured and, therefore, unphysical. See next subsection. } at the interaction vertex. This ``conservation'' rule
helps us to assign 4-momentum labels to all lines: The external
lines correspond to real observable particles, so their
energy-momenta are considered given ($p_0, \mathbf{p}$) and ($p'_0,
\mathbf{p}'$) and they are always ``on the mass shell.''
\index{mass shell} I.e., they satisfy conditions

\begin{eqnarray}
p_0 &=& \omega_{\mathbf{p}} = \sqrt{m^2c^4 + p^2c^2} \label{eq:p0omega} \\
p'_0 &=& \omega_{\mathbf{p}'} = \sqrt{m^2c^4 + (p')^2c^2} \label{eq:p'0omega}
\end{eqnarray}

We can arbitrarily assign directions of the momentum flow through
internal lines. For each independent loop we should also introduce
an additional 4-momentum, which, as we will see, is a dummy
integration variable. Then, following the above ``conservation'' rule, we can assign a unique
4-momentum label to each line in the diagram.

\begin{table}[h]
\caption{The correspondence between elements in a Feynman diagram
and factors in the corresponding  scattering
amplitude.}
\begin{tabular*}{\textwidth}{@{\extracolsep{\fill}}lll}
\hline
 Diagram element    & Factor  & Physical interpretation   \cr
\hline incoming electron line & $\sqrt{\frac{mc^2}{(2 \pi
\hbar)^3\omega_{\mathbf{p}}}}\overline{u}_a(\mathbf{p}, \sigma)$ &
electron in the state $| \mathbf{p}, \sigma \rangle$ \cr
 & &  at $t=-\infty$\cr
outgoing electron line & $\sqrt{\frac{mc^2}{(2 \pi
\hbar)^3\omega_{\mathbf{p}}}}u_a(\mathbf{p}, \sigma)$ & electron in
the state $| \mathbf{p}, \sigma \rangle$ \cr
 & & at $t=+\infty$
\cr incoming photon line & $\frac{\hbar\sqrt{c}}{\sqrt{(2 \pi
\hbar)^32p}}e^*_{\mu}(\mathbf{p}, \tau)$ & photon in the state $|
\mathbf{p}, \tau \rangle$ \cr
 & & at $t=-\infty$
\cr outgoing photon line & $\frac{\hbar \sqrt{c}}{\sqrt{(2 \pi
\hbar)^32p}}e_{\mu}(\mathbf{p}, \tau)$ & photon in the state $|
\mathbf{p}, \tau \rangle$ \cr
 & & at $t=+\infty$
\cr internal electron line &
$\frac{\left(\cross{p}+mc^2\right)_{ab}}{(2 \pi i)(2 \pi
\hbar)^3(\tilde{p}^2 -m^2c^4)}$ & no interpretation \cr carrying 4-momentum $p^{\mu}$ & &
\cr
 & & \cr
internal photon line &  $\frac{\hbar^2 c^2g_{\mu \nu}}{(2
\pi i)(2 \pi \hbar)^3 \tilde{p}^{2}}$ & no interpretation \cr
carrying 4-momentum $p^{\mu}$ & & \cr
 & & \cr
interaction vertex &  $\frac{i(2 \pi \hbar)^4e}{\hbar}\gamma^{\mu}_{ab}$ & no
interpretation \cr \hline
\end{tabular*}
\label{table:8.1}
\end{table}

In addition to $d^4x$ integrations discussed above we also have
4-momentum integrals associated with each external line and each
internal line (propagator). As we discussed in subsection
\ref{ss:diagrams}, these integrations are sufficient to kill all
4-momentum delta functions, leaving just one delta function
expressing the conservation of the overall momentum and energy in
the scattering process. The number of integrals left is equal to the
number of independent loops in the diagram.

To summarize, we have the following rules for writing a
$\mathcal{V}$-order matrix element of the scattering operator

\begin{enumerate}
\item Draw a Feynman diagram with $\mathcal{V}$ vertices, $\mathcal{I}$ internal lines, and $\mathcal{L} = \mathcal{I} - \mathcal{V} +1$
independent loops. Each vertex in the diagram should be connected to
two electron (or proton) lines (either external or internal) and to
one photon line (either external or internal). External incoming
(outgoing) lines correspond to initial (final) configuration of
particles in the considered scattering event. Momenta and spins of
particles in these asymptotic states are assumed to be given.
\item Assign arbitrary 4-momentum labels to $\mathcal{L}$ internal
loop lines.
\item Following the 4-momentum ``conservation'' rule  at each vertex,
 assign
4-momentum labels to all remaining internal lines.
\item The integrand is now obtained by putting together numerical
factors corresponding to all lines and vertices in the diagram as
shown in Table \ref{table:8.1}.
\item Integrate the obtained expression on all loop 4-momenta.
\item Multiply  by $1/\mathcal{V}!$ (which comes
from formula (\ref{eq:F-D})) and by an appropriate \emph{symmetry
factor}.\footnote{For example, in our calculation (\ref{eq:t-order})
the symmetry factor of 2 is needed due to the appearance of two
identical expressions under the $t$-ordering sign. This symmetry
factor canceled the $1/2!$ multiplier exactly. In all S-matrix
calculations in this book such cancelation of the symmetry factor
and the $1/\mathcal{V}!$ multiplier also occurs.}
\item Multiply  by a 4D delta function that expresses the conservation
of the total energy-momentum in the scattering process.
\end{enumerate}

\subsection{Compton scattering}
\label{ss:compton}

\begin{figure}
\centering
\includegraphics{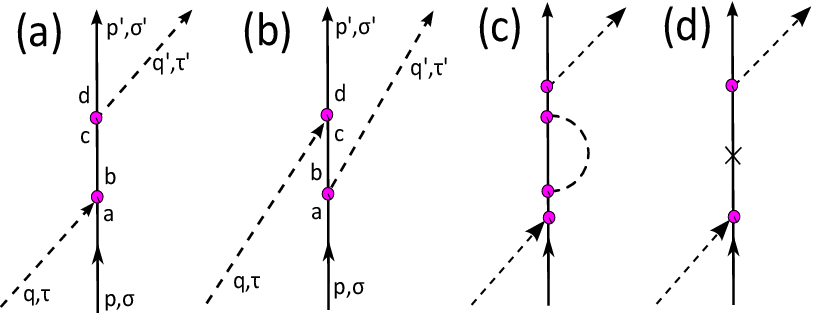} \caption{Feynman diagrams for
electron-photon (Compton) scattering: (a) and (b) - 2nd order terms, (c) and (d) - selected 4th order terms (see section \ref{sc:renormalization-in-QED}).  }
\label{fig:compton}
\end{figure}

As an example of the above rules, consider  the \emph{Compton scattering} \index{Compton scattering} electron+photon $\to$ electron+photon. The two 2nd order diagrams describing this process are shown in Figs. \ref{fig:compton}(a)-(b). According to our rules,  the corresponding scattering amplitude is

\begin{eqnarray*}
&\ &\langle 0 | a_{\mathbf{p}, \sigma } c_{\mathbf{q}, \tau}
S_2[a^{\dag}c^{\dag}ac] a^{\dag}_{\mathbf{p'}, \sigma'}
c^{\dag}_{\mathbf{q'}, \tau'} |0
\rangle \\
&=& \left[\sqrt{\frac{mc^2}{(2 \pi \hbar)^3 \omega_{\mathbf{p}}}} \overline{u}_a(\mathbf{p}, \sigma)\right] \left[ \frac{i(2 \pi \hbar)^4 e \gamma^{\mu}_{ab}}{\hbar} \right] \left[ \frac{\hbar \sqrt{c}e^*_{\mu}(\mathbf{q}, \tau)}{\sqrt{(2 \pi \hbar)^3 2q}} \right]  \left[\frac{\left(\cross{p}+ \cross{q}+mc^2\right)_{bc}}{(2 \pi i)(2 \pi
\hbar)^3((\tilde{p}+ \tilde{q})^2 -m^2c^4)}\right] \\
 &\ & \left[\frac{i(2 \pi \hbar)^4 e \gamma^{\nu}_{cd}}{\hbar} \right]\left[ \frac{\hbar \sqrt{c}e_{\nu}(\mathbf{q}', \tau')}{\sqrt{(2 \pi \hbar)^3 2q'}} \right]  \left[\sqrt{\frac{mc^2}{(2 \pi \hbar)^3 \omega_{\mathbf{p}'}}} u_d(\mathbf{p}', \sigma')\right] \delta^4(\tilde{p} + \tilde{q} - \tilde{p}' - \tilde{q}') \\
 &+& \left[\sqrt{\frac{mc^2}{(2 \pi \hbar)^3 \omega_{\mathbf{p}}}} \overline{u}_a(\mathbf{p}, \sigma)\right] \left[ \frac{i(2 \pi \hbar)^4 e \gamma^{\nu}_{ab}}{\hbar} \right] \left[ \frac{\hbar \sqrt{c}e_{\nu}(\mathbf{q}', \tau')}{\sqrt{(2 \pi \hbar)^3 2q'}} \right]  \left[\frac{\left(\cross{p}- \cross{q}'+mc^2\right)_{bc}}{(2 \pi i)(2 \pi
\hbar)^3((\tilde{p}- \tilde{q}')^2 -m^2c^4)}\right] \\
 &\ & \left[\frac{i(2 \pi \hbar)^4 e \gamma^{\mu}_{cd}}{\hbar} \right] \left[ \frac{\hbar \sqrt{c}e^*_{\mu}(\mathbf{q}, \tau)}{\sqrt{(2 \pi \hbar)^3 2q}} \right] \left[\sqrt{\frac{mc^2}{(2 \pi \hbar)^3 \omega_{\mathbf{p}'}}} u_d(\mathbf{p}', \sigma')\right] \delta^4(\tilde{p} + \tilde{q} - \tilde{p}' - \tilde{q}')
\end{eqnarray*}

\noindent From this expression one can obtain the cross section for the elastic electron-photon scattering. We are not going to reproduce this standard calculation,\footnote{See, for example section 5.5 in \cite{Peskin} and section 8.7 in \cite{book}.} but only note that in the limit of low photon energy one gets exactly the \emph{Thomson cross section formula} \index{Thomson formula} familiar from classical electrodynamics. Since the Thomson formula was well tested experimentally, this suggests that all higher order contributions to our low-energy result should vanish. We will find this observation useful in our discussion of the charge renormalization condition in the next chapter.

\subsection{Virtual particles?}
\label{ss:virtual_particles}

As we mentioned above, external lines of Feynman diagrams represent
asymptotic states of real particles. The momentum and spin labels
attached to these lines directly correspond to the values of
observables that can be measured in these states.

In some QFT textbooks one can also read about interpretation,
according to which internal lines in Feynman diagrams correspond to
so-called ``virtual particles.'' Thus, internal photon lines
correspond to virtual photons and internal electron (positron) lines
correspond to virtual electrons (positrons). Clearly,
energy-momentum labels attached to internal lines do not satisfy the
basic energy-momentum relationships (\ref{eq:p0omega}) - (\ref{eq:p'0omega}) characteristic for
real particles.\footnote{In the example
shown in diagram \ref{fig:9.6}, the ``momentum'' of the virtual
photon is $\mathbf{q}-\mathbf{q}'$ and its ``energy'' is
$\omega_{\mathbf{q}} - \omega_{\mathbf{q}'}$. It is easy to show
that in a general case the usual massless relationship
$c|\mathbf{q}-\mathbf{q}'| = \omega_{\mathbf{q}} -
\omega_{\mathbf{q}'}$ is not satisfied.} This is usually described
as virtual particles being ``out of the mass shell.'' In spite of
their strange properties, virtual particles are often regarded as
real objects and their ``exchange'' is claimed to be the origin of
interactions. For example, diagram \ref{fig:9.6} is interpreted as a
depiction of a process in which a virtual photon is ``exchanged''
between electron and proton, thus leading to their attraction.

These interpretations are not supported by evidence. They are rather misleading. In
fact, Feynman diagrams are not depictions of particle trajectories or real events.
Lines and vertices in Feynman diagrams are simply graphical representations
of certain factors in integrals in scattering amplitudes. Quantum
theory does not provide any mechanistic description of interactions
(like ``exchange'' of virtual particles). The only reliable
information about interactions is contained in the interaction
Hamiltonian, which does not suggest that some invisible virtual
particles are emitted, absorbed, and exchanged during particle collisions.

\chapter{RENORMALIZATION} \label{ch:renormalization}

\begin{quote}
\textit{There is no great thing that would not be surmounted by a
still greater thing. There is no thing so small that no smaller
thing could fit into it.}

\small
\hspace{1in} Kozma Prutkov
\normalsize
\end{quote}

\vspace{0.5in}

In the preceding chapter we calculated 2nd order contributions to the $S$-operator. These were the 2nd and 3rd terms in the perturbation expansion (\ref{eq:11.34}). It can be demonstrated that the obtained result (\ref{eq:S_2}) is in a pretty good agreement with experiments on electron-proton scattering. Similarly, one can obtain rather accurate 2nd order
results for other scattering events, such as the Compton scattering or the
electron-positron annihilation. Can we then expect to get even better agreement by evaluating higher order terms in  (\ref{eq:S_2})? Sadly, the answer to this question is \emph{no}.  As we are going to see in this chapter, many high-order terms in (\ref{eq:S_2}) are not just inaccurate - they are infinite!

\begin{figure}
\centering
\includegraphics {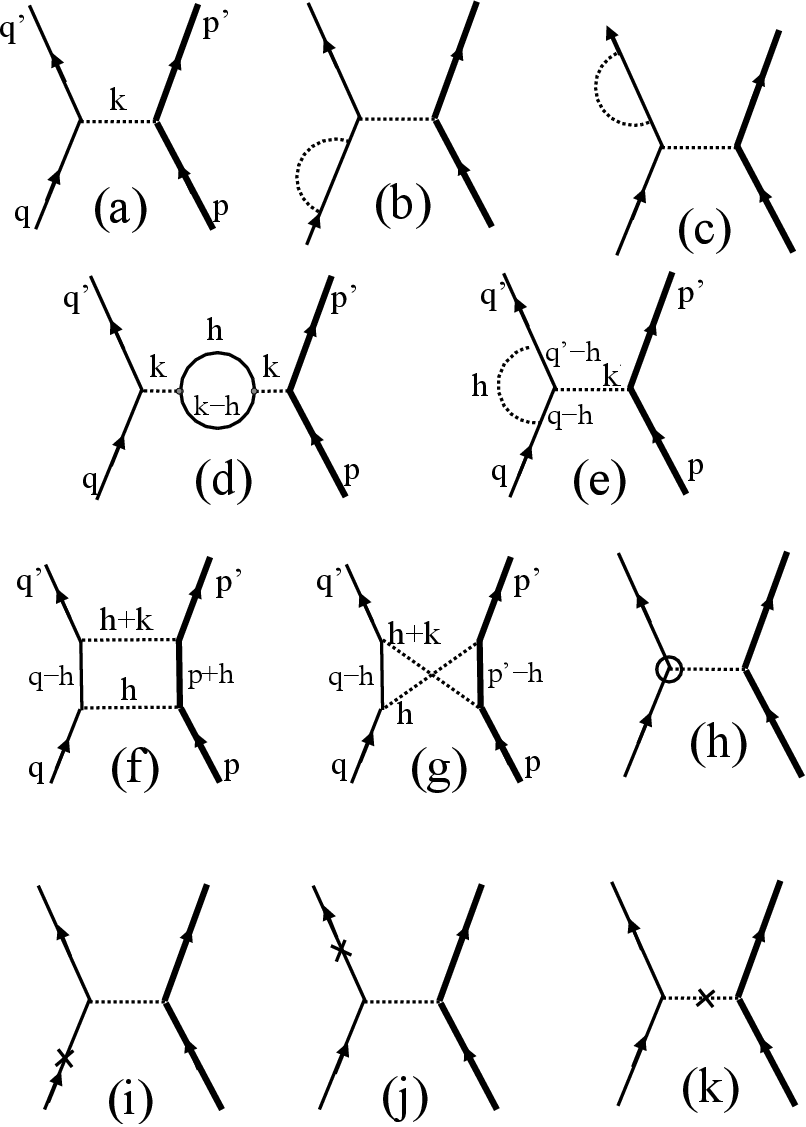} \caption{Feynman diagrams for the
electron-proton scattering up to the 4th perturbation order. Thick full lines - protons, thin full lines - electrons, dotted lines - photons. A small circle in (h) and crosses in (i) - (k) denote counterterms that will be discussed in section \ref{sc:renormalization-in-QED}.}
\label{fig:9.10}
\end{figure}

To get a better understanding of this remarkable failure, in Fig.
\ref{fig:9.10} we show all 2nd and 4th order Feynman diagrams that are relevant to the
electron-proton scattering.\footnote{Diagrams \ref{fig:9.10}(h-k) are
obtained from renormalization counterterms that will be discussed in
section \ref{sc:renormalization-in-QED}.} Generally, all diagrams can be divided into two broad groups: \emph{tree diagrams} and \emph{loop diagrams}. \index{tree diagram} \index{loop diagram} In our example there is only one tree diagram \ref{fig:9.10}(a). This is the same diagram as
\ref{fig:9.6} evaluated in the preceding chapter.   In general, all amplitudes whose
Feynman diagrams are tree-like (i.e., do not contain loops), come
out fairly accurate.

Serious troubles are associated with loop diagrams, such as \ref{fig:9.10}(b-g) in the 4th perturbation order.\footnote{ Here we show only diagrams, in which
loops are associated with electron lines. For a complete
treatment, one should also take into account loops, similar to \ref{fig:9.10}(b), (c), (g), but associated with
proton lines. However, it can be shown that their contributions to scattering amplitudes is much smaller due to the inequality $m \ll M$. So, we will omit them in our analysis.} As we saw in section \ref{ss:diagrams-general}, the appearance of loops is inevitable in high order calculations, and such loops lead to potentially divergent integrals. There are two types of divergences associated with loops in QED. First, it can
be shown\footnote{see Appendix \ref{ss:b-integrals}} that loop
integrals in diagrams \ref{fig:9.10}(b-c) and (e-g) diverge due to singularity
at zero loop momentum. These are \emph{infrared divergences}
\index{infrared divergences} \cite{book}, whose cancelation will be discussed in Chapter \ref{ss:hydrogen}.

Another problem is the divergence of loop integrals \ref{fig:9.10}(b-e)
 at high loop momenta.\footnote{see subsection
\ref{ss:convergence}} These are also known as
 \emph{ultraviolet divergences}. \index{ultraviolet divergences}
The way to fix this problem is provided by the
\emph{renormalization} \index{renormalization} theory developed by
Tomonaga,
 Schwinger, and Feynman in the late 1940's.
 Basically, this approach says that the QED interaction
operator (\ref{eq:11.6}) - (\ref{eq:11.7})

\begin{eqnarray}
V  &=& V_1  + V_2  \label{eq:V1-2}
\end{eqnarray}

\noindent is not complete. It must be corrected by adding certain
\emph{counterterms}. The counterterms are formally infinite
operators. However, if they are carefully selected,  then one can cancel the
infinities occurring in loop integrals, so that only some residual
finite contributions (\emph{radiative corrections} \index{radiative
corrections}) remain in each perturbation order.   These small radiative corrections are exactly what is needed to obtain scattering cross sections, energies of bound states,  and some other properties in remarkable agreement with
experiments. The procedure of adding counterterms to the interaction operator is called \emph{renormalization}.\footnote{We should note, however, that the brief story of renormalization presented above is different from what can be found in most textbooks. The usual explanation of renormalization involves discussion of the difference between \emph{bare} and \emph{physical} particles. While former  have infinite masses and charges, the latter acquire observable masses and charges due to formation of \emph{virtual clouds}. See section \ref{sc:troubles}. In our approach we deal exclusively with physical particles. We formulate the renormalization program as a modification of the full Hamiltonian.} \index{renormalization}

\section{Renormalization conditions}
\label{sc:renormalization-in-QED2}

In this section we will be interested in  rather general physical principles that underlie the renormalization approach. We will summarize these principles in the form of two \emph{renormalization conditions} \index{renormalization conditions} that are called the \emph{no-self-scattering condition} and the \emph{charge renormalization condition}.

\subsection{Regularization}
\label{ss:regularization}

As we mentioned above, loop integrals in QED are ultraviolet-divergent and/or infrared-divergent and it is difficult to do calculations
with infinite quantities. To make things easier, it is convenient to
perform
 \emph{regularization}. \index{regularization} The idea is
 to modify the theory by hand in such a way that all loop integrals are
forced to be finite, so that intermediate manipulations do not
involve infinities.  The simplest regularization approach adopted in
Appendix \ref{ss:b-integrals} is to introduce momentum cutoffs
 in all loop integrals. The cutoffs depend on two parameters having the dimensionality of mass:
the ultraviolet cutoff  $\Lambda$ \index{ultraviolet cutoff} limits integrals at high loop momenta and
infrared cutoff $\lambda$ \index{infrared cutoff} controls integrations at low momenta singularities.
 Of course, the theory with such
truncated integrals cannot be exact. To obtain final results, at the end of calculations the ultraviolet cutoff
momentum should be set to infinity $\Lambda \to \infty$ and the infrared cutoff should be set to zero $\lambda \to 0$.\footnote{In this chapter we will keep $\lambda$ non-zero. The limit $\lambda \to 0$ and associated infrared divergences will be discussed in chapter \ref{ss:hydrogen}.} If counterterms in the Hamiltonian are chosen  correctly, then in these limits the $S$-matrix elements should remain finite, accurate, and cutoff-independent.

\subsection{No-self-scattering renormalization condition} \label{ss:mass-renorm}

First we would like to  note that the divergence of loop
integrals is not the biggest problem that we face in QED. Even if
all loop integrals were convergent,
the $S$-operator might still contain nasty infinities. Let us now consider
in more detail how these infinities appear and what we can do about
them.

Recall that QED interaction operator (\ref{eq:V1-2}) has \emph{unphys}, \emph{phys}, and
\emph{renorm}  terms. So the corresponding scattering phase operator $F$\footnote{see equation (\ref{eq:11.34})}
must contain terms of the same types. Then we can write the most general expression for the $S$-operator as

\begin{eqnarray}
S &=& \exp(\underbrace{F}) = \exp(\underbrace{F^{unp}} + \underbrace{F^{ph}} +
 \underbrace{F^{ren}} )  \nonumber\\
&=& \exp(\underbrace{F^{ph}} + \underbrace{F^{ren}})
\label{eq:11.41}
\end{eqnarray}

\noindent where we noticed that \emph{unphys} terms in $F$ do not
contribute to the $S$-operator due to equation (\ref{eq:9.61}). Let us
now apply the scattering operator (\ref{eq:11.41}) to an one-electron
state $ a^{\dag}_{\mathbf{p}, \sigma}|0 \rangle$. It follows from
Lemma \ref{Lemma9.4} that  \emph{phys} operators yield zero when acting on
one-particle states. \emph{Renorm}  operators do not change the number of
particles. Therefore, we can write

\begin{eqnarray}
&\ & S a^{\dag}_{\mathbf{p}, \sigma}|0 \rangle \nonumber \\
&=&
\exp(\underbrace{F^{ph}} + \underbrace{F^{ren}})
a^{\dag}_{\mathbf{p},
\sigma}|0 \rangle  \nonumber \\
&=& \left(1 +  \underbrace{F^{ph}} + \underbrace{F^{ren}} + \frac{1}{2!}
(\underbrace{F^{ph}} + \underbrace{F^{ren}})^2 + \ldots \right)
a^{\dag}_{\mathbf{p},
\sigma}|0 \rangle  \nonumber \\
&=& \left(1 +  \underbrace{F^{ph}} + \underbrace{F^{ren}} +\frac{1}{2!}
(\underbrace{F^{ph}})^2 + \frac{1}{2!} \underbrace{F^{ph}}
\underbrace{F^{ren}} + \frac{1}{2!} \underbrace{F^{ren}}
\underbrace{F^{ph}} + \frac{1}{2!}  (\underbrace{F^{ren}})^2 +
\ldots \right) a^{\dag}_{\mathbf{p},
\sigma}|0 \rangle  \nonumber \\
&=& \left(1  + \underbrace{F^{ren}} + \frac{1}{2!}
(\underbrace{F^{ren}})^2 + \ldots \right) a^{\dag}_{\mathbf{p},
\sigma}| 0 \rangle  \nonumber \\
&=& \exp(\underbrace{F^{ren}})   a^{\dag}_{\mathbf{p}, \sigma}|0
\rangle \label{eq:9.2aa}
\end{eqnarray}

\noindent  A similar derivation can be performed for a one-photon
state $c^{\dag}_{\mathbf{p}, \tau}|0 \rangle$ and the vacuum vector

\begin{eqnarray}
S c^{\dag}_{\mathbf{p}, \tau}|0 \rangle &=&
\exp(\underbrace{F^{ren}}) c^{\dag}_{\mathbf{p}, \tau}|0 \rangle \label{eq:9.2bb} \\
S
| 0 \rangle &=& \exp(\underbrace{F^{ren}})   |0 \rangle \label{eq:9.2cc}
\end{eqnarray}

\noindent So, the ``scattering'' in these states depends only on
the \emph{renorm}  part of $F$. We know from (\ref{eq:9.58}) that the
$t$-integral $\underbrace{F^{ren}}$ is infinite, even if $F^{ren}$ is itself finite. Therefore, if $F^{ren} \neq 0$ then the $S$-operator multiplies 0- and 1-particle states by infinite phase factors. This divergence is unacceptable.

Intuitively, we expect single particle states and the vacuum to evolve freely during the entire time interval from $t=-\infty$ to $t=+\infty$. This means that there can be no non-trivial scattering in such states. This also means that the $S$-operator must be equal to the identity operator when acting on such states. But this condition is not satisfied in the QED theory presented so far.

We have two options to deal with this problem. One option is to claim (as advised in many textbooks) that $|0 \rangle $ is not the true (physical) vacuum state of the system and that $a^{\dag}_{\mathbf{p}, \sigma}|0 \rangle$, $c^{\dag}_{\mathbf{p}, \tau}|0 \rangle$ are not true (physical) one-particle states. These are examples of the so-called ``bare'' states.\footnote{Some say that the bare electron is surrounded by a cloud of virtual photons and particle-antiparticle pairs.} The real physical 0-particle and 1-particle states should be obtained as linear combinations of the bare states for which the self-scattering is absent. Then scattering theory of such physical particles would not have divergences and paradoxes.

In this book we will adopt a different\footnote{but, in some respect, equivalent} point of view. We will maintain that  $|0 \rangle $, $a^{\dag}_{\mathbf{p}, \sigma}|0 \rangle$, $c^{\dag}_{\mathbf{p}, \tau}|0 \rangle, \ldots$ are true representatives of real physical 0-particle and 1-particle states. Then our explanation for the divergent results (\ref{eq:9.2aa}) - (\ref{eq:9.2cc}) is that we have started to develop our theory from a wrong interaction operator (\ref{eq:V1-2}). We insist that this operator must be modified or \emph{renormalized}, so that the theory is forced to be finite. In particular, we will demand that the new renormalized interaction satisfies the condition

 \begin{eqnarray}
F^{ren} = 0 \label{eq:11.42}
\end{eqnarray}

\noindent This implies that operator $\underbrace{F}$ is
purely \emph{phys}

 \begin{eqnarray*}
\underbrace{F} = \underbrace{F^{ph}} \label{eq:11.43}
\end{eqnarray*}

\noindent If this condition is satisfied then (\ref{eq:9.2aa}) - (\ref{eq:9.2cc}) take physically acceptable forms

\begin{eqnarray*}
S a^{\dag}_{\mathbf{p}, \sigma}|0 \rangle &=& a^{\dag}_{\mathbf{p}, \sigma}|0 \rangle \\
S c^{\dag}_{\mathbf{p}, \tau}|0 \rangle &=& c^{\dag}_{\mathbf{p}, \tau}|0 \rangle \\
S | 0 \rangle &=& | 0 \rangle
\end{eqnarray*}

\noindent Taking into account the perturbation expansion $S= 1 + S_2 + S_3 + \ldots$, we can also write in each perturbation order of our theory

\begin{eqnarray}
S_n a^{\dag}_{\mathbf{p}, \sigma}|0 \rangle &=& 0 \label{eq:S_na} \\
S_n c^{\dag}_{\mathbf{p}, \tau}|0 \rangle &=& 0 \label{eq:S_nc} \\
S_n | 0 \rangle &=& 0 \label{eq:S_n0}
\end{eqnarray}

\noindent where $n=2,3, \ldots$. And for elements of the $S$-matrix we have

\begin{eqnarray}
\langle 0| S_n |0 \rangle &=& 0 \label{eq:scatter-ampl1}\\
\langle 0|a_{\mathbf{p}, \sigma} S_n
a^{\dag}_{\mathbf{p}', \sigma'}|0 \rangle &=& 0 \label{eq:scatter-ampl}\\
\langle 0|c_{\mathbf{p}, \sigma} S_n c^{\dag}_{\mathbf{p}',
\sigma'}|0 \rangle &=& 0 \label{eq:scatter-ampl2}
\end{eqnarray}

The above  conditions can be summarized  as the following

\begin{statement} [no-self-scattering renormalization condition] \label{statementU}
There should be no
(self-)scattering in the vacuum and one-particle states.\footnote{Note that our no-self-scattering condition is actually equivalent to the more traditional ``mass renormalization'' condition.   For example, in section \ref{sc:renormalization-in-QED} we will see that our 2nd order renormalization counterterms are exactly the same as electron and photon self-energy counterterms in textbook QED.}
\end{statement}

\noindent The physical interpretation is obvious: scattering is expected to occur only when
there are at least two particles which interact with each other. One
particle has nothing to collide with, so it cannot experience scattering. Similarly, no scattering can happen in the
no-particle vacuum state.

\subsection{Charge renormalization condition} \label{ss:crc}

The no-self-scattering renormalization condition sets strict requirements
(\ref{eq:scatter-ampl1}) - (\ref{eq:scatter-ampl2}) on matrix
elements of the $S$-operator between 0-particle and 1-particle
states. However, this condition alone is not sufficient to guarantee cancelation of ultraviolet divergences in scattering calculations. On physical grounds we can derive another necessary renormalization
condition.

Recall that the 2nd order electron-proton scattering amplitude
(\ref{eq:2nd-order}) has a singularity $\propto e^2/\tilde{k}^2$ at
zero transferred momentum $\tilde{k} = \tilde{q}' - \tilde{q} =0$.
As shown in subsection \ref{ss:effective}, in the position space this singularity gets Fourier-transformed into the
long-range Coulomb potential $-e^2/(4 \pi r)$. From classical physics and experiment
we also know that the Coulomb potential is a very accurate description of the interaction of charges at large distances and low energies. Similarly, in subsection \ref{ss:compton} we have established that the 2nd perturbation order describes the low-energy electron-photon scattering very accurately. So, we should not expect any high-order corrections to this result. We are now raising these observations to the level of a fundamental physical principle

\begin{postulate}
 [charge renormalization condition]
Charge-charge and charge-photon elastic scattering cross sections at large distances and low energies
are described \emph{exactly} by the 2nd order term $S_2$ in the
$S$-operator. All higher order contributions to these results should vanish. \label{postulateV}
\end{postulate}

\noindent When applied to charge-charge scattering,  the charge renormalization condition implies
that in orders higher than 2nd, scattering amplitudes should not be singular at low values of the transfer
momentum $\tilde{k}$. Suppose that this is not true, and that the 4th order
electron-proton scattering amplitude has a singularity $\propto
e^4/\tilde{k}^2$. Then the long-range electron-proton
potential would obtain an unacceptable form

\begin{eqnarray*}
V(r) \approx -\frac{e^2 + Ce^4}{4 \pi r}
\end{eqnarray*}

\noindent with a non-zero constant $C$, but from experiments we know that $C=0$ to a high degree of precision.

\subsection{Renormalization in Feynman-Dyson theory}
\label{ss:ren-fd}

In the next section we will see that the no-self-scattering and charge
renormalization conditions are not satisfied in QED with interaction
Hamiltonian (\ref{eq:V1-2}). This can be seen already from the fact
that interaction operator $V_1$ (\ref{eq:11.6}) contains \emph{unphys} terms like $a^{\dag}c^{\dag}a + a^{\dag}ac$.
Commutators of two \emph{unphys} terms may contain \emph{renorm}
parts.\footnote{see Table \ref{table:7.2}} So, there is nothing to
prevent the appearance of \emph{renorm}  terms in the scattering phase
operator (\ref{eq:7.63b})

\begin{eqnarray*}
F   &=&  V_1  - \frac{1}{2 }[\underline{V_1 },V_1 ] +\ldots
\end{eqnarray*}

\noindent in violation of the condition
(\ref{eq:11.42}). We will see that interaction (\ref{eq:V1-2}) violates the charge
renormalization condition as well. These two problems are very serious even though
they are not directly related to divergences in loop diagrams.
The presence of such divergences is just another argument that QED
interaction $V_1+V_2 $ must be modified.

The idea of the renormalization approach is to switch from the
interaction Hamiltonian $V_1+V_2$ to another
interaction $V^c $ by addition of
\emph{renormalization counterterms}, \index{counterterms} which will
be denoted collectively by $Q $

\begin{eqnarray}
V^c  &=& V_1  + V_2  +Q \label{eq:V1-2x}
\end{eqnarray}

\noindent The form of $Q $ must be chosen such that  the no-self-scattering
and charge
renormalization conditions are satisfied. In particular, the new scattering phase operator

\begin{eqnarray}
F^c   &=&  V^c  - \frac{1}{2 }[\underline{V^c },V^c ]
+\ldots
\end{eqnarray}

\noindent should not contain \emph{renorm}  terms. Moreover, high-order contributions $F^c_n \ \   (n>2)$ to the scattering of charged particles should be nonsingular at $\tilde{k} = 0$.

Unfortunately, the program outlined above is difficult to implement.
The reason is that scattering calculations with the QED Hamiltonian
(\ref{eq:V1-2}) are rather cumbersome.  As we discussed in subsection
\ref{ss:Fey}, it is much easier to use the Feynman-Dyson approach,
in which the 2nd order interaction operator $V_2$ is omitted, and the momentum space
photon propagator is taken simply as $\propto g_{\mu
\nu}/\tilde{p}^2$. This is the standard way to calculate the
$S$-matrix in QED, and we will adopt this approach in the present
chapter. The general idea of renormalization remains the same. We
are looking for certain counterterms $Q^{FD} $ that can be added to the original
interaction operator

\begin{eqnarray}
 V_1(t)
 &=& -e  \int d\mathbf{x}  \overline{\psi}(\tilde{x}) \gamma_{\mu}
 \psi (\tilde{x}) A^{\mu} (\tilde{x}) \label{eq:basic-int}
\end{eqnarray}

\noindent to obtain renormalized interaction

\begin{eqnarray}
V_{FD}^c  &=& V_1  +Q^{FD} \label{eq:V1-2y}
\end{eqnarray}

\noindent with which the Feynman-Dyson perturbation expansion
(\ref{eq:F-D})

\begin{eqnarray}
 S^c = 1 - \frac{i}{\hbar}\int \limits_{-\infty }^{+\infty} dt_1 V_{FD}^c(t_1)
 - \frac{1}{2!\hbar^2} \int \limits_{-\infty }^{+\infty} dt_1 dt_2 T[V_{FD}^c(t_1)
 V_{FD}^c(t_2)] \ldots \label{eq:9.11x}
\end{eqnarray}

\noindent becomes finite, and our two renormalization
conditions are satisfied.

\section{Counterterms}
\label{sc:renormalization-in-QED}

Next we would like to  see how the program outlined above works in
practice.  In this section we are going to derive explicit expressions for counterterms $Q^{FD}$ in the 2nd and 3rd order.

\subsection{Electron self-scattering} \label{ss:self-en}

 Let us first
discuss the electron$\to$electron scattering and see how exactly the condition (\ref{eq:scatter-ampl}) is violated.

\begin{figure}
\centering
\includegraphics {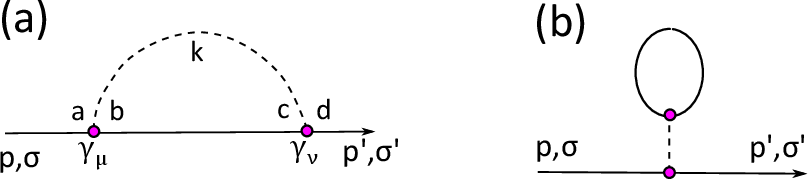} \caption{Feynman diagrams for the
scattering electron$\to$ electron in the 2nd perturbation order. }
\label{fig:9.5}
\end{figure}

There are only two connected diagrams that contribute to the
electron$\to$electron scattering in the 2nd order. They are shown in Fig.
\ref{fig:9.5}. First we will investigate the effect of \ref{fig:9.5}(a). Applying  Feynman rules from Table
\ref{table:8.1} to this diagram, we obtain

\begin{eqnarray}
&\ & \langle 0|a_{\mathbf{p}, \sigma} S_{2a}^{FD}
a^{\dag}_{\mathbf{p}', \sigma'} | 0 \rangle \nonumber \\
&=& -\frac{me^2c^4 \delta^4 (\tilde{p}-\tilde{p}') }{(2 \pi i)^2 (2
\pi \hbar) \omega_{\mathbf{p}}}
 \overline{u}_a(\mathbf{p}, \sigma) \left[\int d^4k
\gamma^{ab}_{\mu} \frac{(\cross{p} - \cross{k} + mc^2)_{bc}}{(\tilde{p}
- \tilde{k})^2 - m^2 c^4} \cdot \frac{g^{\mu \nu}}{\tilde{k}^2 }
\gamma^{cd}_{\nu} \right]
u_b(\mathbf{p}', \sigma') \nonumber \\
\label{eq:square}\\
 &=& \frac{me^2c^4
\delta^4(\tilde{p}-\tilde{p}')}{(2 \pi)^2 (2 \pi \hbar)
\omega_{\mathbf{p}}}\overline{u}_a(\mathbf{p}, \sigma)\Bigl( C_0
\delta_{ab} + C_1(\cross{p} - mc^2)_{ab} + R_{ab}(\cross{p}) \Bigr)
u_b(\mathbf{p}, \sigma') \nonumber \\
 \label{eq:plog}
\end{eqnarray}

\noindent where (divergent) constants $C_0$ and $C_1$ are
calculated in (\ref{eq:CC}) and (\ref{eq:coeff-a}), respectively.\footnote{ Quantities $C_0$ and $R$ have the dimension
of $\langle m \rangle \langle c^{-1} \rangle$ and $\langle C_1 \rangle
= \langle c^{-3} \rangle$. Therefore, the dimension of the scattering amplitude
(\ref{eq:plog}) is $\langle p^{-3} \rangle$, as expected from (\ref{eq:a-dimension}).}

\begin{eqnarray*}
C_0 &=&  \frac{3 \pi^2mc^2}{2i c^3} \left( 4 \ln \frac{\Lambda}{m}
+1 \right)  \\
C_1
&=& -\frac{2\pi^2}{i c^3} \left(\ln \frac{
\Lambda}{ m }   +2  \ln
\frac{\lambda}{m} + \frac{9}{4}\right)
\end{eqnarray*}

\noindent The finite quantity  $R(\cross{p})$ includes terms quadratic, cubic, and higher order in $\cross{p} - mc^2$

\begin{eqnarray}
R(\cross{p}) = C_2 (\cross{p} -mc^2)^2 + C_3 (\cross{p} -mc^2)^3 + \ldots \label{eq:notp}
\end{eqnarray}

 By stripping away factors corresponding to external electron
 lines
and the delta function in (\ref{eq:plog}), we obtain the
contribution from the loop itself and from two attached vertices

\begin{eqnarray}
\mathcal{D}_{loop}(\cross{p})
 &=& \hbar^2 e^2 c^2 \Bigl(C_0  + C_1(\cross{p} - mc^2) +
 R(\cross{p}) \Bigr) \label{eq:e2c2}
\end{eqnarray}

\noindent This non-vanishing result contradicts the no-self-scattering
condition (\ref{eq:scatter-ampl}). Moreover, this expression is clearly infinite in the limit $\Lambda
\to \infty$. So, we are dealing with an ultraviolet divergence here.

Now let us consider an arbitrary Feynman diagram, which can contain
electron-photon loops in external and/or internal lines. If the loops shown in Figure \ref{fig:9.5}(a) is inserted in an external electron line, then the 4-momentum $\tilde{p}$ is on the
mass shell,\footnote{This is true for loops in diagrams
\ref{fig:9.10}(b-c) and also for the diagram \ref{fig:9.5}(a).} and
only the constant term in (\ref{eq:e2c2}) survives\footnote{Here we
formally write the mass shell condition ($\tilde{p}^2 = m^2c^4$) as $\cross{p}=mc^2$ due to (\ref{eq:J.45a}).}

\begin{eqnarray}
\mathcal{D}_{loop}(\cross{p}=mc^2)
 &=& \hbar^2 e^2 c^2 C_0 = 3ie^2 \pi^2 mc\left(2 \ln \frac{\Lambda}{m} + \frac{1}{2}\right) \label{eq:e2c2x}
\end{eqnarray}

\noindent
For loops in internal electron lines, the 4-momentum $\tilde{p}$ is
not necessarily on the mass shell, so the full factor
(\ref{eq:e2c2}) should be taken into account.

Next consider the other electron self-scattering diagram \ref{fig:9.5}(b).\footnote{The divergent denominator $1/(\tilde{p} -\tilde{p}')^2$ can be regularized by prescription (\ref{eq:ck2}).}

\begin{eqnarray*}
&\mbox{ }& \langle 0|a_{\mathbf{p}, \sigma} S_{2b}^{FD}
a^{\dag}_{\mathbf{p}', \sigma'} | 0 \rangle  \\
 &=& \frac{mc^2}{(2 \pi \hbar)^3 \omega_{\mathbf{p}}} \overline{u}_a(\mathbf{p}, \sigma)\gamma_{ab}^{\mu} u_b(\mathbf{p}', \sigma') \left(-\frac{(2 \pi \hbar)^8e^2}{\hbar^2}\right) \frac{\hbar^2 c^2 g_{\mu \nu} \delta^4(\tilde{p} - \tilde{p}')}{(2 \pi i) (2 \pi \hbar)^3 (\tilde{p} -\tilde{p}')^2} \times \\
&\ & \frac{1}{(2 \pi i) (2 \pi \hbar)^3 }\int d^4k \frac{(\cross{k} +mc^2)_{cd} \gamma^{\nu}_{cd}}{\tilde{k}^2 -m^2c^4}
\end{eqnarray*}

\noindent The integral on $\tilde{k}$ vanishes due to (\ref{eq:trace-gamma}) and (\ref{eq:trace-gamma2})

\begin{eqnarray*}
 \int d^4k \frac{Tr(\gamma^{\nu} \gamma^{\rho}k_{\rho} + \gamma^{\nu} mc^2 )}{ \tilde{k}^2 -m^2c^4} = \int d^4k \frac{4k^{\nu} }{ \tilde{k}^2 -m^2c^4} = 0
\end{eqnarray*}

\noindent so, diagrams like \ref{fig:9.5}(b) make no  contributions in both external and internal electron lines.

\subsection{Electron self-energy counterterm}
\label{ss:elec-mass}

From subsection \ref{ss:ren-fd} we know that the above
divergences (\ref{eq:e2c2}) and (\ref{eq:e2c2x}) should be canceled by a 2nd order
counterterm.
Of course, that counterterm cannot be chosen arbitrarily. It becomes a part  of the interaction operator, so it should obey the  conditions formulated  for such operators in subsection \ref{ss:weinberg}. In particular, our addition of the counterterm should  not affect the relativistic invariance of the theory, so the condition (\ref{eq:as-a-scalar}) is essential.
Taking these considerations into account, let us choose the following electron self-energy counterterm

\begin{eqnarray}
Q^{FD}_{2el}(t) = \delta m_2 \int d
\mathbf{x}\overline{\psi}(\tilde{x})\psi(\tilde{x}) + (Z_2-1)_2 \int
d \mathbf{x} \overline{\psi}(\tilde{x})(-i \hbar
c\gamma_{\mu}\partial^{\mu} + mc^2)\psi(\tilde{x}) \nonumber \\
 \label{eq:deltamc2}
\end{eqnarray}

\noindent where the 4-gradient $\partial^{\mu}$ is defined  as

\begin{eqnarray}
\partial^{\mu} \equiv \left(-\frac{1}{c}\frac{\partial}{\partial t},
\frac{\partial}{\partial x}, \frac{\partial}{\partial y},
\frac{\partial}{\partial z}\right) \label{eq:partial-mu}
\end{eqnarray}

\noindent and parameters $\delta m_2$ and $(Z_2-1)_2$ must be
adjusted to satisfy renormalization conditions.\footnote{$\delta m_2$
has the dimension of energy and $(Z_2-1)_2$ is dimensionless.
Both of them are second-order quantities, as indicated by the
subscripts. Here the numerical factors $\delta m_2$ and $(Z_2)_2$ are regarded as unknown constants whose values need to be adjusted to cancel out divergent terms in (\ref{eq:e2c2}). Eventually, we will see that these factors coincide with renormalization parameters - the mass shift and the electron wave function renormalization factor - in usual approaches. Note that in our philosophy of renormalization we are not following the usual logic. In particular, we are not ``shifting'' the electron's mass and do not multiply the electron field by a factor, as in \cite{book}. } The
2nd order contribution to the electron$\to$electron scattering
amplitude resulting from interaction (\ref{eq:deltamc2}) is\footnote{This formula is obtained
by inserting $Q^{FD}_{2el}(t)$ instead of $V_{FD}^c(t)$ in
the second term on the right hand side of (\ref{eq:9.11x}).}

\begin{eqnarray}
&\ & \langle 0|a_{\mathbf{p}, \sigma} S^{count}_2
a^{\dag}_{\mathbf{p}', \sigma'}|0 \rangle \nonumber \\
&= & -\frac{i\delta m_2}{ \hbar} \langle 0|a_{\mathbf{p}, \sigma}\int
d^4x   \overline{\psi}(\tilde{x})\psi(\tilde{x}) a^{\dag}_{\mathbf{p}', \sigma'}|0 \rangle \nonumber \\
&\ &-\frac{i(Z_2-1)_2}{\hbar} \langle 0|a_{\mathbf{p}, \sigma}\int
d^4x \overline{\psi}(\tilde{x})(-i \hbar c\gamma_{\mu}\partial^{\mu}
+ mc^2)\psi(\tilde{x})
a^{\dag}_{\mathbf{p}', \sigma'}|0 \rangle \nonumber  \\
&= & -\frac{i\delta m_2}{\hbar} \int d^4x \frac{1}{(2 \pi \hbar)^{3}}
\frac{mc^2}{\sqrt{\omega_{\mathbf{p}}\omega_{\mathbf{p}'}}}
 e^{\frac{i}{\hbar}\tilde{p}\cdot
\tilde{x}}e^{-\frac{i}{\hbar}\tilde{p}'\cdot
\tilde{x}}\overline{u}_a(\mathbf{p}, \sigma)
u_a(\mathbf{p}', \sigma') \nonumber   \\
&\ &-\frac{i(Z_2-1)_2}{ \hbar} \int d^4x
\frac{mc^2}{(2 \pi \hbar)^{3}\sqrt{\omega_{\mathbf{p}}\omega_{\mathbf{p}'}}}
 e^{\frac{i}{\hbar}\tilde{p}\cdot
\tilde{x}}e^{-\frac{i}{\hbar}\tilde{p}'\cdot
\tilde{x}}\overline{u}_a(\mathbf{p}, \sigma)
(-\cross{p} + mc^2) u_a(\mathbf{p}', \sigma') \nonumber   \\
&= & -
\frac{2 \pi i(\delta m_2)   mc^2 \delta^4(\tilde{p}-\tilde{p}')}{\omega_{\mathbf{p}}}
 \overline{u}_a(\mathbf{p}, \sigma)
u_a(\mathbf{p}', \sigma') \nonumber  \\
&\ &-
\frac{2\pi i(Z_2-1)_2 mc^2   \delta^4(\tilde{p}-\tilde{p}')}{  \omega_{\mathbf{p}}}
 \overline{u}_a(\mathbf{p},
\sigma)(-\cross{p} + mc^2) u_a(\mathbf{p}', \sigma') \label{eq:asa}
\end{eqnarray}

\noindent Dropping factors corresponding to external electron
lines and the delta function, we
obtain the pure counterterm contribution

\begin{eqnarray}
\mathcal{D}_{count}(\cross{p}) &=& -\frac{i(2 \pi
\hbar)^4}{\hbar}\delta m_2  +\frac{i(2 \pi \hbar)^4}{\hbar}(Z_2-1)_2
(\cross{p} - mc^2) \label{eq:asa2}
\end{eqnarray}

\noindent The interaction potential (\ref{eq:deltamc2}) will be represented by a new interaction vertex,
which will be denoted by a cross placed on electron lines, as  in figs. \ref{fig:9.10}(i-j).
Such vertices can be placed on electron lines in Feynman diagrams of arbitrary
topological shape thus increasing the order of the diagram by 2. If the counterterm vertex is placed on an
external electron line, then the 4-momentum $\tilde{p}$ is on the
mass shell where the 2nd term in (\ref{eq:asa2}) vanishes. So, in this case

\begin{eqnarray*}
\mathcal{D}_{count}(\cross{p}=mc^2) &= & -\frac{i(2 \pi
\hbar)^4}{\hbar}\delta m_2
\end{eqnarray*}

\noindent This contribution should be added to the loop term (\ref{eq:e2c2x}). Thus we conclude
that loops in external electron lines will be  canceled
exactly\footnote{In the language of traditional mass renormalization theory this cancelation comes from the requirement that the renormalized electron
propagator  has a pole at $\cross{p}=mc^2$, where
$m$ is the physical mass of the electron.} if we
choose\footnote{This expression for the ``mass shift'' can be
compared with formula for $\delta m$ in (21) \cite{Feynman}, with expression right
after equation (8.42) in \cite{Bjorken1} and with second equation
on page 523 in \cite{Schweber}.}

\begin{eqnarray}
\delta m_2 &=& -\frac{ie^2 c^2 C_0}{(2 \pi)^4 \hbar} = -\frac{ 3  mc^3 e^2
}{16 \pi^2 \hbar} \left( \frac{1}{2} + 2 \ln
\frac{\Lambda}{m}\right)\label{eq:delta-2}
\end{eqnarray}

\noindent  In other words, when doing calculations in the
renormalized theory the self-energy loops in external electron lines and contributions from the  counterterm (\ref{eq:deltamc2})
can be simply ignored. In our study this means that diagrams \ref{fig:9.10}(b-c) and (i-j) can be omitted.

 Electron-photon loops and ``cross'' vertices can also appear in
internal electron lines whose 4-momentum $\tilde{p}$ is not
necessarily on the mass shell. Then  the loop contribution (\ref{eq:e2c2}) has a non-vanishing divergent term proportional to $C_1$. In
order to cancel this divergence, it is sufficient to choose the
other renormalization factor in (\ref{eq:deltamc2})\footnote{Compare this result with
(8.43) in \cite{Bjorken1} and with equation (94b) in section 15 in
\cite{Schweber}. Note that the finiteness requirement does not
specify this factor uniquely. One can replace $C_1$ with $C_1+\delta$, where $\delta$ is any finite
constant, and still have a finite result. Usually, the correct choice $\delta=0$ is justified by the
requirement that the residue of the renormalized electron propagator
is equal to 1. In our approach this requirement is covered by the charge renormalization condition \ref{postulateV}. For example, if $\delta \neq 0$ then the 4th order diagrams \ref{fig:compton}(c) and (d)  would not cancel each other at low energies. This would result in a (finite) 4th order correction the low-energy charge-photon scattering, which is inconsistent with the classical Thomson formula. Compare with discussion of equation (\ref{eq:s4d+k}).}

\begin{eqnarray}
(Z_2-1)_2 &=& \frac{ie^2c^2 C_1}{(2 \pi)^4 \hbar} = -\frac{e^2  }{8
\pi^2 \hbar c}\left( \ln \frac{\Lambda}{m} + 2\ln
\frac{\lambda}{m} + \frac{9}{4}\right) \label{eq:L2-ren}
\end{eqnarray}

\noindent Then infinite contributions from the loop and the
counterterm cancel each other, and  only a finite and harmless $R$-correction
remains

\begin{eqnarray*}
\mathcal{D}_{loop}(\cross{p}) + \mathcal{D}_{count}({\cross{p}}) &= &
\hbar^2 e^2 c^2 R(\cross{p})
\end{eqnarray*}

\noindent It is responsible for so-called \emph{electron
self-energy} radiative corrections \index{electron self-energy} to
scattering amplitudes. Such corrections do not play any role in processes discussed in this book, so we will not discuss them any further.

\subsection{Photon self-scattering}
\label{ss:vacuum-polarization}

\begin{figure}
\centering
\includegraphics {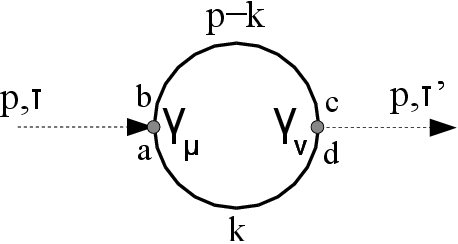} \caption{Feynman diagram for the
scattering photon$\to$photon in the 2nd perturbation order. }
\label{fig:9.7}
\end{figure}

\noindent  The amplitude of scattering photon$\to$photon in the
second perturbation order is obtained from the diagram
\ref{fig:9.7}\footnote{We omit calculation of the integral in square brackets. This calculation can be found, e.g., in  section 11.2 of \cite{book}, in section 7.5 of
\cite{Peskin} and in section 8.2 of \cite{Bjorken1}.}

\begin{eqnarray}
&\ & \langle 0|c_{\mathbf{p}, \tau} S_2^{FD}
c^{\dag}_{\mathbf{p}', \tau'}|0 \rangle \nonumber \\
&=& -\frac{c  e^2}{(2 \pi \hbar) 2p (2 \pi i)^2} \delta^4(\tilde{p}'
- \tilde{p}) e_{\mu}^*(\mathbf{p}, \tau) \times \nonumber \\
&\ & \Bigl[\int d^4k
\frac{(\cross{k} + mc^2)_{ac}}{\tilde{k}^2 - m^2c^4}\gamma_{ab}^{\mu}
\frac{(\cross{p} -\cross{k} + mc^2)_{bd}}{(\tilde{p}-\tilde{k})^2 -
m^2c^4}
\gamma_{cd}^{\nu} \Bigr] e_{\nu} (\mathbf{p}, \tau') \nonumber \\
&=& \frac{c  e^2}{(2 \pi \hbar) 2p (2 \pi)^2} \delta^4(\tilde{p}'
- \tilde{p}) e_{\mu}^*(\mathbf{p}, \tau) (\tilde{p}^2 g^{\mu \nu} -
p^{\mu} p^{\nu}) \Pi(\tilde{p}^2) e_{\nu} (\mathbf{p}, \tau') \nonumber \\
\label{eq:vac-pol}
\end{eqnarray}

\noindent where $\Pi(\tilde{p}^2)$ is a divergent
function. It is
convenient to write $\Pi(\tilde{p}^2)$ as a sum of its (infinite)
value $\Pi(0)$ on the photon's mass shell ($\tilde{p}^2=0$) plus a
finite remainder $\eta(\tilde{p}^2)$:

\begin{eqnarray*}
\Pi(\tilde{p}^2) &=& \Pi(0) + \eta(\tilde{p}^2)
\end{eqnarray*}

\noindent Function $\eta(\tilde{p}^2)$ can be represented by integral (11.2.22) in \cite{book}, which takes the following form at low values of energy-momentum $\tilde{p}$

\begin{eqnarray}
\eta(\tilde{p}^2)= -\frac{(2 \pi)^4 }{2 \pi^2 i c^3} \int_0^{1} x(1-x) \ln \left( 1 + \frac{\tilde{p}^2x(1-x)}{m^2c^4}\right)\approx \frac{i(2 \pi)^4 \tilde{p}^2}{60 \pi^2  m^2c^7}  \label{eq:pik2}
 \end{eqnarray}

\noindent In particular, this function vanishes on the photon's mass shell

\begin{eqnarray}
 \eta(0)&=&0
\label{eq:tildepi}
\end{eqnarray}

\noindent  The factor in (\ref{eq:vac-pol}) associated only with the
loop and two attached vertices (no contributions from external
lines) is

\begin{eqnarray}
\mathcal{D}_{loop}(\tilde{p}) &= & e^2 \Pi(\tilde{p}^2) (\tilde{p}^2
g^{\mu \nu} - p^{\mu} p^{\nu}) \label{eq:Dloop}
\end{eqnarray}

\noindent In equation (\ref{eq:vac-pol}), the 4-momentum $\tilde{p}$ is
on the mass shell, therefore
 the loop contribution (\ref{eq:Dloop})
vanishes there\footnote{The second term in parentheses in (\ref{eq:Dloop}) is not
contributing due to the property $\tilde{p} \cdot e (\mathbf{p},
\tau) = 0$ proved in (\ref{eq:emupmu}).} and the no-self-scattering renormalization condition (\ref{eq:scatter-ampl2}) is satisfied without extra effort. The
same can be said for loops in external photon legs of any diagram:
these loops can be simply ignored. However, we cannot ignore loop
contributions in internal photon lines.\footnote{See, for example, the graph \ref{fig:9.10}(d).} In such cases the 4-momentum
$\tilde{p}$ is not necessarily on the mass shall, and factor
(\ref{eq:Dloop}) is divergent.

\subsection{Photon self-energy counterterm} \label{ss:photon-se}

 Similar to the electron self-energy renormalization described above, we are going to cancel this infinity by adding to the interaction operator of QED a new
counterterm\footnote{From the dimension (\ref{eq:dim-phot}) of the
photon quantum field, it is easy to show that this operator has the
required dimension of energy if $(Z_3-1)_2$ is dimensionless. Moreover, this operator explicitly satisfies the Poincar\'e invariance condition (\ref{eq:as-a-scalar}).}

\begin{eqnarray}
Q^{FD}_{2ph}(t) = -\frac{(Z_3-1)_2}{4} \int d \mathbf{x}
 F^{\mu \nu}(\tilde{x}) F_{\mu \nu}(\tilde{x}) \label{eq:r2pht}
\end{eqnarray}

\noindent  where we denoted

\begin{eqnarray*}
F_{\mu \nu} &\equiv& \partial_{\mu} A_{\nu} - \partial_{\nu} A_{\mu}
\\
F^{\mu \nu}F_{\mu \nu} &=& (\partial^{\mu} A^{\nu} - \partial^{\nu}
A^{\mu})(\partial_{\mu} A_{\nu} - \partial_{\nu} A_{\mu}) \\
&=& \partial^{\mu} A^{\nu} \partial_{\mu} A_{\nu} - \partial^{\mu}
A^{\nu} \partial_{\nu} A_{\mu} - \partial^{\nu}
A^{\mu}\partial_{\mu} A_{\nu} + \partial^{\nu}
A^{\mu}\partial_{\nu} A_{\mu} \\
&=& 2\partial^{\mu} A^{\nu} \partial_{\mu} A_{\nu} - 2\partial^{\mu}
A^{\nu} \partial_{\nu} A_{\mu}
\end{eqnarray*}

\noindent and $(Z_3-1)_2$ is yet unspecified 2nd order renormalization factor. Let us now evaluate the effect of the counterterm
(\ref{eq:r2pht}) on the photon$\to$photon scattering amplitude. From
definitions (\ref{eq:10.26}) and (\ref{eq:partial-mu}) it follows\footnote{Here symbol $\to$ denotes the part that is relevant for calculation of the matrix element (\ref{eq:xi-count}).}

\begin{eqnarray*}
 \partial_{\nu} A_{\mu}(\mathbf{x},t) &=& \frac{i \sqrt{c} }{(2\pi \hbar)^{3/2}} \int
\frac{d\mathbf{q}}{\sqrt{2q}}\frac{ q_{\nu}}{c}\sum_{ \tau}
[-e^{-\frac{i}{\hbar}\tilde{q} \cdot \tilde{x}} e  _{\mu }
(\mathbf{q}, \tau)  c _{\mathbf{q},\tau} + e^{ \frac{i}{\hbar}
\tilde{q}\cdot \tilde{x}} e^*_{\mu }(\mathbf{q}, \tau)
 c^{\dag}_{\mathbf{q},\tau} ] \\
\langle 0|c_{\mathbf{p}, \tau} \partial_{\nu}
A_{\mu}(\mathbf{x},t) &\to& \langle 0| \frac{i  }{(2\pi
\hbar)^{3/2}\sqrt{2pc}}
 e^{\frac{i}{\hbar}\tilde{p} \cdot \tilde{x}} p_{\nu}e^*_{\mu }(\mathbf{p}, \tau) \\
  \partial_{\nu} A_{\mu}(\mathbf{x},t)c^{\dag}_{\mathbf{p}', \tau'}|0 \rangle &\to&
    -\frac{i  }{(2\pi \hbar)^{3/2}\sqrt{2p'c}}
 e^{-\frac{i}{\hbar}\tilde{p}' \cdot \tilde{x}}p'_{\nu} e_{\mu }(\mathbf{p}', \tau') |0 \rangle \\
\end{eqnarray*}

\noindent Then the $S$-matrix contribution from (\ref{eq:r2pht}) has a form
similar to
(\ref{eq:vac-pol})

\begin{eqnarray}
&\ &\langle 0|c_{\mathbf{p}, \tau}
 S_2^{count} c^{\dag}_{\mathbf{p}', \tau'}|0 \rangle \nonumber \\
 &= & -\frac{i(Z_3-1)_2}{2 \hbar} \langle 0|c_{\mathbf{p}, \tau} \int d^4x
\Bigr(\partial^{\lambda} A^{\kappa}(\tilde{x})\partial_{\lambda}
A_{\kappa}(\tilde{x}) - \partial^{\lambda}
A^{\kappa}(\tilde{x})\partial_{\kappa} A_{\lambda}(\tilde{x})\Bigl)
c^{\dag}_{\mathbf{p}', \tau'}|0 \rangle \nonumber \\
 &=& -\frac{i(Z_3-1)_2}{2 \hbar c} \int d^4x \frac{ 1 }{(2\pi
\hbar)^{3}\sqrt{4pp'}}
 e^{\frac{i}{\hbar}(\tilde{p}-\tilde{p}') \cdot \tilde{x}} p^{\lambda}
 p'_{\lambda} e^{*\kappa }(\mathbf{p}, \tau) e_{\kappa }(\mathbf{p}', \tau) \nonumber \\
 &\ &+\frac{i(Z_3-1)_2}{4 \hbar c} \int d^4x \frac{ 1 }{(2\pi
\hbar)^{3}\sqrt{4pp'}}
 e^{\frac{i}{\hbar}(\tilde{p}-\tilde{p}') \cdot x} p^{\lambda}
 p'_{\kappa} e^{*\kappa }(\mathbf{p}, \tau) e_{\lambda }(\mathbf{p}', \tau) \nonumber \\
 &=& -\frac{i(Z_3-1)_2\pi  \delta^4(\tilde{p}-\tilde{p}')}{2  pc}
  [p^{\lambda}p_{\lambda} e^{*\kappa }(\mathbf{p}, \tau) e_{\kappa
}(\mathbf{p}, \tau)
 - p^{\lambda} p_{\kappa} e^{*\kappa }(\mathbf{p}, \tau) e_{\lambda }(\mathbf{p}, \tau)
 ] \nonumber \\
 &=&  -\frac{i(Z_3-1)_2\pi  \delta^4(\tilde{p}-\tilde{p}')}{2 pc}
e^{*}_{\mu }(\mathbf{p}, \tau) [ \tilde{p}^2 g^{\mu \nu} - p^{\mu}
p^{\nu} ]
 e_{\nu}(\mathbf{p}, \tau) \label{eq:xi-count}
\end{eqnarray}

\noindent In the Feynman diagram notation the counterterm (\ref{eq:r2pht}) gives rise to a new type
of vertex,\footnote{ denoted by a cross placed on photon lines, as in Fig. \ref{fig:9.10}(k)} which corresponds to
the factor\footnote{This factor is obtained from (\ref{eq:xi-count}) by
stripping off the 4-momentum delta function as well as factors $\hbar c^{1/2} (2 \pi \hbar)^{-3/2}(2p)^{-1/2}
e_{\nu}(\mathbf{p}, \tau)$ and $\hbar c^{1/2} (2 \pi
\hbar)^{-3/2}(2p)^{-1/2} e^{*}_{\mu }(\mathbf{p}, \tau)$ associated
with external photon lines. }

\begin{eqnarray}
\mathcal{D}_{count} (\tilde{p})
 &=&-\frac{8 i(Z_3-1)_2\pi^4 \hbar  }{c^2}
 ( \tilde{p}^2 g^{\mu \nu} - p^{\mu} p^{\nu} )
 \label{eq:D'n+2}
\end{eqnarray}

\noindent The constant $(Z_3-1)_2$ should be chosen such that expression (\ref{eq:D'n+2}) exactly cancels the loop factor (\ref{eq:Dloop}) when $\tilde{p}^2 = 0$, i.e., for external photon lines

\begin{eqnarray}
(Z_3-1)_2 = -\frac{ie^2c^2\Pi(0)}{8 \pi^4 \hbar} \label{eq:xi2}
\end{eqnarray}

\noindent Then for internal photon lines\footnote{i.e., when $\tilde{p}^2 \neq 0$} the sum of the loop (\ref{eq:Dloop}) and the
counterterm (\ref{eq:D'n+2}) is finite

\begin{eqnarray}
\mathcal{D}_{loop}(\tilde{p})+\mathcal{D}_{count} (\tilde{p})&=&   e^2 \eta(\tilde{p}^2)
(\tilde{p}^2 g^{\mu \nu} - p^{\mu} p^{\nu}) \label{eq:ph-finite}
\end{eqnarray}

\noindent This means that an electron-positron loop and a photon-line cross taken together\footnote{e.g., the sum of diagrams \ref{fig:9.10}(d) and \ref{fig:9.10}(k)} result in a finite correction to the scattering amplitude. This is the so-called \emph{vacuum
polarization} radiative correction. \index{vacuum polarization}

\subsection{Charge renormalization} \label{ss:charge-renorm}

If our only goal is to make the perturbation theory expansion
finite, then the choice of the renormalization parameter
(\ref{eq:xi2}) is not unique. Indeed, we could add to $\Pi(0)$ in
(\ref{eq:xi2}) an arbitrary finite number $\delta$, so that

\begin{eqnarray*}
(Z_3-1)_2 = -\frac{ie^2c^2(\Pi(0) +\delta)}{8 \pi^4 \hbar}
\end{eqnarray*}

\noindent and (\ref{eq:ph-finite}) would remain finite

\begin{eqnarray}
\mathcal{D}_{loop}(\tilde{p}) + \mathcal{D}_{count}(\tilde{p}) &=&   e^2 (\eta(\tilde{p}^2) -
\delta)(\tilde{p}^2 g^{\mu \nu} - p^{\mu} p^{\nu})
\label{eq:ph-finite2}
\end{eqnarray}

\noindent Why don't we do that? The answer is that such an addition
would be inconsistent with the charge renormalization condition in Postulate
\ref{postulateV}.

To see that, let us evaluate the contribution to the electron-proton
scattering from diagrams \ref{fig:9.10}(d) and (k). Using Feynman's
rules and (\ref{eq:ph-finite2}) we obtain\footnote{Here we denoted $\tilde{k} \equiv \tilde{q}' - \tilde{q} = \tilde{p} - \tilde{p}'$, used non-relativistic approximations from Appendix \ref{ss:non-rel} and formulas (\ref{eq:kk}) - (\ref{eq:kk2}): $U^{\mu} g_{\mu \nu}k^{\nu}k^{\lambda} g_{\lambda \kappa} W^{\kappa} = U^{\mu}k_{\mu}k_{\kappa}  W^{\kappa} = 0$.}

\begin{eqnarray}
&\ &\langle 0|a_{\mathbf{q}, \sigma} d_{\mathbf{p}, \tau} S_4^{(d)+(k)} d^{\dag}_{\mathbf{p}', \tau'} a^{\dag}_{\mathbf{q}', \sigma'} |0 \rangle \nonumber \\
&=& -\frac{e^2 mMc^4 \hbar^4 c^4}{(2 \pi i)^2(2 \pi \hbar)^4  \sqrt{\omega_{\mathbf{q}} \omega_{\mathbf{q'}}
\Omega_{\mathbf{p}} \Omega_{\mathbf{p'}} } } \delta^4(\tilde{q} -
\tilde{q}' -\tilde{p}'+\tilde{p}) \times \nonumber \\
&\ &\frac{  e^2 (\eta(\tilde{k}^2) - \delta)}{ \hbar^2}
U^{\mu}(\mathbf{q}, \sigma;\mathbf{q'}, \sigma')  \frac{ g_{\mu
\nu}}{ \tilde{k}^2}(\tilde{k}^2 g^{\nu \lambda} - k^{\nu}
k^{\lambda}) \frac{ g_{\lambda \kappa}}{ \tilde{k}^2}
  W^{\kappa}(\mathbf{p}, \tau; \mathbf{p'}, \tau') \nonumber \\
&=& \frac{e^4mMc^4 c^4}{\hbar^2 (2 \pi)^6 \sqrt{\omega_{\mathbf{q}} \omega_{\mathbf{q'}}
\Omega_{\mathbf{p}} \Omega_{\mathbf{p'}} } } \delta^4(\tilde{q} -
\tilde{q}' -\tilde{p}'+\tilde{p}) \times \nonumber \\
&\ &\frac{  \eta(\tilde{k}^2) - \delta}{\tilde{k}^2}
U_{\mu}(\mathbf{q}, \sigma;\mathbf{q'}, \sigma')
  W^{\mu}(\mathbf{p}, \tau; \mathbf{p'}, \tau') \nonumber \\
&\approx& \frac{e^4 c^4}{(2 \pi)^{6} \hbar^2  } \delta^4(\tilde{q} -
\tilde{q}' -\tilde{p}'+\tilde{p}) \frac{ \eta(\tilde{k}^2) - \delta}{\tilde{k}^2}
 \delta_{\sigma, \sigma'}\delta_{\tau, \tau'} \label{eq:s4d+k}
 \end{eqnarray}

\noindent Taking into account equation (\ref{eq:tildepi}), we conclude that if $\delta \neq 0$, then this matrix element has a
singularity $\propto - \delta/\tilde{k}^2$ at small values of $\tilde{k}$.  This singularity leads to a 4th order correction to the long-range scattering of charged particles and, therefore, violates
the charge renormalization condition \ref{postulateV}. The only way to satisfy this
condition is to set $\delta=0$.

\subsection{Vertex renormalization}
\label{ss:approximation}

Our renormalization task is not completed yet. One more type of
counterterm is required in order to make QED calculations finite and
accurate.

Let us evaluate diagram \ref{fig:9.10}(e) using Feynman rules

\begin{eqnarray}
&\ & \langle 0|a_{\mathbf{q}, \sigma} d_{\mathbf{p}, \tau} S_4^{(e)} d^{\dag}_{\mathbf{p}', \tau'} a^{\dag}_{\mathbf{q}', \sigma'} |0 \rangle \nonumber \\
&\approx&       \frac{e^4c^4mMc^4}{(2 \pi i)^4(2 \pi \hbar)^2
\sqrt{\omega_{\mathbf{q}} \omega_{\mathbf{q'}} \Omega_{\mathbf{p}}
\Omega_{\mathbf{p'}} } } \delta^4(\tilde{q} - \tilde{q}'
-\tilde{p}'+\tilde{p}) \overline{u}(\mathbf{q}, \sigma) \times
 \nonumber \\
&\mbox{ }&  \left[\int d^4h \gamma_{\mu} \frac{-\cross{h} + \cross{q} +
mc^2}{(\tilde{h}-\tilde{q})^2 - m^2c^4}  \gamma_{\kappa}
\frac{-\cross{h} + \cross{q}' + mc^2}{(\tilde{h}-\tilde{q}')^2 - m^2c^4}
\gamma^{\mu}\frac{1}{\tilde{h}^2} \right] \times \nonumber \\
&\mbox{ } & u(\mathbf{q'}, \sigma')
\frac{1}{(\tilde{q}'-\tilde{q})^2}
  W^{\kappa}(\mathbf{p}, \tau; \mathbf{p'}, \tau') \label{eq:s4-1112a}
\end{eqnarray}

\noindent  The integral in square brackets $I^{\kappa}(\tilde{q},
\tilde{q}')$ is calculated in (\ref{eq:I-kappa-qq})

\begin{eqnarray}
&\ &I^{\kappa}(\tilde{q}, \tilde{q}') \nonumber \\
&=&\frac{\pi^2\gamma^{\kappa}}{ic^3}\Bigr(\frac{8 \theta }{ \tan (2
\theta)}  \ln \frac{ \lambda}{m }  + \frac{8 }{\tan (2 \theta)}
\int \limits_{0}^{\theta} \alpha \tan \alpha d\alpha + \frac{1}{2}
+6\theta
\cot \theta + 2 \ln \frac{\Lambda }{m}\Bigl) \nonumber \\
&\ &-
  \frac{2 \pi^2 \theta ( q +q')^{\kappa}}{i m c^5  \sin(2
\theta)} \label{eq:I-kappa-qqx}
\end{eqnarray}

 To apply the
charge renormalization condition, we should consider this expression
at small values of the transferred momentum (i.e., when $\tilde{q} \approx
\tilde{q}'$, $\theta \approx 0$). In this case we can use equation (\ref{gamma-mu})

\begin{eqnarray*}
0 &=& \overline{u}(\mathbf{q}, \sigma)(\gamma^{\mu}(\cross{q}-mc^2) +
(\cross{q}-mc^2)\gamma^{\mu}) u(\mathbf{q}, \sigma') \\
&=& \overline{u}(\mathbf{q}, \sigma)(\gamma^{\mu}\gamma^{\nu}q_{\nu}
+ \gamma^{\nu}q_{\nu}\gamma^{\mu} -2 \gamma^{\mu}mc^2) u(\mathbf{q},
\sigma')
\\
&=& \overline{u}(\mathbf{q}, \sigma)(2g^{\mu \nu}q_{\nu}  -2
\gamma^{\mu}mc^2) u(\mathbf{q}, \sigma') \\
&=& 2\overline{u}(\mathbf{q}, \sigma)(q^{\mu}  -
\gamma^{\mu}mc^2) u(\mathbf{q}, \sigma')
\end{eqnarray*}

\noindent to see that when sandwiched between $\overline{u}$ and $u$, the 4-vector $q^{\mu}$ can be replaced by $\gamma^{\mu} mc^2$. Therefore, in (\ref{eq:I-kappa-qqx}) we will set $q^{\kappa}
\approx (q')^{\kappa} \approx \gamma^{\kappa}mc^2 $
 and obtain\footnote{In our notation (\ref{eq:M.35a}), (\ref{eq:k2sinth}) $
\theta \equiv \sin^{-1}\left(
\frac{\sqrt{\tilde{k}^2}}{2mc^2}\right)$.}

\begin{eqnarray*}
\lim_{\tilde{k} \to 0, \tilde{q} \to 0}I^{\kappa}(\tilde{q},
 \tilde{q}')
 &= & \frac{F \pi^2 \gamma^{\kappa}}{ic^3}
\end{eqnarray*}

\noindent where we introduced an ultraviolet-divergent and infrared-divergent
constant\footnote{Compare with equation (23) in \cite{Feynman}.}

\begin{eqnarray*}
F\equiv  4  \ln \frac{ \lambda}{m} + 2\ln \frac{\Lambda
}{m} + \frac{9}{2}
\end{eqnarray*}

\noindent Then at small values of
$\tilde{k}$ the scattering amplitude (\ref{eq:s4-1112a})

\begin{eqnarray}
&\ & \langle 0|a_{\mathbf{q}, \sigma} d_{\mathbf{p}, \tau} S_4^{(e)} d^{\dag}_{\mathbf{p}', \tau'} a^{\dag}_{\mathbf{q}', \sigma'} |0 \rangle \nonumber \\
&=&       -\frac{ie^4c^4F \pi^2 mMc^4\delta^4(\tilde{q} -
\tilde{q}' -\tilde{p}'+\tilde{p})}{c^3(2 \pi i)^4(2 \pi \hbar)^2
\sqrt{\omega_{\mathbf{q}} \omega_{\mathbf{q'}} \Omega_{\mathbf{p}}
\Omega_{\mathbf{p'}} } (\tilde{q}'-\tilde{q})^2} U_{\kappa}(\mathbf{q},
\sigma;\mathbf{q'}, \sigma')
  W^{\kappa}(\mathbf{p}, \tau; \mathbf{p'}, \tau') \nonumber \\
\label{eq:s4loop}
\end{eqnarray}

\noindent  has a singularity $\propto F/\tilde{k}^2$. This means that in
disagreement with our Postulate \ref{postulateV}, the 4th
perturbation order makes a non-trivial contribution to the
long-distance electron-proton scattering.  Even more disturbing is the
fact that this effect is infinite in the limit $\Lambda \to \infty$.

This unacceptable situation can be fixed by adding one more (\emph{vertex}) renormalization counterterm to the QED interaction

\begin{eqnarray}
Q^{FD}_3(t) =-e(Z_1-1)_2 \int d \mathbf{x}
\overline{\psi}(\tilde{x})\gamma_{\mu}\psi(\tilde{x})
A^{\mu}(\tilde{x}) \label{eq:qfd3t}
\end{eqnarray}

\noindent In
Feynman diagrams we will denote the corresponding three-leg vertex by a
circle,
as shown in diagram \ref{fig:9.10}(h). The renormalization constant $(Z_1-1)_2$ is of the 2nd perturbation order, so the order of the counterterm (\ref{eq:qfd3t}) is 3. It has the
same form as the basic QED interaction (\ref{eq:basic-int}), so the
diagram \ref{fig:9.10}(h) is easily evaluated\footnote{compare with
equation (\ref{eq:8.31a})}

\begin{eqnarray*}
&\ & \langle 0|a_{\mathbf{q}, \sigma} d_{\mathbf{p}, \tau} S_4^{(h)} d^{\dag}_{\mathbf{p}', \tau'} a^{\dag}_{\mathbf{q}', \sigma'} |0 \rangle \nonumber \\
&=&       \frac{ie^2c^2(Z_1-1)_2 mMc^4\delta^4(\tilde{q}
+\tilde{p} - \tilde{q}' -\tilde{p}')}{4 \pi^2 \hbar
\sqrt{\omega_{\mathbf{q}} \omega_{\mathbf{q'}} \Omega_{\mathbf{p}}
\Omega_{\mathbf{p'}} } (\tilde{q}'-\tilde{q})^2}
U_{\kappa}(\mathbf{q}, \sigma;\mathbf{q'}, \sigma')
  W^{\kappa}(\mathbf{p}, \tau; \mathbf{p'}, \tau')
\end{eqnarray*}

\noindent Our requirement to cancel the infinite/singular term
(\ref{eq:s4loop}) tells us that we need to choose our
renormalization constant as\footnote{The
equality with the electron-photon loop renormalization factor
$(Z_2-1)_2$ in (\ref{eq:L2-ren}) is not accidental. It is explained in
section 8.6 in \cite{Bjorken1}. }

\begin{eqnarray*}
(Z_1-1)_2 &=& \frac{e^2 F}{16 \pi^2 c\hbar} = -(Z_2-1)_2
\end{eqnarray*}

After adding all three renormalization counterterms
(\ref{eq:deltamc2}), (\ref{eq:r2pht}), and (\ref{eq:qfd3t}) the full
Feynman-Dyson interaction operator (\ref{eq:V1-2y}) takes the form

\begin{eqnarray}
&\ & V^c_{FD}(t) \nonumber \\
&=& V_1(t) +Q^{FD}_{2el}(t) + Q^{FD}_{2ph}(t)
+Q^{FD}_{3}(t) + \ldots \nonumber \\
&=&-e\int d \mathbf{x}
\overline{\psi}(\tilde{x})\gamma_{\mu}\psi(\tilde{x})
A^{\mu}(\tilde{x}) + e \int d \mathbf{x}
\overline{\Psi}(\tilde{x})\gamma_{\mu}\Psi(\tilde{x})
A^{\mu}(\tilde{x}) \nonumber \\
&\ &+ \delta m_2  \int d
\mathbf{x}\overline{\psi}(\tilde{x})\psi(\tilde{x})
 +(Z_2-1)_2 \int d \mathbf{x} \overline{\psi}(\tilde{x})(-i \hbar
c\gamma_{\mu}\partial_{\mu} + mc^2)\psi(\tilde{x}) \nonumber \\
&\ &-\frac{(Z_3-1)_2}{4}
\int d \mathbf{x}
 F^{\mu \nu}(\tilde{x}) F_{\mu \nu}(\tilde{x})  - e (Z_1-1)_2 \int d \mathbf{x}
\overline{\psi}(\tilde{x})\gamma_{\mu}\psi(\tilde{x})
A^{\mu}(\tilde{x}) + \ldots \nonumber \\
\label{eq:11.61}
\end{eqnarray}

\section{Renormalized $S$-matrix}
\label{ss:ren-s-matrix}

Equation (\ref{eq:11.61}) is the renormalized QED interaction operator that is accurate up to the 3rd perturbation order. Our claim was that inserting this interaction in the usual formula (\ref{eq:9.11x}) for the $S$-operator we can obtain ultraviolet-finite and accurate scattering amplitudes. Let us now support this claim with explicit calculation of all 4th order diagrams in Fig. \ref{fig:9.10}. As we already know, diagrams \ref{fig:9.10}(b), (c), (i), (j) cancel out exactly. In this section we are going to calculate six other diagrams that we arrange in four coefficient functions $s_4$ on the right hand side of

\begin{eqnarray}
\langle 0|a_{\mathbf{q}, \sigma} d_{\mathbf{p}, \tau} S_4^{c} d^{\dag}_{\mathbf{p}', \tau'} a^{\dag}_{\mathbf{q}', \sigma'} |0 \rangle = \left(s_4^{(d)+(k)} + s_4^{(e)+(h)} + s_4^{(f)} + s_4^{(g)}\right) \delta^4(\tilde{q}
+\tilde{p} - \tilde{q}' -\tilde{p}') \nonumber \\
\label{eq:10.46}
\end{eqnarray}

\noindent For our purposes it will be sufficient to work in the limit of low momenta of particles\footnote{see Appendix \ref{ss:non-rel}} and small transferred momentum $\tilde{k}$.

\subsection{Vacuum polarization diagrams}
\label{ss:910d+k}

Inserting (\ref{eq:pik2}) in (\ref{eq:s4d+k}) and setting $\delta=0$ we find that in our approximation the $S$-matrix elements described by diagrams \ref{fig:9.10}(d) and (k) do not depend on particle momenta and spins\footnote{Here  $\alpha \equiv e^2/(4 \pi \hbar c) \approx 1/137$ is the \emph{fine structure constant}. \index{fine structure constant}}

\begin{eqnarray}
s_{4}^{(d)+(k)} \approx \frac{ie^4  }{\hbar^2 (2 \pi)^2 60 \pi^2 m^2c^3} \delta_{\sigma, \sigma'}\delta_{\tau, \tau'} = \frac{i \alpha^2  }{15 \pi^2 m^2c} \delta_{\sigma, \sigma'}\delta_{\tau, \tau'} \label{eq:s4-vac-pol}
 \end{eqnarray}

\subsection{Vertex diagrams}
\label{ss:910e+h}

 The full electron vertex contribution in Figs.
\ref{fig:9.10}(e) and (h)  is given by equation
(\ref{eq:s4-1112a}), where the square bracket should be replaced by
the ultraviolet-finite expression

\begin{eqnarray*}
&\ & I^{\kappa}(\tilde{q}, \tilde{q}') - \frac{F \pi^2}{ic^3} \gamma^{\kappa} \\
 &=&\frac{ \pi^2 \gamma^{\kappa}}{ic^3}\Bigr(\frac{8 \theta }{ \tan (2 \theta)}
\ln\frac{ \lambda}{m}   + \frac{8 }{ \tan (2 \theta)} \int
\limits_{0}^{\theta}
\alpha \tan \alpha d\alpha + \frac{1}{2} + 6\theta \cot \theta + 2 \ln \frac{\Lambda }{m} \Bigl) \\
&\ &-
  \frac{2 \pi^2 \theta ( \tilde{q} +\tilde{q}')^{\kappa}}{i m c^5  \sin(2
\theta)} -\frac{\pi^2 \gamma^{\kappa}}{ic^3} \Bigl(4
\ln\frac{\lambda}{m} + 2 \ln \frac{\Lambda }{m} +\frac{9}{2} \Bigr) \\
 &=&\frac{\pi^2 \gamma^{\kappa}}{ic^3}\Bigr(
 \left(\frac{8 \theta }{ \tan (2 \theta)} -4 \right)
\ln\frac{ \lambda}{m}   + \frac{8 }{ \tan (2 \theta)} \int
\limits_{0}^{\theta} \alpha \tan \alpha d\alpha -4 + 6\theta \cot
\theta \Bigl) \\
&\ &-
  \frac{2 \pi^2 \theta ( \tilde{q} +\tilde{q}')^{\kappa}}{i m c^5  \sin(2
\theta)}
\end{eqnarray*}

\noindent In the limit of small momentum transfer ($\theta \approx 0$) this formula simplifies

\begin{eqnarray*}
\theta &\approx&  \frac{\sqrt{\tilde{k}^2}}{2mc^2} \\
\frac{8 \theta }{ \tan (2 \theta)} &\approx& \frac{4|\tilde{k}|} {mc^2 \tan \left(  \frac{|\tilde{k}|}{mc^2}\right)} \approx \frac{4|\tilde{k}|}{mc^2 \left(  \frac{|\tilde{k}|}{mc^2} + \frac{|\tilde{k}|^3}{3m^3c^6}\right)} \approx 4 - \frac{4 \tilde{k}^2}{3 m^2c^4} \\
\frac{8}{\tan(2 \theta)} \int
\limits_{0}^{\theta} \alpha \tan \alpha d\alpha &\approx& \frac{4}{\theta} \int
\limits_{0}^{\theta} \alpha^2 d\alpha = \frac{4}{\theta} \cdot \frac{\theta^3}{3} \approx \frac{\tilde{k}^2}{3m^2c^4} \\
6 \theta \cot \theta &\approx& 6 \theta \left(\frac{1}{\theta} - \frac{\theta}{3}  \right) = 6 -2 \theta^2 \approx 6 - \frac{\tilde{k}^2}{2m^2c^4} \\
\frac{2  \theta }{ \sin(2
\theta)} &\approx&  \frac{2 \theta }{2
\theta - (4/3) \theta^3} \approx  1 + \frac{2 \theta^2}{3} \approx
 1 + \frac{\tilde{k}^2}{6m^2c^4}
\end{eqnarray*}

\begin{eqnarray*}
 I^{\kappa}(\tilde{q}, \tilde{q}') - \frac{F \pi^2}{ic^3} \gamma^{\kappa}
 &\approx&\frac{2 \pi^2 \gamma^{\kappa}}{ic^3}\left(1 - \frac{\tilde{k}^2}{12m^2c^4} \right) -
  \frac{\pi^2  ( \tilde{q} +\tilde{q}')^{\kappa}}{i m c^5}\left( 1 + \frac{\tilde{k}^2}{6m^2c^4}\right) \\
    &-& \frac{\pi^2 \gamma^{\kappa}}{ic^3} \frac{4 \tilde{k}^2}{3m^2c^4} \ln\frac{ \lambda}{m}
\end{eqnarray*}

\noindent so that  diagrams \ref{fig:9.10}(e) + (h) become

\begin{eqnarray}
&\ & s_4^{(e)+(h)} \approx       \frac{-ic^3 \alpha^2}{4 \pi^2 \tilde{k}^2}
\frac{Mmc^4}{\sqrt{\omega_{\mathbf{q}} \omega_{\mathbf{q'}} \Omega_{\mathbf{p}} \Omega_{\mathbf{p'}} } }
 W^{\kappa}(\mathbf{p}, \tau; \mathbf{p'}, \tau') \times \nonumber \\
&\mbox{ }&
 \overline{u}(\mathbf{q}, \sigma) \left[2\gamma^{\kappa}\left(1 - \frac{\tilde{k}^2}{12m^2c^4} \right) -\frac{( \tilde{q} +\tilde{q}')^{\kappa} }{m c^2}\left(1 + \frac{\tilde{k}^2}{6m^2c^4}\right) - \frac{4  \gamma^{\kappa} \tilde{k}^2}{3m^2c^4} \ln\frac{ \lambda}{m} \right]
u(\mathbf{q'}, \sigma') \nonumber \\
  \label{eq:s4-1112y}
\end{eqnarray}

\noindent This expression can be divided into two parts

\begin{eqnarray*}
&\ & s_4^{(e)+(h)} = s_4^{(e)+(h)AMM} + s_4^{(e)+(h)div}
\end{eqnarray*}

\noindent where $s_4^{(e)+(h)AMM}$ remains finite in the infrared limit $\lambda \to 0$.\footnote{In chapter \ref{ss:hydrogen} we will see that this expression is related to the electron's anomalous magnetic moment (AMM).} and $s_4^{(e)+(h)div}$ contains the infrared-divergent logarithm $\ln(\lambda / m)$.
 Let us now introduce the vector of transferred momentum $\mathbf{k} = \mathbf{q'} - \mathbf{q} = \mathbf{p} - \mathbf{p}'$, apply the limit $M \to \infty $, and  $(v/c)^2$ approximation\footnote{see Appendix \ref{ss:non-rel}. For example, in this limit we can replace $W^0 \approx \delta_{\tau, \tau'}$ and $\mathbf{W} \approx 0$.} to the infrared-finite part of the  amplitude

\begin{eqnarray}
&\ & s_4^{(e)+(h)AMM} =   \frac{i  c \alpha^2}{4 \pi^2 k^2} \frac{ Mmc^4}{\sqrt{ \Omega_{\mathbf{p-k}}
\Omega_{\mathbf{p}}  \omega_{\mathbf{q+k}} \omega_{\mathbf{q}} }} \times \nonumber \\
&\mbox{ }& \overline{u}(\mathbf{q+k}, \sigma)
 \left[2  \gamma^{\kappa} + \frac{ \gamma^{\kappa} k^2 }{6m^2 c^2} -
  \frac{  (q+k)^{\kappa} + q^{\kappa}}{m c^2} \left(1 - \frac{k^2}{6m^2c^2}\right) \right]  u(\mathbf{q}, \sigma') \times \nonumber \\
&\ &  W^{\kappa}(\mathbf{p-k}, \tau; \mathbf{p}, \tau') \label{eq:s4ehAMM}  \\
&\approx& \frac{i  c \alpha^2 \delta_{ \tau, \tau'}}{4 \pi^2 k^2} \frac{ Mmc^4 }{\sqrt{ \Omega_{\mathbf{p-k}}
\Omega_{\mathbf{p}}  \omega_{\mathbf{q+k}} \omega_{\mathbf{q}} }} \times \nonumber \\
&\ &\Bigl[ 2\left(1 +\frac{ k^2 }{12m^2 c^2} \right)U^{0}(\mathbf{q+k}, \sigma; \mathbf{q}, \sigma') - \left(1 - \frac{k^2}{6m^2c^2}\right) \frac{ \omega_{\mathbf{q+k}} + \omega_{\mathbf{q}}}{mc^2} \overline{u}(\mathbf{q+k}, \sigma) u(\mathbf{q}, \sigma')   \Bigr] \nonumber
\end{eqnarray}

\noindent We use formulas from Appendices \ref{ss:explicit-u}, \ref{ss:non-rel}, and (\ref{eq:A.69c}) to further simplify both parts of this expression

\begin{eqnarray}
&\ & \frac{2i  c \alpha^2 \delta_{ \tau, \tau'}}{4 \pi^2 k^2} \cdot \frac{Mmc^4} {\sqrt{\Omega_{\mathbf{p-k}}
\omega_{\mathbf{q+k}} \Omega_{\mathbf{p}}
\omega_{\mathbf{q}}}} \left(1 +\frac{ k^2 }{12m^2 c^2} \right)
 U^0(\mathbf{q}+ \mathbf{k},\epsilon;
\mathbf{q},\epsilon') \nonumber \\
&\approx & \frac{i  c \alpha^2 \delta_{ \tau, \tau'}}{2\pi^2 k^2}    \left(1   -
\frac{q^2}{2m^2c^2} - \frac{\mathbf{q}\mathbf{k}}{2m^2c^2}
- \frac{k^2}{4m^2c^2}\right) \left(1 +\frac{ k^2 }{12m^2 c^2} \right)\times \nonumber \\
&\ & \chi^{\dag}_{\sigma} \left(1 + \frac{(2 \mathbf{q} +
\mathbf{k})^2 + 2i \vec{\sigma}_{el} \cdot [\mathbf{k}\times
\mathbf{q}]}{8m^2c^2} \right)\chi_{\sigma'}    \nonumber  \\
&\approx& \frac{i  c \alpha^2 \delta_{ \tau, \tau'}}{2\pi^2 k^2} \chi^{\dag}_{\sigma}
 \Bigl(1  -
\frac{q^2}{2m^2c^2} - \frac{\mathbf{q}\mathbf{k}}{2m^2c^2}
- \frac{k^2}{4m^2c^2} +\frac{ k^2 }{12m^2 c^2} + \frac{q^2}{2m^2 c^2}
+\frac{\mathbf{q}\mathbf{k}}{2m^2c^2} +\frac{k^2}{8m^2 c^2} \nonumber \\
 &+& i
\frac{\vec{\sigma}_{el} [\mathbf{k} \times \mathbf{q}]}{4m^2 c^2}
\Bigr) \chi_{\sigma'}
\nonumber  \\
& \approx& \frac{i  c \alpha^2 \delta_{ \tau, \tau'}}{2\pi^2 } \chi^{\dag}_{\sigma}   \left(\frac{1}{k^2}
 - \frac{1}{24m^2c^2}  +i
\frac{\vec{\sigma}_{el} [\mathbf{k} \times \mathbf{q}]}{4m^2 c^2k^2}\right) \chi_{\sigma'}   \label{eq:u0w0x}
\end{eqnarray}

\begin{eqnarray}
&\ & -\frac{i  c \alpha^2 \delta_{ \tau, \tau'}}{4 \pi^2 k^2} \cdot \frac{Mmc^4} {\sqrt{\Omega_{\mathbf{p-k}}
\omega_{\mathbf{q+k}} \Omega_{\mathbf{p}}
\omega_{\mathbf{q}}}k^2} \cdot \left(1 - \frac{k^2}{6m^2c^2}\right) \frac{ \omega_{\mathbf{q+k}} + \omega_{\mathbf{q}}}{mc^2} \overline{u}(\mathbf{q+k}, \sigma) u(\mathbf{q}, \sigma') \nonumber  \\
&=& -\frac{i  c \alpha^2 \delta_{ \tau, \tau'}}{2 \pi^2 k^2} \left(1   -
\frac{q^2}{2m^2c^2} - \frac{\mathbf{q}\mathbf{k}}{2m^2c^2}
- \frac{k^2}{4m^2c^2} - \frac{k^2}{6m^2c^2} + \frac{q^2 + k^2 +  2\mathbf{qk}}{4m^2c^2} + \frac{q^2}{4m^2c^2} \right) \times \nonumber   \\
&\ &\chi^{\dag}_{\sigma}  \left[\sqrt{\omega_{\mathbf{q+k}} + mc^2},
 \sqrt{\omega_{\mathbf{q+k}} - mc^2} \left(\frac{\mathbf{q+k}}{|\mathbf{q+k}|} \cdot
\vec{\sigma}_{el}\right)  \right] \times \nonumber \\
&\ & \left[ \begin{array}{c}
 \sqrt{\omega_{\mathbf{q}} + mc^2}   \nonumber \\
 -\sqrt{\omega_{\mathbf{q}} - mc^2} \left(\frac{\mathbf{q}}{q} \cdot
\vec{\sigma}_{el}\right)
\end{array} \right]   \frac{\chi_{\sigma'}}{2mc^2} \nonumber   \\
&=& -\frac{i  c \alpha^2 \delta_{ \tau, \tau'}}{2 \pi^2 k^2} \left(1   -
 \frac{k^2}{6m^2c^2}  \right)  \chi^{\dag}_{\sigma}  \Bigl(\frac{\sqrt{\omega_{\mathbf{q+k}} + mc^2} \sqrt{\omega_{\mathbf{q}} + mc^2}}{2mc^2} \nonumber \\
&-&
 \frac{\sqrt{\omega_{\mathbf{q+k}} - mc^2} \sqrt{\omega_{\mathbf{q}} - mc^2}}{2mc^2} \left(\frac{\mathbf{q+k}}{|\mathbf{q+k}|} \cdot
\vec{\sigma}_{el}\right)   \left(\frac{\mathbf{q}}{q} \cdot
\vec{\sigma}_{el}\right) \Bigr)   \chi_{\sigma'} \nonumber   \\
&\approx& -\frac{i  c \alpha^2 \delta_{ \tau, \tau'}}{2 \pi^2 k^2} \left(1   - \frac{k^2}{6m^2c^2} \right) \times \nonumber  \\
&\ &\chi^{\dag}_{\sigma}  \left(  \left(1 + \frac{(\mathbf{q+k})^2}{8m^2c^2}\right) \left(1 + \frac{q^2}{8m^2c^2}\right) - \frac{|\mathbf{q+k}|q}{4m^2c^2} \left(\frac{\mathbf{q+k}}{|\mathbf{q+k}|} \cdot
\vec{\sigma}_{el}\right)   \left(\frac{\mathbf{q}}{q} \cdot
\vec{\sigma}_{el}\right) \right)   \chi_{\sigma'} \nonumber  \\
&=& \frac{i  c \alpha^2 \delta_{ \tau, \tau'}}{2 \pi^2 }\chi^{\dag}_{\sigma}  \left( -\frac{1}{k^2}  + \frac{1}{24m^2c^2}
+\frac{i \vec{\sigma}_{el} [\mathbf{k} \times \mathbf{q}]}{4m^2c^2k^2} \right)   \chi_{\sigma'} \label{eq:10.48a}
\end{eqnarray}

\noindent Putting (\ref{eq:u0w0x}) and (\ref{eq:10.48a}) together, we obtain\footnote{As expected from the charge renormalization condition, Coulomb-like terms $\propto 1/k^2$ have canceled out.}

\begin{eqnarray}
 s_{4}^{(e)+(h)AMM} =   - \frac{  \alpha^2  \delta_{\tau, \tau'} }{4 \pi^2 m^2 c} \chi^{\dag}_{\sigma}
\frac{(\vec{\sigma}_{el} \cdot [\mathbf{k} \times \mathbf{q}])}{k^2}  \chi_{\sigma'}
\label{eq:s4pqk}
\end{eqnarray}

\noindent For the $\lambda$-dependent part of  (\ref{eq:s4-1112y}) we use the non-relativistic approximation

\begin{eqnarray}
s_4^{(e)+(h)div}
&=&  \frac{i \alpha^2 Mmc^4}{3 \pi^2 m^2c
\sqrt{\Omega_{\mathbf{p-k}}
\omega_{\mathbf{q+k}} \Omega_{\mathbf{p}}
\omega_{\mathbf{q}}} }
   \ln \left(\frac{ \lambda}{m} \right)
  (\tilde{U} \cdot \tilde{W}) \nonumber \\
&\approx& \frac{i \alpha^2 }{3 \pi^2 m^2c}
   \ln \left(\frac{ \lambda}{m } \right) \delta_{\sigma, \sigma'} \delta_{\tau, \tau'}\label{eq:s4vertex}
\end{eqnarray}

\subsection{Ladder diagram}
\label{ss:ladder-diagram}

Let us investigate the contribution in $S_4^c$ corresponding to the ladder diagram \ref{fig:9.10}(f). According to Feynman  rules, it is given by the integral\footnote{Here we dropped spin indices, as in our approximation the spin dependence will be lost anyway.}

\begin{eqnarray*}
s_4^{(f)} &=&  \frac{e^4 (2 \pi \hbar)^{16}}{\hbar^4} \frac{-1}{(2 \pi)^2 (2 \pi
\hbar)^6} \frac{-\hbar^4 c^4}{(2 \pi)^2 (2 \pi
\hbar)^6} \frac{Mmc^4 }{(2 \pi
\hbar)^6
\sqrt{\omega_{\mathbf{q}}\omega_{\mathbf{q'}}\Omega_{\mathbf{p}}\Omega_{\mathbf{p'}}}} \nonumber \\
&\mbox{ } & \int d^4h \frac{\overline{u}(\mathbf{q})\gamma_{\mu} (
\cross{q}-\cross{h} + mc^2) \gamma_{\nu} u(\mathbf{q'})}{(\tilde{q}-\tilde{h})^2
-m^2c^4 } \cdot \frac{\overline{w}(\mathbf{p})\gamma_{\mu}(\cross{p}+\cross{h} + Mc^2) \gamma_{\nu}
w(\mathbf{p'})}{(\tilde{p}+\tilde{h})^2 -M^2c^4 } \nonumber \\
&\mbox{ }&  \frac{1}{[\tilde{h}^2  - \lambda^2c^4][(\tilde{h}+\tilde{k})^2  - \lambda^2c^4]}
\end{eqnarray*}

\noindent  We use Dirac equations (\ref{gamma-mu1}) - (\ref{gamma-mu2}) for  functions
$u(\mathbf{q})$ and $w(\mathbf{p})$
and the anticommutator relationship (\ref{gamma-mu}) for gamma matrices to rewrite the numerator

\begin{eqnarray*}
&\mbox { }& [\overline{u}(\mathbf{q})\gamma_{\mu} (
\cross{q}-\cross{h} + mc^2) \gamma_{\nu} u(\mathbf{q'})] \cdot
[\overline{w}(\mathbf{p})\gamma_{\mu}(\cross{p}+\cross{h} +
Mc^2) \gamma_{\nu} w(\mathbf{p'})] \\
&=& [\overline{u}(\mathbf{q})\gamma_{\mu} ( \cross{q} + mc^2)
\gamma_{\nu} u(\mathbf{q'}) - \overline{u}(\mathbf{q})\gamma_{\mu} \cross{h}  \gamma_{\nu} u(\mathbf{q'})] \times
\\
&\ & [\overline{w}(\mathbf{p})\gamma_{\mu}(\cross{p} +
Mc^2) \gamma_{\nu} w(\mathbf{p'}) +
\overline{w}(\mathbf{p})\gamma_{\mu}\cross{h} \gamma_{\nu} w(\mathbf{p'})] \\
&=& [2\overline{u}(\mathbf{q}) q^{\mu}  \gamma_{\nu}
u(\mathbf{q'}) - \overline{u}(\mathbf{q})\gamma_{\mu}
\cross{h}  \gamma_{\nu} u(\mathbf{q'})]  [2\overline{w}(\mathbf{p}) p^{\mu} \gamma_{\nu}
w(\mathbf{p'}) + \overline{w}(\mathbf{p})\gamma_{\mu}\cross{h} \gamma_{\nu} w(\mathbf{p'})] \\
&=& 4 (\tilde{q} \cdot \tilde{p}) \overline{u}(\mathbf{q})   \gamma_{\nu}
u(\mathbf{q'}) \overline{w}(\mathbf{p}) \gamma_{\nu}
w(\mathbf{p'}) + 2\overline{u}(\mathbf{q})
\gamma_{\nu} u(\mathbf{q'}) \overline{w}(\mathbf{p})\cross{q}\gamma_{\alpha} \gamma_{\nu} w(\mathbf{p'}) h^{\alpha} \\
&-& 2 \overline{u}(\mathbf{q}) \cross{p} \gamma_{\alpha}
 \gamma_{\nu} u(\mathbf{q'}) \overline{w}(\mathbf{p})  \gamma_{\nu} w(\mathbf{p'}) h^{\alpha} -
\overline{u}(\mathbf{q})\gamma_{\mu} \gamma_{\alpha}
\gamma_{\nu} u(\mathbf{q'}) \overline{w}(\mathbf{p})\gamma_{\mu} \gamma_{\beta} \gamma_{\nu} w(\mathbf{p'}) h^{\alpha} h^{\beta}
\end{eqnarray*}

\noindent In the denominators we use $\tilde{q}^2  = m^2c^4$ and $\tilde{p}^2  = M^2c^4$ to write

\begin{eqnarray*}
(\tilde{q}-\tilde{h})^2 -m^2c^4  &=& \tilde{h}^2 - 2 (\tilde{q} \cdot \tilde{h}) \\
(\tilde{p}+\tilde{h})^2 -M^2c^4 &=&  \tilde{h}^2 + 2 (\tilde{p} \cdot \tilde{h})
\end{eqnarray*}

\noindent Using non-relativistic approximation (\ref{eq:J.70a}) we then obtain

\begin{eqnarray}
s_4^{(f)} &\approx&    \frac{e^4 c^4 }{(2 \pi)^4 (2 \pi
\hbar)^2} \times \nonumber \\
&\mbox{ } & [4 (\tilde{q} \cdot \tilde{p}) \overline{u}(\mathbf{q})   \gamma_{\nu}
u(\mathbf{q'}) \overline{w}(\mathbf{p}) \gamma_{\nu}
w(\mathbf{p'}) b (\mathbf{p, q, k}) \nonumber \\
&+& 2\overline{u}(\mathbf{q}) \gamma_{\nu} u(\mathbf{q'}) \overline{w}(\mathbf{p})\cross{q}\gamma_{\alpha}
\gamma_{\nu} w(\mathbf{p'}) b^{\alpha} (\mathbf{p, q, k})
\nonumber \\
&-& 2 \overline{u}(\mathbf{q}) \cross{p} \gamma_{\alpha}
 \gamma_{\nu} u(\mathbf{q'}) \overline{w}(\mathbf{p})  \gamma_{\nu} w(\mathbf{p'}) b^{\alpha} (\mathbf{p,
q, k}) \nonumber \\
&-& \overline{u}(\mathbf{q})\gamma_{\mu} \gamma_{\alpha}
\gamma_{\nu} u(\mathbf{q'}) \overline{w}(\mathbf{p})\gamma_{\mu} \gamma_{\beta} \gamma_{\nu} w(\mathbf{p'}) b^{\alpha \beta} (\mathbf{p, q, k})] \label{eq:I1}
\end{eqnarray}

\noindent where

\begin{eqnarray}
b(\mathbf{p, q, k}) &=&   \int   \frac{d^4h}{[\tilde{h}^2 - 2(\tilde{q} \cdot \tilde{h}][\tilde{h}^2 +
2(\tilde{p} \cdot \tilde{h})][\tilde{h}^2  - \lambda^2c^4][(\tilde{h}+\tilde{k})^2  - \lambda^2c^4]} \nonumber \\
\label{eq:bpqk} \\
b^{\alpha}(\mathbf{p, q, k}) &=&   \int   \frac{d^4hh^{\alpha}}{[\tilde{h}^2 - 2(\tilde{q} \cdot \tilde{h})][\tilde{h}^2 +
2(\tilde{p} \cdot \tilde{h})][\tilde{h}^2  - \lambda^2c^4][(\tilde{h}+\tilde{k})^2  - \lambda^2c^4]} \nonumber \\
b^{\alpha \beta}(\mathbf{p, q, k}) &=&   \int  \frac{d^4h h^{\alpha}
h^{\beta}}{[\tilde{h}^2 - 2(\tilde{q} \cdot \tilde{h})][\tilde{h}^2 +
2(\tilde{p} \cdot \tilde{h})][\tilde{h}^2  - \lambda^2c^4][(\tilde{h}+\tilde{k})^2  - \lambda^2c^4]} \nonumber
\end{eqnarray}

In our calculations we are interested only in leading infrared-divergent terms. They come from those regions in the 4D space of the integration variable $\tilde{h}$, where integrand's denominators vanish in the limit $\lambda \to 0$. These are regions $\tilde{h} \approx 0$ and $\tilde{h} \approx -\tilde{k}$. Using these approximations in the numerators, we get

\begin{eqnarray}
b^{\alpha}(\mathbf{p, q, k}) &\approx& - k^{\alpha} b(\mathbf{p, q, k})\nonumber \\
b^{\alpha \beta}(\mathbf{p, q, k}) &\approx&    k^{\alpha} k^{\beta}b(\mathbf{p, q, k})\nonumber
\end{eqnarray}

\noindent Now substitute these results in (\ref{eq:I1}) and use definitions (\ref{eq:uz}) - (\ref{eq:vz}) of functions $U^{\mu}$ and $W^{\mu}$

\begin{eqnarray*}
s_4^{(f)}
&=& \frac{e^4 c^4 }{(2 \pi)^4 (2 \pi
\hbar)^2} b(\mathbf{p, q, k}) \times \nonumber \\
&\mbox{ } &  [4 (\tilde{q} \cdot \tilde{p})( \overline{u}(\mathbf{q})   \gamma_{\nu}
u(\mathbf{q'}) \overline{w}(\mathbf{p}) \gamma_{\nu}
w(\mathbf{p'})  - 2 \overline{u}(\mathbf{q}) \gamma_{\nu} u(\mathbf{q'}) \overline{w}(\mathbf{p})\cross{q}\cross{k}
\gamma_{\nu} w(\mathbf{p'})
\nonumber \\
&+& 2 \overline{u}(\mathbf{q}) \cross{p} \cross{k}
 \gamma_{\nu} u(\mathbf{q'}) \overline{w}(\mathbf{p})  \gamma_{\nu} w(\mathbf{p'})  -   \overline{u}(\mathbf{q})\gamma_{\mu} \cross{k}
\gamma_{\nu} u(\mathbf{q'}) \overline{w}(\mathbf{p})\gamma_{\mu} \cross{k} \gamma_{\nu} w(\mathbf{p'})]  \\
&=&   \frac{e^4 c^4 }{(2 \pi)^4 (2 \pi
\hbar)^2} b(\mathbf{p, q, k}) [4 (\tilde{q} \cdot \tilde{p}) (\tilde{U} \cdot \tilde{W})  - 2U_{\nu} \overline{w}(\mathbf{p})\cross{q}\cross{k}
\gamma_{\nu} w(\mathbf{p'}) \nonumber \\
&+& 2 \overline{u}(\mathbf{q}) \cross{p} \cross{k}
 \gamma_{\nu} u(\mathbf{q'}) W_{\nu}  -   \overline{u}(\mathbf{q})\gamma_{\mu} \cross{k}
\gamma_{\nu} u(\mathbf{q'}) \overline{w}(\mathbf{p})\gamma_{\mu} \cross{k} \gamma_{\nu} w(\mathbf{p'})]
\end{eqnarray*}

\noindent Next we need to simplify  separate pieces of this expression

\begin{eqnarray}
&\ &  \overline{w}(\mathbf{p})\cross{q}\cross{k}
\gamma_{\nu} w(\mathbf{p'}) = \overline{w}(\mathbf{p})\cross{q}\cross{p}
\gamma_{\nu} w(\mathbf{p'}) - \overline{w}(\mathbf{p})\cross{q}\cross{p}'
\gamma_{\nu} w(\mathbf{p'}) \nonumber \\
&=& \overline{w}(\mathbf{p})\cross{q}\cross{p}
\gamma_{\nu} w(\mathbf{p'}) - \overline{w}(\mathbf{p})\cross{q}(2(p')^{\nu} - mc^2
\gamma_{\nu}) w(\mathbf{p'})
\nonumber \\
&=& \overline{w}(\mathbf{p})(-\cross{p}\cross{q} +2 (\tilde{q} \cdot \tilde{p}))
\gamma_{\nu} w(\mathbf{p'}) - 2(p')^{\nu} (\tilde{q} \cdot \tilde{W}) + mc^2\overline{w}(\mathbf{p})\cross{q}
\gamma_{\nu} w(\mathbf{p'})
\nonumber \\
&=& \overline{w}(\mathbf{p})(-mc^2\cross{q} +2 (\tilde{q} \cdot \tilde{p}))
\gamma_{\nu} w(\mathbf{p'}) - 2(p')^{\nu} (\tilde{q} \cdot \tilde{W}) + mc^2\overline{w}(\mathbf{p})\cross{q}
\gamma_{\nu} w(\mathbf{p'})
\nonumber \\
&=& -mc^2 \overline{w}(\mathbf{p})\cross{q}
\gamma_{\nu} w(\mathbf{p'}) + 2 (\tilde{q} \cdot \tilde{p}) W_{\nu} - 2(p')^{\nu} (\tilde{q} \cdot \tilde{W}) \nonumber \\
&+& mc^2\overline{w}(\mathbf{p})\cross{q}
\gamma_{\nu} w(\mathbf{p'})
=  2 (\tilde{q} \cdot \tilde{p}) W_{\nu} - 2(p')^{\nu} (\tilde{q} \cdot \tilde{W}) \label{eq:wqkg}
\end{eqnarray}

\begin{eqnarray*}
&\ &  \overline{u}(\mathbf{q}) \cross{p} \cross{k}
 \gamma_{\nu} u(\mathbf{q'}) = \overline{u}(\mathbf{q}) \cross{p} (\cross{q}'- \cross{q})
 \gamma_{\nu} u(\mathbf{q'}) \\
 &=& \overline{u}(\mathbf{q}) \cross{p} \cross{q}'
 \gamma_{\nu} u(\mathbf{q'}) - \overline{u}(\mathbf{q}) \cross{p}  \cross{q}
 \gamma_{\nu} u(\mathbf{q'}) \\
  &=& \overline{u}(\mathbf{q}) \cross{p} (-\gamma_{\nu} \cross{q}' + 2 (q')^{\nu}
) u(\mathbf{q'}) - \overline{u}(\mathbf{q}) (- \cross{q}\cross{p} + 2 (\tilde{q} \cdot \tilde{p}))
 \gamma_{\nu} u(\mathbf{q'}) \\
&=& -\overline{u}(\mathbf{q}) \cross{p} \gamma_{\nu} \cross{q}'  u(\mathbf{q'}) + 2\overline{u}(\mathbf{q}) \cross{p} (q')^{\nu} u(\mathbf{q'}) + \overline{u}(\mathbf{q}) \cross{q}\cross{p}\gamma_{\nu} u(\mathbf{q'}) \\
&-& 2 (\tilde{q} \cdot \tilde{p}) U_{\nu} \\
&=& -mc^2\overline{u}(\mathbf{q}) \cross{p} \gamma_{\nu}   u(\mathbf{q'}) + 2(q')^{\nu}(\tilde{p} \cdot \tilde{U}) + mc^2\overline{u}(\mathbf{q}) \cross{p}\gamma_{\nu} u(\mathbf{q'}) - 2 (\tilde{q} \cdot \tilde{p}) U_{\nu} \\
&=&  2(q')^{\nu}(\tilde{p} \cdot \tilde{U})  - 2 (\tilde{q} \cdot \tilde{p})U_{\nu}
\end{eqnarray*}

\begin{eqnarray}
&\ &  \overline{u}(\mathbf{q})\gamma_{\mu} \cross{k}
\gamma_{\nu} u(\mathbf{q'})  \nonumber \\
&=& \overline{u}(\mathbf{q})\gamma_{\mu} \cross{q}'
\gamma_{\nu} u(\mathbf{q'}) -\overline{u}(\mathbf{q})\gamma_{\mu} \cross{q}
\gamma_{\nu} u(\mathbf{q'}) \nonumber \\
&=& \overline{u}(\mathbf{q})\gamma_{\mu} (-
\gamma_{\nu}\cross{q}' + 2(q')^{\nu}) u(\mathbf{q'}) -\overline{u}(\mathbf{q})(-\cross{q}\gamma_{\mu} + 2 q^{\mu})
\gamma_{\nu} u(\mathbf{q'}) \nonumber \\
&=& -\overline{u}(\mathbf{q})\gamma_{\mu}
\gamma_{\nu}\cross{q}'  u(\mathbf{q'}) + 2\overline{u}(\mathbf{q})\gamma_{\mu} (q')^{\nu} u(\mathbf{q'})  +\overline{u}(\mathbf{q})\cross{q}\gamma_{\mu}
\gamma_{\nu} u(\mathbf{q'}) \nonumber \\
&\ & -2\overline{u}(\mathbf{q}) q^{\mu}
\gamma_{\nu} u(\mathbf{q'}) \nonumber \\
&=& -mc^2 \overline{u}(\mathbf{q})\gamma_{\mu}
\gamma_{\nu}  u(\mathbf{q'}) + 2(q')^{\nu} U_{\mu}  +mc^2\overline{u}(\mathbf{q})\gamma_{\mu}
\gamma_{\nu} u(\mathbf{q'}) -2q^{\mu} U_{\nu} \nonumber \\
&=&  2(q')^{\nu} U_{\mu}  -2q^{\mu} U_{\nu} \label{eq:qUqU}
\end{eqnarray}

\begin{eqnarray}
\overline{w}(\mathbf{p})\gamma_{\mu} \cross{k} \gamma_{\nu} w(\mathbf{p'})
&=& -2(p')^{\nu} W_{\mu}  +2p^{\mu} W_{\nu} \label{eq:pWpW}
\end{eqnarray}

\noindent Then we use equalities (\ref{eq:M.50}), (\ref{eq:M.51}) and non-relativistic approximations $(\tilde{q}' \cdot \tilde{p}') \approx Mmc^4$, $(\tilde{U} \cdot \tilde{W}) \approx 1$ to write

\begin{eqnarray}
&\ &s_4^{(f)}  = \frac{e^4 c^4 }{(2 \pi)^4 (2 \pi
\hbar)^2} b(\mathbf{p, q, k}) \times \nonumber \\
&\ & [4 (\tilde{q} \cdot \tilde{p}) (\tilde{U} \cdot \tilde{W})  - 2U_{\nu} (2 (\tilde{q} \cdot \tilde{p}) W_{\nu} - 2(p')^{\nu} (\tilde{q} \cdot \tilde{W}))
\nonumber \\
&+& 2(2(q')^{\nu}(p \cdot U)  - 2 (\tilde{q} \cdot \tilde{p})U_{\nu}) W_{\nu}  -   (2(q')^{\nu} U_{\mu}  -2q^{\mu} U_{\nu}) (-2(p')^{\nu} W_{\mu}  +2p^{\mu} W_{\nu})] \nonumber  \\
&=&  \frac{4e^4 c^4 }{(2 \pi)^4 (2 \pi
\hbar)^2} b(\mathbf{p, q, k}) \times \nonumber \\
&\ &  [(\tilde{q} \cdot \tilde{p}) (\tilde{U} \cdot \tilde{W})  -  (\tilde{q} \cdot \tilde{p})(\tilde{U} \cdot \tilde{W}) +(\tilde{p}' \cdot \tilde{U}) (\tilde{q} \cdot \tilde{W})
+ (\tilde{q}' \cdot \tilde{W})(\tilde{p} \cdot \tilde{U})  -  (\tilde{q} \cdot \tilde{p})(\tilde{U} \cdot \tilde{W})  \nonumber \\
&+&
(\tilde{q}' \cdot \tilde{p}')(\tilde{U} \cdot \tilde{W}) - (\tilde{q}' \cdot \tilde{W})(\tilde{p} \cdot \tilde{U}) - (\tilde{q} \cdot \tilde{W}) (\tilde{p}' \cdot \tilde{U})  + (\tilde{q} \cdot \tilde{p}) (\tilde{U} \cdot \tilde{W}) ] \nonumber \\
&=&  \frac{4e^4 c^4 }{(2 \pi)^4 (2 \pi
\hbar)^2} b(\mathbf{p, q, k})  (\tilde{q}' \cdot \tilde{p}')(\tilde{U} \cdot \tilde{W}) \nonumber \\
&\approx&  \frac{4e^4 Mmc^8 }{(2 \pi)^4 (2 \pi
\hbar)^2} b(\mathbf{p, q, k}) \label{eq:15.12a}
\end{eqnarray}

\noindent Function $b(\mathbf{p, q, k})$ is evaluated in  (\ref{eq:int-py2})

\begin{eqnarray}
b(\mathbf{p, q, k}) &=& \frac{ \pi^2 }{ic^3 \tilde{k}^2} \ln \left( \frac{\tilde{k}^2 }{ \lambda^2 c^4}\right) \int \limits_0^1 \frac{dy}{(\tilde{p} + \tilde{q})^2y^2 - 2 \tilde{p}(\tilde{p} + \tilde{q}) y + \tilde{p}^2} \label{eq:15.12b}
\end{eqnarray}

\noindent This integral is of the table form

\begin{eqnarray*}
\int \frac{dy}{ay^2 +by+c} = \frac{2}{\sqrt{4ac - b^2}}\tan^{-1}\left( \frac{2ax+b}{\sqrt{4ac - b^2}}  \right) + const
\end{eqnarray*}

\noindent So, we get

\begin{eqnarray}
&\ &  \int \limits_0^1 \frac{dy}{(\tilde{p} + \tilde{q})^2y^2 - 2 \tilde{p}(\tilde{p} + \tilde{q}) y + \tilde{p}^2} \nonumber \\
&=& 2B^{-1} \tan^{-1} \left( \frac{2(\tilde{p} + \tilde{q})^2y - 2\tilde{p}(\tilde{p} + \tilde{q})}{B }  \right) \Bigl|_{y=0}^{y=1} \nonumber \\
&=& 2B^{-1} \left[ \tan^{-1} \left( \frac{ 2\tilde{q}^2 +2(\tilde{p} \cdot \tilde{q})}{B } \right)
  + \tan^{-1} \left( \frac{ 2\tilde{p}^2 +2(\tilde{p} \cdot \tilde{q})}{B }\right) \right] \nonumber \\
&\approx& \frac{1}{iMc^3q} \left[ \tan^{-1} \left( \frac{ mc}{iq} \right)
  + \tan^{-1} \left( \frac{ Mc}{iq}\right) \right] \label{eq:15.12c}
\end{eqnarray}

\noindent  where we used the inequality $M \gg m$ and denoted

\begin{eqnarray*}
B &\equiv& \sqrt{4 (\tilde{p} + \tilde{q})^2\tilde{p}^2 - 4(\tilde{p}^2 + (\tilde{p}\cdot \tilde{q}))^2} = 2\sqrt{\tilde{p}^2 \tilde{q}^2 - (\tilde{p}\cdot \tilde{q})^2} = 2\sqrt{M^2m^2c^8 - (\tilde{p}\cdot \tilde{q})^2} \\
&\approx& 2\sqrt{M^2m^2c^8 - \left[\left(Mc^2 + \frac{p^2}{2M}\right)\left(mc^2 + \frac{q^2}{2m}\right)-c^2(\mathbf{pq})\right]^2} \\
&\approx& 2\sqrt{-M^2c^6q^2} = 2iMc^3q
\end{eqnarray*}

\noindent Putting  results (\ref{eq:15.12a}) - (\ref{eq:15.12c}) together and using $\tilde{k}^2 \approx -c^2k^2$, we finally obtain

\begin{eqnarray}
 s_4^{(f)}
&\approx&  \frac{\alpha^2 mc^2 }{\pi^2 q k^2} \left[ \tan^{-1} \left( \frac{ mc}{iq} \right)
  + \tan^{-1} \left( \frac{ Mc}{iq}\right) \right] \ln\left(-\frac{k^2}{\lambda^2c^2} \right) \nonumber \\
\label{eq:ladder}
\end{eqnarray}

\noindent We will not elaborate this result further as we expect some cancelations with the crossed ladder diagram  evaluated in the next subsection.

\subsection{Cross-ladder diagram}
\label{ss:ladder-diagram2}

Similar to the above ladder diagram we calculate the cross-ladder diagram shown in Fig. \ref{fig:9.10}(g)

\begin{eqnarray*}
s_4^{(g)}&\approx&   \frac{e^4 c^4 }{(2 \pi)^4 (2 \pi
\hbar)^2}  \int d^4h  \frac{
\overline{u}(\mathbf{q}) \gamma_{\mu}(\cross{q}- \cross{h} +
mc^2)\gamma_{\nu} u(\mathbf{q'}) }{(\tilde{q}-\tilde{h})^2 -m^2c^4 + i
\sigma} \times \\
&\ &  \frac{\overline{w}(\mathbf{p}) \gamma_{\nu}
(\cross{p}'-\cross{h} + Mc^2) \gamma_{\mu} w(\mathbf{p'})}{(\tilde{p}'-\tilde{h})^2 -M^2c^4 }  \frac{1}{[\tilde{h}^2 - \lambda^2c^4][(\tilde{h}+\tilde{k})^2 - \lambda^2c^4]}
\end{eqnarray*}

\noindent In the numerator we use (\ref{gamma-mu}), (\ref{gamma-mu1}) - (\ref{gamma-mu2})  to write

\begin{eqnarray*}
&\mbox { }& [\overline{u}(\mathbf{q}) \gamma_{\mu}(\cross{q}-
\cross{h} + mc^2)\gamma_{\nu} u(\mathbf{q'})] \cdot
[\overline{w}(\mathbf{p}) \gamma_{\nu} (\cross{p}'-\cross{h} +
Mc^2) \gamma_{\mu} w(\mathbf{p'})] \\
&=& [\overline{u}(\mathbf{q})\gamma_{\mu} ( \cross{q} + mc^2)
\gamma_{\nu} u(\mathbf{q'}) - \overline{u}(\mathbf{q})\gamma_{\mu} \cross{h}  \gamma_{\nu} u(\mathbf{q'})]
\\
&\times& [\overline{w}(\mathbf{p})\gamma_{\mu}(\cross{p}' +
Mc^2) \gamma_{\nu} w(\mathbf{p'}) -
\overline{w}(\mathbf{p})\gamma_{\mu}\cross{h} \gamma_{\nu} w(\mathbf{p'})] \\
&=& [2\overline{u}(\mathbf{q}) q^{\mu}  \gamma_{\nu}
u(\mathbf{q'}) - \overline{u}(\mathbf{q})\gamma_{\mu}
\cross{h}  \gamma_{\nu} u(\mathbf{q'})] \\
&\times& [2\overline{w}(\mathbf{p}) (p')^{\nu} \gamma_{\mu}
w(\mathbf{p'}) - \overline{w}(\mathbf{p})\gamma_{\mu}\cross{h} \gamma_{\nu} w(\mathbf{p'})] \\
&=& 4  \overline{u}(\mathbf{q})   \cross{p}' u(\mathbf{q'}) \overline{w}(\mathbf{p}) \cross{q} w(\mathbf{p'}) - 2\overline{u}(\mathbf{q}) \gamma_{\nu}
u(\mathbf{q'}) \overline{w}(\mathbf{p})\cross{q}\gamma_{\alpha} \gamma_{\nu} w(\mathbf{p'}) h^{\alpha} \\
&-& 2 \overline{u}(\mathbf{q})  \gamma_{\mu} \gamma_{\alpha}
\cross{p}' u(\mathbf{q'}) \overline{w}(\mathbf{p})
\gamma_{\mu} w(\mathbf{p'}) h^{\alpha} \\
&+& \overline{u}(\mathbf{q})\gamma_{\mu} \gamma_{\alpha}
\gamma_{\nu} u(\mathbf{q'}) \overline{w}(\mathbf{p})\gamma_{\mu} \gamma_{\beta} \gamma_{\nu} w(\mathbf{p'}) h^{\alpha} h^{\beta}
\end{eqnarray*}

\noindent and

\begin{eqnarray}
&\ & s_4^{(g)} =  \frac{e^4 c^4}{(2 \pi)^4 (2 \pi
\hbar)^2} \times \nonumber \\
&\mbox{ } & [4  \overline{u}(\mathbf{q})   \cross{p}'
u(\mathbf{q'}) \overline{w}(\mathbf{p}) \cross{q}
w(\mathbf{p'}) b (-\mathbf{p', q, k}) \nonumber \\
&-& 2\overline{u}(\mathbf{q}) \gamma_{\nu} u(\mathbf{q'}) \overline{w}(\mathbf{p})\cross{q}\gamma_{\alpha}
\gamma_{\nu} w(\mathbf{p'}) b^{\alpha} (-\mathbf{p', q, k})
\nonumber \\
&-& 2 \overline{u}(\mathbf{q})  \gamma_{\mu} \gamma_{\alpha}
\cross{p}' u(\mathbf{q'}) \overline{w}(\mathbf{p})
\gamma_{\mu} w(\mathbf{p'}) b^{\alpha} (-\mathbf{p',
q, k}) \nonumber \\
&+& \overline{u}(\mathbf{q})\gamma_{\mu} \gamma_{\alpha}
\gamma_{\nu} u(\mathbf{q'}) \overline{w}(\mathbf{p})\gamma_{\mu} \gamma_{\beta} \gamma_{\nu} w(\mathbf{p'}) b^{\alpha \beta} (-\mathbf{p', q, k})] \label{eq:15.8a}
\end{eqnarray}

\noindent Here we notice that  integral

\begin{eqnarray*}
b(\mathbf{-p', q, k}) \equiv \int   \frac{d^4h}{[\tilde{h}^2 - 2(\tilde{q} \cdot \tilde{h})][\tilde{h}^2 -
2(\tilde{p}' \cdot \tilde{h})][\tilde{h}^2 - \lambda^2c^4][(\tilde{h}+\tilde{k})^2 - \lambda^2c^4]}  \label{eq:bpqk2}
\end{eqnarray*}

\noindent can be obtained from (\ref{eq:bpqk}) by replacing $\tilde{p} \to -\tilde{p}'$.
Using the same assumptions as in the preceding subsection, two other integrals can be expressed in terms of $b(\mathbf{-p', q, k})$

\begin{eqnarray*}
b^{\alpha}(\mathbf{-p', q, k}) &\equiv&   \int   \frac{d^4hh^{\alpha}}{[\tilde{h}^2 - 2(\tilde{q} \cdot \tilde{h})][\tilde{h}^2 -
2(\tilde{p}' \cdot \tilde{h})][\tilde{h}^2 - \lambda^2c^4][(\tilde{h}+\tilde{k})^2 - \lambda^2c^4]} \\
&\approx& -k^{\alpha} b(\mathbf{-p', q, k})\nonumber \\
b^{\alpha \beta}(\mathbf{-p', q, k}) &\equiv &   \int  \frac{d^4h h^{\alpha}
h^{\beta}}{[\tilde{h}^2 - 2(\tilde{q} \cdot \tilde{h})][\tilde{h}^2 -
2(\tilde{p}' \cdot \tilde{h})][\tilde{h}^2 - \lambda^2c^4][(\tilde{h}+\tilde{k})^2 - \lambda^2c^4]} \\
&\approx& k^{\alpha} k^{\beta}b(\mathbf{-p', q, k})\nonumber
\end{eqnarray*}

\noindent Then in  (\ref{eq:15.8a}) we can use (\ref{eq:wqkg}), (\ref{eq:qUqU}), (\ref{eq:pWpW}), and

\begin{eqnarray*}
&\ &  \overline{u}(\mathbf{q})\gamma_{\mu} \cross{k} \cross{p'}
 u(\mathbf{q'}) = \overline{u}(\mathbf{q})  \gamma_{\mu} (\cross{q}'- \cross{q})
 \cross{p}' u(\mathbf{q'}) \\
 &=& \overline{u}(\mathbf{q}) \gamma_{\mu} \cross{q}' \cross{p}'
  u(\mathbf{q'}) - \overline{u}(\mathbf{q}) \gamma_{\mu} \cross{q} \cross{p}'
  u(\mathbf{q'}) \\
 &=& -\overline{u}(\mathbf{q}) \gamma_{\mu} \cross{p}'  \cross{q}' u(\mathbf{q'})
 + 2\overline{u}(\mathbf{q}) \gamma_{\mu} (q'p') u(\mathbf{q'})+ \overline{u}(\mathbf{q}) \cross{q} \gamma_{\mu}  \cross{p}' u(\mathbf{q'}) \\
 &\ & -2\overline{u}(\mathbf{q}) q^{\mu} \cross{p}' u(\mathbf{q'}) = 2\overline{u}(\mathbf{q}) \gamma_{\mu} (q'\cdot p') u(\mathbf{q'})
 - 2\overline{u}(\mathbf{q}) q^{\mu} \cross{p}' u(\mathbf{q'}) \\
&=& 2(\tilde{q}' \cdot \tilde{p}') U_{\mu} - 2 (\tilde{p}' \cdot \tilde{U}) q^{\mu}
\end{eqnarray*}

\noindent to obtain

\begin{eqnarray*}
&\ & s_4^{(g)} = \frac{e^4 c^4 }{(2 \pi)^4 (2 \pi
\hbar)^2} b(\mathbf{-p', q, k}) \times\\
&\mbox{ } & [ 4(\tilde{p}'\cdot \tilde{U})(\tilde{q} \cdot \tilde{W})  + 2U_{\nu}(2 (\tilde{q} \cdot \tilde{p}) W_{\nu} - 2(p')^{\nu} (q \cdot W))
\nonumber \\
&+& 2(2 (\tilde{q}' \cdot \tilde{p}') U_{\mu} - 2(\tilde{p}'\cdot \tilde{U})q^{\mu} ) W_{\mu}  +  (2(q')^{\nu} U_{\mu}  -2q^{\mu} U_{\nu})(-2(p')^{\nu} W_{\mu}  +2p^{\mu} W_{\nu}) ]\\
&=& \frac{4e^4 c^4 }{(2 \pi)^4 (2 \pi
\hbar)^2} b(\mathbf{-p', q, k}) \times\\
&\mbox{ } &[  (\tilde{p}'\cdot \tilde{U})(\tilde{q} \cdot \tilde{W})  +  (\tilde{q} \cdot \tilde{p}) (\tilde{U} \cdot \tilde{W}) - (\tilde{p}'\cdot \tilde{U}) (\tilde{q} \cdot \tilde{W}) +  (\tilde{q}' \cdot \tilde{p}')(\tilde{U} \cdot \tilde{W})
\nonumber \\
&-& (\tilde{p}'\cdot \tilde{U})(\tilde{q} \cdot \tilde{W}) - (\tilde{q}' \cdot \tilde{p}')(\tilde{U} \cdot \tilde{W}) + (\tilde{q} \cdot \tilde{W})(\tilde{p}'\cdot \tilde{U}) + (\tilde{q}' \cdot \tilde{W})(\tilde{p}\cdot \tilde{U}) \\
&-&(\tilde{q} \cdot \tilde{p})(\tilde{U} \cdot \tilde{W})]
\\
&=&   \frac{4e^4 c^4 }{(2 \pi)^4 (2 \pi
\hbar)^2} b(\mathbf{-p', q, k})(\tilde{q}' \cdot \tilde{W})(\tilde{p}\cdot \tilde{U}) \\
&\approx& \frac{4e^4 Mmc^8 }{(2 \pi)^4 (2 \pi
\hbar)^2} b(\mathbf{-p', q, k})
\end{eqnarray*}

\noindent For the integral

\begin{eqnarray*}
b(\mathbf{-p', q, k}) &=& \frac{\pi^2}{ic^3\tilde{k}^2} \ln\left(\frac{\tilde{k}^2}{\lambda^2c^4} \right)\int \limits_0^1 \frac{dy}{(-\tilde{p}' + \tilde{q})^2y^2 + 2 \tilde{p}'(-\tilde{p}' + \tilde{q}) y + \tilde{p}^2}
\end{eqnarray*}

\noindent we use the same method as in (\ref{eq:15.12c}). This time in our non-relativistic approximation $(\tilde{p}' \cdot \tilde{q}) \approx Mmc^4$

\begin{eqnarray*}
B' &\equiv& \sqrt{4(\tilde{q} - \tilde{p}')^2 (\tilde{p}')^2 - 4 ((\tilde{p}')^2 - (\tilde{p}' \cdot \tilde{q}))^2} \approx B = 2iMc^3q \\ \\
&\ & \int \limits_0^1 \frac{dy}{(-\tilde{p}' + \tilde{q})^2y^2 + 2 \tilde{p}'(-\tilde{p}' + \tilde{q}) y + \tilde{p}^2} \\
&\approx& 2 B^{-1} \tan^{-1} \left( \frac{2(-\tilde{p}' + \tilde{q})^2y + 2\tilde{p}'(-\tilde{p}' + \tilde{q})}{B }  \right) \Bigl|_{y=0}^{y=1} \\
&=& 2B^{-1} \left[ \tan^{-1} \left( \frac{ 2\tilde{q}^2 -2(\tilde{p}' \cdot \tilde{q})}{B } \right)
  + \tan^{-1} \left( \frac{ 2\tilde{p}^2 -2(\tilde{p}' \cdot \tilde{q})}{B }\right) \right] \\
&\approx& \frac{1}{iMc^3q} \left[ -\tan^{-1} \left( \frac{ mc}{iq} \right)
  + \tan^{-1} \left( \frac{ Mc}{iq}\right) \right]
\end{eqnarray*}

\noindent and

\begin{eqnarray*}
&\ & s_4^{(g)} \approx  \frac{\alpha^2 mc^2 }{\pi^2 qk^2}   \left[ -\tan^{-1} \left( \frac{ mc}{iq} \right)
  + \tan^{-1} \left( \frac{ Mc}{iq}\right) \right] \ln\left(-\frac{k^2}{\lambda^2c^2} \right)
\end{eqnarray*}

\noindent Adding this result to (\ref{eq:ladder}) and using approximation $\tan^{-1}(Mc/(iq)) \approx -\pi/2$ we obtain the joint contribution of the ladder and crossed ladder diagrams

\begin{eqnarray}
s_4^{(f)+(g)} \approx -\frac{\alpha^2  mc^2  }{\pi q k^2}  \ln\left(-\frac{k^2}{\lambda^2c^2} \right)
\label{eq:joint}
\end{eqnarray}

\subsection{Renormalizability}
\label{ss:qed-renorm}

Combining results (\ref{eq:s4-vac-pol}), (\ref{eq:s4pqk}), (\ref{eq:s4vertex}), (\ref{eq:joint})  we get the following $\Lambda$-independent 4th order amplitude for the electron-proton scattering

\begin{eqnarray}
&\ & \langle 0|a_{\mathbf{q}, \sigma} d_{\mathbf{p}, \tau} S_4^c d^{\dag}_{\mathbf{p}', \tau'} a^{\dag}_{\mathbf{q}', \sigma'} |0 \rangle \approx   \delta^4(\tilde{q} - \tilde{q}'
-\tilde{p}'+\tilde{p})  \delta_{\tau \tau'} \times \nonumber \\
&\ &\Bigl[\frac{i \alpha^2 }{15 \pi^2  m^2c} \delta_{\sigma \sigma'}  + \frac{i \alpha^2}{3 \pi^2 m^2c} \ln\left( \frac{\lambda}{m} \right) \delta_{\sigma \sigma'} -\frac{  mc^2 \alpha^2 }{\pi q k^2}  \ln\left(-\frac{k^2}{\lambda^2c^2} \right) \delta_{\sigma \sigma'}\nonumber \\
&-& \frac{\alpha^2\chi_{\sigma}^{\dag} (\vec{\sigma}_{el} \cdot [\mathbf{k} \times \mathbf{q}]) \chi_{\sigma'}}{4 \pi^2 m^2 c k^2}\Bigr] \label{eq:S4-renorm}
\end{eqnarray}

\noindent As expected, this result is finite and does not depend on the cutoff parameter $\Lambda$. In other words, this result is ultraviolet-safe, so that we fulfilled the promise of the renormalization approach. Unfortunately, the amplitude (\ref{eq:S4-renorm}) still contains unpleasant infrared-divergent logarithms. Their physical origin is related to the vanishing photon mass. Any collision involving charged particles\footnote{in particular, the $e^--p^+$ collision considered here} is inevitably  accompanied by the emission of a large (even infinite) number of low-energy (soft) photons. In most cases these soft photons escape experimental detection, but in a rigorous theoretical treatment one must take them all into account in order to obtain scattering cross-sections in good agreement with experiments. This would require rather involved non-perturbative calculations \cite{book, Peskin} that are beyond the scope of this book. The cancelation of infrared divergences in calculations of the hydrogen energy spectrum (the Lamb shift) will be discussed in chapter \ref{ss:hydrogen}.

So, we conclude that our renormalization approach has achieved its goal: ultraviolet divergences in loop integrals have been canceled, and accurate description of scattering is within reach. Can we get even better accuracy by extending renormalization  to higher perturbation orders?
Yes, but then we would need to add higher-order ultraviolet-divergent counterterms to our interaction (\ref{eq:11.61}), so that the no-self-scattering and charge renormalization conditions are enforced in each perturbation order. It is remarkable that all these
higher-order counterterms will have exactly the same functional form as those
already discussed.  In other words, the complete infinite-order interaction operator
of the renormalized QED will have the same form as our low-order expression (\ref{eq:11.61}). Only
the values of exact renormalization constants $\delta m, Z_2-1, Z_3-1$, and
$Z_1-1$ will get more complicated forms than our 2nd order expressions $\delta m_2, (Z_2-1)_2, (Z_3-1)_2, (Z_1-1)_2$. This fact is referred to as the
\emph{renormalizability} \index{renormalizability} of QED.

We have applied renormalization only to the potential
energy operator $V$ in QED. In order to have a relativistic theory one
also need to find appropriate counterterms for the potential boost
operator $\mathbf{Z}$, so that the ``renormalized boost'' satisfies appropriate
Poincar\'e commutation relations with the ``renormalized energy.''
As far as I know, there were no attempts to extend the
renormalization theory to boosts. Nevertheless, we will assume that such a construction is possible, and that renormalized QED is a fully relativistic theory.

\section{Troubles with renormalized QED}
\label{sc:troubles}

Great successes of the renormalized QED are well known. In this section we are going to
focus on its weak points. The most obvious problem of QED is related
to extremely weird properties of its fundamental ingredients - bare
particles.  The masses and charges of bare electrons and protons are
infinite, and the Hamiltonian of QED is formally
infinite as well. More precisely, coefficient functions of interaction terms in $H^c_{FD} = H_0 + V^c_{FD}$ written in the bare-particle representation (\ref{eq:11.61}) diverge as  the ultraviolet cutoff momentum $\Lambda c$ is sent to infinity.  From results obtained in this chapter, we know that this operator can be used successfully in $S$-matrix calculations, because all divergences cancel out. However, its use for bound state or time evolution studies seems problematic.

\subsection{Renormalization in QED revisited} \label{ss:unable3}

Let us now review the material of this chapter and recall the
logic which led us from the original Feynman-Dyson interaction Hamiltonian $V_1$ in
(\ref{eq:basic-int}) to the interaction with
counterterms $V^c_{FD} $ in (\ref{eq:11.61}).\footnote{Everything said about Feynman-Dyson interaction $V^c_{FD} $ in this subsection applies also to the conventional renormalized interaction operator $V^c$ in (\ref{eq:V1-2x}).}

One distinctive feature of the operator $V_1$  is its \emph{unphys} ($U$)  type.
In order to obtain the scattering phase operator $F$  one needs to
calculate multiple commutators of $V_1$ as in (\ref{eq:7.63b}). It is clear
from Table \ref{table:7.2} that these commutators will give rise to
\emph{renorm}  terms\footnote{in commutators $[U,U']$} in each
perturbation order of $F$. However, according to equation
(\ref{eq:11.42}) and Statement \ref{statementU} (the no-self-scattering renormalization condition), there should be absolutely no \emph{renorm}  terms in the
operator $F$ of any sensible theory.
So, we have a contradiction.

The renormalization approach presented in this chapter suggested the following resolution of this
paradox: change the interaction operator from $V_1$ to $V^c_{FD}$ by adding
infinite counterterms (\ref{eq:11.61}). Two conditions were used to select the counterterms. The first (no-self-scattering renormalization) condition required cancelation of all \emph{renorm}  terms in the
scattering phase operator $F^c$ calculated with $V^c_{FD}$. This requirement was
satisfied by making sure there was a certain balance of \emph{unphys} and \emph{renorm}
(counter)terms in $V_{FD}^c$, so that all \emph{renorm}  terms in $F^c$ canceled out.  The second (charge renormalization) condition demanded
a consistency with classical electrodynamics  in the low
energy regime. These two conditions were sufficient to specify all counterterms in $V_{FD}^c$.
Somewhat miraculously, the $S$-matrix obtained with thus modified Hamiltonian
$H^c_{FD} =H_0 +V^c_{FD}$ agreed with experiment at all energies and for all
scattering processes.

Frequently one can meet interpretations of the renormalization approach, which say
that  infinities in the Hamiltonian $H^c_{FD}$ (\ref{eq:11.61}) have a
real physical meaning. The usual interpretation is that renormalization (=the addition of counterterms) is equivalent to redefinition of parameters (masses and charges of particles) in the Lagrangian. In fact, these parameters become infinite after the renormalization. Thus, it is declared that particle operators specified in section \ref{sc:annih-operators} refer not to real (or physical) particles, but to so-called \emph{bare} particles. One can hear also that bare
\index{bare particle} electrons and protons \emph{really} have
infinite masses and infinite charges.\footnote{It is also common to hypothesize that
these bare parameters may be actually very large rather than
infinite. The idea is that the ``granularity'' of space-time or other
yet unknown Planck-scale effect sets a natural momentum cutoff. This
``effective field theory'' approach assumes that QED is just a low
energy approximation to some unknown divergence-free truly
fundamental theory operating at the Planck scale. Speculations of
this kind are not needed for the dressed particle approach that will be developed
in the second part of this book. The dressed particle Hamiltonian
and the corresponding $S$-operator remain finite, even for infinite cutoff momentum.} The fact that such particles were never observed
in nature is then explained as follows: Bare particles
 are not eigenstates of the total Hamiltonian
$H^c_{FD}$.  The ``physical'' electrons and protons observed in experiments are
complex linear combinations of multiparticle bare states. These
linear combinations  \emph{are} eigenstates of the total Hamiltonian, and they do have correct (finite) measurable masses $m$
and $M$ and charges $\pm e$. This situation is often described as
bare particles being surrounded by ``clouds''
of \emph{virtual particles}, \index{virtual particle} thus forming
\emph{physical} or \emph{dressed particles}. \index{physical
particle} \index{dressed particle} The virtual cloud modifies
 the mass  of  the bare particle by an infinite amount, so that
the resulting mass is exactly the one measured in experiments. The
cloud also ``shields'' the (infinite) charge of the bare particle,
so that the effective charge becomes $\pm e$ \cite{Huang}.

Even if we accept this weird description of physical reality, it is
clear that the renormalization  program did not solve the problem of
ultraviolet divergences in quantum field theory. The divergences
were removed from the $S$-operator, but they reappeared in the
Hamiltonian $H^c_{FD}$ in the form of infinite counterterms. This
introduction of counterterms just shifted the problem of infinities
 from one place  to another.
Inconsistencies of the renormalization approach concerned many
prominent scientists, such as Dirac and Landau.  For example,
Rohrlich wrote

\begin{quote}
\emph{Thus, present quantum electrodynamics is one of the strangest
achievements of the human mind. No theory has been confirmed by
experiment to higher precision; and no theory has been plagued by
greater mathematical difficulties which have withstood repeated
attempts at their elimination. There can be no doubt that the
present agreement with experiments is not fortuitous. Nevertheless,
the renormalization procedure can only be regarded as a temporary
crutch which holds up the present framework. It should be noted
that, even if the renormalization constants were not infinite, the
theory would still be unsatisfactory, as long as the unphysical
concept of ``bare particle'' plays a dominant role. } F. Rohrlich
\cite{Rohrlich}
\end{quote}

In our interpretation of renormalization we do not distinguish bare and physical particles. We claim that $|0 \rangle$, $a^{\dag}_{\mathbf{p}} | 0 \rangle$ and $c^{\dag}_{\mathbf{p}} | 0 \rangle$ represent real physical 0-particle and 1-particle states with finite masses and charges. The only effect of renormalization is to add certain  counterterms to the original QED interaction $V_1$, as shown in (\ref{eq:11.61}). The counterterms are formally divergent, thus rendering the QED Hamiltonian unusable for most quantum-mechanical calculations.  In particular, time evolution studies become virtually impossible in this approach, as we will see in the next subsection.

\subsection{Time evolution in QED } \label{ss:unable}

 Let us
forget for a moment that interaction (counter)terms in $H^c$ are infinite and apply the time evolution operator $U^c(t \leftarrow 0) = \exp(
-\frac{i}{\hbar}H^ct)$ to the vacuum (no-particle) state. Expressing
$V^c$ in terms of creation and annihilation operators  of particles
 we obtain\footnote{Here we are concerned only with the
presence of $a^{\dag}b^{\dag}c^{\dag}$ and
$d^{\dag}f^{\dag}c^{\dag}$ interaction terms in (\ref{eq:11.32}).
All other terms are omitted. We also omit factors $i$, $\hbar$ and
coefficient functions which are not relevant in this context.}

\begin{eqnarray}
| 0 (t) \rangle &=& e^{ -\frac{i}{\hbar}H^ct} |0 \rangle
= (1 - \frac{it}{\hbar}(H_0 + V^c) + \ldots) |0 \rangle \nonumber \\
&\propto&  |0 \rangle  + ta^{\dag}b^{\dag}c^{\dag} |0 \rangle
+ td^{\dag}f^{\dag}c^{\dag} |0 \rangle + \ldots \nonumber \\
&\propto& |0 \rangle  +   t|abc \rangle +  t|dfc \rangle +\ldots
\label{eq:12.1}
\end{eqnarray}

\noindent  We see that various multiparticle states ($|abc \rangle,
|dfc \rangle,$ etc.) are
 created from the vacuum during time evolution.
 The physical vacuum in QED is not just an empty state
without particles. It is more like a boiling ``soup'' of bare particles,
antiparticles, and photons.

Similarly, the time
evolution of bare one-electron states is accompanied by appearing and disappearing virtual
particles.  Such behaviors have not
been seen in experiments. Obviously, if a theory cannot get right
the time evolutions of simplest zero-particle and one-particle
states, there is no hope of predicting the time evolution in more
complex multiparticle states.

The reason for these unphysical time evolutions  is the presence of
\emph{unphys} (e.g., $a^{\dag}b^{\dag}c^{\dag} + d^{\dag}f^{\dag}c^{\dag}$)
and \emph{renorm}  ($a^{\dag}a +b^{\dag}b +d^{\dag}d + f^{\dag}f$) terms in
the interaction operator $V^c$ of the renormalized QED.  How is it possible that such an unrealistic Hamiltonian
leads to exceptionally accurate experimental predictions?\footnote{Note that in the traditional renormalized QED  $S$-matrix elements are calculated on \emph{bare} particle states. This appears to be in contradiction with the absence of well-defined time evolution of such states and with the general understanding that bare particles are not physical.}

The important point is that \emph{unphys} and \emph{renorm}  interaction terms in
$H^c$  are absolutely harmless when the time evolution in the
infinite time range (from $-\infty$ to $ \infty$) is considered. As
we saw in equation (\ref{eq:8.65a}), such time evolution is represented
exactly by the product of the non-interacting time evolution
operator and the $S$-operator

\begin{eqnarray*}
U^c(\infty \gets -\infty) &=&   S^c U_0(\infty \gets -\infty)  =
U_0(\infty \gets -\infty) S^c
\end{eqnarray*}

\noindent  The factor $U_0$ in this product leaves invariant
no-particle and one-particle states. The factor $S^c$ has the same
property due to the cancelation of \emph{unphys} and \emph{renorm}  terms in
$F^c$, as discussed in subsection \ref{ss:mass-renorm}. So, in spite
of ill-defined operators $H^c$ and $\exp( -\frac{i}{\hbar}H^ct)$, the
renormalized QED is perfectly capable of describing scattering.

Luckily for QED, current experiments with elementary particles are
not designed to measure  time-dependent dynamics in the interaction
region. These experiments are,
basically, limited to measurements of scattering cross-sections as well as
energies and lifetimes of bound states, i.e., properties encoded in the
$S$-matrix. In collision experiments, interaction processes occur almost instantaneously, so their detailed time evolution is beyond reach. Measured scattering cross sections do reflect this time evolution, but only in an averaged, integrated form. In bound states (atoms, nuclei, etc.), the interaction is present all the time, but the time evolution is trivial: stationary wave function acquire simple time-dependent phase factors $\exp\left( - \frac{i}{\hbar}E_n t \right)$. Experimental measurements of such bound states are usually limited to their energies $E_n$, while accessing their wave functions is virtually impossible. As we know from subsection \ref{ss:s-matr-bound}, $E_n$ can be obtained as positions of poles of the $S$-matrix $S(E)$ on the complex energy plane. Thus, for description of most experiments it is sufficient to know the $S$-operator; the knowledge of the Hamiltonian is not required.\footnote{It is also important to note that even if we had complete information about the $S$-matrix from our experiments, it would not allow us to reconstruct the Hamiltonian in an unique way. According to subsection \ref{ss:scatt-equiv}, there is infinite number of (unitary equivalent) Hamiltonians producing the same given $S$-matrix.}

So, in the present experimental situation, the lack of a well-defined Hamiltonian and the inability of renormalized QFT to
describe time evolution
 can be tolerated.  But there is no doubt that time-dependent
processes in high energy physics will be eventually accessible to
more advanced experimental techniques. See, for example, recent measurements of the time-dependent dynamics of atomic wave functions with attosecond resolution \cite{Dahlstrom}. Moreover, the time evolution is
clearly observable in everyday ``macroscopic'' life. So, a consistent and comprehensive subatomic theory must describe time-dependent phenomena. The renormalized QED cannot do that, so the scope of this theory is limited.

\subsection{Unphys and renorm operators in QED } \label{ss:unable2}

In the preceding subsection we saw that the presence of \emph{unphys} and
\emph{renorm}  interaction operators in $V^c$  was responsible for unphysical
time evolution of bare particles. It is not difficult to see that the presence of such
questionable interaction terms is inevitable in any local quantum field theory where interaction
Hamiltonian is constructed as a polynomial of quantum fields \cite{Shirokov-Haag}. We saw
in (\ref{eq:10.10}) and (\ref{eq:10.26}) that quantum fields of both
massive and massless particles always have the form of a sum
(creation operator + annihilation operator)

\begin{eqnarray*}
\psi \propto \alpha^{\dag} + \alpha
\end{eqnarray*}

\noindent Therefore, if we constructed interaction as a product (or
polynomial) of fields,\footnote{as prescribed by general QFT rules from subsection \ref{ss:weinberg}.} we would necessarily have \emph{unphys} and \emph{renorm}
terms there. For example, converting a product
of four fields to the normally ordered form

\begin{eqnarray}
V &\propto& \psi^4 \nonumber\\
&=& (\alpha^{\dag} + \alpha)(\alpha^{\dag} + \alpha)(\alpha^{\dag} +
\alpha)(\alpha^{\dag} + \alpha) \nonumber \\
&=& \alpha^{\dag}\alpha^{\dag}\alpha^{\dag}\alpha^{\dag} +
\alpha^{\dag}\alpha^{\dag}\alpha^{\dag}\alpha +
 \alpha^{\dag}\alpha\alpha
\alpha + \alpha\alpha\alpha \alpha + \alpha^{\dag}\alpha^{\dag} + \alpha\alpha \label{eq:unph4}\\
&\ & +\alpha^{\dag}\alpha +
 C \label{eq:renorm4} \\
&\ & +\alpha^{\dag}\alpha^{\dag}\alpha\alpha \label{eq:phys4}
\end{eqnarray}

\noindent we obtain \emph{unphys} terms (\ref{eq:unph4}) together with
\emph{renorm}  terms (\ref{eq:renorm4}) and one \emph{phys} term (\ref{eq:phys4}).
The presence of \emph{unphys} terms is an indication that bare states created by operators $\alpha^{\dag}$ cannot be sensibly associated with true physical particles.

This analysis suggests that  \emph{any} quantum field theory is destined to
suffer from renormalization difficulties. Do we have any
alternative? We are going to show that with certain modifications, quantum field theories can be salvaged.
It \emph{is} possible to construct a satisfactory relativistic
quantum approach where the Hamiltonian is well-defined, renormalization problems are absent, and time evolution can be described in a natural way. This
approach will be discussed in the second part of this book.

\part{QUANTUM THEORY OF PARTICLES}

In the first part of this book we presented a fairly traditional
view on relativistic quantum field theory. This well-established
approach had great successes in many important areas of high energy
physics, in particular, in the description of scattering events.
However, it also had a few troubling spots. The first one is the problem of
ultraviolet divergences. The idea of self-interacting bare particles with
infinite masses and charges seems completely unphysical. Moreover,
QFT is not suitable for the description of time evolution of
particle observables and wave functions. In this second part
of the book, we suggest how to solve these problems by
abandoning the idea of quantum fields as basic ingredients of nature
and returning to the old (going back to Newton) concept of particles
interacting via direct forces. This reformulation of QFT will be achieved
by applying the ``dressed particle'' approach first developed by
Greenberg and Schweber \cite{GS}.

Instantaneous forces acting between dressed particles imply the real possibility of
sending superluminal signals. Then we find ourselves in
contradiction with special relativity, where faster-than-light
signaling is strictly forbidden (see Appendix
\ref{ss:super-signal}). This paradox forces us to take a second look on
derivations of basic results in special relativity, such as Lorentz
transformations for space and time coordinates of events. We find
that previous theories missed one important point. Specifically,
they ignored the fact that in interacting systems generators of
boost transformations are interaction-dependent. A proper
recognition of this fact will allow us to reconcile instantaneous action-at-a-distance with
the principle of causality in all reference frames and to build a consistent
relativistic theory of interacting quantum particles.

\chapter{DRESSED PARTICLE APPROACH}
\label{ch:rqd}

\begin{quote}
\textit{The first principle is that you must not fool yourself -- and
you are the easiest person to
                     fool.}

\small
\hspace{1in} Richard Feynman
\normalsize
\end{quote}

\vspace{0.5in}

In this chapter we will continue our discussion of quantum
electrodynamics - the theory of interacting charged particles
(electrons, protons, etc.) and photons.  We are going to demonstrate that the
formalism of  QED can be significantly improved by removing
ultraviolet-divergent terms from the Hamiltonian and abandoning the
ideas of non-observable virtual and bare particles. In particular, in sections \ref{sc:dressed} - \ref{ss:3dressed}
 we will find a finite ``dressed''  particle  Hamiltonian
$H^d$ which, in addition to  accurate scattering operators, also provides a good description of the time evolution and
bound states. We will call this approach the \emph{relativistic
quantum dynamics} \index{relativistic quantum dynamics} (RQD). The word ``dynamics'' is used here
because, unlike the traditional quantum field theory concerned with
calculations of time-independent $S$-matrices, RQD
 emphasizes the dynamical, i.e., time-dependent, nature of
interacting processes.

\section{Dressing transformation}
\label{sc:dressed}

In section \ref{sc:renormalization-in-QED} we established the presence of \emph{unphys} and \emph{renorm} terms (as well as their divergence) in the Hamiltonian $H^c = H_0 + V^c$ of the renormalized QED. The viewpoint adopted in this book is that such terms are not acceptable. Any realistic interaction operator must be finite and purely physical. Therefore, we conclude
that the Tomonaga-Schwinger-Feynman renormalization program was just
a first step in the process of elimination of infinities from
quantum field theory. In this section we are going to propose how to
make a second step in this direction: remove infinite contributions
from the Hamiltonian $H^c$ and solve the paradox of ultraviolet
divergences in QED.

In subsection \ref{ss:origins} we will see that there are no compelling  reasons\footnote{except, possibly, historical} for using traditional Hamiltonians of QED. So, we should not be afraid of trying other Hamiltonians $H^d = H_0 + V^d$ if we can show that they reproduce existing experimental data. Our starting idea is that QED interaction $V^c$ is not good, and we are going to use a completely different interaction operator $V^d$ for our version of quantum electrodynamics.

Our solution for $H^d$ will be based on the  \emph{dressed particle} \index{dressed
particle}\index{particle dressed} approach which has a long history.
Initial ideas about ``persistent interactions'' in QFT were
expressed by van Hove \cite{Van_Hove, Van_Hove2}. First clear
formulation of the dressed particle concept
 and its applications to model quantum field
theories are contained in a brilliant paper by Greenberg and
Schweber \cite{GS}. This formalism was further applied to various
quantum field models including the scalar-field model \cite{Walter},
the Lee model \cite{Lee, Lee1, Lee2, Lee3, Arefeva}, and the Ruijgrok-Van
Hove model \cite{Ruij, Lopuszanski}.
 The   way to construct  the dressed particle
Hamiltonian  as a perturbation series in a general QFT theory was
suggested by   Faddeev \cite{Faddeev} (see also \cite{Tani, Sato,
Fateev}). Shirokov with coworkers \cite{Shirokov-72, Vishinesku,
Shirokov, Shirokov2,
 Shirokov4}
  further developed these ideas and, in particular,
demonstrated how ultraviolet divergences can be removed from the
dressed Hamiltonian (see also \cite{Kobayashi, Kruger3, Shebeko}).

An interesting and somewhat related approach to particle
interactions in QFT was recently developed  by Weber and co-authors
\cite{Weber, Weber2, Weber3, Weber4}.

In this chapter we will present a general theory of the unitary dressing transformation. In chapter \ref{sc:coulomb} we are going to use this theory to construct the dressed Hamiltonian $H^d$ in the 2nd perturbation order. Unfortunately, in higher orders computations become very difficult. This challenge will be addressed in chapter \ref{ch:hydrogen-revisited} where we will introduce a trick allowing us to go beyond 2nd order results.

\subsection{On the origins of QED interaction} \label{ss:origins}

In subsection \ref{ss:interaction-qed} we simply postulated the QED interaction operator (\ref{eq:11.5}) - (\ref{eq:11.7}) or its renormalized form (\ref{eq:11.61}). What are the physical origins of these expressions? Are there deeper fundamental principles that demand this particular form of electromagnetic interactions? The standard textbook answer is  that the true reason for interactions between charged particles and photons is the principle of \emph{local gauge invariance}. \index{local gauge invariance} It is usually postulated that the Lagrangian of electromagnetic theory must be invariant with respect to certain simultaneous ``gauge'' transformations of the fermion fields $\psi(\mathbf{x},t)$, $\Psi(\mathbf{x},t)$, and the photon field $A^{\mu}( \mathbf{x},t)$. Then it appears that the free field Lagrangian does not satisfy this requirement and that the local gauge invariance can be ensured only after addition of ``minimal'' interaction terms there. This idea is explained in all modern textbooks on field theory, so we will not dwell on it here. It is sufficient to say that the same principle of local gauge invariance has been used to derive interaction Lagrangians for both electro-weak theory and quantum chromodynamics.

In spite of its wide theoretical use, the physical meaning of the gauge invariance  remains obscure. For example, the original idea of gauge freedom comes from Maxwell's electrodynamics. However, in chapter \ref{ch:theories} we will see that this theory can be replaced by a direct interaction approach in which electromagnetic fields, potentials, and gauges do not play any role at all. Moreover, the physical meaning of quantum fields themselves is not clear, as will be discussed in section \ref{ss:are-fields-meas}. For these reasons, in our book we do not accept the usual claim about the fundamental physical importance of fields and gauges. We maintain that the local gauge invariance should be considered only as an heuristic principle, whose remarkable effectiveness still awaits its proper explanation.

Perhaps, a more convincing justification of interactions (\ref{eq:11.5}) - (\ref{eq:11.7}) was explained by Weinberg in section 8.1 of his book \cite{book}. He assumed that interaction must be a polynomial function of field components. Then he advanced an experiment-based argument that this polynomial must be linear in $A^{\mu}( \mathbf{x},t)$. Another requirement is the invariance of the interaction polynomial with respect to the non-interacting representation of the Poincar\'e group.\footnote{see condition (II) in step \textbf{2.} on page \pageref{eq:as-a-scalar}} If $A^{\mu}$ transformed as a 4-vector with respect to the Lorentz group, then the latter condition could be satisfied if one chooses interaction in the form $ \propto A^{\mu}J_{\mu}$, where $J_{\mu}$ is any 4-vector composed of fermion fields. However, Lorentz transformations of $A^{\mu}$ are different from the 4-vector law by the presence of an additional term.\footnote{see equation (\ref{eq:10.35a})} This difficulty can be overcome if for $J_{\mu}$ one chooses a \emph{conserved} fermionic 4-vector. Then $A^{\mu}J_{\mu}$ is a Lorentz scalar despite the non-4-vector character of $A^{\mu}$. The simplest choice for $J_{\mu}$ is the fermion current density (\ref{eq:11.1}). This line of arguments leads one to the electromagnetic interaction operator (\ref{eq:11.5}) - (\ref{eq:11.7}) that is consistent with usual gauge-based derivations.

However, neither gauge-based nor Weinberg's arguments appear very convincing, especially if one takes into account the need for adding divergent renormalization counterterms to the resulting interaction Hamiltonian. The apparent success of the renormalization program seems mysterious and accidental. It is puzzling how infinite counterterms (almost) cancel divergences in the $S$-matrix expansion and how the tiny residual radiative corrections come out in perfect agreement with measurements.\footnote{see chapter \ref{ss:hydrogen}}

So, we just have to accept that the present formulation of QED lacks solid theoretical foundation. Quantum fields and gauges seem to be heuristic devices, and the whole construction is supported by agreement with experiments more than by reliance on well-tested physical principles. Bearing this weakness in mind, in this second part of our book we will attempt to reformulate the QED formalism. Instead of fields and gauges, our approach will be based on the ideas of point particles and instantaneous interaction potentials.

\subsection{No-self-interaction condition}
\label{ss:dressed-ham}

In section \ref{sc:troubles} we saw that the presence of unphys and renorm interactions is the immediate cause of many QFT problems, such as the need for renormalization and the absence of a well-defined time evolution.
The simplest way to avoid these problems is to demand that
the true (dressed) interaction Hamiltonian $V^d$ does not contain \emph{unphys} and \emph{renorm}
terms. So, we are now going to postulate that our desired interaction operator $V^d$ has \emph{phys} terms\footnote{
Recall that decay and oscillation operators are not present in QED. So, we will not consider them here.}

\begin{eqnarray}
V^d &=& \alpha^{\dag}\alpha^{\dag}\alpha\alpha +
\alpha^{\dag}\alpha^{\dag}\alpha\alpha\alpha +
\alpha^{\dag}\alpha^{\dag}\alpha^{\dag}\alpha\alpha + \ldots
\label{eq:physV}
\end{eqnarray}

\noindent According to Table \ref{table:7.2} in subsection
\ref{ss:products}, commutators of \emph{phys} terms can be only \emph{phys}.
Therefore, when the scattering operator $F^d$ is calculated from $V^d$ via equation (\ref{eq:7.63b}) only
\emph{phys} terms can appear there in each perturbation order.  Then both $V^d$ and $F^d$ yield zero when acting on
zero-particle and one-particle states

\begin{eqnarray*}
V^d |0 \rangle &=& V^d \alpha^{\dag}|0 \rangle = 0 \\
F^d |0 \rangle &=& F^d \alpha^{\dag}|0 \rangle = 0
\end{eqnarray*}

\noindent  as required by the no-self-scattering renormalization
condition \ref{statementU}.
 Moreover, time evolutions of the vacuum and
one-particle states are not different from their free time evolutions

\begin{eqnarray*}
| 0(t) \rangle &=& e^{ -\frac{i}{\hbar}H^dt} |0 \rangle
=  \left(1 - \frac{it}{\hbar}(H_0 + (V^d)^{ph}) + \ldots \right) |0 \rangle \nonumber \\
&=&  \left(1 - \frac{it}{\hbar}H_0  + \ldots \right) |0 \rangle
= e^{ -\frac{i}{\hbar}H_0t} |0 \rangle \\
| \alpha(t) \rangle &=& e^{ -\frac{i}{\hbar}H^dt} \alpha^{\dag}|0 \rangle
= \left(1 - \frac{it}{\hbar}(H_0 + (V^d)^{ph}) + \ldots \right) \alpha^{\dag}|0\rangle \nonumber \\
&=&  \left(1 - \frac{it}{\hbar}H_0  + \ldots \right)
\alpha^{\dag}|0\rangle = e^{ -\frac{i}{\hbar}H_0t} |\alpha
\rangle
\end{eqnarray*}

\noindent as they should be. Physically this means that, in addition to
forbidding self-scattering in zero-particle and one-particle states,\footnote{
 i.e., satisfying Statement \ref{statementU}} our \emph{ansatz} (\ref{eq:physV}) also forbids any \emph{self-interaction} in
these states. So, our search for a better QED interaction will be based on the following

\begin{postulate}
[stability of vacuum and one-particle states]  There is no
(self-)interaction in the vacuum and one-particle states, i.e., the
time evolution of these states is not affected by interaction and
 is governed by the non-interacting Hamiltonian $H_0$.
Mathematically, this means that the type of the interaction Hamiltonian $V^d$ is
\emph{phys}. \label{postulateW}
\end{postulate}

\noindent Summarizing discussions from various parts of this book,
we can put together a list of conditions that should be satisfied by
any realistic interaction

\begin{itemize}
\item [(A)] Poincar\'e invariance (Statement \ref{statementO});
\item [(B)] instant form of dynamics (Postulate \ref{postulateR});
\item [(C)] cluster separability (Postulate \ref{postulateS});
\item [(D)] no self-interactions = \emph{phys} character of $V^d$ (Postulate \ref{postulateW});
\item [(E)] finiteness of coefficient functions of interaction potentials;
\item [(F)] coefficient functions should rapidly tend to zero at large values of momenta \footnote{According
to Theorem \ref{Theorem9.10}, this condition guarantees convergence of all
loop integrals involving vertices $V^d$ and, therefore, the
finiteness of the corresponding scattering operator $S^d$. }
\end{itemize}

As we saw in subsection \ref{ss:unable2}, requirement (D)
practically excludes all usual field-theoretical Hamiltonians. The
question is whether there are non-trivial ``good'' interactions that have all
the properties (A) - (F)? And the answer is ``yes.''

One set of examples of allowed interacting theories is provided by
``direct interaction'' models.\footnote{Some of them were discussed
in section \ref{sc:instant-form}.} Two-particle models of this kind
were first constructed by Bakamjian and Thomas
\cite{Bakamjian_Thomas}. Sokolov \cite{Sokolov_Shatnii,
Sokolov_Shatnii2}, Coester and Polyzou \cite{Coester_Polyzou} showed
how this approach can be extended to cover multi-particle systems.
There are recent attempts \cite{Polyzou-production} to extend this
formalism to include description of systems with variable number of
particles. In spite of these achievements, the ``direct
interaction'' approach is currently applicable only to model
systems. One of the reasons is that conditions for satisfying the
cluster separability are very cumbersome. This mathematical
complexity was evident even in the simplest 3-particle
case discussed in subsection \ref{ss:3-particle}.

In the ``direct interaction'' approach, interactions are expressed
as functions of (relative) particle observables, e.g., relative distances and
momenta. However, it appears more
convenient to write interactions as polynomials in particle creation
and annihilation operators  (\ref{eq:9.49}). We saw in Statement
\ref{statementT} that in this case the cluster separability
condition (C) is trivially satisfied if coefficient functions
have smooth dependence on particle momenta. The no-self-interaction
condition (D) simply means that all interaction terms are \emph{phys}. The instant
form condition (B) means that generators of space translations
$\mathbf{P}= \mathbf{P}_0$ and rotations $\mathbf{J}= \mathbf{J}_0$
are interaction-free and that interaction $V$ commutes with
$\mathbf{P}_0$ and $\mathbf{J}_0$. The most difficult part is to
ensure the relativistic invariance (condition (A)), i.e.,
commutation relations of the Poincar\'e group. One way to solve this
problem is to fix the operator structure of interaction terms and
then try to find the momentum dependence of coefficient functions by
solving a set of differential equations resulting from Poincar\'e
commutators (\ref{eq:8.17}) - (\ref{eq:8.21}) \cite{Kazes, Kita, Kita1, Kita2, Kita3,
Kita4, Kita5, Shebeko11}. Kita and Kazes demonstrated that there is an infinite number of
solutions for these equations and provided some non-trivial examples. This means that the above conditions (A) - (F) are not restrictive enough. Ideally, we would like to formulate additional physical principles that would single out
the unique theory of interacting particles that agrees with
all experimental observations. Unfortunately, these additional
principles are not known at this moment.

\subsection{Main idea of the dressed particle approach}
\label{ss:dressed-ham2}

The Kita-Kazes approach  is difficult to apply to realistic particle
interactions, so, currently, it cannot compete with QFT. Thus, it might be
more promising to abandon the idea to build relativistic
interactions from scratch and, instead, try to modify traditional field theories to make them
consistent with our requirements (D), (E), and (F). One idea how to make this possible is to note that the $S$-matrix of the usual
renormalized QED agrees with experiments very well. So, we may add
the following requirement to the above list (A) - (F):

\begin{itemize}
\item [(G)] the scattering operator $S^d$ in our ``dressed'' theory is exactly the same (in each perturbation order) as
the operator $S^c$ in renormalized QED.
\end{itemize}

Previously we have denoted the desired \emph{phys} interaction operator by $V^d$.
Then, condition (G) means that the dressed Hamiltonian $H^d = H_0 + V^d$ is
scattering equivalent to the renormalized QED Hamiltonian $H^c = H_0 + V^c$ from which the accurate scattering operators $S^c$ is usually calculated.
According to our discussion in subsection \ref{ss:scatt-equiv}, this means that
 $H^d$ and $H^c$ are related by  a unitary
transformation

\begin{eqnarray}
H^d &=& H_0 + V^d = e^{i \Phi} H^c e^{-i \Phi} \label{eq:12.14} \\
    &=&  e^{i\Phi} (H_0 + V^c) e^{-i\Phi} \nonumber \\
    &=& (H_0 + V^c) + i[\Phi,(H_0 + V^c)] - \frac{1}{2!}[\Phi,[\Phi,(H_0
+ V^c)]] + \ldots \label{eq:un_tr_hd}
\end{eqnarray}

\noindent where Hermitian operator $\Phi$  satisfies condition
(\ref{eq:8.75}). Transformation $e^{i \Phi}$ will be called the
\emph{unitary dressing transformation}. \index{dressing transformation}

\subsection{Unitary dressing transformation}
\label{ss:scatt-ham}

Now our goal is to find a unitary transformation $e^{i \Phi}$, which
ensures that the dressed particle Hamiltonian $H^d$ satisfies all
properties (A) - (G). In this study we will need the following
useful results

\begin{theorem} [transformations preserving the $S$-operator]
A unitary transformation of the Hamiltonian

\begin{eqnarray*}
H' &=& e^{i\Phi} H e^{-i\Phi}
\end{eqnarray*}

\noindent preserves the $S$-operator if the Hermitian operator
$\Phi$ has the form (\ref{eq:9.48}) - (\ref{eq:9.49}) where all
terms $\Phi_{NM}$ are either \emph{phys} or \emph{unphys} and have smooth coefficient functions.
\label{theorem:non-singular}
\end{theorem}
\begin{proof} [Idea of the proof] Assume that operator $\Phi$ has the standard
form (\ref{eq:9.48}) - (\ref{eq:9.49})

\begin{eqnarray*}
 \Phi &=& \sum_{N=0}^{\infty} \sum_{M=0}^{\infty} \Phi_{NM} \\
 \Phi_{NM}     &=&
 \sum_{\{\eta, \eta' \}}\int   d\mathbf{q}'_1 \ldots d\mathbf{q}'_N
d\mathbf{q}_1 \ldots d\mathbf{q}_M
   D_{NM}(\mathbf{q}'_1 \eta'_1;  \ldots ;\mathbf{q}'_N \eta'_N;
\mathbf{q}_1 \eta_1; \ldots ;\mathbf{q}_M \eta_M)
  \times \nonumber \\
    &\mbox{ }&  \delta
\left(\sum_{i=1}^{N} \mathbf{q'}_i - \sum_{j=1}^{M} \mathbf{q}_j
\right) \alpha^{\dag}_{\mathbf{q}'_1, \eta'_1} \ldots
\alpha^{\dag}_{\mathbf{q}'_N, \eta'_N} \alpha_{ \mathbf{q}_1,
\eta_1} \ldots \alpha_{\mathbf{q}_M, \eta_M}
\end{eqnarray*}

\noindent Then the left hand side of the scattering equivalence
condition (\ref{eq:8.75}) for each term $\Phi_{NM}$ is

\begin{eqnarray}
&\mbox{ } &\lim _{t \to  \pm \infty}  e^{\frac{i}{\hbar}H_0 t} \Phi_{NM}  e^{-\frac{i}{\hbar}H_0t} \nonumber \\
   &=& \lim _{t \to  \pm \infty}
 \sum_{\{\eta, \eta' \}}\int   d\mathbf{q}'_1 \ldots d\mathbf{q}'_N
d\mathbf{q}_1 \ldots d\mathbf{q}_M
   D_{NM}(\mathbf{q}'_1 \eta'_1;  \ldots ;\mathbf{q}'_N \eta'_N;
\mathbf{q}_1 \eta_1; \ldots ;\mathbf{q}_M \eta_M)
  \times \nonumber \\
    &\mbox{ }&  \delta
\left(\sum_{i=1}^{N} \mathbf{q'}_i - \sum_{j=1}^{M} \mathbf{q}_j \right)
e^{\frac{i}{\hbar}E_{NM} t} \alpha^{\dag}_{\mathbf{q}'_1, \eta'_1}
\ldots \alpha^{\dag}_{\mathbf{q}'_N, \eta'_N} \alpha_{ \mathbf{q}_1,
\eta_1} \ldots \alpha_{\mathbf{q}_M, \eta_M} \label{eq:limphi}
\end{eqnarray}

\noindent where $E_{NM}$ is the energy function of this term. In the
limits $t \to \pm \infty$  momentum integrals tend to zero by
Riemann-Lebesgue lemma \ref{lemma:Rim-Leb}, because the coefficient
function $D_{NM}$ is smooth, while the factor
$e^{-\frac{i}{\hbar}E_{NM} t}$ oscillates rapidly in the momentum
space. Therefore, according to (\ref{eq:6.101a}), Hamiltonians $H$
and $H'$ are scattering-equivalent. This theorem does not apply to \emph{renorm} operators $\Phi_{NM}$, because for them the energy functions $E_{NM}$ are identically zero, products $e^{\frac{i}{\hbar}H_0 t} \Phi_{NM}  e^{-\frac{i}{\hbar}H_0t}$ do not depend on $t$, and the scattering equivalence condition (\ref{eq:8.75}) is violated.
\end{proof}

\bigskip

\begin{lemma}
Potential $\underline{B}$\footnote{For definition of an underlined
operator symbol see (\ref{eq:underline}) and (\ref{eq:underline2}).} is smooth\footnote{i.e., it has a
smooth coefficient function} if $B$ is either \emph{unphys} with arbitrary
smooth coefficient function or \emph{phys} with a smooth coefficient
function, which is identically zero on the energy shell.
\label{lemma:non-singular}
\end{lemma}
\begin{proof} The only possible source of singularity in
$\underline{B}$ is the energy denominator $E_B^{-1}$, which is singular on the energy shell. However, for
operators satisfying conditions of this Lemma, either the energy
shell does not exist, or the coefficient function vanishes there.
In both cases $\underline{B}$ is not singular on the energy shell.
\end{proof}

\bigskip

 We will assume that all relevant
operators can be written as expansions in powers of the coupling
constant, and that all series converge

\begin{eqnarray}
  H^c &=& H_0 + V_1^c + V_2^c + \ldots
\label{eq:hc}\\
  H^d &=& H_0 + V_1^d + V_2^d + \ldots
\label{eq:hd} \\
 \Phi &=& \Phi_1 + \Phi_2 + \ldots
\label{eq:phi}
\end{eqnarray}

\noindent As usual, the subscript denotes the power of $e$ (= the perturbation order).

Next, following the plan outlined in  subsection
\ref{ss:regularization}, we introduce regularization cutoffs $\lambda$ and $\Lambda$, which ensure that interactions and counterterms $V_i^c$
in the Hamiltonian of QED are non-singular and finite in all perturbation orders. Moreover, with these cutoffs all loop integrals
involved in calculations of products and commutators of
$V_i^c$ become convergent. In this section we are going to prove that in this
regulated theory the operator $\Phi$
 can be chosen so that  conditions (A) - (G) are satisfied in all
perturbation orders. Of course, to get accurate results, in the end of
calculations the regularization cutoffs $\lambda$ and $\Lambda$ should be lifted. Only
those quantities may have physical meaning, which remain finite in
the limits $\lambda \to 0$ and $\Lambda \to \infty$. The latter limit will be considered in subsection \ref{ss:lim-inf-cutoff}. For the infrared limit $\lambda \to 0$ see section \ref{ss:hydrogen}.

Using expansions (\ref{eq:hc}) - (\ref{eq:phi}) in (\ref{eq:un_tr_hd}) and collecting together terms of equal order
 we obtain an infinite set of  equations

\begin{eqnarray}
  V_1^d  &=& V_1^c  + i[\Phi_1 , H_0] \label{eq:1st-order-un} \\
  V_2^d  &=& V_2^c  + i[\Phi_2 , H_0] + i[\Phi_1 , V_1^c ] -
\frac{1}{2!}
[\Phi_1 , [\Phi_1 , H_0]] \label{eq:2nd-order-un}\\
  V_3^d  &=& V_3^c  + i[\Phi_3 , H_0] + i[\Phi_2 , V_1^c ] +
i[\Phi_1 ,
V_2^c ] \nonumber \\
&\ & -\frac{1}{2!} [\Phi_2 , [\Phi_1 , H_0]] - \frac{1}{2!}
[\Phi_1 , [\Phi_2 , H_0]] - \frac{1}{2!}
[\Phi_1 , [\Phi_1 , V_1^c ]] \nonumber \\
&\ & -\frac{i}{3!} [\Phi_1 ,[\Phi_1 , [\Phi_1 ,
H_0]]]  \ldots \label{eq:3rd-order-un} \\
&\ldots& \nonumber
\end{eqnarray}

\noindent Now we need to solve these equations order-by-order. This
means that we need to choose appropriate operators $\Phi_i = \Phi_i^{ph} + \Phi_i^{unp} + \Phi_i^{ren} $, so
that interaction terms $V_i^d$ on  left hand sides satisfy above
conditions (B) - (G).\footnote{We will discuss condition (A)
separately in subsection \ref{ss:relat-invar-dressed}.} Let us start
with equation (\ref{eq:1st-order-un}).

\subsection{Dressing in the first perturbation order}
\label{ss:uni-1st-order}

In renormalized QED the 1st order interaction operator  $V_1^c  =
V_1 $ is \emph{unphys}.\footnote{see equation (\ref{eq:11.32})} According to condition (D), this term should be canceled exactly. This can be achieved if we choose\footnote{More generally, we can also choose $\Phi_1^{ph}$ to be any \emph{phys} operator whose coefficient function vanishes on the energy shell. See next subsection.} $\Phi_1^{ph}   =
\Phi_1^{ren} =0$ and use (\ref{eq:soluti}) to solve the commutator equation

\begin{eqnarray}
 i [\Phi_1^{unp}, H_0]  &=&- V_1 \nonumber \\
 \Phi_1^{unp}  &=&  i \underline{V_1 } \label{eq:v1ev1}
\end{eqnarray}

\noindent for the unphys part of $\Phi_1$. This choice ensures that not only the \emph{unphys} part of $V_1^d $ is zero, but the entire first order dressed interaction potential vanishes
  $ V_1^d  = 0 $, so that conditions (B) - (F)
 are trivially satisfied in this order.
The coefficient function of $V_1 $ is non-singular.  By Lemma \ref{lemma:non-singular}
this implies that $\Phi_1^{unp} $ in equation (\ref{eq:v1ev1}) is
smooth. By Theorem \ref{theorem:non-singular}, the presence of this
term in the dressing transformation $e^{i \Phi}$ does not affect the
$S$-operator in agreement with our condition (G). So, we managed to satisfy all necessary conditions in the first perturbation order.

\subsection{Dressing in the second perturbation order}
\label{ss:uni-2nd-order}

Now we can substitute the operator $\Phi_1 $ found above into equation
(\ref{eq:2nd-order-un}) and obtain expression for the 2nd order
dressed potential

\begin{eqnarray}
  V_2^d  &=& V_2^c  +i [\Phi_2, H_0]
- [\underline{V_1 }, V_1 ] + \frac{1}{2!}
[\underline{V_1 }, V_1 ] \nonumber \\
&=& V_2^c  +i [\Phi_2, H_0] - \frac{1}{2 }
[\underline{V_1 }, V_1 ] \label{eq:2nd-order-un2}
\end{eqnarray}

\noindent It is convenient to write separately \emph{unphys}, \emph{phys}, and
\emph{renorm}  parts of this equation and take into account that
$[\Phi_2^{ren}, H_0] = 0$

\begin{eqnarray}
  (V_2^d)^{unp}
&=& (V_2^c)^{unp}  +i [\Phi_2^{unp}, H_0] -
\frac{1}{2 } [\underline{V_1 }, V_1 ]^{unp}
\label{eq:2nd-order-un2_unp} \\
(V_2^d)^{ph}  &=& (V_2^c)^{ph}  +i [\Phi_2^{ph}, H_0]
 - \frac{1}{2 } [\underline{V_1 }, V_1 ]^{ph}
\label{eq:2nd-order-un2_ph} \\
(V_2^d)^{ren} &=& (V_2^c)^{ren} - \frac{1}{2 } [\underline{V_1 },
V_1 ]^{ren} \label{eq:2nd-order-un2_ren}
\end{eqnarray}

\noindent All components on the right hand sides of (\ref{eq:2nd-order-un2_unp}) - (\ref{eq:2nd-order-un2_ren}) (except commutators of $\Phi_2^{unp}$ and $\Phi_2^{ph}$) are, basically, known to us already: In QED, $(V_2^c)^{ph}$ is the same as  $V_2^{ph}$ in (\ref{eq:v2-final-ph}); operator $(V_2^c)^{ren}$ is coming from electron and photon self-energy counterterms discussed in subsections \ref{ss:elec-mass} and \ref{ss:photon-se}, respectively; $(V_2^c)^{unp}$ takes contributions from $V_2^{unp}$ in (\ref{eq:v2-final-unp}) as well as from self-energy counterterms; and calculation of commutators $[\underline{V_1 }, V_1 ]^{unp}$ and $[\underline{V_1 }, V_1 ]^{ren}$ should follow the same steps as calculation of  $[\underline{V_1 }, V_1 ]^{unp}$ in subsection \ref{ss:2-nd-order}. Now our goal is to choose operators $\Phi_2^{ph}$ and $\Phi_2^{unp}$, so that dressed interactions on the left hand sides of (\ref{eq:2nd-order-un2_unp}) - (\ref{eq:2nd-order-un2_ren}) satisfy our conditions (B) - (G).

From the condition (D) it follows that $(V_2^d)^{unp} $
must vanish.To achieve that, we can choose

\begin{eqnarray}
\Phi_2^{unp}  &=& i \underline{V_2^{unp} } -\frac{i}{2}
\underline{[\underline{V_1 }, V_1 ]^{unp} } \label{eq:phi2unp}
\end{eqnarray}

\noindent Operators $V_1 $ and $V_2^{unp}$ are smooth.
Then, by Lemma \ref{lemma:non-singular}, the operator
$\underline{V_1 }$ is also smooth and by Lemma
\ref{theorem:linked-comm} the commutator $[\underline{V_1 },
V_1 ]^{unp}$ is smooth as well. Using Lemma
\ref{lemma:non-singular} again, we see that operator
$\Phi_2^{unp} $ is smooth, and by Theorem
\ref{theorem:non-singular} its presence in the transformation $e^{i
\Phi}$ does not affect the $S$-operator. This is exactly what we need.

Let us now turn to equation (\ref{eq:2nd-order-un2_ph}) for the \emph{phys} part
of the dressed particle interaction $V_2^d$. What are the conditions for selecting $\Phi_2^{ph}$?
For example, we cannot simply choose $\Phi_2^{ph}  = 0$, because
in this case the dressed particle interaction would acquire the form

\begin{eqnarray*}
(V_2^d)^{ph}  &=& V_2^{ph}   - \frac{1}{2 }
[\underline{V_1 }, V_1 ]^{ph}
\end{eqnarray*}

\noindent and there is absolutely no guarantee that the coefficient
function of $(V_2^d)^{ph} $ rapidly tends to zero at large values
of particle momenta (condition (F)). In order to have this
guarantee, we are going to choose $\Phi_2^{ph}  $ such that
the right hand side of (\ref{eq:2nd-order-un2_ph}) rapidly tends to zero when momenta are
far from the energy shell. In addition, we will require that
$\Phi_2^{ph}  $ is non-singular.\footnote{This is needed to obey the charge renormalization postulate from subsection \ref{ss:crc}.} Both conditions can be satisfied
by choosing\footnote{ Note that this part of our
dressing transformation closely resembles the ``similarity
renormalization'' procedure suggested by G\l azek and Wilson
\cite{Glazek, Glazek2}. }

\begin{eqnarray}
\Phi_2^{ph}  &=& \underline{\left(i V_2^{ph}  - \frac{i}{2}
[\underline{V_1 }, V_1 ]^{ph} \right) \circ (1 - \zeta_2)}
\label{eq:phi2ph}
\end{eqnarray}

\noindent where $\zeta_2$ is a real function,\footnote{The arguments of $\zeta_2$ (particle momenta and spin projections) should be the same as
arguments of coefficient functions in $V_2^{ph}$ and
$[\underline{V_1}, V_1]^{ph}$. The ``small circle'' notation was defined in subsection \ref{ss:inter-oper}} such that

\begin{itemize}
\item[(I)] $\zeta_2$ is equal to 1 on the energy shell;
\label{cond:I}
\item[(II)] $\zeta_2$ depends on rotationally invariant
combinations of momenta (to make sure that $V_2^d$ commutes with
$\mathbf{P}_0$ and $\mathbf{J}_0$);
\item[(III)] $\zeta_2$ is smooth;
\item[(IV)] $\zeta_2$ rapidly tends to zero
when the arguments move away from the energy shell.\footnote{For
example, we can choose $\zeta_2 = e^{-\alpha E^2}$ where $\alpha$ is
a positive constant and $E$ is the energy function of the
 operator on the right hand side of (\ref{eq:phi2ph}). Actually, it may happen that loop integrals involving $V_2^d$ converge even without involvement of convergency factors $\zeta_2$. For example, in subsection \ref{ss:comm-term} we will see that in QED the loop integral in the product $V_2^d\underline{V_2^d}$ converges even if $\zeta_2=1$ everywhere.}
\end{itemize}

\noindent With the choice (\ref{eq:phi2ph}) we obtain

\begin{eqnarray}
(V_2^d)^{ph}  &=& \left(V_2^{ph}  - \frac{1}{2}
[\underline{V_1 }, V_1 ]^{ph} \right) \circ \zeta_2
\label{eq:v2dphys}
\end{eqnarray}

\noindent so that $(V_2^d)^{ph} $ rapidly tends to zero when
momenta of particles move away from the energy shell in agreement
with condition (F). Moreover, property (I) guarantees that expression under the $t$-integral in (\ref{eq:phi2ph}) vanishes on the energy shell. Therefore, this $t$-integral\footnote{calculated by formula (\ref{eq:underline2})} is non-singular and, according to Theorem \ref{theorem:non-singular}, $\Phi_2^{ph}$ does not modify the $S$-operator, i.e., condition (G) is satisfied.

In (\ref{eq:v2dphys}) $V_1 $ and $\underline{V_1 }$ are smooth operators, so, according to Theorem \ref{theorem:linked-comm}, their commutator is also smooth. Operator $V_2^{ph}$ and function $\zeta_2$ are smooth as well. So, due to Statement \ref{statementT}, we conclude that the second-order dressed interaction $(V_2^d)^{ph}$ is separable in accordance with our requirement (C).

Finally, we need to choose $\Phi_2^{ren}$. Note that this operator is not present in the system of equations (\ref{eq:2nd-order-un2_unp}) - (\ref{eq:2nd-order-un2_ren}), so it should be derived from other considerations. Apparently, the only sensible choice is

\begin{eqnarray}
\Phi_2^{ren} &=& 0 \label{eq:phi2ren}
\end{eqnarray}

\noindent because, according to Theorem \ref{theorem:non-singular}, a non-zero renorm part of $\Phi_2$ would destroy the scattering equivalence (condition (G)). we can satisfy all our conditions. Let us analyze the only non-obvious condition (D). How can we be sure that our choice (\ref{eq:phi2ren}) satisfies the condition (D), in particular, that $(V_2^d)^{ren} = 0$? We already know that (\ref{eq:phi2ren}) guarantees the scattering equivalence of the dressed theory. This means that  the $S$-operator obtained with the
transformed interaction $V_2^d $ agrees with the $S$-operator
$S^c$ up to the second perturbation order (condition (G)). In particular, $F_2^d = F_2^c $. This
would be impossible if $V_2^d $ contained a non-zero \emph{renorm}
term. Indeed, $(V_2^d)^{ren} \neq 0$  would imply that operator $F_2^d $ and, therefore,
$F_2^c $  have non-zero \emph{renorm}  terms in disagreement with equation
(\ref{eq:11.42}). Thus we must conclude that $(V_2^d)^{ren} =
0$, and that two terms on the right hand side of
(\ref{eq:2nd-order-un2_ren}) cancel each other. This cancelation can be verified by a direct calculation as well \cite{Kruger3}.

\subsection{Dressing in arbitrary order}
\label{ss:uni-arb-order}

For any higher perturbation order $i > 2$, the selection of $\Phi_i $ and
proofs of the conditions (B) - (G) are similar to those described above for the 2nd
order.
 The defining equation for $V_i^d $
 can be written in
a general form\footnote{compare with (\ref{eq:2nd-order-un2}) }

\begin{eqnarray}
  V_i^d  &=& V_i^c  +i [\Phi_i, H_0] + \Xi_i
\label{eq:vid_in_arb_order}
\end{eqnarray}

\noindent where $\Xi_i $ is a sum of multiple commutators
involving $V_j^c $ from lower orders $(1 \leq j < i)$ and their
$t$-integrals (= ``underlines''). This equation is solved by

\begin{eqnarray}
\Phi_i^{ren} &=& 0 \nonumber \\
\Phi_i^{unp}  &=& i \underline{\Xi_i^{unp} } +i
\underline{(V_i^c)^{unp} }, \nonumber \\
\Phi_i^{ph}  &=& i \underline{(\Xi_i^{ph}  + (V_i^c)^{ph} ) \circ
(1-\zeta_i)} \label{eq:phiiph}
\end{eqnarray}

\noindent where functions $\zeta_i$ have properties (I) - (IV) from the preceding subsection.
Similar to the 2nd order discussed above, one then demonstrates that
 $\Phi_i $ is smooth, so
that condition (G) is satisfied in the $i$-th order and that $(V_i^d)^{ren} = (V_i^d)^{unp} = 0$.

Solving  equations (\ref{eq:vid_in_arb_order}) order-by-order we
obtain the  dressed particle Hamiltonian

\begin{eqnarray}
H^d  &=& e^{i \Phi } H^c  e^{-i \Phi } = H_0 +  V^d_2  + V^d_3  + V^d_4  + \ldots
\label{eq:12.3}
\end{eqnarray}

\noindent which has all required properties (B) - (G), as promised.

\subsection{Infinite momentum cutoff limit}
\label{ss:lim-inf-cutoff}

So far our calculations of operators $\Phi$ and $V^d$ were performed under the assumption of a finite momentum cutoff $\Lambda c$. This permitted us to avoid ultraviolet divergences in our formulas. In the complete and final theory we must, obviously, take the limit $\Lambda \to \infty$. Our approach can be viable only if we can prove that all physically relevant dressed operators remain finite in this limit.\footnote{Note that operator $\Phi$ providing the link (\ref{eq:12.3}) between Hamiltonians $H^c$ and $H^d$ does not correspond to any observable property, so it is OK if $\Phi$ does not converge in the large cutoff limit.}

 It seems rather obvious that conditions (B)  - (D) and (F) are independent on the momentum cutoff $\Lambda c$.
Therefore, they also remain valid in the limit $\Lambda \to \infty$.
Let us now demonstrate that condition (E) is satisfied in this
limit as well. To do that,  we note than on one hand the traditional QED gives us a perturbation series for the $S$-operator

\begin{eqnarray}
  S^c &=& 1+   S^c_2  + S^c_3   + S^c_4  + \ldots \nonumber \\
  &=& 1 + \underbrace{\Sigma_2^c} + \underbrace{\Sigma_3^c} + \underbrace{\Sigma_4^c} + \ldots \label{eq:ScSigma}
\end{eqnarray}

\noindent On the other hand, in the dressed particle approach with Hamiltonian (\ref{eq:12.3}), the $S$-operator can be written using formulas
(\ref{eq:7.63a}) and (\ref{eq:7.63c})

\begin{eqnarray*}
  S^d
&=& 1+ \underbrace {V^d_2 }
    + \underbrace {V^d_3 } +
\underbrace {V^d_4 } + \underbrace {V^d_2 \underline{V^d_2 }}  \ldots  \nonumber
\end{eqnarray*}

\noindent According to our condition (G), these two operators should be equal order-by-order. Thus we obtain the
following set of relations between $V_i^d $ and $S_i^c $ on the
energy shell\footnote{Recall that the $S$-operator is defined only on the energy shell. Moreover the ``underbrace'' symbol was defined in (\ref{eq:9.52}) as $\underbrace{V} = 2 \pi i V \circ \delta(E_V)$. So, $\underbrace{V}$ is non-zero only on the energy shell of the operator $V$.}

\begin{eqnarray}
\underbrace{V_2^d } &=& S_2^c = \underbrace{\Sigma_2^c }
\label{eq:12.4} \\
\underbrace{V_3^d } &=& S_3^c = \underbrace{\Sigma_3^c }
\label{eq:12.5} \\
\underbrace{V_4^d } &=& S_4^c  -
\underbrace {V^d_2 \underline{V^d_2 }} = \underbrace{\Sigma_4^c } -
\underbrace {V^d_2 \underline{V^d_2 }}
\label{eq:12.6}\\
\underbrace{V_i^d } &=& S_i^c  +
\underbrace{Y_i }, \mbox{        } i > 4 \label{eq:12.7}
\end{eqnarray}

\noindent where  $Y_i $  stands for a  sum of certain products of
$V_j^d $ from lower orders ($2 \leq j \leq i-2$) with
$t$-integrations (``underlines''). The relations (\ref{eq:12.4}) - (\ref{eq:12.7}) are independent on the cutoff
$\Lambda$, so they remain valid when $\Lambda \to \infty$. In this
limit  operators $S_i^c $ and $\Sigma_i^c $ are finite and assumed to be known on
the energy shell from the standard renormalized QED theory. This
immediately implies that $V^d_2 $ and $V^d_3 $ are finite on the
energy shell and, due to our condition (F), they are finite for all
momenta even outside the energy shell.

Can we say that operator $V^d_4 $ is finite too? The part $\underbrace{\Sigma_4^c}$ is definitely finite, but how can we be sure that the term

\begin{eqnarray}
-\underbrace {V^d_2 \underline{V^d_2 }}
\label{eq:12.10}
\end{eqnarray}

 is finite
on the energy shell? This is where the yet undefined factor $\zeta_2$ comes into play. According to our discussion in subsection \ref{ss:uni-arb-order}, this factor can be chosen to decay sufficiently rapidly at large values of arguments (momenta).  Then all loop integrals present in the product  (\ref{eq:12.10}) are guaranteed to converge.\footnote{see Theorem
\ref{Theorem9.10}}  Consequently,  operator
(\ref{eq:12.10}) is finite on the energy shell, and $V_4^d$ in (\ref{eq:12.6}) is
also finite on the energy shell and everywhere else. These arguments can
be repeated in all higher orders, thus proving that the dressed
particle Hamiltonian $H^d$ is free of ultraviolet divergences.

\subsection{Poincar\'e invariance of the dressed
particle approach} \label{ss:relat-invar-dressed}

The next question is whether our theory with the transformed
Hamiltonian $H^d$ is Poincar\'e invariant (condition (A))? In other
words, whether there exists a boost operator $\mathbf{K}^d$
such that the set of generators $\{\mathbf{P}_0, \mathbf{J}_0,
\mathbf{K}^d, H^d \}$ satisfies Poincar\'e commutators? With the
 dressing operator $\exp(i \Phi)$ constructed above, this
problem has a simple solution. If we define $\mathbf{K}^d =
e^{i\Phi}
 \mathbf{K}^c e^{-i\Phi}$, then we can obtain a full set of dressed
generators via
 unitary transformation of the old
 generators\footnote{Note that operator $\exp(i
\Phi)$ commutes with $\mathbf{P}_0$ and $ \mathbf{J}_0$ by
construction.}

\begin{eqnarray*}
\{ \mathbf{P}_0, \mathbf{J}_0, \mathbf{K}^d, H^d \}= e^{i\Phi}
\{\mathbf{P}_0, \mathbf{J}_0, \mathbf{K}^c, H^c \}e^{-i\Phi}
\end{eqnarray*}

\noindent The dressing transformation $e^{i\Phi}$ is unitary and,
therefore, preserves commutators. Since old operators  obey the
Poincar\'e commutators, the same is true for the new generators.
This proves that the transformed
 theory is Poincar\'e invariant and belongs to the  instant form
of dynamics \cite{Shebeko-Shirokov}.

\section{Dressed interactions between particles}
\label{ss:3dressed}

\subsection{General properties of dressed potentials}
\label{ss:3-rd-order-dressed}

\noindent One may notice that even after conditions (I) - (IV) on page \pageref{cond:I} are
satisfied for functions $\zeta_i$, there is a great deal of
ambiguity in choosing their behavior  outside the energy shell.
Therefore, the dressing transformation $e^{i \Phi}$ is not unique, and there is an infinite set of dressed particle Hamiltonians that
satisfy our requirements (A) - (G). Which dressed Hamiltonian should
we choose? Before trying to answer this question, we can notice that
all  Hamiltonians satisfying conditions (A) - (G) have some
important common properties, which will be described here.

Note that electromagnetic interactions are rather weak. In most
situations the (expectation value) of the interaction potential
energy is much less than the electron's rest energy ($mc^2$).\footnote{For example, the ratio of the hydrogen's binding energy (13.6 eV) and the electron's rest energy (511 keV) is only 2.6$\times 10^{-5}$.}  To
describe such situations it is sufficient to know the coefficient
functions of the interaction only near the energy shell where we can
use condition (I) and set approximately $\zeta_i \approx 1$ for each perturbation order
$i$. This observation immediately
allows us to obtain a good approximation for the second-order
interaction from equation (\ref{eq:v2dphys}) by setting $\zeta_2 \approx
1$ there

\begin{eqnarray}
 V_2^d    &\approx& V_2^{ph}
 - \frac{1}{2 } [\underline{V_1 }, V_1 ]^{ph}
\label{eq:12.9}
\end{eqnarray}

\noindent Operator $V_2^{ph}$ can be taken from formula (\ref{eq:v2-final-ph}), and calculations involved in $[\underline{V_1 }, V_1 ]^{ph}$ have been explained in subsection \ref{ss:2-nd-order}. So, obtaining the full operator $V_2^d$ is not that difficult.

In higher perturbation orders commutator formulas (\ref{eq:vid_in_arb_order}) become rather complicated and the method of unitary dressing transformation becomes impractical. Fortunately, there is an equivalent, but a much simpler alternative approach: One can fit the desired dressed interaction operators  $V_i^d$ directly to the renormalized $S$-operator (or, more precisely, to its components $\Sigma_i^c$ in each perturbation order) of traditional QED, as described in subsection \ref{ss:lim-inf-cutoff}.

For example, in the 2nd order we obtain from (\ref{eq:12.4}) and our assumption $\zeta_i \approx 1$

\begin{eqnarray*}
 V_2^d    &\approx& (\Sigma_2^c)^{ph}
\end{eqnarray*}

\noindent which is consistent with (\ref{eq:12.9}). Let us now briefly describe the structure of this operator. Some examples of  potentials present in $V_2^d$ are shown
in Table \ref{table:12.1}. We can  classify them into two groups:
\emph{elastic} potentials and \emph{inelastic} potentials.
\index{inelastic potential} Elastic  \index{elastic potential}
potentials do not change the particle content of the system: they
 have equal number of annihilation and
creation operators of the same particle types.  As shown in subsection
\ref{ss:2-particle}, elastic potentials correspond to particle
interactions familiar from ordinary quantum mechanics and classical
physics.
Inelastic potentials change the number and/or types of particles.
Among inelastic 2nd order potentials in RQD  there are potentials for \emph{pair creation},
\index{pair creation} \emph{pair annihilation}, \index{pair annihilation} and
\emph{pair conversion}. \index{pair conversion}

Similarly to the 2nd order discussed above,  the third-order
interaction $V_3^d$ can be unambiguously obtained near the energy shell
by setting $\zeta_3 \approx 1$ in (\ref{eq:12.5})

\begin{eqnarray}
 V_3^d    &\approx& (\Sigma_3^c)^{ph} \label{eq:V3d-approx}
\end{eqnarray}

\noindent All 3rd order potentials are
inelastic. Two of them are shown in Table \ref{table:12.1}:
 The term
$d^{\dag}a^{\dag}c^{\dag}da$ (\emph{bremsstrahlung})
\index{bremsstrahlung} describes creation of a photon in a
proton-electron collision.\footnote{A more detailed discussion of this effect can be found in section \ref{sc:spontaneous}.} In the  language of classical
electrodynamics, this can be interpreted as emission of radiation due to
acceleration of charged particles and is also related to the
\emph{radiation reaction} \index{radiation reaction} force. The
Hermitian-conjugated term $d^{\dag}a^{\dag}dac$ describes absorption
of a photon by a colliding pair of charged particles.

\begin{table}[h]
\caption{Examples of interaction potentials in RQD.
 Bold numbers in the third
column  indicate perturbation orders  in which explicit interaction
operators can be unambiguously obtained near the energy shell
  as described in the text.}
\begin{tabular*}{\textwidth}{@{\extracolsep{\fill}}lll}
\hline

 Operator     &Physical meaning        &    Perturbation     \cr
                 &               &   Orders      \cr
\hline
   & \bf{Elastic potentials} & \cr
$a^{\dag}a^{\dag}aa$ &$e^- - e^-$ potential  & $\mathbf{2},4,6,
\ldots$ \cr $d^{\dag}a^{\dag}da$ &$e^- - p^+$ potential  &
$\mathbf{2},4,6, \ldots$ \cr $a^{\dag}c^{\dag}ac$  &$e^- - \gamma$
potential (Compton scattering)  & $\mathbf{2},4,6,\ldots$ \cr
$a^{\dag}a^{\dag}a^{\dag}aaa$ & $e^- - e^- - e^-$ potential  &
$4,6,\ldots$ \cr
 & \bf{Inelastic potentials} & \cr
$a^{\dag}b^{\dag}cc$  & $e^- - e^+$  pair creation   &
$\mathbf{2},4,6,\ldots$ \cr $c^{\dag}c^{\dag}ab$  & $e^- - e^+$
annihilation   & $\mathbf{2},4,6,\ldots$ \cr $d^{\dag}f^{\dag}ab$
&conversion of $e^- - e^+$ pair to $p^- - p^+$ pair   &
$\mathbf{2},4,6,\ldots$ \cr $d^{\dag}a^{\dag}c^{\dag}da$ & $e^- -
p^+$ bremsstrahlung & $\mathbf{3},5, \ldots $\cr
$d^{\dag}a^{\dag}dac$ & photon absorption in $e^- - p^+$ collision &
$\mathbf{3},5, \ldots $ \cr $a^{\dag}a^{\dag}a^{\dag}b^{\dag} aa$ &
pair creation  in $e^- - e^-$ collision & $\mathbf{4},6, \ldots $
\cr \hline
\end{tabular*}
\label{table:12.1}
\end{table}

The situation is less certain for the 4th and higher order dressed
particle interactions.
 Near the energy
shell   we can again set $\zeta_4 \approx 1$ in equation
(\ref{eq:12.6})

\begin{eqnarray}
V^d_4  &\approx& (\Sigma^c_4)^{ph}  - (V^d_2 \underline
{V^d_2 })^{ph}  \label{eq:12.11}
\end{eqnarray}

\noindent The operator $V^d_4$ obtained by this formula is a sum of
various interaction potentials (some of them are shown in Table
\ref{table:12.1}; see also Chapter \ref{ss:hydrogen})

\begin{eqnarray}
V^d_4 = d^{\dag} a^{\dag} da + a^{\dag} a^{\dag}a^{\dag} b^{\dag} aa
+ \ldots \label{eq:12.12}
\end{eqnarray}

\noindent The contribution  $ (\Sigma^c_4)^{ph}$ in equation (\ref{eq:12.11})
is well-defined near the energy shell, because we assume exact
knowledge of the $S$-operator of renormalized QED in all
perturbation orders. However, there is less clarity about the contribution
$(V^d_2 \underline
{V^d_2 })^{ph} $. As explained in section \ref{ss:diagrams-general}, the diagram for the product  $(V^d_2 \underline
{V^d_2 })^{ph} $ should be constructed from diagrams $V^d_2 $ and $\underline
{V^d_2 } $ by coupling some of their external lines, thus transforming them into internal lines and loops. Loop integration momenta are not limited, so the product $(V^d_2 \underline
{V^d_2 })^{ph} $ generally
 depends on the behavior of the factors
everywhere in the  momentum space, even outside their energy shells. So, (\ref{eq:12.11}) depends on our global choice
$\zeta_2$ outside the energy shell. The function $\zeta_2$ satisfies
conditions (I) - (IV), but still there is a great freedom in choosing the out-of-shell behavior of $\zeta_2$. This freedom is
reflected in the uncertainty of $V_4^d$ even on the energy shell.
Therefore, we have two possibilities depending on the operator
structure of the 4th order potential we are interested in.

First, there are potentials contained only in the term
$(\Sigma^c_4)^{ph}$ in (\ref{eq:12.11}) and not present in the product $(V^d_2 \underline
{V^d_2 })^{ph} $. For example, this product does not
contain operator $a^{\dag} a^{\dag}a^{\dag}b^{\dag} aa$
responsible for the creation of an electron-positron pair in
two-electron collisions. For such potentials, their 4th order
expression near the energy shell can be explicitly obtained from
formula (\ref{eq:12.11}).\footnote{This certainty is stressed by the bold
\textbf{4} in the last row of Table \ref{table:12.1}.}

Second, there are potentials $V_4^d$ whose contributions
 come from both two terms on the right hand side of (\ref{eq:12.11}).
For such potentials, the second term on the right hand side of
this equation is dependent on the particular choice of function $\zeta_2$ and, therefore, remains uncertain. One example  is the 4th order
contribution to the electron-proton interaction
$d^{\dag}a^{\dag}da$, which is responsible for the famous \emph{Lamb
shift}. \index{Lamb shift} See subsection \ref{sc:Lamb-shift}.

The uncertainty of high order interactions $V_i^d$ is perfectly
understandable: It simply reflects the one-to-many correspondence
between the $S$-operator and Hamiltonians. It means that there is a
broad class of finite \emph{phys} interactions $ \{V^d\} $ all of which  can be
used for $S$-matrix calculations without encountering divergent
integrals. Then which member of the class $\{V^d\}$ is the unique
\emph{correct} interaction Hamiltonian $V^d$? As we are not aware of
any theoretical condition allowing to determine the off-energy-shell
behavior of functions $\zeta_i$, this question should be deferred to
experiments. There seems to be no other way but to fit functions
$\zeta_i$ to experimental measurements. Such experiments are bound
to be rather challenging because they should go beyond usual
information contained in the $S$-operator (scattering
cross-sections, energies and lifetimes of bound states, etc.) and
should be capable of measuring  radiative corrections to wave
functions and time evolution of observables in the region of
interaction.
  Modern experiments do not have sufficient resolution to meet this
challenge.

 To summarize the above discusion, let us now mention a few important differences between the original QED interaction $V^c$ and the dressed particle potential energy $V^d$. First, we see that in interaction operators $V^d_i$ of higher perturbation orders
there are more and more terms with increasing complexity. In
contrast to QED Hamiltonians $H$ and $H^c$, there seems to be no way to
write $H^d$ in a closed form. However, to the credit of RQD,
 all these high order terms
directly reflect real interactions and processes observable in
nature. Unfortunately, the above construction of the dressed
particle Hamiltonian does not allow us to obtain full information
about $V^d$: The off-energy-shell behavior of potentials is fairly
arbitrary and the on-energy-shell behavior\footnote{which is the most
relevant for comparison with experiments} can be determined theoretically only
for lowest order terms. Uncovering the dressed interaction potentials in the entire momentum space would require additional information that can be made available only from sophisticated experiments.

The idea of defining ``effective'' particle interactions, which reproduce scattering amplitudes obtained from quantum field theory and satisfy equations like (\ref{eq:12.4}) - (\ref{eq:12.7}) has a long history. Approaches based on this idea can be found in a number of works \cite{Holstein, PinSot1, PinSot2, Gupta, GupRepSuc, FeiSuc}. The important difference of our dressed approach is somewhat philosophical: In contrast to previous works, we do not consider quantum fields as fundamental physical entities, and we do not regard effective potentials as mere approximations to the ``rigorous'' field based description. For us particles and their direct dressed interactions $V^d$ are the ultimate ingredients of nature.

\subsection{Energy spectrum of the dressed theory}

\begin{figure}
\centering
\includegraphics {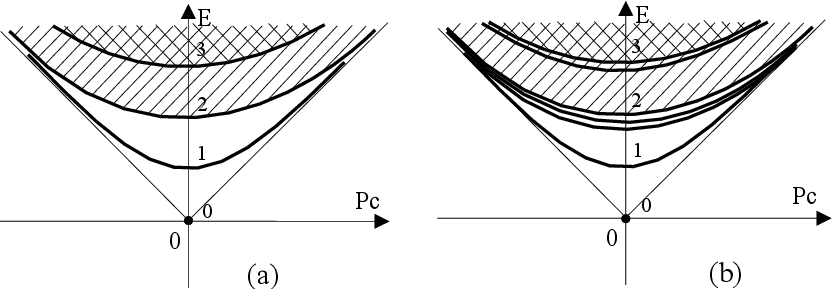} \caption{Typical momentum-energy spectrum
of (a) non-interacting  and (b) interacting  dressed particle theory. }
\label{fig:10.3}
\end{figure}

Properties of interactions between dressed particles discussed in the preceding subsection allow us to analyze some general features of the energy spectrum of our theory. In Fig. \ref{fig:10.3}(a) we show the energy spectrum of a non-interacting theory with one (massive) particle type.\footnote{compare with Fig. \ref{fig:8.1}(a)} The 0-particle state (vacuum) has vanishing energy and momentum. The 1-particle state has energy $E = \sqrt{m^2c^4 + P^2c^2}$. Energy-momenta of 2-particle states form a dense (hatched) region limited from below by the hyperboloid $E = \sqrt{(2mc^2)^2 + P^2c^2}$. Energy-momenta of 3-particle states form a (double-hatched) region limited from below by the hyperboloid $E = \sqrt{(3mc^2)^2 + P^2c^2}$, etc.

We know that dressed interaction does not affect 0-particle and 1-particle states, so the corresponding energies remain exactly the same as in the non-interacting case.\footnote{See Fig. \ref{fig:10.3}(b) and compare with Fig. \ref{fig:8.1}(b).} Dressed interaction does perturb states with two or more particles. In particular, if inter-particle potentials in $V^d$ are relatively weak and attractive, one can expect formation of hyperboloids of bound states, as shown in Fig. \ref{fig:10.3}(b). In the next section we will illustrate the description of bound states in RQD using the hydrogen atom as an example.

Traditional renormalized quantum field theories also make similar statements about the energy-momentum spectrum of multiparticle states.\footnote{See, for example, Fig. 17.4 in \cite{Schweber}, Fig. 16.1 in \cite{Bjorken2}, and Fig. 7.1 in \cite{Peskin}.} However, in field theories these statements are not obvious. They cannot be deduced directly from the renormalized Hamiltonian $H^c$. The main reason is that the interacting part of $H^c$ cannot be regarded as weak, because it includes divergent renormalization counterterms.

\subsection{Comparison with other dressed particle
approaches} \label{ss:comparison}

In this subsection, we would like to discuss another point of view
on the dressing transformation. This point of view is
philosophically different but mathematically equivalent to ours. It
is exemplified by the works of
 Shirokov and coauthors \cite{Shirokov, Shirokov2, Shirokov4}.
 In contrast to our approach in which the dressing transformation $e^{i
\Phi}$ was applied to the field-theoretical Hamiltonian $H^c$ of
QED while (bare) particle creation and annihilation operators were not
affected, Shirokov et al. kept the  $H^c$ intact, but applied the (inverse) dressing
transformation $e^{-i \Phi}$ to creation and annihilation operators
of particles

\begin{eqnarray*}
\alpha_d^{\dag} &=& e^{-i \Phi} \alpha^{\dag} e^{i \Phi} \\
\alpha_d &=& e^{-i \Phi} \alpha e^{i \Phi}
\end{eqnarray*}

\noindent to the vacuum state

\begin{eqnarray*}
| 0 \rangle_d &=& e^{-i \Phi} | 0 \rangle
\end{eqnarray*}

\noindent and to particle observables. Physically, this means that
instead of bare particles \index{bare particle} (created and
annihilated by $\alpha^{\dag}$ and $\alpha$, respectively) the
theory is formulated in terms of fully dressed particles (created
and annihilated by operators $\alpha_d^{\dag}$ and $\alpha_d$,
respectively), \index{dressed particle} i.e., particles together
with their virtual clouds. Within this approach the Hamiltonian
$H^c$ must be expressed as a function of the new particle operators
$H^c = \mathcal{F} (\alpha_d^{\dag}, \alpha_d)$. Apparently, the
same function $ \mathcal{F}$ expresses the Hamiltonian $H^d$ of our
approach through original (bare) particle operators: $H^d =
\mathcal{F} (\alpha^{\dag}, \alpha)$. Indeed, from equation
(\ref{eq:12.14}) we can write

\begin{eqnarray*}
H^c  &=& e^{-i \Phi}H^d e^{i \Phi} = e^{-i \Phi}\mathcal{F}
(\alpha^{\dag}, \alpha) e^{i \Phi} = \mathcal{F} (e^{-i \Phi}
\alpha^{\dag} e^{-i \Phi},
e^{-i \Phi} \alpha e^{i \Phi})  \\
&=& \mathcal{F} ( \alpha_d^{\dag} ,
 \alpha_d )  \\
\end{eqnarray*}

\noindent So, mathematically, these two approaches are equivalent.
Let us demonstrate this equivalence on a simple example. Suppose we
want to calculate a trajectory (=the time dependence of the expectation value of the
position operator) of the electron  in a 2-particle
system (electron + proton). In our approach the initial state of the
system has two particles

 \begin{eqnarray*}
| \Psi \rangle = a^{\dag}d^{\dag}| 0\rangle
\end{eqnarray*}

\noindent and the expectation value of the electron's position is
given by formula

\begin{eqnarray}
 \mathbf{r}(t)  &=& \langle \Psi | \mathbf{R}(t) | \Psi \rangle
\nonumber \\
&=& \langle 0 | da e^{\frac{i}{\hbar} H^dt}\mathbf{R}
e^{-\frac{i}{\hbar} H^dt} a^{\dag}d^{\dag}| 0\rangle \label{eq:12.15}
\end{eqnarray}

\noindent where $\mathbf{R}$ is the position operator for the
bare electron, and the time evolution is governed by the dressed Hamiltonian $H^d$. However, we can rewrite this expression in the following
form characteristic for the Shirokov's approach

\begin{eqnarray}
 \mathbf{r}(t)
 &=& \langle 0 | da (e^{i \Phi} e^{\frac{i}{\hbar} H^ct} e^{-i \Phi}
)\mathbf{R} (e^{i \Phi} e^{-\frac{i}{\hbar} H^ct} e^{-i \Phi})
a^{\dag}d^{\dag}| 0\rangle \nonumber \\
&=& \langle 0 |e^{i \Phi} (e^{-i \Phi} da e^{i \Phi})
e^{\frac{i}{\hbar} H^ct} (e^{-i \Phi}\mathbf{R} e^{i \Phi})
e^{-\frac{i}{\hbar} H^ct}( e^{-i \Phi} a^{\dag}d^{\dag}
e^{i \Phi} )e^{-i \Phi}| 0\rangle \nonumber\\
&=& _d\langle 0 |  d_d a_d  e^{\frac{i}{\hbar} H^ct} \mathbf{R}^d
 e^{-\frac{i}{\hbar} H^ct} a_d^{\dag}d_d^{\dag}
| 0\rangle_d  \label{eq:12.16}
\end{eqnarray}

\noindent where the time evolution is generated by the original
Hamiltonian $H^c$, while  ``dressed'' definitions are used for the
vacuum state, particle operators, and the position observable

\begin{eqnarray*}
| 0\rangle_d   &=&  e^{-i \Phi}| 0\rangle  \\
\{a_d,d_d,a_d^{\dag}, d_d^{\dag} \} &=& e^{-i \Phi}\{a,d,a^{\dag},
d^{\dag} \}
e^{i \Phi} \\
\mathbf{R}^d &=& e^{-i \Phi}  \mathbf{R} e^{i \Phi}
\end{eqnarray*}

\noindent In spite of different formalisms, physical predictions of
both theories, e.g., expectation values of observables
(\ref{eq:12.15}) and (\ref{eq:12.16}), are exactly the same.

\chapter{COULOMB POTENTIAL AND BEYOND}
\label{sc:coulomb}

\begin{quote}
\textit{This work contains many things which are new and interesting. Unfortunately, everything that is new is not interesting, and everything which is interesting, is not new.}

\small
\hspace{1in} Lev D. Landau
\normalsize
\end{quote}

\vspace{0.5in}

In the preceding chapter we have derived basic formulas of the dressed particle approach. Our next goal is to demonstrate that this method is really useful in concrete calculations. This task will occupy us throughout the next three chapters. Their plan is as follows: In the present chapter we will derive the dressed electron-proton interaction potential in the 2nd perturbation order. Then we will use it to calculate the energy spectrum of the hydrogen atom. The next step is to extend this theory to higher perturbation orders. In the 3rd order the hydrogen spectrum is disturbed by spontaneous photon emission from excited energy levels. To analyze this effect we will introduce a relativistic quantum theory of unstable systems in chapter \ref{ch:decays}. In chapter \ref{ch:hydrogen-revisited} we are going to use this theory as well as 4th order radiative corrections to the electron-proton potential and derive two classic QED results - the anomalous magnetic moment of the electron and the Lamb shift in hydrogen atoms.

In the preceding chapter we obtained formulas (\ref{eq:12.9}) and (\ref{eq:12.11}) for the dressed
particle Hamiltonian $H^d$ in a rather abstract form. In this chapter we would
like to demonstrate how this Hamiltonian can be cast
 into a form suitable for calculations,
i.e., expressed  through creation and annihilation operators
 of electrons,
protons, photons, etc.  Here we
will focus on pair interactions between electrons and protons in the
lowest (second) order of the perturbation theory. In section \ref{ss:darwin-breit} we will use the $(v/c)^2$
approximation to obtain what is commonly known as the
\emph{Darwin-Breit potential} between charged particles with spin 1/2. \index{Darwin-Breit potential} The major part of
this interaction is the usual Coulomb potential. In addition, there
are relativistic corrections responsible for magnetic, contact,
spin-orbit, spin-spin, and other interactions which are routinely
used in relativistic calculations of atomic and molecular systems.
This derivation demonstrates how formulas familiar from ordinary
quantum mechanics and classical electrodynamics follow naturally
from RQD.  In section \ref{ss:hydrogen-nonr} we will solve the stationary Schr\"odinger equation for the hydrogen atom and obtain its energy spectrum  with relativistic corrections.

\section{Darwin-Breit Hamiltonian}
\label{ss:darwin-breit}

\subsection{Electron-proton potential in the momentum space} \label{ss:low-energy}

Note that the second-order dressed interaction near the energy shell
(\ref{eq:12.9}) is given by the same formula as $F_2$ in (\ref{eq:11.29}). So, for the electron-proton potential we can simply reuse our result (\ref{eq:F2xy})\footnote{Note that we could also choose a different expression (\ref{eq:s-oper-2nd}) for $F_2$, thus obtaining a different formula for $V_2^d$. The two operators (\ref{eq:F2xy}) and (\ref{eq:s-oper-2nd}) coincide on the energy shell, so they result in identical 2nd order $S$-matrices and energy spectra. However, their corresponding dressed Hamiltonians are different, as they have different off-energy-shell behaviors.  As shown in chapter \ref{ch:theories}, our decision to use (\ref{eq:F2xy}) leads to a good agreement with Maxwell's electrodynamics. However, it is not clear how much this macroscopic theory would be affected by a different choice of $V^d_2$ outside the energy shell. This question requires a deeper investigation.}

\begin{eqnarray}
 V_2^d [d^{\dag}a^{\dag} da]  &=& V_{2A}^d + V_{2B}^d +
V_{2C}^d \label{eq:12.18} \\
  V_{2A}^d
&=&
 -\frac{e^2 \hbar^2}{(2 \pi \hbar)^3} \int
\frac{d\mathbf{k} d\mathbf{q} d\mathbf{p} Mm c^4}
{\sqrt{\omega_{\mathbf{q}}\omega_{\mathbf{q}+ \mathbf{k}}
\Omega_{\mathbf{p}}\Omega_{\mathbf{p}- \mathbf{k}}}}
\frac{1}{  k^2} \times \nonumber \\
&\mbox{ } & \sum_{\epsilon \epsilon' \pi \pi'} W^0(\mathbf{p}-
\mathbf{k},\pi;\mathbf{p},\pi') U^0(\mathbf{q}+ \mathbf{k},\epsilon;
\mathbf{q},\epsilon') \times \nonumber
\\
&\mbox{ }&   d^{\dag}_{\mathbf{p}- \mathbf{k},\pi}
a^{\dag}_{\mathbf{q}+ \mathbf{k},\epsilon} d_{\mathbf{p},\pi'}
a_{\mathbf{q},\epsilon'}
\label{eq:12.19}\\
  V_{2B}^d &=& -\frac{ e^2  \hbar^2 c^2}{ (2 \pi \hbar)^3} \int
\frac{d\mathbf{k} d\mathbf{q} d\mathbf{p} Mm
c^4}{\sqrt{\omega_{\mathbf{q}}\omega_{\mathbf{q}+
\mathbf{k}}\Omega_{\mathbf{p}}\Omega_{\mathbf{p}- \mathbf{k}}}}
  \frac{1}{(\tilde{q}+\tilde{k} \div \tilde{q} )^2} \times \nonumber\\
&\mbox{ } & \sum_{\epsilon \epsilon' \pi \pi'}
  \mathbf{W}
(\mathbf{p}- \mathbf{k},\pi;\mathbf{p},\pi') \cdot
\mathbf{U}(\mathbf{q}+ \mathbf{k},\epsilon; \mathbf{q},\epsilon')
\times
\nonumber \\
&\mbox{ }&  d^{\dag}_{\mathbf{p}- \mathbf{k},\pi}
a^{\dag}_{\mathbf{q}+ \mathbf{k},\epsilon} d_{\mathbf{p},\pi'}
a_{\mathbf{q},\epsilon'}
\label{eq:12.20}\\
  V_{2C}^d
&=& \frac{ e^2   \hbar^2c^2}{ (2 \pi \hbar)^3} \int
\frac{d\mathbf{k} d\mathbf{q} d\mathbf{p} Mm
c^4}{\sqrt{\omega_{\mathbf{q}}\omega_{\mathbf{q}+
\mathbf{k}}\Omega_{\mathbf{p}}\Omega_{\mathbf{p}- \mathbf{k}}}}
  \frac{1}{(\tilde{q}+\tilde{k} \div \tilde{q} )^2k^2} \times \nonumber \\
&\mbox { }& \sum_{\epsilon \epsilon' \pi \pi'}
 (\mathbf{k} \cdot \mathbf{W}(\mathbf{p}-
\mathbf{k},\pi;\mathbf{p},\pi'))(\mathbf{k} \cdot \mathbf{U}
(\mathbf{q}+ \mathbf{k},\epsilon; \mathbf{q},\epsilon')) \times
\nonumber
\\
&\mbox{ }&  d^{\dag}_{\mathbf{p}- \mathbf{k},\pi}
a^{\dag}_{\mathbf{q}+ \mathbf{k},\epsilon} d_{\mathbf{p},\pi'}
a_{\mathbf{q},\epsilon'} \label{eq:12.21}
\end{eqnarray}

\noindent In these formulas we integrate over the electron
($\mathbf{q}$), proton ($\mathbf{p}$) and transferred
($\mathbf{k}$) momenta and sum over spin projections of the ``incoming'' and ``outgoing''
particles $\epsilon, \epsilon', \pi'$ and $\epsilon, \pi$.

Operator (\ref{eq:12.18}) has non-trivial action in all sectors of
the Fock space which contain at least one electron and one proton.
However, for simplicity, we  will limit our attention to the ``1
proton + 1 electron'' subspace $\mathcal{H}_{pe}$ in the Fock space.
If $\Psi(\mathbf{p}, \pi; \mathbf{q}, \epsilon)$ is the wave
function of a two-particle  state in $\mathcal{H}_{pe}$, then the interaction Hamiltonian
(\ref{eq:12.18}) will transform it to\footnote{See
subsection \ref{ss:2-particle}.}

\begin{eqnarray}
&\ & \Psi'(\mathbf{p}, \pi,  \mathbf{q}, \epsilon) = V_2^d[d^{\dag}
a^{\dag} da] \Psi(\mathbf{p}, \pi;  \mathbf{q}, \epsilon) \nonumber \\
&=& \sum_{\pi' \epsilon'} \int  d \mathbf{k}
 v_2
(\mathbf{p},\mathbf{q},\mathbf{k}; \pi, \epsilon, \pi', \epsilon')
 \Psi (\mathbf{p-k}, \pi'; \mathbf{q+k}, \epsilon')
\label{eq:12.22}
\end{eqnarray}

\noindent where $v_2$ is the coefficient function of the interaction
operator $V_2^d [d^{\dag}a^{\dag} da]$. We are going to write our
formulas with the accuracy of $(v/c)^2$. So, we  use
(\ref{eq:qkdivq}) - (\ref{eq:8.48d})
 to obtain the coefficient function in (\ref{eq:12.22}) as a sum of three terms

\begin{eqnarray}
v_2 = v_{2A} + v_{2B} + v_{2C}
\label{eq:12.23}
\end{eqnarray}

\noindent where\footnote{We used properties of Pauli matrices
from Appendix \ref{ss:pauli-matrices} and formulas from Appendix \ref{ss:non-rel}. Our calculations in this section can be compared with \S 83 in ref. \cite{BLP} and with
\cite{Itoh}. }

\begin{eqnarray*}
&\ & v_{2A} \nonumber \\
&=&-\frac{e^2 \hbar^2}{(2 \pi \hbar)^3} \chi^{(el) \dag}_{\epsilon}\chi^{(pr) \dag}_{\pi} \left(\frac{1}{k^2}
- \frac{1}{8M^2c^2} - \frac{1}{8m^2c^2} -i \frac{\vec{\sigma}_{pr}
[\mathbf{k} \times \mathbf{p}]}{4M^2 c^2k^2} +i
\frac{\vec{\sigma}_{el} [\mathbf{k} \times \mathbf{q}]}{4m^2 c^2k^2} \right) \chi^{(el)}_{\epsilon'}
\chi^{(pr)}_{\pi'}\nonumber
\end{eqnarray*}

\begin{eqnarray*}
v_{2B}
&=& \frac{e^2 \hbar^2}{(2 \pi \hbar)^3} \chi^{(el) \dag}_{\epsilon}\chi^{(pr) \dag}_{\pi}\Bigl( \frac{\mathbf{pq}}{ Mmc^2k^2} -
\frac{\mathbf{kq}}{2 Mmc^2k^2}
+ \frac{\mathbf{pk} }{2 Mmc^2k^2}  -\frac{1}{4 Mmc^2} \nonumber \\
&\ & - \frac{i [\vec{\sigma}_{pr} \times
\mathbf{k}] \cdot \mathbf{q}}{2 Mmc^2k^2}
 +
\frac{i \mathbf{p } \cdot [\vec{\sigma}_{el} \times \mathbf{k}]}{2
Mmc^2k^2} +\frac{ (\vec{\sigma}_{el} \cdot \vec{\sigma}_{pr}) }{4 Mmc^2} -\frac{
(\vec{\sigma}_{pr} \cdot \mathbf{k}) (\vec{\sigma}_{el}
\cdot\mathbf{k} )}{4 Mmc^2k^2} \Bigr)\chi^{(el)}_{\epsilon'}
\chi^{(pr)}_{\pi'}
\end{eqnarray*}

\begin{eqnarray*}
 v_{2C}
&=&  -\frac{ e^2 \hbar^2 }{ (2  \pi \hbar)^3 k^4}
 \frac{1} { 4Mmc^2}
 \chi^{(el) \dag}_{\epsilon}\chi^{(pr) \dag}_{\pi} (2\mathbf{p}\mathbf{k} -k^2)
(2\mathbf{q}\mathbf{k} +k^2) \chi^{(el)}_{\epsilon'}
\chi^{(pr)}_{\pi'} \\
&=&  -\frac{ e^2 \hbar^2}{ (2  \pi \hbar)^3 }
 \chi^{(el) \dag}_{\epsilon}\chi^{(pr) \dag}_{\pi}\left(\frac{(\mathbf{p}\mathbf{k})(
\mathbf{q}\mathbf{k})}{ Mmc^2k^4} - \frac{ \mathbf{q}\mathbf{k}}{
2Mmc^2k^2} + \frac{ \mathbf{p}\mathbf{k}}{ 2Mmc^2k^2} - \frac{1}{
4Mmc^2}\right) \chi^{(el)}_{\epsilon'}
\chi^{(pr)}_{\pi'}
\end{eqnarray*}

\noindent Putting these three terms together we finally rewrite  (\ref{eq:12.23}) in the form

\begin{eqnarray}
&\ & v_2
(\mathbf{p},\mathbf{q},\mathbf{k}; \pi, \epsilon, \pi', \epsilon') \nonumber \\
&=& \frac{e^2 \hbar^2 }{(2 \pi \hbar)^3} \chi^{(el) \dag}_{\epsilon}\chi^{(pr) \dag}_{\pi}\Bigl(-\frac{1} { k^2} +
\frac{1}{8M^2 c^2} + \frac{1}{8m^2c^2} + \frac{ \mathbf{p}\mathbf{q}
}{Mmc^2k^2} \nonumber\\
&\ &- \frac{ (\mathbf{p} \mathbf{k})(\mathbf{q} \mathbf{k})}
{Mmc^2k^4} +i \frac{ \vec{\sigma}_{pr} [\mathbf{k} \times \mathbf{p}]}{4M^2
c^2k^2} -i \frac{ \vec{\sigma}_{el} [\mathbf{k} \times
\mathbf{q}]}{4m^2 c^2k^2} - i \frac{\vec{\sigma}_{pr} [
\mathbf{k}\times \mathbf{q}]}{2Mmc^2k^2}
 +i  \frac{\vec{\sigma}_{el} [\mathbf{k} \times \mathbf{p}
]}{2Mmc^2k^2} \nonumber\\
&\ & + \frac{ (\vec{\sigma}_{el} \cdot \vec{\sigma}_{pr})} {4Mmc^2} -
 \frac{(\vec{\sigma}_{pr} \cdot \mathbf{k})
(\vec{\sigma}_{el} \cdot\mathbf{k} )}{4Mmc^2 k^2} \Bigr)  \chi^{(el)}_{\epsilon'}
\chi^{(pr)}_{\pi'}
\label{eq:12.24}
\end{eqnarray}

\noindent In a reasonable approximation one can assume that the proton is infinitely heavy ($M \gg m$) and skip terms with $M$ in denominators. Then

\begin{eqnarray}
 v_2
(\mathbf{p},\mathbf{q},\mathbf{k}; \pi, \epsilon, \pi', \epsilon') \approx \frac{e^2 \hbar^2 }{(2 \pi \hbar)^3} \delta_{\pi, \pi'}\chi^{(el) \dag}_{\epsilon}\Bigl(-\frac{1} { k^2} +
 \frac{1}{8m^2c^2}   -i \frac{ \vec{\sigma}_{el} [\mathbf{k} \times
\mathbf{q}]}{4m^2 c^2k^2}  \Bigr)  \chi^{(el)}_{\epsilon'} \nonumber \\
\label{eq:12.24z}
\end{eqnarray}

\subsection{Position representation} \label{ss:position-breit}

The physical meaning of interaction (\ref{eq:12.24}) is more
transparent in the position representation, which is derived by
replacing variables $\mathbf{p}$ and $\mathbf{q}$
 with differential operators $\hat{\mathbf{p}} = -i\hbar (d/d\mathbf{x})$ and
$\hat{\mathbf{q}} = -i\hbar (d/d\mathbf{y})$ and taking the Fourier
transform\footnote{see subsection \ref{ss:2-particle}; $\mathbf{x}$ is the proton's position
and $\mathbf{y}$ is the electron's position}

\begin{eqnarray*}
&\mbox{ } & \hat{V}_2^d[d^{\dag}a^{\dag}da] \Psi(\mathbf{x}, \pi;
\mathbf{y},
\epsilon) \\
 &=& \sum_{\pi', \epsilon'} \int   d \mathbf{k}
e^{\frac{i}{\hbar} \mathbf{k}(\mathbf{x-y})} v_2(\hat{\mathbf{p}},
\hat{\mathbf{q}}, \mathbf{k}; \pi, \epsilon, \pi', \epsilon')
\Psi(\mathbf{x}, \pi'; \mathbf{y},
\epsilon') \\
&=& \frac{e^2 \hbar^2}{(2 \pi \hbar)^3} \int   d \mathbf{k}
e^{\frac{i}{\hbar} \mathbf{k}(\mathbf{x-y})} \Big( -\frac{1} {k^2} +
\frac{1}{8M^2 c^2}+ \frac{1}{8m^2c^2} + \frac{\hat{\mathbf{p}}
\hat{\mathbf{q}} }{Mmc^2k^2} - \frac{(\hat{\mathbf{p}}
\mathbf{k})(\hat{\mathbf{q}} \mathbf{k})}
{Mmc^2k^4}\\
&\ &+i  \frac{\vec{\sigma}_{pr} [\mathbf{k} \times \hat{\mathbf{p}}]}
{4M^2 c^2k^2} - i \frac{\vec{\sigma}_{el} [\mathbf{k} \times
\hat{\mathbf{q}}]} {4m^2 c^2k^2} - i \frac{\vec{\sigma}_{pr} [
\mathbf{k}\times \hat{\mathbf{q}}]} {2Mmc^2k^2}
 +i  \frac{\vec{\sigma}_{el} [\mathbf{k} \times \hat{\mathbf{p}} ]}
{2Mmc^2k^2} \\
&\ & + \frac{ (\vec{\sigma}_{el} \cdot \vec{\sigma}_{pr})} {4Mmc^2} -
  \frac{(\vec{\sigma}_{pr} \cdot \mathbf{k})
(\vec{\sigma}_{el} \cdot\mathbf{k} )}{4Mmc^2 k^2} \Big)
\Psi(\mathbf{x}, \pi; \mathbf{y}, \epsilon)
\end{eqnarray*}

\noindent Using integral formulas (\ref{eq:A.90}) - (\ref{eq:A.94})
 we obtain the following position-space representation of
this potential (where $\mathbf{r} \equiv \mathbf{x-y}$)\footnote{Some of these interaction terms are non-Hermitian due to the non-commutativity of operators $\mathbf{r}$ and $\mathbf{p}, \mathbf{q}$. This minor problem can be solved by symmetrizing non-commutative products, e.g., by replacing $AB \to (AB+BA)/2$.}

 \begin{eqnarray}
\hat{V}_2^d[d^{\dag}a^{\dag}da] &=& -\frac{e^2}{4 \pi r} + \frac{ e^2
\hbar^2}{8c^2} \Big(\frac{ 1}{M^2} + \frac{1 }{m^2} \Big)
\delta(\mathbf{r}) + \frac{e^2}{8 \pi  M m c^2 r}
\Big[\hat{\mathbf{p}} \cdot \hat{\mathbf{q}}+ \frac{(\mathbf{r}
\cdot \hat{\mathbf{q}}) (\mathbf{r} \cdot \hat{\mathbf{p}})
}{r^2}\Big] \nonumber \\
&\ & -\frac{e^2 \hbar [\mathbf{r} \times \hat{\mathbf{p}}] \cdot
\vec{\sigma}_{pr}}{16 \pi  M^2 c^2 r^3} + \frac{e^2 \hbar
[\mathbf{r} \times \hat{\mathbf{q}}] \cdot \vec{\sigma}_{el}}{16 \pi
m^2 c^2 r^3}  +\frac{e^2 \hbar [\mathbf{r} \times \hat{\mathbf{q}}]
\cdot \vec{\sigma}_{pr}}{8 \pi   M m c^2 r^3} -\frac{e^2 \hbar
[\mathbf{r} \times \hat{\mathbf{p}}]
\cdot \vec{\sigma}_{el}}{8 \pi   M m c^2 r^3} \nonumber \\
&\ & +\frac{e^2 \hbar^2 }{4 M m c^2 } \Big( -\frac{\vec{\sigma}_{pr}
\cdot \vec{\sigma}_{el}} {4 \pi r^3} +3 \frac{(\vec{\sigma}_{pr}
\cdot \mathbf{r}) (\vec{\sigma}_{el} \cdot \mathbf{r})}{ 4 \pi r^5}
+ \frac{2}{3} (\vec{\sigma}_{pr} \cdot \vec{\sigma}_{el})
\delta(\mathbf{r})\Big) \nonumber \\
\label{eq:V2d}
\end{eqnarray}

\noindent With the accuracy of $(v/c)^2$ the free Hamiltonian $H_0$
can be written as

\begin{eqnarray*}
\hat{H}_0 &=& \sqrt{M^2 c^4 + \hat{p}^2 c^2} + \sqrt{m^2 c^4 + \hat{q}^2 c^2}
\\
&=& M c^2 + m c^2 + \frac {\hat{p}^2}{2M} +  \frac {\hat{q}^2}{2m} -
\frac{\hat{p}^4}{8M^3 c^2}  - \frac{\hat{q}^4}{8m^3 c^2} +
\ldots
\end{eqnarray*}

\noindent where the rest energies of particles $M c^2$ and $m c^2$
are simply constants, which can be eliminated by a proper choice of
zero on the energy scale. Note also that Pauli matrices
 are proportional to particle spin
operators (\ref{eq:si-sigma}): $\hat{\mathbf{S}}_{el} =
\frac{\hbar}{2} \vec{\sigma}_{el}$, $\hat{\mathbf{S}}_{pr} =
\frac{\hbar}{2} \vec{\sigma}_{pr}$. So, finally, the QED Hamiltonian
responsible for the electron-proton interaction in the 2nd order is
obtained in the form of \emph{Darwin-Breit
potential} \index{Darwin-Breit potential}

\begin{eqnarray}
&\ &\hat{H}^d = \hat{H}_0 + \hat{V}_2^d(\hat{\mathbf{p}}, \hat{\mathbf{q}},
\mathbf{r},
\hat{\mathbf{S}}_{el}, \hat{\mathbf{S}}_{pr}) + \ldots\nonumber\\
&=&  \frac {\hat{p}^2}{2M} +  \frac
{\hat{q}^2}{2m} + \hat{V}_{Coulomb} +  \hat{V}_{orbit} + \hat{V}_{spin-orbit}
+ \hat{V}_{spin-spin} + \ldots \label{eq:12.25}
\end{eqnarray}

\noindent This form is similar to the familiar non-relativistic
Hamiltonian in which $\hat{p}^2/(2M) +
\hat{q}^2/(2m)$ is treated as the kinetic energy operator
and \index{kinetic energy}  $\hat{V}_{Coulomb}$ is the usual Coulomb
interaction \index{Coulomb potential}
between two charged particles

\begin{eqnarray}
 \hat{V}_{Coulomb} &=& -\frac{e^2}{4 \pi r}
\label{eq:12.26}
\end{eqnarray}

\noindent  This is the only interaction term which survives in the
non-relativistic limit $c \to \infty$. $\hat{V}_{orbit}$  is a spin-independent
relativistic correction to the Coulomb interaction

\begin{eqnarray}
 \hat{V}_{orbit} &=& - \frac{\hat{p}^4}{8M^3 c^2}
- \frac{\hat{q}^4}{8m^3 c^2} + \frac{ e^2 \hbar^2}{8 c^2}
\left(\frac{ 1}{M^2}
+ \frac{1 }{m^2} \right) \delta(\mathbf{r})\nonumber \\
&\ & +\frac{e^2}{8 \pi M m c^2 r} \Big[\hat{\mathbf{p}} \cdot
\hat{\mathbf{q}} + \frac{(\mathbf{r} \cdot
\hat{\mathbf{q}})(\mathbf{r} \cdot \hat{\mathbf{p}}) }{r^2} \Big]
\label{eq:12.27}
\end{eqnarray}

\noindent The first two terms do not depend on relative variables,
so they can be regarded as relativistic corrections to energies of
single particles. The \emph{contact interaction} \index{contact
interaction} (proportional to $\hbar^2 \delta(\mathbf{r})$) can be
neglected in the classical limit $\hbar \to 0$. Keeping the
$(v/c)^2$ accuracy and substituting $\hat{\mathbf{p}}/M \to \hat{\mathbf{v}}_{pr}$ and $\hat{\mathbf{q}}/m \to
\hat{\mathbf{v}}_{el}$, the remaining terms can be rewritten in a more familiar form
of the \emph{Darwin potential} \cite{Breitenberger} \index{Darwin
potential}

\begin{eqnarray}
\hat{V}_{Darwin} &=&  \frac{e^2}{8 \pi c^2 r} \Big[\hat{\mathbf{v}}_{el} \cdot
\hat{\mathbf{v}}_{pr} + \frac{(\mathbf{r} \cdot
\hat{\mathbf{v}}_{pr})(\mathbf{r} \cdot \hat{\mathbf{v}}_{el}) }{r^2} \Big]
\label{eq:12.28}
\end{eqnarray}

\noindent which describes velocity-dependent (magnetic) interaction between charged
 particles.

  Two other
terms $\hat{V}_{spin-orbit}$ \index{spin-orbit potential} and $\hat{V}_{spin-spin}$ \index{spin-spin
potential} in (\ref{eq:12.25}) depend on
particle spins

\begin{eqnarray}
 \hat{V}_{spin-orbit}
&=&  -\frac{e^2 [\mathbf{r} \times \hat{\mathbf{p}}] \cdot
\hat{\mathbf{S}}_{pr}}{8 \pi M^2 c^2 r^3} +
 \frac{e^2  [\mathbf{r} \times \hat{\mathbf{q}}]
\cdot \hat{\mathbf{S}}_{el}}{8 \pi m^2 c^2 r^3} + \frac{e^2 [\mathbf{r}
\times \hat{\mathbf{q}}] \cdot \hat{\mathbf{S}}_{pr}}{4 \pi M m c^2 r^3} -
\frac{e^2 [\mathbf{r} \times \hat{\mathbf{p}}] \cdot
\hat{\mathbf{S}}_{el}}{4 \pi  M m c^2 r^3} \nonumber \\ \label{eq:12.29} \\
 \hat{V}_{spin-spin}
&=& \frac{e^2 }{M m c^2 } \Big(- \frac{\hat{\mathbf{S}}_{pr} \cdot
\hat{\mathbf{S}}_{el}} {4 \pi r^3} + 3 \frac{(\hat{\mathbf{S}}_{pr} \cdot
\mathbf{r}) (\hat{\mathbf{S}}_{el} \cdot \mathbf{r})}{4 \pi r^5} +
\frac{2}{3} (\hat{\mathbf{S}}_{pr} \cdot \hat{\mathbf{S}}_{el})
\delta(\mathbf{r})\Big) \nonumber
\\ \label{eq:12.30}
\end{eqnarray}

Since our dressing transformation preserved commutation
relations of the Poincar\'e Lie algebra,\footnote{See subsection \ref{ss:relat-invar-dressed}.} we can be confident that the
Darwin-Breit Hamiltonian is relativistically invariant, at least up
to the order $(v/c)^2$. In Appendix \ref{sc:proof-comm} we
additionally verify this important fact by a direct
calculation.\footnote{see also \cite{Coleman, Close-Osborn,
Krajcik-Foldy}}

  The Darwin-Breit Hamiltonian was
successfully applied to various electromagnetic problems, such as
the fine structure of atomic spectra \cite{BLP,
Breitenberger}, superconductivity, and properties of plasma
\cite{Essen, Essen3, Essen1, Essen2, Essen4}. In chapter \ref{ch:theories}
we will see that in the classical limit this Hamiltonian reproduces
correctly all major results of classical electrodynamics.
In chapter \ref{ss:hydrogen} we will calculate small radiative corrections to the Darwin-Breit potential.

\section{Hydrogen atom}
\label{ss:hydrogen-nonr}

Having derived the electron-proton interaction potential,  now we can study the bound state of these two particles - the hydrogen atom. \index{hydrogen atom} We are interested in energies and wave functions of its stationary states. In the dressed particle approach, this task is accomplished simply by diagonalizing the dressed particle Hamiltonian in the ``electron+proton'' sector $ \mathcal{H}_{pe}$ of the Fock space. In other words, the stationary Schr\"odinger equation needs to be solved. In this section, we will study this solution with the Hamiltonian (\ref{eq:12.25}) including interaction terms up to the 2nd perturbation order. Higher order corrections will be considered in chapter \ref{ss:hydrogen}.

\subsection{Non-relativistic Schr\"odinger equation}
\label{ss:nonr-schr}

We can use the fact that Hamiltonian (\ref{eq:12.25}) commutes with
the operator of total momentum $\mathbf{P} = \mathbf{p} +
\mathbf{q}$. Therefore this Hamiltonian leaves invariant
the zero-total-momentum subspace of $ \mathcal{H}_{pe}$. Working in
this subspace we can set $\hat{\mathbf{Q}} \equiv - \hat{\mathbf{q}}
= \hat{\mathbf{p}}$ in equations (\ref{eq:12.27}) and (\ref{eq:12.29}) and
consider $\hat{\mathbf{Q}}$ as operator of differentiation with
respect to $\mathbf{r}$

\begin{eqnarray*}
\hat{\mathbf{Q}}= i \hbar \frac{\partial}{\partial \mathbf{r}}
\end{eqnarray*}

\noindent If we make these substitutions in  (\ref{eq:12.25}),
 then the energies $\varepsilon$ and wave
functions $\Psi_{\varepsilon} (\mathbf{r}, \pi, \epsilon)$ of
stationary states of the hydrogen atom  at rest can be found as
solutions of the stationary \index{stationary Schr\"odinger
equation} Schr\"odinger equation

 \begin{eqnarray}
\hat{H}^d \left(i \hbar \frac{\partial}{\partial \mathbf{r}}, \mathbf{r},
\mathbf{S}_{el}, \mathbf{S}_{pr}  \right) \Psi_{\varepsilon}
(\mathbf{r}, \pi, \epsilon) = \varepsilon \Psi_{\varepsilon}
(\mathbf{r}, \pi, \epsilon) \label{eq:12.31}
\end{eqnarray}

\noindent  Analytical solution of equation (\ref{eq:12.31}) is not
possible. Realistically, one can first
 solve equation (\ref{eq:12.31}) leaving just the Coulomb interaction
term (\ref{eq:12.26}) there and rewriting the first two terms in (\ref{eq:12.25}) as

\begin{eqnarray*}
 \frac {\hat{p}^2}{2M} +  \frac
{\hat{q}^2}{2m} = \frac{(m+M) \hat{Q}^2}{2mM} = \frac{ \hat{Q}^2}{2 \mu}
\end{eqnarray*}

\noindent where $\mu = mM/(m+M) \approx m$ is the \emph{reduced mass}. \index{reduced mass} In this approximation equation (\ref{eq:12.31}) takes the form

 \begin{eqnarray}
\left(\frac{\hat{Q}^2}{2 \mu} - \frac{e^2}{4 \pi r} \right) \Psi_{\varepsilon}
(\mathbf{r}, \pi, \epsilon) = \varepsilon \Psi_{\varepsilon}
(\mathbf{r}, \pi, \epsilon) \label{eq:10.54a}
\end{eqnarray}

\noindent It does not depend on spin variables $\pi, \epsilon$, so the solution can be written as a product of orbital and spin parts

\begin{eqnarray*}
 \Psi_{\varepsilon}
(\mathbf{r}, \pi, \epsilon) =  \psi_{\varepsilon}
(\mathbf{r}) \chi (\pi, \epsilon)
\end{eqnarray*}

\noindent The energy $\varepsilon$ is independent on the spin part $\chi (\pi, \epsilon)$, which can be chosen as an arbitrary set of four complex numbers, satisfying the normalization condition

\begin{eqnarray*}
|\chi (1/2, 1/2)|^2 + |\chi (-1/2, 1/2)|^2 + |\chi (1/2, -1/2)|^2 + |\chi (-1/2, -1/2)|^2 = 1
\end{eqnarray*}

\noindent The orbital parts and their energy eigenvalues  satisfy the differential equation

\begin{eqnarray}
\left(-\frac{\hbar^2}{2 \mu} \frac{\partial^2}{\partial \mathbf{r}^2} - \frac{e^2}{4 \pi r} \right) \psi_{\varepsilon}
(\mathbf{r}) = \varepsilon \psi_{\varepsilon} (\mathbf{r}) \label{eq:10.54b}
\end{eqnarray}

\noindent or in spherical coordinates

\begin{eqnarray*}
&\ & \left(-\frac{\hbar^2}{2 \mu} \left(\frac{1}{r^2}\frac{\partial}{\partial r}\left[r^2 \frac{\partial}{\partial r}\right] + \frac{1}{r^2 \sin \theta}\frac{\partial}{\partial \theta}\left[\sin \theta \frac{\partial}{\partial \theta}\right] + \frac{1}{r^2 \sin^2 \theta} \frac{\partial^2}{ \partial \phi^2}\right) - \frac{e^2}{4 \pi r} \right) \psi_{\varepsilon}
(r, \theta, \phi) \\
&=& \varepsilon \psi_{\varepsilon} (r, \theta, \phi)
\end{eqnarray*}

\noindent This is the familiar non-relativistic
problem with a well-known analytical solution, which can be found in
any textbook on quantum mechanics, e.g., \cite{Ballentine, Landau}. Eigenstates will be labeled by the \emph{principal} ($n$), \emph{orbital} ($l$), and \emph{magnetic} ($m$) \emph{quantum numbers}. \index{principal quantum number} \index{orbital quantum number} \index{magnetic quantum number} Energy eigenvalues are degenerate with respect to $l$ and $m$

\begin{eqnarray}
 \varepsilon (n,l,m) = -\frac{\mu c^2 \alpha^2}{2n^2} \label{enlm}
\end{eqnarray}

\noindent  Few low energy solutions are shown in table \ref{table:wavefunctions}, where $a_0 \equiv 4 \pi \hbar^2/(\mu e^2) \approx \hbar/(\alpha m c)$ denotes the \emph{Bohr radius}, \index{Bohr radius} and $\alpha \equiv e^2/(4 \pi \hbar c) \approx 1/137$ is the fine structure constant.

\begin{table}[h]
\caption{Normalized low energy solutions for non-relativistic hydrogen atom}
\begin{tabular*}{\textwidth}{@{\extracolsep{\fill}}c|cc}
\hline
 State($n,l,m$) &  Wave function $\psi_{\varepsilon}
(r, \theta, \phi)$ & Energy  ($\varepsilon$)    \cr
\hline
 $1S(1,0,0)$ & $ \frac{1}{\sqrt{\pi a_0^3}}e^{-r/a_0}$   & $-\mu c^2 \alpha^2/2$ = -13.6 eV\cr
 $2S(2,0,0)$ & $ \frac{1}{4 \sqrt{2\pi a_0^3}}(2 - \frac{r}{a_0})e^{-r/(2a_0)}$   & $-\mu c^2 \alpha^2/8$ = -3.4 eV \cr
 $2P(2,1,0)$ & $\frac{r}{4 a_0\sqrt{2\pi a_0^3}} e^{-r/(2a_0)} \cos \theta$   & $-\mu c^2 \alpha^2/8$ = -3.4 eV \cr
 $2P(2,1,-1)$ & $ \frac{r}{8 a_0\sqrt{\pi a_0^3}} e^{-r/(2a_0)} \sin \theta e^{-i \phi}$   & $-\mu c^2 \alpha^2/8$ = -3.4 eV \cr
 $2P(2,1,1)$ &  $ \frac{r}{8 a_0\sqrt{\pi a_0^3}} e^{-r/(2a_0)} \sin \theta e^{i \phi}$  & $-\mu c^2 \alpha^2/8$ = -3.4 eV \cr
 \hline
\end{tabular*}
\label{table:wavefunctions}
\end{table}

For further calculations we will need expectation values for inverse powers of $r$ in different eigenstates. For example

\begin{eqnarray*}
\langle r^{-1}\rangle(2S) &\equiv& \int d \mathbf{r} \psi_{2S}^*(\mathbf{r}) \frac{1}{r} \psi_{2S}(\mathbf{r})  =  \frac{1}{8  a_0^3}\int_0^{\infty} dr r \left(2 - \frac{r}{a_0} \right)^2 e^{-r/a_0} \\
&=&  \frac{ 1}{8 a_0^3} \left( 4\int_0^{\infty} dr r  e^{-r/a_0} - \frac{4}{a_0} \int_0^{\infty} dr r^2 e^{-r/a_0} + \frac{1}{a_0^2}\int_0^{\infty} dr r^3 e^{-r/a_0} \right) \\
&=&  \frac{1}{8 a_0^3} \left( 4a_0^2 - 8a_0^2 + 6a_0^2 \right)  = \frac{1}{4 a_0}
\end{eqnarray*}

\noindent These results are shown in Table \ref{table:powersr} along with probability densities at the origin $|\psi(0)|^2$.

\begin{table}[h]
\caption{Properties of low energy solutions for non-relativistic hydrogen atom}
\begin{tabular*}{\textwidth}{@{\extracolsep{\fill}}c|cccc}
\hline
 State &  $|\psi(0)|^2$ & $\langle r^{-1}\rangle$ & $\langle r^{-2}\rangle$ & $\langle r^{-3}\rangle$    \cr
\hline
 $1S$ & $ \frac{1}{\pi a_0^3}$   & $ \frac{1}{a_0}$ & $ \frac{2}{a_0^2}$  & \cr
 $2S$ & $ \frac{1}{8\pi a_0^3}$  & $ \frac{1}{4a_0}$ & $ \frac{1}{4a_0^2}$ &  \cr
 $2P$ & 0 &  $ \frac{1}{4a_0}$ & $ \frac{1}{12a_0^2}$ & $ \frac{1}{24a_0^3}$ \cr
 \hline
\end{tabular*}
\label{table:powersr}
\end{table}

\subsection{Relativistic energy corrections (orbital)}
\label{ss:hydrogen-relat}

In the preceding subsection we obtained energies $\varepsilon$ and wave functions $\Psi_{\varepsilon} $ for a simple model of the hydrogen atom in which the electron-proton interaction is approximated by the Coulomb potential $-e^2/(4 \pi r)$.  We can consider these results  as a zero-order approximation for the full exact solution.
Then  other interaction terms in (\ref{eq:12.25}) can be treated as
a perturbation $V_{pert}\equiv V_{orbit} + V_{spin-orbit}
+ V_{spin-spin}$. In the first approximation, this perturbation does
not affect the wave functions but shift energies \cite{Ballentine}. The resulting energy correction for the state $|\Psi_{\varepsilon} \rangle$  is given by the matrix element

\begin{eqnarray}
\Delta \varepsilon = \langle \Psi_{\varepsilon}| V_{pert}|\Psi_{\varepsilon} \rangle
\end{eqnarray}

\begin{figure}
\centering
\includegraphics[width=10cm,height=6cm] {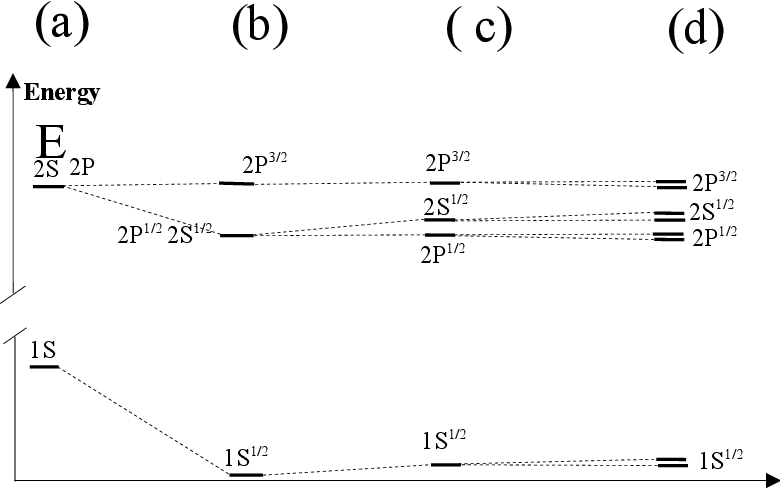} \caption{Low energy states of the
hydrogen atom: (a) the non-relativistic approximation (with the
Coulomb potential (\ref{eq:12.26})); (b) with the fine structure
(due to the \emph{orbit} (\ref{eq:12.27}) and \emph{spin-orbit}
(\ref{eq:12.29}) corrections); (c) with Lamb
 shifts (due to the 4th and higher order
radiative corrections); (d) with the hyperfine structure (due to the
\emph{spin-spin} corrections (\ref{eq:12.30})). Not to scale.} \label{fig:12.1}
\end{figure}

\noindent Then perturbations  $V_{orbit}$
and $V_{spin-orbit}$ are responsible for the \emph{fine structure}
\index{fine structure} of the hydrogen atom and
 $V_{spin-spin}$ is responsible for
the \emph{hyperfine structure} \index{hyperfine structure} (see Fig.
\ref{fig:12.1}). More details can be
found in \S 84 of ref. \cite{BLP}.

Let us first calculate energy level corrections due to the perturbation $V_{orbit}$.
We will take into account that $M \gg m$, thus ignoring terms proportional to inverse powers of $M$ in (\ref{eq:12.27})\footnote{For example, we see that the last term in (\ref{eq:12.27}) - which is Darwin's magnetic electron-proton potential - is negligibly small in this approximation.} and assuming that $\mu = m$. The energy correction due to the second term in (\ref{eq:12.27}) is

\begin{eqnarray}
\Delta \varepsilon_{relat} = -\frac{1}{8m^3c^2} \int d \mathbf{r} \psi^* \hat{Q}^4 \psi \label{eq:deltae2}
\end{eqnarray}

\noindent If $\psi_{\varepsilon}$ is an eigenfunction of $H$ with eigenvalue $\varepsilon$, then\footnote{Here we used (\ref{eq:10.54b}) and (\ref{eq:lapl-of-delta}).}

\begin{eqnarray*}
\hat{Q}^4 \psi_{\varepsilon} &=& \hat{Q}^2 \hat{Q}^2 \psi_{\varepsilon} \\
 &=& 2m \hat{Q}^2 \left( \varepsilon  + \frac{e^2}{4 \pi r} \right) \psi_{\varepsilon} \\
 &=& - 2m \hbar^2 \frac{\partial}{\partial \mathbf{r}} \left[ \varepsilon \frac{\partial \psi_{\varepsilon}}{\partial \mathbf{r}}  + \frac{\partial}{\partial \mathbf{r}}\left( \frac{e^2}{4 \pi r}\psi_{\varepsilon} \right)\right] \\
&=& -2m \hbar^2 \left[ \varepsilon \frac{\partial^2 \psi_{\varepsilon}}{\partial \mathbf{r}^2} + \frac{e^2}{4 \pi} \left(\frac{\partial^2}{\partial \mathbf{r}^2} \frac{1}{r} \right)\psi_{\varepsilon} + \frac{e^2}{2 \pi }\left( \frac{\partial}{\partial \mathbf{r}} \frac{1}{r} \cdot \frac{\partial \psi_{\varepsilon} }{\partial \mathbf{r}} \right)
+ \frac{e^2}{4 \pi r } \frac{\partial^2 \psi_{\varepsilon}}{\partial \mathbf{r}^2}   \right] \\
&=& -2m \hbar^2 \left[ - \frac{2m}{\hbar^2}\left(\varepsilon + \frac{e^2}{4 \pi r } \right)^2  \psi_{\varepsilon} - e^2 \delta(\mathbf{r})\psi_{\varepsilon} + \frac{e^2}{2 \pi }\left( \frac{\partial}{\partial \mathbf{r}} \frac{1}{r} \cdot \frac{\partial \psi_{\varepsilon}}{\partial \mathbf{r}}  \right)  \right]
\end{eqnarray*}

\noindent Using expression for the gradient in spherical coordinates\footnote{Here $\hat{\mathbf{r}}\equiv \mathbf{r}/r, \hat{\theta}, \hat{\phi} $ are unit vectors directed along directions of growth of the corresponding coordinates.}

\begin{eqnarray*}
\frac{\partial f(r, \theta, \phi)}{\partial \mathbf{r}} = \frac{\partial f}{\partial r} \hat{\mathbf{r}} + \frac{\hat{\theta}}{r \sin \phi} \frac{\partial f}{\partial \theta}  + \frac{\hat{\phi}}{r} \frac{\partial f}{\partial \phi} \label{eq:gradient}
\end{eqnarray*}

\noindent and inserting the resulting expression for $\hat{Q}^4 \psi_{\varepsilon}$ in (\ref{eq:deltae2}) we obtain

\begin{eqnarray*}
\Delta \varepsilon_{relat} = \frac{\hbar^2}{4m^2c^2 } \int d \mathbf{r} \psi_{\varepsilon}^* \left[ - \frac{2m}{\hbar^2}\left(\varepsilon + \frac{e^2}{4 \pi r } \right)^2 \psi_{\varepsilon} - e^2 \delta(\mathbf{r})\psi_{\varepsilon} - \frac{e^2}{2 \pi r^2} \frac{\partial \psi_{\varepsilon}}{\partial r}   \right]
\end{eqnarray*}

\noindent The last term in square brackets can be evaluated as\footnote{we take into account that $\psi_{\varepsilon}(r, \theta, \phi) \to 0$ as $r \to \infty$}

\begin{eqnarray}
&\ & -\frac{e^2}{2 \pi} \int d \mathbf{r} \psi_{\varepsilon}^*  \frac{1}{r^2} \frac{\partial \psi_{\varepsilon}}{\partial r} = -\frac{e^2}{2 \pi} \int_{0}^{2 \pi} d \phi \int_{0}^{\pi}\sin \theta d \theta \int_{0}^{\infty} dr  \psi_{\varepsilon}^*  \frac{\partial \psi_{\varepsilon}}{\partial r} \nonumber \\
&=& -\frac{e^2}{4 \pi} \int_{0}^{2 \pi} d \phi \int_{0}^{\pi}\sin \theta d \theta \int_{0}^{\infty} dr    \frac{\partial |\psi_{\varepsilon}|^2}{\partial r} = \frac{e^2}{4 \pi} \int_{0}^{2 \pi} d \phi \int_{0}^{\pi}\sin \theta d \theta |\psi_{\varepsilon}(0)|^2 \nonumber \\
&=& e^2 |\psi_{\varepsilon}(0)|^2 \label{eq:psirpsi}
\end{eqnarray}

\noindent The second term in square brackets is

\begin{eqnarray*}
&\ &  -e^2 \int d \mathbf{r} \psi_{\varepsilon}^*   \delta(\mathbf{r})\psi_{\varepsilon} = -e^2 |\psi_{\varepsilon}(0)|^2
\end{eqnarray*}

\noindent so it cancels with (\ref{eq:psirpsi}) and

\begin{eqnarray*}
\Delta \varepsilon_{relat} &=& -\frac{1}{2mc^2 } \int d \mathbf{r} \psi_{\varepsilon}^*  \left(\varepsilon^2 + \frac{e^2\varepsilon }{2 \pi r }  + \frac{e^4}{16 \pi^2 r^2 } \right) \psi_{\varepsilon}    \\
&=& -\frac{1}{2mc^2} \left(\varepsilon^2 + \frac{e^2\varepsilon }{2 \pi  } \langle r^{-1}\rangle + \frac{e^4}{16 \pi^2  } \langle r^{-2}\rangle\right) \label{eq:10.58a}
\end{eqnarray*}

\noindent Energy correction due to the third term in (\ref{eq:12.27}) is

\begin{eqnarray}
\Delta \varepsilon_{contact} &=& \frac{e^2 \hbar^2}{8 m^2 c^2} \int d \mathbf{r} \delta(\mathbf{r} )|\psi_{\varepsilon}(\mathbf{r})|^2 = \frac{e^2 \hbar^2}{8 m^2 c^2} |\psi_{\varepsilon}(0)|^2 \label{eq:contact}
\end{eqnarray}

\noindent Using data from Tables \ref{table:wavefunctions} and \ref{table:powersr} we obtain orbital energy corrections for individual states $\psi_{nlm}$ as shown in the 2nd and 3rd rows of Table \ref{table:corrections2}.

\begin{table}[h]
\caption{2nd order perturbative relativistic energy corrections to low-lying states of the hydrogen atom. }
\begin{tabular*}{\textwidth}{@{\extracolsep{\fill}}l|cccc}
\hline
  &  $1S^{1/2}$ & $2S^{1/2}$ & $2P^{1/2}$ & $2P^{3/2}$    \cr
\hline
non-relativistic energy (\ref{enlm}) & $-\frac{mc^2 \alpha^2}{2}$  & $-\frac{mc^2 \alpha^2}{8}$ & $-\frac{mc^2 \alpha^2}{8}$ & $-\frac{mc^2 \alpha^2}{8}$ \cr \hline
 Energy corrections: & &  &  &     \cr
\hline
relativistic   (\ref{eq:deltae2}) $-\frac{1}{8 m^3c^2} \langle Q^4 \rangle$ & $ -\frac{5mc^2 \alpha^4}{8}$   & $ -\frac{13mc^2 \alpha^4}{128}$ & $ -\frac{7mc^2 \alpha^4}{384}$ & $-\frac{7mc^2 \alpha^4}{384}$ \cr \cr
contact (\ref{eq:contact}) $\frac{e^2 \hbar^2}{8 m^2c^2} \langle \delta(\mathbf{r}) \rangle$ & $ \frac{mc^2 \alpha^4}{2}$   & $ \frac{mc^2 \alpha^4}{16}$ & 0 & 0 \cr \cr
spin-orbit (\ref{eq:spin-orbit}) $\frac{e^2 }{8 \pi m^2c^2}\langle \frac{\mathbf{L} \cdot \mathbf{S}_{el}}{r^3} \rangle$ & 0   & 0 & $ -\frac{mc^2 \alpha^4}{48}$& $ \frac{mc^2 \alpha^4}{48}$ \cr \cr
Total correction   & $ -\frac{mc^2 \alpha^4}{8}$  & $ -\frac{5mc^2 \alpha^4}{128}$ & $ -\frac{5mc^2 \alpha^4}{128}$ & $ \frac{mc^2 \alpha^4}{384}$ \cr \cr
\hline
\end{tabular*}
\label{table:corrections2}
\end{table}

\subsection{Relativistic energy corrections (spin-orbital)}
\label{ss:hydrogen-spin-orb}

Let us now consider the effect of the spin-orbit interaction (\ref{eq:12.29})

\begin{eqnarray}
 \hat{V}_{spin-orbit}
&\approx&
 \frac{e^2  [\mathbf{r} \times \hat{\mathbf{q}}]
\cdot \hat{\mathbf{S}}_{el}}{8 \pi m^2 c^2 r^3} = \frac{e^2  \mathbf{L}
\cdot \hat{\mathbf{S}}_{el}}{8 \pi m^2 c^2 r^3} \label{eq:spin-orbit}
\end{eqnarray}

\noindent where $\mathbf{L} = [\mathbf{r} \times \hat{\mathbf{Q}}]$ is the orbital angular momentum of the atom. This interaction does not act on states with $l=0$, which are eigenvectors of the orbital angular momentum operator $L^2$ with eigenvalue 0. So, we need to consider only  $2P$-states, where $l=1$.

Totally, there are 6 different substates in $2P$: those with different combinations of $l= -1, 0, 1$ and $s=-1/2, 1/2$. In these substates the total angular momentum\footnote{Here we ignore the proton's spin $ \mathbf{S}_{pr} $ whose contribution to the energy can be ignored in our approximation.} $\mathbf{J} = \mathbf{L} + \mathbf{S}_{el} $ can be either $j \hbar =(1-(1/2))\hbar=\hbar/2$ or $j \hbar=(1+(1/2))\hbar=3 \hbar/2 $. So, the 6 substates separate into two groups. One group of two states corresponds to $j=1/2$. It is denoted by $2P^{1/2}$. The other group of four states corresponds to $j=3/2$ and is denoted by $2P^{3/2}$. The non-perturbed Hamiltonian

\begin{eqnarray}
H_{e-p} = \frac{Q^2}{2 m} - \frac{e^2}{4 \pi r} \label{eq:Hdcoulomb}
\end{eqnarray}

\noindent commutes with the orbital angular momentum operator $\mathbf{L}$, with the electron spin operator $\mathbf{S}_{el} $ and with the total angular momentum operator $\mathbf{J}$. So, all six substates are degenerate with respect to (\ref{eq:Hdcoulomb}).

On the other hand, the total Hamiltonian (\ref{eq:12.25}) commutes only with the total angular momentum $\mathbf{J}$ and it does not commute with $\mathbf{L} $ and $\mathbf{S}_{el} $ separately. Therefore, the total energies of the two groups $2P^{1/2}$ and $2P^{3/2}$ may be different. Let us demonstrate the effect of perturbation (\ref{eq:spin-orbit}) on the state $2P^{1/2}$. We use formula

\begin{eqnarray*}
J^2 = (\mathbf{L} + \mathbf{S}_{el})^2 = L^2 + S_{el}^2 + 2 (\mathbf{L} \cdot \mathbf{S}_{el})
\end{eqnarray*}

\noindent Then

\begin{eqnarray*}
J^2 \psi_{2P^{1/2}} &=& \hbar^2 j(j+1) \psi_{2P^{1/2}} = 3/4 \hbar^2  \psi_{2P^{1/2}} \\
L^2 \psi_{2P^{1/2}} &=& \hbar^2 l(l+1) \psi_{2P^{1/2}} = 2 \hbar^2  \psi_{2P^{1/2}} \\
S_{el}^2 \psi_{2P^{1/2}} &=& \hbar^2 s(s+1) \psi_{2P^{1/2}} = 3/4 \hbar^2  \psi_{2P^{1/2}} \\
(\mathbf{L} \cdot \mathbf{S}_{el}) \psi_{2P^{1/2}} &=& \frac{(J^2 - L^2 - S_{el}^2)}{2}  \psi_{2P^{1/2}} = \frac{(3/4 - 2 - 3/4)\hbar^2}{2}   \psi_{2P^{1/2}} \\
&=& -\hbar^2  \psi_{2P^{1/2}} \\
 \Delta \varepsilon_{spin-orbit} (2P^{1/2})
&=&    -\frac{e^2 \hbar^2 }{8 \pi m^2 c^2} \langle r^{-3}\rangle =  -\frac{m c^2 \alpha^4 }{48 }
\end{eqnarray*}

\noindent A similar calculation gives us the spin-orbit correction to the energy of $2P^{3/2}$

\begin{eqnarray*}
 \Delta \varepsilon_{spin-orbit} (2P^{3/2})
&=&   \frac{m c^2 \alpha^4 }{48 }
\end{eqnarray*}

One can see from Table \ref{table:corrections2} that the total 2nd order  energy corrections to states $2S^{1/2}$ and $2P^{1/2}$ are the same. So, these two states remain degenerate in our approximation

\begin{eqnarray}
\varepsilon(2S^{1/2}) - \varepsilon(2P^{1/2})  = 0 \label{eq:2p12}
\end{eqnarray}

\noindent In chapter \ref{ch:hydrogen-revisited} we will take into account higher perturbation orders of the dressed theory and find a small energy gap between the $2S^{1/2}$ and $2P^{1/2}$ levels, which is known as  the \emph{Lamb shift}.

The description of the hydrogen atom presented here is by no means new or original. Similar calculations can be found in many textbooks on quantum mechanics. However, these textbook calculations always assume the Coulomb potential $-e^2/(4 \pi r)$ between the proton and the electron as given, and this assumption is never properly justified. Usually, the introduction of the Coulomb potential is explained by a reference to classical electrodynamics, but we do not want to base our fundamental relativistic quantum theory on such a shaky foundation as classical Maxwell's theory. Unfortunately, the interaction between the two charges cannot be extracted from the basic formalism of QED as well.  It is impossible to recognize traces of the Coulomb potential among  components\footnote{some of them even divergent} of the QED field-based interaction operator (\ref{eq:11.61}).

The true value of the dressed particle approach presented here is that it shows a clear and rigorous path from the abstract and divergent QED Hamiltonian to the intuitive and physically transparent picture of particles interacting via familiar action-at-a-distance forces.

\chapter{DECAYS} \label{ch:decays}

\begin{quote}
 \emph{Many things are incomprehensible to us not because our
 comprehension is weak, but because those things
are not
 within the frames of our comprehension.}

\small
\hspace{1in} Kozma Prutkov
\normalsize
\end{quote}

\vspace {0.5in}

Our formulation of quantum theory in the Fock space with undetermined particle numbers gives us an opportunity to describe not just
inter-particle interactions, but also processes of particle creation and
absorption. The simplest example of such a process is
the decay of an unstable particle.  This is the topic of the present chapter.

Unstable particles are interesting for several
reasons. First, an unstable particle  is a rare example of a quantum
interacting system whose time evolution can be observed relatively
easily.  This time evolution is especially simple, because in many cases it can be
described by just one parameter - the non-decay probability
$\omega(t)$. Second, a rigorous description of the decay is possible in a small
portion of the Fock space that contains only states of the particle and its
decay products, so a rather accurate solution of this time-dependent problem can be obtained in a closed form.

In sections \ref{sc:general-decay} - \ref{ss:breit-wigner} we will discuss the decay law of
an unstable system at rest.
Most of this material is rather traditional. Less familiar ideas will be presented in the last two sections of this chapter.
In section \ref{sc:formalism} we will be interested in the decay law
observed from a moving reference frame.
In section \ref{sc:discus} we are going to show that the famous Einstein's time dilation
formula is not exactly applicable to such decays.

\section{Unstable system at rest} \label{sc:general-decay}

In this section we will pursue two goals. The first goal is to provide a
preliminary material for our discussion of decays of moving
particles in sections \ref{sc:formalism} and \ref{sc:discus}. The
second goal is to derive a beautiful result, due to Breit and
Wigner, which explains why the time dependence of particle decays is
(almost) always exponential.

\subsection{Quantum mechanics  of particle decays}
\label{ss:general}

The decay of unstable particles is described mathematically by the
\emph{non-decay probability} \index{non-decay probability} which has
the following definition.
 Suppose that we have a piece of
radioactive material with
 $N$ unstable nuclei prepared simultaneously at time $t=0$
 and denote
$N_u(t)$  the number of nuclei that remain undecayed at time $t>0$.
So, at each time point the piece of radioactive material can be
characterized by the ratio $N_u(t)/N$.

In the spirit of quantum mechanics, we will treat $N$
unstable nuclei or particles as an ensemble of identically prepared systems
and consider the ratio $N_u(t)/N$ as a property of a single particle -- the probability of finding this particle in the
undecayed state. Then the non-decay probability $\omega(t)$  is
defined as a large $N$ limit

\begin{eqnarray}
\omega(t) =\lim_{N \to \infty} N_u(t)/N \label{eq:dec_law}
\end{eqnarray}

\noindent Function $\omega(t)$ will be
called the \emph{decay law} of the particle. \index{decay law}

 Let us now turn to the description of an isolated unstable system from the
point of view of quantum theory.
 We will consider a model theory with three
 particles
$a$, $b$, and $c$, so that particle $a$ is massive and unstable. In order to simplify calculations and avoid being
distracted by issues that are not relevant to the problem at hand we
assume that particle $a$ is spinless and has only one decay channel.
 The \emph{decay products} \index{decay products} $b$ and $c$
are assumed to be stable, and their
 masses satisfy the inequality

\begin{eqnarray}
m_a > m_b + m_c \label{eq:masses}
\end{eqnarray}

\noindent which makes the  decay $a \to b + c $ energetically
possible.  Decays of elementary particles are forbidden in quantum electrodynamics, because, as we discussed in subsection \ref{ss:five-types}, there are no decay type interactions\footnote{like (\ref{eq:13.1})} in the Hamiltonian of QED. So, our analysis in this chapter is more relevant to decays governed by weak nuclear interactions. However, even in QED decays may occur in compound systems, e.g., in the hydrogen atom. In section \ref{sc:spontaneous} we will consider a specific example in which $a$ and $b$ are stationary states of the atom ($0< m_b < m_a$) and $c$ is the photon ($m_c = 0$).  Most result from the present chapter will be applicable there without modifications.

Observations performed on the unstable system may result in only two basic
outcomes. One can find either a non-decayed particle $a$ intact or its
decay products $b+c$. Thus it is appropriate to describe  states
of this system in just two sectors of the Fock space\footnote{In
principle, a rigorous description of systems involving these three types
of particles must be formulated in the full \emph{Fock space}  where particle numbers $N_a, N_b$, and $ N_c$ are allowed to
take any values from zero to infinity. However, in most cases the interaction between decay products $b$ and $c$ in
the final state can be ignored. Creation of additional particles due
to this interaction can be ignored too. So, limiting our attention  only to the
Fock subspace (\ref{eq:tensor-product}) is a
reasonable approximation.}

\begin{eqnarray}
\mathcal{H} = \mathcal{H}_ {a} \oplus \mathcal{H}_ {bc}
\label{eq:tensor-product}
\end{eqnarray}

\noindent  where $\mathcal{H}_ {a}$ is the subspace of states of the
unstable particle $a$ and  $\mathcal{H}_ {bc} \equiv \mathcal{H}_ {b} \otimes \mathcal{H}_
{c}$ is the orthogonal subspace of the decay products.

Now we can introduce a Hermitian operator $T$  that corresponds to
the experimental proposition ``particle $a$ exists.''  The operator $T$ can be
fully defined by its eigensubspaces and eigenvalues. When a
measurement performed on the unstable system finds it in a state
corresponding to the particle $a$, then the value of $T$ is 1. When
the decay products $b+c$ are observed, the value of $T$ is 0.
  Apparently,  $T$
is the projection operator on the subspace $\mathcal{H}_a$.
 For each normalized state vector $| \Phi \rangle \in \mathcal{H}$, the
probability of finding the unstable particle $a$  is given by the
expectation value of this projection

\begin{eqnarray}
\omega = \langle \Phi | T | \Phi \rangle \label{eq:exp-val-non-dec}
\end{eqnarray}

\noindent Alternatively, one can say that $\omega$ is a square of
the norm of the projection $ T | \Phi \rangle$

\begin{eqnarray}
\omega &=& \langle \Phi | TT | \Phi \rangle = \Vert T | \Phi \rangle
\Vert^2 \label{square}
\end{eqnarray}

\noindent where we used property $T^2 = T$ from Theorem
\ref{TheoremA.13}.

Any  vector $
 |\Psi\rangle \in \mathcal{H}_a
$ describes  a state in which the unstable particle $a$ is
found with 100\% certainty. We will assume that the unstable system was prepared in
such a state $| \Psi(0) \rangle$ at time $t=0$

\begin{eqnarray}
\omega( 0) &=& \langle \Psi(0)| T | \Psi(0) \rangle = 1 \label{eq:3f}
\end{eqnarray}

To study the time evolution (= decay) of this state, we need to know the full Hamiltonian $H = H_0 + V$ in trhe Hilbert space $\mathcal{H}$. The non-interacting part $H_0$ can be easily constructed by usual rules from subsection \ref{ss:non-int-rep}

For the interaction part we choose the simplest operator of the decay type (\ref{eq:9.49}) that can be
responsible for the process $a \to b+c$ is\footnote{Note that in order
to have a Hermitian Hamiltonian we need to include in the
interaction both the term $b^{\dag}c^{\dag}a$ responsible for the
decay and the term $a^{\dag} bc$ responsible for the inverse process
$b+c \to a$. Due to the relation (\ref{eq:masses}), these two terms
have non-empty energy shells, so, according to our classification in
subsection \ref{ss:five-types}, they belong to the ``decay'' type.}

\begin{eqnarray}
V = \int d \mathbf{p} d \mathbf{q} \Bigl(G(\mathbf{p},
\mathbf{q})a^{\dag}_{\mathbf{p+q}} b_{\mathbf{p}}c_{\mathbf{q}} +
G^*(\mathbf{p},
\mathbf{q})b^{\dag}_{\mathbf{p}}c^{\dag}_{\mathbf{q}}
a_{\mathbf{p+q}} \Bigr) \label{eq:13.1}
\end{eqnarray}

\noindent As expected, the interaction operator (\ref{eq:13.1})
leaves invariant the sector $\mathcal{H} = \mathcal{H}_ {a} \oplus
\mathcal{H}_ {bc} $ of the total Fock space.

Then the time evolution of
the initial state $|\Psi(0) \rangle$ is given by formula (\ref{eq:psi-time})\footnote{In this chapter we are working in the Schr\"odinger picture.}

\begin{eqnarray}
| \Psi (t)  \rangle &=& e^{-\frac{i}{\hbar}Ht} | \Psi (0)\rangle
\label{eq:3g}
\end{eqnarray}

\noindent and the decay law  is

\begin{eqnarray}
\omega( t) &=& \langle \Psi(0)| e^{\frac{i}{\hbar}Ht} T
e^{-\frac{i}{\hbar}Ht} | \Psi (0)\rangle \label{eq:3c}
\end{eqnarray}

\noindent It is clear that our Hamiltonian $H$
does not commute with the
projection operator $T$

\begin{eqnarray}
 [H, T] \neq 0
\label{eq:44b}
\end{eqnarray}

\begin{figure}
\centering
\includegraphics {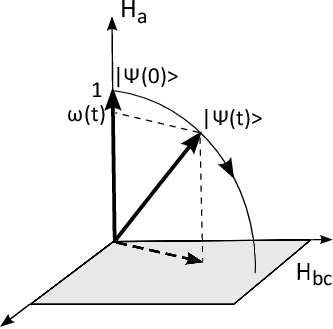} \caption{Time evolution of the state
vector of an unstable system.} \label{fig:13.1}
\end{figure}

\noindent Therefore, the subspace
$\mathcal{H}_a$ of states of the particle $a$ is not invariant
with respect to time translations, so that the decay law (\ref{eq:3c}) is a nontrivial function of time.

Now we can suggest a schematic visual representation of the decay
process in the Hilbert space. In Fig. \ref{fig:13.1} we show the
full Hilbert space $\mathcal{H}_ {a} \oplus \mathcal{H}_ {bc} $ as a
sum of two orthogonal subspaces $\mathcal{H}_ {a}$ and $\mathcal{H}_
{bc} $. We assume that  the initial normalized state vector $|\Psi(0)
\rangle$ at time $t=0$ lies entirely in the subspace
$\mathcal{H}_a$. So that the non-decay probability $\omega(0)$  is
equal to 1 as in equation (\ref{eq:3f}).  At later times $t > 0$ the state vector
$| \Psi(t) \rangle = e^{ -\frac{i}{\hbar}Ht} | \Psi(0)  \rangle$
develops a component\footnote{shown by a broken-line arrow in the figure} lying in the subspace of decay products
$\mathcal{H}_ {bc} $. As we will see later, in our model the state vector gradually moves from the subspace of unstable particle
$\mathcal{H}_ {a}$ to the subspace of decay products $\mathcal{H}_
{bc}$,  so that the non-decay probability $\omega(t)$ decreases with
time monotonically, while the probability $(1 - \omega(t))$ of finding the decay products grows.  Moreover, we are going to see that under very broad range of conditions, the decay law takes the unversal exponential form  $\omega(t) \approx e^{-\Gamma t}$, where the decay rate $\Gamma$ is controlled by the strength of the decay interaction $V$.

Before calculating the decay law (\ref{eq:3c}) we will need to do
some preparatory work first. In subsections \ref{sc:non-int} - \ref{ss:instant} we are going to construct two useful bases. One is
the basis  $|\mathbf{p} \rangle$ of eigenvectors of the total momentum
operator $\mathbf{P}_0$ in $\mathcal{H}_a$. Another one is the basis
$|\mathbf{p},m \rangle$ of common eigenvectors of $\mathbf{P}_0$ and
the interacting mass operator $M = \sqrt{H^2 - P_0^2c^2}/c^2$ in $\mathcal{H}$.

\subsection{Non-interacting representation of the Poincar\'e group}
\label{sc:non-int}

Let us first consider a simple case when the interaction responsible
for the decay is ``turned off,'' e.g., by setting $G(\mathbf{p}, \mathbf{q})$ in (\ref{eq:13.1}). This means that dynamics of the
system is governed by  the \emph{non-interacting} representation
\index{non-interacting representation}
 of the Poincar\'e group $U_g^0$
in $\mathcal{H}$.\footnote{see subsection \ref{ss:non-int-fock}}  This representation is constructed in accordance
with the structure of the Hilbert space (\ref{eq:tensor-product})
as

\begin{eqnarray}
U_g^0 \equiv U_g^a \oplus (U_g^b \otimes U_g^c) \label{eq:non-int}
\end{eqnarray}

\noindent  where $U_g^{a}$, $U_g^{b}$, and $U_g^{c}$ are unitary
irreducible representations
 of the Poincar\'e group
corresponding to the particles $a$, $b$, and $c$, respectively. Generators of this representation will be denoted by
$\mathbf{P}_0$, $\mathbf{J}_0$, $H_0$, and $\mathbf{K}_0$. According
to (\ref{eq:masses}), the operator of the non-interacting \emph{mass}
\index{mass}

\begin{eqnarray*}
M_0 = +\frac{1}{c^2} \sqrt{H_0^2 - P_0^2 c^2}
\end{eqnarray*}

\noindent  has a continuous spectrum in the interval $[m_b+m_c,
\infty)$ and a discrete point $m_a$ embedded in this interval.

From definition (\ref{eq:non-int}) it is clear that  the subspaces
$\mathcal{H}_ {a}$ and $\mathcal{H}_ {bc}$ are separately invariant with
respect to   $U_g^0$. Moreover,  the projection operator $T$
commutes with non-interacting generators

\begin{eqnarray}
[T,\mathbf{P}_0] = [T,\mathbf{J}_0] = [T,\mathbf{K}_0] =  [T,H_0]=0
\label{eq:T-comm}
\end{eqnarray}

Exactly as we did in subsection \ref{ss:momentum-basis}, we can use
the non-interacting representation $U_g^0$  to build a basis
 $|\mathbf{p} \rangle$ of eigenvectors of the momentum
operator $\mathbf{P}_0$ in the subspace $\mathcal{H}_a$. Then any
state $| \Psi \rangle \in \mathcal{H}_a$ can be represented by a
linear combination of these basis vectors

\begin{eqnarray}
| \Psi \rangle &=&  \int d\mathbf{p} \psi(\mathbf{p}) |\mathbf{p}
\rangle \label{eq:expansion}
\end{eqnarray}

\noindent and the projection operator $T$ can be written as\emph{compare with equation (\ref{eq:7.14})}

\begin{eqnarray}
T &=& \int d\mathbf{p} |\mathbf{p} \rangle \langle \mathbf{p} |
\label{eq:projector}
\end{eqnarray}

\subsection{Normalized eigenvectors of momentum}
\label{ss:normal-mom}

Improper basis vectors $|\mathbf{p} \rangle$ are convenient for writing
arbitrary states $|\Psi \rangle \in \mathcal{H}_a$ as linear
combinations (\ref{eq:expansion}). However vectors $|\mathbf{p}
\rangle$ themselves are not good representatives of quantum states,
because they are not normalized.\footnote{see subsection \ref{ss:spectral}} For example, the momentum space
``wave function'' of the basis vector $|\mathbf{q} \rangle$ is a
delta function

\begin{eqnarray}
 \psi_{\mathbf{q}} (\mathbf{p}) &=& \langle \mathbf{p} |\mathbf{q} \rangle
 = \delta(\mathbf{p} -\mathbf{q}) \label{eq:psi-delta}
\end{eqnarray}

\noindent and the corresponding  ``probability'' of finding the
particle is infinite

\begin{eqnarray*}
 \int d \mathbf{p} |\psi_{\mathbf{q}} (\mathbf{p})|^2 &=&
 \int d \mathbf{p} |\delta(\mathbf{p} -\mathbf{q})|^2 = \infty
\end{eqnarray*}

\noindent Therefore, improper states like $|\mathbf{q} \rangle$ cannot be used in
formula (\ref{eq:exp-val-non-dec}) for calculating the decay law of states with definite  (or almost definite) momentum
$\mathbf{p}_0$. For such calculations we should use other (proper) state
vectors  $|\mathbf{p}_0)$
whose normalized momentum-space wave functions are sharply localized near
$\mathbf{p}_0$. In order to satisfy the normalization condition

\begin{eqnarray*}
 \int d \mathbf{p} |\psi (\mathbf{p})|^2 &=& 1
\end{eqnarray*}

\noindent  the wave function of $|\mathbf{p}_0)$ may be formally
represented as a square root of the Dirac's delta
function\footnote{Another way to achieve the same goal would be to
keep the delta-function representation (\ref{eq:psi-delta}) of
definite-momentum states, but use (formally vanishing) normalization
factors, like $N=(\int d \mathbf{p} |\psi (\mathbf{p})|^2)^{-1/2}$.
Perhaps such manipulations with infinitely large and infinitely
small numbers can be justified within \emph{non-standard analysis}
\cite{Friedman}. }

\begin{eqnarray}
\psi_{\mathbf{p}_0}(\mathbf{p})=\sqrt{\delta (\mathbf{p} - \mathbf{p}_0)}
\label{eq:square-delta}
\end{eqnarray}

\subsection{Interacting representation of the Poincar\'e group}
\label{ss:instant}

In order to study dynamics of the unstable system we need to define
an interacting unitary representation of the Poincar\'e group in the
Hilbert space $\mathcal{H}$. This
 representation will allow us to relate
results of measurements in different reference frames. In this
section we will take the point of view of the observer at rest. We
will discuss particle decay from the point of view of a moving
observer in sections \ref{sc:formalism} and \ref{sc:discus}.

Let us now ``turn on'' the interaction (\ref{eq:13.1}) responsible for the decay and
discuss the interacting representation $U_g$ of the Poincar\'e group
in $\mathcal{H}$ with generators $\mathbf{P}$, $\mathbf{J}$,
$\mathbf{K}$, and $H$. As usual, we prefer to work in the Dirac's
instant form of dynamics. Then the generators of space translations
and rotations are interaction-free

\begin{eqnarray*}
\mathbf{P}&=&\mathbf{P}_0 \\
\mathbf{J} &=& \mathbf{J}_0
\end{eqnarray*}

\noindent while generators of time translations (the Hamiltonian
$H$) and boosts contain  interaction-dependent
terms

\begin{eqnarray*}
 H &=& H_0+V \\
 \mathbf{K} &=& \mathbf{K}_0+\mathbf{Z}
\end{eqnarray*}

\noindent We will further assume that the interacting representation
$U_g$ belongs to the Bakamjian-Thomas form of dynamics\footnote{For the Bakamjian-Thomas theory see subsection \ref{ss:bakamjian}. The possibilities for
the decay interaction to be in other forms of dynamics will be
discussed in subsection \ref{sc:different2}.} in which the
interacting operator of mass $M$ commutes with the Newton-Wigner
position operator (\ref{eq:6.26})

\begin{eqnarray}
M &\equiv& c^{-2}\sqrt{H^2 - \mathbf{P}_0^2 c^2} \nonumber \\
\mbox{ } [\mathbf{R}_0, M] &=& 0 \label{eq:nw-comm}
\end{eqnarray}

Our next goal is to define the basis of common eigenvectors of
commuting operators $\mathbf{P}_0$ and $M$ in
$\mathcal{H}$.\footnote{In addition to these two operators, whose
eigenvalues are used for labeling eigenvectors $| \mathbf{p},m
\rangle$, there are other independent operators in the mutually
commuting set containing $\mathbf{P}_0$ and $M$. These are, for
example, the operators of the square of the total angular momentum
 $J_0^2$ and the projection of $\mathbf{J}_0$ on the $z$-axis $J_{0z}$. Therefore a unique
characterization of any basis vector requires specification of all
corresponding quantum numbers as $| \mathbf{p},m , j^2, j_z, \ldots
\rangle$. However these additional quantum numbers  are not relevant for our
discussion, and we omit them.} These  eigenvectors must satisfy
conditions

\begin{eqnarray}
\mathbf{P}_0 | \mathbf{p},m \rangle &=&
 \mathbf{p} | \mathbf{p},m \rangle
\label{eq:mom-eigen} \\
 M| \mathbf{p}, m \rangle &=&   m| \mathbf{p}, m \rangle
\label{eq:mass-eigen}
\end{eqnarray}

\noindent They are also eigenvectors of the interacting Hamiltonian
$H = \sqrt{M^2 c^4 + \mathbf{P}_0^2 c^2}$

\begin{eqnarray*}
 H| \mathbf{p}, m \rangle &=&
\omega_{ \mathbf{p}}| \mathbf{p}, m \rangle
\end{eqnarray*}

\noindent where $\omega_{ \mathbf{p}} \equiv \sqrt{m^2c^4 +
c^2p^2}$.
In the zero-momentum eigensubspace of the momentum operator
$\mathbf{P}_0$ we can introduce a basis $| \mathbf{0},m \rangle$ of
eigenvectors of the
 interacting mass $M$

\begin{eqnarray*}
\mathbf{P}_0 | \mathbf{0}, m \rangle &=& \mathbf{0} \\
M | \mathbf{0}, m \rangle &=& m| \mathbf{0}, m \rangle
\end{eqnarray*}

\noindent Then the basis $| \mathbf{p}, m \rangle$ in the entire
Hilbert space $\mathcal{H}$ can be built by formula\footnote{Compare
with (\ref{eq:7.3}) and (\ref{eq:7.16}). Here $\mathbf{K}$ is the interacting boost operator whose explicit form will not be required in our derivation.}

\begin{eqnarray*}
| \mathbf{p}, m \rangle =
\sqrt{\frac{mc^2}{\omega_{\mathbf{p}}}}e^{-\frac{ic}{\hbar}\mathbf{K}
\vec{\theta}} | \mathbf{0}, m \rangle
\end{eqnarray*}

\noindent where vector $\vec{\theta}$ is related to momentum by formula $\mathbf{p} = mc \vec{\theta} \theta^{-1} \sinh
\theta$. These improper eigenvectors are normalized to delta functions

\begin{eqnarray}
\langle \mathbf{q},m | \mathbf{p}, m' \rangle &=& \delta
(\mathbf{q}- \mathbf{p}) \delta (m - m') \label{eq:28a}
\end{eqnarray}

\noindent The actions of inertial transformations on these states
are found by the same method as in section \ref{sc:massive}. In
particular, for boosts along the $x$-axis and for time translations we
obtain\footnote{compare with equation (\ref{eq:7.17})}

\begin{eqnarray}
  e^{-\frac{ic}{\hbar}K_x \theta}|\mathbf{p}, m
\rangle &=&  \sqrt{\frac{\omega_{\Lambda
\mathbf{p}}}{\omega_{\mathbf{p}}}}
|\Lambda \mathbf{p},m \rangle \label{eq:29a} \\
e^{\frac{i}{\hbar}Ht} |\mathbf{p}, m \rangle &=& e^{\frac{i}{\hbar}
\omega_{ \mathbf{p}}t} | \mathbf{p},m \rangle \label{eq:29b} \\
\Lambda \mathbf{p}&=& \left(p_x \cosh \theta +
\frac{\omega_{\mathbf{p}}}{c}
 \sinh
\theta, p_y, p_z \right) \label{eq:lambdaz}
\end{eqnarray}

\noindent Next we notice that due to equations (\ref{eq:6.21}) and
(\ref{eq:nw-comm}) vectors $e^{\frac{i}{\hbar} \mathbf{R}_0 \cdot
\mathbf{p}}|\mathbf{0} , m \rangle$ also satisfy eigenvector
equations (\ref{eq:mom-eigen}) - (\ref{eq:mass-eigen}), so they must
be proportional to the basis vectors $|\mathbf{p}, m \rangle$

\begin{eqnarray}
  e^{\frac{i}{\hbar}
\mathbf{R}_0 \cdot \mathbf{p}}|\mathbf{0} , m \rangle = \gamma(\mathbf{p},m) |\mathbf{p}, m \rangle \label{eq:xxx}
\end{eqnarray}

\noindent where $\gamma(\mathbf{p},m)$ is a unimodular factor.
Unlike in (\ref{eq:alpha}), we cannot conclude that
$\gamma(\mathbf{p},m) = 1$. However, if the interaction is not
pathological we can assume that the factor $\gamma(\mathbf{p},m)$ is
smooth, i.e., without rapid oscillations.

 Obviously,  vector  $| \mathbf{0}
\rangle$ from the basis  (\ref{eq:alpha}) can be expressed as a
linear combination of
 zero-momentum basis vectors
$| \mathbf{0}, m \rangle$, so we can write\footnote{Here we assume
that interaction responsible for the decay  does not change the
spectrum of mass. In particular, we will neglect the possibility of
existence of bound states of particles $b$ and $c$, i.e., discrete
eigenvalues of $M$ below $m_b+m_c$. Then the spectrum of $M$
(similar to the spectrum of $M_0$) is continuous in the interval
$[m_b+m_c, \infty)$, and integration in (\ref{eq:integral}) should
be performed from $m_b + m_c$ to infinity. Note that this assumption does not hold in the example considered in subsection \ref{ss:eigenfunc}, where one discrete eigenvalue of the mass operator $M$ appears below the threshold $m_b + m_c$, as shown in Figs. \ref{fig:13.2} and \ref{fig:13.3}.}

\begin{eqnarray}
|\mathbf{0} \rangle &=& \int \limits_{m_b + m_c}^{\infty} dm \mu(m)
| \mathbf{0},m \rangle \label{eq:integral}
\end{eqnarray}

\noindent where $\mu(m)$ is yet unknown function, which depends on the choice of the
interaction Hamiltonian $V$ and satisfies equations

\begin{eqnarray}
\mu(m) &=& \langle \mathbf{0},m | \mathbf{0} \rangle \label{eq:7.135a} \\
\int \limits_{m_b + m_c}^{\infty} dm |\mu(m)|^2 &=& 1 \nonumber
\end{eqnarray}

\noindent  The physical meaning of $\mu(m)$ is the probability
amplitude for finding the value $m$ of the interacting mass $M$ in the initial unstable state
$| \mathbf{0} \rangle \in \mathcal{H}_a$. It will be referred to as the \emph{mass distribution} \index{mass distribution} of the unstable particle. We now use equations (\ref{eq:alpha}) and
(\ref{eq:integral}) to expand vectors $|\mathbf{p} \rangle \in \mathcal{H}_a$ in the
basis $| \mathbf{p},m \rangle$

\begin{eqnarray}
|\mathbf{p} \rangle &=&  e^{\frac{i}{\hbar} \mathbf{R}_0 \mathbf{p}}
|\mathbf{0} \rangle =    e^{\frac{i}{\hbar} \mathbf{R}_0 \mathbf{p}}
\int \limits_{m_b + m_c}^{\infty} dm \mu(m) | \mathbf{0},m
\rangle \nonumber \\
  &=&
\int \limits_{m_b + m_c}^{\infty} dm \mu(m) \gamma(\mathbf{p}, m) |
\mathbf{p},m \rangle \label{eq:p-expansion}
\end{eqnarray}

\noindent Then any state vector from the subspace $\mathcal{H}_a$
can be written as

\begin{eqnarray}
| \Psi \rangle &=& \int d\mathbf{p} \psi(\mathbf{p}) |\mathbf{p}
\rangle \label{eq:expansion3} \\
&=& \int d\mathbf{p} \int \limits_{m_b + m_c}^{\infty} dm \mu(m)
\gamma(\mathbf{p}, m) \psi(\mathbf{p}) | \mathbf{p},m \rangle
\label{eq:expansion2}
\end{eqnarray}

\noindent From (\ref{eq:28a}) we also obtain a useful formula

\begin{eqnarray}
\langle \mathbf{q} |\mathbf{p},m \rangle &=&
 \int \limits_{m_b + m_c}^{\infty} dm' \mu^*(m') \gamma^*(\mathbf{q}, m')
\langle \mathbf{q},m'
|\mathbf{p},m \rangle \nonumber \\
&=&
 \gamma^*(\mathbf{p}, m) \mu^*(m) \delta(\mathbf{q-p})
\label{eq:35a}
\end{eqnarray}

\subsection{Decay law}
\label{ss:the-non-decay}

We are now fully equipped to find the time evolution of the state vector
(\ref{eq:expansion3}) prepared within the subspace $\mathcal{H}_a$ at
time $t=0$. We apply equations
(\ref{eq:3g}),  (\ref{eq:29b}), and (\ref{eq:p-expansion})

\begin{eqnarray*}
| \Psi(t) \rangle & =& \int
 d\mathbf{p}  \psi (\mathbf{p})
e^{-\frac{i}{\hbar}Ht}
 |\mathbf{p}\rangle \\
& =& \int
 d\mathbf{p}  \psi (\mathbf{p}) \int \limits_{m_b + m_c}^{\infty} dm \mu(m)
\gamma(\mathbf{p}, m) e^{-\frac{i}{\hbar}Ht}
 |\mathbf{p},m \rangle \\
& =& \int
 d\mathbf{p}  \psi (\mathbf{p})
\int \limits_{m_b + m_c}^{\infty} dm \mu(m) \gamma(\mathbf{p}, m)
e^{-\frac{i}{\hbar}\omega_{\mathbf{p}}t}
 |\mathbf{p},m \rangle
\end{eqnarray*}

\noindent The inner product of this vector with $| \mathbf{q}
\rangle $ is found by using (\ref{eq:35a})

\begin{eqnarray}
&\mbox{ }& \langle \mathbf{q} |  \Psi( t) \rangle \nonumber
\\
&=&  \int d\mathbf{p}  \psi (\mathbf{p}) \int \limits_{m_b +
m_c}^{\infty} dm \mu(m)\gamma(\mathbf{p}, m)
e^{-\frac{i}{\hbar}\omega_{\mathbf{p}}t} \langle \mathbf{q}
|\mathbf{p} ,m \rangle  \nonumber
\\
&=& \int  d\mathbf{p} \psi (\mathbf{p})  \int \limits_{m_b +
m_c}^{\infty} dm  |\mu(m)|^2
 \gamma(\mathbf{p}, m)\gamma^*(\mathbf{p}, m)
e^{-\frac{i}{\hbar}\omega_{\mathbf{p}}t} \delta(\mathbf{q} -
\mathbf{p}) \nonumber \\
&=& \psi (\mathbf{q})  \int \limits_{m_b + m_c}^{\infty} dm
|\mu(m)|^2 e^{-\frac{i}{\hbar}\omega_{\mathbf{q}}t}
 \label{eq:36a}
\end{eqnarray}

\noindent
 The decay law is then obtained
by substituting (\ref{eq:projector}) in equation (\ref{eq:3c}) and using
(\ref{eq:36a})

\begin{eqnarray}
\omega( t) &=& \int  d\mathbf{q} \langle \Psi(t)
 |\mathbf{q} \rangle\langle \mathbf{q}   | \Psi( t)\rangle
= \int  d\mathbf{q} |\langle \mathbf{q}
  |\Psi ( t)\rangle\ |^2  \nonumber \\
&=& \int  d\mathbf{q} |\psi(\mathbf{q})|^2 \left| \int \limits_{m_b
+ m_c}^{\infty} dm
   \vert \mu(m)\vert^2 e^{-\frac{i}{\hbar}\omega_{
\mathbf{q}}t} \right|^2 \label{eq:50}
\end{eqnarray}

\noindent This formula is valid for the decay law of any state $|
\Psi \rangle \in \mathcal{H}_a$.  In the particular case of the
normalized state $|\mathbf{0} )$ whose wave function $\psi(\mathbf{q})$ is
well-localized in the momentum space near zero momentum,  we can set
approximately

\begin{eqnarray*}
\psi(\mathbf{q}) &\approx& \sqrt{\delta (\mathbf{q})}
\\
\int d\mathbf{q} |\psi(\mathbf{q})|^2 &\approx& \int d\mathbf{q} \delta (\mathbf{q}) = 1
\end{eqnarray*}

\noindent and\footnote{compare, for example, with equation (3.8) in
\cite{Fonda}}

\begin{eqnarray}
\omega_{|\mathbf{0})}( t)  &\approx& \left| \int \limits_{m_b +
m_c}^{\infty} dm
   \vert \mu(m)\vert^2 e^{-\frac{i}{\hbar}mc^2 t} \right|^2 \label{eq:50a}
\end{eqnarray}

\noindent  This result demonstrates that the decay law is fully
determined by the \emph{mass distribution} \index{mass distribution}  $ |\mu(m) |^2 $.
In the next section we will consider an exactly solvable
decay model in which the mass distribution and the decay law can be
 calculated explicitly.

\section{Breit-Wigner formula}
\label{ss:breit-wigner}

\subsection{Schr\"odinger equation}
\label{ss:eigenvalue}

In this section we are discussing decay of a particle at rest.
Therefore, it is sufficient to consider the subspace $\mathcal{H}_0
\subseteq \mathcal{H}$ of states having zero total momentum. The
subspace $\mathcal{H}_{\mathbf{0}}$ can be further decomposed into the direct
sum

\begin{eqnarray*}
\mathcal{H}_0 = \mathcal{H}_{a\mathbf{0}} \oplus \mathcal{H}_{(bc)\mathbf{0}}
\end{eqnarray*}

\noindent where\footnote{Recall that symbol $\cap$ denotes intersection of two subspaces in the Hilbert space.}

\begin{eqnarray*}
\mathcal{H}_{a\mathbf{0}} &=& \mathcal{H}_{\mathbf{0}} \cap \mathcal{H}_a \\
\mathcal{H}_{(bc)\mathbf{0}} &=& \mathcal{H}_{\mathbf{0}} \cap (\mathcal{H}_{b} \otimes
\mathcal{H}_{c})
\end{eqnarray*}

\noindent $\mathcal{H}_{a\mathbf{0}}$ is, of course, the one-dimensional
subspace spanning the zero-momentum vector $| \mathbf{0} \rangle$ of
the particle $a$. In the subspace $\mathcal{H}_{(bc)\mathbf{0}}$ of decay
products the total momentum is zero
$\mathbf{P}=\mathbf{p}_b+\mathbf{p}_c = 0$. Then 2-particle
basis states $| \vec{\rho} \rangle $ there can be labeled by eigenvectors
of the relative momentum operator

\begin{eqnarray*}
\vec{\rho} = \mathbf{p}_b = -\mathbf{p}_c
\end{eqnarray*}

\noindent and each state $| \Psi \rangle$ in $\mathcal{H}_{\mathbf{0}}$ can be written as
an expansion in the above basis $(| \mathbf{0} \rangle, |
\vec{\rho} \rangle) $

\begin{eqnarray*}
| \Psi \rangle = \xi | \mathbf{0} \rangle + \int d \vec{\rho}
\zeta(\vec{\rho} ) | \vec{\rho} \rangle
\end{eqnarray*}

\noindent The coefficients of this expansion can be represented
as an infinite column vector

\begin{eqnarray*}
| \Psi \rangle = \left[ \begin{array}{c}
\xi \\
\zeta(\vec{\rho}_1 ) \\
\zeta(\vec{\rho}_2 ) \\
\zeta(\vec{\rho}_3 ) \\
\ldots
\end{array} \right]
\end{eqnarray*}

\noindent whose first component is a complex number $\xi \equiv \langle \mathbf{0} | \Psi \rangle$. All other components are values of the
complex function $\zeta(\vec{\rho} )$ at different momenta
$\vec{\rho}$.\footnote{These are projections of $| \Psi \rangle$ on the relative momentum
eigenvectors $| \vec{\rho} \rangle$. Of course, the spectrum of $\vec{\rho}$ is
continuous and, strictly speaking, cannot be represented by a set
of discrete values $\vec{\rho}_i$. However, we can justify our approximation by the usual trick of placing the system in a finite box
(then the momentum spectrum becomes discrete) and then taking the
limit, in which the size of the box goes to infinity.  \label{discret}}
For brevity, we will use the following notation

\begin{eqnarray}
| \Psi \rangle = \left[ \begin{array}{c}
\xi \\
\zeta(\vec{\rho} )
\end{array} \right]
\label{eq:13.5}
\end{eqnarray}

\noindent  The vector $ | \Psi \rangle$ should be normalized, hence
its wave function satisfies the normalization condition

\begin{eqnarray}
|\xi|^2 + \int d \vec{\rho} |\zeta(\vec{\rho} )|^2  = 1
\label{eq:7.145b}
\end{eqnarray}

\noindent The probability of finding the unstable particle in the
state $| \Psi \rangle$ is

\begin{eqnarray*}
\omega = |\xi|^2
\end{eqnarray*}

\noindent and in the initial state

\begin{eqnarray}
|\mathbf{0} \rangle = \left[ \begin{array}{c}
1 \\
0
\end{array} \right] \label{eq:0-rangle}
\end{eqnarray}

\noindent the unstable particle is found with 100\% probability.

We can now find representations of various operators in the basis
 ($|\mathbf{0} \rangle, |\vec{\rho} \rangle $). The matrix of the free Hamiltonian  is diagonal

\begin{eqnarray*}
H_0 &=& \left[ \begin{array}{ccccc}
m_a c^2& 0 & 0 & 0 & \ldots \\
0 & \eta_{\vec{\rho}_1} c^2 & 0 & 0 & \ldots \\
0 & 0  & \eta_{\vec{\rho}_2} c^2& 0 & \ldots \\
0 & 0  & 0  & \eta_{\vec{\rho}_3} c^2& \ldots \\
0 & 0  & 0  & 0  & \ldots
\end{array} \right]
\equiv \left[ \begin{array}{cc}
m_a c^2& 0 \\
0 & \eta_{\vec{\rho}}c^2
\end{array} \right]
\end{eqnarray*}

\noindent where

\begin{eqnarray}
\eta_{\vec{\rho}} = \frac{1}{c^2}\left(\sqrt{m_b^2c^4 + c^2 \rho^2}
+ \sqrt{m_c^2c^4 + c^2 \rho^2}\right) \label{eq:13.6}
\end{eqnarray}

\noindent is the mass of the two-particle ($b+c$) system expressed
as a function of the relative momentum. In the subspace $\mathcal{H}_0$
interaction operator (\ref{eq:13.1}) takes the form

\begin{eqnarray*}
V &=& \int d \vec{\rho}  \Bigl(G(\vec{\rho},
-\vec{\rho})a^{\dag}_{\mathbf{0}} b_{\vec{\rho}}c_{-\vec{\rho}} +
G^*(\vec{\rho},
-\vec{\rho})b^{\dag}_{\vec{\rho}}c^{\dag}_{-\vec{\rho}}
a_{\mathbf{0}} \Bigr) \\
&\equiv& \int d \vec{\rho}  \Bigl(g(\vec{\rho})a^{\dag}_{\mathbf{0}}
b_{\vec{\rho}}c_{-\vec{\rho}} +
g^*(\vec{\rho})b^{\dag}_{\vec{\rho}}c^{\dag}_{-\vec{\rho}}
a_{\mathbf{0}} \Bigr)
\end{eqnarray*}

\noindent Its matrix representation is\footnote{Here symbol
$\int d \mathbf{q} g(\mathbf{q}) \ldots$ denotes a linear operator,
which produces a number $\int d \mathbf{q}   g(\mathbf{q})
\zeta(\mathbf{q})$ when acting on an arbitrary test function
$\zeta(\mathbf{q})$. The function $g(\vec{\rho})$ coincides with
$G(\mathbf{p}, \mathbf{q})$ on the zero-momentum subspace
$g(\vec{\rho}) \equiv G(\vec{\rho}, -\vec{\rho})$.}

\begin{eqnarray*}
V  &=& \left[ \begin{array}{ccccc}
0 & g(\vec{\rho}_1) & g(\vec{\rho}_2) & g(\vec{\rho}_3) & \ldots \\
g^*(\vec{\rho}_1) & 0 & 0 & 0 & \ldots \\
g^*(\vec{\rho}_2) & 0  & 0 & 0 & \ldots \\
g^*(\vec{\rho}_3) & 0  & 0  & 0 & \ldots \\
\ldots & 0  & 0  & 0  & \ldots
\end{array} \right] \equiv \left[ \begin{array}{cc}
0 & \int d \mathbf{q}   g(\mathbf{q}) \ldots \\
g^*(\vec{\rho})  & 0
\end{array} \right]
\end{eqnarray*}

\noindent  where $g(\vec{\rho})$ is the matrix element of the
interaction operator between states $| \mathbf{0} \rangle$ and  $|
\vec{\rho} \rangle$

\begin{eqnarray}
g(\vec{\rho})  = \langle \mathbf{0} | V | \vec{\rho} \rangle
\label{eq:gvecpi}
\end{eqnarray}

\noindent Then the action of the full  Hamiltonian $H = H_0 + V$ on
 vectors (\ref{eq:13.5}) is

\begin{eqnarray*}
H \left[ \begin{array}{c}
\xi  \\
\zeta(\vec{\rho} )
\end{array} \right]
&=& \left[ \begin{array}{ccccc}
m_ac^2 & g(\vec{\rho}_1) & g(\vec{\rho}_2) & g(\vec{\rho}_3) & \ldots \\
g^*(\vec{\rho}_1) & \eta_{\vec{\rho}_1}c^2 & 0 & 0 & \ldots \\
g^*(\vec{\rho}_2) & 0  & \eta_{\vec{\rho}_2}c^2 & 0 & \ldots \\
g^*(\vec{\rho}_3) & 0  & 0  & \eta_{\vec{\rho}_3}c^2 & \ldots \\
\ldots & 0  & 0  & 0  & \ldots
\end{array} \right] \left[ \begin{array}{c}
\xi \\
\zeta(\vec{\rho}_1 ) \\
\zeta(\vec{\rho}_2 ) \\
\zeta(\vec{\rho}_3 ) \\
\ldots
\end{array} \right]\\
&=& \left[ \begin{array}{c}
m_a c^2 \xi + \int d \mathbf{q}   g(\mathbf{q}) \zeta(\mathbf{q}) \\
g^*(\vec{\rho}_1)\xi + \eta_{\vec{\rho}_1}\zeta(\vec{\rho}_1 )c^2 \\
g^*(\vec{\rho}_2)\xi + \eta_{\vec{\rho}_2}\zeta(\vec{\rho}_2 )c^2 \\
g^*(\vec{\rho}_3)\xi + \eta_{\vec{\rho}_3}\zeta(\vec{\rho}_3 )c^2 \\
\ldots
\end{array} \right]  \equiv \left[ \begin{array}{c}
m_a c^2\xi + \int d \mathbf{q}   g(\mathbf{q}) \zeta(\mathbf{q}) \\
g^*(\vec{\rho}) \xi  + \eta_{\vec{\rho}} \zeta(\vec{\rho} )c^2
\end{array} \right]
\end{eqnarray*}

The next step is to find  eigenvalues (which we denote $mc^2$) and
eigenvectors

\begin{eqnarray}
| \mathbf{0}, m\rangle \equiv \left[ \begin{array}{c}
\mu^*(m)  \\
\zeta_m(\vec{\rho} )
\end{array} \right] \label{eq:7.145a}
\end{eqnarray}

\noindent of the Hamiltonian $H$.\footnote{Note that the function
$\mu(m)$ in (\ref{eq:7.145a}) is the same as in (\ref{eq:7.135a}).
So, in order to calculate the decay law (\ref{eq:50a}), all we need
to know is $|\mu(m)|^2$.} This task
 is equivalent to the solution of the following  system of linear
equations:

\begin{eqnarray}
m_ac^2 \mu^*(m) + \int  d \mathbf{q} g(\mathbf{\mathbf{q}})
\zeta_{m}(\mathbf{q})  &=& mc^2 \mu^*(m)
\label{eq:13.7}\\
g^*(\vec{\rho}) \mu^*(m) + \eta_{\vec{\rho}} c^2 \zeta_{m}(\vec{\rho})
&=& mc^2 \zeta_{m}(\vec{\rho}) \label{eq:13.8}
\end{eqnarray}

\noindent From  Equation (\ref{eq:13.8}) we obtain

\begin{eqnarray}
\zeta_{m}(\vec{\rho}) = \frac{g^*(\vec{\rho})\mu^*(m)}{mc^2 -
\eta_{\vec{\rho}} c^2} \label{eq:13.9}
\end{eqnarray}

\noindent Substituting this result to (\ref{eq:13.7}), we get
a non-linear equation determining the spectrum of eigenvalues $m$

\begin{eqnarray}
m - m_a =  \frac{1}{c^4}\int d \mathbf{q} \frac{|g(\mathbf{q})|^2}{m
- \eta_{\mathbf{q}}} \label{eq:13.10}
\end{eqnarray}

\noindent To comply with the law of conservation of the angular
momentum,\footnote{for a spinless particle $a$}  the function $|g(\mathbf{q})| $ must depend only on the
absolute value  $q \equiv |\mathbf{q}|$ of its argument. Therefore,
we can rewrite equation (\ref{eq:13.10}) in the form

\begin{eqnarray}
m - m_a &=&  F(m) \label{eq:13.11}
\end{eqnarray}

\noindent where

\begin{eqnarray}
F(m)  &\equiv& \int \limits _0 ^{\infty}  d q \frac{G(q)}{m - \eta_{q}}
\label{eq:13.12} \\
G(q) &\equiv& \frac{4 \pi q^2}{c^4} |g(q)|^2 \label{eq:Gq}
\end{eqnarray}

\noindent From the normalization condition
(\ref{eq:7.145b})

\begin{eqnarray}
|\mu(m)|^2 +\int  d \mathbf{q} |\zeta_{m}(\mathbf{q})|^2 &=& 1
\label{eq:m-norm}
\end{eqnarray}

\noindent and equation (\ref{eq:13.9}) we finally obtain a formula for the mass distribution

\begin{eqnarray}
|\mu(m)|^2 \Bigl(1  + \int  d q \frac{G(q) }{(m - \eta_q)^2}\Bigr)
&=& 1 \nonumber \\
|\mu(m)|^2 &=& \frac{1}{1- F'(m)} \label{eq:13.13}
\end{eqnarray}

\noindent where $F'(m)$ is the derivative of  $F(m)$.  So, in order
to calculate the decay law (\ref{eq:50a}) we just need to know the
derivative of $F(m)$ at points $m$ of the spectrum of the
interacting mass operator.\footnote{These points are solution of the equation (\ref{eq:13.10}).} The following subsection will detail such
a calculation.

\subsection{Finding function $\mu(m)$}
 \label{ss:eigenfunc}

\begin{figure}
\centering
\includegraphics {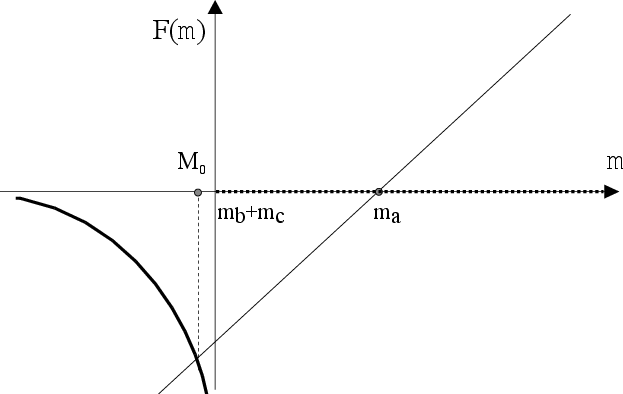} \caption{A graphical solution of equation
(\ref{eq:13.11}) for $m < m_b + m_c$. The thick dashed line indicates the continuous
part of the spectrum of the non-interacting mass operator. The
thick full line shows function $F(m)$ at $m < m_b+m_c$.}
\label{fig:13.2}
\end{figure}

Function $\eta_{\vec{\rho}}$ in (\ref{eq:13.6}) expresses
dependence of the total mass of the two decay products on their
relative momentum. This function has minimum value $\eta_{\mathbf{0}} = m_b +
m_c$ at $\vec{\rho}=\mathbf{0}$ and grows to infinity with increasing $\rho$. Then
the solution of equation (\ref{eq:13.11}) for values of $m$ in the interval $
[-\infty, m_b + m_c]$ is rather straightforward. In this region the
denominator in the integrand of (\ref{eq:13.12}) does not vanish, and $F(m)$ is a well-defined continuous function which tends to zero
at $m = -\infty$ and decreases monotonically  as $m$ grows. A
graphical solution of equation (\ref{eq:13.11}) in the
interval $[-\infty, m_b + m_c]$ can be obtained  as an intersection of the line $m -
m_a$ and the graph of the function $F(m)$ (point $M_0$ in Fig. \ref{fig:13.2}).
The corresponding value $m = M_0 < m_b + m_c$ is a discrete eigenvalue of the interacting
mass operator and the corresponding eigenstate is a superposition
of the unstable particle $a$ and its decay products $b+c$.

\begin{figure}
\centering
\includegraphics {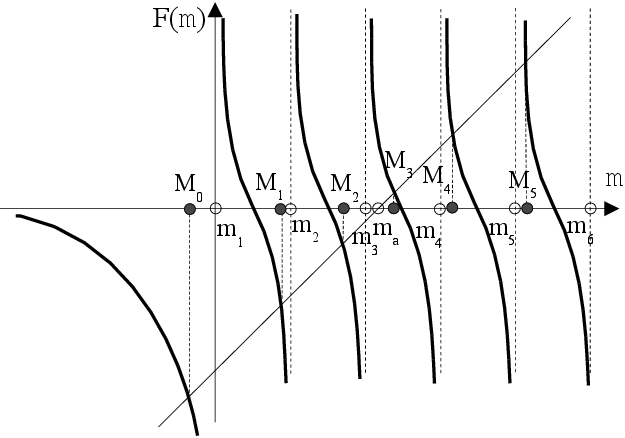} \caption{Spectra of the free (opened
circles, $H_0$) and interacting (full circles, $H=H_0+V$)
Hamiltonians.} \label{fig:13.3}
\end{figure}

Finding the spectrum of the interacting mass in the region $[m_b +
m_c, \infty]$ is more tricky due to a singularity in the integrand
of (\ref{eq:13.12}). Let us first discuss our approach
qualitatively, using graphical representation in Fig.
\ref{fig:13.3} and the discrete approximation for the spectrum of $\vec{\rho}$. Then equation (\ref{eq:13.11}) takes the form

\begin{eqnarray}
m - m_a = \frac{1}{c^2} \sum_{i=1}^{\infty}
\frac{|g(\vec{\rho}_i)|^2}{m - m_i} = F(m) \label{eq:13.14}
\end{eqnarray}

\noindent where $m_i \equiv \eta_{\vec{\rho}_i}$ are eigenvalues of the
non-interacting mass of the 2-particle system $b+c$ and the lowest
eigenvalue is $m_1 = m_b + m_c$. The function on the right hand side
of equation (\ref{eq:13.14}) is a superposition of functions
$c^{-2}|g(\vec{\rho}_i)|^2(m - m_i)^{-1} $ for all values of $i = 1,2,3,
\ldots$. These functions have singularities at points $m_i$.
Positions of these singularities are shown as open circles and
dashed vertical lines in Fig. \ref{fig:13.3}. The overall shape of
the function $F(m)$ in this approximation is shown by the thick full
line. According to equation (\ref{eq:13.14}), the spectrum of the
interacting mass operator can be found at points where the line $m -
m_a$ intersects with the graph  $F(m)$. These points $M_i$ are
shown by full circles in Fig. \ref{fig:13.3}. So, the derivatives
required in equation (\ref{eq:13.13}) are graphically represented as
slopes of the function $F(m)$ at points $M_1, M_2, M_3, \ldots$. The
difficulty is that in the continuous limit the distances
between points $m_i$ tend to zero, function $F(m)$ wildly
oscillates, and its derivative tends to infinity everywhere.

To overcome this difficulty we will use the following idea \cite{Livshitz}.
 Let us first change the integration variable in (\ref{eq:13.12})

\begin{eqnarray*}
z &=& \eta_{\rho}
\end{eqnarray*}

\noindent so that inverse function

\begin{eqnarray*}
 \rho  &=& \eta^{-1}(z)
\end{eqnarray*}

\noindent expresses the relative momentum $\rho$ as a function of
the total mass $z$ of the decay products. Then denoting

\begin{eqnarray}
\Gamma(z) \equiv 2 \pi \frac{d \eta^{-1}(z)}{dz} G(\eta^{-1}(z))
 \label{eq:gammam}
\end{eqnarray}

\noindent we obtain

\begin{eqnarray}
F(m) &=& \int \limits_{m_b + m_c}^{\infty} dz \frac{\Gamma(z)}{2
\pi(m - z) } = \int \limits_{m_b + m_c}^{m-\Delta} dz
\frac{\Gamma(z)}{2\pi(m - z) } + \int \limits_{m -
\Delta}^{m+\Delta} dz \frac{
\Gamma(z)}{2\pi(m - z) } \nonumber \\
&+& \int \limits_{m +
\Delta}^{\infty} dz \frac{ \Gamma(z)}{2\pi(m - z) } \label{eq:13.15}
\end{eqnarray}

\noindent Here we split the integration interval $[m_b+m_a, +\infty)$ into three segments.  When $\Delta \to 0$, the
first and third terms on the right hand side of (\ref{eq:13.15})
give the \emph{principal value} integral (denoted by $P \int$) \index{principal value integral}

\begin{eqnarray}
\int \limits_{m_b + m_c }^{m-\Delta} dz \frac{ \Gamma(z)}{2\pi(m -
z) } + \int \limits_{m + \Delta}^{\infty} dz \frac{\Gamma(z)}{2\pi(m
- z) } \longrightarrow P \int \limits_{m_b + m_c}^{\infty} dz \frac{
\Gamma(z)}{2\pi(m - z) } &\equiv&
\mathcal{P}(m) \nonumber \\
 \label{eq:13.16}
\end{eqnarray}

\noindent Let us now look more closely at the second integral on the
right hand side of (\ref{eq:13.15}).    If $\Delta$ is sufficiently small,\footnote{but still much larger than the distance between adjacent points $m_j$ and $m_{j+1}$} then function
$\Gamma(z)$ may be considered constant $\Gamma(z) = \Gamma(m)$.      Moreover, in our discrete approximation, the density of points $m_j$ in the interval $[m-\Delta, m+
\Delta]$ is almost constant. So, this interval can be divided into $2N$ small equal segments

\begin{eqnarray*}
m_j = m_0 + j \frac{\Delta}{N}
\end{eqnarray*}

\noindent where $m_0 = m$, integer $j$ runs from $-N$ to $N$, and
the integral is represented by a \emph{partial sum}

\begin{eqnarray}
\int \limits_{m - \Delta}^{m+\Delta} dz \frac{ \Gamma(z)}{2\pi(m -
z) } &\approx& \frac{\Gamma(m_0)}{2\pi} \int \limits_{m -
\Delta}^{m+\Delta} dz \frac{1}{m - z } \approx
\frac{\Gamma(m_0)}{2\pi} \sum_{j= -N}^N \frac{\Delta/N}{m - m_0 - j
\frac{\Delta}{N}} \nonumber \\
\label{eq:analyt}
\end{eqnarray}

\noindent Next we assume that $N \to \infty$ and index $j$ runs from
$-\infty$ to $\infty$. Then the  right hand side of equation
(\ref{eq:analyt}) defines an analytical function  with poles at
points

\begin{eqnarray}
m_j = m_0 + j \frac{\Delta}{N} \label{eq:13.17}
\end{eqnarray}

\noindent and with residues $\Gamma(m_0) \Delta/(2 \pi N)$. As any
analytical function is uniquely determined by the positions of its
poles and the values of its residues, we conclude that integral
(\ref{eq:analyt}) has the following representation

\begin{eqnarray}
\int \limits_{m - \Delta}^{m+\Delta} dz \frac{ \Gamma(z)}{2\pi(m -
z) } \approx \frac{\Gamma(m_0)}{2}  \cot \left(\frac{\pi N}{ \Delta}
(m-m_0) \right) \label{eq:13.18}
\end{eqnarray}

\noindent Indeed, the $\cot$  function on the right hand side of
(\ref{eq:13.18}) also has  poles at points (\ref{eq:13.17}). The
residues of this function are exactly as required too.\footnote{For example,
near the point $m_0$ (where $j=0$) the right hand side of
(\ref{eq:13.18}) can be approximated as

\begin{eqnarray*}
\frac{\Gamma(m_0)}{2}  \cot\left(\frac{\pi N}{ \Delta}
(m-m_0)\right)
 &\approx&   \frac{ \Gamma(m_0) \Delta /N}{ 2\pi (m-m_0)}
\end{eqnarray*}

\noindent which agrees with (\ref{eq:analyt}).} Now we can put equations (\ref{eq:13.16}) and (\ref{eq:13.18})
together and write

\begin{eqnarray*}
F(m) &=& \mathcal{P}(m) + \frac{\Gamma(m)}{2} \cot \left(\frac{\pi N
m} {\Delta} \right)
\end{eqnarray*}

\noindent  Then,  using

\begin{eqnarray*}
\cot(ax)'
  = -a(1 + \cot^2(ax))
\end{eqnarray*}

\noindent and ignoring derivatives of smooth functions $\mathcal{P}(m)$ and $\Gamma(m)$ we obtain

\begin{eqnarray}
F'(m) &=&  - \frac{\pi \Gamma(m) N}{2\Delta} (1 + \cot ^2 (\pi N
\Delta^{-1} m )) \label{eq:13.19}
\end{eqnarray}

\noindent For formula (\ref{eq:13.13}) we need values of $F'(m)$ at the discrete set of solutions
of the equation

\begin{eqnarray*}
 F(m)= m - m_a
\end{eqnarray*}

\noindent At these points we can write

\begin{eqnarray*}
m - m_a &=&  \mathcal{P}(m) - \frac{\Gamma(m)}{2} \cot (\pi N \Delta^{-1} m ) \\
\cot (\pi N \Delta^{-1} m ) &=& -\frac{2(m - m_a -\mathcal{P}(m))}{
\Gamma(m)}
\\
\cot^2 (\pi N \Delta^{-1} m ) &=& \frac{4(m - m_a
-\mathcal{P}(m))^2}{\Gamma^2(m)}
\end{eqnarray*}

\noindent Substituting this to (\ref{eq:13.19}) and (\ref{eq:13.13})
we obtain the desired result

\begin{eqnarray}
F'(m) &=&  -\pi  \frac{\Gamma(m) N}{2\Delta} \left(1 + \frac{4(m -
m_a
-\mathcal{P}(m))^2}{ \Gamma^2(m)} \right) \nonumber \\
 |\mu(m)|^2 &=& \frac{1}{1+
\pi \Gamma(m) N /(2\Delta) \left(1 + \frac{4(m_a + \mathcal{P}(m)-
m)^2)}{ \Gamma^2(m)} \right)} \label{eq:mu-m} \\
&\approx& \frac{\Gamma(m) \Delta /(2 \pi N)}{  \Gamma^2(m)/4  + (m_a
+ \mathcal{P}(m) - m)^2  } \label{eq:mu-m2}
\end{eqnarray}

\noindent where we neglected the unity in the denominator of
(\ref{eq:mu-m}) as compared to the large factor $\propto N
\Delta^{-1}$. Formula (\ref{eq:mu-m2}) gives the probability for
finding particle $a$ at each point of the discrete spectrum $M_1,
M_2, M_3, \ldots$. This probability naturally tends to zero as the density of
points $N \Delta^{-1}$ tends to infinity. However, when approaching
the continuous spectrum in the limit $N \to \infty$ we do not need
the probability at each spectrum point. We, actually, need the
\emph{probability density} \index{probability density} which can be
obtained by multiplying the right hand side of equation (\ref{eq:mu-m2})
by the number of points per unit interval $N \Delta^{-1} $. Then
the mass distribution for the unstable particle takes the famous
\emph{Breit-Wigner} form \index{Breit-Wigner distribution}

\begin{eqnarray}
|\mu(m)|^2 = \frac{\Gamma(m) /(2 \pi)}{\Gamma^2(m)/4  + (m_a +
\mathcal{P}(m) - m)^2} \label{eq:13.21}
\end{eqnarray}

\begin{figure}
\centering
\includegraphics {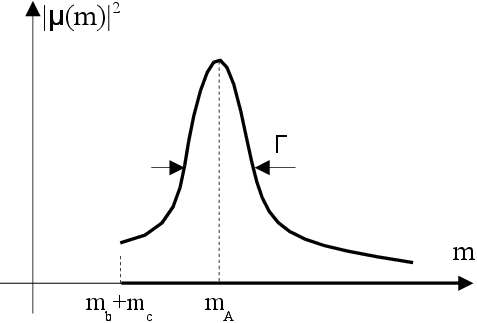} \caption{ Mass distribution of a typical
unstable particle.} \label{fig:13.4}
\end{figure}

\noindent  This resonance mass distribution describes an unstable
particle with the expectation value of mass\footnote{the center of
the resonance} $m_A =m_a + \mathcal{P}(m_A)$ and the width of
$\Delta m \approx \Gamma(m_A)$ (see Fig. \ref{fig:13.4}).

For unstable systems whose decays are slow enough to be observed in
time-resolved experiments, the resonance shown in Fig. \ref{fig:13.4} is very
narrow, so that instead of functions $\Gamma(m)$ and
$\mathcal{P}(m)$ we can use their values (constants) at $m=m_A$:
$\Gamma \equiv \Gamma(m_A)$ and  $\mathcal{P} \equiv
\mathcal{P}(m_A)$. Moreover, we will assume that the instability of
the particle $a$ does not have a large effect on its mass, i.e.,
that $\mathcal{P} \ll m_a$ and $m_A \approx m_a$.\footnote{This assumption is not trivial, as will be seen from subsections \ref{ss:instab} and \ref{sc:Lamb-shift}.} We also neglect
a small contribution from the isolated point $M_0$ of the mass spectrum
discussed in the beginning of this subsection. Then

\begin{eqnarray}
|\mu(m)|^2 \approx \frac{\Gamma /(2 \pi)}{\Gamma^2/4  + (m_a - m)^2} \label{eq:13.21a}
\end{eqnarray}

\subsection{Exponential decay law} \label{ss:decay_law}

To complete our discussion of the unstable system at rest we are now going to calculate its decay law and the  time-dependent wave function. For the initial state vector at $t=0$ we choose state (\ref{eq:0-rangle}) of the particle $a$. Its time dependence is described by the time evolution operator

\begin{eqnarray*}
| \mathbf{0},t \rangle = e^{-\frac{i}{\hbar}Ht} | \mathbf{0} \rangle
\end{eqnarray*}

\noindent To evaluate this expression it is convenient to represent $| \mathbf{0} \rangle$ as an expansion (\ref{eq:integral}) in the basis of eigenvectors of the Hamiltonian $H$. Then, using (\ref{eq:7.145a}) and (\ref{eq:13.9})  we obtain

\begin{eqnarray}
e^{-\frac{i}{\hbar}Ht}|\mathbf{0} \rangle &=& \int \limits_{m_b + m_c}^{\infty} dm \mu(m)
e^{-\frac{i}{\hbar}Ht}| \mathbf{0},m \rangle = \int \limits_{m_b + m_c}^{\infty} dm \mu(m)
e^{-\frac{i}{\hbar}mc^2t}| \mathbf{0},m \rangle \nonumber \\
&=& \int \limits_{m_b + m_c}^{\infty} dm
e^{-\frac{i}{\hbar}mc^2t} |\mu(m)|^2 \left[ \begin{array}{c}
1  \\
g^*(\vec{\rho})/ (mc^2 - \eta_{\vec{\rho}}c^2)
\end{array} \right] \equiv \left[ \begin{array}{c}
I(t)  \\
J(\vec{\rho}, t)
\end{array} \right] \nonumber \\
\label{eq:decay}
\end{eqnarray}

\noindent The first integral $I(t)$ determines the decay law
for the particle at rest.
Substituting (\ref{eq:13.21a}) in the integrand, we obtain

\begin{eqnarray}
\omega(t) = |I(t)|^2 \approx \frac{1}{4 \pi^2} \left| \int_{m_b + m_c}^{\infty}
dm \frac{\Gamma e^{-\frac{i}{\hbar}mc^2t}}{\Gamma ^2/4  + (m_a  -
m)^2} \right|^2 \label{eq:7.164}
\end{eqnarray}

\noindent For most unstable systems

\begin{equation}
\Gamma \ll m_a - (m_b + m_c) \label{eq:gamma}
\end{equation}

\noindent  so the integrand is well localized around the value $m \approx m_a$,  and we can introduce further approximation by setting the
lower integration limit in (\ref{eq:7.164}) to $-\infty$. Then the
decay law obtains a familiar exponential form

\begin{eqnarray}
\omega(t) &\approx& \frac{1}{4 \pi^2} \left| 2 \pi e^{-\frac{i}{\hbar}m_ac^2t}\exp \left(-\frac{\Gamma c^2 t}{2 \hbar}\right) \right|^2 =  \exp\left(- \frac{\Gamma c^2t}{\hbar}\right) \nonumber \\
&=&
\exp \left(- \frac{ t}{\tau_0} \right) \label{eq:13.23}
\end{eqnarray}

\noindent where

\begin{eqnarray}
\tau_0 = \frac{\hbar}{\Gamma c^2} \label{eq:lifetime}
\end{eqnarray}

\noindent is the \emph{lifetime} of the unstable particle. The nondecay probability decreases from 1 to $1/e$ during the
lifetime interval. \index{lifetime}

Using formulas (\ref{eq:gvecpi}), (\ref{eq:Gq}) and (\ref{eq:gammam}) we can also see
that the \emph{decay rate} \index{decay rate}
$\Gamma$\footnote{Actually,  parameter $\Gamma$ has the
dimensionality of mass, so the true decay rate is $1/\tau_0 = \Gamma
c^2/\hbar$ (Hz). The momentum $\rho$ in (\ref{eq:Gamma2}) should be
calculated as $\rho = \eta^{-1}(m_a)$.}

\begin{eqnarray}
\frac{1}{\tau_0 } = \frac{\Gamma c^2}{\hbar}  &=&  \frac{2 \pi c^2}{\hbar} G(\eta^{-1}(m_a)) \frac{d \eta^{-1}(z)}{dz} \Bigl|_{z=m_a} \nonumber \\
&=& \frac{8 \pi^2 \rho^2}{c^2 \hbar} |\langle \mathbf{0}|V|\rho \rangle|^2
\frac{d \eta^{-1}(z)}{dz} \Bigl|_{z=m_a} \label{eq:Gamma2}
\end{eqnarray}

\noindent is proportional to the square of the matrix element of the
perturbation $V$  the initial and final states of the system.
It is also proportional to the ``kinematical'' factor $d
\eta^{-1}(z)/dz|_{z=m_a}$, which is fully determined by the three
involved masses $m_a, m_b$, and $m_c$.\footnote{See equation (\ref{eq:13.6}).}

The importance of formulas (\ref{eq:13.21a}), (\ref{eq:13.23}), and
(\ref{eq:Gamma2}) is that they were derived from very general
assumptions. We have not used the perturbation theory. Actually, the only
significant approximation was the weakness of the interaction
responsible for the decay, i.e., the narrow width $\Gamma$ of the
resonance (\ref{eq:gamma}). This condition is satisfied for all
known decays.\footnote{Approximation (\ref{eq:gamma}) may be not
accurate for particles (or \emph{resonances}) \index{resonance}
decaying due to strong nuclear forces. However, their lifetime is
very short $\tau_0 \approx 10^{-23}$s, so the time dependence of
their decays cannot be observed experimentally.} Therefore, the
exponential decay law is expected to be universally valid. This
prediction is confirmed by experiment: so far no deviations from the
exponential decay law (\ref{eq:13.23}) were observed.

\subsection{Wave function of decay products} \label{sc:time-dependent}

The second integral $J(\vec{\rho}, t)$ in (\ref{eq:decay}) describes the wave function of decay product $b$ and $c$. We note that function $|\mu(m)|^2$ has poles at $m_a - i \Gamma/2$ and $m_a + i \Gamma/2$. Then

\begin{eqnarray}
J(\vec{\rho}, t) = \frac{\Gamma g^*(\vec{\rho})}{2 \pi c^2} \int \limits_{-\infty}^{\infty}
\frac{ e^{-\frac{i}{\hbar}mc^2t} dm}{(m - \eta_{\vec{\rho}})(m-m_a - i \Gamma/2)(m-m_a + i \Gamma/2)} \label{eq:complex}
\end{eqnarray}

\begin{figure}
\centering
 \includegraphics {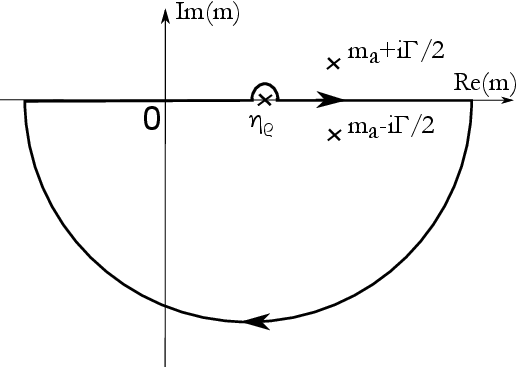} \caption{Complex plane integration contour for integral (\ref{eq:complex}).}
\label{fig:13x}
\end{figure}

\noindent The integration contour should be closed as shown in Fig. \ref{fig:13x}, because then the integral along the large semi-circle in the lower half-plane can be ignored, as the factor $e^{-\frac{i}{\hbar}mc^2t}$ tends to zero there. So we obtain

\begin{eqnarray}
J(\vec{\rho}, t) =   \frac{ g^*(\vec{\rho})}{ c^2(m_a - \eta_{\vec{\rho}} -i \Gamma/2)} \left(
 e^{-\frac{i}{\hbar}(m_a-i \Gamma/2)c^2t} -
\frac{i\Gamma e^{-\frac{i}{\hbar}\eta_{\vec{\rho}} c^2t}}{2(m_a - \eta_{\vec{\rho}}) +i \Gamma} \right) \label{eq:it}
\end{eqnarray}

\noindent To be consistent with the initial condition (\ref{eq:0-rangle}), our solution must satisfy $J(\vec{\rho}, 0) = 0$. This is, indeed, true as within our approximations we can set $\eta_{\vec{\rho}} = m_a$ in the second term in the parentheses, so that the whole expression vanishes at $t=0$.

The first term in the parentheses is significant only at short times comparable with the particle's lifetime $\tau_0$. In the limit $t \to \infty$ only the second term contributes, and we can write the wave function in the position representation (\ref{eq:7.27})

\begin{eqnarray}
J(\mathbf{r}, t) \approx -\frac{i\Gamma}{2c^2(2 \pi \hbar)^{3/2}} \int d \vec{\rho} e^{\frac{i}{\hbar}\vec{\rho} \mathbf{r}}  g^*(\vec{\rho})
\frac{ e^{-\frac{i}{\hbar}\eta_{\vec{\rho}} c^2t}}{(m_a - \eta_{\vec{\rho}})^2 +\Gamma^2/4} \label{eq:position-rep}
\end{eqnarray}

\noindent where $\mathbf{r} \equiv \mathbf{r}_b - \mathbf{r}_c$ is the relative position of the two decay products, which is an observable conjugate to $\vec{\rho}$. In our model, the interaction strength $|g(\vec{\rho})|$ is spherically symmetric. We will also assume that function $g(\vec{\rho})$ is real $g^*(\vec{\rho}) = |g(\rho)|$. The largest contribution to the integral (\ref{eq:position-rep}) comes from a thin spherical layer with momenta $|\vec{\rho}| \approx \rho_0$, such that $\eta_{\rho_0} \approx m_a$. We can assume that within this layer the modulus of the interaction function stays nearly constant $|g(\rho)| \approx |g(\rho_0)|$. Similarly, we can approximate equation (\ref{eq:13.6}) as

\begin{eqnarray*}
\eta_{\vec{\rho}} &\approx& \eta_{\rho_0} + \frac{1}{c^2}\left(\frac{c^2 \rho_0}{\sqrt{m_b^2c^4 + c^2\rho_0^2}}+ \frac{c^2 \rho_0}{\sqrt{m_c^2c^4 + c^2\rho_0^2}} \right) (\rho - \rho_0) \\
&=& \eta_{\rho_0} + \frac{1}{c^2}\left(v_b + v_c \right) (\rho - \rho_0)
\end{eqnarray*}

\noindent where $v_b$ $v_c$ are the average speeds with which the two decay products leave the region of their creation. With these approximations we rewrite equation (\ref{eq:position-rep})

\begin{eqnarray}
J(\mathbf{r}, t) \approx \frac{C}{(2 \pi \hbar)^3} \int d \vec{\rho} e^{\frac{i}{\hbar}\vec{\rho} \mathbf{r}}   e^{-\frac{i}{\hbar}\left(v_b + v_c \right) \rho t}
\end{eqnarray}

\noindent where $C$ is a constant whose value is not important to us here. Evaluating this integral in spherical coordinates we obtain

\begin{eqnarray*}
J(\mathbf{r}, t) &\approx& \frac{2 \pi C}{(2 \pi \hbar)^3}\int \limits_{0}^{\pi} \sin \theta d \theta \int \limits_{0}^{\infty} \rho^2 d \rho e^{\frac{i}{\hbar} \rho r \cos \theta}   e^{-\frac{i}{\hbar}\left(v_b + v_c \right) \rho t} \\
&\approx& \frac{2 \pi \hbar C}{i r(2 \pi \hbar)^3}  \int \limits_{-\infty}^{\infty} \rho d \rho (e^{\frac{i}{\hbar} \rho r } - e^{-\frac{i}{\hbar} \rho r })  e^{-\frac{i}{\hbar}\left(v_b + v_c \right) \rho t} \\
&\propto& \frac{1}{r}[\delta(r  - (v_b+v_c)t) + \delta(r + (v_b+v_c)t)]
\end{eqnarray*}

\noindent The second delta function in square brackets can be ignored, because its argument never turns to zero as $r, v_b, v_c, t$ are positive quantities and  the limit $t \to +\infty$ is taken. Then

\begin{eqnarray*}
J(\mathbf{r}, t)
&\propto& \frac{1}{r} \delta(r  - (v_b+v_c)t)
\end{eqnarray*}

\noindent which means that after all approximations adopted here the wave function of the decay products has the form of a spherical shell expanding around the decay point with a constant speed. The separation between two decay products changes as

\begin{eqnarray}
r =  (v_b+v_c)t \label{r-gamma}
\end{eqnarray}

\noindent This indicates that particles $b$ and $c$ are not interacting in the asymptotic regime: they move apart with constant velocities, as expected.

\section{Decay law for moving particles}
\label{sc:formalism}

Equation (\ref{eq:50}) is the decay law $\omega(0,t)$ observed from the reference frame $O$ at rest. In the present section we will derive
a formula for the decay law $\omega(\theta,t)$ in a moving frame $O'$. Particular
cases of this formula relevant to unstable particles with sharply
defined momenta or velocities
 will be considered in subsections \ref{sc:mov}  and \ref{sc:general},
respectively.

\subsection{General formula for the decay law}
\label{sc:moving-frame}

Suppose that observer $O$ describes the initial state (at $t=0$) by
the state vector $| \Psi \rangle$. Then moving  observer $O'$
describes the same state\footnote{at $t'=t=0$, where $t'$ is time measured
by the observer's $O'$ clock}
 by the vector

\begin{eqnarray*}
| \Psi (\theta, 0)  \rangle =  e^{\frac{ic}{\hbar}K_x \theta} | \Psi
\rangle \label{eq:44}
\end{eqnarray*}

\noindent The time dependence of this state  is

\begin{eqnarray}
| \Psi(\theta, t') \rangle &=&
e^{-\frac{i}{\hbar}Ht'}e^{\frac{ic}{\hbar}K_x \theta} | \Psi \rangle
\label{eq:44a}
\end{eqnarray}

\noindent According to the general formula (\ref{square}), the
decay law from the point of view of $O'$ is

\begin{eqnarray}
\omega(\theta, t') &=& \langle \Psi(\theta, t') | T | \Psi (\theta,
t') \rangle
\label{eq:3b} \\
&=& \Vert T | \Psi (\theta, t') \rangle \Vert^2 \label{eq:3d}
\end{eqnarray}

Let us use the basis set decomposition (\ref{eq:expansion2}) of the
state vector $| \Psi \rangle$.  Then, applying equations (\ref{eq:44a}),
(\ref{eq:29a}), and (\ref{eq:29b}) we obtain

\begin{eqnarray*}
| \Psi(\theta, t') \rangle & =& \int
 d\mathbf{p}  \psi (\mathbf{p})
e^{-\frac{i}{\hbar}Ht'} e^{\frac{ic}{\hbar}K_x \theta}
 |\mathbf{p}\rangle \\
& =& \int
 d\mathbf{p}  \psi (\mathbf{p}) \int \limits_{m_b + m_c}^{\infty} dm \mu(m)
\gamma(\mathbf{p}, m) e^{-\frac{i}{\hbar}Ht'} e^{\frac{ic}{\hbar}K_x
\theta}
 |\mathbf{p},m \rangle \\
& =& \int
 d\mathbf{p}  \psi (\mathbf{p})
\int \limits_{m_b + m_c}^{\infty} dm \mu(m) \gamma(\mathbf{p}, m)
e^{-\frac{i}{\hbar}\omega_{\Lambda \mathbf{p}}t'}
\sqrt{\frac{\omega_{\Lambda \mathbf{p}}}{\omega_{\mathbf{p}}}}
 |\Lambda\mathbf{p},m \rangle
\end{eqnarray*}

\noindent The inner product of this vector with $| \mathbf{q}
\rangle $ can be found with the help of (\ref{eq:35a}) and new integration variables $ \mathbf{r} =
\Lambda \mathbf{p}$

\begin{eqnarray*}
&\mbox{ }& \langle \mathbf{q} |  \Psi(\theta, t') \rangle \nonumber
\\
&=&  \int d\mathbf{p}  \psi (\mathbf{p}) \int \limits_{m_b +
m_c}^{\infty} dm \mu(m)\gamma(\mathbf{p}, m)
e^{-\frac{i}{\hbar}\omega_{\Lambda \mathbf{p}}t'} \langle \mathbf{q}
|\Lambda \mathbf{p} ,m \rangle \sqrt{\frac{\omega_{\Lambda
\mathbf{p}}}{\omega_{\mathbf{p}}}} \nonumber
\\
&=& \int  d\mathbf{p} \psi (\mathbf{p})  \int \limits_{m_b +
m_c}^{\infty} dm  |\mu(m)|^2
 \gamma(\mathbf{p}, m)\gamma^*(\Lambda \mathbf{p}, m)
e^{-\frac{i}{\hbar}\omega_{\Lambda \mathbf{p}}t'} \delta(\mathbf{q}
- \Lambda \mathbf{p}) \sqrt{\frac{\omega_{\Lambda
\mathbf{p}}}{\omega_{\mathbf{p}}}}
\label{eq:36ab} \\
&=& \int \limits_{m_b + m_c}^{\infty} dm \int
 d \mathbf{r}
\frac{\omega_{\Lambda^{-1} \mathbf{r}}} {\omega_{\mathbf{r}}}
\sqrt{\frac{\omega_{\mathbf{r}}}{\omega_{\Lambda^{-1} \mathbf{r}}}}
\psi(\Lambda^{-1} \mathbf{r}) \gamma(\Lambda^{-1} \mathbf{r})
\gamma^*( \mathbf{r}) \vert \mu(m)\vert^2
e^{-\frac{i}{\hbar}\omega_{ \mathbf{r}}t'}
    \delta(\mathbf{q} - \mathbf{r}) \nonumber \\
&=& \int \limits_{m_b + m_c}^{\infty} dm
\sqrt{\frac{\omega_{\Lambda^{-1} \mathbf{q}}} {\omega_{\mathbf{q}}}}
\psi(\Lambda^{-1} \mathbf{q}) \gamma(\Lambda^{-1} \mathbf{q}, m)
\gamma^*( \mathbf{q}, m)  \vert \mu(m)\vert^2
e^{-\frac{i}{\hbar}\omega_{ \mathbf{q}}t'}
\end{eqnarray*}

\noindent
 The non-decay probability in the reference frame $O'$ for all values of $\theta$
and $t'$ is then found
by substituting (\ref{eq:projector}) in equation (\ref{eq:3b})

\begin{eqnarray}
&\mbox{ }& \omega(\theta, t') \nonumber \\ &=& \int  d\mathbf{q}
\langle \Psi(\theta, t')
 |\mathbf{q} \rangle\langle \mathbf{q}   | \Psi(\theta, t')\rangle
= \int  d\mathbf{q} |\langle \mathbf{q}
  |\Psi (\theta, t')\rangle\ |^2  \nonumber \\
&=& \int  d\mathbf{q} \left| \int \limits_{m_b + m_c}^{\infty} dm
\sqrt{\frac{\omega_{\Lambda^{-1} \mathbf{q}}} {\omega_{\mathbf{q}}}}
\psi(\Lambda^{-1} \mathbf{q}) \gamma(\Lambda^{-1} \mathbf{q}, m)
\gamma^*( \mathbf{q}, m)   \vert \mu(m)\vert^2
e^{-\frac{i}{\hbar}\omega_{
\mathbf{q}}t'} \right|^2 \nonumber \\
&\mbox{ }& \label{eq:50x}
\end{eqnarray}

This general formula is not very convenient for calculations. So, in the following subsections we will consider some specific situations in which (\ref{eq:50x}) can be simplified and fully evaluated.

\subsection{Decays of states with definite momentum}
\label{sc:mov}

In the reference frame at rest ($\theta = 0$), formula
(\ref{eq:50x}) coincides exactly with our earlier result
(\ref{eq:50})

\begin{eqnarray}
\omega(0, t) = \int d \mathbf{q} |\psi(\mathbf{q})|^2 \left|\int
\limits_{m_b + m_c}^{\infty} dm |\mu(m)|^2
  e^{-\frac{i}{\hbar} \omega_{\mathbf{q}} t} \right|^2 \label{eq:omega-0-t}
\end{eqnarray}

\noindent In section \ref{sc:general-decay} we applied this formula
to calculate the decay law of a particle with zero momentum. Here we
will consider the case when the unstable particle has a
non-zero momentum $\mathbf{p}$, i.e., the state is described by a normalized
vector $| \mathbf{p})$ whose wave function is (\ref{eq:square-delta})

\begin{eqnarray}
\psi(\mathbf{q}) = \sqrt{\delta(\mathbf{ q-p})} \label{eq:sqrt}
\end{eqnarray}

\noindent From equation (\ref{eq:omega-0-t}) the decay law for such a
state is

\begin{eqnarray}
\omega_{ |\mathbf{p})}(0,t) &=& \left|\int \limits_{m_b +
m_c}^{\infty}dm |\mu(m)|^2
  e^{-\frac{i}{\hbar}\omega_{\mathbf{p}} t} \right|^2
\label{eq:35}
\end{eqnarray}

\noindent In a number of works \cite{Stefanovich_decay, Khalfin,
Shirokov_decay, Urbanowski} it was noticed that this result disagrees with
Einstein's time dilation formula (\ref{eq:1b}). Indeed, if  one
interprets the state $|\mathbf{p})$ as a state of unstable particle
moving with definite speed

\begin{eqnarray*}
v &=& \frac{c^2 p}{\sqrt{m_a^2c^4 + p^2c^2}} = c \tanh \theta
\end{eqnarray*}

\noindent  then the decay law (\ref{eq:35}) \emph{cannot} be
connected with the decay law of the particle at rest (\ref{eq:50a})
by Einstein's formula (\ref{eq:1b})\footnote{In subsection \ref{sc:numerical} we will illustrate this inequality with numerical calculations.}

\begin{eqnarray}
\omega_{ |\mathbf{p})}(0,t) &\neq& \omega_{ |\mathbf{0})}(0, t/
\cosh \theta) \label{eq:35b}
\end{eqnarray}

\noindent This observation prompted authors of
\cite{Stefanovich_decay, Khalfin, Shirokov_decay} to question the
applicability of special relativity to particle decays. However, at
a closer inspection it appears that this result by itself does not challenge
the special-relativistic time dilation (\ref{eq:1b}) directly.
Formula (\ref{eq:35b}) is comparing decay laws of two different
momentum eigenstates $|\mathbf{0})$ and $|\mathbf{p})$ viewed
from the same reference frame. This is not exactly the same as
(\ref{eq:1b}) which compares observations made on the same particle
from two frames of reference moving with respect to each other.  If
from the point of view of observer $O$ the particle is described by
the state vector $|\mathbf{0}) $ which has zero momentum and zero
velocity, then from the point of view of $O'$ this particle is
described by the state

\begin{eqnarray}
 e^{\frac{ic}{\hbar}\mathbf{K}  \vec{\theta}} | \mathbf{0})
\label{eq:velo-state}
\end{eqnarray}

\noindent which is \emph{not} an eigenstate of the momentum operator
$\mathbf{P}_0$. So, strictly speaking, formula (\ref{eq:35}) is not
applicable to this state. However, it is not difficult to see that
(\ref{eq:velo-state})
 is an eigenstate of
the velocity operator \cite{Shirokov-osc}. Indeed, taking into
account that $V_x | \mathbf{0}) = \mathbf{0}$ and equations (\ref{eq:6.2}) -
(\ref{eq:6.3}), we obtain

\begin{eqnarray}
V_x e^{\frac{ic}{\hbar}K_x  \theta} | \mathbf{0}) &=&
e^{\frac{ic}{\hbar}K_x  \theta} e^{-\frac{ic}{\hbar}K_x  \theta} V_x
e^{\frac{ic}{\hbar}K_x  \theta} | \mathbf{0} ) =
e^{\frac{ic}{\hbar}K_x  \theta} \frac{V_x - c \tanh \theta}{ 1 -
\frac{V_x \tanh \theta}{c} } | \mathbf{0}
) \nonumber \\
&\approx&  -c \tanh \theta e^{\frac{ic}{\hbar}K_x \theta} |
\mathbf{0} ) \label{eq:velocity-eig}
\end{eqnarray}

 \noindent   Thus, a fair comparison with the time
dilation formula (\ref{eq:1b}) requires consideration of unstable
states having definite values of velocity for both observers. This will be done in subsection \ref{sc:def-vel}.

\subsection{Decay law in the moving reference frame}
\label{sc:general}

Before addresing the decay law seen from a moving frame, let us first introduce a few realistic approximations and simplify our general formula (\ref{eq:50x}) a little bit.
First, we may notice that in all realistic cases the initial state $|\Psi \rangle \in \mathcal{H}_a$ is not an exact eigenstate of the
total momentum operator: the wave function of the unstable particle
is never localized at one point in the momentum space (as was assumed,
for example, in (\ref{eq:sqrt})) but has a spread (or uncertainty)
of momentum $|\Delta \mathbf{p}|$ and, correspondingly, an
uncertainty of position $|\Delta \mathbf{r}| \approx \hbar/ |\Delta
\mathbf{p}|$. Second, the state $|\Psi \rangle \in \mathcal{H}_a$ is
not an eigenstate of the mass operator $M$. The initial state $|\Psi \rangle$ is
characterized by the uncertainty of mass $\Gamma$ (see Fig.
\ref{fig:13.4}) that is related to the particle's lifetime ($
\tau_0$)  by formula (\ref{eq:lifetime}). It is important to note that
in all cases of practical interest the mentioned uncertainties are
related by inequalities

 \begin{eqnarray}
|\Delta \mathbf{p}| &\gg& \Gamma c
\label{eq:gg}\\
|\Delta \mathbf{r}| &\ll&  c \tau_0 \label{eq:gg2}
\end{eqnarray}

\noindent In particular, the latter inequality means that the
uncertainty of position is mush less than the distance passed by
light during the lifetime of the particle. For example, in the case
of muon $\tau_0 \approx 2.2 \cdot 10^{-6}$s and, according to
(\ref{eq:gg2}), the spread of the wave function in the position
space must be much less than 600m, which is a reasonable assumption.
 Therefore, we can safely assume that the
 factor  $|\mu(m)|^2$ in (\ref{eq:50x})
has a very sharp peak near the value $m = m_a$. Then we can move  the
value of the smooth\footnote{see discussion after equation
(\ref{eq:xxx})} function $\sqrt{\frac{\omega_{\Lambda \mathbf{q}}}
{\omega_{\mathbf{q}}}} \psi(\Lambda \mathbf{q}) \gamma(\Lambda
\mathbf{q},m) \gamma^*(\mathbf{q},m)$ at $m=m_a$ outside the
integral on $m$

\begin{eqnarray}
&\mbox{ }& \omega(\theta, t') \nonumber \\
&\approx&  \int  d\mathbf{q} \left| \sqrt{\frac{\Omega_{\Lambda^{-1}
\mathbf{q}}} {\Omega_{\mathbf{q}}}} \psi(\Lambda^{-1} \mathbf{q})
\gamma(\Lambda^{-1} \mathbf{q}, m_a) \gamma^*( \mathbf{q}, m_a)
\right| ^2 \left| \int \limits_{m_b + m_c}^{\infty} dm
  \vert \mu(m)\vert^2
e^{-\frac{i}{\hbar}\omega_{ \mathbf{q}}t'}    \right|^2 \nonumber \\
&=& \int  d\mathbf{q} \frac{\Omega_{L^{-1} \mathbf{q}}} {\Omega_{
\mathbf{q}}} |\psi(L^{-1} \mathbf{q}) |^2 \left| \int \limits_{m_b +
m_c}^{\infty} dm
 \vert \mu(m)\vert^2
e^{-\frac{i}{\hbar}\omega_{ \mathbf{q}}t'}    \right|^2 \nonumber \\
&=& \int  d\mathbf{p} |\psi( \mathbf{p})|^2 \left| \int \limits_{m_b
+ m_c}^{\infty} dm
 \vert \mu(m)\vert^2
e^{-\frac{i}{\hbar}\omega_{ L \mathbf{p}}t'}    \right|^2
\label{eq:appr_dec}
\end{eqnarray}

\noindent Here $\Omega_{\mathbf{p}} = \sqrt{m_a^2 c^4 + p^2 c^2}$, $\Lambda \mathbf{p}$ is given by equation
(\ref{eq:lambdaz}) and $ L \mathbf{p}= (p_x \cosh \theta +
\frac{\Omega_{\mathbf{p}}}{c}
 \sinh
\theta, p_y, p_z)$.

\subsection{Decays of states with definite velocity} \label{sc:def-vel}

Next we consider an initial state which has zero velocity from the point of
view of observer $O$. The wave function of this state is localized
near zero momentum $\mathbf{p} = \mathbf{0}$.
So, we can set in equation (\ref{eq:appr_dec})\footnote{As we mentioned
in the preceding subsection, in reality this state is not exactly an
eigenstate of momentum (velocity) $|\mathbf{0})$. However, its wave
function is still much better localized in the $\mathbf{p}$-space
than the slowly varying second factor under the integral in
(\ref{eq:appr_dec}), so, approximation (\ref{eq:psi2}) is
justified.}

\begin{eqnarray}
|\psi( \mathbf{p})|^2 \approx  \delta(\mathbf{p}) \label{eq:psi2}
\end{eqnarray}

\noindent and obtain the decay law seen by the moving observer

\begin{eqnarray}
\omega_{|\mathbf{0})}(\theta, t') &\approx& \left| \int \limits_{m_b
+ m_c}^{\infty} dm
 \vert \mu(m)\vert^2
e^{-\frac{it'}{\hbar}\sqrt{m^2 c^4 + m_a^2 c^4\sinh^2 \theta} }
\right|^2 \label{eq:omega-0-t2}
\end{eqnarray}

\noindent If we approximately identify $m_a c \sinh \theta$ with the
momentum $p$ of the particle $a$ from the point of view of the
moving observer $O'$\footnote{From the point of view of this
observer, particle's velocity is $c \tanh \theta$.} then

\begin{eqnarray}
\omega_{|\mathbf{0})}(\theta, t') &\approx& \left|\int \limits_{m_b
+ m_c}^{\infty} dm |\mu(m)|^2
  e^{-\frac{i}{\hbar}\omega_{\mathbf{p}} t'}\right|^2
\label{eq:62}
\end{eqnarray}

\noindent So, in this approximation the decay law (\ref{eq:62}) in
the frame of reference $O'$ moving with the speed $c \tanh \theta$
takes the same form as the decay law (\ref{eq:35}) of a particle
moving with momentum $m_a c \sinh \theta$ with respect to the
stationary observer $O$.\footnote{Note the disagreement between
 ourresult and conclusions of ref. \cite{Shirokov-osc}.} So, the violation of the Einstein's time dilation formula mentioned in subsection \ref{sc:mov} is a real effect that warrants a more in-depth study. In the next section we will evaluate (\ref{eq:62}) numerically.

\section{``Time dilation'' in decays}
\label{sc:discus}

In this section we will present a specific example, in which predictions of our RQD approach deviate from special relativity. In particular, we will demonstrate the approximate character of the Einstein's ``time dilation''
formula (\ref{eq:1b}) for decays of fast moving particles.

\subsection{Numerical results} \label{sc:numerical}

In this subsection we will calculate the difference between the
accurate quantum mechanical result (\ref{eq:62})\footnote{For similar calculations of the decay law of fast moving particles see \cite{Urbanowski}.} and the
special-relativistic formula (\ref{eq:1b})

\begin{eqnarray}
\omega_{|\mathbf{0})}^{SR}(\theta, t) =
\omega_{|\mathbf{0})}\left(0, \frac{t}{\cosh \theta} \right)
\label{eq:time-dilation2}
\end{eqnarray}

\noindent In this calculation we assume that the mass
distribution $|\mu(m)|^2$ of the unstable particle has the
Breit-Wigner form\footnote{see equation (\ref{eq:13.21a}) and Fig.
\ref{fig:13.4}}

\begin{eqnarray}
|\mu(m)|^2  = \left \{
\begin{array}{c}
 \frac{\alpha \Gamma /2 \pi}{\Gamma^2/4  + (m -
m_a )^2}, \mbox{
  } if \mbox{ } m \geq m_b + m_c  \\
  \mbox{ } \\
0, \mbox{
  } \ \ \ \ \ \ \ \ \ \ \ if \mbox{ }m < m_b + m_c
\end{array} \right.  \label{eq:64}
\end{eqnarray}

\noindent where  parameter $\alpha$ is a factor that ensures the
normalization to unity

\begin{eqnarray*}
\int \limits_{m_b + m_c}^{\infty} |\mu(m)|^2 =1
\end{eqnarray*}

\noindent The following parameters of this distribution have been chosen: The mass of the unstable particle was $m_a=
1000$ MeV/$c^2$, the total mass of the decay products
 was $m_b + m_c = 900$ MeV/$c^2$, and the width of the mass distribution was $\Gamma$= 20 MeV/$c^2$.
 These values do not
correspond to any real particle, but they are typical for strongly
decaying baryon resonances.

It is convenient to measure time in units of the lifetime $\tau _0
\cosh \theta$.
 Denoting $\chi \equiv t/(\tau _0 \cosh \theta)$, we find that
special-relativistic decay laws (\ref{eq:time-dilation2}) for any
rapidity $\theta$ are given by the same universal  function
$\omega^{SR}(\chi)$.  This function   was evaluated
 for values of $\chi$ in the interval
from 0 to 6
 with the step of 0.1.   Calculations were performed by direct numerical
integration of equation (\ref{eq:35}) using the \emph{Mathematica}
program shown below

\begin{verbatim}
gamma = 20

mass = 1000

theta = 0.0

Do[Print[(1/0.9375349) Abs[NIntegrate [gamma/(2 Pi) / (gamma^2/4 +(x
- mass)^2) Exp[ I t Sqrt [x^2 + mass^2 (Sinh [theta])^2] Cosh
[theta] / gamma], {x, 900, 1010, 1100, 300000}, MinRecursion -> 3,
MaxRecursion -> 16, PrecisionGoal -> 8, WorkingPrecision -> 18]]^2],
{t, 0.0, 6.0, 0.1}]
\end{verbatim}

\begin{figure}
\centering
\includegraphics {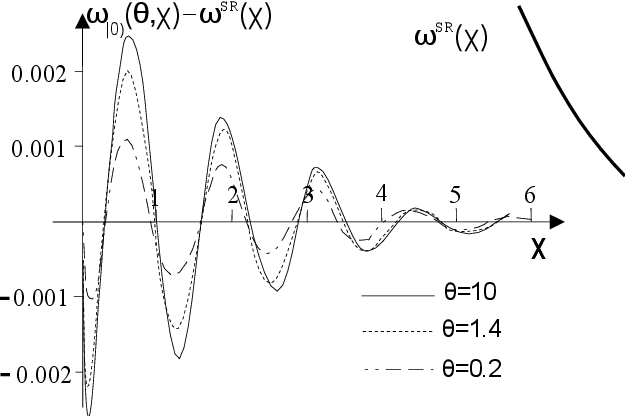} \caption{ Corrections to the Einstein's
``time dilation'' formula (\ref{eq:1b}) for the decay law of
unstable particle
 moving with the speed $v = c \tanh \theta$.
Parameter $\chi$ is time measured in units of $\tau_0 \cosh
\theta$.} \label{fig:13.5}
\end{figure}

\noindent As expected, function $\omega^{SR}(\chi)$ (shown by the
thick solid line in Fig. \ref{fig:13.5}) is very close to the
exponent $e^{- \chi}$.

Next we used equation (\ref{eq:35})\footnote{or, what is essentially the same, equation (\ref{eq:62})} to calculate decay laws
$\omega_{|\mathbf{0})}(\theta, \chi)  $ of moving particles. These calculations were done by the same Mathematica code, only instead of $\theta=0$ we used
three values
 of the rapidity parameter $\theta$ (=\verb/theta/), namely $\theta =$ 0.2, 1.4, and
10.0. These rapidities corresponded to moving frame velocities of 0.197c, 0.885c, and 0.999999995c, respectively.  These calculations revealed important differences between acurate quantum mechanical result (\ref{eq:35}) and the special-relativistic approximation (\ref{eq:time-dilation2}). These differences $
\omega_{|\mathbf{0})}(\theta, \chi)  - \omega ^{SR}(\chi) $
are plotted as thin lines  in Fig. \ref{fig:13.5}.  Deviations from the Einstein's time dilation formula are as high as 0.3\% in the example considered here.  Is it possible to observe these deviations in experiments with moving unstable particles?

The lifetime of the particle $a$ considered in our example ($ \tau_0
\approx 2 \times 10^{-22}$  s) is too short to be observed
experimentally. Unstable baryon resonances are identified
experimentally by the resonance behavior of the scattering
cross-section as a function of the collision energy, rather than by
direct measurements of the decay law.  So, calculated corrections
to the Einstein's time dilation law have only illustrative value.
However, from these data we can estimate the magnitude of
corrections for particles whose time-dependent decay laws can be
measured in a laboratory, e.g., for muons. Taking into account that
the magnitude of corrections is roughly proportional to the ratio
$\Gamma/m_a$ \cite{Stefanovich_decay, Shirokov_decay} and that in
our example $\Gamma/m_a= 0.02$,
 we can
expect that for  muons ($\Gamma \approx 2 \times 10^{-9} eV/c^2$,
$m_a \approx 105 MeV/c^2$, $\Gamma/m_a \approx 0.02 \times
10^{-15}$) the maximum magnitude of the correction should be about
$2 \times 10^{-18}$, which is much smaller than the precision of
modern experiments.\footnote{Most accurate experiments with decaying muons confirm
Einstein's time dilation formula with the precision of only
$10^{-3}$ \cite{muons, muons2}. }

 Our results indicate that all physical processes\footnote{not just particle decays considered here} viewed from a moving
reference frame do not go exactly $\cosh \theta$ slower, as special
relativity would predict. The exact slowdown pattern depends on
the physical makeup of the process and on interactions responsible for it.  For more discussions on how our RQD approach compares with special relativity and experiments see chapter \ref{sc:obs-interact}.

\subsection{Decays caused by boosts} \label{sc:dec-boost}

Recall that in subsection \ref{ss:dirac-forms} we discussed two
classes of inertial transformations of observers -
\emph{kinematical} and \emph{dynamical}. \index{kinematical inertial
transformation} \index{dynamical inertial transformation} According
to our Postulate \ref{postulateR}, we are working in the instant form of Dirac's dynamics, so space translations and rotations
are kinematical, while time translations and boosts are dynamical.
 Kinematical transformations
impose only trivial changes on the external appearance of the object and do
not influence its internal state. The description of kinematical
space translations and rotations  is a purely geometrical exercise
which does not require intricate knowledge of interactions in the
physical system. This conclusion is supported by observations of
unstable particles:   For two observers in different places or with
different orientations, the non-decay probability of the particle
has exactly the same value.

On the other hand, dynamical transformations depend on interaction
and directly affect the internal structure of the observed
system.\footnote{see subsection \ref{ss:dynamical-relativity}} The
dynamical effect of time translations on the unstable particle is
obvious - the particle decays with time. But what about boost transformations? Does the non-decay probability depend on the observer's velocity? Special relativity answers: ``No, there is no such dependence.''\footnote{see Appendix
\ref{ss:relativity}} And this answer is often believed to be
self-evident in discussions of relativistic effects. For example,
Polishchuk writes\footnote{This assertion does not hold in our RQD theory, as illustrated in Fig. \ref{fig:10.2}.}

\begin{quote}
\emph{Any event that is ``seen'' in one inertial system is ``seen''
in all others. For example if observer in one system ``sees'' an
explosion on a rocket then so do all other observers.} R. Polishchuk
\cite{Polishchuk}
\end{quote}

\noindent Applying this statement to decaying particles, we would
expect that the non-decay probability does not depend on the observer's velocity.
In particular, this would mean that at time $t=0$ we should have

\begin{eqnarray}
\omega(\theta, 0) =1 \label{eq:omega-theta-1}
\end{eqnarray}

\noindent for all $\theta$. Here we are going to prove that this
expectation is incorrect.

 Suppose that  special-relativistic
equation (\ref{eq:omega-theta-1}) is valid, i.e., for any $|\Psi \rangle
\in \mathcal{H}_a$ and any $\theta
> 0$, boost transformations of the observer keep the initial state vector within the unstable particle's subspace

\begin{eqnarray*}
e^{\frac{ic}{\hbar}K_x \theta} | \Psi \rangle &\in& \mathcal{H}_a
\end{eqnarray*}

\noindent Then the subspace $\mathcal{H}_a$  is invariant under
action of boosts $e^{\frac{ic}{\hbar}K_x \theta}$, which means that
 operator $K_x$
commutes with the projection $T$ on the subspace $\mathcal{H}_a$.
Then from Poincar\'e commutator (\ref{eq:5.55}) and $[T,
P_{0x}]= 0$ it follows by Jacobi identity that

\begin{eqnarray*}
[T,H] &=& \frac{ic^2}{\hbar}[T,[K_x,P_{0x}]] = \frac{ic^2}{\hbar}
[K_x,[T, P_{0x}]] - \frac{ic^2}{\hbar}[P_{0x},[T,K_x]] \\ &=& 0
\end{eqnarray*}

\noindent which contradicts the fundamental property  (\ref{eq:44b}) of unstable
systems. This contradiction implies that, in fact, the state
$e^{\frac{ic}{\hbar}K_x \theta} | \Psi \rangle$ does not correspond
to the particle $a$ with 100\% probability. This state must contain
contributions from decay products even at $t=0$

\begin{eqnarray}
e^{\frac{ic}{\hbar}K_x \theta} | \Psi \rangle &\notin& \mathcal{H}_a
\label{eq:not-in} \\
\omega(\theta, 0) &<& 1, \mbox{  } for \; \; \theta \neq 0
\label{eq:not-in2}
\end{eqnarray}

\noindent This is the ``decay caused by boost,'' which means that
special-relativistic equations (\ref{eq:1b}) and
(\ref{eq:omega-theta-1}) are not accurate and that boosts of the
observer have a non-trivial effect on the internal state of the
observed unstable system.

The presence of ``decays caused by boosts'' means that particle composition of systems involving unstable states is not a relativistic invariant. For example, one should be careful when making assertions like this one:

\begin{quote}
\emph{Flavor is the quantum number that distinguishes the different
types of quarks and leptons. It is a Lorentz invariant quantity. For
example, an electron is seen as an electron by any observer, never
as a muon.} C. Giunti and M. Lavender \cite{oscillations}
\end{quote}

\noindent  Although this statement about the electron is correct
(because the electron is a stable particle), it is not true about
the muon. An unstable muon can be
seen as a single particle by the observer at rest and, at the same time, it will be perceived as a group of
three decay products (an electron, a neutrino $\nu_{\mu}$, and an
antineutrino $\tilde{\nu}_{e}$) by a moving observer.

In spite of its fundamental importance, the effect of boosts on the
non-decay probability is very small. For example, our rather
accurate approximation (\ref{eq:appr_dec}) failed to ``catch'' this
effect. Indeed, for $t=0$ this formula predicted

\begin{eqnarray*}
\omega(\theta, 0) &=& \int  d\mathbf{p} |\psi( \mathbf{p})|^2 \left|
\int \limits_{m_b + m_c}^{\infty} dm
 \vert \mu(m)\vert^2    \right|^2 = 1
\end{eqnarray*}

\noindent instead of the expected $\omega(\theta, 0) < 1$.

\subsection{Particle decays in different forms of dynamics}
\label{sc:different2}

Throughout this section we assumed that interaction responsible for
the decay belongs to the Bakamjian-Thomas instant form of dynamics.
However, as we saw in subsection \ref{ss:non-separability}, the
Bakamjian-Thomas form does not allow separable interactions, so,
most likely, this is not the form preferred by nature. Therefore, it
would be interesting to calculate decay laws  in non-Bakamjian-Thomas
instant forms of dynamics as well. Although no such calculations
have been done yet, one can say with certainty that there is no form
of interaction in which the special-relativistic time dilation formula
(\ref{eq:time-dilation2}) is exactly valid. This follows from the
fact that in any instant form of dynamics  boost operators contain
interaction terms, so the ``decays caused by boosts'' - which contradict equation
(\ref{eq:time-dilation2}) - are always present.

What if the interaction responsible for the decay has a non-instant
form? Is it possible that there is a form of dynamics in which
Einstein's time dilation formula (\ref{eq:time-dilation2}) is exactly true? Our
answer to this question is \emph{No}. Let us consider, for example,
the point form of dynamics.\footnote{see subsection
\ref{bakam-point}} In this case the subspace $\mathcal{H}_a$ of the
unstable particle is invariant with respect to boosts, $[K_{0x}, T]
= 0$, so there can be no boost-induced decays (\ref{eq:not-in}).
However, we obtain a rather surprising relationship between decay
laws of the same particle viewed from the moving reference frame
$\omega(\theta, t)$ and from the frame at rest $\omega(0,
t)$\footnote{In this derivation we assumed that the state of the
particle at rest $|\Psi \rangle$ is an eigenvector of the
interacting momentum operator $\mathbf{P} |\Psi \rangle = 0$. }

\begin{eqnarray*}
\omega(\theta, t) &=& \langle \mathbf{0} |
e^{-\frac{ic}{\hbar}K_{0x}  \theta} e^{\frac{i}{\hbar}Ht}T
e^{-\frac{i}{\hbar}Ht}
e^{\frac{ic}{\hbar}K_{0x}  \theta} | \mathbf{0} \rangle \\
&=& \langle \mathbf{0} | e^{-\frac{ic}{\hbar}K_{0x}  \theta}
e^{\frac{i}{\hbar}Ht} e^{\frac{ic}{\hbar}K_{0x}  \theta}
 e^{-\frac{ic}{\hbar}K_{0x}  \theta}T
e^{\frac{ic}{\hbar}K_{0x}  \theta}e^{-\frac{ic}{\hbar}K_{0x}
\theta}e^{-\frac{i}{\hbar}Ht}
e^{\frac{ic}{\hbar}K_{0x}  \theta} | \mathbf{0} \rangle \\
&=& \langle \mathbf{0} | e^{-\frac{ic}{\hbar}K_{0x}  \theta}
e^{\frac{i}{\hbar}Ht} e^{\frac{ic}{\hbar}K_{0x}  \theta} T
e^{-\frac{ic}{\hbar}K_{0x}  \theta}e^{-\frac{i}{\hbar}Ht}
e^{\frac{ic}{\hbar}K_{0x}  \theta} | \mathbf{0} \rangle \\
&=& \langle \mathbf{0} | e^{\frac{it}{\hbar} (H \cosh \theta  + cP_x
\sinh \theta)} T e^{-\frac{it}{\hbar}(H \cosh \theta  + cP_x \sinh
\theta)}
 | \mathbf{0} \rangle \\
&=& \langle \mathbf{0} | e^{\frac{it}{\hbar} H \cosh \theta } T
e^{-\frac{it}{\hbar}H \cosh \theta }
 | \mathbf{0} \rangle \\
&=& \omega(0, t \cosh \theta)
\end{eqnarray*}

\noindent where the last equality follows from comparison with the decay law at rest (\ref{eq:3c}). This means that the decay rate in the moving frame is
$\cosh \theta$ times \emph{faster} than in the rest frame. This
is in direct contradiction with experiments.

The point form of dynamics is not acceptable for the description of
decays for yet another reason. Due to the interaction-dependence of
the total momentum operator (\ref{eq:inter-p-point}), one should
expect decays induced by space translations

\begin{eqnarray}
e^{\frac{i}{\hbar}P_x a} | \Psi \rangle &\notin& \mathcal{H}_a,
\mbox{  } for \; \; a\neq 0 \label{eq:not-in-tr}
\end{eqnarray}

\noindent Translation-induced and/or rotation-induced decays  are expected
in all forms of dynamics (except the instant form). This
contradicts  our experience, which suggests that the
composition of an unstable particle is not affected by these
kinematical transformations. Therefore only the instant form of
dynamics is appropriate for the description of particle decays. This conclusion supports our Postulate \ref{postulateR}.

\chapter{RQD IN HIGHER ORDERS} \label{ch:hydrogen-revisited}

\begin{quote}
\textit{There must be no barriers for freedom of inquiry.
There is no place for dogma in science. The scientist
is free and must be free to ask any question,
to doubt any assertion, to seek for any evidence,
to correct any errors.}

\small
\hspace{1in} J. Robert Oppenheimer
 \normalsize
\end{quote}

\vspace{0.5in}

In section \ref{sc:dressed} we derived the 2nd order dressed particle Hamiltonian $H_0 + V_2^d$ by applying a unitary dressing transformation to the field-based Hamiltonian of QED. However, this approach is hardly applicable to higher perturbation orders because dressing  requires separate processing of terms of different types (\emph{unphys}, \emph{renorm},  \emph{phys}). To achieve that, instead of using the compact field representation, one needs to keep all terms expressed through creation and annihilation operators as in (\ref{eq:9.48}) - (\ref{eq:9.49}). In this representation even the original QED Hamiltonian takes rather inconvenient cumbersome form shown in Appendix \ref{ss:int-part-oper}. The task of dressing transformation is further complicated by the necessity to calculate multiple commutators of these long expressions.

Fortunately, there is a much simpler approach, which leads to the same desired result for the dressed Hamiltonian $H^d$. This approach is based on formulas from subsection \ref{ss:lim-inf-cutoff}. Suppose, for example, that we want to find the 3rd and 4th order contributions $V_3^d$ and $V_4^d$ to the electron-proton dressed interaction. To do that we can  use equations (\ref{eq:V3d-approx}) - (\ref{eq:12.11}) and write

\begin{eqnarray}
V_3^d    &\approx& (\Sigma_3^c)^{ph} \label{eq:V3x} \\
V_4^d &\approx&
  (\Sigma_4^c)^{ph} -
V_2^d \underline{V_2^d} \label{eq:V4}
\end{eqnarray}

\noindent All terms on the right hand sides can be found relatively easily. All warnings from subsection \ref{ss:3-rd-order-dressed} regarding the non-uniqueness of the off-energy-shell behavior of the potentials remain valid here. Indeed, according to (\ref{eq:ScSigma}), $\Sigma$-operators are directly related to the scattering matrix $S^c$

\begin{eqnarray}
 \underbrace{(\Sigma_3^c)^{ph}} &=& S_3^c \label{eq:sigma_3}\\
\underbrace{(\Sigma_4^c)^{ph}} &=& S_4^c \label{eq:sigma_4}
\end{eqnarray}

\noindent whose calculation is the primary goal of the QED formalism, as described in all textbooks.  In particular, the 4th order scattering matrix $S_4^c$ has been calculated in (\ref{eq:S4-renorm}). The 3rd order contribution $S_3^c$ will be found in subsection \ref{ss:brems}.

We will use formula (\ref{eq:V3x}) to calculate the 3rd order dressed interaction $V_3^d$  in section \ref{sc:spontaneous}. This interaction potential will help us to calculate lifetimes and energy shifts of unstable levels in the hydrogen atom. In section \ref{ss:hydrogen} we will use (\ref{eq:V4}) to derive  the 4th order electron-proton interaction $V_4^d$, whose experimental manifestations include the electron's anomalous magnetic moment and the Lamb shift.

\section{Spontaneous radiative transitions} \label{sc:spontaneous}

In  chapter \ref{ch:decays} we discussed rather general properties of
the decay process. In particular, we did not specify the exact
type of the unstable system and the form of the decay interaction
operator $V$. So, we left our formula for the decay rate in an
unprocessed form (\ref{eq:Gamma2}). In this section we would like to
fill this gap and perform a complete calculation of the decay rate
for a realistic system  -- an excited state of the hydrogen atom.

The simplest interaction responsible for the spontaneous photon emission from hydrogen excited states has the structure $d^{\dag}a^{\dag}c^{\dag}da$. In the dressed particle Hamiltonian such terms first appear in the 3rd perturbation order.\footnote{see Table \ref{table:12.1}} So, our plan in this section is as follows:
In subsection \ref{ss:brems} we are going  to find the 3rd order contribution
to the $S$-operator having the desired structure
$d^{\dag}a^{\dag}c^{\dag}da$. Then in subsection \ref{ss:perturbation2} we will use
the correspondence (\ref{eq:V3x}), (\ref{eq:sigma_3}) between the $S$-operator and the
dressed particle Hamiltonian in order to obtain
$V_3^d[d^{\dag}a^{\dag}c^{\dag}da]$ near the energy shell. Next, in subsections \ref{sc:instability}-\ref{ss:instab}, we will use approach developed in
chapter \ref{ch:decays} to calculate the radiative transition rate between two
states of the hydrogen atom and associated  energy shifts.

\subsection{Bremsstrahlung scattering amplitude}
\label{ss:brems}

To find the (bremsstrahlung) $d^{\dag}a^{\dag}c^{\dag}da$ part of the scattering
operator in the 3rd order, we use the Feynman-Dyson perturbation
theory with interaction operator  $V_1$  from (\ref{eq:11.6a}). Then

\begin{eqnarray}
 S_3 &=&  \frac{i}{3! \hbar^3} \int \limits_{-\infty }^{+\infty}
 dt_1 dt_2 dt_3 T[V_1(t_1) V_1(t_2) V_1(t_3)] \nonumber \\
 &=&
 \frac{i}{3! \hbar^3} \int d^4x_1 d^4x_2 d^4x_3 T[V_1(\tilde{x}_1) V_1(\tilde{x}_2) V_1(\tilde{x}_3)] \\
  &=& \frac{i}{3! \hbar^3} \int d^4x_1 d^4x_2 d^4x_3  \times \nonumber \\
&\mbox{ }&   T[(J_{\mu}(\tilde{x}_1)A^{\mu} (\tilde{x}_1) +
\mathcal{J}_{\mu}(\tilde{x}_1)A^{\mu} (\tilde{x}_1)) \times \nonumber \\
&\mbox{ }&  (J_{\nu}(\tilde{x}_2)A^{\nu} (\tilde{x}_2) +
\mathcal{J}_{\nu}(\tilde{x}_2)A^{\nu} (\tilde{x}_2)) \times \nonumber \\
&\mbox{ }&  (J_{\lambda}(\tilde{x}_3)A^{\lambda} (\tilde{x}_3) +
\mathcal{J}_{\lambda}(\tilde{x}_3)A^{\lambda} (\tilde{x}_3))] \label{eq:S3x}
\end{eqnarray}

\noindent where $J_{\mu}$ and $\mathcal{J}_{\mu}$ are electron-positron and proton-antiproton current operators, respectively, as defined in Appendix \ref{sc:current-dens}. Expanding the three parentheses we get 8 terms under the integral sign. The term
of the type $JJJ$ cannot contribute to the electron-proton
bremsstrahlung, because it lacks the proton component. Similarly the term
$\mathcal{J}\mathcal{J}\mathcal{J}$ does not contribute and should
be omitted as well. Let us first consider the three terms $JJ\mathcal{J} +
J\mathcal{J}J + \mathcal{J}JJ$. As the order of factors under the
time-ordering sign is irrelevant, these three terms are equal. So,
the corresponding contribution to the coefficient function of the
$S$-operator is\footnote{Summation is performed on repeating indices $\mu, \nu, \lambda = 0,1,2,3$.}

\begin{eqnarray*}
&\mbox{ }& S^{JJ\mathcal{J}}_3 (\mathbf{p},\mathbf{q},\mathbf{p}',
\mathbf{q}',\mathbf{s}; \sigma, \tau, \sigma ', \tau ', \kappa  ) \\
  &=& \frac{i}{2 \hbar^3 c^3} \int d^4x_1 d^4x_2 d^4x_3 \times  \\
&\mbox{ }&  \langle 0 | a_{\mathbf{q}, \tau } d_{\mathbf{p}, \sigma}
T[J_{\mu}(\tilde{x}_1)A^{\mu} (\tilde{x}_1)
J_{\nu}(\tilde{x}_2)A^{\nu} (\tilde{x}_2)
\mathcal{J}_{\lambda}(\tilde{x}_3)A^{\lambda} (\tilde{x}_3)]
d^{\dag}_{\mathbf{p'}, \sigma'} a^{\dag}_{\mathbf{q'}, \tau'}
c^{\dag}_{\mathbf{s}, \kappa }
|0 \rangle \\
  &=& \frac{ie^3 }{2 \hbar^3} \int d^4x_1 d^4x_2 d^4x_3 \times  \\
&\mbox{ }&  \langle 0 | a_{\mathbf{q}, \tau } d_{\mathbf{p}, \sigma}
T[(\overline{\psi}(\tilde{x}_1) \gamma_{\mu} \psi
(\tilde{x}_1)A^{\mu} (\tilde{x}_1)) (\overline{\psi}(\tilde{x}_2)
\gamma_{\nu} \psi (\tilde{x}_2) A^{\nu} (\tilde{x}_2)) \times
\\
&\mbox{ } & (\overline{\Psi} (\tilde{x}_3) \gamma_{\lambda} \Psi
(\tilde{x}_3) A^{\lambda} (\tilde{x}_3))] d^{\dag}_{\mathbf{p'},
\sigma'} a^{\dag}_{\mathbf{q'}, \tau'} c^{\dag}_{\mathbf{s}, \kappa
} |0 \rangle
\end{eqnarray*}

\begin{figure}
\centering
\includegraphics {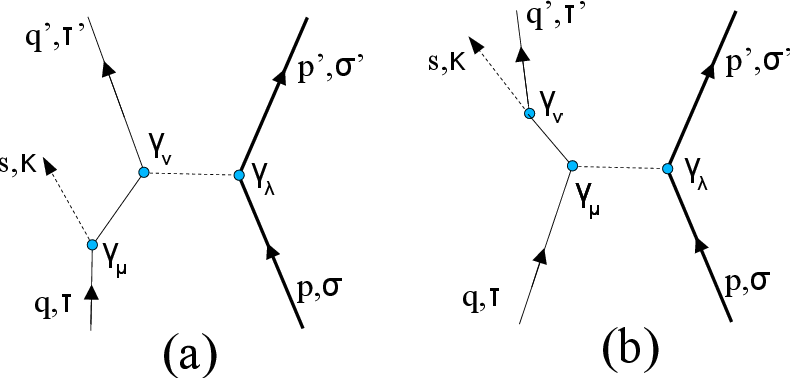} \caption{ 3rd order Feynman diagrams for
the photon emission in electron-proton collisions.} \label{fig:14.1}
\end{figure}

\noindent This function can be evaluated  by
drawing two Feynman diagrams shown in Fig. \ref{fig:14.1} and
processing them according to Feynman rules from subsection
\ref{ss:Fdiagrams}.

\begin{eqnarray*}
&\mbox{ }& S^{JJ\mathcal{J}}_3 (\mathbf{p},\mathbf{q},\mathbf{p}',
\mathbf{q}', \mathbf{s}; \sigma, \tau, \sigma ', \tau ', \kappa) \\
   &=& -\frac{ie^3c^{3/2}}{4 \pi^2}\frac{mMc^4 }{
\sqrt{\omega_{\mathbf{q}}\omega_{\mathbf{q'}}\Omega_{\mathbf{p}}\Omega_{\mathbf{p'}}}}
\frac{c }{(2 \pi \hbar)^{3/2}\sqrt{2s}} \times
\\
&\mbox{ }& \delta^4(-\tilde{s} +\tilde{q}
+\tilde{p}-\tilde{p}'-\tilde{q}') \frac{1}{(\tilde{p}-\tilde{p}')^2
} \times
 \\ &\mbox{ } &
 \overline{u}_a(\mathbf{q}, \tau) \Bigl(
\cross{e}^{ab}(\mathbf{s}, \kappa) \frac{(\cross{q} -\cross{s}
+mc^2)_{bc}}{(\tilde{q}-\tilde{s})^2 - m^2c^4 }
\cross{W}^{cd}(\mathbf{p}, \sigma;\mathbf{p'}, \sigma') \\
&\ & +\cross{W}^{ab}(\mathbf{p}, \sigma;\mathbf{p'}, \sigma')
\frac{(\cross{s} + \cross{q}' +mc^2)_{bc}}{(\tilde{s}+\tilde{q}')^2 -
m^2c^4 }
 \cross{e}^{cd}(\mathbf{s}, \kappa) \Bigr) u_d(\mathbf{q'}, \tau')
\end{eqnarray*}

Now let us assume that the electron and the proton are
non-relativistic and simplify the above expression. According to
approximations derived in Appendix \ref{ss:non-rel} and using

\begin{eqnarray*}
\tilde{p}^2 &=& (\tilde{p}')^2 = m^2c^4 \\
\tilde{s}^2 &=& 0 \\
(\tilde{p}-\tilde{p}')^2 &\approx& -c^2(\mathbf{p}-\mathbf{p}')^2 \\
(\tilde{s}+\tilde{q}')^2 - m^2c^4 &=& 2 \tilde{s} \cdot \tilde{q}'\\
(\tilde{q}-\tilde{s})^2 - m^2c^4 &=& -2 \tilde{s} \cdot \tilde{q} \\
 W^0 (\mathbf{p}, \sigma,\mathbf{p'},
\sigma') &\approx& \delta_{\sigma,
\sigma'} \\
\mathbf{W}(\mathbf{p}, \sigma,\mathbf{p'},
\sigma') &\approx& 0 \\
\sqrt{\omega_{\mathbf{q}}\omega_{\mathbf{q'}}\Omega_{\mathbf{p}}\Omega_{\mathbf{p'}}}
&\approx& Mmc^4
\end{eqnarray*}

\noindent we obtain

\begin{eqnarray*}
&\mbox{ }& S^{JJ\mathcal{J}}_3 (\mathbf{p},\mathbf{q},\mathbf{p}',
\mathbf{q}', \mathbf{s}; \sigma, \tau, \sigma ', \tau ', \kappa) \\
    &\approx&  ie^3 c^{5/2}\frac{1 }{ 4 \pi^2}
    \frac{1}{(2 \pi \hbar)^{3/2}\sqrt{2s}} \frac{\delta^4(-\tilde{s}
    +\tilde{q}
+\tilde{p}-\tilde{p}'-\tilde{q}')\delta_{ \sigma
\sigma'}}{c^2(\mathbf{p}-\mathbf{p}')^2 } \times
 \\ &\mbox{ } &
 \overline{u}_a(\mathbf{q}, \tau) \Bigl(
-\cross{e}^{ab}(\mathbf{s}, \kappa) \frac{(\cross{q} -\cross{s}
+mc^2)_{bc}}{2\tilde{q} \cdot \tilde{s}}
\gamma_0^{cd}  \\
&+&  \gamma_0^{ab} \frac{(\cross{s} + \cross{q}'
+mc^2)_{bc}}{2\tilde{q}' \cdot \tilde{s}}
 \cross{e}^{cd}(\mathbf{s}, \kappa ) \Bigr) u_d(\mathbf{q'}, \tau')
\end{eqnarray*}

Next we  assume that the energy and momentum of the
emitted photon is much less than energies and momenta of charged
particles. We also use Dirac equations (\ref{gamma-mu1}),
(\ref{gamma-mu2}) and the non-relativistic approximation
(\ref{eq:U0}) to write\footnote{Here we used the tilde to write
$\tilde{e}$ in order to stress the 4-component nature of this quantity despite the fact that it does not transform as a 4-vector.}

\begin{eqnarray*}
&\ & \overline{u}(\mathbf{q}, \tau) \cross{e} (\cross{q} -\cross{s}
+mc^2) \gamma_0 u(\mathbf{q'}, \tau') \\
&\approx& \overline{u}(\mathbf{q}, \tau) e_{\mu} \gamma^{\mu}
(q_{\nu}
\gamma^{\nu}  +mc^2)\gamma_0 u(\mathbf{q'}, \tau')\\
&=& \overline{u}(\mathbf{q}, \tau)((- q_{\nu} \gamma^{\nu}
+mc^2)e_{\mu} \gamma^{\mu}  + 2 g^{\mu \nu} q_{\nu} e_{\mu})\gamma_0 u(\mathbf{q'}, \tau')\\
&=& \overline{u}(\mathbf{q}, \tau)((- \cross{q}
+mc^2)e_{\mu} \gamma^{\mu}  + 2 \tilde{q} \cdot \tilde{e})\gamma_0 u(\mathbf{q'}, \tau')\\
&=& 2\overline{u}(\mathbf{q}, \tau) (\tilde{q} \cdot \tilde{e}) \gamma_0 u(\mathbf{q'}, \tau')
= 2U^0(\mathbf{q}, \tau; \mathbf{q'}, \tau') (\tilde{q} \cdot \tilde{e}) \\
&\approx& 2 \delta_{\tau,  \tau'} (\tilde{q} \cdot \tilde{e}) \\ \\
&\ & \overline{u}(\mathbf{q}, \tau) \gamma_0 (\cross{s} + \cross{q}'
+mc^2) \cross{e}  u(\mathbf{q'}, \tau') \\
&=& 2U^0(\mathbf{q}, \tau; \mathbf{q'}, \tau') (\tilde{q}' \cdot \tilde{e})
\approx 2 \delta_{\tau,  \tau'} (\tilde{q}' \cdot \tilde{e}) \\
\end{eqnarray*}

\noindent Further approximations yield

\begin{eqnarray*}
 \tilde{s} \cdot \tilde{q}'
&=& cs \omega_{\mathbf{q}'} -c^2 (\mathbf{s} \cdot
\mathbf{q}') \approx mc^3s \\
\tilde{s} \cdot \tilde{q} &\approx& mc^3s \\
\tilde{q} \cdot \tilde{e} &=& -c (\mathbf{q} \cdot \mathbf{e}) \\
\tilde{q}' \cdot \tilde{e} &=& -c (\mathbf{q}' \cdot \mathbf{e})
\end{eqnarray*}

\noindent Therefore\footnote{Our result in (\ref{eq:s_3^112a}) can
be compared with equations (7.57) - (7.58) in \cite{Bjorken1}.}

\begin{eqnarray}
&\mbox{ }& S^{JJ\mathcal{J}}_3 (\mathbf{p},\mathbf{q},\mathbf{p}',
\mathbf{q}', \mathbf{s}; \sigma, \tau, \sigma ', \tau ', \kappa) \nonumber \\
    &\approx&   \frac{ie^3\sqrt{c} \delta^4(-\tilde{s} +\tilde{q}  +\tilde{p}-\tilde{p}'-\tilde{q}') }{4 \pi^2(2 \pi \hbar)^{3/2}\sqrt{2s}} \cdot
  \frac{\delta_{ \sigma \sigma'}
\delta_{\tau,  \tau'}}{(\mathbf{p}-\mathbf{p}')^2}
\left(\frac{\tilde{q}'
\cdot \tilde{e}}{\tilde{q}' \cdot \tilde{s} } - \frac{\tilde{q} \cdot \tilde{e}}
{\tilde{q} \cdot \tilde{s} } \right) \nonumber  \\
\label{eq:s_3^112a} \\
    &\approx&  - \frac{ie^3 \delta^4(\tilde{q}
+\tilde{p}-\tilde{p}'-\tilde{q}') }{4\pi^2m(2 \pi
\hbar)^{3/2}\sqrt{2(cs)^3}} \cdot \frac{\delta_{ \sigma \sigma'}
\delta_{\tau,  \tau'} (\mathbf{q}' - \mathbf{q})
\cdot \mathbf{e}(\mathbf{s}, \kappa)}{(\mathbf{q}'-\mathbf{q})^2} \nonumber \\
\label{eq:s_3^112}
\end{eqnarray}

\noindent This is our final expression for the terms $JJ\mathcal{J}$ in the
scattering operator (\ref{eq:S3x}). The contribution from $J\mathcal{J}\mathcal{J} + \mathcal{J}J\mathcal{J} + \mathcal{J}\mathcal{J}J$
terms can be obtained simply by replacing the electron's mass $m$ in
(\ref{eq:s_3^112}) by the proton's mass $M$. So, this contribution
is much smaller and will be neglected.

\subsection{3rd order perturbation Hamiltonian}
\label{ss:perturbation2}

From results in the preceding subsection we can find the 3rd order
contribution $V_3^d[d^{\dag}a^{\dag}c^{\dag}da]$ to the dressed particle interaction Hamiltonian. The relationship between the dressed Hamiltonian and the scattering operator in the 3rd order is given by equations (\ref{eq:V3x}) and (\ref{eq:sigma_3}). So\footnote{For brevity we drop the label $[d^{\dag}a^{\dag}c^{\dag}da]$ from the $V_3^d$ operator symbol.}

\begin{eqnarray}
&\mbox{ }& V_3^d \nonumber \\
&=& \sum_{\sigma \tau \sigma ' \tau ' \kappa} \int
d\mathbf{p} d\mathbf{q} d\mathbf{p}' d\mathbf{q}'d\mathbf{s}
V^{JJ\mathcal{J}}_3(\mathbf{p},\mathbf{q},\mathbf{p}', \mathbf{q}',
\mathbf{s}; \sigma, \tau, \sigma ', \tau ', \kappa)
a^{\dag}_{\mathbf{q}', \tau'} d^{\dag}_{\mathbf{p}', \sigma'}
c^{\dag}_{\mathbf{s}, \kappa} a_{\mathbf{q}, \tau} d_{\mathbf{p},
\sigma} \nonumber \\
 \label{eq:v1123}
\end{eqnarray}

\noindent whose coefficient function is\footnote{This formula is obtained simply by dividing scattering amplitude (\ref{eq:s_3^112}) by the factor $(-2 \pi i)$ and omitting the energy delta function $\delta(\omega_q + \Omega_p - \omega_{q'} - \Omega_{p'})$. }

\begin{eqnarray}
&\mbox{ }& V^{JJ\mathcal{J}}_3 (\mathbf{p},\mathbf{q},\mathbf{p}',
\mathbf{q}', \mathbf{s}; \sigma, \tau, \sigma ', \tau ', \kappa) \nonumber \\
    &\approx&  \frac{ e^3 }{8 \pi^3 m(2 \pi
\hbar)^{3/2}\sqrt{2(cs)^3}}  \delta(\mathbf{q  +p-p'-q'})
\frac{\delta_{ \sigma \sigma'} \delta_{\tau,  \tau'} (\mathbf{q}' -
\mathbf{q}) \cdot \mathbf{e}(\mathbf{s},
\kappa)}{(\mathbf{q}'-\mathbf{q})^2} \nonumber \\
\label{eq:14.9a}
\end{eqnarray}

\noindent The action of the operator (\ref{eq:v1123}) on a two particle
(electron+proton) initial state

\begin{eqnarray*}
|\Psi_i \rangle \equiv \sum_{\lambda \nu} \int d\mathbf{p}''
d\mathbf{q}''
 \Psi(\mathbf{p}'',\mathbf{q}''; \lambda, \nu)
a^{\dag}_{\mathbf{q}'', \nu} d^{\dag}_{\mathbf{p}'', \lambda} |0
\rangle
\end{eqnarray*}

\noindent is

\begin{eqnarray*}
&\mbox{ }& V_3^d |\Psi_i \rangle = \sum_{\sigma \tau \sigma ' \tau '
\kappa} \sum_{\lambda \nu} \int d\mathbf{p}'' d\mathbf{q}'' \int
d\mathbf{p} d\mathbf{q} d\mathbf{p}' d\mathbf{q}'d\mathbf{s}
V_3^{JJ\mathcal{J}}(\mathbf{p},\mathbf{q},\mathbf{p}', \mathbf{q}',
\mathbf{s}; \sigma, \tau, \sigma ', \tau ', \kappa) \times  \\
&\ & \Psi(\mathbf{p}'',\mathbf{q}''; \lambda, \nu)
a^{\dag}_{\mathbf{q}', \tau'} d^{\dag}_{\mathbf{p}', \sigma'}
c^{\dag}_{\mathbf{s}, \kappa} a_{\mathbf{q}, \tau} d_{\mathbf{p},
\sigma} a^{\dag}_{\mathbf{q}'', \nu} d^{\dag}_{\mathbf{p}'',
\lambda} |0 \rangle \\
&=& \sum_{\sigma \tau \sigma ' \tau ' \kappa} \sum_{\lambda \nu}
\int d\mathbf{p}'' d\mathbf{q}'' \int d\mathbf{p} d\mathbf{q}
d\mathbf{p}' d\mathbf{q}'d\mathbf{s}
V_3^{JJ\mathcal{J}}(\mathbf{p},\mathbf{q},\mathbf{p}', \mathbf{q}',
\mathbf{s}; \sigma, \tau, \sigma ', \tau ', \kappa ) \times  \\
&\ & \Psi(\mathbf{p}'',\mathbf{q}''; \lambda, \nu)
a^{\dag}_{\mathbf{q}', \tau'} d^{\dag}_{\mathbf{p}', \sigma'}
c^{\dag}_{\mathbf{s}, \kappa} \delta(\mathbf{q} -
\mathbf{q}'')\delta_{\tau \nu}\delta(\mathbf{p}- \mathbf{p}'')
\delta_{\sigma \lambda} |0 \rangle \\
&=& \sum_{\sigma \tau \sigma ' \tau ' \kappa}  \int d\mathbf{p}
d\mathbf{q} d\mathbf{p}' d\mathbf{q}'d\mathbf{s}
V_3^{JJ\mathcal{J}}(\mathbf{p},\mathbf{q},\mathbf{p}', \mathbf{q}',
\mathbf{s}; \sigma, \tau, \sigma ', \tau ', \kappa) \times \\
&\ & \Psi(\mathbf{p},\mathbf{q}; \sigma, \tau)
a^{\dag}_{\mathbf{q}', \tau'} d^{\dag}_{\mathbf{p}', \sigma'}
c^{\dag}_{\mathbf{s}, \kappa}  |0 \rangle \\
&=& \sum_{\sigma ' \tau ' \kappa} \int d\mathbf{p}'
d\mathbf{q}'d\mathbf{s} \Bigl( \sum_{\sigma \tau} \int d\mathbf{p}
d\mathbf{q} V_3^{JJ\mathcal{J}}(\mathbf{p},\mathbf{q},\mathbf{p}', \mathbf{q}',
\mathbf{s}; \sigma, \tau, \sigma ', \tau ', \kappa) \times \\
&\ & \Psi(\mathbf{p},\mathbf{q}; \sigma, \tau) \Bigr)
a^{\dag}_{\mathbf{q}', \tau'} d^{\dag}_{\mathbf{p}', \sigma'}
c^{\dag}_{\mathbf{s}, \kappa}  |0 \rangle
\end{eqnarray*}

\noindent where expression in big parentheses is the transformed
wave function of the 3-particle system (electron+proton+photon).
Using (\ref{eq:14.9a}), this wave function can be written as

\begin{eqnarray*}
&\ &\Psi'(\mathbf{p}', \mathbf{q}',\mathbf{s}; \sigma ' \tau '
\kappa) \\
&=&
 \sum_{\sigma \tau} \int d\mathbf{p}
d\mathbf{q}  \frac{e^3 \delta(\mathbf{q  +p-p'-q'}) \delta_{ \sigma
\sigma'} \delta_{\tau,  \tau'} (\mathbf{q}' - \mathbf{q}) \cdot
\mathbf{e}(\mathbf{s}, \kappa)}{8 \pi^3m(2 \pi \hbar)^{3/2}\sqrt{2(cs)^3}(\mathbf{q}'-\mathbf{q})^2} \Psi(\mathbf{p},\mathbf{q}; \sigma, \tau) \\
&=& \frac{e^3  }{8 \pi^3 m(2 \pi \hbar)^{3/2}\sqrt{2(cs)^3}}
 \int d\mathbf{k}    \frac{
\mathbf{k} \cdot \mathbf{e}(\mathbf{s}, \kappa)}{k^2}
\Psi(\mathbf{p}' + \mathbf{k},\mathbf{q}'-\mathbf{k}; \sigma',
\tau')
\end{eqnarray*}

By taking a Fourier transform and using (\ref{eq:psixy}), (\ref{eq:A.91}) we can switch to the position
representation for fermions\footnote{$\mathbf{x}$ and $\mathbf{y}$ are position vectors of the proton and the electron, respectively.}

\begin{eqnarray*}
&\ &\Psi'(\mathbf{x}, \mathbf{y},\mathbf{s}; \sigma ' \tau '
\kappa) \\
&=& \frac{e^3  }{8 \pi^3 m(2 \pi \hbar)^{9/2}\sqrt{2(cs)^3}}\int d\mathbf{p}'d\mathbf{q}'
e^{\frac{i}{\hbar}\mathbf{p}'\mathbf{x} +
\frac{i}{\hbar}\mathbf{q}'\mathbf{y}} \int d\mathbf{k} \frac{ \mathbf{k} \cdot \mathbf{e}(\mathbf{s},
\kappa) }{k^2}  \times \\
&\ &\Psi(\mathbf{p}' + \mathbf{k},\mathbf{q}'-\mathbf{k}; \sigma',
\tau') \\
&=&  \frac{e^3  }{8 \pi^3 m(2 \pi \hbar)^{9/2}\sqrt{2(cs)^3}}\int d\mathbf{p}'d\mathbf{q}'
e^{\frac{i}{\hbar}(\mathbf{p}' - \mathbf{k})\mathbf{x} +
\frac{i}{\hbar}(\mathbf{q}' + \mathbf{k})\mathbf{y}} \int
d\mathbf{k}  \frac{  \mathbf{k} \cdot
\mathbf{e}(\mathbf{s},
\kappa) }{k^2}  \times \\
&\ &\Psi(\mathbf{p}',\mathbf{q}'; \sigma',
\tau') \\
&=&  \frac{ \hbar^{3/2} e^3   }{(2 \pi \hbar)^3 m(2 \pi)^{3/2}\sqrt{2(cs)^3}}
\int d\mathbf{k} e^{\frac{i}{\hbar}\mathbf{k}(\mathbf{y}
-\mathbf{x})}   \frac{ \mathbf{k} \cdot \mathbf{e}(\mathbf{s},
\kappa)}{k^2} \times \\
&\ &\left(\frac{1}{(2 \pi \hbar)^3} \int d\mathbf{p}'d\mathbf{q}'
e^{\frac{i}{\hbar}\mathbf{p}' \mathbf{x} +
\frac{i}{\hbar}\mathbf{q}'\mathbf{y}} \Psi(\mathbf{p}',\mathbf{q}';
\sigma',\tau') \right) \\
&=& \frac{ e^3  \hbar^{1/2} }{m  (2 \pi)^{3/2}\sqrt{2(cs)^3}}
\frac{i(\mathbf{y-x})\cdot
\mathbf{e}(\mathbf{s}, \kappa)}{4 \pi |\mathbf{y-x}|^3}  \Psi(\mathbf{x},\mathbf{y}; \sigma',\tau')
\end{eqnarray*}

\noindent This means that the 3rd order position-space bremsstrahlung potential
between two charged particles
 is

\begin{eqnarray}
V_3^d(\mathbf{r}, \mathbf{s}, \kappa) &=& \frac{ i\hbar^{1/2}  e^3
 \mathbf{r} \cdot
\mathbf{e}(\mathbf{s}, \kappa)}{4 \pi m \sqrt{2(2 \pi cs)^3}r^3}  \label{eq:ham-posit}
\end{eqnarray}

\noindent  In contrast to 2nd order interactions discussed in chapter \ref{sc:coulomb}, this
potential does not conserve the number of particles. It is
responsible for the emission of photons with momentum $\mathbf{s}$ and helicity $\kappa$ by an electron moving in the
field of a heavy proton.\footnote{To maintain the Hermiticity of the
full Hamiltonian it must contain also a term, which is a Hermitian
conjugate of $V_3^d$. Apparently, this term is
responsible for the absorption of photons by the interacting system
electron+proton. }  We can expect that the radiation emission rate should be
proportional to the square of the matrix element of this operator
between appropriate initial and final states. One can notice that potential (\ref{eq:ham-posit}) is proportional to the electron's
acceleration in the Coulomb field\footnote{see equation (\ref{eq:d2r2/dt2})}

\begin{eqnarray*}
\mathbf{a} &\approx& \frac{e^2\mathbf{r}}{4 \pi m r^3}
\end{eqnarray*}

\noindent Thus we conclude that
the total radiated power should depend on the square of electron's
acceleration $a^2$. This is in agreement with the well-know
\emph{Larmor's formula} \index{Larmor's formula} of classical
electrodynamics. Thus the 3rd order bremsstrahlung interaction $V_3^d$ has direct relevance to the ``radiation reaction'' effect
\cite{McDonald, Parrott1, Parrott2}.

\subsection{Instability of excited atomic states}
\label{sc:instability}

The (bremsstrahlung) perturbation $V_3$ derived in the preceding subsection is also responsible for \emph{radiative
transitions} \index{radiative transitions} between energy levels in
atoms and other bound systems. As an example, let us consider two stationary states $2P^{1/2}$ and $1S^{1/2}$ of the hydrogen atom. The corresponding state vectors will be denoted $| \Psi_i \rangle$
and $| \Psi_f \rangle$, respectively. They are eigenvectors of the two-particle electron-proton Hamiltonian\footnote{For simplicity, here we ignore relativistic corrections. $\mathbf{p}_e $ and $\mathbf{p}_p $ are the electron's and proton's momentum operators, respectively, and
$\mathbf{r} \equiv \mathbf{r}_e - \mathbf{r}_p$.}

\begin{eqnarray}
H_{e-p} = \frac{p_e^2}{2m} + \frac{p_p^2}{2M} -\frac{e^2}{4 \pi r}
\label{eq:h'}
\end{eqnarray}

\noindent with eigenvalues $E_i$ and $E_f$

\begin{eqnarray*}
H_{e-p} | \Psi_i \rangle &=& E_i | \Psi_i \rangle \\
H_{e-p} | \Psi_f \rangle &=& E_f | \Psi_f \rangle \\
E_i &>& E_f
\end{eqnarray*}

 If we add interaction potential $V_3^d + (V_3^d)^{\dag}$ to the Hamiltonian $H_{e-p}$ then the state $| \Psi_i
\rangle$ is no longer stationary. The
two-particle subspace $\mathcal{H}_{pe}$ is not invariant with
respect to this potential. Operator  $V_3^d$ has a non-zero
matrix element between the stationary state $2P^{1/2}$ of the
hydrogen atom and the state $1S^{1/2} + \gamma$ which contains the
ground state of the atom and one emitted photon $\gamma$. Thus, an
atom initially  prepared in the high-energy state $| \Psi_i \rangle$ decays over time into two decay products: the atom in the state $| \Psi_f
\rangle$ plus a photon. This is exactly the
situation discussed in the preceding chapter: particles $a, b, c$ from section
\ref{sc:general-decay} are analogous to our states $| \Psi_i \rangle$,
$| \Psi_f \rangle$, and $\gamma$, respectively. Thus, according to arguments in subsection \ref{ss:eigenfunc}, the presence of the perturbation $V_3^d$ must result in energy shifts of $E_i$ and $E_f$ and in broadening of the level $2P^{1/2}$. Note that the level broadening does not apply to the
lowest-energy ground state $1S^{1/2}$. This state  cannot
decay spontaneously, simply because
there are no any lower energy states to which it can decay. Thus only the
ground state $| 1S^{1/2} \rangle$ is the true sharp-energy stationary
state of the RQD Hamiltonian $H_{e-p} + V_3^d + (V_3^d)^{\dag}$.

\subsection{Transition rate}

From formula
(\ref{eq:Gamma2}) we can obtain the probability density for the radiative transition between two
stationary atomic states $| \Psi_f \rangle$ and $| \Psi_i \rangle$ with the emission of one photon

\begin{eqnarray}
\Gamma(\mathbf{s}, \kappa) &=& \frac{8 \pi^2 s^2}{c^4} |\langle \Psi_i|
V_3^d(\mathbf{r}, \mathbf{s}, \kappa) | \Psi_f \rangle|^2 \frac{d
\eta^{-1}(z)}{dz} \Bigl|_{z=m_a} \label{eq:Gamma-new}
\end{eqnarray}

\noindent  Let us now simplify this expression a bit. Using equality

\begin{eqnarray}
[\mathbf{p}_e, H_{e-p}] &=& \left[\mathbf{p}_e, \frac{e^2}{4\pi r}
\right]= i \hbar e^2\frac{\mathbf{r} }{4 \pi r^3} \label{eq:peHep}
\end{eqnarray}

\noindent and denoting $E \equiv E_i -E_f = cs$ the energy of the
emitted photon we obtain for the matrix element

\begin{eqnarray*}
\langle \Psi_i| V_3^d(\mathbf{r}, \mathbf{s}, \kappa) | \Psi_f
\rangle &=& \frac{ i\hbar^{1/2}  e^3  }{ m \sqrt{2(2 \pi cs)^3}}
\left\langle \Psi_i \left| \frac{\mathbf{r} \cdot
\mathbf{e}(\mathbf{s}, \kappa)}{4 \pi r^3} \right|
\Psi_f \right\rangle \nonumber \\
&=& \frac{  e }{  m \sqrt{2\hbar(2 \pi E)^3}} \langle
\Psi_i| (\mathbf{p}_e\cdot \mathbf{e}) H_{e-p} -H_{e-p}(\mathbf{p}_e\cdot
\mathbf{e}) |
\Psi_f \rangle \nonumber \\
&=& \frac{   e  }{  m \sqrt{2 \hbar(2 \pi E)^3}} E\langle
\Psi_i| (\mathbf{p}_e\cdot \mathbf{e})  |\Psi_f \rangle \nonumber
\end{eqnarray*}

\noindent Next we use

\begin{eqnarray*}
-\frac{im}{ \hbar}[\mathbf{r}, H_{e-p}] &=& - \frac{im}{
\hbar}[\mathbf{r}_e, H_{e-p}] +\frac{im}{ \hbar}[\mathbf{r}_p, H_{e-p}]
\approx  \mathbf{p}_e - \frac{m}{M}\mathbf{p}_p
\approx \mathbf{p}_e
\end{eqnarray*}

\noindent to obtain

\begin{eqnarray}
\langle \Psi_i| V_3^d(\mathbf{r}, \mathbf{s}, \kappa) | \Psi_f
\rangle &=& -\frac{  ie  }{  \sqrt{2(2 \pi \hbar  E)^3}} E
\langle \Psi_i| (\mathbf{r}\cdot \mathbf{e})H_{e-p} - H_{e-p}
(\mathbf{r}\cdot \mathbf{e})
 |\Psi_f \rangle \nonumber \\
&=& \frac{ i   e  }{ \sqrt{2(2  \pi \hbar E)^3}} E^2 \langle
\Psi_i| (\mathbf{r}\cdot \mathbf{e})  |\Psi_f \rangle \nonumber \\
&=& \frac{ i e \sqrt{E} }{  \sqrt{2(2  \pi \hbar )^3}} \langle
\Psi_i| (\mathbf{r}\cdot \mathbf{e}) |\Psi_f \rangle \label{V-psii}
\end{eqnarray}

\noindent In order to find function $\eta^{-1}(z)$ in (\ref{eq:Gamma-new}) we turn to the definition (\ref{eq:13.6}). In the case of atomic radiative
transitions considered here, one of the decay products (the photon) is massless
($m_c =0$), and its energy is much smaller
than rest energies of the atomic states\footnote{In the case of the hydrogen atom we can assume, for example, $a=2P^{1/2}$ and $b = 1S^{1/2}$.} $cs \ll m_ac^2 \approx
m_bc^2$

\begin{eqnarray*}
z &=&\eta_{s} = \frac{1}{c^2}\left(\sqrt{m_b^2c^4 + c^2s^2} + cs \right) \approx \frac{1}{c^2}(m_bc^2 +cs) \\
s &=& \eta^{-1}(z) \approx (z-m_b)c \\
 \frac{d
\eta^{-1}(z)}{dz}\Bigl|_{z=m_a} &\approx& c \label{eq:eta-1z}
\end{eqnarray*}

\noindent Putting these results in (\ref{eq:Gamma-new}) we obtain

\begin{eqnarray*}
\Gamma(\mathbf{s}, \kappa) &=& \frac{8 \pi^2 E^2}{c^5}
\left(\frac{e \sqrt{E} }{ \sqrt{2(2 \pi \hbar)^3}}\right)^2 |
\langle \Psi_i| (\mathbf{r}\cdot
\mathbf{e}) |\Psi_f \rangle|^2 \nonumber \\
 &=&  \frac{   E^3}{2 \hbar^3 c^5} |
\langle \Psi_i| (\mathbf{d}\cdot \mathbf{e}(\mathbf{s}, \kappa)) |\Psi_f \rangle|^2  \ \ \ \label{eq:decay-rate}
\end{eqnarray*}

\noindent where $\mathbf{d} \equiv -e \mathbf{r}$ is the atom's
dipole moment operator.

The full transition probability\footnote{This probability is related to the brightness of the corresponding
spectral line. See also section 19.5 in \cite{Ballentine} and section 45 in \cite{BLP}.} should be obtained by summing  (\ref{eq:Gamma-new}) over two photon polarizations $\kappa = \pm 1$ and integrating over all possible directions $\mathbf{s}/s$ of the emitted photon\footnote{Here $\int d \Omega$ denotes the integral over orientations of $\mathbf{s}$, and $\mathbf{d}_{if} \equiv \langle \Psi_i| \mathbf{d} |\Psi_f \rangle$ is the matrix element of the dipole moment operator calculated on eigenfunctions of atomic states $| \Psi_i \rangle$ and $| \Psi_f \rangle$.}

\begin{eqnarray*}
\frac{1}{\tau_0} &=& \frac{c^2}{\hbar} \sum_{\kappa = -1}^{+1} \int d \Omega \Gamma(\mathbf{s}, \kappa) =  \frac{   E^3}{2 \pi \hbar^4 c^3} \sum_{\kappa = -1}^{+1} \int d \Omega |
\langle \Psi_i| (\mathbf{d}\cdot \mathbf{e}(\mathbf{s}, \kappa)) |\Psi_f \rangle|^2   \\
&=& \frac{   E^3}{2 \pi \hbar^4 c^3} \sum_{\kappa = -1}^{+1} \int d \Omega | (\mathbf{d}_{if}\cdot \mathbf{e}(\mathbf{s}, \kappa))|^2
\end{eqnarray*}

\noindent For simplicity, we assume that vector $\mathbf{d}_{if}$ is directed along the $z$-axis. Then, with the help of (\ref{eq:eptau}) we obtain

\begin{eqnarray*}
&\ & \sum_{\kappa = -1}^{+1} | (\mathbf{d}_{if}\cdot \mathbf{e}(\mathbf{s}, \kappa))|^2 = \left|\frac{ d_{if}(-s_x +is_y)}{\sqrt{2}s}  \right|^2
+\left|\frac{ d_{if}(-s_x -is_y)}{\sqrt{2}s}  \right|^2 = \frac{|d_{if}|^2(s_x^2 + s_y^2)}{s^2}
\end{eqnarray*}

\noindent and the total transition rate (measured in Hz) is given by the well-known formula

\begin{eqnarray*}
\frac{1}{\tau_0} &=& \frac{ |d_{if}|^2  E^3}{2 \pi \hbar^4 c^3} \int \limits_0^{\pi} \sin \theta d \theta \int \limits_0^{2 \pi} d \phi \frac{(s_x^2 + s_y^2)}{s^2} = \frac{ |d_{if}|^2  E^3}{\hbar^4 c^3} \int \limits_0^{\pi} \sin^3 \theta d \theta = \frac{4|d_{if}|^2  E^3}{3\hbar^4 c^3}
\end{eqnarray*}

\subsection{Energy correction due to level instability}
\label{ss:instab}

Let us now assume that the photon has a small mass $\lambda$. This can be modeled by adding a ``rest energy'' term $\lambda^2c^4$  to the (squared) photon's energy $c^2s^2$. Then interaction (\ref{eq:ham-posit}) can be rewritten as

\begin{eqnarray}
V_3^d(\mathbf{r}, \mathbf{s}, \kappa) &=& -\frac{ i\hbar^{1/2}  e^3
 \mathbf{r} \cdot
\mathbf{e}(\mathbf{s}, \kappa)}{4 \pi m \sqrt{2(2 \pi )^3} (\lambda^2c^4 + c^2s^2)^{3/4} r^3}  \label{eq:ham-posit3}
\end{eqnarray}

\noindent Consider state $|n \rangle$ of the hydrogen atom with energy $E_n$. Perturbation (\ref{eq:ham-posit3}) changes the energy of this state by the amount $\Delta E_n$, which can be calculated by the perturbation theory formula\footnote{See equation (10.70) in \cite{Ballentine}. This energy shift can be compared with the mass shift $\mathcal{P}(m_A)$ in formula (\ref{eq:13.21}).}

\begin{eqnarray*}
\Delta E_n &=& \int d \mathbf{s} \sum_l \sum_{\kappa} \frac{\langle n | V_3^d | l; \mathbf{s}, \kappa \rangle \langle l; \mathbf{s}, \kappa | V_3^d | n \rangle}{E_n - E_l - cs} \\
&=& \frac{ \hbar  e^6}{ 2 m^2 (2 \pi )^3 } \sum_l \sum_{\kappa=-1}^{+1} \int d \mathbf{s} \left\langle n \left| \frac{(\mathbf{r} \cdot \mathbf{e}(\mathbf{s}, \kappa))}{4 \pi r^3} \right| l; \mathbf{s}, \kappa  \right\rangle \left\langle l; \mathbf{s}, \kappa  \left| \frac{(\mathbf{r} \cdot \mathbf{e}(\mathbf{s}, \kappa))}{4 \pi r^3} \right| n \right\rangle \times \\
 &\ &\frac{ 1}{  (\lambda^2c^4 + c^2 s^2)^{3/2} (E_n - E_l - cs)}
\end{eqnarray*}

\noindent where $| l; \mathbf{s}, \kappa \rangle \equiv |l \rangle | \mathbf{s}, \kappa \rangle$ is a basis state, which has the atom in a stationary state $|l\rangle$\footnote{$|l\rangle$ is an eigenstate of the non-perturbed 2nd order Hamiltonian $H_{e-p}$ (\ref{eq:h'}) with eigenvalue $E_l$.}  and a free photon in the state $| \mathbf{s}, \kappa \rangle$ with momentum $\mathbf{s}$ and helicity $\kappa$. From equality (\ref{eq:10.29a})

\begin{eqnarray*}
 \sum_{\kappa=-1}^{+1}  e_i(\mathbf{s}, \kappa) e_j(\mathbf{s}, \kappa) &=& \delta_{ij} - \frac{s_i s_j}{s^2}
\end{eqnarray*}

\noindent we obtain

\begin{eqnarray*}
\Delta E_n &=& \frac{ \hbar  e^6}{ 2 m^2 (2 \pi )^3 } \sum_{lij}  \int d \mathbf{s} \left(\delta_{ij} - \frac{s_i s_j}{s^2}\right) \left\langle n \left| \frac{r_i }{4 \pi r^3} \right| l  \right\rangle \left\langle l \left| \frac{r_j}{4 \pi r^3} \right| n \right\rangle \times \\
 &\ &\frac{ 1}{  (\lambda^2c^4 + c^2 s^2)^{3/2} (E_n - E_l - cs)}
\end{eqnarray*}

\noindent Let us now set $E_{nl} \equiv E_n - E_l$ and

\begin{eqnarray*}
I_{nl} &\equiv& \int \limits_{0}^{\infty}  \frac{ s^2ds }{  (\lambda^2c^4 + c^2 s^2)^{3/2} (E_{nl} - cs)}
\end{eqnarray*}

\noindent Then

\begin{eqnarray*}
&\ & \int d \mathbf{s}\frac{ s_is_j}{  s^2(\lambda^2c^4 + c^2 s^2)^{3/2} (E_{nl} - cs)} = \delta_{ij}\int d \mathbf{s}\frac{ s_z^2}{  s^2(\lambda^2c^4 + c^2 s^2)^{3/2} (E_{nl} - cs)} \\
&=& 2 \pi \delta_{ij}\int \limits_0^{\pi} \sin \theta d \theta \int\limits_0^{\infty} ds\frac{ s^2 \cos^2 \theta}{  (\lambda^2c^4 + c^2 s^2)^{3/2} (E_{nl} - cs)} \\
&=& 2 \pi \delta_{ij}\int \limits_{-1}^{1} dt \int\limits_0^{\infty} ds\frac{ s^2 t^2}{  (\lambda^2c^4 + c^2 s^2)^{3/2} (E_{nl} - cs)} = \frac{4\pi \delta_{ij}}{3} I_{nl}
\end{eqnarray*}

\noindent and we can write

\begin{eqnarray*}
\Delta E_n
&\approx& \frac{ \hbar  e^6}{ 2 m^2 (2 \pi )^3 }  \sum_{lij}  \left\langle n \left| \frac{r_i }{4 \pi r^3} \right| l  \right\rangle \left\langle l  \left| \frac{r_j}{4 \pi r^3} \right| n \right\rangle  \left( 4 \pi \delta_{ij} I_{nl}
-  \frac{4\pi \delta_{ij}}{3} I_{nl} \right) \\
&=& \frac{4 \hbar  e^6}{ 3 m^2 (2 \pi )^3 }  \sum_{li}  \left\langle n \left| \frac{r_i }{4 \pi r^3} \right| l  \right\rangle \left\langle l  \left| \frac{r_i}{4 \pi r^3} \right| n \right\rangle  I_{nl}
\end{eqnarray*}

\noindent Integral $I_{nl}$ is calculated as follows

\begin{eqnarray*}
&\ &I_{nl} \\
&=& \frac{1}{c^3} \Bigl[\frac{\lambda^2c^4 -E_{nl}cs}{(\lambda^2c^4 +E_{nl}^2)\sqrt{\lambda^2c^4 + c^2s^2}} +\frac{E_{nl}^2}{(\lambda^2c^4 +E_{nl}^2)^{3/2}}  \times \\ &\ & \ln \left( \frac{\sqrt{\lambda^2c^4 +E_{nl}^2}\sqrt{\lambda^2c^4 + c^2s^2} + \lambda^2 c^2 + E_{nl}cs}{E_{nl} - cs} \right) \Bigr] \Bigl|_{s=0}^{s=\infty} \\
&\approx& \frac{1}{c^3} \Bigl[-\frac{ 1}{E_{nl}} +\frac{1}{|E_{nl}|}  \ln \left( -(|E_{nl}| + \lambda^2c^4 |E_{nl}|^{-1}/2)  - E_{nl} \right)  - \frac{1}{|E_{nl}|}  \ln \left(\frac{ |E_{nl}|\lambda c^2 }{E_{nl}} \right)\Bigr]  \\
&\approx&  \frac{1}{c^3|E_{nl}|}  \ln \left( \frac{-[|E_{nl}| + \lambda^2c^4 |E_{nl}|^{-1}/2 + E_{nl}]E_{nl}}{|E_{nl}|\lambda c^2} \right)
\end{eqnarray*}

\noindent If $E_{nl} > 0$

\begin{eqnarray*}
I_{nl} &=&   \frac{1}{c^3E_{nl}}  \ln \left( \frac{-2E_{nl}}{\lambda c^2} \right)
\end{eqnarray*}

\noindent If $E_{nl} < 0$

\begin{eqnarray*}
I_{nl} &=& -  \frac{1}{c^3E_{nl}}  \ln \left( \frac{[\lambda^2c^4 E_{nl}^{-1}/2 ]E_{nl}}{E_{nl}\lambda c^2} \right) =   \frac{1}{c^3E_{nl}}  \ln \left( \frac{2E_{nl}}{\lambda c^2  } \right)
\end{eqnarray*}

\noindent So, taking into account that $\ln(-1) \ll \ln \left( 2|E_{nl}|/(\lambda c^2) \right)$, we can write for all values of $E_{nl}$

\begin{eqnarray*}
I_{nl} &\approx&   \frac{1}{c^3E_{nl}}  \ln \left( \frac{2|E_{nl}|}{\lambda c^2} \right)
\end{eqnarray*}

\noindent Then

\begin{eqnarray*}
\Delta E_n &\approx&   \frac{4 \pi \hbar  e^6}{ 3 m^2 (2 \pi )^3 c^3}  \sum_{li}  \left|\left\langle n \left| \frac{r_i }{4 \pi r^3} \right| l  \right\rangle \right|^2  \frac{1}{E_{nl}}  \ln \left( \frac{2|E_{nl}|}{\lambda c^2}\right)
\end{eqnarray*}

\noindent Next we use equation  (\ref{eq:peHep})

\begin{eqnarray*}
\frac{r_i}{4 \pi r^3} &=& -\frac{1}{i \hbar e^2} [p_i, H_{e-p}] \\
\left|\left\langle n \left| \frac{r_i }{4 \pi r^3} \right| l  \right\rangle \right|^2 &=& -\frac{1}{\hbar^2 e^4} \left\langle n \left| (p_iH_{e-p} - H_{e-p}p_i) \right| l  \right\rangle \left\langle l  \left| (p_iH_{e-p} - H_{e-p}p_i)  \right| n \right\rangle \\
&=& \frac{E_{nl}^2}{\hbar^2 e^4} \left\langle n \left| p_i \right| l  \right\rangle \left\langle l  \left| p_i \right| n \right\rangle
\end{eqnarray*}

\noindent   to obtain

\begin{eqnarray*}
\Delta E_n
&\approx& -\frac{ e^2}{ 6 m^2 \pi^2 \hbar c^3 }  \sum_{li} \left| \langle n | p_i | l  \rangle \right|^2    E_{nl}  \ln \left( \frac{\lambda c^2}{2|E_{nl}|}\right)
\end{eqnarray*}

\noindent Our next step is to use the so-called \emph{Bethe logarithm} \index{Bethe logarithm} $\overline{E}_n$ defined for $s$-states as\footnote{See, e.g., formulas (8.87) in \cite{Bjorken1} and (14.3.51) in \cite{book}. }

\begin{eqnarray*}
\sum_{li}  \left| \langle n | p_i | l  \rangle \right|^2   E_{nl}  \ln \left( \frac{\lambda c^2}{2|E_{nl}|}\right) \equiv \ln \left( \frac{\lambda c^2}{2\overline{E}_n}\right) \sum_{li}  \left| \langle n | p_i | l  \rangle \right|^2   E_{nl}
\end{eqnarray*}

\noindent Then\footnote{here we used equation (\ref{eq:lapl-of-delta})}

\begin{eqnarray}
\Delta E_n
&\approx& -\frac{ 2  \alpha}{ 3\pi m^2c^2 }  \ln \left( \frac{\lambda c^2}{2\overline{E}_n}\right) \sum_l  \left| \langle n | p_i | l  \rangle \right|^2    E_{nl} \nonumber \\
&=& -\frac{   \alpha}{3\pi m^2 c^2 }  \ln \left( \frac{\lambda c^2}{2\overline{E}_n}\right) \sum_l \Bigl( \langle n | (H_{e-p}p_i - p_i H_{e-p})| l  \rangle \langle l | p_i | n  \rangle \nonumber  \\
&\ &  - \langle n | p_i | l  \rangle \langle l | (H_{e-p}p_i - p_i H_{e-p}) | n  \rangle \Bigr)\nonumber \\
&=& -\frac{   \alpha}{3\pi m^2 c^2 }  \ln \left( \frac{\lambda c^2}{2\overline{E}_n}\right)  ( \langle n | (H_{e-p}p_i - p_i H_{e-p}) p_i | n  \rangle  \nonumber \\
 &\ &  - \langle n | p_i  (H_{e-p}p_i - p_i H_{e-p}) | n  \rangle )\nonumber \\
&=& -\frac{   \alpha}{3\pi m^2 c^2 }  \ln \left( \frac{\lambda c^2}{2\overline{E}_n}\right)  \langle n | ([H_{e-p},p_i] p_i - p_i [H_{e-p},p_i])| n  \rangle \nonumber   \\
&=& -\frac{   \alpha}{3 \pi m^2 c^2 }  \ln \left( \frac{\lambda c^2}{2\overline{E}_n}\right)  \langle n | [p_i,[p_i,H_{e-p}]]| n  \rangle \nonumber   \\
&=& -\frac{ \hbar^2  \alpha}{3\pi m^2 c^2 } \ln \left( \frac{\lambda c^2}{2\overline{E}_n}\right)  \left\langle n \left| \frac{\partial^2}{\partial \mathbf{r}^2} \frac{e^2}{4 \pi r} \right| n  \right\rangle \nonumber \\
&=& \frac{ 4\hbar^3  \alpha^2}{3 m^2 c } \ln \left( \frac{\lambda c^2}{2\overline{E}_n}\right)  \langle n | \delta(\mathbf{r})| n  \rangle \label{eq:lamb}
\end{eqnarray}

This energy correction affects only spherically-symmetric $S$-states of the atom. From formula (\ref{eq:lamb}) and the well-known result\footnote{See  section 14.3 in \cite{book}.} $2\overline{E}_{2S^{1/2}}= 16.64 mc^2 \alpha^2$ we can calculate the  effect of spontaneous emission on the hydrogen $2S^{1/2}$ state's energy

\begin{eqnarray}
\Delta E^{se}_{2S^{1/2}} &=&     \frac{ 4\hbar^3  \alpha^2}{3 m^2 c} \ln \left( \frac{\lambda }{16.64 m \alpha^2}\right) |\psi_{2S^{1/2}}(0)|^2 = \frac{ mc^2  \alpha^5}{6 \pi} \ln \left(\frac{\lambda }{16.64 m \alpha^2} \right) \nonumber \\
\label{eq:rad-en-corr}
\end{eqnarray}

It is rather shocking that in the limit of zero photon mass ($\lambda \to 0$) this energy shift becomes infinite. This is an example of \emph{infrared divergence} that will be discussed in greater detail in the next section.

\section{Radiative corrections}
\label{ss:hydrogen}

In the preceding section we got a disturbing result: the instability of atomic levels with respect to spontaneous photon emission has resulted in divergent energy shifts (\ref{eq:rad-en-corr}) in the 3rd perturbation order. This problem can be solved by taking into account even higher perturbation orders.
In this section
we are going to derive 4th order \emph{radiative corrections}  to the electron-proton interaction potential by using formula (\ref{eq:V4}).   In particular, we will see that this potential contains an infrared divergence, which cancels the infinite level shift (\ref{eq:rad-en-corr}) and leads to a finite energy spectrum of the hydrogen atom, which compares favorably with experiment. Here, within our dressed particle approach, we are going to reproduce such classic results of renormalized QED as the electron's \emph{anomalous magnetic moment} and the \emph{Lamb shifts}.

\subsection{Product term in (\ref{eq:V4})}
\label{ss:comm-term}

In this subsection we would like to calculate the 4th order electron-proton interaction potential $V_4^d$ via formula (\ref{eq:V4}). The first term on the right hand side of this formula  can be obtained from (\ref{eq:S4-renorm}) by dropping the energy delta function and dividing by $-2 \pi i$

\begin{eqnarray}
(\Sigma_4^c)^{ph} &\approx&  \frac{ \delta^4(\tilde{q} - \tilde{q}'
-\tilde{p}'+\tilde{p})  \delta_{\tau \tau'}}{-2 \pi i} \times \nonumber \\
&\ &\Bigl[-\frac{\alpha^2 }{30 \pi^3  m^2c} \delta_{\sigma \sigma'}  + -\frac{ \alpha^2}{6 \pi^3 m^2c} \ln\left( \frac{\lambda}{m} \right) \delta_{\sigma \sigma'} -\frac{  imc^2 \alpha^2 }{2\pi^2 q k^2}  \ln\left(-\frac{k^2}{\lambda^2c^2} \right) \delta_{\sigma \sigma'}\nonumber \\
&+& \frac{i\alpha^2\chi_{\sigma}^{\dag} (\vec{\sigma}_{el} \cdot [\mathbf{k} \times \mathbf{q}]) \chi_{\sigma'}}{8 \pi^3 m^2 c k^2}\Bigr] \label{eq:14.18a}
\end{eqnarray}

\noindent The only yet unknown part on the right hand side of (\ref{eq:V4}) is the product term $-V_2^d \underline{V_2^d}$. Here $V_2^d$ is the usual non-relativistic Coulomb potential\footnote{This is the leading term in the momentum representation of the Darwin-Breit interaction (\ref{eq:12.24}) with a screening provided by the photon mass $\lambda$. Here we are interested only in the dominant infrared-divergent term in the product $V_2^d \underline{V_2^d}$, so we will work in the non-relativistic approximation from Appendix \ref{ss:non-rel} and thus assume that momenta of interacting particles are much less that $mc$. Accordingly, we will omit all terms containing positive powers of $\tilde{p}$, $\tilde{q}$ and/or $\tilde{k}$. In this approximation the product $V_2^d \underline{V_2^d}$ becomes spin-independent. So, we disregard spins and drop spin labels of particle states. The same approximations will be employed for other infrared-divergent terms in this chapter.}

\begin{eqnarray*}
&\ & V_2^d(t) \\
&=& -\frac{e^2 \hbar^2 }{(2 \pi \hbar)^3} \int d\mathbf{p} d\mathbf{q} d\mathbf{p}' d\mathbf{q}' \frac{\delta(\mathbf{p} +\mathbf{q}
-\mathbf{p'} -\mathbf{q'} ) e^{\frac{i}{\hbar}(\Omega_p + \omega_q - \Omega_{p'} - \omega_{q'})t}}{(\mathbf{q} - \mathbf{q}')^2 + \lambda^2c^2}
d^{\dag}_{\mathbf{p}} a^{\dag}_{\mathbf{q}}
d_{\mathbf{p}'} a_{\mathbf{q'}}
\end{eqnarray*}

\noindent and $\underline{V_2^d}$ is obtained with the help of (\ref{eq:underline2})

\begin{eqnarray*}
&\ & \underline{V_2^d(t)} \\
&=& \frac{e^2 \hbar^2 }{(2 \pi \hbar)^3} \int d\mathbf{t} d\mathbf{s} d\mathbf{t}' d \mathbf{s}'\frac{\delta(\mathbf{t} +\mathbf{s}
-\mathbf{t'} -\mathbf{s'} )e^{\frac{i}{\hbar}(\Omega_t + \omega_s - \Omega_{t'} - \omega_{s'})t}}{[(\mathbf{s} - \mathbf{s}')^2 + \lambda^2c^2][\omega_{\mathbf{s}} - \omega_{\mathbf{s'}}+
\Omega_{\mathbf{t}}- \Omega_{\mathbf{t'}}]}
 d^{\dag}_{\mathbf{t}} a^{\dag}_{\mathbf{s}}
d_{\mathbf{t}'} a_{\mathbf{s'}}
\end{eqnarray*}

\noindent  Using

\begin{eqnarray}
&\ &d^{\dag}_{\mathbf{p}}a^{\dag}_{\mathbf{q}}
d_{\mathbf{p}'}a_{\mathbf{q}'}
d^{\dag}_{\mathbf{t}}a^{\dag}_{\mathbf{s}} d_{\mathbf{t}'}a_{\mathbf{s}'} \nonumber \\
&=&
(d^{\dag}_{\mathbf{p}}
d_{\mathbf{p}'}
d^{\dag}_{\mathbf{t}} d_{\mathbf{t}'})( a^{\dag}_{\mathbf{q}} a_{\mathbf{q}'} a^{\dag}_{\mathbf{s}} a_{\mathbf{s}'}) \nonumber \\
&=& (-d^{\dag}_{\mathbf{p}} d^{\dag}_{\mathbf{t}}
d_{\mathbf{p}'}
 d_{\mathbf{t}'} + d^{\dag}_{\mathbf{p}}
 d_{\mathbf{t}'}\delta(\mathbf{t}-\mathbf{p}'))( -a^{\dag}_{\mathbf{q}} a^{\dag}_{\mathbf{s}} a_{\mathbf{q}'} a_{\mathbf{s}'} + a^{\dag}_{\mathbf{q}}  a_{\mathbf{s}'} \delta(\mathbf{s}-\mathbf{q}'))  \nonumber \\
 &=& d^{\dag}_{\mathbf{p}}
 a^{\dag}_{\mathbf{q}} d_{\mathbf{t}'} a_{\mathbf{s}'}  \delta(\mathbf{t}-\mathbf{p}') \delta(\mathbf{s}-\mathbf{q}') + \ldots \label{eq:antic}
\end{eqnarray}

\noindent we obtain\footnote{For definition of the \emph{underbrace} sign see (\ref{eq:9.52}). As we have mentioned in subsection \ref{ss:3-rd-order-dressed}, operators $V_2^d$ and $\underline{V_2^d}$ are well-defined only on the energy shell. However, momentum integrations in (\ref{eq:15.66a}) include regions outside the energy shell, where the integrands are not known precisely.  Nevertheless, this uncertainty does not seem to be significant, because here we are interested only in the leading infrared-divergent contribution, which comes from integration in a small region near the $\mathbf{k} \equiv \mathbf{q}' - \mathbf{q}=0$ singularity located on the energy shell.}

\begin{eqnarray}
&\ & - V_2^d(t) \underline{V_2^d(t)} \nonumber \\
&=& \frac{  e^4 \hbar^4}{(2 \pi \hbar)^6}\int d\mathbf{p}
d\mathbf{q}d\mathbf{p'} d\mathbf{q'} d\mathbf{s} d\mathbf{t}
d\mathbf{s'} d\mathbf{t'} d^{\dag}_{\mathbf{p}} a^{\dag}_{\mathbf{q}}
d_{\mathbf{p}'} a_{\mathbf{q'}} d^{\dag}_{\mathbf{t}} a^{\dag}_{\mathbf{s}} d_{\mathbf{t'}} a_{\mathbf{s'}} \times \nonumber
\\
&\mbox{ }& \frac{\delta(\mathbf{p} +\mathbf{q}
-\mathbf{p'} -\mathbf{q'} ) \delta(\mathbf{s} +\mathbf{t} -\mathbf{s'}
-\mathbf{t'} )}{[(\mathbf{q-q'})^2
+ \lambda^2 c^2][\omega_{\mathbf{s}} - \omega_{\mathbf{s'}}+
\Omega_{\mathbf{t}}- \Omega_{\mathbf{t'}} ][(\mathbf{s-s}')^2 + \lambda^2c^2]}
 \label{eq:15.66a} \\
&=& \frac{ e^4 \hbar^4}{(2 \pi \hbar)^6}\int d\mathbf{p}
d\mathbf{q}d\mathbf{p'} d\mathbf{q'} d\mathbf{s} d\mathbf{t}
d\mathbf{s'} d\mathbf{t'} d^{\dag}_{\mathbf{p}}a^{\dag}_{\mathbf{q}}
d_{\mathbf{t}'}  a_{\mathbf{s'}} \times \nonumber \\
&\mbox{ }& \frac{\delta(\mathbf{p} +\mathbf{q}
-\mathbf{p'} -\mathbf{q'} )\delta(\mathbf{s} +\mathbf{t} -\mathbf{s'}
-\mathbf{t'} )\delta(\mathbf{q'-s}) \delta(\mathbf{p'}-\mathbf{t})}{[(\mathbf{q-q'})^2
+ \lambda^2 c^2][\omega_{\mathbf{s}} - \omega_{\mathbf{s'}}+
\Omega_{\mathbf{t}}- \Omega_{\mathbf{t'}} ][(\mathbf{s-s}')^2 + \lambda^2c^2]} \nonumber \\
&=& \frac{ e^4 \hbar^4}{(2 \pi \hbar)^6}\int
d\mathbf{q} d\mathbf{p} d\mathbf{p'} d\mathbf{q'} \delta(\mathbf{q} +\mathbf{p}
-\mathbf{p'} -\mathbf{q'} ) d^{\dag}_{\mathbf{p}} a^{\dag}_{\mathbf{q}}
 d_{\mathbf{p'}} a_{\mathbf{q'}} \times \nonumber \\
&\ &\int
 \frac{d\mathbf{s}}{[(\mathbf{q-s})^2 + \lambda^2c^2][\omega_{\mathbf{s}} - \omega_{\mathbf{q'}}+
\Omega_{\mathbf{q} +\mathbf{p} -\mathbf{s}}- \Omega_{\mathbf{p'}} ][(\mathbf{s-q'})^2
+ \lambda^2 c^2]}  \nonumber
\end{eqnarray}

\noindent Next we introduce non-relativistic approximations $\omega_{\mathbf{q}} \approx mc^2 + q^2/(2m)$ and $\Omega_{\mathbf{p}} \approx \Omega_{\mathbf{p}'} \approx Mc^2$. We also choose the center-of-mass frame, where the heavy proton remains motionless, and the electron scattering is elastic: $q=q'$.\footnote{Here $\mu$ is a small positive constant, which should be taken to 0 at the end of calculations.}

\begin{eqnarray*}
&\mbox{ }&  \frac{ e^4 \hbar^4}{(2 \pi \hbar)^6} \int
 \frac{d\mathbf{s}}{[(\mathbf{q-s})^2 + \lambda^2c^2][(\mathbf{s-q'})^2
+ \lambda^2 c^2]} \cdot   \frac{1}{\omega_{\mathbf{s}} - \omega_{\mathbf{q'}}}  \nonumber  \\
&\approx&  \frac{2 e^4 \hbar^4 m}{(2 \pi \hbar)^6}\int
\frac{d\mathbf{s}}{[(\mathbf{q-s})^2 + \lambda^2c^2][s^2 - q^2 -i \mu][(\mathbf{q'-s})^2 + \lambda^2c^2]}
\end{eqnarray*}

\noindent This integral  is calculated in equation (\ref{eq:D1D2}) in Appendix \ref{sc:integral-comm}. So, the coefficient function of the operator $-V_2^d(t) \underline{V_2^d(t)}$  is

\begin{eqnarray}
 \frac{\alpha^2  mc^2}{(-2 \pi i) \pi q k^2} \ln \left(\frac{k^2}{\lambda^2 c^2} \right)
\label{eq:commu}
\end{eqnarray}

\noindent Adding this term to (\ref{eq:14.18a}), we see that it cancels the third term in square brackets there, i.e., the contribution from ladder and crossed ladder diagrams.\footnote{This observation justifies the omission of contributions (\ref{eq:commu}) and (\ref{eq:joint}) in most textbook calculations of the Lamb shift.} Then for the left hand side of (\ref{eq:V4}) we obtain

\begin{eqnarray}
&\ & \langle 0|a_{\mathbf{q}, \sigma} d_{\mathbf{p}, \tau} V^d_4 d^{\dag}_{\mathbf{p}', \tau'} a^{\dag}_{\mathbf{q}', \sigma'} |0 \rangle \nonumber \\
&\approx&   \frac{ \delta_{\tau \tau'}}{-2 \pi i}
\Bigl[\frac{i \alpha^2 }{ 15 \pi^2  m^2c} \delta_{\sigma \sigma'}   - \frac{\alpha^2\chi_{\sigma}^{\dag} (\vec{\sigma}_{el} \cdot [\mathbf{k} \times \mathbf{q}]) \chi_{\sigma'}}{4 \pi^2 m^2 c k^2} + \frac{i \alpha^2}{3 \pi^2 m^2c} \ln\left( \frac{\lambda}{m} \right) \delta_{\sigma \sigma'}\Bigr] \nonumber \\
\label{eq:S4-renormx}
\end{eqnarray}

\subsection{Radiative corrections to the Coulomb potential}
\label{ss:uehling}

Equation (\ref{eq:S4-renormx}) gives us the 4th order interaction $V_4^d$ only on the
energy shell. Outside the energy shell we can adopt the usual assumption about the near constancy of the coefficient function.\footnote{See condition $\zeta_i \approx 1$ in subsections \ref{ss:uni-2nd-order} - \ref{ss:uni-arb-order} and discussion in subsection \ref{ss:3-rd-order-dressed}.} Then the momentum-space coefficient function of the operator $V_4^d$ has three components

\begin{eqnarray*}
&\ & v_4^d(\mathbf{p}, \mathbf{q}, \mathbf{k}; \tau,\sigma, \tau'  \sigma') \\
  &=& -\frac{ \alpha^2 \delta_{\sigma, \sigma'} \delta_{\tau, \tau'} }{30 \pi^3  m^2c } - \left(\frac{\alpha}{ \pi} \right)\frac{e^2 \hbar^2}{(2 \pi \hbar)^3} \cdot \chi^{\dag}_{\sigma}
 \frac{i \vec{\sigma}_{el} \cdot[\mathbf{k} \times \mathbf{q}]}{4m^2c^2 k^2}   \chi_{\sigma'} \delta_{\tau, \tau'}  -\frac{\alpha^2  \delta_{\sigma, \sigma'} \delta_{\tau, \tau'} }{6  \pi^3   m^2c }
   \ln \left(\frac{ \lambda}{m } \right) \label{eq:v4vertex}
\end{eqnarray*}

\noindent The corresponding position-space potential is obtained using the Fourier transform with respect to the transferred momentum $\mathbf{k}$, formulas from Appendix \ref{sc:delta}, and $\mathbf{S}_{el} = \hbar \vec{\sigma}_{el}/2$.

\begin{eqnarray}
V_{4}^{d}(\mathbf{q},
\mathbf{r}, \mathbf{S}_{el})
&=& -\frac{8\hbar^3  \alpha^2  }{ 30  m^2c } \delta(\mathbf{r})  +
 \frac{e^2  [\mathbf{r} \times \mathbf{q}] \cdot \mathbf{S}_{el}}{8 \pi m^2c^2r^3} \left(\frac{\alpha}{\pi} \right)-\frac{4 \hbar^3 \alpha^2  }{3    m^2c }
   \ln \left(\frac{ \lambda}{m} \right)  \delta(\mathbf{r}) \nonumber \\
   \label{eq:Vd24z}
\end{eqnarray}

\noindent  In a theory of electron-proton interaction valid to the 4th perturbation order this expression must be added to the 2nd order interaction operator  (\ref{eq:V2d}), thus leading to the following complete potential that depends on the position, momentum, and spin of the electron

\begin{eqnarray}
V^d_{2+4}(\mathbf{q},
\mathbf{r},
\mathbf{S}_{el}) &=& -\frac{e^2}{4 \pi r} +
 \frac{e^2  [\mathbf{r} \times \mathbf{q}]
\cdot \mathbf{S}_{el}}{8 \pi m^2 c^2 r^3}\left(1 + \frac{\alpha}{ \pi}\right)
 +  \frac{ e^2 \hbar^2}{8 c^2m^2} \left(1 - \frac{8\alpha}{15\pi}\right) \delta(\mathbf{r}) \nonumber \\
&-& \frac{4 \hbar^3 \alpha^2  }{3    m^2c }
   \ln \left(\frac{ \lambda}{m} \right)  \delta(\mathbf{r})   \label{eq:Vd24}
\end{eqnarray}

\noindent The 1st term is the usual Coulomb potential. The 2nd term describes the spin-orbit interaction of the electron's spin  with its own momentum. The 3rd and 4th terms are contact potentials. The latter one is rather troubling: it diverges in the limit of zero photon mass $\lambda \to 0$. From the point of view of classical electrodynamics,\footnote{See chapter \ref{ch:theories}.} this is not a reason for concern, because such short-range potentials do not affect macroscopic dynamics of classical point-like charges. However, this potential has infinite effect on eigenstates and eigenvalues of quantum compound systems, e.g., the hydrogen atom. How are we going to solve this problem?

\subsection{Lamb shift}
\label{sc:Lamb-shift}

As we saw in (\ref{eq:2p12}), in the 2nd order theory hydrogen \index{hydrogen atom} levels $2S^{1/2}$ and  $2P^{1/2}$ have exactly the same energies.  However, in 1947 Lamb and Retherford  found a small gap between the two levels, which is now known as the \emph{Lamb shift}. \index{Lamb shift} The modern experimental value for the Lamb shift is \cite{Lamb_shift}

\begin{eqnarray}
\varepsilon_{2S^{1/2}} - \varepsilon_{2P^{1/2}} &=& 4.37 \times 10^{-6} \mbox{  } eV \nonumber  \\
&=& 2 \pi \hbar \times 1057.8 \ \  MHz \label{eq:lamb2}
\end{eqnarray}

\noindent The presence of the Lamb shift was completely unexpected from the point of view of the quantum theory available at the time. Attempts to explain this effect played an important role in the development of quantum electrodynamics. Successful calculation of the shift value (\ref{eq:lamb2})  was a major triumph of QED.
Here we are going to calculate the Lamb shift within our dressed particle approach.

The effects of the 4th order  potentials (\ref{eq:Vd24z}) on energies of the $2S^{1/2}$ and  $2P^{1/2}$ hydrogen states  can be calculated using perturbation theory, as in section \ref{ss:hydrogen-nonr}. The results are collected in Table \ref{table:corrections3}.

The short-range contact potential in the 1st term in (\ref{eq:Vd24z}) shifts only the $s$-state,  whose wave function is non-zero at the origin

\begin{eqnarray*}
\Delta \varepsilon^{contact}_{2S^{1/2}} &=&   -\frac{8\hbar^3 \alpha^2}{ 30m^2c  } |\psi_{2S}(0)|^2 =  -\frac{ mc^2 \alpha^5 }{30 \pi}
\end{eqnarray*}

\begin{table}[h]
\caption{4th order perturbative energy corrections to $2S^{1/2}$ and $2P^{1/2}$  energy levels of the hydrogen atom. }
\begin{tabular*}{\textwidth}{@{\extracolsep{\fill}}ll|cc}
\hline
contribution & potential   &   $\Delta \varepsilon_{2S^{1/2}}$ & $\Delta \varepsilon_{2P^{1/2}}$    \cr
\hline
contact (\ref{eq:Vd24z}) & $-\frac{e^2 \hbar^2 \alpha}{30 m^2c^2 \pi} \delta(\mathbf{r})$ & $ -\frac{mc^2 \alpha^5}{30 \pi}$   & 0 \cr &&& \cr
vertex (\ref{eq:Vd24z}) & $-\frac{4 \hbar^3 \alpha^2  }{3    m^2c }
   \ln \left(\frac{ \lambda}{m} \right)  \delta(\mathbf{r})$  & $ -\frac{mc^2 \alpha^5}{6 \pi} \ln \left(\frac{\lambda}{m}\right)$ & 0 \cr &&& \cr
emission (\ref{eq:rad-en-corr}) &  & $ \frac{mc^2 \alpha^5}{6 \pi} \ln \left(\frac{\lambda}{17.6 \alpha^2 m}\right)$  & 0  \cr &&&  \cr
spin-orbit (\ref{eq:Vd24z}) & $-\frac{e^2 \alpha [\mathbf{r} \times \mathbf{q}]
\cdot \mathbf{S}_{el}}{16\pi^2 m^2 c^2 r^3} $ & 0  & $-\frac{mc^2 \alpha^5}{48 \pi}$  \cr &&& \cr \hline
Total correction &  & $ \frac{mc^2 \alpha^5}{6\pi } \left[ -\frac{1}{5} -  \ln (17.6 \alpha^2)\right]$ & $-\frac{mc^2 \alpha^5}{48 \pi}$ \cr
\hline
\end{tabular*}
\label{table:corrections3}
\end{table}

\noindent As we mentioned in the preceding subsection, the energy correction due to the last term in  (\ref{eq:Vd24z})

\begin{eqnarray}
\Delta \varepsilon_{2S^{1/2}}^{vertex} = -\frac{4 \hbar^3 \alpha^2  }{3    m^2c }
   \ln \left(\frac{ \lambda}{m} \right)  |\psi_{2S}(0)|^2 = - \frac{mc^2 \alpha^5}{6 \pi} \ln \left(\frac{\lambda}{m} \right) \label{eq:E2sep}
\end{eqnarray}

\noindent diverges in the limit $\lambda \to 0$, which seems to be unphysical.
Luckily,  there is another interaction that cancels this divergence.
This is the 3rd order bremsstrahlung potential\footnote{This interaction couples the ``electron+proton'' subspace of the Fock space with the ``electron+proton+photon'' subspace. So, strictly speaking, this is not a true  electron-proton potential.} (\ref{eq:ham-posit}), which induces the energy shift (\ref{eq:rad-en-corr})

\begin{eqnarray*}
\Delta \varepsilon^{emission}_{2S^{1/2}} &=&    \frac{ mc^2  \alpha^5}{6 \pi} \ln \left(\frac{\lambda}{16.64  \alpha^2 m}\right)
\end{eqnarray*}

\noindent So, the total energy correction becomes finite. The only interaction affecting the $2P^{1/2}$ level is the 2nd term in (\ref{eq:Vd24z}). The corresponding energy shift can be evaluated by the method from subsection \ref{ss:hydrogen-spin-orb}.

Finally, within our approximations, the full 4th order contribution to the Lamb shift

\begin{eqnarray*}
\varepsilon_{2S^{1/2}} - \varepsilon_{2P^{1/2}}&=& \Delta \varepsilon^{contact}_{2S^{1/2}} + \Delta \varepsilon^{emission}_{2S^{1/2}} + \Delta \varepsilon^{vertex}_{2S^{1/2}} - \Delta \varepsilon^{spin-orbit}_{2P^{1/2}} \\
&=& \frac{mc^2 \alpha^5}{6\pi } \left[ -\frac{3}{40} -  \ln (16.64 \alpha^2)\right]= 3.91 \times 10^{-6} eV \\
&=& \ 2 \pi \hbar \times 945 \ MHz
\end{eqnarray*}

\noindent is in a  good agreement with the experimental value (\ref{eq:lamb2}). \index{Lamb shift} Relative positions of lowest energy levels of the hydrogen atom are shown in Fig. \ref{fig:12.1}.

Let us stress once again that in RQD we do not assume the existence of virtual particles or non-trivial vacuum. Therefore, we explain the high-order effects entirely in terms of small corrections to inter-particle potentials without any reference to ``virtual particle exchanges'', ``vacuum polarization'', and other  field-theoretical terminology. In the literature one can find a number of similar ``effective'' particle approaches \cite{Holstein, PinSot1, PinSot2, Gupta, GupRepSuc, FeiSuc}, which use inter-particle potentials with radiative corrections.

\subsection{Electron's anomalous magnetic moment}
\label{sc:AMM}

As we discussed in chapter \ref{ch:renormalization}, renormalization has no effect on the electron's charge, because this is forbidden by the charge renormalization condition postulate \ref{postulateV}. However, there is another electron's property - the \emph{magnetic moment} \index{magnetic moment} - which is not restricted by any postulate. The effect of renormalization on the electron's magnetic moment was first calculated by Schwinger in 1948. This was a major triumph of the renormalized QED. In this subsection we are going to reproduce this result within our dressed particle approach.

The electron's magnetic moment manifests itself by the electron's dynamics in external ``magnetic fields''.\footnote{Experimental manifestations of particle magnetic moments will be discussed in more detail in chapter \ref{ch:theories}.} In our electron-proton system, in principle, the proton can play the role of a source of such a ``magnetic field''. Unfortunately, so far we assumed that the proton's mass is  infinite and that this particle is motionless. Thus, in our approximation  we have lost the effect of the proton's magnetic field. In order to have a model of electron-magnet interaction, let us consider the effect of a finite proton mass $M < \infty$. In the 2nd perturbation order the relevant potential was obtained in equation (\ref{eq:12.29})

\begin{eqnarray}
 V_{s-o}
&=&   -\frac{e^2 [\mathbf{r} \times \mathbf{p}] \cdot
\mathbf{S}_{el}}{4 \pi  M m c^2 r^3} \label{eq:V-spin-orb1}
\end{eqnarray}

\noindent where $\mathbf{p}$ is the proton's momentum. It is customary to define the electron's magnetic moment and its interaction with a moving charge $e$ by formulas\footnote{see equation (11.100) in
\cite{Jackson}}

\begin{eqnarray*}
\vec{\mu}_{el}
&\equiv&   -\frac{g e \mathbf{S}_{el}}{2mc} \\
 V_{s-o}
&=&   \frac{e [\mathbf{r} \times \mathbf{p}] \cdot
\vec{\mu}_{el}}{4 \pi  M  c r^3}  \label{eq:14.26a}
\end{eqnarray*}

\noindent where $g$ is the so-called \emph{gyromagnetic ratio} \index{gyromagnetic ratio} or simply the $g$-\emph{factor}. \index{g-factor} Thus, comparing (\ref{eq:V-spin-orb1}) and (\ref{eq:14.26a}), we conclude that  in the 2nd perturbation order $g=2$. In higher orders we expect some corrections to this value. Here we will be interested in the 4th order correction $\beta_4$, such that

\begin{eqnarray}
g = 2(1 + \beta_4) \label{eq:g2beta}
\end{eqnarray}

Recall that interaction (\ref{eq:V-spin-orb1}) has resulted from the momentum-space coefficient function in (\ref{eq:12.24})

\begin{eqnarray}
 v_{2s-o}
(\mathbf{p},\mathbf{q},\mathbf{k}; \pi, \epsilon, \pi', \epsilon') = \frac{ie^2 \hbar^2 \delta_{\pi, \pi'}}{(2 \pi \hbar)^3} \chi^{(el) \dag}_{\epsilon}
 \left( \frac{\vec{\sigma}_{el} \cdot [\mathbf{k} \times \mathbf{p}
]}{2Mmc^2k^2} \right)  \chi^{(el)}_{\epsilon'}  \label{eq:v2s-o}
\end{eqnarray}

\noindent Now our plan is to find a 4th order terms with the similar structure.  One can easily see that the only relevant $S$-matrix contribution is (\ref{eq:s4ehAMM}). Retaining only the leading first term in the square bracket there, we obtain

\begin{eqnarray*}
 s_{4s-o}
(\mathbf{p},\mathbf{q},\mathbf{k}; \pi, \epsilon, \pi', \epsilon') \approx -\frac{ic \alpha^2 }{2 \pi^2 k^2} \mathbf{U}(\mathbf{q+k}, \epsilon; \mathbf{q}, \epsilon') \cdot  \mathbf{W}(\mathbf{p-k}, \pi; \mathbf{p}, \pi')
\end{eqnarray*}

\noindent To obtain the coefficient function of the corresponding potential we just need to use (\ref{eq:8.48c}), (\ref{eq:8.48d}), and multiply the result by the usual factor $1/(-2 \pi i)$

\begin{eqnarray*}
v_{4s-o}
 &=& \frac{ic \alpha^2 \delta_{\pi, \pi'}}{(2 \pi i) 2 \pi^2 k^2} \chi^{(el) \dag}_{\epsilon}
\left( \frac{i [\vec{\sigma}_{el} \times \mathbf{k}] }{2mc}\right) \cdot \left(  \frac{2 \mathbf{p} - \mathbf{k} }{2Mc} \right)  \chi^{(el)}_{\epsilon'}  \nonumber \\
&=& \frac{ie^2 \hbar^2 \delta_{\pi, \pi'}}{(2 \pi \hbar)^3} \left(\frac{\alpha}{2 \pi} \right) \chi^{(el) \dag}_{\epsilon}
\left( \frac{ \vec{\sigma}_{el} \cdot [\mathbf{k} \times \mathbf{p}] }{2Mmc^2 k^2}\right)   \chi^{(el)}_{\epsilon'} \label{eq:v4s-o}
\end{eqnarray*}

\noindent This 4th order matrix element  differs from the similar 2nd order matrix element (\ref{eq:v2s-o}) only by the factor $\alpha/(2 \pi)$, which is the desired value of $\beta_4$ in (\ref{eq:g2beta}). So, in our approximation, the $g$-factor is

\begin{eqnarray}
g = 2\left(1 + \frac{\alpha}{2 \pi}\right) \approx 2.0023
\end{eqnarray}

\noindent which is the standard 4th order QED result.

\chapter{CLASSICAL ELECTRODYNAMICS} \label{ch:theories}

\begin{quote}
\textit{All of physics is either impossible or trivial. It is
impossible until you understand it and then it becomes trivial.}

\small
\hspace{1in} Ernest Rutherford
\normalsize
\end{quote}

\vspace{0.5in}

So far in this book we were concerned with quantum mechanical description of electromagnetic phenomena. We focused on scattering, decays, and bound states of systems of charges and achieved a good agreement with experiments in the 4th order of perturbation theory. We can expect even better accuracy by extending our dressed particle approach to higher perturbation orders.  We are not going to do that in our bok. Instead, we will consider another broad class of electromagnetic phenomena, namely the dynamics of macroscopic charges in the classical limit.

Classical theory of electromagnetic phenomena was formulated
a century-and-a-half ago. It was based on a set of equations, which
were designed by Maxwell as a theoretical generalization for a large
number of experiments, in particular those performed by Faraday. By and large, this theory enjoyed a good agreement with experiment. However, in section \ref{sc:maxwell} and in chapter \ref{ch:support} we will meet a number of paradoxes and experiments that cannot be explained within the Maxwell's theory.

In the present
chapter we will apply the direct interaction RQD approach to
classical electrodynamics. Our goal is to show that this is a plausible alternative to the standard
theory based on Maxwell's equations.

We would like to emphasise that our classical electromagnetic theory is obtained by a straightforward application of the $\hbar \to 0 $ limit to RQD equations. There is no such a direct connection between the traditional QED and the Maxwell's theory.

\section{Hamiltonian formulation}
\label{ss:classical-ed}

The central idea of Maxwell's electrodynamics is that charged
particles interact with each other indirectly via electric and
magnetic fields and that electromagnetic radiation is an
electromagnetic field varying in time and space. In this book we are
challenging these universally accepted points of view. Our main concern is that such primary ingredients of the classical theory as Maxwell fields, Li\'enard-Wiechert potentials, and the Lorentz force law cannot be expressed in the language of Poincar\'e group generators, e.g., the Hamiltonian. In our opinion, this is the universal language in which all physical theories ought to be formulated. In particular, the Poincar\'e invariant Hamiltonian dynamics is the most natural way to ensuring conservation laws and correct transformations of observables between different reference frames. In Maxwell's approach the validity of these important requirements is not obvious at all.

We argue  that
all results of conventional classical electromagnetic theory can be
equally well (or even better) explained from the viewpoint of Hamiltonian
dynamics of charged particles with direct interactions, where
``fields'' are not involved at all. In our approach light is
described as a flow of massless particles -- photons, rather than the so-called transverse ``electromagnetic wave''.

In section \ref{sc:coulomb} we already derived the Darwin-Breit
Hamiltonian (\ref{eq:12.25}) for charged particles as an approximation to the full-fledged
RQD. This Hamiltonian was obtained in the 2nd order perturbation
theory within the $(v/c)^2$ approximation. Our goal now is to
demonstrate that this Hamiltonian can be used successfully even in classical (non-quantum) approximation. Then it provides a reasonably  accurate description
of electromagnetic processes in which acceleration of charged
particles is low, so that one can neglect the emission of
electromagnetic radiation (photons).
The Darwin-Breit approach adopted here is fundamentally different
from the generally accepted Maxwell's theory. In the Darwin-Breit approach charged particles interact
via instantaneous potentials; there are no electromagnetic fields and no specific ``field energy'' associated with them. In spite of these
differences, we will see that in many cases it is very difficult to
distinguish these two approaches experimentally, as both of them
lead to very similar predictions. We will also find situations\footnote{See section \ref{sc:maxwell} and chapter \ref{sc:fast-charge}.} in which the traditional Maxwell's theory leads to contradictions and paradoxes. These paradoxes find their resolution in the Darwin-Breit electrodynamics.

In this chapter, we will be working in the classical approximation: ignoring all quantum effects,\footnote{The only exception is  our discussion of the Aharonov-Bohm effect in section \ref{sc:a-b-effect}.}  not
paying attention to the order of dynamical variables in their products and using Poisson brackets $[\ldots, \ldots]_P$ instead of
quantum commutators $(-i/\hbar)[\ldots, \ldots]$. We will also represent all quantities as series in powers of
$v/c$ and leave only terms whose order is not higher than $(v/c)^{2}$.

\subsection{Darwin-Breit Hamiltonian}
\label{sc:poinc-lie}

The Darwin-Breit Hamiltonian $H = H_0 + V$ for a system of two charges $q_1$ and $q_2$  consists of the free part

\begin{eqnarray}
H_0 &=& h_1 + h_2 \nonumber
\\
&=& \sqrt{m_1^2c^4 + p_1^2c^2} + \sqrt{m_2^2c^4 + p_2^2c^2} \nonumber \\
&\approx& m_1c^2 + m_2c^2 + \frac{p_1^2}{2m_1} + \frac{p_2^2}{2m_2}-
\frac{p_1^4}{8m_1^3c^2}   - \frac{p_2^4}{8m_2^3c^2}\label{eq:mc2}
\end{eqnarray}

\noindent and  the potential energy $V$
(\ref{eq:12.26}) - (\ref{eq:12.30})\footnote{We denote $\mathbf{r} \equiv \mathbf{r}_1 -
\mathbf{r}_2$ throughout this chapter. In the case of electron-proton system, the charges are $q_1 = -e$ and $q_2 = +e$. Contact terms proportional
to $\delta(\mathbf{r})$ are not relevant for classical mechanics and
 are omitted. We also omitted 3rd and 4th order corrections to this Hamiltonian that were derived in sections \ref{sc:spontaneous} and \ref{ss:hydrogen}. In Appendix \ref{sc:proof-comm} we verified that with a properly chosen boost generator $\mathbf{K} = \mathbf{K}_0 + \mathbf{Z}$ the Hamiltonian $H=H_0+V$ satisfies all Poincar\'e Lie algebra relationships within the $(v/c)^2$ approximation.}

\begin{eqnarray}
 V &\approx&  \frac{q_1q_2}{4 \pi r}   - \frac{q_1q_2}{8 \pi m_1m_2c^2 r}
\left((\mathbf{p_1} \cdot \mathbf{p}_2) + \frac{(\mathbf{p_1} \cdot
\mathbf{r})(\mathbf{p_2}
\cdot \mathbf{r})}{r^2} \right) \nonumber \\
&\ & -\frac{q_1q_2 [\mathbf{r} \times \mathbf{p}_1] \cdot
\mathbf{s}_{1}}{8 \pi m_1^2 c^2 r^3} +
 \frac{q_1q_2  [\mathbf{r} \times \mathbf{p}_2]
\cdot \mathbf{s}_{2}}{8 \pi m_2^2 c^2 r^3} + \frac{q_1q_2
[\mathbf{r} \times \mathbf{p}_2] \cdot \mathbf{s}_{1}}{4 \pi m_1m_2
c^2 r^3}
\nonumber \\
&\ & -\frac{q_1q_2 [\mathbf{r} \times \mathbf{p}_1] \cdot
\mathbf{s}_{2}}{4 \pi  m_1m_2 c^2 r^3} + \frac{q_1q_2(\mathbf{s}_1
\cdot \mathbf{s}_2)} {4 \pi m_1m_2c^2
 r^3} - \frac{3q_1q_2(\mathbf{s}_1 \cdot \mathbf{r}) (\mathbf{s}_2 \cdot
\mathbf{r})}{4 \pi m_1m_2 c^2 r^5} \nonumber
\\
\label{eq:H-spin-spin}
\end{eqnarray}

In order to use this Hamiltonian  in practical calculations, we introduce few
adjustments. First, we  omit the rest energies of the two particles,
because they have no effect on dynamics. Second, we notice that
particle spins $\mathbf{s}_i$ are not easily measurable in classical
experiments. It is more convenient to replace them with magnetic
moments $\vec{\mu}_i$, which are known to be proportional to spins.
This dependence includes anomalous contributions, i.e., those not described by the classical formula $\vec{\mu}_i = q_i\mathbf{s}_i/(m_i c)$. The electron's anomalous magnetic moment has been discussed in subsection \ref{sc:AMM}. We will not dwell on this issue here and simply postulate that  the full Hamiltonian for two charged spinning classical
particles takes the form

\begin{eqnarray}
H &=&  \frac {p_1^2}{2m_1} +  \frac {p_2^2}{2m_2}
  -  \frac{p_1^4}{8m_1^3 c^2}
- \frac{p_2^4}{8m_2^3 c^2} +\frac{q_1q_2}{4 \pi r} \nonumber \\
&\ & -\frac{q_1q_2}{8 \pi m_1m_2 c^2 r} \left(\mathbf{p}_1 \cdot
\mathbf{p}_2 + \frac{(\mathbf{r} \cdot
\mathbf{p}_2)(\mathbf{r} \cdot \mathbf{p}_1) }{r^2} \right) \nonumber \\
&\ & -\frac{q_1 [\mathbf{r} \times \mathbf{p}_1] \cdot
\vec{\mu}_{2}}{4 \pi  m_1 c r^3}  +
 \frac{q_1  [\mathbf{r} \times \mathbf{p}_2]
\cdot \vec{\mu}_{2}}{8 \pi m_2 c r^3} -\frac{q_2 [\mathbf{r} \times \mathbf{p}_1] \cdot
\vec{\mu}_{1}}{8 \pi m_1 c r^3}
\nonumber \\
&\ &   + \frac{q_2 [\mathbf{r} \times
\mathbf{p}_2] \cdot \vec{\mu}_{1}}{4 \pi m_2 c r^3}+  \frac{(\vec{\mu}_1 \cdot
\vec{\mu}_2)} {4 \pi
 r^3} - \frac{3(\vec{\mu}_1 \cdot \mathbf{r}) (\vec{\mu}_2 \cdot
\mathbf{r})}{4 \pi  r^5} \label{eq:full-H}
\end{eqnarray}

\subsection{Two charges} \label{sc:two-charges}

Let us consider a system of two spinless charged particles. The full
Hamiltonian of this system (which is called the \emph{Darwin
Hamiltonian}) \index{Darwin Hamiltonian} is obtained by dropping
spin-dependent terms in the Darwin-Breit Hamiltonian
(\ref{eq:full-H})

\begin{eqnarray}
 H &=& \frac{p_1^2}{2m_1} + \frac{p_2^2}{2m_2} -
\frac{p_1^4}{8m_1^3c^2}   -
\frac{p_2^4}{8m_2^3c^2} + \frac{q_1q_2}{4 \pi r} \nonumber \\
&\ & -\frac{q_1q_2}{8 \pi m_1m_2c^2 r} \left((\mathbf{p_1} \cdot
\mathbf{p}_2) + \frac{(\mathbf{p_1} \cdot \mathbf{r})(\mathbf{p_2}
\cdot \mathbf{r})}{r^2} \right) \label{eq:h(1)}
\end{eqnarray}

\noindent This Hamiltonian fully determines the dynamics in the system via Hamilton's equations of motion (\ref{eq:7.40}) - (\ref{eq:7.39}) and Poisson brackets (\ref{eq:7.37}). The time derivative of the first particle's momentum can be obtained
from the first Hamilton's equation

\begin{eqnarray}
 \frac{d \mathbf{p}_1}{d t} &=& [\mathbf{p}_1, H]_P = -\frac{\partial H}{\partial \mathbf{r}_1} \nonumber \\
&=& \frac{q_1q_2 \mathbf{r}}{4 \pi r^3} -\frac{q_1q_2(\mathbf{p_1}
\cdot \mathbf{p}_2) \mathbf{r}}{8 \pi m_1m_2c^2 r^3}
 +  \frac{q_1q_2\mathbf{p_1} (\mathbf{p_2} \cdot
\mathbf{r})}{8 \pi m_1m_2c^2 r^3} +    \frac{q_1q_2(\mathbf{p_1}
\cdot \mathbf{r})\mathbf{p_2} }{8 \pi m_1m_2c^2 r^3} \nonumber \\
&\ & -\frac{3q_1q_2(\mathbf{p_1} \cdot \mathbf{r})(\mathbf{p_2} \cdot
\mathbf{r}) \mathbf{r}}{8 \pi m_1m_2c^2 r^5} \label{eq:dp1/dtx}
\end{eqnarray}

\noindent Since Hamiltonian (\ref{eq:h(1)}) is symmetric with
respect to permutations of the two particles, we can obtain the time
derivative of the second particle's momentum by replacing indices $1
\leftrightarrow 2$ in (\ref{eq:dp1/dtx})

\begin{eqnarray}
 \frac{d \mathbf{p}_2}{dt} &=&  -\frac{d
\mathbf{p}_1}{dt} \label{eq:dp2/dtx}
\end{eqnarray}

\noindent Velocities of particles 1 and 2 are obtained from the second
Hamilton's equation\footnote{This relationship between velocity and
momentum is interaction-dependent because interaction energy in
(\ref{eq:h(1)}) is momentum-dependent. }

\begin{eqnarray}
\mathbf{v}_1 &\equiv& \frac{d \mathbf{r}_1}{d t} = [\mathbf{r}_1,
H]_P =
\frac{\partial H}{\partial \mathbf{p}_1} \nonumber \\
&=& \frac{\mathbf{p}_1}{m_1} - \frac{p_1^2 \mathbf{p}_1}{2m_1^3c^2}
- \frac{q_1q_2\mathbf{p}_2}{8 \pi m_1m_2c^2 r}  -
\frac{q_1q_2(\mathbf{p_2} \cdot \mathbf{r}) \mathbf{r}}{8 \pi
m_1m_2c^2 r^3} \label{eq:dr1/dt} \\
 \mathbf{v}_2 &\equiv&\frac{d
\mathbf{r}_2}{d t} = \frac{\mathbf{p}_2}{m_2} - \frac{p_2^2
\mathbf{p}_2}{2m_2^3c^2} - \frac{q_1q_2\mathbf{p}_1}{8 \pi m_1m_2c^2
r}  - \frac{q_1q_2(\mathbf{p_1} \cdot \mathbf{r}) \mathbf{r}}{8 \pi
m_1m_2c^2 r^3} \label{eq:dr2/dt}
\end{eqnarray}

\noindent   From these results we can calculate second time derivatives   of
particle positions (=accelerations)\footnote{In this derivation we
omitted terms proportional to $q_1^2q_2^2$ due to their smallness
(formally they belong to the 4th perturbation order). Also keeping
the accuracy of $(v/c)^2$ we can set $ \dot{\mathbf{r}} = \frac{d
\mathbf{r}_1}{d t} - \frac{d \mathbf{r}_2}{d t} \equiv \mathbf{v}_1
- \mathbf{v}_2 \approx \frac{\mathbf{p}_1}{m_1} -
\frac{\mathbf{p}_2}{m_2} $ in those terms that already have the factor
$(1/c)^2$. }

\begin{eqnarray}
 \frac{d^2 \mathbf{r}_1}{d t^2}
&=& \frac{\dot{\mathbf{p}}_1}{m_1}- \frac{p_1^2 \dot{\mathbf{p}}_1}{2m_1^3c^2} -
\frac{2(\mathbf{p}_1 \cdot \dot{\mathbf{p}}_1)
\mathbf{p}_1}{2m_1^3c^2}  + \frac{q_1q_2\mathbf{p}_2 (\mathbf{r}
\cdot \dot{\mathbf{r}})}{8 \pi m_1
m_2c^2 r^3}  \nonumber \\
&\ & -\frac{q_1q_2(\mathbf{p_2} \cdot \dot{\mathbf{r}}) \mathbf{r}}{8
\pi m_1m_2c^2 r^3} + \frac{3q_1q_2(\mathbf{p_2} \cdot \mathbf{r})
\mathbf{r}(\mathbf{r} \cdot \dot{\mathbf{r}})}{8 \pi m_1m_2c^2 r^5}
- \frac{q_1q_2(\mathbf{p_2} \cdot \mathbf{r}) \dot{\mathbf{r}}}{8
\pi m_1m_2c^2 r^3} \nonumber \\
&\approx& \frac{q_1q_2 \mathbf{r}}{4 \pi m_1 r^3}
-\frac{q_1q_2(\mathbf{p_1} \cdot \mathbf{p}_2) \mathbf{r}}{8 \pi
m_1^2m_2c^2 r^3}
 +  \frac{q_1q_2\mathbf{p_1} (\mathbf{p_2} \cdot
\mathbf{r})}{8 \pi m_1^2m_2c^2 r^3} +    \frac{q_1q_2(\mathbf{p_1}
\cdot \mathbf{r})\mathbf{p_2} }{8 \pi m_1^2m_2c^2 r^3} \nonumber \\
&\ & -\frac{3q_1q_2(\mathbf{p_1} \cdot \mathbf{r})(\mathbf{p_2} \cdot
\mathbf{r}) \mathbf{r}}{8 \pi m_1^2m_2c^2 r^5}
 - \frac{p_1^2 }{2m_1^2c^2}\frac{q_1q_2 \mathbf{r}}{4 \pi m_1r^3} -
\frac{2q_1q_2(\mathbf{p}_1 \cdot \mathbf{r}) \mathbf{p}_1}{8 \pi
m_1^3c^2 r^3}  + \frac{q_1q_2\mathbf{p}_2 (\mathbf{r} \cdot
\mathbf{p}_1)}{8 \pi m_1^2m_2c^2 r^3}  \nonumber \\
&\ & -\frac{q_1q_2\mathbf{p}_2 (\mathbf{r} \cdot \mathbf{p}_2)}{8 \pi
m_1m_2^2c^2 r^3} - \frac{q_1q_2(\mathbf{p_2} \cdot \mathbf{p}_1)
\mathbf{r}}{8 \pi m_1^2m_2c^2 r^3} +  \frac{q_1q_2(\mathbf{p_2}
\cdot \mathbf{p}_2) \mathbf{r}}{8 \pi m_1m_2^2c^2 r^3} +
\frac{3q_1q_2(\mathbf{p_2} \cdot \mathbf{r}) \mathbf{r}(\mathbf{r}
\cdot \mathbf{p}_1)}{8 \pi
m_1^2m_2c^2 r^5} \nonumber \\
&\ & -\frac{3q_1q_2(\mathbf{p_2} \cdot \mathbf{r})
\mathbf{r}(\mathbf{r} \cdot \mathbf{p}_2)}{8 \pi m_1m_2^2c^2 r^5} -
\frac{q_1q_2(\mathbf{p_2} \cdot \mathbf{r}) \mathbf{p}_1}{8 \pi
m_1^2m_2c^2 r^3} + \frac{q_1q_2(\mathbf{p_2} \cdot \mathbf{r})
\mathbf{p}_2}{8 \pi
m_1m_2^2c^2 r^3} \nonumber \\
&=& \frac{q_1q_2 \mathbf{r}}{4 \pi m_1 r^3}
+\frac{q_1q_2(\mathbf{v}_1 - \mathbf{v}_2)^2 \mathbf{r}}{8 \pi
m_1c^2 r^3} -\frac{q_1q_2v_1^2 \mathbf{r}}{4 \pi m_1c^2 r^3} +
\frac{q_1q_2(\mathbf{v}_1 \cdot \mathbf{r}) (\mathbf{v}_2 -
\mathbf{v}_1)}{4 \pi m_1c^2 r^3} \nonumber \\
&\ & -\frac{3q_1q_2(\mathbf{v_2} \cdot \mathbf{r})^2
\mathbf{r}}{8 \pi m_1c^2 r^5}  \label{eq:d2r1/dt2} \\
\frac{d^2 \mathbf{r}_2}{d t^2} &\approx& -\frac{q_1q_2 \mathbf{r}}{4
\pi m_2r^3} -\frac{q_1q_2(\mathbf{v}_1 - \mathbf{v}_2)^2
\mathbf{r}}{8 \pi m_2c^2 r^3} +\frac{q_1q_2v_2^2 \mathbf{r}}{4 \pi
m_2c^2 r^3} - \frac{q_1q_2(\mathbf{v}_2 \cdot \mathbf{r})
(\mathbf{v}_1 -
\mathbf{v}_2)}{4 \pi m_2c^2 r^3} \nonumber \\
&\ & +\frac{3q_1q_2(\mathbf{v_1} \cdot \mathbf{r})^2 \mathbf{r}}{8 \pi
m_2 c^2 r^5} \label{eq:d2r2/dt2}
\end{eqnarray}

\subsection{Definition of force}
\label{sc:force-def}

There are two definitions of \emph{force} \index{force} commonly
used in classical mechanics. In one definition the force acting on a
particle is identified with the time derivative of that particle's
momentum

\begin{eqnarray}
\mathbf{f}_i &\equiv& \frac{d\mathbf{p}_i}{d t} \label{eq:force22}
\end{eqnarray}

\noindent In another definition \cite{Coleman} the force is a
product of the particle's rest mass and its acceleration\footnote{This is equivalent to the second Newton's law of motion.}

\begin{eqnarray}
\mathbf{f}_i &\equiv& m_i \frac{d^2 \mathbf{r}_i}{d t^2}
\label{eq:force2}
\end{eqnarray}

\noindent These two definitions are identical only for
not-so-interesting potentials that do not depend on momenta (or
velocities) of particles. In the Darwin-Breit electrodynamics we are dealing with momentum-dependent potentials, so we need to decide which
definition of force we are going to use.

The usual definition (\ref{eq:force22}) has the advantage that the third
Newton's law \index{Newton's third law} of motion (the law of action
and reaction) in a two-body system has a simple formulation\footnote{see (\ref{eq:dp2/dtx})}

\begin{eqnarray}
\mathbf{f}_1 = -\mathbf{f}_2 \label{eq:balance}
\end{eqnarray}

\noindent This is a trivial consequence of the law of
conservation of the total momentum. It follows immediately from
the vanishing Poisson bracket $[\mathbf{P}_0, H]_P = 0$ in the instant form of relativistic dynamics\footnote{Note that in the traditional Maxwell's theory the proof of the validity of the third Newton's law is rather non-trivial. This ``proof'' requires introduction of  such dubious notions as
``hidden momentum''  and/or momentum of the electromagnetic fields  \cite{Keller, Page, Shockley,
Jefimenko-99}.}

\begin{eqnarray}
 \frac{d \mathbf{p}_2}{dt} = [\mathbf{p}_2,H]_P = [\mathbf{P}_0
-\mathbf{p}_1,H ]_P = -[\mathbf{p}_1,H]_P = -\frac{d
\mathbf{p}_1}{dt} \label{eq:dp2/dt}
\end{eqnarray}

 Contrary to the
usual practice, in this book we will use an alternative definition of force
(\ref{eq:force2}).
\index{Newton's second law} Although this definition does not
imply the \emph{balance of forces} (\ref{eq:balance}),\footnote{This can be
seen from comparing (\ref{eq:d2r1/dt2}) and (\ref{eq:d2r2/dt2}): $\mathbf{f}_1 \neq - \mathbf{f}_2$.} it is preferable for several reasons. First, definition
(\ref{eq:force2}) is consistent with the standard notion that
equilibrium (or zero acceleration $d^2\mathbf{r}/dt^2=0$) is
achieved when the force vanishes.\footnote{This is the first Newton's law of motion.}
\index{Newton's first law} Second, definition (\ref{eq:force22}) is less convenient because it is rather difficult to measure
momenta of particles and their time derivatives in experiments. It
is much easier to measure velocities and accelerations of particles,
 e.g., by time-of-flight techniques. For
example, by measuring current in a wire we actually
measure the amount of charge passing through the cross-section of
the wire in a unit of time. This quantity is directly related to the
velocity of electrons, while it has no direct connection to
electrons' momenta.

\subsection{Wire with current}
\label{sc:wire}

Experimentally, it is very difficult to isolate two charged
particles and measure their trajectories with the precision sufficient
to verify theoretical predictions (\ref{eq:d2r1/dt2}) - (\ref{eq:d2r2/dt2}). In many cases it is more convenient to study
behavior of electrons whose movement is confined inside wires made of
conducting materials. In this subsection we will consider forces
acting between electrons in wires and outside charges.

Let us consider the force exerted by a metal wire on a test charge $q_1$
located at point $\mathbf{r}_1$ outside the wire and moving with velocity
$\mathbf{v}_1$.\footnote{Here we ignore the magnetic moment of the test particle: $\vec{\mu}_1 = 0$.}  There are two kinds of charges in the wire:
fixed positive ions of the lattice and mobile negatively charged
electrons. In most cases the total charge of the ions compensates
exactly the total charge of the electrons, so that the wire is
electrically neutral. We assume that the spins (or magnetic moments $\vec{\mu}_2$) of ions and that electrons in the wire
are oriented randomly. Therefore, all $\vec{\mu}_2$-dependent terms in (\ref{eq:full-H}) vanish after averaging over angles. If the wire
moves as a whole with velocity $\mathbf{w}$, then $\mathbf{w}$-dependent Darwin interactions\footnote{the
second line in equation (\ref{eq:full-H})} of the charge 1 with electrons
and ions in the wire cancel each other. So, the force acting on the charge 1 does not depend on the wire's velocity,  and we can assume
that the wire remains stationary and that only electrons in the
wire are moving with velocity $\mathbf{v}_2$. Electrons in the
wire participate in two kinds of movements: \emph{thermal} and \emph{drift} movements. The
velocities of the thermal movement are rather high, but their
orientations are distributed randomly. The drift velocity is
directed along the applied voltage, and its magnitude is very
small ($\approx$mm/sec).

Let us first see the effect of thermally agitated
electrons\footnote{They are marked
 by the index 2 in this derivation.} on the external charge 1. In this case  we can
omit terms that do not depend on $\mathbf{v}_2$ in equation (\ref{eq:d2r1/dt2}), because these terms are canceled by forces from positively charged lattice
ions. We can also neglect terms having linear dependence on
$\mathbf{v}_2$, because they average out to zero due to the
isotropy of the thermal movement. So, the force acting on the charge 1 due to the thermal chaotic movement of electrons 2 is
proportional to $v_2^2$

\begin{eqnarray}
\mathbf{f}_1&=& \frac{q_1q_2 v_2^2 \mathbf{r}}{8 \pi c^2 r^3} -
\frac{3q_1q_2(\mathbf{v_2} \cdot \mathbf{r})^2 \mathbf{r}}{8 \pi c^2
r^5} \label{eq:thermal}
\end{eqnarray}

\begin{figure}
\centering
 \includegraphics {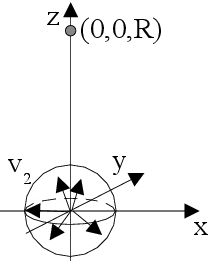} \caption{Interaction of a charge 1 at $(0,0,R)$
 with a piece of conductor placed in the origin. It is assumed that conductor's electrons have thermal velocities with absolute value $v_2$ and
 random orientations.} \label{fig:11.17}
\end{figure}

\noindent Now consider a small piece of conductor located in the origin
and the charge 1 at the point $(0,0,R)$ on the $z$-axis (see Fig.
\ref{fig:11.17}). Our goal is to show that the total force\footnote{or the average of forces (\ref{eq:thermal}) over
different values of $\mathbf{v}_2$} acting on the
charge 1 is zero. To prove this fact it
is sufficient to show that expression (\ref{eq:thermal}) yields zero
when averaged over directions of $\mathbf{v}_2$ with the
absolute value $v_2$ kept fixed. The $x$- and $y$-components of
this average are zero by symmetry, and the $z$-component is
given by the integral on the surface of a sphere of radius
$v_2$\footnote{Here we  use spherical coordinates with angles
$\varphi \in [0, 2 \pi)$ and $\theta \in [0, \pi]$, so that
$(\mathbf{v_2} \cdot \mathbf{r}) = v_2 R \cos \theta$.}

\begin{eqnarray*}
I_z&=& \frac{q_1q_2}{8 \pi}\int \limits_{0}^{\pi} \sin \theta d \theta \int
\limits_{0}^{2\pi} d \varphi \left( \frac{ v_2^2 }{R^2} - \frac{3
v_2^2  \cos^2 \theta }{ R^2} \right) \\
&=& \frac{q_1q_2}{8 \pi} \left( \frac{  4 \pi v_2^2 }{R^2} + \frac{6 \pi v_2^2}{R^2} \int
\limits_{1}^{-1}   t^2 dt\right)  = 0
\end{eqnarray*}

\noindent By similar arguments one can show that the reciprocal force
 exerted by the charge on the
conductor without current vanishes as well. So, we conclude that the thermal movement
of electrons can be ignored in conductor-charge calculations.

\begin{figure}
\centering
 \includegraphics {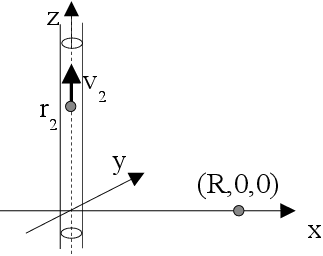} \caption{Interaction of a charge at $(R,0,0)$
 with an infinite straight vertical wire with current.} \label{fig:11.18}
\end{figure}

Let us now consider the charge 1 and an infinite straight wire with a
non-zero drift velocity of electrons $\mathbf{v}_2$ in the geometry shown in
Fig. \ref{fig:11.18}.  The linear density of
conduction electrons in the wire is denoted by $\rho_2$. First we would like to calculate the force acting on
the charge 1 from a small portion $dr_{2z}$
of the wire. Then we use formula (\ref{eq:d2r1/dt2}), keeping only terms dependent on $\mathbf{v}_2$

\begin{eqnarray}
d\mathbf{f}_1 &=& q_1 \rho_2 dr_{2z} \left(-\frac{(\mathbf{v}_1
\cdot \mathbf{v}_2) \mathbf{r}}{4 \pi c^2 r^3} + \frac{v_2^2
\mathbf{r}}{8 \pi c^2 r^3}   + \frac{(\mathbf{v}_1 \cdot \mathbf{r})
\mathbf{v}_2 }{4 \pi c^2 r^3} - \frac{3(\mathbf{v_2} \cdot
\mathbf{r})^2 \mathbf{r}}{8 \pi
c^2 r^5} \right) \nonumber \\
&=& q_1 \rho_2 dr_{2z} \left( \frac{[\mathbf{v}_1 \times[
\mathbf{v}_2 \times \mathbf{r}]]}{4 \pi c^2 r^3} + \frac{v_2^2
\mathbf{r}}{8 \pi c^2 r^3}  - \frac{3(\mathbf{v_2} \cdot
\mathbf{r})^2 \mathbf{r}}{8 \pi c^2 r^5} \right) \label{eq:wire}
\end{eqnarray}

\noindent The full force is obtained by integrating
(\ref{eq:wire}) on the full length of the wire. Let us first show that
the integral of the 2nd and 3rd term vanishes. The $y$- and $z$-components of this integral are zero due to symmetry. For the
$x$-component we obtain

\begin{eqnarray*}
I_x &=& \frac{q_1 \rho_2 v_2^2}{8 \pi m_1c^2} \int
\limits_{-\infty}^{\infty} dr_{2z} \left(
 \frac{ R}{ (R^2 + r_{2z}^2)^{3/2}}  - \frac{3 r_{2z}^2
R}{ (R^2 + r_{2z}^2)^{5/2}} \right) \\
&=& 0
\end{eqnarray*}

\noindent This result means, in particular, that a neutral
superconducting (=zero resistance) wire with current does not create
$v_2^2$-dependent electrostatic potential in the surrounding space. In other words,
a straight wire with current does not act on charges at rest. The observation of
such a potential was erroneously reported in \cite{Edwards}.
Subsequent more accurate measurements \cite{Lemon, Shishkin} did not
confirm that report.

So, the full force acting on the charge 1 is obtained by integration
of the first term in (\ref{eq:wire}) on the length of the wire

\begin{eqnarray}
\mathbf{F}_1 &=& \frac{q_1 \rho_2}{4 \pi c^2} \int
\limits_{-\infty}^{\infty} dr_{2z} \frac{[\mathbf{v}_1 \times[
\mathbf{v}_2 \times \mathbf{r}]]}{r^3} \label{eq:Biot}
\end{eqnarray}

\noindent In this expression one easily recognizes the Biot-Savart
force law \index{Biot-Savart force law} of the traditional Maxwell's
theory. This means that all results of Maxwell's theory referring to
magnetic properties of wires with currents remain valid in our
approach.

\subsection{Charge and current loop} \label{ch:curr-loop}

\begin{figure}
\centering
 \includegraphics {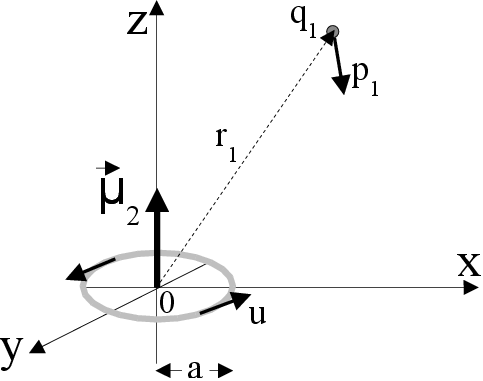} \caption{Interaction between a current loop and a charge. The
charge $q_1$ is located at a general point in space $\mathbf{r}_1 =
(r_{1x},r_{1y},r_{1z})$ and has an arbitrary momentum $\mathbf{p}_1
= (p_{1x},p_{1y},p_{1z})$.}
\label{fig:1}
\end{figure}

  Let
us use the Darwin Hamiltonian (\ref{eq:h(1)}) to calculate the interaction
energy between a
 neutral circular current-carrying wire of a small radius $a$  and a
point charge in the geometry shown in Fig. \ref{fig:1}. As we saw in the preceding subsection,
 the movement of the wire as a whole does not have any effect on its
 interaction with the charge. So, we will assume that the current loop
 is fixed in the origin. We need to take into account only the
velocity-dependent interaction between the charge 1 and negative
charges of conduction electrons having linear density $ \rho_2  $ and drift velocity $\mathbf{v}_2 \approx
\mathbf{p}_2/m_2$, whose tangential component is $u$, as shown in
Fig. \ref{fig:1}. Then the potential energy of interaction between
the charge 1 and the loop element $dl$ is given by the Darwin's
formula

\begin{eqnarray*}
 V_{dl2-q1}
&\approx& -\frac{q_1\rho_2 dl}{8 \pi m_1 c^2}
\left(\frac{(\mathbf{p}_1 \cdot \mathbf{v}_2)}{r} +
\frac{(\mathbf{p}_1 \cdot \mathbf{r})( \mathbf{v}_2 \cdot \mathbf{r}
) }{r^3} \right) \nonumber
\end{eqnarray*}

\noindent In the coordinate system shown in Fig. \ref{fig:1} the
line element in the loop is $dl = a d \theta$ and $\mathbf{v}_2 =
(-u \sin \theta, u \cos \theta, 0)$. In the limit $a \to 0$ we can
approximate

\begin{eqnarray}
\frac{1}{r} &\equiv& \frac{1}{|\mathbf{r}_1 - \mathbf{r}_2|}
\approx  \frac{1}{r_1} + \frac{a (r_{1x}\cos \theta + r_{1y}\sin \theta)}{r_1^3} \label{eq:r-1} \\
\frac{1}{r^3} &\equiv& \frac{1}{|\mathbf{r}_1 - \mathbf{r}_2|^3} \approx \frac{1}{r_1^3} + \frac{3a (r_{1x}\cos \theta +
r_{1y}\sin \theta)}{r_1^5} \label{eq:r-3}
\end{eqnarray}

\noindent The full interaction between the charge and the loop is
obtained by integrating $V_{dl2-q1}$ on $\theta$ from 0 to $2 \pi$
and neglecting small terms proportional to $a^3$

\begin{eqnarray}
&\ & V_{loop2-q1} \nonumber \\
&\approx&  - \frac{aq_1 \rho_2}{8 \pi m_1 c^2} \int \limits _{0}^{2
\pi} d \theta \Bigl[(-up_{1x} \sin \theta+ u p_{1y}\cos \theta)
 \left(\frac{1}{r_1} + \frac{a (r_{1x}\cos \theta + r_{1y}\sin \theta)}{r_1^3} \right) \nonumber \\
 &\ & +(-ur_{1x} \sin \theta+ u r_{1y}\cos \theta)(
(\mathbf{p}_1 \cdot \mathbf{r}_1) - p_{1x}a \cos \theta- p_{1y} a
\sin \theta ) \times \nonumber \\
&\ & \left(\frac{1}{r_1^3} + \frac{3a (r_{1x}\cos \theta +
r_{1y}\sin \theta )}{r_1^5} \right) \Bigr]
\nonumber \\
&\approx&  -\frac{a^2uq_1 \rho_2 [\mathbf{r}_1 \times
\mathbf{p}_1]_z}{4 m_1 c^2 r_1^3} \label{eq:vloop-q1}
\end{eqnarray}

\noindent  Taking into account the usual definition of the loop's
magnetic moment\footnote{see equation (5.42) in
\cite{Jackson}} as a vector $ \vec{\mu}_2 $ whose length is $ \mu_2 = \pi a^2 \rho_2 u/c$ and whose direction is orthogonal to the plane of the
loop, we can generalize (\ref{eq:vloop-q1}) for arbitrary position and orientation of the
loop

\begin{eqnarray}
V_{loop2-q1} &\approx&  -
 \frac{ q_1[\vec{\mu}_2 \times \mathbf{r}]
\cdot \mathbf{p}_1}{4 \pi m_1 c r^3} \label{eq:vloopq1}
\end{eqnarray}

\noindent  So, the full Hamiltonian for
the system of charge 1 and current loop 2 is

\begin{eqnarray}
 H
&=& \frac{p_1^2}{2m_1} + \frac{p_2^2}{2m_2} -
\frac{p_1^4}{8m_1^3c^2}   - \frac{p_2^4}{8m_2^3c^2} - \frac{q_1
[\vec{\mu}_2 \times \mathbf{r}] \cdot \mathbf{p}_1}{4 \pi m_1c r^3}
\label{eq:h-charge-loop}
\end{eqnarray}

Now we can use this Hamiltonian to obtain the dynamics in the ``loop+charge'' system. The time derivative of the particle's momentum can be obtained
from the Hamilton's equation of motion (\ref{eq:7.40})

\begin{eqnarray*}
\frac{d \mathbf{p}_1}{dt} &=&  - \frac{\partial H}{\partial
\mathbf{r}_1} =
 \frac{q_1[ \mathbf{p}_1 \times \vec{\mu}_2] }{4 \pi m_1 cr^3}
 -
 \frac{3q_1 ([ \mathbf{p}_1 \times \vec{\mu}_2] \cdot \mathbf{r})\mathbf{r} }{4\pi m_1 cr^5}
\end{eqnarray*}

\noindent The time derivative  of the loop's momentum
follows from the momentum conservation law
(\ref{eq:dp2/dt})

\begin{eqnarray*}
 \frac{d \mathbf{p}_2}{dt} &=& -\frac{d
\mathbf{p}_1}{dt}
\end{eqnarray*}

\noindent The velocity of the charge 1  is obtained from
the 2nd Hamilton's equation of motion

\begin{eqnarray}
\mathbf{v}_1 &=& \frac{\partial H}{\partial \mathbf{p}_1} =
\frac{\mathbf{p}_1}{m_1}- \frac{p_1^2\mathbf{p}_1}{2m_1^3 c}  -
\frac{q_1 [\vec{\mu}_2 \times \mathbf{r}] }{4 \pi m_1c r^3}
\label{eq:v1-ch}
\end{eqnarray}

\noindent Acceleration of this particle  is obtained as a time
derivative of (\ref{eq:v1-ch})\footnote{Here we noticed that
 $\dot{\mathbf{p}}_1 \propto (v/c)^2$, therefore the
time derivative of the second term on the right hand side of
(\ref{eq:v1-ch}) is $\propto (v/c)^3$, so it can be ignored. We also neglected the time derivative of the magnetic moment $\dot{\vec{\mu}}_2 = [\vec{\mu}_2, H]_P$, because, due to (\ref{eq:sisjsk}), the Poisson bracket $[\mu_{2i}, \mu_{2j}]_P = q_2/(m_2c)\sum_k \epsilon_{ijk} \mu_{2k}$ has an extra factor of $c$ in the denominator, which means that terms proportional to $\dot{\vec{\mu}}_2$ are much smaller than other terms in (\ref{eq:lorentz-forcea}). Vector identities
(\ref{eq:A.17}) and (\ref{eq:A.18}) were used in the
derivation of (\ref{eq:lorentz-forcea}). \label{pg:mu2}}

\begin{eqnarray}
\mathbf{a}_1 &\equiv& \frac{d \mathbf{v}_1}{dt} \nonumber \\
 &\approx& \frac{\dot{\mathbf{p}}_1}{m_1}-
 \frac{q_1[  \vec{\mu}_2 \times \dot{\mathbf{r}}] }{4 \pi m_1c r^3} +
 \frac{3q_1[  \vec{\mu}_2 \times \mathbf{r}] (\mathbf{r} \cdot \dot{\mathbf{r}})}{4 \pi m_1c r^5}
 \nonumber \\
 &=& \frac{q_1[ \mathbf{p}_1 \times \vec{\mu}_2] }{2 \pi m_1^2 cr^3}
 -
 \frac{3q_1 ([ \vec{\mu}_2 \times \mathbf{r}] \cdot \mathbf{p}_1)\mathbf{r} }{4 \pi m_1^2 cr^5}
 +
 \frac{3q_1[  \vec{\mu}_2 \times \mathbf{r}] (\mathbf{r} \cdot \mathbf{p}_1)}{4 \pi m_1^2 cr^5}
 \nonumber \\
 &\ & +\frac{q_1[ \vec{\mu}_2 \times  \mathbf{p}_2] }{4 \pi m_1m_2 cr^3}
 -
 \frac{3q_1[  \vec{\mu}_2 \times \mathbf{r}] (\mathbf{r} \cdot \mathbf{p}_2)}{4 \pi m_1 m_2 cr^5}
 \nonumber \\
  &=& \frac{q_1 [\mathbf{p}_1 \times \vec{\mu}_2]}{2
\pi m_1^2 cr^3} -
 \frac{3q_1 [ \mathbf{p}_1 \times [\mathbf{r} \times [\vec{\mu}_2 \times \mathbf{r}]]]
 }{4 \pi m_1^2c  r^5} - \left(\frac{d}{dt}\right)_2
 \frac{q_1  [\vec{\mu}_2 \times \mathbf{r}] }{4 \pi m_1 c  r^3}
 \nonumber \\
 &=& -\frac{q_1 [\mathbf{p}_1 \times \vec{\mu}_2]}{4 \pi
m_1^2 cr^3} +
 \frac{3q_1 [ \mathbf{p}_1 \times \mathbf{r}](\vec{\mu}_2 \cdot
 \mathbf{r})
 }{4 \pi m_1^2c  r^5} -  \left(\frac{d}{dt} \right)_2
 \frac{q_1  [\vec{\mu}_2 \times \mathbf{r}] }{4 \pi m_1 c  r^3}
  \label{eq:lorentz-forcea}
\end{eqnarray}

\noindent The notation $(\frac{d}{dt})_2$ means the time derivative
(of $\mathbf{r}$) when only particle 2 (the loop) is allowed to
move. For example

\begin{eqnarray*}
\left(\frac{d}{dt} \right)_2 \mathbf{r} = -\mathbf{v}_2 \approx
-\frac{\mathbf{p}_2}{m_2}
\end{eqnarray*}

\subsection{Charge and spin's magnetic moment}
\label{sc:two-particle}

Let us now consider the system of a spinless charged particle 1 and
a spin's magnetic moment 2. The relevant Hamiltonian is obtained from
(\ref{eq:full-H}) by assuming $\vec{\mu}_1=0$ and
$q_2=0$ and dropping the corresponding terms\footnote{Note that if the spin's magnetic moment is not moving ($\mathbf{p}_2 = 0$) then the interaction energy of ``charge + moment'' (the last term in (\ref{eq:h-charge-spin})) is exactly the same as the interaction energy of ``charge + current loop'' (\ref{eq:h-charge-loop}). For a moving spin the interaction energy has an additional term (the last term in (\ref{eq:h-charge-spin})) which is absent in (\ref{eq:h-charge-loop}).}

\begin{eqnarray}
 H
&=& \frac{p_1^2}{2m_1} + \frac{p_2^2}{2m_2} -
\frac{p_1^4}{8m_1^3c^2}   - \frac{p_2^4}{8m_2^3c^2} + \frac{q_1 [\vec{\mu}_2 \times \mathbf{r}] \cdot
\mathbf{p}_2}{8\pi m_2 c r^3} - \frac{q_1 [\vec{\mu}_2 \times
\mathbf{r}] \cdot \mathbf{p}_1}{4 \pi m_1c r^3} \nonumber \\
\label{eq:h-charge-spin}
\end{eqnarray}

\noindent As usual, we employ Hamilton's equations of motion to
calculate the time derivative of the momentum, the velocity, and the
acceleration

\begin{eqnarray}
\frac{d \mathbf{p}_1}{dt} &=& [\mathbf{p}_1,H]_P = - \frac{\partial
H}{\partial
\mathbf{r}_1} \nonumber \\
&=&
 \frac{q_1[ \mathbf{p}_1 \times \vec{\mu}_2] }{4 \pi m_1 cr^3}
 -
 \frac{3q_1 ([ \mathbf{p}_1 \times \vec{\mu}_2] \cdot \mathbf{r})\mathbf{r} }{4\pi m_1 cr^5}
 - \frac{q_1
[\mathbf{p}_2 \times \vec{\mu}_2]  }{8\pi m_2 c r^3} +  \frac{3q_1
([\mathbf{p}_2 \times \vec{\mu}_2] \cdot \mathbf{r}) \mathbf{r}
}{8\pi m_2 c r^5} \nonumber \\
 \frac{d \mathbf{p}_2}{dt} &=& -\frac{d
\mathbf{p}_1}{dt} \nonumber \\
\mathbf{v}_1 &\equiv& \frac{d \mathbf{r}_1}{dt} = [\mathbf{r}_1,H]_P = \frac{\partial H}{\partial \mathbf{p}_1} =
\frac{\mathbf{p}_1}{m_1}- \frac{p_1^2\mathbf{p}_1}{2m_1^3 c}  -
\frac{q_1 [\vec{\mu}_2 \times \mathbf{r}] }{4 \pi m_1c r^3}
\nonumber \\
\mathbf{a}_1 &\equiv& \frac{d \mathbf{v}_1}{dt} \nonumber  \\
 &\approx& \frac{\dot{\mathbf{p}}_1}{m_1}-
 \frac{q_1[  \vec{\mu}_2 \times \dot{\mathbf{r}}] }{4 \pi m_1c r^3} +
 \frac{3q_1[  \vec{\mu}_2 \times \mathbf{r}] (\mathbf{r} \cdot \dot{\mathbf{r}})}{4 \pi m_1c r^5}
 \nonumber \\
 &=& \frac{q_1[ \mathbf{p}_1 \times \vec{\mu}_2] }{2 \pi m_1^2 cr^3}
 -
 \frac{3q_1 ([ \mathbf{p}_1 \times \vec{\mu}_2] \cdot \mathbf{r})\mathbf{r} }{4 \pi m_1^2 cr^5}
 +
 \frac{3q_1[  \vec{\mu}_2 \times \mathbf{r}] (\mathbf{r} \cdot \mathbf{p}_1)}{4 \pi m_1^2 cr^5}
 \nonumber \\
 &\ & -\frac{3q_1[ \mathbf{v}_2 \times \vec{\mu}_2] }{8 \pi m_1 cr^3}
 +
 \frac{3q_1 ([\vec{\mu}_2 \times \mathbf{r}] \cdot \mathbf{v}_2)\mathbf{r} }{8 \pi m_1 cr^5}
 -
 \frac{3q_1[  \vec{\mu}_2 \times \mathbf{r}] (\mathbf{r} \cdot \mathbf{v}_2)}{4 \pi m_1 cr^5}
 \nonumber \\
  &=& \frac{q_1 [\mathbf{p}_1 \times \vec{\mu}_2]}{2
\pi m_1^2 cr^3} -
 \frac{3q_1 [ \mathbf{p}_1 \times [\mathbf{r} \times [\vec{\mu}_2 \times \mathbf{r}]]]
 }{4 \pi m_1^2c  r^5} \nonumber \\
  &\ & -\frac{q_1[ \mathbf{v}_2 \times \vec{\mu}_2] }{8 \pi m_1 cr^3}
 +
 \frac{3q_1 ([\vec{\mu}_2 \times \mathbf{r}] \cdot \mathbf{v}_2)\mathbf{r} }{8 \pi
m_1 cr^5} +\frac{q_1[ \vec{\mu}_2 \times \mathbf{v}_2  ] }{4 \pi m_1 cr^3} -
 \frac{3q_1[  \vec{\mu}_2 \times \mathbf{r}] (\mathbf{r} \cdot \mathbf{v}_2)}{4 \pi m_1 cr^5}
 \nonumber \\
 &=& -\frac{q_1 [\mathbf{p}_1 \times \vec{\mu}_2]}{4 \pi
m_1^2 cr^3} +
 \frac{3q_1 [ \mathbf{p}_1 \times \mathbf{r}](\vec{\mu}_2 \cdot
 \mathbf{r})
 }{4 \pi m_1^2c  r^5} -\left(\frac{d}{dt} \right)_2
 \frac{q_1  [\vec{\mu}_2 \times \mathbf{r}] }{4 \pi m_1 c  r^3}
   \nonumber \\
&\ &    - \frac{d}{d\mathbf{r}_1} \frac{q_1 ([\mathbf{v}_2 \times \vec{\mu}_2] \cdot
  \mathbf{r})}{8
\pi m_1 cr^3}
  \label{eq:lorentz-force}
\end{eqnarray}

\noindent This means that acceleration of the charge 1 in the field
of the spin's magnetic moment  is basically the same as in the field
of a current loop  (\ref{eq:lorentz-forcea}). The only difference is
the presence of an additional gradient term - the last term on the right hand side of  (\ref{eq:lorentz-force}).  This difference will be discussed
in subsections \ref{ch:induction} - \ref{ch:rot-sol} in greater
detail.

\subsection{Two types of magnets}
\label{sc:magnets}

Let us now consider the system ``moving charge 1 + magnetic moment 2 at
rest.'' As we discussed above, the magnetic moment can be produced either by a spinning particle or by a small current loop. The Hamiltonian\footnote{which is the same in both cases} is obtained either from
(\ref{eq:h-charge-loop}) or from (\ref{eq:h-charge-spin}) by setting
$\mathbf{p}_2=0$

\begin{eqnarray}
H &=&  \frac {p_1^2}{2m_1} +  \frac {p_2^2}{2m_2}
  -  \frac{p_1^4}{8m_1^3 c^2}
- \frac{p_2^4}{8m_2^3 c^2}  -
 \frac{q_1 [\mathbf{r} \times
\mathbf{p}_1] \cdot \vec{\mu}_{2}}{4 \pi  m_1 c r^3}
\label{eq:full-H2}
\end{eqnarray}

\noindent The force acting on the charge 1 is given by formula
(\ref{eq:lorentz-forcea})

\begin{eqnarray}
\mathbf{f}_1 &=& m_1\mathbf{a}_1 = -\frac{q_1 [\mathbf{p}_1 \times \vec{\mu}_2]}{4 \pi m_1 cr^3} +
 \frac{3q_1 [ \mathbf{p}_1 \times \mathbf{r}](\vec{\mu}_2 \cdot
 \mathbf{r})
 }{4 \pi m_1c  r^5} \nonumber \\
 & \approx&
\frac{q_1}{c}[\mathbf{v}_1 \times \mathbf{b}_1] \label{eq:fieldy}
\end{eqnarray}

\noindent which is the standard definition of the magnetic part of
the \emph{Lorentz force} if another standard expression

\begin{eqnarray}
 \mathbf{b}_1 = - \frac{\vec{\mu}_2}{4 \pi   r^3}
 + \frac{3(\vec{\mu}_2 \cdot \mathbf{r}) \mathbf{r}}{4 \pi
r^5} \label{eq:fieldx}
\end{eqnarray}

\noindent  is used for the ``magnetic field'' of the magnetic
moment $\vec{\mu}_2$ at point
$\mathbf{r}_1$.\footnote{ See equation (5.56) in \cite{Jackson}.} There
are, however, important differences between our formulas and the standard
approach. First, in the usual Lorentz force equation\footnote{See, e.g., equation
(11.124) in \cite{Jackson}.} the force is identified with the time
derivative of momentum (\ref{eq:force22}). In our case, the
force is ``mass times acceleration.'' Second, in
our approach, there are no fields (electric or magnetic) having
independent existence at each space point. There are only direct
inter-particle forces. This is why we put ``magnetic field'' in
quotes.

\begin{figure}
\centering
 \includegraphics {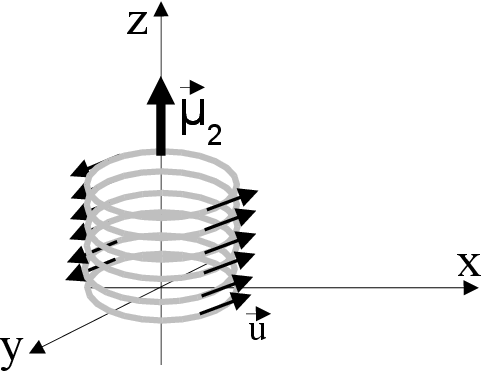} \caption{A thin solenoid can be represented
 as a stack of small current loops. The magnetization vector $\vec{\mu}_2$ is directed
 along the solenoid's axis.} \label{fig:11.7}
\end{figure}

For comparison with experiment it is not sufficient to discuss point
magnetic moments. We need to apply the above results to macroscopic
magnets as well. It is important to mention that there are two origins of
magnetization in materials. The first origin is due to the orbital motion of
electrons. The second one is due to spin magnetic moments of
electrons\footnote{the contribution from nuclear spins is much
weaker}. In permanent magnets both components play roles. The
relative strength of the ``orbital'' and ``spin'' magnetizations
varies among different types of magnetic materials. However, in most
cases the dominant contribution is due to  electron spins
\cite{Reck}. The full magnetization can be described by summing
up total magnetic moments over all atoms in the body, and the full
``magnetic field'' of the macroscopic magnet is obtained by adding
up contributions like (\ref{eq:fieldx}).

The above discussion referred to permanent bar magnets. However,
there is an alternative way to produce ``magnetic field'' by means
of electromagnets -- solenoids with current. In solenoids only the
orbital component of magnetization (due to electrons moving in wires) is present.  For
example, a straight thin solenoid can be
represented as a collection of small current loops\footnote{see subsection \ref{ch:curr-loop}} stacked on top of
each other (see Fig. \ref{fig:11.7}). The ``magnetic field'' of such
a stack can be obtained by integrating (\ref{eq:fieldx}) along the
length of the stack.

\begin{figure}
\centering
 \includegraphics {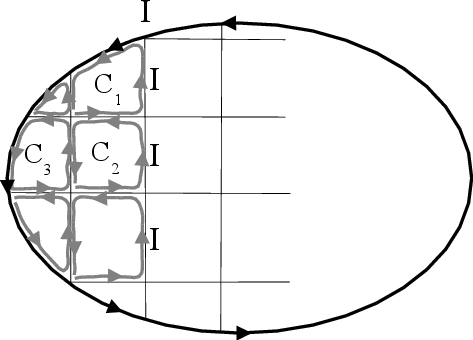} \caption{A wire coil (black thick line) with current $I$
 can be represented as
a superposition of infinitesimally small wire loops $C_1, C_2, C_3,
\ldots$ (grey lines) with the same current $I$. All (imaginary) inside currents cancel each other, so that only the (real) peripheral current remains.} \label{fig:11.12}
\end{figure}

This result can be also generalized for macroscopic solenoids with non-vanishing cross-sections. It is
easy to see that each current-carrying coil in such a solenoid can
be represented as a superposition of infinitely small loops (see
Fig. \ref{fig:11.12}). Then a macroscopic thick cylindrical solenoid
can be represented as a set of parallel thin solenoids joined
together.

\subsection{Longitudinal forces in conductors}
\label{sc:longitudinal}

According to classical electrodynamics, the magnetic force
(\ref{eq:fieldy}) is always perpendicular to the particle's
velocity. Consequently, there can be no magnetic force between two
electrons moving inside the same straight thin wire with a steady current. Indeed, if we
substitute $\mathbf{v}_1 = \mathbf{v}_2 \equiv \mathbf{v}$ and
$\mathbf{r}
\parallel \mathbf{v}$ in the standard Biot-Savart force law
(\ref{eq:biot1}) - (\ref{eq:biot2}), we obtain $\mathbf{f}_1 =
\mathbf{f}_2 = 0$. However, this result does not hold in our
approach. Similar substitutions in our formulas (\ref{eq:d2r1/dt2})
- (\ref{eq:d2r2/dt2}) yield\footnote{Here we ignore the Coulomb
force components, which are shielded in metal conductors. $q$
denotes electron's charge.}

\begin{eqnarray*}
 \mathbf{f}_1 &=&
 -\frac{q^2v^2 \mathbf{r}}{4 \pi c^2 r^3} -
\frac{3q^2(\mathbf{v} \cdot \mathbf{r})^2 \mathbf{r}}{8 \pi c^2 r^5}
=
 -\frac{5q^2v^2 \mathbf{r}}{ 8\pi c^2 r^3}   \\
 \mathbf{f}_2 &=&  \frac{5q^2v^2 \mathbf{r}}{ 8\pi  c^2 r^3}
\end{eqnarray*}

\noindent which indicates the presence of an (longitudinal) attractive
force parallel to the electrons velocity vectors. As discussed in
\cite{Essen, Essen1, Essen3}, this magnetic attraction of conduction
electrons may contribute to superconductivity at low temperatures.

It is interesting to note that the issue of longitudinal
interactions in conductors was discussed ever since Amp\`ere
suggested his charge interaction law in the early 19th
century.\footnote{for a good review see \cite{Johansson}} However,
in contrast to our predicted attraction, the Amp\`ere's formula predicted longitudinal
repulsion between two electrons in the same wire.
Numerous experiments attempting to detect such a repulsion did not
yield conclusive results. A recent study \cite{Graneau}
declared a confirmation of the Amp\`ere's repulsion. However, this
conclusion was challenged in \cite{Cavalleri-03}. So,
experimentally, the presence of longitudinal forces in conductors and their signs
(i.e., attractive or repulsive) remains an unsettled issue.

\section{Experiments and paradoxes} \label{sc:maxwell}

In this section we will discuss a number of real or thought
electromagnetic experiments, whose description in classical
Maxwell's electrodynamics is inadequate or paradoxical. We will also
consider these experiments from the point of view of the RQD direct interaction approach
developed in the preceding section. Our goal is to demonstrate that
in all cases the RQD description is more logical and consistent.

\subsection{Conservation laws in Maxwell's theory}
 \label{ss:newton}

 One important class
of difficulties characteristic to Maxwell's electrodynamics is
related to the apparent non-conservation of total observables
(energy, momentum, angular momentum) in systems of
interacting charges. Indeed, in Maxwell's
equations there is no built-in guarantee that total observables are conserved
and that the total energy and momentum form a 4-vector quantity.  Various electromagnetic paradoxes were formulated on the basis of Maxwell's equations and the Lorentz force law
\cite{Butler, Rohrlich-60,
Furry, Aharonov-88, Comay-hidden, Comay-hidden2, Kholmetskii-2004, Kholmetskii-hidden,
Hnizdo, Spavieri, Teukolsky, Jackson-torque, Onoochin, Kholmetskii-2007,
Kholmetskii-2007a,
Kholmetskii-2008, Tuval, Mansuripur, Babson, Caprez09}. The suggested ``solutions'' of these paradoxes involved such \emph{ad hoc} constructions as
``hidden momentum,'' the energy and momentum of ''electromagnetic
fields,'' ``Poincar\'e stresses,'' alternative non-Lorentz  force laws, etc.

The simplest example of a paradox in Maxwell's theory refers to an isolated system of two charges 1 and 2, which are free to move without
influence of external forces. By
applying the standard Biot-Savart force law \index{Biot-Savart force
law}

\begin{eqnarray}
\mathbf{f}_1 &=& \frac{q_1 q_2}{4 \pi c^2}  \frac{[\mathbf{v}_1
\times[ \mathbf{v}_2 \times \mathbf{r}]]}{r^3} \label{eq:biot1} \\
\mathbf{f}_2 &=& -\frac{q_1 q_2}{4 \pi c^2}  \frac{[\mathbf{v}_2
\times[ \mathbf{v}_1 \times \mathbf{r}]]}{r^3} \label{eq:biot2}
\end{eqnarray}

\noindent and the traditional force definition $\mathbf{f} = d \mathbf{p}/dt$ it is easy to see that the Newton's third
law ($\mathbf{f}_1 = -\mathbf{f}_2$) is not satisfied for most
geometries \cite{Howe}. As we discussed in subsection
\ref{sc:force-def}, this means that the total momentum of particles
$\mathbf{P}^p$ is not conserved. The usual explanation
\cite{Keller, Page} of this paradox is that the two charges alone do
not constitute a closed physical system. In order to restore the
momentum conservation one needs to take into account the momentum
contained in the electromagnetic field surrounding the charges.

 According to Maxwell's theory, electric and magnetic fields $\mathbf{E}(\mathbf{r}),
 \mathbf{B}(\mathbf{r})$  have momentum
and energy given by integrals over entire space

\begin{eqnarray}
\mathbf{P}^f &=& \frac{1}{4 \pi c} \int d\mathbf{r}
[\mathbf{E}(\mathbf{r}) \times \mathbf{B}(\mathbf{r})] \label{eq:field-mom}\\
H^f &=& \frac{1}{8 \pi} \int d\mathbf{r} (E^2(\mathbf{r}) +
B^2(\mathbf{r})) \label{eq:field-en}
\end{eqnarray}

\noindent So, the idea of  the standard explanation is that the total
momentum of ``particles + fields'' ($\mathbf{P}^p + \mathbf{P}^f$) is
conserved in all circumstances.

From the point of view of RQD, it is understandable when Maxwell's
theory associates momentum and energy with transverse time-varying
electromagnetic fields in free space. As we will discuss in subsection
\ref{ss:way-forward}, these fields can be accepted as rough models
of electromagnetic radiation. For free propagating fields, equations
(\ref{eq:field-mom}) and (\ref{eq:field-en}) are supposed to be
equivalent to the sums of momenta and energies of photons,
respectively.

However, Maxwell's theory goes even farther and claims that bound\footnote{stationary, non-radiating} electromagnetic fields
surrounding charges or magnets
also have non-zero momentum and energy. If this were true then one
could easily imagine stationary systems (e.g., a charged magnet) where nothing is moving and where fields
 $\mathbf{E}, \mathbf{B}$ would possess a
non-zero momentum.\footnote{See, e.g., subsection \ref{ch:toroid}.
The angular momentum of static electromagnetic fields in Maxwell's
theory was discussed in \cite{Romer}.} However, this ``electromagnetic field energy''
idea does not seem attractive for a couple of reasons.

First, the ``electromagnetic energy'' integral (\ref{eq:field-en})
for the electric field $\mathbf{E}= q \mathbf{r}/(4 \pi r^3)$
associated with a stationary point charge (e.g., an electron) is
infinite.\footnote{See also \cite{Franklin} for discussion of other
difficulties related to the idea of energy and momentum contained in
the electromagnetic field. An interesting critical review of
Maxwell's electrodynamics and Minkowski space-time picture can be
found in section 1 of \cite{Gill}.} To avoid this difficulty,
various ``classical models'' of the electron were suggested, the
simplest of which is a charged sphere of a small but finite radius.
However, these models led to other problems. One of them is the
famous ``4/3 paradox'': It can be shown that the momentum of the
electromagnetic field associated with a finite-radius electron
does not form a 4-vector quantity together with its electromagnetic
energy \cite{Rohrlich-60, Butler, Comay}. This violation of
relativistic invariance can be ``fixed'' if one introduces an extra
factor of 4/3 in the formula for the field momentum. To justify this
extra factor the \emph{ad hoc} idea of \emph{Poincar\'e stresses}
\index{Poincar\'e stress} is sometimes introduced.\footnote{see
sections 16.4 - 16.6 in \cite{Jackson}}

\subsection{Conservation laws in RQD}
 \label{ss:conservation-RQD}

On the other hand, if one adopts the RQD ``no-fields'' approach, then total
observables of particle systems will be conserved, and their correct relativistic transformation laws
 will hold exactly without any \emph{ad hoc}
assumptions. No ``electromagnetic field'' contributions to these quantities need to be taken into acount. Indeed, in relativistic Hamiltonian dynamics (which is the
basis of our RQD approach to electrodynamics) the conservation laws
and transformation properties of observables are direct
consequences of the Poincar\'e group structure. The Poisson bracket
of any observable $F$ with the Hamiltonian $H$ determines
 the time evolution of this observable

\begin{eqnarray*}
\frac{dF(t)}{dt}  &=&   [F, H]_P
\end{eqnarray*}

\noindent Then the conservation of observables $H, \mathbf{P}$, and
$\mathbf{J}$ follows automatically from their vanishing Poisson
brackets with $H$.\footnote{For example, the explicit conservation of the total momentum $\mathbf{P}$ in our theory guarantees the resolution of paradoxes 6 and 7 in \cite{Kholmetskii-paradox}.} Similarly, boost transformations of $F$ can be obtained as solutions of (\ref{eq:icKF})

\begin{eqnarray*}
\frac{dF(\vec{\theta})}{d\vec{\theta}}  &=&   - c[F, \mathbf{K}]_P
\end{eqnarray*}

\noindent In the case of total momentum-energy $(\mathbf{P}, H)$, the commutators (\ref{eq:5.55}) - (\ref{eq:5.56}) hold independent on the strength of interaction. Then the 4-vector transformation formulas  (\ref{eq:6.2}) - (\ref{eq:6.3}) follow. So, in RQD the conservation of total observables of isolated systems and their correct transformation laws are embedded in the formalism at the most fundamental level and can be never violated. In particular, the total momentum  $\mathbf{P} = \mathbf{p}_1 + \mathbf{p}_2$ of the two-charge system described in the beginning of subsection \ref{ss:newton} is conserved automatically.

\subsection{Trouton-Noble ``paradox''}
\label{sc:trouton-noble}

To be more specific, let us now discuss the conservation of the total angular momentum in the two competing theories.

In RQD the total angular momentum $\mathbf{J}$ of any isolated
system of interacting particles is conserved. In other words, there can be no torque\footnote{The torque
is defined here as the time derivative of the total angular
momentum of the system.} in any isolated system of charges. This follows directly
from the following Poisson bracket in the Poincar\'e Lie algebra

\begin{eqnarray*}
\frac{d \mathbf{J}}{dt} &=& [\mathbf{J}, H ]_P = 0
\end{eqnarray*}

\noindent
This result should hold in any inertial frame of reference. For example, in a moving frame the relevant dynamical variables are \footnote{in quantum notation}

\begin{eqnarray*}
\mathbf{J}(\vec{\theta}) &=& e^{\frac{ic}{\hbar} \mathbf{K} \cdot \vec{\theta}} \mathbf{J} e^{-\frac{ic}{\hbar} \mathbf{K} \cdot \vec{\theta}} \\
H(\vec{\theta}) &=& e^{\frac{ic}{\hbar} \mathbf{K} \cdot \vec{\theta}} H e^{-\frac{ic}{\hbar} \mathbf{K} \cdot \vec{\theta}}
\end{eqnarray*}

\noindent and the equation of motion for the total angular momentum is\footnote{Here $t'$ is time measured by the moving observer. Note that this result is valid only if $\mathbf{K}$ is the full interaction-dependent boost (\ref{eq:k-z}) - (\ref{eq:boost-spin-orb}).}

\begin{eqnarray*}
\frac{d \mathbf{J}(\vec{\theta})}{dt'} &=& [\mathbf{J} (\vec{\theta}), H (\vec{\theta})]_P =
[ e^{\frac{ic}{\hbar} \mathbf{K} \cdot \vec{\theta}} \mathbf{J} e^{-\frac{ic}{\hbar} \mathbf{K} \cdot \vec{\theta}},e^{\frac{ic}{\hbar} \mathbf{K} \cdot \vec{\theta}} H e^{-\frac{ic}{\hbar} \mathbf{K} \cdot \vec{\theta}}]_P \\
&=& e^{\frac{ic}{\hbar} \mathbf{K} \cdot \vec{\theta}}[  \mathbf{J} , H ]_P e^{-\frac{ic}{\hbar} \mathbf{K} \cdot \vec{\theta}} = 0
\end{eqnarray*}

\noindent

Maxwell's
classical electrodynamics cannot make such a clear statement about
the conservation of the total angular momentum and the absence of
torque in all frames. This failure is in the center of the ``Trouton-Noble
paradox'' \index{Trouton-Noble paradox} which haunted Maxwell's
theory for more than a century \cite{Trouton-Noble, Page, Furry,
Spavieri, Jackson-torque, Butler, Teukolsky, Jefimenko-99a}.

\begin{figure}
\centering
 \includegraphics {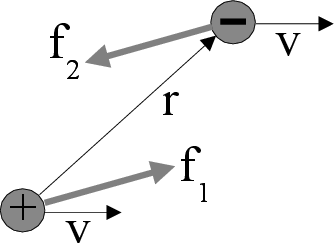} \caption{The Trouton-Noble ``paradox'': two charges
 moving with the same velocity $\mathbf{v}$. The forces $\mathbf{f}_1$ and $\mathbf{f}_2$
produce a non-zero torque.} \label{fig:11.11}
\end{figure}

Imagine two charges moving with the same velocity vector
$\mathbf{v}$, which makes an angle\footnote{The angle between $\mathbf{r}$ and $\mathbf{v}$ should be different from 0
and $90^{\circ}$. Note that in the original Trouton-Noble experiment
\cite{Trouton-Noble}, two charged capacitor plates were used instead
of point charges, but this difference has no significant effect on our
theoretical analysis. } with the vector $\mathbf{r} = \mathbf{r}_1 -
\mathbf{r}_2$ connecting positions of the charges (see Fig.
\ref{fig:11.11}). A calculation using the standard Biot-Savart force
formulas (\ref{eq:biot1}) - (\ref{eq:biot2})  predicts that there
should be a non-zero torque, which tries to turn vector $\mathbf{r}$
until it is perpendicular to the direction of motion $\mathbf{v}$
\cite{Sard}. This result is paradoxical for two reasons. First, as
we said earlier, one should expect zero torque from the conservation
of the total angular momentum. Second, there is no torque in the
reference frame that moves together with the charges,\footnote{In
this reference frame velocities of both charges are zero. So,
only the Coulomb force remains, which is directed along the vector
$\mathbf{r}$, thus causing no torque. } so the presence of the torque in the reference frame
at rest violates the principle of relativity. Numerous attempts to
explain this paradox within Maxwell's theory \cite{Page, Furry,
Spavieri, Jackson-torque, Butler, Teukolsky, Jefimenko-99a}  do not
look convincing.

\section{Electromagnetic induction} \label{ch:elinduction}

From equation (\ref{eq:fieldy})  it is clear that if both the charge 1 and the magnet 2 are at rest, then the force between them vanishes. Classical theory describes this situation as ``a magnet at rest does not create electric field''. One of greatest Faraday's discoveries was the realization that  a varying magnetic field does produce electric field, i.e., it acts on stationary charges. This phenomenon is called \emph{electromagnetic induction}. \index{electromagnetic induction} Magnetic field variation can result either from changing magnetic moment $\vec{\mu}_2$ or from changing its position in space $\mathbf{r}_2$. In this section we will consider the latter source of electromagnetic induction and some of its experimental manifestations.

\subsection{Moving magnets} \label{ch:induction}

 \begin{figure}
 \centering
 \includegraphics {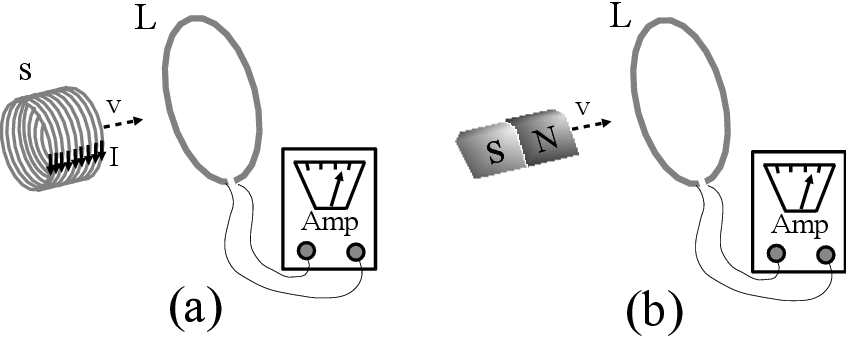} \caption{The electromagnetic
  induction. Current in the wire loop $L$ can be
 induced by (a) a moving solenoid with current; (b) a moving permanent magnet.}
\label{fig:11.13}
\end{figure}

 In this subsection we are going to consider
the force acting on a charge at rest ($\mathbf{p}_1=0$) from a
moving magnet $\vec{\mu}_2$. In the traditional Maxwell's theory, a moving bar magnet creates qualitatively the same fields and forces as a moving solenoid. However, this is not so in our approach. If the magnetic moment $\vec{\mu}_2$ is created by a particle with spin, then
the force on the charge 1 at rest is given by (\ref{eq:lorentz-force})\footnote{Recall that
$\left(\frac{d}{dt}\right)_2$ denotes the time derivative when
$\mathbf{r}_1$ is kept fixed.}

\begin{eqnarray}
\mathbf{f}_1^{spin} &=& - \frac{d}{d\mathbf{r}_1} \frac{q_1
([\mathbf{v}_2 \times \vec{\mu}_2] \cdot
  \mathbf{r})}{8
\pi cr^3} - \left(\frac{d}{dt} \right)_2
 \frac{q_1  [\vec{\mu}_2 \times \mathbf{r}] }{4 \pi c  r^3}
  \label{eq:11.10a}
\end{eqnarray}

\noindent If the magnetic moment is created by a small current loop, then we should use
(\ref{eq:lorentz-forcea})

\begin{eqnarray}
\mathbf{f}_1^{orb} &=& -\left(\frac{d}{dt} \right)_2
 \frac{q_1  [\vec{\mu}_2 \times \mathbf{r}] }{4 \pi c  r^3}
 \label{eq:11.11a}
\end{eqnarray}

\noindent  In other words, the
force produced by a moving spin has two components, the first of
which is conservative and the second is non-conservative\footnote{
The force is defined as \emph{conservative} if it can be represented
as a gradient of a scalar function (an example is given by the first
term on the right hand side of (\ref{eq:11.10a})).
\index{conservative force} Otherwise the force is called
\emph{non-conservative}. The integral of a conservative force
vector along any closed loop is zero. Therefore, conservative
forces on electrons cannot be detected by measuring a current in a
closed circuit. \index{non-conservative force}}

\begin{eqnarray}
\mathbf{f}_1^{spin} &=& \mathbf{f}_1^{cons} +
\mathbf{f}_1^{non-cons} \label{eq:11.18z}
\end{eqnarray}

\noindent The force produced by the current loop has only a
non-conservative component

\begin{eqnarray}
\mathbf{f}_1^{orb} &=&  \mathbf{f}_1^{non-cons} \label{eq:11.19z}
\end{eqnarray}

\noindent Let us first focus on the non-conservative force component
$\mathbf{f}_1^{non-cons}$, which is common for both spin and orbital
magnetic moments. We will return to the conservative force component
in subsection \ref{ch:rot-sol}.

 For macroscopic magnets the infinitesimal
quantities considered thus far  should be integrated
 on the magnet's volume $V$, e.g., the full non-conservative force
 exerted by a macroscopic magnet on the charge at rest $q_1$ is

\begin{eqnarray}
\mathbf{F}_1^{non-cons}
 &=&  -\left(\frac{d}{dt} \right)_2 \int_V
\frac{q_1[  \vec{\mu}_2 \times  \mathbf{r}] }{4 \pi  c r^3}
d\mathbf{r}_2
 \label{eq:induction2}
\end{eqnarray}

 \noindent   This means that the magnet,\footnote{either a permanent magnet or a solenoid with
current} moving near a wire loop $L$, induces a current in the loop
as shown in Fig. \ref{fig:11.13}(a) and (b).

Let us now show that this prediction agrees quantitatively with
Maxwell's electrodynamics. We denote by symbol
$\mathbf{e}_1$ the force with which a
microscopic magnetic moment acts on a unit charge\footnote{In Maxwell's electrodynamics this is the
definition of the \emph{electric field} $\mathbf{e}_1$. \index{electric field}}

\begin{eqnarray*}
\mathbf{e}_1 \equiv \mathbf{f}_1^{non-cons}/q_1
\end{eqnarray*}

\noindent  If we take \emph{curl} of this quantity, we obtain

\begin{eqnarray}
\left[\frac{\partial}{\partial \mathbf{r}_1} \times \mathbf{e}_1
\right]
 &=&  -\frac{1}{4 \pi  c} \left(\frac{d}{dt} \right)_2
\left[\frac{\partial}{\partial \mathbf{r}_1} \times \frac{[
\vec{\mu}_2
\times  \mathbf{r}] }{r^3} \right] \nonumber \\
 &=& -\frac{1}{4 \pi c} \left(\frac{d}{dt} \right)_2 \left(
-\frac{\vec{\mu}_2}{r^3} + \frac{3 (\vec{\mu}_2 \cdot \mathbf{r})
\mathbf{r}}{r^5} \right)
\nonumber \\
 &=&  -\frac{1}{c} \left(\frac{d}{dt} \right)_2 \mathbf{b}_1
 \label{eq:induction}
\end{eqnarray}

\noindent where $\mathbf{b}_1$ is the ``magnetic field''
(\ref{eq:fieldx}) of the magnetic moment. After integrating  both
sides of equation (\ref{eq:induction}) on the magnet's volume we obtain
exactly the Maxwell's equation

\begin{eqnarray*}
\left[\frac{\partial}{\partial \mathbf{r}_1} \times \mathbf{E}_1
\right]
 &=&  -\frac{1}{c} \left(\frac{d}{dt} \right)_2 \mathbf{B}_1
\end{eqnarray*}

\noindent which expresses the Faraday's law of induction.
\index{Faraday's law of induction}

It is important to stress that the origin of electromagnetic induction proposed in our work is fundamentally different from that adopted in Maxwell's theory. The traditional explanation is that electromagnetic induction results from inter-dependence of time-varying electric and magnetic fields. In our approach, the electromagnetic induction is the consequence of velocity-dependent interactions between magnetic dipoles and charges.

\subsection{Homopolar induction: non-conservative forces} \label{ch:homopolar}
\index{homopolar generator}

 \begin{figure}
 \centering
 \includegraphics {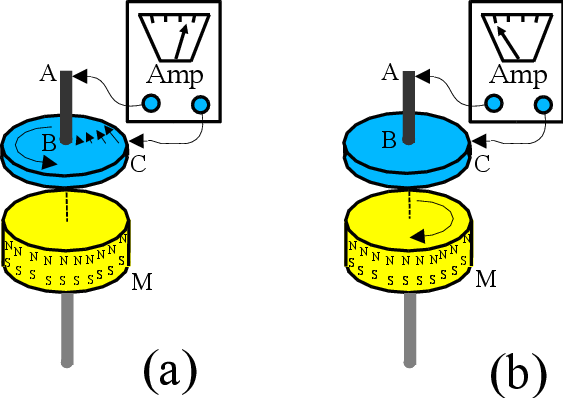} \caption{Homopolar generator. (a) the conducting disk $C$ rotates;
 (b) the magnet $M$ rotates.}
\label{fig:7}
\end{figure}

One interesting application of the electromagnetic induction law is
the homopolar generator shown in Fig. \ref{fig:7}. This device consists of a
conducting disk $C$ and a cylindrical magnet $M$. Both the
conducting disk and the magnet are rigidly attached to their own
shafts, and both can independently rotate about their common vertical axis.
The magnetization vector $\vec{\mu}_2$ of each small volume element of the
magnet is directed along the axis, so the total magnetic moment is
time-independent for both stationary and rotating magnets. The shaft
$AB$ is conducting. Points $A$ and $C$ are connected to sliding
contacts (shown by arrows), and the circuit is closed through the
galvanometer.

There are two modes of operation of this device. In the first mode
(see Fig. \ref{fig:7}(a)) the magnet is stationary while the
conducting disk rotates about its axis. The galvanometer detects a
current in the circuit. This has a simple explanation: The
force acting on electrons in the metal can be obtained by
integrating formula (\ref{eq:fieldy}),\footnote{As we saw in
subsection \ref{ch:induction}, this formula is applicable to both
``orbital'' and ``spin'' magnets at rest.} on the magnet's volume
$V$

\begin{eqnarray}
 \mathbf{F}_1 (\mathbf{r}_1, \mathbf{v}_1)
 &=&  \int_V \frac{q_1}{c}[\mathbf{v}_1 \times \mathbf{b}_1(\mathbf{r}_1)] d
 \mathbf{r}_2
 \label{eq:lorentz-force3}
\end{eqnarray}

\noindent The full electromotive force in the circuit is obtained by
integrating expression (\ref{eq:lorentz-force3}) on the variable $\mathbf{r}_1$
 along the closed contour $A \to B \to C \to$ galvanometer
$\to A$. The
velocity $\mathbf{v}_1$ is non-zero only on the segment $B \to
C$,\footnote{Velocities of electrons in the rotating conductor are
shown by small arrows in Fig. \ref{fig:7}(a).} where the force
$\mathbf{F}_1$ is directed radially. The integral is non-zero and
the galvanometer must show a non-vanishing current in agreement with
experiments.

In the second operation mode (see Fig. \ref{fig:7}(b)) the disk $C$
is fixed, and the magnet rotates. It was established by careful
experiments \cite{DasGupta, Then} that there is no current in this
case. If both the magnet and the disk rotate, then the current is
the same as in the ``first mode,'' i.e., with a fixed magnet. This
means that rotation of the magnet has no effect on the produced
current. This experimental result looks somewhat surprising,
because from the principle of relativity one could expect that the
physical outcome (the current) should depend only on the relative
movement of the magnet and the disk. However, this conclusion is
incorrect, because the principle of relativity is applicable only to
inertial movements. It cannot be applied to rotational movements
without contradictions.

Let us now analyze the rotating magnet case shown in Fig.
\ref{fig:7}(b) from the point of view of the Darwin-Breit
electromagnetic theory. We need to know the integral of the force
acting on electrons along the closed circuit $A \to B \to C \to$
galvanometer $\to A$. The conservative portion of the force
$\mathbf{f}_1^{cons}$ does not contribute to this integral. Since
here we have a cylindrical magnet rotating about its axis of
magnetization, the volume integral in the expression
(\ref{eq:induction2}) for the non-conservative force is
time-independent, and the total non-conservative force acting on
electrons is zero. This agrees with the observed absence of the
current.

\subsection{Homopolar induction: conservative forces} \label{ch:rot-sol}

 \begin{figure}
 \centering
 \includegraphics {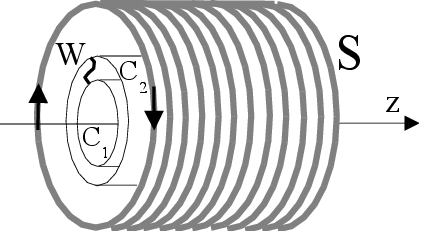} \caption{The Barnett's experiment.}
\label{fig:11.10}
\end{figure}

So far in our discussion of homopolar induction we considered only
non-conservative forces, mainly because they can be detected rather
easily by measuring induced currents in closed
circuits. In the beginning of the 20th century Barnett and Kennard
performed experiments \cite{Barnett, Kennard}\footnote{see also
\cite{Kholmetskii2}} with the specific purpose to detect the
conservative part of forces from moving magnets. Barnett's experimental setup
\index{Barnett experiment} (shown schematically in Fig.
\ref{fig:11.10}) resembled the homopolar generator discussed above.
Its main parts were a cylindrical solenoid $S$ with current and two
conducting cylinders $C_1$ and $C_2$ placed inside the solenoid. All
three cylinders shared the same rotation axis $z$. Conductors $C_1$
and $C_2$ formed a cylindrical capacitor. Initially they were
connected by a conducting wire $W$. Note that in contrast to the
homopolar generator experiment, where a current in a closed circuit
was measured, the system $C_1 - W - C_2$ in the Barnett's setup did
not form a closed circuit. So, the capacitor would obtain a non-zero
charge even if the force acting on electrons in the wire $W$ was
conservative.

Similar to the homopolar generator discussed above, this apparatus
could operate in two different modes. In the first mode the
cylindrical capacitor spun about its axis. Due to the presence of
the magnetic field inside $S$ a current ran through the wire $W$,
and the capacitor $C_1 - C_2$ became charged. Then the wire was
disconnected, capacitor's rotation stopped and the capacitor's
charge measured. As expected, the measured charge was consistent
with the standard Lorentz force formula (\ref{eq:lorentz-force3}).

In the second operation mode, the capacitor was fixed while the
solenoid rotated about its axis. No charge on the capacitor was
registered in this case. This result is consistent with our theory,
because, just as in the case of homopolar generator, the
non-conservative force (\ref{eq:induction2}) vanishes due to the
cylindrical symmetry of the setup, and the conservative force is
absent in the case of a moving solenoid (\ref{eq:11.19z}).  So, the
null result of the Barnett's experiment confirms our earlier
conclusion that moving solenoids with current do not exert
conservative forces on nearby charges.

A different result is expected in the case of a rotating permanent
magnet. In this case the conservative force component is non-zero.
It can be obtained by integrating the first term on the right hand
side of (\ref{eq:11.10a}) on the volume $V$ of the magnet

\begin{eqnarray}
\mathbf{F}_1^{cons}
 &=&  -\frac{d}{d \mathbf{r}_1}  \int_V
\frac{q_1([ \mathbf{v}_2 \times \vec{\mu}_2] \cdot  \mathbf{r}) }{8
\pi c r^3} d\mathbf{r}_2
 \label{eq:induction23}
\end{eqnarray}

\noindent So, rotating cylindrical permanent magnet should induce a
non-zero charge in a stationary capacitor. A relevant experiment was performed in 1913 by Wilson and Wilson
\index{Wilson-Wilson experiment}   \cite{Wilson}. This experiment was
repeated again in 2001 with improved accuracy \cite{Hertzberg}.
For theoretical discussion of the Wilson-Wilson experiment from the
point of view of Maxwell's electrodynamics see \cite{McD-Wilson,
Pellegrini}.

 \begin{figure}
 \centering
 \includegraphics {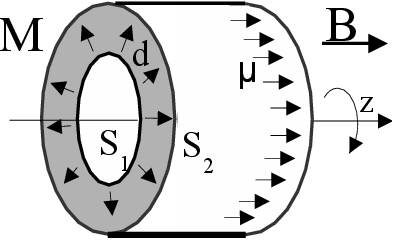} \caption{Schematic of the Wilson-Wilson experiment.}
\label{fig:11.16}
\end{figure}

Schematic representation of the Wilson-Wilson experiment is shown in
Fig. \ref{fig:11.16}. A hollow cylinder $M$ made of magnetic
dielectric (non-conducting material) was placed in a constant
magnetic field $\mathbf{B}$  parallel to the axis $z$. The inner and
outer surfaces of the cylinder ($S_1$ and $S_2$, respectively) were
covered by metal, and the electrostatic potential between the two
surfaces was measured. When the cylinder was at rest, no potential
was recorded, as expected. However, when the cylinder was rotated a
non-zero potential difference was observed. This potential is a
result of electric dipoles $d$ created in the bulk of the magnet.
There are two physical mechanisms for the appearance of these
dipoles. First, molecules of the dielectric material moving in the
magnetic field $\mathbf{B}$ get polarized (the Lorentz forces act in
opposite directions on positive and negative charges in the
molecules). Second, moving induced magnetic moments $\vec{\mu}$ create
``electric fields'' similar to the ``field'' of an electric dipole.
Indeed, if we compare expression (\ref{eq:induction23}) with the
force exerted on the charge $q_1$ by an electric dipole $\mathbf{d}$

\begin{eqnarray*}
 \mathbf{f}_1^{dipole}
 &=&  -\frac{\partial}{\partial \mathbf{r}_1} \frac{q_1(\mathbf{d} \cdot \mathbf{r})}
 {4 \pi r^3}
\end{eqnarray*}

\noindent we see that a magnetic moment $\vec{\mu}$ moving with
velocity $\mathbf{v}$ acquires an electric dipole of the
magnitude\footnote{The presence of the dipole electric field near
moving magnetic moment is predicted in the traditional
special-relativistic theory as well \cite{Einstein-Laub, Rosser},
however this prediction
\begin{eqnarray*}
 \mathbf{d}^{SR}
 =  [ \mathbf{v} \times \vec{\mu}]/c \label{eq:dsr}
\end{eqnarray*}
 is twice larger than our result.}

\begin{eqnarray}
 \mathbf{d}
 &=&   \frac{[ \mathbf{v} \times \vec{\mu}]} {2c} \label{eq:dour}
\end{eqnarray}

\noindent Thus, the Wilson-Wilson experiment clearly demonstrated that
both kinds of dipole moments (those due to dielectric polarization
and those due to moving dipole moments) are present in the rotating
magnet. This confirms qualitatively our conclusion about the
presence of conservative forces (\ref{eq:induction23}) near moving
permanent magnets. A quantitative description of this experiment
would require calculation of the polarization and magnetization of
bodies moving in an external magnetic field. This is beyond the
scope of the theory developed here.

\section{Aharonov-Bohm effect}
\label{sc:a-b-effect}

The central idea of our approach to classical electrodynamics is the rejection of electric and magnetic fields. This also means that we reject the notion of electromagnetic potentials $A^{\mu}(\mathbf{x},t)$. In Maxwell's electrodynamics these potentials are assumed to be non-observable. However, there exists a class of experiments, which allegedly proves the reality of electromagnetic potentials. The oldest and the most famous representative in this class is the Aharonov-Bohm effect. In its traditional interpretation a claim is made that this effect is a manifestation of non-vanishing electromagnetic potentials in regions of space where both electric and magnetic fields are zero. If this interpretation were true, then our particle-only theory would be in trouble. Our goal in this section is to show that there is no reason for concern. We are going to demonstrate that the Aharonov-Bohm effect can be easily explained in terms of particles interacting via Darwin-Breit potentials. This explanation relies also on quantum properties of particles, in particular, on how the interaction potential affects phases of quasiclassical wave packets, as described in subsection \ref{ss:time-wave}.

\subsection{Infinitely long solenoids or magnets}
\label{sc:infinitely}

It is not difficult to show that the ``magnetic field'' outside an
infinitely long thin solenoid vanishes. Assuming that the solenoid
is oriented along the $z$-axis with $x=y=0$\footnote{i.e., solenoid's points have coordinates $\mathbf{r}_2 =
(0,0,z)$} and that the observation point is at $\mathbf{r}_1 = (x_1,
y_1, 0)$, we obtain\footnote{Here we integrate equation (\ref{eq:fieldx}) on the (infinite) length of the solenoid. This time $\mu_2$ should be understood
as magnetization per unit length of the solenoid.}

\begin{eqnarray}
 \mathbf{B}_{long}(\mathbf{r}_1) &=& \int \limits_{-\infty}^{\infty} dz
 \Big(- \frac{(0,0,\mu_2)}{4 \pi (x_1^2 + y_1^2 + z^2)^{3/2}}
 - \frac{3 \mu_2 z (x_1, y_1, -z)}{4 \pi
(x_1^2 + y_1^2 + z^2)^{5/2}} \Big) \nonumber \\
&=& 0 \label{eq:inf-long}
\end{eqnarray}

A solenoid with
arbitrary cross-section can be represented as a bunch of parallel thin solenoids.\footnote{see Fig. \ref{fig:11.12}} If the observation point $\mathbf{r}_1$ is outside the
solenoid's volume, then equation (\ref{eq:inf-long}) holds for each thin
segment, and the total ``magnetic field'' at point $\mathbf{r}_1$ also vanishes. The same analysis applies to infinitely long bar magnets of arbitrary cross-section. Thus we conclude that the force acting on a
moving charge outside infinitely long magnet (either permanent
magnet or solenoid with current) is zero. This conclusion agrees with
calculations based on Maxwell's equations. See, for example, Problem
5.2(a) in \cite{Jackson}.

However, the vanishing force does not mean that the potential energy
of the charge-solenoid interaction is zero as well. In the case of a thin solenoid, the potential
energy can be found by
integrating the last term in equation (\ref{eq:full-H2}) along the length
of the solenoid and noticing that the mixed product $([\vec{\mu}_2
\times \mathbf{v}_1 ] \cdot \mathbf{r}_1)$ is independent of
$z$. Denoting $r \equiv (x_1^2+y_1^2)^{1/2}$ the particle-solenoid distance, we obtain

\begin{eqnarray}
V_{long} &=&  \int \limits_{-\infty}^{\infty} dz \frac{q_1
([\vec{\mu}_2 \times \mathbf{v}_1 ] \cdot \mathbf{r}_1)}{4 \pi  c
(x_1^2+y_1^2+z^2)^{3/2}} = \frac{q_1 ([\vec{\mu}_2 \times
\mathbf{v}_1 ] \cdot \mathbf{r}_1)}{2 \pi c r^2}
\label{eq:inf-solen}
\end{eqnarray}

\noindent The acceleration of the moving charge is found, as usual,
by application of the Hamilton's equations of motion

\begin{eqnarray}
\frac{d \mathbf{p}_1}{dt} &=& [\mathbf{p}_1,H]_P = - \frac{\partial
V_{long}}{\partial
\mathbf{r}_1} =
 \frac{q_1[ \mathbf{p}_1 \times \vec{\mu}_2] }{2 \pi m_1 cr^2}
 +
 \frac{q_1 ([ \vec{\mu}_2 \times \mathbf{p}_1] \cdot \mathbf{r}_1)\mathbf{r}_1 }{\pi m_1
 cr^4} \nonumber \\
\frac{d \mathbf{r}_1}{dt} &=& [\mathbf{r}_1, H]_P = \frac{\partial
H}{\partial \mathbf{p}_1} = \frac{\mathbf{p}_1}{m_1}-
\frac{p_1^2\mathbf{p}_1}{2m_1^3 c}  + \frac{q_1 [\mathbf{r}_1 \times
\vec{\mu}_2] }{2 \pi m_1c r^2}
\nonumber \\
\frac{d^2 \mathbf{r}_1}{dt^2}
 &\approx& \frac{\dot{\mathbf{p}}_1}{m_1}+
 \frac{q_1[ \mathbf{p}_1 \times \vec{\mu}_2 ] r_1^2}{2 \pi m_1^2c r^4}
 -
 \frac{q_1[ \mathbf{r}_1 \times \vec{\mu}_2 ] (\mathbf{r}_1 \cdot \mathbf{p}_1)}{\pi m_1^2c r^4}
 \nonumber \\
&=& \frac{q_1}{\pi m_1^2c r^4} ( [ \mathbf{p}_1 \times \vec{\mu}_2 ]
r_1^2-
 ([ \mathbf{p}_1 \times \vec{\mu}_2 ] \cdot \mathbf{r}_1) \mathbf{r}_1
 -[ \mathbf{r}_1 \times \vec{\mu}_2 ] (\mathbf{r}_1 \cdot \mathbf{p}_1))
 \nonumber \\
 &=& \frac{q_1}{\pi m_1^2c r^4} ( -[\mathbf{r}_1 \times[ \mathbf{r}_1 \times
 [\mathbf{p}_1 \times \vec{\mu}_2 ]]]
 -[ \mathbf{r}_1 \times \vec{\mu}_2 ] (\mathbf{r}_1 \cdot \mathbf{p}_1))
 \nonumber \\
 &=& \frac{q_1}{\pi m_1^2c r^4} ( -[\mathbf{r}_1 \times  \mathbf{p}_1] ( \mathbf{r}_1
 \cdot
  \vec{\mu}_2 )
 +[\mathbf{r}_1 \times \vec{\mu}_2 ] (\mathbf{r}_1 \cdot \mathbf{p}_1)
 -[ \mathbf{r}_1 \times \vec{\mu}_2 ] (\mathbf{r}_1 \cdot \mathbf{p}_1))
 \nonumber \\
 &=& 0 \label{eq:d2rdt2ab}
\end{eqnarray}

\noindent where we took into account that $(\mathbf{r}_1 \cdot
\vec{\mu}_2) = 0$. This agrees with the vanishing ``magnetic field''
found earlier and presents a curious example of a non-vanishing
potential, which does not produce any force on charges. Experimental
manifestations of such potentials will be discussed in this section.

\subsection{Aharonov-Bohm experiment}
\label{sc:aharonov}

In the preceding subsection  we concluded that charges do not
experience any force (acceleration) when they move in the vicinity
of a straight infinite magnetized solenoid or a permanent bar magnet.
However, the absence of force does not mean that charges do not
``feel'' the presence of the magnet. In spite of zero
magnetic (and electric) ``field'', infinite solenoids/rods have a
non-zero effect on particle wave functions and their interference. This effect was first
predicted by Aharonov and Bohm \index{Aharonov-Bohm effect}
\cite{Aharonov} and later confirmed in experiments \cite{Chambers,
Tonomura-86, Tonomura-86a}. Experimentally it was found  that the interference of the
two wave packets at point $B$ depends on the magnetization of the
solenoid/rod, in spite of zero force acting on the
electrons.  The explanation proposed by Aharonov and Bohm was based on electromagnetic potentials in the
multiply-connected topology of space induced by the presence of the solenoid. There exist attempts to explain the
Aharonov-Bohm effect as a result of classical electromagnetic force
that creates a ``time lag'' between wave packets moving on different
sides of the solenoid \cite{Rubio, Boyer-Darwin, Boyer-Ahar,
Boyer-2007-08, Boyer-2007}. However, this approach seems to be in
contradiction with recent measurements, which failed to detect such
a ``time lag'' \cite{Caprez}.  Several other non-conventional
explanations of the Aharonov-Bohm effect were also suggested in the
literature \cite{Spavieri-92, Wesley, Pinheiro, Hegerfeldt-08}.   Here we suggest a different explanation \cite{Ahar}.

\begin{figure}
\centering
 \includegraphics {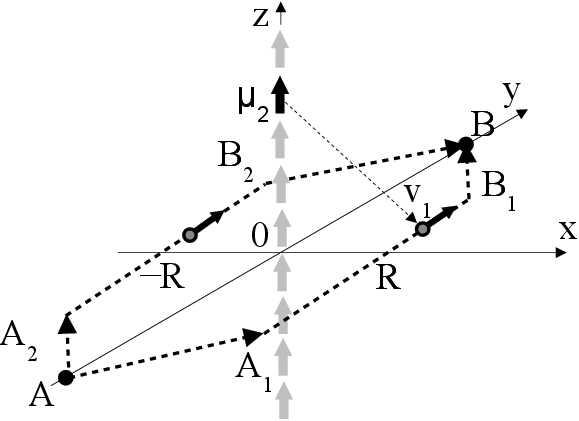} \caption{The Aharonov-Bohm experiment. The vertical
 infinite thin magnetized rod with linear magnetization density $\mu_2$ is
 shown by grey arrows.} \label{fig:4}
\end{figure}

 Let us consider the
idealized version of the Aharonov-Bohm experiment shown in Fig.
\ref{fig:4}: An infinite solenoid or
 ferromagnetic fiber with a negligible cross-section and linear magnetization density $\mu_2$
 is erected vertically in the origin. The electron wave packet is split
 into two parts (e.g., by using a
double-slit)
 at point $A$. The subpackets travel on both sides of the
 solenoid/bar
 with constant velocity $\mathbf{v}_1$, and the distance of the
 closest approach is $R$.
 The subpackets  rejoin at point $B$, where the interference is measured.
 (The two trajectories $AA_1B_1B$ and $AA_2B_2B$ are denoted by dashed lines.)
 The distance $AB$ is sufficiently large, so that the two paths
can be regarded as parallel to the $y$-axis everywhere

\begin{eqnarray}
\mathbf{r}_1(t) = (\pm R, v_{1y}t, 0) \label{eq:rt}
\end{eqnarray}

To estimate the solenoid's effect on the interference, we need
to turn to the quasiclassical representation of particle dynamics
from subsection \ref{ss:time-wave}. We have established there that the center
of the wave packet is moving in accordance with Heisenberg's
equations of motion. In our case, no force is acting on the
electrons, so their trajectories (\ref{eq:rt}) are independent on
magnetization. We also established in \ref{ss:time-wave} that the
overall phase factor of the wave packet changes in time as
 $\exp(\frac{i}{\hbar}\phi(t)) $, where the \emph{action integral}
 \index{action integral} $\phi(t)$
is given by

\begin{eqnarray}
\phi(t) &\equiv& \int \limits_{t_0}^{t} \left(\frac{ m_1
v_{1y}^2(t')}{2} - V_{long}(t') \right) dt' \label{eq:phase-factor}
\end{eqnarray}

\noindent and $V_{long}(t)$ is the time dependence of the potential (\ref{eq:inf-solen})
experienced by the electron. In the Aharonov-Bohm experiment the
electron's wave packet separates into two subpackets that travel
along different paths $AA_1B_1B$ and $AA_2B_2B$. Therefore, the
phase factors accumulated by the two subpackets are generally
different, and the interference of the ``left'' and ``right'' wave
packets at point $B$ will depend on this phase difference

\begin{eqnarray*}
\Delta \phi &=& \frac{1}{\hbar}(\phi_{left} - \phi_{right})
\label{eq:right}
\end{eqnarray*}

Let us now calculate the relative phase difference in the geometry of
Fig. \ref{fig:4}. The kinetic energy term in (\ref{eq:phase-factor})
does not contribute, because velocity remains constant and equal for both
paths due to (\ref{eq:d2rdt2ab}). However, the potential energy of the charge 1 is different for
the two paths. For all points on the ``right'' path the numerator of
the expression (\ref{eq:inf-solen}) is $ - q_1\mu_2 v_{1y} R$ and
for the ``left'' path the numerator is $q_1\mu_2 v_{1y} R$. Then the
total phase difference

\begin{eqnarray}
\Delta \phi &=& \frac{1}{\hbar} \int \limits_{-\infty}^{\infty}
\frac{ q_1\mu_2 Rv_{1y}}{ \pi  c (R^2 + v_1^2t^2)} dt  =
\frac{e\mu_2 }{\hbar c} \label{eq:phase-shift}
\end{eqnarray}

\noindent does not depend on the electron's
velocity and on the value of $R$. This phase difference is proportional to the
solenoid's magnetization
 $\mu_2$.   So, all essential properties of the
Aharonov-Bohm effect are fully reproduced within our approach.\footnote{Our result was derived for
 thin ferromagnetic rods and solenoids, however the same arguments
apply to infinite cylindrical rods and solenoids of any
cross-section.}

It is interesting that the presence of the phase difference is not
specific to line magnets of infinite length. This effect was also
seen in experiments with short magnetized nanowires
\cite{Matteucci}. This observation presents a challenge for the traditional
explanation, which must apply one logic (electromagnetic potential
in the space with ``multiple-connected topology'') for infinitely
long magnets and another logic (the presence of the magnetic field)
for finite bar magnets. Our description of the Aharonov-Bohm effect is
more economical, as it applies the same logic independent on whether
the magnet is infinite or finite.\footnote{To find the potential
energy in the case of a finite linear magnet one should simply use
finite integral limits in (\ref{eq:inf-solen}).} In both cases there
is a difference between action integrals for electron's paths
passing the line magnet on the right and on the left.

\subsection{Toroidal magnet and moving charge} \label{ch:toroid}

The system consisting of a toroidal magnet and a moving charge is
interesting for two reasons.  First, toroidal permanent magnets were
used in Tonomura's experiments \cite{Tonomura-86, Tonomura-86a},
which are regarded as the best evidence for the Aharonov-Bohm effect.
Second, classical Maxwell's electrodynamics has a serious trouble in
explaining how the total momentum is conserved in this system. This
is known as the ``Cullwick's paradox'' \index{Cullwick paradox}
\cite{Cullwick, Aguirregabiria, McDonald-06}.

In an attempt to explain this paradox, let us apply Maxwell's
theory to a charge moving along the symmetry axis through the center
of a magnetized torus (see Fig. \ref{fig:11.14}). As we will see below, there is no
``magnetic field'' outside the toroidal magnet, so the force
acting on the charge is zero. However, the moving charge creates its
own ``magnetic field'' which does act on the torus with a non-zero
force. So, the Newton's third law is
apparently violated. According to McDonald \cite{McDonald-06}, the
balance of force can be restored if one takes into account the
hypothetical ``momentum of the electromagnetic field.''\footnote{Note that in the McDonald's treatment the force is identified with the time derivative of
momentum, while in our approach the force is defined as (mass)$\times$(acceleration).}  However, this is not the whole story
yet. The field momentum turns out to be non-zero even in the case
when both the magnet and the charge are at rest. This leads to the
absurd conclusion that the linear momentum of the system does not
vanish even if nothing moves. The problem is allegedly fixed by assuming
the existence of the ``hidden momentum'' in the magnet.
\index{hidden momentum}  However this explanation does not seem
satisfactory, and here we would like to suggest a different version
of events.

\begin{figure}
\centering
 \includegraphics {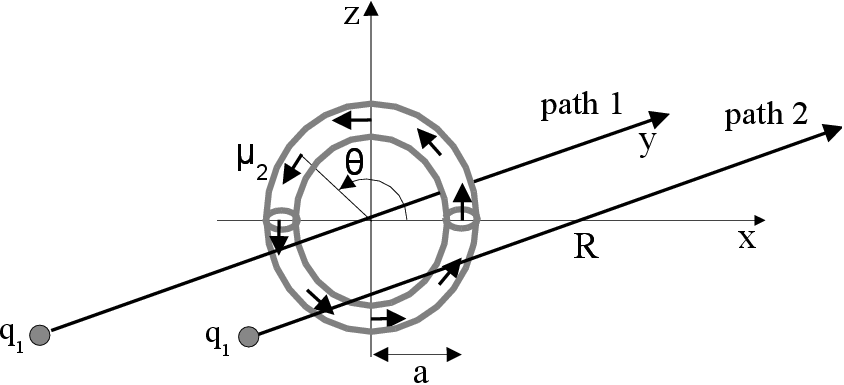} \caption{Toroidal magnet  and moving charge.
 ``path 1'' passes through the center of the torus; ``path 2'' is outside the torus.}
\label{fig:11.14}
\end{figure}

First we need to derive the Hamiltonian describing dynamics of the
system ``charge 1 + toroidal magnet 2.'' We introduce a Cartesian
coordinate system shown in fig \ref{fig:11.14}. Assume that the
torus has radius $a$ and linear magnetization density $\mu_2$ and
that  particle 1 moves straight through the center of the torus
with momentum $\mathbf{p}_1 = (0,p_{1y}, 0)$ (path 1 in Fig. \ref{fig:11.14}). Then we can use symmetry arguments to
disregard $x$ and $z$ components of forces and write\footnote{ We do not
assume that the magnet is stationary. It can move along the
$y$-axis. The $y$-component of its velocity is denoted $V_{2y}\approx P_{2y}/M_2$, where  $M_2$ is the full mass of the magnet and $\mathbf{P}_2 = (0, P_{2y}, 0)$ is the magnet's
momentum.}

\begin{eqnarray*}
\mathbf{r}_2 &=& (a \cos \theta, 0, a \sin \theta) \\
\mathbf{r} &=& \mathbf{r}_1 - \mathbf{r}_2= (-a \cos \theta, r_{1y}, -a \sin \theta) \\
\vec{\mu}_2 &=& ( -\mu_2 \sin \theta, 0, \mu_2 \cos \theta) \\
\ [\vec{\mu}_2 \times \mathbf{r} ]_y &=& a \mu_2 \sin^2 \theta + a
\mu_2
\cos^2 \theta = a \mu_2 \\
\ [\vec{\mu}_2 \times \mathbf{r} ] \cdot \mathbf{p}_1 &=& a \mu_2
p_{1y} \\
\ [\vec{\mu}_2 \times \mathbf{r} ] \cdot \mathbf{P}_2 &=& a \mu_2
M_2 V_{2y}
\end{eqnarray*}

\noindent Then the potential energy of interaction between the
charge and the magnet is obtained by integrating the potential
energy in (\ref{eq:h-charge-spin}) on the length of the
torus\footnote{ $\mathbf{R}_2$ is the center of mass of the toroid. Here we assume that we are dealing with a permanent
toroidal magnet. For a toroidal solenoid one should integrate the
potential energy expression (\ref{eq:vloopq1}). Then the second term
on the right hand side of (\ref{eq:pot-toroid}) would be absent.
}

\begin{eqnarray}
V
&=&    \int \limits _0^{2 \pi} d \theta  \Bigl( -\frac{q_1 a^2
\mu_2 p_{1y}}{4 \pi m_1 c (a^2 \cos^2 \theta + (r_{1y}-R_{2y})^2 +
a^2\sin^2
 \theta)^{3/2}} \nonumber \\
 &\ & +\frac{q_1 a^2 \mu_2
V_{2y}}{8 \pi c (a^2 \cos^2 \theta + (r_{1y}-R_{2y})^2 + a^2\sin^2
 \theta)^{3/2}} \Bigr) \nonumber \\
&=&    -\frac{q_1 a^2 \mu_2 p_{1y}}{2 m_1 c (a^2  +
(r_{1y}-R_{2y})^2)^{3/2}} + \frac{q_1 a^2 \mu_2 P_{2y}}{4 M_2 c (a^2
+ (r_{1y}-R_{2y})^2)^{3/2}} \nonumber \\
\label{eq:pot-toroid}
\end{eqnarray}

\noindent and the full Hamiltonian can be written as

\begin{eqnarray*}
H &=&  \frac {p_{1y}^2}{2m_1} +  \frac {P_{2y}^2}{2M_2}
  -  \frac{p_{1y}^4}{8m_1^3 c^2}
- \frac{P_{2y}^4}{8M_2^3 c^2}  -\frac{q_1 a^2 \mu_2 p_{1y}}{2 m_1 c
(a^2  + (r_{1y}-R_{2y})^2)^{3/2}} \\
&\ & +\frac{q_1 a^2 \mu_2 P_{2y}}{4 M_2 c (a^2  +
(r_{1y}-R_{2y})^2)^{3/2}}
\end{eqnarray*}

\noindent  We will now switch to the quasiclassical approximation in which the light particle 1 (presumably, an electron) is described by a localized wave packet , whose center moves according to the laws of classical mechanics. The first Hamilton's equation of motion leads to the following
results

\begin{eqnarray*}
\frac{d p_{1y}}{dt} &=& -\frac{\partial V}{\partial r_{1y}} \\
&=& -\frac{3q_1 a^2 \mu_2 p_{1y}(r_{1y}-R_{2y})}{2 m_1 c (a^2  +
(r_{1y}-R_{2y})^2)^{5/2}} +  \frac{3q_1 a^2 \mu_2
P_{2y}(r_{1y}-R_{2y})}{4 M_2 c (a^2  +
(r_{1y}-R_{2y})^2)^{5/2}} \\
\frac{d P_{2y}}{dt} &=& - \frac{d p_{1y}}{dt}
\end{eqnarray*}

\noindent So, unlike in Maxwell's theory, the rate of change of the
1st particle's momentum is non-zero and the 3rd Newton's law is
satisfied without involvement of the ``electromagnetic field
momentum.'' Acceleration of the charge 1 is calculated as follows

\begin{eqnarray*}
\frac{dr_{1y}}{dt} &=& \frac{p_{1y}}{m_1} - \frac{p_{1y}^3}{2m_1^3
c} + \frac{\partial V}{\partial p_{1y}} \nonumber \\
&=& \frac{p_{1y}}{m_1} - \frac{p_{1y}^3}{2m_1^3
c} -\frac{q_1 a^2 \mu_2 }{2 m_1 c (a^2  + (r_{1y}-R_{2y})^2)^{3/2}} \nonumber \\
\frac{d^2r_{1y}}{dt^2} &=& \frac{\dot{p}_{1y}}{m_1} +\frac{3q_1 a^2
\mu_2 (r_{1y}-R_{2y})(v_{1y}-V_{2y})}{2 m_1 c (a^2 +
(r_{1y}-R_{2y})^2)^{5/2}}  \nonumber \\
&\approx& -  \frac{3q_1 a^2 \mu_2 V_{2y}(r_{1y}-R_{2y})}{4 m_1 c
(a^2  + (r_{1y}-R_{2y})^2)^{5/2}}
\end{eqnarray*}

\noindent When the magnet is at rest ($V_{2y}=0$) this expression
vanishes, so there is no force (acceleration) on the particle 1, as
expected.\footnote{This result holds also for a toroidal solenoid.} The force (acceleration) acting on the magnet is found by
the following steps

\begin{eqnarray*}
\frac{dR_{2y}}{dt} &=& \frac{P_{2y}}{M_2} - \frac{P_{2y}^3}{2M_2^3
c} + \frac{\partial V}{\partial P_{2y}} \nonumber \\
&=& \frac{P_{2y}}{M_2} - \frac{P_{2y}^3}{2M_2^3 c} + \frac{q_1 a^2
\mu_2 }{4
M_2 c (a^2  + (r_{1y}-R_{2y})^2)^{3/2}}\nonumber  \\
 \frac{d^2R_{2y}}{dt^2}
 &=& \frac{\dot{p}_{2y}}{M_2} - \frac{3q_1 a^2 \mu_2 (r_{1y}-R_{2y})(v_{1y}-V_{2y})}{4 M_2 c
(a^2 + (r_{1y}-R_{2y})^2)^{5/2}} \\
& \approx& \frac{3q_1 a^2 \mu_2 v_{1y}(r_{1y}-R_{2y})}{4 M_2 c (a^2
+ (r_{1y}-R_{2y})^2)^{5/2}}
\end{eqnarray*}

\noindent So, the magnet's acceleration does not vanish even if $V_{2y}=0$. This is an example of the  situation described in subsection \ref{sc:force-def}: the forces are not balanced despite exact conservation of the total momentum.

To complete consideration of the quasiclassical wave packet passing
through the center of the stationary torus we need to calculate the
action integral (\ref{eq:phase}). Essentiall, we are going to integrate on time the
potential (\ref{eq:pot-toroid})

\begin{eqnarray}
\phi_0 &=&  -\int \limits_{-\infty}^{\infty} dt \frac{q_1 a^2 \mu_2
v_{1y}}{2  c (a^2  + v_{1y}^2t^2)^{3/2}} =   \frac{q_1 \mu_2 }{ c }
\label{eq:ph-shift}
\end{eqnarray}

\noindent Here we set $R_{2y}=P_{2y}=0$,
$p_{1y} \approx m_1v_{1y}, r_{1y} = v_{1y}t$.

Now let us consider a charge whose trajectory passes outside the
stationary torus (path 2 in Fig. \ref{fig:11.14}). The force acting on the charge vanishes, so we can
assume that the wave packet travels with constant velocity along
straight line

\begin{eqnarray}
\mathbf{r}(t)\approx \mathbf{r}_1(t)  = (R, v_{1y}t, 0)
\label{eq:rr1t}
\end{eqnarray}

\noindent To calculate the action integral we repeat our earlier
derivation of the potential energy (\ref{eq:pot-toroid}), this time
taking into account $x$- and $z$-components of vectors. We will
assume that the torus is small, so that at all times $r \gg a$,
$\mathbf{r}_1 \approx \mathbf{r}$ and

\begin{eqnarray*}
\ [\vec{\mu}_2 \times \mathbf{r} ] &=& (-\mu_2 r_{1y} \cos \theta,
\mu_2 r_{1z} \sin \theta + \mu_2 r_{1x} \cos \theta -\mu_2 a,
- r_{1y} \sin \theta) \\
\ [\vec{\mu}_2 \times \mathbf{r} ] \cdot \mathbf{p}_1 &=& \mu_2
(-p_{1x} r_{1y} \cos \theta + p_{1y} r_{1z} \sin \theta + p_{1y}
r_{1x} \cos
\theta - p_{1y} a - p_{1z} r_{1y} \sin \theta) \\
&=& - \frac{1}{a}(\mathbf{N}_2 \cdot \mathbf{p}_1) + \mu_2
[\mathbf{r}_1 \times \mathbf{p}_1]_z \cos \theta -
\mu_2[\mathbf{r}_1 \times \mathbf{p}_1]_x \sin \theta
\end{eqnarray*}

\noindent  Here we characterized magnetic properties of the toroidal magnet
by the vector $\mathbf{N}_2 = (0, \mu_2 a^2, 0)$ which is
perpendicular to the plane of the torus and whose length is $\mu_2
a^2$. Then using approximation (\ref{eq:r-3}), setting $\mathbf{p}_2=0$ and integrating the potential energy in (\ref{eq:h-charge-spin}) on the
length of the torus, we obtain

\begin{eqnarray}
V &=& -\int \limits_{0}^{2 \pi} d\theta \frac{q_1 a[\vec{\mu}_2
\times
\mathbf{r}] \cdot \mathbf{p}_1}{4 \pi m_1 cr^3} \nonumber \\
&\approx& -\frac{q_1}{4 \pi m_1 c}\int \limits_{0}^{2 \pi} d\theta
\left(- (\mathbf{N}_2 \cdot \mathbf{p}_1) + \mu_2a [\mathbf{r}
\times \mathbf{p}_1]_z \cos \theta - \mu_2a[\mathbf{r} \times
\mathbf{p}_1]_x
\sin \theta \right) \times \nonumber \\
&\ & \left( \frac{1}{r^3} + \frac{3a(r_{x} \cos \theta + r_{z}
\sin \theta)}{r^5} \right) \nonumber \\
&=&   \frac{q_1 (\mathbf{N}_2 \cdot \mathbf{p}_1) }{2m_1 c r^3} +
\frac{3q_1 \mu_2 a^2}{4 \pi m_1 c r^5} \int \limits_{0}^{2 \pi}
d\theta \left([\mathbf{r} \times \mathbf{p}_1]_zr_{x}\cos^2 \theta -
[\mathbf{r} \times
\mathbf{p}_1]_x r_{z} \sin^2 \theta \right) \nonumber \\
&=&   \frac{q_1 (\mathbf{N}_2 \cdot \mathbf{p}_1) }{2m_1 c r^3} +
\frac{3q_1 \mu_2 a^2}{4  m_1 c r^5}  \left([\mathbf{r} \times
\mathbf{p}_1]_zr_{x} - [\mathbf{r} \times
\mathbf{p}_1]_x r_{z}  \right) \nonumber \\
&=& \frac{q_1 (\mathbf{N}_2 \cdot \mathbf{p}_1) }{2m_1 c r^3} +
\frac{3q_1 }{4  m_1 c r^5} ([[\mathbf{p}_1
\times \mathbf{r}]\times \mathbf{r}] \cdot \mathbf{N}_2)  \nonumber \\
&=& \frac{q_1 (\mathbf{N}_2 \cdot \mathbf{p}_1) }{2m_1 c r^3} -
\frac{3q_1 }{4  m_1 c r^5} ((\mathbf{p}_1 \cdot \mathbf{N}_2)r^2 -
(\mathbf{r} \cdot \mathbf{N}_2)
(\mathbf{p}_1 \cdot \mathbf{r})) \nonumber   \\
&=&   -\frac{q_1 (\mathbf{N}_2 \cdot \mathbf{p}_1)}{4 m_1 c r^3} +
\frac{3q_1 (\mathbf{r} \cdot \mathbf{N}_2) (\mathbf{p}_1 \cdot
\mathbf{r})}{4  m_1 c r^5} \label{eq:vtoroid}
\end{eqnarray}

\noindent The time dependence of this potential energy is obtained by
substitution of (\ref{eq:rr1t}) in (\ref{eq:vtoroid}). As expected, the corresponding
action integral vanishes

\begin{eqnarray*}
\phi_R &=& \int \limits_{-\infty}^{\infty} V(t) dt = \int
\limits_{-\infty}^{\infty}  dt \left(-\frac{q_1 N_2 v_{1y}}{4 c (R^2
+ v_{1y}^2t^2 )^{3/2}} + \frac{3q_1 N_2 v_{1y}^3t^2 }{4 c (R^2 +
v_{1y}^2t^2 )^{5/2}}\right) \\
&=& 0
\end{eqnarray*}

\noindent Comparing this result with (\ref{eq:ph-shift}) we see
that the phase difference for the two paths (inside and outside the torus) is

\begin{eqnarray*}
\Delta \phi = \frac{1}{\hbar}(\phi_0 - \phi_R) = \frac{e
\mu}{\hbar c}
\end{eqnarray*}

\noindent  This is the same result as in the case of infinite linear solenoid
(\ref{eq:phase-shift}). Note that this phase shift does not depend
on the radius of the magnet $a$ and on the charge's velocity
$v_{1y}$. This is in full agreement with Tonomura's experiments
\cite{Tonomura-86, Tonomura-86a}.

\section{Fast moving charges and radiation}
\label{sc:em-radiation}

In subsection \ref{sc:two-charges} we calculated forces (\ref{eq:d2r1/dt2}) - (\ref{eq:d2r2/dt2}) acting between two charges in relative motion. These formulas were approximate as they included only terms of order $(v/c)^2$ and lower. So, they could not be applied to situations in which charges move with high velocities comparable to the speed of light.
Moreover, we tacitly assumed that accelerations of our charges were low, so that the 3rd order interaction (\ref{eq:ham-posit}) could be ignored. By doing so, we neglected the possibility of the photon emission by the interacting charges.

In this section we will try to fill these gaps and discuss (albeit only qualitatively) RQD effects associated with high velocities and accelerations of charges. We will compare these effects with those predicted by the standard Maxwell's theory. In the next chapter we will see how these differences can be observed in experiments.

\begin{figure}
\centering
 \includegraphics {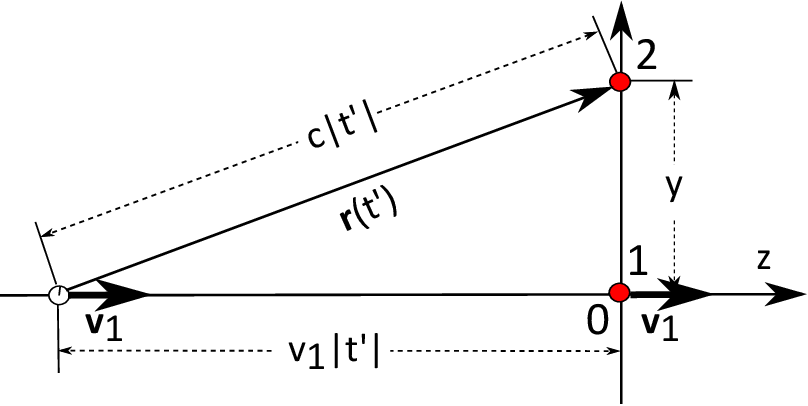} \caption{For calculation of RQD ``fields'' in (\ref{eq:exmy}) - (\ref{eq:ezmy}) and  Li\'enard-Wiechert fields in (\ref{eq:lienard}). Full circles mark positions of particles 1 and 2 at time $t=0$. The open circle marks position of the particle 1 at an earlier time $t'$.} \label{fig:lienard}
\end{figure}

\subsection{Fast moving charge in RQD}
\label{ss:fast-charge-RQD}

Let us now derive the RQD inter-particle potential beyond the $(v/c)^2$ approximation. We will be interested in a specific setup in which charge $q_1$ is moving with a high constant velocity $v_1 \approx c$ and  momentum $p_1 \gg m_1c$ along the $z$-axis, while the charge 2 is resting at the distance $y$ from the beam line, as shown in fig. \ref{fig:lienard}. For simplicity, we will choose our axes in such a way that the point $z_1=0$ on the beam line corresponds to the closest approach between the two charges. Likewise, $t=0$ is the time when particle 1 passes through this point.
We will assume that charge 2 is very small ($q_2 \ll q_1$), so that its presence has no visible effect on the straight-line movement of $q_1$. Furthermore, the mass $m_2$ is taken to be infinitely large, so that in the course of our thought experiment this particle does not move ($v_2=0$). Our goal is to calculate the force $\mathbf{f}_2$ experienced by the test charge $q_2$. More precisely, we are interested in the ratio

\begin{eqnarray}
 \mathbf{e} &\equiv& \mathbf{f}_2/q_2 \label{eq:el_field}
\end{eqnarray}

\noindent which in Maxwell's electrodynamics goes by the name ``electric field''.

Let us start from evaluating the interaction energy between charges 1 and 2 beyond the $(v_1/c)^2$ approximation.  Near the energy shell we can use formula  (\ref{eq:s-oper-2nd}) for the interaction operator\footnote{We ignored spins of the two particles, took the limit $m_2 \to \infty$ and used formula (\ref{eq:12.4}), which tells us that on the energy shell $V_2^d = \Sigma_2^c = F_2^c$.}

\begin{eqnarray}
&\ &  V_{2}^d  \nonumber \\
&\approx&
 -\frac{q_1q_2\hbar^2c^2}{(2 \pi \hbar)^3} \int
\frac{d\mathbf{k} d\mathbf{p}_2 d\mathbf{p}_1 m_1 c^2}
{\sqrt{\omega_{\mathbf{p}_1}\omega_{\mathbf{p}_1+ \mathbf{k}}}} \frac{W_{\mu}(\mathbf{p}_2-\mathbf{k}; \mathbf{p}_2) U^{\mu}(\mathbf{p}_1+ \mathbf{k};\mathbf{p}_1)}{(\omega_{\mathbf{p}_1} - \omega_{\mathbf{p}_1+ \mathbf{k}})^2 -  c^2k^2} \times \nonumber \\
&\mbox{ } &      d^{\dag}_{\mathbf{p}_2- \mathbf{k}}
a^{\dag}_{\mathbf{p}_1+ \mathbf{k}} d_{\mathbf{p}_2}
a_{\mathbf{p}_1} \nonumber \\
&\approx&
 -\frac{q_1q_2\hbar^2c^2}{(2 \pi \hbar)^3} \int
\frac{d\mathbf{k} d\mathbf{p}_2 d\mathbf{p}_1 m_1 c^2}
{\sqrt{\omega_{\mathbf{p}_1}\omega_{\mathbf{p}_1+ \mathbf{k}}}} \frac{ U^{0}(\mathbf{p}_1+ \mathbf{k};\mathbf{p}_1)}{(\omega_{\mathbf{p}_1} - \omega_{\mathbf{p}_1+ \mathbf{k}})^2 -  c^2k^2}       d^{\dag}_{\mathbf{p}_2- \mathbf{k}}
a^{\dag}_{\mathbf{p}_1+ \mathbf{k}} d_{\mathbf{p}_2}
a_{\mathbf{p}_1} \nonumber \\
&\equiv&  \int d\mathbf{k} d\mathbf{p}_2 d\mathbf{p}_1
w_{2}^d(\mathbf{p}_1+ \mathbf{k}, \mathbf{p}_1, \mathbf{k})       d^{\dag}_{\mathbf{p}_2- \mathbf{k}}
a^{\dag}_{\mathbf{p}_1+ \mathbf{k}} d_{\mathbf{p}_2}
a_{\mathbf{p}_1}
\label{eq:12.19a}
\end{eqnarray}

\noindent According to subsection \ref{ss:2-particle}, the position-space representation of this potential can be obtained by Fourier-transforming the coefficient function $w_{2}^d(\mathbf{p}_1+ \mathbf{k}, \mathbf{p}_1, \mathbf{k})$ in (\ref{eq:12.19a}). Since we are interested only in the long-range component of our interaction, the relevant integration range is around $|\mathbf{k}|=0$. So, we will assume $k \ll p_1$ and $\omega_{\mathbf{p}} \approx cp$.
Then from (\ref{eq:8.48b}) and (\ref{eq:A.69c}) we obtain

\begin{eqnarray}
&\mbox{ }& \frac{ m_1 c^2 }
{\sqrt{\omega_{\mathbf{p}_1}\omega_{\mathbf{p}_1+ \mathbf{k}}}} \approx \frac{m_1c}{p_1}   \nonumber \\ \nonumber \\
&\mbox{ }& U^0(\mathbf{p}_1+ \mathbf{k}; \mathbf{p}_1)   = \Bigl(\sqrt{\omega_{\mathbf{p}_1+ \mathbf{k}} + m_1c^2}
\sqrt{\omega_{\mathbf{p}_1} + m_1c^2} \nonumber \\
&+& \sqrt{\omega_{\mathbf{p}_1+ \mathbf{k}} - m_1c^2} \sqrt{\omega_{\mathbf{p}_1} - m_1c^2} \frac{(\mathbf{p}_1+ \mathbf{k}) \cdot \mathbf{p}_1 }{|\mathbf{p}_1+ \mathbf{k}|p_1} \Bigr)  \frac{1}{2 m_1 c^2}  \approx  \frac{p_1}{m_1c} \nonumber \\ \nonumber \\
&\ & w_{2}^d(\mathbf{p}_1+ \mathbf{k}, \mathbf{p}_1, \mathbf{k}) \approx
- \frac{q_1q_2\hbar^2}{(2 \pi \hbar)^3} \cdot \frac{c^2}{(\omega_{\mathbf{p}_1} - \omega_{\mathbf{p}_1+ \mathbf{k}})^2 -  c^2k_x^2-c^2k_y^2 -c^2k_z^2}  \nonumber \\
 \label{eq:v2ap1k}
\end{eqnarray}

\noindent The non-negative expression $\Omega(k_x, k_y, k_z) \equiv (\omega_{\mathbf{p}_1} - \omega_{\mathbf{p}_1+ \mathbf{k}})^2$ in the denominator is a function, which vanishes at $k_x= k_y= k_z=0$ and has zero first derivatives there\footnote{We took into account that $p_{1x} = p_{1y} = 0$.}

\begin{eqnarray*}
\frac{\partial \Omega}{\partial k_y} \Bigr|_{\mathbf{k} = 0} &=& -2 (\omega_{\mathbf{p}_1} - \omega_{\mathbf{p}_1+ \mathbf{k}})\frac{c^2 k_y}{\omega_{\mathbf{p}_1+ \mathbf{k}}}\Bigr|_{\mathbf{k} = 0} = 0 \\
\frac{\partial \Omega}{\partial k_z} \Bigr|_{\mathbf{k} = 0} &=& -2 (\omega_{\mathbf{p}_1} - \omega_{\mathbf{p}_1+ \mathbf{k}})\frac{c^2 (p_{1z} -k_z)}{\omega_{\mathbf{p}_1+ \mathbf{k}}} \Bigr|_{\mathbf{k} = 0} = 0
\end{eqnarray*}

\noindent For second derivatives we obtain

\begin{eqnarray*}
\frac{\partial^2 \Omega}{\partial k_z^2} \Bigr|_{\mathbf{k} = 0} &=& -2 \frac{c^2 (p_{1z} -k_z)}{\omega_{\mathbf{p}_1+ \mathbf{k}}} \frac{\partial }{\partial k_z}(\omega_{\mathbf{p}_1} - \omega_{\mathbf{p}_1+ \mathbf{k}})\Bigr|_{\mathbf{k} = 0}  \\
&=& 2 \frac{c^2 (p_{1z} -k_z)}{\omega_{\mathbf{p}_1+ \mathbf{k}}}\frac{c^2 (p_{1z} -k_z)}{\omega_{\mathbf{p}_1+ \mathbf{k}}}\Bigr|_{\mathbf{k} = 0} = \frac{2c^4 p_{1z}^2}{\omega^2_{\mathbf{p}_1}} = 2v_{1z}^2 \\
\frac{\partial^2 \Omega}{\partial k_y^2} \Bigr|_{\mathbf{k} = 0} &=&
\frac{\partial^2 \Omega}{\partial k_y \partial k_z} \Bigr|_{\mathbf{k} = 0} =  0
\end{eqnarray*}

\noindent Then the Taylor expansion around $\mathbf{k} = 0$ yields
$\Omega(k_x, k_y, k_z) \approx k_z^2 v_{1}^2 $. Substituting this expression in (\ref{eq:v2ap1k}), we obtain

\begin{eqnarray*}
w_{2}^d(\mathbf{p}_1+ \mathbf{k}, \mathbf{p}_1, \mathbf{k}) &\approx&
 \frac{q_1q_2\hbar^2}{(2 \pi \hbar)^3} \cdot \frac{1}
{k_x^2 + k_y^2 + k_z^2(1 - v_{1}^2/c^2 )}
\end{eqnarray*}

\noindent and the position-space potential is

\begin{eqnarray}
w_{2}^d(\mathbf{p}_1, \mathbf{r}) &= & \int d \mathbf{k} w_{2}(\mathbf{p}_1+ \mathbf{k}, \mathbf{p}_1, \mathbf{k}) e^{\frac{i}{\hbar}\mathbf{kr}} = \frac{q_1q_2\hbar^2}{(2 \pi \hbar)^3} \int d \mathbf{k} \frac{e^{\frac{i}{\hbar}\mathbf{kr}}}
{k_x^2 + k_y^2 + k_z^2/ \gamma^2 } \nonumber \\
&=&
\frac{q_1q_2 \gamma }{4 \pi \sqrt{x^2 + y^2 + \gamma^2 z^2}} \label{eq:v2pot}
\end{eqnarray}

\noindent where we defined $\gamma \equiv 1/\sqrt{1 - v_1^2/c^2} \gg 1 $ and $\mathbf{r} \equiv \mathbf{r}_2 - \mathbf{r}_1$.

Formula (\ref{eq:v2pot}) is the potential energy of interaction between charges 1 and 2. Within our approximations, the full Hamiltonian can be written as

\begin{eqnarray*}
H \approx m_2c^2 + cp_1 + \frac{q_1q_2 \gamma }{4 \pi \sqrt{x^2 + y^2 + \gamma^2 z^2}}
\end{eqnarray*}

\noindent The force acting on the particle 2 is

\begin{eqnarray*}
\mathbf{f}_2 \equiv m_2 \frac{d \mathbf{r}_2}{d t^2} =  \frac{d \mathbf{p}_2}{dt} = -\frac{\partial H}{ \partial \mathbf{r}_2}
\end{eqnarray*}

\noindent and the ``electric field'' (\ref{eq:el_field}) at $t=0$ can be obtained as the gradient of the potential (\ref{eq:v2pot})

\begin{eqnarray}
e_x^{(\gamma)}(x,y,z) &=& -\frac{1}{q_2} \cdot \frac{ \partial w_{2}^d}{\partial x_2}
=  \frac{q_1 \gamma x}{4 \pi (x^2 + y^2 + \gamma^2 z^2)^{3/2}} \label{eq:exmy} \\
e_y^{(\gamma)}(x,y,z) &=& -\frac{1}{q_2} \cdot \frac{ \partial w_{2}^d}{\partial y_2}
= \frac{q_1 \gamma y}{4 \pi (x^2 + y^2 + \gamma^2 z^2)^{3/2}} \label{eq:eymy} \\
e_z^{(\gamma)}(x,y,z) &=& -\frac{1}{q_2} \cdot \frac{ \partial w_{2}^d}{\partial z_2}
= \frac{q_1 \gamma^3 z}{4 \pi (x^2 + y^2 + \gamma^2 z^2)^{3/2}} \label{eq:ezmy}
\end{eqnarray}

\noindent It is interesting to compare this field with the one produced by a charge at rest ($\gamma = 1$)

\begin{eqnarray}
\mathbf{e}^{(\gamma = 1)}(x,y,z) &=&
\frac{q_1 \mathbf{r}}{4 \pi(x^2 + y^2 + z^2)^{3/2}} \label{eq:e-zero}
\end{eqnarray}

\begin{figure}
\centering
 \includegraphics {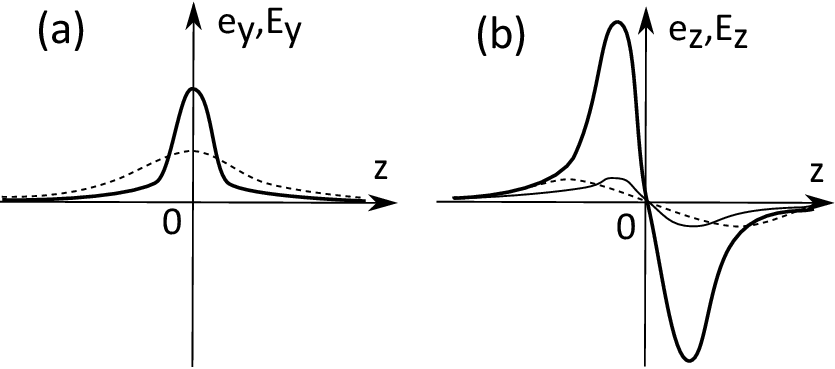} \caption{Schematic ``electric field'' profiles along $z$-direction (with coordinates $x=0, y>0$ fixed) for a charge moving with velocity $v=c \sqrt{\gamma^2-1}/\gamma$ along the $z$-axis. The profiles are taken at time $t=0$ when the charge is located in the origin (see Fig. \ref{fig:lienard}). Broken line - charge at rest ($\gamma = 0$); thick line - RQD ``electric field'' $\mathbf{e}$ for a moving charge ($\gamma = 2$); thin full line - field $\mathbf{E}$ for the moving charge ($\gamma = 2$) in Maxwell's theory: (a) transversal field components $e_y$ and $E_y$ coincide for all $\gamma$; (b) longitudinal field components $|e_z| > |E_z|$.} \label{fig:field}
\end{figure}

\noindent  For $x=0$ and fixed value $y>0$ we plotted $e_y$- and $e_z$-components of (\ref{eq:e-zero}) as functions of $z$ in Fig. \ref{fig:field}. They are shown by broken lines. Field components for the moving charge\footnote{See equations (\ref{eq:eymy}) and (\ref{eq:ezmy}), where the $\gamma = 2$ was chosen as an example.} are shown there by thick full lines. The effect of the charge's velocity is twofold: First, the field profile gets squeezed towards the charge's position ($z_1=0$). Second, the peak magnitude of the field increases. This means that the electric field configuration around a fast-moving charge 2 is concentrated in a narrow disk perpendicular to the direction of motion. This disk moves together with the charge as if it was rigidly attached to the instantaneous charge's position.

\subsection{Fast moving charge in Maxwell's electrodynamics}
\label{ss:fast-charge-CED}

In the preceding subsection we used our RQD formalism to find the ``electric field'' generated by a fast-moving charge. Let us now see how the same problem is solved in classical Maxwell's electrodynamics.

The standard derivation\footnote{see \cite{Ibison, Carlip} and section 14.1 in \cite{Jackson}} involves the concept of retarded Li\'enard-Wiechert fields. \index{Li\'enard-Wiechert fields} The idea is that electric (and magnetic) fields  are not rigidly attached to the moving charge. They radially spread around the charge with the speed equal to the speed of light. So, the total field around the charge is not determined by the charge's instantaneous position. It is rather a function of previous locations of the particle. The formula for the Li\'enard-Wiechert field produced by the uniformly moving charge 1 at the point 2 at time $t=0$  is\footnote{See equation (14.14) in \cite{Jackson}. The Li\'enard-Wiechert field is denoted by the capital $\mathbf{E}$ in order to distinguish it from the ``electric field'' $\mathbf{e}$ (\ref{eq:exmy}) - (\ref{eq:ezmy}) predicted in our theory.}

\begin{eqnarray}
\mathbf{E}(\mathbf{r}_2) = \frac{q_1}{4 \pi}\cdot \frac{\mathbf{r}(t') - r(t')\mathbf{v}_1/c}{\gamma^2[r(t') - (\mathbf{r}(t') \cdot \mathbf{v}_1)/c]^3} \label{eq:lienard}
\end{eqnarray}

\noindent Various components in this formula are shown in Fig. \ref{fig:lienard}. In particular,  $\mathbf{r}(t') = \mathbf{r}_2 - \mathbf{r}_1(t')$ is the vector connecting the two charges at an earlier time $t' =  - r(t')/c$. Now, let us find the time $t'$ and the charge's position $\mathbf{r}_1(t') =(0,0,z_1(t'))$ at which the Li\'enard-Wiechert field was ``emitted'', such that it reached the test particle 2 at time $t=0$. As the field propagates with the speed of light, we can write

\begin{eqnarray}
-ct' = \sqrt{y^2 + z_1^2(t')} \label{eq:1}
\end{eqnarray}

\noindent On the other hand, in the time interval $[t', 0]$ particle 1 has traveled along the $z$-axis from $z=z_1(t')$ to $z=0$. This condition yields

\begin{eqnarray}
v_1t' = z_1(t') \label{eq:2}
\end{eqnarray}

\noindent Solving the system of  equations (\ref{eq:1}) - (\ref{eq:2}) we obtain

\begin{eqnarray}
t' &=& -\frac{y}{\sqrt{c^2-v_1^2}} = -\frac{y \gamma}{c} \label{eq:tprime} \\
\mathbf{r}(t') &\equiv& \mathbf{r}_2 - \mathbf{r}_1(t) = \left(0, y, y \gamma\frac{v_1}{c} \right) \nonumber \\
r(t') &=& y \sqrt{1 + \gamma^2 \frac{v_1^2}{c^2}} =  y \gamma \label{eq:rtprime}
\end{eqnarray}

\noindent Using these results in the Li\'enard-Wiechert formula (\ref{eq:lienard}) we find electric field components at time $t=0$\footnote{For a detailed derivation see references quoted in \cite{Calcaterra}, e.g.,  section 14.1 in \cite{Jackson}. Note that this is the same result as the one obtained by  Lorentz-transforming field components (\ref{eq:e-zero}) to the moving frame (see section 11.10 in \cite{Jackson}). However, the method of Lorentz transformations is questionable, because, as we explained in section \ref{ss:fields-disc}, standard special-relativistic formulas for such transformations are valid only in the absence of interactions and cannot be used for transforming forces between particles (=electric fields).}

\begin{eqnarray}
E_x(x,y,z) &=&
\frac{q_1\gamma x }{4 \pi(x^2  + y^2 + \gamma^2 z^2)^{3/2}} \label{eq:Ex} \\
E_y(x,y,z) &=&
\frac{q_1\gamma y}{4 \pi(x^2 + y^2 + \gamma^2 z^2)^{3/2}} \label{eq:Ey} \\
E_z(x,y,z) &=&
\frac{q_1\gamma z }{4 \pi(x^2  + y^2 + \gamma^2 z^2)^{3/2}} \label{eq:Ez}
\end{eqnarray}

\noindent Field components $E_x$ and $E_y$ perpendicular to the direction of motion are exactly the same as our results (\ref{eq:exmy}) - (\ref{eq:eymy}). But the parallel component  (\ref{eq:Ez}) is $\gamma^2$ time smaller than (\ref{eq:ezmy}). This component is shown by the thin full line in Fig. \ref{fig:field}(b).

\subsection{Kislev-Vaidman ``paradox''} \label{ss:kislev}

RQD predicts that electric and magnetic forces between charged particles propagate instantaneously. In chapter \ref{ch:support} we will discuss a number of experiments indicating that this is more accurate representation of electromagnetic interactions than the traditional retarded Li\'enard-Wiechert potentials. In section \ref{ss:fields-disc} we will see that action-at-a-distance can be consistent with the principle of causality.

\begin{figure}
\centering
 \includegraphics {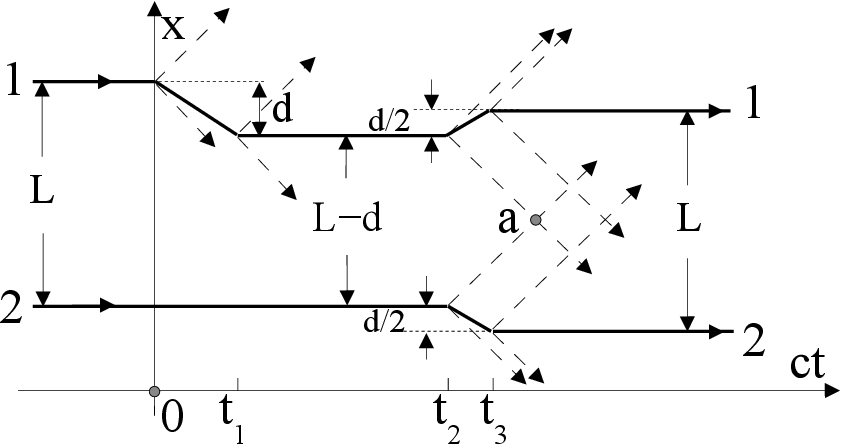} \caption{ Movements of two charged particles (full bold lines)
 in the Kislev-Vaidman paradox plotted on the $t-x$ plane. The time on the horizontal axis is multiplied
 by $c$, so that photon trajectories (dashed arrows) are at $45^{\circ}$ angles.}
\label{fig:8}
\end{figure}

Additional support for these ideas is provided by the remarkable paradox \cite{Kislev} associated with the
assumption of retarded interactions in standard Maxwell's
electrodynamics.\footnote{A number of related paradoxes were
discussed also in \cite{Engelhardt2, Kholmetskii-2004, Kholmetskii-hidden,
Kholmetskii-2006}.} Consider two particles 1 and 2 both having the
unit charge. Let us assume that their electromagnetic interaction is
transmitted by retarded potentials and that the movement of both
particles is confined on the $x$ axis. Let us now force the two particles to move along certain prescribed paths  plotted in Fig. \ref{fig:8} by full thick lines. Initially (at times
$t<0$) both particles are kept at rest with the distance $L$ between
them. The Coulomb interaction energy is $1/(4 \pi L)$. At time $t=0$
we apply external force which displaces particle 1 by the distance
$d < L$ toward the particle 2. The work performed by this force will
be denoted $W_1$

\begin{eqnarray*}
4 \pi W_1 = \frac{1}{L-d} - \frac{1}{L}
\end{eqnarray*}

\noindent  Then we wait\footnote{The displacement of the charge 1
and its acceleration results in emission of electromagnetic
radiation (indicated by dashed arrows in Fig. \ref{fig:8}). So, we would need to wait for a sufficiently long time until the emitted
photons propagated far enough, so that they do not have any effect on our two-charge
system anymore. } until time $t_2$ and move both particles
simultaneously by the distance $d/2$ away from each other. If we make
this move rapidly during a short time interval $(t_3-t_2) < (L-d)/c$,
then the retarded field of the particle 2 in the vicinity of the
particle 1 remains unperturbed as if the particle 2 has not been
moved at all. The same is true for the field of the particle 1 in
the vicinity of the particle 2. Therefore the work performed by such
a move is

\begin{eqnarray*}
4 \pi W_2 &=& 2\left(\frac{1}{L-d/2} - \frac{1}{L-d} \right)
\end{eqnarray*}

\noindent The total work performed in these two steps is nonzero

\begin{eqnarray}
&\ &4 \pi(W_1 + W_2) \nonumber \\
&=& \frac{1}{L-d} - \frac{1}{L} +
\frac{2}{L-d/2} -
\frac{2}{L-d} \nonumber \\
&\approx&  \frac{1}{L(1-d/L)} - \frac{1}{L} + \frac{2}{2(1-d/(2L))}
-\frac{2}{L(1-d/L)} \nonumber \\
&\approx& \frac{1}{L} \left(1+\frac{d}{L} + \frac{d^2}{L^2}\right) -
\frac{1}{L} + \frac{2}{L}\left(1 + \frac{d}{2L} +
\frac{d^2}{4L^2}\right) -
\frac{2}{L}\left(1+\frac{d}{L} + \frac{d^2}{L^2}\right) \nonumber \\
&=& \frac{1}{L}+\frac{d}{L^2} + \frac{d^2}{L^3} - \frac{1}{L} +
\frac{2}{L} + \frac{d}{L^2} + \frac{d^2}{2L^3} -
\frac{2}{L}-\frac{2d}{L^2} - \frac{2d^2}{L^3} \nonumber \\
&=&    - \frac{d^2}{2L^3} \label{eq:gain}
\end{eqnarray}

\noindent This means that after the shifts are completed we find both
charges in the same configuration as before (at rest and separated
by the distance $L$), however we gained some amount of energy
(\ref{eq:gain}). Of course, the balance of energy (\ref{eq:gain}) is
not complete. It does not include the energy of photons emitted by
accelerated charges.\footnote{According to
Larmor's formula, the energy of emitted photons is proportional to the square of acceleration of the charges. See subsection \ref{ss:perturbation2}.}  However, one could, in principle, recapture
this emitted energy by surrounding the pair of particles by
appropriate photon absorbers and redirect the captured energy to
perform the work of moving the charges again. Then, it would become
possible to build a \emph{perpetuum mobile} machine in which the two
steps described above are repeated indefinitely and each time some amount of
energy (\ref{eq:gain}) is gained.

The following explanation of this paradox was suggested by Kislev
and Vaidman \cite{Kislev}: They claim that there is another energy
term missed in the above analysis, which is related to the
interference of electromagnetic waves emitted by the two
particles\footnote{For example, in Fig. \ref{fig:8} electromagnetic
waves emitted by the two charges meet at point $a$, and the
interference of the waves proceeds from that time on.} and which restores the
energy balance. This explanation does not look plausible, because
there is actually no interaction energy associated with
\index{interference} the interference of light waves: The interference
results in a redistribution of the wave's amplitude (formation of
minima and maxima) and the wave's local energy in space, while the total
energy of the wave remains unchanged \cite{Gauthier}. In other
words, there is no interaction between photons.\footnote{QED
predicts a very weak photon-photon interaction in the 4th
perturbation order, however it is negligibly small in the situation
considered here.}

The correct explanation of the Kislev-Vaidman ``paradox'' is provided
by the Darwin-Breit instantaneous action-at-a-distance theory. In the absence of
retardation of the Coulomb potential, it is easy to show that
$W_1+W_2 = 0$, and the total work performed by moving the
charges is equal to the energy of the emitted radiation.

Another reason to doubt the validity of retarded Li\'enard-Wiechert potentials is the paradoxical prediction \cite{Cornish, Griffiths2, Steane} that an isolated electric dipole can move with a constant acceleration in the absence of any external force. This paradox does not appear when charges interact instantaneously, as in RQD.

\subsection{Accelerated charges}
\label{ss:accelerate}

As we saw above, RQD ``fields''  depend on the source charge's velocity, but not on its acceleration. However, this does not mean that accelerations do not play any role in electromagnetic interactions. As we have shown in subsection \ref{ss:perturbation2}, an accelerated charge emits photons. One result of this effect is that a part of the charge's kinetic energy gets lost to radiation. This is called the \emph{radiation reaction} \index{radiation reaction} or radiation braking.

The other result is the appearance of an additional indirect interaction between charges:
An accelerated charge emits a large number of photons, which spread around with the speed of light. Some photons reach the other (receiving) charge and interact with it either by absorption\footnote{if the receiving charge is accelerating, then it can absorb photons in a process that is reverse to the bremsstrahlung emission.} or by the Compton scattering,\footnote{see subsection \ref{ss:compton}} thus causing the receiving charge to accelerate. \index{Compton scattering} This indirect transmission of interaction occurs with the speed of light, and is responsible for TV, radio, and cell-phone signal propagation through air.

\subsection{Electromagnetic fields vs. photons}
\label{ss:way-forward}

So far in this book we presented many arguments  suggesting that electrodynamics (both quantum and classical) does not need electromagnetic potentials (scalar and vector) and/or electromagnetic fields ($\mathbf{E}$ and $\mathbf{B}$). Instantaneous inter-particle forces and the laws of quantum mechanics are quite sufficient for the description of all electromagnetic phenomena.

The same conclusion remains true with respect to properties of the free electromagnetic radiation.
In RQD we claim that electromagnetic radiation is simply a flow of a large number of point-like massless particles -- photons. The wave properties of light are manifestations of the quantum nature of individual particles, and the Huygens-Maxwell wave theory of
light is, in fact, an attempt to approximate quantum wave functions
of billions of photons by two surrogate functions
$\mathbf{E}(\mathbf{x},t)$ and $\mathbf{B}(\mathbf{x},t)$
\cite{Field-photons, Torre3, Carroll}.
As we saw in section \ref{sc:thought}, ordinary quantum theory of
photons (particles) can also explain the diffraction and
interference phenomena without involving the classical waves $\mathbf{E}$ and
$\mathbf{B}$. Moreover, our particle-based explanation works much better than the classical wave model in the limit of
low-intensity light and for the photo-electric
effect.\footnote{see subsection \ref{ss:low-intensity}} So, the
representation of electromagnetic radiation as a flow of discrete
countable quantum particles better agrees with experiment and is
more general than the field (or wave) theory of light
\cite{Field-photons, Field-photons2, Torre1, Torre2, Torre3}. This
is an invitation to reconsider the status of Maxwell's
electrodynamics:

\begin{quote}
\emph{Finally, the remark may be made, as previously pointed out by
Feynman \cite{Feynman-QED} and other authors adopting a similar
approach \cite{Quantics}, that the so called `classical wave theory
of light' developed in the early part of the 19th century by Young,
Fresnel and others \textbf{is} QM as it applies to photons
interacting with matter. Similarly, Maxwell's theory of CEM
\emph{[=Classical ElectroMagnetism]} is most economically regarded as
simply the limit of QM when the number of photons involved in a
physical measurement becomes very large. \emph{[...]} Thus
experiments performed by physicists during the last century and
even earlier, \textbf{were} QM experiments, now interpreted via the
wavefunctions of QM, but then in terms of `light waves'.
\emph{[...]} The essential and mysterious aspects of QM, as embodied
in the wavefunction (superposition, interference) were already well
known, in full mathematical detail, almost a hundred years earlier!}
J. H. Field \cite{Field-photons}
\end{quote}

\chapter{EXPERIMENTAL SUPPORT FOR RQD}
\label{ch:support}

\begin{quote}
\textit{Let a hundred flowers bloom, let a hundred schools of thought contend.}

\small
\hspace{1in} Mao Zedong
\normalsize
\end{quote}

\vspace{0.5in}

Our discussion in subsection \ref{ss:accelerate} suggests that there are two distinct kinds of forces between charged particles: One is the direct \emph{bound} \index{bound field} Coulomb or magnetic force, which is dominant when the charges are at rest or move with low accelerations.  In RQD these force fields are rigidly and permanently attached to the source charges and react immediately to any perturbation of the charges' trajectories. This is equivalent to saying that the speed of propagation of these interactions is infinite. On the other hand, in Maxwell electrodynamics the attachment of the bound fields to the charge is not rigid: The electric and magnetic forces are described by Li\'enard-Wiechert fields, which propagate with the speed of light.

The second type of force is the indirect \emph{radiation} \index{radiation field} interaction. In RQD this interaction is transmitted by photons traveling with the speed of light. In Maxwell's theory, the radiation field is represented by the familiar transverse electromagnetic wave, in which mutually perpendicular $\mathbf{E}$ and $\mathbf{B}$ vectors oscillate and travel in space with the light's speed.

Thus, in RQD the total force field produced by a group of moving charges can be written symbolically as a superposition of instantaneous and retarded components\footnote{For brevity here we consider only the ``electric'' part of the field, omitting ``magnetic'' interactions that act only on moving charges. For
similar ideas about electromagnetic interactions being composed of
both instantaneous and retarded parts see \cite{Chubykalo, Field,
Field-06, Kholmetskii-2006, Kholmetskii-102}.}

\begin{eqnarray}
\mathbf{e} = \mathbf{e}^{inst}_{bound} + \mathbf{e}^{ret}_{radiation}
\end{eqnarray}

\noindent The Li\'enard-Wiechert electric field of the Maxwell's theory is fully retarded

\begin{eqnarray}
\mathbf{E} = \mathbf{E}^{ret}_{bound} + \mathbf{E}^{ret}_{radiation}
\end{eqnarray}

\noindent In this book we will not derive the explicit form of the $ \mathbf{e}^{ret}_{radiation}$ component\footnote{As explained in subsection \ref{ss:accelerate}, its action on a test charge is a combination of the photon absorption and the Compton scattering.} We will simply assume that $ \mathbf{e}^{ret}_{radiation}=\mathbf{E}^{ret}_{radiation} $ and focus on the more interesting difference between bound fields $ \mathbf{e}^{inst}_{bound}$ and $\mathbf{E}^{ret}_{bound} $.

The infinite propagation speed of the bound electromagnetic fields is, perhaps, the most controversial prediction of our RQD approach. In this chapter we will be interested in experimental techniques that can be used to verify this prediction. As we have mentioned already, the bound fields are easily observable in static or quasistatic situations when accelerations of charges are low. However, if we want to measure field velocities, then we need to disrupt these static configurations, thus introducing charge accelerations and, inevitably, the emission of radiation fields. So, the experimental challenge is to somehow minimize the effect of the radiation fields, so that dynamical properties of the bound fields can be studied in their pure form. Several experimental approaches that can meet this challenge will be described below. For other relevant experiments see review articles \cite{Recami2009,systemized}.

\section{Relativistic electron bunches}
\label{sc:fast-charge}

In subsections \ref{ss:fast-charge-RQD} and \ref{ss:fast-charge-CED} we have calculated bound electric fields produced by a fast moving charge in RQD (\ref{eq:exmy}) - (\ref{eq:ezmy}) and in the Maxwell-Li\'enard-Wiechert theory (\ref{eq:Ex}) - (\ref{eq:Ez}), respectively. In both cases the fields have a form of a thin ``pancake'' perpendicular to the charge's velocity. However, there is one crucial difference in the field dynamics (time evolution) predicted by these two theories. Remarkably, this difference is so significant that it can be seen even in unsophisticated experimental setups.  Let us now discuss this experimental opportunity in more detail.

Fast-moving charges are routinely available as electron beams in accelerators.  For example, a bunch of 500 MeV electrons has the factor $\gamma$ as large as  $10^3$.  Then Li\'enard-Wiechert equations (\ref{eq:tprime}) - (\ref{eq:rtprime}) suggest  a peculiar electric field dynamics when such an electron bunch leaves the accelerator's pipe and enters the open space. Immediately after the bunch's emergence there is no electric field around it. The disk-shaped field builds up gradually,\footnote{Sometimes such a buildup is described as a ``field recovery'' of the ``semi-bare electron'' \cite{Naumenko1,Naumenko2}.} starting from low values of $y$ and extending to larger $y$ as the time progresses. Thus the electric field grows ``older'' as the observation point moves away from the beam line, i.e., as $y$ increases. For example, the peak electric field at the distance of $y=50$ cm from the beam's axis becomes fully formed only after the bunch has traveled  $\gamma y =500$ m away from the pipe exit point.

Unlike in the Li\'enard-Wiechert theory, in our RQD approach there is no ``field recovery'' dynamics as described above. The electric field (\ref{eq:exmy}) - (\ref{eq:ezmy}) of the bunch emerging from the accelerator is fully formed in the entire space without delay. Here we will have a chance to compare these two competing predictions with the experiment performed recently by the group of prof. Pizzella at the Frascati National Laboratory in Italy \cite{Calcaterra}.

\subsection{Experiment at Frascati}
\label{ss:frascati}

Frascati experiment used 500 MeV electron bunches ($\gamma \approx 10^3$) with $(0.5 - 5.0) \times 10^8$ electrons per pulse. The transverse $y$-component of the electric field was measured at distances $y=3, 5, 10, 20, 40$, and 55 cm from the beam line. This experiment confirmed theoretical results (\ref{eq:eymy}) or (\ref{eq:Ey}) within experimental errors.

As we discussed above, the traditional result (\ref{eq:Ey}) relied on the requirement that the electron bunch was moving uniformly long before the measurement was done. For example, for the validity of (\ref{eq:Ey}) at $y=55$ cm and $t=0$,\footnote{Here we use the same notation and assumptions as in subsections \ref{ss:fast-charge-RQD} - \ref{ss:fast-charge-CED}. In particular, $t=0$ is the time when the sensor (=charge 2) registers the maximum $e_y$ field strength.} the earlier trajectory of the bunch should be linear, uniform, and unconstrained  for at least  $(55 \  cm) \cdot \gamma \approx 550 \ m$ from the point of observation. However, this essential condition was badly violated in the experiment! In particular, a fully formed electric field was detected at the distance of only 92 $cm$ from the beam pipe exit flange.

Thus the Pizzella group's measurements were totally inconsistent with the gradual build-up of the bunch's electric field predicted by the classical Maxwell-Li\'enard-Wiechert theory. However, they were in full agreement with our RQD explanation, which maintains that the field is rigidly attached to the instantaneous position of the moving charge.

\subsection{Proposal for modified experiment}
\label{ss:frascati2}

Let us  suggest two simple modifications of the Frascati experimental setup, which may provide even more spectacular validation of RQD. First, one can change orientation of the electric field sensors so as to measure the longitudinal $z-$component of the field and compare it with our prediction (\ref{eq:eymy}). As we mentioned earlier, this quantity is expected to be $\gamma^2 \approx 10^6$ times greater than the Li\'enard-Wiechert's prediction (\ref{eq:Ez}).

\begin{figure}
\centering
 \includegraphics {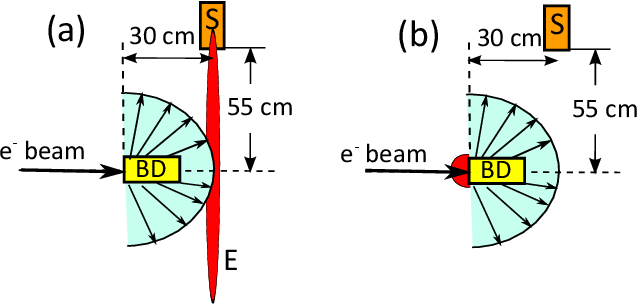} \caption{Field configurations at $t=0$, i.e., after the electron bunch was stopped by the beam dump ($BD$): (a) Maxwell's theory in which the  disk-shaped electric field $E$ of the beam has reached the sensor ($S$); (b) RQD in which the runaway disk-shaped electric field is absent (a much weaker stationary field surrounds the stopped electron bunch), and the photons emitted from the collision point have not reached the sensor yet.}  \label{fig:beam}
\end{figure}

Second, we propose to check whether the field at large transverse distances $y$ follows instantaneous positions of the moving charge, as predicted by RQD in subsection \ref{ss:fast-charge-RQD}. In order to verify this prediction, the experimentalists can stop the electron bunch abruptly by placing a lead brick (beam dump) on the beam's path, and investigate the time evolution of the electric field after such an interruption of the beam. An example of the proposed setup is shown in Fig. \ref{fig:beam}.\footnote{Coordinate axes and beam properties are the same as in Fig. \ref{fig:lienard}.} The electric field sensor\footnote{which was earlier modeled by the fixed charge 2} is placed at $(x=0, y=55 \ cm, z=0)$. The lead brick is positioned at $(x=0, y=0, z=-30 \  cm)$, so that the beam stoppage occurs at time $t_1=- 1 \ ns$, i.e., 1 ns earlier than the field maximum is supposed to reach the sensor. Next consider response of the sensor in the two theories discussed above.

In the traditional Li\'enard-Wiechert approach, the electric field of the beam will continue its motion with velocity $(0, 0, v_1)$  even after the electron beam has been interrupted. So, the sensor will register the onset of the field pulse at time $t=0$, as if the lead brick was not there.\footnote{It is also important to note that in the Li\'enard-Wiechert theory the amplitude of the $t=0$ signal must be the same independent on the presence or absence of the beam dump.  This prediction was \emph{not} confirmed by the Frascati experiment, which showed a significant reduction of the signal in the presence of the beam dump. See Fig. 15 in \cite{Calcaterra}.} The beam's collision with the lead brick will also result in formation of a burst of electromagnetic radiation $\mathbf{E}^{ret}_{radiation}$, which will propagate radially with the speed of light, as shown in the figure. The distance between the sensor and the collision point is  $R = \sqrt{y^2 + z^2} = \sqrt{(55 \ cm)^2 + (30 \ cm)^2} \approx 64 \ cm$. Therefore, the electromagnetic pulse will reach the sensor  at time $t_2 = t_1 + R/c = -1 \  ns + (64 \  cm)/c \approx 1 \ ns$, i.e., 1 ns later than the signal onset.

In our RQD approach, the field configuration does not depend on the previous history of the beam.  So, after the electron bunch is stopped, its field suddenly transforms into a spherically symmetric shape (\ref{eq:e-zero}), characteristic for a charge at rest. At the same time, the field strength reduces by the factor of $\gamma \approx 1000$, i.e., it weakens below detector's sensitivity. So, in contrast to the traditional theory described above, we are not expecting to see any sensor response at $t=0$. Formation of the bremsstrahlung photon pulse $\mathbf{e}^{ret}_{radiation}$ upon the beam-brick collision will proceed as described above, and this signal will reach the sensor at time $t_2 \approx 1 \ ns$.

In short, the two theories predict quite different timings of initial sensor responses: In Maxwell's theory, the signal onset will occur at $t=0$, while in our approach the first signal will reach the sensor at $t\approx 1$ ns. The expected time difference of about 1 ns should be easily detectable by the available experimental equipment.

\section{Radiation and bound fields}
\label{ss:near-field}

Experimental studies of the electromagnetic field propagation normally involve two antennas: the \emph{emitter} \index{emitting antenna} and the \emph{receiver}. \index{receiving antenna} The signal in the receiver is a combined effect of both the bound and radiation fields produced by the emitter. Separation of these two effects is a challenging experimental task, but it is still possible due to their different physical properties.

According to RQD, the radiation signal must be proportional to the number of photons reaching the receiver.  In the simplest spherically symmetric case, this number drops with the distance as $e_{radiation}^{ret} \propto r^{-2}$. A different distance dependence can be expected for the bound field component $\mathbf{e}^{inst}_{bound}$.
Regarded as a source of the bound field, a simple antenna can be approximated by a time-varying electric dipole. Then at large distances we should expect the electric field to behave like $e_{bound}^{inst} \propto r^{-3}$. This  suggests that at large distances  the bound field effect is overwhelmed by the radiation effect. So, in order to detect superluminal bound fields one should either look very close to the emitter, e.g., in its near-field zone, or try to reshape the photon flow, so that certain regions of space are free from radiation,  and the weaker bound fields can be detected there.

A few experiments, which succeeded in measuring the bound field velocity, will be discussed in this section.

\subsection{Near field studies}
\label{ss:near-field2}

A remarkable study of the electromagnetic field propagation was performed by Kholmetskii and coworkers \cite{Kholmetskii, Kholmetskii-102, Kholmetskii-2011}. They used the classical Hertz's setup with two antennas. The emitter ($E$) antenna produced a short pulse of electromagnetic radiation in the surrounding space. The receiver ($R$) antenna was placed at a variable distance $r$ from $E$, and time-resolved voltage waveforms were recorded at $R$. The authors have demonstrated that at large separations (the distance $r$ was up to 3m) the signal in $R$ was dominated by the radiation field with the distance dependence $\propto r^{-2}$. In the near-field zone ($r<50 cm$) the signals from both bound and radiation fields were mixed with the prevalence of the bound ($\propto r^{-3}$) field.  Assuming that the radiation field propagated with the constant speed $c$ within the entire range of $r$ values, the authors were able to subtract the radiation field contribution from the total signal for all $r$, thus obtaining pure bound field waveforms. Analyzing these waveforms, they estimated the propagation speed of the bound component. In the near-field zone this value appeared much higher than the speed of light (or even infinite).\footnote{These issues have been discussed also in \cite{superluminal5, Budko, BudkoPRL}.} This result fully agreed with the RQD idea about the instantaneous propagation of  bound fields.

\subsection{Microwave horn antennas}
\label{ss:evanescent-micro}

\begin{figure}
\centering
\includegraphics {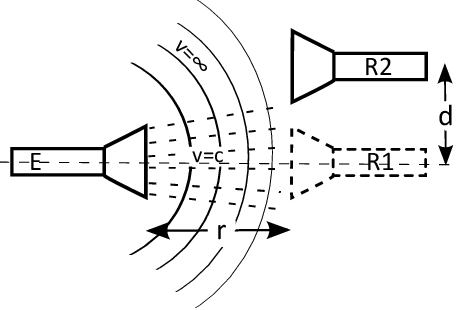} \caption {Sketch of the experiment with microwave horn antennas and its interpretation. Broken lines indicate the flux of photons concentrated along the emitter's ($E$) axis . Half-circles represent bound electric and magnetic fields that propagate instantaneously.} \label{fig:horn}
\end{figure}

Perhaps the first convincing experimental observation of the superluminal character of bound electromagnetic fields was performed by Giakos and Ishii in 1991 \cite{Giakos, Giakos-91}. They studied the propagation of microwave pulses between two horn antennas arranged as shown in Fig. \ref{fig:horn}. In the first run of this experiment the emitter ($E$) and the receiver ($R1$) antennas were placed face to face and separated by a distance of $r=$71.5 cm. The signal crossed this air gap in 2.378 ns, which corresponded to the propagation velocity of 3.01 $\times 10^8$ m/s, i.e., the speed of light, as expected. In the second run the receiver ($R2$) was shifted away from the emitter's axis (with or without tilting towards the emitter). For shift distances as large as $d=$34 cm the transit time remained nearly constant despite substantial increase of the $E-R2$ separation. Thus, signal propagation velocities were observed as high as  3.32 $\times 10^8$ m/s, i.e., 10\% higher than the speed of light.

These results were later confirmed in a set of experiments performed by Ranfagni and coworkers \cite{Ranfagni, superluminal3, Mugnai}. Wide ranges of microwave frequencies and $E-R$ separations were explored.

These observations are fully consistent with the following RQD narrative: The emitting horn antenna generates both bound and radiation electromagnetic fields. Due to the specific $E$ antenna horn shape, the beam of photons (=the radiation field) is concentrated near the antenna axis. On the other hand, the bound field is more diffuse and short-range. When the receiver is placed on the emitter's axis, the signal is dominated by microwave photons, and the apparent signal velocity is close to $c$. When the receiver is displaced away from the axis, the photon contribution decreases, and a more prominent role is played by the bound instantaneous field. Thus the effective signal propagation speed tends to increase. The radiation field still dominates the signal, so the effective speed exceeds $c$ by only few percents. If it were possible to shut off the radiation field completely, we would see an infinite propagation speed.

Such an elimination of the radiation component is, actually, possible in another kind of experiment that we would like to discuss in the next subsection.

\subsection{Frustrated total internal reflection}
\label{ss:evanescent}

\begin{figure}
\centering
\includegraphics {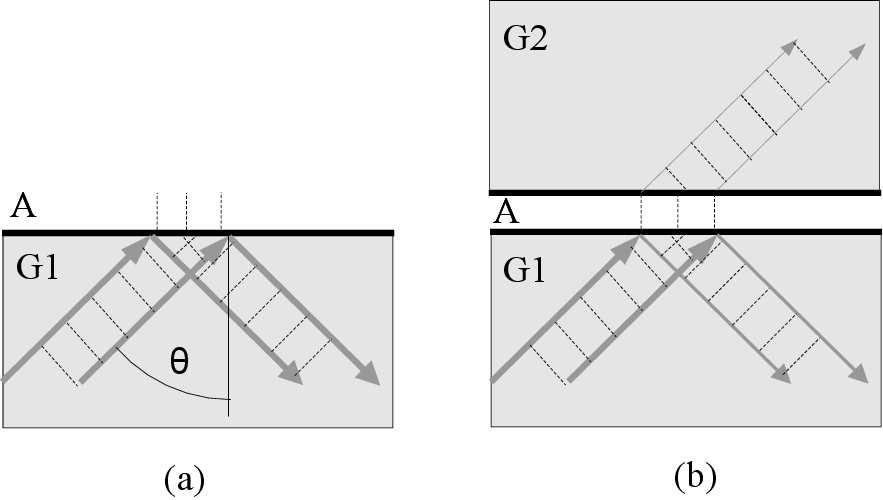} \caption {A beam of light impinging on
the glass-air interface. (a) If the incidence angle $\theta$ is
greater than the Brewster angle, then all light is reflected at the
surface. The region of evanescent light is shown by vertical dashed
lines. (b) If a second piece of glass is placed near the interface,
then ``evanescent light'' is converted to the ``normal light'' propagating into
the second piece of glass (G2).} \label{fig:12.4}
\end{figure}

 Consider a beam of light directed
from the glass side on the interface between glass (G1) and air (A)
(see Fig. \ref{fig:12.4}(a)). The \emph{total internal reflection}
\index{total internal reflection} occurs when the incidence angle
$\theta$ is greater than the \emph{Brewster angle}. \index{Brewster
angle} In such cases, all light is reflected at the interface, and no
radiation leaks into the air. The radiation field is blocked by the interface completely.

However, a rather surprising effect occurs if another piece
of glass (G2) is placed close to the interface (see Fig.
\ref{fig:12.4}(b)). In this case, some part of the initial light beam does cross the air gap and penetrates the piece of glass G2.  At the same time the intensity of the reflected light decreases. The total internal reflection becomes
\emph{frustrated}, and the phenomenon described above is called the
\emph{frustrated total internal reflection} (FTIR).
\index{frustrated total internal reflection}

 In recent experiments
 \cite{superluminal1, FTIR, Balcou} the speed of light's propagation
across the air gap (A) was investigated, and there were strong indications that
this speed may be superluminal.\footnote{There are discrepancies in
data interpretations by different groups (see,
for example, \cite{Reiten1, Reiten2}) and the
question remains open whether or not the speed of the signal crossing the gap may
exceed $c$.}

The usual explanation of this effect is that the air gap serves as a barrier for the photon propagation, and the gap crossing is an example of the well-known quantum tunneling effect, which is widely believed to be superluminal.\footnote{This is known as the \emph{Hartman effect} \index{Hartman effect} \cite{Hartman}. For a recent observation of superluminal electron tunneling see \cite{Eckle}.} We suggest another (perhaps, not an alternative but a complementary) explanation:
Being excited by the light wave, the
charged particles (electrons and nuclei) at the interface G1 - A
oscillate. These oscillations give rise to variable dipole moments at the
interface. The dipoles generate bound electric and magnetic fields in the gap,\footnote{they are also called \emph{evanescent waves} \index{evanescent waves}} which follow the dipole dynamics instantaneously. The amplitude of the bound fields decreases exponentially with the distance from the G1 - A interface. So, when the size of the gap is sufficiently small, these fields
 affect charged particles on the other side of the gap forcing them to oscillate and to emit
photons. These newly created photons propagate inside the piece of class G2 in the form of
a ``normal'' light beam. In our interpretation, the evanescent wave
in the gap is equivalent to instantaneous Coulomb and magnetic
forces acting between oscillating charges on the two interfaces. Although no real photons cross the air gap A, the apparent travel time of the light pulse does not depend on the size of the gap \cite{Nimtz,Aichmann}, i.e., the  signal transmission across the gap occurs superluminaly.

\chapter{PARTICLES AND RELATIVITY}
\label{sc:obs-interact}

\begin{quote}
\textit{How often have I said to you that when you have eliminated the
impossible, whatever remains, \emph{however improbable}, must be the
truth?}

\small
\hspace{1in} Sherlock Holmes
\normalsize
\end{quote}

\vspace{0.5in}

In the preceding chapters of this book we constructed a dressed particle version of
quantum electrodynamics which we called \emph{relativistic quantum
dynamics}  or RQD. One
important property of RQD was that this theory reproduces exactly the
$S$-matrix of the standard renormalized quantum electrodynamics.
Therefore, RQD can describe existing experiments (e.g., scattering
cross-sections, bound state energies, and lifetimes) just as well as QED.
However RQD is fundamentally different from QED. The main
ingredients of RQD are particles (not fields) that interact with
each other via instantaneous potentials.
The usual attitude toward
such a theory is that it cannot be mathematically and physically
consistent \cite{Strocchi, Halvorson, Wallace2, Wilczek, Hobson}.

In this chapter we are going to delve into more philosophical aspects of RQD. In particular, we will try to reassure the reader that this particle-based theory with instantaneous interactions still obeys the principles of relativity and causality. We will also argue that our theory provides a more accurate representation of relativistic phenomena than the traditional Minkowski space-time model.
One type
of objections against particle-based theories is related to the
alleged incompatibility between the existence of localized particle states and
principles of relativity and causality. We will analyze these
objections in section \ref{sc:localizability} and demonstrate that
there is no reason for concern: the Newton-Wigner position operator
and sharply localized particle states do not contradict any
fundamental physical principle. In particular, we will analyze the well-known
paradox of superluminal spreading of localized wave packets.

In
section \ref{sc:new-approach}  we will define the notion of a
localized physical \emph{event} and attempt to derive transformations of space-time coordinates of such events between different reference frames. We will notice that spatial
translations and rotations induce kinematical transformations of
observables, but translations in time are always
dynamical (i.e., they depend on interactions). Then
 boost transformations must be dynamical as well.
This implies first that interactions are governed by the instant
form of dynamics, and second that the connection between space and
time coordinates of events in different moving reference frames are
generally different from Lorentz transformations of special
relativity.

In section \ref{ss:fields-disc} we will conclude that
Minkowski space-time picture is not an accurate representation of
the principle of relativity. We will also dispel another misconception about the alleged incompatibility between instantaneous action-at-a-distance and causality. We will see that in some cases superluminal effects may not violate causality and may be physically acceptable.

Section
\ref{ss:are-fields-meas} is devoted to more philosophical
speculations on the role of quantum fields and their interpretation.

\section{Localizability of particles}
\label{sc:localizability}

In section \ref{sc:spin} we found that in relativistic quantum
theory particle positions are described by the Newton-Wigner operator.
However, this idea is often regarded as controversial. There are at
least three arguments that are usually cited to ``explain'' why
there can be no position operator and localized states in
relativistic quantum theory, in particular, in QFT:

\begin{itemize}
\item Single particle localization is impossible, because it requires an unlimited
amount of energy (due to the Heisenberg's uncertainty relation) and
leads to creation of extra particles \cite{BLP}:

\begin{quote}
\emph{In quantum field theory, where the particle propagators do not
allow acausal effects, it is impossible to define a
position operator, whose measurement will leave the particle in a
sharply defined spot, even though the interaction between the
fields is local. The argument is always that, to
localize the electric charge on a particle with an accuracy better
than the Compton wavelength of the electron, so much energy should
be put in, that electron-positron pairs would be formed. This would
make the concept of position meaningless.} Th. W. Ruijgrok \cite{Ruijgrok98}
\end{quote}

\item Newton-Wigner particle localization is relative, i.e., different moving
observers may disagree on whether the particle is localized or not.
\item Perfectly localized wave packets spread out with superluminal
speeds, which contradicts the principle of causality
\cite{Hegerfeldt}:

\begin{quote}
\emph{The 'elementary particles' of particle physics are generally
understood as pointlike objects, which would seem to imply the
existence of position operators for such particles. However, if we
add the requirement that such operators are covariant (so that, for
instance, a particle localized at the origin in one Lorentz frame
remains so localized in another), or the requirement that the
wave-functions of the particles do not spread out faster than light,
then it can be shown that no such position operator exists. (See
Halvorson and Clifton (2001) \emph{\cite{Halvorson}} and references
therein, for details.)} D. Wallace \cite{Wallace2}
\end{quote}
\end{itemize}

\noindent In the present section we are going to show that
relativistic localized states of particles have a well-defined and non-controversial meaning
in spite of these arguments.

\subsection{Measurements of position} \label{ss:localizable}

Let us first consider the idea that precise measurements of position
disturb the number of particles in the system.

It is true that due to the Heisenberg's uncertainty relation
(\ref{eq:heisenberg}), sharply localized 1-particle states do not
have well-defined momentum and energy. For a sufficiently localized
state, the energy uncertainty can be made greater than the energy
required to create a particle-antiparticle pair. However, large
uncertainty in energy does not immediately imply any uncertainty in
the number of particles, and sharp localization does not necessarily
require pair creation. The number of particles in a localized state
would be uncertain if the particle number operator did not commute
with position operators of particles. However, this is not true. One
can easily demonstrate that Newton-Wigner particle position
operators do commute with particle number operators. This follows
directly from the definition of particle observables in the Fock space.\footnote{see subsection
\ref{ss:sectors} } By their construction, all 1-particle
observables (position, momentum, spin, etc.) commute with
projections on $n$-particle sectors in the Fock space. Therefore
these 1-particle observables commute with particle number operators.
So, one can measure position of any particle without disturbing the
number of particles in the system. This conclusion is valid for both
non-interacting and interacting particle systems, because the Fock
space structure and definitions of one-particle observables do not
depend on interaction.

\subsection{Localized states in a
moving reference frame} \label{ss:moving-frame}

In this subsection we will discuss the second objection against the
use of localized states in relativistic quantum theories, i.e., the
non-invariance of the particle localization.

The position-space wave function of a single massive spinless
particle in a state  sharply localized in the origin
is\footnote{This is a non-normalizable state that we called
``improper'' in section \ref{sc:representations}.  Similar arguments
apply to normalized localized wave functions, like $\sqrt{\delta
(\mathbf{r})}$.}

\begin{eqnarray}
\psi(\mathbf{r}) = \delta (\mathbf{r}) \label{eq:7.28}
\end{eqnarray}

\noindent The  corresponding momentum space wave function
 is (\ref{eq:7.25})

\begin{eqnarray}
 \psi(\mathbf{p}) &=&
(2\pi \hbar )^{-3/2} \label{eq:7.29}
\end{eqnarray}

\noindent Let us now find the wave function of this  state  from the
point of view of a moving observer $O'$. By applying a boost
transformation to (\ref{eq:7.29})\footnote{see equation (\ref{eq:x})}

\begin{eqnarray*}
e^{-\frac{ic}{\hbar}\hat{K}_x \theta} \psi(\mathbf{p}) &=& (2\pi \hbar
)^{-3/2} \sqrt{\frac{\omega_{\mathbf{p}} \cosh \theta -
 c p_x \sinh \theta}{\omega_{\mathbf{p}}}
 }
\end{eqnarray*}

\noindent and transforming  back to the position representation
via (\ref{eq:7.27}) we obtain

\begin{eqnarray}
e^{-\frac{ic}{\hbar}\hat{K}_x \theta} \psi(\mathbf{r}) &=& (2\pi \hbar
)^{-3} \int d\mathbf{p}
\sqrt{ \cosh \theta - (c p_x/ \omega_{\mathbf{p}})\sinh \theta}
e^{\frac{i}{\hbar}\mathbf{p}\mathbf{r}} \label{eq:7.30}
\end{eqnarray}

\noindent We are not going to calculate this integral explicitly,
but one property of the function (\ref{eq:7.30}) must be clear: for
non-zero $\theta$ this function is non-vanishing for all values of
$\mathbf{r}$.\footnote{This property follows from the
non-analyticity of the square root in the integrand
\cite{Strocchi}.\label{non-analyticity} } Therefore, the moving observer $O'$ would
not agree with $O$ that the particle is localized. Observer $O'$ can
find the particle anywhere in space. This means that the notion of localization is
relative: a state which looks localized to the observer $O$ does not
look localized to the moving observer $O'$.

The non-invariant nature of localization is a property not familiar
in classical physics. Although this property has not been observed
in experiments yet, it does not contradict any postulates of
relativistic quantum theory and does not constitute a sufficient
reason to reject the notion of localizability.

\subsection{Spreading of well-localized states}
\label{ss:spreading}

\begin{figure}
\centering
 \includegraphics {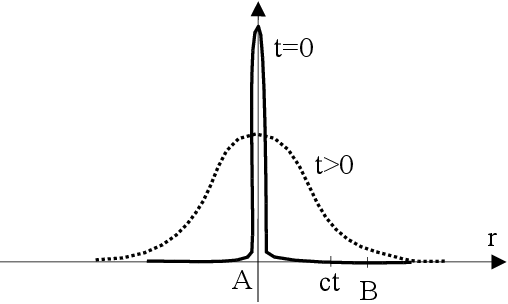} \caption{Spreading of the
probability distribution of a localized wave function. Full line: at
time $t=0$; dashed line: at time $t > 0$ (the distance between
points $A$ and $B$ is greater than $ct$).} \label{fig:7.3}
\end{figure}

Here we are going to discuss the wide-spread opinion that
superluminal spreading of particle wave functions violates the
principle of causality \cite{Halvorson, Wallace2, Malament,
Hegerfeldt, Buchholz}. \index{causality}

In the preceding subsection we found how a localized state
(\ref{eq:7.28}) looks from the point of view of a moving observer.
Now, let us find  the appearance of this state from the point of
view of an observer displaced in time. Again, we first make a detour
to the momentum space (\ref{eq:7.29}),
 apply the time translation operator

\begin{eqnarray*}
  \psi (\mathbf{p}, t) &=& e^{-\frac{i}{\hbar}\hat{H}t} \psi (\mathbf{p}, 0)
= (2 \pi \hbar)^{-3/2} e^{-\frac{it}{\hbar}\sqrt{m^2c^4 + p^2c^2} }
\end{eqnarray*}

\noindent and then use equation (\ref{eq:7.27}) to find the
position-space wave function at non-zero $t$

\begin{eqnarray*}
  \psi (\mathbf{r}, t)
&=& (2 \pi  \hbar)^{-3/2}\int d\mathbf{p}
 \psi (\mathbf{p}, t)
  e^{\frac{i}{\hbar} \mathbf{p} \mathbf{r}}
  = (2 \pi  \hbar)^{-3}\int  d\mathbf{p} e^{-\frac{it}{\hbar}
\sqrt{m^2c^4 + p^2c^2}}
  e^{\frac{i}{\hbar} \mathbf{p} \mathbf{r}}
\end{eqnarray*}

\noindent This integral can be calculated analytically
\cite{Ruijgrok}. However, for us the most important result is that
the wave function  is non-zero at distances larger than $ct$ from the initial
point $A$ ($r >
ct$), i.e., outside the ``light cone''.\footnote{This fact can be justified by the same analyticity
argument as in footnote on page \pageref{non-analyticity}. See also \cite{Wagner} and
section 2.1 in \cite{Peskin}.} The corresponding probability density
$|\psi (\mathbf{r}, t)|^2$ is shown schematically by the dashed line
in Fig. \ref{fig:7.3}. Although the  probability density outside the
light cone is very small, there is still a non-zero chance that the
particle propagates faster than the speed of light.

Note that superluminal propagation of the particle's wave function
in the position space does not mean that particle's speed is greater
than $c$. As we have established in subsection \ref{ss:momentum-basis}, for a free
massive particle,
eigenvalues of the quantum-mechanical operator of speed  are less than $c$. So, the possibility of wave
functions propagating faster than $c$ is a purely quantum effect
associated with the non-commutativity of operators $\mathbf{R}$ and
$\mathbf{V}$.

\subsection{Superluminal spreading and causality}
\label{ss:spreading-sup}

The superluminal spreading of localized wave packets described in
the preceding subsection holds under very general assumptions in
relativistic quantum theory \cite{Hegerfeldt}. It is usually
regarded as a sign of a serious trouble \cite{Halvorson, Wallace2,
Malament, Buchholz, Zuben}, because the superluminal propagation of
any signal is strictly forbidden in special relativity.\footnote{see
Appendix \ref{ss:super-signal}} This contradiction is often claimed
to be the major obstacle for the particle interpretation of
relativistic quantum theories. Since particle interpretation is the
major aspect of our approach, we definitely need to resolve this
controversy. This is what we are going to do in this subsection.

\begin{figure}
\centering
\includegraphics {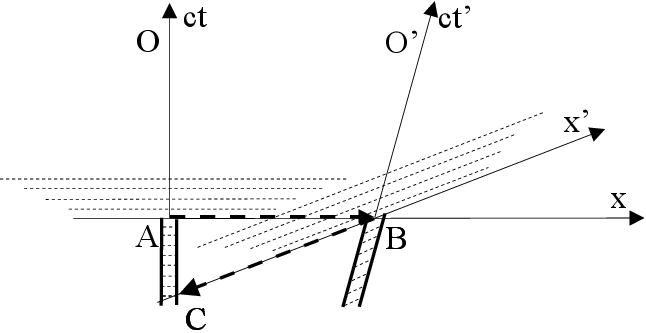} \caption{Space-time diagram
demonstrating the alleged causality paradox associated with superluminal
spreading of wave functions. Observers $O$ and $O'$ have coordinate
systems with space-time axes $(x,ct)$ and $(x',ct')$, respectively.
Observers $O$ and $O'$ send superluminal signals to each other by
opening boxes with localized quantum particles. See text for more
details.} \label{fig:10.1}
\end{figure}

Let us first describe the reason why the superluminal spreading of
wave functions is claimed to be unacceptable in the traditional
approach. One idea is that this phenomenon can be used to build a
device which would violate the principle of causality, as discussed
in Appendix \ref{ss:super-signal}. In that discussion we have not
specified the mechanism by which the instantaneous signals were sent
between observers $O$ and $O'$. Let us now assume that these signals
are transmitted by spreading quantum wave packets. More
specifically, suppose that the signaling device used by the observer
$O$ is simply a small impenetrable box containing massive spinless quantum
particles. Before time $t=0$ (point $A$ in Fig. \ref{fig:10.1}) the
box is tightly closed, so that wave functions of the particles are
well-localized inside it. The walls of the closed box at $t<0$ are
shown by two thick vertical parallel lines on the space-time diagram \ref{fig:10.1}. At time
$t=0$ observer $O$  sends a signal to the moving observer
$O'$ by opening the box. The wave function of spreading particles at
$t>0$ is shown schematically in Fig. \ref{fig:10.1} by thin dashed
lines parallel to the $x$-axis. Due to the superluminal spreading of the wave
function, there is, indeed, a non-zero probability of finding
particles at the location of the moving observer $O'$ (point $B$)
immediately after the box was opened.

The observer $O'$ has
a similar closed box with particles. Upon receiving the signal from
$O$ (point $B$) she opens her box. It is clear that the wave packet
of her particles $\psi'(\mathbf{r}',t')$ spreads instantaneously in
her own reference frame. The question is how this spreading will be
perceived by the stationary observer $O$? The traditional answer is
that the wave function $\psi(\mathbf{r},t)$ from the point of view
of $O$ should be obtained by applying Lorentz transformations
(\ref{eq:lorentz-transform-t}) - (\ref{eq:lorentz-transform-comp})
to the arguments of $\psi'$

\begin{eqnarray}
  \psi (x,y,z, t) = \psi'(x \cosh \theta + ct \sinh \theta, y, z, t \cosh \theta +
  (x/c) \sinh \theta) = \psi'(\Lambda \tilde{x}) \nonumber \\
  \label{eq:wave-pack-transf}
\end{eqnarray}

\noindent This wave function is shown schematically in Fig.
\ref{fig:10.1} by inclined parallel thin dashed lines. Then we see that
there is a non-zero probability of finding particles emitted by $O'$
at point $C$. This means that the response signal sent by $O'$
arrives to $O$ \emph{earlier} than the initial signal $O \to O'$ was
sent (at point $A$). This is clearly a violation of causality.

Actually, the time evolution of the wave packet
(\ref{eq:wave-pack-transf}) looks totally absurd from the point of
view of $O$. The particles do not look as emitted from $B$ at all.
In fact, the wave function approaches observer $O$ (point $C$) from
the opposite side (from the side of negative $x$) and moves in the positive $x$ direction. So, one
cannot even talk about the ``signal'' being sent from $O'$ to $O$!

What is wrong with this picture? The traditional answer is that this
weird behavior is the consequence of the superluminal propagation of
the wave function. The usual conclusion is that sanity can be
restored by forbidding such superluminal effects. However, this
would go against the entire theory developed in this book. Could
there be a different answer?

We would like to suggest the following explanation: Apparently, the
crucial step in the above derivation is the use of the wave function
transformation law (\ref{eq:wave-pack-transf}). However, there are
serious reasons to doubt that this formula is applicable even
approximately. First, here we are dealing with a system (particles
confined in a box) where interactions play a significant role. In
such a system the boost operator is interaction-dependent, and boost
transformations of wave functions should depend on the details of
interaction potentials.\footnote{This dependence will be discussed
in the next section in greater detail.} Therefore, it is obvious
that the wave function transformation cannot be described by the
universal interaction-independent formula
(\ref{eq:wave-pack-transf}). Moreover, our system is not isolated.
It is described by a time-dependent Hamiltonian (the box is opened
at some point in time). This makes Poincar\'e group arguments
unapplicable and further complicates the analysis of boost
transformations. Even if we assume that interaction-dependence of
boost transformations can be neglected in some approximation, there
is still no justification for equation (\ref{eq:wave-pack-transf}). This
formula cannot be used to transform even wave functions of free
particles. For example, this formula contradicts the transformation
law (\ref{eq:7.30}) derived earlier for localized particle
states. So, there is absolutely no evidence that the wave function of
particles emitted by observer $O'$ will behave as shown in Fig.
\ref{fig:10.1} from the point of view of $O$. In particular, there
is no evidence that the signal sent by $O'$ arrives to $O$ at point
$C$ in violation of the causality law.

It is plausible that using the correct boost transformation law one
would obtain that  the wave function of particles released by $O'$
at point $B$ propagates superluminally in the reference frame $O$
as well.\footnote{this means that thin dashed lines around point $B$ in
Fig. \ref{fig:10.1} should be drawn parallel to the axis $x$} This
conclusion can be supported by the following argument. From the
point of view of observer $O$, the particle emitted at point $B$
($t=0$) is in a localized state with definite position. Such states
do not have any definite velocity (or momentum), so their
free\footnote{At times $t>0$ the box $B$ remains opened, and the particle's evolution
is described by the non-interacting Hamiltonian $H_0$.} time evolution is determined only by the value
of the position characteristic for the initial state. Therefore
particles emitted by boxes at rest (e.g., the box $A$) and by moving
boxes (e.g., the box $B$) are described by essentially the same
time-dependent wave functions at $t>0$. The only difference being a
relative shift along the $x$-axis. Then instead of the acausal
response signal $B \to C$\footnote{as predicted incorrectly in
the traditional approach based on equation (\ref{eq:wave-pack-transf})} one
would have a signal $B \to A$, which, in spite of being
instantaneous, does not violate the principle of causality.

\section{Inertial transformations in multiparticle systems}
\label{sc:new-approach}

One of the goals of physics declared in Introduction\footnote{see page
\pageref{goals}} includes finding transformations of observables
between different inertial reference frames. In chapter
\ref{ch:operators} and in subsection \ref{ss:total-observables} we
discussed inertial transformations of \emph{total} observables in a
multiparticle system and we found that these transformations have
universal forms, which do not depend on the system's composition and
interactions acting there. In this section we will be interested in
establishing inertial transformations for observables of \emph{individual
particles} within an interacting multiparticle system. Our goal is to
compare these predictions of RQD  with Lorentz transformations for time and position of
\emph{events} in special relativity

\begin{eqnarray}
 t' &=&  t \cosh \theta - (x/c) \sinh \theta \label{eq:lor-transform-t} \\
 x' &=& x \cosh \theta - ct \sinh \theta \label{eq:lor-transform-x} \\
 y' &=& y \\
 z' &=& z
 \label{eq:lor-transform-comp}
 \end{eqnarray}

\noindent  Here we will reach a surprising
conclusion that formulas of special relativity may be not accurate.

\subsection{Events and observables} \label{sc:events-obs}

One of the most fundamental concepts in physics is the concept of an
\emph{event}. \index{event} Generally, event is some physical
process or phenomenon occurring in a small volume of space in a
short interval of time. So, each event can be characterized by four
numbers: its time $t$ and its position $\mathbf{r} = (x,y,z)$. These numbers
 are referred to as \emph{space-time coordinates} $(t, \mathbf{r})$ of the event.
For the event to be observable, there should be some material
particles present at time $t$ at the point
$\mathbf{r}$. The simplest example of an event is an
intersection of trajectories (a collision) of two particles. We will define $t$ as the reading of the clock belonging to the observer witnessing the particles' collision and $ \mathbf{r}$ as the
(expectation) value of the positions of particles present in the
event's volume.

In this section we would like to derive the relationship
between event's space-time coordinates $(t, \mathbf{r})$ measured in the
reference frame at rest $O$ and space-time coordinates $(t',
\mathbf{r}')$ measured in the moving reference frame $O'$. Since we
just identified event's position with the expectation value of particles' position operators, finding
boost transformations $\mathbf{r} \to \mathbf{r}'$ is just an exercise in
straightforward application of the general rule for transformations of operators of
observables between different reference frames.\footnote{see subsections \ref{ss:lorentz-position} and \ref{ss:inertial-obs}} By following this rule we should be
able to derive analogs of Lorentz transformations (\ref{eq:lor-transform-t}) - (\ref{eq:lor-transform-comp}) without artificial assumptions from Appendix
\ref{sc:lorentz-time-pos}, so we should be able to tell whether
Lorentz transformations formulas are exact or approximate. This is
the plan of our presentation  in this section.

For simplicity, here we will consider a system of two
massive spinless particles described in the Hilbert space
$\mathcal{H} = \mathcal{H}_1 \otimes \mathcal{H}_2 $, where
one-particle observables (position, momentum, velocity, angular momentum,
spin, energy,...) are denoted by lowercase letters:

\begin{eqnarray}
&\ &\mathbf{r}_1, \mathbf{p}_1, \mathbf{v}_1, \mathbf{j}_1,
\mathbf{s}_1, h_1, \ldots \label{eq:part-obs1} \\
&\ &\mathbf{r}_2, \mathbf{p}_2, \mathbf{v}_2, \mathbf{j}_2,
\mathbf{s}_2, h_2, \ldots \label{eq:part-obs2}
\end{eqnarray}

 Transformations of these observables between reference frames $O$
and $O'$ should be found by the general rule outlined in subsection
\ref{ss:fundamental}. Suppose that observers $O$ and $O'$ are
related by an inertial transformation, which is
generated by the Hermitian operator $F$ and parameter $b$. If $g$
is an observable (a Hermitian operator) of one particle in the
reference frame $O$, and $g(b)$ is the same observable in the
reference frame $O'$ then we use equations (\ref{eq:5.59}) and
(\ref{eq:A.39}) to obtain

\begin{eqnarray}
g(b) &=& e^{-\frac{i}{\hbar}Fb} g e^{\frac{i}{\hbar}Fb} = g -
\frac{ib}{\hbar} [F,g] - \frac{b^2}{2! \hbar} [F, [F,g]] +
     \ldots \label{eq:comm-poiss}
\end{eqnarray}

\noindent Application of this formula to event's position is not straightforward, because
particle localization does not have  absolute meaning in quantum
mechanics. If observer $O$ registers a localized event (or locallized particles
constituting this event), then other observers may disagree that the
event is localized or that it has occurred at all. Examples of such
a behavior are common in quantum mechanics. Some of them were discussed in subsections \ref{ss:wave-packets},
\ref{ss:moving-frame}, and \ref{ss:spreading}. Thus we are going to apply boost
transformations only to expectation values of
positions.
In other words, in the rest of this chapter we will work in the
classical limit, where the spreading wave packet  can be ignored, and particle trajectories can be
unambiguously defined. Then we will interpret (\ref{eq:part-obs1})
and (\ref{eq:part-obs2}) as numerical (expectation)  values of
observables in quasiclassical states, and instead of quantum operator
equation (\ref{eq:comm-poiss}) with commutators we will use its
classical analog involving Poisson brackets (\ref{eq:poiss-br})

\begin{eqnarray*}
g(b)  &\approx& g + b[F,g]_P + \frac{b^2}{2!} [F, [F,g]_P]_P +
\ldots
\end{eqnarray*}

\noindent In order to perform calculations with this formula one
needs two major things. First, one needs to know expressions for
Poincar\'e generators $F$ in terms of one-particle observables
(\ref{eq:part-obs1}) - (\ref{eq:part-obs2}). This is equivalent to
having a full dynamical description of the system. Such a description can be easily obtained in the case of a non-interacting particle system. However, for interacting particles this is a rather non-trivial problem that can be solved only approximately. Second, one needs to know
Poisson brackets between all one-particle observables
(\ref{eq:part-obs1}) - (\ref{eq:part-obs2}). This is an easier
task, which has been accomplished already in chapters
\ref{ch:operators}, \ref{ch:single}, and in section
\ref{sc:many-particle}. In particular, we found there that
observables of different particles always have vanishing Poisson
brackets. The Poisson brackets  for observables referring to the
same particle are ($i,j,k = 1,2,3$)

\begin{eqnarray}
\ [r_i, r_j]_P &=& [p_i, p_j]_P = [r_i, s_j]_P = [p_i, s_j]_P = 0 \label{eq:rirj} \\
\ [r_i, p_j]_P &=&  \delta_{ij} \label{eq:ripj}  \\
\ [s_i, s_j]_P &=& \sum_{k=1}^3 \epsilon_{ijk} s_k \label{eq:sisjsk}\\
\ [\mathbf{p}, h]_P &=& [\mathbf{s}, h]_P = 0 \label{eq:pihj} \\
\ [\mathbf{r}, h]_P &=& \frac{\mathbf{p} c^2}{h} \label{eq:rih}
\end{eqnarray}

\subsection{Non-interacting particles} \label{ss:lorentz-free}

 First we assume that the two particles 1 and 2 are non-interacting,
so that generators of inertial transformations in the Hilbert space
$\mathcal{H}$ are

\begin{eqnarray}
H_0 &=& h_1 + h_2 \label{eq:h-non-inter} \\
\mathbf{P}_0 &=& \mathbf{p}_1 + \mathbf{p}_2 \\
\mathbf{J}_0 &=& \mathbf{j}_1 + \mathbf{j}_2 \\
\mathbf{K}_0 &=& \mathbf{k}_1 + \mathbf{k}_2 \label{eq:k-non-inter}
\end{eqnarray}

\noindent The trajectory of the particle 1 in the reference frame
$O$ is obtained from the usual formula (\ref{eq:6.47})

\begin{eqnarray}
\mathbf{r}_1(t) &=& e^{\frac{i}{\hbar}H_0t} \mathbf{r}_1 e^{-\frac{i}{\hbar}H_0t}
= e^{\frac{i}{\hbar}(h_1 + h_2)t} \mathbf{r}_1 e^{-\frac{i}{\hbar}(h_1 + h_2)t}
= e^{\frac{i}{\hbar}h_1t} \mathbf{r}_1 e^{-\frac{i}{\hbar}h_1 t} \nonumber \\
&\approx& \mathbf{r}_1 + t [h_1, \mathbf{r}_1]_P + \frac{t^2}{2!}
[h_1, [h_1, \mathbf{r}_1]_P]_P + \ldots =  \mathbf{r}_1 +
\mathbf{v}_1 t \label{eq:r1t-non-int}
 \end{eqnarray}

\noindent Applying boost transformations to (\ref{eq:r1t-non-int}) and taking
into account (\ref{eq:6.4}) -
(\ref{eq:6.6}), (\ref{eq:7.46}) - (\ref{eq:7.47}), and (\ref{eq:7.43})
we find the trajectory of this particle in the reference frame $O'$
moving with the speed $v = c \tanh \theta$ along the
$x$-axis\footnote{If we set $t'=0$ then these formulas coincide with
(23) - (24) in ref. \cite{Monahan}. By setting also $\mathbf{v}_1=0 $
we obtain the usual Lorentz \emph{length contraction} \index{length
contraction} formulas $r_{1x}(\theta, 0) = r_{1x}/(\cosh \theta)$,
$r_{1y}(\theta, 0) = r_{1y}$, $r_{1z}(\theta, 0) = r_{1z}$. Compare with equation
(\ref{eq:length-contraction}).}

\begin{eqnarray}
r_{1x}(\theta, t') &=& \beta \left(\frac{r_{1x}}{\cosh \theta}
+ (v_{1x} -v)t' \right) \label{eq:r-x-theta-t'}\\
r_{1y}(\theta,t') &=& \beta \left( r_{1y} + \frac{j_{1z}v}{h_1}
+ \frac{v_{1y} t'}{\cosh \theta} \right) \nonumber \\
&=& r_{1y} + \beta \left( \frac{r_{1x}v_{1y}v }{c^2} + \frac{v_{1y}
t'}{\cosh \theta} \right) \\
 r_{1z}(\theta,t') &=& \beta \left(
r_{1z} + \frac{j_{1y}v}{h_1} + \frac{v_{1z} t'}{\cosh \theta}\right)
\nonumber \\
&=& r_{1z} + \beta \left( \frac{r_{1x}v_{1z}v }{c^2} + \frac{v_{1z}
t'}{\cosh \theta} \right) \label{eq:r-z-theta-t'}
\end{eqnarray}

\noindent where we denoted $\beta \equiv  (1-v_{1x} v c^{-2})^{-1}
$. Similar formulas are valid for the particle 2.

The important feature of these formulas is that inertial
transformations for particle observables are completely independent
on the presence of other particles in the system, e.g. formulas for
$\mathbf{r}_1(\theta, t')$ do not depend on observables of the
particle 2. This is hardly surprising, since the two particles were
assumed to be non-interacting.

\subsection{Lorentz transformations for non-interacting particles}
\label{ss:trajectory}

Now, let us consider a localized event associated with the intersection of particle
trajectories. Suppose that from the point of view of the observer $O$ this event has space-time coordinates $(t, \mathbf{r})$. This means that

\begin{eqnarray*}
x &\equiv& r_{1x}(t) = r_{2x}(t) \\
y &\equiv& r_{1y}(t) = r_{2y}(t) \\
z &\equiv& r_{1z}(t) = r_{2z}(t)
 \end{eqnarray*}

\noindent Apparently, these two trajectories intersect from the point
of view of the moving observer $O'$ as well. So $O'$ also sees the
event. Now, the question is: \emph{what are the space-time
coordinates of the event seen by $O'$?} The answer to this question
is given by the following theorem.

\begin{theorem} [Lorentz transformations for time and position]
For events defined as intersections of trajectories of
non-interacting particles, the Lorentz transformations for time and
position (\ref{eq:lor-transform-t}) -
(\ref{eq:lor-transform-comp}) are exactly valid.
 \label{theorem:lorentz-tr} \index{Lorentz transformations}
\end{theorem}
\begin{proof}
Let us first prove that Lorentz formulas
(\ref{eq:lor-transform-t}) -
(\ref{eq:lor-transform-comp})
are correct transformations for the trajectory of the particle 1
between reference frames $O$ and $O'$. For simplicity, we will
consider only the case in which this particle is moving along the
$x$-axis: $r_{1y}(t)=r_{1z}(t)=v_{1y}=v_{1z}=0$. (More general
situations can be analyzed similarly.) Then we can ignore the $y$-
and $z$-coordinates in our proof. So, we need to prove that\footnote{Here the left hand side is the Newton-Wigner position of the particle 1 seen from the reference frame $O'$ at time $t'$. This is formula (\ref{eq:r-x-theta-t'}). The right hand side is Lorentz-transformed position as in equation (\ref{eq:lor-transform-x}).}

\begin{eqnarray}
r_{1x}(\theta, t') &=& r_{1x}(0,t) \cosh \theta - ct \sinh \theta \nonumber \\
&=& ( r_{1x} + v_{1x}t) \cosh \theta - ct \sinh \theta
\label{eq:7.48}
\end{eqnarray}

\noindent where

\begin{eqnarray}
 t' &=& t \cosh \theta -\frac{r_{1x}(t)}{c}  \sinh
\theta \label{eq:7.49}
\end{eqnarray}

\noindent To do that, we calculate the difference between the right
hand sides of equations (\ref{eq:r-x-theta-t'}) and (\ref{eq:7.48}) with
$t'$ taken from (\ref{eq:7.49}) and using $v = c \tanh \theta$

\begin{eqnarray*}
&\mbox{ }& \frac{\beta r_{1x}}{\cosh \theta} + (v_{1x} - v)\beta (t
\cosh \theta - c^{-1}(r_{1x} +v_{1x}t)
\sinh \theta) \\
 &\ &-
(r_{1x}+v_{1x}t) \cosh \theta + c t \sinh \theta \\
&=& \frac{\beta }{\cosh \theta} [ r_{1x} + v_{1x} t \cosh^2 \theta
-v t \cosh^2 \theta  - (v_{1x}r_{1x}/c) \sinh \theta \cosh \theta
\\
&\ & +(vr_{1x}/c) \sinh \theta \cosh \theta - (v^2_{1x}/c) t\sinh
\theta \cosh \theta + (vv_{1x}/c) t\sinh \theta \cosh \theta \\
 &\ & -r_{1x} \cosh^2 \theta + r_{1x}  (v_{1x}v/ c^{2}) \cosh^2 \theta - v_{1x} t\cosh^2 \theta +
 (v^2_{1x} v /c^{2}) t\cosh^2 \theta
+c t \sinh \theta \cosh \theta \\
&\ & -(v_{1x}v /c) t\sinh \theta \cosh \theta]\\
&=& \frac{\beta }{\cosh \theta} [ r_{1x} -v t \cosh^2 \theta  -
(v_{1x}r_{1x}/c) \sinh \theta \cosh \theta
\\
&\ & +(vr_{1x}/c) \sinh \theta \cosh \theta - (v^2_{1x}/c)t \sinh
\theta \cosh \theta  \\
 &\ & -r_{1x} \cosh^2 \theta + r_{1x}  (v_{1x}v /c^{2}) \cosh^2 \theta  +
 (v^2_{1x} v /c^{2})t \cosh^2 \theta
+c t \sinh \theta \cosh \theta] \\
&=& \frac{\beta }{\cosh \theta} [ r_{1x} -c t \sinh \theta \cosh
\theta - (v_{1x}r_{1x}/c) \sinh \theta \cosh \theta
\\
&\ & +r_{1x} \sinh^2 \theta  - (v^2_{1x}/c)t \sinh
\theta \cosh \theta  \\
 &\ & -r_{1x} \cosh^2 \theta + (r_{1x}  v_{1x}/c) \sinh \theta \cosh \theta  +
 (v^2_{1x} /c)t \sinh \theta \cosh \theta
+c t \sinh \theta \cosh \theta] \\
&=& 0
\end{eqnarray*}

\noindent This proves equation (\ref{eq:7.48}) boost-transformed trajectory (\ref{eq:r-x-theta-t'}) of the particle 1
 is consistent with Lorentz formulas
(\ref{eq:lor-transform-t}) and (\ref{eq:lor-transform-x}).
The same is true for the  particle 2. This implies that times and
positions of intersections of the two trajectories
also undergo Lorentz transformations
(\ref{eq:lor-transform-t}) - (\ref{eq:lor-transform-comp})
when the reference frame is boosted.
\end{proof}

\bigskip

\subsection{Interacting particles} \label{ss:new_approach}

This time we will assume that the two-particle system is
interacting. This means that the unitary representation $U_g$ of the
Poincar\'e group in $\mathcal{H}$ is different from the
non-interacting representation $U_g^0$ with generators
(\ref{eq:h-non-inter}) - (\ref{eq:k-non-inter}). Generally, we can
write generators of $U_g$ as\footnote{see equations (\ref{eq:8.11}) - (\ref{eq:8.14})}

\begin{eqnarray}
H &=& h_1 + h_2 + V(\mathbf{r}_1, \mathbf{p}_1, \mathbf{r}_2,
\mathbf{p}_2) \label{eq:inter-h}\\
\mathbf{P} &=& \mathbf{p}_1 + \mathbf{p}_2 +
\mathbf{U}(\mathbf{r}_1, \mathbf{p}_1, \mathbf{r}_2,
\mathbf{p}_2)\\
\mathbf{J} &=& \mathbf{j}_1 + \mathbf{j}_2 +
\mathbf{Y}(\mathbf{r}_1, \mathbf{p}_1, \mathbf{r}_2,
\mathbf{p}_2) \label{eq:inter-j}\\
\mathbf{K} &=& \mathbf{k}_1 + \mathbf{k}_2 +
\mathbf{Z}(\mathbf{r}_1, \mathbf{p}_1, \mathbf{r}_2, \mathbf{p}_2)
\label{eq:inter-k}
\end{eqnarray}

\noindent where $V, \mathbf{U}, \mathbf{Y}$, and $\mathbf{Z}$ are
interaction operators that are functions of one-particle observables.
One goal of this section is to find out more about the interaction
terms $V, \mathbf{U}, \mathbf{Y}$, and $\mathbf{Z}$, e.g., to see if
some of these terms can be set to zero. In other words, we would like to
understand if one can find an observational evidence about  the
relativistic form of dynamics in nature.

\subsection{Time translations in interacting systems}
\label{ss:inequivalence2}

The most obvious effect of interaction is modification of the time evolution
of the system as compared to the non-interacting time evolution. We
estimate the strength of interaction between particles by how much
their trajectories deviate from the uniform straight-line movement
(\ref{eq:r1t-non-int}). Therefore in any realistic form of dynamics,
the Hamiltonian - the generator of time translations - should
contain a non-vanishing interaction $V $, and we can discard as
unphysical any form of dynamics in which $V=0$. Then the time
evolution of the position of particle 1 is

\begin{eqnarray}
\mathbf{r}_1(t) &=& e^{\frac{i}{\hbar}Ht} \mathbf{r}_1 e^{-\frac{i}{\hbar}Ht}
= e^{\frac{i}{\hbar}(h_1 + h_2 + V)t} \mathbf{r}_1 e^{-\frac{i}{\hbar}(h_1 + h_2 + V)t} \nonumber \\
&=& \mathbf{r}_1 + t[h_1 + V, \mathbf{r}_1]_P + \frac{t^2}{2}
[(h_1 + h_2 + V), [h_1 + V, \mathbf{r}_1]_P]_P
+ \ldots \nonumber \\
&=& \mathbf{r}_1 + \mathbf{v}_1 t + t [V, \mathbf{r}_1]_P +
\frac{t^2}{2} [V, \mathbf{v}_1 ]_P + \frac{t^2}{2}[(h_1 + h_2), [V,
\mathbf{r}_1]_P]_P \nonumber \\
&\mbox{ }& + \frac{t^2}{2} [V, [V, \mathbf{r}_1]_P]_P + \ldots
\label{eq:int-corr}
 \end{eqnarray}

\noindent In the simplest case when interaction  $V$ commutes with
particle positions and in the non-relativistic
approximation $\mathbf{v}_1 \approx
\mathbf{p}_1/m_1$, this formula simplifies

\begin{eqnarray*}
\mathbf{r}_1(t) &\approx& \mathbf{r}_1 + \mathbf{v}_1 t  -
\frac{t^2}{2 m_1} \frac{ \partial V}{ \partial \mathbf{r}_1 } +
\ldots = \mathbf{r}_1 + \mathbf{v}_1 t  + \frac{ \mathbf{f}_1 t^2}{2
m_1}
+ \ldots \\
&=& \mathbf{r}_1 + \mathbf{v}_1 t  + \frac{ \mathbf{a}_1 t^2}{2} +
\ldots
 \end{eqnarray*}

\noindent where we denoted

\begin{eqnarray*}
\mathbf{f}_1 (\mathbf{r}_1, \mathbf{p}_1, \mathbf{r}_2,
\mathbf{p}_2) \equiv - \frac{ \partial V(\mathbf{r}_1, \mathbf{p}_1,
\mathbf{r}_2, \mathbf{p}_2)}{ \partial \mathbf{r}_1 }
 \end{eqnarray*}

\noindent the \emph{force} \index{force} with which particle 2 acts
on the particle 1. The vector $\mathbf{a}_1 \equiv \mathbf{f}_1/m_1$
can be interpreted as \emph{acceleration} \index{acceleration} of
the particle 1 in agreement with the Newton's second law of
mechanics. The trajectory $\mathbf{r}_1(t)$ of the particle 1
depends in a non-trivial way on the trajectory $\mathbf{r}_2(t)$ of
the particle 2 and \emph{vice versa}. Curved trajectories of particles 1
and 2 are definitely observable in macroscopic
experiments. However, this
interacting time evolution, by itself, cannot tell us which form of
relativistic dynamics is responsible for the interaction. Other
types of inertial transformations should be examined in order to
make this determination.

As an example, in  this section we will explain which experimental
measurements should be performed to tell apart two popular forms of
dynamics: the instant form

\begin{eqnarray}
H &=& h_1 + h_2 + V \label{eq:inter-h-instant}\\
\mathbf{P} &=& \mathbf{p}_1 + \mathbf{p}_2 \\
\mathbf{J} &=& \mathbf{j}_1 + \mathbf{j}_2 \\
\mathbf{K} &=& \mathbf{k}_1 + \mathbf{k}_2 + \mathbf{Z}
\label{eq:inter-k-instant}
\end{eqnarray}

\noindent and the point form

\begin{eqnarray}
H &=& h_1 + h_2 + V \label{eq:inter-h-point}\\
\mathbf{P} &=& \mathbf{p}_1 + \mathbf{p}_2 + \mathbf{U} \label{eq:inter-p-point}\\
\mathbf{J} &=& \mathbf{j}_1 + \mathbf{j}_2 \\
\mathbf{K} &=& \mathbf{k}_1 + \mathbf{k}_2  \label{eq:inter-k-point}
\end{eqnarray}

\subsection{Boost transformations in interacting systems}
\label{ss:inequivalence1}

Similar to the above analysis of time translations, we can examine
boost transformations. For interactions in the point form
(\ref{eq:inter-h-point}) - (\ref{eq:inter-k-point}), the potential
boost $\mathbf{Z}$ is zero, so boost transformations of the position and
velocity are the same as in the non-interacting case\footnote{For
simplicity, we consider only $x$-components here. For the  general
case, see (\ref{eq:6.4}) - (\ref{eq:6.6}) and
(\ref{eq:r-x-theta-t'}) - (\ref{eq:r-z-theta-t'}). As usual, $v \equiv c \tanh \theta$.}

\begin{eqnarray}
r_{1x}(\theta) &=& e^{-\frac{ic}{\hbar} K_{0x}  \theta}
 r_{1x}
e^{\frac{ic}{\hbar} K_{0x}  \theta} = e^{-\frac{ic}{\hbar} k_{1x}
\theta}
 r_{1x}
e^{\frac{ic}{\hbar} k_{1x}  \theta} \approx
 r_{1x} - c \theta [k_{1x}, r_{1x}]_P + \ldots  \nonumber  \\
&=& \frac{r_{1x}}{\cosh \theta (1-v_{1x} v c^{-2})}
 \label{eq:r-1x} \\
 v_{1x}(\theta) &=& e^{-\frac{ic}{\hbar} K_{0x}\theta}
 v_{1x}
e^{\frac{ic}{\hbar} K_{0x} \theta} = e^{-\frac{ic}{\hbar} k_{1x}
\theta}
 v_{1x}
e^{\frac{ic}{\hbar} k_{1x}  \theta} \approx
 v_{1x} - c\theta [k_{1x}, v_{1x}]_P + \ldots  \nonumber  \\
&=& \frac{v_{1x} -v}{ 1- v_{1x} v c^{-2} }
 \label{eq:v-1x}
\end{eqnarray}

On the other hand, in the instant form,  generators of
boosts  (\ref{eq:inter-k-instant}) are dynamical, and transformation formulas are different. For example, the
boost transformation of position is

\begin{eqnarray}
r_{1x}(\theta) &=& e^{-\frac{ic}{\hbar} K _x\theta}
 r_{1x}
e^{\frac{ic}{\hbar} K_x \theta} \nonumber \\
&=& e^{-\frac{ic}{\hbar} (K_{0x} + Z_x) \theta}
 r_{1x}
e^{\frac{ic}{\hbar} (K_{0x} + Z_x) \theta} \nonumber  \\
&\approx&
 r_{1x} - c \theta[k_{1x}, r_{1x}]_P
- c \theta [Z _x, r_{1x}]_P + \ldots  \nonumber  \\
&=& \frac{r_{1x}}{\cosh \theta (1-v_{1x} v c^{-2})}- c \theta [Z _x,
r_{1x}]_P + \ldots
 \label{eq:r-1x2}
\end{eqnarray}

\noindent The first term on the right hand side is the same
interaction-independent term as in (\ref{eq:r-1x}). This term is
responsible for the well-known relativistic effect of length
contraction (\ref{eq:length-contraction}).  The second term in
(\ref{eq:r-1x2}) is a correction due to interaction with the
particle 2. This correction depends on observables of both particles
1 and 2, and it makes boost transformations of positions
dependent in a non-trivial way on the state of the system  and on
interactions acting there. So, in the instant form of
dynamics, there is a strong analogy between time translations and
boosts of particle observables. They are both
interaction-dependent, i.e., dynamical.

In order to observe the dynamical effect of boosts described
above, one would need to use measuring devices moving with very high speeds
comparable to the speed of light. This presents enormous technical
difficulties. So, boost
 transformations of particle positions  have not been directly observed
 with an accuracy sufficient
 to detect the kinematical relativistic effect (\ref{eq:r-1x}), let alone
 the deviation  $[Z _x, r_{1x}]_P$ due to interactions.

Similarly, we can consider boost transformations of velocity in the
instant form of dynamics

\begin{eqnarray*}
v_{1x}(\theta) &=& e^{-\frac{ic}{\hbar} K _x\theta}
 v_{1x}
e^{\frac{ic}{\hbar} K_x \theta} = e^{-\frac{ic}{\hbar} (K_{0x} +
Z_x) \theta}
 v_{1x}
e^{\frac{ic}{\hbar} (K_{0x} + Z_x) \theta} \nonumber  \\
&=& v_{1x} - c \theta [k_{1x}, v_{1x}]_P - c \theta[Z _x, v_{1x}]_P + \ldots =
\frac{v_{1x} -v}{ 1- v_{1x} vc^{-2} } - c \theta [Z _x, v_{1x}]_P +
\ldots \nonumber \\
&=& (v_{1x} -v) + \frac{v_{1x} v (v_{1x} -v)}{c^2} - c \theta [Z _x,
v_{1x}]_P + \ldots
 \label{eq:v-1x2}
\end{eqnarray*}

\noindent The terms on the right hand side have clear physical
meaning: The first term $v_{1x} -v$ is the usual non-relativistic
change of velocity  in the moving reference frame.
This is the most obvious effect of boosts that is visible in our
everyday life. The second term is a relativistic correction that is
valid for both interacting and non-interacting particles. This
correction is a contribution of the order $c^{-2}$ to the
relativistic law of addition of velocities (\ref{eq:6.4}). Currently, there is abundant experimental evidence
for the validity of this law.\footnote{see subsection
\ref{ss:exp-special}} The third term is a correction due to the
interaction between particles 1 and 2. This effect has not been seen
experimentally, because it is very difficult to perform accurate
measurements of observables of interacting particles from fast moving reference
frames, as we mentioned above.

In conclusion, detailed measurements of boost transformations of
particle observables are very difficult, and with the present level
of experimental  precision they cannot help us to decide which form
of dynamics is active in any given physical system. Let us now turn
to space translations and rotations.

\subsection{Spatial translations and rotations}
\label{ss:inequivalence}

In both  instant and point forms of dynamics, rotations are
interaction-independent, so the term $\mathbf{Y}$ in the generator
of rotations (\ref{eq:inter-j}) is zero, and  rotation
transformations of particle positions (and other observables) are
exactly the same as in the non-interacting case, e.g.,\footnote{Rotation matrix $R_{\vec{\phi}}$ has been defined in (\ref{eq:r-vec-phi}).}

\begin{eqnarray}
\mathbf{r}_1(\vec{\phi}) &=& e^{-\frac{i}{\hbar} \mathbf{J} \cdot
\vec{\phi}}
 \mathbf{r}_1
e^{\frac{i}{\hbar} \mathbf{J} \cdot \vec{\phi}} =
e^{-\frac{i}{\hbar} \mathbf{j}_1 \cdot \vec{\phi}}
 \mathbf{r}_1
e^{\frac{i}{\hbar} \mathbf{j}_1 \cdot \vec{\phi}} =
R_{\vec{\phi}}\mathbf{r}_1 \label{eq:rotation-position}
\end{eqnarray}

\noindent This is in full agreement with experimental observations.

In the instant form of dynamics, space translations are
interaction-independent as well

\begin{eqnarray*}
\mathbf{r}_1(\mathbf{a}) &=& e^{-\frac{i}{\hbar} \mathbf{P} \cdot
\mathbf{a}}
 \mathbf{r}_1
e^{\frac{i}{\hbar} \mathbf{P} \cdot \mathbf{a}} =
e^{-\frac{i}{\hbar} (\mathbf{p}_1 + \mathbf{p}_2) \cdot \mathbf{a}}
 \mathbf{r}_1
e^{\frac{i}{\hbar} (\mathbf{p}_1 + \mathbf{p}_2) \cdot \mathbf{a}} =
e^{-\frac{i}{\hbar} \mathbf{p}_1 \cdot \mathbf{a}}
 \mathbf{r}_1
e^{\frac{i}{\hbar} \mathbf{p}_1 \cdot \mathbf{a}} \nonumber \\
&=& \mathbf{r}_1 - \mathbf{a}
\end{eqnarray*}

\noindent Again this result is supported by experimental
observations and by our common experience in various physical systems
and in a wide range of values of the transformation parameter
$\mathbf{a}$.

However, the point-form generator of space translations  (\ref{eq:inter-p-point}) does depend
on interaction. Thus translations of the
observer have a non-trivial effect on measured positions of
interacting particles. For example, the action of a translation
along the $x$-axis on the $x$-component of position of the particle
1 is

\begin{eqnarray}
r_{1x}(a) &=& e^{-\frac{i}{\hbar} P_x a}
 r_{1x} e^{\frac{i}{\hbar} P_x a} = e^{-\frac{i}{\hbar} (p_{1x}+ p_{2x} + U_x) a}
 r_{1x}
e^{\frac{i}{\hbar} (p_{1x}+ p_{2x} + U_x) a} \nonumber \\
&\approx& r_{1x}- a [(p_{1x}+  U_x),
 r_{1x}]_P + \ldots \nonumber \\
 &=& r_{1x} - a - a [U_x, r_{1x}]_P + \ldots\label{eq:int-x}
\end{eqnarray}

\noindent where the last term on the right hand side  is the interaction correction. Such a correction has
not been seen in experiments in spite of the fact that there is no difficulty in
arranging observations from reference frames displaced by large
distances $a$. So, there is a good reason to believe that
interaction dependence (\ref{eq:int-x}) has not been seen because it
is non-existent.

Thus we conclude that the effect of space translations and rotations
must be independent on interactions in the system. This means that these
transformations are kinematical  as in the instant form\footnote{It follows immediately from (\ref{eq:p=p0}) that boosts ought to be dynamical. Indeed,
suppose that boosts are kinematical, i.e., $\mathbf{K} =
\mathbf{K}_0$. Then from commutator (\ref{eq:5.55}) we obtain

\begin{eqnarray*}
H
  &=& c^2[K_x, P_x]_P= c^2[K_{0x}, P_{0x}]_P = H_0
\end{eqnarray*}

\noindent which means that $V=0$, and the system is non-interacting
in disagreement with our initial assumption.}

\begin{eqnarray}
 \mathbf{P} &=& \mathbf{P}_0 \label{eq:p=p0}\\
 \mathbf{J} &=& \mathbf{J}_0 \label{eq:j=j0}
\end{eqnarray}

\noindent  Therefore available experimental data imply
that

\begin{postulate} [instant form of dynamics] The unitary
representation of the Poincar\'e group acting in the Hilbert space
of any interacting physical system belongs to the instant form of
dynamics. \label{postulateR}
\end{postulate}

Throughout this book we assumed (without much discussion) that interactions
belong to the instant form.
Now we see that this was the correct choice.

Our arguments in this section used the assumption that one can observe particle
trajectories while interaction takes place. In order to make such observations, the
range of interaction should be larger than the spatial resolution of
instruments. This condition is certainly true for particles
interacting via long-range electromagnetic forces. Their non-trivial interacting dynamics (\ref{eq:int-corr}) is
directly observed. See chapters \ref{ch:theories} and \ref{ch:support}. Therefore, for such systems one should use the  instant form of relativistic dynamics with interacting boost operators (\ref{eq:boost-spin-orb}). In chapter \ref{ch:decays}, from the analysis of particle decays we
showed that Postulate \ref{postulateR} must be valid also for short-range
weak nuclear forces.  In the case of
systems governed by short-range  strong nuclear
forces, neither interacting trajectories nor time-dependent decay laws can be observed.\footnote{The presence of interaction
becomes evident only through scattering effects or through
formation of bound states, which are insensitive to the form of dynamics, as shown in subsection \ref{ss:form-equiv-point}.} Thus, the form of dynamics governing strong nuclear interactions remains an open issue.

\subsection{Physical inequivalence of forms of dynamics }
\label{ss:inequivalence3}

Postulate \ref{postulateR} contradicts a widely shared belief that
different forms of dynamics are physically equivalent. In the
literature one can find examples of calculations performed in the
instant, point, and front forms. The common assumption is that one
can freely choose the form of dynamics which is more convenient.
Where does this idea come from? There are two sources. The first
source is the fact\footnote{ explained in subsection
\ref{ss:form-equiv-point}} that
 different forms of dynamics are scattering equivalent. The second source is the questionable
assumption that all physically relevant information can be obtained
from the $S$-matrix:

\begin{quote}
\emph{If one adopts the point of view, first expressed by
Heisenberg, that all experimental information about the physical
world is ultimately deduced from scattering experiments and reduces
to knowledge of certain elements of the scattering matrix (or the
analogous classical quantity), then different dynamical theories
which lead to the same $S$-matrix must be regarded as physically
equivalent.} S. N. Sokolov and A. N. Shatnii \cite{Sokolov_Shatnii2}
\end{quote}

\noindent We already discussed in chapter \ref{sc:scattering} that
having exact knowledge of the on-shell $S$-matrix one can easily calculate
scattering cross-sections. Moreover, the energy levels and lifetimes
of bound states are encoded in positions of poles of the $S$-matrix
on the complex energy plane. It is true that in modern high energy
physics experiments it is very difficult to measure anything beyond
these data. This is the reason why scattering-theoretical methods
play such an important role in particle physics. It is also true
that in order to describe these data, one can choose any convenient
form of dynamics and a wide range of scattering-equivalent
expressions for the Hamiltonian.

However, it is definitely not true that the $S$-matrix provides a
complete description of everything that can be observed. For
example, the time evolution and other inertial transformations of
particle observables discussed earlier in this section, cannot be
described within the $S$-matrix formalism. A theoretical description
of these phenomena requires exact knowledge of generators of the
Poincar\'e group.
 Two scattering-equivalent forms of
dynamics  may yield very
different transformations of states with respect to space translations, rotations and/or boosts. These differences can be measured in experiments: For example, the interaction-independence of spatial translations rules out the point form of dynamics, as we established in subsection \ref{ss:inequivalence}.

\subsection{``No interaction'' theorem}
\label{ss:no-interaction}

The fact that boost generators are interaction-dependent has very
important implications for relativistic effects in interacting
systems. For example, consider a system of two interacting
particles. The arguments used to prove Theorem
\ref{theorem:lorentz-tr} are no longer valid in this case. Boost
transformations of particle positions (\ref{eq:r-1x2}) contain
interaction-dependent terms. This means that Lorentz transformations (\ref{eq:lor-transform-t}) - (\ref{eq:lor-transform-comp}) are no longer applicable to trajectories of individual particles and associated events.

The contradiction between the usually assumed ``invariant world
lines'' and relativistic interactions was noticed first by Thomas
\cite{Thomas}. Currie, Jordan, and Sudarshan analyzed this problem
in greater detail \cite{CJS}  and proved their famous theorem

\bigskip

\begin{theorem}  [Currie, Jordan, and Sudarshan]
In a two-particle system,\footnote{This theorem can be proven for
many-particle systems as well. We limit ourselves to two particles
in order to simplify the proof.} trajectories of particles obey
Lorentz transformation formulas (\ref{eq:lor-transform-t}) -
(\ref{eq:lor-transform-comp}) if and only if the particles do
not interact\footnote{In our version of the proof we actually demonstrate the impossibility of cluster-separable interactions. For a more general
proof see original paper \cite{CJS}.} with each other. \label{Theorem12.1}
\end{theorem}
\begin{proof} We have demonstrated in Theorem \ref{theorem:lorentz-tr} that
trajectories of non-interacting particles do transform by Lorentz
formulas. So, we only need to prove the reverse statement: Any physical system whose trajectories transform by Lorentz formulas, is interaction-free.

 In our proof we will need to study inertial transformations
of particle observables (position $\mathbf{r}$ and momentum
$\mathbf{p}$), with respect to time translations and boosts. In
particular, given observables $\mathbf{r}(0, t)$ and
$\mathbf{p}(0,t)$ at time $t$ in the reference frame $O$, we would
like to find observables $\mathbf{r}(\theta, t')$ and
$\mathbf{p}(\theta, t')$ in the moving reference frame $O'$, where
time $t'$ is measured by its own clock. As before, we will assume that
$O'$ is moving relative to $O$ with velocity $v = c \tanh \theta$
along the $x$-axis.

Our plan is similar to the proof of Theorem
\ref{theorem:lorentz-tr}. We will compare formulas for
$\mathbf{r}(\theta, t')$ and $\mathbf{p}(\theta, t')$ obtained by
two methods. In the first method, we will use Lorentz
transformations of special relativity. In the second method, we will
apply interacting unitary operators of time translation and boost to
$\mathbf{r}$ and $\mathbf{p}$. Our goal is to
show that these two methods give different results. It is sufficient
to demonstrate that differences occur already in terms linear
with respect to $t'$ and $\theta$. So, we will work in this
approximation.

Let us apply the first method (i.e., traditional Lorentz formulas).
From equations (\ref{eq:lor-transform-t}) -
(\ref{eq:lor-transform-comp}) and (\ref{eq:6.2}) we obtain the
following transformations for the position and momentum of the
particle 1 (formulas for the particle 2 are similar) \index{Lorentz
transformations}

\begin{eqnarray}
r_{1x}(\theta, t') &\approx& r_{1x} (0,t)  - ct  \theta
\label{eq:12.35}\\
r_{1y}(\theta,t') &=& r_{1y} (0,t)  \\
r_{1z}(\theta,t') &=& r_{1z} (0,t)  \\
p_{1x}(\theta,t') &\approx& p_{1x} (0,t)  - \frac{1}{c} h_1(0,t)
\theta
\label{eq:12.36}\\
p_{1y}(\theta,t') &=& p_{1y} (0,t)  \\
p_{1z}(\theta,t') &=& p_{1z}  (0,t)  \\
t' &\approx& t  - \frac{\theta}{c} r_{1x}(0,t) \label{eq:12.37}
\end{eqnarray}

\noindent We can rewrite equation (\ref{eq:12.35}) without affecting the first order
accuracy level in $t'$ and $\theta$

\begin{eqnarray}
r_{1x}(\theta,t')
 &=& r_{1x}\left(0,t' + \frac{r_{1x}(0,t)}{c}  \theta\right) - \left(t' +
\frac{r_{1x}(t)}{c}  \theta\right) c \theta \nonumber\\
&\approx& r_{1x}(0,t') +  \frac{dr_{1x}(0,t')}{c dt'}
r_{1x}(0,t)
\theta -   c \theta t' \nonumber\\
&\approx& r_{1x}(0,t') +  \frac{dr_{1x}(0,t')}{c dt'}
r_{1x}(0,t') \theta -   c \theta t' \label{eq:12.38}
\end{eqnarray}

Next we use the second method (i.e., the direct application of interacting time
translations and boosts)

\begin{eqnarray}
r_{1x}(\theta,t') &=& e^{-\frac{ic}{\hbar} K_x \theta} e^{
\frac{i}{\hbar}Ht'} e^{\frac{ic}{\hbar} K_x \theta}
e^{-\frac{ic}{\hbar} K_x \theta}r_{1x}(0,0) e^{\frac{ic}{\hbar} K_x
\theta} e^{-\frac{ic}{\hbar} K_x \theta}  e^{
-\frac{i}{\hbar} Ht'} e^{\frac{ic}{\hbar} K_x  \theta} \nonumber\\
&=&  e^{ \frac{i}{\hbar}Ht' \cosh
 \theta}
e^{- \frac{ic}{\hbar}P_x t' \sinh  \theta}
 e^{-\frac{ic}{\hbar} K_x \theta}r_{1x}(0,0)
e^{\frac{ic}{\hbar} K_x \theta} e^{ \frac{ic}{\hbar}P_x t' \sinh
\theta} e^{ -\frac{i}{\hbar}Ht' \cosh
\theta}\nonumber \\
&\approx&  e^{ \frac{i}{\hbar}Ht' } e^{- \frac{ic}{\hbar}P_x t'
\theta}
 (r_{1x}(0,0) - c \theta [r_{1x}(0,0),K_x]_P)
e^{ \frac{ic}{\hbar}P_x t' \theta} e^{ -\frac{i}{\hbar}Ht' } \nonumber\\
&\approx&  e^{ \frac{i}{\hbar}Ht' }
 (r_{1x}(0,0) - c \theta t' - c \theta
[r_{1x}(0,0),K_x]_P )
e^{ -\frac{i}{\hbar}Ht' }\nonumber \\
& =&
 r_{1x}(0,t') -c \theta [r_{1x}(0,t'),K_x(t')]_P
- c \theta t' \label{eq:12.39} \\
p_{1x}(\theta, t') &=&  e^{ \frac{i}{\hbar}Ht' \cosh
 \theta}
e^{- \frac{ic}{\hbar}P_x t' \sinh c \theta}
 e^{-\frac{ic}{\hbar} K_x \theta}p_{1x} (0,0)
e^{\frac{ic}{\hbar} K_x \theta} e^{ \frac{ic}{\hbar}P_x t' \sinh c
\theta} e^{ -\frac{i}{\hbar}Ht' \cosh
\theta} \nonumber \\
&\approx&  e^{ \frac{i}{\hbar}Ht' } e^{- \frac{ic}{\hbar}P_x t'
\theta}
 (p_{1x}(0,0) -c \theta [p_{1x}(0,0),K_x]_P)
e^{ \frac{ic}{\hbar}P_x t' \theta} e^{ -\frac{i}{\hbar}Ht' } \nonumber \\
&\approx&  e^{ \frac{i}{\hbar}Ht' }
 (p_{1x}(0,0) -c \theta [p_{1x}(0,0),K_x ]_P )
e^{ -\frac{i}{\hbar}Ht' } \nonumber \\
& =&
 p_{1x}(0,t') -c \theta
[p_{1x}(0,t'),K_x(t')]_P \label{eq:12.39b}
\end{eqnarray}

\noindent If Lorentz transformations were exact, then results of both
methods would be identical and comparing (\ref{eq:12.38}) and
(\ref{eq:12.39}) we would obtain

\begin{eqnarray*}
\frac{1}{c} \frac{dr_{1x}(0,t)}{dt}  r_{1x}(0,t)  \theta = - c
\theta
 [r_{1x}(0,t),K_x(t)]_P
\end{eqnarray*}

\noindent or using $\frac{dr_{1x}}{dt} =  [r_{1x},H]_P =
\frac{\partial H}{\partial p_{1x}} $ and
 $ [r_{1x},K_x]_P = \frac{\partial K_x}{\partial p_{1x}}$

\begin{eqnarray*}
c^2 \frac{\partial K_x}{\partial p_{1x}} = - r_x \frac{\partial
H}{\partial p_{1x}}
\end{eqnarray*}

\noindent Similar arguments lead us to the general case
($i,j=1,2,3$)

\begin{eqnarray}
c^2 \frac{\partial K_j}{\partial p_{1i}} = -r_j \frac{\partial
H}{\partial p_{1i}} \label{eq:dkdp}
\end{eqnarray}

\noindent Comparing equations (\ref{eq:12.39b}) and
(\ref{eq:12.36}) we  get

\begin{eqnarray*}
p_{1x}(0,t) - \frac{h_1(0,t) \theta}{c} &=&
 p_{1x}\left(0,t - \frac{r_{1x}(0,t)}{c}  \theta \right)
-c \theta
[p_{1x}(0,t'),K_x(t')]_P \\
&\approx& p_{1x}(0,t) - \frac{r_{1x}(0,t) \theta}{c} \frac{\partial
p_{1x}(0,t)}{\partial t} -c \theta
[p_{1x}(0,t'),K_x(t')]_P \\
&\approx&
 p_{1x}(0,t) - \frac{r_{1x}(0,t) \theta}{c } [ p_{1x}(0,t), H]_P
-c \theta [p_{1x}(0,t),K_x(t)]_P
\end{eqnarray*}

\noindent and

\begin{eqnarray*}
c^2 [p_{1x},K_x]_P = -r_{1x} [ p_{1x}, H]_P +  h_1
\end{eqnarray*}

\noindent In the general case ($i,j = 1,2,3$) we have

\begin{eqnarray}
c^2 \frac{\partial K_j}{\partial r_{1i}} &=&- r_{1j} \frac{\partial
H}{\partial r_{1i}}  +  \delta_{ij}h_1 \label{eq:dkdr}
\end{eqnarray}

Putting equations (\ref{eq:dkdp}) - (\ref{eq:dkdr}) together, we conclude
that if trajectories of interacting particles transform by Lorentz,
then the following equations must be valid

\begin{eqnarray}
c^2\frac{\partial K_k}{ \partial \mathbf{p}_1} &=& -r_{1k}
\frac{\partial H}{ \partial \mathbf{p}_1}
\label{eq:12.40}\\
c^2\frac{\partial K_k}{ \partial \mathbf{p}_2} &=&- r_{2k}
\frac{\partial H}{ \partial \mathbf{p}_2}
\label{eq:12.41}\\
c^2\frac{\partial K_k}{ \partial r_{1i}} &=& -r_{1k} \frac{\partial
H}{ \partial r_{1i}} +  \delta_{ik}h_1
\label{eq:12.42}\\
c^2\frac{\partial K_k}{ \partial r_{2i}} &=& -r_{2k} \frac{\partial
H}{ \partial r_{2i}} +  \delta_{ik}h_2 \label{eq:12.43}
\end{eqnarray}

\noindent Our next goal is to show that these equations lead to a
contradiction. Taking derivatives of (\ref{eq:12.40}) by
$\mathbf{p}_2$ and  (\ref{eq:12.41}) by $\mathbf{p}_1$ and
subtracting them we obtain

\begin{eqnarray*}
\frac{\partial^2 H}{ \partial \mathbf{p}_2 \partial \mathbf{p}_1} =
0
\end{eqnarray*}

\noindent In a similar way we get\footnote{Here we used $\partial h_2/\partial \mathbf{r}_1 = \partial \sqrt{m_2c^4 + p_2^2c^2}/\partial \mathbf{r}_1 = 0$ and $\partial h_2/\partial \mathbf{p}_1 = \partial h_1/\partial \mathbf{r}_2 = \partial h_1/\partial \mathbf{p}_2 = 0$.}

\begin{eqnarray*}
\frac{\partial^2 H}{ \partial \mathbf{r}_2 \partial \mathbf{r}_1}
&=& \frac{\partial^2 H}{ \partial \mathbf{r}_2 \partial \mathbf{p}_1}
=\frac{\partial^2 H}{ \partial \mathbf{p}_2 \partial \mathbf{r}_1}
= 0
\\
\end{eqnarray*}

\noindent The only non-zero  cross-derivatives are

\begin{eqnarray*}
\frac{\partial^2 H}{ \partial \mathbf{p}_1 \partial \mathbf{r}_1}
&\neq& 0
\\
\frac{\partial^2 H}{ \partial \mathbf{p}_2 \partial \mathbf{r}_2}
&\neq& 0
\end{eqnarray*}

\noindent Therefore, only pairs of arguments $(\mathbf{p}_1,
\mathbf{r}_1)$ and $(\mathbf{p}_2, \mathbf{r}_2)$ are allowed to be
together in $H$, and we can represent the full Hamiltonian in the
form

\begin{eqnarray*}
H = H_1 (\mathbf{p}_1, \mathbf{r}_1) +  H_2 (\mathbf{p}_2,
\mathbf{r}_2 )
\end{eqnarray*}

\noindent This result means that the force acting on the particle 1 does not depend on the state (position and momentum) of the particle 2.

\begin{eqnarray*}
\mathbf{f}_1 = \frac{\partial \mathbf{p}_1}{\partial t} = [\mathbf{p}_1, H]_P =
[\mathbf{p}_1, H_1(\mathbf{p}_1, \mathbf{r}_1)]_P
\end{eqnarray*}

\noindent and \emph{vice versa}. Therefore, both particles move independently, i.e., there is no interaction. This already proves the statement of the Theorem. A stronger result can be obtained if we disregard the possibility of non-cluster separable (or long-range) interactions. From the Poisson bracket with the total momentum we obtain

\begin{eqnarray}
0 &=& [\mathbf{P}, H]_P = [\mathbf{p}_1 + \mathbf{p}_2, H]_P =
-\frac{\partial H_1 (\mathbf{p}_1, \mathbf{r}_1)}{\partial
\mathbf{r}_1} -  \frac{\partial H_2 (\mathbf{p}_2, \mathbf{r}_2)}
{\partial \mathbf{r}_2} \label{eq:12.44}
\end{eqnarray}

\noindent Since two terms on the right hand side of (\ref{eq:12.44})
depend on different variables, we must have

\begin{eqnarray*}
 \frac{\partial H_1 (\mathbf{p}_1, \mathbf{r}_1)}{\partial
\mathbf{r}_1} = -  \frac{\partial H_2 (\mathbf{p}_2, \mathbf{r}_2)}
{\partial \mathbf{r}_2} = \mathbf{C}
\end{eqnarray*}

\noindent where $\mathbf{C}$ is a constant vector. Then the
Hamiltonian can be written in the form

\begin{eqnarray*}
H =  H_1 (\mathbf{p}_1) +  H_2 (\mathbf{p}_2) +
 \mathbf{C} \cdot (\mathbf{r}_1 - \mathbf{r}_2)
\end{eqnarray*}

\noindent To ensure the cluster separability (=short range) of the interaction we
must set $\mathbf{C} = 0$. Then the resulting form of the Hamiltonian $H =  H_1
(\mathbf{p}_1) +  H_2 (\mathbf{p}_2)$ implies that the force acting on
the particle 1 vanishes

\begin{eqnarray*}
\mathbf{f}_1 &=& \frac{\partial \mathbf{p}_1}{ \partial t} =  [
\mathbf{p}_1, H]_P =  [ \mathbf{p}_1, H_1 (\mathbf{p}_1)]_P = 0
\end{eqnarray*}

\noindent The same is true for the force acting on the particle 2.
\end{proof}
\bigskip

The above theorem shows us that if particles have Lorentz-invariant
``worldlines,'' then they are not interacting. In special
relativity, Lorentz transformations are assumed to be exactly and
universally valid (see Assertion \ref{lorentz-trnsf2}). Then the
theorem leads to the conclusion that inter-particle
interactions are simply impossible. This explains why this theorm is commonly referred to as
``the no-interaction theorem.''
 Of course, it is absurd to think that there
are no interactions in nature. So, in current literature there are
two interpretations of this result. One interpretation is that the
Hamiltonian dynamics cannot properly describe interactions. Then a
variety of non-Hamiltonian versions of dynamics were suggested
\cite{Keister, Van_Dam, covariant}. Another view is that variables
$\mathbf{r}$ and $\mathbf{p}$ do not describe real observables of
particle positions and momenta, or even that the notion of particles
themselves becomes irrelevant in quantum field theory. Quite often
the Currie-Jordan-Sudarshan theorem is considered as an evidence that particle-based description of nature is not adequate, and one should seek a field-based approach \cite{Boyer-08}.

 However, we reject both these explanations. Non-Hamiltonian versions
 of particle dynamics contradict fundamental postulates of relativistic quantum
theory, which were formulated and analyzed throughout this book. We
also would like to stick to the idea that physical world is
described by particles with well-defined positions, momenta, spins,
etc. We will also assume that these physical particles interact via instantaneous potentials
obtained in the dressed particle approach to QFT. So, for us the
only way out of the paradox is to admit that Lorentz transformations
of special relativity are not applicable to observables of
interacting particles. Then
from our point of view, it is more appropriate to call Theorem
\ref{Theorem12.1} the ``no-Lorentz-transformation theorem.''  In
contrast to the special-relativistic Assertion
\ref{lorentz-trnsf2}, boost transformations of observables of
individual particles should depend on the observed system, its
state and on interactions acting in the system. So, we conclude that boost
transformations are dynamical.

\section{Comparison with special relativity} \label{ss:fields-disc}

In this section we would like to discuss the physical significance
of our conclusion about the dynamical character of boosts and how this contradicts  Einstein's special
relativity. In subsection \ref{ss:fields-disc2} we will analyze
existing proofs of Lorentz transformations and show that these
proofs do not apply to transformations of observables of interacting particles. In
subsection \ref{ss:exp-special} we are going to discuss experimental
verifications of special relativity. We will see that in  most cases
these experiments are dealing with free particles, for which our theory makes the same predictions as the standard special relativity. When genuine interacting systems are observed (such as decaying particles), the measurements are not accurate enough to register our predicted deviations from Einstein's theory. So, it is not surprising that this theory has withstood all experimental tests so far. In subsections
\ref{ss:poincare-manifest} - \ref{ss:dynamical-relativity} we will suggest
that  such fundamental assertions of relativistic theories as the
manifest covariance and the 4-dimensional  Minkowski space-time continuum should be rejected.

\subsection{On ``derivations'' of Lorentz transformations}
\label{ss:fields-disc2} \index{Lorentz transformations}

 Einstein based his special relativity
\cite{Einstein_1905} on two postulates. One of them was the
principle of relativity. The other was the independence of the speed
of light on the velocity of the source and/or observer. Both these
statements remain true in our theory as well (see our Postulate
\ref{postulateA} and Statement \ref{statementB}). Then Einstein
discussed a series of thought experiments with measuring rods,
clocks, and light rays, which demonstrated the relativity of
simultaneity, the length contraction of moving rods, and the
slowing-down of moving clocks. These observations were
formalized in Lorentz formulas (\ref{eq:lor-transform-t}) -
(\ref{eq:lor-transform-comp}),
 which supposedly connected  times and
positions of a localized event in different moving reference frames.
As we demonstrated in Theorem \ref{theorem:lorentz-tr}, our approach
leads to exactly the same transformation laws for events
associated with non-interacting particles.  So far our approach and
special relativity are in complete agreement.

Note that although Einstein's relativity postulate had
universal applicability to all kinds of events and processes, his
``invariance of the speed of light'' postulate is only relevant to
freely propagating light pulses. So, strictly speaking, all
conclusions made in \cite{Einstein_1905} can be applied only to
space and time coordinates of events (such as intersections of light
pulses) related in some way to the propagation of light. Nevertheless, in his work Einstein tacitly assumed\footnote{and this assumption was being repeated in all relativity textbooks ever since} that
the same conclusions should be extended to all other events independent on
their physical nature and on involved interactions.\footnote{see Assertion
\ref{lorentz-trnsf2}}

 There is
a large number of publications  \cite{Schwartz, Field, Lee-Kalotas, Levy-Leblond, Sardelis, Polishchuk}, which claim that Lorentz
transformation formulas (\ref{eq:lor-transform-t}) -
(\ref{eq:lor-transform-comp})
 can be derived even without using the Einstein's second postulate. However, these works do not look conclusive.
There are two common features in these derivations, which we find
troublesome. First, they often assume an abstract (i.e., independent on
real physical processes and interactions) nature of events occurring in
 space-time points $(t,x,y,z)$. Second, they often postulate the isotropy
and homogeneity of space around these points \cite{Schwartz, Field, Lee-Kalotas, Levy-Leblond}.  The main problem with these approaches is that in physics we
should be interested in transformations of observables of real
interacting particles, rather than abstract space-time points. One cannot
make an assumption that transformations of these observables  are
completely independent of what occurs in the space surrounding the
particle and what are interactions of this particle with the rest of
the physical system.

One can reasonably assume that all directions
in space are exactly equivalent for a single isolated particle
\cite{Sardelis}, but this is not at all obvious when the particle
participates in interactions with other particles.
Suppose that we have two interacting particles 1 and 2 at some
distance from each other. Suppose that we want to derive boost
transformations for observables of the particle 1. Clearly, for this
particle different directions in space are not equivalent: For
example, the direction pointing to the particle 2 is different from
other directions. So, the assumption of spatial isotropy cannot
be applied in this case.

Sometimes the following argument is presented in order to justify the applicability of (\ref{eq:lor-transform-t}) - (\ref{eq:lor-transform-comp}) even for interacting systems. Suppose that we have two events $A$ and $B$ having the same coordinates $(\mathbf{r}, t)$ in the frame at rest. Suppose also that event $A$ is related to light pulses (therefore, Lorentz formulas are exactly applicable to it), but event $B$ is associated with some interacting system. If space-time coordinates of $A$ and $B$ transformed by different formulas, then we would have a seemingly intolerable situation in which events $A$ and $B$ coincide in the frame at rest, but they occur at different space-time points if observed from a moving frame. Therefore, the argument goes, all events, independent on their physical nature, must transform by the same universal formulas (\ref{eq:lor-transform-t}) - (\ref{eq:lor-transform-comp}).
Though seemingly reasonable, the above argument is not convincing. There is absolutely no experimental or theoretical support for the above ``coincidence postulate'' (i.e., that events, overlapping in one frame, overlap in all other frames as well).

Thus we conclude that  there are no compelling theoretical
reasons to believe in the universal validity of Lorentz transformations (\ref{eq:lor-transform-t}) - (\ref{eq:lor-transform-comp}). Note that special relativity first postulates these transformations and then tries to formulate dynamical (interacting) theories, which conform with this assumption. Our approach to relativistic dynamics is fundamentally different, in fact, opposite. We start with formulation of relativistic (=Poincar\'e invariant) interacting theory. Then we derive boost
transformations of particle observables using
standard formulas of quantum theory and see\footnote{in section
\ref{sc:new-approach}} that they are different from universal Lorentz formulas
(\ref{eq:lor-transform-t}) - (\ref{eq:lor-transform-comp}).
Correct boost transformations
depend on the state of the observed multiparticle system and on
interactions acting there.

 We also see\footnote{ in subsection
\ref{ss:inequivalence}} that ``geometric'' universality of boosts  contradicts
 the (well-established) dynamical character of time
translations.  A theory, in which time translations are dynamical
while space translations, rotations, and boost are kinematical,
cannot be invariant with respect to the Poincar\'e group. So,
ironically, the assumptions of kinematical boosts, universal Lorentz
transformations, and ``invariant worldlines'' are in conflict with
the principle of relativity. This contradiction is the main reason
for the ``no interaction'' Theorem \ref{Theorem12.1}.

\subsection{On experimental tests of special relativity}
\label{ss:exp-special}

Supporters of special relativity usually invoke an argument that
predictions of this theory were confirmed by experiments with
astonishing precision. This is, indeed, true. However, at a closer
inspection it appears that existing experiments cannot distinguish
between special relativity and the approach presented in this book. In some cases, this is because two theories really agree. In other cases, the disagreement is so small that the required precision is out of reach for modern technology.

From the preceding discussion it should be clear that Lorentz formulas of special relativity are exactly applicable to observables of non-interacting
particles and to \emph{total} observables of any
physical system,  whether interacting or
not. It appears that
almost all experimental tests of special relativity are concerned with these kinds of measurements: they either look at non-interacting (free)
particles or at total observables in a compound system. Below we
briefly discuss several major classes of such experiments
\cite{experimentSR, MacArthur, schleif, Roberts}.

 One class of experiments is related to measurements of the
frequency (energy) of light and its dependence on the movement of the
source and/or observer.\footnote{See subsections \ref{ss:doppler} and \ref{ss:doppler-effect}.} These Doppler effect experiments \index{Doppler effect}
\cite{Ives-Stilwell, Ives-Stilwell2, Kaivola, Hasselkamp}
 can be formulated either as measurements of the photon's
energy dependence on the velocity of the source (or observer) or as
velocity dependence of the energy level separation in the source.
These two interpretations were discussed in subsections
\ref{ss:doppler} and \ref{ss:doppler-effect}, respectively. In the
former interpretation, one is measuring the energy (or frequency) of
a free particle - the photon. In the latter interpretation,
measurements of the total energy (differences) in an interacting
system are performed. In both these formulations, predictions of our
theory exactly coincide with special relativity.

Another class of experiments is concerned with measuring
the speed of light and confirming its independence on
the movement of the source and/or observer. This class includes
interference experiments of \emph{Michelson-Morley}
\index{Michelson-Morley experiment} and \emph{Kennedy-Thorndike}
\index{Kennedy-Thorndike experiment} as well as direct speed measurements
\cite{Alvager}. These experiments are
performed with free photons or light rays, so, again, our
theory and special relativity make exactly the same predictions for
such non-interacting systems. The same is true for tests of relativistic kinematics, which
include relationships between velocities, momenta, and energies of
free massive particles as well as changes of these parameters after
particle collisions or decays.

An exceptional type of experiment where one \emph{can}, at least in
principle, observe the differences between our theory and special
relativity is the decay of fast moving unstable particles. In this
case we are dealing with a physical system in which the interaction
acts during a long time interval (of the order of particle's
lifetime), and there is a clearly observable time-dependent process
(the decay) which is controlled by the strength of the interaction.
It was established in chapter \ref{ch:decays} that experiments measuring decays of fast moving particles are not accurate enough to see the small difference between predictions of special relativity and RQD.

\subsection{Poincar\'e invariance vs. manifest covariance}
\label{ss:poincare-manifest}

From our above discussion it should be clear that there are two
rather different approaches to constructing relativistic theories.
One is the traditional approach pioneered by Einstein and Minkowski
and used in theoretical physics ever since.
This approach accepts without proof the validity of Assertion
\ref{lorentz-trnsf2} (the universality of Lorentz transformations)
and its various consequences, like Assertions \ref{assertionEE} (no
superluminal signaling) and \ref{manifest} (manifest covariance). It
also assumes the existence of space-time, its 4-dimensional
geometry, and universal tensor transformations of space-time
coordinates of events. The distinguishing feature  of this \emph{manifestly
covariant} approach
is that boost transformations of observables are
assumed to be interaction-independent and universal.  \index{manifest covariance}

In this book we take a  different viewpoint on relativity.
We would like to call it a \emph{Poincar\'e invariant} approach.
\index{Poincar\'e invariance}  It is built on two
fundamental Postulates: the principle of relativity (Postulate
\ref{postulateA}) and the laws of quantum mechanics from sections
\ref{sc:quant-mech} and \ref{ss:quant-obs}. From these Postulates we
found that

\begin{statement} [Poincar\'e invariance]
 Descriptions of the system in
different inertial reference frames are related by transformations
which furnish a unitary representation of the Poincar\'e group.
\label{rel-inv}
\end{statement}

\noindent For example, if $F$ is an operator of observable in the reference frame at rest, then in the moving frame the same observable is represented by the transformed operator

\begin{eqnarray*}
F' = e^{-\frac{ic}{\hbar}\mathbf{K} \vec{\theta}} F e^{\frac{ic}{\hbar}\mathbf{K} \vec{\theta}}
\end{eqnarray*}

Most textbooks in relativistic quantum theory tacitly assume that
the Poincar\'e invariance and the manifest covariance do not
contradict each other, in fact, they are often assumed to be
equivalent. However, it is important to realize that there is no
convincing proof of such an equivalence. For example, Foldy wrote

\begin{quote}
\emph{To begin our discussion of relativistic covariance, we would
like first to make clear that we are not in the least concerned with
appropriate tensor or spinor equations, or with ``manifest
covariance'' or with any other mathematical apparatus which is
intended to exploit  the space-time symmetry of relativity, useful
as such may be. We are instead concerned with the \emph{group} of
inhomogeneous Lorentz transformations as expressing the
inter-relationship of physical phenomena as viewed by different
equivalent observers in un-accelerated reference frames. That this
group has its basis in the symmetry properties of an underlying
space-time continuum is interesting, important, but not directly
relevant to the considerations we have in mind.} L. Foldy
\cite{Foldy}
\end{quote}

\noindent This issue was also discussed by H. Bacry  who came to a
similar conclusion

\begin{quote}
\emph{The \emph{Minkowski manifest covariance} cannot be present in
quantum theory but we want to preserve the \emph{Poincar\'e
covariance}.} H. Bacry \cite{Bacry}
\end{quote}

\subsection{Is there an observable of time?} \label{ss:fields-time}

Special relativity and the manifestly covariant approach to
relativistic physics adopt a ``geometrical'' viewpoint on Lorentz
transformations.\footnote{see Appendix \ref{sc:lorentz-time-pos}} In
these theories,  time and position are unified as components of one
4-vector, and they are treated on equal footing. Such an
unification implies that there should be certain similarity between
space and time coordinates. However, in quantum mechanics\footnote{and in our everyday experience} there is a
significant physical difference between space and time. Space
coordinates $x, y, z$ are attributes (observables) of a material
physical system -- a collection of particles. In the formalism of
quantum mechanics these coordinates are represented by (expectation)
values of the position operator $\mathbf{R}$. There are position operators for each particle in the system as well as the center-of-mass position operator. On the other hand,
time is not an observable in quantum mechanics.

In order to better understand this difference between $\mathbf{r}$ and $t$,
recall our definitions of measurements, clocks, and observables from
Introduction. The measurements yield values of observables, such
as positions, momenta, energies, etc, and these values depend on the nature of the measured physical system and on the state in which the system finds itself. ``Time'' is not in the list of
physical observables. Time is just a numerical label attached to each
measurement according to the reading of the clock at the instant of
the measurement. The clock is separate from the observed physical system. Clock readings do not depend on the kind of
system being measured and on its state. We can record time even
if we do not measure anything, even if there is no physical system
to observe in our laboratory.

The clock is the necessary component of any reference frame or
observer, but this component is different from any measuring apparatus.
In order to ``measure'' time the observer needs to look at hour and
minute hands or at the digital display of her laboratory clock. In practical
applications clocks are macroscopic classical systems, such that
there are no quantum uncertainties in the hand's positions, or these uncertainties are reduced to a minimum. Of
course, there is a certain logical controversy here. We know that
all systems (including clocks) obey the laws of quantum mechanics.
When we look at clock's hands we basically measure their positions,
while assuming that their velocities are exactly zero. This situation is
explicitly forbidden by the Heisenberg's uncertainty principle. So,
there should be some uncertainty associated with the measurement of the clock hand's position, which implies an uncertainty associated with
``measurements of time''. Does it mean that some ``quantum nature''
of time should be taken into account? The answer is \textbf{no}.
Only those systems which produce well-defined countable periodic
``ticks'' without any (or with negligible) quantum uncertainty are
suitable as good clocks. So, if our laboratory clock exhibits some annoying quantum fuzziness, then this is simply a bad clock that should be replaced by a more accurate and stable one. Similarly, in order to measure positions
one needs to have heavy macroscopic sticks as rulers whose length is
not subject to quantum fluctuations. The existence of such ideal
clocks and rulers is questionable from the formal theoretical point
of view. But there is no doubt that distances and time intervals can
be measured with very high precision in practice. So, for theoretical purposes, it is reasonable to assume the availability of ideal clocks and rulers, whose performance is not affected by quantum effects.

The clock and the observed physical system are two separate objects, and  time ``measurements'' do not involve any interaction between the
physical system and the measuring apparatus. Therefore, in quantum
mechanics there can be no ``operator of time'', \index{operator of
time} such that $t$ is the expectation value (or eigenvalue) of this
operator. All attempts to introduce time operator in quantum
mechanics were not successful.

There were numerous attempts to introduce the ``time of arrival''
observable (and a corresponding Hermitian operator) in quantum
mechanics, see, e.g., \cite{Oppenheim, Galapon, Wang, Grot} and
references cited therein. For example, one can mark a certain space
point $(X,Y,Z)$ and ask ``at what time the particle arrives at this
point?'' Observations can yield a specific value for this time
$T$, and this value depends on the particle's state. Of course,
these are important attributes of an observable. However, they are
not sufficient to justify the introduction of the ``time of arrival''
observable. According to our definitions from Introduction, any
observable is an attribute of the system that can be measured by
\emph{all} observers. The time of arrival is a different kind of
attribute. For those observers whose time label\footnote{Recall that
in our definition (see Introduction) observers are
``instantaneous'': each observer is characterized by a definite time
label. } is different from $T$ the particle is not at the point
$(X,Y,Z)$, so the time of arrival value is completely undefined. So,
one cannot associate the time of arrival with any true observable.
It is more correct to say that the ``time of arrival'' is a time
label of a particular inertial observer (or observers) for whom the
measured value of the particle's position coincides with the
pre-determined point $(X,Y,Z)$.

An alternative proposal to introduce the time operator was presented
in \cite{Nikolic}. The author suggested to define the action of the
time operator $\hat{T}$ on particle wave functions
$\psi(\mathbf{r},t)$ as

\begin{eqnarray*}
\hat{T}\psi(\mathbf{r},t) = t \psi(\mathbf{r},t)
\end{eqnarray*}

\noindent According to our postulate \ref{postulateJ}, the (assumed) existence
of such an observable implies existence of states (eigenstates of
$\hat{T}$) in which time acquires a definite fixed value $t_0$. The
wave function of the particle in such a state is zero at all times
except (small neighborhood of the) time $t_0$. Physically this means
that the particle was created spontaneously out of nothing, existed
for a short time interval around $t_0$ and then disappeared. Such
states violate all kinds of conservation laws, and they are clearly
unphysical.

\subsection{Is geometry 4-dimensional?} \label{ss:fields-maxw}

Our position in this book is that there is no ``symmetry'' between
space and time coordinates. So, there is no need for a 4-dimensional
``background'' continuum of special relativity. All we care about
(in both experiment and in theory) are particle observables (e.g.,
positions and momenta) and how they transform with respect to inertial
transformations (e.g., time translations and boosts) of observers.
Particle observables are given by Hermitian operators in the Hilbert
space of the system. Inertial transformations enter the theory
through the unitary representation of the Poincar\'e group in the
same Hilbert space. Once these ingredients are known, one can
calculate the effect of any transformation on any observable. To do
that, there is no need to make assumptions about the ``symmetry''
between space and time coordinates and to introduce a 4-dimensional
spacetime geometry. The clear evidence for non-universal,
non-geometrical and interaction-dependent character of boost
transformations was obtained in section \ref{sc:new-approach}. So,
we suggest that 4D Minkowski space-time should not be used in
physical theories at all.

Likewise, there is no basis for the special-relitivistic classification of physical quantities into 4-scalars, 4-vectors, 4-tensors, etc. In reality, boost transformations of observables in interacting systems  may have quite complicated interaction-dependent forms that deviate from universal linear Lorentz formulas.

Historical and philosophical discussion of the idea that
relativistic effects (such as length contraction and time dilation)
result from the dynamical behavior of individual physical systems rather
than from kinematical properties of the universal ``space-time
continuum'' can be found in the book \cite{Brown-book}. In our work
we go further and claim that the difference between ``dynamical''
and ``kinematical'' approaches is not just philosophical one. It has
real observable consequences. We have shown in section
\ref{sc:new-approach} that boost transformations are
interaction-dependent and that they cannot be reduced to simple
universal Lorentz formulas or
``pseudo-rotations'' in the Minkowski space-time. Then the ideas of the universal pseudo-Euclidean
space-time continuum and of the manifest covariance of physical laws\footnote{see Assertions \ref{lorentz-trnsf2} and \ref{manifest}} can be
accepted only as approximations. Additional physical arguments
against the notion of the Minkowski space-time can be found in
\cite{Bacry-04}.

We cannot deny that the Minkowski space-time idea turned out to be very
fruitful in the formalism of quantum field theory. However, in
section \ref{ss:are-fields-meas} we will
take a viewpoint that quantum fields are just formal
mathematical objects and that the 4-dimensional manifold on which
the fields are defined has nothing to do with real physical space
and time.

\subsection{``Dynamical'' relativity}
\label{ss:dynamical-relativity}

From the dynamical character of boosts advocated in this book one can predict some curious
effects which, nevertheless, do not contradict any experimental
observations. For example, our approach implies that two measuring
rods made from different materials (e.g., wood and tungsten) may
contract in slightly different ways when viewed from a moving frame
of reference. Another consequence is that  Einstein's time dilation formula
(\ref{eq:time-dilation}) may be not accurate. See
section \ref{sc:formalism}.

\begin{figure}
\centering
\includegraphics {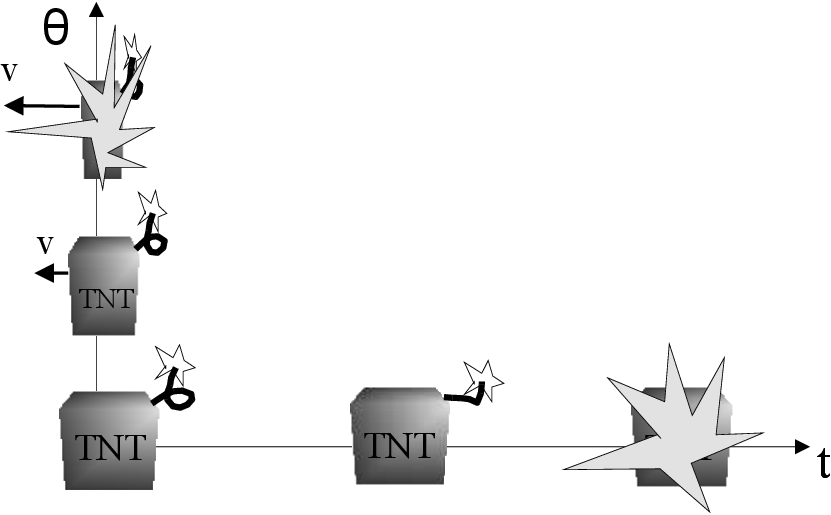} \caption{Non-trivial dynamics of an
isolated interacting system as a function of time ($t$) and boost
parameter ($\theta$). Three images on the $t$-axis illustrate the
usual time sequence of events associated with a piece of explosive
and an attached burning fuse. Three images on the $\theta$ axis show
how the unexploded device is perceived by a moving observer. If the
observer's velocity ($v = c \tanh \theta$) is low, then the observer
sees a moving (unexploded) device whose length is contracted along
the velocity's direction. This trivial (kinematical) change is
predicted also by special relativity. At higher speeds the moving
observer may notice more significant (dynamical) changes, e.g., the
device may be seen as exploded. Such non-trivial changes result from
the interaction-dependence of boost generators predictions by RQD.} \label{fig:10.2}
\end{figure}

The most significant difference between our approach and special
relativity concerns the effect of boosts on interacting systems. Let us
see how an isolated system is seen by time-translated and boosted
observers. In our cartoon \ref{fig:10.2} we placed images of the same
system on the plane $t-\theta$, where $t$ is the time parameter of
the observer and $\theta$ is its rapidity. Our approach and special
relativity agree about the effect of time translations: As the time
parameter increases, the system may undergo some dramatic changes
(e.g., an explosion) caused by internal forces acting in the system.
These changes result from the presence of interaction $V$ in the
Hamiltonian (the generator of time translations $H=H_0 + V$)  that
describes the system.

The fundamental disagreement is about the effects of boosts. From the point of view of special relativity, the boosted observer can see
only simple kinematical changes in the system. They include the change of the
system's velocity and relativistic contraction. These effects also
take place in our approach. However, in addition to them, we expect non-trivial changes, which result from the presence of the
interaction  $\mathbf{Z}$ in the generator of boost
transformations $\mathbf{K} = \mathbf{K}_0+\mathbf{Z}$. For example,
it is quite possible that for sufficiently high boost parameters
$\theta$ the system may look completely different, e.g., exploded
(the image in the upper left corner of Fig. \ref{fig:10.2}). For
this reason our approach can be characterized as \emph{dynamical
relativity} in contrast to \emph{kinematical relativity} of
Einstein's special theory.

\subsection{Does action-at-a-distance violate causality?}
\label{ss:charged-rqd3x}

 We saw in chapters \ref{sc:coulomb} and \ref{ch:theories}
that RQD describes interactions between particles in terms of
instantaneous potentials. However, textbooks teach us that
interactions cannot propagate faster than light:

\begin{quote}
\emph{In non-relativistic quantum mechanics, it is straightforward
to construct Hamiltonians which describe particles interacting via
long-range forces (for a simple example, consider two charged
particles interacting via a Coulomb force). However, the concept of
a long-range interaction \emph{prima facie} requires some sort of
preferred reference frame, which seems to cast doubt upon the
possibility of constructing such an interaction in a
relativistically covariant way.} D. Wallace \cite{Wallace2}
\end{quote}

\noindent The traditional viewpoint is that interactions between
particles ought to be retarded, i.e., they should propagate with the
speed of light. The usual argument in favor of this hypothesis
is the observation that faster-than-light interactions violate the
special relativistic ban on superluminal signals.\footnote{see
Appendix \ref{ss:super-signal}} If one accepts the validity of this
ban, then logically there is no other choice, but to accept also a
field-based approach, rather than the picture of directly
interacting particles advocated in this book. Indeed, interactions
are always accompanied by redistribution of the momentum and energy
between particles. If we assume that interactions are retarded, then
the transferred momentum-energy must exist in some form while
\emph{en route} from one particle to another. This implies existence
of some \emph{interaction carriers}  and
corresponding degrees of freedom not directly related to particle
observables. These degrees of freedom are usually associated with
fields, e.g., the electromagnetic field of Maxwell's theory. In
other words

\begin{quote}
\emph{...the interaction is a result of energy momentum exchanges
between the particles through the field, which propagates energy and
momentum and can transfer them to the particles by contact.} F.
Strocchi \cite{Strocchi}
\end{quote}

\begin{quote}
\emph{The field concept came to dominate physics starting with the
work of Faraday in the mid-nineteenth century. Its conceptual
advantage over the earlier Newtonian program of physics, to
formulate the fundamental laws in terms of forces among atomic
particles, emerges when we take into account the circumstance,
unknown to Newton (or, for that matter, Faraday) but fundamental in
special relativity, that influences travel no farther than a finite
limiting speed. For then the force on a given particle at a given
time cannot be deduced from the positions of other particles at that
time, but must be deduced in a complicated way from their previous
positions. Faraday's intuition that the fundamental laws of
electromagnetism could be expressed most simply in terms of fields
filling space and time was of course brilliantly vindicated in
Maxwell's mathematical theory.} F. Wilczek \cite{Wilczek}
\end{quote}

\noindent In this subsection we will argue against this traditional logic. Our point is that if the dynamical character of boosts is
properly taken into account, then instantaneous action-at-a-distance
does not contradict the principle of causality in all reference
frames.

Let us now consider two  particles interacting via instantaneous action-at-a-distance potentials. By definition, the  momentum is being exchanged between the two particles. In RQD there are no interaction carriers or intermediate fields or extra degrees of freedom where the transferred momentum could
be stored. Therefore,
when particle 1 loses some part of its momentum, particle 2
instantaneously acquires the same amount of momentum. Otherwise the momentum conservation law would be violated. So, in RQD the instantaneous character of interactions is not an approximation, but a necessity.

The dressed particle Hamiltonian in the 2-particle sector of the Fock
space is a function of positions and momenta of the two particles
$H^d = H^d(\mathbf{r}_1, \mathbf{p}_1, \mathbf{r}_2, \mathbf{p}_2)$. Particle
 trajectories can be obtained from equation (\ref{eq:int-corr})

\begin{eqnarray*}
\mathbf{r}_1(t) &=& e^{\frac{i}{\hbar}H^dt}\mathbf{r}_1
e^{-\frac{i}{\hbar}H^dt} \\
\mathbf{p}_1(t) &=&  e^{\frac{i}{\hbar}H^dt}\mathbf{p}_1
e^{-\frac{i}{\hbar}H^dt} \\
\mathbf{r}_2(t) &=&  e^{\frac{i}{\hbar}H^dt}\mathbf{r}_2
e^{-\frac{i}{\hbar}H^dt}\\
\mathbf{p}_2(t) &=&  e^{\frac{i}{\hbar}H^dt}\mathbf{p}_2
e^{-\frac{i}{\hbar}H^dt}
\end{eqnarray*}

\noindent  and the force\footnote{The same conclusions can be reached with the alternative definition of force $\mathbf{f} = m \frac{d^2\mathbf{r}}{dt^2}$, as in   subsection \ref{sc:force-def}.} acting on the particle 2

\begin{eqnarray}
\mathbf{f}_2 (t) &=& \frac{d}{dt} \mathbf{p}_2(t) = -\frac{i}{\hbar}[\mathbf{p}_2(t), H^d ] \nonumber \\
&\equiv& \mathbf{f}_2(\mathbf{r}_1(t), \mathbf{p}_1 ( t); \mathbf{r}_2
(t), \mathbf{p}_2(t)) \label{eq:f1}
\end{eqnarray}

\noindent depends on positions and momenta of both particles at the
same time instant $t$. Thus, interaction propagates instantaneously in the reference frame $O$.

 The
impossibility of superluminal signals is usually ``proven''  by
applying Lorentz transformations to space-time coordinates of two
causally related events and claiming that there exists a moving frame in which the temporal order of these events is reversed.\footnote{see Appendix
\ref{ss:super-signal}} However, we know from subsection
\ref{ss:no-interaction} that for systems with interactions Lorentz
transformations are no longer exact. So, the
special-relativistic ban on superluminal propagation of interactions may not be valid as
well.

Now consider  the above two-particle system  from the point of view of
a moving reference frame $O'$. Trajectories of particles 1 and 2 in
this frame are\footnote{Here $t'$
is time measured by the clock of observer $O'$; $\theta$ is the
rapidity of this observer. $\mathbf{K}^d$ is the dressed boost operator introduced in subsection \ref{ss:relat-invar-dressed}. }

\begin{eqnarray*}
\mathbf{r}_1(\theta, t') &=&e^{-\frac{ic}{\hbar} \mathbf{K}^d
\vec{\theta}} e^{\frac{i}{\hbar}H^dt'}\mathbf{r}_1
e^{-\frac{i}{\hbar}H^dt'}
e^{\frac{ic}{\hbar} \mathbf{K}^d \vec{\theta}} \\
\mathbf{p}_1(\theta, t') &=& e^{-\frac{ic}{\hbar} \mathbf{K}^d
\vec{\theta}} e^{\frac{i}{\hbar}H^dt'}\mathbf{p}_1
e^{-\frac{i}{\hbar}H^dt'} e^{\frac{ic}{\hbar} \mathbf{K}^d \vec{\theta}}\\
\mathbf{r}_2(\theta, t') &=& e^{-\frac{ic}{\hbar} \mathbf{K}^d
\vec{\theta}} e^{\frac{i}{\hbar}H^dt'}\mathbf{r}_2
e^{-\frac{i}{\hbar}H^dt'}
e^{\frac{ic}{\hbar} \mathbf{K}^d \vec{\theta}}\\
\mathbf{p}_2(\theta, t') &=& e^{-\frac{ic}{\hbar} \mathbf{K}^d
\vec{\theta}} e^{\frac{i}{\hbar}H^dt'}\mathbf{p}_2
e^{-\frac{ic}{\hbar}H^dt'} e^{\frac{i}{\hbar} \mathbf{K}^d
\vec{\theta}}
\end{eqnarray*}

\noindent The Hamiltonian in the reference frame $O'$ is

\begin{eqnarray}
H^d(\theta) &=&e^{-\frac{ic}{\hbar} \mathbf{K}^d \vec{\theta}} H^d
e^{\frac{ic}{\hbar} \mathbf{K}^d \vec{\theta}} \label{eq:mov-frame-h}
\end{eqnarray}

\noindent therefore the force acting on the particle 2 in this
frame

\begin{eqnarray}
\mathbf{f}_2 (\theta,t') &=& \frac{d}{dt'} \mathbf{p}_2(\theta, t') = - \frac{i}{\hbar}[\mathbf{p}_2(\theta, t'), H^d(\theta)] \nonumber \\
&=& - \frac{ic}{\hbar}[e^{-\frac{i}{\hbar} \mathbf{K}^d \vec{\theta}}
e^{\frac{i}{\hbar}H^dt'}\mathbf{p}_2 e^{-\frac{i}{\hbar}H^dt'}
e^{\frac{ic}{\hbar} \mathbf{K}^d \vec{\theta}}, e^{-\frac{i}{\hbar}
\mathbf{K}^d c\vec{\theta}} H^d
e^{\frac{ic}{\hbar} \mathbf{K}^d \vec{\theta}} ] \nonumber \\
&=& - \frac{i}{\hbar}e^{-\frac{ic}{\hbar} \mathbf{K}^d \vec{\theta}}[
e^{\frac{i}{\hbar}H^dt'}\mathbf{p}_2 e^{-\frac{i}{\hbar}H^dt'} ,
 H^d
] e^{\frac{ic}{\hbar} \mathbf{K}^d \vec{\theta}} \nonumber \\
&=& - \frac{i}{\hbar}e^{-\frac{ic}{\hbar} \mathbf{K}^d \vec{\theta}}[
\mathbf{p}_2 (0,t') ,
 H^d
] e^{\frac{ic}{\hbar} \mathbf{K}^d \vec{\theta}} =e^{-\frac{ic}{\hbar} \mathbf{K}^d \vec{\theta}}\mathbf{f}_2 (0,t')
e^{\frac{ic}{\hbar} \mathbf{K}^d \vec{\theta}} \nonumber \\
&=& e^{-\frac{ic}{\hbar} \mathbf{K}^d \vec{\theta}}
\mathbf{f}_2(\mathbf{r}_1(0,t'), \mathbf{p}_1 (0,t'); \mathbf{r}_2 (0,
t'), \mathbf{p}_2(0,t')) e^{\frac{ic}{\hbar} \mathbf{K}^d
\vec{\theta}} \nonumber \\
&=& \mathbf{f}_2(\mathbf{r}_1(\theta,t'), \mathbf{p}_1 (\theta,t');
\mathbf{r}_2 (\theta,t'), \mathbf{p}_2(\theta,t')) \label{eq:f1a}
\end{eqnarray}

\noindent  is a function of positions and momenta of both particles
at the same time instant $t'$. Moreover, in agreement with
the principle of relativity, this function $\mathbf{f}_2$ has
exactly the same form as in the reference frame at rest
(\ref{eq:f1}).
   Therefore, for the moving observer $O'$ the
interaction propagates instantaneously as for the observer at rest $O$.

If inter-particle potential is used to transmit information between two events $A$ (the cause) and $B$ (the effect), then in the reference frame at rest these two events are simultaneous. Then, due to the interaction dependence of $\mathbf{K}^d$,  these two events remain
simultaneous in all frames, so instantaneous potentials do not
contradict causality.  \index{causality}

The above arguments remain valid for any system of $N$ particles interacting via Poincar\'e invariant action-at-a-distance potentials.

\section{Are quantum fields necessary? } \label{ss:are-fields-meas}

 The general idea of RQD is that particles are the
most fundamental ingredients of nature and that everything we know
in physics can be explained as manifestations of quantum behavior of
particles interacting with each other at a distance. If this idea is correct, then the notion of fields
becomes redundant. On the other hand, it is also true that (quantum)
fields are in the center of all modern relativistic quantum
theories, and we actually started our formulation of RQD from the
quantum field version of QED in section \ref{sc:inter-qed}. This
surely looks like a contradiction. Then we are pressed to answer the
following question: \emph{what is the role of quantum fields in
relativistic quantum theory?}

\subsection{Dressing transformation in a nutshell}
\label{ss:dressing-nut}

\begin{figure}
\centering
\includegraphics {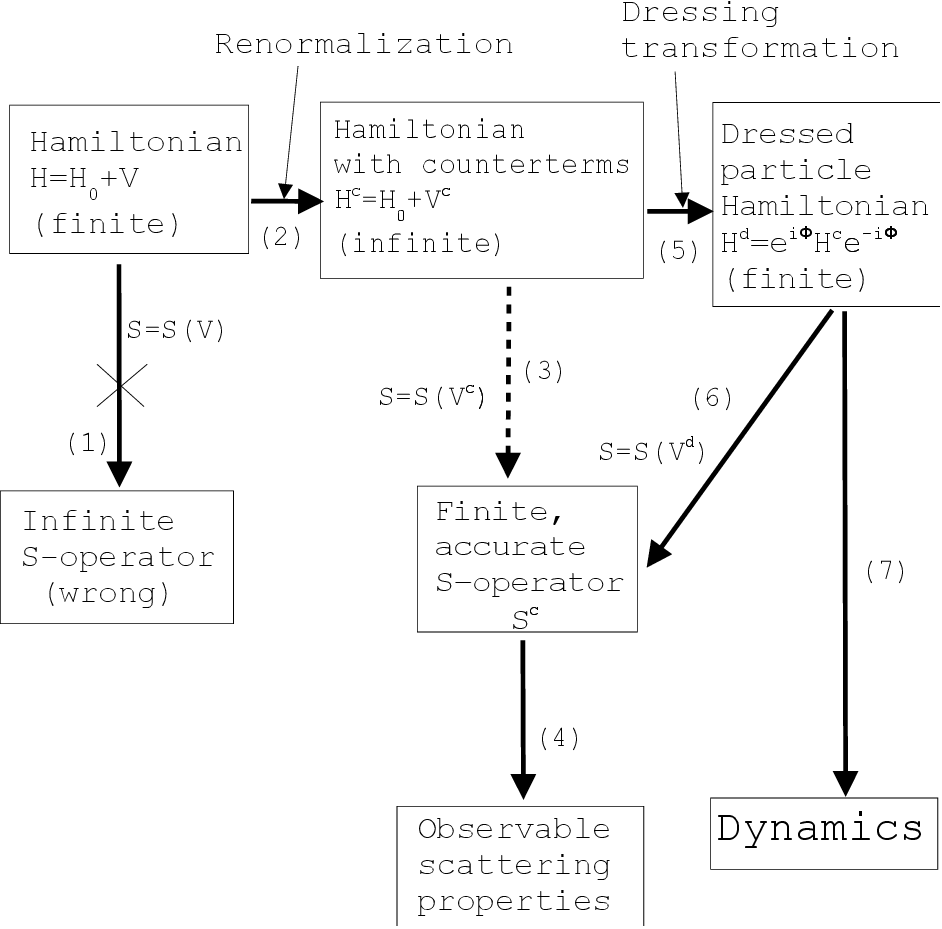} \caption{The logic of construction of
the dressed particle Hamiltonian $H^d = H_0 + V^d$.  $S(V)$ is the
perturbation formula (\ref{eq:7.63a}) that allows one to calculate
the $S$-operator from the known interaction Hamiltonian $V$.}
\label{fig:12.2}
\end{figure}

Before discussing the meaning of quantum fields, let us now review
the process by which we arrived to the finite dressed particle
Hamiltonian $H^d = H_0 + V^d$ in sections \ref{ss:darwin-breit} and \ref{ss:hydrogen}. We started
with the QED Hamiltonian $H = H_0 + V$
 in subsection \ref{ss:interaction-qed} (the upper
left box in Fig. \ref{fig:12.2})
 and
demonstrated some of its good properties, such as the Poincar\'e
invariance and the cluster separability.
 However, when we used this Hamiltonian to calculate the $S$-operator
beyond the lowest
non-vanishing perturbation order (arrow (1) in Fig. \ref{fig:12.2})  we
obtained meaningless infinite results. The solution to this
problem was given by  renormalization theory in chapter
\ref{ch:renormalization} (arrow (2)): infinite counterterms were
added to the Hamiltonian $H$, and a new Hamiltonian was obtained $H^c
= H_0 + V^c$.  Although the Hamiltonian $H^c$ was infinite, these
infinities canceled in the process of calculation of the
$S$-operator (dashed arrow (3)), and very accurate values for
observable scattering cross-sections and energies of bound states
were obtained (arrow (4)). As a result of the renormalization
procedure, the divergences were ``swept under the rug,'' and this
rug was the Hamiltonian $H^c$. This Hamiltonian was not
satisfactory: First, in the limit of infinite cutoff  matrix
elements of  $H^c$ on bare particle states were infinite. Second, the Hamiltonian $H^c$
contained \emph{unphys} terms like $a^{\dag} b^{\dag} c^{\dag}$ and
$a^{\dag}  c^{\dag} a$, which implied that in the course of time
evolution the (bare) vacuum state and (bare) one-electron states
rapidly dissociated into complex linear combinations of
multiparticle states.\footnote{Although the divergences in the Hamiltonian
$H^c$ can be avoided by the ``similarity renormalization'' approach
\cite{Glazek, Glazek2, Walhout}, the problem of unphysical time
evolution (=the instability of bare particles)
persists in all current formulations of QED that do not use
dressing.} Therefore, $H^c$ could not be used to describe dynamics
of interacting particles. To solve this problem, we applied a
unitary dressing transformation to the Hamiltonian $H^c$ (arrow (5))
and obtained a new ``dressed particle'' Hamiltonian

\begin{eqnarray}
H^d = e^{i \Phi} H^c e^{-i \Phi} \label{eq:hsphi}
\end{eqnarray}

\noindent We managed to select the unitary transformation $e^{i \Phi}$ so that all infinities from
$H^c$  were canceled out.\footnote{One can say that our approach has swept
the divergences under another rug. This time the rug is the phase
$\Phi$ of the transformation operator $e^{i \Phi}$. This
operator has no physical meaning, so there is no harm in choosing it
infinite.} In addition, the Poincar\'e invariance and cluster
separability of the theory remained intact, and the $S$-operator
computed with the dressed particle Hamiltonian $H^d$ was exactly the
same as the accurate $S$-operator of the renormalized QED (arrow (6)).

 The Hamiltonian $H^d$ of RQD  has a number of advantages over the Hamiltonian  $H^c$ of QED.
Unlike ``trilinear'' interactions in $H^c$, all terms in
$H^d$ have very clear and direct physical meaning and correspond to
real observable physical processes (see Table \ref{table:12.1}).
Both Hamiltonians $H^c$ and $H^d$ can be used to calculate
scattering amplitudes and energies of bound states.
 However, only with $H^d$ one can do that without regularization,
 renormalization, and other tricks. Only $H^d$ can describe the time evolution
in a simple and straightforward way (arrow (7)). It is also important that our ``quantum theory of
dressed particles'' (which is based on the Hamiltonian $H^d$) is conceptually much simpler than the ``quantum
theory of fields'' (which is based on the Hamiltonian $H^c$). RQD is similar to the ordinary quantum
mechanics: states are described by normalized wave functions, the
time evolution and scattering amplitudes are governed by a finite well-defined Hamiltonian, the
stationary states and their energies can be found by diagonalizing
this Hamiltonian.
 The only significant difference between RQD and conventional
 quantum mechanics is that in RQD the
number of particles is not conserved: particle creation and
annihilation can be adequately described. \index{annihilation}

The above derivation of the dressed particle
Hamiltonian $H^d$  involved a sequence of dubious
steps: ``canonical gauge field quantization $\to$ renormalization
$\to$ dressing''. Are these steps
inevitable ingredients of a realistic physical theory? Is nature meant to be that
complicated? Our answer to
these questions is ``no.'' Apparently, the ``first principles'' used
in constructions of traditional relativistic quantum field theories
(local fields, gauge invariance, etc.) are not fundamental.\footnote{see subsection \ref{ss:origins}}
Otherwise, we would not need such a painful procedure, involving
infinities and their cancelations, to derive a satisfactory dressed
particle Hamiltonian. We believe that it should be possible to build
a fully consistent relativistic quantum theory without ever invoking quantum fields. Unfortunately, this goal has not been
achieved yet, and we must rely on quantum fields and on the messy
renormalization and dressing procedures to arrive to an
acceptable theory of physical particles.

\subsection{What was the reason for having quantum fields? }
\label{ss:derivation}

In a nutshell, the traditional idea of quantum fields is that
particles that we observe in experiments -- photons, electrons,
protons, etc. -- are not the fundamental ingredients of nature. Allegedly, the
most fundamental ingredients are fields. For each kind of particle,
there exists a corresponding field -- a continuous all-penetrating
``substance'' that extends all over the universe. Dyson called it
\emph{``a single fluid which fills the whole of space-time''}
\cite{Dyson}. The fields are present even in situations when there
are no particles, i.e., in the vacuum. The fields cannot be measured
or observed by themselves. We can only see their excitations in the
form of small bundles of energy and momentum that we recognize as
particles. Photons are excitations of the photon field; electrons
and positrons are two kinds of excitations of the Dirac
electron-positron field, etc.

In this book we adopted a different attitude toward quantum fields. Our
viewpoint is that quantum fields are not the fundamental ingredients
of nature.  They are just formal mathematical objects (linear combinations of particle
creation and annihilation operators) which just happen to be
rather helpful in constructing relativistic quantum theories of
interacting particles.   However, it is not necessary to assign any physical
significance to quantum fields themselves.\footnote{It should be noted that in
non-relativistic (e.g., condensed matter) physics, quantum fields
may have perfectly valid physical meaning. However, in these cases
the field description is approximate and works only in the
low-energy long-distance limit. For example, the quantum field
description of crystal vibrations is applicable when the wavelength
is much greater than the inter-atomic distance. The excitations of
the crystal elastic field give rise to (pseudo-)particles called
\emph{phonons}. \index{phonon} The concept of renormalization also
makes a perfect sense in these systems. For example, the
\emph{polaron} \index{polaron} (a conduction band electron
interacting with lattice vibrations) has renormalized mass that is
different from the effective mass of the ``free'' conduction band
electron in a ``frozen'' lattice. In this book we are discussing
only fundamental relativistic quantum fields for which the above
relationships between quantum fields and underlying small-scale
physics do not apply.}

If (as usually suggested) fields are important ingredients of
physical reality, then we should be able to measure them. However,
the things that are measured in physical experiments are intimately
related to particles and their properties, not to fields. For
example, we can measure (expectation values of) positions, momenta,
velocities, angular momenta, and energies of particles as functions
of time (= trajectories).  In interacting systems of particles one
can probe the energies of bound states and their wave functions. A
wealth of information can be obtained by studying the connections
between values of particle observables before and after their
collisions (the $S$-matrix).
 All these measurements have a transparent and natural description in the language
of particles and operators of their observables.

On the other hand, field properties (field values at points, their
space and time derivatives, etc.) are not directly observable.
Fermion quantum fields are not Hermitian operators, so that, even
formally, they cannot correspond to quantum mechanical observables.
Even for the electric and magnetic fields of classical
electrodynamics their direct measurability is very
questionable. When we say that we have ``measured the electric
field'' at a certain point in space, we  have actually placed a test
charge at that point and measured the force  exerted on this
charge by surrounding charges. Nobody has ever measured electric and
magnetic fields themselves.

\subsection{Quantum fields and space-time}

The formal character of quantum fields is clear also from the fact
that their arguments  $t$ and $\mathbf{x}$ have no relationship to
measurable times and positions. The variable $t$ is a parameter,
which we used in (\ref{eq:9.50}) to describe the ``$t$-dependence''
of regular operators generated by the non-interacting Hamiltonian
$H_0$. As we explained in subsection \ref{ss:perturbation},
this $t$-dependence has no relationship to the observable time
dependence of physical quantities, but is rather added as a help in
calculations. Three variables $\mathbf{x}$ are just coordinates in
an abstract Minkowski space-time, and they should not be confused with physical positions of particles.

Arguments $\mathbf{x}$ of the fields should not be regarded as
eigenvalues of the Newton-Wigner  position operator. This can be seen from the simplest example of the scalar field taken at time $t=0$\footnote{See section 5.2 in \cite{book}.}

\begin{eqnarray*}
\psi(\mathbf{x}, 0) &=& \psi^+(\mathbf{x}, 0) + \psi^-(\mathbf{x}, 0) \\
&=& \int \frac{d \mathbf{p}}{\sqrt{2(2 \pi \hbar)\omega_{\mathbf{p}}}}e^{\frac{i}{\hbar}(\mathbf{p} \cdot \mathbf{x})} \alpha_{\mathbf{p}} + \int \frac{d \mathbf{p}}{\sqrt{2 (2 \pi \hbar )\omega_{\mathbf{p}}}}e^{-\frac{i}{\hbar}(\mathbf{p} \cdot \mathbf{x})} \alpha^{\dag}_{\mathbf{p}}
\end{eqnarray*}

\noindent The annihilation part $\psi^+(\mathbf{x}, 0)$ of this expression  cannot be regarded as an operator annihilating a particle at the space point $\mathbf{x}$, and operator $\psi^-(\mathbf{x}, 0)$ does not create the particle at point $\mathbf{x}$. The correct expressions for operators creating and annihilating one electron with Newton-Wigner position $\mathbf{x}$ and spin projection $\sigma$ can be obtained using formulas  from  subsection \ref{ss:position-representation}\footnote{Note the absence of the  denominator $\sqrt{\omega_{\mathbf{p}}}$ under the integral.}

\begin{eqnarray}
\alpha_{\mathbf{x}}
&=&\frac{1}{(2 \pi \hbar)^{3/2}} \int d \mathbf{p} e^{\frac{i}{\hbar}(\mathbf{p} \cdot \mathbf{x})} \alpha_{\mathbf{p}} \label{eq:alphax} \\
\alpha^{\dag}_{\mathbf{x}}
&=&\frac{1}{(2 \pi \hbar)^{3/2}} \int d \mathbf{p} e^{-\frac{i}{\hbar}(\mathbf{p} \cdot \mathbf{x})} \alpha^{\dag}_{\mathbf{p}} \label{eq:alphaxx}
\end{eqnarray}

\noindent Likewise, the product $\psi^-(\mathbf{x}, 0) \psi^+(\mathbf{x}, 0)$ cannot be interpreted as the spatial density of particles, but the product $\alpha^{\dag}_{\mathbf{x}} \alpha_{\mathbf{x}}$ has exactly this interpretation \cite{Wagner}.

As can be seen from formulas for scattering operators in subsection \ref{ss:perturbation} and from equations
(\ref{eq:10.7}) - (\ref{eq:10.8}), the parameters $t$ and
$\mathbf{x}$ are just integration variables, and they are not
present in the final expression for the fundamental measurable
quantity calculated in QFT - the $S$-matrix.

We certainly agree with the following two quotes:

\begin{quote}
\emph{Every physicist would easily convince himself that all quantum
calculations are made in the energy-momentum space and that the
Minkowski $x_{\mu}$ are just dummy variables without physical
meaning (although almost all textbooks insist on the fact that these
variables are not related with position, they use them to express
\emph{locality} of interactions!)} H. Bacry \cite{Bacry}
\end{quote}

\begin{quote}
\emph{It is important to note that the $\mathbf{x}$ and $t$ that
appear in the quantized field $A(\mathbf{x}, t)$ are not
quantum-mechanical variables but just parameters on which the field
operator depends. In particular, $\mathbf{x}$ and $t$ should not be
regarded as the space-time coordinates of the photon.} J. Sakurai
\cite{Sakurai}
\end{quote}

So, we arrive to the conclusion that quantum fields
$\psi(\mathbf{x},t)$ are simply formal linear combinations of
particle creation and annihilation operators.  Their arguments $t$
and $\mathbf{x}$ are some dummy variables, which are not related to
temporal and spatial properties of the physical system. Quantum
fields should not be regarded as ``generalized'' or ``second
quantized'' versions of wave functions. Their role is more technical
than fundamental: They provide convenient ``building blocks'' for
the construction of Poincar\'e invariant operators of potential
energy $V$ (\ref{eq:11.6}) - (\ref{eq:11.7}) and potential boost
$\mathbf{Z}$ (\ref{eq:11.8}) in the Fock space. That's all there is to quantum fields.

It seems appropriate to end this section with the following quote
from Mermin

\begin{quote}
\emph{But what is the ontological status of those quantum fields
that quantum field theory describes? Does reality consist of a
four-dimensional spacetime at every point of which there is a
collection of operators on an infinite-dimensional Hilbert space?
... But I hope you will agree that \emph{you} are not a continuous
field of operators on an infinite-dimensional Hilbert space. Nor,
for that matter, is the page you are reading or the chair you are
sitting in. Quantum fields are useful mathematical tools. They
enable us to calculate things.} N. D. Mermin \cite{Mermin}
\end{quote}

\chapter{CONCLUSIONS} \label{ch:summary}

\begin{quote}
\emph{Don't worry about people stealing your ideas. If your ideas
are any good, you'll have to ram them down people's throats.}

\small
\hspace{1in} Howard Aiken
\normalsize
\end{quote}

\vspace {0.5in}

In this book we presented a new relativistic quantum theory of interactions. Our approach is based on two claims that disagree with traditional textbook theories:

\begin{enumerate}
\item The primary constituents of matter are particles. These particles (electrons, protons, photons, etc.) obey the rules of quantum mechanics and interact with each other via position- and velocity-dependent instantaneous potentials. Potentials that change the number of particles are allowed as well.
\item The dynamical character of boosts. Perception of the system by a moving observer is different from that predicted by Einstein's special relativity. In addition to universal special-relativitic effects, such as length contraction and time dilation, we predict other phenomena whose exact nature and magnitude depend on the composition of the observed system and on interactions acting there.
\end{enumerate}

Our first claim about the primary role of particles contradicts the fundamental assumptions of such field-based approaches as quantum field theory and Maxwell's electrodynamics. We agree that quantum fields are useful mathematical constructs for building invariant interaction operators and calculating scattering amplitudes. However, for solving more general problems that include the time evolution and bound state properties, one is advised to switch to the dressed particle representation, which, incidentally, solves the problem of ultraviolet divergences. In the classical limit, the Hamiltonian theory of particles interacting via instantaneous potentials is a viable alternative to the traditional Maxwell's electrodynamics.

In the majority of experimental situations, predictions of our theory are either the same as in old approaches or the differences are too small to be measurable by modern techniques. So, experimental confirmation of RQD is rather challenging. The most compelling experimental evidence in favor of QED is observation of superluminal propagation of electromagnetic forces discussed in chapter \ref{ch:support}.

The most common argument against instantaneous interactions uses the special-relativistic ban on superluminal signal propagation. We explain this apparent contradiction by invoking our second claim that boost transformations are dynamical or interaction-dependent. This interaction-dependence of boosts follows naturally from the well-understood invariance of physical laws with respect to the Poincar\'e group. It is well-known that space translations and rotations of observers are purely kinematical and independent on interactions. On the other hand, it is also well-known that time translations induce highly non-trivial interaction-dependent (dynamical) changes in the observed system. Then, the Poincar\'e group structure demands that boosts have a non-trivial interaction-dependent effect as well. This simple observation has far-reaching consequences. In particular, it implies that universal Lorentz transformations of special relativity can be rigorously applied only to non-interacting systems. In the interacting case, the boost transformations should involve small, but crucially important, system-dependent and interaction-dependent corrections. Thus, in our approach, the Minkowski space-time is a non-rigorous, approximate concept.

The validity of special relativity is usually supported by reference to numerous experiments. However, at closer inspection, it appears that the majority of these measurements refer either to total observables of compound systems or to non-interacting particles. In these cases, predictions of our theory and special relativity are exactly the same. When truly interacting systems are observed (as in the case of ``time dilation'' in decays of moving particles), the differences between the two approaches ar extremely small.

\noindent \textbf{Summary:}

\begin{itemize}
\item Lorentz transformations of special relativity are not exact. Correct boost transformation laws must depend on the state
of the observed system and on interactions acting there.
\item The equivalence between space and time coordinates postulated
in special (and general) relativity is neither exact nor fundamental. The
4-dimensional Minkowski space-time formalism should not be used for
describing interacting relativistic systems.
\item Interactions between particles propagate  instantaneously.  This does not violate the principle of
causality.
\item Fields (either quantum or classical) should not be considered
as fundamental constituents of physical reality. Quantum fields are
formal mathematical constructs, which cannot be observed or
measured.
\item Classical electrodynamics can be formulated as a theory of
directly interacting particles, where electromagnetic fields (as
well as their momentum and energy) do not play any role.
\item The most direct way to confirm RQD experimentally is to measure the superluminal speed of propagation of bound (evanescent) electric and/or magnetic ``fields''.
\end{itemize}

\appendix

\part {MATHEMATICAL APPENDICES} \label{ch:appendix}

\chapter{Groups and vector spaces} \label{sc:sets}

\section{Groups} \label{ss:groups}

\emph{Group} \index{group} is a set where a \emph{product}
\index{group product} $ab$ of any two elements $a$ and $b$ is
defined. This product is also an element of the group, and the
following conditions are satisfied:

1. associativity: \index{associativity}

\begin{eqnarray}
(ab)c = a(bc)
\label{eq:A.1}
\end{eqnarray}

2. there is a unique \emph{unit element} \index{unit element} $e$
such that for any $a$

\begin{eqnarray}
ea = ae =a \label{eq:A.2}
\end{eqnarray}

3. for each element $a$ there is a unique \emph{inverse} element
\index{inverse element} $a^{-1}$ such that

\begin{eqnarray}
aa^{-1} = a^{-1}a = e
\label{eq:A.3}
\end{eqnarray}

\begin{figure}
\centering
\includegraphics {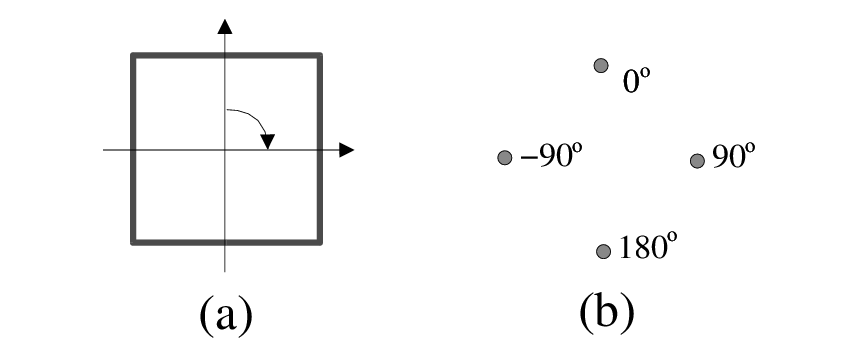} \caption{(a) Square; (b) the group of (rotational) symmetries
of the square.} \label{fig:A.4}
\end{figure}

In many cases a group can be described as a set of transformations
preserving certain symmetries. Consider, for example, a square shown
in Fig. \ref{fig:A.4}(a) and the set of rotations around its center. There are
four special rotations (by the angles $0^{\circ}, 90^{\circ},
180^{\circ}, -90^{\circ}$) which transform the square into itself.
This set of four elements (see Fig. \ref{fig:A.4}(b)) is the group of symmetries
of the square.  Apparently, $0^{\circ}$ is the unit element of
the group. The composition law of rotations leads us to the
 \emph{multiplication table} \ref{table:C.x} and the \emph{inversion table} \ref{table:C.y}
 for this simple group.
\index{group multiplication} \index{group inversion}

\begin{table}[h]
\caption{Multiplication table for the symmetry group of the square}
\begin{tabular*}{\textwidth}{@{\extracolsep{\fill}}c|cccc}
\hline
  &           $0^{\circ}$ & $90^{\circ}$ & $180^{\circ}$ & $-90^{\circ}$ \cr
\hline
$0^{\circ}$  & $0^{\circ}$ & $90^{\circ}$ & $180^{\circ}$ & $-90^{\circ}$ \cr
$90^{\circ}$ & $90^{\circ}$ & $180^{\circ}$ & $-90^{\circ}$ & $0^{\circ}$\cr
$180^{\circ}$ & $180^{\circ}$ & $-90^{\circ}$ & $0^{\circ}$ & $90^{\circ}$\cr
$-90^{\circ}$ & $-90^{\circ}$ & $0^{\circ}$ & $90^{\circ}$ & $180^{\circ}$ \cr
 \hline
\end{tabular*}
\label{table:C.x}
\end{table}

\begin{table}[h c]
\caption{Inversion table for the symmetry group of the square}
\begin{tabular*}{3in}{@{\extracolsep{\fill}}c|c}
\hline
  \ \ \ \ \ \ \ \ \ \ \ \ \ \ \  element & inverse element \cr
\hline
$0^{\circ}$  & $0^{\circ}$ \cr
$90^{\circ}$ & $-90^{\circ}$ \cr
$180^{\circ}$ & $180^{\circ}$ \cr
$-90^{\circ}$ & $90^{\circ}$ \cr
 \hline
\end{tabular*}
\label{table:C.y}
\end{table}

The group considered above is commutative (or \emph{Abelian}),
\index{Abelian group} because $ab = ba$ for any two elements $a$ and $b$
in the group. However, this property is not required for a  general group. For example, it is easy to
see that the group of rotational symmetries of a cube is not Abelian: A $90^{\circ}$ rotation of the cube about its $x$-axis followed by a $90^{\circ}$ rotation about the $y$-axis is a transformation that is different from these two rotations performed in the reverse order.

\section{Vector  spaces} \label{ss:vector-space}

A vector space $H$  is a set of objects (called \emph{vectors}
\index{vector} and further denoted by boldface letters $\mathbf{x}$)
with two operations: addition of two vectors and multiplication of a
vector by \emph{scalars}. \index{scalar} In this book we are
interested only in vector spaces whose scalars are either complex ($\mathbb{C}$) or
real ($\mathbb{R}$) numbers. If $\mathbf{x}$ and $\mathbf{y}$ are two vectors and
$a$ and $b$ are two scalars, then

\begin{eqnarray*}
a\mathbf{x} + b\mathbf{y}
\end{eqnarray*}

\noindent is also a vector. A vector space forms an Abelian group
\index{Abelian group} with respect to vector additions. This means
associativity \index{associativity}

\begin{eqnarray*}
\mathbf{(x+y) +z} = \mathbf{x+(y +z)},
\end{eqnarray*}

\noindent  existence of the group unity (denoted by $\mathbf{0}$ and
called \emph{zero vector}) \index{zero vector}

\begin{eqnarray*}
\mathbf{x+0} = \mathbf{0+x} = \mathbf{x}
\end{eqnarray*}

\noindent  and existence of the opposite
(additive inverse) element denoted by $-\mathbf{x}$

\begin{eqnarray*}
\mathbf{x+(-x)} = \mathbf{0},
\end{eqnarray*}

In addition, the following properties are postulated in the vector
space: The associativity \index{associativity} of scalar
multiplication

 \begin{eqnarray*}
a(b\mathbf{x}) = (a b)\mathbf{x}
\end{eqnarray*}

\noindent The distributivity of scalar sums:

 \begin{eqnarray*}
(a+b)\mathbf{x} = a\mathbf{x} + b\mathbf{x}
\end{eqnarray*}

\noindent The distributivity of vector sums:

 \begin{eqnarray*}
a (\mathbf{x} + \mathbf{y}) = a\mathbf{x} + a\mathbf{y}
\end{eqnarray*}

\noindent The scalar multiplication identity:

 \begin{eqnarray*}
1\mathbf{x} = \mathbf{x}
\end{eqnarray*}

\noindent We leave it to the reader to prove  from these axioms the
following useful results for an arbitrary scalar $a$ and a vector
$\mathbf{x}$

\begin{eqnarray*}
0\mathbf{x} &=& a\mathbf{0} = \mathbf{0} \\
(-a)\mathbf{x} &=& a (-\mathbf{x}) = - (a\mathbf{x}) \\
a\mathbf{x} &=& \mathbf{0}  \Rightarrow a = 0 \mbox{ or } \mathbf{x} =
\mathbf{0}
\end{eqnarray*}

An example of a vector space is the set of all columns of $n$
numbers\footnote{If $x_i$ are real (complex) numbers then this
vector space is denoted by $\mathbb{R}^n$ ($\mathbb{C}^n$).}

\begin{eqnarray*}
 \left[ \begin{array}{c}
x_1   \\
 x_2     \\
\vdots    \\
x_n
\end{array} \right]
\end{eqnarray*}

\noindent The sum of two columns is

\begin{eqnarray*}
 \left[ \begin{array}{c}
x_1   \\
 x_2     \\
\vdots    \\
x_n
\end{array} \right] +
 \left[ \begin{array}{c}
y_1   \\
 y_2     \\
\vdots    \\
y_n
\end{array} \right] =
 \left[ \begin{array}{c}
x_1+y_1   \\
 x_2+y_2     \\
\vdots    \\
x_n+y_n
\end{array} \right]
\end{eqnarray*}

\noindent The multiplication of a column by a scalar $\lambda$ is

\begin{eqnarray*}
\lambda \left[ \begin{array}{c}
x_1   \\
 x_2     \\
\vdots    \\
x_n
\end{array} \right] =
 \left[ \begin{array}{c}
\lambda x_1   \\
\lambda x_2     \\
\vdots    \\
\lambda x_n
\end{array} \right]
\end{eqnarray*}

A set of nonzero vectors $\{\mathbf{x}_i\}$ is called \emph{linearly
independent} \index{linear independence} if from

\begin{eqnarray*}
\sum_i a_i \mathbf{x}_i = \mathbf{0}
\end{eqnarray*}

\noindent it follows that $a_i = 0$ for each  $i$. A set of linearly
independent vectors $\mathbf{x}_i$ is called \emph{basis}
\index{basis} if by adding arbitrary nonzero vector $\mathbf{y}$ to
this set
 it is no longer linearly
independent. If $\mathbf{x}_i$ is a basis and $\mathbf{y}$ is an arbitrary nonzero vector, then equation

\begin{eqnarray*}
a_0 \mathbf{y}  + \sum_i a_i \mathbf{x}_i  = \mathbf{0}
\end{eqnarray*}

\noindent has a solution in which  $a_0 \neq 0$.\footnote{because otherwise we would
have $a_i = 0$ for all $i$, meaning that the full set $\{\mathbf{x}_i, \mathbf{y} \}$ is linearly independent in disagreement with our assumption.} This means that we can express an
arbitrary vector $\mathbf{y} $ as a linear combination of basis
vectors

\begin{eqnarray}
\mathbf{y} = - \sum_i\frac{ a_i}{a_0}  \mathbf{x}_i
= \sum_i y_i \mathbf{x}_i
\label{eq:A.4}
\end{eqnarray}

\noindent Note that any vector $\mathbf{y}$ has unique
\emph{components} \index{vector components} $y_i$ with respect to
the basis $\mathbf{x}_i$. Indeed, suppose we found another set of
components $y_i'$, so that

\begin{eqnarray}
\mathbf{y} =  \sum_i y_i' \mathbf{x}_i
\label{eq:A.5}
\end{eqnarray}

\noindent Then subtracting (\ref{eq:A.5}) from (\ref{eq:A.4}) we obtain

\begin{eqnarray*}
\mathbf{0} =  \sum_i (y_i' - y_i) \mathbf{x}_i
\end{eqnarray*}

\noindent and $y_i' = y_i$ since $\mathbf{x}_i $ are linearly independent.

One can choose many different bases in the same vector space. However, the number of vectors in any basis is the same,   and this number
is called the \emph{dimension} \index{dimension} of the vector space
$V$ (denoted $\dim V$). The dimension of the space of $n$-member
columns is $n$. An example of a basis set in this space is given by
$n$ vectors

\begin{eqnarray*}
 \left[ \begin{array}{c}
1   \\
 0     \\
\vdots    \\
0
\end{array} \right] ,
 \left[ \begin{array}{c}
0   \\
1     \\
\vdots    \\
0
\end{array} \right], \ldots,
 \left[ \begin{array}{c}
0   \\
0     \\
\vdots    \\
1
\end{array} \right]
\end{eqnarray*}

A \emph{linear subspace} \index{linear subspace} is a subset of
vectors in $H$ which is closed with respect to addition and
multiplication by scalars. For any set of vectors $\mathbf{x}_1,
\mathbf{x}_2, \ldots$ there is a \emph{spanning subspace} (or simply
\emph{span}) \index{span}  $Sp(\mathbf{x}_1, \mathbf{x}_2, \ldots)$
which is the set of all linear combinations $\sum_i a_i
\mathbf{x}_i$ with arbitrary coefficients $a_i$. The span of a
single non-zero vector $Sp(\mathbf{x})$ is also called a \emph{ray}.
\index{ray}

\chapter{Delta function and  useful integrals} \label{sc:delta}

Dirac's delta function $\delta(x)$ \index{delta function} is defined
by the property of the integral

\begin{eqnarray*}
\int \limits_{-a}^a f(x) \delta(x) dx &=& f(0)
\end{eqnarray*}

\noindent where $f(x)$ is any smooth function, and $a > 0$. The
delta function can be also defined by its integral representation

\begin{eqnarray*}
\frac{1}{2 \pi \hbar}\int \limits_{-\infty}^{\infty}
e^{\frac{i}{\hbar}ax} da &=& \delta(x)
\end{eqnarray*}

\noindent Another useful property is

\begin{eqnarray*}
\delta(ax) &=& \frac{1}{a} \delta(x)
\end{eqnarray*}

\noindent The delta function of a vector argument $\mathbf{r} = (x, y, z)$ is defined as

\begin{eqnarray*}
\delta(\mathbf{r}) &=& \delta(x) \delta(y) \delta(z)
\end{eqnarray*}

\noindent or

\begin{eqnarray}
\frac{1}{(2 \pi \hbar)^3}  \int e^{\frac{i}{\hbar} \mathbf{k} \cdot
\mathbf{r}} d \mathbf{k} = \delta(\mathbf{r}) \label{eq:delta-rep}
\end{eqnarray}

\noindent It has the property

\begin{eqnarray}
\frac{\partial^2}{\partial \mathbf{r}^2} \frac{1}{4 \pi r} &=& - \delta(\mathbf{r}) \label{eq:lapl-of-delta}
\end{eqnarray}

The \emph{step function} $\theta(t)$ \index{step function} is
defined as

\begin{eqnarray}
\theta(t) \equiv  \left \{
\begin{array}{c}
1, \mbox{  } if \mbox{  } t \geq 0 \\
0, \mbox{  } otherwise
\end{array}\right. \label{eq:theta-function}
\end{eqnarray}

\noindent  It has the following integral representation

\begin{eqnarray}
\theta(t) = -\frac{1}{2 \pi i}\int \limits_{- \infty}^{\infty} ds
\frac{e^{-ist}}{s+ i \epsilon} \label{eq:step}
\end{eqnarray}

Consider integral\footnote{In this derivation one can set
$\cos(\infty) = 0$ because in applications the
plane wave $e^{\frac{i}{\hbar}\mathbf{p}\mathbf{r}}$ in the
integrand does not have infinite extension. Typically it has a
smooth damping factor that makes it tend to zero at large values of
$\mathbf{r}$, so that $\cos(\infty)$ can be effectively taken as
zero. }

\begin{eqnarray}
&\ & \int \frac{d\mathbf{r}} { r}
e^{\frac{i}{\hbar}\mathbf{p}\mathbf{r}} = \int\limits_0^{\pi}
\sin\theta d\theta \int\limits_0^{2\pi} d \phi
\int\limits_0^{\infty} r^2 dr \frac{e^{\frac{i}{\hbar}pr \cos
\theta}}{r} = 2 \pi  \int\limits_{-1}^{1} d z
\int\limits_0^{\infty} dr r e^{\frac{i}{\hbar}prz} \nonumber \\
&=& 2 \pi \hbar \int\limits_0^{\infty} r dr
\frac{e^{\frac{i}{\hbar}pr}- e^{-\frac{i}{\hbar} pr} }{ipr} =
\frac{4 \pi \hbar}{p} \int\limits_0^{\infty}
dr \sin \left(\frac{pr}{\hbar} \right) \nonumber \\
& =& \frac{4 \pi \hbar^2}{p^2} \int\limits_0^{\infty} d\rho
\sin(\rho) = -\frac{4 \pi \hbar^2}{p^2} (\cos (\infty) - \cos(0)) =
\frac{4 \pi \hbar^2 }{p^2} \label{eq:A.89}
\end{eqnarray}

\noindent Next consider integral

\begin{eqnarray*}
K &=&   \int d\mathbf{x} d\mathbf{y}
\frac{e^{\frac{i}{\hbar}(\mathbf{p} \cdot \mathbf{x} + \mathbf{q}
\cdot \mathbf{y})}} {|\mathbf{x} - \mathbf{y}|}
\end{eqnarray*}

\noindent First we change the integration  variables

\begin{eqnarray*}
\mathbf{x} &=& \frac{1}{2}(\mathbf{z} + \mathbf{t}) \\
\mathbf{y} &=& \frac{1}{2}(\mathbf{z} - \mathbf{t}) \\
\mathbf{x-y} &=& \mathbf{t} \\
\mathbf{x+y} &=& \mathbf{z}
\end{eqnarray*}

\noindent The Jacobian of this transformation  is

\begin{eqnarray*}
J &\equiv& \det \left|\frac{\partial (\mathbf{x}, \mathbf{y})} {\partial
(\mathbf{z}, \mathbf{t})} \right| =1/8
\end{eqnarray*}

\noindent  Then, using integrals (\ref{eq:delta-rep}) and
(\ref{eq:A.89}), we obtain

\begin{eqnarray}
K&=&  \frac{1}{8}\int d\mathbf{t} d\mathbf{z}
\frac{e^{\frac{i}{2\hbar}(\mathbf{p} \cdot (\mathbf{z+t}) +
\mathbf{q} \cdot (\mathbf{z-t}))}} {t} = \frac{1}{8}\int d\mathbf{t}
d\mathbf{z} \frac{e^{\frac{i}{2\hbar}(\mathbf{z} \cdot
(\mathbf{p+q}) + \mathbf{t} \cdot (\mathbf{p-q}))}}
{t} \nonumber \\
&=& (2 \pi \hbar)^3 \delta(\mathbf{p+q}) \int d\mathbf{t}
\frac{e^{\frac{i}{2\hbar} \mathbf{t} \cdot (\mathbf{p-q})}} {t} =
\frac{(2 \pi \hbar)^6}{2\pi^2 \hbar} \frac{
\delta(\mathbf{p+q})}{p^2} \label{eq:int1/x}
\end{eqnarray}

\noindent Other useful integrals are

\begin{eqnarray}
 \int  \frac{ d \mathbf{k}}{k^2}
e^{\frac{i}{\hbar} \mathbf{k}\mathbf{r}}   &=& \frac{(2 \pi \hbar )^3}{4 \pi
\hbar^2 r} \label{eq:A.90} \\
\int \frac{ d \mathbf{k}  \mathbf{k}}{k^2} e^{\frac{i}{\hbar}
\mathbf{k}\mathbf{r}} &=& -i \hbar
\frac{\partial}{\partial \mathbf{r}}  \int   \frac{d
\mathbf{k}}{k^2} e^{\frac{i}{\hbar} \mathbf{k}\mathbf{r}} =
-\frac{i(2 \pi \hbar )^3}{4 \pi \hbar} \frac{\partial}{\partial
\mathbf{r}}\left(\frac{1}{ r}\right) = \frac{i (2 \pi \hbar )^3 \mathbf{r}}{4 \pi \hbar r^3} \nonumber \\
\label{eq:A.91} \\
 \int  \frac{ d \mathbf{k} \mathbf{q}\cdot
[\mathbf{k} \times \mathbf{p}]}{k^2} e^{\frac{i}{\hbar}
\mathbf{k}\mathbf{r}}
 &=&
\frac{ i(2 \pi \hbar )^3\mathbf{q} \cdot [\mathbf{r} \times \mathbf{p}]}{4 \pi \hbar
r^3} \label{eq:A.92}
\end{eqnarray}

\begin{eqnarray}
 \int  \frac{ d \mathbf{k}
 (\mathbf{q} \cdot \mathbf{k})(\mathbf{p} \cdot \mathbf{k})}{k^4}
e^{\frac{i}{\hbar} \mathbf{k}\mathbf{r}} &=& \frac{(2 \pi \hbar )^3}{8 \pi \hbar^2
r} \left[(\mathbf{q} \cdot \mathbf{p}) - \frac{(\mathbf{q} \cdot
\mathbf{r}) (\mathbf{p} \cdot
\mathbf{r})}{r^2} \right] \label{eq:A.93} \\
\int  \frac{ d \mathbf{k} (\mathbf{p}
 \cdot \mathbf{k})(\mathbf{q} \cdot \mathbf{k})}{k^2}
e^{\frac{i}{\hbar} \mathbf{k}\mathbf{r}} &=& \frac{(2 \pi \hbar )^3}{4 \pi r^3}
\left[(\mathbf{p} \cdot \mathbf{q}) - 3 \frac{(\mathbf{p} \cdot
\mathbf{r}) (\mathbf{q} \cdot \mathbf{r})}{r^2}\right] + \frac{1}{3}
(\mathbf{p} \cdot
\mathbf{q}) \delta(\mathbf{r}) \nonumber \\
&\mbox{ }& \label{eq:A.94} \\
 \int  \frac{ d \mathbf{k}
 }{k^4} e^{\frac{i}{\hbar}
\mathbf{k}\mathbf{r}}  &=& \mathcal{E} - \frac{(2 \pi \hbar )^3r}{8 \pi \hbar^4}
\label{eq:A.95}
\end{eqnarray}

\noindent where $\mathcal{E}$ is an infinite constant (see
\cite{Weinberg_1049}).

\begin{eqnarray}
 \int  d \mathbf{r} e^{-a \mathbf{r}^2 + \mathbf{b} \mathbf{r}}
= (\pi/a)^{3/2}e^{\mathbf{b}^2/(4a)} \label{eq:A.96}
\end{eqnarray}

\begin{lemma} [Riemann-Lebesgue \cite{Riemann}]
Fourier image of a smooth function tends to zero at
infinity.\index{Riemann-Lebesgue lemma}\label{lemma:Rim-Leb}
\end{lemma}

\noindent When talking about \emph{smooth functions} \index{smooth function} in this book we will
presume that these functions are continuous, can be differentiated
as many times as needed and do not contain singularities.

\chapter{Some lemmas for orthocomplemented lattices.}
\label{ss:theorems}

From axioms of orthocomplemented lattices\footnote{They are
summarized in Table \ref{table:2.1} as statements \ref{lemmaK1} - \ref{postulateM}.} one can prove a variety of
useful results

\bigskip

\begin{lemma} \label{Lemma4.1}

\begin{eqnarray}
 z \leq x \wedge y &\Rightarrow&  z \leq x
\label{eq:lemma4.1}
\end{eqnarray}
\end{lemma}
\begin{proof}   From Postulate \ref{postulateK5a} we have $x \wedge y \leq x$, hence
$z \leq x \wedge y \leq x$ and by the transitivity Lemma
\ref{lemmaK3} we obtain $z \leq x$.
\end{proof}
\bigskip

\begin{lemma} \label{Lemma4.2}

\begin{eqnarray}
x \leq y \Leftrightarrow x \wedge y = x \label{eq:lemma4.2}
\end{eqnarray}
\end{lemma}
\begin{proof}
   From $x \leq y$ and $x \leq x$ it follows by Postulate \ref{postulateK6a} that
$x \leq x \wedge y$. On the other hand, $x \wedge y \leq x$
(\ref{postulateK5a}). Lemma \ref{lemmaK2} then  implies $x \wedge y
= x$. The reverse statement follows from Postulate \ref{postulateK5a} written
in the form

\begin{eqnarray}
x \wedge y \leq y \label{eq:(4.9)}
\end{eqnarray}

\noindent  If $x \wedge y = x$, then we can replace the left hand
side of (\ref{eq:(4.9)}) with $x$ and obtain the left hand side of
 (\ref{eq:lemma4.2})
\end{proof}
\bigskip

\begin{lemma} \label{Lemma4.3}
For any proposition $z$

\begin{eqnarray}
 x \leq y  &\Rightarrow& x \wedge  z \leq  y \wedge z
\label{eq:lemma-4.3}
\end{eqnarray}
\end{lemma}
\begin{proof}   This follows from $x \wedge  z \leq
x \leq  y $ and $ x \wedge  z \leq z $ by using Postulate
\ref{postulateK6a}.
\end{proof}
\bigskip

\noindent One can also prove equations

\begin{eqnarray}
x \wedge x &=& x  \\
\emptyset \wedge x &=& \emptyset  \\
\mathcal{I} \wedge x &=& x  \\
\emptyset^{\perp} &=& \mathcal{I}  \label{eq:(4.14)}
\end{eqnarray}

\noindent which are left as an exercise for the reader.

Proofs of lemmas and theorems for orthocomplemented lattices are
facilitated by the following observation:
 Given an expression composed of lattice elements
we can form a \emph{dual} expression \index{duality} by the
following rules:

\begin{itemize}
\item 1) change places of $\wedge$ and $\vee$ signs;
\item 2) change the direction of the implication signs $\leq $;
\item 3) change $\emptyset$ to $\mathcal{I}$ and change $\mathcal{I}$ to $\emptyset$.
\end{itemize}

\noindent Then it is easy to see that all axioms in Table \ref{table:2.1} have
the property of \emph{duality}: \index{duality} Each axiom is either
self-dual or its dual is also a valid axiom. Therefore, for each
logical (in)equality, its dual is also a valid (in)equality. For
example, by duality we have from (\ref{eq:lemma4.1}),
(\ref{eq:lemma4.2}) and (\ref{eq:lemma-4.3}) - (\ref{eq:(4.14)})

\begin{eqnarray*}
x \vee y \leq z &\Rightarrow&  x \leq z \\
x \leq y &\Leftrightarrow& x \vee y = y \\
y \leq x &\Rightarrow& y \vee z \leq x \vee z \\
x \vee x &=& x \\
\mathcal{I} \vee x &=& \mathcal{I} \\
\emptyset \vee x &=& x \\
\mathcal{I}^{\perp} &=& \emptyset
\end{eqnarray*}

\chapter{Rotation group } \label{sc:rotations}

\section{Basics of the 3D space} \label{ss:3d-space}

\begin{figure}
\centering
\includegraphics {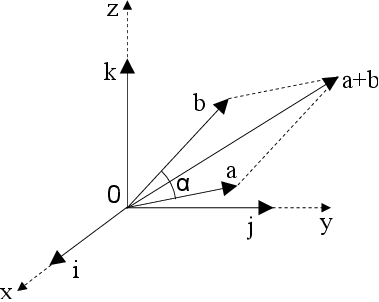} \caption{Some objects in the vector space
$\mathbb{R}^3$: the origin $\mathbf{0}$, the basis vectors
$\mathbf{i,j,k} $, a sum of two vectors $\mathbf{a}+\mathbf{b} $ via
the parallelogram rule.} \label{fig:A.1}
\end{figure}

Let us now consider the familiar 3D position space.
 This  space  consists of \emph{points}. We can arbitrarily
select one such point $\mathbf{0}$ and call it the \emph{origin}.
\index{origin} Then
 we can draw a \emph{vector} \index{vector} $\mathbf{a}$ from the origin to any other
point in space. We can also  define a  sum of two vectors (by the
parallelogram rule as shown in Fig. \ref{fig:A.1}) and the multiplication
of a vector by a real scalar. There is a natural definition of the
length of a vector $|\mathbf{a}|$ (also denoted by $a$) and the
angle $\alpha(\mathbf{a}, \mathbf{b})$ between two vectors
$\mathbf{a}$ and $\mathbf{b}$. Then the \emph{dot product}
\index{dot product} (or \emph{scalar product}
\index{scalar product}) of two vectors is defined by formula

\begin{eqnarray}
\mathbf{a} \cdot \mathbf{b} = \mathbf{b} \cdot \mathbf{a} = ab \cos \alpha
(\mathbf{a}, \mathbf{b})
\label{eq:A.6}
\end{eqnarray}

\noindent Two non-zero vectors are called \emph{perpendicular} or
\emph{orthogonal} \index{orthogonal vectors} if their dot product is
zero.

We can build an \emph{orthonormal basis}
\index{orthonormal basis} of 3 mutually perpendicular  vectors of
unit length $\mathbf{i}$, $\mathbf{j}$, and $\mathbf{k} $ along $x$,
$y$, and $z$ axes respectively.\footnote{  Let us agree that the
triple of basis vectors $(\mathbf{i},\mathbf{j}, \mathbf{k})$ forms
a \emph{right-handed system} \index{right-handed coordinate system}
as shown in Fig. \ref{fig:A.1}. Such a system is easy to recognize
by the following rule of thumb: If we point a corkscrew in the
direction of $\mathbf{k} $ and rotate it in the clockwise direction
(from $\mathbf{i} $ to $\mathbf{j} $), then the corkscrew will move
in the direction of vector $\mathbf{k}$.} Then each vector
$\mathbf{a}$ can be represented as a linear combination

\begin{eqnarray*}
\mathbf{a} &=& a_x \mathbf{i} + a_y \mathbf{j} + a_z \mathbf{k}
\end{eqnarray*}

\noindent or as a column  of its components or \index{coordinates}
\emph{coordinates}\footnote{So, physical space can be identified
with the vector space $\mathbb{R}^3$ of all triples of real numbers
(see subsection \ref{ss:vector-space}). We will mark vector indices
either by letters $x,y,z$ or by numbers 1,2,3, as convenient.}

\begin{eqnarray*}
 \mathbf{a} =  \left[ \begin{array}{c}
 a_x  \\
 a_y \\
 a_z \\
\end{array} \right]
\end{eqnarray*}

\noindent The \emph{transposed} \index{transposed vector} vector can
be represented as a row

\begin{eqnarray*}
\mathbf{a}^T = [a_x, a_y, a_z]
\end{eqnarray*}

\noindent One can easily verify that the dot product (\ref{eq:A.6}) can be
written in several equivalent forms

\begin{eqnarray*}
\mathbf{b} \cdot \mathbf{a} = \sum_{i=1}^3 b_i a_i = b_x a_x  + b_y a_y  + b_z
a_z = [b_x, b_y, b_z] \left[ \begin{array}{c}
 a_x  \\
 a_y \\
 a_z \\
\end{array} \right] =
\mathbf{b}^T \mathbf{a}
\end{eqnarray*}

\noindent where $\mathbf{b}^T \mathbf{a}$ denotes the usual ``row by column''
product
of the row $\mathbf{b}^T $ and  column $\mathbf{a} $.

The length of the vector $\mathbf{a}$ can be written  as
$ a \equiv |\mathbf{a}| = \sqrt{\mathbf{a} \cdot
 \mathbf{a}} \equiv \sqrt{a ^2}$, and the distance between two points (or vectors) $\mathbf{a}$
and $\mathbf{b} $ is defined as $d = |\mathbf{a} - \mathbf{b}| $.

\section{Scalars and vectors} \label{ss:scalars}

 There are two approaches to rotations, as well
as to any inertial transformation: \emph{active} and \emph{passive}.
\index{active transformation} An active rotation \index{active
rotation} rotates all objects around the origin while keeping the
orientation of basis vectors. A \emph{passive} rotation
\index{passive transformation} simply changes the directions of the
basis vectors and thus affects only components of ``real'' vectors but not the physical vectors themselves. Unless noted otherwise, we will use the passive
representation of rotations.

We call a quantity $\mathcal{A}$ a \emph{3-scalar} \index{scalar} if
it is not affected by rotations.  Examples of scalars are distances and angles.

Let us now find how rotations change the coordinates of vectors in
$\mathbb{R}^3$. By definition, rotations preserve the origin and
linear combinations of vectors, so the action of a rotation on a
column vector can be represented as multiplication by a $3 \times 3$
matrix $R$

\begin{eqnarray}
a'_i = \sum_{j=1}^3 R_{ij}a_j
\label{eq:A.7}
\end{eqnarray}

\noindent or in the matrix form

\begin{eqnarray}
\mathbf{a}' &=& R \mathbf{a} \label{eq:rot-matrix}\\
\mathbf{b'}^T &=& (R\mathbf{b})^T = \mathbf{b}^T R^T
\end{eqnarray}

\noindent where $R^T$ denotes the \emph{transposed matrix}.
\index{transposed matrix}

\section{Orthogonal matrices} \label{ss:orthogonal-m}

Since rotations preserve
distances and angles,  they also preserve the dot product:

\begin{eqnarray}
\mathbf{b} \cdot \mathbf{a} &=& \mathbf{b}^T \mathbf{a} =
(R\mathbf{b})^T (R \mathbf{a}) =  \mathbf{b}^T R^T R \mathbf{a}
\label{eq:A.8}
\end{eqnarray}

\noindent
The validity of  equation (\ref{eq:A.8}) for any
$\mathbf{a}$ and $\mathbf{b}$ implies that rotation matrices satisfy the
condition

\begin{eqnarray}
R^T R = I
\label{eq:A.9}
\end{eqnarray}

\noindent where $I$ denotes the unit matrix

\begin{eqnarray*}
I = \left[ \begin{array}{ccc}
 1 & 0 & 0  \\
 0 & 1 & 0 \\
 0 & 0 & 1  \\
\end{array} \right]
\end{eqnarray*}

\noindent Multiplying by the inverse matrix $R^{-1}$ from the right,
equation (\ref{eq:A.9})  can be also written as

\begin{eqnarray}
R^T =  R^{-1}
\label{eq:A.10}
\end{eqnarray}

\noindent This implies a useful property

\begin{eqnarray}
R\mathbf{b} \cdot \mathbf{a} &=& \mathbf{b}^T R^T \mathbf{a} =
\mathbf{b}^T R^{-1} \mathbf{a}
=  \mathbf{b} \cdot R^{-1}\mathbf{a} \label{eq:Rba}
\end{eqnarray}

In the coordinate notation, condition (\ref{eq:A.9}) takes the form

\begin{eqnarray}
\sum_{j = 1}^3 R_{ij}^TR_{jk} = \sum_{j = 1}^3 R_{ji}R_{jk} =
\delta_{ik}
\label{eq:A.11}
\end{eqnarray}

\noindent where $\delta_{ij}$ is  the   \emph{Kronecker delta
symbol} \index{Kronecker delta symbol}

\begin{eqnarray}
\delta_{ij} =  \left \{
\begin{array}{c}
1, \mbox{  } if \mbox{  } i = j \\
0, \mbox{  } if \mbox {  } i \neq j
\end{array}\right.       \label{eq:Kron}
\end{eqnarray}

\noindent Matrices satisfying condition (\ref{eq:A.10}) are called
\emph{orthogonal}. \index{orthogonal matrix}  Thus, any rotation has
a unique representative in the set of orthogonal matrices.

However,  not every orthogonal matrix $R$
corresponds to a rotation. To see that, we can write

\begin{eqnarray*}
1 &=& \det(I) =\det(R^TR) = \det(R^T) \det(R) = (\det(R))^2
\end{eqnarray*}

\noindent which implies  that if $R$ is orthogonal then $\det(R) = \pm 1$. Any
rotation can be connected by
a continuous path with the trivial rotation which is represented, of
course, by the unit matrix with unit determinant. Since continuous
transformations cannot abruptly change the determinant from 1 to -1,
 only matrices
with

\begin{eqnarray}
\det(R) = 1
\label{eq:A.12}
\end{eqnarray}

\noindent  have a chance to represent
rotations.\footnote{Matrices with
$\det(R) = -1$
describe rotations coupled with inversion (see subsection
\ref{ss:inversions}).} We conclude that rotations are
in one-to-one correspondence with
orthogonal matrices having a unit determinant.

The notion of a vector is more general
than just an arrow directed to a point in space. We will call any
triple of quantities $\vec{\mathcal{A}} = (\mathcal{A}_x,
\mathcal{A}_y, \mathcal{A}_z)$ a  \emph{3-vector} \index{3-vector}
if it transforms under rotations in the same way as vector arrows (\ref{eq:A.7}).

Let us now derive explicit forms  of rotation matrices.  Any rotation around the $z$-axis
does not change $z$-components of 3-vectors. The most general
 matrix satisfying this property can be
written as

\begin{eqnarray*}
R_z = \left[ \begin{array}{ccc}
  a & b & 0  \\
  c & d & 0  \\
  0 & 0 & 1
\end{array} \right]
\end{eqnarray*}

\noindent and condition (\ref{eq:A.12})  translates into $ad -  bc =
1$. One can verify directly that the inverse matrix  is

\begin{eqnarray*}
R_z^{-1} = \left[ \begin{array}{ccc}
  d & -b & 0  \\
  -c & a & 0  \\
  0 & 0 & 1
\end{array} \right]
\end{eqnarray*}

\noindent According to the property (\ref{eq:A.10}) we must have

\begin{eqnarray*}
a &=& d \\
b &=& -c
\end{eqnarray*}

\noindent therefore

\begin{eqnarray*}
R_z = \left[ \begin{array}{ccc}
  a & b & 0  \\
  -b & a & 0  \\
  0 & 0 & 1
\end{array} \right]
\end{eqnarray*}

\noindent The condition $\det(R_z) = a^2 + b^2 = 1$ implies that matrix $R_z$
depends on
 one parameter $\phi$ such that
$a =  \cos \phi$ and $b =  \sin \phi$

\begin{eqnarray}
R_z = \left[ \begin{array}{ccc}
  \cos \phi & \sin \phi & 0  \\
  -\sin \phi & \cos \phi & 0  \\
  0 & 0 & 1
\end{array} \right]
\label{eq:A.13}
\end{eqnarray}

\noindent Obviously, parameter $\phi$ is just the rotation
angle.\footnote{ Note that positive values of $\phi$ correspond to a
clockwise rotation (from $\mathbf{i}$ to $\mathbf{j}$) of the basis
 vectors which drives the corkscrew in the positive
$z$-direction.} The matrices for  rotations around the $x$- and $y$-axes are

\begin{eqnarray}
R_x = \left[ \begin{array}{ccc}
 1 &  0 &  0 \\
 0 &  \cos \phi & \sin \phi  \\
 0 &  -\sin \phi & \cos \phi
\end{array} \right]
\label{eq:A.14}
\end{eqnarray}

\noindent and

\begin{eqnarray}
R_y = \left[ \begin{array}{ccc}
  \cos \phi & 0 &  -\sin \phi  \\
  0  & 1 & 0 \\
  \sin \phi & 0 & \cos \phi
\end{array} \right]
\label{eq:A.15}
\end{eqnarray}

\noindent respectively.

\section{Invariant tensors} \label{ss:invariant}

\emph{Tensor} \index{tensor} of the second rank\footnote{Scalars and
vectors are sometimes called tensors of rank 0 and 1, respectively.}
$ \mathcal{A}_{ij}$ is defined as a set of 9 quantities, which
 depend on two indices and transform
as a vector with respect to each index

\begin{eqnarray}
 \mathcal{A}'_{ij}
= \sum_{kl=1}^3 R_{ik}R_{jl} \mathcal{A}_{kl} \label{eq:tensor-tr}
\end{eqnarray}

\noindent   Similarly, one can also define
 tensors of higher
\emph{rank}, \index{rank} e.g., $\mathcal{A}_{ijk}$.

There are two \emph{invariant tensors} \index{invariant tensor}
which play a special role
 because they  have the same components
independent on the orientation of the basis vectors. The first
invariant tensor is the \emph{Kronecker delta} \index{Kronecker delta
symbol} $\delta_{ij}$.\footnote{see equation (\ref{eq:Kron})} Its
invariance follows from the orthogonality of $R$-matrices
(\ref{eq:A.11}).

\begin{eqnarray*}
\delta'_{ij} = \sum_{kl=1}^3 R_{ik} R_{jl}\delta_{kl} = \sum_{k=1}^3 R_{ik}
R_{jk} = \delta_{ij}
\end{eqnarray*}

\noindent Another invariant tensor is the \emph{Levi-Civita symbol}
\index{Levi-Civita symbol} $\epsilon_{ijk}$, which is defined as
$\epsilon_{xyz}=\epsilon_{zxy}=\epsilon_{yzx}=-\epsilon_{xzy}=
-\epsilon_{yxz}=-\epsilon_{zyx}=1$, and all other components of
$\epsilon_{ijk}$ are zero. We show its invariance by applying an
arbitrary rotation $R$ to $\epsilon_{ijk}$.  Then

\begin{eqnarray}
 \epsilon'_{ijk}
&=& \sum_{lmn=1}^3 R_{il} R_{jm} R_{kn}\epsilon_{lmn} = R_{i1} R_{j2} R_{k3} + R_{i3} R_{j1} R_{k2} + R_{i2} R_{j3} R_{k1}
\nonumber \\
&\ & -R_{i2} R_{j1} R_{k3} - R_{i3} R_{j2} R_{k1} - R_{i1} R_{j3} R_{k2}
\label{eq:A.16}
\end{eqnarray}

\noindent The right hand side has the following properties:

\begin{itemize}
\item[1.] it is equal to zero if any two indices coincide: $i=j$ or $i=k$ or
$j=k$;
\item[2.]it does not change after cyclic permutation of indices $ijk$.
\item[3.]  $\epsilon'_{123} = \det (R) = 1$.
\end{itemize}

\noindent  These are the same properties as those used to define the Levi-Civita symbol above. So, the right hand side of (\ref{eq:A.16}) must have the same components
as $\epsilon_{ijk}$

\begin{eqnarray*}
 \epsilon'_{ijk} = \epsilon_{ijk}
\end{eqnarray*}

Using invariant tensors $\delta_{ij}$ and $\epsilon_{ijk}$ we can
convert between scalar, vector and tensor quantities, as shown in
Table \ref{table:C.1}. For example, any antisymmetric 3-tensor has 3 independent components, so it can be always represented as

\begin{eqnarray*}
 \mathcal{A}_{ij} = \sum_{k=1}^3 \epsilon_{ijk} V_k
\end{eqnarray*}

\noindent where $V_k$ are components of some 3-vector.

\begin{table}[h]
\caption{Converting between quantities of different rank using
invariant tensors}
\begin{tabular*}{\textwidth}{@{\extracolsep{\fill}}rcl}
\hline
Scalar $S$           &          $\to $   &   $  S\delta_{ij}$  (tensor)
\cr
Scalar $S$           &          $\to $   &   $S\epsilon_{ijk}$  (antisymmetric tensor)
\cr
Vector $V_i$         &          $\to$    & $\sum \limits_{k=1}^3
\epsilon_{ijk} V_{k}$ (antisymmetric tensor)\cr
Tensor $T_{ij}$      & $\to $&  $\sum \limits_{ij=1}^3 \delta_{ij}
T_{ji}$  (scalar) \\
Tensor $T_{ij}$      & $\to $   & $\sum \limits_{jk=1}^3\epsilon_{ijk}
T_{kj}$  (vector) \cr
 \hline
\end{tabular*}
\label{table:C.1}
\end{table}

Using invariant tensors one can also build a scalar or a vector from
two independent vectors $\mathbf{A}$ and $\mathbf{B}$. The scalar
is constructed by using the Kronecker delta

\begin{eqnarray*}
\mathbf{A} \cdot \mathbf{B} = \sum _{ij=1}^3 \delta_{ij} A_i B_j
\equiv A_1B_1 +A_2B_2+A_3B_3
\end{eqnarray*}

\noindent This is the usual dot product (\ref{eq:A.6}). \index{dot
product} The vector can be constructed using the Levi-Civita tensor

\begin{eqnarray*}
[\mathbf{A} \times \mathbf{B}]_i = \sum_{jk=1}^3 \epsilon_{ijk} A_j B_k
\end{eqnarray*}

\noindent This vector is called the \emph{cross product}
\index{cross product} (or \emph{vector product}
\index{vector product}) of $\mathbf{A}$ and $\mathbf{B}$. It has the
following components

\begin{eqnarray*}
[\mathbf{A} \times \mathbf{B}]_x = A_y B_z - A_z B_y \\ \mbox{ }
[\mathbf{A} \times \mathbf{B}]_y = A_z B_x - A_x B_z
\\ \mbox{ }
[\mathbf{A} \times \mathbf{B}]_z = A_x B_y - A_y B_x
\end{eqnarray*}

\noindent and   properties

\begin{eqnarray}
[\mathbf{A} \times \mathbf{B}] &=& -[\mathbf{B} \times \mathbf{A}] \nonumber \\
\ [\mathbf{A} \times [\mathbf{B} \times \mathbf{C}]] &=& \mathbf{B}
(\mathbf{A} \cdot \mathbf{C}) - \mathbf{C} (\mathbf{A} \cdot
\mathbf{B}) \label{eq:A.17}
\end{eqnarray}

\noindent The \emph{mixed product} \index{mixed product}  is a
scalar which can be build from three vectors with the help of the
Levi-Civita invariant tensor

\begin{eqnarray*}
[\mathbf{A} \times \mathbf{B}] \cdot \mathbf{C} &=&
\sum_{ijk=1}^3 \epsilon_{ijk} A_i B_j C_k
\end{eqnarray*}

\noindent Its properties are

\begin{eqnarray}
[\mathbf{A} \times \mathbf{B}] \cdot \mathbf{C} &=& [\mathbf{B}
\times \mathbf{C}] \cdot \mathbf{A} = [\mathbf{C} \times \mathbf{A}]
\cdot \mathbf{B} \label{eq:A.18} \\
\mbox{ } [\mathbf{A} \times \mathbf{B}] \cdot \mathbf{B} &=& 0
\nonumber
\end{eqnarray}

\section{Vector parameterization of rotations}
\label{ss:parameterization}

The matrix representation of rotations (\ref{eq:A.7}) is useful for
describing transformations of vector and tensor components. However,
sometimes it is more convenient to characterize rotation in a more
physical way by the rotation  axis and the rotation angle. In other
words, a rotation can be described by a single vector $\vec{\phi} =
\phi_x \mathbf{i} + \phi_y \mathbf{j}  + \phi_z \mathbf{k} $, such
that its direction represents the axis of the rotation and its
length $\phi \equiv |\vec{\phi}|$ represents the angle of the
rotation. So we can characterize any rotation by three real
numbers
 $\{ \vec{\phi} \} = \{ \phi_x, \phi_y, \phi_z\}$.\footnote{This
characterization is not unique: there are many vectors describing the
same rotation (see Appendix \ref{ss:double-valued}).}

\begin{figure}
\centering
\includegraphics {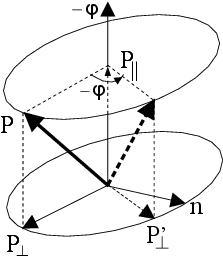}
\caption{Transformation of vector components under active rotation
 through the angle $-\phi$.} \label{fig:A.2}
\end{figure}

Let us now make a link between the matrix and
vector representations of rotations. First, we find the matrix $R_{\vec{\phi}}$
corresponding to the passive rotation $\{ \vec{\phi} \}$. Here it will be
convenient to consider the equivalent \emph{active} rotation
\index{active rotation} by the angle $\{ -\vec{\phi} \}$. Each
vector $\mathbf{P}$ in $\mathbb{R}^3$ can be decomposed into two parts: $\mathbf{P} = \mathbf{P}_{\parallel} + \mathbf{P}_{\perp}$
The first part $\mathbf{P}_{\parallel} \equiv (\mathbf{P} \cdot
\frac{\vec{\phi}}{\phi}) \frac{\vec{\phi}}{\phi}$ is parallel to the
rotation axis, and the second part $\mathbf{P}_{\perp} = \mathbf{P} -
\mathbf{P}_{\parallel}$ is perpendicular to the rotation axis (see Fig.
\ref{fig:A.2}).
 Rotation does not affect  the
parallel part of the vector, so after rotation

\begin{eqnarray}
\mathbf{P}_{\parallel}' = \mathbf{P}_{\parallel}
\label{eq:A.19}
\end{eqnarray}

\noindent  If  $\mathbf{P}_{\perp} = 0$  then rotation does
not change the vector $\mathbf{P}$ at all.
If $\mathbf{P}_{\perp} \neq 0$, we denote

\begin{eqnarray*}
\mathbf{n} = -\frac{[\mathbf{P}_{\perp} \times \vec{\phi}]}{\phi}
\end{eqnarray*}

\noindent the vector which is orthogonal to both $\vec{\phi}$ and
$\mathbf{P}_{\perp}$ and is equal to the latter in length.
Note  that the triple $(\mathbf{P}_{\perp}, \mathbf{n},
\vec{\phi})$ forms a right-handed system, just like vectors
$(\mathbf{i}, \mathbf{j},\mathbf{k})$.
 Then
the result of the passive rotation through the angle $\vec{\phi}$
 in the plane spanned by
vectors $\mathbf{P}_{\perp}$ and $\mathbf{n}$ is the same as a rotation
about the axis $\mathbf{k}$ in the plane spanned by vectors $\mathbf{i}$
and $\mathbf{j}$, i.e., it is given by the matrix
(\ref{eq:A.13})

\begin{eqnarray}
\mathbf{P}_{\perp}' &=& \mathbf{P}_{\perp} \cos \phi + \mathbf{n}
\sin \phi
\label{eq:A.20}
\end{eqnarray}

\noindent Combining equations (\ref{eq:A.19}) and (\ref{eq:A.20}) we obtain

\begin{eqnarray}
\mathbf{P}' = \mathbf{P}'_{\parallel} + \mathbf{P}_{\perp}' =
\left(\mathbf{P} \cdot \frac{\vec{\phi}}{\phi}\right) \frac{\vec{\phi}}{\phi}
(1 - \cos \phi) + \mathbf{P} \cos \phi - \left[\mathbf{P} \times
\frac{\vec{\phi}}{\phi}\right] \sin \phi \label{eq:A.21}
\end{eqnarray}

\noindent or in the component notation

\begin{eqnarray*}
P'_x = (P_x \phi_x + P_y \phi_y + P_z \phi_z) \frac{\phi_x}{\phi^2} (1 - \cos
\phi) +
P_x \cos \phi - (P_y \phi_z - P_z \phi_y) \frac{\sin \phi}{\phi} \\
P'_y = (P_x \phi_x + P_y \phi_y + P_z \phi_z) \frac{\phi_y}{\phi^2} (1 - \cos
\phi) +
P_y \cos \phi - (P_z \phi_x - P_x \phi_z) \frac{\sin \phi}{\phi} \\
P'_z = (P_x \phi_x + P_y \phi_y + P_z \phi_z) \frac{\phi_z}{\phi^2} (1 - \cos
\phi) +
P_z \cos \phi - (P_x \phi_y - P_y \phi_x) \frac{\sin \phi}{\phi}
\end{eqnarray*}

\noindent This transformation  can be also
represented in a matrix form.

\begin{eqnarray*}
 \mathbf{P}' =
R_{\vec{\phi}}^{-1} \mathbf{P} = R_{-\vec{\phi}} \mathbf{P}
\end{eqnarray*}

\noindent where the orthogonal  matrix $R_{\vec{\phi}}$  has the
following matrix elements

\begin{eqnarray*}
(R_{\vec{\phi}}) _{ij} = \cos \phi \delta_{ij} +  \sum_{k=1}^3 \phi_k
\epsilon_{ijk} \frac{ \sin \phi}{\phi} + \phi_i \phi_j \frac{1 - \cos
\phi}{\phi^2}
\end{eqnarray*}

\begin{eqnarray}
&\mbox{ }&R_{\vec{\phi}} \nonumber \\
&=& \left[ \begin{array}{ccc}
  \cos \phi + m_x^2 (1-\cos \phi) & m_x m_y (1-\cos \phi) - m_z \sin
\phi & m_xm_z (1-\cos \phi) + m_y \sin \phi  \\
 m_x m_y (1-\cos \phi) + m_z \sin \phi & \cos \phi + m_y^2 (1 - \cos
\phi) & m_y m_z(1- \cos \phi) -m_x \sin \phi  \\
  m_xm_z (1-\cos \phi) -m_y \sin \phi & m_ym_z (1-\cos \phi) +m_x \sin
\phi & \cos \phi + m_z^2 (1-\cos \phi)
\end{array} \right] \nonumber \\
\label{eq:r-vec-phi}
\end{eqnarray}

\noindent and $\mathbf{m} \equiv \vec{\phi}/\phi$.

Inversely, let us start from an arbitrary orthogonal matrix
$R_{\vec{\phi}}$ with $\det(R_{\vec{\phi}}) = 1$ and try to find the corresponding rotation vector
$\vec{\phi}$. Obviously, this vector is not changed by the
transformation $R_{\vec{\phi}}$, so

\begin{eqnarray*}
R_{\vec{\phi}} \vec{\phi}  = \vec{\phi}
\end{eqnarray*}

\noindent which means that $\vec{\phi}$ is eigenvector of the matrix
$R_{\vec{\phi}}$ with eigenvalue 1. Each orthogonal $3 \times 3$
matrix with unit determinant has eigenvalues $(1, e^{i\phi}, e^{-i\phi})$,\footnote{One
can check this result by using the explicit representation
(\ref{eq:r-vec-phi})} so that eigenvalue 1 is not degenerate. Then
the direction of the vector $\vec{\phi}$ is uniquely specified. Now
we need to find the length of this vector, i.e., the rotation angle
$\phi$. The trace of the matrix $R_{\vec{\phi}}$ is given by the sum
of its eigenvalues

\begin{eqnarray*}
Tr(R_{\vec{\phi}}) &=& 1 + e^{i\phi} + e^{-i\phi} = 1 + 2 \cos \phi
\end{eqnarray*}

\noindent Therefore, we can define  the function $\vec{\Phi}
(R_{\vec{\phi}}) = \vec{\phi}$ (which maps from the set of rotation
matrices to corresponding rotation angles) by the following rules:

\begin{itemize}
\item the direction of the rotation vector $\vec{\phi}$
coincides with the direction of the eigenvector of $R_{\vec{\phi}}$
with eigenvalue 1;
\item the length of the
 rotation angle $\phi$ is given by

\begin{eqnarray}
\phi = \cos ^{-1}\frac{Tr(R_{\vec{\phi}}) - 1}{2} \label{eq:A.23}
\end{eqnarray}
\end{itemize}

\noindent As expected, this formula is basis-independent, because the trace of a matrix does not depend on the basis (see Lemma
\ref{LemmaA.8}).

\section{Group properties of rotations} \label{ss:rotation-group}

One can see that rotations form a group. If we perform a rotation
$\{ \vec{\phi}_1\} $ followed by a rotation $\{ \vec{\phi}_2\} $,
then the resulting
transformation preserves the origin, the linear combinations of vectors and
their dot product, so it is another rotation.

The identity element in the rotation group  is the  rotation through
zero angle $\{ \vec{0} \}$, which leaves all vectors intact and is
represented
 by the unit matrix $R_{\vec{0}} = I$.
For each rotation  $ \{\vec{\phi} \}$ there exists an opposite (or
inverse) rotation
$\{ -\vec{\phi} \}$ such that

\begin{eqnarray*}
\{ -\vec{\phi} \} \{  \vec{\phi} \}  =  \{\vec{0}\}
\end{eqnarray*}

\noindent The inverse rotation is represented by the inverse matrix
$R_{-\vec{\phi}} = R_{\vec{\phi}}^{-1}= R_{\vec{\phi}}^{T} $. The
\emph{associativity law} \index{associativity}

\begin{eqnarray*}
\{\vec{\phi}_1 \}(
\{ \vec{\phi}_2 \} \{  \vec{\phi}_3 \}) =
(\{\vec{\phi}_1 \}
\{ \vec{\phi}_2 \} )\{  \vec{\phi}_3 \}
\end{eqnarray*}

\noindent follows from the associativity of the matrix product.

Rotations about
different axes do not commute. However, two rotations $\{\phi
\vec{n}\}$ and $\{\psi \vec{n}\}$
about the same
axis\footnote{here $\vec{n}$ is a unit vector.} do commute. Moreover,
our choice of the vector parameterization of rotations leads to the following
important
relationship

\begin{eqnarray}
R_{\phi \vec{n}} R_{\psi \vec{n}} = R_{\psi \vec{n}} R_{\phi
\vec{n}} = R_{(\phi + \psi) \vec{n}}
\label{eq:A.24}
\end{eqnarray}

\noindent For example, considering two rotations around the $z$-axis
 we can write

\begin{eqnarray*}
R_{(0,0,\phi)} R_{(0,0,\psi)} &=& \left[ \begin{array}{ccc}
  \cos \phi & \sin \phi & 0  \\
  -\sin \phi & \cos \phi & 0  \\
  0 & 0 & 1
\end{array} \right] \left[ \begin{array}{ccc}
  \cos \psi & \sin \psi & 0  \\
  -\sin \psi & \cos \psi & 0  \\
  0 & 0 & 1
\end{array} \right] \\
& =& \left[ \begin{array}{ccc}
  \cos (\phi + \psi) & \sin (\phi + \psi) & 0  \\
  -\sin (\phi + \psi) & \cos (\phi + \psi) & 0  \\
  0 & 0 & 1
\end{array} \right] \\
&=& R_{(0,0,\phi+ \psi)}
\end{eqnarray*}

\noindent We will say that rotations about the same axis $\mathbf{n}$ form an
\emph{one-parameter subgroup} \index{one-parameter subgroup} of the
rotation group. \label{rotation-end}

\section{Generators of rotations} \label{ss:rotation-generators}

Rotations in the vicinity of the unit element, can be represented as a
Taylor series\footnote{Here we denote $1 \equiv  \{ \vec{0} \}$ the
identity element of the group. }

\begin{eqnarray*}
\{ \vec{\theta} \} = 1 + \sum_{i=1}^3 \theta^i t_i
+ \frac{1}{2} \sum_{ij=1}^3\theta^i \theta^j t_{ij} + \ldots
\end{eqnarray*}

\noindent At small values of $\theta$ we have simply

\begin{eqnarray*}
 \{ \vec{\theta} \} \approx 1 + \sum_{i=1}^3 \theta^i t_i
\end{eqnarray*}

\noindent Quantities $ t_i$  are called \emph{generators}
\index{generator} or \emph{infinitesimal rotations}.
\index{infinitesimal rotation} Generators can be formally
represented as derivatives of elements in one-parameter subgroups
with respect to parameters $\theta_i$, e.g.,

\begin{eqnarray*}
t_i = \lim _{\vec{\theta} \to \mathbf{0}} \frac{d}{d \theta^i} \{ \vec{\theta}
\}
\end{eqnarray*}

\noindent For example, in the matrix notation, the generator of
rotations around the $z$-axis is given by the matrix

\begin{eqnarray}
\mathcal{J}_z = \lim_{\phi \to 0} \frac{d}{d \phi} R_z (\phi) = \lim_{\phi
\to
0} \frac{d}{d \phi}
\left[ \begin{array}{ccc}
  \cos \phi & \sin \phi & 0  \\
  -\sin \phi & \cos \phi & 0  \\
  0 & 0 & 1
\end{array} \right] = \left[ \begin{array}{ccc}
  0 & 1 & 0  \\
  -1 & 0 & 0  \\
  0 & 0 & 0
\end{array} \right]
\label{eq:A.25}
\end{eqnarray}

\noindent Similarly, for generators of rotations around $x$- and $y$-axes we
obtain
from (\ref{eq:A.14}) and (\ref{eq:A.15})

\begin{eqnarray}
\mathcal{J}_x  = \left[ \begin{array}{ccc}
  0 & 0 & 0  \\
  0 & 0 & 1  \\
  0 & -1 & 0
\end{array} \right], \mbox{\ \ \ \ \ \ }
 \mathcal{J}_y  = \left[ \begin{array}{ccc}
  0 & 0 & -1  \\
  0 & 0 & 0  \\
  1 & 0 & 0
\end{array} \right]   \label{eq:A.26}
\end{eqnarray}

Using the additivity property (\ref{eq:A.24}) we can express general
rotation
$\{
\vec{\theta} \}$ as exponential function of generators

\begin{eqnarray}
\{ \vec{\theta} \} &=& \lim_{N \to \infty} \left\{ N
\frac{\vec{\theta}}{N} \right\}  = \lim_{N \to \infty} \left\{
\frac{\vec{\theta}}{N} \right\}^N  = \lim_{N \to \infty} \left(1 + \sum_{i=1}^3
\frac{\theta^i}{N} t_i \right)^N \nonumber \\
&=& \exp\left(\sum_{i=1}^3 \theta^i t_i \right) \label{gener-expon}
\end{eqnarray}

\noindent Let us verify this formula in the case of a rotation
around the $z$-axis

\begin{eqnarray*}
 e^{\mathcal{J}_z \phi}
&=& 1 + \phi \mathcal{J}_z + \frac{1}{2!}\phi^2 \mathcal{J}_z^2 + \ldots \\
 &=& \left[ \begin{array}{ccc}
  1 & 0 & 0  \\
  0 & 1 & 0  \\
  0 & 0 & 1
\end{array} \right] +
\left[ \begin{array}{ccc}
  0 & \phi & 0  \\
  -\phi & 0 & 0  \\
  0 & 0 & 0
\end{array} \right] +
\left[ \begin{array}{ccc}
  -\frac{\phi^2}{2} & 0 & 0  \\
 0 &  -\frac{\phi^2}{2} & 0  \\
  0 & 0 & 0
\end{array} \right] + \ldots \\
 &=& \left[ \begin{array}{ccc}
  1-\frac{\phi^2}{2} +  \ldots & \phi + \ldots & 0  \\
  -\phi + \ldots & 1-\frac{\phi^2}{2} +  \ldots & 0  \\
  0 & 0 & 1
\end{array} \right] =  \left[ \begin{array}{ccc}
  \cos \phi & \sin \phi & 0  \\
  -\sin \phi & \cos \phi & 0  \\
  0 & 0 & 1
\end{array} \right] \\
&=& R_z = \{0, 0, \phi \}
\end{eqnarray*}

\noindent Exponent of any linear combination of generators $t_i$
also results in an orthogonal matrix with unit determinant, i.e.,
represents a rotation. Therefore, objects $t_i$ form a basis
 in the vector space of generators of the rotation
group. This vector space is referred to as the \emph{Lie algebra}
\index{Lie algebra} of the rotation group. General properties of Lie
algebras will be discussed in Appendix \ref{ss:lie-algebras}.

\chapter{Lie groups and Lie algebras} \label{sc:lie-groups}

\section{Lie groups} \label{ss:lie}

In general, a group\footnote{see subsection \ref{ss:groups}} can be
thought of as a set of points (elements) with a multiplication law such that
the ``product'' of any two points gives you a third element in the set. In addition,
there is an inversion law that map each point to an ``inverse''
point. For some groups the corresponding sets of points are
discrete.\footnote{See example in Appendix \ref{ss:groups}.} Here we would like to discuss a special
class of groups that are called \emph{Lie groups}.\footnote{  Lie
groups and algebras were named after Norwegian mathematician Sophus
Lie who first developed their theory.} \index{Lie group} The
characteristic feature of a Lie group is that its set of points is
continuous and smooth and that multiplication and inversion laws are
described by smooth functions. This set of points can be visualized
as a multi-dimensional ``hypersurface'', which is called the
\emph{group manifold}. \index{group manifold}

We saw in the Appendix \ref{ss:parameterization}\footnote{ see also Appendix
\ref{ss:double-valued}} that elements of the rotation group are in
isomorphic correspondence with points $\vec{\phi}$ in a certain smooth manifold. The multiplication and
inversion laws define two smooth mappings between points in this
manifold. Thus, the rotation group is an example of a Lie group.
Similar to the rotation group, elements in a general \emph{Lie
group}  can be parameterized by $n$ continuous parameters $
\theta_i$, where $n$ is the dimension of the Lie group. We will join
these parameters in one $n$-dimensional ``vector'' $\vec{\theta}$
and denote a general group element  as $ \{\vec{\theta} \} = \{
\theta_1, \theta_2, \ldots \theta_n\}$, so that the group
multiplication and inversion laws are smooth functions of these
parameters.

It appears  that similar to the rotation group, in a general Lie
group it is also possible to choose a parameterization $\{
\theta_1, \theta_2, \ldots \theta_n\}$ such that the following
properties are satisfied

\begin{itemize}
\item the unit
element has parameters (0,0,...,0);
\item $\{ \vec{\theta} \}^{-1} = \{ -\vec{\theta} \}$;
\item if elements
$\{  \vec{\psi}\}$ and $ \{  \vec{\phi}\}$ belong to the same
one-parameter subgroup, then

\begin{eqnarray*}
\{  \vec{\psi}\} \{  \vec{\phi}\} = \{ \vec{\psi} + \vec{\phi} \}
\end{eqnarray*}
\end{itemize}

\noindent We will always assume that group parameters satisfy these
properties. Then, similar to subsection
\ref{ss:rotation-generators}, we can
introduce \emph{infinitesimal transformations} \index{infinitesimal
transformation} or \emph{generators} \index{generator} $t_a$ ($a =
1,2, \ldots, n$) for a general Lie group and express group elements
in the vicinity of the unit element as exponential functions of
generators

\begin{eqnarray}
\{ \vec{\theta} \} &=&  \exp \left(\sum_{a=1}^n \theta^a t_a \right) = 1 +
\sum_{a=1}^n \theta^a t_a + \frac{1}{2!} \sum_{bc=1}^n \theta^b
\theta^c t_{bc} + \ldots \label{eq:A.28}
\end{eqnarray}

\noindent Let us
 introduce function $\vec{g}(\vec{\zeta}, \vec{\xi})$ which associates with
two points $\vec{\zeta}$ and
$\vec{\xi}$ in the group manifold a third point $ \vec{g}(\vec{\zeta},
\vec{\xi}) $ according to the group multiplication law, i.e.,

\begin{eqnarray}
\{\vec{\zeta} \} \{ \vec{\xi} \} = \{ \vec{g}(\vec{\zeta},
\vec{\xi}) \}
\label{eq:A.29}
\end{eqnarray}

\noindent Function $\vec{g}(\vec{\zeta}, \vec{\xi})$ must
satisfy   conditions

\begin{eqnarray}
\vec{g} (\vec{0},\vec{\theta}) &=& \vec{g} (\vec{\theta}, \vec{0}) =
\vec{\theta}
\label{eq:A.30}\\
\vec{g} (\vec{\theta},-\vec{\theta}) &=& \vec{0} \nonumber
\end{eqnarray}

\noindent which follow from the group properties (\ref{eq:A.2})
 and (\ref{eq:A.3}), respectively.
To ensure
agreement with  equation (\ref{eq:A.30}), the Taylor expansion of $\vec{g}$ up to
the 2nd order
in parameters must look like

\begin{eqnarray}
g^a (\vec{\zeta}, \vec{\xi}) &=& \zeta^a + \xi^a +
\sum _{bc=1}^n f^a_{bc} \xi^b \zeta^c + \ldots
\label{eq:A.31}
\end{eqnarray}

\noindent where $f^a_{bc}$ are real coefficients. Now we substitute
expansions (\ref{eq:A.28}) and (\ref{eq:A.31}) into (\ref{eq:A.29})

\begin{eqnarray*}
&\mbox{ }& \left(1 + \sum_{a=1}^n \xi^a t_a + \frac{1}{2} \sum_{bc=1}^n
\xi^b
\xi^c t_{bc} + \ldots \right)\left(1 + \sum_{a=1}^n \zeta^a t_a + \frac{1}{2}
\sum_{bc=1}^n \zeta^b  \zeta^c t_{bc} + \ldots \right) \\ &=&
1 + \sum_{a=1}^n \left(\zeta^a + \xi^a +
\sum_{bc=1}^n f^a_{bc} \xi^b \zeta^c + \ldots \right)t_a + \frac{1}{2}
\sum_{ab=1}^n (\zeta^a + \xi^a + \ldots) (\zeta^b +
\xi^b + \ldots) t_{ab} + \ldots
\end{eqnarray*}

\noindent Factors multiplying  $1, \zeta, \xi, \zeta^2, \xi^2$ are
exactly the same on both sides of this equation, but the factor in front of
$\xi \zeta$ produces a non-trivial condition

\begin{eqnarray*}
 \frac{1}{2}(t_{bc} + t_{cb}) = t_b t_c - \sum_{a=1}^n f^a_{bc}t_a
\end{eqnarray*}

\noindent The left hand side is symmetric with respect to the interchange of
indices $b$ and $c$. Therefore the right hand side must by symmetric as
well

\begin{eqnarray}
  t_b t_c - \sum_{a=1}^n f^a_{bc}t_a -
t_c t_b + \sum_{a=1}^n f^a_{cb}t_a = 0
\label{eq:A.32}
\end{eqnarray}

\noindent If we define the \emph{commutator} \index{commutator} of
two generators by formula \index{$[\ldots, \ldots]$ commutator}

\begin{eqnarray*}
[t_b, t_c] \equiv  t_b t_c - t_c t_b
\end{eqnarray*}

\noindent then, according to (\ref{eq:A.32}), this commutator is a linear
combination of generators

\begin{eqnarray}
[t_b, t_c]   = \sum_{a=1}^n  C^a_{bc} t_a
\label{eq:A.33}
\end{eqnarray}

\noindent where real parameters $C^a_{bc} = f^a_{bc} - f^a_{cb}$
 are called \emph{structure constants} \index{structure constants}
 of the Lie group.

\bigskip

\begin{theorem} \label{TheoremA.1} Generators of a Lie group satisfy the
Jacobi identity

\begin{eqnarray}
[t_a,[t_b, t_c]] + [t_b,[t_c, t_a]] + [t_c,[t_a, t_b]] = 0
\label{eq:A.34}
\end{eqnarray}
\end{theorem}
\begin{proof}    Let us first write the associativity
law (\ref{eq:A.1}) in the form\footnote{The burden of writing
summation signs becomes unbearable at this point, so we will adopt
here the Einstein's summation rule which allows us to drop the
summation signs and assume that the sums are performed over all
pairs of repeating indices. Moreover, we keep only 2nd order terms in the expansion (\ref{eq:A.31}).}

\begin{eqnarray*}
0 &=& g^a(\vec{\zeta}, \vec{g}(\vec{\xi}, \vec{\eta})) -
g^a(\vec{g}(\vec{\zeta}, \vec{\xi}),
\vec{\eta}) \\
 &\approx& \zeta^a +  g^a(\vec{\xi}, \vec{\eta}) + f^a_{bc} \zeta^b
g^c(\vec{\xi}, \vec {\eta})
 - g^a(\vec{\zeta}, \vec{\xi}) -  \eta^a - f^a_{bc}  g^b(\vec{\zeta},
\vec{\xi}) \eta^c \\
 &\approx& \zeta^a +  \xi^a +  \eta^a + f^a_{bc} \xi^b
\eta^c + f^a_{bc} \zeta^b
(\xi^c +  \eta^c + f^c_{xy} \xi^x  \eta^y  ) \\
 &\ & -\zeta^a -  \xi^a - f^a_{xy} \zeta^x  \xi^y -
\eta^a
- f^a_{bc}  (\zeta^b + \xi^b + f^b_{xy} \zeta^x \xi^y)
\eta^c \\
 &=& f^a_{bc} \xi^b
\eta^c +
f^a_{bc} \zeta^b\xi^c +  f^a_{bc} \zeta^b\eta^c +
f^a_{bc} f^c_{xy} \zeta^b \xi^x  \eta^y \\
 &\ & -  f^a_{xy} \zeta^x  \xi^y
  - f^a_{bc} \eta^c\zeta^b - f^a_{bc} \eta^c \xi^b -
f^a_{bc} f^b_{xy}\eta^c  \zeta^x \xi^y \\
 &=&
f^a_{bc} f^c_{xy} \zeta^b \xi^x  \eta^y   -
f^a_{bc} f^b_{xy} \zeta^x \xi^y \eta^c   \\
 &=&
(f^a_{bc} f^c_{xy}    - f^a_{cy} f^c_{bx}) \zeta^b \xi^x \eta^y
\end{eqnarray*}

\noindent Since elements  $\{\vec{\zeta} \}$, $\{\vec{\xi} \}$ and
$\{\vec{\eta} \}$
are arbitrary, this implies

\begin{eqnarray}
f^c_{kl}  f^a_{bc} -
f^c_{bk} f^a_{cl} = 0
\label{eq:A.35}
\end{eqnarray}

\noindent Now let us turn to the left hand side of the Jacobi identity
(\ref{eq:A.34})

\begin{eqnarray*}
&\mbox{ }& [t_a,[t_b, t_c]] + [t_b,[t_c, t_a]] + [t_c,[t_a, t_b]] \\
 &=& [t_a, C_{bc}^xt_x] + [t_b,C_{ca}^xt_x] + [t_c,C_{ab}^xt_x] \\
 &=& (C_{bc}^x C_{ax}^y + C_{ca}^x C_{bx}^y + C_{ab}^x
C_{cx}^y) t_y
\end{eqnarray*}

\noindent The expression in parentheses is

\begin{eqnarray}
&\mbox{ } &  (f_{bc}^x  - f_{cb}^x) (f_{ax}^y  - f_{xa}^y) +
(f_{ca}^x  - f_{ac}^x) (f_{bx}^y  - f_{xb}^y) +
(f_{ab}^x  - f_{ba}^x) (f_{cx}^y  - f_{xc}^y) \nonumber \\
&=&  f_{bc}^xf_{ax}^y - f_{bc}^x f_{xa}^y - f_{cb}^x f_{ax}^y +
f_{cb}^x f_{xa}^y +
f_{ca}^x f_{bx}^y - f_{ca}^x f_{xb}^y \nonumber\\
&\ & -f_{ac}^x f_{bx}^y + f_{ac}^x
f_{xb}^y   +
f_{ab}^x  f_{cx}^y - f_{ab}^x f_{xc}^y - f_{ba}^x f_{cx}^y +  f_{ba}^x
f_{xc}^y \nonumber \\
&=&  (f_{bc}^xf_{ax}^y - f_{ab}^x f_{xc}^y) +
(f_{ca}^x f_{bx}^y - f_{bc}^x f_{xa}^y ) -
(f_{cb}^x f_{ax}^y - f_{ac}^x f_{xb}^y) \nonumber\\
&\ &-
(f_{ba}^x f_{cx}^y-f_{cb}^x f_{xa}^y  ) +
(f_{ab}^x  f_{cx}^y - f_{ca}^x f_{xb}^y) -
(f_{ac}^x f_{bx}^y     -  f_{ba}^x f_{xc}^y)
\label{eq:A.36}
\end{eqnarray}

\noindent According to (\ref{eq:A.35}) all terms in parentheses on the
right hand side of (\ref{eq:A.36})
 are zero, which proves the theorem.
\end{proof}
\bigskip

 \section{Lie algebras}
\label{ss:lie-algebras}

\emph{Lie algebra} \index{Lie algebra} is a vector space over real
numbers $\mathbb{R}$
 with the additional operation called the  \emph{Lie bracket}.
\index{Lie bracket} \index{$[\ldots , \ldots]$ Lie bracket} This operation is denoted $[A,B]$ and it maps
two vectors $A$ and $B$ to a third vector. The Lie bracket is postulated to satisfy
the following set of conditions\footnote{equation (\ref{eq:jacobi}) is called the \emph{Jacobi identity} \index{Jacobi identity}}

\begin{eqnarray}
[A,B] &=& -[B,A] \nonumber \\ \mbox{ }
 [A, B+C] &=& [A,B] + [A,C] \nonumber
\\ \mbox{ }
[A, \lambda B] &=& [\lambda A,  B] = \lambda [A,B], \mbox{ } for
\mbox { } any \mbox { } \lambda \in \mathbb{R} \nonumber
\\ \mbox{ }
0 &=& [A, [B,C]] + [B, [C,A]] + [C, [A,B]] \label{eq:jacobi}
\end{eqnarray}

From our discussion in the preceding section it is clear that
generators of a Lie group form a Lie algebra, in which the role of the Lie bracket is played by the commutator of generators. Consider, for example,
the group of rotations. In the matrix representation, the generators
are  linear combinations of matrices (\ref{eq:A.25}) -
(\ref{eq:A.26}), i.e., they are arbitrary antisymmetric matrices
\index{antisymmetric tensor} satisfying $A^T = -A$. The commutator
is represented by\footnote{Note that this representation of the Lie
bracket as a difference of two products can be used only when the
generators are identified with matrices. This formula (as well as
(\ref{eq:A.37})) does not apply to abstract Lie algebras, because
the product of two elements $AB$ is not defined there.}

\begin{eqnarray*}
[A,B] &=& AB - BA
\end{eqnarray*}

\noindent which is also an antisymmetric matrix, because

\begin{eqnarray*}
(AB-BA)^T &=& B^T A^T - A^T B^T = BA - AB = - (AB-BA)
\end{eqnarray*}

We will frequently use the following property of commutators in the
matrix representation

\begin{eqnarray}
[A, BC] &=& ABC -BCA = ABC -BAC +BAC - BCA \nonumber\\
&=& (AB - BA)C + B(AC-CA) = [A,B] C + B[A,C] \label{eq:A.37}
\end{eqnarray}

The structure constants of the Lie algebra of the rotation group can
be obtained by direct calculation from explicit expressions
(\ref{eq:A.25}) - (\ref{eq:A.26})

\begin{eqnarray*}
[\mathcal{J}_x, \mathcal{J}_y] &=& \mathcal{J}_z \\ \mbox{ }
[\mathcal{J}_x, \mathcal{J}_z] &=& -\mathcal{J}_y
\\ \mbox{ }
[\mathcal{J}_y, \mathcal{J}_z] &=& \mathcal{J}_x
\end{eqnarray*}

\noindent This can be written more compactly as

\begin{eqnarray*}
[\mathcal{J}_i, \mathcal{J}_j] &=& \sum_{k=1}^3 \epsilon_{ijk} \mathcal{J}_k
\end{eqnarray*}

In the vicinity of the unit element, any Lie group element can be
represented as exponent $\exp(x)$ of a Lie algebra element $x$ (see
equation (\ref{eq:A.28})). As product of two group elements is another
group element, we must have for any two Lie algebra elements $x$ and $y$

\begin{eqnarray}
\exp(x) \exp(y) = \exp(z)
\label{eq:A.38}
\end{eqnarray}

\noindent where $z$ is also an element from the Lie algebra. Then there should
exist a
mapping in the Lie algebra which associates with any two elements $x$
and $y$ a third element $z$, such that equation (\ref{eq:A.38}) is satisfied.
The Baker-Campbell-Hausdorff theorem  \cite{Weiss} gives us the explicit
form of this mapping

 \begin{eqnarray*}
z &=& x + y + \frac{1}{2} [x,y] + \frac{1}{12} [[x,y], y]  + \frac{1}{12}
[[y,x], x] \\
&\ & +\frac{1}{24}  [[[y,x], x],y] - \frac{1}{720} [[[[x,y],y], y],y] +
\frac{1}{360}  [[[[x,y],y], y], x] \\
&\ & +\frac{1}{360} [[[[y,x],x], x],y]  -  \frac{1}{120}  [[[[x,y],y], x],y]
 -  \frac{1}{120}  [[[[y,x],x], y],x] \ldots
\end{eqnarray*}

\noindent This means that Lie bracket relations in the Lie algebra
contain full information about the group multiplication law in the
vicinity of the unit element.
 In many cases, it is much
easier to deal with generators and their Lie brackets than directly
with group elements and their multiplication law.

In applications one often finds useful the following identity

\begin{eqnarray}
 \exp(ax) y \exp(-ax) = y +
 a[x,y] +\frac{a^2}{2!}[x,[x,y]] + \frac{a^3}{3!}[x,[x,[x,y]]] \ldots
\label{eq:A.39}
\end{eqnarray}

\noindent where $a \in \mathbb{R}$. This formula can be proved by
noticing that both sides are solutions of the same differential
operator equation

\begin{eqnarray*}
 \frac{dy(a)}{da} = [x,y(a)]
\end{eqnarray*}

\noindent with the same initial condition $y(a) = y$.

There is a unique Lie algebra $A_G$ corresponding to each Lie group
$G$. However, there are many Lie groups with the same Lie algebra. These
groups have the same structure in the vicinity of the unit element,
but their global topological properties can be different.

A Lie \emph{subalgebra} \index{subalgebra} $B$ of a Lie algebra $A$
is a subspace in $A$ which is closed with respect to the Lie bracket,
i.e., if $x,y \in B$, then $[x,y] \in B$. If $H$ is a subgroup of a
Lie group $G$, then its Lie algebra $A_H$ is a Lie subalgebra of $A_G$.

\chapter{Hilbert space} \label{sc:hilbert-space}

\section{Inner product} \label{sc:inner-product}

 An \emph{inner product  space} \index{inner product space} $H$ is defined as a complex vector
space\footnote{See Appendix \ref{ss:vector-space}.   Vectors in $H$
will be denoted by $|x \rangle$.}
 which
has a mapping from ordered pairs of vectors to complex numbers. This mapping is
called the  \emph{inner  product} \index{inner product} $(| y \rangle,
|x \rangle)$ and it satisfies the following properties

\begin{eqnarray}
(| x \rangle, |y \rangle) &=& (| y \rangle, | x \rangle)^*
\label{eq:A.40} \\ \mbox{ }
 (| z \rangle, \alpha |  x \rangle +
\beta | y \rangle) &=& \alpha (| z \rangle, | x \rangle) + \beta (|
z \rangle, |  y \rangle) \label{eq:A.41}
\\ \mbox{ }
(| x \rangle, | x \rangle) &\in& \mathbb{R} \label{eq:A.42}
\\ \mbox{ }
 (| x \rangle, | x \rangle) &\geq& 0
\label{eq:A.43}
\\ \mbox{ }
 (| x \rangle, | x \rangle) &=& 0
\Leftrightarrow |x\rangle = \mathbf{0}
\label{eq:A.44}
\end{eqnarray}

\noindent where  $\alpha$ and $\beta$ are complex numbers. Given inner
product we can define the \emph{distance}  between
two vectors by formula $d(|x \rangle, |y \rangle) \equiv \sqrt{(|
x-y \rangle, | x-y \rangle)}$.

 The inner product space $H$ is called \emph{ complete} \index{complete inner product space} if
any Cauchy sequence\footnote{Cauchy sequence is an infinite sequence of
vectors
$| x_i \rangle$ in which the distance between two vectors $| x_n
\rangle$ and $| x_m \rangle$ tends to zero
 when their indices tend to infinity $n,m \to \infty$.}
 of vectors in $H$  converges to a vector
in $H$. Analogously, a subspace in $H$ is called a \emph{closed
subspace} \index{closed subspace} if any Cauchy sequence of vectors
belonging to the subspace converges to a vector in this subspace.
The \emph{Hilbert space} \index{Hilbert space} is simply a complete
inner product space.\footnote{The notions of completeness and
closedness are rather technical. Finite dimensional inner product
spaces are always complete, and their subspaces are always closed.
Although in quantum mechanics we normally deal with
infinite-dimensional spaces, most properties having relevance to
physics do not depend on the number of dimensions. So, we will
ignore the difference between finite- and infinite-dimensional
spaces and freely use finite $n$-dimensional examples in our proofs
and demonstrations. In particular, we will tacitly assume that every
subspace $A$ is closed or forced to be closed by adding all vectors
which are limits of Cauchy sequences in $A$.}

\section{Orthonormal bases} \label{ss:orthonormal}

Two vectors $|x \rangle$ and $|y \rangle$ are called
\emph{orthogonal} \index{orthogonal vectors} if $(| x \rangle, |y
\rangle) = 0$. Vector $|x \rangle$ is called \emph{unimodular}
\index{unimodular vector} if $(| x \rangle, |x \rangle) = 1$. In
Hilbert space we can consider \emph{orthonormal} \index{orthonormal
basis} bases consisting of mutually orthogonal
unimodular vectors $|e_i \rangle$ which satisfy

\begin{eqnarray*}
(| e_i \rangle, | e_j \rangle)  = \left \{
\begin{array}{c}
 1, \mbox{ if } i = j  \\
0, \mbox{ if } i \neq j
\end{array} \right.
\end{eqnarray*}

\noindent or, using the Kronecker delta symbol

\begin{eqnarray}
(| e_i \rangle, | e_j \rangle) = \delta_{ij}
\label{eq:A.45}
\end{eqnarray}

Suppose that  vectors $|x \rangle$ and $|y \rangle$ have
components $x_i$ and $y_i$, respectively, in this basis

\begin{eqnarray*}
|x \rangle &=& x_1| e_1 \rangle + x_2| e_2 \rangle + \ldots + x_n| e_n
\rangle \\
|y \rangle &=& y_1| e_1 \rangle + y_2| e_2 \rangle + \ldots + y_n| e_n
\rangle
\end{eqnarray*}

\noindent Then using (\ref{eq:A.40}), (\ref{eq:A.41}) and (\ref{eq:A.45}) we can  express  the inner product through
vector components

\begin{eqnarray*}
 (| x \rangle, |y \rangle) &=&
( x_1 |e_1 \rangle  + x_2 |e_2 \rangle  + \ldots + x_n |e_n \rangle,
y_1 |  e_1 \rangle  + y_2 |e_2 \rangle + \ldots + y_n |e_n
 \rangle) \nonumber\\
&=& x_1^* y_1 + x_2^* y_2 + \ldots + x_n^* y_n = \sum_i x^*_i y_i
\end{eqnarray*}

\section{Bra and ket vectors} \label{ss:bra-ket}

The notation $(|x \rangle, |y \rangle)$ for the inner product is
rather cumbersome. We will use instead a more convenient
\emph{bra-ket} formalism \index{bra-ket formalism} suggested by
Dirac, which greatly simplifies manipulations with objects in the
Hilbert space. Let us call vectors in the Hilbert space \emph{ket}
\index{ket vector} vectors. We define a \emph{linear functional}
\index{linear functional} $ \langle f|: H \to \mathbb{C}$ as a
function (denoted by $\langle f| x \rangle$) which maps each ket
vector $|x \rangle$ in $H$ to complex numbers in such a way  that
linearity is preserved, i.e.

\begin{eqnarray*}
\langle f|( \alpha |x\rangle  + \beta |y \rangle) = \alpha \langle f|x \rangle + \beta
\langle f|y \rangle
\end{eqnarray*}

\noindent  Since any
linear combination $\alpha \langle f| + \beta \langle g|$ of
functionals $\langle f|$ and $\langle g|$ is again a functional,
then all  functionals form a vector space (denoted $H^*$). The
vectors in $H^*$ will be called \emph{bra} vectors. \index{bra
vector} We can define an inner product in $H^*$ so that it becomes a
Hilbert space. To do that, let us choose an orthonormal basis
 $|e_i \rangle$ in $H$. Then each functional $\langle f
| $ defines a set of complex numbers $f_i$ which are values of this
functional on the basis vectors

\begin{eqnarray*}
 f_i = \langle f|e_i \rangle
\end{eqnarray*}

\noindent These numbers define the functional uniquely, i.e., if two
functionals
 $\langle f|$ and $\langle g|$ are different, then  their
values are different for at least one basis vector $|e_k \rangle$:
$f_k \neq g_k$.\footnote{ Otherwise, using linearity we would be
able to prove that the values of functionals $\langle f|$ and
$\langle g|$ are equal on all vectors in $H$, i.e., $\langle f| =
\langle g|$.} Now we can define the inner product of bra vectors
$\langle f |$ and $\langle g |$ by formula

\begin{eqnarray*}
 (\langle f|, \langle g|) = \sum_i f_i g^*_i
\label{eq:A.46}
\end{eqnarray*}

\noindent and verify that it satisfies all properties of the inner product
listed
 in (\ref{eq:A.40}) - (\ref{eq:A.44}). The Hilbert space $H^*$ is
called a \emph{dual} \index{dual Hilbert space} of the Hilbert space
$H$. Note that each vector $|y \rangle$ in $H$  defines a unique
linear functional $\langle y|$ in $H^*$  by formula

\begin{eqnarray}
 \langle y|x \rangle \equiv (|y \rangle, |x \rangle) \label{eq:F.7a}
\end{eqnarray}

\noindent for each $|x \rangle \in H$. This bra vector $\langle y |$
will be called the dual \index{dual vector} of the ket vector $|y
\rangle$. Equation (\ref{eq:F.7a}) tells us that in order to calculate
the inner product of $|y \rangle$ and $|x \rangle$ we should find
the bra vector (functional) dual to $|y \rangle$ and then find its
value on $|x \rangle$. So, the inner product is obtained
 by coupling bra and ket vectors $ \langle x |y \rangle$,
thus forming a closed \emph{bra}(c)\emph{ket} expression, which is a complex
number.

Clearly, if $|x \rangle$ and $|y \rangle$ are different kets, then
their dual bras $\langle x|$ and $\langle y|$ are different as well.
We may notice that just like vectors in $H^*$ define linear
functionals on vectors in $H$, any vector  $ |x \rangle \in H$ also
defines an  \emph{antilinear} \index{antilinear functional}
functional on bra vectors by formula $\langle y | x \rangle $, i.e.,

\begin{eqnarray*}
( \alpha \langle y| +  \beta \langle z|) | x \rangle = \alpha^* \langle y|x \rangle
+ \beta ^*\langle z|x \rangle
\end{eqnarray*}

 \noindent Then
applying the same arguments as above, we see that
 if $\langle y |$ is a bra vector, then
there is a unique ket $|y \rangle$ such that for any $\langle
x | \in H^*$ we have

\begin{eqnarray}
\langle x | y \rangle = (\langle x |, \langle y
|)
\label{eq:A.47}
\end{eqnarray}

\noindent  Thus we established an isomorphism of two Hilbert spaces
$H$ and $H^*$.  This statement is known as the \emph{Riesz theorem}. \index{Riesz theorem}

\bigskip

\begin{lemma} \label{LemmaA.2} If kets $|e_i \rangle $ form an orthonormal
 basis
in $H$, then dual bras $\langle e_i |$ also form an orthonormal
basis in $H^*$.
\end{lemma}
\begin{proof}
Suppose that $\langle e_i|$ do not form a basis. Then there is a
nonzero vector  $\langle z | \in H^*$ which is orthogonal to all
$\langle e_i|$, and the values of the functional
 $\langle z |$ on all basis vectors $| e_i \rangle$ are zero,
so  $\langle z | = \mathbf{0}$.
The orthonormality of $\langle e_i|$ follows from equations (\ref{eq:A.47}) and
\ref{eq:A.45})

\begin{eqnarray*}
(\langle e_i |, \langle e_j |) = \langle e_i | e_j \rangle = (| e_i
\rangle, | e_j \rangle) = \delta_{ij}
\end{eqnarray*}
\end{proof}
\bigskip

\noindent The components $x_i$ of a vector $|x \rangle$ in the basis
 $|e_i \rangle$ are conveniently represented in the
bra-ket notation as

\begin{eqnarray*}
\langle e_i | x \rangle  &=& \langle e_i |(x_1 |e_1\rangle + x_2
|e_2\rangle + \ldots + x_n |e_n\rangle) = x_i
\end{eqnarray*}

\noindent So we can write

\begin{eqnarray}
| x \rangle &=& \sum \limits_i |e_i \rangle x_i = \sum \limits_i
|e_i\rangle \langle e_i |x\rangle \label{eq:A.48}
\end{eqnarray}

\noindent The bra vector $\langle y |$ dual to the ket $|y \rangle$
has complex conjugate components in the dual basis

\begin{eqnarray}
 \langle y | =  \sum \limits_i y_i^* \langle e_i|
\label{eq:A.49}
\end{eqnarray}

\noindent This can be verified by checking that  the functional
on the right hand side being applied to any vector $|x
\rangle \in H$ yields

\begin{eqnarray*}
   \sum \limits_i y_i^* \langle e_i| x \rangle &=&  \sum \limits_i y_i^*
x_i=  (|y \rangle, |x \rangle) =  \langle y |x \rangle
\end{eqnarray*}

\label{bra-end}

\section{Tensor product of Hilbert spaces} \label{ss:tens-prod}

Given two Hilbert spaces $\mathcal{H}_1$ and $\mathcal{H}_2$ one can
construct a third Hilbert space $\mathcal{H}$  which is called the
\emph{tensor product} \index{tensor product} of $\mathcal{H}_1$ and
$\mathcal{H}_2$ and denoted by $\mathcal{H} = \mathcal{H}_1 \otimes
\mathcal{H}_2$. For each pair of basis  ket vectors $|i \rangle \in
\mathcal{H}_1 $ and $|j \rangle \in \mathcal{H}_2 $ there is exactly
one basis ket in $\mathcal{H}$, which is denoted by $|i \rangle
\otimes |j \rangle$. Other vectors in $\mathcal{H}$ are linear
products of the basis kets $|i \rangle \otimes |j \rangle$ with
complex coefficients.

 The inner product of two basis vectors $|a_1 \rangle \otimes
|a_2 \rangle \in \mathcal{H}$ and $|b_1 \rangle \otimes |b_2 \rangle
\in \mathcal{H}$ is defined as $ \langle a_1 | b_1 \rangle \langle
a_2 | b_2 \rangle $. This inner product is extended to linear
combinations of basis vectors by linearity.

\section{Linear operators} \label{ss:operators}

Linear transformations of vectors in the Hilbert space (also called
\emph{operators}) \index{operator} play a very important role in
quantum formalism. Such transformations

\begin{eqnarray*}
T |x \rangle = | x' \rangle
\end{eqnarray*}

\noindent have the  property

\begin{eqnarray*}
T ( \alpha |x \rangle + \beta |y \rangle) &=& \alpha T | x\rangle +
\beta T | y\rangle
\end{eqnarray*}

\noindent for any two complex numbers $\alpha$ and $\beta$ and any two vectors $|x \rangle$ and $|y \rangle$. Given an operator $T$ we can find images of basis
 vectors

\begin{eqnarray*}
T  |e_i \rangle &=&  | e_i'\rangle
\end{eqnarray*}

\noindent and find the expansion of these images in the original basis
$|e_i \rangle$

\begin{eqnarray*}
  | e_i'\rangle  = \sum \limits_j t_{ij} |e_j \rangle
\end{eqnarray*}

\noindent Coefficients $t_{ij}$ of this expansion are called the
\emph{matrix elements} \index{matrix element} of the operator  $T$
in the  basis  $| e_i \rangle$. In the bra-ket notation we can find
a convenient expression for the matrix elements

\begin{eqnarray*}
\langle e_j| (T |e_i \rangle) &=& \langle e_j| e_i' \rangle =
\langle e_j | \sum \limits_k t_{ik} |e_k \rangle
= \sum \limits_k t_{ik} \langle e_j   |e_k \rangle= \sum \limits_k t_{ik} \delta_{jk}\\
&=& t_{ij}
\end{eqnarray*}

\noindent Knowing  matrix elements of the operator $T$ and
components of  vector $| x \rangle $ in the basis $|e_i \rangle$ one
can always find the components of the transformed vector $| x'
\rangle = T |x \rangle $

\begin{eqnarray}
 x'_i &=&  \langle e_i | x' \rangle  = \langle e_i |(T | x \rangle)
 =  \langle e_i |\sum \limits_j  (T |  e_j \rangle) x_j
 = \sum \limits_{jk} \langle e_i |  e_k \rangle t_{kj} x_j \nonumber \\
      &=& \sum \limits_{jk} \delta_{ik} t_{kj} x_j= \sum \limits_jt_{ij} x_j \label{eq:xi'}
\end{eqnarray}

\noindent In the bra-ket notation, the operator
$T$ has the form

\begin{eqnarray}
T= \sum \limits_{ij} |e_i \rangle t_{ij} \langle e_j|
\label{eq:A.50}
\end{eqnarray}

\noindent Indeed,  by applying the right hand side of equation (\ref{eq:A.50}) to
 arbitrary vector $|x \rangle $ we obtain

\begin{eqnarray*}
 \sum \limits_{ij} |e_i \rangle t_{ij} \langle e_j|x \rangle
&=& \sum \limits_{ij} |e_i \rangle t_{ij} x_j = \sum \limits_i x'_i
|e_i \rangle = |x' \rangle = T|x \rangle
\end{eqnarray*}

\section{Matrices and operators} \label{ss:matrices}

Sometimes it is convenient to represent vectors and operators in the
Hilbert space $\mathcal{H}$ in a matrix notation. Let us fix an
orthonormal basis $|e_i \rangle \in \mathcal{H}$  and
 represent each ket vector $|y \rangle$ by a column of
its components

\begin{eqnarray*}
 |y \rangle =
\left[ \begin{array}{c}
y_1  \\
y_2   \\
 \vdots     \\
y_n
\end{array} \right]
\end{eqnarray*}

\noindent  The bra vector
$\langle x|$  will be
represented by a row

\begin{eqnarray*}
 \langle x | = [x_1^*, x_2^*, \ldots, x_n^*]
\end{eqnarray*}

\noindent of complex conjugate components in the dual basis
 $\langle e_i |$. Then the inner product  is obtained
by the usual ``row by column'' matrix multiplication rule.

\begin{eqnarray*}
 \langle x |y \rangle = [x_1^*, x_2^*, \ldots, x_n^*]
\left[ \begin{array}{c}
y_1  \\
y_2   \\
 \vdots     \\
y_n
\end{array} \right] = \sum_i x^*_i y_i
\end{eqnarray*}

\noindent Matrix elements of the operator $T$ in (\ref{eq:A.50})
can be conveniently arranged in the \emph{matrix} \index{matrix}

\begin{eqnarray*}
T = \left[ \begin{array}{cccc}
t_{11} & t_{12}  & \ldots  & t_{1n}   \\
 t_{21}  & t_{22}  & \ldots  &  t_{2n}     \\
\vdots  & \vdots  & \ddots &  \vdots     \\
t_{n1}  & t_{n2}  & \ldots  &  t_{nn}
\end{array} \right]
\end{eqnarray*}

\noindent Then the action of the operator $T$ on a vector  $|x' \rangle = T |x
\rangle $ can be represented as a product of the
matrix corresponding to $T$ and the column vector $| x \rangle$\footnote{compare with (\ref{eq:xi'})}

\begin{eqnarray*}
\left[ \begin{array}{c}
x'_{1}  \\
 x'_{2}      \\
\vdots     \\
x'_{n}
\end{array} \right] = \left[ \begin{array}{cccc}
t_{11} & t_{12}  & \ldots  & t_{1n}   \\
 t_{21}  & t_{22}  & \ldots  &  t_{2n}     \\
\vdots  & \vdots  & \ddots &  \vdots     \\
t_{n1}  & t_{n2}  & \ldots  &  t_{nn}
\end{array} \right]
\left[ \begin{array}{c}
x_{1}  \\
 x_{2}      \\
\vdots     \\
x_{n}
\end{array} \right]
= \left[ \begin{array}{c}
\sum_j t_{1j}x_{j}  \\
\sum_j t_{2j}x_{j}    \\
\vdots     \\
\sum_j t_{nj}x_{j}
\end{array} \right]
\end{eqnarray*}

\noindent So, each operator has a unique matrix and each $n \times
n$ matrix  defines a unique linear operator. This establishes an
isomorphism between operators and  matrices. In what follows we will
often use the terms \emph{operator} and \emph{matrix}
interchangeably.

The matrix corresponding to the \emph{identity operator}
\index{identity matrix} is $\delta_{ij}$, i.e., the unit matrix

\begin{eqnarray*}
I = \left[ \begin{array}{cccc}
1 & 0  & \ldots  & 0   \\
 0  &1 & \ldots  &  0    \\
\vdots  & \vdots  & \ddots &  \vdots     \\
0  &0  & \ldots  &  1
\end{array} \right]
\end{eqnarray*}

\noindent A \emph{diagonal operator} has diagonal matrix $d_i
\delta_{ij}$ \index{diagonal matrix}

\begin{eqnarray*}
D = \left[ \begin{array}{cccc}
d_1 & 0  & \ldots  & 0   \\
 0  & d_2 & \ldots  &  0    \\
\vdots  & \vdots  & \ddots &  \vdots     \\
0  &0  & \ldots  &  d_n
\end{array} \right]
\end{eqnarray*}

\noindent The action of operators in the dual space $\mathcal{H}^*$ will be
denoted by multiplying
bra row by the  operator matrix from the right

\begin{eqnarray*}
[y'_{1}, y'_{2}, \ldots, y'_{n}]  =
[y_{1}, y_{2}, \ldots, y_{n}]
  \left[ \begin{array}{cccc}
s_{11} & s_{12}  & \ldots  & s_{1n}   \\
 s_{21}  & s_{22}  & \ldots  &  s_{2n}     \\
\vdots  & \vdots  & \ddots &  \vdots     \\
s_{n1}  & s_{n2}  & \ldots  &  s_{nn}
\end{array} \right]
\end{eqnarray*}

\noindent or symbolically

\begin{eqnarray*}
y'_i &=& \sum \limits_j y_j s_{ji}  \\
\langle y'| &=& \langle y| S
\end{eqnarray*}

Suppose that operator $T$ with matrix $t_{ij}$ in the ket space $H$ transforms
vector
$|x\rangle$ to $|y \rangle$, i.e.,

\begin{eqnarray}
y_i = \sum \limits_j t_{ij} x_j
\label{eq:A.51}
\end{eqnarray}

\noindent What is the matrix of operator $S$ in the bra
space $H^*$ which connects corresponding dual vectors $\langle x |$ and $
\langle
y|$? As $\langle x |$ and $\langle y |$ have
components complex conjugate to those of $| x \rangle$ and $| y
\rangle$, and $S$ acts on bra vectors from the right, we can write

\begin{eqnarray}
y^*_i =  \sum \limits_j x^*_j s_{ji}
\label{eq:A.52}
\end{eqnarray}

\noindent On the other hand, taking complex conjugate of equation (\ref{eq:A.51})
we obtain

\begin{eqnarray*}
y^*_i =\sum \limits_j  t^*_{ij} x^*_j
\end{eqnarray*}

\noindent Comparing this with (\ref{eq:A.52}) we have

\begin{eqnarray*}
s_{ij} = t^*_{ji}
\end{eqnarray*}

\noindent This means  that the matrix
 representing the action of the operator $T$ in the dual space $H^*$,
is different from the  matrix $T$ in that rows are substituted by
columns,\footnote{This is equivalent to the  reflection of the matrix
with respect to the main diagonal. Such matrix operation is called
\emph{transposition}. \index{transposition}} and matrix elements are complex-conjugated. This combined operation ``transposition +
complex conjugation''
 is called \emph{Hermitian conjugation}. \index{Hermitian conjugation} Hermitian conjugate
(or \emph{adjoint}) \index{adjoint operator} of
 operator $T$ is denoted $T^{\dag}$. In particular, we can write

\begin{eqnarray}
\langle x | (T | y \rangle) &=& (\langle x | T^{\dag}) | y \rangle
\label{eq:A.53} \\
det(T^{\dag}) &=& (det(T))^* \label{eq:A.53x}
\end{eqnarray}

\section{Functions of operators} \label{ss:functions}

The sum of two operators $A$ and $B$ and the multiplication of an
operator $A$ by a complex number $\lambda$ are easily expressed in
terms of matrix elements

\begin{eqnarray*}
(A+B)_{ij} &=& a_{ij} + b_{ij} \\
(\lambda A)_{ij} &=& \lambda a_{ij}
\end{eqnarray*}

\noindent We can define the product $AB$ of two operators  as the
transformation obtained by a sequential application of $B$ and then
$A$. This product is also a linear transformation, i.e., an
operator. The matrix of the product $AB$ is the ``row-by-column''
product of their matrices $a_{ij}$ and $b_{ij}$

\begin{eqnarray*}
(AB)_{ij} = \sum \limits_k a_{ik}b_{kj}
\end{eqnarray*}

\begin{lemma}
\label{LemmaA.3} Adjoint of a product of operators is equal to the
product of adjoint operators \index{adjoint operator} in the opposite order.

\begin{eqnarray*}
 (AB)^{\dag} = B^{\dag} A^{\dag}
\end{eqnarray*}
\end{lemma}
\begin{proof}

\begin{eqnarray*}
 (AB)^{\dag}_{ij} &=& (AB)^*_{ji}
 = \sum \limits_k a^*_{jk} b^*_{ki}
 = \sum \limits_k b^*_{ki} a^*_{jk}
 = \sum \limits_k (B^{\dag})_{ik} (A^{\dag})_{kj}
 = (B^{\dag} A^{\dag})_{ij}
\end{eqnarray*}
\end{proof}
\bigskip

\noindent The  \emph{inverse operator} \index{inverse operator}
$A^{-1}$ is defined by its two properties

\begin{eqnarray*}
A^{-1} A = A A^{-1} = I
\end{eqnarray*}

\noindent The corresponding matrix is the inverse of the matrix $A$.
\index{inverse matrix}

Using the basic operations of addition, multiplication, and inversion
 we can define various functions $f(A)$ of the
operator $A$. For example, the exponential
function is defined by its Taylor series

\begin{eqnarray}
e^{F} = 1 +F + \frac{1}{2!}F^2 + \ldots
\label{eq:A.54}
\end{eqnarray}

\noindent For any two operators $A$ and $B$ the expression

\begin{eqnarray}
[A,B] \equiv AB - BA
\label{eq:A.55}
\end{eqnarray}

\noindent is called the \emph{commutator}. \index{commutator}
\index{$[\ldots, \ldots]$ commutator} We say
that two operators $A$ and $B$ \emph{commute} with each other if
$[A,B] = 0$. Clearly, any two powers of $A$ commute: $[A^n, A^m] =
0$, and $[A, A^{-1}] = 0$. Consequently, any two functions of $A$
commute as well: $[f(A), g(A)] = 0$.

 \emph{Trace} \index{trace} of a matrix is defined as a sum of its diagonal
elements

\begin{eqnarray*}
Tr(A) = \sum_i A_{ii}
\end{eqnarray*}

\begin{lemma} \label{LemmaA.4} Trace of a product of operators is invariant
with respect to any cyclic permutation of factors.
\end{lemma}
\begin{proof}    Take for example a trace of the
product of three operators

\begin{eqnarray*}
Tr(ABC) = \sum \limits_{ijk} A_{ij} B_{jk} C_{ki}
\end{eqnarray*}

\noindent Then

\begin{eqnarray*}
Tr(BCA) &=& \sum \limits_{ijk} B_{ij} C_{jk} A_{ki}
\end{eqnarray*}

\noindent Changing in this expression summation indices $k \to i$, $i \to j$, and $j
\to k$, we obtain

\begin{eqnarray*}
Tr(BCA) &=& \sum \limits_{ijk} B_{jk} C_{ki} A_{ij} = Tr(ABC)
\end{eqnarray*}
\end{proof}
\bigskip

We can define two classes of operators (and their matrices) which
play important roles in quantum mechanics (see Table
\ref{table:A.2}). These are Hermitian and unitary operators. We call
operator $T$ \emph{Hermitian} \index{Hermitian operator} or
\emph{self-adjoint} if \index{self-adjoint operator}

\begin{eqnarray}
T = T^{\dag} \label{eq:A.56}
\end{eqnarray}

\noindent For a Hermitian $T$  we can write

\begin{eqnarray}
t_{ii} = t^*_{ii} \label{eq:F.19a} \\
t_{ij} = t^*_{ji}  \nonumber
\end{eqnarray}

\begin{table}[h]
\caption{Actions on operators and types of linear operators in the
Hilbert space}
\begin{tabular*}{\textwidth}{@{\extracolsep{\fill}}ccc}
 \hline
                          & Symbolic    &Condition on matrix elements
\cr
                          &             & or eigenvalues  \cr
\hline
 & \textbf{Action on operators}             &     \cr
Complex conjugation                  & $A \to A^*$ & $(A^*)_{ij} = A^*_{ij}
$ \cr Transposition                 & $A \to A^T$    &
$(A^T)_{ij} = A_{ji} $  \cr Hermitian conjugation & $A \to A^{\dag}
= (A^*)^T $ & $(A^{\dag})_{ij} = A^*_{ji}$ \cr Inversion & $A \to
A^{-1}$        & inverse eigenvalues \cr Determinant & $\det(A) $ &
product of eigenvalues \cr Trace & $Tr(A) $ & $\sum_i A_{ii}$ \cr
                  &                      &      \cr
 & \textbf{Types of operators}        &           \cr
Identity          &  $I$          &  $I_{ij} = \delta_{ij}$ \cr
Diagonal          &  $D$          &  $D_{ij} = d_i\delta_{ij}$ \cr
Hermitian         & $A  = A^{\dag}$  & $A_{ij} = A^*_{ji}$ \cr
AntiHermitian         & $A  = -A^{\dag}$  & $A_{ij} = -A^*_{ji}$ \cr
Unitary           & $A^{-1}  = A^{\dag}$ & unimodular eigenvalues
\cr Projection        & $A  = A^{\dag}$, $A^2 = A$ & eigenvalues 0
and 1 only \cr
 \hline
\end{tabular*}
\label{table:A.2}
\end{table}

\noindent i.e., diagonal matrix elements are real, and non-diagonal
matrix elements symmetrical with respect to the main diagonal are
complex conjugates of each other. Moreover, from equations (\ref{eq:A.53})
and (\ref{eq:A.56}) we can calculate the inner product of vectors
$\langle x |$ and $T | y \rangle$ with a Hermitian $T$

\begin{eqnarray*}
\langle x | (T | y \rangle) &=& (\langle x | T^{\dag}) | y \rangle =
(\langle x | T) | y \rangle \equiv \langle x | T | y \rangle
\end{eqnarray*}

\noindent From this symmetric notation it is clear that a Hermitian
$T$ can act either to the right (on $|y \rangle$) or to the left (on
$\langle x |$)

Operator $U$ is called \emph{unitary} \index{unitary operator} if

\begin{eqnarray*}
U^{-1} = U^{\dag}
\end{eqnarray*}

\noindent or, equivalently

\begin{eqnarray*}
U^{\dag}U = UU^{\dag} = I
\end{eqnarray*}

\noindent A unitary operator preserves the inner product of vectors, i.e.,

\begin{eqnarray}
\langle Ua| Ub \rangle &\equiv& (\langle a|U^{\dag}) (U |b \rangle)
= \langle a|U^{-1}U |b \rangle = \langle a|I |b \rangle = \langle
a|b \rangle \label{eq:preserve}
\end{eqnarray}

\begin{lemma} \label{LemmaA.5}
If $F$ is an Hermitian operator then $U = e^{iF}$ is unitary.
\end{lemma}
\begin{proof}

\begin{eqnarray*}
U^*U &=& (e^{iF})^{\dag} (e^{iF})  = e^{-iF^{\dag}}e^{iF}  =
e^{-iF}e^{iF}  = e^{-iF+iF}  = e^{0}  = I
\end{eqnarray*}

\end{proof}
\bigskip

\begin{lemma} \label{LemmaA.6} Determinant of a unitary matrix $U$ is
unimodular.
\end{lemma}
\begin{proof} We use equation (\ref{eq:A.53x}) to write

\begin{eqnarray*}
|\det(U)|^2 = \det(U) (\det(U))^* = \det(U) \det(U^{\dag}) =
  \det(UU^{\dag}) = \det(I) = 1
\end{eqnarray*}
\end{proof}
\bigskip

Operator $A$ is called \emph{antilinear} \index{antilinear operator}
if $ A ( \alpha |x \rangle + \beta |y \rangle) = \alpha^* A |
x\rangle + \beta^* A | y\rangle $ for any complex $\alpha$ and
$\beta$. An antilinear operator with the property $\langle Ay | Ax
\rangle = \langle y | x \rangle^*$ is called \emph{antiunitary.}
\index{antiunitary operator}

\section{Linear operators in different orthonormal bases}
\label{ss:bases}

So far, we have been working with matrix elements of operators in a
fixed orthonormal basis $|e_i \rangle$. However,  in a
different basis the operator is represented by a different matrix.
Nevertheless, we are going to show that properties of operators defined above remain
valid in all orthonormal basis sets. In other words, we would like
to demonstrate that above operator properties are basis-independent.

\bigskip

\begin{theorem} \label{TheoremA.7} $|e_i \rangle$ and $|e'_i \rangle$ are two
orthonormal bases if and only if there exists a unitary operator $U$
such that

\begin{eqnarray}
U |e_i \rangle = |e'_i \rangle
\label{eq:A.57}
\end{eqnarray}
\end{theorem}
\begin{proof}
 The basis  $|e'_i \rangle$ obtained by
applying a unitary transformation $U$ to the orthonormal  basis
$|e_i \rangle$ is  orthonormal, because unitary transformations preserve inner products of vectors (\ref{eq:preserve}). To prove the reverse statement let us form a matrix

\begin{eqnarray*}
 \left[ \begin{array}{cccc}
\langle e_1  |e'_1 \rangle  & \langle e_1  |e'_2 \rangle  & \ldots  & \langle
e_1  |e'_n \rangle   \\
\langle e_2  |e'_1 \rangle  & \langle e_2  |e'_2 \rangle  & \ldots  & \langle
e_2  |e'_n \rangle   \\
\vdots  & \vdots  & \ddots &  \vdots     \\
\langle e_n  |e'_1 \rangle  & \langle e_n  |e'_2 \rangle  & \ldots  & \langle
e_n  |e'_n \rangle
\end{array} \right]
\end{eqnarray*}

\noindent with matrix elements

\begin{eqnarray*}
 u_{ji} = \langle e_j  |e'_i \rangle
\end{eqnarray*}

\noindent The operator $U$ corresponding to this matrix can be written as

\begin{eqnarray*}
U  &=&  \sum \limits_{jk} | e_j \rangle u_{jk} \langle e_k|=  \sum
\limits_{jk} | e_j \rangle \langle e_j  |e'_k \rangle \langle e_k |
\end{eqnarray*}

\noindent So, acting on the vector $|e_i \rangle$

\begin{eqnarray*}
U |e_i \rangle &=& \sum \limits_{jk} | e_j \rangle \langle e_j |e'_k
\rangle \langle e_k | e_i \rangle
= \sum \limits_{jk} | e_j \rangle
\langle e_j  |e'_k \rangle \delta_{ki}
= \sum \limits_j | e_j \rangle \langle e_j  |e'_i \rangle  \\
&=& |e'_i \rangle
\end{eqnarray*}

\noindent it makes vector $|e'_i \rangle$ as required. Moreover, this operator
is unitary because\footnote{Here we use the following representation
of the identity operator \index{identity operator}

\begin{eqnarray}
I = \sum_i |e_i \rangle \langle e_i |
\label{eq:A.58}
\end{eqnarray}

\noindent which is valid in each orthonormal basis $|e_i \rangle$.
}

\begin{eqnarray*}
 (UU^{\dag})_{ij} &=& \sum \limits_k u_{ik} u_{jk}^*
 = \sum \limits_k \langle e'_i  |e_k \rangle \langle e'_j  |e_k
\rangle ^* = \sum \limits_k \langle e'_i  |e_k \rangle \langle e_k
|e'_j \rangle = \langle e'_i   |e'_j \rangle \\
&=& \delta_{ij}  = I_{ij}
\end{eqnarray*}

\end{proof}
\bigskip

If $F$ is operator with matrix elements $f_{ij}$ in the basis
 $| e_k \rangle$, then its matrix elements $f'_{ij}$ in
the basis $| e'_k \rangle = U | e_k \rangle$ can be obtained by formula

\begin{eqnarray}
f'_{ij} &=&  \langle e'_i | F | e'_j \rangle  =  (\langle e_i |
U^{\dag}) F (U | e_j  \rangle) =  \langle e_i | U^{\dag}F U | e_j
\rangle \nonumber \\
&=&  \langle e_i | U^{-1}F U | e_j \rangle \label{eq:A.59}
\end{eqnarray}

\noindent Equation (\ref{eq:A.59}) can be viewed from two different but equivalent
perspectives. One can regard (\ref{eq:A.59}) either as matrix
elements of $F$ in the new basis set $U|e_i \rangle$ (a
\emph{passive} view) or as matrix elements of the transformed
operator $U^{-1}FU$ in the original basis set $|e_i \rangle$ (an
\emph{active} view).

When the basis is changing, the matrix of the operator changes as well, but the operator's type
remains the same. If operator
$F$ is Hermitian, then in the new basis\footnote{adopting the active view
and omitting symbols for basis vectors}

\begin{eqnarray*}
(F')^{\dag} &=& (U^{-1} F U)^{\dag} = U^{\dag} F^{\dag}
(U^{-1})^{\dag}
       = U^{-1} F U = F'
\end{eqnarray*}

\noindent it is Hermitian as well.

If operator $V$ is unitary, then for the transformed operator $V'$ we have

\begin{eqnarray*}
(V')^{\dag}V' &=& (U^{-1} V U)^{\dag}V'
= U^{\dag} V^{\dag} (U^{-1})^{\dag}V'
= U^{-1} V^{\dag} U V'
= U^{-1} V^{\dag} U U^{-1} V U \\
       &=& U^{-1} V^{\dag} V U = U^{-1}  U = I
\end{eqnarray*}

\noindent so, $V'$ is also unitary.

\bigskip

\begin{lemma} \label{LemmaA.8} The trace  of an operator is basis-independent.
\end{lemma}
\begin{proof}    From Lemma \ref{LemmaA.4} we obtain

\begin{eqnarray*}
Tr(U^{-1} A U) &=&  Tr( A U U^{-1}) =Tr(A)
\end{eqnarray*}
\end{proof}
\bigskip

\section{Diagonalization of Hermitian and unitary matrices}
\label{ss:diagonalization}

We see that the choice of basis in the Hilbert space is a matter of
convenience. So, when performing calculations it is always a good
idea  to choose a basis in which operators have the simplest form,
e.g., diagonal. It appears that Hermitian and unitary operators can
always be made diagonal by an appropriate choice of basis. Suppose
that vector $|x \rangle$ satisfies equation

\begin{eqnarray*}
F  |x \rangle  = \lambda | x \rangle
\end{eqnarray*}

\noindent where $\lambda$ is a complex number called
\emph{eigenvalue} \index{eigenvalue} of the operator $F$. Then $|x
\rangle $ is called \emph{eigenvector} \index{eigenvector} of the
operator $F$.

\bigskip

\begin{theorem} [spectral theorem] \label{spectral-th} \index{spectral theorem}   For
any Hermitian or unitary operator $F$ there is an orthonormal basis
 $| e_i \rangle$ such that

\begin{eqnarray}
F  |e_i \rangle  = f_i | e_i \rangle
\label{eq:A.60}
\end{eqnarray}

\noindent where $f_i$ are complex numbers.
\end{theorem}

\noindent For the proof of this theorem see ref.
\cite{spectral_theorem}.

Equation (\ref{eq:A.60}) means that
 the matrix of the operator $F$ is diagonal in the basis $|e_i \rangle$

\begin{eqnarray}
F =  \left[ \begin{array}{cccc}
f_1 & 0  & \ldots  & 0   \\
 0  & f_2  & \ldots  &  0     \\
\vdots  & \vdots  & \ddots &  \vdots     \\
0  & 0  & \ldots  &  f_n
\end{array} \right]  \label{eq:F.24a}
\end{eqnarray}

\noindent and according to (\ref{eq:A.50}) each Hermitian or unitary operator
can be
expressed through its eigenvectors and eigenvalues

\begin{eqnarray}
F = \sum_{i} | e_i \rangle f_i\langle  e_i | \label{eq:A.61}
\end{eqnarray}

\begin{lemma} \label{LemmaA.10} Eigenvalues of a Hermitian operator are
real.
\end{lemma}
\begin{proof}    Any Hermitian operator can be brought to the diagonal form (\ref{eq:F.24a}) with eigenvalues on the diagonal. It follows from (\ref{eq:F.19a}) that these
diagonal matrix elements are real.
\end{proof}
\bigskip

\begin{lemma} \label{LemmaA.11} Eigenvalues of an unitary operator are
unimodular.
\end{lemma}
\begin{proof}   Using representation
(\ref{eq:A.61}) we can write

\begin{eqnarray*}
I &=& U U^{\dag} =\left( \sum \limits_i |e_i \rangle f_i \langle e_i | \right) \left(
\sum \limits_j |e_j \rangle f^*_j \langle e_j
| \right) \\
  &=& \sum \limits_{ij} f_i f^*_j |e_i \rangle \langle e_i    |e_j \rangle
\langle e_j | = \sum \limits_{ij} f_i f^*_j |e_i \rangle \delta_{ij}
\langle e_j | = \sum \limits_i |f_i|^2 |e_i \rangle  \langle e_i |
\end{eqnarray*}

\noindent Since all eigenvalues of the identity operator are 1, we
obtain $ |f_i|^2 = 1$.
\end{proof}
\bigskip

One benefit of diagonalization is that
 functions of operators
are easily defined in the diagonal form. If operator $A$ has a
diagonal form

\begin{eqnarray*}
A =  \left[ \begin{array}{cccc}
a_1 & 0  & \ldots  & 0   \\
 0  & a_2  & \ldots  &  0     \\
\vdots  & \vdots  & \ddots &  \vdots     \\
0  & 0  & \ldots  &  a_n
\end{array} \right]
\end{eqnarray*}

\noindent then operator $f(A)$ (in the same basis)  has the form

\begin{eqnarray*}
f(A) = \left[ \begin{array}{cccc}
f(a_1) & 0  & \ldots  & 0   \\
 0  & f(a_2)  & \ldots  &  0     \\
\vdots  & \vdots  & \ddots &  \vdots     \\
0  & 0  & \ldots  &  f(a_n)
\end{array} \right]
\end{eqnarray*}

\noindent For example, the matrix of the inverse operator is\footnote{Note
that inverse operator $A^{-1}$ is defined only if all
eigenvalues of $A$ are nonzero.}

\begin{eqnarray*}
A^{-1}  =
\left[ \begin{array}{cccc}
a_1^{-1} & 0  & \ldots  & 0   \\
 0  & a_2^{-1}  & \ldots  &  0     \\
\vdots  & \vdots  & \ddots &  \vdots     \\
0  & 0  & \ldots  &  a_n^{-1}
\end{array} \right]
\end{eqnarray*}

\noindent From Lemma \ref{LemmaA.11}, there is a basis in which the
matrix of unitary operator $U$ is diagonal

\begin{eqnarray*}
U =
\left[ \begin{array}{cccc}
e^{if_1} & 0  & \ldots  & 0   \\
 0  & e^{if_2}  & \ldots  &  0     \\
\vdots  & \vdots  & \ddots &  \vdots     \\
0  & 0  & \ldots  &  e^{if_n}
\end{array} \right]
\end{eqnarray*}

\noindent with real $f_i$. It then follows that each unitary operator can be
represented as

\begin{eqnarray*}
U = e^{iF}
\end{eqnarray*}

\noindent where $F$ is Hermitian

\begin{eqnarray*}
F =
\left[ \begin{array}{cccc}
f_1 & 0  & \ldots  & 0   \\
 0  & f_2  & \ldots  &  0     \\
\vdots  & \vdots  & \ddots &  \vdots     \\
0  & 0  & \ldots  &  f_n
\end{array} \right]
\end{eqnarray*}

\noindent Together with Lemma \ref{LemmaA.5} this establishes an
isomorphism between the sets of Hermitian and unitary operators.

\bigskip

\begin{lemma} \label{LemmaA.12} Unitary transformation of a Hermitian or
unitary operator does not change the spectrum of its eigenvalues.
\end{lemma}
\begin{proof}     If $ | \psi_k \rangle$ is
eigenvector of $M$ with eigenvalue $m_k$

\begin{eqnarray*}
M | \psi_k \rangle = m_k | \psi_k \rangle
\end{eqnarray*}

\noindent then vector $ | U \psi_k \rangle$ is eigenvector of the unitarily
transformed operator $M' = U M U^{-1}$ with the same eigenvalue

\begin{eqnarray*}
M' (U | \psi_k \rangle) &=&  U M U^{-1} (U | \psi_k \rangle )= U M |
\psi_k \rangle = U m_k | \psi_k \rangle =   m_k (U | \psi_k \rangle
)
\end{eqnarray*}
\end{proof}
\bigskip

\chapter{Subspaces and projection operators} \label{sc:subspaces}

\section{Projections} \label{sc:projections}

 Two subspaces $A$ and $B$ in the Hilbert space $\mathcal{H}$ are called
\emph{orthogonal} \index{orthogonal subspaces} (denoted $A \perp B$)
if any vector from $A$ is orthogonal to any vector from $B$. The
span of all vectors which are orthogonal to $A$ is called the
\emph{orthogonal complement} \index{orthogonal complement} to the
subspace $A$ and denoted $A'$.

For a
subspace $A$ (with  $\dim(A) = m$) in the Hilbert space $\mathcal{H}$
(with  $\dim(\mathcal{H}) = n > m$)
 we can select an
orthonormal basis  $|e_i \rangle$ such that first $m$ vectors with
indices $i = 1,2, \ldots, m$ belong to $A$ and vectors with indices
$i = m+1, m+2, \ldots, n$ belong to the orthogonal complement $A'$.
Then for each vector $|y \rangle$ we can write

\begin{eqnarray*}
|y \rangle &=& \sum \limits_i^n |e_i \rangle \langle e_i | y \rangle
= \sum \limits_{i=1}^{m} |e_i \rangle \langle e_i | y \rangle +
      \sum \limits_{i=m+1}^{n} |e_i \rangle \langle e_i | y \rangle
\end{eqnarray*}

\noindent The first sum lies entirely in $A$ and is denoted by
$|y_{\parallel}
\rangle$. The second sum lies in $A'$ and is
denoted $|y_{\perp} \rangle$. This means that we can
 always make a
decomposition of  $|y \rangle$ into two uniquely defined mutually orthogonal components  $|y_{\parallel}
\rangle$ and  $|y_{\perp} \rangle$\footnote{We will also say that
Hilbert space $H$ is represented as a \emph{direct sum}
\index{direct sum} ( $H = A \oplus A'$) of \emph{orthogonal
subspaces} $A$ and $A'$.}

\begin{eqnarray*}
|y \rangle &=& |y_{\parallel} \rangle + |y_{\perp}
\rangle \\
|y_{\parallel} \rangle &\in& A \\
|y_{\perp} \rangle &\in& A'
\end{eqnarray*}

\noindent Then we can define a linear operator $P_A$ called
\emph{projection} \index{projection operator} on the subspace $A$
which associates with any vector $|y \rangle$ its component in the
subspace $A$

\begin{eqnarray*}
P_A |y \rangle = |y_{\parallel}\rangle
\end{eqnarray*}

\noindent The subspace $A$ is called the \emph{range} \index{range}
of the projection $P_A$. In the bra-ket notation we can also write

\begin{eqnarray*}
P_A &=& \sum_{i=1}^{m} |e_i \rangle \langle e_i |
\end{eqnarray*}

\noindent  so that in the above basis $|e_i \rangle$ the operator
$P_A$ has diagonal matrix with first $m$ diagonal entries equal to 1,
and all  others equal to 0. From this, it immediately follows that

\begin{eqnarray*}
P_{A'} = 1 - P_A
\end{eqnarray*}

A set of projections $P_{\alpha} $ on mutually orthogonal subspaces
$H_{\alpha} $ is called \emph{decomposition of unity}
\index{decomposition of unity} if

\begin{eqnarray*}
1 = \sum_{\alpha} P_{\alpha}
\end{eqnarray*}

\noindent or, equivalently

\begin{eqnarray*}
H = \oplus_{\alpha} H_{\alpha}
\end{eqnarray*}

\noindent Thus $P_{A}$ and  $P_{A'}$ provide an example of the decomposition
of
unity.

\bigskip

\begin{theorem} \label{TheoremA.13} Operator $P$ is a projection if and only
if
 $P$  is Hermitian and $P^2 = P$.
 \end{theorem}
\begin{proof}     For Hermitian  $P$, there is a
basis $|e_i \rangle$ in which this operator is diagonal.

\begin{eqnarray*}
P = \sum \limits_i |e_i \rangle p_i \langle e_i |
\end{eqnarray*}

\noindent Then

\begin{eqnarray*}
0 &=& P^2 -P  = \left(\sum \limits_i |e_i \rangle p_i \langle e_i
|\right)\left(\sum \limits_j |e_j \rangle p_j \langle e_j |\right)
- \sum \limits_i |e_i \rangle p_i \langle e_i | \\
  & =& \sum \limits_{ij} |e_i \rangle p_i  p_j \delta_{ij}\langle e_j |
- \sum \limits_i |e_i \rangle p_i \langle e_i | = \sum \limits_i
|e_i \rangle \left(p_i^2 - p_i \right) \langle e_i |
\end{eqnarray*}

\noindent Therefore $p_i^2 - p_i = 0$ and  either $p_i = 0$ or $ p_i = 1$.
From
this we conclude that  $P$ is
a projection on the  subspace spanning
eigenvectors with eigenvalue 1.

To prove the inverse statement we note that any projection operator
is Hermitian because it has real eigenvalues 1 and 0. Furthermore,
for any vector $|y \rangle$

\begin{eqnarray*}
P^2 |y\rangle &=& P |y_{\parallel} \rangle =  |y_{\parallel} \rangle
= P |y \rangle
\end{eqnarray*}

\noindent which proves that $P^2 = P$.
\end{proof}
\bigskip

\section{Commuting operators} \label{ss:commuting}

\begin{lemma} \label{LemmaA.14} Subspaces $A$ and $B$ are orthogonal if and
only if $P_AP_B = P_B P_A = 0$.
\end{lemma}
\begin{proof}    Assume that

\begin{eqnarray}
P_AP_B = P_B P_A = 0 \label{eq:papb}
\end{eqnarray}

\noindent and suppose that there is vector $|y \rangle \in B$ such
that $|y \rangle $ is not orthogonal to $A$. Then $P_A |y \rangle =
|y_A \rangle \neq \mathbf{0}$. From these properties we obtain

\begin{eqnarray*}
P_A P_B |y\rangle &=& P_A  |y\rangle  =  |y_A \rangle = P_A
|y_A\rangle \\
P_B P_A |y\rangle &=& P_B  |y_A\rangle
\end{eqnarray*}

\noindent From the commutativity of $P_A $ and $P_B$ we obtain

\begin{eqnarray*}
 P_A |y_A\rangle &=& P_A P_B |y\rangle  = P_B P_A |y\rangle  =P_B  |y_A\rangle \\
 P_A  P_B  |y_A\rangle  &=&  P_A  P_A  |y_A\rangle  = P_A    |y_A\rangle   \neq \mathbf{0}
\end{eqnarray*}

\noindent So, we found a vector $|y_A\rangle$ for which $P_A  P_B
|y_A\rangle \neq \mathbf{0}$ in disagreement with our original
assumption (\ref{eq:papb}).

The inverse statement is proven as follows. For each vector $|x
\rangle$, the  projection
$P_A |x \rangle$ is in the subspace $A$. If $A$ and $B$ are
orthogonal, then the second projection $P_B P_A |x
\rangle$ yields zero vector. The same arguments show that
$P_A P_B |x
\rangle = \mathbf{0}$, and $P_AP_B = P_B P_A$.
\end{proof}
\bigskip

\begin{lemma} \label{LemmaA.15} If $A \perp B$ then $P_A + P_B$ is the
projection on the direct sum $A \oplus B$.
\end{lemma}
\begin{proof}    If we build an orthonormal  basis
$|e_i \rangle$ in  $A \oplus B$ such that first $\dim (A)$ vectors
belong to $A$, and next $\dim (B)$ vectors belong to $B$, then

\begin{eqnarray*}
 P_A + P_B &=& \sum_{i=1}^{\dim(A)} |e_i \rangle \langle e_i|
+ \sum_{j=1}^{\dim(B)} |e_j \rangle \langle e_j| = P_{A \oplus B}
\end{eqnarray*}
\end{proof}
\bigskip

\begin{lemma} \label{LemmaA.16} If $A \subseteq B$ ($A$ is a subspace of $B$) then

\begin{eqnarray*}
 P_A  P_B = P_B P_A = P_A
\end{eqnarray*}
\end{lemma}
\begin{proof}    If $A \subseteq B$ then there exists
a subspace $C$ in $B$ such that $C \perp A$ and $B = A \oplus
C$.\footnote{This subspace is composed of vectors in $B$, which
are orthogonal to $A$.} According to Lemmas \ref{LemmaA.14} and
\ref{LemmaA.15}

\begin{eqnarray*}
  P_A P_C &=& P_C P_A = 0 \\
  P_B &=&  P_A + P_C \\
  P_A P_B &=& P_A( P_A + P_C) = P_A^2 = P_A \\
  P_B P_A &=& ( P_A + P_C)P_A = P_A
\end{eqnarray*}
\end{proof}
\bigskip

 If there exist three mutually orthogonal subspaces
$X$, $Y$, and $Z$, such that $A = X \oplus Y$ and $B = X \oplus Z$,
then subspaces $A$ and $B$ (and projections $P_A$ and $P_B$) are
called \emph{compatible}. \label{compat-sub} \index{compatible
subspaces}

\bigskip

\begin{lemma} \label{LemmaA.17} Subspaces $A$ and $B$ are compatible if and
only if their corresponding projections commute
\begin{eqnarray*}
[P_A, P_B ] = 0
\end{eqnarray*}
\end{lemma}
\begin{proof}    Let us first show that if $[P_A,P_B]
= 0$ then $P_AP_B = P_B P_A = P_{A \cap B}$ is the projection on the
intersection of subspaces $A$ and $B$.

First we find that

\begin{eqnarray*}
(P_A P_B)^2 &=& P_A P_B P_A P_B =   P_A^2 P_B^2 =   P_A P_B
\end{eqnarray*}

\noindent and that operator $P_AP_B$ is Hermitian, because

\begin{eqnarray*}
(P_A P_B)^{\dag} &=& P_B^{\dag} P_A^{\dag}  =   P_B P_A =   P_A P_B
\end{eqnarray*}

\noindent Therefore, $P_AP_B$ is  a projection by Theorem
\ref{TheoremA.13}. If $A \perp B$, then the direct statement of the
Lemma follows from Lemma \ref{LemmaA.14}. Suppose that $A$ and $B$
are not orthogonal and denote $C = A \cap B$ ($C$ can be empty, of
course). We can always represent $A = C \oplus X$ and $B = C \oplus
Y$, therefore

\begin{eqnarray*}
P_A &=& P_C + P_X \\
P_B &=& P_C + P_Y \\ \mbox{ }
 [P_C, P_X ] &=& 0
\\ \mbox{ }
[P_C, P_Y ] &=& 0
\end{eqnarray*}

\noindent We are left to show that $X$ and $Y$ are orthogonal. This follows
from the  commutator

\begin{eqnarray*}
0 &=& [P_A, P_B] = [P_C + P_X , P_C + P_Y] \\
     &=& [P_C , P_C] + [P_C , P_Y] + [P_X , P_C] + [P_X ,  P_Y] =  [P_X ,  P_Y]
\end{eqnarray*}

Let us now prove the inverse statement.
From the compatibility of $A$ and $B$ it follows that

\begin{eqnarray*}
P_A  &=&  P_X + P_Y \\
P_B  &=&   P_X + P_Z \\
P_X P_Y &=& P_X P_Z = P_Y P_Z = 0 \\
\  [P_A, P_B]   &=&  [P_X + P_Y,    P_X + P_Z] = 0
\end{eqnarray*}
\end{proof}

\bigskip

\begin{lemma} \label{LemmaG.6} If projection $P$ is compatible with
all other projections in the Hilbert space, then either $P=0$ or $P=1$.
\end{lemma}
\begin{proof}
Suppose that $P \neq 0$ and $P\neq 1$. Then $P$ has a non-empty
range $A$, which is different from $\mathcal{H}$. So, the orthogonal
complement $A'$ is not empty as well. Choose an arbitrary vector $y$
with non-zero components $| y_{\parallel} \rangle$ and $|y_{\perp}
\rangle$ with respect to $A$. Then it is easy to show that
projection on $|y \rangle$ does not commute with $P$. Therefore, by
Lemma \ref{LemmaA.17} this projection is not compatible with $P$.
\end{proof}
\bigskip

 Note that two or more eigenvectors of a Hermitian operator $F$ may
correspond to  the same eigenvalue (such an eigenvalue is called
\emph{degenerate}). \index{degenerate eigenvalue} Then any linear
combination of these eigenvectors is again an eigenvector with the
same eigenvalue. The span of all eigenvectors with the same
eigenvalue $f$ is called the \emph{ eigensubspace}
\index{eigensubspace} of the operator $F$, and  one can associate a
projection $P_{f}$ on this
 subspace with eigenvalue $f$. Then Hermitian
operator $F$ can be written as

\begin{eqnarray}
F = \sum \limits_{f} f P_{f} \label{eq:A.62}
\end{eqnarray}

\noindent where index $f$ now runs over all distinct eigenvalues of
$F$ and $P_{f}$ are referred to as \emph{spectral projections}
\index{spectral projection} of $F$. This means that each Hermitian
operator defines an unique decomposition of unity $I = \sum _{f} P_{f}$. Inversely, if
$P_{f}$ is a decomposition of unity and $f$ are real numbers then
equation (\ref{eq:A.62}) defines an unique Hermitian operator.

\bigskip

\begin{lemma} \label{LemmaA.18} If two Hermitian operators $F$ and $G$
commute then all spectral projections of $F$ commute with $G$.
\end{lemma}
\begin{proof}
  Consider operator $P$
which is a spectral projection of $F$. Take any vector $|x \rangle $
in the range of $P$,
i.e.,

\begin{eqnarray*}
P |x \rangle &=&  |x \rangle \\
F |x \rangle &=& f |x \rangle
\end{eqnarray*}

\noindent for some real $f$. Let us first prove that  the vector
$G|x \rangle$ also lies in the range of $P$. Indeed, using the
commutativity of $F$ and $G$ we obtain

\begin{eqnarray*}
F G |x \rangle &=& G F |x \rangle = G f |x \rangle =f  G  |x \rangle
\end{eqnarray*}

\noindent This means that operator $G$ leaves all eigensubspaces of $F$
invariant.
Then for any vector $|x \rangle$ the vectors $P |x \rangle$ and $G P |x
\rangle$ lie in the range of $P$. Therefore

\begin{eqnarray}
P G P = G P
\label{eq:A.63}
\end{eqnarray}

\noindent Taking adjoint of both sides we obtain

\begin{eqnarray}
P G P = P G
\label{eq:A.64}
\end{eqnarray}

\noindent Now subtracting (\ref{eq:A.64}) from (\ref{eq:A.63}) we obtain

\begin{eqnarray*}
[G, P] = G P - P G = 0
\end{eqnarray*}
\end{proof}
\bigskip

\begin{theorem} \label{TheoremA.19} Two Hermitian operators $F$ and $G$
commute if and only if  all their spectral projections commute.
\end{theorem}
\begin{proof}    We write

\begin{eqnarray}
F &=& \sum \limits_i f_i P_i \label{eq:FfP}\\
G &=& \sum \limits_j g_j Q_j \label{eq:GgQ}
\end{eqnarray}

\noindent If $[P_i, Q_j] = 0$ for all $i,j$, then obviously $[F,G]=0$. To prove
the reverse statement we notice that from Lemma \ref{LemmaA.18} each
spectral projection $P_i$ commutes with $G$. From the same Lemma it
follows that each spectral projection of $G$ commutes with $P_i$.
\end{proof}
\bigskip

\begin{theorem} \label{TheoremA.20} If two Hermitian operators $F$ in (\ref{eq:FfP}) and $G$ in (\ref{eq:GgQ})
commute then there is a basis $|e_i \rangle$ in which both $F$ and
$G$ are diagonal, i.e., $|e_i \rangle$  are common eigenvectors of
$F$ and $G$.
\end{theorem}
\begin{proof} The identity operator can be written in three different ways

\begin{eqnarray*}
I &=& \sum_i P_i \\
I &=& \sum_j Q_j \\
I &=& I \cdot I =\left(\sum_i P_i\right) \left(\sum_j Q_j\right) =
\sum_{ij} P_i Q_j
\end{eqnarray*}

\noindent where $P_i$ and $Q_j$ are spectral projections of operators $F$ and
$G$, respectively.
Since $F$ and $G$ commute, the operators $P_i Q_j$ with different $i$
and/or $j$ are projections on
mutually orthogonal subspaces. So, these projections form a spectral
decomposition of unity, and the desired basis is obtained
by coupling bases in the subspaces $P_i Q_j$.
\end{proof}
\bigskip

\chapter{Representations of groups} \label{sc:group-reps}

A \emph{representation} \index{representation of group} of a group
$G$ is a \emph{homomorphism}\footnote{homomorphism = a mapping that preserves group operations}
 between the group $G$ and the  group of  linear
transformations in a vector space. In other words, to each group
element $g$ there corresponds a   matrix $U_g$ with non-zero
determinant.\footnote{Matrices with zero determinant cannot be inverted, so they cannot represent group elements.} The group multiplication is represented by the matrix
product and the following conditions are satisfied

\begin{eqnarray*}
U_{g_1} U_{g_2}  &=&  U_{g_1g_2} \\
U_{g^{-1}}  &=& U_{g}^{-1} \\
U_e &=& I
\end{eqnarray*}

\noindent Each group has a \emph{trivial} \index{trivial
representation} representation in which each group element is
represented by the identity operator. If the linear space of the
representation is a Hilbert space $\mathcal{H}$, then we can define
a particularly useful class of \emph{unitary} representations.
\index{unitary representation} These representations are made of
unitary operators.

\section{Unitary representations of groups}
\label{ss:unitary-reps}

 Two representations $U_g$ and $U_g'$ of a group $G$ in the Hilbert space
$\mathcal{H}$ are called \emph{unitarily equivalent} \index{unitary
equivalent representations} if there exists a unitary operator $V$
such that for each $g \in G$

\begin{eqnarray}
U_g' = VU_g V^{-1} \label{eq:un-tr}
\end{eqnarray}

\noindent Having two representations $U_g$ and $V_g$ in Hilbert spaces $H_1$
and
$H_2$ respectively, we can always
build another representation $W_g$ in the Hilbert space $H = H_1
\oplus H_2$ by joining two matrices in the block diagonal form.

\begin{eqnarray}
W_g =   \left[ \begin{array}{cc}
 U_g & 0 \\
 0   & V_g
\end{array} \right]
\label{eq:A.65}
\end{eqnarray}

\noindent This is called the \emph{direct sum} \index{direct sum} of
two representations. The direct sum is denoted by the sign $\oplus$

\begin{eqnarray*}
W_g =   U_g \oplus  V_g
\end{eqnarray*}

A representation is called \emph{reducible} \index{reducible
representation} if there is a unitary transformation (\ref{eq:un-tr}) that brings
representation matrices to the block diagonal form (\ref{eq:A.65}) for all $g$. Otherwise, the
representation is called \emph{irreducible}.  \index{irreducible
representation}

 \emph{Casimir operators} are operators which commute with all
representatives of group elements. \index{Casimir operator}

\bigskip
\label{schur}
\begin{lemma} [Schur's first lemma  \cite{Schur}]
\label{Lemma.schur} Casimir operators   of an unitary irreducible
representation of any group are constant multiples of the unit
matrix.
\end{lemma}

\bigskip

\noindent From Appendix \ref{ss:lie} we know that elements of
any Lie group in the vicinity of the unit element can be represented
as

\begin{eqnarray*}
g =   e^{A}
\end{eqnarray*}

\noindent where $A$ is an element from the Lie algebra of the
group. Correspondingly,  any matrix of the unitary group
representation in $\mathcal{H}$ can be written as

\begin{eqnarray*}
U_g =   e^{-\frac{i}{\hbar}F_A}
\end{eqnarray*}

\noindent where $F_A$ is a Hermitian operator and $\hbar$ is a real
constant.\footnote{Here we use the Planck's constant, but any other
nonzero real constant will do as well.} Operators $F_A$ form a
\emph{representation of the Lie algebra} \index{representation of
Lie algebra} in the Hilbert space $\mathcal{H}$. If the Lie bracket
of two Lie algebra elements is $[A,B] = C$, then the commutator of
their Hermitian representatives is

\begin{eqnarray*}
[F_A, F_B] \equiv F_A F_B - F_B F_A = i \hbar F_c
\end{eqnarray*}

\section{Stone's theorem}
\label{ss:stone-theorem}

\index{Stone's theorem} Stone's theorem provides a valuable
information about unitary representations of 1-dimensional Lie
groups. Such groups are called also one-parameter Lie groups,
because all their elements $g(z)$ can be parameterized with one real
parameter $z \in \mathbb{R}$, so that

\begin{eqnarray*}
g(0) &=& e \\
g(z_1) g(z_2) &=& g(z_1+z_2) \\
g(z)^{-1}&=&  g(-z)
\end{eqnarray*}

\begin{theorem} [Stone \cite{Stones}]
\label{Theorem.Stone's} If $U_g$ is a unitary representation of a
1-dimensional Lie group in the Hilbert space $\mathcal{H}$, then
there exists an Hermitian operator $T$ in $\mathcal{H}$, such that

\begin{eqnarray}
U_{g(z)}&=&  e^{-\frac{i}{\hbar}Tz} \label{eq:ugz=}
\end{eqnarray}

\end{theorem}

This theorem is useful not only for 1-dimensional Lie groups, but
also for Lie groups of arbitrary dimension. The reason is that in
any Lie group one can find multiple \emph{one-parameter subgroups},
\index{one-parameter subgroup} for which the theorem can be applied.\footnote{See Appendix \ref{ss:lie}}
For example consider an arbitrary Lie group $G$ and a basis vector
$\vec{t}$ from its Lie algebra. Consider a set of group elements of
the form

\begin{eqnarray}
g(z) &=& e^{z\vec{t} } \label{eq:gz}
\end{eqnarray}

\noindent where parameter $z$ runs through all real numbers $z \in
\mathbb{R}$. It is easy to see that the set (\ref{eq:gz}) forms a
one-parameter subgroup in $G$. Indeed, this set contains the unit
element (when $z = 0$); the group product is defined as

\begin{eqnarray*}
g(z_1) g(z_2)&=& e^{z_1 \vec{t} } e^{z_2 \vec{t} } = e^{(z_1+z_2)
\vec{t} } = g(z_1+z_2)
\end{eqnarray*}

\noindent and the inverse element is

\begin{eqnarray*}
g(z)^{-1}&=& e^{-z \vec{t} }  = g(-z)
\end{eqnarray*}

From the Stone's theorem we can then conclude that in any unitary
representation of $G$ representatives of $g(z)$ have the form
(\ref{eq:ugz=}) with some fixed Hermitian operator $T$.

\section{Heisenberg Lie algebra}
\label{ss:heisenberg-lie}

The \emph{Heisenberg Lie algebra} \index{Heisenberg Lie algebra}
$h_{2n}$ of dimension $2n$ has basis  elements $\mathcal{P}_i$ and
$\mathcal{R}_i$ ($i=1,2, \ldots, n$) with Lie brackets

\begin{eqnarray*}
[\mathcal{P}_i, \mathcal{P}_j] &=& [\mathcal{R}_i, \mathcal{R}_j] =
0 \\
\  [\mathcal{R}_i, \mathcal{P}_j] &=& \delta_{ij}
\end{eqnarray*}

\noindent  The following theorem is applicable

\bigskip
\label{stone}
\begin{theorem} [Stone-von Neumann \cite{Stone}]
\label{Theorem.Stone}   If $(P_i, R_i)$ ($i=1,2, \ldots, n$) is a
Hermitian representation\footnote{This means that Hermitian
operators $P_i$ and  $R_i$ satisfy commutation relations
\begin{eqnarray*}
\ [P_i, P_j] &=& [R_i, R_j] = 0 \\
\  [R_i, P_j] &=& i \hbar\delta_{ij}
\end{eqnarray*}
\noindent where $\hbar$ is a real constant.}
 of the Heisenberg Lie algebra $h_{2n}$ in the Hilbert space $\mathcal{H}$,
then

\begin{enumerate}
\item  representatives $P_i$ and $R_i$ have continuous spectra from $- \infty$
to
$\infty$.
\item any irreducible representation of $h_{2n}$ is unitary equivalent to the
so-called Schr\"odinger representation. In the physically relevant
case $n=3$, the Schr\"odinger representation is the one described in
subsection \ref{ss:position-representation}: Vectors in the Hilbert
space are represented by complex functions on $\mathbb{R}^3$;
operator $\mathbf{R}$ multiplies these functions by $\mathbf{r}$;
operator $\mathbf{P}$ is differentiation $-i \hbar d/d\mathbf{r}$.
\end{enumerate}
\end{theorem}

\bigskip

\section{Double-valued representations of the rotation group}
\label{ss:double-valued}

\begin{figure}
\centering
 \includegraphics {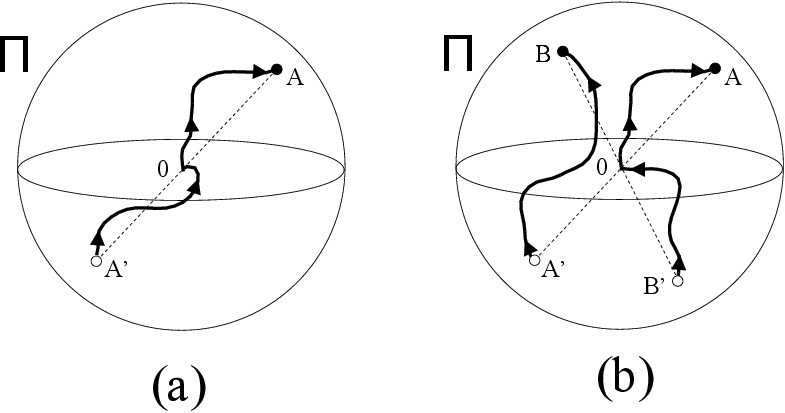} \caption{The space of
parameters of the rotation group is not simply connected: (a) a loop
which starts from the center of the ball $\{ \vec{0}\}$, reaches the
surface of the sphere $\Pi$ at point $A$ and then continues from
the opposite point $A'$ back to $\{ \vec{0}\}$; this loop  cannot be
continuously collapsed to $\{ \vec{0}\}$, because it crosses the
surface an odd  number of times (1); (b) a loop $\{ \vec{0}\} \to A
\to A' \to B \to B' \to \{ \vec{0}\}$ which crosses the surface of
the sphere $\Pi$ twice can be deformed to the point $\{ \vec{0}\}$.
This can be achieved by moving the points $A'$ and $B$ (and,
correspondingly the points $A$ and $B'$) close to each other, so
that the segment $A' \to B$ of the path disappears.} \label{fig:5.1}
\end{figure}

The rotation group\footnote{see Appendix \ref{sc:rotations}} has a peculiar non-trivial topology: Results of two
rotations around the same axis by angles $\phi + 2 \pi n$
(with different integer $n$) are physically indistinguishable. Then the region
of independent rotation vectors\footnote{Recall from Appendix
\ref{ss:parameterization} that direction of the rotation vector
$\vec{\phi}$ coincides with the axis of rotation, and its length
$\phi$ is the rotation angle.} in $\mathbb{R}^3$ can be described as
the interior of the sphere of radius $\pi$ with opposite points on
the surface of the sphere being equivalent.  This set of points will be
referred to as the ball $\Pi$ (see Fig. \ref{fig:5.1}). The unit element
$\{ \vec{0}\}$ is in the center of the ball. We will be interested
in one-parameter families of group elements\footnote{They are not necessarily one-parameter subgroups.} which form continuous
curves in the group manifold $\Pi$. Since the opposite points on the
surface of the ball are identical in our topology, any continuous
path that crosses the surface must reappear on the opposite side of
the sphere (see Fig. \ref{fig:5.1}(a)).

A topological space is \emph{simply connected} \index{simply
connected space} if every loop can be continuously deformed to a
single point. An example of a simply connected topological space is
the surface of a sphere. However, the manifold $\Pi$ of the rotation
parameters is not simply connected. The loop shown in Fig.
\ref{fig:5.1}(a) crosses the sphere once and  can not be shrunk to a
single point. However, the loop shown in Fig. \ref{fig:5.1}(b) can
be continuously deformed to a point, because it crosses the sphere twice. It appears that for any
rotation $R$ there are two classes of paths from the group's unit
element $\{ \vec{0}\}$ to $R$. They are also called the
\emph{homotopy classes}. \index{homotopy class} These two classes
consist of paths that cross the surface of the sphere $\Pi$ even and
odd number of times, respectively. Two paths from different classes
cannot be continuously deformed to each other.

If we build a projective representation of the rotation group, then, similar to our discussion of the Poincar\'e group in subsection \ref{ss:central-charges}, central charges
can be eliminated by a proper
choice of  numerical constants added to generators. Then a unitary
representation of the rotation group can be constructed in which
 the identity rotation is represented by the identity operator and
by traveling a small loop in the group manifold from the identity
element $\{ \vec{0}\}$ back to $\{ \vec{0}\}$ we will  end up with
the identity operator $I$ again. However, if we travel the long path
$\{ \vec{0}\} \to A \to A' \to \{ \vec{0}\}$ in Fig.
\ref{fig:5.1}(a), there is no guarantee that in the end we will find
the same representative $I$ of the identity transformation. We can get
some other equivalent unitary operator from the ray containing $I$,
so the representative of $\{ \vec{0}\}$ may acquire
 a phase factor
$e^{i\phi} $ after travel along such a loop.  On the other hand,
making two passes on the loop $\{ \vec{0}\} \to A \to A' \to \{
\vec{0}\} \to A \to A' \to \{ \vec{0}\}$  we obtain a loop which
crosses the surface of the sphere twice and hence can be deformed
to a point. Therefore $e^{2i\phi}  = 1$ and $e^{i\phi}  = \pm 1$.
This demonstrates that
 there are two
types of unitary  representations of the rotation group:
single-valued and double-valued representations. For \emph{single-valued}
\index{single-valued representation} representations, the
representative of the identity rotation is always $I$. For
\emph{double-valued} \index{double-valued representation}
representations, the identity rotation has two representatives $I$
and $-I$ and the product of two operators in (\ref{eq:5.15}) may
have a non-trivial sign factor

\begin{eqnarray*}
U_{g_1}U_{g_2} = \pm U_{g_1 g_2}
\end{eqnarray*}

\noindent For irreducible representations of the rotation group (both single-valued and double-valued) see Appendix \ref{ss:rotation-reps}.

\section{Unitary irreducible representations of the
rotation group} \label{ss:rotation-reps}

There is an infinite number of unitary irreducible representation
$D^s$ of the rotation group which are characterized by the value of
spin $s = 0, 1/2, 1, \ldots$. These representations are thoroughly
discussed in a number of good textbooks, see, e.g., ref.
\cite{rotation_group}. In Table \ref{table:A.3} we just provide a
summary of these results: the dimension of the representation space,
the value of the Casimir operator $\mathbf{S}^2$, the spectrum of
each component of the spin operator\footnote{We denote $S_x, S_y, S_z$ Hermitian representatives of the Lie algebra basis vectors $\mathcal{J}_x, \mathcal{J}_y, \mathcal{J}_z$. See Appendix \ref{ss:rotation-generators}.} and an explicit form of the
three generators of the representation. \index{representations of
rotation group}

\begin{table}[h]
\caption{Unitary irreducible representations of SU(2)}
\begin{tabular*}{\textwidth}{@{\extracolsep{\fill}}ccccc}
\hline
 Spin:   &  $s=0$         &   $s=1/2$       &  $s=1$     &  $s=3/2,2, \ldots$
  \cr
 \hline
dimension  & 1  & 2 &  3  & $ 2s+1$ \cr $<\mathbf{S}^2>$  & 0  &
$\frac{3}{4} \hbar^2 $     & $2 \hbar^2$ & $\hbar^2 s(s+1)$   \cr
$s_x $ or $s_y$ or $s_z$ &  0 & $-\hbar/2, \hbar/2$ & $-\hbar, 0,
\hbar$ & $-\hbar s, \hbar(-s+1), \ldots,$ \cr
 & & & & $\hbar(s-1), \hbar s$ \cr
 $S_x$ & 0  & $\left[ \begin{array}{cc}
  0 &  \hbar/2   \\
  \hbar/2 & 0
\end{array} \right]$       &  $\left[ \begin{array}{ccc}
  0 & 0 & 0 \\
  0 & 0 & -i \hbar   \\
 0 &  i\hbar & 0
\end{array} \right] $ &      \cr
 $S_y$ & 0      & $\left[ \begin{array}{cc}
  0 &  -i\hbar/2   \\
  i\hbar/2 & 0
\end{array} \right]$         &  $\left[ \begin{array}{ccc}
  0 & 0 & i \hbar   \\
  0 & 0 & 0 \\
  -i\hbar & 0 & 0
\end{array} \right] $ &  see, e.g., ref. \cite{rotation_group}   \cr
$S_z$  & 0   &  $ \left[ \begin{array}{cc}
  \hbar/2 & 0  \\
 0 & -\hbar/2
\end{array} \right]$      &  $\left[ \begin{array}{ccc}
  0 & -i \hbar & 0  \\
  i\hbar & 0 & 0  \\
  0 & 0 & 0
\end{array} \right] $ &    \cr
 \hline
\end{tabular*}
\label{table:A.3}
\end{table}

Representations characterized by integer spin $s$ are single-valued.
Half-integer spin representations are double-valued.\footnote{see
Appendix \ref{ss:double-valued}} For example, in the 2-dimensional
representation ($s = 1/2$), the rotation through the angle $2
\pi$ around the $z$-axis is represented by negative unity

\begin{eqnarray*}
e^{- \frac{i}{\hbar} S_z 2 \pi} &=& \exp \left(- \frac{2 \pi
i}{\hbar}\left[
\begin{array}{cc}
  \hbar/2 & 0  \\
 0 & -\hbar/2
\end{array} \right] \right) = \left[ \begin{array}{cc}
  e^{i \pi} & 0  \\
 0 & e^{-i \pi}
\end{array} \right]  \\
&=& \left[ \begin{array}{cc}
 -1 & 0  \\
 0 & -1
\end{array} \right] = - I
\end{eqnarray*}

\noindent while a $4 \pi$ rotation is represented by the unit matrix

\begin{eqnarray*}
e^{- \frac{i}{\hbar} S_z 4 \pi} &=&  \left[ \begin{array}{cc}
 1 & 0  \\
 0 & 1
\end{array} \right] =  I
\end{eqnarray*}

\section{Pauli matrices} \label{ss:pauli-matrices}

Generators of the spin 1/2 representation of the rotation group (see Table \ref{table:A.3}) can be conveniently
expressed through \emph{Pauli matrices} \index{Pauli matrices}
$\sigma_i$  $(i=x,y,z)$

\begin{eqnarray}
S_i = \frac{\hbar}{2} \sigma_i \ \ \ \ \label{eq:si-sigma}
\end{eqnarray}

\noindent where

\begin{eqnarray*}
 \sigma_x \equiv \sigma_1 &=&  \left[
\begin{array}{cc}
 0 & 1  \\
 1 & 0 \\
\end{array} \right]
\label{eq:A.67} \\
 \sigma_y  \equiv \sigma_2 &=&  \left[
\begin{array}{cc}
 0 & -i  \\
 i & 0 \\
\end{array} \right]
\label{eq:A.68} \\
 \sigma_z  \equiv \sigma_3 &=&  \left[
\begin{array}{cc}
 1 & 0  \\
 0 & -1 \\
\end{array} \right]
\label{eq:A.69}
\end{eqnarray*}

\noindent Sometimes it is convenient to define a fourth Pauli matrix

\begin{eqnarray*}
\sigma_t \equiv \sigma_0 &=&  \left[ \begin{array}{cc}
 1 & 0  \\
 0 & 1 \\
\end{array} \right]
\label{eq:A.66}
\end{eqnarray*}

\noindent For reference we list here some properties of the Pauli
matrices

\begin{eqnarray*}
[\sigma_i, \sigma_j] &=& 2i\sum_{i=1}^3 \epsilon_{ijk}  \sigma_k \\
\{ \sigma_i ,\sigma_j \} &=& 2 \delta_{ij} \\
\sigma_i^2 &=& 1
\end{eqnarray*}

\noindent For arbitrary numerical 3-vectors $\mathbf{a}$ and $\mathbf{b}$ we have

\begin{eqnarray}
(\vec{\sigma}\cdot \mathbf{a}) \vec{\sigma} &=&\mathbf{a} \sigma_0 +
i[\vec{\sigma} \times \mathbf{a}] \label{eq:A.69a}\\
\vec{\sigma}(\vec{\sigma}\cdot \mathbf{a})  &=& \mathbf{a} \sigma_0 -
i[\vec{\sigma} \times \mathbf{a}] \label{eq:A.69b} \\
(\vec{\sigma}\cdot \mathbf{a}) (\vec{\sigma}\cdot \mathbf{b}) &=&
(\mathbf{a} \cdot \mathbf{b}) \sigma_0 + i \vec{\sigma} \cdot [\mathbf{a}
\times \mathbf{b}]
\label{eq:A.69c} \\
\ [\vec{\sigma}_{pr} \times \mathbf{a}]\cdot [\vec{\sigma}_{el} \times
\mathbf{a}] &=& [[\vec{\sigma}_{el}
\times \mathbf{a}] \times \vec{\sigma}_{pr}] \cdot \mathbf{a} \nonumber \\
&=& (\mathbf{a} (\vec{\sigma}_{el} \cdot \vec{\sigma}_{pr}) -
\vec{\sigma}_{pr}
(\vec{\sigma}_{el} \cdot\mathbf{a} ) ) \cdot \mathbf{a} \nonumber \\
&=& a^2 (\vec{\sigma}_{el} \cdot \vec{\sigma}_{pr}) -
(\vec{\sigma}_{pr} \cdot \mathbf{a}) (\vec{\sigma}_{el}
\cdot\mathbf{a} ) \nonumber
\end{eqnarray}

\chapter{Special relativity} \label{sc:lorentz-time-pos}
\index{special relativity}

In this Appendix we present major assertions of Einstein's special
relativity \cite{Einstein_1905}. In chapter \ref{sc:obs-interact} we
argued that this theory is approximate. We also suggested there an alternative
rigorous approach, which  ensures the validity of the relativity principle in interacting systems.

\section{4-vector representation of the Lorentz group}
\label{ss:4-dim-rep}

The Lorentz group is a 6-dimensional subgroup of the Poincar\'e
group, which is formed by rotations and boosts. Linear (tensor)
representations of the Lorentz group play a significant role in many
physical problems.

The 4-vector \index{4-vector} representation of the Lorentz group
forms the mathematical framework of special relativity discussed in
this Appendix. This representation resembles the 3-vector
 representation of the rotation group.\footnote{see Appendix
\ref{ss:scalars}} Let us first define the vector space where this representation is acting. This is a 4-dimensional real
vector space $\mathcal{M}$ whose vectors are denoted by\footnote{Here $c$ is the
speed of light.  Also in this book we always denote 4-vectors by the
tilde. By the way, here we do not bestow the Minkowski space $\mathcal{M}$ with any physical meaning. For us $\mathcal{M}$ is just an abstract vector space, unrelated to the physical space and time.} \index{position-time 4-vector} \index{$\tilde{\tau}$
4-vector}

 \begin{eqnarray*}
\tilde{\tau}  = \left[ \begin{array}{c}
  ct   \\
  x   \\
  y  \\
  z  \\
\end{array} \right]
\end{eqnarray*}

\noindent  and the \emph{pseudoscalar product}
\index{pseudoscalar product} of any two 4-vectors  $\tilde{\tau}_1$ and $\tilde{\tau}_2$ can be written in a number of equivalent
forms\footnote{Here indices $\mu$ and $\nu$ run from 0 to 3: $\tau_0
= ct, \tau_1 = x, \tau_2=y, \tau_3=z$.}

\begin{eqnarray}
\tilde{\tau}_1 \cdot \tilde{\tau}_2 &\equiv&  c^2t_1 t_2  -x_1 x_2 -
y_1 y_2 - z_1 z_2
=  \sum_{\mu \nu = 0}^3 (\tau_1)_{\mu} g^{\mu \nu} (\tau_2)_{\nu} \nonumber\\
&=&  [ct_1, x_1, y_1, z_1] \left[ \begin{array}{cccc}
 1 & 0 & 0 & 0 \\
 0 & -1 & 0 & 0 \\
 0 &  0 & -1 & 0 \\
 0 &  0 &  0 & -1 \\
\end{array} \right]
\left[ \begin{array}{c}
 ct_2 \\
 x_2  \\
 y_2 \\
 z_2  \\
\end{array} \right] \nonumber \\
& =& \tilde{\tau}_1^T g \tilde{\tau}_2 \label{eq:A.71}
\end{eqnarray}

\noindent where $g^{\mu \nu}$ are matrix elements of the so-called \emph{metric tensor}. \index{metric
tensor}

\begin{eqnarray*}
g=  \left[ \begin{array}{cccc}
 1 & 0 & 0 & 0 \\
 0 & -1 & 0 & 0 \\
 0 &  0 & -1 & 0 \\
 0 &  0 &  0 & -1 \\
\end{array} \right]
\end{eqnarray*}

For compact notation it is convenient to define a vector with a
``raised index'' and to adopt the Einstein's convention about summation over repeated
indices

\begin{eqnarray*}
\tau^{\mu} &\equiv&  \sum_{\nu = 0}^3 g^{\mu \nu} \tau_{\nu} \equiv
g^{\mu \nu} \tau_{\nu}
= (ct, -x , -y, -z)
\end{eqnarray*}

\noindent Then, the pseudoscalar product can be rewritten as

\begin{eqnarray}
\tilde{\tau}_1 \cdot \tilde{\tau}_2 &\equiv& (\tau_1)_{\mu}
(\tau_2)^{\mu} = (\tau_1)^{\mu}
(\tau_2)_{\mu} \label{eq:A.71a}
\end{eqnarray}

\noindent The tilde notation allows us to distinguish the
\emph{pseudoscalar square} (or \emph{4-square}) \index{4-square} of the 4-vector $\tilde{\tau}$

\begin{eqnarray*}
\tilde{\tau}^2 \equiv \tilde{\tau} \cdot \tilde{\tau} = \tau_{\mu} \tau^{\mu} = \tau_0^2 - \tau_1^2 -
\tau_2^2 - \tau_3^2 = \tau_0^2 - \tau^2
\end{eqnarray*}

\noindent from the square of its 3-vector part

\begin{eqnarray*}
\tau^2 \equiv  (\vec{\tau} \cdot \vec{\tau}) =  \tau_1^2 + \tau_2^2
+ \tau_3^2
\end{eqnarray*}

\noindent A 4-vector $(\tau_0, \vec{\tau})$ is called \emph{space-like} \index{space-like 4-vector} if
$\tau^2 > \tau_0^2$. \emph{Time-like} 4-vectors \index{time-like 4-vector} have $\tau^2 < \tau_0^2$, and for \emph{null} 4-vectors \index{null 4-vector} the condition is $\tau^2 = \tau_0^2$.

The 4-vector representation of the Lorentz group is defined as a representation  by linear transformations in the vector space $\mathcal{M}$ that conserve the pseudoscalar product of 4-vectors. In other words, representation matrices $\Lambda$ must satisfy

\begin{eqnarray*}
\tilde{\tau}_1' \cdot \tilde{\tau}_2' &\equiv& \Lambda
\tilde{\tau}_1 \cdot \Lambda \tilde{\tau}_2
= \tilde{\tau}_1^T \Lambda^T g \Lambda \tilde{\tau}_2
=  \tilde{\tau}_1^T  g  \tilde{\tau}_2
= \tilde{\tau}_1 \cdot \tilde{\tau}_2
\end{eqnarray*}

\noindent which means that matrices $\Lambda$ must have the property

\begin{eqnarray}
 g &=&  \Lambda^T g \Lambda
\label{eq:A.74}
\end{eqnarray}

\noindent One useful implication of this result is

\begin{eqnarray}
\Lambda \tilde{\tau}_1 \cdot  \tilde{\tau}_2 &=& \tilde{\tau}_1^T \Lambda^T g  \tilde{\tau}_2 = \tilde{\tau}_1^T g \Lambda^{-1} \tilde{\tau}_2 = \tilde{\tau}_1
\cdot \Lambda^{-1} \tilde{\tau}_2
 \label{eq:A.73}
\end{eqnarray}

 Another property of $\Lambda$ can be obtained by taking the determinant of both sides of (\ref{eq:A.74})

\begin{eqnarray*}
-1 &=& \det(g) =  \det(\Lambda^T g \Lambda) =  \det(\Lambda^T) \det(
g) \det( \Lambda) = - \det(\Lambda)^2
\end{eqnarray*}

\noindent which implies $ \det( \Lambda) = \pm 1$. Writing equation
(\ref{eq:A.74}) for the $g_{00}$ component we also get

\begin{eqnarray*}
1 &=& g_{00} = \sum_{\alpha', \beta' = 0}^3 \Lambda_{\alpha' 0}
g_{\alpha' \beta'} \Lambda_{\beta' 0} =   -\Lambda_{10}^2 -
\Lambda_{20}^2 - \Lambda_{30}^2 + \Lambda_{00}^2
\end{eqnarray*}

\noindent It then follows that $\Lambda_{00}^2 \geq 1$, which means
that either  $\Lambda_{00} \geq 1$ or $\Lambda_{00} \leq -1$.

 The unit element of the group is represented by the identity
transformation $I$, which obviously has $\det(I) = 1$ and $I_{00} =
1$. As we are interested only in rotations and boosts which can be
continuously connected to the unit element, we must choose

\begin{eqnarray}
\det(\Lambda) &=& 1
\label{eq:A.75}\\
\Lambda_{00} &\geq& 1 \label{eq:A.76}
\end{eqnarray}

\noindent The matrices satisfying equation (\ref{eq:A.74}) with
additional conditions (\ref{eq:A.75}) - (\ref{eq:A.76}) will be
called \emph{pseudoorthogonal}. \index{pseudoorthogonal matrix} Thus
we can say that $4 \times 4$ pseudoorthogonal matrices form a
representation of the Lorentz group.

 Boost transformations can be written as

\begin{eqnarray}
  \left[ \begin{array}{c}
   ct'  \\
   x' \\
   y' \\
   z'
\end{array} \right] = B(\vec{\theta})  \left[ \begin{array}{c}
   ct \\
   x \\
  y \\
   z
\end{array} \right]
\label{eq:A.70}
\end{eqnarray}

\noindent where general pseudoorthogonal matrix of boost is\footnote{compare with equations (\ref{eq:boost-momentum}) and (\ref{eq:boost-energy})}

\begin{eqnarray}
B(\vec{\theta}) =  \left[ \begin{array}{cccc}
  \cosh \theta &  -\frac {\theta_x}{\theta} \sinh \theta&
-\frac {\theta_y}{\theta} \sinh \theta &  -\frac {\theta_z}{\theta}
\sinh
\theta \\
   -\frac {\theta_x}{\theta} \sinh \theta & 1 + \chi \theta_x^2 &
\chi \theta_x \theta_y & \chi \theta_x \theta_z \\
  -\frac {\theta_y}{\theta} \sinh \theta & \chi \theta_x \theta_y &
1 + \chi \theta_y^2 & \chi \theta_y \theta_z \\
   -\frac {\theta_z}{\theta} \sinh \theta & \chi \theta_x \theta_z &
\chi \theta_y \theta_z & 1 + \chi \theta_z^2
\end{array} \right]
\label{eq:boost-matrix}
\end{eqnarray}

\noindent where we denoted   $\chi = (\cosh \theta -1)\theta^{-2}$.
In particular,
 boosts along  $x$, $y$, and $z$ axes are
represented by the following $4 \times 4$ matrices

\begin{eqnarray}
B(\theta, 0, 0) &=& \left[ \begin{array}{cccc}
  \cosh \theta & -\sinh \theta & 0 & 0 \\
    -\sinh \theta &\cosh \theta & 0 & 0 \\
  0 & 0 & 1 & 0 \\
  0 & 0 & 0 & 1
\end{array} \right]
\label{eq:boost-matrix-x} \\
 B(0, \theta, 0) &=&  \left[
\begin{array}{cccc}
  \cosh \theta & 0 &  -\sinh \theta & 0 \\
  0 & 1 & 0 & 0 \\
   -\sinh \theta & 0 & \cosh \theta & 0 \\
  0 & 0 & 0 & 1
\end{array} \right]
\label{eq:boost-matrix-y} \\
 B(0,0, \theta) &=& \left[
\begin{array}{cccc}
  \cosh \theta & 0 & 0 &  -\sinh \theta  \\
  0 & 1 & 0 & 0 \\
  0 & 0 & 1 & 0 \\
   -\sinh \theta & 0 & 0 & \cosh \theta
\end{array} \right]
\label{eq:boost-matrix-z}
\end{eqnarray}

\noindent Conservation of the pseudoscalar product by these transformations can be easily verified.

 Rotations are represented  by $4 \times 4$ matrices

\begin{eqnarray*}
R(\vec{\phi}) = \left[ \begin{array}{cc}
 1 & 0 \\
 0 & R_{\vec{\phi}}
\end{array} \right].
\end{eqnarray*}

\noindent where $R_{\vec{\phi}}$ is a  $3 \times 3$ rotation matrix (\ref{eq:r-vec-phi}).   A general element of the Lorentz group
\index{Lorentz group}
 can be represented as (rotation) $\times$ (boost),\footnote{This order of factors agrees with our convention (\ref{eq:poincare_exp}).} so its matrix

\begin{eqnarray}
\Lambda = R(\vec{\phi}) B(\vec{\theta}) \label{eq:LRB}
\end{eqnarray}

\noindent preserves the pseudoscalar product just as $R(\vec{\phi})$ and  $B(\vec{\theta})$ do.

So far we discussed the matrix representation of finite Lorentz transformations. Let us now find the matrix representation of the corresponding Lie
algebra. According to our discussion in
Appendix \ref{ss:unitary-reps}, the matrix of a general Lorentz
group element can be represented in the exponential form

\begin{eqnarray*}
  \Lambda &=&   e^{aF}
\end{eqnarray*}

\noindent where $F$ is an element of the Lie algebra and $a$ is a
real constant. Condition (\ref{eq:A.74}) then can be rewritten as

\begin{eqnarray*}
 0 &=& \Lambda^T g \Lambda -    g  = e^{aF^T} g e^{aF} -    g
 = (1 + aF^T + \ldots) g (1 + aF + \ldots) -    g  \\
&=&  a(F^T g -  g F) + \ldots
\end{eqnarray*}

\noindent where the ellipsis indicates terms proportional to $a^2$,
$a^3$, etc. This sets the following restriction on the matrices $F$

\begin{eqnarray*}
F^T g -  g F = 0.
\end{eqnarray*}

\noindent We can easily find 6 linearly independent $4 \times 4$
matrices satisfying this condition. Three generators of
rotations are\footnote{Note that matrices (\ref{eq:A.25})
- (\ref{eq:A.26}) are 3$\times$3 submatrices of (\ref{eq:A.77}).}

\begin{eqnarray}
\mathcal{J}_{x}=  \left[ \begin{array}{cccc}
 0 & 0 & 0 & 0 \\
 0 & 0 & 0 & 0 \\
 0 &  0 & 0 & 1 \\
 0 &  0 &  -1 & 0 \\
\end{array} \right],
\mathcal{J}_{y}=  \left[ \begin{array}{cccc}
 0 & 0 & 0 & 0 \\
 0 & 0 & 0 & -1 \\
 0 &  0 & 0 & 0 \\
 0 &  1 &  0 & 0 \\
\end{array} \right],
\mathcal{J}_{z}=  \left[ \begin{array}{cccc}
 0 & 0 & 0 & 0 \\
 0 & 0 & 1 & 0 \\
 0 &  -1 & 0 & 0 \\
 0 &  0 &  0 & 0 \\
\end{array} \right]
\label{eq:A.77}
\end{eqnarray}

\noindent Three generators of boosts  can be
obtained by differentiating explicit representation of boosts
(\ref{eq:boost-matrix-x}) - (\ref{eq:boost-matrix-z})

\begin{eqnarray}
\mathcal{K}_{x}= \frac{1}{c} \left[ \begin{array}{cccc}
 0 & -1 & 0 & 0 \\
 -1 & 0 & 0 & 0 \\
 0 &  0 & 0 & 0 \\
 0 &  0 &  0 & 0 \\
\end{array} \right],
\mathcal{K}_{y}=  \frac{1}{c}\left[ \begin{array}{cccc}
 0 & 0 & -1 & 0 \\
 0 & 0 & 0 & 0 \\
 -1 &  0 & 0 & 0 \\
 0 &  0 &  0 & 0 \\
\end{array} \right],
\mathcal{K}_{z}= \frac{1}{c} \left[ \begin{array}{cccc}
 0 & 0 & 0 & -1 \\
 0 & 0 & 0 & 0 \\
 0 &  0 & 0 & 0 \\
 -1 &  0 &  0 & 0 \\
\end{array} \right]
\label{eq:A.78}
\end{eqnarray}

\noindent  These six matrices represent basis elements of the Lie algebra, so that for representatives of finite transformations we have

\begin{eqnarray*}
R(\vec{\theta}) &=& e^{\vec{\mathcal{J}} \vec{\theta}} \\
B(\vec{\theta}) &=& e^{c\vec{\mathcal{K}} \vec{ \theta}}
\end{eqnarray*}

\section{Lorentz transformations for time and position}
\label{sc:lorentz-time-pos2}

The fundamental idea of special relativity is that Minkowski space-time $\mathcal{M}$ is a faithful representation of the physical space and time. In particular, coordinates $(ct, x, y, z)$ can be interpreted as space-time coordinates of real physical events\footnote{For definition of an \emph{event} see subsection \ref{sc:events-obs}.} localized in space and time. Moreover, it is claimed that $4 \times 4$ matrices (\ref{eq:LRB}) accurately describe transformations of these coordinates to reference frames affected by rotations and/or boosts. Suppose that observer $O'$ moves with respect to $O$ with
rapidity $\vec{\theta}$. Suppose also that $(t, \mathbf{x})$ are
space-time coordinates of an event viewed by observer $O$. Then,
according to special relativity, the space-time coordinates $(t',
\mathbf{x}')$ of this event from the point of view of $O'$ are given
by formula (\ref{eq:A.70}), which is called \emph{the Lorentz
transformation for time and position of the event}. \index{Lorentz
transformations} In particular, if observer $O'$ moves with the
speed $v = c \tanh \theta$ along the $x$-axis, then the matrix
$B(\vec{\theta})$ is (\ref{eq:boost-matrix-x})

\begin{eqnarray}
B(\theta, 0, 0) = \left[ \begin{array}{cccc}
  \cosh \theta & -\sinh \theta & 0 & 0 \\
    -\sinh \theta &\cosh \theta & 0 & 0 \\
  0 & 0 & 1 & 0 \\
  0 & 0 & 0 & 1
\end{array} \right]
\label{eq:lorentz-lambda}
\end{eqnarray}

\noindent and  Lorentz transformation (\ref{eq:A.70}) can be written
in a more familiar form

\begin{eqnarray}
 t' &=&  t \cosh \theta - (x/c) \sinh \theta \label{eq:lorentz-transform-t} \\
 x' &=& x \cosh \theta - ct \sinh \theta \label{eq:lorentz-transform-x} \\
 y' &=& y \\
 z' &=& z
 \label{eq:lorentz-transform-comp}
 \end{eqnarray}

It is important to note that special relativity makes the following
assertion

\begin{assertion} [the universality of Lorentz transformations]
 \label{lorentz-trnsf2} Lorentz transformations (\ref{eq:lorentz-transform-t}) -
(\ref{eq:lorentz-transform-comp}) are exact and universal: they  are
valid for all kinds of events in any physical system; they do not depend  on the
composition of the physical system and on interactions acting there.
\end{assertion}

In the main body of this book\footnote{See, especially, chapter \ref{sc:obs-interact}.} we explain why Assertion \ref{lorentz-trnsf2} does not hold in relativistic theory (RQD) developed here. The key difference between our approach and the standard logic of special relativity is that in RQD boost transformations of space-time coordinates of events involving interacting particles have a more complicated form, which depends on the interaction and on the state of the physical system. So, from our standpoint all
consequences of the Assertion \ref{lorentz-trnsf2} described in the rest of this Appendix are neither rigorous nor accurate.

\section{Minkowski space-time and manifest covariance}
\label{ss:manifest-covar}

An important consequence of the Assertion \ref{lorentz-trnsf2} is
the idea of the Minkowski 4-dimensional space-time. It wouldn't be
an exaggeration to say that this idea is the foundation of the
entire mathematical formalism  of modern relativistic physics.

The logic of introducing the Minkowski space-time was as follows:
According to Assertion \ref{lorentz-trnsf2}, Lorentz transformations
(\ref{eq:lorentz-transform-t}) - (\ref{eq:lorentz-transform-comp})
are universal and interaction-independent. These transformations
coincide with the abstract 4-vector representation of the Lorentz
group introduced in Appendix \ref{ss:4-dim-rep}.  It is then
natural to assume that the abstract 4-dimensional vector space $\mathcal{M}$ with
pseudo-scalar product defined in Appendix \ref{ss:4-dim-rep}
 can be identified with the space-time arena where all real physical processes occur. Then space and time
coordinates of any event become unified as different components of
the same time-position 4-vector, and the real geometry of the world
becomes  4-dimensional one.
Space and time of the old physics become unified as the
\emph{Minkowski space-time} endowed with
\emph{pseudo-Euclidean metric}. \index{pseudo-Euclidean metric}
\index{Minkowski space-time} Minkowski described this space and time
unification in following words:

\begin{quote}
\emph{From henceforth, space by itself and time by itself, have
vanished into the merest shadows and only a kind of blend of the two
exists in its own right.} H. Minkowski
\end{quote}

In analogy with familiar 3D scalars, vectors, and tensors (see
Appendix \ref{sc:rotations}), special relativity of Einstein and
Minkowski requires that physical  quantities  transform in a linear
``manifestly covariant'' way, i.e., as 4-scalars, \index{4-scalar}
 or 4-vectors, \index{4-vector} or
4-tensors, etc.

\begin{assertion} [manifest covariance of physical laws \cite{Einstein_1916}]
\index{manifest covariance} \label{manifest} Every general law of
nature must be so constituted that it is transformed into a law of
exactly the same form when, instead of the space-time variables $t$,
$x$, $y$, $z$ of the original coordinate system $K$, we introduce
new space-time variables $t'$, $x'$, $y'$, $z'$ of a coordinate
system $K'$. In this connection the relation between the ordinary
and the accented magnitudes is given by the Lorentz transformation.
Or in brief: General laws of nature are co-variant with respect to
Lorentz transformations.
\end{assertion}
\label{einstein-end}

From Assertions \ref{lorentz-trnsf2} and \ref{manifest} one can
immediately obtain many important physical predictions of special
relativity. One consequence of  Lorentz transformations is that the
length of a measuring rod reduces by a universal factor
\index{length contraction}

\begin{eqnarray}
l' = l/\cosh \theta \label{eq:length-contraction}
\end{eqnarray}

\noindent from the point of view of a moving reference frame.
Another well-known result is that the duration of time intervals
between any two events increases by the same factor $\cosh \theta$

\begin{eqnarray}
\Delta t' = \Delta t\cosh \theta \label{eq:time-dilation}
\end{eqnarray}

\noindent One experimentally verifiable consequence of this \emph{time
dilation}  formula will be discussed in the next section.

\section{Decay of moving particles in special relativity}
\label{ss:relativity}

Suppose that from the viewpoint of observer  $O$ the unstable
particle is prepared at rest in the origin $x=y=z=0$ at time $t=0$
in the  non-decayed state, so that $\omega(0,0) = 1$.\footnote{Here
we follow notation from chapter \ref{ch:decays} by writing
$\omega(\theta,t)$ the non-decay probability observed from the
reference frame $O'$ moving with respect to $O$ with rapidity
$\theta$ at time $t$ (measured by a clock attached to $O'$).} Then
observer $O$ may associate the space-time point

\begin{eqnarray}
(t, x,y,z)_{prep} = (0,0,0,0) \label{eq:event1}
\end{eqnarray}

\noindent with the event of preparation. We know that
 the non-decay probability decreases with time by (almost) exponential
\emph{decay law}\footnote{Actually, as we saw in subsection
\ref{ss:decay_law}, the decay law is not exactly exponential, but
this is not important for our derivation of equation (\ref{eq:1b}) here.}
\index{decay law}

 \begin{eqnarray}
\omega(0,t) &\approx& \exp \left(- \frac{ t}{\tau_0} \right) \label{eq:expo}
\end{eqnarray}

\noindent  At time  $t = \tau_{0}$
 the non-decay probability is exactly $\omega(0,\tau_{0}) = e^{-1}$.
 This ``one lifetime'' event  has
space-time coordinates

\begin{eqnarray}
(t,x,y,z)_{life} = (\tau_{0},0,0,0) \label{eq:event2}
\end{eqnarray}

\noindent according to the observer $O$.

Let us now take the point of view of the moving observer $O'$.
According to special relativity, this observer will also see the
``preparation'' and the ``one lifetime'' events, when the non-decay
probabilities are 1 and $e^{-1}$, respectively. However, observer
$O'$ may disagree with $O$ about the space-time coordinates of these
events. Substituting (\ref{eq:event1})
 and (\ref{eq:event2}) in (\ref{eq:lorentz-transform-t}) -
 (\ref{eq:lorentz-transform-comp})
we see  that from the point of view of $O'$, the ``preparation''
event  has coordinates $(0,0,0,0)$, and the ``lifetime'' event  has
coordinates $(\tau_{0}\cosh\theta,  - c \tau_0 \sinh\theta, 0, 0)$.
Therefore, the time elapsed between these two events  is
$\cosh\theta$ times longer than in the reference frame $O$. This
also means that the decay law is exactly $\cosh\theta$ slower from
the point of view of the moving observer $O'$. This finding is
summarized in the famous Einstein's ``time dilation'' formula \index{time dilation}

\begin{eqnarray}
\omega(\theta,t) = \omega\left(0, \frac{t}{\cosh \theta}\right) \label{eq:1b}
\end{eqnarray}

\noindent which was confirmed in numerous experiments
\cite{dilation, dilation1, dilation2}, most accurately for muons
accelerated
 to relativistic speeds in a
cyclotron \cite{muons, muons2}. These experiments were certainly a
triumph of Einstein's theory.
 However, as we see from the above discussion,
 equation (\ref{eq:1b}) can be derived only under assumption \ref{lorentz-trnsf2},
 which lacks proper justification. Therefore, a question remains
whether equation (\ref{eq:1b}) is a fundamental exact result or simply an
approximation that can be disproved by more accurate measurements?
 This question is addressed in chapter
\ref{ch:decays}.

 \section{Ban on superluminal signaling}
\label{ss:super-signal}

\begin{figure}
\centering
 \includegraphics {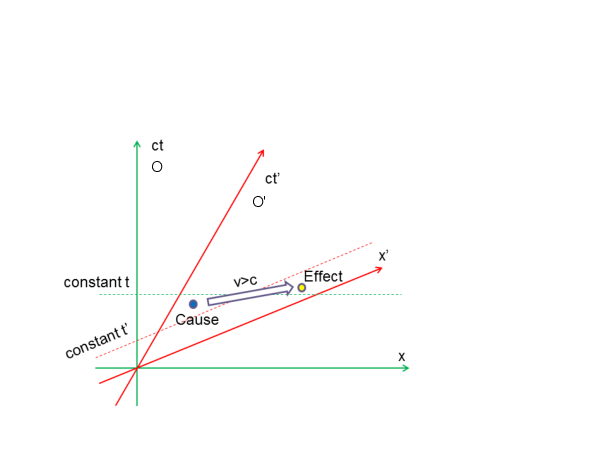} \caption{Illustration to the special-relativistic ``proof'' that superluminal signals violate the principle of causality.
} \label{fig:1y}
\end{figure}

Perhaps, the most famous assertion of special relativity  is

\begin{assertion} [no superluminal signaling] \label{assertionEE} No signal may
propagate faster than the speed of light.
\end{assertion}

\noindent The ``proof'' of this Assertion \cite{Russo} relies on the \emph{principle of causality}, \index{principle of causality} which says that \emph{the cause
precedes the effect in all reference frames}. Suppose that two events ``Cause'' and ``Effect'' are causally related, while separated by a space-like interval\footnote{This means that the signal has propagated from the ``Cause'' to the ``Effect'' superluminally.} in the reference frame $O$ with coordinate axes $(t, \mathbf{x})$, as in Fig. \ref{fig:1y}. In special relativity, we obtain coordinates of the two events in the moving reference frame $O'$ from Lorentz formulas (\ref{eq:lorentz-transform-t}) - (\ref{eq:lorentz-transform-comp}). This transformation can be represented graphically as a \emph{pseudorotation} of the coordinate axes shown in the figure. If the speed of  $O'$ is high enough, then this observer will find that the ``Effect'' happens earlier than the ``Cause,'' which clearly violates  the principle of causality. Thus, within special relativity all superluminal signals are forbidden.  In chapter \ref{ch:support} we will discuss experiments that challenge this conclusion.

\chapter{Quantum fields for fermions} \label{sc:fermions}

According to our interpretation of quantum field theory, quantum
fields are not fundamental ingredients of the material world. They
are just convenient mathematical expressions, which simplify the
construction of relativistic and cluster-separable interaction operators. For this reason, discussion of quantum fields is
placed in this Appendix rather than in the main body of the book. Here we will discuss quantum fields for
spin 1/2 fermions (electrons, protons, neutrinos, and their antiparticles). In the
next Appendix we will consider the photon's quantum field.

\section{Dirac's gamma matrices}
\label{ss:spinor-rep}

Let us introduce the following $4\times 4$ \emph{Dirac
gamma matrices}.\footnote{On the right hand sides each $2 \times 2$
block is expressed in terms of Pauli matrices from Appendix
\ref{ss:pauli-matrices}} \index{gamma matrices}

\begin{eqnarray}
\gamma^0 &=& \left[ \begin{array}{cccc}
 1 & 0 & 0 & 0   \\
 0 & 1 & 0 & 0 \\
 0 & 0 & -1 & 0 \\
 0 &  0 & 0 & -1
\end{array} \right] = \left[ \begin{array}{cc}
\sigma_0 & 0  \\
0 & -\sigma_0\\
\end{array} \right]  = \left[ \begin{array}{cc}
1 & 0  \\
0 & -1\\
\end{array} \right] \label{eq:A.81}\\
\gamma^x &=& \left[ \begin{array}{cccc}
 0 & 0 & 0 & 1   \\
 0 & 0 & 1 & 0 \\
 0 & -1 & 0 & 0 \\
 -1 &  0 & 0 & 0
\end{array} \right]  = \left[ \begin{array}{cc}
0 & \sigma_x  \\
-\sigma_x & 0\end{array} \right] \nonumber \\
\gamma^y &=& \left[ \begin{array}{cccc}
 0 & 0 & 0 & -i   \\
 0 & 0 & i & 0 \\
 0 & i & 0 & 0 \\
 -i &  0 & 0 & 0
\end{array} \right] = \left[ \begin{array}{cc}
0 & \sigma_y  \\
-\sigma_y & 0\end{array} \right] \nonumber \\
\gamma^z &=& \left[ \begin{array}{cccc}
 0 & 0 & 1 & 0   \\
 0 & 0 & 0 & -1 \\
 -1 & 0 & 0 & 0 \\
 0 &  1 & 0 & 0
\end{array} \right] = \left[ \begin{array}{cc}
0 & \sigma_z  \\
-\sigma_z & 0\end{array} \right] \nonumber \\
 \vec{\gamma} &=& \left[ \begin{array}{cc}
0 & \vec{\sigma}  \\
- \vec{\sigma} & 0\\
\end{array} \right]
\label{eq:A.82}
\end{eqnarray}

\noindent These matrices have the following
properties\footnote{The indices take values $\mu, \nu = 0,1,2,3$, $i
= 1,2,3$.}

\begin{eqnarray}
\gamma^0 \vec{\gamma} &=& \vec{\gamma}^{\dag} \gamma^0 = -\vec{\gamma} \gamma^0 \label{eq:H.17a}\\
 \gamma^{\mu}
\gamma^{\nu} + \gamma^{\nu} \gamma^{\mu} &=& 2g^{\mu
\nu} \label{gamma-mu}  \\
\gamma^0 \gamma^0 &=& 1 \label{eq:gamma0} \\
\gamma^i \gamma^i &=&  -1 \label{eq:gammaxyz} \\
Tr(\gamma^{\mu}) &=& 0 \label{eq:trace-gamma} \\
Tr(\gamma^{\mu} \gamma^{\nu}) &=& 4 g_{\mu \nu}
\label{eq:trace-gamma2} \\
\gamma_{\mu}\gamma^{\mu} &=& -\gamma^x \gamma^x - \gamma^y \gamma^y
-
\gamma^z \gamma^z + \gamma^0 \gamma^0 = 4 \label{eq:gamma-mu-gamma-mu}\\
\gamma_{\mu}\gamma_{\nu}\gamma^{\mu} &=& -
\gamma_{\nu}\gamma_{\mu}\gamma^{\mu} + 2g_{\mu \nu} \gamma^{\mu} =
-4 \gamma_{\nu} +2 \gamma_{\nu} = -2 \gamma_{\nu}
\label{eq:gamma-munu-gamma-mu}
\end{eqnarray}

\noindent If $A,B,C$ are any linear combinations of gamma-matrices,
then

\begin{eqnarray}
\gamma_{\mu}A\gamma^{\mu} &=& -2A \label{eq:gammaA} \\
\gamma_{\mu}AB\gamma^{\mu} &=& 2(AB+BA) \label{eq:gammaAB}\\
\gamma_{\mu}ABC\gamma^{\mu} &=& -2CBA \label{eq:gammaABC}
\end{eqnarray}

\section{Bispinor representation of the Lorentz group}
\label{ss:bispinor-rep}

In this section, we would like to build the \emph{bispinor
representation} \index{bispinor representation} $\mathcal{D}(\Lambda)$ of the Lorentz group.  Similar to the 4-vector representation from  Appendix \ref{ss:4-dim-rep}, the bispinor representation is realized by $4 \times 4$ matrices.

The boost and rotation generators of the bispinor representation of
the Lorentz group are defined through commutators of gamma matrices

\begin{eqnarray}
\vec{\mathcal{K}} &=& \frac{i\hbar}{4c} [\gamma^0, \vec{\gamma}] =
\frac{i \hbar}{2c} \left[ \begin{array}{cc}
 0 & \vec{\sigma}  \\
 \vec{\sigma} & 0 \\
\end{array} \right]
\label{eq:A.83} \\
 \mathcal{J}_x &=& \frac{i\hbar}{4} [\gamma_y ,
\gamma_z]  =  \frac{\hbar}{2} \left[
\begin{array}{cc}
 \sigma_x & 0   \\
 0 & \sigma_x
\end{array} \right]
\label{eq:A.84}\\
\mathcal{J}_y &=& \frac{i\hbar}{4} [\gamma_z , \gamma_x]  =
\frac{\hbar}{2} \left[
\begin{array}{cc}
 \sigma_y & 0   \\
 0 & \sigma_y
\end{array} \right]
\label{eq:A.85}\\
\mathcal{J}_z &=& \frac{i\hbar}{4} [\gamma_x , \gamma_y]  =
\frac{\hbar}{2} \left[
\begin{array}{cc}
 \sigma_z & 0   \\
 0 & \sigma_z
\end{array} \right]
\label{eq:A.86}
\end{eqnarray}

\noindent  Using properties of Pauli matrices from Appendix \ref{ss:pauli-matrices}, it is not difficult to verify that these generators
indeed satisfy commutation relations of the Lorentz algebra
(\ref{eq:5.51}), (\ref{eq:5.52}), and (\ref{eq:5.54}). For example,

\begin{eqnarray*}
[\mathcal{J}_x, \mathcal{J}_y]  &=&  \frac{\hbar^2}{4} \left[
\begin{array}{cc}
 [\sigma_x, \sigma_y] & 0   \\
 0 & [\sigma_x, \sigma_y]
\end{array} \right] =  \frac{i\hbar^2 }{2} \left[ \begin{array}{cc}
 \sigma_z & 0   \\
 0 & \sigma_z
\end{array} \right] = i\hbar \mathcal{J}_z \\
\mbox{ } [\mathcal{J}_x, \mathcal{K}_y]  &=& \frac{i\hbar^2}{4c}
\left( \left[
\begin{array}{cc}
  \sigma_x & 0  \\
 0 & \sigma_x
\end{array} \right] \left[ \begin{array}{cc}
 0 & \sigma_y  \\
 \sigma_y & 0
\end{array} \right] - \left[ \begin{array}{cc}
 0 & \sigma_y  \\
 \sigma_y & 0
\end{array} \right] \left[ \begin{array}{cc}
  \sigma_x  & 0\\
 0 & \sigma_x
\end{array} \right] \right)\\
 &=&  -\frac{\hbar^2}{2c}  \left[ \begin{array}{cc}
 0 & \sigma_z \\
\sigma_z & 0
\end{array} \right]=  i\hbar \mathcal{K}_z \\
\mbox{ } [\mathcal{K}_x, \mathcal{K}_y]  &=&
-\frac{\hbar^2}{4c^2}\left( \left[
\begin{array}{cc}
 0 & \sigma_x  \\
 \sigma_x & 0
\end{array} \right] \left[ \begin{array}{cc}
 0 & \sigma_y  \\
 \sigma_y & 0
\end{array} \right] - \left[ \begin{array}{cc}
 0 & \sigma_y  \\
 \sigma_y & 0
\end{array} \right] \left[ \begin{array}{cc}
 0 & \sigma_x  \\
 \sigma_x & 0
\end{array} \right] \right)\\
&=&  -\frac{\hbar^2}{4c^2} \left[ \begin{array}{cc}
 [\sigma_x, \sigma_y] & 0  \\
0 & [\sigma_x, \sigma_y]
\end{array} \right]  =  -\frac{i\hbar^2}{2c^2} \left[ \begin{array}{cc}
 \sigma_z & 0  \\
0 & \sigma_z
\end{array} \right]  = -\frac{i\hbar}{c^2} \mathcal{J}_z
\end{eqnarray*}

\noindent  We also get the following representation of finite boosts\footnote{Note that this representation is not unitary.}

\begin{eqnarray}
 \mathcal{D}_{ij}(e^{-\frac{ic}{\hbar}\mathbf{K} \cdot \vec{ \theta}})
&=& \exp \left( \frac{1}{2} \left[ \begin{array}{cc}
 0 & \vec{\sigma} \cdot \vec{ \theta}  \\
\vec{\sigma} \cdot \vec{ \theta} & 0 \\
\end{array} \right]\right) \nonumber \\
&=& 1 +  \frac{1}{2} \left[ \begin{array}{cc}
 0 & \vec{\sigma} \cdot \vec{ \theta}  \\
\vec{\sigma} \cdot \vec{ \theta} & 0 \\
\end{array} \right] + \frac{1}{2!} \left(\frac{\theta}{2} \right)^2
\left[ \begin{array}{cc}
 1& 0  \\
 0 & 1 \\
\end{array} \right] + \ldots \nonumber \\
&=& I \cosh \frac{\theta}{2} + \frac{2c}{i \hbar} \vec{\mathcal{K}}
\cdot \frac{\vec{ \theta}}{\theta} \sinh \frac{\theta}{2}
\label{eq:A.87}
\end{eqnarray}

\noindent This equation allows us to prove another important
property of gamma matrices

\begin{eqnarray}
\mathcal{D}^{-1}(\Lambda) \gamma^{\mu} \mathcal{D}(\Lambda) =
\sum_{\nu} \Lambda_{\mu \nu} \gamma^{\nu} \label{eq:A.88}
\end{eqnarray}

\noindent where $\Lambda$ is any Lorentz transformation, and  $\Lambda_{\mu \nu}$ is a $4 \times 4$ matrix (\ref{eq:LRB}) realizing the 4-vector representation of the Lorentz group. Indeed, let us consider a particular case of this formula
with $\mu = 0$ and $\Lambda$ being a boost with rapidity $\theta$
along the $x$-axis.\label{spinor-rep-end} Then

\begin{eqnarray*}
&\mbox{ }& \mathcal{D}^{-1}(\Lambda) \gamma^{0} \mathcal{D}(\Lambda)
\nonumber
\\ &=& \left(I \cosh \frac{\theta}{2} - \frac{2c}{i \hbar} \mathcal{K}_x  \sinh
\frac{\theta}{2}\right) \gamma^{0} \left(I \cosh \frac{\theta}{2} +
\frac{2c}{i
\hbar} \mathcal{K}_x  \sinh \frac{\theta}{2} \right) \nonumber \\
&=& \left(\cosh \frac{\theta}{2} \left[ \begin{array}{cc}
 1& 0  \\
 0 & 1 \\
\end{array} \right] -   \sinh
\frac{\theta}{2}\left[ \begin{array}{cc}
 0& \sigma_x  \\
 \sigma_x  & 0 \\
\end{array} \right]\right) \left[ \begin{array}{cc}
 1& 0  \\
 0  & -1 \\
\end{array} \right] \times \nonumber \\
&\ &\left(\cosh \frac{\theta}{2} \left[ \begin{array}{cc}
 1& 0  \\
 0 & 1 \\
\end{array} \right] +
  \sinh \frac{\theta}{2} \left[ \begin{array}{cc}
 0& \sigma_x  \\
 \sigma_x  & 0 \\
\end{array} \right] \right) \nonumber \\
&=& \cosh^2 \frac{\theta}{2} \left[ \begin{array}{cc}
 1& 0  \\
 0 & -1 \\
\end{array} \right] -2   \sinh
\frac{\theta}{2} \cosh \frac{\theta}{2}\left[ \begin{array}{cc}
 0& -\sigma_x  \\
 \sigma_x  & 0 \\
\end{array} \right] +
  \sinh^2 \frac{\theta}{2} \left[ \begin{array}{cc}
 1 & 0  \\
 0 & -1 \\
\end{array} \right] \nonumber \\
&=& \gamma^0 \cosh \theta  + \gamma^x \sinh \theta
\end{eqnarray*}

\noindent In agreement with formula for the boost matrix
$\Lambda_{\mu \nu}$ (\ref{eq:boost-matrix-x}).

One can also check for pure boosts

\begin{eqnarray*}
\gamma^0  \mathcal{D}(e^{-\frac{ic}{\hbar}\mathbf{K} \cdot
\vec{ \theta}}) \gamma^0 &=& 1 +  \frac{1}{2} \gamma^0 \left[
\begin{array}{cc}
 0 & \vec{\sigma} \cdot \vec{ \theta}  \\
\vec{\sigma} \cdot \vec{ \theta} & 0 \\
\end{array} \right] \gamma^0 + \frac{1}{2!} \left(\frac{\theta}{2} \right)^2
\left[ \begin{array}{cc}
 1& 0  \\
 0 & 1 \\
\end{array} \right] + \ldots \nonumber \\
&=& 1 -  \frac{1}{2}  \left[
\begin{array}{cc}
 0 & \vec{\sigma} \cdot \vec{ \theta}  \\
\vec{\sigma} \cdot \vec{ \theta} & 0 \\
\end{array} \right]  + \frac{1}{2!} \left(\frac{\theta}{2}\right)^2
\left[ \begin{array}{cc}
 1& 0  \\
 0 & 1 \\
\end{array} \right] + \ldots \nonumber \\
&=& \mathcal{D}\left(e^{\frac{ic}{\hbar}\mathbf{K} \cdot
\vec{ \theta}}\right) =
\mathcal{D}^{-1}\left(e^{-\frac{ic}{\hbar}\mathbf{K} \cdot
\vec{ \theta}}\right)
\end{eqnarray*}

\noindent A similar calculation for rotations should convince us
that for a general  transformation $\Lambda$ from the Lorentz
group

\begin{eqnarray}
\gamma^0  \mathcal{D}(\Lambda) \gamma^0 = \mathcal{D}^{-1}(\Lambda)
\label{eq:d-lambda}
\end{eqnarray}

\noindent Another useful formula is

\begin{eqnarray}
 D(\Lambda) \gamma^0  D(\Lambda) &=& D(\Lambda) \gamma^0  D(\Lambda) \gamma^0
 \gamma^0 = D(\Lambda)  D^{-1}(\Lambda)  \gamma^0 =\gamma^0 \label{eq:d-gamma-d}
\end{eqnarray}

 It will be convenient to  introduce a \emph{slash}
\index{$\cross{k}$ slash notation} notation for pseudoscalar  products of $\gamma^{\mu}$ with 4-vectors $\tilde{k}$

\begin{eqnarray}
\cross{k} &\equiv& k_{\mu}\gamma^{\mu} \equiv \gamma^0 k_0 - \vec{\gamma} \cdot \mathbf{k} \label{eq:slash} \\
\cross{k}^2 &=&
\gamma^{\mu}k_{\mu}\gamma^{\nu}k_{\nu} =
1/2(\gamma^{\mu}\gamma^{\nu} + \gamma^{\nu}\gamma^{\mu})
k_{\mu}k_{\nu} = g^{\mu \nu} k_{\mu}k_{\nu} \nonumber \\
&=& \tilde{k}^2 \label{eq:J.45a} \\
(\cross{k} -mc^2)(\cross{k} +mc^2) &=& \cross{k}\cross{k} -m^2c^4  = \tilde{k}^2 -m^2c^4
\label{eq:slashk} \\
\gamma_{\mu} \cross{k} + \cross{k} \gamma_{\mu} &=& 2 k_{\mu}
\end{eqnarray}

\section{Construction of the Dirac field} \label{ss:cons-ferm}

According to the \textbf{Step 1} in subsection \ref{ss:weinberg}, in order to
construct relativistic interaction operators, we need to associate
with each particle type a finite-dimensional representation of the
Lorentz group and a quantum field. In this section we are going to
build the quantum field for electrons and positrons.
 We postulate
that this  \emph{Dirac field} \index{Dirac field} has 4 components that transform by means of the representation $\mathcal{D}(\Lambda)$ constructed above. The explicit formula for the field is\footnote{This form (apart from the overall
normalization of the field) can be uniquely established \cite{book} from the
properties (I) - (IV) in \textbf{Step 1} of subsection \ref{ss:weinberg}. The bispinor index $\alpha$ takes values 1,2,3,4.}

\begin{eqnarray}
\psi_{\alpha}(\tilde{x}) &\equiv& \psi_{\alpha}(\mathbf{x},t) \nonumber \\
&=& \int \frac{d\mathbf{p}}{(2\pi \hbar)^{3/2}}
\sqrt{\frac{mc^2}{\omega_{\mathbf{p}}}} \sum_{ \sigma} \left
(e^{-\frac{i}{\hbar}\tilde{p} \cdot \tilde{x}}u_{\alpha}(\mathbf{p},
\sigma)
 a _{\mathbf{p},\sigma}
+ e^{\frac{i}{\hbar}\tilde{p} \cdot \tilde{x}}v_{\alpha }
(\mathbf{p}, \sigma)b^{\dag}_{\mathbf{p},\sigma} \right) \nonumber
\\
\label{eq:10.10}
\end{eqnarray}

\noindent Here $a _{\mathbf{p},\sigma}$ is the electron annihilation
operator, and $b^{\dag}_{\mathbf{p},\sigma}$ is the positron creation
operator.  For brevity,   we  denote $\tilde{p} \equiv
(\omega_{\mathbf{p}}, cp_x, cp_y, cp_z)$ the energy-momentum
4-vector and $\tilde{x} \equiv (t, x/c, y/c, z/c)$ the 4-vector in the
Minkowski space-time.\footnote{As discussed in section
\ref{ss:are-fields-meas}, the only purpose for introducing quantum
fields is to build interaction operators as in (\ref{eq:11.6}) -
(\ref{eq:11.7}). In these formulas field arguments $x,y,z$ are
integration variables. Therefore, they should not be  identified with
positions in the physical space.  Moreover, in applications bispinor labels $\alpha$ serve as dummy
summation indices, so no physical meaning should be assigned to them
as well. } The pseudo-scalar product of the 4-vectors is denoted by a
dot:
 $\tilde{p} \cdot \tilde{x} \equiv p_{\mu}x^{\mu} \equiv \mathbf{p}
\mathbf{x} - \omega_{\mathbf{p}} t$ and $\omega_{\mathbf{p}} \equiv
\sqrt{m^2c^4 + p^2c^2}$. Numerical factors $u_{\alpha }(\mathbf{p},
\sigma)$ and $v_{\alpha }(\mathbf{p}, \sigma)$ will be discussed in
Appendix \ref{ss:factors}. Note that according to equations
(\ref{eq:9.36}) and (\ref{eq:9.37})

\begin{eqnarray}
\psi_{\alpha}(\mathbf{x},t) &=&  e^{-\frac{i}{\hbar} H_0t }
\psi_{\alpha}(\mathbf{x},0) e^{\frac{i}{\hbar} H_0t }
\label{eq:t-evolu}
\end{eqnarray}

\noindent so, the $t$-dependence demanded by equation (\ref{eq:9.50}) for
regular operators  is satisfied in our definition (\ref{eq:10.10}).

The Dirac field can be represented by a 4-component column of
operator functions

\begin{eqnarray*}
\psi (\tilde{x}) &=& \left[ \begin{array}{c}
\psi_1 (\tilde{x})  \\
 \psi_2 (\tilde{x})     \\
\psi_3 (\tilde{x})   \\
\psi_4 (\tilde{x})
\end{array} \right]
\end{eqnarray*}

\noindent We will also need the \emph{conjugate} field
\index{conjugate field}

\begin{eqnarray*}
\psi_{\alpha}^{\dag}(\tilde{x}) &=&  \int \frac{d\mathbf{p}}{(2\pi
\hbar)^{3/2}} \sqrt{\frac{mc^2}{\omega_{\mathbf{p}}}} \sum_{ \sigma}
\left(e^{\frac{i}{\hbar}\tilde{p} \cdot \tilde{x}}u^{\dag}_{\alpha
}(\mathbf{p}, \sigma) a^{\dag} _{\mathbf{p},\sigma} \nonumber +
e^{-\frac{i}{\hbar}\tilde{p} \cdot \tilde{x}}v^{\dag}_{\alpha }
(\mathbf{p}, \sigma)b_{\mathbf{p},\sigma} \right) \\
\end{eqnarray*}

\noindent which is usually represented as a row

\begin{eqnarray*}
\psi^{\dag} &=& [ \psi^*_1,    \psi^*_2,    \psi^*_3,    \psi^*_4]
\end{eqnarray*}

 \noindent The \emph{adjoint} field \index{adjoint field}

 \begin{eqnarray}
 \overline{\psi}_{\alpha}( \tilde{x}) &\equiv&
 \sum_{\beta} \psi^{\dag}_{\beta}(\tilde{x})
 \gamma^0_{\beta \alpha} \label{eq:adjoint-field}
 \end{eqnarray}

 \noindent is also represented as a row

 \begin{eqnarray*}
 \overline{\psi} &\equiv&
 \psi^{\dag} \gamma^0 = [ \psi^*_1,    \psi^*_2,    \psi^*_3,
 \psi^*_4]
 \left[ \begin{array}{cccc}
 1 & 0 & 0 & 0 \\
 0 & 1 & 0 & 0 \\
 0 & 0 & -1 & 0\\
 0 & 0 & 0 & -1
 \end{array} \right] \\
 &=& [ \psi^*_1,    \psi^*_2,    -\psi^*_3,
 -\psi^*_4]
 \end{eqnarray*}

The quantum field for the proton-antiproton system is built
similarly to (\ref{eq:10.10})

\begin{eqnarray}
\Psi(\tilde{x}) &=&  \int \frac{d\mathbf{p} } {(2\pi \hbar)^{3/2}}
\sqrt{ \frac{Mc^2} {\Omega_{\mathbf{p}}} }\sum_{ \sigma}
\left(e^{-\frac{i}{\hbar}\tilde{p} \cdot \tilde{x}}w(\mathbf{p},
\sigma) d _{\mathbf{p},\sigma} \nonumber +
e^{\frac{i}{\hbar}\tilde{p} \cdot \tilde{x}}
s (\mathbf{p}, \sigma)f^{\dag}_{\mathbf{p},\sigma} \right) \\
\label{eq:10.10a}
\end{eqnarray}

\noindent where $\Omega_{\mathbf{p}} = \sqrt{M^2 c^4 + p^2 c^2}$,
$M$ is the proton mass, $\tilde{P} \cdot \tilde{x} \equiv \mathbf{px
} - \Omega_{\mathbf{p}} t $, and coefficient functions $w(\mathbf{p}, \sigma)$
and $s(\mathbf{p}, \sigma)$ are the same as $u(\mathbf{p}, \sigma)$
and $v(\mathbf{p}, \sigma)$ but with the electron mass $m$ replaced
by the proton mass $M$.

\section{Properties of factors $u$ and $v$} \label{ss:factors}

The key components of the quantum field formula (\ref{eq:10.10})
are numerical functions $u_{\alpha} (\mathbf{p}, \sigma)$ and
$v_{\alpha } (\mathbf{p}, \sigma)$. We can represent them as  $4
\times 2$ matrices with the (bispinor) index $\alpha =1,2,3,4$ enumerating
rows and the (spin projection) index $\sigma = -1/2, 1/2$ enumerating
columns. Let us first postulate the following form of these matrices
at zero momentum

\begin{eqnarray*}
u(\mathbf{0}) =   \left[ \begin{array}{cc}
 0 & 1   \\
 1 &  0  \\
 0  & 0 \\
 0  & 0 \\
\end{array} \right], \mbox{   } v(\mathbf{0}) =   \left[ \begin{array}{cc}
 0  & 0 \\
 0  & 0 \\
 0 & 1   \\
 1 &  0  \\
\end{array} \right]
\end{eqnarray*}

\noindent Sometimes it is convenient  to represent these matrices as
four vectors-columns

\begin{eqnarray}
u(\mathbf{0}, -1/2) &=& \left[ \begin{array}{c}
 0   \\
 1   \\
 0   \\
 0
\end{array} \right]
\label{eq:10.12}\\
u(\mathbf{0}, 1/2) &=& \left[ \begin{array}{c}
 1   \\
 0   \\
 0   \\
 0
\end{array} \right]
\label{eq:10.13}\\
v(\mathbf{0}, -1/2) &=& \left[ \begin{array}{c}
 0   \\
 0   \\
 0   \\
 1
\end{array} \right]
\label{eq:10.14}\\
v(\mathbf{0}, 1/2) &=& \left[ \begin{array}{c}
 0   \\
 0   \\
 1   \\
 0
\end{array} \right]
\label{eq:10.15}
\end{eqnarray}

\noindent We will get more compact formulas if we introduce
2-component quantities

\begin{eqnarray}
\chi_{1/2} = \left[ \begin{array}{c}
 1 \\
 0 \\
\end{array} \right], \mbox {    }
\chi_{-1/2} = \left[ \begin{array}{c}
 0 \\
 1 \\
\end{array} \right] , \mbox {    } \chi^{\dag}_{1/2} = (1,0), \mbox {    }
\chi^{\dag}_{-1/2} = (0,1) \label{eq:chi}
\end{eqnarray}

\noindent Then we can write

\begin{eqnarray*}
u(\mathbf{0}, \sigma) = \left[ \begin{array}{c}
 \chi_{\sigma} \\
 0 \\
\end{array} \right], \mbox {    } v(\mathbf{0}, \sigma) = \left[
\begin{array}{c}
 0 \\
 \chi_{\sigma} \\
\end{array} \right]
\end{eqnarray*}

\noindent Let us verify that matrix $u(\mathbf{0})$ has the
following property

\begin{eqnarray}
 \sum _{\beta}    \mathcal{D}_{\alpha \beta} (R) u_{\beta } (\mathbf{0},
\sigma)
    =
\sum _{\tau}  u_{\alpha} (\mathbf{0}, \tau) D^{1/2}_{\tau \sigma}
(R) \label{eq:10.16}
\end{eqnarray}

\noindent where $\mathcal{D}$ is the bispinor representation of the
Lorentz group,\footnote{see Appendix \ref{ss:spinor-rep}} $D^{1/2}$
is the 2-dimensional unitary irreducible representation of the
rotation group,\footnote{see Table \ref{table:A.3}} and $R$ is any rotation.
By denoting $\mathcal{J}_k$ the generators of rotations in the
representation $\mathcal{D}_{\alpha \beta}(R)$ and $S_k$ the
generators of rotations in the representation $D^{1/2}_{\sigma
\sigma'}(R)$ we can write equation (\ref{eq:10.16}) in an equivalent
differential form

\begin{eqnarray*}
 \sum _{\beta}    (\mathcal{J}_k)_{\alpha \beta} (R) u_{\beta } (\mathbf{0},
\sigma)
    =
\sum _{\tau}  u_{\alpha } (\mathbf{0}, \tau) (S_k)_{\tau \sigma} (R)
\end{eqnarray*}

\noindent Let us check, for example,  that this equation is satisfied for rotations
around the $x$-axis. Acting with the $4 \times 4$  matrix
(\ref{eq:A.84})

\begin{eqnarray*}
\mathcal{J}_x =  \frac{\hbar}{2}\left[ \begin{array}{cccc}
 0 & 1 & 0 & 0  \\
 1 & 0 & 0 & 0 \\
 0 & 0 & 0 & 1 \\
 0 & 0 & 1 & 0
\end{array} \right]
\end{eqnarray*}

\noindent on the index $\beta$ in $u_{\beta } (\mathbf{0}, \sigma)$
we obtain

\begin{eqnarray*}
\mathcal{J}_x u (\mathbf{0})  &=&
 \frac{\hbar}{2}\left[ \begin{array}{cccc}
 0 & 1 & 0 & 0  \\
 1 & 0 & 0 & 0 \\
 0 & 0 & 0 & 1 \\
 0 & 0 & 1 & 0
\end{array} \right]
 \left[ \begin{array}{cc}
 0 & 1   \\
 1 &  0  \\
 0  & 0 \\
 0  & 0 \\
\end{array} \right]
= \frac{\hbar}{2} \left[ \begin{array}{cc}
 1 & 0   \\
 0 &  1  \\
 0  & 0 \\
 0  & 0 \\
\end{array} \right]
\end{eqnarray*}

\noindent This has the same  effect as acting with $2 \times 2$
matrix (see Table \ref{table:A.3})

\begin{eqnarray*}
S_x = \frac{\hbar}{2} \left[ \begin{array}{cc}
 0 & 1  \\
 1 & 0
\end{array} \right]
\end{eqnarray*}

\noindent  on the index $\tau$ in $u_{\alpha} (\mathbf{0}, \tau)$

\begin{eqnarray*}
u (\mathbf{0}) J_x
 &=&
 \frac{\hbar}{2} \left[ \begin{array}{cc}
 0 & 1   \\
 1 &  0  \\
 0  & 0 \\
 0  & 0 \\
\end{array} \right]  \left[ \begin{array}{cc}
  0 & 1  \\
  1 & 0
\end{array} \right]
=\frac{\hbar}{2} \left[ \begin{array}{cc}
 1 & 0   \\
 0 &  1  \\
 0  & 0 \\
 0  & 0 \\
\end{array} \right]
\end{eqnarray*}

\noindent This proves equation (\ref{eq:10.16}). Similarly, one can show

\begin{eqnarray*}
 \sum _{\beta}    \mathcal{D}_{\alpha \beta} (R) v_{\beta } (\mathbf{0},
\sigma)
    =
\sum _{\tau}  v_{\alpha} (\mathbf{0}, \tau) D^{* 1/2}_{\tau \sigma}
(R)
\end{eqnarray*}

\noindent The corresponding formula for the adjoint factor
$\overline{u}$ is obtained as follows: take the Hermitian
conjugate of (\ref{eq:10.16}), multiply it by $\gamma^0$ from the
right and take into account equations (\ref{eq:gamma0}) and
(\ref{eq:d-lambda})

\begin{eqnarray}
 u^{\dag}(\mathbf{0}, \sigma) \gamma^0 \gamma^0 \mathcal{D}^{\dag} (R)
 \gamma^0   &=&
\sum _{\tau}  u^{\dag} (\mathbf{0}, \tau) \gamma^0 D^{1/2}_{\tau
\sigma}
(R) \nonumber \\
 \overline{u}(\mathbf{0}, \sigma)  \mathcal{D}^{\dag} (-R)
  &=&
\sum _{\tau}  \overline{u} (\mathbf{0}, \tau)  D^{1/2}_{\tau \sigma}
(R) \label{eq:overline-u}
\end{eqnarray}

So far, we have discussed zero-momentum values of functions $u$ and $v$. The values of  $u_{\alpha }(\mathbf{p}, \sigma)$  and $v_{\alpha
}(\mathbf{p}, \sigma)$ at arbitrary momentum $\mathbf{p}$ are defined  by applying the bispinor representation matrix (\ref{eq:A.87}) of
the standard boost $\lambda_{\mathbf{p}}$ (\ref{eq:unique-pure}) to
zero-momentum values

\begin{eqnarray}
u_{\alpha }(\mathbf{p}, \sigma) &\equiv& \sum_{\beta} \mathcal{D}_{\alpha
\beta}(\lambda_{\mathbf{p}})
 u_{\beta }(\mathbf{0}, \sigma)
 \label{eq:u-alpha} \\
v_{\alpha }(\mathbf{p}, \sigma) &\equiv& \sum_{\beta} \mathcal{D}_{\alpha
\beta}(\lambda_{\mathbf{p}})
 v_{\beta }(\mathbf{0}, \sigma) \label{eq:v-beta}
\end{eqnarray}

\noindent Taking a Hermitian conjugate of (\ref{eq:u-alpha}) and
multiplying by $\gamma^0$ from the right we obtain factors in adjoint fields

\begin{eqnarray}
\overline{u}(\mathbf{p}, \sigma)  &\equiv& u^{\dag}(\mathbf{p},
\sigma) \gamma^0 =
 u^{\dag}(\mathbf{0}, \sigma) \mathcal{D}^{\dag}(\lambda_{\mathbf{p}})
\gamma^0 =
 u^{\dag}(\mathbf{0}, \sigma) \gamma^0 \gamma^0 \mathcal{D}(\lambda_{\mathbf{p}})
\gamma^0 \nonumber \\
 &=&
 u^{\dag}(\mathbf{0}, \sigma) \gamma^0  \mathcal{D}^{-1}(\lambda_{\mathbf{p}})
 =
 \overline{u}(\mathbf{0}, \sigma)  \mathcal{D}^{-1}(\lambda_{\mathbf{p}})
\label{eq:u-alpha2} \\
\overline{v}(\mathbf{p}, \sigma)
 &=&
 \overline{v}(\mathbf{0}, \sigma)  \mathcal{D}^{-1}(\lambda_{\mathbf{p}}) \nonumber
\end{eqnarray}

\section{Explicit formulas for $u$ and $v$} \label{ss:explicit-u}

Now let us find explicit expressions for factors $u, v, \overline{u}
$, and $\overline{v}$ for all momenta. Using formulas (\ref{eq:unique-pure}), (\ref{eq:A.87}),
(\ref{eq:A.83}) and

\begin{eqnarray*}
\theta &=& \tanh^{-1}(v/c) \\
 \tanh \frac{\theta}{2} &=& \frac{\tanh \theta}{1 + \sqrt{1 - \tanh^2
\theta}} = \frac{v/c}{1 + \sqrt{1- v^2/c^2}} = \frac{pc}{\omega_{\mathbf{p}} +
mc^2} \\
\cosh \frac{\theta}{2} &=& \frac{1}{\sqrt{1 - \tanh ^2
\frac{\theta}{2}}}
= \sqrt{\frac{\omega_{\mathbf{p}} + mc^2}{2mc^2}} \\
\sinh \frac{\theta}{2} &=& \tanh \frac{\theta}{2} \cosh
\frac{\theta}{2}
\end{eqnarray*}

\noindent  we obtain

\begin{eqnarray*}
\mathcal{D}(\lambda_{\mathbf{p}}) &=& e^{- \frac{ic}{\hbar}
\vec{\mathcal{K}} \cdot \frac{\mathbf{p}}{p}  \theta_\mathbf{p}}= I
\cosh \frac{\theta_\mathbf{p}}{2} + \frac{2c}{i
\hbar}\frac{\vec{\mathcal{K}} \cdot
\mathbf{p}}{p} \sinh \frac{\theta_\mathbf{p}}{2} \\
&=&  \cosh \frac{\theta_\mathbf{p}}{2}\left[ \begin{array}{cc}
 1 & 0  \\
0 & 1 \\
\end{array} \right] + \sinh
\frac{\theta_\mathbf{p}}{2} \left[ \begin{array}{cc}
 0 & \frac{\vec{\sigma} \cdot \mathbf{p}}{p}  \\
\frac{\vec{\sigma} \cdot \mathbf{p}}{p} & 0 \\
\end{array} \right] \\
 &=&
 \cosh \frac{\theta_\mathbf{p}}{2}\left(1 + \tanh
\frac{\theta_\mathbf{p}}{2} \left[ \begin{array}{cc}
 0 & \frac{\vec{\sigma} \cdot \mathbf{p}}{p}  \\
\frac{\vec{\sigma} \cdot \mathbf{p}}{p} & 0 \\
\end{array} \right] \right)\\
&=& \sqrt{\frac{\omega_{\mathbf{p}} + mc^2}{2mc^2}}\left(1 +
\frac{pc}{\omega_{\mathbf{p}} + mc^2}  \left[
\begin{array}{cc}
 0 & \frac{\vec{\sigma} \cdot
\mathbf{p}}{p} \\
\frac{\vec{\sigma} \cdot \mathbf{p}}{p} & 0
\end{array} \right] \right)  \\
&=& \sqrt{\frac{\omega_\mathbf{p} + mc^2}{2mc^2}} \left[
\begin{array}{cc}
 1 & \frac{\vec{\sigma} \cdot \mathbf{p} c}{\omega_\mathbf{p} + mc^2}   \\
 \frac{\vec{\sigma} \cdot \mathbf{p} c}{\omega_\mathbf{p} + mc^2} & 1
\end{array} \right]
\end{eqnarray*}

\noindent Then, inserting this result in (\ref{eq:u-alpha}) we
obtain

\begin{eqnarray}
 u(\mathbf{p}, \sigma ) &=& \sqrt{\frac{\omega_\mathbf{p} + mc^2}{2mc^2}}
\left[ \begin{array}{cc}
 1 & \frac{\vec{\sigma} \cdot \mathbf{p} c}{\omega_\mathbf{p} + mc^2}   \\
 \frac{\vec{\sigma} \cdot \mathbf{p} c}{\omega_\mathbf{p} + mc^2} & 1
\end{array} \right]  \left[ \begin{array}{c}
\chi_{\sigma}   \\
0
\end{array} \right]   \nonumber \\
&=&\left[ \begin{array}{c}
 \sqrt{\omega_{\mathbf{p}} + mc^2}   \\
 \sqrt{\omega_{\mathbf{p}} - mc^2}
\left(\vec{\sigma} \cdot \frac{\mathbf{p}}{p}  \right)  \\
\end{array} \right] \frac{\chi_{\sigma}}{\sqrt{2mc^2}}
\label{eq:10.17}
\end{eqnarray}

\noindent Similarly, the explicit expressions for $v $,
 $ \overline{u}$, and $ \overline{v}$ are

\begin{eqnarray}
 v (\mathbf{p}, \sigma) &=& \left[ \begin{array}{c}
 \sqrt{\omega_{\mathbf{p}} - mc^2} (\vec{\sigma} \cdot
\frac{\mathbf{p}}{p}  )
  \\
 \sqrt{\omega_{\mathbf{p}} + mc^2}   \\
\end{array} \right] \frac{\chi_{\sigma}}{\sqrt{2mc^2}}
\label{eq:10.18} \\
 \overline{u}(\mathbf{p}, \sigma) &=&
\frac{\chi^{\dag}_{\sigma}}{\sqrt{2mc^2}} \left[
\sqrt{\omega_{\mathbf{p}} + mc^2} ,
  -\sqrt{\omega_{\mathbf{p}} - mc^2}
\left(\vec{\sigma} \cdot \frac{\mathbf{p}}{p} \right) \right] \label{eq:10.19} \\
\overline{v} (\mathbf{p}, \sigma) &=&
\frac{\chi^{\dag}_{\sigma}}{\sqrt{2mc^2}}
 \left[ \sqrt{\omega_{\mathbf{p}} - mc^2}  \left(\vec{\sigma} \cdot \frac{\mathbf{p}}{p}
 \right)  ,
  \sqrt{\omega_{\mathbf{p}} + mc^2} \right]
\label{eq:10.20}
\end{eqnarray}

\noindent These functions are normalized to unity in the sense that\footnote{Here we used (\ref{eq:A.69c}).}

\begin{eqnarray}
&\mbox{ }&
 \overline{u}(\mathbf{p}, \sigma)
u(\mathbf{p}, \sigma') \nonumber \\
&=& \chi^{\dag}_{\sigma}  \left[\sqrt{\omega_{\mathbf{p}} + mc^2},
 \sqrt{\omega_{\mathbf{p}} - mc^2} \left(\frac{\mathbf{p}}{p} \cdot
\vec{\sigma}\right)  \right] \left[ \begin{array}{c}
 \sqrt{\omega_{\mathbf{p}} + mc^2}   \nonumber \\
 -\sqrt{\omega_{\mathbf{p}} - mc^2} \left(\frac{\mathbf{p}}{p} \cdot
\vec{\sigma}\right)
\end{array} \right]  \chi_{\sigma'}
\frac{1}{2 m c^2} \nonumber \\
&=& \chi^{\dag}_{\sigma}\left(\omega_{\mathbf{p}} + mc^2
 -( \omega_{\mathbf{p}} -
mc^2) \frac{(\mathbf{p} \cdot \vec{\sigma}) (\mathbf{p} \cdot
\vec{\sigma}) }{p^2} \right) \chi_{\sigma'}
\frac{1}{2 m c^2}
= \chi^{\dag}_{\sigma} \chi_{\sigma'} \nonumber  \\
&=& \delta_{\sigma,\sigma'} \label{eq:8.48cx}
\end{eqnarray}

\noindent Let us also calculate the sum $
   \sum_{\sigma
= -1/2}^{1/2} u(\mathbf{p}, \sigma)u^{\dag}(\mathbf{p}, \sigma) $.
At zero momentum
 we can use the explicit representation (\ref{eq:10.12}) - (\ref{eq:10.15})

\begin{eqnarray*}
   \sum_{\sigma  -1/2}^{1/2}
u(\mathbf{0}, \sigma) u^{\dag}(\mathbf{0}, \sigma) &=& \left[
\begin{array}{cccc}
 1 & 0 & 0 & 0   \\
 0 & 0 & 0 & 0 \\
 0 & 0 & 0 & 0 \\
 0 &  0 & 0 & 0
\end{array} \right] + \left[ \begin{array}{cccc}
 0 & 0 & 0 & 0   \\
 0 & 1 & 0 & 0 \\
 0 & 0 & 0 & 0 \\
 0 &  0 & 0 & 0
\end{array} \right] = \left[ \begin{array}{cccc}
 1 & 0 & 0 & 0   \\
 0 & 1 & 0 & 0 \\
 0 & 0 & 0 & 0 \\
 0 &  0 & 0 & 0
\end{array} \right] \\
&=& \frac{1}{2} \left(1 + \gamma^0 \right)
\end{eqnarray*}

\noindent To generalize this formula for arbitrary momentum, we use (\ref{eq:u-alpha}),
(\ref{eq:u-alpha2}), the Hermiticity of the matrix $\mathcal{D}
(\lambda_{\mathbf{p}})$ and
properties (\ref{eq:gamma0}), (\ref{eq:A.88}) - (\ref{eq:d-gamma-d})

\begin{eqnarray}
&\ & \sum_{\sigma = -1/2}^{1/2} u(\mathbf{p}, \sigma)
u^{\dag}(\mathbf{p}, \sigma) = \mathcal{D}(\lambda_{\mathbf{p}})
\left(\sum_{\sigma = -1/2}^{1/2} u(\mathbf{0},
\sigma)u^{\dag}(\mathbf{0}, \sigma) \right) \mathcal{D}^{\dag}
(\lambda_{\mathbf{p}}) \nonumber \\
&=& \frac{1}{2} \mathcal{D}(\lambda_{\mathbf{p}}) \left(1 +\gamma^0
\right) \mathcal{D} (\lambda_{\mathbf{p}}) = \frac{1}{2}
\left(\mathcal{D}(\lambda_{\mathbf{p}})\mathcal{D}(\lambda_{\mathbf{p}})
   + \gamma^0 \right) = \frac{1}{2} \left(\mathcal{D}(\lambda_{\mathbf{p}})\gamma^0
 \gamma^0\mathcal{D}(\lambda_{\mathbf{p}})\gamma^0
\gamma^0
 + \gamma^0 \right) \nonumber \\
 &=& \frac{1}{2} \left (\mathcal{D}(\lambda_{\mathbf{p}})\gamma^0
 \mathcal{D}^{-1}(\lambda_{\mathbf{p}})
\gamma^0
 + \gamma^0 \right) \nonumber \\
 &=& \frac{1}{2} \left(\mathcal{D}(\lambda_{\mathbf{p}}) \gamma^0
\mathcal{D}^{-1}(\lambda_{\mathbf{p}}) + 1 \right) \gamma^0 =
\frac{1}{2} \left( \gamma^0 \cosh \theta +
\vec{\gamma}\frac{\vec{\theta}}{\theta} \sinh \theta  + 1 \right)
\gamma^0 \nonumber \\
&=& \frac{1}{2mc^2} \left( \gamma^0 \omega_{\mathbf{p}} -
\vec{\gamma} \mathbf{p}c  + mc^2 \right) \gamma^0 = \frac{1}{2mc^2}
\left(\cross{p}  + mc^2 \right) \gamma^0\label{eq:needed}
\end{eqnarray}

\noindent Similarly we can derive a number of useful formulas\footnote{Here we used the facts that the
trace of any gamma-matrix is zero, and that the trace of the unit
4$\times$4 matrix is 4.}

\begin{eqnarray}
 \sum_{\sigma
= -1/2}^{1/2} u(\mathbf{p}, \sigma) \overline{u}(\mathbf{p}, \sigma)
&=&  \frac{1}{2mc^2} \left(\cross{p}  + mc^2 \right) \label{eq:J.43a} \\
\sum_{\alpha} \sum_{\sigma = -1/2}^{1/2} u_{\alpha}(\mathbf{p},
\sigma) \overline{u}_{\alpha}(\mathbf{p}, \sigma) &=&
\frac{1}{2mc^2} Tr \left[ \gamma^0 \omega_{\mathbf{p}} -
c\vec{\gamma}
\mathbf{p}  + mc^2 \right] = 2 \nonumber \\
 \sum_{\sigma
= -1/2}^{1/2} v(\mathbf{p}, \sigma)v^{\dag}(\mathbf{p}, \sigma) &=&
\frac{1}{2mc^2} \left( \cross{p}  - mc^2 \right) \gamma^0 \label{eq:needed2} \\
 \sum_{\sigma
= -1/2}^{1/2} v(\mathbf{p}, \sigma)\overline{v}(\mathbf{p}, \sigma)
&=& \frac{1}{2mc^2} \left( \cross{p} - mc^2 \right) \label{eq:J.44a}   \\
\sum_{\alpha} \sum_{\sigma = -1/2}^{1/2} v_{\alpha}(\mathbf{p},
\sigma) \overline{v}_{\alpha}(\mathbf{p}, \sigma) &=& -2 \nonumber
\end{eqnarray}

\section{Convenient notation}
\label{sc:convenient}

To simplify QED calculations we  introduce the following
combinations of particle operators

\begin{eqnarray}
  A_{\alpha}(\mathbf{p}) &=& \sqrt{\frac{mc^2}{\omega_{\mathbf{p}}}}
\sum_{\sigma} u_{\alpha}(\mathbf{p},\sigma) a _{\mathbf{p}, \sigma}
\label{eq:11.25}  \\
 \overline{A}^{\dag}_{\alpha}(\mathbf{p}) &=& \sqrt{\frac{mc^2}{\omega_{\mathbf{p}}}}
\sum_{\sigma} \overline{u}_{\alpha}(\mathbf{p},\sigma) a^{\dag}
_{\mathbf{p}, \sigma}
\label{eq:11.25a}  \\
  B^{\dag}_{\alpha}(\mathbf{p}) &=&
\sqrt{\frac{mc^2}{\omega_{\mathbf{p}}}}\sum_{\sigma}
v_{\alpha}(\mathbf{p},\sigma) b^{\dag} _{\mathbf{p}, \sigma}
\label{eq:11.26}  \\
 \overline{B}_{\alpha}(\mathbf{p}) &=&
\sqrt{\frac{mc^2}{\omega_{\mathbf{p}}}}\sum_{\sigma}
\overline{v}_{\alpha}(\mathbf{p},\sigma) b _{\mathbf{p}, \sigma}
\label{eq:11.26a}  \\
  D_{\alpha}(\mathbf{p}) &=& \sqrt{\frac{Mc^2}{\Omega_{\mathbf{p}}}}\sum_{\sigma}
w_{\alpha}(\mathbf{p},\sigma) d _{\mathbf{p}, \sigma}
\label{eq:11.27} \\
  \overline{D}^{\dag}_{\alpha}(\mathbf{p}) &=& \sqrt{\frac{Mc^2}{\Omega_{\mathbf{p}}}}\sum_{\sigma}
\overline{w}_{\alpha}(\mathbf{p},\sigma) d^{\dag} _{\mathbf{p},
\sigma}
\label{eq:11.27a} \\
  F^{\dag}_{\alpha}(\mathbf{p}) &=& \sqrt{\frac{Mc^2}{\Omega_{\mathbf{p}}}}
\sum_{\sigma} s_{\alpha}(\mathbf{p},\sigma) f^{\dag} _{\mathbf{p},
\sigma}
\label{eq:11.28}\\
  \overline{F}_{\alpha}(\mathbf{p}) &=& \sqrt{\frac{Mc^2}{\Omega_{\mathbf{p}}}}
\sum_{\sigma} \overline{s}_{\alpha}(\mathbf{p},\sigma) f
_{\mathbf{p}, \sigma} \label{eq:11.28a}
\end{eqnarray}

\noindent In this notation, indices $\alpha, \beta = 1,2,3,4$ are
those corresponding to the bispinor representation of the Lorentz
group, and index
$\sigma = \pm 1/2$ enumerates two spin projections of fermions.

 With the above conventions, the
electron/positron  and
proton/antiproton quantum fields can be written compactly

\begin{eqnarray}
\psi_{ \alpha}(\tilde{x}) &=& (2\pi \hbar)^{-3/2} \int d\mathbf{p}
 \left[e^{-\frac{i}{\hbar} \tilde{p} \cdot \tilde{x}} A _{
\alpha}(\mathbf{p}) + e^{\frac{i}{\hbar}\tilde{p} \cdot
\tilde{x}}B^{\dag}_{ \alpha}(\mathbf{p}) \right]
\label{eq:J.55a}\\
\Psi_{ \alpha}(\tilde{x}) &=& (2\pi \hbar)^{-3/2} \int d\mathbf{p}
 \left[e^{-\frac{i}{\hbar}\tilde{P} \cdot \tilde{x}}D _{
\alpha}(\mathbf{p}) + e^{\frac{i}{\hbar}\tilde{P} \cdot
\tilde{x}}F^{\dag}_{ \alpha}(\mathbf{p}) \right] \label{eq:J.55b}
\end{eqnarray}

\section{Transformation laws} \label{ss:transformations}

Operators (\ref{eq:11.25})-(\ref{eq:11.28a}) have simple boost
transformation laws. For example, we can use (\ref{eq:5.57}),
(\ref{eq:9.37}), (\ref{eq:7.9x}), and (\ref{eq:10.16}) to obtain

\begin{eqnarray}
U_0 (\Lambda; 0) A(\mathbf{p}) U_0^{-1} (\Lambda; 0) &=&
e^{-\frac{ic}{\hbar}\mathbf{K}_0 \vec{\theta}}A(\mathbf{p})
e^{\frac{ic}{\hbar}\mathbf{K}_0 \vec{\theta}} \nonumber \\
& =& \sqrt{\frac{mc^2}{\omega_{\mathbf{p}}}} \sum_{\sigma} u (
\mathbf{p}, \sigma) e^{\frac{ic}{\hbar}\mathbf{K}_0 \vec{\theta}} a
_{\mathbf{p}, \sigma}
e^{-\frac{ic}{\hbar}\mathbf{K}_0 \vec{\theta}} \nonumber \\
& =&  \sqrt{\frac{mc^2}{\omega_{\mathbf{p}}}}
\sqrt{\frac{\omega_{\Lambda \mathbf{p}}}{\omega_{\mathbf{p}}}}
\sum_{\sigma} u ( \mathbf{p}, \sigma) \sum_{\sigma'} D^{1/2}_{\sigma
\sigma'}(-\vec{\phi}_W) a_{\Lambda\mathbf{p}, \sigma'} \nonumber
\\
& =&  \sqrt{\frac{mc^2}{\omega_{\mathbf{p}}}}
\sqrt{\frac{\omega_{\Lambda \mathbf{p}}}{\omega_{\mathbf{p}}}}
 \mathcal{D}(\lambda_{\mathbf{p}}) \sum_{\sigma} u(\mathbf{0}, \sigma)
\sum_{\sigma'}
D^{1/2}_{\sigma \sigma'}(-\vec{\phi}_W) a_{\Lambda\mathbf{p}, \sigma'} \nonumber  \\
& =&  \sqrt{\frac{mc^2}{\omega_{\mathbf{p}}}}
\sqrt{\frac{\omega_{\Lambda \mathbf{p}}}{\omega_{\mathbf{p}}}}
 \mathcal{D}(\lambda_{\mathbf{p}})
 \mathcal D(-\vec{\phi}_W) \sum_{\sigma} u(\mathbf{0}, \sigma)
 a_{\Lambda\mathbf{p}, \sigma} \nonumber  \\
& =&  \sqrt{\frac{mc^2}{\omega_{\mathbf{p}}}}
\sqrt{\frac{\omega_{\Lambda \mathbf{p}}}{\omega_{\mathbf{p}}}}
 \mathcal{D}(\lambda_{\mathbf{p}})
 \mathcal D(\lambda^{-1}_{\mathbf{p}} \Lambda^{-1}
\lambda_{\Lambda \mathbf{p}}) \sum_{\sigma} u(\mathbf{0}, \sigma)
 a_{\Lambda\mathbf{p}, \sigma} \nonumber  \\
& =&  \sqrt{\frac{mc^2}{\omega_{\mathbf{p}}}}
\sqrt{\frac{\omega_{\Lambda \mathbf{p}}}{\omega_{\mathbf{p}}}}
\mathcal {D}( \Lambda^{-1})   \mathcal D ( \lambda_{\Lambda
\mathbf{p}}) \sum_{\sigma} u(\mathbf{0}, \sigma)
 a_{\Lambda\mathbf{p}, \sigma} \nonumber \\
& =& \sqrt{\frac{\omega_{\Lambda \mathbf{p}}}{\omega_{\mathbf{p}}}}
 \sqrt{\frac{mc^2}{\omega_{\mathbf{p}}}}
\mathcal {D}( \Lambda^{-1}) \sum_{\sigma}   u(\Lambda \mathbf{p},
\sigma)
 a_{\Lambda\mathbf{p}, \sigma} \nonumber \\
& =& \frac{\omega_{\Lambda \mathbf{p}}}{\omega_{\mathbf{p}}}
\mathcal {D}( \Lambda^{-1}) A(\Lambda \mathbf{p}) \label{eq:J.54a}
\end{eqnarray}

\noindent Similarly, using (\ref{eq:overline-u})

\begin{eqnarray}
U_0 (\Lambda; 0) \overline{A}^{\dag}(\mathbf{p}) U_0^{-1} (\Lambda;
0) & =& \sqrt{\frac{mc^2}{\omega_{\mathbf{p}}}} \sum_{\sigma}
\overline{u} ( \mathbf{p}, \sigma) e^{-\frac{ic}{\hbar}\mathbf{K}_0
\vec{\theta}} a^{\dag} _{\mathbf{p}, \sigma}
e^{\frac{ic}{\hbar}\mathbf{K}_0 \vec{\theta}} \nonumber \\
& =&  \sqrt{\frac{mc^2}{\omega_{\mathbf{p}}}}
\sqrt{\frac{\omega_{\Lambda \mathbf{p}}}{\omega_{\mathbf{p}}}}
\sum_{\sigma} \overline{u} ( \mathbf{p}, \sigma) \sum_{\sigma'}
(D^{1/2})^*_{\sigma \sigma'}(-\vec{\phi}_W)
a^{\dag}_{\Lambda\mathbf{p}, \sigma'} \nonumber
\\
& =&  \sqrt{\frac{mc^2}{\omega_{\mathbf{p}}}}
\sqrt{\frac{\omega_{\Lambda \mathbf{p}}}{\omega_{\mathbf{p}}}}
\sum_{\sigma} \sum_{\sigma'} \overline{u}(\mathbf{0}, \sigma')
 (D^{1/2})^*_{\sigma \sigma'}(-\vec{\phi}_W)
\mathcal{D}^{-1}(\lambda_{\mathbf{p}})
a^{\dag}_{\Lambda\mathbf{p}, \sigma'} \nonumber \\
&=&  \sqrt{\frac{mc^2}{\omega_{\mathbf{p}}}}
\sqrt{\frac{\omega_{\Lambda \mathbf{p}}}{\omega_{\mathbf{p}}}}
\sum_{\sigma} \overline{u}(\mathbf{0}, \sigma)
  \mathcal {D}(\lambda^{-1}_{\Lambda \mathbf{p}}  \Lambda
\lambda_{\mathbf{p}}) \mathcal{D}^{-1}(\lambda_{\mathbf{p}})
 a^{\dag}_{\Lambda\mathbf{p}, \sigma} \nonumber \\
&=&  \sqrt{\frac{mc^2}{\omega_{\mathbf{p}}}}
\sqrt{\frac{\omega_{\Lambda \mathbf{p}}}{\omega_{\mathbf{p}}}}
\sum_{\sigma} \overline{u}(\Lambda \mathbf{p}, \sigma)
  \mathcal {D}(  \Lambda)
 a_{\Lambda\mathbf{p}, \sigma} \nonumber \\
&=& \frac{\omega_{\Lambda \mathbf{p}}}{\omega_{\mathbf{p}}}
\overline{A}^{\dag}(\Lambda \mathbf{p})  \mathcal {D}(
\Lambda) \label{eq:J.54b}
\end{eqnarray}

Let us show that quantum field $\psi_{\alpha}(\tilde{x})$ has the required
covariant transformation law (\ref{eq:10.1})

\begin{eqnarray}
U_0(\Lambda; \tilde{a}) \psi_{\alpha}(\tilde{x}) U_0^{-1}(\Lambda; \tilde{a}) =
\sum_j \mathcal{D}_{\alpha \beta}(\Lambda^{-1}) \psi_{\beta}(\Lambda
(\tilde{x} + \tilde{a})) \label{eq:10.21}
\end{eqnarray}

\noindent  Transformations with respect to translations are

\begin{eqnarray*}
&\mbox{ }& U_0(1; \tilde{a}) \psi_{\alpha}(\tilde{x})  U_0^{-1} (1; \tilde{a}) \\
&=&
 \int \frac{d\mathbf{p}}{(2\pi \hbar)^{3/2}}
\left( e^{-\frac{i}{\hbar} \tilde{p} \cdot \tilde{x}} U_0(1;
\tilde{a}) A_{\alpha}(\mathbf{p}) U_0^{-1} (1; \tilde{a}) +
e^{\frac{i}{\hbar} \tilde{p} \cdot
\tilde{x}}U_0 (1; \tilde{a}) B^{\dag}_{\alpha}(\mathbf{p})U_0^{-1} (1; \tilde{a}) \right) \\
&=&  \int \frac{d\mathbf{p}}{(2\pi \hbar)^{3/2}}
\left(e^{-\frac{i}{\hbar} \tilde{p} \cdot (\tilde{x} + \tilde{a})}
A_{\alpha } ( \mathbf{p}) + e^{\frac{i}{\hbar} \tilde{p} \cdot
(\tilde{x} + \tilde{a})} B^{\dag}_{\alpha } ( \mathbf{p}) \right)=
\psi_{\alpha}( \tilde{x} + \tilde{a})
\end{eqnarray*}

\noindent For transformations with respect to boosts we use equations
(\ref{eq:J.54a}), (\ref{eq:J.54b}),
(\ref{eq:7.15a}), and (\ref{eq:A.73})

\begin{eqnarray*}
&\mbox{ }& U_0 (\Lambda; 0) \psi(\tilde{x}) U_0^{-1} (\Lambda; 0)
\\&=& (2\pi \hbar)^{-3/2} \int d\mathbf{p}
 \left(e^{-\frac{i}{\hbar} \tilde{p} \cdot \tilde{x}} U_0 (\Lambda; 0) A (\mathbf{p})
 U_0^{-1} (\Lambda; 0)
 + e^{\frac{i}{\hbar}\tilde{p} \cdot \tilde{x} }U_0
(\Lambda; 0)B^{\dag}(\mathbf{p})U_0^{-1} (\Lambda; 0) \right) \\
&=& (2\pi \hbar)^{-3/2} \mathcal{D}( \Lambda^{-1}) \int d\mathbf{p}
\frac{\omega_{\Lambda \mathbf{p}}}{\omega_{\mathbf{p}}}
 \Bigl(e^{-\frac{i}{\hbar} \tilde{p} \cdot \tilde{x}}  A (\Lambda \mathbf{p})
 + e^{\frac{i}{\hbar}\tilde{p} \cdot \tilde{x} }B^{\dag}(\Lambda \mathbf{p}) \Bigr) \\
 &=& (2\pi \hbar)^{-3/2} \mathcal{D}( \Lambda^{-1}) \int d\mathbf{q}
 \Bigl(e^{-\frac{i}{\hbar} (\Lambda^{-1} \tilde{q} \cdot \tilde{x})}  A (\mathbf{q})
 + e^{\frac{i}{\hbar} (\Lambda^{-1} \tilde{q} \cdot \tilde{x}) }B^{\dag}(\mathbf{q}) \Bigr) \\
  &=& (2\pi \hbar)^{-3/2} \mathcal{D}( \Lambda^{-1}) \int d\mathbf{q}
 \Bigl(e^{-\frac{i}{\hbar} ( \tilde{q} \cdot \Lambda \tilde{x})}  A (\mathbf{q})
 + e^{\frac{i}{\hbar} (\tilde{q} \cdot \Lambda \tilde{x}) }B^{\dag}(\mathbf{q}) \Bigr) \\
 &=&  \mathcal{D}( \Lambda^{-1})
\psi(\Lambda \tilde{x})
\end{eqnarray*}

\noindent  We leave to the
reader the proof of equation (\ref{eq:10.21}) in the case of rotations.

Thus we conclude that in agreement with \textbf{Step 1}(II) in subsection \ref{ss:weinberg}, the Dirac field transforms according to the 4D bispinor representation of the Lorentz
group.

\section{Functions $U_{\mu}$ and $W_{\mu}$.}
\label{ss:Umuwmu}

In QED calculations one often meets products like
$\overline{u}\gamma^{\mu}u$ and  $\overline{w}\gamma^{\mu}w$. It is
convenient to introduce special symbols for them

\begin{eqnarray}
U^{\mu} (\mathbf{p}, \sigma; \mathbf{p}', \sigma') &\equiv& \overline{u}
(\mathbf{p}, \sigma)  \gamma^{\mu} u( \mathbf{p}', \sigma') \label{eq:uz}\\
W^{\mu} (\mathbf{p}, \sigma; \mathbf{p}', \sigma') &\equiv& \overline{w}
(\mathbf{p}, \sigma)  \gamma^{\mu} w( \mathbf{p}', \sigma')
\label{eq:vz}
 \end{eqnarray}

\noindent Note that quantities $U^{\mu}$ and $W^{\mu}$ are four-vectors with
respect to Lorentz transformations of their momentum and spin
labels.\footnote{Such a transformation acts by matrix $\Lambda$
(rotation$\times$boost) on momentum
arguments and by the corresponding Wigner rotation $R$ on spin components. See subsection \ref{ss:Lor-S}.} For
example, using (\ref{eq:10.16}), (\ref{eq:overline-u}), (\ref{eq:u-alpha}), (\ref{eq:u-alpha2}),  and
(\ref{eq:A.88}), we obtain\footnote{Here $R(\mathbf{p}, \Lambda)$ is the Wigner rotation defined in (\ref{eq:7.9x}).}

\begin{eqnarray*}
&\mbox{ }& U^{\mu} \left(\Lambda \mathbf{p}, R^{-1}(\mathbf{p},
\Lambda)\sigma; \Lambda \mathbf{p}', R(\mathbf{p'}, \Lambda) \sigma' \right) \\
&\equiv& \overline{u} \left(\Lambda \mathbf{p}, R^{-1}(\mathbf{p},
\Lambda) \sigma \right)  \gamma^{\mu} u \left( \Lambda \mathbf{p}',
R(\mathbf{p'},
\Lambda) \sigma' \right) \\
&=& \overline{u} \left( \mathbf{0}, R^{-1}(\mathbf{p}, \Lambda)
\sigma \right) \mathcal{D}^{-1}(\lambda_{\Lambda \mathbf{p}})
\gamma^{\mu} \mathcal{D}(\lambda_{\Lambda \mathbf{p'}}) u\left(
\mathbf{0}, R(\mathbf{p'},
\Lambda) \sigma' \right) \\
&=& \overline{u} \left(\mathbf{0}, R^{-1}(\mathbf{p}, \Lambda)
\sigma \right) \mathcal{D}^{-1} \left(\Lambda \lambda_{\mathbf{p}}
R^{-1}(\mathbf{p}, \Lambda) \right) \gamma^{\mu}
\mathcal{D}\left(\Lambda \lambda_{\mathbf{p'}} R^{-1}(\mathbf{p'},
\Lambda)\right) u \left(\mathbf{0}, R(\mathbf{p'},
\Lambda) \sigma' \right) \\
&=& \overline{u} \left( \mathbf{0}, R^{-1}(\mathbf{p}, \Lambda)
\sigma \right) \mathcal{D} \left(R^{-1}(\mathbf{p}, \Lambda)
\right)\mathcal{D}^{-1} \left(\lambda_{\mathbf{p}}
 \right)\mathcal{D}^{-1}(\Lambda) \gamma^{\mu} \mathcal{D}(\Lambda)
\mathcal{D} \left( \lambda_{\mathbf{p'}} \right)
\mathcal{D}\left(R^{-1}(\mathbf{p'}, \Lambda) \right) \times \\
&\ & u\left(  \mathbf{0}, R(\mathbf{p'},
\Lambda) \sigma' \right) \\
&=& \overline{u} ( \mathbf{0},  \sigma) \mathcal{D}^{-1}
(\lambda_{\mathbf{p}} )\mathcal{D}^{-1}(\Lambda) \gamma^{\mu}
\mathcal{D}(\Lambda) \mathcal{D}( \lambda_{\mathbf{p'}})
 u(  \mathbf{0}, \sigma') \\
 &=& \overline{u} ( \mathbf{p},  \sigma) \mathcal{D}^{-1}(\Lambda) \gamma^{\mu}
\mathcal{D}(\Lambda)
 u(  \mathbf{p'}, \sigma') \\
 &=& \overline{u} ( \mathbf{p},  \sigma) \left(\Lambda_{\mu \nu}
 \gamma^{\nu} \right)
 u(  \mathbf{p'}, \sigma') \\
 &=& \Lambda_{\mu \nu} U^{\nu} (\mathbf{p}, \sigma; \mathbf{p}', \sigma')
\end{eqnarray*}

\section{$(v/c)^2$ approximation}
\label{ss:non-rel}

Often it is useful to obtain QED results in a weakly-relativistic or
non-relativistic case, when momenta of electrons are much less than
$mc$ and momenta of protons are much less than $Mc$. In these cases,
with reasonable accuracy we can represent all quantities as series
in powers of $v/c$ and leave only terms having orders not higher than
$(v/c)^{2}$. First, we can use (\ref{eq:omega-non-rel}) to write

\begin{eqnarray*}
 \sqrt
{\omega_{\mathbf{p}} + mc^2} &\approx& \sqrt {mc^2 + \frac{p^2}{2m}
+ mc^2}= \sqrt {2mc^2 + \frac{p^2}{2m}} \nonumber \\
&=& \sqrt {2mc^2} \sqrt{1 + \frac{p^2}{4m^2c^2}} \approx
\sqrt{2mc^2}\left(1 + \frac{p^2}{8m^2c^2} \right) \nonumber \\
\sqrt
{\omega_{\mathbf{p}} - mc^2} &\approx& \sqrt {mc^2 + \frac{p^2}{2m}
- mc^2} =
\frac{p}{\sqrt{2m}}
\end{eqnarray*}

\begin{eqnarray}
&\ & (q+k \div q)^2 =
(\omega_{\mathbf{q+k}}-\omega_{\mathbf{q}})^2 - c^2k^2
\approx-c^2k^2  \label{eq:qkdivq}
\end{eqnarray}

\begin{eqnarray}
&\mbox{}& \frac{Mmc^4} {\sqrt{\Omega_{\mathbf{p-k}}
\omega_{\mathbf{q+k}} \Omega_{\mathbf{p}}
\omega_{\mathbf{q}}}} \nonumber \\
&\approx& \frac{1}{\sqrt{1 + \frac{(\mathbf{p-k})^2}{2 M^2 c^2}}}
 \frac{1}{\sqrt {1 + \frac{p^2}{2 M^2 c^2}}}
 \frac{1}{\sqrt{1 + \frac{(\mathbf{q+k})^2}{2 m^2
c^2}}}
 \frac{1}{\sqrt{ 1 + \frac{q^2}{2 m^2 c^2}}} \nonumber \\
&\approx& 1 - \frac{(\mathbf{p-k})^2}{4M^2c^2} - \frac{p^2}{4M^2c^2}
 - \frac{(\mathbf{q+k})^2}{4m^2c^2} - \frac{q^2}{4m^2c^2} \nonumber
\\
&=& 1 - \frac{p^2}{2M^2c^2} + \frac{\mathbf{p}\mathbf{k}}{2M^2c^2} -
\frac{k^2}{4M^2c^2} - \frac{q^2}{2m^2c^2} -
\frac{\mathbf{q}\mathbf{k}}{2m^2c^2} - \frac{k^2}{4m^2c^2} \label{eq:exp-den}
\end{eqnarray}

\noindent To obtain the  $(v/c)^2$ approximation for expressions
(\ref{eq:uz}), (\ref{eq:vz}) we use equations (\ref{eq:10.17}) -
(\ref{eq:10.20}) and (\ref{eq:A.69a}) - (\ref{eq:A.69c})

\begin{eqnarray}
&\mbox{ }& U^0(\mathbf{p}, \sigma;  \mathbf{p}', \sigma') =
 \overline{u}(\mathbf{p}, \sigma)   \gamma^0
u(\mathbf{p}', \sigma') =
 u^{\dag}(\mathbf{p}, \sigma)  u(\mathbf{p}', \sigma') \nonumber \\
&=& \chi^{\dag}_{\sigma}  \left[\sqrt{\omega_{\mathbf{p}} + mc^2},
 \sqrt{\omega_{\mathbf{p}} - mc^2} \left(\frac{\mathbf{p}}{p} \cdot
\vec{\sigma}\right)  \right] \left[ \begin{array}{c}
 \sqrt{\omega_{\mathbf{p'}} + mc^2}   \nonumber \\
 \sqrt{\omega_{\mathbf{p'}} - mc^2} (\frac{\mathbf{p'}}{p'} \cdot
\vec{\sigma})
\end{array} \right]  \chi_{\sigma'}
\frac{1}{2 m c^2} \nonumber \\
&=& \chi^{\dag}_{\sigma}\left(\sqrt{\omega_{\mathbf{p}} + mc^2}
\sqrt{\omega_{\mathbf{p'}} + mc^2} + \sqrt{\omega_{\mathbf{p}} -
mc^2} \sqrt{\omega_{\mathbf{p'}} - mc^2} \frac{(\mathbf{p} \cdot
\vec{\sigma}) (\mathbf{p'} \cdot \vec{\sigma}) }{pp'} \right)
\chi_{\sigma'}
\frac{1}{2 m c^2}\nonumber  \\
 &\approx&  \chi^{\dag}_{\sigma}
\left(\left(1 + \frac{p^2}{8m^2c^2} \right)\left(1 +
\frac{(p')^2}{8m^2c^2} \right) + \frac{pp'}{4m^2c^2}
\frac{(\mathbf{p} \cdot
\vec{\sigma})( \mathbf{p'} \cdot \vec{\sigma}) }{pp'} \right) \chi_{\sigma'} \nonumber \\
&=&  \chi^{\dag}_{\sigma} \left(1 + \frac{p^2 + (p')^2 + 2
\mathbf{p} \cdot \mathbf{p}'+ 2i \vec{\sigma} \cdot  [\mathbf{p}
\times
\mathbf{p'}]}{8m^2c^2} \right) \chi_{\sigma'}  \nonumber \\
&=&  \chi^{\dag}_{\sigma} \left(1 + \frac{(\mathbf{p} +
\mathbf{p}')^2 + 2i \vec{\sigma} \cdot  [\mathbf{p} \times
\mathbf{p'}]}{8m^2c^2} \right) \chi_{\sigma'} \label{eq:8.48a} \\ \nonumber  \\
&\ &W^0(\mathbf{p}, \sigma; \mathbf{p}', \sigma')  =
 \overline{w}(\mathbf{p}, \sigma)  \gamma^0
w(\mathbf{p}', \sigma') \approx
 \chi^{\dag}_{\sigma} \left(1 + \frac{(\mathbf{p} + \mathbf{p'})^2+ 2i
\vec{\sigma} \cdot  [\mathbf{p} \times \mathbf{p'}]}{8M^2c^2}
\right) \chi_{\sigma'} \nonumber \\
\label{eq:8.48b} \\
&\mbox{ }& \mathbf{U}(\mathbf{p}, \sigma;  \mathbf{p}', \sigma') =
\overline{u} (\mathbf{p}, \sigma) \vec{\gamma}u(\mathbf{p}', \sigma') \nonumber \\
&=& \chi^{\dag}_{\sigma}\left[ \sqrt{\omega_{\mathbf{p}} + mc^2},
-\sqrt{\omega_{\mathbf{p}} - mc^2} \frac{\mathbf{p} \cdot
\vec{\sigma}}{p} \right]  \left[ \begin{array}{cc}
0 & \vec{\sigma}     \\
-\vec{\sigma} & 0
\end{array} \right]
\left[ \begin{array}{c}
\sqrt{\omega_{\mathbf{p'}} + mc^2}    \\
 \sqrt{\omega_{\mathbf{p'}} - mc^2} \frac{\mathbf{p'} \cdot \vec{\sigma}}{p'}
\end{array} \right] \chi_{\sigma'}
 \frac{1}{2 mc^2} \nonumber \\
&=& \chi^{\dag}_{\sigma} \left[ \sqrt{\omega_{\mathbf{p}} + mc^2}, -
\sqrt{\omega_{\mathbf{p}} - mc^2} \frac{\mathbf{p} \cdot
\vec{\sigma}}{p} \right] \left[ \begin{array}{c}
 \sqrt{\omega_{\mathbf{p'}} - mc^2} \frac{\vec{\sigma} (\mathbf{p'} \cdot
\vec{\sigma})}{p'} \\
-\vec{\sigma} \sqrt{\omega_{\mathbf{p'}} + mc^2}
\end{array} \right] \chi_{\sigma'}
\frac{1}{2 mc^2}
\nonumber \\
&=&  \chi^{\dag}_{\sigma} \left(\sqrt{\omega_{\mathbf{p}} + mc^2}
\sqrt{\omega_{\mathbf{p'}} - mc^2} \frac{\vec{\sigma} (\mathbf{p'}
\cdot \vec{\sigma})}{p'} +  \sqrt{\omega_{\mathbf{p}} - mc^2}
\sqrt{\omega_{\mathbf{p'}} + mc^2} \frac{ (\mathbf{p} \cdot
\vec{\sigma}) \vec{\sigma}}{p} \right)
\chi_{\sigma'} \frac{1}{2 mc^2} \nonumber  \\
&\approx& \chi^{\dag}_{\sigma} \left(\sqrt{2mc^2}
\frac{p'}{\sqrt{2m}} \frac{\vec{\sigma} (\mathbf{p}' \cdot
\vec{\sigma})}{p'} + \sqrt{2mc^2} \frac{p}{\sqrt{2m}} \frac{
(\mathbf{p} \cdot \vec{\sigma}) \vec{\sigma}}{p} \right)
\chi_{\sigma'} \frac{1}{2
m c^2} \nonumber  \\
&=&  \chi^{\dag}_{\sigma} \left((\vec{\sigma}\cdot \mathbf{p})
\vec{\sigma}  + \vec{\sigma}(\vec{\sigma}\cdot \mathbf{p}') \right)
\chi_{\sigma'}
\frac{1}{2 mc} \nonumber  \\
 &=&
\chi^{\dag}_{\sigma}\left(\mathbf{p} + i[\vec{\sigma} \times
\mathbf{p}] + \mathbf{p}' - i[\vec{\sigma} \times
\mathbf{p}']\right) \chi_{\sigma'} \frac{1}{2 mc} \nonumber
\\ &=&\chi^{\dag}_{\sigma} \left(\mathbf{p+p'} + i[\vec{\sigma} \times
(\mathbf{p-p'})]\right) \chi_{\sigma'} \frac{1}{2 mc}
\label{eq:8.48c} \\
\nonumber \\
&\ &\mathbf{W}(\mathbf{p}, \sigma; \mathbf{p}', \sigma') \approx
\chi^{\dag}_{\sigma}\left(\mathbf{p+p'} + i[\vec{\sigma} \times
(\mathbf{p-p'})]\right) \chi_{\sigma'} \frac{1}{2 Mc}
\label{eq:8.48d}
\end{eqnarray}

\noindent In the non-relativistic limit $c \to \infty$, all formulas
are further simplified

\begin{eqnarray}
\lim_{c \to \infty} \omega_{\mathbf{p}} &=& mc^2 \nonumber \\
\lim_{c \to \infty} \Omega_{\mathbf{p}} &=& Mc^2 \nonumber \\
\lim_{c \to \infty} \frac{Mmc^4} {\sqrt{\Omega_{\mathbf{p-k}}
\omega_{\mathbf{q+k}} \Omega_{\mathbf{p}}
\omega_{\mathbf{q}}}} &=& 1 \label{eq:J.70a} \\
\lim_{c \to \infty} U_0(\mathbf{p}, \sigma; \mathbf{p}', \sigma')
&=&  \chi^{\dag}_{\sigma}  \chi_{\sigma'}  =  \delta_{\sigma,\sigma'} \label{eq:U0} \\
 \lim_{c \to \infty} W_0(\mathbf{p}, \sigma; \mathbf{p}',
\sigma') &=& \delta_{\sigma,\sigma'} \nonumber \\
\lim_{c \to \infty} \mathbf{U}(\mathbf{p}, \sigma; \mathbf{p}',
\sigma') &=& 0 \nonumber \\
\lim_{c \to \infty} \mathbf{W}(\mathbf{p}, \sigma; \mathbf{p}',
\sigma') &=& 0 \nonumber
\end{eqnarray}

\section{Anticommutation relations} \label{ss:anticomm}

To check the anticommutation relations (\ref{eq:10.3}) we calculate,
for example,\footnote{Here we used (\ref{eq:needed}) and
(\ref{eq:needed2}).}

\begin{eqnarray}
&\mbox{ }& \{ \psi_{\alpha} ( \mathbf{x}, 0), \psi^{\dag}_{\beta} (
\mathbf{y}, 0) \} \nonumber\\
&=&
 \int \frac{d\mathbf{p}}{(2\pi \hbar)^{3/2}}
\sqrt{\frac{mc^2}{\omega_{\mathbf{p}}}}
 \frac{d\mathbf{p}'}{(2\pi \hbar)^{3/2}}
\sqrt{\frac{mc^2}{\omega_{\mathbf{p}'}}} \sum_{\sigma, \sigma'
= -1/2}^{1/2}  \nonumber \\
&\ & \{ \left(e^{-\frac{i}{\hbar}\mathbf{p}
\mathbf{x}}u_{\alpha}(\mathbf{p}, \sigma)
 a_{\mathbf{p}, \sigma} +
e^{\frac{i}{\hbar}\mathbf{p} \mathbf{x}}v_{\alpha } (\mathbf{p},
\sigma)b^{\dag}_{\mathbf{p}, \sigma} \right), \nonumber \\
&\ & \left(e^{\frac{i}{\hbar}\mathbf{p}' \mathbf{y}}u^{\dag}_{\beta }
(\mathbf{p}', \sigma') a^{\dag}_{\mathbf{p}', \sigma'} +
e^{-\frac{i}{\hbar}\mathbf{p}' \mathbf{y}}v^{\dag}_{\beta }
(\mathbf{p}', \sigma')b_{\mathbf{p}', \sigma'} \right)\} \nonumber\\
&=&
 \int \frac{d\mathbf{p}d\mathbf{p}'}{(2\pi \hbar)^{3}}
\frac{mc^2}{\sqrt{\omega_{\mathbf{p}}\omega_{\mathbf{p}'}}}
 \sum_{\sigma, \sigma'
= -1/2}^{1/2}  \Bigl(e^{-\frac{i}{\hbar}\mathbf{p}
\mathbf{x}+\frac{i}{\hbar}\mathbf{p}' \mathbf{y}}u_{\alpha
}(\mathbf{p}, \sigma)u^{\dag}_{\beta}(\mathbf{p}', \sigma') \{
a_{\mathbf{p}, \sigma} , a^{\dag}_{\mathbf{p}', \sigma'}\} \nonumber \\
&\ &+
e^{\frac{i}{\hbar}\mathbf{p} \mathbf{x}-\frac{i}{\hbar}\mathbf{p}'
\mathbf{y}}v_{\alpha }(\mathbf{p},
\sigma)v^{\dag}_{\beta}(\mathbf{p}', \sigma')
\{b^{\dag}_{\mathbf{p}, \sigma},
b_{\mathbf{p}', \sigma'}\} \Bigr) \nonumber\\
&=&  \int \frac{ d \mathbf{p}d \mathbf{p}'
mc^2}{(2\pi \hbar)^{3}\omega_{\mathbf{p}}}
  \sum_{\sigma, \sigma'
= -1/2}^{1/2} \Bigl(e^{-\frac{i}{\hbar}
\mathbf{p}(\mathbf{x-y})}u_{\alpha}(\mathbf{p}, \sigma)
u^{\dag}_{\beta}(\mathbf{p}', \sigma') \delta (\mathbf{p}-
\mathbf{p}') \delta _{\sigma, \sigma'} \nonumber\\
 &\ &+
e^{\frac{i}{\hbar}\mathbf{p}(\mathbf{x-y})}v_{\alpha }(\mathbf{p},
\sigma) v^{\dag}_{\beta}(\mathbf{p}', \sigma')
\delta (\mathbf{p}- \mathbf{p}') \delta _{\sigma, \sigma'}\Bigr) \nonumber  \\
&=& (2\pi \hbar)^{-3} \int \frac{ d \mathbf{p}
mc^2}{\omega_{\mathbf{p}}}
  \sum_{\sigma
= -1/2}^{1/2} \nonumber \\
&\ &\Bigl(e^{-\frac{i}{\hbar}\mathbf{p}(\mathbf{x-y})}
u_{\alpha}(\mathbf{p}, \sigma)u^{\dag}_{\beta}(\mathbf{p}, \sigma)
 +
e^{\frac{i}{\hbar}\mathbf{p}(\mathbf{x-y})}v_{\alpha }(\mathbf{p},
\sigma) v^{\dag}_{\beta}(\mathbf{p}, \sigma) \Bigr) \nonumber \\
&=&  \int \frac{ d \mathbf{p}
mc^2}{(2\pi \hbar)^{3}\omega_{\mathbf{p}}}
e^{-\frac{i}{\hbar}\mathbf{p}(\mathbf{x-y})}
  \sum_{\sigma
= -1/2}^{1/2} \left( u_{\alpha}(\mathbf{p},
\sigma)u^{\dag}_{\beta}(\mathbf{p}, \sigma)
 + v_{\alpha }(-\mathbf{p},
\sigma) v^{\dag}_{\beta}(-\mathbf{p}, \sigma) \right) \nonumber \\
&=&  \int \frac{ d \mathbf{p}
mc^2}{(2\pi \hbar)^{3} \omega_{\mathbf{p}}}
e^{-\frac{i}{\hbar}\mathbf{p}(\mathbf{x-y})}
\frac{\omega_{\mathbf{p}} }{mc^2} ( \gamma^0 \gamma^0)_{\alpha
\beta} \nonumber \\
&=& \delta(\mathbf{x-y}) \delta_{\alpha \beta} \label{eq:10.23}
\end{eqnarray}

We will also find useful the following anticommutators

\begin{eqnarray}
\{A_{\alpha}(\mathbf{p}),\overline{ A}^{\dag}_{\beta}(\mathbf{p}')\}
&=& \frac{mc^2}{\omega_{\mathbf{p}}} \sum_{\sigma \sigma'}
u_{\alpha}(\mathbf{p}, \sigma)
\overline{u}^{\dag}_{\beta}(\mathbf{p}', \sigma')  \{a_{\mathbf{p},
\sigma}, a^{\dag}_{\mathbf{p}', \sigma'}\} \nonumber \\
 &=&
\frac{mc^2}{\omega_{\mathbf{p}}} \left(\sum_{\sigma}
u_{\alpha}(\mathbf{p}, \sigma) \overline{u}_{\beta}(\mathbf{p},
\sigma) \right) \delta(\mathbf{p}- \mathbf{p}') \nonumber \\
&=& \frac{1}{2\omega_{\mathbf{p}}} (\gamma^0 \omega_{\mathbf{p}} -
\vec{\gamma} \mathbf{p} c + mc^2 )_{\alpha \beta} \delta(\mathbf{p}-
\mathbf{p}') \label{eq:A-over-A} \\
 \sum_{\alpha}
\{A^{\dag}_{\alpha}(\mathbf{p}),
\overline{A}_{\alpha}(\mathbf{p}')\} &=& 2 \delta(\mathbf{p}-
\mathbf{p}') \label{eq:anticomm-A} \\
 \{B_{\alpha}(\mathbf{p}),
\overline{B}^{\dag}_{\beta}(\mathbf{p}')\} &=&
\frac{1}{2\omega_{\mathbf{p}}} (\gamma^0 \omega_{\mathbf{p}}-
\vec{\gamma} \mathbf{p} c - mc^2 )_{\alpha \beta} \delta(\mathbf{p}-
\mathbf{p}') \label{eq:B-over-B} \\
 \sum_{\alpha}
\{B^{\dag}_{\alpha}(\mathbf{p}),
\overline{B}_{\alpha}(\mathbf{p}')\} &=& 2 \delta(\mathbf{p}-
\mathbf{p}') \label{eq:anticomm-B}
\end{eqnarray}

\section{Dirac equation} \label{ss:dirac-equation}

  We can write the electron-positron quantum
field (\ref{eq:10.10}) as a sum of two terms

\begin{eqnarray*}
\psi_{\alpha}(\tilde{x})   &=& \psi_{\alpha}^+(\tilde{x})  +
\psi_{\alpha}^-(\tilde{x}) \\
\psi_{\alpha}^+(\tilde{x})   &\equiv& \sum_{\sigma}\int
\frac{d\mathbf{p}}{(2\pi \hbar)^{3/2}}
\sqrt{\frac{mc^2}{\omega_{\mathbf{p}}}} e^{
-\frac{i}{\hbar}\tilde{p}
\cdot \tilde{x}} u_{\alpha } ( \mathbf{p}, \sigma) a _{\mathbf{p}, \sigma}  \\
\psi_{\alpha}^-(\tilde{x})   &\equiv& \sum_{\sigma}\int
\frac{d\mathbf{p}}{(2\pi \hbar)^{3/2}}
\sqrt{\frac{mc^2}{\omega_{\mathbf{p}}}} e^{ \frac{i}{\hbar}\tilde{p}
\cdot \tilde{x}}
 v_{\alpha } (\mathbf{p}, \sigma) b^{\dag} _{\mathbf{p}, \sigma}
\end{eqnarray*}

\noindent Let us now act on the component $\psi^+(\tilde{x})$
by operator in parentheses\footnote{Here
we use explicit definitions of gamma matrices from (\ref{eq:A.81})
and (\ref{eq:A.82}) as well as equation (\ref{eq:10.17}).}

\begin{eqnarray*}
&\mbox { }& \left(\gamma^0 \frac{\partial}{\partial t}  +
c\vec{\gamma} \frac{\partial}{\partial \mathbf{x}}
-\frac{imc^2}{\hbar} \right)\psi ^+(\tilde{x})
\nonumber\\
&=& \left(\gamma^0 \frac{\partial}{\partial t}  +  c\vec{\gamma}
\frac{\partial}{\partial \mathbf{x}} -\frac{imc^2}{\hbar} \right)
\sum_{\sigma} \int \frac{d\mathbf{p}}{(2\pi \hbar)^{3/2}}
\sqrt{\frac{mc^2}{\omega_{\mathbf{p}}}} e^{ -\frac{i}{\hbar}
\mathbf{px} +  \frac{i}{\hbar} \omega_{\mathbf{p}}t} u( \mathbf{p},
\sigma)
a _{\mathbf{p}, \sigma}  \nonumber\\
&=& \frac{i}{\hbar} \sum_{\sigma} \int \frac{d\mathbf{p}}{(2\pi
\hbar)^{3/2}} \sqrt{\frac{mc^2}{\omega_{\mathbf{p}}}} (\gamma^0
\omega_{\mathbf{p}}
 -  c\vec{\gamma} \cdot \mathbf{p}
 -mc^2  ) u( \mathbf{p}, \sigma)  e^{
-\frac{i}{\hbar} \mathbf{px} +  \frac{i}{\hbar}
\omega_{\mathbf{p}}t} a _{\mathbf{p}, \sigma}  \nonumber
\end{eqnarray*}

\noindent For the product on the right hand side we obtain

\begin{eqnarray}
&\mbox { }&  (\gamma^0 \omega_{\mathbf{p}}
 -  c\vec{\gamma} \cdot \mathbf{p}
 -mc^2  ) u( \mathbf{p}, \sigma)   \nonumber\\
&=&  \omega_{\mathbf{p}} \left[ \begin{array}{c}
 \sqrt{\omega_{\mathbf{p}} + mc^2}   \\
 -\sqrt{\omega_{\mathbf{p}} - mc^2}
(\vec{\sigma} \cdot \frac{\mathbf{p}}{p}  )  \\
\end{array} \right] \frac{\chi_{\sigma}}{\sqrt{2mc^2}} -   \left[ \begin{array}{c}
 \sqrt{\omega_{\mathbf{p}} - mc^2} p \\
- \sqrt{\omega_{\mathbf{p}} + mc^2} (\vec{\sigma} \cdot \mathbf{p})  \\
\end{array} \right] \frac{\chi_{\sigma}}{\sqrt{2mc^2}}
 -mc^2 u( \mathbf{p}, \sigma)   \nonumber \\
&=&  \left[ \begin{array}{c} \omega_{\mathbf{p}}
\sqrt{\omega_{\mathbf{p}} + mc^2} - (\omega_{\mathbf{p}} - mc^2)
\sqrt{\omega_{\mathbf{p}} + mc^2} -
mc^2 \sqrt{\omega_{\mathbf{p}} + mc^2} \\
 -\omega_{\mathbf{p}}  \sqrt{\omega_{\mathbf{p}} - mc^2}
(\vec{\sigma} \cdot \frac{\mathbf{p}}{p}  ) +
\sqrt{\omega_{\mathbf{p}} + mc^2} (\vec{\sigma} \cdot
\frac{\mathbf{p}}{p}) pc - mc^2 \sqrt{\omega_{\mathbf{p}} - mc^2}
(\vec{\sigma} \cdot \frac{\mathbf{p}}{p})\\
\end{array} \right] \frac{\chi_{\sigma}}{\sqrt{2mc^2}}  \nonumber  \\
&=&   \left[ \begin{array}{c} (\omega_{\mathbf{p}} -mc^2)
\sqrt{\omega_{\mathbf{p}} + mc^2}
- (\omega_{\mathbf{p}} - mc^2) \sqrt{\omega_{\mathbf{p}} + mc^2}  \\
( -(\omega_{\mathbf{p}} +mc^2) \sqrt{\omega_{\mathbf{p}} - mc^2}
 +  (\omega_{\mathbf{p}} + mc^2)\sqrt{\omega_{\mathbf{p}} - mc^2} )
(\vec{\sigma} \cdot \frac{\mathbf{p}}{p})\\
\end{array} \right] \frac{\chi_{\sigma}}{\sqrt{2mc^2}}   \nonumber \\
&=& 0 \label{eq:dirac-eq}
\end{eqnarray}

\noindent This leads to the \emph{Dirac equation} \index{Dirac equation} for the field component $\psi^+ (\tilde{x})$

\begin{eqnarray}
 \left(\gamma^0 \frac{\partial}{\partial t}  +  c\vec{\gamma}
\frac{\partial}{\partial \mathbf{x}} -\frac{imc^2}{\hbar}
\right)\psi^+ (\tilde{x}) = 0 \label{eq:10.24x}
\end{eqnarray}

\noindent The same equation is satisfied by the component
$\psi^-(\tilde{x})$. So, the Dirac equation for the full field is

\begin{eqnarray}
 \left(\gamma^0 \frac{\partial}{\partial t}  +  c\vec{\gamma}
\frac{\partial}{\partial \mathbf{x}} -\frac{imc^2}{\hbar}\right)\psi
(x) = 0 \label{eq:10.24}
\end{eqnarray}

\noindent The  equation conjugate to (\ref{eq:dirac-eq}) is

\begin{eqnarray}
0&=& u^{\dag}( \mathbf{p}, \sigma)  \left((\gamma^0)^{\dag}
\omega_{\mathbf{p}}
 -  c(\vec{\gamma})^{\dag} \cdot \mathbf{p}
 -mc^2  \right)   = u^{\dag}( \mathbf{p}, \sigma)  \left(\gamma^0 \omega_{\mathbf{p}}
 +  c\vec{\gamma} \cdot \mathbf{p}
 -mc^2  \right)   \nonumber\\
 &=& u^{\dag}( \mathbf{p}, \sigma) \gamma^0\gamma^0 \left(\gamma^0
\omega_{\mathbf{p}}
 +  c\vec{\gamma} \cdot \mathbf{p}
 -mc^2  \right)   = \overline{u}( \mathbf{p}, \sigma) \gamma^0 \left(\gamma^0
\omega_{\mathbf{p}}
 +  c\vec{\gamma} \cdot \mathbf{p}
 -mc^2  \right)   \nonumber\\
 &=& \overline{u}( \mathbf{p}, \sigma)  \left(\gamma^0
\omega_{\mathbf{p}}
 -  c\vec{\gamma} \cdot \mathbf{p}
 -mc^2  \right) \gamma^0  \label{eq:dirac-eq2}
\end{eqnarray}

\noindent Therefore, the equation satisfied by the conjugated field
is

\begin{eqnarray*}
  \frac{\partial}{\partial t}\psi ^{\dag}(\tilde{x})\gamma^0  -
c\frac{\partial}{\partial \mathbf{x}}\psi ^{\dag}(\tilde{x})
\vec{\gamma}^{\dag}
 +\frac{imc^2}{\hbar}\psi ^{\dag}(\tilde{x})  = 0
\end{eqnarray*}

\noindent or multiplying from the right by $\gamma^0$ and using
(\ref{eq:H.17a}), we obtain

\begin{eqnarray}
  \frac{\partial}{\partial t}\overline{\psi}(\tilde{x})\gamma^0  +
c\frac{\partial}{\partial \mathbf{x}}\overline{\psi}(\tilde{x})
\vec{\gamma}
 +\frac{imc^2}{\hbar} \overline{\psi}(\tilde{x})  = 0
\label{eq:10.25}
\end{eqnarray}

It should be emphasized that in our approach to QFT Dirac equation
appears as a rather unremarkable property of the electron-positron
quantum field $\psi(\tilde{x})$ . This equation does not play a
fundamental role assigned to it in many textbooks. Definitely, Dirac
equation cannot be regarded as a ``relativistic analog of the
Schr\"odinger equation for electrons''.\footnote{A point of view similar to ours is adopted also in textbook \cite{book}.} The
correct electron wave functions and corresponding relativistic
Schr\"odinger equations should be constructed by using Wigner-Dirac
theory of unitary representations of the Poincar\'e group. For free
electrons such derivations are performed in chapter \ref{ch:single}.
The relativistic analog of the Schr\"odinger equation for an
interacting electron-proton system is constructed in chapter \ref{sc:coulomb}.

In the slash notation (\ref{eq:slash}) the momentum-space Dirac
equations (\ref{eq:dirac-eq}) and (\ref{eq:dirac-eq2}) take compact
forms

\begin{eqnarray}
 (\cross{p} - mc^2) u(\mathbf{p}, \tau) &=& 0 \label{gamma-mu2} \\
\overline{u}(\mathbf{p}, \tau) (\cross{p} - mc^2) &=& 0 \label{gamma-mu1}
\end{eqnarray}

\noindent If we denote $\cross{k} \equiv \cross{p}' - \cross{p}$, then it follows from (\ref{gamma-mu2}) - (\ref{gamma-mu1})

\begin{eqnarray}
&\ & U^{\mu} (\mathbf{p}, \sigma; \mathbf{p}', \sigma') k_{\mu} = \overline{u}(\mathbf{p}, \sigma) \cross{k}  u(\mathbf{p}', \sigma') = \overline{u}(\mathbf{p}, \sigma) [\cross{p}' u(\mathbf{p}', \sigma')] - [\overline{u}(\mathbf{p}, \sigma) \cross{p}] u(\mathbf{p}', \sigma') \nonumber \\
&=& (mc^2 - mc^2) \overline{u}(\mathbf{p}, \sigma)   u(\mathbf{p}', \sigma') = 0 \label{eq:kk} \\
&\ & W^{\mu} (\mathbf{p}, \sigma; \mathbf{p}', \sigma') k_{\mu} = 0 \label{eq:kk2}
\end{eqnarray}

\noindent We will also need the \emph{Gordon identity}\footnote{See Problem 3.2 in \cite{Peskin}.} \index{Gordon identity}

\begin{eqnarray}
&\ & \overline{u}(\mathbf{p}, \sigma)(\gamma_{\kappa} \cross{k} -
\cross{k} \gamma_{\kappa} ) u(\mathbf{p}', \sigma') =
\overline{u}(\mathbf{p}, \sigma)(\gamma_{\kappa} (\cross{p}' -
\cross{p})- (\cross{p}' - \cross{p}) \gamma_{\kappa} ) u(\mathbf{p}',
\sigma') \nonumber \\
&=& \overline{u}(\mathbf{p}, \sigma)(\gamma_{\kappa} (mc^2 -\cross{p})
- (\cross{p}' -mc^2 ) \gamma_{\kappa} ) u(\mathbf{p}', \sigma') \nonumber \\
&=& \overline{u}(\mathbf{p}, \sigma)(2\gamma_{\kappa} mc^2
-\gamma_{\kappa}\cross{p} -\cross{p}' \gamma_{\kappa} ) u(\mathbf{p}', \sigma') \nonumber \\
&=& \overline{u}(\mathbf{p}, \sigma)(2\gamma_{\kappa} mc^2
+\cross{p}\gamma_{\kappa} - 2p_{\kappa} +\gamma_{\kappa}\cross{p}' -
2p'_{\kappa}) u(\mathbf{p}', \sigma') \nonumber \\
&=& \overline{u}(\mathbf{p}, \sigma)(2\gamma_{\kappa} mc^2
+mc^2\gamma_{\kappa} - 2p_{\kappa} + mc^2\gamma_{\kappa} -
2p'_{\kappa}) u(\mathbf{p}', \sigma') \nonumber \\
&=& \overline{u}(\mathbf{p}, \sigma)(4 mc^2\gamma_{\kappa}
 - 2p_{\kappa}  - 2p'_{\kappa}) u(\mathbf{p}', \sigma') \label{eq:gkk-kgk}
\end{eqnarray}

\section{Fermion propagator} \label{ss:fermion-prop}

Let us calculate the  \emph{electron propagator}, \index{electron
propagator} which is frequently used in Feynman-Dyson perturbation
theory

\begin{eqnarray*}
\mathcal{D}_{a b}(\tilde{x}_1, \tilde{x}_2) \equiv \langle 0
|T(\psi_{a} (\tilde{x}_1) \overline{\psi}_{b}(\tilde{x}_2)) | 0
\rangle
\end{eqnarray*}

\noindent if $t_1 > t_2$ we can omit the time ordering \index{time ordering} sign and use
(\ref{eq:J.43a})

\begin{eqnarray*}
\mathcal{D}_{a b}(\tilde{x}_1, \tilde{x}_2) &=& \langle 0 |\psi_{a}
(\tilde{x}_1) \overline{\psi}_{b}(\tilde{x}_2) | 0 \rangle
\propto \langle 0 |(a + b^{\dag})(a^{\dag} + b)| 0 \rangle
\propto \langle 0 |a a^{\dag} | 0 \rangle \\
&=& \langle 0 | \left(\int \frac{d\mathbf{p}}{(2\pi \hbar)^{3/2}}
\sqrt{\frac{mc^2}{\omega_{\mathbf{p}}}} \sum_{ \sigma}
e^{-\frac{i}{\hbar}\tilde{p} \cdot \tilde{x}_1}u_{a}(\mathbf{p},
\sigma)
 a _{\mathbf{p},\sigma} \right) \times \\
 &\ & \left(\int
\frac{d\mathbf{q}}{(2\pi \hbar)^{3/2}}
\sqrt{\frac{mc^2}{\omega_{\mathbf{q}}}} \sum_{ \tau}
e^{\frac{i}{\hbar}\tilde{q} \cdot \tilde{x}_2}u^{\dag}_{b
}(\mathbf{q}, \tau) a^{\dag} _{\mathbf{q},\tau} \right) \gamma^0
 | 0 \rangle \\
 &=&  \int \frac{d\mathbf{p}d\mathbf{q}}{(2\pi \hbar)^{3}}
\frac{mc^2}{\sqrt{\omega_{\mathbf{p}}\omega_{\mathbf{q}}}} \sum_{
\sigma \tau} e^{-\frac{i}{\hbar}\tilde{p} \cdot
\tilde{x}_1}u_{a}(\mathbf{p}, \sigma)  e^{\frac{i}{\hbar}\tilde{q}
\cdot \tilde{x}_2}\overline{u}_{b }(\mathbf{q}, \tau)
\delta(\mathbf{p-q}) \delta_{\sigma\tau}
  \\
  &=&  \int \frac{d\mathbf{p}}{(2\pi \hbar)^{3}}
\frac{mc^2}{\omega_{\mathbf{p}}} e^{\frac{i}{\hbar}\tilde{p}
\cdot(\tilde{x}_2 - \tilde{x}_1)} \sum_{ \sigma}u_{a}(\mathbf{p},
\sigma) \overline{u}_{b
}(\mathbf{p}, \sigma) \\
&=& \int \frac{d\mathbf{p}}{(2\pi \hbar)^{3}}
e^{\frac{i}{\hbar}(\omega_{\mathbf{p}}(t_2 - t_1) -
\mathbf{p}(\mathbf{x}_2 -
\mathbf{x}_1))}\frac{1}{2\omega_{\mathbf{p}}} \left(\gamma^0
\omega_{\mathbf{p}} - \vec{\gamma} \mathbf{p} c + mc^2
\right)_{a b}
\end{eqnarray*}

\noindent if $t_1 < t_2$ we use (\ref{eq:J.44a}) to obtain\footnote{Note that for the anticommuting fermion field the definition of the time ordered product involves a change of sign (compare with (\ref{eq:time-order}))

\begin{eqnarray}
T[\psi_{a}(\tilde{x}_1)\overline{\psi}_{b}(\tilde{x}_2)]  =  \left \{
\begin{array}{c}
\psi_{a}(\tilde{x}_1)\overline{\psi}_{b}(\tilde{x}_2), \mbox{  } if \mbox{  } t_1 > t_2 \\
-\overline{\psi}_{b}(\tilde{x}_2)\psi_{a}
(\tilde{x}_1), \mbox{  } if \mbox {  } t_1 < t_2
\end{array}\right.  \label{eq:J.79}
\end{eqnarray}
}

\begin{eqnarray*}
\mathcal{D}_{a b}(\tilde{x}_1, \tilde{x}_2) &=& -\langle 0 |
\overline{\psi}_{b}(\tilde{x}_2)\psi_{a}
(\tilde{x}_1) | 0 \rangle
\propto -\langle 0 |(a^{\dag} + b)(a + b^{\dag})| 0 \rangle
\propto -\langle 0 |b b^{\dag} | 0 \rangle \\
&=& -\langle 0 |(2\pi \hbar)^{-3/2} \int d\mathbf{p}
e^{-\frac{i}{\hbar}\tilde{p} \cdot \tilde{x}_2 }B_{b}(\mathbf{p})
2\pi \hbar)^{-3/2} \int d\mathbf{q}
  e^{\frac{i}{\hbar}\tilde{q} \cdot \tilde{x}_1 }B^{\dag}_{
a}(\mathbf{q})  | 0 \rangle \\
&=& -(2\pi \hbar)^{-3} \int d\mathbf{p}
\frac{mc^2}{\omega_{\mathbf{p}}}e^{\frac{i}{\hbar}\tilde{p} \cdot
(\tilde{x}_1 -\tilde{x}_2) }\sum_{\sigma} v_b(\mathbf{p}, \sigma)\overline{v}_a(\mathbf{p}, \sigma)  \\
&=& -\int \frac{d\mathbf{p}}{(2\pi \hbar)^{3}}
e^{\frac{i}{\hbar}(\omega_{\mathbf{p}}(t_1 - t_2) -
\mathbf{p}(\mathbf{x}_1 -
\mathbf{x}_2))}\frac{1}{2\omega_{\mathbf{p}}} \left(\gamma^0
\omega_{\mathbf{p}} - \vec{\gamma} \mathbf{p} c - mc^2
\right)_{a b}
\end{eqnarray*}

\noindent The sum of these two terms gives

\begin{eqnarray}
\mathcal{D}_{a b}(\tilde{x}_1, \tilde{x}_2) &=& \theta(t_1 - t_2) \int
\frac{d\mathbf{p}}{(2\pi \hbar)^{3}}
e^{\frac{i}{\hbar}(\omega_{\mathbf{p}}(t_2 - t_1) -
\mathbf{p}(\mathbf{x}_2 -
\mathbf{x}_1))}\frac{1}{2\omega_{\mathbf{p}}}P_{a b}
(\mathbf{p}, \omega_{\mathbf{p}}) \nonumber \\
&+& \theta(t_2 - t_1) \int \frac{d\mathbf{p}}{(2\pi \hbar)^{3}}
e^{\frac{i}{\hbar}(\omega_{\mathbf{p}}(t_1 - t_2) -
\mathbf{p}(\mathbf{x}_1 -
\mathbf{x}_2))}\frac{1}{2\omega_{\mathbf{p}}}P_{a b}
(-\mathbf{p}, -\omega_{\mathbf{p}}) \nonumber \\
\label{eq:Dab}
\end{eqnarray}

\noindent where  we denoted

\begin{eqnarray*}
P_{a b} (\mathbf{p}, \omega_{\mathbf{p}}) &=&
\left(\gamma^0 \omega_{\mathbf{p}} - \vec{\gamma} \mathbf{p} c +
mc^2 \right)_{a b}
\end{eqnarray*}

\noindent and $\theta(t)$ is the step function defined in
(\ref{eq:theta-function}). Our next goal is to rewrite equation
(\ref{eq:Dab}) so that integration goes by 4 independent components
of the 4-vector of momentum $(p_0, p_x, p_y, p_z)$. We use an
integral representation (\ref{eq:step}) for the step function
 to obtain

\begin{eqnarray}
&\ &\mathcal{D}_{a b}(\tilde{x}_1, \tilde{x}_2) \nonumber \\
&=& -\frac{1}{2 \pi i} \int \frac{d\mathbf{p}}{(2\pi \hbar)^{3}}
\int ds \frac{e^{-is(t_1 - t_2)}}{s+ i \epsilon}
e^{-\frac{i}{\hbar}(\omega_{\mathbf{p}}(t_1 - t_2) -
\mathbf{p}(\mathbf{x}_1 -
\mathbf{x}_2))}\frac{1}{2\omega_{\mathbf{p}}}P_{a b}
(\mathbf{p}, \omega_{\mathbf{p}}) \nonumber \\
&-& \frac{1}{2 \pi i} \int \frac{d\mathbf{p}}{(2\pi \hbar)^{3}} \int
ds \frac{e^{is(t_1 - t_2)}}{s+ i \epsilon}
e^{\frac{i}{\hbar}(\omega_{\mathbf{p}}(t_1 - t_2) -
\mathbf{p}(\mathbf{x}_1 -
\mathbf{x}_2))}\frac{1}{2\omega_{\mathbf{p}}}P_{a b}(-\mathbf{p}, -\omega_{\mathbf{p}}) \nonumber \\
&=& -\frac{1}{2 \pi i} \int \frac{d\mathbf{p}}{(2\pi \hbar)^{3}}
\int ds \frac{1}{s+ i
\epsilon} \frac{1}{2\omega_{\mathbf{p}}} \times \nonumber \\
&\mbox{ } & \Bigl[e^{-\frac{i}{\hbar}((\omega_{\mathbf{p} + \hbar
s})(t_1 - t_2) - \mathbf{p}(\mathbf{x}_1 - \mathbf{x}_2))}P_{a b}
(\mathbf{p}, \omega_{\mathbf{p}}) +
e^{\frac{i}{\hbar}((\omega_{\mathbf{p}} + \hbar s)(t_1 - t_2) -
\mathbf{p}(\mathbf{x}_1 -
\mathbf{x}_2))}P_{a b}(-\mathbf{p}, -\omega_{\mathbf{p}})\Bigl] \nonumber \\
&=& -\frac{1}{2 \pi i} \int \frac{d\mathbf{p}}{(2\pi \hbar)^{3}}
\int dp_0 \frac{1}{p_0 - \omega_{\mathbf{p}} + i
\epsilon} \frac{1}{2\omega_{\mathbf{p}}} \times \nonumber \\
&\mbox{ } & \left[e^{-\frac{i}{\hbar}(p_0(t_1 - t_2) -
\mathbf{p}(\mathbf{x}_1 - \mathbf{x}_2))}P_{a b} (\mathbf{p},
\omega_{\mathbf{p}}) + e^{\frac{i}{\hbar}(p_0(t_1 - t_2) -
\mathbf{p}(\mathbf{x}_1 - \mathbf{x}_2))}P_{a b}(-\mathbf{p},
-\omega_{\mathbf{p}}) \right]
\nonumber \\
&=& -\frac{1}{2 \pi i} \int \frac{d\mathbf{p}}{(2\pi \hbar)^{3}}
\int dp_0 \frac{1}{p_0 - \omega_{\mathbf{p}} + i
\epsilon} \frac{1}{2\omega_{\mathbf{p}}} \times \nonumber \\
&\mbox{ } & \left[e^{-\frac{i}{\hbar}(p_0(t_1 - t_2) -
\mathbf{p}(\mathbf{x}_1 - \mathbf{x}_2))}P_{a b} (\mathbf{p}, p_0) +
e^{-\frac{i}{\hbar}(-p_0(t_1 - t_2) + \mathbf{p}(\mathbf{x}_1
- \mathbf{x}_2))}P_{a b}(-\mathbf{p}, -p_0) \right] \nonumber \\
&=& -\frac{1}{2 \pi i} \int \frac{d\mathbf{p}}{(2\pi \hbar)^{3}}
\int dp_0 e^{-\frac{i}{\hbar}(p_0(t_1 - t_2) -
\mathbf{p}(\mathbf{x}_1 -
\mathbf{x}_2))} \frac{1}{2\omega_{\mathbf{p}}}  \left[ \frac{P_{a b} (\mathbf{p}, p_0)}{p_0 - \omega_{\mathbf{p}} + i \epsilon}
 + \frac{P_{a b}(\mathbf{p}, p_0)}{-p_0 - \omega_{\mathbf{p}} + i
\epsilon} \right]
\nonumber \\
&=&  \frac{1}{2 \pi i}\int \frac{d\mathbf{p}}{(2\pi \hbar)^{3}} \int
dp_0 e^{-\frac{i}{\hbar}(p_0(t_1 - t_2) - \mathbf{p}(\mathbf{x}_1 -
\mathbf{x}_2))} \frac{1}{2\omega_{\mathbf{p}}} P_{a b}(\mathbf{p},
p_0)
 \frac{2 \omega_{\mathbf{p}}}{p_0^2 - \omega^2_{\mathbf{p}} + i
 \epsilon} \nonumber \\
 &=& \frac{1}{2 \pi i(2\pi \hbar)^{3}} \int d^4p
e^{-\frac{i}{\hbar}(\tilde{p}\cdot \tilde{x})}
 \frac{P_{a b}(\mathbf{p}, p_0)}{p_0^2 - c^2\mathbf{p}^2 - m^2c^4 + i
 \epsilon} \nonumber \\
 &=&   \frac{1}{2 \pi i(2\pi \hbar)^{3}} \int d^4p
e^{-\frac{i}{\hbar}(\tilde{p}\cdot \tilde{x})}
 \frac{(\gamma^0
p_0 - \vec{\gamma} \mathbf{p} c + mc^2 )_{a b}}{\tilde{p}^2c^2 -
m^2c^4 + i \epsilon} \nonumber \\
&=&   \frac{1}{2 \pi i(2\pi \hbar)^{3}} \int d^4p
e^{-\frac{i}{\hbar}(\tilde{p}\cdot \tilde{x})}
 \frac{(\cross{p} + mc^2 )_{a b}}{\tilde{p}^2c^2 -
m^2c^4 + i \epsilon} \label{eq:elec-prop}
\end{eqnarray}

\chapter{Quantum field for photons} \label{sc:photons}

\section{Construction of the photon's quantum field}
\label{ss:construction}

Let us now  construct a quantum field based on creation ($c^{\dag}_{\mathbf{p}, \tau}$) and annihilation ($c_{\mathbf{p}, \tau}$) operators for photons.\footnote{See subsection \ref{ss:discrete-momentum-phot}.} Our goal is to satisfy conditions listed in \textbf{Step 1.} in subsection \ref{ss:weinberg}.

We will postulate that Lorentz transformations (\ref{eq:10.1}) of the photon
field $A_{\mu}(\tilde{x})$ are associated with the 4-dimensional representation of the Lorentz group from subsection
\ref{ss:4-dim-rep}

\begin{eqnarray}
U_0(\Lambda; \tilde{a}) A_{\mu}(\tilde{x}) U_0^{-1}(\Lambda; \tilde{a}) = \sum_{\nu}\Lambda_{\mu \nu}^{-1} A^{\nu}(\Lambda(\tilde{x} + \tilde{a})) \label{eq:9.1x}
\end{eqnarray}

\noindent with indices $\mu$ and $\nu$ taking values 0,1,2,3. Then we attempt to define a 4-component quantum field for photons as\footnote{In equation (\ref{eq:10.35a}) we will see that, actually, this field does not satisfy our requirement (\ref{eq:9.1x}) for boosts.}

\begin{eqnarray}
&\ &A_{\mu}(\tilde{x}) \equiv   A_{\mu}(\mathbf{x},t) \nonumber \\
 &=& \frac{\hbar \sqrt{c} }{(2\pi \hbar)^{3/2}} \int \frac{d\mathbf{p}}{\sqrt{2p}}
\sum_{ \tau}
 \left[e^{-\frac{i}{\hbar}\tilde{p} \cdot \tilde{x}}e  _{\mu }
(\mathbf{p}, \tau)  c _{\mathbf{p},\tau} + e^{ \frac{i}{\hbar}
\tilde{p}\cdot \tilde{x}}e^*_{\mu }(\mathbf{p}, \tau)
 c^{\dag}_{\mathbf{p},\tau} \right]
\label{eq:10.26}
\end{eqnarray}

\noindent where $\tilde{p} \cdot \tilde{x} \equiv
\mathbf{p}\mathbf{x}- cpt$ and the coefficient functions $e  _{\mu } (\mathbf{p},
\tau)$ should be chosen such that (\ref{eq:9.1x}) is satisfied. Following the recipe from subsection \ref{ss:factors}, we first choose the value of the coefficient function at the standard momentum $\mathbf{k} =
(0,0,1)$\footnote{see equation (\ref{eq:7.57})} appropriate for massless photons

\begin{eqnarray}
e_{\mu  }(\mathbf{k}, \tau) =
 \frac{1}{\sqrt{2}}  \left[ \begin{array}{c}
0  \\
1 \\
i \tau\\
0
\end{array} \right] \label{eq:e-mu}
\end{eqnarray}

\noindent For all other photon momenta $\mathbf{p}$ we define\footnote{This is similar in spirit to  the massive
case (\ref{eq:u-alpha}) - (\ref{eq:v-beta})}

\begin{eqnarray}
e( \mathbf{p}, \tau) &=& \lambda_{\mathbf{p}}   e(\mathbf{k}, \tau)
\label{eq:10.27}\\
e^{\dag}(\mathbf{p}, \tau) &=&  e^{\dag}(\mathbf{k}, \tau)
\lambda_{\mathbf{p}} \label{eq:10.28}
\end{eqnarray}

\noindent where $ \lambda_{\mathbf{p}}$ is a boost transformation
which takes the particle from the standard momentum \index{standard
momentum} $\mathbf{k}$ to an arbitrary $\mathbf{p}$.

\begin{eqnarray}
 \lambda_{\mathbf{p}} = R_{\mathbf{p}} B_{\mathbf{p}} \label{eq:10.27a}
\end{eqnarray}

\noindent where  $B_{\mathbf{p}}$ is a boost along the $z$-axis and
$R_{\mathbf{p}}$ is a pure rotation, as in equation (\ref{eq:7.60}).

\section{Explicit formula for $e_{\mu} ( \mathbf{p}, \tau)$}
\label{ss:explicite2}

Note that the boost $B_{\mathbf{p}}$ in equation
(\ref{eq:7.60a}) has no effect on the 4-vector (\ref{eq:e-mu}). The 0-th component of this vector is not affected by rotations
$R_{\mathbf{p}}$ as well. Therefore, we conclude that for all $\mathbf{p}$ and $\mathbf{x}$

\begin{eqnarray}
e_{0 } ( \mathbf{p}, \tau) &=&0 \label{eq:10.33} \\
A_0 (\mathbf{x}, t) &=& 0 \label{eq:K13.a}
\end{eqnarray}

\noindent Let us now find the 3-vector part of $e_{\mu } (
\mathbf{p}, \tau)$, which we denote by $\mathbf{e} ( \mathbf{p}, \tau)$. From (\ref{eq:e-mu}), (\ref{eq:10.27}), (\ref{eq:10.27a}), and
(\ref{eq:r-vec-phi}) we obtain

\begin{eqnarray*}
&\ & \sqrt{2} \mathbf{e} ( \mathbf{p}, \tau) \\
&=&   \left[
\begin{array}{ccc}
  \cos \phi + n_x^2 (1-\cos \phi) & n_x n_y (1-\cos \phi) - n_z \sin
\phi & n_xn_z (1-\cos \phi) + n_y \sin \phi  \\
 n_x n_y (1-\cos \phi) + n_z \sin \phi & \cos \phi + n_y^2 (1 - \cos
\phi) & n_y n_z(1- \cos \phi) -n_x \sin \phi  \\
  n_xn_z (1-\cos \phi) -n_y \sin \phi & n_yn_z (1-\cos \phi) +n_x \sin
\phi & \cos \phi + n_z^2 (1-\cos \phi)
\end{array} \right] \times \\
&\ &    \left[ \begin{array}{c}
1 \\
i \tau\\
0
\end{array} \right] \\
&=&   \left[
\begin{array}{ccc}
  \frac{p_z}{p} + \frac{p_y^2(p-p_z)}{p(p_x^2+p_y^2)} & -\frac{p_xp_y(p-p_z)}{p(p_x^2+p_y^2)}
  & \frac{p_x}{p}  \\
 -\frac{p_xp_y(p-p_z)}{p(p_x^2+p_y^2)} & \frac{p_z}{p} + \frac{p_x^2(p-p_z)}{p(p_x^2+p_y^2)} &
 \frac{p_y}{p}   \\
  -\frac{p_x}{p}  & -\frac{p_y}{p}  & \frac{p_z}{p}
\end{array} \right]
   \left[ \begin{array}{c}
1 \\
i \tau\\
0
\end{array} \right] \\
&=&   \left[
\begin{array}{ccc}
   \frac{p_z p_x^2 + pp_y^2}{p(p_x^2+p_y^2)} & -\frac{p_xp_y(p-p_z)}{p(p_x^2+p_y^2)}
  & \frac{p_x}{p}  \\
 -\frac{p_xp_y(p-p_z)}{p(p_x^2+p_y^2)} &  \frac{p_zp_y^2 + pp_x^2}{p(p_x^2+p_y^2)} &
 \frac{p_y}{p}   \\
  -\frac{p_x}{p}  & -\frac{p_y}{p}  & \frac{p_z}{p}
\end{array} \right]
   \left[ \begin{array}{c}
1 \\
i \tau\\
0
\end{array} \right] \\
&=& \frac{1}{p(p_x^2+p_y^2)}
   \left[ \begin{array}{c}
p_z p_x^2 + pp_y^2 - i \tau p_xp_y(p-p_z)\\
-p_xp_y(p-p_z) + i \tau (p_z p_y^2 + pp_x^2)\\
-p_x(p_x^2+p_y^2) - i \tau p_y (p_x^2+p_y^2)
\end{array} \right]
\end{eqnarray*}

\noindent Therefore

\begin{eqnarray}
e ( \mathbf{p}, \tau) &=& \frac{1}{\sqrt{2}p(p_x^2+p_y^2)}
   \left[ \begin{array}{c}
   0 \\
p_z p_x^2 + pp_y^2 - i \tau p_xp_y(p-p_z)\\
-p_xp_y(p-p_z) + i \tau (p_z p_y^2 + pp_x^2)\\
-p_x(p_x^2+p_y^2) - i \tau p_y (p_x^2+p_y^2)
\end{array} \right] \label{eq:eptau}
\end{eqnarray}

\noindent One can easily see that $\mathbf{e} (
\mathbf{p}, \tau)$ is orthogonal to the momentum vector
$\mathbf{p}=(p_x, p_y, p_z)$ and that

\begin{eqnarray}
e_{\mu} ( \mathbf{p}, \tau)p^{\mu} &=& 0 \label{eq:emupmu}
\end{eqnarray}

\section{Useful commutator}
\label{ss:photon-propagator2}

For our derivations in subsection \ref{ss:2-nd-order} we need the following expression

\begin{eqnarray}
  C_{\alpha \beta}(\mathbf{p}) &=& \frac{\hbar \sqrt{c} }{\sqrt{2p}}
 \gamma^{\mu}_{ \alpha \beta}
\sum_{\tau} e_{\mu}(\mathbf{p},\tau)
  c _{\mathbf{p}, \tau} \label{eq:K.6}
\end{eqnarray}

\noindent and the  commutator

\begin{eqnarray}
\left[C^{\dag}_{\alpha \beta} (\mathbf{p}), C_{\gamma
\delta}(\mathbf{p}') \right] &=& \frac{\hbar^2 c}{2\sqrt{pp'}}
\sum_{\tau \tau'} \gamma^{\mu}_{ \alpha \beta}
 \gamma^{\nu}_{\gamma \delta} e^{\dag}_{\mu} (\mathbf{p}, \tau )e_{\nu}
(\mathbf{p}', \tau' )
[c^{\dag}_{\mathbf{p}, \tau}, c_{\mathbf{p}', \tau'}] \nonumber \\
&=& -\frac{\hbar^2 c}{2p} \sum_{\tau \tau'}\gamma^{\mu}_{ \alpha
\beta}
 \gamma^{\nu}_{\gamma \delta} e^{\dag}_{\mu} (\mathbf{p}, \tau )e_{\nu}
(\mathbf{p}', \tau' )
\delta(\mathbf{p}-\mathbf{p}') \delta_{\tau, \tau'} \nonumber \\
&=& -\frac{\hbar^2 c}{2p} \sum_{\tau } \gamma^{\mu}_{ \alpha \beta}
 \gamma^{\nu}_{\gamma \delta} e^{\dag}_{\mu} (\mathbf{p}, \tau )e_{\nu}
(\mathbf{p}, \tau )
\delta(\mathbf{p}-\mathbf{p}')  \nonumber \\
&=& -\frac{\hbar^2 c}{2p}  \gamma^{\mu}_{ \alpha \beta}
 \gamma^{\nu}_{\gamma \delta} h_{\mu \nu}(\mathbf{p})
\delta(\mathbf{p}-\mathbf{p}') \label{eq:K.5a}
\end{eqnarray}

\noindent where

\begin{eqnarray*}
h_{\mu \nu} (\mathbf{p}) &\equiv&\sum_{\tau} e _{\mu }(\mathbf{p},
\tau) e^{\dag} _{\nu }(\mathbf{p}, \tau)
\end{eqnarray*}

\noindent is a sum frequently appearing in calculations. First we calculate this sum at
the standard momentum $\mathbf{k}= (0,0,1)$ with the help of (\ref{eq:e-mu})

\begin{eqnarray*}
h_{\mu \nu} (\mathbf{k})&=&
 \frac{1}{2}  \left[ \begin{array}{c}
0  \\
1 \\
i\\
0
\end{array} \right] \left[ \begin{array}{cccc}
0 & 1 & -i & 0
\end{array} \right] + \frac{1}{2}\left[ \begin{array}{c}
0  \\
1 \\
-i\\
0
\end{array} \right] \left[ \begin{array}{cccc}
0 & 1 & i & 0
\end{array} \right]  \\
&=& \frac{1}{2} \left[ \begin{array}{cccc}
0 & 0 & 0 & 0 \\
0& 1 & -i & 0 \\
0 & i & 1 & 0\\
0 & 0 & 0 & 0
\end{array} \right] + \frac{1}{2} \left[ \begin{array}{cccc}
0 & 0 & 0 & 0 \\
0& 1 & i & 0 \\
0 & -i & 1 & 0\\
0 & 0 & 0 & 0
\end{array} \right] \\
&=&
  \left[ \begin{array}{cccc}
0 & 0 & 0 & 0 \\
0 & 1 & 0 & 0 \\
0 & 0 & 1 & 0\\
0 & 0 & 0 & 0
\end{array} \right]
\end{eqnarray*}

\noindent which can be also expressed in terms of components of
the standard vector $\mathbf{k} $

\begin{eqnarray*}
h_{0 \mu} (\mathbf{k}) &=& h_{ \mu 0} (\mathbf{k}) = 0 \\
h_{ij} (\mathbf{k}) &=&  \delta_{ij} - \frac{k_i k_j}{k^2}
\end{eqnarray*}

\noindent At arbitrary momentum $\mathbf{p}$ we use formulas
(\ref{eq:10.27}), (\ref{eq:10.28}), and (\ref{eq:10.27a})

\begin{eqnarray*}
h_{\mu \nu} (\mathbf{p}) &=&\sum_{\tau} e _{\mu }(\mathbf{p}, \tau)
e^{\dag} _{\nu
}(\mathbf{p}, \tau) \label{eq:hmunu2} \\
&=&
 R_{\mathbf{p}} B_{\mathbf{p}} \left[ \begin{array}{cccc}
0 & 0 & 0 & 0 \\
0 & 1 & 0 & 0 \\
0 & 0 & 1 & 0\\
0 & 0 & 0 & 0
\end{array} \right] B^{-1}_{\mathbf{p}} R^{-1}_{\mathbf{p}} \nonumber \\
&=&  R_{\mathbf{p}}  \left[ \begin{array}{cccc}
0 & 0 & 0 & 0 \\
0 & 1 & 0 & 0 \\
0 & 0 & 1 & 0\\
0 & 0 & 0 & 0
\end{array} \right]  R^{-1}_{\mathbf{p}} \nonumber
\end{eqnarray*}

\noindent   It then follows  that $h_{0
\mu}(\mathbf{p}) = h_{\mu 0} (\mathbf{p}) = 0$, that the $3
\times 3$ submatrix is

\begin{eqnarray}
h_{i j} (\mathbf{p}) &=&
 R_{\mathbf{p}}  \left[\delta_{ij} - \frac{k_i k_j}{k^2} \right]
R^{-1}_{\mathbf{p}} = \delta_{ij} - \frac{p_i p_j}{p^2}
\label{eq:10.29a}
\end{eqnarray}

\noindent and the final formula for $h_{\mu \nu} (\mathbf{p})$ is

\begin{eqnarray}
 h_{\mu \nu} (\mathbf{p})&=&
  \left[ \begin{array}{cccc}
0 & 0 & 0 & 0 \\
0 & 1- \frac{p_x^2}{p^2} & - \frac{p_xp_y}{p^2} & - \frac{p_xp_z}{p^2} \\
0 & - \frac{p_xp_y}{p^2} & 1- \frac{p_y^2}{p^2} & - \frac{p_yp_z}{p^2}\\
0 & - \frac{p_xp_z}{p^2} & - \frac{p_zp_y}{p^2} & 1-
\frac{p_z^2}{p^2}
\end{array} \right]
\label{eq:10.29b}
\end{eqnarray}

\section{Equal time commutator of photon fields}
\label{ss:eq-time-comms}

The photon quantum field  (\ref{eq:10.26}) commutes with itself at
space-like intervals ($\mathbf{x} \neq \mathbf{y}$), as required in
equation (\ref{eq:10.4})

\begin{eqnarray}
&\mbox{ } & [ A_{\mu}(\mathbf{x},0), A_{\nu}^{\dag} (\mathbf{y},0)]
\nonumber \\
& =& \frac{\hbar^2 c}{2(2\pi \hbar)^{3}} \int \frac{d\mathbf{p}
d\mathbf{p}'}{\sqrt{p p'}} \sum_{ \tau \tau'}
 \Bigl[\left(e^{-\frac{i}{\hbar}\mathbf{p}\mathbf{x}}e  _{\mu }
(\mathbf{p}, \tau)  c _{\mathbf{p},\tau} +
e^{\frac{i}{\hbar}\mathbf{p}\mathbf{x}}e^*_{\mu }(\mathbf{p}, \tau)
 c^{\dag}_{\mathbf{p},\tau} \right), \nonumber \\
&\mbox { }& \left(e^{ \frac{i}{\hbar}\mathbf{p}'\mathbf{y}}e^*
_{\nu } (\mathbf{p}', \tau')  c^{\dag} _{\mathbf{p}',\tau'} +
e^{-\frac{i}{\hbar}\mathbf{p}'\mathbf{y}}e_{\nu }(\mathbf{p}',
\tau')
 c_{\mathbf{p}',\tau'} \right) \Bigl] \nonumber\\
& =& \frac{\hbar^2 c}{2(2\pi \hbar)^{3}} \int \frac{d\mathbf{p}
d\mathbf{p}'}{\sqrt{pp'}} \sum_{ \tau \tau'}
 \Bigl(e^{-\frac{i}{\hbar}\mathbf{p}\mathbf{x}}e^{
\frac{i}{\hbar}\mathbf{p}'\mathbf{y}} e  _{\mu } (\mathbf{p}, \tau)
e^{\dag}  _{\nu } (\mathbf{p}', \tau')[ c _{\mathbf{p},\tau},
c^{\dag}
_{\mathbf{p}',\tau'} ] \nonumber \\
&\ &+
e^{\frac{i}{\hbar}\mathbf{p}\mathbf{x}}e^{-\frac{i}{\hbar}\mathbf{p}'\mathbf{x}}
e^*_{\mu }(\mathbf{p}, \tau) e^{* \dag}_{\nu }(\mathbf{p}', \tau')
 [c^{\dag}_{\mathbf{p},\tau} ,
 c_{\mathbf{p}',\tau'} ] \Bigr)\nonumber \\
& =& \frac{\hbar^2 c}{2(2\pi \hbar)^{3}} \int \frac{d\mathbf{p}
d\mathbf{p}'}{p}
\delta(\mathbf{p} - \mathbf{p}') \times \nonumber \\
&\mbox { }& \sum_{ \tau \tau'} \delta_{\tau, \tau'}
 \left(e^{-\frac{i}{\hbar}\mathbf{p}(\mathbf{x-y})}e  _{\mu}
(\mathbf{p}, \tau) e^{\dag}  _{\nu } (\mathbf{p}', \tau') -
e^{\frac{i}{\hbar}\mathbf{p}(\mathbf{x-y})}e^*_{\mu }(\mathbf{p},
\tau) e_{\nu }^{* \dag}(\mathbf{p}', \tau') \right)\nonumber
 \\
& =& \frac{\hbar^2 c}{2(2\pi \hbar)^{3}} \int \frac{d\mathbf{p}}{p}
 \sum_{
\tau }
 \left(e^{-\frac{i}{\hbar}\mathbf{p}(\mathbf{x-y})}e  _{\mu }
(\mathbf{p}, \tau) e^{\dag}  _{\nu } (\mathbf{p}, \tau) -
e^{\frac{i}{\hbar}\mathbf{p}(\mathbf{x-y})}e^*_{\mu }(\mathbf{p},
\tau) e_{\nu }^{*\dag}(\mathbf{p}, \tau) \right)\nonumber
 \\
& =& \frac{\hbar^2 c}{2(2\pi \hbar)^{3}} \int \frac{d\mathbf{p}}{p}
 \left(e^{-\frac{i}{\hbar}\mathbf{p}(\mathbf{x-y})} -
e^{\frac{i}{\hbar}\mathbf{p}(\mathbf{x-y})} \right)h  _{\mu \nu}
(\mathbf{p})
\nonumber \\
& =& - \frac{i \hbar^2 c}{2(2\pi \hbar)^{3}} \int
\frac{d\mathbf{p}}{p} \sin(\mathbf{p}(\mathbf{x-y})) h _{\mu \nu}
(\mathbf{p})
\label{eq:10.29} \\
&=& 0 \nonumber
\end{eqnarray}

\noindent because the integrand in (\ref{eq:10.29}) is an odd
function of $\mathbf{p}$.

\section{Photon propagator}
\label{ss:photon-propagator}

Next we need to calculate the photon propagator. We use
the integral representation (\ref{eq:step}) of the step function
 to write

\begin{eqnarray*}
&\mbox{ }& \langle 0| T[A_{\mu}(\tilde{x}_1) A_{\nu}(\tilde{x}_2)] | 0 \rangle  \\
&=& \hbar^2 c\int \frac{ d\mathbf{p}}{2 (2 \pi \hbar)^3 p} h_{\mu
\nu} (\mathbf{p}) \left[e^{\frac{i}{\hbar} \tilde{p} \cdot
(\tilde{x}_1-\tilde{x}_2)} \theta (t_1 - t_2) +
e^{\frac{i}{\hbar} \tilde{p} \cdot (\tilde{x}_2 - \tilde{x}_1)} \theta (t_2 - t_1)\right] \\
&=& -\frac{\hbar^2 c}{2 \pi i}\int \limits_{- \infty}^{\infty} ds
\int \frac{d\mathbf{p}}{2 (2 \pi \hbar)^3 p} h_{\mu \nu}
(\mathbf{p}) \left[e^{\frac{i}{\hbar} \tilde{p} \cdot (\tilde{x}_1 -
\tilde{x}_2)} \frac{e^{-is(t_1 - t_2)}}{s+ i \epsilon} +
e^{\frac{i}{\hbar} \tilde{p} \cdot (\tilde{x}_2-\tilde{x}_1)}
\frac{e^{is(t_1-t_2)}}{s+ i \epsilon} \right] \\
&=& -\frac{\hbar^2 c}{2 \pi i}\int \limits_{- \infty}^{\infty} ds
\int \frac{ d\mathbf{p}}{2 (2 \pi
\hbar)^3 p} h_{\mu \nu} (\mathbf{p}) \frac{1}{s+ i \epsilon} \times \\
&\mbox{ }& \left[e^{\frac{i}{\hbar} (cp  (t_1 - t_2) -
\mathbf{p} (\mathbf{x}_1 - \mathbf{x}_2))} e^{-is(t_1 - t_2)} +
e^{\frac{i}{\hbar} (-cp  (t_1 - t_2)
+ \mathbf{p}(\mathbf{x}_1 - \mathbf{x}_2))} e^{is(t_1 - t_2)} \right] \\
&=& -\frac{\hbar^2 c}{2 \pi i}\int \limits_{- \infty}^{\infty} ds
\int \frac{ d\mathbf{p}}{2 (2 \pi \hbar)^3 p} h_{\mu \nu}
(\mathbf{p}) e^{-\frac{i}{\hbar} \mathbf{p} (\mathbf{x}_1 -
\mathbf{x}_2)} \left[\frac{e^{\frac{i}{\hbar} (cp - \hbar s) \cdot
(t_1 - t_2) } }{s+ i \epsilon} + \frac{e^{-\frac{i}{\hbar} (cp-
\hbar s) \cdot (t_1 - t_2) }}{s+ i \epsilon} \right]
\end{eqnarray*}

\noindent Next we change variables: in the first integral $p_0 = cp -
\hbar s$; in the second integral $p_0 = -cp + \hbar s$

\begin{eqnarray}
&\mbox{ }& \langle 0| T[A_{\mu}(\tilde{x}_1) A_{\nu}(\tilde{x}_2)] | 0 \rangle \nonumber  \\
 &=& -\frac{\hbar^2 c}{2 \pi i}\int
\limits_{- \infty}^{\infty} dp_0  \int \frac{ d\mathbf{p}}{2 (2 \pi
\hbar)^3 p} h_{\mu \nu} (\mathbf{p}) e^{-\frac{i}{\hbar} \mathbf{p}
(\mathbf{x}_1 - \mathbf{x}_2)} \left[\frac{ e^{\frac{i}{\hbar} p_0
(t_1 - t_2) } }{cp - p_0+ i \epsilon} +
\frac{ e^{\frac{i}{\hbar} p_0 (t_1 - t_2) }}{cp + p_0+ i \epsilon} \right] \nonumber \\
&=& \frac{\hbar^2 c^2}{2 \pi i}\int \limits_{- \infty}^{\infty} dp_0
\int \frac{ d\mathbf{p}}{(2 \pi \hbar)^3 } h_{\mu \nu} (\mathbf{p})
e^{\frac{i}{\hbar} p_0 (t_1 - t_2) } e^{-\frac{ic}{\hbar} \mathbf{p}
(\mathbf{x}_1 - \mathbf{x}_2)} \frac{1}{\tilde{p}^2+ i \epsilon} \nonumber \\
&=& \frac{\hbar^2 c^2}{2 \pi i} \int \frac{ d^4p}{(2 \pi \hbar)^3 }
h_{\mu \nu} (\mathbf{p})
 e^{\frac{i}{\hbar} \tilde{p}  (\tilde{x}_1 - \tilde{x}_2) }  \frac{1 }{\tilde{p}^2+ i \epsilon} \label{eq:K.17a}
\end{eqnarray}

\noindent where we denoted $d^4p \equiv dp_0 d\mathbf{p}$.

The matrix $h_{\mu \nu} (\mathbf{p})$ has been calculated in (\ref{eq:10.29b}). However, as explained in subsection \ref{ss:Fey}, it is more convenient to use the Feynman-Dyson approach where this matrix is replaced by the metric tensor
$h_{\mu \nu} (\mathbf{p})= g_{\mu \nu}$. Then  we obtain our final propagator formula

\begin{eqnarray}
 \langle 0| T[A_{\mu}(\tilde{x}_1) A_{\nu}(\tilde{x}_2)] | 0 \rangle &=& \frac{\hbar^2 c^2}{2 \pi i}\int
\frac{ d^4p}{(2 \pi \hbar)^3 }
 e^{\frac{i}{\hbar} \tilde{p}  (\tilde{x}_1 - \tilde{x}_2) }
  \frac{g_{\mu \nu} }{\tilde{p}^2+ i \epsilon}
 \label{eq:photon-prop}
\end{eqnarray}

\index{photon propagator}

\section{Poincar\'e transformations of the photon field}
\label{ss:lorentz-photon}

Now we need to determine transformations of the photon field with
respect to the non-interacting representation of the Poincar\'e
group. Note that we have defined coefficient functions $e _{\mu}(\mathbf{p},\tau)$ in subsection \ref{ss:explicite2} in the hope to achieve the transformation law (\ref{eq:9.1x}) for the photon field. This approach was successful in the case of electron-positron field in Appendix \ref{ss:transformations}. However, for massless photons the situation is more complicated. The actions of translations and rotations do agree with our
condition (\ref{eq:9.1x})

\begin{eqnarray}
 U_0 (R; 0) A_{0}(\mathbf{x},t) U_0^{-1} (R; 0)  &=& A_{0}(R
\mathbf{x},t) \nonumber \\
 U_0 (R; 0)  \mathbf{A}(\mathbf{x},t) U_0^{-1} (R; 0)&=&
R^{-1}\mathbf{A}(R \mathbf{x},t) \nonumber \\
 U_0 (1; \mathbf{r}, \tau)  A_{\mu}(\mathbf{x},t) U_0^{-1} (1; \mathbf{r},
\tau)
 &=& A_{\mu}(
\mathbf{x+r},t+ \tau) \label{eq:K18a}
\end{eqnarray}

\noindent However, transformations with respect to boosts disagree with our expectation \cite{Weinberg_1049}

\begin{eqnarray}
U_0 (\Lambda; 0)  A_{\mu}(\tilde{x}) U_0^{-1} (\Lambda; 0) &=&
\sum_{\nu } \Lambda^{-1}_{\mu \nu}  A^{\nu}(\Lambda \tilde{x}) \label{eq:ULA}
\end{eqnarray}

\noindent To demonstrate this disagreement we first use equations (\ref{eq:9.38}) and
(\ref{eq:9.39}) to write

\begin{eqnarray}
&\mbox{ } &  U_0 (\Lambda; 0)  A_{\mu}(\tilde{x}) U_0^{-1}
(\Lambda; 0)  \nonumber \\
&=& \frac{\hbar \sqrt{c}}{(2\pi \hbar)^{3/2} }\int
\frac{d\mathbf{p}} {\sqrt{2p}} \sum_{ \tau}
 \Bigl(e^{-\frac{i}{\hbar}\tilde{p} \cdot \tilde{x}} e _{\mu}(\mathbf{p},\tau)
 U_0 (\Lambda; 0) c _{\mathbf{p},\tau} U_0^{-1} (\Lambda; 0)
\nonumber \\
&\ & +e^{\frac{i}{\hbar}\tilde{p} \cdot
\tilde{x}}e^*_{\mu}(\mathbf{p},\tau) U_0 (\Lambda; 0) c^{\dag}
_{\mathbf{p},\tau} U_0^{-1} (\Lambda; 0) \Bigr)
\nonumber \\
 &=& \frac{\hbar \sqrt{c}}{(2\pi \hbar)^{3/2} } \int \frac{d\mathbf{p}}
{\sqrt{2p}}  \sqrt{\frac{|\Lambda \mathbf{p}|}{p}}\sum_{ \tau}
\Bigl(e^{-\frac{i}{\hbar}\tilde{p} \cdot \tilde{x}}
e_{\mu}(\mathbf{p},\tau) e^{-i \tau \phi_W (\mathbf{p}, \Lambda)} c
_{\Lambda \mathbf{p},\tau} \nonumber \\
&\ & +
e^{\frac{i}{\hbar}\tilde{p} \cdot \tilde{x}}
e^*_{\mu}(\mathbf{p},\tau) e^{i \tau \phi_W (\mathbf{p}, \Lambda)}
c^{\dag} _{\Lambda \mathbf{p},\tau} \Bigr) \label{eq:10.30}
\end{eqnarray}

\noindent Next we take equation (\ref{eq:10.27}) for vector $\Lambda \mathbf{p}$

\begin{eqnarray*}
 e(\Lambda \mathbf{p},\tau)
= \lambda_{\Lambda \mathbf{p}}  e( \mathbf{k},\tau)
\end{eqnarray*}

\noindent   and multiply both sides from the left by $\Lambda^{-1}$

\begin{eqnarray*}
\Lambda^{-1}  e (\Lambda \mathbf{p}, \tau) &=& \lambda_{\mathbf{p}}
(\lambda^{-1}_{\mathbf{p}}\Lambda^{-1} \lambda_{\Lambda \mathbf{p}})
e (\mathbf{k}, \tau)
\end{eqnarray*}

\noindent The term in parentheses
is a member of the little group\footnote{see subsection
\ref{ss:little-group}} which corresponds to a Wigner rotation
through the angle $-\phi_W$, so we can use representation
(\ref{eq:7.58})

\begin{eqnarray*}
&\mbox{ }& \lambda^{-1}_{\mathbf{p}}\Lambda^{-1} \lambda_{\Lambda
\mathbf{p}} e (\mathbf{k}, \tau) =
 S(X_1, X_2, -\phi_{W}) e (\mathbf{k}, \tau) \\
&=& \left[ \begin{array}{cccc}
1+ (X_1^2 + X_2^2)/2 & X_1 & X_2 & - (X_1^2 + X_2^2)/2 \\
X_1 \cos \phi_W - X_2 \sin \phi_W &\cos \phi_W
 & -\sin \phi_W & -X_1
\cos \phi_W + X_2 \sin \phi_W  \\
 X_1 \sin \phi_W +X_2 \cos \phi_W &  \sin \phi_W &  \cos \phi_W  &
-X_1
\sin \phi_W -X_2\cos \phi_W  \\
(X_1^2 + X_2^2)/2 & X_1 & X_2 & 1-(X_1^2 + X_2^2)/2
\end{array} \right]
\left[ \begin{array}{c}
0 \\
1 \\
i \tau\\
 0
\end{array} \right] \\ &=& e^{ -i \tau \phi_W (\mathbf{p}, \Lambda)}
\left[ \begin{array}{c}
0 \\
1 \\
 i \tau\\
 0
\end{array} \right] + (X_1 + i \tau X_2)
\left[ \begin{array}{c}
1 \\
0 \\
 0  \\
 1
\end{array} \right] \\
&=& e^{ -i \tau \phi_W(\mathbf{p}, \Lambda)}  e (\mathbf{k},
\tau) + \frac{X_1 + i \tau X_2}{c}
 \tilde{k}
\end{eqnarray*}

\noindent where $k_{\mu} = (c, 0,0,c)$ and  $X_1$, $X_2$ are certain
functions of $\Lambda$ and $\mathbf{p}$. Our next goal is to eliminate these unknown functions from our formulas. Denoting

\begin{eqnarray}
X_{\tau} (\mathbf{p}, \Lambda)= \frac{X_1 + i \tau X_2}{c}
\label{eq:10.31}
\end{eqnarray}

\noindent we obtain

\begin{eqnarray}
\sum_{\nu = 0}^3\Lambda ^{-1}_{\mu \nu} e^{\nu} (\Lambda \mathbf{p},
\tau) &=&
 e^{ -i \tau \phi_W (\mathbf{p}, \Lambda)}
\lambda_{\mathbf{p}} e_{\mu} ( \mathbf{k}, \tau) +  X_{\tau} (\mathbf{p},
\Lambda)\lambda_{\mathbf{p}} k_{\mu}
\nonumber  \\
&=& e^{ -i \tau \phi_W (\mathbf{p}, \Lambda)} e_{\mu} ( \mathbf{p},
\tau) + X_{\tau} (\mathbf{p}, \Lambda) \frac{p_{\mu}}{p}
\label{eq:10.32}
\end{eqnarray}

\noindent where $p_{\mu} = (p, p_x, p_y, p_z)$ is the
energy-momentum 4-vector corresponding to the 3-momentum $\mathbf{p}$.
 By letting $\mu = 0$
 and taking
into account (\ref{eq:10.33}) we also obtain

\begin{eqnarray*}
 \sum_{\nu = 0}^3 \Lambda ^{-1}_{0 \nu}
 e^{\nu} (\Lambda \mathbf{p}, \tau) &=&
e^{ -i \tau \phi_W (\mathbf{p}, \Lambda)} e_{0 } ( \mathbf{p}, \tau)
+ X_{\tau} (\mathbf{p}, \Lambda) \frac{p_{0}}{p} =
 X_{\tau} (\mathbf{p}, \Lambda) \nonumber \\
e^{ -i \tau \phi_W (\mathbf{p}, \Lambda)} e_{\mu } ( \mathbf{p},
\tau) &=& \sum_{\nu  =0}^3  \Lambda ^{-1}_{\mu \nu} e^{\nu }
(\Lambda \mathbf{p}, \tau)
 - X_{\tau} (\mathbf{p}, \Lambda) \frac{p_{\mu}}{p} \nonumber  \\
&=& \sum_{\nu  =0}^3  \Lambda ^{-1}_{\mu \nu}  e^{\nu } (\Lambda
\mathbf{p}, \tau)
 - \frac{ p_{\mu}}{p}\sum_{\nu =0}^3
\Lambda ^{-1}_{0 \nu}
e^{\nu } (\Lambda \mathbf{p}, \tau) \nonumber \\
&=& \sum_{\nu =0}^3  \left(\Lambda ^{-1}_{\mu \nu} - \Lambda
^{-1}_{0 \nu} \frac{p_{\mu}}{p} \right)  e^{\nu } (\Lambda
\mathbf{p}, \tau)
\end{eqnarray*}

\noindent The complex conjugate of this equation  is

\begin{eqnarray*}
e^{ i \tau \phi_W (\mathbf{p}, \Lambda)} e^*_{\mu } ( \mathbf{p},
\tau) = \sum_{\nu =0}^3  \left(\Lambda ^{-1}_{\mu \nu} - \Lambda
^{-1}_{0 \nu} \frac{p_{\mu}}{p} \right)  e^{* \nu } (\Lambda
\mathbf{p}, \tau)
\end{eqnarray*}

\noindent Then using (\ref{eq:7.15x}) and (\ref{eq:A.73}) we can
rewrite equation (\ref{eq:10.30}) as

\begin{eqnarray}
&\mbox{ } & U_0(\Lambda; 0)  A_{\mu}(\tilde{x}) U_0^{-1}(\Lambda; 0)  \nonumber \\
 &=& \frac{\hbar \sqrt{c}}{\sqrt{2}(2\pi \hbar)^{3/2} } \int
\frac{d\mathbf{p}}{\sqrt{p}} \sqrt{\frac{|\Lambda \mathbf{p}|}{p}}
\sum_{ \tau = -1}^1 \sum_{ \nu =0}^3
 \Bigl(e^{-\frac{i}{\hbar}\tilde{p} \cdot \tilde{x}}
 \left(\Lambda^{-1}_{\mu \nu} - \Lambda^{-1}_{0
\nu} \frac{p_{\mu}}{p} \right)
 e^{\nu}(\Lambda \mathbf{p}, \tau)  c _{\Lambda \mathbf{p},\tau} \nonumber
\\
&\ &+
e^{\frac{i}{\hbar}\tilde{p} \cdot \tilde{x}} \left(\Lambda^{-1}_{\mu
\nu} - \Lambda^{-1}_{0 \nu} \frac{p_{\mu}}{p}\right) e^{* \nu
}(\Lambda \mathbf{p}, \tau)
c^{\dag} _{\Lambda \mathbf{p},\tau} \Bigr) \nonumber \\
 &=& \frac{\hbar \sqrt{c}}{\sqrt{2}(2\pi \hbar)^{3/2} } \sum_{\nu =0}^3
\Lambda^{-1}_{\mu \nu} \int \frac{d (\Lambda \mathbf{p})}{ |\Lambda
\mathbf{p}|} \sqrt{|\Lambda \mathbf{p}|} \sum_{ \tau = -1}^1
 \left(e^{-\frac{i}{\hbar}\tilde{p} \cdot \tilde{x}}
 e^{\nu }(\Lambda \mathbf{p}, \tau)  c _{\Lambda \mathbf{p},\tau}
+ e^{\frac{i}{\hbar}\tilde{p} \cdot \tilde{x}}  e^{* \nu }(\Lambda
\mathbf{p},
\tau)  c^{\dag}  _{\Lambda \mathbf{p},\tau}\right) \nonumber \\
 &\ & -\frac{\hbar \sqrt{c}}{(2\pi \hbar)^{3/2} }  \int \frac{d
(\Lambda \mathbf{p})}{ |\Lambda \mathbf{p}|} \sqrt{|\Lambda
\mathbf{p}|} p^{\mu} \sum_{ \tau = -1}^1 \sum_{ \nu = 0}^3
\frac{\Lambda^{-1}_{0 \nu} }{p}
 \left(e^{-\frac{i}{\hbar}\tilde{p} \cdot \tilde{x}}
 e^{\nu }(\Lambda \mathbf{p}, \tau)  c _{\Lambda \mathbf{p},\tau} +
e^{\frac{i}{\hbar}\tilde{p} \cdot \tilde{x} }  e^{* \nu }(\Lambda
\mathbf{p}, \tau) c^{\dag}  _{\Lambda
\mathbf{p},\tau}\right) \nonumber \\
&=&  \sum_{ \nu =
0}^3 \Lambda^{-1} _{\mu \nu} \left[ \frac{\hbar \sqrt{c}}{\sqrt{2}(2\pi \hbar)^{3/2} } \int \frac{d
 \mathbf{p}}{\sqrt{p}}   \sum_{ \tau = -1}^1
 \left(e^{-\frac{i}{\hbar}\tilde{p} \cdot  \Lambda \tilde{x}}
 e^{\nu  }( \mathbf{p}, \tau)  c  _{ \mathbf{p},\tau} +
e^{\frac{i}{\hbar}\tilde{p}\cdot \Lambda \tilde{x}}  e^{* \nu
}(\mathbf{p}, \tau)
c^{\dag} _{ \mathbf{p},\tau}\right) \right] \nonumber \\
 &\ & -\frac{\hbar \sqrt{c}}{(2\pi \hbar)^{3/2} }
\int \frac{d
 \mathbf{p}}{\sqrt{p}}  \sum_{ \tau = -1}^1
\frac{(\Lambda^{-1} p)_{\mu}}{|\Lambda^{-1} \mathbf{p}|} \sum_{ \nu
= 0}^3 \Lambda^{-1} _{0 \nu}  \left(e^{-\frac{i}{\hbar} \Lambda^{-1}
\tilde{p} \cdot \tilde{x}}
 e^{\nu  }( \mathbf{p}, \tau)  c _{ \mathbf{p},\tau} +
e^{\frac{i}{\hbar} \Lambda^{-1} \tilde{p} \cdot \tilde{x}}  e^{* \nu
}(\mathbf{p}, \tau)  c^{\dag}
_{ \mathbf{p},\tau} \right) \nonumber \\
&=& \sum_{\nu = 0}^3 \Lambda_{\mu \nu}^{-1} A^{\nu}(\Lambda
\tilde{x}) +
 \Omega_{\mu} (\tilde{x}, \Lambda)
\label{eq:10.35a}
\end{eqnarray}

\noindent Thus we see that property (\ref{eq:ULA}) is not satisfied.
In addition to the desired covariant transformation $\Lambda^{-1} A(\Lambda \tilde{x})$,  there is an
 extra term

\begin{eqnarray}
\Omega_{\mu} (\tilde{x}, \Lambda) &\equiv& -\frac{\hbar \sqrt{c}}{(2\pi
\hbar)^{3/2} }   \int \frac{d
 \mathbf{p}}{\sqrt{2p}}  \sum_{
\tau = -1}^1 \frac{(\Lambda^{-1}p)_{\mu}}{|\Lambda^{-1} \mathbf{p}|}
 \sum_{ \nu = 0}^3 \Lambda_{0 \nu}^{-1} \times \nonumber \\
&\mbox{ } & \left[e^{-\frac{i}{\hbar} \Lambda^{-1} \tilde{p} \cdot
\tilde{x}}
 e^{\nu }( \mathbf{p}, \tau)  c _{ \mathbf{p},\tau} +
e^{\frac{i}{\hbar} \Lambda^{-1} \tilde{p} \cdot  \tilde{x}} e^{* \nu
}(\mathbf{p}, \tau)  c^{\dag} _ {\mathbf{p},\tau} \right]
\label{eq:10.35}
\end{eqnarray}

\noindent in the boost transformation law. The presence of this extra term is the reason why QED with massless photons cannot be formulated via simple steps outlined in subsection \ref{ss:weinberg}. A more elaborate construction is required in order to maintain the relativistic invariance of QED as detailed in subsection \ref{ss:interaction-qed} and in Appendix \ref{ss:relat-invar}.

 From

\begin{eqnarray}
\lim_{\theta \to 0 } \sum_{\rho=0}^3 \Lambda^{-1}_{0 \rho}
 e^{\rho}( \mathbf{p}, \tau)  &=& \sum_{\rho=0}^3 \delta _{0 \rho}
 e_{\rho}( \mathbf{p}, \tau) =
 e_{0 }( \mathbf{p}, \tau)= 0 \label{eq:11.14}
\end{eqnarray}

\noindent we obtain the following useful property

\begin{eqnarray}
\Omega(\tilde{x}, 1) = 0 \label{eq:11.14a}
\end{eqnarray}


\chapter{QED interaction in terms of particle operators}
\label{ss:int-part-oper}

\section{ Current density }
\label{sc:current-dens}

In QED an important role is played by the operator of \emph{current
density} \index{current density} which is defined as a sum of the
electron/positron $J^{\mu}(\tilde{x})$ and proton/antiproton
$\mathcal{J}^{\mu}(\tilde{x})$ current densities

\begin{eqnarray}
j^{\mu}(\tilde{x}) &=& J^{\mu}(\tilde{x}) + \mathcal{J}^{\mu}(\tilde{x}) \nonumber \\
&\equiv&  -e c \overline{\psi}(\tilde{x}) \gamma^{\mu}
\psi(\tilde{x}) + e c \overline{\Psi}(\tilde{x})  \gamma^{\mu} \Psi
(\tilde{x}) \label{eq:11.1}
\end{eqnarray}

\noindent where $e$ is the absolute value of the electron charge, gamma
matrices $\gamma^{\mu}$ are defined in equations (\ref{eq:A.81}) -
(\ref{eq:A.82}) and quantum fields $\psi(\tilde{x})$,
$\overline{\psi}(\tilde{x})$, $\Psi(\tilde{x})$,
$\overline{\Psi}(\tilde{x})$ are defined in Appendix
\ref{ss:cons-ferm}.\footnote{Note that $\psi(\tilde{x})$ is a
4-component bispinor-column and $\overline{\psi}(\tilde{x})$  is a
4-component bispinor-row.
So, the product $\overline{\psi}(\tilde{x}) \gamma^{\mu}
\psi(\tilde{x})$ is a scalar in the bispinor space.} Let us consider
the electron/positron part $J^{\mu}(\tilde{x})$ of the current
density and derive three important properties of this
operator.\footnote{Properties of the proton/antiproton part $\mathcal{J}^{\mu}(\tilde{x})$ are
similar. } First, with the help of (\ref{eq:A.88}),
(\ref{eq:d-gamma-d}), and (\ref{eq:10.21}) we can find that the
current operator (\ref{eq:11.1}) transforms as a 4-vector function
on the Minkowski space-time

\begin{eqnarray}
&\ &U_0(\Lambda; 0) J^{\mu}(\tilde{x}) U_0^{-1}(\Lambda; 0) \nonumber \\
& =& -e c U_0(\Lambda; 0) \psi^{\dag}(\tilde{x})  \gamma^0
\gamma^{\mu} \psi(\tilde{x}) U_0^{-1}(\Lambda; 0) \nonumber \\
& =& -e c U_0(\Lambda; 0) \psi^{\dag}(\tilde{x}) U_0^{-1}(\Lambda;
0) \gamma^0 \gamma^{\mu}
U_0(\Lambda; 0) \psi(\tilde{x}) U_0^{-1}(\Lambda; 0) \nonumber \\
& =& -ec \psi^{\dag}(\Lambda \tilde{x}) \mathcal{D}^{\dag}
(\Lambda^{-1}) \gamma^0 \gamma^{\mu}
\mathcal{D} (\Lambda^{-1}) \psi (\Lambda \tilde{x}) \nonumber \\
& =& -ec  \psi^{\dag}(\Lambda \tilde{x}) \mathcal{D}(\Lambda^{-1})
\gamma^0 \mathcal{D}(\Lambda^{-1})
\mathcal{D}(\Lambda)\gamma^{\mu}\mathcal{D} (\Lambda^{-1}) \psi
(\Lambda \tilde{x}) \nonumber
\\
& =& -e c \psi^{\dag}(\Lambda \tilde{x})  \gamma^0
\mathcal{D}(\Lambda)\gamma^{\mu}\mathcal{D} (\Lambda^{-1}) \psi
(\Lambda \tilde{x})
\nonumber \\
& =& -e c \sum_{\nu=0}^3 \psi^{\dag} (\Lambda \tilde{x}) \gamma^0
(\Lambda^{-1})_{\mu}^ {\nu } \gamma^{\nu}
\psi (\Lambda \tilde{x}) \nonumber \\
& =&  \sum_{\nu=0}^3 (\Lambda^{-1})_{\mu}^ {\nu } J_{\nu} (\Lambda
\tilde{x}) \label{eq:curr-trans}
\end{eqnarray}

\noindent
 From this we
obtain a useful commutator

\begin{eqnarray}
&\mbox{ } &[K_{0z}, J_0(\tilde{x})] \nonumber \\
&=& -\frac{i \hbar}{c} \lim_{\theta \to 0} \frac{d}{d \theta}
e^{\frac{ic}{\hbar} K_{0z}  \theta}
J_0(\tilde{x}) e^{-\frac{ic}{\hbar} K_{0z}  \theta} \nonumber \\
&=& -\frac{i \hbar}{c} \lim_{\theta \to 0} \frac{d}{d \theta}
\Bigl[J_0 \left(x , y, z \cosh \theta - ct \sinh \theta, t \cosh
\theta - \frac{z}{c} \sinh \theta \right)
 \cosh \theta \nonumber \\
&\ & +J_z\left( x, y, z \cosh \theta - ct \sinh \theta, t \cosh \theta
- \frac{z}{c} \sinh \theta\right)   \sinh
\theta \Bigr] \nonumber \\
&=& i \hbar \left(\frac{z}{c^2} \frac{d}{dt} + t \frac{d}{dz}
\right)J_0(\tilde{x}) - \frac{i \hbar}{c}J_z(\tilde{x})
\label{eq:11.2}
\end{eqnarray}

\noindent Space-time translations act by shifting the argument of
the current

\begin{eqnarray}
U_0(0; \tilde{a}) J^{\mu}(\tilde{x}) U_0^{-1}(0; \tilde{a}) & =&
J^{\mu}(\tilde{x} + \tilde{a}) \label{eq:cur-trans}
\end{eqnarray}

 \noindent
Second, the current density satisfies the \emph{continuity equation}
\index{continuity equation} which can be proven by using Dirac
equations (\ref{eq:10.24}), (\ref{eq:10.25}), and property
(\ref{eq:H.17a})

\begin{eqnarray}
\frac{\partial}{\partial t} J^0(\tilde{x}) &=& -e
c\frac{\partial}{\partial t} (\overline{\psi}(\tilde{x})
\gamma^0 \psi(\tilde{x}) ) \nonumber \\
&=& -e c \left(\frac{\partial}{\partial t}
\overline{\psi}(\tilde{x}) \right) \gamma^0 \psi(\tilde{x}) +
\overline{\psi}(\tilde{x}) \left( \gamma^0 \frac{\partial}{\partial
t}
\psi(\tilde{x}) \right) \nonumber \\
&=& e c \left(c\frac{\partial}{\partial \mathbf{x}}
\psi^{\dag}(\tilde{x})\vec{\gamma}^{\dag} + \frac{i}{\hbar} mc^2
\psi^{\dag}(\tilde{x}) \right) \gamma^0 \psi(\tilde{x}) \nonumber \\
&\ & +ec \overline{\psi}(\tilde{x}) \left( c \vec{\gamma}
\frac{\partial}{
\partial \mathbf{x} } \psi(\tilde{x}) - \frac{i}{\hbar}mc^2
\psi(\tilde{x}) \right) \nonumber \\
&=& ec^2 \frac{\partial}{\partial \mathbf{x}}
\overline{\psi}(\tilde{x}) \vec{\gamma}
 \psi(\tilde{x}) +  ec^2 \overline{\psi}(\tilde{x})
 \vec{\gamma} \frac{\partial}{ \partial \mathbf{x} }  \psi(\tilde{x}) \nonumber  \\
&=& ec^2 \frac{\partial}{\partial \mathbf{x}}
(\overline{\psi}(\tilde{x}) \vec{\gamma}
 \psi(\tilde{x})) \nonumber  \\
&=& -c \frac{\partial}{\partial \mathbf{x}} \mathbf{J} (\tilde{x})
\label{eq:11.3}
\end{eqnarray}

\noindent Third, from equation (\ref{eq:10.23}) it follows that current
components commute at spacelike separations

\begin{eqnarray*}
[ j^{\mu} ( \mathbf{x}, t), j^{\nu} ( \mathbf{y}, t)] = 0, \mbox{ }
if  \mbox{   }\mathbf{x} \neq \mathbf{y}
\end{eqnarray*}

 Using expressions for fields
(\ref{eq:J.55a}) and (\ref{eq:J.55b}), we can also write the
current density operator (\ref{eq:11.1}) in the normally ordered
form\footnote{Summation on bispinor indices $\alpha$ and $\beta$ is
assumed.}

\begin{eqnarray*}
 j^{\mu}(\tilde{x})& =&   -e c\overline{\psi}(\tilde{x}) \gamma^{\mu} \psi (\tilde{x})
 + e c\overline{\Psi} (\tilde{x}) \gamma^{\mu} \Psi (\tilde{x})\\
&=& e c (2\pi \hbar)^{-3}   \int
d\mathbf{p} d\mathbf{p}'  \times \\
\Bigl(&-&[e^{\frac{i}{\hbar} \tilde{p} \cdot \tilde{x}}
\overline{A}^{\dag} _{ \alpha}(\mathbf{p}) +
e^{-\frac{i}{\hbar}\tilde{p} \cdot \tilde{x} }\overline{B}_{
\alpha}(\mathbf{p}) ] \gamma^{\mu} _{\alpha \beta}
[e^{-\frac{i}{\hbar} \tilde{p}' \cdot \tilde{x}} A _{
\beta}(\mathbf{p}') +
e^{\frac{i}{\hbar}\tilde{p}' \cdot \tilde{x} }B^{\dag}_{ \beta}(\mathbf{p}') ] \\
&+&  [e^{\frac{i}{\hbar}\tilde{P} \cdot
\tilde{x}}\overline{D}^{\dag} _{ \alpha}(\mathbf{p}) +
e^{-\frac{i}{\hbar}\tilde{P} \cdot \tilde{x}}\overline{F}_{
\alpha}(\mathbf{p}) ] \gamma^{\mu} _{\alpha \beta}
[e^{-\frac{i}{\hbar}\tilde{P}' \cdot \tilde{x}}D _{
\beta}(\mathbf{p}') + e^{\frac{i}{\hbar}\tilde{P}' \cdot
\tilde{x}}F^{\dag}_{
\beta}(\mathbf{p}') ] \Bigr) \\
&=& e c (2\pi \hbar)^{-3}   \int d\mathbf{p} d\mathbf{p}'
\gamma^{\mu} _{\alpha \beta} \times
 \\
\Bigl(&-& \overline{A}^{\dag}_{ \alpha}(\mathbf{p})  A _{
\beta}(\mathbf{p}') e^{-\frac{i}{\hbar}(\tilde{p}'-\tilde{p}) \cdot
\tilde{x}} - \overline{A}^{\dag}_{ \alpha}(\mathbf{p})  B^{\dag}_{
\beta}(\mathbf{p}') e^{\frac{i}{\hbar}(\tilde{p}'+\tilde{p}) \cdot \tilde{x}} \\
&-& \overline{B}_{ \alpha}(\mathbf{p})  A _{ \beta}(\mathbf{p}')
e^{-\frac{i}{\hbar}(\tilde{p}'+\tilde{p}) \cdot \tilde{x}} -
\overline{B}_{ \alpha}(\mathbf{p})  B^{\dag}_{
\beta}(\mathbf{p}') e^{\frac{i}{\hbar}(\tilde{p}'-\tilde{p}) \cdot \tilde{x}} \\
&+& \overline{D}^{\dag}_{ \alpha}(\mathbf{p})  D _{
\beta}(\mathbf{p}') e^{-\frac{i}{\hbar}(\tilde{P}'-\tilde{P}) \cdot
\tilde{x}} + \overline{D}^{\dag}_{ \alpha}(\mathbf{p})  F^{\dag} _{
\beta}(\mathbf{p}') e^{+\frac{i}{\hbar}(\tilde{P}'+\tilde{P}) \cdot \tilde{x}} \\
&+& \overline{F}_{ \alpha}(\mathbf{p}) D _{ \beta}(\mathbf{p}')
e^{-\frac{i}{\hbar}(\tilde{P}'+\tilde{P}) \cdot \tilde{x}} +
\overline{F}_{ \alpha}(\mathbf{p})  F^{\dag} _{
\beta}(\mathbf{p}') e^{\frac{i}{\hbar}(\tilde{P}'-\tilde{P}) \cdot \tilde{x}}\Bigr) \\
&=& e c(2\pi \hbar)^{-3}   \int d\mathbf{p} d\mathbf{p}'
\gamma^{\mu} _{\alpha
\beta} \times \\
\Bigl(&-& \overline{A}^{\dag}_{ \alpha}(\mathbf{p})  A _{
\beta}(\mathbf{p}') e^{-\frac{i}{\hbar}(\tilde{p}'-\tilde{p}) \cdot
\tilde{x}} - \overline{A}^{\dag}_{ \alpha}(\mathbf{p})  B^{\dag}_{
\beta}(\mathbf{p}') e^{\frac{i}{\hbar}(\tilde{p}'+\tilde{p}) \cdot \tilde{x}} \\
&-& \overline{B}_{ \alpha}(\mathbf{p})  A _{ \beta}(\mathbf{p}')
e^{-\frac{i}{\hbar}(\tilde{p}'+\tilde{p}) \cdot \tilde{x}} +
B^{\dag}_{ \beta}(\mathbf{p}')\overline{B}_{ \alpha}(\mathbf{p})
 e^{\frac{i}{\hbar}(\tilde{p}'-\tilde{p}) \cdot \tilde{x}} \\
&+& \overline{D}^{\dag}_{ \alpha}(\mathbf{p})  D _{
\beta}(\mathbf{p}') e^{-\frac{i}{\hbar}(\tilde{P}'-\tilde{P}) \cdot
\tilde{x}} + \overline{D}^{\dag}_{ \alpha}(\mathbf{p})  F^{\dag} _{
\beta}(\mathbf{p}') e^{+\frac{i}{\hbar}(\tilde{P}'+\tilde{P}) \cdot \tilde{x}} \\
&+& \overline{F}_{ \alpha}(\mathbf{p}) D _{ \beta}(\mathbf{p}')
e^{-\frac{i}{\hbar}(\tilde{P}'+\tilde{P}) \cdot \tilde{x}} -
F^{\dag} _{ \beta}(\mathbf{p}') \overline{F}_{ \alpha}(\mathbf{p})
 e^{\frac{i}{\hbar}(\tilde{P}'-\tilde{P}) \cdot \tilde{x}} \\
&-& \{\overline{B}_{ \alpha}(\mathbf{p}),  B^{\dag}_{
\beta}(\mathbf{p}')\} e^{\frac{i}{\hbar}(\tilde{p}'-\tilde{p}) \cdot
\tilde{x}} + \{\overline{F}_{ \alpha}(\mathbf{p}),  F^{\dag} _{
\beta}(\mathbf{p}')\} e^{\frac{i}{\hbar}(\tilde{P}'-\tilde{P}) \cdot
\tilde{x}} \Bigr)
\end{eqnarray*}

\noindent Let us show that the two last terms vanish. We
 use anticommutator (\ref{eq:B-over-B}) and properties of gamma
matrices to rewrite these two terms as

\begin{eqnarray}
&\mbox{ } & e c (2\pi \hbar)^{-3}   \int d\mathbf{p} d\mathbf{p}'
\gamma^{\mu} _{\alpha \beta} \times \nonumber
\\
\Bigl(&-& \frac{1}{2\omega_{\mathbf{p}}} (\gamma^0
\omega_{\mathbf{p}} + \vec{\gamma} \mathbf{p} c - mc^2)_{\beta
\alpha}
\delta(\mathbf{p}- \mathbf{p}') e^{\frac{i}{\hbar}(\tilde{p}'-\tilde{p}) \cdot x} \nonumber \\
&+& \frac{1}{2\Omega_{\mathbf{p}}} (\gamma^0 \Omega_{\mathbf{p}}  +
\vec{\gamma} \mathbf{p} c - Mc^2 )_{\beta \alpha}
\delta(\mathbf{p}- \mathbf{p}') e^{\frac{i}{\hbar}(\tilde{P}'-\tilde{P}) \cdot x}\Bigr) \nonumber \\
&=& e c (2\pi \hbar)^{-3}   \int d\mathbf{p} \gamma^{\mu} _{\alpha
\alpha}   \left( \frac{mc^2}{2\omega_{\mathbf{p}}}
- \frac{Mc^2}{2\Omega_{\mathbf{p}}} \right) \nonumber \\
&+& e c (2\pi \hbar)^{-3}   \int d\mathbf{p} (\gamma^{\mu} \gamma^0)
_{\alpha
\alpha}   \left( -\frac{1}{2} + \frac{1}{2} \right) \label{eq:half} \\
&+& e c (2\pi \hbar)^{-3}   \int d\mathbf{p} (\gamma^{\mu}
\vec{\gamma}) _{\alpha \alpha}   \left(
-\frac{\mathbf{p}c}{2\omega_{\mathbf{p}}} +
\frac{\mathbf{p}c}{2\Omega_{\mathbf{p}}} \right) \nonumber \\
&=& e c (2\pi \hbar)^{-3}  Tr(\gamma^{\mu}) \int d\mathbf{p}
\left( \frac{mc^2}{2\omega_{\mathbf{p}}}
- \frac{Mc^2}{2\Omega_{\mathbf{p}}} \right) \nonumber \\
 &+& ec^2 (2\pi \hbar)^{-3} Tr(\gamma^{\mu}
\vec{\gamma})  \int d\mathbf{p}  \mathbf{p}  \left(
-\frac{1}{2\omega_{\mathbf{p}}} + \frac{1}{2\Omega_{\mathbf{p}}}
\right ) \nonumber
\end{eqnarray}

\noindent The first term vanishes due to the property
(\ref{eq:trace-gamma}). The second integral is zero, because the
integrand is an odd function of $\mathbf{p}$.\footnote{Note that
cancelation in (\ref{eq:half}) was possible only because our theory
contains two particle types (electrons and protons) with opposite
electric charges.} So, finally, the normally ordered form of the
current density is

\begin{eqnarray}
 j^{\mu}(\tilde{x}) &=&
e c (2\pi \hbar)^{-3}   \int d\mathbf{p} d\mathbf{p}' \gamma^{\mu}
_{\alpha \beta} \times
 \nonumber \\
\Bigl(&-& \overline{A}^{\dag}_{ \alpha}(\mathbf{p})  A _{
\beta}(\mathbf{p}') e^{-\frac{i}{\hbar}(\tilde{p}'-\tilde{p}) \cdot
\tilde{x}} - \overline{A}^{\dag}_{ \alpha}(\mathbf{p})  B^{\dag}_{
\beta}(\mathbf{p}') e^{\frac{i}{\hbar}(\tilde{p}'+\tilde{p}) \cdot \tilde{x}} \nonumber \\
&-& \overline{B}_{ \alpha}(\mathbf{p})  A _{ \beta}(\mathbf{p}')
e^{-\frac{i}{\hbar}(\tilde{p}'+\tilde{p}) \cdot \tilde{x}} + B^{\dag}_{
\beta}(\mathbf{p}')\overline{B}_{ \alpha}(\mathbf{p})
 e^{\frac{i}{\hbar}(\tilde{p}'-\tilde{p}) \cdot \tilde{x}} \nonumber \\
&+& \overline{D}^{\dag}_{ \alpha}(\mathbf{p})  D _{
\beta}(\mathbf{p}') e^{-\frac{i}{\hbar}(\tilde{P}'-\tilde{P}) \cdot
\tilde{x}} + \overline{D}^{\dag}_{ \alpha}(\mathbf{p})  F^{\dag} _{
\beta}(\mathbf{p}') e^{+\frac{i}{\hbar}(\tilde{P}'+\tilde{P}) \cdot \tilde{x}} \nonumber \\
&+& \overline{F}_{ \alpha}(\mathbf{p}) D _{ \beta}(\mathbf{p}')
e^{-\frac{i}{\hbar}(\tilde{P}'+\tilde{P}) \cdot \tilde{x}} -
F^{\dag} _{ \beta}(\mathbf{p}') \overline{F}_{ \alpha}(\mathbf{p})
 e^{\frac{i}{\hbar}(\tilde{P}'-\tilde{P}) \cdot \tilde{x}}\Bigr) \label{eq:L.1a}
\end{eqnarray}

\section{ First-order interaction in QED}

Inserting (\ref{eq:L.1a}) and (\ref{eq:K.6}) in (\ref{eq:11.6}) we obtain
 the 1st order interaction expressed in terms of
  creation and
annihilation operators

\begin{eqnarray}
 V_1
&=&
 \frac{e }{ (2\pi \hbar)^{9/2} }  \int
d\mathbf{x} d\mathbf{p} d\mathbf{p}' d\mathbf{k} \left( -
\overline{A}^{\dag}_{ \alpha}(\mathbf{p})   A _{ \beta}(\mathbf{p}')
e^{-\frac{i}{\hbar}(\mathbf{p}'-\mathbf{p}) \cdot \mathbf{x}} + \ldots \right) \times \nonumber\\
&\mbox{ } & \left(e^{-\frac{i}{\hbar}\mathbf{kx}}C _{\alpha
\beta}(\mathbf{k})
+ e^{\frac{i}{\hbar}\mathbf{kx}}C^{\dag}_{\alpha \beta}(\mathbf{k}) \right) \nonumber\\
&=& \frac{e }{ (2\pi \hbar)^{3/2} } \int
 d\mathbf{k}
d\mathbf{p} \times  \nonumber \\
\Bigl(&-&  \overline{A}^{\dag}_{ \alpha}(\mathbf{p+k}) A _{
\beta}(\mathbf{p})C_{\alpha \beta} (\mathbf{k}) -
\overline{A}^{\dag}_{ \alpha}(\mathbf{p-k}) A _{
\beta}(\mathbf{p})C^{\dag}_{\alpha \beta}(\mathbf{k}) \nonumber\\
&+& \overline{D}^{\dag}_{ \alpha}(\mathbf{p+k}) D _{
\beta}(\mathbf{p})C_{\alpha \beta} (\mathbf{k}) +
\overline{D}^{\dag}_{ \alpha}(\mathbf{p-k}) D _{
\beta}(\mathbf{p})C^{\dag}_{\alpha \beta}(\mathbf{k}) \nonumber \\
&+&     B^{\dag}_{ \beta}(\mathbf{p+k}) \overline{B} _{
\alpha}(\mathbf{p})C_{\alpha \beta} (\mathbf{k})   + B^{\dag}_{
\beta}(\mathbf{p-k}) \overline{B}_{
\alpha}(\mathbf{p})C^{\dag}_{\alpha \beta}(\mathbf{k}) \nonumber\\
&-& F^{\dag}_{ \beta}(\mathbf{p+k}) \overline{F} _{
\alpha}(\mathbf{p})C_{\alpha \beta} (\mathbf{k})  -  F^{\dag}_{
\beta}(\mathbf{p-k}) \overline{F} _{
\alpha}(\mathbf{p})C^{\dag}_{\alpha \beta}(\mathbf{k}) \nonumber \\
&-&    \overline{A}^{\dag}_{ \alpha}(\mathbf{p+k}) B^{\dag}_{
\beta}(\mathbf{p})C_{\alpha \beta} (\mathbf{k})  -
\overline{A}^{\dag}_{ \alpha}(\mathbf{p-k}) B^{\dag} _{
\beta}(\mathbf{p})C^{\dag}_{\alpha \beta}(\mathbf{k}) \nonumber\\
&-& A_{ \beta}(\mathbf{p+k}) \overline{B} _{
\alpha}(\mathbf{p})C_{\alpha \beta} (\mathbf{k})  -  A_{
\beta}(\mathbf{p-k}) \overline{B} _{
\alpha}(\mathbf{p})C^{\dag}_{\alpha \beta}(\mathbf{k})\nonumber \\
&+&    \overline{D}^{\dag}_{ \alpha}(\mathbf{p+k}) F^{\dag} _{
\beta}(\mathbf{p})C_{\alpha \beta} (\mathbf{k})  +
\overline{D}^{\dag}_{ \alpha}(\mathbf{p-k}) F^{\dag} _{
\beta}(\mathbf{p})C^{\dag}_{\alpha \beta}(\mathbf{k}) \nonumber\\
&+& D_{ \beta}(\mathbf{p+k}) \overline{F} _{
\alpha}(\mathbf{p})C_{\alpha \beta} (\mathbf{k})  +  D_{
\beta}(\mathbf{p-k}) \overline{F} _{
\alpha}(\mathbf{p})C^{\dag}_{\alpha \beta}(\mathbf{k}) \Bigr)
\label{eq:11.32}
\end{eqnarray}

\noindent This operator is of the pure \emph{unphys} type.

\section{ Second-order interaction in QED} \label{ss:2nd-order-int}

The second order interaction Hamiltonian  (\ref{eq:11.7}) has rather
long expression in terms of particle operators

\begin{eqnarray}
 V_2
&=& \frac{1}{c^2} \int d\mathbf{x} d\mathbf{y} j_0(\mathbf{x} , 0)
 \frac{1}{8 \pi |\mathbf{x} - \mathbf{y}|}
          j_0(\mathbf{y}, 0) \nonumber \\
&=& e^2  (2\pi \hbar)^{-6}  \sum_{\alpha \beta \gamma \delta}
\gamma^0_{\alpha \beta} \gamma^0_{\gamma \delta}
 \int d\mathbf{x} d\mathbf{y}
\int d\mathbf{p} d\mathbf{p}' d\mathbf{q}
d\mathbf{q}' \frac{1}{8 \pi |\mathbf{x} - \mathbf{y}|}  \times  \nonumber \\
\Bigl(&-& \overline{A}^{\dag}_{ \alpha}(\mathbf{p})  A _{
\beta}(\mathbf{p}') e^{-\frac{i}{\hbar}(\mathbf{p}'-\mathbf{p})
\cdot \mathbf{x}} - \overline{A}^{\dag}_{ \alpha}(\mathbf{p})
B^{\dag}_{
\beta}(\mathbf{p}') e^{\frac{i}{\hbar}(\mathbf{p}'+\mathbf{p}) \cdot \mathbf{x}} \nonumber \\
&-& \overline{B}_{ \alpha}(\mathbf{p})  A _{ \beta}(\mathbf{p}')
e^{-\frac{i}{\hbar}(\mathbf{p}'+\mathbf{p}) \cdot \mathbf{x}} -
\overline{B}_{ \alpha}(\mathbf{p})  B^{\dag}_{
\beta}(\mathbf{p}') e^{\frac{i}{\hbar}(\mathbf{p}'-\mathbf{p}) \cdot \mathbf{x}} \nonumber \\
&+& \overline{D}^{\dag}_{ \alpha}(\mathbf{p})  D _{
\beta}(\mathbf{p}') e^{-\frac{i}{\hbar}(\mathbf{p}'-\mathbf{p})
\cdot \mathbf{x}} + \overline{D}^{\dag}_{ \alpha}(\mathbf{p})
F^{\dag} _{
\beta}(\mathbf{p}') e^{+\frac{i}{\hbar}(\mathbf{p}'+\mathbf{p}) \cdot \mathbf{x}} \nonumber \\
&+& \overline{F}_{ \alpha}(\mathbf{p}) D _{ \beta}(\mathbf{p}')
e^{-\frac{i}{\hbar}(\mathbf{p}'+\mathbf{p}) \cdot \mathbf{x}} +
\overline{F}_{ \alpha}(\mathbf{p})  F^{\dag} _{
\beta}(\mathbf{p}') e^{\frac{i}{\hbar}(\mathbf{p}'-\mathbf{p}) \cdot \mathbf{x}}\Bigr)  \times \nonumber \\
\Bigl(&-& \overline{A}^{\dag}_{ \gamma}(\mathbf{q})  A _{
\delta}(\mathbf{q}') e^{-\frac{i}{\hbar}(\mathbf{q}'-\mathbf{q})
\cdot \mathbf{y}} - \overline{A}^{\dag}_{ \gamma}(\mathbf{q})
B^{\dag}_{
\delta}(\mathbf{q}') e^{\frac{i}{\hbar}(\mathbf{q}'+\mathbf{q}) \cdot \mathbf{y}} \nonumber \\
&-& \overline{B}_{ \gamma}(\mathbf{q})  A _{ \delta}(\mathbf{q}')
e^{-\frac{i}{\hbar}(\mathbf{q}'+\mathbf{q}) \cdot \mathbf{y}} -
\overline{B}_{ \gamma}(\mathbf{q})  B^{\dag}_{
\delta}(\mathbf{q}') e^{\frac{i}{\hbar}(\mathbf{q}'-\mathbf{q}) \cdot \mathbf{y}} \nonumber \\
&+& \overline{D}^{\dag}_{ \gamma}(\mathbf{q})  D _{
\delta}(\mathbf{q}') e^{-\frac{i}{\hbar}(\mathbf{q}'-\mathbf{q})
\cdot \mathbf{y}} +\overline{D}^{\dag}_{ \gamma}(\mathbf{q})
F^{\dag} _{
\delta}(\mathbf{q}') e^{+\frac{i}{\hbar}(\mathbf{q}'+\mathbf{q}) \cdot \mathbf{y}} \nonumber \\
&+& \overline{F}_{ \gamma}(\mathbf{q}) D _{ \delta}(\mathbf{q}')
e^{-\frac{i}{\hbar}(\mathbf{q}'+\mathbf{q}) \cdot \mathbf{y}} +
\overline{F}_{ \gamma}(\mathbf{q})  F^{\dag} _{ \delta}(\mathbf{q}')
e^{\frac{i}{\hbar}(\mathbf{q}'-\mathbf{q}) \cdot
\mathbf{y}}\Bigr) \nonumber \\
&=& e^2  (2\pi \hbar)^{-6}  \sum_{\alpha \beta \gamma \delta}  \int
d\mathbf{x} d\mathbf{y} \int d\mathbf{p} d\mathbf{p}' d\mathbf{q}
d\mathbf{q}' \frac{\gamma^0_{\alpha \beta} \gamma^0_{\gamma \delta} }{8 \pi |\mathbf{x} - \mathbf{y}|}  \times  \nonumber \\
\Bigl(&+& \overline{A}^{\dag}_{ \alpha}(\mathbf{p})  A _{
\beta}(\mathbf{p}') \overline{A}^{\dag}_{ \gamma}(\mathbf{q})  A _{
\delta}(\mathbf{q}') e^{-\frac{i}{\hbar}(\mathbf{q}'-\mathbf{q})
\cdot \mathbf{y}} e^{-\frac{i}{\hbar}(\mathbf{p}'-\mathbf{p}) \cdot
\mathbf{x}} \nonumber \nonumber \\
&+& \overline{A}^{\dag}_{ \alpha}(\mathbf{p}) A _{
\beta}(\mathbf{p}')\overline{A}^{\dag}_{ \gamma}(\mathbf{q})
B^{\dag}_{ \delta}(\mathbf{q}')
e^{\frac{i}{\hbar}(\mathbf{q}'+\mathbf{q}) \cdot \mathbf{y}}
e^{-\frac{i}{\hbar}(\mathbf{p}'-\mathbf{p}) \cdot \mathbf{x}}\nonumber \\
&+& \overline{A}^{\dag}_{ \alpha}(\mathbf{p})  A _{
\beta}(\mathbf{p}')\overline{B}_{ \gamma}(\mathbf{q})  A _{
\delta}(\mathbf{q}') e^{-\frac{i}{\hbar}(\mathbf{q}'+\mathbf{q})
\cdot \mathbf{y}} e^{-\frac{i}{\hbar}(\mathbf{p}'-\mathbf{p}) \cdot
\mathbf{x}} \nonumber \nonumber \\
&+& \overline{A}^{\dag}_{ \alpha}(\mathbf{p})  A _{
\beta}(\mathbf{p}')\overline{B}_{ \gamma}(\mathbf{q})  B^{\dag}_{
\delta}(\mathbf{q}') e^{\frac{i}{\hbar}(\mathbf{q}'-\mathbf{q})
\cdot \mathbf{y}}
e^{-\frac{i}{\hbar}(\mathbf{p}'-\mathbf{p}) \cdot \mathbf{x}}\nonumber \\
&-& \overline{A}^{\dag}_{ \alpha}(\mathbf{p})  A _{
\beta}(\mathbf{p}')\overline{D}^{\dag}_{ \gamma}(\mathbf{q})  D _{
\delta}(\mathbf{q}') e^{-\frac{i}{\hbar}(\mathbf{q}'-\mathbf{q})
\cdot \mathbf{y}} e^{-\frac{i}{\hbar}(\mathbf{p}'-\mathbf{p}) \cdot
\mathbf{x}} \nonumber \nonumber \\
&-& \overline{A}^{\dag}_{ \alpha}(\mathbf{p})  A _{
\beta}(\mathbf{p}')\overline{D}^{\dag}_{ \gamma}(\mathbf{q})
F^{\dag} _{ \delta}(\mathbf{q}')
e^{+\frac{i}{\hbar}(\mathbf{q}'+\mathbf{q}) \cdot \mathbf{y}}
e^{-\frac{i}{\hbar}(\mathbf{p}'-\mathbf{p}) \cdot \mathbf{x}}\nonumber \\
&-& \overline{A}^{\dag}_{ \alpha}(\mathbf{p})  A _{
\beta}(\mathbf{p}')\overline{F}_{ \gamma}(\mathbf{q}) D _{
\delta}(\mathbf{q}') e^{-\frac{i}{\hbar}(\mathbf{q}'+\mathbf{q})
\cdot \mathbf{y}} e^{-\frac{i}{\hbar}(\mathbf{p}'-\mathbf{p}) \cdot
\mathbf{x}} \nonumber \nonumber \\
&-&
 \overline{A}^{\dag}_{ \alpha}(\mathbf{p})  A _{
\beta}(\mathbf{p}') \overline{F}_{ \gamma}(\mathbf{q})  F^{\dag} _{
\delta}(\mathbf{q}') e^{\frac{i}{\hbar}(\mathbf{q}'-\mathbf{q})
\cdot \mathbf{y}}
e^{-\frac{i}{\hbar}(\mathbf{p}'-\mathbf{p}) \cdot \mathbf{x}} \nonumber \\
&+& \overline{A}^{\dag}_{ \alpha}(\mathbf{p})  B^{\dag}_{
\beta}(\mathbf{p}')\overline{A}^{\dag}_{ \gamma}(\mathbf{q})  A _{
\delta}(\mathbf{q}') e^{-\frac{i}{\hbar}(\mathbf{q}'-\mathbf{q})
\cdot \mathbf{y}} e^{\frac{i}{\hbar}(\mathbf{p}'+\mathbf{p}) \cdot
\mathbf{x}} \nonumber \nonumber \\
&+& \overline{A}^{\dag}_{ \alpha}(\mathbf{p})  B^{\dag}_{
\beta}(\mathbf{p}')\overline{A}^{\dag}_{ \gamma}(\mathbf{q})
B^{\dag}_{ \delta}(\mathbf{q}')
e^{\frac{i}{\hbar}(\mathbf{q}'+\mathbf{q}) \cdot \mathbf{y}}
e^{\frac{i}{\hbar}(\mathbf{p}'+\mathbf{p}) \cdot \mathbf{x}}\nonumber \\
&+& \overline{A}^{\dag}_{ \alpha}(\mathbf{p})  B^{\dag}_{
\beta}(\mathbf{p}')\overline{B}_{ \gamma}(\mathbf{q})  A _{
\delta}(\mathbf{q}') e^{-\frac{i}{\hbar}(\mathbf{q}'+\mathbf{q})
\cdot \mathbf{y}} e^{\frac{i}{\hbar}(\mathbf{p}'+\mathbf{p}) \cdot
\mathbf{x}} \nonumber \nonumber \\
&+& \overline{A}^{\dag}_{ \alpha}(\mathbf{p})  B^{\dag}_{
\beta}(\mathbf{p}')\overline{B}_{ \gamma}(\mathbf{q})  B^{\dag}_{
\delta}(\mathbf{q}') e^{\frac{i}{\hbar}(\mathbf{q}'-\mathbf{q})
\cdot \mathbf{y}}
e^{\frac{i}{\hbar}(\mathbf{p}'+\mathbf{p}) \cdot \mathbf{x}}\nonumber \\
&-& \overline{A}^{\dag}_{ \alpha}(\mathbf{p})  B^{\dag}_{
\beta}(\mathbf{p}')\overline{D}^{\dag}_{ \gamma}(\mathbf{q})  D _{
\delta}(\mathbf{q}') e^{-\frac{i}{\hbar}(\mathbf{q}'-\mathbf{q})
\cdot \mathbf{y}} e^{\frac{i}{\hbar}(\mathbf{p}'+\mathbf{p}) \cdot
\mathbf{x}} \nonumber \nonumber \\
&-& \overline{A}^{\dag}_{ \alpha}(\mathbf{p}) B^{\dag}_{
\beta}(\mathbf{p}')\overline{D}^{\dag}_{ \gamma}(\mathbf{q})
F^{\dag} _{ \delta}(\mathbf{q}')
e^{+\frac{i}{\hbar}(\mathbf{q}'+\mathbf{q}) \cdot \mathbf{y}}
e^{\frac{i}{\hbar}(\mathbf{p}'+\mathbf{p}) \cdot \mathbf{x}}\nonumber \\
&-& \overline{A}^{\dag}_{ \alpha}(\mathbf{p})  B^{\dag}_{
\beta}(\mathbf{p}')\overline{F}_{ \gamma}(\mathbf{q}) D _{
\delta}(\mathbf{q}') e^{-\frac{i}{\hbar}(\mathbf{q}'+\mathbf{q})
\cdot \mathbf{y}} e^{\frac{i}{\hbar}(\mathbf{p}'+\mathbf{p}) \cdot
\mathbf{x}} \nonumber \nonumber \\ &-& \overline{A}^{\dag}_{
\alpha}(\mathbf{p})  B^{\dag}_{ \beta}(\mathbf{p}') \overline{F}_{
\gamma}(\mathbf{q})  F^{\dag} _{ \delta}(\mathbf{q}')
e^{\frac{i}{\hbar}(\mathbf{q}'-\mathbf{q}) \cdot \mathbf{y}}
e^{\frac{i}{\hbar}(\mathbf{p}'+\mathbf{p}) \cdot \mathbf{x}} \nonumber \\
&+& \overline{B}_{ \alpha}(\mathbf{p})  A _{
\beta}(\mathbf{p}')\overline{A}^{\dag}_{ \gamma}(\mathbf{q})  A _{
\delta}(\mathbf{q}') e^{-\frac{i}{\hbar}(\mathbf{q}'-\mathbf{q})
\cdot \mathbf{y}} e^{-\frac{i}{\hbar}(\mathbf{p}'+\mathbf{p}) \cdot
\mathbf{x}} \nonumber \nonumber \\
&+& \overline{B}_{ \alpha}(\mathbf{p})  A _{
\beta}(\mathbf{p}')\overline{A}^{\dag}_{ \gamma}(\mathbf{q})
B^{\dag}_{ \delta}(\mathbf{q}')
e^{\frac{i}{\hbar}(\mathbf{q}'+\mathbf{q}) \cdot \mathbf{y}}
e^{-\frac{i}{\hbar}(\mathbf{p}'+\mathbf{p}) \cdot \mathbf{x}}\nonumber \\
&+& \overline{B}_{ \alpha}(\mathbf{p})  A _{
\beta}(\mathbf{p}')\overline{B}_{ \gamma}(\mathbf{q})  A _{
\delta}(\mathbf{q}') e^{-\frac{i}{\hbar}(\mathbf{q}'+\mathbf{q})
\cdot \mathbf{y}} e^{-\frac{i}{\hbar}(\mathbf{p}'+\mathbf{p}) \cdot
\mathbf{x}} \nonumber \nonumber \\
&+& \overline{B}_{ \alpha}(\mathbf{p})  A _{
\beta}(\mathbf{p}')\overline{B}_{ \gamma}(\mathbf{q})  B^{\dag}_{
\delta}(\mathbf{q}') e^{\frac{i}{\hbar}(\mathbf{q}'-\mathbf{q})
\cdot \mathbf{y}}
e^{-\frac{i}{\hbar}(\mathbf{p}'+\mathbf{p}) \cdot \mathbf{x}} \nonumber \\
&-& \overline{B}_{ \alpha}(\mathbf{p})  A _{
\beta}(\mathbf{p}')\overline{D}^{\dag}_{ \gamma}(\mathbf{q})  D _{
\delta}(\mathbf{q}') e^{-\frac{i}{\hbar}(\mathbf{q}'-\mathbf{q})
\cdot \mathbf{y}} e^{-\frac{i}{\hbar}(\mathbf{p}'+\mathbf{p}) \cdot
\mathbf{x}} \nonumber \nonumber \\
&-& \overline{B}_{ \alpha}(\mathbf{p})  A _{
\beta}(\mathbf{p}')\overline{D}^{\dag}_{ \gamma}(\mathbf{q})
F^{\dag} _{ \delta}(\mathbf{q}')
e^{+\frac{i}{\hbar}(\mathbf{q}'+\mathbf{q}) \cdot \mathbf{y}}
e^{-\frac{i}{\hbar}(\mathbf{p}'+\mathbf{p}) \cdot \mathbf{x}}\nonumber \\
&-& \overline{B}_{ \alpha}(\mathbf{p})  A _{
\beta}(\mathbf{p}')\overline{F}_{ \gamma}(\mathbf{q}) D _{
\delta}(\mathbf{q}') e^{-\frac{i}{\hbar}(\mathbf{q}'+\mathbf{q})
\cdot \mathbf{y}} e^{-\frac{i}{\hbar}(\mathbf{p}'+\mathbf{p}) \cdot
\mathbf{x}} \nonumber \nonumber \\
&-& \overline{B}_{ \alpha}(\mathbf{p})  A _{
\beta}(\mathbf{p}')\overline{F}_{ \gamma}(\mathbf{q})  F^{\dag} _{
\delta}(\mathbf{q}') e^{\frac{i}{\hbar}(\mathbf{q}'-\mathbf{q})
\cdot \mathbf{y}}
e^{-\frac{i}{\hbar}(\mathbf{p}'+\mathbf{p}) \cdot \mathbf{x}} \nonumber \\
&+& \overline{B}_{ \alpha}(\mathbf{p})  B^{\dag}_{
\beta}(\mathbf{p}')\overline{A}^{\dag}_{ \gamma}(\mathbf{q})  A _{
\delta}(\mathbf{q}') e^{-\frac{i}{\hbar}(\mathbf{q}'-\mathbf{q})
\cdot \mathbf{y}} e^{\frac{i}{\hbar}(\mathbf{p}'-\mathbf{p}) \cdot
\mathbf{x}} \nonumber \nonumber \\
&+& \overline{B}_{ \alpha}(\mathbf{p}) B^{\dag}_{
\beta}(\mathbf{p}')\overline{A}^{\dag}_{ \gamma}(\mathbf{q})
B^{\dag}_{ \delta}(\mathbf{q}')
e^{\frac{i}{\hbar}(\mathbf{q}'+\mathbf{q}) \cdot \mathbf{y}}
e^{\frac{i}{\hbar}(\mathbf{p}'-\mathbf{p}) \cdot \mathbf{x}} \nonumber \\
&+& \overline{B}_{ \alpha}(\mathbf{p})  B^{\dag}_{
\beta}(\mathbf{p}')\overline{B}_{ \gamma}(\mathbf{q})  A _{
\delta}(\mathbf{q}') e^{-\frac{i}{\hbar}(\mathbf{q}'+\mathbf{q})
\cdot \mathbf{y}} e^{\frac{i}{\hbar}(\mathbf{p}'-\mathbf{p}) \cdot
\mathbf{x}} \nonumber \nonumber \\
&+& \overline{B}_{ \alpha}(\mathbf{p}) B^{\dag}_{
\beta}(\mathbf{p}')\overline{B}_{ \gamma}(\mathbf{q}) B^{\dag}_{
\delta}(\mathbf{q}') e^{\frac{i}{\hbar}(\mathbf{q}'-\mathbf{q})
\cdot \mathbf{y}}
e^{\frac{i}{\hbar}(\mathbf{p}'-\mathbf{p}) \cdot \mathbf{x}} \nonumber \\
&-& \overline{B}_{ \alpha}(\mathbf{p})  B^{\dag}_{
\beta}(\mathbf{p}')\overline{D}^{\dag}_{ \gamma}(\mathbf{q})  D _{
\delta}(\mathbf{q}') e^{-\frac{i}{\hbar}(\mathbf{q}'-\mathbf{q})
\cdot \mathbf{y}} e^{\frac{i}{\hbar}(\mathbf{p}'-\mathbf{p}) \cdot
\mathbf{x}} \nonumber \nonumber \\
&-& \overline{B}_{ \alpha}(\mathbf{p}) B^{\dag}_{
\beta}(\mathbf{p}')\overline{D}^{\dag}_{ \gamma}(\mathbf{q})
F^{\dag} _{ \delta}(\mathbf{q}')
e^{+\frac{i}{\hbar}(\mathbf{q}'+\mathbf{q}) \cdot \mathbf{y}}
e^{\frac{i}{\hbar}(\mathbf{p}'-\mathbf{p}) \cdot \mathbf{x}} \nonumber \\
&-& \overline{B}_{ \alpha}(\mathbf{p})  B^{\dag}_{
\beta}(\mathbf{p}')\overline{F}_{ \gamma}(\mathbf{q}) D _{
\delta}(\mathbf{q}') e^{-\frac{i}{\hbar}(\mathbf{q}'+\mathbf{q})
\cdot \mathbf{y}} e^{\frac{i}{\hbar}(\mathbf{p}'-\mathbf{p}) \cdot
\mathbf{x}} \nonumber \nonumber \\
&-& \overline{B}_{ \alpha}(\mathbf{p}) B^{\dag}_{
\beta}(\mathbf{p}')\overline{F}_{ \gamma}(\mathbf{q}) F^{\dag} _{
\delta}(\mathbf{q}') e^{\frac{i}{\hbar}(\mathbf{q}'-\mathbf{q})
\cdot \mathbf{y}}
e^{\frac{i}{\hbar}(\mathbf{p}'-\mathbf{p}) \cdot \mathbf{x}} \nonumber \\
&-& \overline{D}^{\dag}_{ \alpha}(\mathbf{p})  D _{
\beta}(\mathbf{p}')\overline{A}^{\dag}_{ \gamma}(\mathbf{q})  A _{
\delta}(\mathbf{q}') e^{-\frac{i}{\hbar}(\mathbf{q}'-\mathbf{q})
\cdot \mathbf{y}} e^{-\frac{i}{\hbar}(\mathbf{p}'-\mathbf{p}) \cdot
\mathbf{x}} \nonumber \nonumber \\
&-& \overline{D}^{\dag}_{ \alpha}(\mathbf{p})  D _{
\beta}(\mathbf{p}')\overline{A}^{\dag}_{ \gamma}(\mathbf{q})
B^{\dag}_{ \delta}(\mathbf{q}')
e^{\frac{i}{\hbar}(\mathbf{q}'+\mathbf{q}) \cdot \mathbf{y}}
e^{-\frac{i}{\hbar}(\mathbf{p}'-\mathbf{p}) \cdot \mathbf{x}} \nonumber \\
&-& \overline{D}^{\dag}_{ \alpha}(\mathbf{p})  D _{
\beta}(\mathbf{p}')\overline{B}_{ \gamma}(\mathbf{q})  A _{
\delta}(\mathbf{q}') e^{-\frac{i}{\hbar}(\mathbf{q}'+\mathbf{q})
\cdot \mathbf{y}} e^{-\frac{i}{\hbar}(\mathbf{p}'-\mathbf{p}) \cdot
\mathbf{x}} \nonumber \nonumber \\
&-& \overline{D}^{\dag}_{ \alpha}(\mathbf{p})  D _{
\beta}(\mathbf{p}')\overline{B}_{ \gamma}(\mathbf{q})  B^{\dag}_{
\delta}(\mathbf{q}') e^{\frac{i}{\hbar}(\mathbf{q}'-\mathbf{q})
\cdot \mathbf{y}}
e^{-\frac{i}{\hbar}(\mathbf{p}'-\mathbf{p}) \cdot \mathbf{x}} \nonumber \\
&+& \overline{D}^{\dag}_{ \alpha}(\mathbf{p})  D _{
\beta}(\mathbf{p}')\overline{D}^{\dag}_{ \gamma}(\mathbf{q})  D _{
\delta}(\mathbf{q}') e^{-\frac{i}{\hbar}(\mathbf{q}'-\mathbf{q})
\cdot \mathbf{y}} e^{-\frac{i}{\hbar}(\mathbf{p}'-\mathbf{p}) \cdot
\mathbf{x}} \nonumber \nonumber \\
&+& \overline{D}^{\dag}_{ \alpha}(\mathbf{p})  D _{
\beta}(\mathbf{p}')\overline{D}^{\dag}_{ \gamma}(\mathbf{q})
F^{\dag} _{ \delta}(\mathbf{q}')
e^{+\frac{i}{\hbar}(\mathbf{q}'+\mathbf{q}) \cdot \mathbf{y}}
e^{-\frac{i}{\hbar}(\mathbf{p}'-\mathbf{p}) \cdot \mathbf{x}} \nonumber \\
&+& \overline{D}^{\dag}_{ \alpha}(\mathbf{p})  D _{
\beta}(\mathbf{p}')\overline{F}_{ \gamma}(\mathbf{q}) D _{
\delta}(\mathbf{q}') e^{-\frac{i}{\hbar}(\mathbf{q}'+\mathbf{q})
\cdot \mathbf{y}} e^{-\frac{i}{\hbar}(\mathbf{p}'-\mathbf{p}) \cdot
\mathbf{x}} \nonumber \nonumber \\
&+& \overline{D}^{\dag}_{ \alpha}(\mathbf{p})  D _{
\beta}(\mathbf{p}')\overline{F}_{ \gamma}(\mathbf{q})  F^{\dag} _{
\delta}(\mathbf{q}') e^{\frac{i}{\hbar}(\mathbf{q}'-\mathbf{q})
\cdot \mathbf{y}}
e^{-\frac{i}{\hbar}(\mathbf{p}'-\mathbf{p}) \cdot \mathbf{x}} \nonumber \\
&-& \overline{D}^{\dag}_{ \alpha}(\mathbf{p})  F^{\dag} _{
\beta}(\mathbf{p}')\overline{A}^{\dag}_{ \gamma}(\mathbf{q})  A _{
\delta}(\mathbf{q}') e^{-\frac{i}{\hbar}(\mathbf{q}'-\mathbf{q})
\cdot \mathbf{y}} e^{+\frac{i}{\hbar}(\mathbf{p}'+\mathbf{p}) \cdot
\mathbf{x}} \nonumber \nonumber \\
&-& \overline{D}^{\dag}_{ \alpha}(\mathbf{p})  F^{\dag} _{
\beta}(\mathbf{p}')\overline{A}^{\dag}_{ \gamma}(\mathbf{q})
B^{\dag}_{ \delta}(\mathbf{q}')
e^{\frac{i}{\hbar}(\mathbf{q}'+\mathbf{q}) \cdot \mathbf{y}}
e^{+\frac{i}{\hbar}(\mathbf{p}'+\mathbf{p}) \cdot \mathbf{x}}\nonumber \\
&-& \overline{D}^{\dag}_{ \alpha}(\mathbf{p})  F^{\dag} _{
\beta}(\mathbf{p}')\overline{B}_{ \gamma}(\mathbf{q})  A _{
\delta}(\mathbf{q}') e^{-\frac{i}{\hbar}(\mathbf{q}'+\mathbf{q})
\cdot \mathbf{y}} e^{+\frac{i}{\hbar}(\mathbf{p}'+\mathbf{p}) \cdot
\mathbf{x}} \nonumber \nonumber \\
&-& \overline{D}^{\dag}_{ \alpha}(\mathbf{p})  F^{\dag} _{
\beta}(\mathbf{p}')\overline{B}_{ \gamma}(\mathbf{q})  B^{\dag}_{
\delta}(\mathbf{q}') e^{\frac{i}{\hbar}(\mathbf{q}'-\mathbf{q})
\cdot \mathbf{y}}
e^{+\frac{i}{\hbar}(\mathbf{p}'+\mathbf{p}) \cdot \mathbf{x}}\nonumber \\
&+& \overline{D}^{\dag}_{ \alpha}(\mathbf{p})  F^{\dag} _{
\beta}(\mathbf{p}')\overline{D}^{\dag}_{ \gamma}(\mathbf{q})  D _{
\delta}(\mathbf{q}') e^{-\frac{i}{\hbar}(\mathbf{q}'-\mathbf{q})
\cdot \mathbf{y}} e^{+\frac{i}{\hbar}(\mathbf{p}'+\mathbf{p}) \cdot
\mathbf{x}} \nonumber \nonumber \\
&+& \overline{D}^{\dag}_{ \alpha}(\mathbf{p})  F^{\dag} _{
\beta}(\mathbf{p}')\overline{D}^{\dag}_{ \gamma}(\mathbf{q})
F^{\dag} _{ \delta}(\mathbf{q}')
e^{+\frac{i}{\hbar}(\mathbf{q}'+\mathbf{q}) \cdot \mathbf{y}}
e^{+\frac{i}{\hbar}(\mathbf{p}'+\mathbf{p}) \cdot \mathbf{x}}\nonumber \\
&+& \overline{D}^{\dag}_{ \alpha}(\mathbf{p})  F^{\dag} _{
\beta}(\mathbf{p}')\overline{F}_{ \gamma}(\mathbf{q}) D _{
\delta}(\mathbf{q}') e^{-\frac{i}{\hbar}(\mathbf{q}'+\mathbf{q})
\cdot \mathbf{y}} e^{+\frac{i}{\hbar}(\mathbf{p}'+\mathbf{p}) \cdot
\mathbf{x}} \nonumber \nonumber \\
&+&\overline{D}^{\dag}_{ \alpha}(\mathbf{p})  F^{\dag} _{
\beta}(\mathbf{p}')\overline{F}_{ \gamma}(\mathbf{q})  F^{\dag} _{
\delta}(\mathbf{q}') e^{\frac{i}{\hbar}(\mathbf{q}'-\mathbf{q})
\cdot \mathbf{y}}
e^{+\frac{i}{\hbar}(\mathbf{p}'+\mathbf{p}) \cdot \mathbf{x}} \nonumber \\
&-& \overline{F}_{ \alpha}(\mathbf{p}) D _{
\beta}(\mathbf{p}')\overline{A}^{\dag}_{ \gamma}(\mathbf{q})  A _{
\delta}(\mathbf{q}') e^{-\frac{i}{\hbar}(\mathbf{q}'-\mathbf{q})
\cdot \mathbf{y}} e^{-\frac{i}{\hbar}(\mathbf{p}'+\mathbf{p}) \cdot
\mathbf{x}} \nonumber \nonumber \\
&-& \overline{F}_{ \alpha}(\mathbf{p}) D _{
\beta}(\mathbf{p}')\overline{A}^{\dag}_{ \gamma}(\mathbf{q})
B^{\dag}_{ \delta}(\mathbf{q}')
e^{\frac{i}{\hbar}(\mathbf{q}'+\mathbf{q}) \cdot \mathbf{y}}
e^{-\frac{i}{\hbar}(\mathbf{p}'+\mathbf{p}) \cdot \mathbf{x}}\nonumber \\
&-& \overline{F}_{ \alpha}(\mathbf{p}) D _{
\beta}(\mathbf{p}')\overline{B}_{ \gamma}(\mathbf{q})  A _{
\delta}(\mathbf{q}') e^{-\frac{i}{\hbar}(\mathbf{q}'+\mathbf{q})
\cdot \mathbf{y}} e^{-\frac{i}{\hbar}(\mathbf{p}'+\mathbf{p}) \cdot
\mathbf{x}} \nonumber \nonumber \\
&-& \overline{F}_{ \alpha}(\mathbf{p}) D _{
\beta}(\mathbf{p}')\overline{B}_{ \gamma}(\mathbf{q})  B^{\dag}_{
\delta}(\mathbf{q}') e^{\frac{i}{\hbar}(\mathbf{q}'-\mathbf{q})
\cdot \mathbf{y}}
e^{-\frac{i}{\hbar}(\mathbf{p}'+\mathbf{p}) \cdot \mathbf{x}} \nonumber \\
&+& \overline{F}_{ \alpha}(\mathbf{p}) D _{
\beta}(\mathbf{p}')\overline{D}^{\dag}_{ \gamma}(\mathbf{q})  D _{
\delta}(\mathbf{q}') e^{-\frac{i}{\hbar}(\mathbf{q}'-\mathbf{q})
\cdot \mathbf{y}} e^{-\frac{i}{\hbar}(\mathbf{p}'+\mathbf{p}) \cdot
\mathbf{x}} \nonumber \nonumber \\
&+& \overline{F}_{ \alpha}(\mathbf{p}) D _{
\beta}(\mathbf{p}')\overline{D}^{\dag}_{ \gamma}(\mathbf{q})
F^{\dag} _{ \delta}(\mathbf{q}')
e^{+\frac{i}{\hbar}(\mathbf{q}'+\mathbf{q}) \cdot \mathbf{y}}
e^{-\frac{i}{\hbar}(\mathbf{p}'+\mathbf{p}) \cdot \mathbf{x}} \nonumber \\
&+& \overline{F}_{ \alpha}(\mathbf{p}) D _{
\beta}(\mathbf{p}')\overline{F}_{ \gamma}(\mathbf{q}) D _{
\delta}(\mathbf{q}') e^{-\frac{i}{\hbar}(\mathbf{q}'+\mathbf{q})
\cdot \mathbf{y}} e^{-\frac{i}{\hbar}(\mathbf{p}'+\mathbf{p}) \cdot
\mathbf{x}} \nonumber \nonumber \\
&+& \overline{F}_{ \alpha}(\mathbf{p}) D _{
\beta}(\mathbf{p}')\overline{F}_{ \gamma}(\mathbf{q})  F^{\dag} _{
\delta}(\mathbf{q}') e^{\frac{i}{\hbar}(\mathbf{q}'-\mathbf{q})
\cdot \mathbf{y}}
e^{-\frac{i}{\hbar}(\mathbf{p}'+\mathbf{p}) \cdot \mathbf{x}} \nonumber \\
&-& \overline{F}_{ \alpha}(\mathbf{p})  F^{\dag} _{
\beta}(\mathbf{p}')\overline{A}^{\dag}_{ \gamma}(\mathbf{q})  A _{
\delta}(\mathbf{q}') e^{-\frac{i}{\hbar}(\mathbf{q}'-\mathbf{q})
\cdot \mathbf{y}} e^{\frac{i}{\hbar}(\mathbf{p}'-\mathbf{p}) \cdot
\mathbf{x}} \nonumber \nonumber \\
&-& \overline{F}_{ \alpha}(\mathbf{p}) F^{\dag} _{
\beta}(\mathbf{p}')\overline{A}^{\dag}_{ \gamma}(\mathbf{q})
B^{\dag}_{ \delta}(\mathbf{q}')
e^{\frac{i}{\hbar}(\mathbf{q}'+\mathbf{q}) \cdot \mathbf{y}}
e^{\frac{i}{\hbar}(\mathbf{p}'-\mathbf{p}) \cdot \mathbf{x}} \nonumber \\
&-& \overline{F}_{ \alpha}(\mathbf{p})  F^{\dag} _{
\beta}(\mathbf{p}')\overline{B}_{ \gamma}(\mathbf{q})  A _{
\delta}(\mathbf{q}') e^{-\frac{i}{\hbar}(\mathbf{q}'+\mathbf{q})
\cdot \mathbf{y}} e^{\frac{i}{\hbar}(\mathbf{p}'-\mathbf{p}) \cdot
\mathbf{x}} \nonumber \nonumber \\
&-& \overline{F}_{ \alpha}(\mathbf{p}) F^{\dag} _{
\beta}(\mathbf{p}')\overline{B}_{ \gamma}(\mathbf{q}) B^{\dag}_{
\delta}(\mathbf{q}') e^{\frac{i}{\hbar}(\mathbf{q}'-\mathbf{q})
\cdot \mathbf{y}}
e^{\frac{i}{\hbar}(\mathbf{p}'-\mathbf{p}) \cdot \mathbf{x}}\nonumber \\
&+& \overline{F}_{ \alpha}(\mathbf{p})  F^{\dag} _{
\beta}(\mathbf{p}')\overline{D}^{\dag}_{ \gamma}(\mathbf{q})  D _{
\delta}(\mathbf{q}') e^{-\frac{i}{\hbar}(\mathbf{q}'-\mathbf{q})
\cdot \mathbf{y}} e^{\frac{i}{\hbar}(\mathbf{p}'-\mathbf{p}) \cdot
\mathbf{x}} \nonumber \nonumber \\
&+& \overline{F}_{ \alpha}(\mathbf{p}) F^{\dag} _{
\beta}(\mathbf{p}')\overline{D}^{\dag}_{ \gamma}(\mathbf{q})
F^{\dag} _{ \delta}(\mathbf{q}')
e^{+\frac{i}{\hbar}(\mathbf{q}'+\mathbf{q}) \cdot \mathbf{y}}
e^{\frac{i}{\hbar}(\mathbf{p}'-\mathbf{p}) \cdot \mathbf{x}} \nonumber \\
&+& \overline{F}_{ \alpha}(\mathbf{p})  F^{\dag} _{
\beta}(\mathbf{p}')\overline{F}_{ \gamma}(\mathbf{q}) D _{
\delta}(\mathbf{q}') e^{-\frac{i}{\hbar}(\mathbf{q}'+\mathbf{q})
\cdot \mathbf{y}} e^{\frac{i}{\hbar}(\mathbf{p}'-\mathbf{p}) \cdot
\mathbf{x}} \nonumber \nonumber \\
&+& \overline{F}_{ \alpha}(\mathbf{p}) F^{\dag} _{
\beta}(\mathbf{p}')\overline{F}_{ \gamma}(\mathbf{q}) F^{\dag} _{
\delta}(\mathbf{q}') e^{\frac{i}{\hbar}(\mathbf{q}'-\mathbf{q})
\cdot \mathbf{y}} e^{\frac{i}{\hbar}(\mathbf{p}'-\mathbf{p}) \cdot
\mathbf{x}} \Bigr) \label{eq:2nd-order-int}
\end{eqnarray}

\noindent We need to convert this expression to the normal order,
i.e., move all creation operators in front of annihilation
operators. After this is done we will obtain a sum of \emph{phys}, \emph{unphys}, and \emph{renorm}  terms. It can be shown that the \emph{renorm}  terms are
infinite. This is an indication of renormalization troubles with
QED. These troubles are discussed in detail in chapter \ref{ch:renormalization}. In
the rest of this section we will simply ignore the \emph{renorm}  part of
interaction.

There are some cancelations among \emph{unphys} terms. To see how they
work, let us convert to the normal order the 12th term in
(\ref{eq:2nd-order-int})

\begin{eqnarray*}
&\mbox{  }& e^2  (2\pi \hbar)^{-6}  \sum_{\alpha \beta \gamma
\delta}  \int d\mathbf{x} d\mathbf{y} \int d\mathbf{p} d\mathbf{p}'
d\mathbf{q}
d\mathbf{q}' \frac{\gamma^0_{\alpha \beta} \gamma^0_{\gamma \delta}}{8 \pi |\mathbf{x} - \mathbf{y}|} \times \\
&\mbox{  }&\overline{ A}^{\dag}_{ \alpha}(\mathbf{p})  B^{\dag}_{
\beta}(\mathbf{p}')\overline{B}_{ \gamma}(\mathbf{q})  B^{\dag}_{
\delta}(\mathbf{q}') e^{\frac{i}{\hbar}(\mathbf{q}'-\mathbf{q})
\cdot \mathbf{y}}
e^{\frac{i}{\hbar}(\mathbf{p}'+\mathbf{p}) \cdot \mathbf{x}} \\
&=& -e^2  (2\pi \hbar)^{-6}  \sum_{\alpha \beta \gamma \delta}  \int
d\mathbf{x} d\mathbf{y} \int d\mathbf{p} d\mathbf{p}' d\mathbf{q}
d\mathbf{q}' \frac{\gamma^0_{\alpha \beta} \gamma^0_{\gamma \delta}}{8 \pi |\mathbf{x} - \mathbf{y}|} \times \\
&\mbox{  }& \overline{A}^{\dag}_{ \alpha}(\mathbf{p})  B^{\dag}_{
\beta}(\mathbf{p}') B^{\dag}_{ \delta}(\mathbf{q}') \overline{B}_{
\gamma}(\mathbf{q})   e^{\frac{i}{\hbar}(\mathbf{q}'-\mathbf{q})
\cdot \mathbf{y}}
e^{\frac{i}{\hbar}(\mathbf{p}'+\mathbf{p}) \cdot \mathbf{x}} \\
&+&e^2  (2\pi \hbar)^{-6}  \sum_{\alpha \beta \gamma \delta}  \int
d\mathbf{x} d\mathbf{y} \int d\mathbf{p} d\mathbf{p}' d\mathbf{q}
d\mathbf{q}' \frac{\gamma^0_{\alpha \beta} \gamma^0_{\gamma \delta}}{8 \pi |\mathbf{x} - \mathbf{y}|} \times \\
&\mbox{  }& \overline{A}^{\dag}_{ \alpha}(\mathbf{p})  B^{\dag}_{
\beta}(\mathbf{p}') \{ \overline{B}_{ \gamma}(\mathbf{q}),
B^{\dag}_{ \delta}(\mathbf{q}')  \}
e^{\frac{i}{\hbar}(\mathbf{q}'-\mathbf{q}) \cdot \mathbf{y}}
e^{\frac{i}{\hbar}(\mathbf{p}'+\mathbf{p}) \cdot \mathbf{x}}
\end{eqnarray*}

\noindent Now we denote the second term on the right hand side
of this expression by $I$ and use (\ref{eq:B-over-B}),  (\ref{eq:trace-gamma}) -
(\ref{eq:trace-gamma2}), and (\ref{eq:int1/x})

\begin{eqnarray}
I &=& e^2  (2\pi \hbar)^{-6}  \sum_{\alpha \beta \gamma \delta} \int
d\mathbf{x} d\mathbf{y} \int d\mathbf{p} d\mathbf{p}' d\mathbf{q}
d\mathbf{q}' \frac{\gamma^0_{\alpha \beta} \gamma^0_{\gamma \delta}}{8 \pi |\mathbf{x} - \mathbf{y}|} \times \nonumber \\
&\mbox{  }& \overline{A}^{\dag}_{ \alpha}(\mathbf{p})  B^{\dag}_{
\beta}(\mathbf{p}') \frac{1}{2 \omega_{\mathbf{q}}}
(\gamma^0\omega_{\mathbf{q}} + \vec{\gamma} \mathbf{q} c -  mc^2
)_{\gamma \delta} \delta(\mathbf{q}'-\mathbf{q})
e^{\frac{i}{\hbar}(\mathbf{q}'-\mathbf{q}) \cdot \mathbf{y}}
e^{\frac{i}{\hbar}(\mathbf{p}'+\mathbf{p}) \cdot \mathbf{x}} \nonumber \\
&=& e^2  (2\pi \hbar)^{-6}  \sum_{\alpha \beta }  \int d\mathbf{x}
d\mathbf{y} \int d\mathbf{p} d\mathbf{p}'
d\mathbf{q} \frac{\gamma^0_{\alpha \beta}}{8 \pi |\mathbf{x} - \mathbf{y}|} \times \nonumber \\
&\mbox{  }& \overline{A}^{\dag}_{ \alpha}(\mathbf{p})  B^{\dag}_{
\beta}(\mathbf{p}') \frac{1}{2 \omega_{\mathbf{q}}}
(\omega_{\mathbf{q}} Tr(\gamma^0 \gamma^0) + \mathbf{q}c Tr
(\gamma^0 \vec{\gamma}) - mc^2 Tr(\gamma^0))
e^{\frac{i}{\hbar}(\mathbf{p}'+\mathbf{p}) \cdot \mathbf{x}} \nonumber \\
&=& 2e^2  (2\pi \hbar)^{-6}  \sum_{\alpha \beta}  \int d\mathbf{x}
d\mathbf{y} \int d\mathbf{p} d\mathbf{p}' d\mathbf{q}
\frac{\gamma^0_{\alpha \beta}}{8 \pi |\mathbf{x} - \mathbf{y}|}
\overline{A}^{\dag}_{ \alpha}(\mathbf{p})  B^{\dag}_{
\beta}(\mathbf{p}')
e^{\frac{i}{\hbar}(\mathbf{p}'+\mathbf{p}) \cdot \mathbf{x}} \nonumber \\
 &=& \frac{2e^2 \hbar^2} { (2\pi \hbar)^3}  \sum_{\alpha \beta}
\int d\mathbf{p} d\mathbf{p}' d\mathbf{q} \gamma^0_{\alpha \beta}
\overline{A}^{\dag}_{ \alpha}(\mathbf{p})  B^{\dag}_{
\beta}(\mathbf{p}')
\frac{\delta(\mathbf{p}'+\mathbf{p})}{(\mathbf{p}'+\mathbf{p})^2}
 \label{eq:inf-term}
\end{eqnarray}

\noindent This term is infinite. However there are three other
infinite terms in (\ref{eq:2nd-order-int}) that arise in a similar
manner from $-A^{\dag}B^{\dag}FF^{\dag} + BB^{\dag}A^{\dag}B^{\dag}
-FF^{\dag} A^{\dag}B^{\dag} $. These terms  cancel  exactly with
(\ref{eq:inf-term}). Similar to (\ref{eq:half}),
this cancelation is possible only because of the condition $q_{electron} + q_{proton}=0$.

Taking into account the above results and using anticommutators like
(\ref{eq:A-over-A}) and (\ref{eq:B-over-B}) we can bring the second
order interaction (\ref{eq:2nd-order-int}) to the normal order

\begin{eqnarray*}
V_2
&=& e^2  (2\pi \hbar)^{-6}  \sum_{\alpha \beta \gamma \delta}  \int
d\mathbf{x} d\mathbf{y} \int d\mathbf{p} d\mathbf{p}' d\mathbf{q}
d\mathbf{q}' \frac{\gamma^0_{\alpha \beta} \gamma^0_{\gamma \delta}}{8 \pi |\mathbf{x} - \mathbf{y}|} \times  \\
(&-& \overline{A}^{\dag}_{ \alpha}(\mathbf{p}) \overline{A}^{\dag}_{
\gamma}(\mathbf{q}) A _{\beta}(\mathbf{p}')   A _{
\delta}(\mathbf{q}') e^{-\frac{i}{\hbar}(\mathbf{q}'-\mathbf{q})
\cdot \mathbf{y}} e^{-\frac{i}{\hbar}(\mathbf{p}'-\mathbf{p}) \cdot
\mathbf{x}} \nonumber \\
&-& \overline{A}^{\dag}_{ \alpha}(\mathbf{p}) \overline{A}^{\dag}_{
\gamma}(\mathbf{q}) A _{ \beta}(\mathbf{p}') B^{\dag}_{
\delta}(\mathbf{q}') e^{\frac{i}{\hbar}(\mathbf{q}'+\mathbf{q})
\cdot \mathbf{y}}
e^{-\frac{i}{\hbar}(\mathbf{p}'-\mathbf{p}) \cdot \mathbf{x}}\\
&+& \overline{A}^{\dag}_{ \alpha}(\mathbf{p})  A
_{\beta}(\mathbf{p}') A _{\delta}(\mathbf{q}') \overline{B}_{
\gamma}(\mathbf{q})   e^{-\frac{i}{\hbar}(\mathbf{q}'+\mathbf{q})
\cdot \mathbf{y}} e^{-\frac{i}{\hbar}(\mathbf{p}'-\mathbf{p}) \cdot
\mathbf{x}} \nonumber \\
&-& \overline{A}^{\dag}_{ \alpha}(\mathbf{p})  A _{
\beta}(\mathbf{p}') B^{\dag}_{ \delta}(\mathbf{q}')\overline{B}_{
\gamma}(\mathbf{q})
  e^{\frac{i}{\hbar}(\mathbf{q}'-\mathbf{q}) \cdot
\mathbf{y}}
e^{-\frac{i}{\hbar}(\mathbf{p}'-\mathbf{p}) \cdot \mathbf{x}}\\
&-& \overline{A}^{\dag}_{ \alpha}(\mathbf{p})  A _{
\beta}(\mathbf{p}')\overline{D}^{\dag}_{ \gamma}(\mathbf{q})  D _{
\delta}(\mathbf{q}') e^{-\frac{i}{\hbar}(\mathbf{q}'-\mathbf{q})
\cdot \mathbf{y}} e^{-\frac{i}{\hbar}(\mathbf{p}'-\mathbf{p}) \cdot
\mathbf{x}} \nonumber \\
&-& \overline{A}^{\dag}_{ \alpha}(\mathbf{p})  A _{
\beta}(\mathbf{p}')\overline{D}^{\dag}_{ \gamma}(\mathbf{q})
F^{\dag} _{ \delta}(\mathbf{q}')
e^{+\frac{i}{\hbar}(\mathbf{q}'+\mathbf{q}) \cdot \mathbf{y}}
e^{-\frac{i}{\hbar}(\mathbf{p}'-\mathbf{p}) \cdot \mathbf{x}}\\
&-& \overline{A}^{\dag}_{ \alpha}(\mathbf{p})  A _{
\beta}(\mathbf{p}') D _{ \delta}(\mathbf{q}')\overline{F}_{
\gamma}(\mathbf{q})
 e^{-\frac{i}{\hbar}(\mathbf{q}'+\mathbf{q}) \cdot
\mathbf{y}} e^{-\frac{i}{\hbar}(\mathbf{p}'-\mathbf{p}) \cdot
\mathbf{x}} \nonumber \\
&+& \overline{A}^{\dag}_{ \alpha}(\mathbf{p})  A _{
\beta}(\mathbf{p}') F^{\dag} _{ \delta}(\mathbf{q}') \overline{F}_{
\gamma}(\mathbf{q})
 e^{\frac{i}{\hbar}(\mathbf{q}'-\mathbf{q}) \cdot
\mathbf{y}}
e^{-\frac{i}{\hbar}(\mathbf{p}'-\mathbf{p}) \cdot \mathbf{x}} \\
&+& \overline{A}^{\dag}_{ \alpha}(\mathbf{p}) \overline{A}^{\dag}_{
\gamma}(\mathbf{q})  A _{ \delta}(\mathbf{q}') B^{\dag}_{
\beta}(\mathbf{p}') e^{-\frac{i}{\hbar}(\mathbf{q}'-\mathbf{q})
\cdot \mathbf{y}} e^{\frac{i}{\hbar}(\mathbf{p}'+\mathbf{p}) \cdot
\mathbf{x}} \nonumber \\
&+& \overline{A}^{\dag}_{ \alpha}(\mathbf{p}) \overline{A}^{\dag}_{
\gamma}(\mathbf{q}) B^{\dag}_{ \beta}(\mathbf{p}') B^{\dag}_{
\delta}(\mathbf{q}') e^{\frac{i}{\hbar}(\mathbf{q}'+\mathbf{q})
\cdot \mathbf{y}}
e^{\frac{i}{\hbar}(\mathbf{p}'+\mathbf{p}) \cdot \mathbf{x}}\\
&+& \overline{A}^{\dag}_{ \alpha}(\mathbf{p}) A
_{\delta}(\mathbf{q}') B^{\dag}_{ \beta}(\mathbf{p}')\overline{B}_{
\gamma}(\mathbf{q})   e^{-\frac{i}{\hbar}(\mathbf{q}'+\mathbf{q})
\cdot \mathbf{y}} e^{\frac{i}{\hbar}(\mathbf{p}'+\mathbf{p}) \cdot
\mathbf{x}} \nonumber \\
&-& \overline{A}^{\dag}_{ \alpha}(\mathbf{p})  B^{\dag}_{
\beta}(\mathbf{p}') B^{\dag}_{ \delta}(\mathbf{q}')\overline{B}_{
\gamma}(\mathbf{q})
 e^{\frac{i}{\hbar}(\mathbf{q}'-\mathbf{q}) \cdot
\mathbf{y}}
e^{\frac{i}{\hbar}(\mathbf{p}'+\mathbf{p}) \cdot \mathbf{x}}\\
&-& \overline{A}^{\dag}_{ \alpha}(\mathbf{p})  B^{\dag}_{
\beta}(\mathbf{p}')\overline{D}^{\dag}_{ \gamma}(\mathbf{q})  D _{
\delta}(\mathbf{q}') e^{-\frac{i}{\hbar}(\mathbf{q}'-\mathbf{q})
\cdot \mathbf{y}} e^{\frac{i}{\hbar}(\mathbf{p}'+\mathbf{p}) \cdot
\mathbf{x}} \nonumber \\
&-& \overline{A}^{\dag}_{ \alpha}(\mathbf{p}) B^{\dag}_{
\beta}(\mathbf{p}')\overline{D}^{\dag}_{ \gamma}(\mathbf{q})
F^{\dag} _{ \delta}(\mathbf{q}')
e^{+\frac{i}{\hbar}(\mathbf{q}'+\mathbf{q}) \cdot \mathbf{y}}
e^{\frac{i}{\hbar}(\mathbf{p}'+\mathbf{p}) \cdot \mathbf{x}}\\
&-& \overline{A}^{\dag}_{ \alpha}(\mathbf{p})  B^{\dag}_{
\beta}(\mathbf{p}')D _{ \delta}(\mathbf{q}')\overline{F}_{
\gamma}(\mathbf{q})  e^{-\frac{i}{\hbar}(\mathbf{q}'+\mathbf{q})
\cdot \mathbf{y}} e^{\frac{i}{\hbar}(\mathbf{p}'+\mathbf{p}) \cdot
\mathbf{x}} \nonumber \\
&+& \overline{A}^{\dag}_{ \alpha}(\mathbf{p}) B^{\dag}_{
\beta}(\mathbf{p}') F^{\dag} _{ \delta}(\mathbf{q}') \overline{F}_{
\gamma}(\mathbf{q})
 e^{\frac{i}{\hbar}(\mathbf{q}'-\mathbf{q}) \cdot
\mathbf{y}}
e^{\frac{i}{\hbar}(\mathbf{p}'+\mathbf{p}) \cdot \mathbf{x}} \\
&-&   \overline{A}^{\dag}_{ \gamma}(\mathbf{q}) A _{
\beta}(\mathbf{p}')  A _{ \delta}(\mathbf{q}') \overline{B}_{
\alpha}(\mathbf{p}) e^{-\frac{i}{\hbar}(\mathbf{q}'-\mathbf{q})
\cdot \mathbf{y}} e^{-\frac{i}{\hbar}(\mathbf{p}'+\mathbf{p}) \cdot
\mathbf{x}} \nonumber \\
&+& \overline{A}^{\dag}_{ \gamma}(\mathbf{q}) A _{
\beta}(\mathbf{p}')    B^{\dag}_{ \delta}(\mathbf{q}')
\overline{B}_{ \alpha}(\mathbf{p})
e^{\frac{i}{\hbar}(\mathbf{q}'+\mathbf{q}) \cdot \mathbf{y}}
e^{-\frac{i}{\hbar}(\mathbf{p}'+\mathbf{p}) \cdot \mathbf{x}}\\
&+& A _{\beta}(\mathbf{p}')A _{\delta}(\mathbf{q}') \overline{B}_{
\alpha}(\mathbf{p})  \overline{B}_{ \gamma}(\mathbf{q})
e^{-\frac{i}{\hbar}(\mathbf{q}'+\mathbf{q}) \cdot \mathbf{y}}
e^{-\frac{i}{\hbar}(\mathbf{p}'+\mathbf{p}) \cdot \mathbf{x}}
\nonumber \\
&+& A _{\beta}(\mathbf{p}')B^{\dag}_{ \delta}(\mathbf{q}')
\overline{B}_{ \alpha}(\mathbf{p}) \overline{B}_{
\gamma}(\mathbf{q}) e^{\frac{i}{\hbar}(\mathbf{q}'-\mathbf{q}) \cdot
\mathbf{y}}
e^{-\frac{i}{\hbar}(\mathbf{p}'+\mathbf{p}) \cdot \mathbf{x}} \\
&-& A _{ \beta}(\mathbf{p}')\overline{B}_{ \alpha}(\mathbf{p})
\overline{D}^{\dag}_{ \gamma}(\mathbf{q})  D _{ \delta}(\mathbf{q}')
e^{-\frac{i}{\hbar}(\mathbf{q}'-\mathbf{q}) \cdot \mathbf{y}}
e^{-\frac{i}{\hbar}(\mathbf{p}'+\mathbf{p}) \cdot \mathbf{x}}
\nonumber \\
&-& A _{\beta}(\mathbf{p}') \overline{B}_{ \alpha}(\mathbf{p})
\overline{D}^{\dag}_{ \gamma}(\mathbf{q}) F^{\dag} _{
\delta}(\mathbf{q}') e^{+\frac{i}{\hbar}(\mathbf{q}'+\mathbf{q})
\cdot \mathbf{y}}
e^{-\frac{i}{\hbar}(\mathbf{p}'+\mathbf{p}) \cdot \mathbf{x}}\\
&-& A _{\beta}(\mathbf{p}')\overline{B}_{ \alpha}(\mathbf{p})  D _{
\delta}(\mathbf{q}')  \overline{F}_{ \gamma}(\mathbf{q})
e^{-\frac{i}{\hbar}(\mathbf{q}'+\mathbf{q}) \cdot \mathbf{y}}
e^{-\frac{i}{\hbar}(\mathbf{p}'+\mathbf{p}) \cdot \mathbf{x}}
\nonumber \\
&+& A _{\beta}(\mathbf{p}')\overline{B}_{ \alpha}(\mathbf{p})
F^{\dag} _{ \delta}(\mathbf{q}') \overline{F}_{ \gamma}(\mathbf{q})
 e^{\frac{i}{\hbar}(\mathbf{q}'-\mathbf{q}) \cdot
\mathbf{y}}
e^{-\frac{i}{\hbar}(\mathbf{p}'+\mathbf{p}) \cdot \mathbf{x}} \\
&-& \overline{A}^{\dag}_{ \gamma}(\mathbf{q})  A _{
\delta}(\mathbf{q}') B^{\dag}_{ \beta}(\mathbf{p}') \overline{B}_{
\alpha}(\mathbf{p})   e^{-\frac{i}{\hbar}(\mathbf{q}'-\mathbf{q})
\cdot \mathbf{y}} e^{\frac{i}{\hbar}(\mathbf{p}'-\mathbf{p}) \cdot
\mathbf{x}} \nonumber \\
&+& \overline{A}^{\dag}_{ \gamma}(\mathbf{q}) B^{\dag}_{
\beta}(\mathbf{p}') B^{\dag}_{ \delta}(\mathbf{q}') \overline{B}_{
\alpha}(\mathbf{p}) e^{\frac{i}{\hbar}(\mathbf{q}'+\mathbf{q}) \cdot
\mathbf{y}}
e^{\frac{i}{\hbar}(\mathbf{p}'-\mathbf{p}) \cdot \mathbf{x}} \\
&-& A _{ \delta}(\mathbf{q}')B^{\dag}_{ \beta}(\mathbf{p}')
\overline{B}_{ \alpha}(\mathbf{p})  \overline{B}_{
\gamma}(\mathbf{q})   e^{-\frac{i}{\hbar}(\mathbf{q}'+\mathbf{q})
\cdot \mathbf{y}} e^{\frac{i}{\hbar}(\mathbf{p}'-\mathbf{p}) \cdot
\mathbf{x}} \nonumber \\
&-& B^{\dag}_{ \beta}(\mathbf{p}') B^{\dag}_{ \delta}(\mathbf{q}')
\overline{B}_{ \alpha}(\mathbf{p}) \overline{B}_{
\gamma}(\mathbf{q}) e^{\frac{i}{\hbar}(\mathbf{q}'-\mathbf{q}) \cdot
\mathbf{y}}
e^{\frac{i}{\hbar}(\mathbf{p}'-\mathbf{p}) \cdot \mathbf{x}} \\
&+& B^{\dag}_{ \beta}(\mathbf{p}') \overline{B}_{
\alpha}(\mathbf{p}) \overline{D}^{\dag}_{ \gamma}(\mathbf{q})  D _{
\delta}(\mathbf{q}') e^{-\frac{i}{\hbar}(\mathbf{q}'-\mathbf{q})
\cdot \mathbf{y}} e^{\frac{i}{\hbar}(\mathbf{p}'-\mathbf{p}) \cdot
\mathbf{x}} \nonumber \\
&+& B^{\dag}_{ \beta}(\mathbf{p}') \overline{B}_{
\alpha}(\mathbf{p})  \overline{D}^{\dag}_{ \gamma}(\mathbf{q})
F^{\dag} _{ \delta}(\mathbf{q}')
e^{+\frac{i}{\hbar}(\mathbf{q}'+\mathbf{q}) \cdot \mathbf{y}}
e^{\frac{i}{\hbar}(\mathbf{p}'-\mathbf{p}) \cdot \mathbf{x}} \\
&+& B^{\dag}_{ \beta}(\mathbf{p}') \overline{B}_{
\alpha}(\mathbf{p})   D _{ \delta}(\mathbf{q}') \overline{F}_{
\gamma}(\mathbf{q}) e^{-\frac{i}{\hbar}(\mathbf{q}'+\mathbf{q})
\cdot \mathbf{y}} e^{\frac{i}{\hbar}(\mathbf{p}'-\mathbf{p}) \cdot
\mathbf{x}} \nonumber \\
&-& B^{\dag}_{ \beta}(\mathbf{p}') \overline{B}_{
\alpha}(\mathbf{p}) F^{\dag} _{ \delta}(\mathbf{q}' \overline{F}_{
\gamma}(\mathbf{q}) e^{\frac{i}{\hbar}(\mathbf{q}'-\mathbf{q}) \cdot
\mathbf{y}}
e^{\frac{i}{\hbar}(\mathbf{p}'-\mathbf{p}) \cdot \mathbf{x}} \\
&-& \overline{A}^{\dag}_{ \gamma}(\mathbf{q})  A _{
\delta}(\mathbf{q}') \overline{D}^{\dag}_{ \alpha}(\mathbf{p})  D _{
\beta}(\mathbf{p}') e^{-\frac{i}{\hbar}(\mathbf{q}'-\mathbf{q})
\cdot \mathbf{y}} e^{-\frac{i}{\hbar}(\mathbf{p}'-\mathbf{p}) \cdot
\mathbf{x}} \nonumber \\
&-& \overline{A}^{\dag}_{ \gamma}(\mathbf{q})  B^{\dag}_{
\delta}(\mathbf{q}') \overline{D}^{\dag}_{ \alpha}(\mathbf{p})  D _{
\beta}(\mathbf{p}') e^{\frac{i}{\hbar}(\mathbf{q}'+\mathbf{q}) \cdot
\mathbf{y}}
e^{-\frac{i}{\hbar}(\mathbf{p}'-\mathbf{p}) \cdot \mathbf{x}} \\
&-& A _{ \delta}(\mathbf{q}') \overline{B}_{ \gamma}(\mathbf{q})
\overline{D}^{\dag}_{ \alpha}(\mathbf{p})  D _{ \beta}(\mathbf{p}')
e^{-\frac{i}{\hbar}(\mathbf{q}'+\mathbf{q}) \cdot \mathbf{y}}
e^{-\frac{i}{\hbar}(\mathbf{p}'-\mathbf{p}) \cdot \mathbf{x}}
\nonumber \\
&+& B^{\dag}_{ \delta}(\mathbf{q}') \overline{B}_{
\gamma}(\mathbf{q}) \overline{D}^{\dag}_{ \alpha}(\mathbf{p})  D _{
\beta}(\mathbf{p}') e^{\frac{i}{\hbar}(\mathbf{q}'-\mathbf{q}) \cdot
\mathbf{y}}
e^{-\frac{i}{\hbar}(\mathbf{p}'-\mathbf{p}) \cdot \mathbf{x}} \\
&-& \overline{D}^{\dag}_{ \alpha}(\mathbf{q}) \overline{D}^{\dag}_{
\gamma}(\mathbf{p})  D _{ \beta}(\mathbf{q}')  D _{
\delta}(\mathbf{p}') e^{-\frac{i}{\hbar}(\mathbf{q}'-\mathbf{q})
\cdot \mathbf{y}} e^{-\frac{i}{\hbar}(\mathbf{p}'-\mathbf{p}) \cdot
\mathbf{x}} \nonumber \\
&-& \overline{D}^{\dag}_{ \alpha}(\mathbf{p}) \overline{D}^{\dag}_{
\gamma}(\mathbf{q}) D _{ \beta}(\mathbf{p}') F^{\dag} _{
\delta}(\mathbf{q}') e^{+\frac{i}{\hbar}(\mathbf{q}'+\mathbf{q})
\cdot \mathbf{y}}
e^{-\frac{i}{\hbar}(\mathbf{p}'-\mathbf{p}) \cdot \mathbf{x}} \\
&+& \overline{D}^{\dag}_{ \alpha}(\mathbf{p})  D _{
\beta}(\mathbf{p}') D _{ \delta}(\mathbf{q}')\overline{F}_{
\gamma}(\mathbf{q})  e^{-\frac{i}{\hbar}(\mathbf{q}'+\mathbf{q})
\cdot \mathbf{y}} e^{-\frac{i}{\hbar}(\mathbf{p}'-\mathbf{p}) \cdot
\mathbf{x}} \nonumber \\
&-&\overline{D}^{\dag}_{ \alpha}(\mathbf{p}) D _{
\beta}(\mathbf{p}') F^{\dag} _{ \delta}(\mathbf{q}') \overline{F}_{
\gamma}(\mathbf{q}) e^{\frac{i}{\hbar}(\mathbf{q}'-\mathbf{q}) \cdot
\mathbf{y}}
e^{-\frac{i}{\hbar}(\mathbf{p}'-\mathbf{p}) \cdot \mathbf{x}} \\
&-& \overline{A}^{\dag}_{ \gamma}(\mathbf{q})  A _{
\delta}(\mathbf{q}')\overline{D}^{\dag}_{ \alpha}(\mathbf{p})
F^{\dag} _{ \beta}(\mathbf{p}')
e^{-\frac{i}{\hbar}(\mathbf{q}'-\mathbf{q}) \cdot \mathbf{y}}
e^{+\frac{i}{\hbar}(\mathbf{p}'+\mathbf{p}) \cdot \mathbf{x}}
\nonumber \\
&-& \overline{A}^{\dag}_{ \gamma}(\mathbf{q}) B^{\dag}_{
\delta}(\mathbf{q}') \overline{D}^{\dag}_{ \alpha}(\mathbf{p})
F^{\dag} _{ \beta}(\mathbf{p}')
e^{\frac{i}{\hbar}(\mathbf{q}'+\mathbf{q}) \cdot \mathbf{y}}
e^{+\frac{i}{\hbar}(\mathbf{p}'+\mathbf{p}) \cdot \mathbf{x}}\\
&-& A _{ \delta}(\mathbf{q}') \overline{B}_{ \gamma}(\mathbf{q})
\overline{D}^{\dag}_{ \alpha}(\mathbf{p})  F^{\dag} _{
\beta}(\mathbf{p}')   e^{-\frac{i}{\hbar}(\mathbf{q}'+\mathbf{q})
\cdot \mathbf{y}} e^{+\frac{i}{\hbar}(\mathbf{p}'+\mathbf{p}) \cdot
\mathbf{x}} \nonumber \\
&+& B^{\dag}_{ \delta}(\mathbf{q}') \overline{B}_{
\gamma}(\mathbf{q}) \overline{D}^{\dag}_{ \alpha}(\mathbf{p})
F^{\dag} _{ \beta}(\mathbf{p}')
e^{\frac{i}{\hbar}(\mathbf{q}'-\mathbf{q}) \cdot \mathbf{y}}
e^{+\frac{i}{\hbar}(\mathbf{p}'+\mathbf{p}) \cdot \mathbf{x}}\\
&+& \overline{D}^{\dag}_{ \alpha}(\mathbf{p})  \overline{D}^{\dag}_{
\gamma}(\mathbf{q})  D _{ \delta}(\mathbf{q}') F^{\dag} _{
\beta}(\mathbf{p}') e^{-\frac{i}{\hbar}(\mathbf{q}'-\mathbf{q})
\cdot \mathbf{y}} e^{+\frac{i}{\hbar}(\mathbf{p}'+\mathbf{p}) \cdot
\mathbf{x}} \nonumber \\
&+& \overline{D}^{\dag}_{ \alpha}(\mathbf{p}) \overline{D}^{\dag}_{
\gamma}(\mathbf{q}) F^{\dag} _{ \beta}(\mathbf{p}') F^{\dag} _{
\delta}(\mathbf{q}') e^{+\frac{i}{\hbar}(\mathbf{q}'+\mathbf{q})
\cdot \mathbf{y}}
e^{+\frac{i}{\hbar}(\mathbf{p}'+\mathbf{p}) \cdot \mathbf{x}}\\
&+& \overline{D}^{\dag}_{ \alpha}(\mathbf{p}) D _{
\delta}(\mathbf{q}') F^{\dag} _{ \beta}(\mathbf{p}')\overline{F}_{
\gamma}(\mathbf{q})  e^{-\frac{i}{\hbar}(\mathbf{q}'+\mathbf{q})
\cdot \mathbf{y}} e^{+\frac{i}{\hbar}(\mathbf{p}'+\mathbf{p}) \cdot
\mathbf{x}} \nonumber \\
&-&\overline{D}^{\dag}_{ \alpha}(\mathbf{p}) F^{\dag} _{
\beta}(\mathbf{p}')F^{\dag} _{ \delta}(\mathbf{q}') \overline{F}_{
\gamma}(\mathbf{q}) e^{\frac{i}{\hbar}(\mathbf{q}'-\mathbf{q}) \cdot
\mathbf{y}}
e^{+\frac{i}{\hbar}(\mathbf{p}'+\mathbf{p}) \cdot \mathbf{x}} \\
&-& \overline{A}^{\dag}_{ \gamma}(\mathbf{q})  A _{
\delta}(\mathbf{q}')  D _{ \beta}(\mathbf{p}') \overline{F}_{
\alpha}(\mathbf{p}) e^{-\frac{i}{\hbar}(\mathbf{q}'-\mathbf{q})
\cdot \mathbf{y}} e^{-\frac{i}{\hbar}(\mathbf{p}'+\mathbf{p}) \cdot
\mathbf{x}} \nonumber \\
&-& \overline{A}^{\dag}_{ \gamma}(\mathbf{q})  B^{\dag}_{
\delta}(\mathbf{q}')  D _{ \beta}(\mathbf{p}') \overline{F}_{
\alpha}(\mathbf{p}) e^{\frac{i}{\hbar}(\mathbf{q}'+\mathbf{q}) \cdot
\mathbf{y}}
e^{-\frac{i}{\hbar}(\mathbf{p}'+\mathbf{p}) \cdot \mathbf{x}}\\
&-& A _{ \delta}(\mathbf{q}') \overline{B}_{ \gamma}(\mathbf{q})  D
_{ \beta}(\mathbf{p}') \overline{F}_{ \alpha}(\mathbf{p})
e^{-\frac{i}{\hbar}(\mathbf{q}'+\mathbf{q}) \cdot \mathbf{y}}
e^{-\frac{i}{\hbar}(\mathbf{p}'+\mathbf{p}) \cdot \mathbf{x}}
\nonumber \\
&+& B^{\dag}_{ \beta}(\mathbf{q}') \overline{B}_{
\gamma}(\mathbf{q})  D _{ \beta}(\mathbf{p}') \overline{F}_{
\alpha}(\mathbf{p}) e^{\frac{i}{\hbar}(\mathbf{q}'-\mathbf{q}) \cdot
\mathbf{y}}
e^{-\frac{i}{\hbar}(\mathbf{p}'+\mathbf{p}) \cdot \mathbf{x}} \\
&-&  \overline{D}^{\dag}_{ \gamma}(\mathbf{q}) D
_{\beta}(\mathbf{p}')  D _{ \delta}(\mathbf{q}') \overline{F}_{
\alpha}(\mathbf{p}) e^{-\frac{i}{\hbar}(\mathbf{q}'-\mathbf{q})
\cdot \mathbf{y}} e^{-\frac{i}{\hbar}(\mathbf{p}'+\mathbf{p}) \cdot
\mathbf{x}} \nonumber \\
&+& \overline{D}^{\dag}_{ \gamma}(\mathbf{q}) D _{
\beta}(\mathbf{p}') F^{\dag} _{ \delta}(\mathbf{q}') \overline{F}_{
\alpha}(\mathbf{p}) e^{+\frac{i}{\hbar}(\mathbf{q}'+\mathbf{q})
\cdot \mathbf{y}}
e^{-\frac{i}{\hbar}(\mathbf{p}'+\mathbf{p}) \cdot \mathbf{x}} \\
&+& D _{ \beta}(\mathbf{p}')D _{ \delta}(\mathbf{q}') \overline{F}_{
\alpha}(\mathbf{p}) \overline{F}_{ \gamma}(\mathbf{q})
e^{-\frac{i}{\hbar}(\mathbf{q}'+\mathbf{q}) \cdot \mathbf{y}}
e^{-\frac{i}{\hbar}(\mathbf{p}'+\mathbf{p}) \cdot \mathbf{x}}
\nonumber \\
&+& D _{ \beta}(\mathbf{p}') F^{\dag} _{ \delta}(\mathbf{q}')
\overline{F}_{ \alpha}(\mathbf{p}) \overline{F}_{
\gamma}(\mathbf{q}) e^{\frac{i}{\hbar}(\mathbf{q}'-\mathbf{q}) \cdot
\mathbf{y}}
e^{-\frac{i}{\hbar}(\mathbf{p}'+\mathbf{p}) \cdot \mathbf{x}} \\
&+& \overline{A}^{\dag}_{ \gamma}(\mathbf{q})  A _{
\delta}(\mathbf{q}') F^{\dag} _{ \beta}(\mathbf{p}') \overline{F}_{
\alpha}(\mathbf{p})   e^{-\frac{i}{\hbar}(\mathbf{q}'-\mathbf{q})
\cdot \mathbf{y}} e^{\frac{i}{\hbar}(\mathbf{p}'-\mathbf{p}) \cdot
\mathbf{x}} \nonumber \\
&+& \overline{A}^{\dag}_{ \gamma}(\mathbf{q})  B^{\dag}_{
\delta}(\mathbf{q}') F^{\dag} _{ \beta}(\mathbf{p}') \overline{F}_{
\alpha}(\mathbf{p}) e^{\frac{i}{\hbar}(\mathbf{q}'+\mathbf{q}) \cdot
\mathbf{y}}
e^{\frac{i}{\hbar}(\mathbf{p}'-\mathbf{p}) \cdot \mathbf{x}} \\
&+& A _{ \delta}(\mathbf{q}') \overline{B}_{ \gamma}(\mathbf{q})
F^{\dag} _{ \beta}(\mathbf{p}') \overline{F}_{ \alpha}(\mathbf{p})
e^{-\frac{i}{\hbar}(\mathbf{q}'+\mathbf{q}) \cdot \mathbf{y}}
e^{\frac{i}{\hbar}(\mathbf{p}'-\mathbf{p}) \cdot \mathbf{x}}
\nonumber \\
&-& B^{\dag}_{ \delta}(\mathbf{q}') \overline{B}_{
\gamma}(\mathbf{q}) F^{\dag} _{ \beta}(\mathbf{p}')\overline{F}_{
\alpha}(\mathbf{p}) e^{\frac{i}{\hbar}(\mathbf{q}'-\mathbf{q}) \cdot
\mathbf{y}}
e^{\frac{i}{\hbar}(\mathbf{p}'-\mathbf{p}) \cdot \mathbf{x}}\\
&-& \overline{D}^{\dag}_{ \gamma}(\mathbf{q})  D _{
\delta}(\mathbf{q}') F^{\dag} _{ \beta}(\mathbf{p}') \overline{F}_{
\alpha}(\mathbf{p})   e^{-\frac{i}{\hbar}(\mathbf{q}'-\mathbf{q})
\cdot \mathbf{y}} e^{\frac{i}{\hbar}(\mathbf{p}'-\mathbf{p}) \cdot
\mathbf{x}} \nonumber \\
&+&  \overline{D}^{\dag}_{ \gamma}(\mathbf{q}) F^{\dag} _{
\beta}(\mathbf{p}')  F^{\dag} _{ \delta}(\mathbf{q}') \overline{F}_{
\alpha}(\mathbf{p}) e^{+\frac{i}{\hbar}(\mathbf{q}'+\mathbf{q})
\cdot \mathbf{y}}
e^{\frac{i}{\hbar}(\mathbf{p}'-\mathbf{p}) \cdot \mathbf{x}} \\
&-& D _{ \delta}(\mathbf{q}') F^{\dag} _{ \beta}(\mathbf{p}')
\overline{F}_{ \alpha}(\mathbf{p})  \overline{F}_{
\gamma}(\mathbf{q})  e^{-\frac{i}{\hbar}(\mathbf{q}'+\mathbf{q})
\cdot \mathbf{y}} e^{\frac{i}{\hbar}(\mathbf{p}'-\mathbf{p}) \cdot
\mathbf{x}} \nonumber \\
&-&F^{\dag} _{ \beta}(\mathbf{p}') F^{\dag} _{
\delta}(\mathbf{q}')\overline{F}_{ \alpha}(\mathbf{p})
\overline{F}_{ \gamma}(\mathbf{q})
 e^{\frac{i}{\hbar}(\mathbf{q}'-\mathbf{q}) \cdot
\mathbf{y}} e^{\frac{i}{\hbar}(\mathbf{p}'-\mathbf{p}) \cdot
\mathbf{x}})
\end{eqnarray*}

Next we switch summation labels $\alpha \leftrightarrow \gamma$ and integration variables $\mathbf{x} \leftrightarrow \mathbf{y}$
and $\mathbf{p} \leftrightarrow \mathbf{q}$  to simplify

\begin{eqnarray*}
V_2
&=& e^2  (2\pi \hbar)^{-6}  \sum_{\alpha \beta \gamma \delta}  \int
d\mathbf{x} d\mathbf{y} \int d\mathbf{p} d\mathbf{p}' d\mathbf{q}
d\mathbf{q}' \frac{\gamma^0_{\alpha \beta} \gamma^0_{\gamma \delta}}{8 \pi |\mathbf{x} - \mathbf{y}|} \times \\
(&-& \overline{A}^{\dag}_{ \alpha}(\mathbf{p}) \overline{A}^{\dag}_{
\gamma}(\mathbf{q}) A _{ \beta}(\mathbf{p}')   A _{
\delta}(\mathbf{q}') e^{-\frac{i}{\hbar}(\mathbf{q}'-\mathbf{q})
\cdot \mathbf{y}}
e^{-\frac{i}{\hbar}(\mathbf{p}'-\mathbf{p}) \cdot \mathbf{x}} \\
&+& 2\overline{A}^{\dag}_{ \alpha}(\mathbf{p})  A
_{\beta}(\mathbf{p}') A _{\delta}(\mathbf{q}') \overline{B}_{
\gamma}(\mathbf{q})   e^{-\frac{i}{\hbar}(\mathbf{q}'+\mathbf{q})
\cdot \mathbf{y}} e^{-\frac{i}{\hbar}(\mathbf{p}'-\mathbf{p}) \cdot
\mathbf{x}} \nonumber \\
&-& 2\overline{A}^{\dag}_{ \alpha}(\mathbf{p})  A _{
\beta}(\mathbf{p}') B^{\dag}_{ \delta}(\mathbf{q}')\overline{B}_{
\gamma}(\mathbf{q})
  e^{\frac{i}{\hbar}(\mathbf{q}'-\mathbf{q}) \cdot
\mathbf{y}}
e^{-\frac{i}{\hbar}(\mathbf{p}'-\mathbf{p}) \cdot \mathbf{x}}\\
&-& 2\overline{A}^{\dag}_{ \alpha}(\mathbf{p})  A _{
\beta}(\mathbf{p}')\overline{D}^{\dag}_{ \gamma}(\mathbf{q})  D _{
\delta}(\mathbf{q}') e^{-\frac{i}{\hbar}(\mathbf{q}'-\mathbf{q})
\cdot \mathbf{y}} e^{-\frac{i}{\hbar}(\mathbf{p}'-\mathbf{p}) \cdot
\mathbf{x}} \nonumber \\
&-& 2\overline{A}^{\dag}_{ \alpha}(\mathbf{p})  A _{
\beta}(\mathbf{p}')\overline{D}^{\dag}_{ \gamma}(\mathbf{q})
F^{\dag} _{ \delta}(\mathbf{q}')
e^{+\frac{i}{\hbar}(\mathbf{q}'+\mathbf{q}) \cdot \mathbf{y}}
e^{-\frac{i}{\hbar}(\mathbf{p}'-\mathbf{p}) \cdot \mathbf{x}}\\
&-& 2\overline{A}^{\dag}_{ \alpha}(\mathbf{p})  A _{
\beta}(\mathbf{p}') D _{ \delta}(\mathbf{q}')\overline{F}_{
\gamma}(\mathbf{q})
 e^{-\frac{i}{\hbar}(\mathbf{q}'+\mathbf{q}) \cdot
\mathbf{y}} e^{-\frac{i}{\hbar}(\mathbf{p}'-\mathbf{p}) \cdot
\mathbf{x}} \nonumber \\
&+& 2\overline{A}^{\dag}_{ \alpha}(\mathbf{p})  A _{
\beta}(\mathbf{p}') F^{\dag} _{ \delta}(\mathbf{q}') \overline{F}_{
\gamma}(\mathbf{q})
 e^{\frac{i}{\hbar}(\mathbf{q}'-\mathbf{q}) \cdot
\mathbf{y}}
e^{-\frac{i}{\hbar}(\mathbf{p}'-\mathbf{p}) \cdot \mathbf{x}} \\
&+& 2\overline{A}^{\dag}_{ \alpha}(\mathbf{p}) \overline{A}^{\dag}_{
\gamma}(\mathbf{q})  A _{ \delta}(\mathbf{q}') B^{\dag}_{
\beta}(\mathbf{p}') e^{-\frac{i}{\hbar}(\mathbf{q}'-\mathbf{q})
\cdot \mathbf{y}} e^{\frac{i}{\hbar}(\mathbf{p}'+\mathbf{p}) \cdot
\mathbf{x}} \nonumber \\
&+& \overline{A}^{\dag}_{ \alpha}(\mathbf{p})  \overline{A}^{\dag}_{
\gamma}(\mathbf{q}) B^{\dag}_{ \beta}(\mathbf{p}') B^{\dag}_{
\delta}(\mathbf{q}') e^{\frac{i}{\hbar}(\mathbf{q}'+\mathbf{q})
\cdot \mathbf{y}}
e^{\frac{i}{\hbar}(\mathbf{p}'+\mathbf{p}) \cdot \mathbf{x}}\\
&+& 2\overline{A}^{\dag}_{ \alpha}(\mathbf{p}) A
_{\delta}(\mathbf{q}') B^{\dag}_{ \beta}(\mathbf{p}')\overline{B}_{
\gamma}(\mathbf{q})   e^{-\frac{i}{\hbar}(\mathbf{q}'+\mathbf{q})
\cdot \mathbf{y}} e^{\frac{i}{\hbar}(\mathbf{p}'+\mathbf{p}) \cdot
\mathbf{x}} \nonumber \\
&-& 2\overline{A}^{\dag}_{ \alpha}(\mathbf{p})  B^{\dag}_{
\beta}(\mathbf{p}') B^{\dag}_{ \delta}(\mathbf{q}')\overline{B}_{
\gamma}(\mathbf{q})
 e^{\frac{i}{\hbar}(\mathbf{q}'-\mathbf{q}) \cdot
\mathbf{y}}
e^{\frac{i}{\hbar}(\mathbf{p}'+\mathbf{p}) \cdot \mathbf{x}}\\
&-& 2\overline{A}^{\dag}_{ \alpha}(\mathbf{p})  B^{\dag}_{
\beta}(\mathbf{p}')\overline{D}^{\dag}_{ \gamma}(\mathbf{q})  D _{
\delta}(\mathbf{q}') e^{-\frac{i}{\hbar}(\mathbf{q}'-\mathbf{q})
\cdot \mathbf{y}} e^{\frac{i}{\hbar}(\mathbf{p}'+\mathbf{p}) \cdot
\mathbf{x}} \nonumber \\
&-& 2\overline{A}^{\dag}_{ \alpha}(\mathbf{p})  B^{\dag}_{
\beta}(\mathbf{p}')\overline{D}^{\dag}_{ \gamma}(\mathbf{q})
F^{\dag} _{ \delta}(\mathbf{q}')
e^{+\frac{i}{\hbar}(\mathbf{q}'+\mathbf{q}) \cdot \mathbf{y}}
e^{\frac{i}{\hbar}(\mathbf{p}'+\mathbf{p}) \cdot \mathbf{x}}\\
&-& 2\overline{A}^{\dag}_{ \alpha}(\mathbf{p})  B^{\dag}_{
\beta}(\mathbf{p}')D _{ \delta}(\mathbf{q}')\overline{F}_{
\gamma}(\mathbf{q})  e^{-\frac{i}{\hbar}(\mathbf{q}'+\mathbf{q})
\cdot \mathbf{y}} e^{\frac{i}{\hbar}(\mathbf{p}'+\mathbf{p}) \cdot
\mathbf{x}} \nonumber \\
&+& 2\overline{A}^{\dag}_{ \alpha}(\mathbf{p})  B^{\dag}_{
\beta}(\mathbf{p}') F^{\dag} _{ \delta}(\mathbf{q}') \overline{F}_{
\gamma}(\mathbf{q})
 e^{\frac{i}{\hbar}(\mathbf{q}'-\mathbf{q}) \cdot
\mathbf{y}}
e^{\frac{i}{\hbar}(\mathbf{p}'+\mathbf{p}) \cdot \mathbf{x}} \\
&+& A _{\beta}(\mathbf{p}')A _{\delta}(\mathbf{q}') \overline{B}_{
\alpha}(\mathbf{p})  \overline{B}_{ \gamma}(\mathbf{q})
e^{-\frac{i}{\hbar}(\mathbf{q}'+\mathbf{q}) \cdot \mathbf{y}}
e^{-\frac{i}{\hbar}(\mathbf{p}'+\mathbf{p}) \cdot \mathbf{x}}
\nonumber \\
&+& 2 A _{\beta}(\mathbf{p}')B^{\dag}_{ \delta}(\mathbf{q}')
\overline{B}_{ \alpha}(\mathbf{p}) \overline{B}_{
\gamma}(\mathbf{q}) e^{\frac{i}{\hbar}(\mathbf{q}'-\mathbf{q}) \cdot
\mathbf{y}}
e^{-\frac{i}{\hbar}(\mathbf{p}'+\mathbf{p}) \cdot \mathbf{x}} \\
&-& 2A _{ \beta}(\mathbf{p}')\overline{B}_{ \alpha}(\mathbf{p})
\overline{D}^{\dag}_{ \gamma}(\mathbf{q})  D _{ \delta}(\mathbf{q}')
e^{-\frac{i}{\hbar}(\mathbf{q}'-\mathbf{q}) \cdot \mathbf{y}}
e^{-\frac{i}{\hbar}(\mathbf{p}'+\mathbf{p}) \cdot \mathbf{x}}
\nonumber \\
&-& 2A _{\beta}(\mathbf{p}') \overline{B}_{ \alpha}(\mathbf{p})
\overline{D}^{\dag}_{ \gamma}(\mathbf{q}) F^{\dag} _{
\delta}(\mathbf{q}') e^{+\frac{i}{\hbar}(\mathbf{q}'+\mathbf{q})
\cdot \mathbf{y}}
e^{-\frac{i}{\hbar}(\mathbf{p}'+\mathbf{p}) \cdot \mathbf{x}}\\
&-& 2\overline{A} _{\beta}(\mathbf{p}')B_{ \alpha}(\mathbf{p})  D _{
\delta}(\mathbf{q}')  \overline{F}_{ \gamma}(\mathbf{q})
e^{-\frac{i}{\hbar}(\mathbf{q}'+\mathbf{q}) \cdot \mathbf{y}}
e^{-\frac{i}{\hbar}(\mathbf{p}'+\mathbf{p}) \cdot \mathbf{x}}
\nonumber \\
&+& 2\overline{A} _{\beta}(\mathbf{p}')B_{ \alpha}(\mathbf{p})
F^{\dag} _{ \delta}(\mathbf{q}') \overline{F}_{ \gamma}(\mathbf{q})
 e^{\frac{i}{\hbar}(\mathbf{q}'-\mathbf{q}) \cdot
\mathbf{y}}
e^{-\frac{i}{\hbar}(\mathbf{p}'+\mathbf{p}) \cdot \mathbf{x}} \\
&-&  B^{\dag}_{ \beta}(\mathbf{p}') B^{\dag}_{ \delta}(\mathbf{q}')
\overline{B}_{ \alpha}(\mathbf{p})  \overline{B}_{
\gamma}(\mathbf{q})   e^{\frac{i}{\hbar}(\mathbf{q}'-\mathbf{q})
\cdot \mathbf{y}}
e^{\frac{i}{\hbar}(\mathbf{p}'-\mathbf{p}) \cdot \mathbf{x}} \\
&+& 2B^{\dag}_{ \beta}(\mathbf{p}') \overline{B}_{
\alpha}(\mathbf{p})  \overline{D}^{\dag}_{ \gamma}(\mathbf{q})  D _{
\delta}(\mathbf{q}') e^{-\frac{i}{\hbar}(\mathbf{q}'-\mathbf{q})
\cdot \mathbf{y}} e^{\frac{i}{\hbar}(\mathbf{p}'-\mathbf{p}) \cdot
\mathbf{x}} \nonumber \\&+& 2B^{\dag}_{ \beta}(\mathbf{p}')
\overline{B}_{ \alpha}(\mathbf{p})  \overline{D}^{\dag}_{
\gamma}(\mathbf{q}) F^{\dag} _{ \delta}(\mathbf{q}')
e^{+\frac{i}{\hbar}(\mathbf{q}'+\mathbf{q}) \cdot \mathbf{y}}
e^{\frac{i}{\hbar}(\mathbf{p}'-\mathbf{p}) \cdot \mathbf{x}} \\
&+& 2B^{\dag}_{ \beta}(\mathbf{p}') \overline{B}_{
\alpha}(\mathbf{p})   D _{ \delta}(\mathbf{q}') \overline{F}_{
\gamma}(\mathbf{q}) e^{-\frac{i}{\hbar}(\mathbf{q}'+\mathbf{q})
\cdot \mathbf{y}} e^{\frac{i}{\hbar}(\mathbf{p}'-\mathbf{p}) \cdot
\mathbf{x}} \nonumber \\
&-& 2B^{\dag}_{ \beta}(\mathbf{p}') \overline{B}_{
\alpha}(\mathbf{p}) F^{\dag} _{ \delta}(\mathbf{q}') \overline{F}_{
\gamma}(\mathbf{q}) e^{\frac{i}{\hbar}(\mathbf{q}'-\mathbf{q}) \cdot
\mathbf{y}}
e^{\frac{i}{\hbar}(\mathbf{p}'-\mathbf{p}) \cdot \mathbf{x}} \\
&-& \overline{D}^{\dag}_{ \alpha}(\mathbf{p}) \overline{D}^{\dag}_{
\gamma}(\mathbf{q})  D _{ \beta}(\mathbf{p}')  D _{
\delta}(\mathbf{q}') e^{-\frac{i}{\hbar}(\mathbf{q}'-\mathbf{q})
\cdot \mathbf{y}} e^{-\frac{i}{\hbar}(\mathbf{p}'-\mathbf{p}) \cdot
\mathbf{x}} \nonumber \\
&-& 2\overline{D}^{\dag}_{ \alpha}(\mathbf{p}) \overline{D}^{\dag}_{
\gamma}(\mathbf{q}) D _{ \beta}(\mathbf{p}') F^{\dag} _{
\delta}(\mathbf{q}') e^{+\frac{i}{\hbar}(\mathbf{q}'+\mathbf{q})
\cdot \mathbf{y}}
e^{-\frac{i}{\hbar}(\mathbf{p}'-\mathbf{p}) \cdot \mathbf{x}} \\
&+& 2\overline{D}^{\dag}_{ \alpha}(\mathbf{p})  D _{
\beta}(\mathbf{p}') D _{ \delta}(\mathbf{q}')\overline{F}_{
\gamma}(\mathbf{q})  e^{-\frac{i}{\hbar}(\mathbf{q}'+\mathbf{q})
\cdot \mathbf{y}} e^{-\frac{i}{\hbar}(\mathbf{p}'-\mathbf{p}) \cdot
\mathbf{x}} \nonumber \\
&-&2 \overline{D}^{\dag}_{ \alpha}(\mathbf{p})  D _{
\beta}(\mathbf{p}') F^{\dag} _{ \gamma}(\mathbf{q}') \overline{F}_{
\delta}(\mathbf{q}) e^{\frac{i}{\hbar}(\mathbf{q}'-\mathbf{q}) \cdot
\mathbf{y}}
e^{-\frac{i}{\hbar}(\mathbf{p}'-\mathbf{p}) \cdot \mathbf{x}} \\
&+& \overline{D}^{\dag}_{ \alpha}(\mathbf{p})  \overline{D}^{\dag}_{
\gamma}(\mathbf{q}) F^{\dag} _{ \beta}(\mathbf{p}') F^{\dag} _{
\delta}(\mathbf{q}') e^{+\frac{i}{\hbar}(\mathbf{q}'+\mathbf{q})
\cdot \mathbf{y}}
e^{+\frac{i}{\hbar}(\mathbf{p}'+\mathbf{p}) \cdot \mathbf{x}}\\
&+& 2\overline{D}^{\dag}_{ \alpha}(\mathbf{p}) D _{
\delta}(\mathbf{q}') F^{\dag} _{ \beta}(\mathbf{p}')\overline{F}_{
\gamma}(\mathbf{q})  e^{-\frac{i}{\hbar}(\mathbf{q}'+\mathbf{q})
\cdot \mathbf{y}} e^{+\frac{i}{\hbar}(\mathbf{p}'+\mathbf{p}) \cdot
\mathbf{x}} \nonumber \\
&-&2\overline{D}^{\dag}_{ \alpha}(\mathbf{p})  F^{\dag} _{
\beta}(\mathbf{p}')F^{\dag} _{ \delta}(\mathbf{q}') \overline{F}_{
\gamma}(\mathbf{q}) e^{\frac{i}{\hbar}(\mathbf{q}'-\mathbf{q}) \cdot
\mathbf{y}}
e^{+\frac{i}{\hbar}(\mathbf{p}'+\mathbf{p}) \cdot \mathbf{x}} \\
&+& D _{ \beta}(\mathbf{p}')D _{ \delta}(\mathbf{q}') \overline{F}_{
\alpha}(\mathbf{p}) \overline{F}_{ \gamma}(\mathbf{q})
e^{-\frac{i}{\hbar}(\mathbf{q}'+\mathbf{q}) \cdot \mathbf{y}}
e^{-\frac{i}{\hbar}(\mathbf{p}'+\mathbf{p}) \cdot \mathbf{x}}
\nonumber \\
&+& 2D _{ \beta}(\mathbf{p}') F^{\dag} _{ \delta}(\mathbf{q}')
\overline{F}_{ \alpha}(\mathbf{p}) \overline{F}_{
\gamma}(\mathbf{q}) e^{\frac{i}{\hbar}(\mathbf{q}'-\mathbf{q}) \cdot
\mathbf{y}}
e^{-\frac{i}{\hbar}(\mathbf{p}'+\mathbf{p}) \cdot \mathbf{x}} \\
&-& F^{\dag} _{ \beta}(\mathbf{p}') F^{\dag} _{
\delta}(\mathbf{q}')\overline{F}_{ \alpha}(\mathbf{p})
\overline{F}_{ \gamma}(\mathbf{q})
 e^{\frac{i}{\hbar}(\mathbf{q}'-\mathbf{q}) \cdot
\mathbf{y}} e^{\frac{i}{\hbar}(\mathbf{p}'-\mathbf{p}) \cdot
\mathbf{x}})
\end{eqnarray*}

Integrals over $\mathbf{x}$ and $\mathbf{y}$ can be evaluated by using
formula (\ref{eq:int1/x})

\begin{eqnarray*}
V_2
&=& \frac{e^2 \hbar^2 }  { 2(2\pi \hbar)^{3}}  \sum_{\alpha \beta
\gamma \delta} \int d\mathbf{p} d\mathbf{p}' d\mathbf{q}
d\mathbf{q}'
\gamma^0_{\alpha \beta} \gamma^0_{\gamma \delta}  \times \\
\Bigl(&-& \overline{A}^{\dag}_{ \alpha}(\mathbf{p}) \overline{A}^{\dag}_{
\gamma}(\mathbf{q}) A _{ \beta}(\mathbf{p}')   A _{
\delta}(\mathbf{q}') \delta (\mathbf{q}'-\mathbf{q} +
\mathbf{p}'-\mathbf{p}) \frac{1}{|\mathbf{q}'-\mathbf{q}|^2} \\
&+& 2\overline{A}^{\dag}_{ \alpha}(\mathbf{p})  A
_{\beta}(\mathbf{p}') A _{\delta}(\mathbf{q}') \overline{B}_{
\gamma}(\mathbf{q}) \delta(\mathbf{q}'+\mathbf{q} +
\mathbf{p}'-\mathbf{p})
\frac{1}{ |\mathbf{q}'+\mathbf{q}|^2} \\
&-& 2\overline{A}^{\dag}_{ \alpha}(\mathbf{p})  A _{
\beta}(\mathbf{p}') B^{\dag}_{ \delta}(\mathbf{q}')\overline{B}_{
\gamma}(\mathbf{q}) \delta (\mathbf{q}'-\mathbf{q} - \mathbf{p}'+
\mathbf{p})
\frac{1}{ |\mathbf{q}'-\mathbf{q}|^2} \\
&-& 2\overline{A}^{\dag}_{ \alpha}(\mathbf{p})  A _{
\beta}(\mathbf{p}')\overline{D}^{\dag}_{ \gamma}(\mathbf{q})  D _{
\delta}(\mathbf{q}') \delta(\mathbf{q}'-\mathbf{q} +
\mathbf{p}'-\mathbf{p}) \frac{1}{ |\mathbf{q}'-\mathbf{q}|^2} \\
&-& 2\overline{A}^{\dag}_{ \alpha}(\mathbf{p})  A _{
\beta}(\mathbf{p}')\overline{D}^{\dag}_{ \gamma}(\mathbf{q})
F^{\dag} _{ \delta}(\mathbf{q}') \delta(\mathbf{q}'+\mathbf{q} -
\mathbf{p}'+\mathbf{p})
\frac{1}{ |\mathbf{q}'+\mathbf{q}|^2}\\
&-& 2\overline{A}^{\dag}_{ \alpha}(\mathbf{p})  A _{
\beta}(\mathbf{p}') D _{ \delta}(\mathbf{q}')\overline{F}_{
\gamma}(\mathbf{q}) \delta(\mathbf{q}'+\mathbf{q} +
\mathbf{p}'-\mathbf{p})
\frac{1}{ |\mathbf{q}'+\mathbf{q}|^2} \\
&+& 2\overline{A}^{\dag}_{ \alpha}(\mathbf{p})  A _{
\beta}(\mathbf{p}') F^{\dag} _{ \delta}(\mathbf{q}') \overline{F}_{
\gamma}(\mathbf{q}) \delta(\mathbf{q}'-\mathbf{q} -
\mathbf{p}'+\mathbf{p})
\frac{1}{ |\mathbf{q}'-\mathbf{q}|^2} \\
&+& 2\overline{A}^{\dag}_{ \alpha}(\mathbf{p}) \overline{A}^{\dag}_{
\gamma}(\mathbf{q})  A _{ \delta}(\mathbf{q}') B^{\dag}_{
\beta}(\mathbf{p}') \delta(\mathbf{q}'-\mathbf{q}-
\mathbf{p}'-\mathbf{p} )
\frac{1}{ |\mathbf{q}'-\mathbf{q}|^2} \\
&+& \overline{A}^{\dag}_{ \alpha}(\mathbf{p})  \overline{A}^{\dag}_{
\gamma}(\mathbf{q}) B^{\dag}_{ \beta}(\mathbf{p}') B^{\dag}_{
\delta}(\mathbf{q}') \delta (\mathbf{q}'+\mathbf{q} +
\mathbf{p}'+\mathbf{p} )
\frac{1}{ |\mathbf{q}'+\mathbf{q}|^2} \\
&+& 2\overline{A}^{\dag}_{ \alpha}(\mathbf{p}) A
_{\delta}(\mathbf{q}') B^{\dag}_{ \beta}(\mathbf{p}')\overline{B}_{
\gamma}(\mathbf{q}) \delta(\mathbf{q}'+\mathbf{q} -
\mathbf{p}'-\mathbf{p})
\frac{1}{ |\mathbf{q}'+\mathbf{q}|^2} \\
&-& 2\overline{A}^{\dag}_{ \alpha}(\mathbf{p})  B^{\dag}_{
\beta}(\mathbf{p}') B^{\dag}_{ \delta}(\mathbf{q}')\overline{B}_{
\gamma}(\mathbf{q}) \delta(\mathbf{q}'-\mathbf{q} +
\mathbf{p}'+\mathbf{p})
\frac{1}{ |\mathbf{q}'-\mathbf{q}|^2} \\
&-& 2\overline{A}^{\dag}_{ \alpha}(\mathbf{p})  B^{\dag}_{
\beta}(\mathbf{p}')\overline{D}^{\dag}_{ \gamma}(\mathbf{q})  D _{
\delta}(\mathbf{q}') \delta(\mathbf{q}'-\mathbf{q} -
\mathbf{p}'-\mathbf{p})
\frac{1}{ |\mathbf{q}'-\mathbf{q}|^2} \\
&-& 2\overline{A}^{\dag}_{ \alpha}(\mathbf{p})  B^{\dag}_{
\beta}(\mathbf{p}')\overline{D}^{\dag}_{ \gamma}(\mathbf{q})
F^{\dag} _{ \delta}(\mathbf{q}') \delta(\mathbf{q}'+\mathbf{q} +
\mathbf{p}'+\mathbf{p})
\frac{1}{ |\mathbf{q}'+\mathbf{q}|^2} \\
&-& 2\overline{A}^{\dag}_{ \alpha}(\mathbf{p})  B^{\dag}_{
\beta}(\mathbf{p}')D _{ \delta}(\mathbf{q}')\overline{F}_{
\gamma}(\mathbf{q}) \delta(\mathbf{q}'+\mathbf{q} -
\mathbf{p}'-\mathbf{p} )
\frac{1}{ |\mathbf{q}'+\mathbf{q}|^2} \\
&+& 2\overline{A}^{\dag}_{ \alpha}(\mathbf{p})  B^{\dag}_{
\beta}(\mathbf{p}') F^{\dag} _{ \delta}(\mathbf{q}') \overline{F}_{
\gamma}(\mathbf{q}) \delta(\mathbf{q}'-\mathbf{q} +
\mathbf{p}'+\mathbf{p})
\frac{1}{ |\mathbf{q}'-\mathbf{q}|^2} \\
&+& A _{\beta}(\mathbf{p}')A _{\delta}(\mathbf{q}') \overline{B}_{
\alpha}(\mathbf{p})  \overline{B}_{ \gamma}(\mathbf{q})
\delta(\mathbf{q}'+\mathbf{q} + \mathbf{p}'+\mathbf{p})
\frac{1}{ |\mathbf{q}'+\mathbf{q}|^2} \\
&+& 2 A _{\beta}(\mathbf{p}')B^{\dag}_{ \delta}(\mathbf{q}')
\overline{B}_{ \alpha}(\mathbf{p}) \overline{B}_{
\gamma}(\mathbf{q}) \delta(\mathbf{q}'-\mathbf{q} -
\mathbf{p}'-\mathbf{p})
\frac{1}{ |\mathbf{q}'-\mathbf{q}|^2} \\
&-& 2A _{ \beta}(\mathbf{p}')\overline{B}_{ \alpha}(\mathbf{p})
\overline{D}^{\dag}_{ \gamma}(\mathbf{q})  D _{ \delta}(\mathbf{q}')
\delta(\mathbf{q}'-\mathbf{q} + \mathbf{p}'+\mathbf{p})
\frac{1}{ |\mathbf{q}'-\mathbf{q}|^2} \\
&-& 2A _{\beta}(\mathbf{p}') \overline{B}_{ \alpha}(\mathbf{p})
\overline{D}^{\dag}_{ \gamma}(\mathbf{q})  F^{\dag} _{
\delta}(\mathbf{q}') \delta(\mathbf{q}'+\mathbf{q} -
\mathbf{p}'-\mathbf{p} )
\frac{1}{ |\mathbf{q}'+\mathbf{q}|^2} \\
&-& 2\overline{A} _{\beta}(\mathbf{p}')B_{ \alpha}(\mathbf{p})  D _{
\delta}(\mathbf{q}')  \overline{F}_{ \gamma}(\mathbf{q})
\delta(\mathbf{q}'+\mathbf{q} + \mathbf{p}'+\mathbf{p})
\frac{1}{ |\mathbf{q}'+\mathbf{q}|^2} \\
&+& 2\overline{A} _{\beta}(\mathbf{p}')B_{ \alpha}(\mathbf{p})
F^{\dag} _{ \delta}(\mathbf{q}') \overline{F}_{ \gamma}(\mathbf{q})
\delta(\mathbf{q}'-\mathbf{q} - \mathbf{p}'-\mathbf{p})
\frac{1}{ |\mathbf{q}'-\mathbf{q}|^2} \\
&-&  B^{\dag}_{ \beta}(\mathbf{p}') B^{\dag}_{ \delta}(\mathbf{q}')
\overline{B}_{ \alpha}(\mathbf{p})  \overline{B}_{
\gamma}(\mathbf{q}) \delta(\mathbf{q}'-\mathbf{q} +
\mathbf{p}'-\mathbf{p})
\frac{1}{ |\mathbf{q}'-\mathbf{q}|^2} \\
&+& 2B^{\dag}_{ \beta}(\mathbf{p}') \overline{B}_{
\alpha}(\mathbf{p})  \overline{D}^{\dag}_{ \gamma}(\mathbf{q})  D _{
\delta}(\mathbf{q}') \delta(\mathbf{q}'-\mathbf{q} -
\mathbf{p}'+\mathbf{p})
\frac{1}{ |\mathbf{q}'-\mathbf{q}|^2} \\
&+& 2B^{\dag}_{ \beta}(\mathbf{p}') \overline{B}_{
\alpha}(\mathbf{p})  \overline{D}^{\dag}_{ \gamma}(\mathbf{q})
F^{\dag} _{ \delta}(\mathbf{q}') \delta(\mathbf{q}'+\mathbf{q} +
\mathbf{p}'-\mathbf{p})
\frac{1}{ |\mathbf{q}'+\mathbf{q}|^2} \\
&+& 2B^{\dag}_{ \beta}(\mathbf{p}') \overline{B}_{
\alpha}(\mathbf{p})   D _{ \delta}(\mathbf{q}') \overline{F}_{
\gamma}(\mathbf{q}) \delta(\mathbf{q}'+\mathbf{q} -
\mathbf{p}'+\mathbf{p})
\frac{1}{ |\mathbf{q}'+\mathbf{q}|^2} \\
&-& 2B^{\dag}_{ \beta}(\mathbf{p}') \overline{B}_{
\alpha}(\mathbf{p}) F^{\dag} _{ \delta}(\mathbf{q}') \overline{F}_{
\gamma}(\mathbf{q}) \delta(\mathbf{q}'-\mathbf{q} +
\mathbf{p}'-\mathbf{p})
\frac{1}{ |\mathbf{q}'-\mathbf{q}|^2} \\
&-& \overline{D}^{\dag}_{ \alpha}(\mathbf{p}) \overline{D}^{\dag}_{
\gamma}(\mathbf{q})  D _{ \beta}(\mathbf{p}')  D _{
\delta}(\mathbf{q}') \delta(\mathbf{q}'-\mathbf{q} +
\mathbf{p}'-\mathbf{p})
\frac{1}{ |\mathbf{q}'-\mathbf{q}|^2} \\
&-& 2\overline{D}^{\dag}_{ \alpha}(\mathbf{p}) \overline{D}^{\dag}_{
\gamma}(\mathbf{q}) D _{ \beta}(\mathbf{p}')  F^{\dag} _{
\delta}(\mathbf{q}') \delta(\mathbf{q}'+\mathbf{q} - \mathbf{p}'+
\mathbf{p})
\frac{1}{ |\mathbf{q}'+\mathbf{q}|^2}  \\
&+& 2\overline{D}^{\dag}_{ \alpha}(\mathbf{p})  D _{
\beta}(\mathbf{p}') D _{ \delta}(\mathbf{q}')\overline{F}_{
\gamma}(\mathbf{q}) \delta(\mathbf{q}'+\mathbf{q} +
\mathbf{p}'-\mathbf{p})
\frac{1}{ |\mathbf{q}'+\mathbf{q}|^2} \\
&-&2 \overline{D}^{\dag}_{ \alpha}(\mathbf{p})  D _{
\beta}(\mathbf{p}') F^{\dag} _{ \gamma}(\mathbf{q}') \overline{F}_{
\delta}(\mathbf{q}) \delta(\mathbf{q}'-\mathbf{q} -
\mathbf{p}'+\mathbf{p})
\frac{1}{ |\mathbf{q}'-\mathbf{q}|^2} \\
&+& \overline{D}^{\dag}_{ \alpha}(\mathbf{p})  \overline{D}^{\dag}_{
\gamma}(\mathbf{q}) F^{\dag} _{ \beta}(\mathbf{p}') F^{\dag} _{
\delta}(\mathbf{q}') \delta(\mathbf{q}'+\mathbf{q} +
\mathbf{p}'+\mathbf{p})
\frac{1}{ |\mathbf{q}'+\mathbf{q}|^2} \\
&+& 2\overline{D}^{\dag}_{ \alpha}(\mathbf{p}) D _{
\delta}(\mathbf{q}') F^{\dag} _{ \beta}(\mathbf{p}')\overline{F}_{
\gamma}(\mathbf{q}) \delta(\mathbf{q}'+\mathbf{q} -
\mathbf{p}'-\mathbf{p})
\frac{1}{ |\mathbf{q}'+\mathbf{q}|^2} \\
&-&2\overline{D}^{\dag}_{ \alpha}(\mathbf{p})  F^{\dag} _{
\beta}(\mathbf{p}')F^{\dag} _{ \delta}(\mathbf{q}') \overline{F}_{
\gamma}(\mathbf{q}) \delta(\mathbf{q}'-\mathbf{q} +
\mathbf{p}'+\mathbf{p})
\frac{1}{ |\mathbf{q}'-\mathbf{q}|^2} \\
&+& D _{ \beta}(\mathbf{p}')D _{ \delta}(\mathbf{q}') \overline{F}_{
\alpha}(\mathbf{p}) \overline{F}_{ \gamma}(\mathbf{q})
\delta(\mathbf{q}'+\mathbf{q} + \mathbf{p}'+\mathbf{p})
\frac{1}{ |\mathbf{q}'+\mathbf{q}|^2} \\
& +& 2D _{ \beta}(\mathbf{p}') F^{\dag} _{ \delta}(\mathbf{q}')
\overline{F}_{ \alpha}(\mathbf{p}) \overline{F}_{
\gamma}(\mathbf{q}) \delta(\mathbf{q}'-\mathbf{q} -
\mathbf{p}'-\mathbf{p})
\frac{1}{ |\mathbf{q}'-\mathbf{q}|^2} \\
&-& F^{\dag} _{ \beta}(\mathbf{p}') F^{\dag} _{
\delta}(\mathbf{q}')\overline{F}_{ \alpha}(\mathbf{p})
\overline{F}_{ \gamma}(\mathbf{q}) \delta(\mathbf{q}'-\mathbf{q} +
\mathbf{p}'-\mathbf{p}) \frac{1}{ |\mathbf{q}'-\mathbf{q}|^2} \Bigr)
\end{eqnarray*}

\noindent Finally, we  integrate this expression on $\mathbf{q}'$ and divide $V_2$ into \emph{phys} and \emph{unphys} parts

\begin{eqnarray*}
 V_2 = V_2^{phys} + V_2^{unphys}
\end{eqnarray*}

\begin{eqnarray}
 V_2^{phys}
&=& \frac{e^2 \hbar^2}  {2(2\pi \hbar)^{3}}    \sum_{\alpha \beta
\gamma \delta} \int d\mathbf{p} d\mathbf{p}' d\mathbf{q}
\gamma^0_{\alpha \beta}
\gamma^0_{\gamma \delta}  \times \nonumber \\
\Bigl(&-& \overline{A}^{\dag}_{ \alpha}(\mathbf{p}) \overline{A}^{\dag}_{
\gamma}(\mathbf{q}) A _{ \beta}(\mathbf{p}')   A _{
\delta}(\mathbf{q} - \mathbf{p}'+\mathbf{p})  \frac{1}{|
\mathbf{p}'-\mathbf{p}|^2} \nonumber \\
&-& 2\overline{A}^{\dag}_{ \alpha}(\mathbf{p})  A _{
\beta}(\mathbf{p}') B^{\dag}_{ \delta}(\mathbf{q} + \mathbf{p}'-
\mathbf{p})\overline{B}_{ \gamma}(\mathbf{q}) \frac{1}{ |
\mathbf{p}'- \mathbf{p}|^2} \nonumber \\
&-& 2\overline{A}^{\dag}_{ \alpha}(\mathbf{p})  A _{
\beta}(\mathbf{p}')\overline{D}^{\dag}_{ \gamma}(\mathbf{q})  D _{
\delta}(\mathbf{q} -
\mathbf{p}'+\mathbf{p})  \frac{1}{ |  \mathbf{p}'-\mathbf{p}|^2} \nonumber \\
&+& 2\overline{A}^{\dag}_{ \alpha}(\mathbf{p})  A _{
\beta}(\mathbf{p}') F^{\dag} _{ \delta}(+\mathbf{q} +
\mathbf{p}'-\mathbf{p}) \overline{F}_{ \gamma}(\mathbf{q}) \frac{1}{
| \mathbf{p}'-\mathbf{p}|^2} \nonumber \\
&+& 2\overline{A}^{\dag}_{ \alpha}(\mathbf{p}) A
_{\delta}(-\mathbf{q} + \mathbf{p}'+\mathbf{p}) B^{\dag}_{
\beta}(\mathbf{p}')\overline{B}_{ \gamma}(\mathbf{q})
\frac{1}{ | \mathbf{p}'+\mathbf{p}|^2} \nonumber \\
&-& 2\overline{A}^{\dag}_{ \alpha}(\mathbf{p})  B^{\dag}_{
\beta}(\mathbf{p}')D _{ \delta}(-\mathbf{q} +
\mathbf{p}'+\mathbf{p})\overline{F}_{ \gamma}(\mathbf{q})
\frac{1}{ | \mathbf{p}'+\mathbf{p}|^2} \nonumber \\
&-& 2A _{\beta}(\mathbf{p}') \overline{B}_{ \alpha}(\mathbf{p})
\overline{D}^{\dag}_{ \gamma}(\mathbf{q})  F^{\dag} _{
\delta}(-\mathbf{q} + \mathbf{p}'+\mathbf{p}) \frac{1}{ |
\mathbf{p}'+ \mathbf{p}|^2} \nonumber \\
&-& 2\overline{A} _{\beta}(\mathbf{p}')B_{ \alpha}(\mathbf{p})  D _{
\delta}(-\mathbf{q} - \mathbf{p}'-\mathbf{p})  \overline{F}_{
\gamma}(\mathbf{q})
\frac{1}{ | \mathbf{p}'+\mathbf{p}|^2} \nonumber \\
&-&  B^{\dag}_{ \beta}(\mathbf{p}') B^{\dag}_{ \delta}(\mathbf{q} -
\mathbf{p}'+\mathbf{p}) \overline{B}_{ \alpha}(\mathbf{p})
\overline{B}_{ \gamma}(\mathbf{q})
\frac{1}{ | \mathbf{p}'-\mathbf{p}|^2} \nonumber \\
&+& 2B^{\dag}_{ \beta}(\mathbf{p}') \overline{B}_{
\alpha}(\mathbf{p})  \overline{D}^{\dag}_{ \gamma}(\mathbf{q})  D _{
\delta}(\mathbf{q} +\mathbf{p}'-\mathbf{p}) \frac{1}{ |
\mathbf{p}'-\mathbf{p}|^2} \nonumber \\
&-& 2B^{\dag}_{ \beta}(\mathbf{p}') \overline{B}_{
\alpha}(\mathbf{p}) F^{\dag} _{ \delta}(\mathbf{q} -
\mathbf{p}'+\mathbf{p}) \overline{F}_{ \gamma}(\mathbf{q})
\frac{1}{ | \mathbf{p}'-\mathbf{p}|^2} \nonumber \\
&-& \overline{D}^{\dag}_{ \alpha}(\mathbf{p}) \overline{D}^{\dag}_{
\gamma}(\mathbf{q})  D _{ \beta}(\mathbf{p}')  D _{
\delta}(\mathbf{q} - \mathbf{p}'+\mathbf{p}) \frac{1}{ |
\mathbf{p}'-\mathbf{p}|^2} \nonumber \\
&-&2 \overline{D}^{\dag}_{ \alpha}(\mathbf{p})  D _{
\beta}(\mathbf{p}') F^{\dag} _{ \gamma}(\mathbf{q} +
\mathbf{p}'-\mathbf{p}) \overline{F}_{ \delta}(\mathbf{q})
\frac{1}{ | \mathbf{p}'-\mathbf{p}|^2} \nonumber \\
&+& 2\overline{D}^{\dag}_{ \alpha}(\mathbf{p}) D _{
\delta}(-\mathbf{q} + \mathbf{p}'+\mathbf{p}) F^{\dag} _{
\beta}(\mathbf{p}')\overline{F}_{ \gamma}(\mathbf{q})
\frac{1}{ | \mathbf{p}'+\mathbf{p}|^2} \nonumber \\
&-& F^{\dag} _{ \beta}(\mathbf{p}') F^{\dag} _{ \delta}(\mathbf{q} -
\mathbf{p}'+\mathbf{p})\overline{F}_{ \alpha}(\mathbf{p})
\overline{F}_{ \gamma}(\mathbf{q}) \frac{1}{ |
\mathbf{p}'-\mathbf{p}|^2} \Bigr) \label{eq:v2-final-ph}
\end{eqnarray}

\begin{eqnarray}
 V_2^{unphys}
&=& \frac{e^2 \hbar^2}  {2(2\pi \hbar)^{3}}    \sum_{\alpha \beta
\gamma \delta} \int d\mathbf{p} d\mathbf{p}' d\mathbf{q}
\gamma^0_{\alpha \beta}
\gamma^0_{\gamma \delta}  \times \nonumber \\
\Bigl(
&+& 2\overline{A}^{\dag}_{ \alpha}(\mathbf{p})  A
_{\beta}(\mathbf{p}') A _{\delta}(-\mathbf{q} -
\mathbf{p}'+\mathbf{p}) \overline{B}_{ \gamma}(\mathbf{q}) \frac{1}{
|
\mathbf{p}'- \mathbf{p}|^2} \nonumber \\
&-& 2\overline{A}^{\dag}_{ \alpha}(\mathbf{p})  A _{
\beta}(\mathbf{p}')\overline{D}^{\dag}_{ \gamma}(\mathbf{q})
F^{\dag} _{ \delta}(-\mathbf{q} + \mathbf{p}'-\mathbf{p}) \frac{1}{
| \mathbf{p}'-\mathbf{p}|^2} \nonumber \\
&-& 2\overline{A}^{\dag}_{ \alpha}(\mathbf{p})  A _{
\beta}(\mathbf{p}') D _{ \delta}(-\mathbf{q} -
\mathbf{p}'+\mathbf{p})\overline{F}_{ \gamma}(\mathbf{q})
\frac{1}{ | \mathbf{p}'- \mathbf{p}|^2} \nonumber \\
&+& 2\overline{A}^{\dag}_{ \alpha}(\mathbf{p}) \overline{A}^{\dag}_{
\gamma}(\mathbf{q})  A _{ \delta}(\mathbf{q}+
\mathbf{p}'+\mathbf{p}) B^{\dag}_{ \beta}(\mathbf{p}')
\frac{1}{ |\mathbf{p}'+\mathbf{p}|^2} \nonumber \\
&+& \overline{A}^{\dag}_{ \alpha}(\mathbf{p})  \overline{A}^{\dag}_{
\gamma}(\mathbf{q}) B^{\dag}_{ \beta}(\mathbf{p}')
B^{\dag}_{\delta}(\mathbf{-q-p'-p}) \frac{1}{ |
\mathbf{p}'+\mathbf{p}|^2} \nonumber \\
&-& 2\overline{A}^{\dag}_{ \alpha}(\mathbf{p})  B^{\dag}_{
\beta}(\mathbf{p}') B^{\dag}_{ \delta}(\mathbf{q} -
\mathbf{p}'-\mathbf{p})\overline{B}_{ \gamma}(\mathbf{q}) \frac{1}{
| \mathbf{p}'+\mathbf{p}|^2} \nonumber \\
&-& 2\overline{A}^{\dag}_{ \alpha}(\mathbf{p})  B^{\dag}_{
\beta}(\mathbf{p}')\overline{D}^{\dag}_{ \gamma}(\mathbf{q})  D _{
\delta}(\mathbf{q}) + \mathbf{p}'+\mathbf{p})
\frac{1}{ | \mathbf{p}'+ \mathbf{p}|^2} \nonumber \\
&-& 2\overline{A}^{\dag}_{ \alpha}(\mathbf{p})  B^{\dag}_{
\beta}(\mathbf{p}')\overline{D}^{\dag}_{ \gamma}(\mathbf{q})
F^{\dag} _{ \delta}(-\mathbf{q} - \mathbf{p}'-\mathbf{p}) \frac{1}{
|\mathbf{p}'+\mathbf{p}|^2} \nonumber \\
&+& 2\overline{A}^{\dag}_{ \alpha}(\mathbf{p})  B^{\dag}_{
\beta}(\mathbf{p}') F^{\dag} _{ \delta}(\mathbf{q} -
\mathbf{p}'-\mathbf{p}) \overline{F}_{ \gamma}(\mathbf{q}) \frac{1}{
| \mathbf{p}'+\mathbf{p}|^2} \nonumber \\
&+& A _{\beta}(\mathbf{p}')A _{\delta}(-\mathbf{q} -
\mathbf{p}'-\mathbf{p}) \overline{B}_{ \alpha}(\mathbf{p})
\overline{B}_{ \gamma}(\mathbf{q})
\frac{1}{ | \mathbf{p}'+\mathbf{p}|^2} \nonumber \\
&+& 2 A _{\beta}(\mathbf{p}')B^{\dag}_{ \delta}(\mathbf{q} +
\mathbf{p}'+\mathbf{p}) \overline{B}_{ \alpha}(\mathbf{p})
\overline{B}_{ \gamma}(\mathbf{q}) \frac{1}{
|\mathbf{p}'+\mathbf{p}|^2} \nonumber \\
&-& 2A _{ \beta}(\mathbf{p}')\overline{B}_{ \alpha}(\mathbf{p})
\overline{D}^{\dag}_{ \gamma}(\mathbf{q})  D _{ \delta}(\mathbf{q} -
\mathbf{p}'-\mathbf{p})
\frac{1}{ | \mathbf{p}'+\mathbf{p}|^2} \nonumber \\
&-& 2\overline{A} _{\beta}(\mathbf{p}')B_{ \alpha}(\mathbf{p})  D _{
\delta}(-\mathbf{q} - \mathbf{p}'-\mathbf{p})  \overline{F}_{
\gamma}(\mathbf{q})
\frac{1}{ | \mathbf{p}'+\mathbf{p}|^2} \nonumber \\
&+& 2\overline{A} _{\beta}(\mathbf{p}')B_{ \alpha}(\mathbf{p})
F^{\dag} _{ \delta}(\mathbf{q} + \mathbf{p}'+\mathbf{p})
\overline{F}_{ \gamma}(\mathbf{q}) \frac{1}{ |
\mathbf{p}'+\mathbf{p}|^2} \nonumber \\
&+& 2B^{\dag}_{ \beta}(\mathbf{p}') \overline{B}_{
\alpha}(\mathbf{p})  \overline{D}^{\dag}_{ \gamma}(\mathbf{q})
F^{\dag} _{ \delta}(-\mathbf{q} - \mathbf{p}'+\mathbf{p})
\frac{1}{ | \mathbf{p}'-\mathbf{p}|^2} \nonumber \\
&+& 2B^{\dag}_{ \beta}(\mathbf{p}') \overline{B}_{
\alpha}(\mathbf{p})   D _{ \delta}(-\mathbf{q} +
\mathbf{p}'-\mathbf{p}) \overline{F}_{ \gamma}(\mathbf{q}) \frac{1}{
| \mathbf{p}'-\mathbf{p}|^2} \nonumber \\
&-& 2\overline{D}^{\dag}_{ \alpha}(\mathbf{p}) \overline{D}^{\dag}_{
\gamma}(\mathbf{q}) D _{ \beta}(\mathbf{p}')  F^{\dag} _{
\delta}(-\mathbf{q} + \mathbf{p}'- \mathbf{p})
\frac{1}{ | \mathbf{p}'- \mathbf{p}|^2}  \nonumber \\
&+& 2\overline{D}^{\dag}_{ \alpha}(\mathbf{p})  D _{
\beta}(\mathbf{p}') D _{ \delta}(-\mathbf{q} -
\mathbf{p}'+\mathbf{p})\overline{F}_{ \gamma}(\mathbf{q}) \frac{1}{
| \mathbf{p}'-\mathbf{p}|^2} \nonumber \\
&+& \overline{D}^{\dag}_{ \alpha}(\mathbf{p})  \overline{D}^{\dag}_{
\gamma}(\mathbf{q}) F^{\dag} _{ \beta}(\mathbf{p}') F^{\dag} _{
\delta}(-\mathbf{q} - \mathbf{p}'-\mathbf{p}) \frac{1}{
|\mathbf{p}'+\mathbf{p}|^2} \nonumber \\
&-&2\overline{D}^{\dag}_{ \alpha}(\mathbf{p})  F^{\dag} _{
\beta}(\mathbf{p}')F^{\dag} _{ \delta}(\mathbf{q} -
\mathbf{p}'-\mathbf{p}) \overline{F}_{ \gamma}(\mathbf{q}) \frac{1}{
| \mathbf{p}'+\mathbf{p}|^2} \nonumber \\
&+& D _{ \beta}(\mathbf{p}')D _{ \delta}(-\mathbf{q} -
\mathbf{p}'-\mathbf{p}) \overline{F}_{ \alpha}(\mathbf{p})
\overline{F}_{ \gamma}(\mathbf{q})
\frac{1}{ | \mathbf{p}'+\mathbf{p}|^2} \nonumber \\
& +& 2D _{ \beta}(\mathbf{p}') F^{\dag} _{ \delta}(\mathbf{q} +
\mathbf{p}'+\mathbf{p}) \overline{F}_{ \alpha}(\mathbf{p})
\overline{F}_{ \gamma}(\mathbf{q}) \frac{1}{ |
\mathbf{p}'+\mathbf{p}|^2}  \Bigr) \label{eq:v2-final-unp}
\end{eqnarray}

\chapter{Loop integrals in QED} \label{ss:b-integrals}

\section{4-dimensional delta function} \label{ss:4dimdellta}

In covariant Feynman-Dyson perturbation theory one often needs
4-dimensional delta function \index{delta function} of 4-momentum
$(p_0,p_x, p_y, p_z)$

\begin{eqnarray}
\delta^4(\tilde{p}) \equiv \delta(p_0) \delta(p_x) \delta (p_y)
\delta(p_z) = \delta(p_0) \delta(\mathbf{p}) \label{eq:4d-delta}
\end{eqnarray}

\noindent which has the following integral representation

\begin{eqnarray}
\frac{1}{(2 \pi \hbar)^4}  \int e^{\frac{i}{\hbar} (\tilde{p} \cdot
\tilde{x})} d^4 x = \delta^4(\tilde{p}) \label{eq:4delta-rep}
\end{eqnarray}

\noindent In our notation

\begin{eqnarray*}
\tilde{x} &=& (t, \mathbf{x}) \\
\tilde{p} &=& (p_0, \mathbf{p}) \\
\tilde{p} \cdot \tilde{x} &=& p_0t - \mathbf{p} \cdot \mathbf{x} \\
d^4x &\equiv& dt d \mathbf{x}
\end{eqnarray*}

\section{Feynman's trick}

In QED loop calculations one often meets integrals on the loop
4-momentum $\tilde{k}$ of expressions like $1/(abc \ldots)$, where
$a, b, c, \ldots$ are certain functions of $\tilde{k}$. Calculations become much simpler if one can replace the integrand
$1/(abc \ldots)$ with an expression in which $a, b, c, \ldots$ are
present in the denominator in a linear form. This can be achieved
using a trick first introduced by Feynman \cite{Feynman}.

The simplest example of such a trick is given by the integral
representation of the product $1/(ab)$

\begin{eqnarray}
 \int \limits_{0}^{1}  \frac{dx}{(ax+ b(1-x))^2}
 &=& \frac{1}{(b-a)(ax+ b(1-x))}\Bigl|_{0}^{1} \nonumber \\
 &=& \frac{1}{(b-a)a} - \frac{1}{(b-a)b} = \frac{1}{ab} \label{eq:ab}
\end{eqnarray}

\noindent The denominator on the left hand side is a square of a
function linear in $a$ and $b$. In spite of adding one more integral (on $x$), the overall integration task is greatly simplified, as we will see in many examples in this Appendix. Using this result, we can convert to
the linear form more complex expressions, e.g.,

\begin{eqnarray}
\frac{1}{a^2b} &=& -\frac{d}{da} \left(\frac{1}{ab} \right)
= -\frac{d}{da} \int \limits_{0}^{1}  \frac{dx}{(ax+ b(1-x))^2} \nonumber  \\
&=& \int \limits_{0}^{1}  \frac{2x dx}{(ax+ b(1-x))^3}
\label{eq:a2b}
\end{eqnarray}

\noindent These two results can be used to get an integral
representation for $1/(abc)$

\begin{eqnarray}
\frac{1}{abc} &=& \left(\frac{1}{bc}\right)\frac{1}{a}
= \left(\int \limits_{0}^{1}  \frac{dy}{(by+ c(1-y))^2} \right)\frac{1}{a} \nonumber  \\
&=& \int \limits_{0}^{1} dy\int \limits_{0}^{1} 2xdx
\frac{1}{[(by+ c(1-y))x + a(1-x)]^3} \nonumber  \\
&=& 2 \int \limits_{0}^{1} xdx \int \limits_{0}^{1}
\frac{dy}{[byx+ cx(1-y) + a(1-x)]^3} \label{eq:abc}
\end{eqnarray}

\noindent Another useful formula is\footnote{equation (131.2) in \cite{BLP}}

\begin{eqnarray}
\frac{1}{abc}
&=& 2 \int \limits_{0}^{1} dx \int \limits_{0}^{1} dy \int \limits_{0}^{1} dz
\frac{\delta(x+y+z-1)}{[ax+ by + cz]^3} \nonumber \\
&=& 2 \int \limits_{0}^{1} dx \int \limits_{0}^{1-x} dy
\frac{1}{[ax+ by + c(1-x-y)]^3} \label{eq:abc2}
\end{eqnarray}

\noindent Next differentiate equation (\ref{eq:a2b}) on $a$

\begin{eqnarray*}
\frac{1}{a^3d} &=& - \frac{1}{2}\frac{d}{da} \left(\frac{1}{a^2d}\right)=-
\frac{d}{da} \int \limits_{0}^{1} \frac{z dz}{[az + d(1-z)]^3} = 3
\int \limits_{0}^{1} \frac{z^2 dz}{[az + d(1-z)]^4}
\end{eqnarray*}

\noindent This results in

\begin{eqnarray}
&\ & \frac{1}{abcd} \nonumber \\
&=&  \left(2 \int \limits_{0}^{1} x dx \int
\limits_{0}^{1}
dy  \frac{1}{[a(1-x) + bxy + cx(1-y)]^3} \right)\frac{1}{d} \nonumber \\
&=& 6 \int \limits_{0}^{1} x dx \int \limits_{0}^{1} dy \int
\limits_{0}^{1} \frac{z^2 dz}{[az(1-x) + bxyz + cxz(1-y) +
d(1-z)]^4} \label{eq:abcd}
\end{eqnarray}

Obviously these calculations can be continued for expressions with
larger numbers of factors in denominators. See, e.g., the last
formula on page 520 of \cite{Schweber} and equation (11.A.1) in
\cite{book}.

\section{Some basic 4D integrals}

In our studies of loop integrals we will follow Feynman's approach
\cite{Feynman} and begin with the following simple integral

\begin{eqnarray}
K &=& \int   \frac{d^4k}{(\tilde{k}^2-L)^3} \equiv \int   \frac{dk_0 d
\mathbf{k}}{(k_0^2- c^2 k^2 - L + i \epsilon)^3} \label{eq:Kd4k}
\end{eqnarray}

\noindent The integral on $k_0$ has two 3rd order poles at $k_0 =
\pm \sqrt{c^2 k^2 + L}$. We can rotate\footnote{This step is known as the \emph{Wick rotation} \index{Wick rotation} \cite{Peskin}.} the integration
contour on $k_0$, so that it goes along the imaginary axis and then
change the integration variables $ik_0 = m_4$ and $c\mathbf{k} =
\mathbf{m}$. Then the integral is

\begin{eqnarray*}
K &=& \frac{1}{ c^3} \int \limits_{-i\infty}^{i\infty} dk_0 \int  \frac{d
\mathbf{m}}{(k_0^2 - m^2 - L)^3} = \frac{i}{ c^3} \int \limits_{-\infty}^{\infty} dm_4 \int  \frac{d
\mathbf{m}}{(-m_4^2 - m^2 - L)^3}
\end{eqnarray*}

\noindent Next we introduce 4-dimensional spherical coordinates
\cite{Blumenson} where $r^2 = m_4^2 + m^2$, and the area of a unit sphere\footnote{See equation (7.81) in \cite{Peskin}.} is $\int d \Omega = 2 \pi^2$

\begin{eqnarray}
K &=&  -\frac{2 \pi^2 i}{c^3}
\int \limits_0^{\infty}  \frac{r^3 dr}{(r^2 + L)^3} = -\frac{ \pi^2
i}{ c^3} \int \limits_L^{\infty}
\frac{(t-L) dt}{t^3} = -\frac{\pi^2}{2 ic^3 L}   \label{eq:k}
\end{eqnarray}

\noindent From symmetry properties we also get

\begin{eqnarray}
 \int d^4k  \frac{k_{\sigma}}{(\tilde{k}^2-L)^3} = 0
\label{eq:k-sigma}
\end{eqnarray}

\noindent Replacing $\tilde{k} \to \tilde{k} -\tilde{p}$ in
(\ref{eq:Kd4k}) and calling $L - \tilde{p}^2 = \Delta$ we get

\begin{eqnarray}
-\frac{\pi^2 }{2i (\tilde{p}^2 + \Delta) c^3} &=& \int
\frac{d^4k}{((\tilde{k}-\tilde{p})^2-L)^3} = \int
\frac{d^4k}{(\tilde{k}^2 - 2\tilde{p}\tilde{k} + \tilde{p}^2 -L)^3} \nonumber  \\
&=& \int  \frac{d^4k}{(\tilde{k}^2 - 2\tilde{p}\tilde{k} -
\Delta)^3} \label{eq:one-den}
\end{eqnarray}

\noindent Making the same substitutions  in (\ref{eq:k-sigma})  we
obtain

\begin{eqnarray*}
0 &=& \int   \frac{d^4k(k_{\sigma} -
p_{\sigma})}{((\tilde{k}-\tilde{p})^2-L)^3} = \int
\frac{d^4k(k_{\sigma} - p_{\sigma})}{(\tilde{k}^2 - 2\tilde{p}\tilde{k} -
\Delta)^3}
\end{eqnarray*}

\noindent Then

\begin{eqnarray}
 \int   \frac{d^4k k_{\sigma} }{(\tilde{k}^2 - 2\tilde{p}\tilde{k} - \Delta)^3}
 &=& \int   \frac{d^4k p_{\sigma}}{(\tilde{k}^2 - 2\tilde{p}\tilde{k} -
 \Delta)^3}= -\frac{ \pi^2 p_{\sigma} }{2i (\tilde{p}^2 + \Delta) c^3} \label{eq:one-den-sigma}
\end{eqnarray}

\noindent Differentiating both sides of (\ref{eq:one-den}) either by
$\Delta$ or by $p_{\sigma}$ we obtain

\begin{eqnarray}
\int  \frac{d^4k}{(\tilde{k}^2 - 2\tilde{p}\tilde{k} - \Delta)^4}
&=& \frac{\pi^2 }{6i (\tilde{p}^2 + \Delta)^2 c^3} \label{eq:k2-pk} \\
\int \frac{d^4k k_{\sigma}}{(\tilde{k}^2 - 2\tilde{p}\tilde{k} -
\Delta)^4} &=& \frac{ \pi^2p_{\sigma} }{6 i (\tilde{p}^2 + \Delta)^2
c^3}
\end{eqnarray}

\noindent Next differentiate both sides of (\ref{eq:one-den-sigma})
by $p_{\tau}$. If $\tau \neq \sigma$ then

\begin{eqnarray}
 \int  \frac{d^4k k_{\sigma} k_{\tau}}{(\tilde{k}^2 - 2\tilde{p}\tilde{k} - \Delta)^4}
 &=& \frac{\pi^2 p_{\sigma} p_{\tau} }{6i (\tilde{p}^2 + \Delta)^2 c^3} \label{eq:k-sigma-tau}
\end{eqnarray}

\noindent If $\tau =\sigma$

\begin{eqnarray}
   \int   \frac{d^4k k_{\sigma} k_{\sigma}}{(\tilde{k}^2 - 2\tilde{p}\tilde{k} - \Delta)^4}
 &=& \frac{ \pi^2 p_{\sigma} p_{\sigma} }{6i (\tilde{p}^2 + \Delta)^2 c^3}
 -\frac{\pi^2 }{12 i (\tilde{p}^2 + \Delta) c^3} \nonumber \\
 \label{eq:k-sigma-sigma}
\end{eqnarray}

\noindent Combining (\ref{eq:k-sigma-tau}) and
(\ref{eq:k-sigma-sigma}) yields

\begin{eqnarray*}
 \int  \frac{d^4k k_{\sigma} k_{\tau}}{(\tilde{k}^2 - 2\tilde{p}\tilde{k} - \Delta)^4}
 &=& \frac{\pi^2 (p_{\sigma} p_{\tau} - \frac{1}{2}\delta_{\sigma \tau }(\tilde{p}^2 + \Delta))}
 {6i (\tilde{p}^2 + \Delta)^2 c^3}
\end{eqnarray*}

\noindent Next we use (\ref{eq:a2b}) and (\ref{eq:one-den}) to calculate

\begin{eqnarray}
 &\ & \int   \frac{d^4k}{(\tilde{k}^2 - 2\tilde{p}_1\tilde{k} - \Delta_1)^2 (\tilde{k}^2
 - 2\tilde{p}_2\tilde{k} -
 \Delta_2)} \nonumber \\
 &=& \int \limits_{0}^{1}  2x dx \int  \frac{d^4k}{[(\tilde{k}^2 - 2\tilde{p}_1\tilde{k}
 - \Delta_1)x + (\tilde{k}^2 - 2\tilde{p}_2\tilde{k} -
 \Delta_2)(1-x)]^3} \nonumber \\
 &=& \int \limits_{0}^{1}  2x dx \int   \frac{d^4k}{[\tilde{k}^2x - 2\tilde{p}_1\tilde{k} x
 - \Delta_1x +
 \tilde{k}^2 - 2\tilde{p}_2\tilde{k} -
 \Delta_2 -\tilde{k}^2x + 2\tilde{p}_2\tilde{k}x +
 \Delta_2x]^3} \nonumber \\
&=& \int \limits_{0}^{1}  2x dx \int   \frac{d^4k}{[\tilde{k}^2 -
2\tilde{p}_x\tilde{k} - \Delta_x ]^3} = - \frac{\pi^2}{i c^3}\int
\limits_{0}^{1}   \frac{ x dx}{ \tilde{p}_x^2 + \Delta_x}
\label{eq:L}
\end{eqnarray}

\noindent where $\tilde{p}_1, \tilde{p}_2$ are two arbitrary
4-vectors, $\Delta_1, \Delta_2$ are numerical constants and

\begin{eqnarray*}
 \tilde{p}_x &=& x\tilde{p}_1 + (1-x) \tilde{p}_2 \\
  \Delta_x &=& x\Delta_1 + (1-x) \Delta_2
\end{eqnarray*}

\noindent Similarly,  we use (\ref{eq:one-den-sigma}) to obtain

\begin{eqnarray}
 &\ & \int   \frac{d^4k k_{\sigma}}{(\tilde{k}^2 - 2\tilde{p}_1\tilde{k}
 - \Delta_1)^2 (\tilde{k}^2 - 2\tilde{p}_2\tilde{k} -
 \Delta_2)} \nonumber \\
&=& \int \limits_{0}^{1}  2x dx \int
\frac{d^4k k_{\sigma}}{[\tilde{k}^2 - 2\tilde{p}_x\tilde{k} - \Delta_x
]^3} = -\frac{\pi^2}{ic^3} \int \limits_{0}^{1} \frac{p_{x \sigma} x
dx}{\tilde{p}_x^2 + \Delta_x } \label{eq:L-sigma}
\end{eqnarray}

\noindent Three more integrals are obtained by differentiating
(\ref{eq:L}) with respect to $\Delta_2$ and $p_{2 \tau}$ and by
differentiating (\ref{eq:L-sigma}) with respect to $p_{2 \tau}$

\begin{eqnarray}
&\ &  \int   \frac{d^4k}{(\tilde{k}^2 - 2\tilde{p}_1\tilde{k} -
\Delta_1)^2 (\tilde{k}^2 - 2\tilde{p}_2\tilde{k} -
 \Delta_2)^2} = \frac{\pi^2}{i c^3}\int \limits_{0}^{1}   \frac{x (1-x)
dx}{(\tilde{p}_x^2 + \Delta_x)^2} \nonumber \\
\label{eq:19b} \\
 &\ &
\int   \frac{d^4k k_{\tau}}{(\tilde{k}^2 - 2\tilde{p}_1\tilde{k} -
\Delta_1)^2 (\tilde{k}^2 - 2\tilde{p}_2\tilde{k} -
 \Delta_2)^2} = \frac{\pi^2}{i c^3}\int \limits_{0}^{1}   \frac{p_{x \tau} x
(1-x) dx}{(\tilde{p}_x^2 + \Delta_x)^2} \nonumber \\
\label{eq:19a} \\
  &\ &  \int
  \frac{d^4k k_{\sigma} k_{\tau}}{(\tilde{k}^2 -
2\tilde{p}_1\tilde{k} - \Delta_1)^2 (\tilde{k}^2 -
2\tilde{p}_2\tilde{k} -
 \Delta_2)^2} \nonumber \\
&=& \frac{\pi^2}{i c^3}\int \limits_{0}^{1}   \frac{(p_{x \sigma }
p_{x \tau} - 1/2 \delta_{\sigma \tau} (\tilde{p}_x^2 + \Delta_x))x
(1-x) dx}{(\tilde{p}_x^2 + \Delta_x)^2} \label{eq:19c}
\end{eqnarray}

\section{Electron self-energy integral}\label{ss:self-energy}

The loop integral in square brackets in (\ref{eq:square}) can be
represented in the form\footnote{Here we used equations
(\ref{eq:gamma-mu-gamma-mu}), (\ref{eq:gamma-munu-gamma-mu}), and
(\ref{eq:gammaA}).}

\begin{eqnarray}
J_{ad}(\cross{p}) &=& \gamma_{\mu} (\cross{p} + mc^2) I \gamma_{\mu} -
 \gamma_{\mu} \gamma^{\kappa}I^{\kappa} \gamma_{\mu} = (-2\cross{p} + 4mc^2) I  + 2 \gamma^{\kappa}I^{\kappa} \nonumber \\
\label{eq:Jadkappa}
\end{eqnarray}

\noindent where

\begin{eqnarray*}
 I &\equiv& \int
 \frac{d^4k}{[(\tilde{p}-\tilde{k})^2 - m^2 c^4]\tilde{k}^2 }  \\
 I^{\kappa} &\equiv& \int
 \frac{d^4k k^{\kappa}}{[(\tilde{p}-\tilde{k})^2 - m^2 c^4]\tilde{k}^2 }
\end{eqnarray*}

\noindent The factor $1/\tilde{k}^2$ in the integrand is a source of
both ultraviolet and infrared divergences. So, the integrals need to
be regularized, as described in subsection \ref{ss:regularization}.
To do that, we introduce two parameters: the \emph{ultraviolet
cutoff} $\Lambda$ and the \emph{infrared cutoff} $\lambda$.\footnote{$\Lambda$ and $\lambda$ have the dimensionality of  (mass)} \index{ultraviolet
cutoff} \index{infrared cutoff} Then we
replace the troublesome factor $1/\tilde{k}^2$ by the integral

\begin{eqnarray}
1/\tilde{k}^2 \to -\int \limits_{\lambda^2 c^4}^{\Lambda^2c^4}
\frac{dL}{(\tilde{k}^2 -L)^2}   \label{eq:ck2}
\end{eqnarray}

\noindent In the end of calculations we should take limits $\Lambda
\to \infty$ and $\lambda \to 0$. In this limit the integral reduces
to $1/\tilde{k}^2$, as expected

\begin{eqnarray*}
 -\int \limits_{0}^{\infty}
\frac{dL}{(\tilde{k}^2 -L)^2} &=& -\int
\limits_{-\tilde{k}^2}^{\infty} \frac{dx}{x^2} =
   \frac{1}{\tilde{k}^2}
\end{eqnarray*}

\noindent  Then we can use (\ref{eq:L}) and (\ref{eq:L-sigma}) with
parameters

\begin{eqnarray}
\Delta_1 = L;  \ \ \tilde{p}_1 = 0; \ \ \Delta_2 = m^2c^4 -
\tilde{p}^2; \ \ \tilde{p}_2 = \tilde{p} \label{eq:delta--1} \\
\tilde{p}_x = (1-x)\tilde{p}; \  \ \Delta_x = xL + (1-x)(m^2c^4 -
\tilde{p}^2) \label{eq:px-deltax}
\end{eqnarray}

\noindent to rewrite our integrals\footnote{we have assumed that
$\Lambda^2 \gg m^2c^4$}

\begin{eqnarray}
 I
 &=& -\int \limits_{\lambda^2c^4}^{\Lambda^2 c^4} dL \int
\frac{d^4k}{(\tilde{k}^2 -2 \tilde{p}\tilde{k} +\tilde{p}^2  - m^2
c^4)(\tilde{k}^2 - L)^2 } \nonumber \\
 &=& \frac{\pi^2}{i c^3}\int \limits_{\lambda^2c^4}^{\Lambda^2 c^4} dL \int \limits_{0}^{1}
  \frac{x dx}{ (\tilde{p}_x^2 + \Delta_x)
} \nonumber \\
&=& \frac{\pi^2}{i c^3}\int \limits_{\lambda^2c^4}^{\Lambda^2 c^4} dL \int
\limits_{0}^{1}
  \frac{x dx}{(1-x)^2 \tilde{p}^2 + xL + (1-x)(m^2c^4 - \tilde{p}^2)} \nonumber \\
&=& \frac{\pi^2}{i c^3}\int \limits_{0}^{1}   dx \ln\Bigl((1-x)^2
\tilde{p}^2 + xL + (1-x)(m^2c^4 - \tilde{p}^2)\Bigr)
\Bigl|_{L=\lambda^2c^4}^{L=\Lambda^2 c^4} \nonumber \\
&=& \frac{\pi^2}{i c^3}\int \limits_{0}^{1} dx  \ln \frac{(1-x)^2
\tilde{p}^2 + x \Lambda^2 c^4+ (1-x)(m^2c^4 - \tilde{p}^2)}{(1-x)^2
\tilde{p}^2  + x \lambda^2c^4 + (1-x)(m^2c^4 - \tilde{p}^2)}  \nonumber \\
&\approx& \frac{\pi^2}{i c^3}\int \limits_{0}^{1} dx  \ln \frac{ x
\Lambda^2 c^4}{(1-x)^2 \tilde{p}^2   + (1-x)(m^2c^4 -
\tilde{p}^2)}  \label{eq:I-offshell}
\end{eqnarray}

\begin{eqnarray}
 I^{\kappa}
 &=& -\int \limits_{\lambda^2 c^4}^{\Lambda^2 c^4} dL \int  d^4k
\frac{k^{\kappa}}{(\tilde{k}^2 - 2\tilde{p}\cdot \tilde{k}
+\tilde{p}^2 -m^2c^4)(\tilde{k}^2 - L)^2 } \nonumber \\
 &=& \frac{\pi^2}{ic^3} \int \limits_{\lambda^2c^4}^{\Lambda^2 c^4 }
 dL \int \limits_{0}^{1}   \frac{p_{\kappa}
 x (1-x) dx}{(1-x)^2 \tilde{p}^2 + x L +(1-x)(m^2c^4-\tilde{p}^2)} \nonumber \\
&=& \frac{\pi^2}{ic^3} \int \limits_{0}^{1} dx  (1-x)p_{\kappa}
\ln\frac{(1-x)^2 \tilde{p}^2 + x \Lambda^2 c^4
+(1-x)(m^2c^4-\tilde{p}^2)}
{(1-x)^2 \tilde{p}^2 + x \lambda^2c^4 +(1-x)(m^2c^4-\tilde{p}^2)} \nonumber \\
&\approx& \frac{\pi^2}{ic^3} \int \limits_{0}^{1} dx  (1-x)p_{\kappa}
\ln\frac{x \Lambda^2 c^4} {(1-x)^2 \tilde{p}^2
+(1-x)(m^2c^4-\tilde{p}^2)} \nonumber \\
 \label{eq:I-kappa-offshell}
\end{eqnarray}

\noindent Inserting  (\ref{eq:I-offshell}) and
(\ref{eq:I-kappa-offshell}) in (\ref{eq:Jadkappa}) we obtain

\begin{eqnarray}
J_{ad}(\cross{p}) &\approx& \frac{\pi^2}{i c^3}(-2\cross{p} + 4mc^2) \int
\limits_{0}^{1} dx  \ln \frac{ x \Lambda^2 c^4}{(1-x)^2 \tilde{p}^2 +
(1-x)(m^2c^4 - \tilde{p}^2)}  \nonumber \\
  &\ & +\frac{2 \pi^2 \cross{p}}{ic^3} \int \limits_{0}^{1}
dx (1-x) \ln\frac{x \Lambda^2 c^4} {(1-x)^2 \tilde{p}^2
+(1-x)(m^2c^4-\tilde{p}^2)} \nonumber \\
&=& \frac{\pi^2}{i c^3} \int \limits_{0}^{1} dx (4mc^2 -2\cross{p}x)
\ln \frac{ x \Lambda^2 c^4}{(1-x)^2 \tilde{p}^2  + (1-x)(m^2c^4 -
\tilde{p}^2)}  \nonumber \\
\label{eq:Jad-p}
\end{eqnarray}

For our discussion in subsections \ref{ss:self-en} and
\ref{ss:elec-mass} it will be convenient to represent $J_{ad}$ in
the form of a Taylor expansion around the on-mass-shell value of the
4-momentum
$\cross{p}=mc^2$

\begin{eqnarray}
J_{ad}(\cross{p}) &=& C_0 \delta_{ad} + C_1(\cross{p} - mc^2)_{ad} +
R(\cross{p}) \label{eq:J-ab-p}
\end{eqnarray}

\noindent where $C_0$ is a constant (independent on $p^{\mu}$) term,
$C_1(\cross{p} - mc^2)_{ad}$ is linear in $\cross{p} - mc^2$, and
$R(\cross{p})$ combines all other terms (quadratic, cubic, etc. in
$\cross{p} - mc^2$).  To calculate $C_0$ we simply set $\tilde{p}^2 = m^2c^4$ in
(\ref{eq:Jad-p}).\footnote{Note that $\tilde{p}^2$
is a function of  $\cross{p}$ due to (\ref{eq:J.45a}).  When doing calculations with slash symbols $\cross{p}$ and $\gamma$-matrices, it is convenient to use properties (\ref{eq:H.17a}) - (\ref{eq:gammaABC})}

\begin{eqnarray}
C_0 &=&   \frac{2 \pi^2 mc^2}{ic^3} \int \limits_{0}^{1} dx (2-x)
\ln\frac{x \Lambda^2 c^4} {(1-x)^2 m^2c^4}
\nonumber \\
&\approx& \frac{2 \pi^2 mc^2}{i c^3} \int \limits_{0}^{1}(2-x) dx
\ln  \frac{\Lambda^2 }{m^2}  +\frac{2 \pi^2 mc^2}{i c^3} \int
\limits_{0}^{1}(2-x) dx  \ln
 \frac{x}{(1-x)^2}  \nonumber \\
&=& \frac{3  \pi^2mc^2}{i c^3}   \ln  \frac{\Lambda^2}{m^2}
 +\frac{3 \pi^2mc^2}{2i c^3}
 =\frac{3 \pi^2mc^2}{2i c^3} \left( 4 \ln \frac{\Lambda}{m}
+1 \right) \label{eq:CC}
\end{eqnarray}

\noindent For the coefficient $C_1$ we obtain\footnote{Here we used
integral $\int \limits_0^1 dx x \ln(1/(1-x)^2) = 5/4$.}

\begin{eqnarray}
C_1 &=& \frac{d J_{ad}}{d \cross{p}}\Bigl|_{\cross{p}=mc^2} \nonumber \\
&=& -\frac{2 \pi^2}{i c^3} \int \limits_{0}^{1}x dx  \ln \frac{ x
\Lambda^2 c^4}{(1-x)^2 \tilde{p}^2  + x \lambda^2c^4 +
(1-x)(m^2c^4 - \tilde{p}^2)} \Bigl|_{\cross{p}=mc^2} \nonumber \\
&\ &-\frac{2 \pi^2}{i c^3} mc^2 \int \limits_{0}^{1} (2-x)dx
\frac{(1-x)^2 \tilde{p}^2 + x \lambda^2c^4 + (1-x)(m^2c^4 -
\tilde{p}^2)}{ x
\Lambda^2 c^4} \times \nonumber \\
&\ &\frac{ x \Lambda^2 c^4(2(1-x)^2
\cross{p} - 2(1-x)\cross{p})}{((1-x)^2 \tilde{p}^2 + x \lambda^2c^4 +
(1-x)(m^2c^4 - \tilde{p}^2))^2} \Bigl|_{\cross{p}=mc^2}\nonumber \\
 &=& -\frac{2 \pi^2}{i c^3} \int \limits_{0}^{1} xdx  \ln \frac{ x
\Lambda^2}{(1-x)^2 m^2 }  \nonumber \\
&\ & -\frac{2 \pi^2 mc^2}{ic^3} \int \limits_{0}^{1} dx (2-x)
\frac{2(1-x)^2
mc^2- 2(1-x)mc^2} {(1-x)^2 m^2c^4 + x \lambda^2c^4} \nonumber \\
&=& -\frac{2 \pi^2}{i c^3} \int \limits_{0}^{1} xdx  \ln \frac{ x
\Lambda^2}{(1-x)^2 m^2 }  -\frac{4 \pi^2}{ic^3} \int
\limits_{0}^{1} dx  \frac{(2-x)(x^2 -x)}{(1-x)^2 + x \lambda^2/m^2} \nonumber \\
&=& -\frac{2 \pi^2}{i c^3} \int \limits_{0}^{1} xdx  \ln \frac{ x
}{(1-x)^2  }  -\frac{ \pi^2}{i c^3}   \ln \frac{ \Lambda^2}{ m^2
}  -\frac{2 \pi^2}{ic^3}
\left(1 + \ln  \frac{\lambda^2}{m^2} \right) \nonumber \\
&=& -\frac{2\pi^2}{i c^3} \left(\ln \frac{
\Lambda}{ m }   +2  \ln
\frac{\lambda}{m} + \frac{9}{4}\right)   \label{eq:coeff-a}
\end{eqnarray}

\noindent Then the residual term

\begin{eqnarray*}
R(\cross{p}) &=& J_{ad}(\cross{p}) - C_0 \delta_{ad} - C_1(\cross{p} -
mc^2)_{ad}
\end{eqnarray*}

\noindent is ultraviolet-finite, because all $\Lambda$-dependent
terms there cancel out

\begin{eqnarray*}
&\ & \frac{\pi^2}{i c^3} \ln\Lambda^2 \int \limits_{0}^{1} dx
(4mc^2 -2\cross{p}x)   - \frac{2 \pi^2 mc^2}{ic^3} \ln\Lambda^2  \int
\limits_{0}^{1} dx (2-x) \\
&+&(\cross{p} - mc^2)\frac{2 \pi^2}{i c^3} \ln  \Lambda^2  \int
\limits_{0}^{1} xdx
 = 0
\end{eqnarray*}

\noindent We see that $C_0$ is ultraviolet-divergent, while $C_1$ has both ultraviolet and infrared divergences. It can be said that $J_{bc}(\cross{p})$, as a function of
$\cross{p}$, is infinite at the point $\cross{p}=mc^2$ and has an
infinite first derivative at this point. However, the 2nd and higher
derivatives are all finite.

\section{Integral for the vertex renormalization}
\label{sc:vertex-ren}

Let us calculate the integral in square brackets in equation
(\ref{eq:s4-1112a})\footnote{We used equation (\ref{eq:ck2}) and took into
account that $\tilde{q}$ and $\tilde{q}'$ are on the mass shell, so
that $\tilde{q}^2 = (\tilde{q}')^2= m^2c^4$.}

\begin{eqnarray*}
I^{\kappa}(\tilde{q}, \tilde{q}') &=& \int d^4h \gamma_{\mu}
\frac{-\cross{h} + \cross{q} + mc^2}{(\tilde{h}-\tilde{q})^2 - m^2c^4}
\gamma_{\kappa} \frac{-\cross{h} + \cross{q}' +
mc^2}{(\tilde{h}-\tilde{q}')^2 - m^2c^4} \gamma_{\mu}
\frac{1}{\tilde{h}^2} \nonumber \\
&\approx& -\int \limits_{\lambda^2c^4}^{\Lambda^2 c^4} dL \int d^4h
\frac{\gamma_{\mu}(-\cross{h} + \cross{q} +
mc^2)\gamma_{\kappa}(-\cross{h} + \cross{q}' +
mc^2)\gamma_{\mu}}{(\tilde{h}^2-2\tilde{q}\tilde{h})(\tilde{h}^2-
2\tilde{q}'\tilde{h})(\tilde{h}^2- L)^2}
\end{eqnarray*}

\noindent  The numerator can be rewritten as

\begin{eqnarray*}
&\ &
 \gamma_{\mu} (-\cross{h} + \cross{q} +
mc^2)  \gamma_{\kappa} (-\cross{h} + \cross{q}' + mc^2) \gamma_{\mu} \\
&=&
 \gamma_{\mu} ( \cross{q} +
mc^2)  \gamma_{\kappa} ( \cross{q}' + mc^2) \gamma_{\mu} -
 \gamma_{\mu} \cross{h}  \gamma_{\kappa} (\cross{q}' + mc^2) \gamma_{\mu} \\
 &\ &-
 \gamma_{\mu} (\cross{q} +
mc^2)  \gamma_{\kappa} \cross{h} \gamma_{\mu} + \gamma_{\mu} \cross{h}
\gamma_{\kappa} \cross{h} \gamma_{\mu}
\end{eqnarray*}

\noindent  Then the desired integral is

\begin{eqnarray}
I^{\kappa}(\tilde{q}, \tilde{q}') &=& \gamma_{\mu} ( \cross{q} + mc^2)
\gamma_{\kappa} ( \cross{q}' +
mc^2) \gamma_{\mu} J \nonumber \\
&\ & - \gamma_{\mu} \gamma_{\sigma} \gamma_{\kappa} (\cross{q}' + mc^2)
\gamma_{\mu}J_{\sigma} -
 \gamma_{\mu} (\cross{q} +
mc^2)  \gamma_{\kappa} \gamma_{\sigma} \gamma_{\mu} J_{\sigma}  \nonumber \\
&\ & +\gamma_{\mu} \gamma_{\sigma} \gamma_{\kappa} \gamma_{\tau}
\gamma_{\mu} J_{\sigma \tau} \label{eq:Ikappa}
\end{eqnarray}

\noindent where\footnote{The denominators were combined using
(\ref{eq:ab})  and $ \tilde{q}_y \equiv  y\tilde{q}  + (1-y)
\tilde{q}'$.}

\begin{eqnarray}
J &=& -\int \limits_0^1 dy \int \limits_{\lambda^2c^4}^{\Lambda^2 c^4} dL
\int
\frac{d^4h }{[\tilde{h}^2 - 2 \tilde{h}\cdot \tilde{q}_y  ]^2[\tilde{h}^2 - L]^2} \label{eq:Jy} \\
J_{\sigma}  &=& -\int \limits_0^1 dy \int
\limits_{\lambda^2c^4}^{\Lambda^2 c^4} dL \int \frac{d^4h
h^{\sigma}}{[\tilde{h}^2 - 2 \tilde{h}\cdot \tilde{q}_y
]^2[\tilde{h}^2 - L]^2}
 \label{eq:J--sigma}\\
J_{\sigma \tau}  &=& -\int \limits_0^1 dy \int
\limits_{\lambda^2c^4}^{\Lambda^2 c^4} dL \int \frac{d^4h
h^{\sigma}h^{\tau}}{[\tilde{h}^2 - 2 \tilde{h}\cdot \tilde{q}_y
]^2[\tilde{h}^2 - L]^2} \label{eq:J--sigma-tau}
\end{eqnarray}

\noindent  These are particular cases of integrals (\ref{eq:19b}) -
(\ref{eq:19c}) with parameters $\tilde{p}_1 = \tilde{q}_y$,
$\Delta_1 = 0$, $ \tilde{p}_2 = 0$, $ \Delta_2 = L$,   $\tilde{p}_x
= x\tilde{q}_y$, and $\Delta_x = (1-x)L$

\begin{eqnarray}
J
 &=&-\frac{\pi^2}{i c^3}\int \limits_0^1 dy \int
\limits_{\lambda^2c^4}^{\Lambda^2 c^4} dL \int \limits_{0}^{1} \frac{x
(1-x) dx}{(x^2\tilde{q}_y^2 + (1-x)L)^2}
\label{eq:JJ}\\
J_{\sigma}
 &=&-\frac{\pi^2}{i c^3}\int \limits_0^1 dy \int
\limits_{\lambda^2c^4}^{\Lambda^2 c^4} dL \int \limits_{0}^{1} \frac{q_{y
\sigma} x^2 (1-x) dx}{(x^2\tilde{q}_y^2 + (1-x)L)^2} \label{eq:JJsigma} \\
J_{\sigma \tau}
 &=&-\frac{\pi^2}{i c^3}\int \limits_0^1 dy \int
\limits_{\lambda^2c^4}^{\Lambda^2 c^4} dL \int \limits_{0}^{1}
\frac{[x^2q_{y \sigma} q_{y \tau} - 1/2 \delta_{\sigma
\tau}(x^2\tilde{q}_y^2 + (1-x)L)]x (1-x) dx}{(x^2\tilde{q}_y^2 +
(1-x)L)^2} \nonumber \\
\label{eq:JJsigmatau}
\end{eqnarray}

\noindent In the limit $\Lambda \to \infty$ we obtain for
(\ref{eq:JJ})

\begin{eqnarray}
J &=& \frac{\pi^2}{i c^3} \int \limits_{0}^{1} dx \int
\limits_{0}^{1} dy \frac{x }{x^2\tilde{q}_y^2 + (1-x)L }
\Bigl|_{L = \lambda^2c^4}^{L=\infty} = -\frac{\pi^2}{i c^3} \int \limits_{0}^{1} dx \int
\limits_{0}^{1} dy
\frac{x }{x^2\tilde{q}_y^2 + (1-x)\lambda^2c^4 } \nonumber \\
&\approx& -\frac{\pi^2}{2i c^3} \int \limits_{0}^{1}
\frac{dy}{\tilde{q}_y^2} \ln(-x^2\tilde{q}_y^2
- (1-x)\lambda^2c^4) \Bigl|_{x=0}^{x=1}= -\frac{\pi^2}{2i c^3} \int \limits_{0}^{1}
\frac{dy}{\tilde{q}_y^2}
\left(\ln(-\tilde{q}_y^2 )  - \ln(-\lambda^2c^4) \right)\nonumber \\
&=& -\frac{\pi^2}{2i c^3} \int \limits_{0}^{1} dy
\frac{1}{\tilde{q}_y^2}\ln \frac{\tilde{q}_y^2 }{\lambda^2c^4}
\label{eq:intJ}
\end{eqnarray}

To proceed further with this integral we introduce the 4-vector of the
transferred momentum

\begin{eqnarray}
\tilde{k} &\equiv& \tilde{q}'-\tilde{q} \label{eq:M.35a}
\end{eqnarray}

\noindent Then from $(\tilde{q}')^2 =  (\tilde{q}+\tilde{k})^2$ and
$(\tilde{q}')^2 = \tilde{q}^2=m^2c^4$ it follows that

\begin{eqnarray*}
 2\tilde{q} \cdot \tilde{k}  &=& - \tilde{k}^2 \\
\tilde{q}_y &=&   \tilde{q} +(1-y)\tilde{k} \\
\tilde{q}_y^2 &=&  m^2c^4 - (1-y)y\tilde{k}^2
\end{eqnarray*}

\noindent Instead of  $\tilde{k}^2$ and integration variable $y$ it
is convenient to introduce two new variables $\theta$ and $\alpha$,
such that\footnote{Note that, by definition, $0 \leq \tilde{k}^2 \leq 4m^2c^4 $.}

\begin{eqnarray}
\tilde{k}^2 &\equiv& 4m^2 c^4\sin^2 \theta \label{eq:k2sinth} \\
y  &=& \frac{1}{2}\left(1 + \frac{\tan \alpha}{\tan \theta}\right) \nonumber \\
1 -y  &=& \frac{1}{2}\left(1 - \frac{\tan \alpha}{\tan \theta}\right) \nonumber \\
\tilde{q}_y^2 &=& m^2c^4 - 4 m^2c^4 \sin^2 \theta \cdot \frac{1}{2}\left(1 +
\frac{\tan \alpha}{\tan \theta}\right) \cdot
\frac{1}{2}\left(1 - \frac{\tan \alpha}{\tan \theta}\right) \nonumber \\
&=& m^2c^4 -  m^2c^4 \cos^2 \theta
(\tan^2 \theta - \tan^2 \alpha) = m^2 c^4\frac{\cos^2 \theta}{\cos ^2 \alpha} \nonumber \\
dy &=& \frac{d \alpha}{2
\tan \theta} \frac{d}{d \alpha} \left(\frac{\sin
\alpha}{\cos \alpha}\right) = \frac{d \alpha}{2
\tan \theta \cos^2 \alpha}  \nonumber \\
 \frac{dy}{\tilde{q}_y^2} &=&  \frac{d \alpha}{2 m^2 c^4 \cos^2 \theta
\tan \theta}  = \frac{d \alpha}{m^2c^4 \sin (2 \theta)} \nonumber
\end{eqnarray}

\noindent Integral $J$ is infrared-divergent\footnote{Here we
took the following integral by parts $ \int \limits_{0}^{\theta} d
\alpha \ln(\cos \alpha ) = \theta \ln (\cos \theta) + \int
\limits_{0}^{\theta} \alpha \tan \alpha d\alpha $}

\begin{eqnarray}
J &=& -\frac{\pi^2}{2i c^3} \int \limits_{- \theta}^{\theta} \frac{d
\alpha}{m^2 c^4 \sin (2 \theta)}\ln \left(\frac{m^2 \cos^2
\theta}{\lambda^2\cos^2
\alpha }\right)  \nonumber \\
&=&  -\frac{ 2 \pi^2 \theta}{i c^3 m^2 c^4 \sin (2
\theta)}\ln\frac{m  }{ \lambda} - \frac{\pi^2}{2i c^3 m^2 c^4
\sin (2 \theta)} \int \limits_{- \theta}^{\theta} d
\alpha\ln\frac{\cos^2 \theta}{\cos^2
\alpha }  \nonumber \\
&=&   -\frac{ 2 \pi^2\theta}{i c^3 m^2 c^4 \sin (2
\theta)}\ln\frac{m }{ \lambda} -  \frac{2 \pi^2}{i c^3 m^2 c^4
\sin (2 \theta)} \int \limits_{0}^{\theta} d \alpha \Bigl(
\ln(\cos\theta) - \ln(\cos
\alpha )\Bigr) \nonumber \\
&=&   -\frac{2\pi^2\theta}{i c^3 m^2 c^4 \sin (2 \theta)}\ln
\frac{m }{ \lambda}  -  \frac{2 \pi^2\theta \ln(\cos\theta)}{i
c^3 m^2 c^4 \sin (2 \theta)} + \frac{2\pi^2 }{i c^3 m^2 c^4 \sin (2
\theta)} \int \limits_{0}^{\theta} d \alpha
\ln(\cos \alpha ) \nonumber \\
 &=&   2A\left[ \theta \ln \frac{m }{ \lambda}   - \int
\limits_{0}^{\theta} \alpha \tan \alpha d\alpha \right] \label{eq:26a}
\end{eqnarray}

\noindent where we defined

\begin{eqnarray*}
A &\equiv& -\frac{\pi^2}{i c^3} \frac{\theta}{m^2 c^4 \sin(2 \theta)}
\end{eqnarray*}

\noindent Next we calculate (\ref{eq:JJsigma}) using variables
$\theta$ and $\alpha$ introduced above. Taking the limits $\lambda
\to 0$, $\Lambda \to \infty$ we obtain a finite result (both
infrared and ultraviolet divergences are absent)

\begin{eqnarray}
 J_{\sigma}
 &=&\frac{\pi^2}{i c^3} \int \limits_{0}^{1} dx \int \limits_{0}^{1} dy \frac{x^2q_{y
\sigma}   }{x^2\tilde{q}_y^2 +
(1-x)L}\Bigl|_{L=\lambda^2c^4}^{L=\Lambda^2 c^4} \approx-\frac{\pi^2}{i
c^3}\int \limits_{0}^{1} dy \frac{q_{y \sigma}
}{\tilde{q}_y^2} \nonumber \\
&=&-\frac{\pi^2}{i c^3}\int \limits_{-\theta}^{\theta} \frac{d
\alpha}{m^2 c^4 \sin(2 \theta)} \left(q_{\sigma}
+ \frac{k_{\sigma}}{2} \left(1 - \frac{\tan \alpha}{ \tan \theta}\right) \right) \nonumber  \\
&=&-\frac{\pi^2}{i c^3}\frac{2 \theta }{m^2 c^4 \sin(2 \theta)}
\left(q_{\sigma} + \frac{k_{\sigma}}{2} \right) + \frac{\pi^2 k_{\sigma}}{2i
c^3 m^2 c^4 \sin(2 \theta) \tan \theta}\int
\limits_{-\theta}^{\theta} d
\alpha \tan \alpha \nonumber  \\
&=& A (q_{\sigma} +q'_{\sigma} ) \label{eq:j-sigma}
\end{eqnarray}

\noindent Next we need to calculate
(\ref{eq:JJsigmatau})\footnote{Here we assumed that $\Lambda^2 c^4\gg
\tilde{q}_y^2 \gg \lambda^2c^4$ and used integral $\int \limits_{0}^{1}
dx
 x \ln \left((1-x)/x^2 \right) = \left[ (x^2/2)
 \ln\left((1-x)/x^2\right) + x^2/4 - x/2
 -1/2\ln(1-x)\right]\Bigl|_{0}^1 = -1/4$.}

\begin{eqnarray*}
 J_{\sigma \tau}
 &=&-\frac{\pi^2}{i c^3} \int \limits_{\lambda^2c^4}^{\Lambda^2 c^4} dL
 \int \limits_{0}^{1} dx \int \limits_{0}^{1} dy
 \frac{x^3 (1-x)q_{y
\sigma} q_{y \tau} }{(x^2\tilde{q}_y^2 + (1-x)L)^2} \nonumber
\\
&\ &+\frac{\pi^2}{2i c^3} \int \limits_{\lambda^2c^4}^{\Lambda^2 c^4} dL \int
\limits_{0}^{1} dx \int \limits_{0}^{1} dy \frac{ \delta_{\sigma
\tau}x (1-x) }{x^2\tilde{q}_y^2 +
(1-x)L} \nonumber \\
&\approx&\frac{\pi^2}{i c^3}  \int \limits_{0}^{1} dx \int
\limits_{0}^{1} dy \left(\frac{x^3 q_{y \sigma} q_{y \tau}
}{x^2\tilde{q}_y^2 + (1-x) \Lambda^2 c^4} - \frac{x q_{y \sigma} q_{y
\tau} }{\tilde{q}_y^2 } \right)
\nonumber \\
&\ &+\frac{\pi^2}{2i c^3}  \int \limits_{0}^{1} dx \int
\limits_{0}^{1} dy
 \delta_{\sigma \tau}x [\ln((1-x) \Lambda^2 c^4) - \ln
 (x^2\tilde{q}_y^2)] \nonumber \\
&\approx&-\frac{\pi^2}{2i c^3}   \int \limits_{0}^{1} dy
 \frac{q_{y \sigma} q_{y \tau} }{\tilde{q}_y^2 }
+\frac{\pi^2 \delta_{\sigma \tau}}{2i c^3}  \int \limits_{0}^{1} dx
\int \limits_{0}^{1} dy
 x \ln \frac{(1-x) \Lambda^2 c^4}{x^2\tilde{q}_y^2}  \\
&=&-\frac{\pi^2}{2i c^3}   \int \limits_{0}^{1} dy
 \frac{q_{y \sigma} q_{y \tau} }{\tilde{q}_y^2 }
+\frac{\pi^2 \delta_{\sigma \tau}}{2i c^3}  \int \limits_{0}^{1} dx
 x \ln \frac{(1-x) }{x^2}
 +\frac{\pi^2 \delta_{\sigma \tau}}{4i c^3}
\int \limits_{0}^{1} dy \ln \frac{ \Lambda^2 c^4}{\tilde{q}_y^2} \nonumber \\
&=&-\frac{\pi^2}{2i c^3}   \int \limits_{0}^{1} dy
 \frac{q_{y \sigma} q_{y \tau} }{q_y^2 }
+\frac{\pi^2 \delta_{\sigma \tau}}{4i c^3}   \int \limits_{0}^{1} dy
  \ln\frac{
  \Lambda^2 c^4}{\tilde{q}_y^2} - \frac{\pi^2 \delta_{\sigma \tau}}{8i
  c^3}
\end{eqnarray*}

\noindent Integrations on $y$ are performed using variables
$\theta$ and $\alpha$\footnote{Here we used integrals $\int \tan^{2}(x) dx= \tan (x) -x +
C$, $\int
\cos^{-2}(x) dx= \tan (x) + C$ and $\int \ln(\cos^2(x))/\cos^2(x) dx = -2x + 2 \tan (x) +
\tan(x) \ln(\cos^2(x)) + C $.}

\begin{eqnarray*}
   \int \limits_{0}^{1} dy
 \frac{q_{y \sigma} q_{y \tau} }{\tilde{q}_y^2 }
&=& \int \limits_{-\theta}^{\theta} \left(q_{\sigma} +
\frac{1}{2}k_{\sigma}  - \frac{k_{\sigma}\tan \alpha}{ 2\tan
\theta}\right) \left(q_{\tau} + \frac{1}{2}k_{\tau}  -
\frac{k_{\tau} \tan
\alpha}{2\tan \theta}\right) \frac{d \alpha}{m^2 c^4 \sin(2 \theta)} \\
&=& \int \limits_{-\theta}^{\theta} \left(q_{\sigma} +
\frac{1}{2}k_{\sigma}\right)\left(q_{\tau} + \frac{1}{2}k_{\tau}\right)\frac{d
\alpha}{m^2 c^4 \sin(2 \theta)}  + \int \limits_{-\theta}^{\theta}
\frac{k_{\sigma} k_{\tau} \tan^2 \alpha}{ 4\tan^2 \theta} \frac{d
\alpha}{m^2 c^4 \sin(2 \theta)} \\
&=& \frac{ \theta}{2 m^2 c^4 \sin(2 \theta)}  (q_{\sigma} +
q'_{\sigma})(q_{\tau} + q'_{\tau})  + \frac{k_{\sigma} k_{\tau} \cos
\theta}{ 4 m^2 c^4\sin^3 \theta   }
 \int \limits_{0}^{\theta}  \tan^2
\alpha d \alpha \\
 &=& \frac{ \theta}{2 m^2 c^4 \sin(2 \theta)}  (q_{\sigma} +
q'_{\sigma})(q_{\tau} + q'_{\tau})  + \frac{k_{\sigma} k_{\tau} }{
\tilde{k}^2 }
 (1 - \theta \cot \theta) \\
   \int \limits_{0}^{1} dy
 \ln\frac{
  \Lambda^2 c^4}{\tilde{q}_y^2}
  &=& \int \limits_{0}^{\theta} \frac{d \alpha}{ \tan \theta \cos^2 \alpha}
  \ln \frac{\Lambda^2 \cos^2 \alpha}{m^2  \cos^2 \theta}  \\
   &=&
  \ln \frac{\Lambda^2 }{m^2  \cos^2 \theta}
  +   \frac{1}{ \tan \theta}
  (- 2 \theta + \ln(\cos^2 \theta) \tan \theta + 2 \tan \theta)\\
        &=&
  2\ln \frac{\Lambda }{m}  +2 (1 -  \theta \cot \theta)
\end{eqnarray*}

\noindent  Then we see that integral $J_{\sigma \tau}$ is
ultraviolet-divergent

\begin{eqnarray}
J_{\sigma \tau}
&=&\frac{A}{4} (q_{\sigma} + q'_{\sigma}) (q_{\tau} + q'_{\tau})
 + D k_{\sigma} k_{\tau} + E \delta_{\sigma \tau} \label{eq:29az} \\
 D &\equiv&  -\frac{\pi^2(1 - \theta \cot \theta)}{2ic^3 \tilde{k}^2} \nonumber \\
 E&\equiv& \frac{\pi^2}{2ic^3 } \left(\ln \frac{\Lambda }{m } + \frac{3}{4}-  \theta \cot
 \theta \right)  \nonumber
\end{eqnarray}

\noindent Using results (\ref{eq:26a}), (\ref{eq:j-sigma}),
(\ref{eq:29az}),  we obtain for the full integral (\ref{eq:Ikappa})

\begin{eqnarray}
I^{\kappa}(\tilde{q}, \tilde{q}')
 &=& J \gamma_{\mu} ( \cross{q} + mc^2)  \gamma_{\kappa} ( \cross{q}' +
mc^2) \gamma_{\mu} \nonumber \\
&\ & -A \gamma_{\mu} (\cross{q} + \cross{q}') \gamma_{\kappa} (\cross{q}' +
mc^2) \gamma_{\mu} -
 A \gamma_{\mu} (\cross{q} +
mc^2)  \gamma_{\kappa} (\cross{q} + \cross{q}') \gamma_{\mu} \nonumber  \\
&\ & +\frac{A}{4}\gamma_{\mu} (\cross{q} + \cross{q}') \gamma_{\kappa} (\cross{q} +
\cross{q}') \gamma_{\mu}
 + D \gamma_{\mu} \cross{k} \gamma_{\kappa} \cross{k}
\gamma_{\mu}
 + 4E  \gamma_{\kappa} \nonumber \\
&=& JT_1 - AT_2 -AT_3 + \frac{A}{4}T_4 + D \gamma_{\mu} \cross{k} \gamma_{\kappa} \cross{k}
\gamma_{\mu}
 + 4E  \gamma_{\kappa} \label{eq:I-kappa2}
\end{eqnarray}

\noindent Let us now use (\ref{eq:gammaA}) -
(\ref{eq:gammaABC}) and process these terms one-by-one

\begin{eqnarray*}
T_1&= & \gamma_{\mu} ( \cross{q} + mc^2)  \gamma_{\kappa} ( \cross{q}' +
mc^2) \gamma_{\mu} \\
 &=& \gamma_{\mu}  \cross{q} \gamma_{\kappa}
\cross{q}' \gamma_{\mu} + mc^2 \gamma_{\mu}  \gamma_{\kappa} \cross{q}'
 \gamma_{\mu} + mc^2\gamma_{\mu}  \cross{q}  \gamma_{\kappa}   \gamma_{\mu} +
 m^2c^4 \gamma_{\mu}
\gamma_{\kappa}   \gamma_{\mu} \\
 &=& -2  \cross{q}' \gamma_{\kappa}
\cross{q} + 2mc^2   \gamma_{\kappa} \cross{q}' + 2mc^2  \cross{q}'
\gamma_{\kappa}
  + 2mc^2  \cross{q}  \gamma_{\kappa} + 2mc^2 \gamma_{\kappa} \cross{q}
  -2 m^2c^4 \gamma_{\kappa}    \\
   &=& -2  (\cross{q} + \cross{k}) \gamma_{\kappa}
(\cross{q}' - \cross{k}) + 2mc^2   \gamma_{\kappa} \cross{q}' + 2mc^2
(\cross{q}+\cross{k}) \gamma_{\kappa}
  + 2mc^2  \cross{q}  \gamma_{\kappa} \\
&\ & +2mc^2 \gamma_{\kappa}
  (\cross{q}' -\cross{k})
  -2 m^2c^4 \gamma_{\kappa}    \\
     &=& -2  \cross{q}  \gamma_{\kappa}\cross{q}'
-2  \cross{k} \gamma_{\kappa} \cross{q}' +2  \cross{q}  \gamma_{\kappa}
\cross{k} +2   \cross{k} \gamma_{\kappa} \cross{k} + 2mc^2 \gamma_{\kappa}
\cross{q}' \\
&\ & +2mc^2 \cross{q} \gamma_{\kappa} + 2mc^2 \cross{k} \gamma_{\kappa}
  + 2mc^2  \cross{q}  \gamma_{\kappa} + 2mc^2 \gamma_{\kappa} \cross{q}'  - 2mc^2 \gamma_{\kappa}
  \cross{k}
  -2 m^2c^4 \gamma_{\kappa}
\end{eqnarray*}

According to (\ref{eq:s4-1112a}), integral $I^{\kappa}(\tilde{q},
\tilde{q}')$ is multiplied by $\overline{u}(\mathbf{q}, \sigma)$
from the left and by $u(\mathbf{q}', \sigma')$ from the right. Then,
due to (\ref{gamma-mu2}) - (\ref{gamma-mu1}), in the above
summands the factor $\cross{q}$ standing on the left and the factor
$\cross{q}'$ standing on the right can both be changed to $mc^2$

\begin{eqnarray*}
 T_1    &=& -2  m^2c^4  \gamma_{\kappa}
-2 mc^2 \cross{k} \gamma_{\kappa}  +2  mc^2 \gamma_{\kappa} \cross{k} +2
\cross{k} \gamma_{\kappa} \cross{k} + 2m^2c^4 \gamma_{\kappa} \\
&\ & +2m^2c^4  \gamma_{\kappa} + 2mc^2 \cross{k} \gamma_{\kappa}
  + 2m^2c^4    \gamma_{\kappa} + 2m^2c^4 \gamma_{\kappa}   - 2mc^2 \gamma_{\kappa}
  \cross{k}
  -2 m^2c^4 \gamma_{\kappa}    \\
&=&     2\cross{k} \gamma_{\kappa} \cross{k} + 4m^2c^4 \gamma_{\kappa} \\
\end{eqnarray*}

\noindent It follows from (\ref{gamma-mu}) and (\ref{eq:J.45a}) that

\begin{eqnarray*}
\cross{k} \gamma_{\kappa} \cross{k}  &=& \gamma_{\mu} \gamma_{\kappa}
\gamma_{\nu} k^{\mu} k^{\nu} = - \gamma_{\kappa}\gamma_{\mu}
\gamma_{\nu} k^{\mu} k^{\nu} + 2
\gamma_{\nu} g_{\mu \kappa} k^{\mu} k^{\nu} \\
&=& - \gamma_{\kappa} \cross{k}^2 + 2 \cross{k} k_{\kappa} = -
\gamma_{\kappa} \tilde{k}^2 + 2 (\cross{q}'-\cross{q}) k_{\kappa}
\end{eqnarray*}

\noindent The last term vanishes when sandwiched between
$\overline{u}(\mathbf{q}, \sigma)$ and $u(\mathbf{q}', \sigma')$.
 So, we can set $ \cross{k} \gamma_{\kappa} \cross{k} = -\tilde{k}^2
 \gamma_{\kappa} $. Then

\begin{eqnarray*}
 T_1
&=&     (-2\tilde{k}^2 +4m^2c^4)\gamma_{\kappa}
\end{eqnarray*}

\noindent We use the same techniques to obtain the 2nd, 3rd, and 4th
terms in (\ref{eq:I-kappa2})

\begin{eqnarray*}
T_2 &=& \gamma_{\mu} (\cross{q} + \cross{q}') \gamma_{\kappa} (\cross{q}'
+ mc^2) \gamma_{\mu} \\
&=& \gamma_{\mu} \cross{q}  \gamma_{\kappa} \cross{q}' \gamma_{\mu}
+\gamma_{\mu}  \cross{q}' \gamma_{\kappa} \cross{q}' \gamma_{\mu}
+mc^2\gamma_{\mu} \cross{q}  \gamma_{\kappa} \gamma_{\mu}
+mc^2\gamma_{\mu} \cross{q}' \gamma_{\kappa}  \gamma_{\mu} \\
&=& -2 \cross{q}'  \gamma_{\kappa} \cross{q} -2  \cross{q}'
\gamma_{\kappa} \cross{q}'  + 2mc^2 \cross{q} \gamma_{\kappa} +2 mc^2
\gamma_{\kappa} \cross{q}
+2 mc^2 \cross{q}' \gamma_{\kappa} +2mc^2 \gamma_{\kappa} \cross{q}'\\
&=& -2 (\cross{q}+\cross{k})  \gamma_{\kappa} (\cross{q}'-\cross{k}) -2
(\cross{q}+ \cross{k}) \gamma_{\kappa} \cross{q}'  + 2mc^2 \cross{q}
\gamma_{\kappa} +2 mc^2 \gamma_{\kappa} (\cross{q}'-\cross{k})
\\
&\ &+2 mc^2 (\cross{q} +\cross{k}) \gamma_{\kappa} +2mc^2 \gamma_{\kappa} \cross{q}'\\
&=& -2 \cross{q}  \gamma_{\kappa} \cross{q}' -2 \cross{k}  \gamma_{\kappa}
\cross{q}' +2 \cross{q}  \gamma_{\kappa} \cross{k} +2 \cross{k}
\gamma_{\kappa} \cross{k} -2 \cross{q} \gamma_{\kappa} \cross{q}' \\
&\ &-2 \cross{k} \gamma_{\kappa} \cross{q}'
 + 2mc^2 \cross{q}
\gamma_{\kappa} +2 mc^2 \gamma_{\kappa} \cross{q}' -2 mc^2
\gamma_{\kappa} \cross{k}
\\
&\ & +2 mc^2 \cross{q}  \gamma_{\kappa} + 2 mc^2 \cross{k} \gamma_{\kappa} +2mc^2 \gamma_{\kappa} \cross{q}'\\
&=& -2 m^2c^4  \gamma_{\kappa}  -2 mc^2\cross{k}  \gamma_{\kappa}
 +2 mc^2  \gamma_{\kappa} \cross{k} +2 \cross{k}
\gamma_{\kappa} \cross{k} -2 m^2c^4 \gamma_{\kappa} \\
&\ &-2 mc^2\cross{k} \gamma_{\kappa}
 + 2m^2c^4
\gamma_{\kappa} +2 m^2c^4 \gamma_{\kappa}  -2 mc^2 \gamma_{\kappa}
\cross{k}
\\
&\ &+2 m^2c^4   \gamma_{\kappa} + 2 mc^2 \cross{k} \gamma_{\kappa} +2m^2c^4 \gamma_{\kappa} \\
&=& 2 \cross{k} \gamma_{\kappa} \cross{k}  -2 mc^2\cross{k}
\gamma_{\kappa}
 + 4m^2c^4
\gamma_{\kappa} \\
&=& (-2 \tilde{k}^2 + 4m^2c^4)\gamma_{\kappa}   -2 mc^2\cross{k}
\gamma_{\kappa} \\
 T_3 &=& \gamma_{\mu} (\cross{q} + mc^2)
\gamma_{\kappa} (\cross{q} +
\cross{q}') \gamma_{\mu} \\
&=& \gamma_{\mu} \cross{q}   \gamma_{\kappa} \cross{q}  \gamma_{\mu}
+\gamma_{\mu}  mc^2  \gamma_{\kappa} \cross{q}  \gamma_{\mu}
+\gamma_{\mu} \cross{q}   \gamma_{\kappa}  \cross{q}' \gamma_{\mu}
+\gamma_{\mu}  mc^2  \gamma_{\kappa}  \cross{q}' \gamma_{\mu} \\
&=& -2 \cross{q}   \gamma_{\kappa} \cross{q}   + 2mc^2 \gamma_{\kappa}
\cross{q} + 2mc^2 \cross{q} \gamma_{\kappa}   -2  \cross{q}'
\gamma_{\kappa} \cross{q} + 2mc^2   \gamma_{\kappa} \cross{q}' + 2mc^2
\cross{q}'\gamma_{\kappa}
\\
&=& -2 \cross{q}   \gamma_{\kappa} (\cross{q}'-\cross{k})   + 2mc^2
\gamma_{\kappa} (\cross{q}'-\cross{k}) + 2mc^2 \cross{q} \gamma_{\kappa}
-2 (\cross{q}+\cross{k}) \gamma_{\kappa} (\cross{q}'-\cross{k}) \\
&\ & +2mc^2 \gamma_{\kappa} \cross{q}' + 2mc^2
(\cross{q}+\cross{k})\gamma_{\kappa}
\\
&=& -2 \cross{q}   \gamma_{\kappa} \cross{q}' +2 \cross{q} \gamma_{\kappa}
\cross{k} + 2mc^2 \gamma_{\kappa} \cross{q}' - 2mc^2 \gamma_{\kappa}
\cross{k}
 + 2mc^2 \cross{q}
\gamma_{\kappa} \\
&\ &-2 \cross{q} \gamma_{\kappa} \cross{q}' -2 \cross{k} \gamma_{\kappa}
\cross{q}' +2 \cross{q} \gamma_{\kappa} \cross{k} +2 \cross{k}
\gamma_{\kappa} \cross{k}
\\
&\ & +2mc^2 \gamma_{\kappa} \cross{q}' + 2mc^2 \cross{q}\gamma_{\kappa} +
2mc^2 \cross{k}\gamma_{\kappa}
\\
&=& -2 m^2c^4   \gamma_{\kappa}  +2 mc^2 \gamma_{\kappa} \cross{k} +
2m^2c^4 \gamma_{\kappa}  - 2mc^2 \gamma_{\kappa} \cross{k}
 + 2m^2c^4
\gamma_{\kappa} \\
&\ & -2 m^2c^4 \gamma_{\kappa}  -2 mc^2\cross{k} \gamma_{\kappa}  +2 mc^2
\gamma_{\kappa} \cross{k} +2 \cross{k} \gamma_{\kappa} \cross{k}
\\
&\ & +2m^2c^4 \gamma_{\kappa}  + 2m^2c^4 \gamma_{\kappa} + 2mc^2
\cross{k}\gamma_{\kappa}
\\
&=& (4m^2c^4 -2\tilde{k}^2) \gamma_{\kappa} +  2 mc^2
\gamma_{\kappa} \cross{k} \\
 T_4 &=& \gamma_{\mu} (\cross{q} + \cross{q}')
\gamma_{\kappa} (\cross{q}
+ \cross{q}') \gamma_{\mu} \\
&=& \gamma_{\mu} \cross{q} \gamma_{\kappa} \cross{q} \gamma_{\mu} +
\gamma_{\mu} \cross{q}' \gamma_{\kappa} \cross{q} \gamma_{\mu} +
\gamma_{\mu} \cross{q} \gamma_{\kappa} \cross{q}' \gamma_{\mu} +
\gamma_{\mu} \cross{q}' \gamma_{\kappa} \cross{q}' \gamma_{\mu} \\
&=& -2 \cross{q} \gamma_{\kappa} \cross{q}  -2 \cross{q} \gamma_{\kappa}
\cross{q}' -2 \cross{q}' \gamma_{\kappa} \cross{q} -2
\cross{q}' \gamma_{\kappa} \cross{q}'  \\
&=& -2 \cross{q} \gamma_{\kappa} (\cross{q}'-\cross{k})  -2 \cross{q}
\gamma_{\kappa} \cross{q}' -2 (\cross{q}+\cross{k}) \gamma_{\kappa}
(\cross{q}'-\cross{k}) -2
(\cross{q}+\cross{k}) \gamma_{\kappa} \cross{q}'  \\
&=& -2 \cross{q} \gamma_{\kappa} \cross{q}' +2 \cross{q} \gamma_{\kappa}
\cross{k} -2 \cross{q} \gamma_{\kappa} \cross{q}' -2 \cross{q}
\gamma_{\kappa} \cross{q}' -2 \cross{k} \gamma_{\kappa} \cross{q}' \\
&\ &+2 \cross{q} \gamma_{\kappa} \cross{k} +2 \cross{k} \gamma_{\kappa}
\cross{k}
 -2
\cross{q} \gamma_{\kappa} \cross{q}' -2
\cross{k} \gamma_{\kappa} \cross{q}'  \\
&=& -2 m^2c^4 \gamma_{\kappa}  +2 mc^2 \gamma_{\kappa} \cross{k} -2
m^2c^4 \gamma_{\kappa} -2 m^2c^4
\gamma_{\kappa}  -2 mc^2\cross{k} \gamma_{\kappa}  \\
&\ &+2 mc^2 \gamma_{\kappa} \cross{k} +2 \cross{k} \gamma_{\kappa} \cross{k}
 -2m^2c^4 \gamma_{\kappa}  -2mc^2
\cross{k} \gamma_{\kappa}  \\
&=& -8 m^2c^4 \gamma_{\kappa}  +4 mc^2 \gamma_{\kappa} \cross{k}   -4
mc^2\cross{k} \gamma_{\kappa}  -2 \tilde{k}^2 \gamma_{\kappa}
\end{eqnarray*}

\noindent Putting all terms together we obtain\footnote{Here we used (\ref{eq:gkk-kgk}) to write $\gamma_{\kappa} \cross{k} - \cross{k} \gamma_{\kappa} = 4
mc^2\gamma_{\kappa}
 - 2(q_{\kappa} +q'_{\kappa})$.}

\begin{eqnarray*}
I^{\kappa}(q, q')
 &=& J (-2\tilde{k}^2 +4m^2c^4)\gamma_{\kappa}  \\
&\ & -A ((-2 \tilde{k}^2 + 4m^2c^4)\gamma_{\kappa}   -2 mc^2\cross{k}
\gamma_{\kappa}) -
 A ((4m^2c^4 -2\tilde{k}^2) \gamma_{\kappa} +  2 mc^2 \gamma_{\kappa}
\cross{k})  \\
&\ & +\frac{A}{4} (-8 m^2c^4 \gamma_{\kappa}  +4 mc^2 \gamma_{\kappa} \cross{k} -4
mc^2\cross{k} \gamma_{\kappa}  -2 \tilde{k}^2 \gamma_{\kappa})
 + 2D \tilde{k}^2 \gamma_{\kappa}
 + 4E  \gamma_{\kappa} \\
 &=& \left(-2J\tilde{k}^2 +2D \tilde{k}^2+4Jm^2c^4 +4E -10Am^2c^4 +\frac{7}{2}
 A\tilde{k}^2\right) \gamma_{\kappa} \\
&\ &- A  mc^2(\gamma_{\kappa} \cross{k}-\cross{k}\gamma_{\kappa}) \\
  &=& \left(-2J\tilde{k}^2 +2D \tilde{k}^2+4Jm^2c^4 +4E -14Am^2c^4
  +\frac{7A \tilde{k}^2}{2} \right)\gamma_{\kappa}  \\
&+& 2A  mc^2( q_{\kappa} +q'_{\kappa})
\end{eqnarray*}

\noindent The coefficient in front of $\gamma_{\kappa}$ is

\begin{eqnarray*}
&\ & -\frac{2 \pi^2(-2\tilde{k}^2 + 4m^2c^4) }{i c^3 m^2 c^4 \sin (2
\theta)}\left[ \theta \ln\frac{m}{ \lambda}  - \int
\limits_{0}^{\theta} \alpha \tan \alpha d\alpha \right]
-  \frac{\pi^2(1 - \theta \cot \theta)}{ic^3 } \\
&\ & +\frac{2 \pi^2}{ic^3 } \left(\ln \frac{\Lambda }{m} + (1 -
\theta \cot
 \theta)- \frac{1}{4} \right) + \frac{14 \pi^2 \theta}{ic^3  \sin(2 \theta)}
- \frac{7 \pi^2\tilde{k}^2\theta}{2ic^3 m^2 c^4 \sin(2 \theta)} \\
&=& -\frac{2 \pi^2(-8m^2c^4\sin^2 \theta + 4m^2c^4) }{i c^3 m^2 c^4
\sin (2 \theta)}\left[ \theta \ln \frac{m}{ \lambda} - \int
\limits_{0}^{\theta}
\alpha \tan \alpha d\alpha \right] \\
&\ & +\frac{ \pi^2(1 - \theta \cot \theta)}{ic^3} + \frac{2
\pi^2}{ic^3 } \ln \frac{\Lambda }{m}  - \frac{\pi^2}{2ic^3 }+
\frac{14 \pi^2 \theta}{ic^3  \sin(2 \theta)}
- \frac{14 \pi^2\theta m^2c^4 \sin^2 \theta}{ic^3 m^2 c^4 \sin(2 \theta)} \\
&=& -\frac{8 \pi^2  }{i c^3  \tan (2 \theta)}\left[ \theta
\ln\frac{m }{ \lambda}  - \int \limits_{0}^{\theta}
\alpha \tan \alpha d\alpha \right] \\
&\ & +\frac{ \pi^2(1 - \theta \cot \theta)}{ic^3} + \frac{2
\pi^2}{ic^3 } \ln \frac{\Lambda }{m}  - \frac{\pi^2}{2ic^3 } +
\frac{7 \pi^2 \theta \cot \theta}{ic^3  } \\
\end{eqnarray*}

\noindent Therefore, finally

\begin{eqnarray}
&\ &I^{\kappa}(\tilde{q}, \tilde{q}') \nonumber \\
&=&\frac{\pi^2\gamma^{\kappa}}{ic^3}\Bigr(-\frac{8 \theta }{ \tan (2
\theta)}  \ln\frac{m }{ \lambda}  + \frac{8 }{\tan (2 \theta)}
\int \limits_{0}^{\theta} \alpha \tan \alpha d\alpha + \frac{1}{2}
+6\theta
\cot \theta + 2 \ln \frac{\Lambda }{m}\Bigl) \nonumber \\
&\ &-
  \frac{2 \pi^2 \theta ( q +q')^{\kappa}}{i m c^5  \sin(2
\theta)} \label{eq:I-kappa-qq}
\end{eqnarray}

\section{Integral for the ladder diagram}
\label{sc:calc-ladder}

For the integral (\ref{eq:bpqk})

\begin{eqnarray*}
b(\mathbf{p, q, k})
 &=&   \int
\frac{ d^4h}{[\tilde{h}^2 + 2(\tilde{q} \cdot \tilde{h})][\tilde{h}^2 - 2(\tilde{p} \cdot \tilde{h})][\tilde{h}^2 - \lambda^2 c^4][\tilde{h}^2 + 2(\tilde{h} \cdot \tilde{k}) + \tilde{k}^2 - \lambda^2 c^4]}
\end{eqnarray*}

\noindent we follow the calculation technique from \cite{Redhead}. First use equation (\ref{eq:abcd}) and notation

\begin{eqnarray*}
a &=& \tilde{h}^2 + 2(\tilde{k} \cdot \tilde{h}) +\tilde{k}^2 - \lambda^2 c^4 \label{eq:a}\\
b &=& \tilde{h}^2 +2 (\tilde{q} \cdot \tilde{h}) \\
c &=& \tilde{h}^2 -2 (\tilde{p} \cdot \tilde{h}) \\
d &=& \tilde{h}^2 - \lambda^2 c^4 \label{eq:d}
\end{eqnarray*}

\noindent to write

\begin{eqnarray*}
&\mbox{ }& b(\mathbf{p, q, k}) \\
&=&   6 \int d^4h \int \limits_{0}^{1}  dx \int \limits_{0}^{1} dy
\int \limits_{0}^{1} xz^2 dz[(\tilde{h}^2 + 2(\tilde{k} \cdot \tilde{h}) +k^2 - \lambda^2 c^4)z(1-x) +
(\tilde{h}^2 +2 (\tilde{q} \cdot \tilde{h}))xyz \\
&+& (\tilde{h}^2 -2 (\tilde{p} \cdot \tilde{h}))xz(1-y) +
(\tilde{h}^2 - \lambda^2 c^4)(1-z)]^{-4} \\
&=&   6 \int d^4h \int \limits_{0}^{1}  dx \int \limits_{0}^{1} dy
\int \limits_{0}^{1} xz^2 dz[\tilde{h}^2
-2\tilde{h} \cdot(-\tilde{k}z(1-x) -\tilde{q}xyz + \tilde{p}xz(1-y)) \\
&+& \tilde{k}^2 z(1-x) + \lambda^2 c^4(zx  -1)  ]^{-4} \\
&=&   6 \int d^4h \int \limits_{0}^{1}  dx \int \limits_{0}^{1} dy
\int \limits_{0}^{1} \frac{xz^2 dz}{[\tilde{h}^2
-2(\tilde{h} \cdot \tilde{p}_x) z - \Delta  ]^4} \\
\end{eqnarray*}

\noindent where

\begin{eqnarray*}
\Delta &\equiv& \lambda^2 c^4(1-zx) -\tilde{k}^2 z(1-x) \\
\tilde{p}_x &=& -\tilde{k}(1-x)  + \tilde{p}x(1-y) -\tilde{q}xy = -\tilde{k}(1-x)  + x \tilde{p}_y \\
\tilde{p}_y &=& \tilde{p}(1-y) -\tilde{q}y
\end{eqnarray*}

\noindent From (\ref{eq:k2-pk}) we obtain

\begin{eqnarray*}
 b(\mathbf{p, q, k}) =  \frac{ \pi^2 }{ic^3}  \int \limits_{0}^{1}  dx \int \limits_{0}^{1}
dy \int
\limits_{0}^{1} \frac{xz^2 dz}{(z^2 \tilde{p}_x^2 + \Delta)^2 } \\
\end{eqnarray*}

\noindent We have $\tilde{q}'=\tilde{q}-\tilde{k}$ and $\tilde{p}'=\tilde{p}+\tilde{k}$. Taking squares of both sides of these equations and using $\tilde{q}^2 = (\tilde{q}')^2 = m^2c^4$ and $\tilde{p}^2 = (\tilde{p}')^2 = M^2c^4$ we obtain

\begin{eqnarray}
(\tilde{q} \cdot \tilde{k}) &=& \tilde{k}^2/2 \label{eq:M.50} \\
(\tilde{p} \cdot \tilde{k}) &=& -\tilde{k}^2/2 \label{eq:M.51}  \\
(\tilde{k} \cdot \tilde{p}_y) &=& (\tilde{k} \cdot \tilde{p})(1-y) - (\tilde{k} \cdot \tilde{q})y = -\frac{\tilde{k}^2}{2} (1-y) -  \frac{\tilde{k}^2}{2}y = -\frac{\tilde{k}^2}{2} \nonumber \\
\tilde{p}_x^2 &=& (x\tilde{p}_y -\tilde{k}(1-x))^2 = x^2 \tilde{p}_y^2 + \tilde{k}^2(1-x)^2 - 2x(1-x)(\tilde{p}_y \cdot \tilde{k}) \nonumber \\
&=& x^2 \tilde{p}_y^2 + \tilde{k}^2 -2\tilde{k}^2x + \tilde{k}^2x^2 + \tilde{k}^2x - \tilde{k}^2x^2 = x^2 \tilde{p}_y^2 + \tilde{k}^2(1-x) \nonumber \\
 b(\mathbf{p, q, k}) &=&   \frac{ \pi^2 }{ic^3}  \int \limits_{0}^{1}  dx \int \limits_{0}^{1}
dy \int
\limits_{0}^{1} \frac{xz^2 dz}{[z^2 (x^2 \tilde{p}_y^2 + \tilde{k}^2(1-x)) + \lambda^2 c^4(1-zx) -\tilde{k}^2 z(1-x)]^2 } \nonumber
\end{eqnarray}

\noindent Even though $\lambda$ is small, the term $\lambda^2 c^4(1-zx)$ cannot be neglected\footnote{because other terms in the denominator can be even smaller} when  $x \to 0, z \to 0$, when  $x \to 1, z \to 0$
and when  $x \to 0, z \to 1$.
Therefore, we are going to break the region of integration on $x$ into three parts $0<x < \epsilon$, $ \epsilon < x < 1 - \delta$ and $ 1 - \delta < x < 1$, where $\epsilon$ and $\delta$ are small, but large enough, so that in the interval $ \epsilon < x < 1 - \delta$ the term $\lambda^2 c^4(1-zx)$ can be neglected. Integrations on $x$ in these three regions split our integral into three parts

\begin{eqnarray*}
 b(\mathbf{p, q, k}) =  L_I + L_{II} + L_{III}
\end{eqnarray*}

In the second region we neglect the $\lambda$-term

\begin{eqnarray*}
 L_{II}
&\approx&   \frac{ \pi^2 }{ic^3}  \int \limits_{\epsilon}^{1- \delta}  dx \int \limits_{0}^{1}
dy \int
\limits_{0}^{1} \frac{x dz}{[z (x^2 \tilde{p}_y^2 + \tilde{k}^2(1-x))  -\tilde{k}^2(1-x)]^2 }
\end{eqnarray*}

\noindent use table integrals

\begin{eqnarray*}
 \int \frac{dz}{(az+b)^2} &=& - \frac{1}{a(ax+b)} + const \\
 \int \frac{dx}{x(1-x)} &=& \ln(x) - \ln(x-1) + const
\end{eqnarray*}

\noindent and obtain

\begin{eqnarray*}
 L_{II}
&=&   -\frac{ \pi^2 }{ic^3}  \int \limits_{\epsilon}^{1- \delta}  dx \int \limits_{0}^{1}
dy \frac{x}{[x^2 \tilde{p}_y^2 + \tilde{k}^2(1-x)][z (x^2 \tilde{p}_y^2 + \tilde{k}^2(1-x))  -\tilde{k}^2(1-x)]} \Big|_{z=0}^{z=1} \\
&=&  - \frac{ \pi^2 }{ic^3}  \int \limits_{\epsilon}^{1- \delta}  dx \int \limits_{0}^{1}
dy \frac{x}{x^2 \tilde{p}_y^2 + \tilde{k}^2(1-x)} \Bigl( \frac{1}{x^2 \tilde{p}_y^2 }
+
\frac{1}{  \tilde{k}^2(1-x)} \Bigr) \\
&=&  -\frac{ \pi^2 }{ic^3 \tilde{k}^2}  \int \limits_{\epsilon}^{1- \delta}  dx \int \limits_{0}^{1}
dy  \frac{1}{x \tilde{p}_y^2 (1-x)} \\
&=&  -\frac{ \pi^2 }{ic^3 \tilde{k}^2}  \int \limits_{0}^{1}
dy  \frac{1}{\tilde{p}_y^2 } \left( \ln(x) - \ln(x-1)  \right)\Bigl|_{x=\epsilon}^{x=1 - \delta} \\
&\approx&  -\frac{ \pi^2 }{ic^3 \tilde{k}^2}  \int \limits_{0}^{1}
dy  \frac{1}{\tilde{p}_y^2 } \left(  - \ln(\delta) -\ln(-1) - \ln(\epsilon) + \ln( - 1) \right) \\
&=& \frac{ \pi^2 \ln(\delta \epsilon) }{ic^3 \tilde{k}^2}  \int \limits_{0}^{1}  \frac{dy}{\tilde{p}_y^2 }
\end{eqnarray*}

\noindent In the third integral we replace $x \to 1-x$

\begin{eqnarray*}
 L_{III}
&=&   -\frac{ \pi^2 }{ic^3}  \int \limits_{\delta}^0  dx \int \limits_{0}^{1}
dy \int
\limits_{0}^{1} \frac{(1-x)z^2 dz}{[z^2 ((1-x)^2 \tilde{p}_y^2 + \tilde{k}^2x) + \lambda^2 c^4(1-z(1-x)) -\tilde{k}^2 zx]^2 } \\
&\approx& -\frac{ \pi^2 }{ic^3}  \int \limits_{\delta}^0  dx \int \limits_{0}^{1}
dy \int
\limits_{0}^{1} \frac{z^2 dz}{[z^2 \tilde{p}_y^2 + z^2  \tilde{k}^2x + \lambda^2 c^4(1-z) -\tilde{k}^2 zx]^2 } \\
&=& \frac{ \pi^2 }{ic^3 k^2}   \int \limits_{0}^{1}
dy \int
\limits_{0}^{1}  \frac{z dz}{(z-1)}
\left(  \frac{1}{z^2 \tilde{p}_y^2 + z^2  \tilde{k}^2x + \lambda^2 c^4(1-z) -\tilde{k}^2 zx} \right)\Bigl|_{x=\delta}^{x=0} \\
&=& \frac{ \pi^2 }{ic^3 \tilde{k}^2}   \int \limits_{0}^{1}
dy \int
\limits_{0}^{1}  \frac{z dz}{(z-1)}
\left( \frac{1}{z^2 \tilde{p}_y^2  + \lambda^2 c^4(1-z)} - \frac{1}{z^2 \tilde{p}_y^2 + z^2  \tilde{k}^2 \delta + \lambda^2 c^4(1-z) -\tilde{k}^2 z \delta} \right) \\
&=& \frac{ \pi^2 }{ic^3 \tilde{k}^2}   \int \limits_{0}^{1}
dy \int
\limits_{0}^{1}  \frac{z dz}{(z-1)} \cdot
 \frac{z^2 \tilde{p}_y^2 + z^2  \tilde{k}^2 \delta + \lambda^2 c^4(1-z) -\tilde{k}^2 z \delta - z^2 \tilde{p}_y^2  -\lambda^2 c^4(1-z)}{(z^2 \tilde{p}_y^2  + \lambda^2 c^4(1-z))(z^2 \tilde{p}_y^2 + z^2  \tilde{k}^2 \delta + \lambda^2 c^4(1-z) -\tilde{k}^2 z \delta)} \\
  &=& \frac{ \pi^2  \delta}{ic^3 }   \int \limits_{0}^{1}
dy \int
\limits_{0}^{1}
 \frac{ z^2  dz }{[z^2 \tilde{p}_y^2  + \lambda^2 c^4(1-z)][z^2 \tilde{p}_y^2 + z^2  \tilde{k}^2 \delta + \lambda^2 c^4(1-z) -\tilde{k}^2 z \delta]}
\end{eqnarray*}

\noindent We now break the $z$ integration into two regions $0 \leq z < z_c$ and $z_c \leq z < 1$, where $z_c$ is chosen such that $\lambda^2 c^4 \ll z_c^2 \tilde{p}_y^2 \ll \tilde{k}^2 z_c \delta $. We also use table integrals

\begin{eqnarray}
 \int \frac{   dz }{   z(az +b)} &=& \frac{1}{b}[\ln(z) - \ln(az+b)] + const \nonumber \\
 \int dz \frac{a+bz}{a+cz^2 } &=& \frac{b}{2c} \ln(a+cz^2) + \sqrt{\frac{a}{c}} \tan^{-1}\left(\frac{\sqrt{c}z}{\sqrt{a}} \right) + const \nonumber \\
 &\approx& \frac{b}{2c} \ln(a+cz^2) + const \label{eq:abzacz} \\
 \int dz \frac{a}{a+cz } &=& \frac{a}{c}\ln(a+cz) + const \nonumber
\end{eqnarray}

\noindent Then

\begin{eqnarray}
L_{III} &=&  L_{IIIa} +  L_{IIIb} \nonumber \\
 L_{IIIb}
  &=& \frac{ \pi^2  \delta}{ic^3 }   \int \limits_{0}^{1}
dy \int
\limits_{z_c}^{1}
 \frac{ z^2  dz }{[z^2 \tilde{p}_y^2  + \lambda^2 c^4(1-z)][z^2 \tilde{p}_y^2 + z^2  \tilde{k}^2 \delta + \lambda^2 c^4(1-z) -\tilde{k}^2 z \delta]} \nonumber \\
   &\approx& \frac{ \pi^2  \delta}{ic^3 }   \int \limits_{0}^{1}
dy \int
\limits_{z_c}^{1}
 \frac{   dz }{ \tilde{p}_y^2  z[z (\tilde{p}_y^2 + \tilde{k}^2 \delta)  -\tilde{k}^2 \delta]} \nonumber \\
   &=& -\frac{ \pi^2  \delta}{ic^3 }   \int \limits_{0}^{1}
\frac{dy}{\tilde{p}_y^2}\frac{1}{\tilde{k}^2 \delta}  \left( \ln(z) - \ln(z (\tilde{p}_y^2 + \tilde{k}^2 \delta)  -\tilde{k}^2 \delta) \right) \Bigl|_{z=z_c}^{z=1} \nonumber \\
   &=& -\frac{ \pi^2  \delta}{ic^3 }   \int \limits_{0}^{1}
\frac{dy}{\tilde{p}_y^2}\frac{1}{\tilde{k}^2 \delta}  \left(  - \ln(\tilde{p}_y^2) - \ln(z_c) + \ln[z_c (\tilde{p}_y^2 + \tilde{k}^2 \delta)  -\tilde{k}^2 \delta] \right) \nonumber \\
  &\approx& -\frac{ \pi^2  }{ic^3 \tilde{k}^2}   \int \limits_{0}^{1}
\frac{dy}{\tilde{p}_y^2} \ln \left( \frac{-\tilde{k}^2 \delta }{\tilde{p}_y^2 z_c} \right) \label{eq:LIIIb}
\end{eqnarray}

\begin{eqnarray}
 L_{IIIa}
  &=& \frac{ \pi^2  \delta}{ic^3 }   \int \limits_{0}^{1}
dy \int
\limits_{0}^{z_c}
 \frac{ z^2  dz }{[z^2 \tilde{p}_y^2  + \lambda^2 c^4(1-z)][z^2 \tilde{p}_y^2 + z^2  \tilde{k}^2 \delta + \lambda^2 c^4(1-z) -\tilde{k}^2 z \delta]} \nonumber \\
  &\approx& \frac{ \pi^2  \delta}{ic^3 }   \int \limits_{0}^{1}
dy \int
\limits_{0}^{z_c}
 \frac{ z^2  dz }{(z^2 \tilde{p}_y^2  + \lambda^2 c^4)( \lambda^2 c^4 -\tilde{k}^2 z \delta)} \nonumber \\
   &=& -\frac{ \pi^2  \delta}{ic^3 }   \int \limits_{0}^{1}
dy \int
\limits_{0}^{z_c}
dz \frac{1}{ \tilde{p}_y^2 \lambda^2 c^4 + \tilde{k}^4 \delta^2} \left(\frac{\lambda^2 c^4 + \tilde{k}^2 z \delta}{z^2 \tilde{p}_y^2  + \lambda^2 c^4} - \frac{\lambda^2 c^4}{ \lambda^2 c^4 -\tilde{k}^2 z \delta}\right) \nonumber \\
   &=& -\frac{ \pi^2  \delta}{ic^3 }   \int \limits_{0}^{1}
dy \frac{1}{ \tilde{p}_y^2 \lambda^2 c^4 + \tilde{k}^4 \delta^2}
 \left( \frac{\tilde{k}^2 \delta}{2 \tilde{p}_y^2} \ln(\lambda^2 c^4 + z^2 \tilde{p}_y^2) + \frac{\lambda^2 c^4}{\tilde{k}^2 \delta}
 \ln(\lambda^2 c^4 -\tilde{k}^2 z \delta)\right)  \Bigl|_{z=0}^{z=z_c} \nonumber \\
    &=& -\frac{ \pi^2  \delta}{ic^3 }   \int \limits_{0}^{1}
dy \frac{1}{ \tilde{p}_y^2 \lambda^2 c^4 + \tilde{k}^4 \delta^2}
 \Bigl( \frac{\tilde{k}^2 \delta}{2 \tilde{p}_y^2} \ln(\lambda^2 c^4 + z_c^2 \tilde{p}_y^2) + \frac{\lambda^2 c^4}{\tilde{k}^2 \delta}
 \ln(\lambda^2 c^4 -\tilde{k}^2 z_c \delta) \nonumber \\
  &-& \frac{k^2 \delta}{2 \tilde{p}_y^2} \ln(\lambda^2 c^4) - \frac{\lambda^2 c^4}{\tilde{k}^2 \delta}
 \ln(\lambda^2 c^4 )\Bigr)  \nonumber \\
&\approx& -\frac{ \pi^2  \delta}{ic^3 }   \int \limits_{0}^{1}
dy \frac{1}{ \tilde{p}_y^2 \lambda^2 c^4 + \tilde{k}^4 \delta^2}
 \Bigl( \frac{\tilde{k}^2 \delta}{2 \tilde{p}_y^2} \ln(z_c^2 \tilde{p}_y^2) + \frac{\lambda^2 c^4}{\tilde{k}^2 \delta}
 \ln( -\tilde{k}^2 z_c \delta) \nonumber \\
  &-& \frac{\tilde{k}^2 \delta}{2 \tilde{p}_y^2} \ln(\lambda^2 c^4) - \frac{\lambda^2 c^4}{\tilde{k}^2 \delta}
 \ln(\lambda^2 c^4 )\Bigr) \nonumber \\
&\approx& -\frac{ \pi^2  }{ic^3  2\tilde{k}^2 }   \int \limits_{0}^{1}
dy \frac{1}{ \tilde{p}_y^2} \ln \left(\frac{z_c^2 \tilde{p}_y^2}{\lambda^2 c^4} \right)  \label{eq:LIIIa}
\end{eqnarray}

Adding together (\ref{eq:LIIIb}) and (\ref{eq:LIIIa}) we obtain

\begin{eqnarray*}
L_{III}&=& L_{IIIa} +  L_{IIIb} = -\frac{ \pi^2  }{ic^3 \tilde{k}^2}   \int \limits_{0}^{1}
\frac{dy}{\tilde{p}_y^2} \ln \left( \frac{-\tilde{k}^2 \delta }{\tilde{p}_y^2 z_c} \right)
- \frac{ \pi^2  }{ic^3  2\tilde{k}^2 }   \int \limits_{0}^{1}
dy \frac{1}{ \tilde{p}_y^2} \ln \left(\frac{z_c^2 \tilde{p}_y^2}{\lambda^2 c^4} \right)  \\
&\approx& -\frac{ \pi^2  }{ic^3 2\tilde{k}^2}   \int \limits_{0}^{1}
\frac{dy}{\tilde{p}_y^2} \ln \left( \frac{\tilde{k}^4 \delta^2 }{\tilde{p}_y^4 z_c^2} \right)
- \frac{ \pi^2  }{ic^3  2\tilde{k}^2 }   \int \limits_{0}^{1}
dy \frac{1}{ \tilde{p}_y^2} \ln \left(\frac{z_c^2 \tilde{p}_y^2}{\lambda^2 c^4} \right) \\
&=& -\frac{ \pi^2  }{ic^3 2\tilde{k}^2}   \int \limits_{0}^{1}
\frac{dy}{\tilde{p}_y^2} \ln \left( \frac{\tilde{k}^4 \delta^2 }{\tilde{p}_y^2 \lambda^2 c^4}\right)
\end{eqnarray*}

\noindent In the integral $L_I$ we replace $z \to 1-z$

\begin{eqnarray*}
&\ & L_{I} =  \frac{ \pi^2 }{ic^3}  \int \limits_{0}^{\epsilon}  dx \int \limits_{0}^{1}
dy \times \\
&\ & \int
\limits_{0}^{1} \frac{x(1-z)^2 dz}{[(1-z)^2 (x^2 \tilde{p}_y^2 + \tilde{k}^2(1-x)) + \lambda^2 c^4(1-(1-z)x) -\tilde{k}^2 (1-z)(1-x)]^2 }
\end{eqnarray*}

\noindent and break $z$-integration into two regions $0 \leq z < z_c$ and $z_c \leq z \leq 1$, where $z_c$ is small, but large enough, so that in the second region we can neglect the $\lambda$-term. Then

\begin{eqnarray*}
 L_{Ia}
&\approx&   \frac{ \pi^2 }{ic^3}  \int \limits_{0}^{\epsilon}  dx \int \limits_{0}^{1}
dy \int
\limits_{0}^{z_c} \frac{x dz}{[(1-2z) (x^2 \tilde{p}_y^2 + \tilde{k}^2(1-x)) + \lambda^2 c^4 -\tilde{k}^2 (1-z)(1-x)]^2 } \\
&=&  \frac{ \pi^2 }{ic^3}  \int \limits_{0}^{\epsilon}  dx \int \limits_{0}^{1}
dy \int
\limits_{0}^{z_c} \frac{x dz}{[(x^2 \tilde{p}_y^2 + \lambda^2 c^4) - (2x^2 \tilde{p}_y^2 + \tilde{k}^2(1-x))z]^2 } \\
&=&   -\frac{ \pi^2 }{ic^3}  \int \limits_{0}^{\epsilon} x dx \int \limits_{0}^{1}
dy  \frac{1}{[2x^2 \tilde{p}_y^2 + \tilde{k}^2(1-x)][-(x^2 \tilde{p}_y^2 + \lambda^2 c^4) + (2x^2 \tilde{p}_y^2 + \tilde{k}^2(1-x))z]} \Bigl|_{z=0}^{z=z_c} \\
&=&   -\frac{ \pi^2 }{ic^3}  \int \limits_{0}^{\epsilon} x dx \int \limits_{0}^{1}
dy \frac{1}{(2x^2 \tilde{p}_y^2 + \tilde{k}^2(1-x))} \Bigl(  \frac{1}{-(x^2 \tilde{p}_y^2 + \lambda^2 c^4) + (2x^2 \tilde{p}_y^2 + \tilde{k}^2(1-x))z_c} \\
&+&  \frac{1}{x^2 \tilde{p}_y^2 + \lambda^2 c^4} \Bigr)  \\
&\approx&  - \frac{ \pi^2 }{ic^3}  \int \limits_{0}^{\epsilon} dx \int \limits_{0}^{1}
dy \frac{x}{\tilde{k}^2(1-x)(x^2 \tilde{p}_y^2 + \lambda^2 c^4)}   \approx   -\frac{ \pi^2 }{ic^3\tilde{k}^2}  \int \limits_{0}^{\epsilon} dx \int \limits_{0}^{1}
dy    \frac{x(1+x)}{x^2 \tilde{p}_y^2 + \lambda^2 c^4}   \\
&\approx&   -\frac{ \pi^2 }{ic^3\tilde{k}^2}  \int \limits_{0}^{\epsilon} dx \int \limits_{0}^{1}
dy  \left(  \frac{x}{x^2 \tilde{p}_y^2 + \lambda^2 c^4} + \frac{1}{\tilde{p}_y^2}  \right)
\end{eqnarray*}

\noindent The last term in parentheses can be neglected when integrated on $x$. Using integral (\ref{eq:abzacz}) with $a=\lambda^2 c^4$, $b=1$, $c=\tilde{p}_y^2$ we  obtain

\begin{eqnarray*}
 L_{Ia}
&\approx&   -\frac{ \pi^2 }{ic^3\tilde{k}^2}   \int \limits_{0}^{1}
\frac{dy }{2\tilde{p}_y^2}  \ln(x^2 \tilde{p}_y^2 + \lambda^2 c^4)  \Bigl|_{x=0}^{x=\epsilon}    \\
&=&   -\frac{ \pi^2 }{ic^3\tilde{k}^2}   \int \limits_{0}^{1}
\frac{dy }{2\tilde{p}_y^2} \ln \left( \frac{\epsilon^2 \tilde{p}_y^2}{ \lambda^2 c^4} \right) \\
\end{eqnarray*}

\noindent In the second part $L_{Ib}$ we neglect the $\lambda$-term

\begin{eqnarray*}
 L_{Ib}
&\approx&   \frac{ \pi^2 }{ic^3}  \int \limits_{0}^{\epsilon}  dx \int \limits_{0}^{1}
dy \int
\limits_{z_c}^{1} \frac{x(1-z)^2 dz}{[(1-z)^2 (x^2 \tilde{p}_y^2 + \tilde{k}^2(1-x))  -\tilde{k}^2 (1-z)(1-x)]^2 } \\
&\approx&   \frac{ \pi^2 }{ic^3}  \int \limits_{0}^{\epsilon}  dx \int \limits_{0}^{1}
dy \int
\limits_{z_c}^{1} \frac{x dz}{[-x^2 \tilde{p}_y^2  + (x^2 \tilde{p}_y^2 +\tilde{k}^2(1-x))z  ]^2 } \\
&=&   -\frac{ \pi^2 }{ic^3}  \int \limits_{0}^{\epsilon} x dx \int \limits_{0}^{1}
dy   \frac{1}{[x^2 \tilde{p}_y^2 +\tilde{k}^2(1-x)][-x^2 \tilde{p}_y^2  + (x^2 p_y^2 +\tilde{k}^2(1-x))z  ]}  \Bigl|_{z =z_c}^{z=1}  \\
&=&   -\frac{ \pi^2 }{ic^3}  \int \limits_{0}^{\epsilon} dx \int \limits_{0}^{1}
dy \frac{x}{x^2 \tilde{p}_y^2 +\tilde{k}^2(1-x)} \Bigl( \frac{1}{\tilde{k}^2(1-x)  } - \frac{1}{-x^2 \tilde{p}_y^2  + (x^2 \tilde{p}_y^2 +\tilde{k}^2(1-x))z_c  } \Bigr)\\
&\approx&   -\frac{ \pi^2 }{ic^3 \tilde{k}^2}  \int \limits_{0}^{\epsilon} dx \int \limits_{0}^{1}
dy x \Bigl( \frac{1}{\tilde{k}^2  } - \frac{1}{ \tilde{k}^2z_c  } \Bigr)\\
&\approx&   0
\end{eqnarray*}

\noindent Collecting all non-vanishing contributions we obtain

\begin{eqnarray}
L &=& L_{Ia} +L_{II} + L_{III} \nonumber \\
&\approx&   -\frac{ \pi^2 }{ic^3\tilde{k}^2}   \int \limits_{0}^{1}
dy \frac{1}{2\tilde{p}_y^2} \ln \left( \frac{\epsilon^2 \tilde{p}_y^2}{ \lambda^2 c^4} \right) + \frac{ \pi^2 \ln(\delta \epsilon) }{ic^3 \tilde{k}^2}  \int \limits_{0}^{1}  \frac{dy}{\tilde{p}_y^2 } - \frac{ \pi^2  }{ic^3 2\tilde{k}^2}   \int \limits_{0}^{1}
\frac{dy}{\tilde{p}_y^2} \ln \left( \frac{\tilde{k}^4 \delta^2 }{\tilde{p}_y^2 \lambda^2 c^4}\right) \nonumber  \\
&=&  -\frac{ \pi^2 }{ic^3\tilde{k}^2} \int \limits_{0}^{1}
dy \frac{1}{2\tilde{p}_y^2}   \ln \left( \frac{\epsilon^2 \tilde{p}_y^2}{ \lambda^2 c^4 \delta^2 \epsilon^2} \cdot \frac{\tilde{k}^4 \delta^2 }{\tilde{p}_y^2 \lambda^2 c^4}\right)  \nonumber    \\
&=&  -\frac{ \pi^2 }{ic^3\tilde{k}^2} \ln \left( \frac{\tilde{k}^2 }{ \lambda^2 c^4}\right) \int \limits_{0}^{1}
dy \frac{1}{\tilde{p}_y^2} \label{eq:int-py2}
\end{eqnarray}

\noindent This is equation (A20) in \cite{Redhead}.

\section{Coulomb scattering in 2nd order}
\label{sc:integral-comm}

Here we will calculate the 3D integral

\begin{eqnarray*}
 D(\mathbf{q},\mathbf{q'})
&=&  \int
\frac{d\mathbf{s}}{[(\mathbf{q-s})^2 + \lambda^2c^2][s^2 - q^2 + i \mu][(\mathbf{s-q'})^2 + \lambda^2c^2]}
\end{eqnarray*}

\noindent in formula (\ref{eq:commu}) for the 4th order commutator term in the electron-proton interaction. We are interested in leading terms surviving in the limits $\lambda \to 0$, $\mu \to 0$. The calculation method was adopted from \S 121 in \cite{BLP}.\footnote{see also \cite{Kacser}}

First we use (\ref{eq:abc2}) and the elastic scattering condition $(q')^2 = q^2$ to write

\begin{eqnarray*}
&\ & D(\mathbf{q},\mathbf{q'}) \\
&=& 2 \int \limits_0^1 dx \int \limits_0^{1-x}dy \int
\frac{d\mathbf{s}}{[((\mathbf{q-s})^2 + \lambda^2c^2)x + ((\mathbf{s-q'})^2 + \lambda^2c^2) y + (s^2 - q^2 - i \mu)(1-x-y)]^3}\\
&=&  2 \int \limits_0^1 dx \int \limits_0^{1-x}dy \int
\frac{d\mathbf{s}}{[s^2 -2(\mathbf{qs})x -2(\mathbf{q's})y   + \lambda^2c^2 (x + y)  + q^2(2x + 2y -1)   - i \mu]^3}
\end{eqnarray*}

\noindent Next we shift the integration variable $\mathbf{s} \to \mathbf{h} \equiv \mathbf{s} - x \mathbf{q}' - y \mathbf{q}$ and take into account that $2 (\mathbf{qq'}) = 2q^2 - k^2$, where the vector of transferred momentum is defined as $\mathbf{k} = \mathbf{q'} - \mathbf{q}$

\begin{eqnarray*}
&\ & D(\mathbf{q},\mathbf{q'}) \\
&=&  2 \int \limits_0^1 dx \int \limits_0^{1-x}dy \int
\frac{d\mathbf{h}}{[h^2 + q^2(-x^2 -y^2 +2x + 2y -1) -2(\mathbf{qq'})xy   + \lambda^2c^2 (x + y)     - i \mu]^3}\\
&=&  2 \int \limits_0^1 dx \int \limits_0^{1-x}dy \int
\frac{d\mathbf{h}}{[h^2 - q^2(x+y -1)^2 + k^2 xy   + \lambda^2c^2 (x + y)     - i \mu]^3} \\
&=& \frac{i \pi^2}{2} \int \limits_0^1 dx \int \limits_0^{1-x}dy
\frac{1}{[q^2(x+y -1)^2 - k^2 xy   - \lambda^2c^2 (x + y)     - i \mu]^{3/2}}
\end{eqnarray*}

\noindent Change integration variables $\xi = x+y$, $\zeta = x-y$

\begin{eqnarray*}
 D(\mathbf{q},\mathbf{q'})
&=& \frac{i \pi^2}{2} \int \limits_0^1 d \xi \int \limits_0^{\xi}d\zeta
\frac{1}{(q^2(\xi -1)^2 - k^2 \xi^2/4 + k^2 \zeta^2/4   - \lambda^2c^2 \xi     - i \mu)^{3/2}} \\
&=& \frac{i \pi^2}{2} \int \limits_0^1
\frac{\xi d \xi}{(q^2(\xi -1)^2 - k^2 \xi^2/4   - \lambda^2c^2 \xi^2  -i \mu)\sqrt{q^2(\xi -1)^2    - \lambda^2c^2 \xi     -i \mu}}
\end{eqnarray*}

\noindent Next we introduce parameter $\delta$, such that $1 \gg \delta \gg \lambda^2 c^2 /q^2$, and split the integration range into two parts

\begin{eqnarray*}
 D(\mathbf{q},\mathbf{q'}) &=& D_1(\mathbf{q},\mathbf{q'}) + D_2(\mathbf{q},\mathbf{q'}) \\
D_1(\mathbf{q},\mathbf{q'}) &=& \frac{i \pi^2}{2} \int \limits_0^{1-\delta} \ldots d \xi \\
D_2(\mathbf{q},\mathbf{q'}) &=& \frac{i \pi^2}{2} \int \limits_{1-\delta}^1 \ldots d \xi
\end{eqnarray*}

\noindent In the first integral we ignore the $\lambda$-term

\begin{eqnarray*}
&\ & D_1(\mathbf{q},\mathbf{q'}) \\
 &\approx& \frac{i \pi^2}{2q^3} \int \limits_0^{1-\delta}
\frac{\xi d \xi}{[(\xi -1)^2 - k^2\xi^2/(4 q^2)     -i \mu](\xi -1)} \\
&=& \frac{i \pi^2}{2q^3} \cdot \frac{2q^2}{k^2}
\ln \left( \frac{-k^2 \xi ^2 /(4q^2) + \xi^2 - 2 \xi + 1}{(1- \xi)^2} \right)\Bigl|_0^{1-\delta} \approx \frac{i \pi^2}{qk^2}\ln \left(\frac{-k^2}{4q^2 \delta^2} \right)
\end{eqnarray*}

\noindent In the second integral we change the integration variable $y = x-1$

\begin{eqnarray*}
&\ & D_2(\mathbf{q},\mathbf{q'}) \\
&\approx& \frac{i \pi^2}{2} \int \limits_{-\delta}^0
\frac{dy (y+1)}{(q^2y^2 - k^2/4) \sqrt{q^2y^2 - \lambda^2 c^2}} \approx \frac{2i \pi^2}{k^2} \int \limits_0^{\delta} \frac{dy }{ \sqrt{q^2y^2 - \lambda^2 c^2}} \\
&=& \frac{2i \pi^2}{qk^2} \ln(q\sqrt{q^2y^2 - \lambda^2 c^2} +q^2y) \Bigl|_0^{\delta} = \frac{2i \pi^2}{qk^2} \ln \left(\frac{q\sqrt{q^2\delta^2 - \lambda^2 c^2} +q^2 \delta}{i q \lambda c} \right)\\
&\approx & \frac{i \pi^2}{qk^2}\ln \left(\frac{-4q^2 \delta^2}{\lambda^2 c^2} \right)
\end{eqnarray*}

\noindent Putting both parts of the integral together we finally obtain

\begin{eqnarray}
 D(\mathbf{q},\mathbf{q'}) &\approx& \frac{i \pi^2}{qk^2}\ln \left(\frac{-k^2}{4q^2 \delta^2} \cdot \frac{-4q^2 \delta^2}{\lambda^2 c^2} \right)= \frac{i \pi^2}{qk^2}\ln \left(\frac{k^2}{\lambda^2 c^2} \right) \label{eq:D1D2}
\end{eqnarray}

\noindent which is equation (121.16) in \cite{BLP}.

\chapter{Relativistic invariance of RQD}

\section{Relativistic invariance of simple QFT} \label{ss:relat-invar-simple}

Here we would like to verify that interacting theory presented in subsection \ref{ss:weinberg} is, indeed, relativistically invariant \cite{book, Weinberg_lecture}. In other words, we are going to prove the
validity of Poincar\'e commutators (\ref{eq:8.17}) -
(\ref{eq:8.21}) for the interacting energy and boost operators

\begin{eqnarray}
V &=&  \int  d\mathbf{x} V(\mathbf{x}, 0) \nonumber \\
 \mathbf{Z}
&=& \frac{1}{c^2}\int d\mathbf{x} \mathbf{x} V(\mathbf{x}, 0)
 \label{eq:11.8ax}
\end{eqnarray}

\noindent in (\ref{eq:10.7}) - (\ref{eq:10.8}).

Equation (\ref{eq:8.17}) follows directly from the property (\ref{eq:as-a-scalar}) in the case of space translations and rotations. The potential boost
$\mathbf{Z}$ in (\ref{eq:11.8ax}) is a 3-vector by
construction, so equation (\ref{eq:8.19}) is valid as well. Let us now
prove the commutator (\ref{eq:8.18})

\begin{eqnarray*}
[P_{0i}, Z_j] &=&   \frac{i \hbar}{c^2} V \delta_{ij}
\end{eqnarray*}

\noindent  Consider the case $i = j = z$. Then, using equation (\ref{eq:as-a-scalar}) with $\Lambda = 1$, we obtain

\begin{eqnarray}
[ P_{0z}, Z_z]
&=&  -\frac{i \hbar}{c^2} \lim _{a \to 0} \frac{d}{d a} \int
d\mathbf{x}e^{\frac{i}{\hbar} P_{0z} a} zV( \mathbf{x},0)
e^{-\frac{i}{\hbar} P_{0z} a} \nonumber\\
&=&  -\frac{i \hbar}{c^2} \lim _{a \to 0} \frac{d}{d a} \int
d\mathbf{x}z V( x , y, z+a,0)\nonumber\\
&=&  -\frac{i \hbar}{c^2} \lim _{a \to 0} \frac{d}{d a} \int
d\mathbf{x}(z-a) V( x, y, z,0) \nonumber \\
&=&   \frac{i \hbar}{c^2}\int d\mathbf{x} V( x, y, z,0) =  \frac{i \hbar}{c^2} V \label{eq:10.9x}
\end{eqnarray}

\noindent which is exactly equation (\ref{eq:8.18}).

The proof of equation (\ref{eq:8.21}) is more challenging. Let us
consider the case $i = z$ and attempt to prove\footnote{In this calculation it is convenient to write
condition (\ref{eq:8.21}) in a $t$-dependent
form, i.e., multiply this equation by
$\exp(\frac{i}{\hbar}H_0t)$ from the left and
$\exp(-\frac{i}{\hbar}H_0t)$ from the right, as in (\ref{eq:8.67}).
At the end of calculations we will set $t= 0$. \label{page:time-depn} }

\begin{eqnarray}
[K_{0z}, V(t)]  + [Z_z(t), H_0] -
[V(t), Z_z(t) ] = 0 \label{eq:11.10x}
\end{eqnarray}

\noindent  For  the first term
on the left
hand side we use  (\ref{eq:as-a-scalar}) and

\begin{eqnarray*}
\lim_{\theta \to 0}  \frac{d}{d \theta} V(\Lambda \tilde{x}) &=&
\lim_{\theta \to 0}  \frac{d}{d \theta} V( x, y, z\cosh \theta - ct
\sinh \theta, t \cosh \theta - \frac{z}{c} \sinh \theta)\nonumber
 \\
&=& \lim_{\theta \to 0} \frac{\partial V} {\partial z} (z\sinh
\theta - ct \cosh \theta) + \frac{\partial V} {\partial t} (t \sinh
\theta - \frac{z}{c} \cosh \theta)\nonumber
 \\
&=& -ct \frac{\partial V} {\partial z} - \frac{z}{c} \frac{\partial
V} {\partial t}
\label{eq:11.12x}
\end{eqnarray*}

\noindent where $\Lambda$ is the boost matrix (\ref{eq:boost-matrix-z}). Then

\begin{eqnarray}
 [K_{0z}, V(t)]
&=&   -\frac{i \hbar}{c} \lim_{\theta \to 0} \frac{d}{d
\theta } e^{\frac{ic}{\hbar}K_{0z} \theta}
 \int d\mathbf{x} V(\tilde{x})  e^{-\frac{ic}{\hbar}K_{0z} \theta} \nonumber\\
&=&  -\frac{i \hbar}{c} \lim_{\theta \to 0} \frac{d}{d
\theta } \int d\mathbf{x}  V(\Lambda \tilde{x})  \nonumber \\
&=& - \frac{i \hbar}{c}
 \int d\mathbf{x} \left(
-ct \frac{\partial V( \mathbf{x}, t)} {\partial z} - \frac{z}{c} \frac{\partial
V( \mathbf{x}, t)} {\partial t}\right)
\label{eq:11.17x}
\end{eqnarray}

 \noindent For the second term we obtain

\begin{eqnarray}
[Z_z(t), H_0] &=& -i \hbar \frac{\partial}{\partial t}  Z_z(  t) = -\frac{i
\hbar}{c} \frac{\partial}{\partial t} \int d\mathbf{x} z
V( \mathbf{x}, t)
\label{eq:11.19x}
\end{eqnarray}

\noindent  The last term in (\ref{eq:11.10x}) vanishes due to (\ref{eq:10.6}).
Now we can set $t=0$ and see that  (\ref{eq:11.17x}) and
(\ref{eq:11.19x}) cancel each other, which  proves (\ref{eq:11.10x}).

 Derivation of the last remaining nontrivial commutation relation

\begin{eqnarray*}
[K_{0i}, Z_j] + [Z_i, K_{0j}] + [Z_i, Z_j]= 0
\end{eqnarray*}

\noindent is left as an exercise for the reader.

\section{Relativistic invariance of QED} \label{ss:relat-invar}

In this Appendix we are going to prove the relativistic invariance
of the field-theoretical formulation of QED presented in subsection
\ref{ss:interaction-qed}. In other words, we are going to prove the
validity of Poincar\'e commutators (\ref{eq:8.17}) -
(\ref{eq:8.21}).\footnote{We write
conditions (\ref{eq:8.17}) - (\ref{eq:8.21}) in a $t$-dependent
form. See footnote on page \pageref{page:time-depn}.} The proof presented
here is taken from Weinberg's works \cite{book, Weinberg_lecture}
and, especially, Appendix B in \cite{ Weinberg_65}.

The interaction operator $V(t)$ in (\ref{eq:11.5}) clearly commutes
with operators of the total momentum and total angular momentum,  so
equation (\ref{eq:8.17}) is easily verified. The potential boost
$\mathbf{Z}$ in (\ref{eq:11.8}) is a 3-vector by
construction, so equation (\ref{eq:8.19}) is valid as well. Let us now
prove the commutator (\ref{eq:8.18})

\begin{eqnarray*}
[P_{0i}, Z_j(t)] &=&   \frac{i \hbar}{c^2} V(t) \delta_{ij}
\end{eqnarray*}

\noindent  Consider the case $i = j = x$ and
denote

\begin{eqnarray*}
V(\mathbf{x}, t) &\equiv&  \frac{\hbar}{\sqrt{c}} \mathbf{j}(
\mathbf{x},t) \mathbf{ A}( \mathbf{x},t) + \frac{1}{2c^2}\int
d\mathbf{y} j_0( \mathbf{x},t) \mathcal{G}(\mathbf{x} - \mathbf{y})
          j_0( \mathbf{y},t) \\
     \mathcal{G}(\mathbf{x}) &\equiv& \frac{1}{4 \pi |\mathbf{x}|}
\end{eqnarray*}

\noindent so that

\begin{eqnarray}
V(t) &=&  \int  d\mathbf{x} V(\mathbf{x}, t) \nonumber \\
 \mathbf{Z}(t)
&=& \frac{1}{c^2}\int d\mathbf{x} \mathbf{x} V(\mathbf{x}, t)
 + \frac{\hbar}{c^{5/2}} \int d\mathbf{x} j_0( \mathbf{x},t)
\mathbf{C} ( \mathbf{x},t) \label{eq:11.8a}
\end{eqnarray}

\noindent where

\begin{eqnarray}
\mathbf{C}(\tilde{x}) \equiv \frac{i \hbar^2 \sqrt{c} }{\sqrt{2(2\pi \hbar)^{3}}} \int
\frac{d\mathbf{p}}{p^{3/2}} \sum_{ \tau}
 \left( e^{ -\frac{i}{\hbar}\tilde{p} \cdot \tilde{x}} \mathbf{e}(\mathbf{p}, \tau)c _{\mathbf{p},\tau} -
e^{ \frac{i}{\hbar} \tilde{p} \cdot \tilde{x}} \mathbf{e}^* (\mathbf{p}, \tau) c^{\dag}
_{\mathbf{p},\tau} \right) \label{eq:11.9}
\end{eqnarray}

\noindent  Then, using equations (\ref{eq:cur-trans}) and
(\ref{eq:9.38}) - (\ref{eq:9.39}) we obtain

\begin{eqnarray}
&\mbox{ }& [ P_{0x}, Z_x(t)] \nonumber \\
&=&  -i \hbar \lim _{a \to
0} \frac{d}{d a}e^{\frac{i}{\hbar} P_{0x} a} Z_x(t)
e^{-\frac{i}{\hbar} P_{0x} a} \nonumber \\
&=&  -\frac{i \hbar}{c^2} \lim _{a \to 0} \frac{d}{d a} \int
d\mathbf{x}e^{\frac{i}{\hbar} P_{0x} a}  \left(xV( \mathbf{x},t) +
\frac{\hbar}{\sqrt{c}}j_0( \mathbf{x},t) C_x ( \mathbf{x},t)\right)
e^{-\frac{i}{\hbar} P_{0x} a} \nonumber\\
&=&  -\frac{i \hbar}{c^2} \lim _{a \to 0} \frac{d}{d a} \int
d\mathbf{x}\left(x V( x +a, y, z,t) + \frac{\hbar}{\sqrt{c}}j_0(
x+a, y, z,t) C_x ( x+a, y, z,t)\right)\nonumber\\
&=&  -\frac{i \hbar}{c^2} \lim _{a \to 0} \frac{d}{d a} \int
d\mathbf{x}\left((x-a) V( x, y, z,t)+ \frac{\hbar}{\sqrt{c}} j_0(
x, y, z,t) C_x ( x, y, z,t)\right) \nonumber \\
&=&   \frac{i \hbar}{c^2}\int d\mathbf{x} V( x, y, z,t) =  \frac{i \hbar}{c^2} V(t) \label{eq:10.9}
\end{eqnarray}

\noindent which is exactly equation (\ref{eq:8.18}).

The proof of equation (\ref{eq:8.21}) is more challenging. Let us
consider the case $i = z$ and attempt to prove

\begin{eqnarray}
[K_{0z}, V_1(t)] + [K_{0z}, V_2(t)] - i \hbar \frac{d}{dt} Z_z(t) -
[V(t), Z_z(t) ] = 0 \label{eq:11.10}
\end{eqnarray}

\noindent where we took into account that $[Z_z(t), H_0] = - i \hbar
\frac{d}{dt} Z_z(t)$. We will calculate all four terms on the left
hand side of (\ref{eq:11.10}) separately. Consider the first term
and use equations (\ref{eq:11.2}), (\ref{eq:10.35a}), (\ref{eq:11.14a}), (\ref{eq:A.73})

\begin{eqnarray}
 &\mbox{ }& [K_{0z}, V_1(t)] \nonumber\\
&=& -\frac{i \hbar}{c} \lim_{\theta \to 0} \frac{d}{d \theta }
 e^{\frac{ic}{\hbar}K_{0z} \theta}
V_1(t) e^{-\frac{ic}{\hbar}K_{0z} \theta} \nonumber\\
&=&   -\frac{i \hbar^2}{c^{3/2}} \lim_{\theta \to 0} \frac{d}{d
\theta } e^{\frac{ic}{\hbar}K_{0z} \theta}
 \int d\mathbf{x} \tilde{j}(\tilde{x}) \cdot
\tilde{A}(\tilde{x}) e^{-\frac{ic}{\hbar}K_{0z} \theta} \nonumber\\
&=&  -\frac{i \hbar^2}{c^{3/2}} \lim_{\theta \to 0} \frac{d}{d
\theta } \int d\mathbf{x} (\Lambda^{-1} \tilde{j}(\Lambda \tilde{x})    \cdot
\Lambda^{-1} \tilde{A}(\Lambda \tilde{x}) + \Lambda^{-1} \tilde{j}(\Lambda \tilde{x}) \cdot  \Omega
(\tilde{x}, \Lambda))\nonumber
\\
&=& - \frac{i \hbar^2}{c^{3/2}} \lim_{\theta \to 0} \frac{d}{d
\theta } \int d\mathbf{x}( \tilde{j}(\Lambda \tilde{x}) \cdot  \tilde{A}(\Lambda \tilde{x}) +
\Lambda^{-1} \tilde{j}(\Lambda \tilde{x}) \cdot  \Omega (\tilde{x}, \Lambda)) \nonumber\\
&=&  -\frac{i \hbar^2}{c^{3/2}} \lim_{\theta \to 0}
 \int d\mathbf{x}\Bigl( \frac{d}{d
\theta } \tilde{j}(\Lambda \tilde{x}) \cdot  \tilde{A}(\tilde{x}) + \tilde{j}(\tilde{x}) \cdot \frac{d}{d \theta }
\tilde{A}(\Lambda \tilde{x}) +
\left(\frac{d}{d \theta} \Lambda^{-1}\right) \tilde{j}(\tilde{x}) \cdot \Omega(\tilde{x},1) \nonumber\\
&\ & +\frac{d}{d \theta } \tilde{j}(\Lambda \tilde{x}) \cdot \Omega (\tilde{x}, 1) + \tilde{j}(\tilde{x})
\cdot \frac{d}{d \theta } \Omega (\tilde{x}, \Lambda)\Bigr) \nonumber \\
&=&  -\frac{i \hbar^2}{c^{3/2}} \lim_{\theta \to 0}
 \int d\mathbf{x}\Bigl( \frac{d}{d
\theta } \tilde{j}(\Lambda \tilde{x}) \cdot  \tilde{A}(\tilde{x}) + \tilde{j}(\tilde{x}) \cdot \frac{d}{d \theta }
\tilde{A}(\Lambda \tilde{x}) +
  \tilde{j}(\tilde{x}) \cdot \frac{d}{d \theta } \Omega (\tilde{x}, \Lambda)\Bigr)
\label{eq:11.11}
\end{eqnarray}

\noindent where $\Omega (\tilde{x}, \Lambda)$ is given by equation
(\ref{eq:10.35}) and $\Lambda$ is matrix (\ref{eq:boost-matrix-z}).
Next we  use the following results

\begin{eqnarray}
\lim_{\theta \to 0}  \frac{d}{d \theta} \tilde{j}(\Lambda \tilde{x}) &=&
\lim_{\theta \to 0}  \frac{d}{d \theta} \tilde{j}( x, y, z\cosh \theta - ct
\sinh \theta, t \cosh \theta - \frac{z}{c} \sinh \theta)\nonumber
 \\
&=& \lim_{\theta \to 0} \frac{\partial \tilde{j}} {\partial z} (z\sinh
\theta - ct \cosh \theta) + \frac{\partial \tilde{j}} {\partial t} (t \sinh
\theta - \frac{z}{c} \cosh \theta)\nonumber
 \\
&=& -ct \frac{\partial \tilde{j}} {\partial z} - \frac{z}{c} \frac{\partial
\tilde{j}} {\partial t}
\label{eq:11.12}\\
\lim_{\theta \to 0}  \frac{d}{d \theta} \tilde{A}(\Lambda \tilde{x}) &=& - ct
\frac{\partial \tilde{A}} {\partial z} - \frac{z}{c} \frac{\partial \tilde{A}}
{\partial t} \label{eq:11.13}
\end{eqnarray}

\noindent Calculation of the $d \Omega/d \theta$ term is more
involved

 \begin{eqnarray} &\mbox{ }& \lim_{\theta \to 0 } \frac{d}{d
\theta }  \Omega_{\mu} ( \tilde{x}, \Lambda)
\nonumber \\
&= & -\frac{\hbar \sqrt{c}}{(2\pi \hbar)^{3/2}}  \lim_{\theta \to 0}
\sum_{ \nu=0}^3 \frac{d}{d \theta }
  \int \frac{d
 \mathbf{p}}{\sqrt{2p}}  \sum_{
\tau= -1}^1
\frac{ (\Lambda^{-1} p)^{\mu}}{|\Lambda^{-1} \mathbf{p}|} \times \nonumber\\
&\mbox{ }& \sum_{ \nu \rho=0}^3 \Lambda_{0 \nu} ^{-1}
\left(e^{-\frac{i}{\hbar} \Lambda^{-1}\tilde{p} \cdot  \tilde{x}}
 e_{\nu }( \mathbf{p}, \tau)  c _{\mathbf{p},\tau} +
e^{\frac{i}{\hbar} \Lambda^{-1} \tilde{p} \cdot  \tilde{x}}  e^*_{\nu }(\mathbf{p},
\tau)
  c^{\dag} _{\mathbf{p},\tau} \right)
\label{eq:11.15}
\end{eqnarray}

\noindent  The only quantities dependent on $\theta$ are $\Lambda$-matrices.
Therefore, taking the derivative on the right hand side of equation
(\ref{eq:11.15}) we will obtain four terms, those containing
$\frac{d}{d \theta } \Lambda^{-1}_{\nu \mu} $, $\frac{d}{d \theta }
\Lambda_{0 \rho}^{-1} $, $\frac{d}{d \theta } |\Lambda^{-1}
\mathbf{p}|^{-1} $ and $\frac{d}{d \theta } \exp(\pm i \Lambda^{-1}
\tilde{p} \cdot  \tilde{x}) $. After taking the derivative we must set $\theta \to
0$. It follows from equation (\ref{eq:11.14}) that the only non-zero term
is that containing

\begin{eqnarray*}
\lim_{\theta \to 0 } \frac{d}{d \theta } \Lambda_{0 \rho}^{-1} &=&
\lim_{\theta \to 0 } \frac{d}{d \theta } (\cosh \theta, 0, 0,
\sinh \theta) \\
 &=& \lim_{\theta \to 0 } (-\sinh \theta, 0, 0,  \cosh \theta) \\
 &=&  (0, 0, 0,  1)
\end{eqnarray*}

\noindent Thus

\begin{eqnarray}
&\mbox{ }& \lim_{\theta \to 0 } \frac{d}{d \theta } \Omega_{\mu} (\tilde{x},
\Lambda) \nonumber \\
 &=&
 - \frac{\hbar \sqrt{c}}{(2\pi \hbar)^{3/2}}  \int \frac{d
 \mathbf{p}  p^{\mu}}{\sqrt{2} p^{3/2}}  \sum_{
\tau= -1}^1
 \left(e^{-\frac{i}{\hbar} \tilde{p} \cdot  \tilde{x}}
  e_{z } ( \mathbf{p}, \tau)   c_{\mathbf{p},\tau} +
e^{\frac{i}{\hbar}  \tilde{p} \cdot  \tilde{x}}
 e^*_{z }(\mathbf{p}, \tau)  c^{\dag} _{\mathbf{p},\tau}\right)\nonumber  \\
& = & - \frac{i \hbar^2 \sqrt{c}}{\sqrt{2 (2\pi \hbar)^{3}}}
\partial_{\mu}  \int \frac{d
 \mathbf{p}}{p^{3/2}}  \sum_{
\tau = -1}^1
 \left(e^{-\frac{i}{\hbar} \tilde{p} \cdot  \tilde{x}}
  e_{ z  }( \mathbf{p}, \tau)   c_{\mathbf{p},\tau} -
e^{\frac{i}{\hbar}  \tilde{p} \cdot  \tilde{x}}  e^*_{z }(\mathbf{p}, \tau) c^{\dag}
_{\mathbf{p},\tau} \right) \nonumber \\
& = & -\partial_{\mu} C_z (\tilde{x}) \label{eq:K.15c}
\end{eqnarray}

\noindent where $\partial_{\mu} \equiv
(-\frac{1}{c}\frac{\partial}{\partial t}, \frac{\partial}{\partial
x}, \frac{\partial}{\partial y}, \frac{\partial}{\partial z})$.
 So, using (\ref{eq:K.15c}) and the continuity equation
(\ref{eq:11.3}), we obtain that the last term on the right hand side
of equation (\ref{eq:11.11}) is\footnote{
 due to the property (\ref{eq:10.2}) all functions $f$ and $g$ of quantum
fields vanish at infinity, therefore we can take integrals by parts
($\xi \equiv (t, x, y, z)$)

\begin{eqnarray*}
&\mbox{ }& \int \limits_{- \infty}^{\infty} dx \left(\frac{d}{dx}
f(\xi)\right) g(\xi) = \int \limits_{- \infty}^{\infty} dx
\frac{d}{dx} (f(\xi) g(\xi)) -
\int \limits_{- \infty}^{\infty} dx  f(\xi) \frac{d}{dx} g(\xi)\\
&=& f(x = \infty)g(x = \infty) - f(x = -\infty)g(x = - \infty) -
\int \limits_{- \infty}^{\infty} dx  f(\xi) \frac{d}{dx} g(\xi) \\
&=&  - \int \limits_{- \infty}^{\infty} dx  f(\xi) \frac{d}{dx}
g(\xi)
\end{eqnarray*}}

\begin{eqnarray}
&\ & -\frac{i \hbar^2}{c^{3/2}} \lim_{\theta \to 0}
 \int d\mathbf{x}  \tilde{j}( \tilde{x}) \cdot \frac{d}{d
\theta }\Omega (\tilde{x}, \Lambda) \nonumber\\
&=& \frac{i \hbar^2}{c^{3/2}} \sum_{\mu \nu} \int d\mathbf{x}
{j}_{\mu} (\mathbf{x}, t) g_{\mu \nu}
\partial_{\nu} C_z (\mathbf{x}, t) \nonumber\\
 &=& \frac{i \hbar^2}{c^{5/2}}\int d\mathbf{x}  j_{0}(\mathbf{x}, t)
\frac{\partial C_z (\mathbf{x}, t)}{\partial t}  +\frac{i
\hbar^2}{c^{3/2}}\int d\mathbf{x}  \mathbf{j}(\mathbf{x}, t)
\frac{\partial C_z (\mathbf{x}, t)}{\partial \mathbf{x}} \nonumber\\
 &=& \frac{i \hbar^2}{c^{5/2}}\int d\mathbf{x}  j_{0}(\mathbf{x}, t)
\frac{\partial C_z (\mathbf{x}, t)}{\partial t}  - \frac{i
\hbar^2}{c^{3/2}} \int d\mathbf{x} \frac{\partial \mathbf{j}(\mathbf{x}, t)}{\partial
\mathbf{x}}
 C_z (\mathbf{x}, t) \nonumber\\
 &=& \frac{i \hbar^2}{c^{5/2}}\int d\mathbf{x}  j_{0}(\mathbf{x}, t)
\frac{\partial C_z (\mathbf{x}, t)}{\partial t}  + \frac{i
\hbar^2}{c^{5/2}} \int d\mathbf{x} \frac{\partial j_0(\mathbf{x}, t)}{\partial t}
 C_z (\mathbf{x}, t) \nonumber\\
&=& \frac{i \hbar^2}{c^{5/2}} \frac{\partial}{\partial t} \int
d\mathbf{x} j_{0}(\mathbf{x}, t)
 C_z (\mathbf{x}, t)
\label{eq:11.16}
\end{eqnarray}

\noindent Substituting results (\ref{eq:11.12}), (\ref{eq:11.13}),
(\ref{eq:K.15c}), and (\ref{eq:11.16}) in equation (\ref{eq:11.11}) and
setting $t=0$ we obtain

\begin{eqnarray}
 &\mbox{ }& [K_{0z}, V_1(t)] \nonumber\\
&=& - \frac{i \hbar^2}{c^{3/2}}
 \int d\mathbf{x} \left(
- \frac{z}{c} \frac{\partial \tilde{j}} {\partial t} \cdot  \tilde{A}(\tilde{x}) - \tilde{j}(\tilde{x})
\cdot
 \frac{z}{c} \frac{\partial \tilde{A}} {\partial t} - \frac{1}{c^2}
 \frac{\partial}{\partial t} (j_{0}(\tilde{x})   C_z (\tilde{x}))\right) \nonumber\\
&=& -\frac{i \hbar^2}{c^{5/2}} \frac{\partial } {\partial t}
 \int d\mathbf{x} \Bigl(
 -z (\tilde{j}(\tilde{x}) \cdot  \tilde{A}(\tilde{x}))  -
 j_{0}(\tilde{x})   C_z (\tilde{x}) \Bigr)
\label{eq:11.17}
\end{eqnarray}

\noindent For the second term on the left hand side of
(\ref{eq:11.10}) we use equation (\ref{eq:11.2})

\begin{eqnarray}
&\mbox{ }&  [K_{0z}, V_2(t)]\nonumber \\
&=& \frac{1}{2c^2}\int d\mathbf{x} d\mathbf{x}' [K_{0z},j_0(\tilde{x})]
\mathcal{G}(\mathbf{x} - \mathbf{x}')
          j_0(\tilde{x}')
+ \frac{1}{2c^2} \int d\mathbf{x} d\mathbf{x}' j_0(\tilde{x})
\mathcal{G}(\mathbf{x} - \mathbf{x}')
          [K_{0z},j_0(\tilde{x}')] \nonumber\\
&=& \frac{1}{c^2}\int d\mathbf{x} d\mathbf{x}' [K_{0z},j_0(\tilde{x})]
\mathcal{G}(\mathbf{x} - \mathbf{x}')
          j_0(\tilde{x}') \nonumber\\
&=&
 \frac{i \hbar}{c^2}\int d\mathbf{x} d\mathbf{x}' \left(\frac{z}{c^2} \frac{\partial
j_0(\tilde{x})}{\partial t}
 -\frac{1}{c} j_z(\tilde{x}) \right) \mathcal{G}(\mathbf{x} - \mathbf{x}')
          j_0(\tilde{x}') \nonumber\\
 &=&    \frac{i \hbar}{2c^4} \int d\mathbf{x} d\mathbf{x}'
\frac{\partial j_0(\tilde{x})}{ \partial t}  (z-z') \mathcal{G}(\mathbf{x} -
\mathbf{x}')
          j_0(\tilde{x}')
 +  \frac{i \hbar}{2c^4} \int d\mathbf{x} d\mathbf{x}'
\frac{\partial j_0(\tilde{x})}{ \partial t}  z \mathcal{G}(\mathbf{x} -
\mathbf{x}')
          j_0(\tilde{x}') \nonumber\\
      &\ & + \frac{i \hbar}{2c^4} \int d\mathbf{x} d\mathbf{x}'
\frac{\partial j_0(\tilde{x})}{ \partial t}  z' \mathcal{G}(\mathbf{x} -
\mathbf{x}')
           j_0(\tilde{x}') - \frac{ i \hbar}{c^3} \int d\mathbf{x} d\mathbf{x}'
j_z(\tilde{x}) \mathcal{G}(\mathbf{x} - \mathbf{x}')
          j_0(\tilde{x}')\nonumber\\
    &=&   - \frac{i \hbar}{2c^3} \int d\mathbf{x} d\mathbf{x}'
\frac{\partial \mathbf{j} (\tilde{x})}{ \partial \mathbf{x}}  (z-z')
\mathcal{G}(\mathbf{x} - \mathbf{x}')
          j_0(\tilde{x}')
 +  \frac{i \hbar}{2c^4} \int d\mathbf{x} d\mathbf{x}'
\frac{\partial j_0(\tilde{x})}{ \partial t}  z \mathcal{G}(\mathbf{x} -
\mathbf{x}')
          j_0(\tilde{x}') \nonumber\\
      &\ & + \frac{i \hbar}{2c^4} \int d\mathbf{x} d\mathbf{x}'
 j_0(\tilde{x}) z \mathcal{G}(\mathbf{x} - \mathbf{x}')
         \frac{\partial j_0(\tilde{x}')}{ \partial t}
- \frac{i \hbar}{c^3} \int d\mathbf{x} d\mathbf{x}' j_z(\tilde{x})
\mathcal{G}(\mathbf{x} - \mathbf{x}')
          j_0(\tilde{x}') \nonumber\\
    &=&    \frac{ i \hbar}{2c^3}  \int d\mathbf{x} d\mathbf{x}'
 \mathbf{j}(\tilde{x})  \frac{\partial ((z-z') \mathcal{G} (\mathbf{x-x'})
)}{\partial \mathbf{x}}
          j_0(\tilde{x}')
 + \frac{i \hbar}{2 c^4}\frac{\partial}{\partial t}  \int d\mathbf{x}
d\mathbf{x}'
 j_0(\tilde{x}) z \mathcal{G}(\mathbf{x} - \mathbf{x}')
          j_0(\tilde{x}') \nonumber\\
&\ & -\frac{ i \hbar}{c^3}\int d\mathbf{x} d\mathbf{x}' j_z(\tilde{x})
\mathcal{G}(\mathbf{x} - \mathbf{x}')
          j_0(\tilde{x}')
\label{eq:11.18}
\end{eqnarray}

 \noindent  Using expression (\ref{eq:11.8a}) for
$\mathbf{Z}(t)$ we obtain for the third term on the left hand side
of equation (\ref{eq:11.10})

\begin{eqnarray}
-i \hbar \frac{\partial}{\partial t}  Z_z(  t) &=& -\frac{i
\hbar^2}{c^{5/2}} \frac{\partial}{\partial t} \int d\mathbf{x} z
\mathbf{j}( \mathbf{x}, t) \mathbf{ A}( \mathbf{x}, t) - \frac{i
\hbar^2 }{c^{5/2}} \frac{\partial}{\partial t}  \int d\mathbf{x}
j_0(
\mathbf{x}, t) C_z ( \mathbf{x}, t) \nonumber  \\
&-&
       \frac{i \hbar }{2c^4} \frac{\partial}{\partial t}
\int d\mathbf{x} d\mathbf{y} j_0( \mathbf{x}, t) z
\mathcal{G}(\mathbf{x} - \mathbf{y})
          j_0( \mathbf{y}, t)
\label{eq:11.19}
\end{eqnarray}

\noindent In order to calculate the last term in (\ref{eq:11.10}),
we notice that the only term in $\mathbf{Z}(t)$ which does not
commute with $V(t)$ is that containing $\mathbf{C}$, therefore

\begin{eqnarray}
-[  V(t), Z_z( t)]
 =  -\frac{\hbar^2}{c^{3}}\int d\mathbf{x} d\mathbf{x}'
\mathbf{j}(\tilde{x}) j_0(\tilde{x}') [\mathbf{A}(\tilde{x}),
        C_z(\tilde{x}')]
\label{eq:11.21}
\end{eqnarray}

\noindent To calculate the commutator, we set $t=0$ and use equation
(\ref{eq:A.95})

\begin{eqnarray}
&\mbox{}&   [A_i(\mathbf{x}, 0),
         C_z( \mathbf{x'}, 0)] \nonumber\\
&=& i \hbar (2\pi \hbar)^{-3}  \int \frac{ d\mathbf{p}d
\mathbf{q}}{2 \sqrt{q^3 p}}
\sum_{\sigma \tau}   \nonumber\\
&\Bigl[&\Bigl( e_i(\mathbf{p}, \sigma) c_{\mathbf{p},
\sigma}e^{\frac{i}{\hbar}\mathbf{px}} + e^*_i(\mathbf{p}, \sigma)
c^{\dag}_{\mathbf{p}, \sigma} e^{-\frac{i}{\hbar}\mathbf{px}}\Bigr),
\Bigl(e_z(\mathbf{q}, \tau) c_{\mathbf{q}, \tau}
e^{\frac{i}{\hbar}\mathbf{qx'}} - e_z(\mathbf{q}, \tau)
c^{\dag}_{\mathbf{q},
\tau} e^{-\frac{i}{\hbar}\mathbf{qx'}}\Bigr) \Bigr] \nonumber\\
&=& i \hbar(2\pi \hbar)^{-3} \int \frac{ d\mathbf{p}d
\mathbf{q}}{2\sqrt{q^3p}} \sum_{\sigma \tau} e_i(\mathbf{p}, \sigma)
e^*_z(\mathbf{q}, \tau) \times
\nonumber \\
&\mbox{ }& \left(- \delta(\mathbf{p} - \mathbf{q}) \delta_{\sigma, \tau}
e^{\frac{i}{\hbar}\mathbf{px} - \frac{i}{\hbar} \mathbf{qx'}}-
\delta(\mathbf{p} - \mathbf{q}) \delta_{\sigma,
\tau}e^{-\frac{i}{\hbar}\mathbf{px} + \frac{i}{\hbar} \mathbf{qx'}} \right)
\nonumber\\
&=& -i \hbar(2\pi \hbar)^{-3} \int \frac{ d\mathbf{p} }{2p^2}
\sum_{\sigma \tau} e_i(\mathbf{p}, \sigma) e^*_z(\mathbf{p}, \tau)
 \delta_{\sigma, \tau} \left(
e^{\frac{i}{\hbar}\mathbf{p}
(\mathbf{x-x'})}+e^{-\frac{i}{\hbar}\mathbf{p}
(\mathbf{x-x'})} \right) \nonumber\\
&=& -i \hbar(2\pi \hbar)^{-3} \int \frac{ d\mathbf{p} }{2p^2}
\left(\delta_{iz} - \frac{p_i p_z}{p^2} \right) \left( e^{\frac{i}{\hbar}\mathbf{p}
(\mathbf{x-x'})}+e^{-\frac{i}{\hbar}\mathbf{p}
(\mathbf{x-x'})} \right) \nonumber\\
&=& - i\hbar(2\pi \hbar)^{-3}  \int \frac{ d\mathbf{p} }{p^2}
\left(\delta_{iz} - \frac{p_i p_z}{p^2} \right)
e^{\frac{i}{\hbar}\mathbf{p} (\mathbf{x-x'})}\nonumber \\
&=& -\frac{i}{ \hbar} \delta_{iz} \mathcal{G}(\mathbf{x} -
\mathbf{x'}) + \frac{i \hbar(-i \hbar)^2}{(2\pi \hbar)^{3}}
 \partial_{x_i} \partial_{z} \int \frac{d
\mathbf{p}}{p^4}
e^{\frac{i}{\hbar}\mathbf{p} (\mathbf{x-x'})}\nonumber\\
&=& - \frac{i} {\hbar}\delta_{iz} \mathcal{G}(\mathbf{x} -
\mathbf{x'}) + i \hbar^3
  \partial_{x_i} \partial_{z}
\frac{ |\mathbf{x-x'}|}
{8 \pi \hbar^4} \nonumber \\
&=& -\frac{i}{ \hbar} \delta_{iz} \mathcal{G}(\mathbf{x} -
\mathbf{x'}) +
 \frac{i}
{2 \hbar} \partial_{x_i}( (z- z') \mathcal{G}(\mathbf{x} -
\mathbf{x'})) \label{eq:11.20}
\end{eqnarray}

\noindent Then

\begin{eqnarray}
&\mbox{ }& -[  V(t), Z_z( t)] \nonumber \\
& =&   \frac{i\hbar}{c^{3}} \sum_{i=1}^3 \int d\mathbf{x} d\mathbf{x}'
j_i(\tilde{x}) \Bigl(\delta_{iz} \mathcal{G}(\mathbf{x} -
\mathbf{x}')j_0(\tilde{x}') - \frac{1}{2} \partial_{ x_i}
[(z-z') \mathcal{G}(\mathbf{x} - \mathbf{x}')]
          j_0(\tilde{x}') \Bigr)\nonumber \\
\label{eq:11.22}
\end{eqnarray}

\noindent Now we can set $t=0$, add four terms (\ref{eq:11.17}),
(\ref{eq:11.18}), (\ref{eq:11.19}), and (\ref{eq:11.22}) together and see that the first two terms in (\ref{eq:11.19}) cancel with the
two terms on the right hand side of (\ref{eq:11.17}); the third term
in (\ref{eq:11.19}) cancels the second term on the right hand side
of (\ref{eq:11.18}); and (\ref{eq:11.22}) exactly cancels the
remaining first and third terms on the right hand side of
(\ref{eq:11.18}). This proves equation (\ref{eq:11.10}).

 The proof of the last remaining commutation relation

\begin{eqnarray}
[K_{0i}, Z_j] + [Z_i, K_{0j}] + [Z_i, Z_j]= 0 \label{eq:11.23}
\end{eqnarray}

\noindent is left as an exercise for the reader.

\section{Relativistic invariance of classical electrodynamics}
\label{sc:proof-comm}

In this Appendix we will prove the relativistic invariance of the
classical limit of RQD  constructed in subsections \ref{ss:position-breit} and \ref{sc:poinc-lie}.

From our derivation in chapter \ref{sc:coulomb} it follows that the
 Darwin-Breit Hamiltonian (\ref{eq:12.25}) is a
part of a relativistically invariant theory in the instant form of
dynamics. This means that there exists an interacting boost operator
$\mathbf{K}$, which satisfies all commutation relations of the
Poincar\'e Lie algebra together with the Darwin-Breit Hamiltonian
$H$. In principle, it should be possible to find the explicit form
of the operator $\mathbf{K}$ by applying the unitary dressing
transformation\footnote{see subsection \ref{ss:relat-invar-dressed}} to the boost operator (\ref{eq:interact-boost}) - (\ref{eq:11.8}) of
QED. However,
here we will choose a different route. Together with \cite{Coleman,
Close-Osborn, Krajcik-Foldy} we will simply postulate the form of
$\mathbf{K}$ and verify that Poincar\'e commutators are, indeed,
satisfied in
the $(v/c)^{2}$ approximation.

Let us first write the non-interacting generators of the Poincar\'e
group for a two-particle system as sums of one-particle
generators\footnote{see equations (\ref{eq:h-non-inter}) - (\ref{eq:k-non-inter})}

\begin{eqnarray}
\mathbf{P}_0 &=& \mathbf{p}_1 + \mathbf{p}_2 \label{eq:P_0} \\
\mathbf{J}_0 &=& [\mathbf{r}_1 \times \mathbf{p}_1]
+ \mathbf{s}_1 + [\mathbf{r}_2 \times \mathbf{p}_2] + \mathbf{s}_2 \label{eq:JJ_0} \\
H_0 &=& h_1 + h_2 \nonumber \\
&\approx& m_1c^2 + m_2c^2 + \frac{p_1^2}{2m_1} + \frac{p_2^2}{2m_2}-
\frac{p_1^4}{8m_1^3c^2}   - \frac{p_2^4}{8m_2^3c^2}\label{eq:mc22} \\
\mathbf{K}_0 &=& -\frac{h_1\mathbf{r}_1}{c^2} - \frac{[\mathbf{p}_1
\times \mathbf{s}_1]}{m_1c^2 + h_1} -\frac{h_2\mathbf{r}_2}{c^2} -
\frac{[\mathbf{p}_2 \times \mathbf{s}_2]}{m_2c^2 + h_2} \nonumber \\
&\approx& -m_1 \mathbf{r}_1 - m_2 \mathbf{r}_2 -\frac{p_1^2
\mathbf{r}_1}{2m_1c^2}  - \frac{p_2^2
\mathbf{r}_2}{2m_2 c^2} \nonumber \\
 &\ & +\frac{1}{2 c^2 }
\left(\frac{[\mathbf{s}_1 \times \mathbf{p}_1 ]}{m_1} +
\frac{[\mathbf{s}_2 \times \mathbf{p}_2 ]}{m_2} \right)
\label{eq:free-boost}
\end{eqnarray}

\noindent The full interacting generators are

\begin{eqnarray}
H &=& H_0 + V \nonumber \\
\mathbf{K} &=& \mathbf{K}_0 + \mathbf{Z} \label{eq:k-z}
\end{eqnarray}

\noindent The potential energy $V$ is given by (\ref{eq:H-spin-spin}), and the potential boost is postulated as \cite{Coleman, Close-Osborn,
Krajcik-Foldy}

\begin{eqnarray}
\mathbf{Z} &\approx&    -  \frac{q_1q_2(\mathbf{r}_1
+ \mathbf{r}_2)}{8 \pi c^2r} \label{eq:boost-spin-orb}
\end{eqnarray}

 The non-trivial Poisson brackets of the
Poincar\'e Lie algebra (\ref{eq:5.50}) - (\ref{eq:5.56}) that need
to be verified are those involving interacting generators
$H$ and $\mathbf{K}$

\begin{eqnarray}
\mbox{ } [J_{0i}, K_j]_P &= & \sum_{k=x,y,z}\epsilon_{ijk} K_k
\label{eq:y5.55}
\\
\mbox{ }  [\mathbf{J}_0,H]_P &=& [\mathbf{P}_0, H]_P = 0
\label{eq:z5.55}
\\
 \mbox{ } [K_i, K_j]_P &=& -\frac{1}{c^2} \sum_{k=x,y,z} \epsilon_{ijk}
 J_{0k}
\label{eq:a5.55}
\\
\mbox{ } [K_i, P_{0j}]_P &=& -\frac{1}{c^2} H \delta_{ij} \label{eq:x5.55}\\
 \mbox{ } [\mathbf{K}, H]_P &=& - \mathbf{P}_0
\label{eq:x5.56}
\end{eqnarray}

\noindent where $i,j,k= (x,y,z)$.

The proof of (\ref{eq:y5.55}) - (\ref{eq:z5.55}) follows easily from
the Poisson brackets of particle observables (\ref{eq:rirj}) -
(\ref{eq:rih}) and formula (\ref{eq:7.37}) for brackets involving
complex expressions. This proof is left as an exercise for the
reader. For the less trivial brackets (\ref{eq:a5.55}) -
(\ref{eq:x5.56}), it will be convenient to write
 $H$ and $\mathbf{K}$ as series in powers of $(v/c)^{2}$ (the superscript in parentheses is
 the power of $(v/c)^{2}$)

\begin{eqnarray*}
H &\approx& H^{(-1)} + H^{(0)} +  H_{orb}^{(1)} +  H_{spin-orb}^{(1)} +  H_{spin-spin}^{(1)}\\
\mathbf{K} &\approx& \mathbf{K}^{(0)} + \mathbf{K}_{orb}^{(1)} +
\mathbf{K}_{spin-orb}^{(1)}
\end{eqnarray*}

\noindent where

\begin{eqnarray*}
H^{(-1)} &=& m_1c^2 + m_2c^2 \\
H^{(0)} &=& \frac{p_1^2}{2m_1} + \frac{p_2^2}{2m_2} + \frac{q_1q_2}{4 \pi r}\\
H_{orb}^{(1)} &=& - \frac{p_1^4}{8m_1^3c^2}   -
\frac{p_2^4}{8m_2^3c^2} - \frac{q_1q_2}{8 \pi m_1m_2c^2 r}
\left((\mathbf{p_1} \cdot \mathbf{p}_2) + \frac{(\mathbf{p_1} \cdot
\mathbf{r})(\mathbf{p_2}
\cdot \mathbf{r})}{r^2} \right) \\
 H_{spin-orb}^{(1)} &=& - \frac{q_1q_2 [\mathbf{r} \times \mathbf{p}_1] \cdot
\mathbf{s}_{1}}{8 \pi m_1^2 c^2 r^3} +
 \frac{q_1q_2  [\mathbf{r} \times \mathbf{p}_2]
\cdot \mathbf{s}_{2}}{8 \pi m_2^2 c^2 r^3} + \frac{q_1q_2
[\mathbf{r} \times \mathbf{p}_2] \cdot \mathbf{s}_{1}}{4 \pi m_1m_2
c^2 r^3}
\nonumber \\
&\ & -\frac{q_1q_2 [\mathbf{r} \times \mathbf{p}_1] \cdot
\mathbf{s}_{2}}{4 \pi  m_1m_2 c^2 r^3} \\
H_{spin-spin}^{(1)} &=& \frac{(\mathbf{s}_1 \cdot \mathbf{s}_2)} {4
\pi m_1m_2c^2
 r^3} - \frac{3(\mathbf{s}_1 \cdot \mathbf{r}) (\mathbf{s}_2 \cdot
\mathbf{r})}{4 \pi m_1m_2 c^2 r^5} \\
\mathbf{K}^{(0)} &=& -m_1 \mathbf{r}_1 - m_2 \mathbf{r}_2 \\
\mathbf{K}_{orb}^{(1)} &=& -\frac{p_1^2 \mathbf{r}_1}{2m_1c^2}  -
\frac{p_2^2 \mathbf{r}_2}{2m_2 c^2} -  \frac{q_1q_2(\mathbf{r}_1 +
\mathbf{r}_2)}{8 \pi c^2r} \\
\mathbf{K}_{spin-orb}^{(1)} &=& \frac{1}{2 c^2 }
\left(\frac{[\mathbf{s}_1 \times \mathbf{p}_1 ]}{m_1} +
\frac{[\mathbf{s}_2 \times \mathbf{p}_2 ]}{m_2} \right)
\end{eqnarray*}

\noindent Then we find that the following relationships need to be
proven

\begin{eqnarray}
\ - \frac{1}{c^2}H^{(-1)} \delta_{ij} &=&[K_i^{(0)}, P_{0j}]_P   \label{eq:21} \\
\ 0 &=& [K_i^{(0)}, H^{(-1)}]_P  \label{eq:23}  \\
\ - P_{0i}  &=& [K_i^{(0)}, H^{(0)}]_P \label{eq:24}  \\
\ 0 &=& [K_i^{(0)}, K_j^{(0)}]_P  \label{eq:26}  \\
\ - \frac{1}{c^2}H^{(0)} \delta_{ij} &=& [K_{i-orb}^{(1)}, P_{0j}]_P + [K_{i-spin-orb}^{(1)}, P_{0j}]_P  \label{eq:22} \\
\ 0 &=& [K_{i-orb}^{(1)}, H^{(0)}]_P + [K_{i-spin-orb}^{(1)}, H^{(0)}]_P
+ [K_i^{(0)}, H_{orb}^{(1)}]_P \nonumber \\
&\ & +[K_i^{(0)}, H_{spin-orb}^{(1)}]_P + [K_i^{(0)}, H_{spin-spin}^{(1)}]_P \label{eq:25}  \\
\ - \frac{1}{c^2} \sum_{k=1}^{3} \epsilon_{ijk} J_{0k} &=&
[K_{i-orb}^{(1)}, K_j^{(0)}]_P + [K_{i-spin-orb}^{(1)}, K_j^{(0)}]_P
+ [K_i^{(0)}, K_{j-orb}^{(1)}]_P \nonumber \\
&\ & +[K_i^{(0)}, K_{j-spin-orb}^{(1)}]_P \label{eq:27}
\end{eqnarray}

\noindent Again, we skip the easy-to-prove (\ref{eq:21}),
(\ref{eq:23}), (\ref{eq:24}), and (\ref{eq:26}). For equation
(\ref{eq:22}) we obtain

\begin{eqnarray*}
&\ &[K_{x-orb}^{(1)} + K_{x-spin-orb}^{(1)}, P_{0x}]_P \nonumber \\
&=& \ -\left[\frac{p_1^2r_{1x}}{2m_1c^2} , p_{1x}\right]_P  -
\left[\frac{p_2^2r_{2x}}{2m_2 c^2} , p_{2x}\right]_P - \ \left[\frac{q_1q_2}{8 \pi
c^2} \frac{r_{1x} + r_{2x}}{r}, p_{1x}\right]_P - \left[\frac{q_1q_2}{8 \pi
c^2}
\frac{r_{1x} + r_{2x}}{r}, p_{2x}\right]_P \nonumber  \\
&=& -\frac{p_1^2}{2m_1c^2}   - \ \frac{p_2^2 }{2m_2 c^2}
- \ \frac{q_1q_2}{4 \pi c^2r} \nonumber \\
&=& - \frac{1}{c^2}H^{(0)}
\end{eqnarray*}

\noindent Individual terms on the right hand side of (\ref{eq:25})
are

\begin{eqnarray}
&\mbox{ }& [K_{x-orb}^{(1)}, H^{(0)}]_P  \nonumber \\
&=& -\frac{p_1^2}{4m_1^2c^2}[ r_{1x}, p_1^2]_P -
\frac{q_1q_2r_{1x}}{8 \pi m_1c^2} \left[p_1^2 , \frac{1}{r}\right]_P -
\frac{p_2^2}{4m_2^2c^2} [ r_{2x}, p_2^2]_P -
\frac{q_1q_2r_{2x}}{8 \pi m_2c^2} \left[p_2^2, \frac{1}{r}\right]_P \nonumber  \\
&\ & -\frac{q_1q_2}{16 \pi m_1 c^2r} [r_{1x}, p_1^2]_P -
\frac{q_1q_2r_{1x}}{16 \pi m_1 c^2 }  \left[\frac{1}{ r} , p_1^2\right]_P -
\frac{q_1q_2r_{2x}}{16 \pi m_1 c^2}\left[\frac{1}{r} , p_1^2\right]_P -
\frac{q_1q_2r_{1x}}{16 \pi m_2 c^2} \left[\frac{1}{r}, p_2^2\right]_P \nonumber  \\
&\ & -\frac{q_1q_2}{16 \pi m_2  c^2r} [r_{2x},p_2^2]_P
- \frac{q_1q_2 r_{2x}}{16 \pi m_2  c^2} \left[\frac{1}{r},p_2^2\right]_P \nonumber  \\
&=& -\frac{p_1^2 p_{1x}}{2m_1^2c^2} -\frac{q_1q_2(r_{1x}-r_{2x})}{8
\pi m_1c^2} \frac{(\mathbf{p}_1 \cdot \mathbf{r})}{r^3} -
\frac{p_2^2 p_{2x}}{2m_2^2c^2}  - \frac{q_1q_2(r_{1x}-r_{2x})}{8 \pi
m_2c^2} \frac{(\mathbf{p}_2 \cdot
\mathbf{r})}{r^3} \nonumber  \\
&\ & -\frac{q_1q_2 p_{1x}}{8 \pi m_1 c^2r}  - \frac{q_1q_2 p_{2x}}{8
\pi m_2 c^2r}  \label{eq:kx0horb} \\ \nonumber  \\
&\mbox{ }&  [K_{x-spin-orb}^{(1)}, H^{(0)}]_P = \frac{1}{2
c^2}\left[\frac{1}{m_1} [\mathbf{s}_1 \times \mathbf{p}_1 ]_x+
\frac{1}{m_2} [\mathbf{s}_2 \times \mathbf{p}_2 ]_x, \frac{q_1q_2}{4
\pi r} \right]_P \nonumber  \\
&=&\frac{q_1q_2[\mathbf{s}_1 \times \mathbf{r} ]_x}{8 \pi m_1c^2
r^3} -
\frac{q_1q_2[\mathbf{s}_2 \times \mathbf{r} ]_x}{8 \pi m_2 c^2 r^3}    \\ \nonumber  \\
&\mbox{ }& [K_x^{(0)}, H_{orb}^{(1)}]_P \nonumber  \\
&=&  \frac{1}{8m_1^2c^2} [r_{1x} , p_1^4]_P +\frac{q_1q_2}{8 \pi
m_2c^2 r} \left[r_{1x}, \left((\mathbf{p_1} \cdot \mathbf{p}_2) +
\frac{(\mathbf{p_1} \cdot
\mathbf{r})(\mathbf{p_2} \cdot \mathbf{r})}{r^2}\right)\right]_P \nonumber  \\
&\ & +\frac{1}{8m_2^2c^2} [  r_{2x}, p_2^4]_P + \frac{q_1q_2}{8 \pi
m_1 c^2 r} \left[r_{2x},
 \left((\mathbf{p_1} \cdot \mathbf{p}_2) +
\frac{(\mathbf{p_1} \cdot
\mathbf{r})(\mathbf{p_2} \cdot \mathbf{r})}{r^2} \right)\right]_P \nonumber  \\
&=&  \frac{p_1^2 p_{1x}}{2m_1^2c^2}    + \frac{q_1q_2}{8 \pi m_2c^2
r}
\left(p_{2x} + \frac{(r_{1x}-r_{2x})(\mathbf{p_2} \cdot \mathbf{r})}{r^2} \right) \nonumber  \\
&\ & +\frac{p_2^2 p_{2x}}{2m_2^2c^2} +  \frac{q_1q_2}{8 \pi m_1 c^2 r}
\left(p_{1x} + \frac{(\mathbf{p_1} \cdot
\mathbf{r})(r_{1x}-r_{2x})}{r^2} \right)   \\  \nonumber   \\
&\mbox{ }&  [K_x^{(0)}, H_{spin-orb}^{(1)}]_P  = \Bigl[-m_1r_{1x}
-m_2r_{2x}, - \frac{q_1q_2 [\mathbf{r} \times \mathbf{p}_1] \cdot
\mathbf{s}_{1}}{8 \pi m_1^2 c^2 r^3} \nonumber \\
&\ & +
 \frac{q_1q_2  [\mathbf{r} \times \mathbf{p}_2]
\cdot \mathbf{s}_{2}}{8 \pi m_2^2 c^2 r^3} + \frac{q_1q_2
[\mathbf{r} \times \mathbf{p}_2] \cdot \mathbf{s}_{1}}{4 \pi m_1m_2
c^2 r^3} - \frac{q_1q_2 [\mathbf{r} \times \mathbf{p}_1] \cdot
\mathbf{s}_{2}}{4 \pi  m_1m_2 c^2 r^3} \Bigr]_P \nonumber  \\
&=& -\frac{q_1q_2 [\mathbf{s}_2 \times \mathbf{r}]_x}{8 \pi m_2 c^2
r^3} +
 \frac{q_1q_2  [\mathbf{s}_1 \times \mathbf{r} ]_x}{8 \pi m_1 c^2 r^3} + \frac{q_1 q_2[\mathbf{s}_2
\times \mathbf{r}]_x}{4 \pi m_2 c^2 r^3} - \frac{q_1q_2
[\mathbf{s}_1 \times \mathbf{r}]_x}{4
\pi m_1  c^2 r^3} \nonumber  \\
&=&
 -\frac{q_1q_2  [\mathbf{s}_1 \times \mathbf{r} ]_x}{8 \pi m_1 c r^3}
  +\frac{q_1 q_2[\mathbf{s}_2 \times \mathbf{r}]_x}{8 \pi m_2 c
r^3}  \\ \nonumber  \\
&\mbox{ }& [K_x^{(0)}, H^{(1)}_{spin-spin}]_P = 0
\label{eq:kx0hspin}
\end{eqnarray}

\noindent Summing up the right hand sides of equations (\ref{eq:kx0horb})
- (\ref{eq:kx0hspin}) we see that equation (\ref{eq:25}) is, indeed,
satisfied. For equation (\ref{eq:27}) we obtain

\begin{eqnarray}
&\ & [K_{x-orb}^{(1)}, K_y^{(0)}]_P + [K_x^{(0)}, K_{y-orb}^{(1)}]_P
+ [K_{x-spin-orb}^{(1)}, K_y^{(0)}]_P + [K_x^{(0)},
K_{y-spin-orb}^{(1)}]_P \nonumber \\
&=& \frac{r_{1x}}{2c^2}[p_1^2, r_{1y}]_P +
\frac{r_{2x}}{2c^2}[p_2^2, r_{2y}]_P + \frac{r_{1y}}{2c^2}[r_{1x},
p_1^2]_P + \frac{r_{2y}}{2c^2}[r_{2x}, p_2^2]_P \nonumber \\
&\ &  -\frac{1}{2 c^2 }\left[-\frac{1}{m_1} s_{1z}p_{1y}- \frac{1}{m_2}
s_{2z}p_{2y}, m_1r_{1y} + m_2r_{2y} \right]_P \nonumber \\
&\ & -\frac{1}{2c^2 }\left[m_1r_{1x} + m_2r_{2x}, \frac{1}{m_1}
s_{1z}p_{1x}+
\frac{1}{m_2} s_{2z}p_{2x}\right]_P \nonumber \\
&=& - \frac{1}{c^2} [\mathbf{r}_1 \times \mathbf{p}_1]_z -
\frac{1}{c^2} [\mathbf{r}_2 \times \mathbf{p}_2]_z   -\frac{1}{ c^2 }( s_{1z} + s_{2z}) = -\frac{1}{c^2}J_{0z} \label{eq:kxky}
\end{eqnarray}

\chapter{Dimensionality checks}
\label{ss:dimensionality}

In our formulas in this book we chose to show explicitly all fundamental constants,
like $c$ and $\hbar$,  rather than adopt the usual
convention $\hbar = c = 1$. This makes our expressions slightly
lengthier, but has the benefit of easier control of dimensions
and checking correctness at each calculation step. In this
subsection we are going to suggest a few rules for such dimension estimates in formulas involving quantum fields.

 From the familiar formula

\begin{eqnarray*}
\int d\mathbf{p} \delta(\mathbf{p}) = 1
\end{eqnarray*}

\noindent it follows that the dimension of the delta function
is\footnote{Angle brackets $\langle A \rangle$ denote the dimension of an observable $A$, as it has been introduced in subsection \ref{sc:poincare-lie}. For example, $\langle p \rangle = \langle m \rangle\langle v\rangle =
\langle E \rangle/\langle v \rangle$ denotes the dimension of
momentum. Note that dimension of the 4D delta function (\ref{eq:4d-delta}) is $\langle E^{-1} \rangle \langle p^{-3} \rangle$.}

\begin{eqnarray*}
\langle \delta(\mathbf{p})\rangle = \frac{1}{\langle p^3 \rangle}
\end{eqnarray*}

\noindent Then (anti)commutation relations of creation and
annihilation operators

\begin{eqnarray*}
\{ a^{\dag}_{\mathbf{p}, \sigma}, a_{\mathbf{p}', \sigma'} \} &=&
\delta(\mathbf{p}- \mathbf{p}') \delta_{\sigma, \sigma'} \\
\ [ c_{\mathbf{p}, \tau}, c^{\dag}_{\mathbf{p}', \tau'} ] &=&
\delta(\mathbf{p}- \mathbf{p}') \delta_{\tau, \tau'}
\end{eqnarray*}

\noindent suggests that dimensions of these operators are

\begin{eqnarray}
\langle a^{\dag}_{\mathbf{p}, \sigma}\rangle =  \langle
a_{\mathbf{p}', \sigma'} \rangle = \langle c_{\mathbf{p}, \tau}
\rangle = \langle c^{\dag}_{\mathbf{p}', \tau'} \rangle &=&
\frac{1}{\langle p^{3/2} \rangle} \label{eq:a-dimension}
\end{eqnarray}

\noindent In the definition of the Dirac's quantum field
(\ref{eq:10.10})

\begin{eqnarray*}
 \psi(\mathbf{x},t) \nonumber =  \int \frac{d\mathbf{p}}{(2\pi \hbar)^{3/2}}
\sqrt{\frac{mc^2}{\omega_{\mathbf{p}}}} \sum_{ \sigma}
\Bigl(e^{-\frac{i}{\hbar}\tilde{p} \cdot \tilde{x}}u(\mathbf{p},
\sigma)
 a _{\mathbf{p},\sigma}
+ e^{\frac{i}{\hbar}\tilde{p} \cdot \tilde{x}}v (\mathbf{p},
\sigma)b^{\dag}_{\mathbf{p},\sigma} \Bigr)
\end{eqnarray*}

\noindent  4-vectors $\tilde{p}$ and
$\tilde{x}$ have dimensions of energy $\langle \tilde{p}
\rangle = \langle E \rangle$ and time $\langle \tilde{x} \rangle =
\langle t \rangle$, respectively. The dimension of the Planck's constant is $ \langle \hbar
\rangle = \langle p\rangle \langle r\rangle = \langle E \rangle
\langle t\rangle $, which implies that arguments $\frac{i}{\hbar}\tilde{p}
\cdot \tilde{x}$ of exponents are dimensionless, as expected. Functions $u$ and $v$ are dimensionless as well.\footnote{see
(\ref{eq:10.17}) - (\ref{eq:10.20})} Then the
dimension of the Dirac quantum field is

\begin{eqnarray*}
\langle \psi \rangle &=&
\frac{\langle p^3 \rangle}{\langle \hbar^{3/2}\rangle \langle p^{3/2} \rangle} = \frac{\langle p^{3/2}
\rangle}{\langle \hbar^{3/2}\rangle}= \frac{1}{\langle
r^{3/2}\rangle}
\end{eqnarray*}

\noindent Similarly, we obtain the dimension of the photon's
quantum field (\ref{eq:10.26})\footnote{In different texts one can
find various definitions of quantum fields, which can differ from definitions
adopted here by
their numerical factors and dimensions. However, as we stress in subsection
\ref{ss:derivation}, quantum fields do not correspond to any
observable quantities. They are just formal mathematical objects,
whose role is to provide convenient ``building blocks'' for
interaction operators (\ref{eq:11.6}), (\ref{eq:11.7}) and
(\ref{eq:11.8}). So, there is a significant freedom in choosing
concrete forms of quantum fields. All these choices should lead to the same
forms of the physically meaningful interaction operators $V_1$, $V_2$, and $\mathbf{Z}$.}

\begin{eqnarray}
\langle A \rangle &=& \frac{\langle \hbar \rangle \langle
c^{1/2}\rangle}{\langle r^{3/2}\rangle \langle p^{1/2}\rangle} =
\frac{\langle p^{1/2} \rangle \langle c^{1/2}\rangle}{\langle
r^{1/2}\rangle} \label{eq:dim-phot}
\end{eqnarray}

\noindent   current density operator (\ref{eq:11.1})

\begin{eqnarray*}
\langle j \rangle &=& \langle ec \overline{\psi} \psi \rangle =
\frac{\langle e\rangle \langle c\rangle}{\langle r^{3}\rangle}
\end{eqnarray*}

\noindent and  potential energy (\ref{eq:11.6})\footnote{This
expression was simplified by using
 $\langle e^2 \rangle = \langle \hbar \rangle
\langle c \rangle$, which follows from the fact that $\alpha \equiv
e^2/(4 \pi \hbar c) \approx 1/137$ is the dimensionless \emph{fine structure constant}.}

\begin{eqnarray*}
\langle V_1 \rangle &=&  \frac{1}{\langle c \rangle} \langle r^3
\rangle  \frac{\langle e\rangle \langle c\rangle}{\langle
r^{3}\rangle} \frac{\langle p^{1/2} \rangle \langle
c^{1/2}\rangle}{\langle r^{1/2}\rangle} \\
 &=&     \frac{\langle
e\rangle \langle p^{1/2} \rangle \langle c^{1/2}\rangle}{\langle
r^{1/2}\rangle} = \frac{\langle e\rangle \langle \hbar^{1/2} \rangle
\langle c^{1/2}\rangle}{\langle r\rangle} = \frac{\langle e^2\rangle
}{\langle r\rangle}
\end{eqnarray*}

\noindent This is exactly the dimension of energy, as one can
expect from the Coulomb law $V = e^2/(4 \pi r)$. The 2nd order QED
potential (\ref{eq:11.7}) also has the dimension of energy

\begin{eqnarray*}
\langle V_2 \rangle &=& \frac{1}{\langle c^{2} \rangle}
\frac{\langle r^3 \rangle \langle r^3 \rangle }{\langle r \rangle}
\frac{\langle e\rangle \langle c\rangle}{\langle r^{3}\rangle}
\frac{\langle e\rangle \langle c\rangle}{\langle r^{3}\rangle} =
\frac{\langle e^2\rangle}{\langle r \rangle}
\end{eqnarray*}

\noindent By following the same rules it is easy to establish that
all three terms in the potential boost (\ref{eq:11.8}) have the
dimension $\langle m\rangle\langle r \rangle$, as expected.

 Let us illustrate the dimensionality checks on the example of the
scattering amplitude (\ref{eq:t-order}). The $S$-operator is a
dimensionless quantity and particle creation-annihilation operators
have the dimension $\langle p^{-3/2} \rangle$. Therefore, the
dimension of the matrix element $\langle 0 | a_{\mathbf{q},
\tau } d_{\mathbf{p}, \sigma} S_2 d^{\dag}_{\mathbf{p'}, \sigma'}
a^{\dag}_{\mathbf{q'}, \tau'} |0 \rangle$ is expected to be $\langle
p^{-6} \rangle$. Turning to the final result (\ref{eq:2nd-order}) we
may note that according to (\ref{eq:en-delta})

\begin{eqnarray*}
\langle \delta^4(p) \rangle &=& \frac{1}{\langle E \rangle \langle p^3
\rangle}
\end{eqnarray*}

\noindent Then the dimension of (\ref{eq:2nd-order})

\begin{eqnarray*}
\frac{\langle e^2 \rangle \langle c^2 \rangle}{\langle \hbar \rangle
\langle E \rangle\langle p^3 \rangle\langle E \rangle} = \frac{
\langle c^3 \rangle}{\langle E^3 \rangle\langle p^3 \rangle} =
\frac{ 1}{\langle p^6 \rangle}
\end{eqnarray*}

\noindent is consistent with expectations.

Note also that $d^4x \equiv dt d\mathbf{x}$ and $d^4p \equiv dE
d\mathbf{p}$, so

\begin{eqnarray*}
\langle d^4x \rangle &=& \langle t \rangle \langle r^3 \rangle \\
\langle d^4p \rangle &=& \langle E \rangle \langle p^3 \rangle
\end{eqnarray*}


\backmatter


\bibliographystyle{alpha}

\newcommand{\etalchar}[1]{$^{#1}$}

\printindex
\input{book.ind}

\end{document}